\documentclass[final,3p,times,nopreprintline]{elsarticle}

\usepackage{amsmath}
\usepackage{amssymb}
\usepackage{booktabs}
\usepackage{braket}
\usepackage[pdftex]{color}
\usepackage{dcolumn}
\usepackage{enumitem}
\usepackage{epsfig}
\usepackage{etoolbox}
\usepackage{graphicx}
\usepackage{hyperref}
\hypersetup{
    colorlinks=true,       
    linkcolor=blue,          
    citecolor=blue,        
    filecolor=blue,      
    urlcolor=blue           
}
\usepackage{latexsym}
\usepackage{longtable}
\usepackage{mathrsfs}
\usepackage{mathtools}
\usepackage{multirow}
\usepackage{rotating}
\usepackage{rotfloat}
\usepackage[squaren,mediumqspace]{SIunits}
\usepackage{slashed}
\usepackage{stackrel}
\usepackage{wrapfig}

\usepackage[utf8]{inputenc}
\usepackage{afterpage}
\usepackage{xspace}
\usepackage{placeins}
\usepackage{natbib}

\usepackage[symbol]{footmisc}

\usepackage[capitalise]{cleveref}

\crefname{section}{Sec.}{Secs.}

\Crefname{section}{Section}{Sections}

\biboptions{sort&compress}

\DeclareGraphicsExtensions{.pdf,.png,.jpg,.eps}

\definecolor{red}{rgb}{0.8,0.0,0.0}
\definecolor{green}{rgb}{0.0,0.6,0.0}
\definecolor{darkblue}{rgb}{0.0,0.1,0.7}
\definecolor{brown}{rgb}{0.6,0.1,0.0}
\definecolor{grey}{rgb}{0.6,0.6,0.6}
\definecolor{darkgreen}{rgb}{0.0, 0.545098, 0.0}
\definecolor{applegreen}{rgb}{0.55, 0.71, 0.0}
\definecolor{purple}{rgb}{0.5,0.0,0.5}
\definecolor{babypink} {rgb}{0.64, 0.44, 0.44}
\definecolor{orange}{rgb}{1.0,0.5,0.0}

\definecolor{DARKBLUE}{rgb}{0.0,0.1,0.7}
\allowdisplaybreaks

\newcommand{\toright}[1]{\hspace*{\fill}{\footnotesize{#1}}}

\newcommand{\ie}{i.e.}

\newcommand{\bi}{\begin{itemize}}
\newcommand{\ei}{\end{itemize}}

\newcommand{\ben}{\begin{enumerate}}
\newcommand{\een}{\end{enumerate}} 

\newcommand{\bt}[1]{\begin{table}[tb]\begin{tabular}{#1} \hline\hline  \\[-1.0em]}
\newcommand{\et}[2]{\hline\hline \end{tabular} \caption{#1} \label{#2} \end{table}}

\newcommand{\be}{\begin{equation}}
\newcommand{\ee}{\end{equation}}

\newcommand{\bea}{\begin{eqnarray}}
\newcommand{\eea}{\end{eqnarray}}

\newcommand{\colrule}{\midrule}
\newcommand{\botrule}{\bottomrule}

\newcommand{\chiS}{\ensuremath{\chi^2}\xspace}

\newcommand{\BR}{\ensuremath{\mathcal{B}}}            
\newcommand{\order}{\ensuremath{\mathcal{O}}} 

\renewcommand{\Im}{\ensuremath{\mathop{\mathrm{Im}}}}

\newcommand{\epem}{\ensuremath{e^+e^-}\xspace}
\newcommand{\pp}{\ensuremath{\pi^+\pi^-}\xspace}
\newcommand{\KK}{\ensuremath{K^+K^-}\xspace}
\newcommand{\piz}{\ensuremath{\pi^0}\xspace}

\newcommand{\amuSM}{\ensuremath{a_\mu^\text{SM}}}

\newcommand{\amuexp}{\ensuremath{a_\mu^\text{exp}}}

\newcommand{\amuHVP}{\ensuremath{a_\mu^\text{HVP}}}

\newcommand{\amuHVPLO}{\ensuremath{a_\mu^\text{HVP, LO}}}
\newcommand{\amuHVPLOpp}{\ensuremath{a_\mu^\text{HVP, LO}[\pi\pi]}}
\newcommand{\amuHVPLOppp}{\ensuremath{a_\mu^\text{HVP, LO}[3\pi]}}

\newcommand{\amuHVPNLO}{\ensuremath{a_\mu^\text{HVP, NLO}}}
\newcommand{\amuHVPNNLO}{\ensuremath{a_\mu^\text{HVP, NNLO}}}

\newcommand{\amuHLbL}{\ensuremath{a_\mu^\text{HLbL}}}
\newcommand{\amuHLbLNLO}{\ensuremath{a_\mu^\text{HLbL, NLO}}}

\newcommand{\amuhad}{\ensuremath{a_\mu^\text{had}}}
\newcommand{\amuQED}{\ensuremath{a_\mu^\text{QED}}}
\newcommand{\amuEW}{\ensuremath{a_\mu^\text{EW}}}

\newcommand{\amuewl}{{a_\mu ^{\text{EW(1)}}}}
\newcommand{\amub}{a_{\mu;\text{bos}}^{\rm EW(2)}}
\newcommand{\amuf}{a_{\mu;\text{ferm}}^{\rm EW(2)}}

\newcommand{\amufrestH}{a_{\mu;\text{f-rest,H}}^{\rm EW(2)}}

\newcommand{\amufrestnoH}{a_{\mu;\text{f-rest,no H}}^{\rm EW(2)}}

\newcommand{\amuHVPLOud}{\ensuremath{a_\mu^\text{HVP, LO}(ud)}}
\newcommand{\amuHVPLOs}{\ensuremath{a_\mu^\text{HVP, LO}(s)}}
\newcommand{\amuHVPLOc}{\ensuremath{a_\mu^\text{HVP, LO}(c)}}
\newcommand{\amuHVPLOb}{\ensuremath{a_\mu^\text{HVP, LO}(b)}}
\newcommand{\amuHVPLOdisc}{\ensuremath{a_{\mu, \text{disc}}^\text{HVP, LO}}}
\newcommand{\amuHVPLOconn}{\ensuremath{a_{\mu, \text{conn}}^\text{HVP, LO}}}
\newcommand{\amuW}{\ensuremath{a_\mu^\text{W}}}
\newcommand{\amuSD}{\ensuremath{a_\mu^\text{SD}}}
\newcommand{\amuLD}{\ensuremath{a_\mu^\text{LD}}}

\newcommand{\Qlow}{\ensuremath{Q^2_\text{low}}}
\newcommand{\Qhigh}{\ensuremath{Q^2_\text{high}}}

\def\gev{\text{GeV}}

\def\mev{\mathrm{Me\kern-0.1em V}}
\def\gev{\mathrm{Ge\kern-0.1em V}}
\def\tev{\mathrm{Te\kern-0.1em V}}
\def\fm{\mathrm{fm}}

\newcommand{\lsim}{ {\
\lower-1.2pt\vbox{\hbox{\rlap{$<$}\lower5pt\vbox{\hbox{$\sim$}}}}\ } }
\newcommand{\gsim}{ {\
\lower-1.2pt\vbox{\hbox{\rlap{$>$}\lower5pt\vbox{\hbox{$\sim$}}}}\ } }

\makeatletter
\DeclareRobustCommand{\text}{%
  \ifmmode\expandafter\text@\else\expandafter\mbox\fi}
\let\nfss@text\text
\def\text@#1{{\mathchoice
  {\textdef@\displaystyle\f@size{#1}}%
  {\textdef@\textstyle\f@size{#1}}%
  {\textdef@\textstyle\sf@size{#1}}%
  {\textdef@\textstyle \ssf@size{#1}}%
  \check@mathfonts
  }%
}
\def\textdef@#1#2#3{\hbox{{%
                    \everymath{#1}%
                    \let\f@size#2\selectfont
                    #3}}}
\makeatother

\newcommand\amu[1]{{a_{\mu}^{#1\textrm{-pole}}}}

\newcommand{\minidiagSize}[2]{\begin{minipage}{#2} \includegraphics[width=#2]{dispHLbL/figures/#1} \end{minipage}}
\newcommand{\nn}{\nonumber\\}
\newcommand{\Imspipi}{\text{Im}_s^{\pi\pi}}
\newcommand{\mpi}{M_{\pi}}
\newcommand{\MeV}{\,\text{MeV}}
\newcommand{\GeV}{\,\text{GeV}}
\newcommand{\keV}{\,\text{keV}}

\newcommand{\Tr}{\ensuremath{\mathop{\mathrm{Tr}}}}
\newcommand{\hyph}{\text{--}}

\def\simge{
    \mathrel{\rlap{\raise 0.511ex
        \hbox{$>$}}{\lower 0.511ex \hbox{$\sim$}}}}
\def\simle{
    \mathrel{\rlap{\raise 0.511ex 
        \hbox{$<$}}{\lower 0.511ex \hbox{$\sim$}}}}

\newcommand{\FF}{F_{\pi^0\gamma^*\gamma^*}}

\DeclareMathOperator{\trace}{Tr}
\newcommand\mcL{\mathcal{L}}
\newcommand\bmcL{\bar{\mathcal{L}}}

\newcommand{\HVPref}{Davier:2017zfy,Keshavarzi:2018mgv,Colangelo:2018mtw,Hoferichter:2019gzf,Davier:2019can,Keshavarzi:2019abf}

\newcommand{\HVPlatticeref}{Chakraborty:2017tqp,Borsanyi:2017zdw,Blum:2018mom,Giusti:2019xct,Shintani:2019wai,Davies:2019efs,Gerardin:2019rua,Aubin:2019usy,Giusti:2019hkz}

\newcommand{\HVPtotalref}{Davier:2017zfy,Keshavarzi:2018mgv,Colangelo:2018mtw,Hoferichter:2019gzf,Davier:2019can,Keshavarzi:2019abf,Kurz:2014wya}

\newcommand{\HLbLlatticeref}{Blum:2019ugy}

\newcommand{\HLbLlatticemethods}{Blum:2014oka,Green:2015mva,Blum:2015gfa,Blum:2016lnc,Asmussen:2016lse,Blum:2017cer,Asmussen:2019act}

\newcommand{\HLbLref}{Melnikov:2003xd,Masjuan:2017tvw,Colangelo:2017fiz,Hoferichter:2018kwz,Gerardin:2019vio,Bijnens:2019ghy,Colangelo:2019uex,Pauk:2014rta,Danilkin:2016hnh,Jegerlehner:2017gek,Knecht:2018sci,Eichmann:2019bqf,Roig:2019reh}

\newcommand{\QEDref}{Aoyama:2012wk,Aoyama:2019ryr}

\newcommand{\EWref}{Czarnecki:2002nt,Gnendiger:2013pva}

\newcommand{\HLbLcombref}{Melnikov:2003xd,Masjuan:2017tvw,Colangelo:2017fiz,Hoferichter:2018kwz,Gerardin:2019vio,Bijnens:2019ghy,Colangelo:2019uex,Pauk:2014rta,Danilkin:2016hnh,Jegerlehner:2017gek,Knecht:2018sci,Eichmann:2019bqf,Roig:2019reh,Blum:2019ugy}

\newcommand{\HLbLtotalref}{Melnikov:2003xd,Masjuan:2017tvw,Colangelo:2017fiz,Hoferichter:2018kwz,Gerardin:2019vio,Bijnens:2019ghy,Colangelo:2019uex,Pauk:2014rta,Danilkin:2016hnh,Jegerlehner:2017gek,Knecht:2018sci,Eichmann:2019bqf,Roig:2019reh,Blum:2019ugy,Colangelo:2014qya}

\newcommand{\SMref}{Aoyama:2012wk,Aoyama:2019ryr,Czarnecki:2002nt,Gnendiger:2013pva,Davier:2017zfy,Keshavarzi:2018mgv,Colangelo:2018mtw,Hoferichter:2019gzf,Davier:2019can,Keshavarzi:2019abf,Kurz:2014wya,Melnikov:2003xd,Masjuan:2017tvw,Colangelo:2017fiz,Hoferichter:2018kwz,Gerardin:2019vio,Bijnens:2019ghy,Colangelo:2019uex,Blum:2019ugy,Colangelo:2014qya}

\newcommand{\HVPexpref}{Bai:1999pk,Akhmetshin:2000ca,Akhmetshin:2000wv,Achasov:2000am,Bai:2001ct,Achasov:2002ud,Akhmetshin:2003zn,Aubert:2004kj,Aubert:2005eg,Aubert:2005cb,Aubert:2006jq,Aulchenko:2006na,Achasov:2006vp,Akhmetshin:2006wh,Akhmetshin:2006bx,Akhmetshin:2006sc,Aubert:2007ur,Aubert:2007ef,Aubert:2007uf,Aubert:2007ym,Akhmetshin:2008gz,Ambrosino:2008aa,Ablikim:2009ad,Aubert:2009ad,Ambrosino:2010bv,Lees:2011zi,Lees:2012cr,Lees:2012cj,Babusci:2012rp,Akhmetshin:2013xc,Lees:2013ebn,Lees:2013uta,Lees:2014xsh,Achasov:2014ncd,Aulchenko:2014vkn,Akhmetshin:2015ifg,Ablikim:2015orh,Shemyakin:2015cba,Anashin:2015woa,Achasov:2016bfr,Achasov:2016lbc,TheBaBar:2017aph,CMD-3:2017tgb,TheBaBar:2017vzo,Kozyrev:2017agm,Anastasi:2017eio,Achasov:2017vaq,Xiao:2017dqv,TheBaBar:2018vvb,Anashin:2018vdo,Achasov:2018ujw,Lees:2018dnv,CMD-3:2019ufp}

\newcommand{\HLbLexpref}{Behrend:1990sr,Gronberg:1997fj,Acciarri:1997yx,Achard:2001uu,Achard:2007hm,Arnaldi:2009aa,Aubert:2009mc,BABAR:2011ad,Berghauser:2011zz,Uehara:2012ag,Babusci:2012ik,Aguar-Bartolome:2013vpw,Ablikim:2015wnx,Masuda:2015yoh,Arnaldi:2016pzu,Adlarson:2016hpp,Adlarson:2016ykr,TheNA62:2016fhr,BaBar:2018zpn,Larin:2020}

\begin{document}

\title{{\footnotesize{FERMILAB-PUB-20-207-T}}\toright{CERN-TH-2020-075}\\[-0.2cm]
{\footnotesize{INT-PUB-20-021}}\toright{IFT-UAM/CSIC-20-74}\\[-0.2cm]
{\footnotesize{KEK Preprint 2020-5}}\toright{LMU-ASC 18/20}\\[-0.2cm]
{\footnotesize{MITP/20-028}}\toright{LTH 1234}\\[-0.2cm]
\toright{LU TP 20-20}\\[-0.2cm]
\toright{MAN/HEP/2020/003}\\[-0.2cm]
\toright{PSI-PR-20-06}\\[-0.2cm]
\toright{UWThPh 2020-14}\\[-0.2cm]
\toright{ZU-TH 18/20}\\[0.5cm]
The anomalous magnetic moment of the muon in the Standard Model}

\renewcommand{\theaffn}{\arabic{affn}}

\renewcommand{\thefootnote}{\fnsymbol{footnote}}

\author[KEK,Nishina,Kobayashi]{T.~Aoyama}
\author[Southampton]{N.~Asmussen}
\author[Sorbonne]{M.~Benayoun}
\author[Lund]{J.~Bijnens} 
\author[Uconn,RIKENBNL]{T.~Blum}
\author[CERN]{M.~Bruno}
\author[Bucharest]{I.~Caprini}
\author[Pavia]{C.~M.~Carloni~Calame}
\author[CERN,Mainz,HIM]{M.~C\`e}
\author[Bern]{G.~Colangelo\footnote[2]{\label{SC}MUON-GM2-THEORY-SC@fnal.gov}}
% ...

\address[KEK]{Institute of Particle and Nuclear Studies, High Energy Accelerator Research Organization (KEK), Tsukuba 305-0801, Japan}
\address[Nishina]{Nishina Center, RIKEN, Wako 351-0198, Japan}
\address[Kobayashi]{Kobayashi--Maskawa Institute for the Origin of Particles and the Universe (KMI), Nagoya University, Nagoya 464-8602, Japan}
\address[Southampton]{School of Physics and Astronomy, University of Southampton, Southampton SO17 1BJ, United Kingdom}
\address[Sorbonne]{LPNHE, Sorbonne Universit\'e, Universit\'e de Paris, CNRS/IN2P3, Paris, France}
\address[Lund]{Department of Astronomy and Theoretical Physics, Lund University, S\"olvegatan 14A, 22362 Lund, Sweden}
\address[Uconn]{Department of Physics, 196 Auditorium Road, Unit 3046, University of Connecticut, Storrs, CT 06269-3046, USA}
\address[RIKENBNL]{RIKEN BNL Research Center,  Brookhaven National Laboratory, Upton, NY 11973, USA}
\address[CERN]{Theoretical Physics Department, CERN, 1211 Geneva 23, Switzerland}
\address[Bucharest]{Horia Hulubei National Institute for Physics and Nuclear Engineering, P.O.B.\ MG-6, 077125 Bucharest-Magurele, Romania}
\address[Pavia]{Istituto Nazionale di Fisica Nucleare (INFN), Sezione di Pavia,
 Via A.\ Bassi 6, 27100 Pavia, Italy}
 \address[Mainz]{PRISMA$^+$ Cluster of Excellence and Institute for Nuclear Physics,
 Johannes Gutenberg University of Mainz, 55099 Mainz, Germany }
 \address[HIM]{Helmholtz Institute Mainz, 55099 Mainz, Germany and GSI Helmholtzzentrum f\"ur Schwerionenforschung, 64291 Darmstadt, Germany }
\address[Bern]{Albert Einstein Center for Fundamental Physics, Institute for Theoretical Physics, University of Bern, Sidlerstrasse 5, 3012 Bern, Switzerland}
% ...

\begin{abstract}
We review the present status of the Standard Model calculation of the anomalous magnetic moment of the muon. 
This is performed in a perturbative expansion in the fine-structure constant $\alpha$ and is broken down into pure QED, electroweak, and hadronic contributions. The pure QED contribution
is by far the largest and has been evaluated up to and including $\mathcal{O}(\alpha^5)$ with negligible numerical uncertainty. 
The electroweak contribution is suppressed by $(m_\mu/M_W)^2$ and only shows up at the level of the seventh significant digit. It has been evaluated up to two loops and is known to better than one percent. 
Hadronic contributions are the most difficult to calculate and are responsible for almost all of the theoretical uncertainty. The leading hadronic contribution appears at $\mathcal{O}(\alpha^2)$ and is due to hadronic vacuum polarization, whereas at $\mathcal{O}(\alpha^3)$ the hadronic light-by-light scattering contribution appears. 
Given the low characteristic scale of this observable, these contributions have to be calculated with nonperturbative methods, in particular, dispersion relations and the lattice approach to QCD. 
The largest part of this review is dedicated to a detailed account of recent efforts to improve the calculation of these two contributions with either a data-driven, dispersive approach, or a first-principle, lattice-QCD approach. The final result reads $\amuSM=116\,591\,810(43)\times 10^{-11}$ and is smaller than the Brookhaven measurement by 3.7$\sigma$.  
The experimental uncertainty will soon be reduced by up to a factor four by the new experiment
currently running at Fermilab, and also by the future J-PARC experiment.
This and the prospects to further reduce the theoretical uncertainty in the near future---which are also discussed here---make this quantity one of the most promising places to look for evidence of new physics.
\end{abstract}

\numberwithin{equation}{section}

\maketitle

\newpage
\tableofcontents
\newpage

\setcounter{footnote}{0}

\setcounter{section}{-1}
\section{Executive Summary} 
\label{sec:execsumm}
\begin{table}[t]
\begin{centering}
\small
	\begin{tabular}{l  l l r l }
	\toprule
	   Contribution & Section & Equation & Value   $\times 10^{11}$ & References\\ \colrule
	   Experiment (E821)     &                   & \cref{amuexp} &$ 116\, 592\, 089(63) $ & Ref.~\cite{Bennett:2006fi}  \\\colrule
	HVP LO ($e^+e^-$)  &  \cref{sec:ConservativeMerging}   & \cref{merging} & $ 6931(40) $  &  Refs.~\cite{Davier:2017zfy,Keshavarzi:2018mgv,Colangelo:2018mtw,Hoferichter:2019gzf,Davier:2019can,Keshavarzi:2019abf}\\
        HVP NLO ($e^+e^-$) &   \cref{sec:higher_orders}   & \cref{HVPNLO} & $-98.3(7) $ & Ref.~\cite{Keshavarzi:2019abf}\\
        HVP NNLO ($e^+e^-$) &   \cref{sec:higher_orders}   & \cref{HVPNNLO} & $ 12.4(1) $ & Ref.~\cite{Kurz:2014wya}\\
        HVP LO (lattice, $udsc$) & \cref{subsec:status}  &\cref{eq:amuhlo_IB_lat}  & $7116 (184)$ & Refs.~\cite{\HVPlatticeref}\\
HLbL (phenomenology) & \cref{sec:HLbL} & \cref{eq:final-estimate2} & $92(19)$ & Refs.~\cite{\HLbLref}\\
HLbL NLO (phenomenology) & \cref{sec:HLbL_NLO} & \cref{HLbL_NLO} & $2(1)$ & Ref.~\cite{Colangelo:2014qya}\\
HLbL (lattice, $uds$) & \cref{sec:LHLBL} & \cref{eqn:hlbllatrbcres} & $79(35)$ & Ref.~\cite{\HLbLlatticeref}\\
HLbL (phenomenology + lattice) & \cref{sec:conclusionsWP} & \cref{eq:HLbL_comb} & $90(17)$ & Refs.~\cite{\HLbLcombref}\\\colrule
QED             &  \cref{sec:QED_sum}  &  \cref{eq:amuQED_Cs}    &  $116\,584\,718.931(104)$  & Refs.~\cite{\QEDref}\\
Electroweak     &  \cref{sec:fullEWresult}  &   \cref{amuEWNew}  & $153.6(1.0)$   & Refs.~\cite{\EWref}\\
HVP ($e^+e^-$, LO + NLO + NNLO) & \cref{sec:conclusionsWP} & \cref{HVPtotal} & $6845(40)$ & Refs.~\cite{\HVPtotalref} \\
HLbL (phenomenology + lattice + NLO) & \cref{sec:conclusionsWP}  & \cref{HLbLtotal} & $92(18)$ & Refs.~\cite{\HLbLtotalref}\\ 
        Total SM Value  & \cref{sec:conclusionsWP}            & \cref{SM_prediction} & $ 116\, 591\, 810(43) $  & Refs.~\cite{\SMref}\\
        Difference:    $\Delta a_\mu:=\amuexp - \amuSM$ & \cref{sec:conclusionsWP}  & \cref{amudiff} & $ 279(76) $ & \\
        \bottomrule
	\end{tabular}
	\caption{Summary of the contributions to $\amuSM$. After the experimental number from E821, the first block gives the main results for the hadronic contributions  from \cref{sec:dataHVP,sec:latticeHVP,sec:dispHLbL,sec:latticeHLbL} as well as the combined result for HLbL scattering from phenomenology and lattice QCD constructed in \cref{sec:conclusionsWP}. 
	The second block summarizes the quantities entering our recommended SM value, in particular, the total HVP contribution, evaluated from $e^+e^-$ data, and the total HLbL number. The construction of the total HVP and HLbL contributions takes into account correlations among the terms at different orders, and the final rounding includes subleading digits at intermediate stages.     
	The HVP evaluation is mainly based on the experimental Refs.~\cite{\HVPexpref}.
	In addition, the HLbL evaluation uses experimental input from Refs.~\cite{\HLbLexpref}.  The lattice QCD calculation of the HLbL contribution builds on crucial methodological advances from Refs.~\cite{\HLbLlatticemethods}. Finally, the QED value uses the fine-structure constant obtained from atom-interferometry measurements of the Cs atom~\cite{Parker:2018vye}.}
\label{tab:summary}
\end{centering}
\end{table}

The current tension between the
experimental and the theoretical value of the muon magnetic anomaly,
$a_\mu \equiv (g-2)_\mu/2$, has generated significant interest in the particle physics
community because it might arise from effects of as yet undiscovered particles contributing through virtual loops.
The final result from the Brookhaven National Laboratory (BNL) experiment E821, published in 2004, has a precision of $0.54$\,ppm.
At that time, the Standard Model (SM) theoretical value of $a_\mu$ that
employed the conventional $e^+e^-$ dispersion relation to determine
hadronic vacuum polarization (HVP), had an uncertainty of 0.7\,ppm, and
$\amuexp$ differed from $\amuSM$ by $2.7\sigma$.  An independent
evaluation of  HVP using hadronic $\tau$ decays, also at 0.7\,ppm
precision, led to a  $1.4\sigma$ discrepancy.    The situation was
interesting, but by no means convincing.  Any enthusiasm for a new-physics
interpretation was further tempered when one considered the variety of
hadronic models used to evaluate higher-order hadronic light-by-light
(HLbL) diagrams, the uncertainties of which were difficult to assess.
A comprehensive experimental effort to produce dedicated, precise, and extensive measurements of $e^+e^-$ cross sections, coupled with the development of sophisticated data combination methods, led to improved SM evaluations that determine a difference between $\amuexp$ and $\amuSM$ of $\approx3\hyph4\sigma$, albeit with concerns over the reliability of the model-dependent HLbL estimates.
On the theoretical side, there was a lot of activity to develop new model-independent approaches, including dispersive methods for HLbL and lattice-QCD methods for both HVP and HLbL. While not mature enough to inform the SM predictions until very recently, they held promise for significant improvements to the reliability and precision of the SM estimates.

This more tantalizing discrepancy is not at the discovery threshold.   Accordingly,  two major initiatives are aimed at resolving whether new physics is being revealed in the precision evaluation of the muon's magnetic moment.  The first is to improve the experimental measurement of $\amuexp$ by a factor of 4.   The Fermilab Muon $g-2$ collaboration is actively taking and analyzing data using proven, but modernized, techniques that largely adopt key features of magic-momenta storage ring efforts at CERN and BNL.  An alternative and novel approach is being designed for J-PARC.  It will feature an ultra-cold, low-momentum muon beam injected into a compact and highly uniform magnet.
The goal of the second effort is to improve the theoretical SM evaluation to a level commensurate with the experimental goals.  To this end, a group was formed---the {\it Muon $g-2$ Theory Initiative}---to holistically evaluate all aspects of the SM and to recommend a single value against which new experimental results should be compared.  This White Paper (WP) is the first product of the Initiative, representing the work of many dozens of authors.

The SM value of $a_\mu$ consists of contributions
from quantum electrodynamics (QED), calculated through fifth order in the fine-structure constant;
the electroweak gauge and Higgs bosons, calculated through second order;
and, from the strong interaction through virtual loops containing hadrons.
The overall uncertainty on the SM value remains dominated by the
strong-interaction contributions, which are the main focus of the Theory Initiative.  
  
In this paper, significant new results are presented, as are re-evaluations and summaries of previous work.
Particularly important advances have been made in distilling the various approaches to obtaining the HVP contribution from the
large number of old and new data sets.  The aim of the Initiative is an inclusive and conservative recommendation.  At this time, HVP
is determined from $e^+e^-$ data; new lattice efforts---while promising---are not yet at the level of precision and consistency to be included in
the overall evaluation.
New here is a data-driven prediction of HLbL based on a recently developed dispersive approach.  Additionally, a lattice-QCD evaluation
has reached the precision  necessary to contribute to the recommended HLbL value.  Together they
replace the older ``Glasgow'' consensus, and reduce the uncertainty on this contribution, while at the same time placing its estimate on 
solid theoretical grounds. 
A compact summary of results is given in \cref{tab:summary}, along with the section and equation
numbers where the detailed discussions are presented. The last column provides for each result the underlying list of references used to obtain it. We strongly recommend that these references be cited in any work that uses the results presented here. 
The Initiative has created a website \cite{MuonInitiative}, which includes links to downloadable bib files and citation commands, to make it easy to add these references to the bibiliography. 
The recommended SM value lies $3.7\sigma$ below the E821 experimental result.

\clearpage

\section{Introduction}
\label{sec:introWP}

The anomalous magnetic moment of the muon\footnote{The muon magnetic moment $\boldsymbol{\mu}$ is a vector along the spin $\boldsymbol{s}$,
$\boldsymbol{\mu} = g({Qe}/{2m_\mu}) \boldsymbol{s}$.  The $g$ factor consists of the Dirac value
of 2 and the factor $a_\mu = (g-2)_\mu/2$, which arises from
radiative corrections. The dimensionless quantity $a_\mu$ is called by
several names in the literature:
``the muon magnetic anomaly,'' the ``muon anomalous magnetic moment,''
and the ``muon anomaly.''  All of these terms are used interchangeably in
this document. }  has, for well over ten years now, provided an enduring hint for new physics, in the form of a tantalizing $3\hyph4\sigma$ tension between SM theory and experiment.  
It is currently measured to a precision of about $0.5$ ppm~\cite{Bennett:2006fi}, commensurate with the theoretical uncertainty in its SM prediction. With a plan to reduce the experimental uncertainty by a factor of four, two new experiments will shed new light on this tension: the E989 experiment at Fermilab \cite{Grange:2015fou}, which started running in 2018, and the E34 experiment at J-PARC, which plans to start its first run in 2024~\cite{Abe:2019thb}. 

However, without improvements on the theoretical side, the discovery potential  of these efforts may be limited.  To leverage the new experimental efforts at Fermilab and J-PARC and hence unambiguously discover whether or not new-physics effects contribute to this quantity, the theory errors must be reduced to the same level as the experimental uncertainties. 
In the SM, $a_\mu$ is calculated from a perturbative expansion in the fine-structure constant $\alpha$, which starts 
with the Schwinger term $\alpha/(2 \pi)$ and has been carried out up to and including ${\mathcal O}(\alpha^5)$. Its uncertainty,  dominated by the unknown ${\mathcal O}(\alpha^6)$ term, is completely negligible. Electroweak corrections have been evaluated at full two-loop order, with dominant three-loop effects estimated from the renormalization group. Their uncertainty, mainly arising from nonperturbative effects in two-loop diagrams involving the light quarks, is still negligible compared to the experimental precision.   
The dominant sources of theory error are by far the hadronic contributions, in particular, the ${\mathcal O}(\alpha^2)$ HVP  term and the ${\mathcal O}(\alpha^3)$ HLbL  term. 
There are a number of complementary theoretical efforts underway to better understand and quantify the hadronic corrections, including using dispersive methods, lattice QCD, and effective field theories, as well as a number of different experimental efforts to provide inputs to dispersive, data-driven evaluations. The {\it Muon $g-2$ Theory Initiative} was created to facilitate interactions among these different groups, as well as between the theoretical and experimental $g-2$ communities.  It builds upon previous community efforts, see, e.g., Refs.~\cite{Blum:2013xva,Benayoun:2014tra}, to improve the SM prediction for $a_\mu$.

The Initiative's activities are being coordinated by a Steering Committee that consists of theorists, experimentalists, and representatives from the Fermilab and J-PARC muon $g-2$ experiments.   
This committee also functions as the Advisory Committee for the  workshops it organizes. Given the precision goals and the potential impact, it is crucially important to have more than one independent method for each of the two hadronic corrections, each with fully quantified uncertainties. Fostering the development of such methods is a prime goal of the Initiative, as this will enable critical cross-checks, and, upon combination, may yield gains in precision, to maximize the impact of E989 and E34. To this end, several workshops were organized in 2017, 2018, and 2019.

The first meeting, held near Fermilab~\cite{FNAL2017}, served to kick-off the Initiative's activities.  All sessions in the workshop were plenary and featured a mix of talks and discussions. Representatives of all major theoretical efforts on the hadronic contributions to the muon were invited to speak about their work, and all theorists working on such calculations were encouraged to participate. Representatives from the $e^+e^-$ experiments, which are performing measurements needed for evaluations of the hadronic corrections to $a_\mu$ based on dispersive methods, also presented invited talks, as did members of the Fermilab and J-PARC experiments. 

The Fermilab workshop's main outcome was a plan to write a WP on the theory status of the SM prediction of the muon $g-2$. Given the high stakes of a possible discovery of new physics the emphasis was on presenting a reliable SM prediction with a conservatively estimated error. The time plan had as a final goal to post the WP before the public release of the Fermilab E989 experiment's  measurement from their run 1 data. For that purpose, two working groups were formed, one on the HVP correction and another on the HLbL correction, and all stakeholders were invited to join them. Each working group held a meeting in early 2018. The HVP workshop was held at KEK \cite{KEK2018}  and the HLbL workshop at the University of Connecticut~\cite{UConn2018}. 

The second plenary meeting of the Initiative was held at the University of Mainz~\cite{Mainz2018} in June 2018. The first four days of the workshop followed the successful format of the Fermilab workshop, while the last day was reserved for editorial meetings for the WP, which produced a detailed outline, including writing assignments. Finally, the most recent meeting took place in September 2019 at the Institute for Nuclear Theory (INT) at the University of Washington in Seattle~\cite{INT2019}. It followed the same format as the previous two workshops, with a mix of talks and extended discussion time. It also included breakout sessions to bring the co-authors of the four main sections together to map out the conclusions of each section and the strategies for finalizing them. The INT meeting was instrumental for setting out the rules and deadlines, collectively referred to as the ``Seattle agreement,'' which are needed for finalizing the WP: 
\begin{itemize}

\item Procedure for obtaining the final estimate.

The consensus reached early on was to aim for a conservative error
estimate, but a concrete implementation of this principle into a detailed
procedure was first worked out and agreed upon during the INT
workshop. Details can be found in~\cref{sec:ConservativeMerging}, the concluding parts of the other sections, as well as \cref{sec:conclusionsWP}.

\item Authorship 

All participants of past Muon $g-2$ Theory Initiative workshops, members of the two working groups, and their collaborators were invited to become co-authors of the WP. The contributions from section authors, defined as members who made significant contributions to the corresponding sections of the WP, are highlighted at the beginning of each section.  

\item Deadline for essential inputs: 31 March 2020

Essential inputs are defined as experimental data used in data-driven, dispersive evaluations, or new theoretical calculations that contribute to the SM prediction of $a_\mu$.  A paper that contains essential inputs must be published by the deadline, in order to be included in the final results and averages. Papers that appear on arXiv, but are not published before the deadline will be mentioned in the WP. 
The original, agreed-upon deadline at the Seattle meeting was earlier (15 October 2019). It was adjusted to the date shown above, to reflect the actual timeline of the WP. 

\end{itemize}

The work of the Muon $g-2$ Theory Initiative will continue, certainly for the duration of the experimental programs at Fermilab and J-PARC. With the focus of the first WP on the consolidation of the  SM prediction, a workshop is planned at KEK \cite{KEK2020} to discuss the next steps towards reducing the theory errors to keep pace with experiment.

The Steering Committee is co-chaired by Aida El-Khadra and Christoph Lehner and includes Gilberto Colangelo, Michel Davier, Simon Eidelman, Tsutomu Mibe, Andreas Nyffeler, Lee Roberts, and  Thomas Teubner. The Steering Committee's tasks are the long-term planning of the Theory Initiative as well as the planning and organization of the workshops that led to the writing of the WP. 
The writing of the WP by the various section authors was coordinated by the WP editorial board, which also performed the final assembly into one document. The WP editorial board included
all the members of the Steering Committee and Martin Hoferichter. 

The remainder of this review is organized as follows. With the focus of this paper on the hadronic corrections, we first discuss the evaluations of HVP, the dominant hadronic contribution, where we summarize the status and prospects of dispersive evaluations in \cref{sec:dataHVP} and 
lattice calculations in \cref{sec:latticeHVP}. The source of the currently second-largest uncertainty, HLbL scattering, is addressed with data-driven and dispersive techniques in \cref{sec:dispHLbL} and with lattice QCD in \cref{sec:latticeHLbL}. The current status of the QED and electroweak contributions is presented in \cref{sec:QED,sec:EW}, respectively. In \cref{sec:conclusionsWP} we summarize the main conclusions and construct our recommendation for the current SM prediction.

\clearpage

\section{Data-driven calculations of HVP}
\label{sec:dataHVP}

\noindent
\begin{flushleft}
 \emph{M.~Benayoun, C.~M.~Carloni Calame, H.~Czy\.z, M.~Davier, S.~I.~Eidelman, M.~Hoferichter, F.~Jegerlehner, A.~Keshavarzi, B.~Malaescu, D.~Nomura, M.~Passera, T.~Teubner, G.~Venanzoni, Z.~Zhang }  
 \end{flushleft}

\subsection{Introduction}
\label{sec:hvp_introduction}

Based on analyticity and unitarity, loop integrals containing
insertions of HVP in photon
propagators can be expressed in the form of dispersion integrals over the
cross section of a virtual photon decaying into hadrons. This cross
section can be determined in $e^+e^-$ annihilation, either in {\em direct
scan} mode, where the beam energy is adjusted to provide measurements at
different center-of-mass (CM) energies, or by relying on the method of
{\em radiative return}, where a collider is operating at a fixed
CM energy. In the latter, the high statistics allow for an
effective scan over different masses of the hadronic system through
the emission of initial-state photons, whose spectrum can be calculated
and, in some cases, measured directly. With the availability of high-luminosity colliders, especially meson factories, this method of
radiative return has become a powerful alternative to the
direct scan experiments. In addition, it is possible to use hadronic $\tau$
decays to determine hadronic spectral functions, which can be related
to the required hadronic cross section. 
As a consequence of the wealth of data from many sources, the hadronic
cross section is now known experimentally with a high precision
over a wide range of energies. This allows one to obtain data-driven determinations of the HVP contributions. 

At leading order (LO), i.e.,
${\cal O}(\alpha^2)$, the dispersion integral 
reads~\cite{Brodsky:1967sr,Lautrup:1969fr}
\be
\amuHVPLO=\, \frac{\alpha^2}{3\pi^2} \int_{\mpi^2}^\infty
\frac{K(s)}{s}R(s)\,ds\,,
\label{eq:amuhvsdisplo}
\ee
with the kernel function 
\be
K(s) = \frac{x^2}{2}(2-x^2) +
\frac{(1+x^2)(1+x)^2}{x^2}\left( \log(1+x)-x+\frac{x^2}{2}\right) +
\frac{1+x}{1-x}x^2\log x\,,
\ee
where $x=\frac{1-\beta_\mu}{1+\beta_\mu}$,
$\beta_\mu=\sqrt{1-4m_\mu^2/s}$. When expressed in the form $\hat K(s) = \frac{3s}{m_\mu^2}K(s)$, the kernel function $\hat K$ is a slowly varying 
monotonic function, rising from $\hat K(4\mpi^2)\approx 0.63$ at the two pion 
threshold to its asymptotic value of 1 in the limit of large $s$. $R(s)$ 
is the so-called (hadronic) $R$-ratio
defined by\footnote{Note that this standard definition of $\sigma_{\rm
    pt}$ does not take into account effects due to the finite electron
  mass, which, for CM energies above the hadronic
  threshold, are completely negligible.}
\be
\label{Rratio}
R(s)=\frac{\sigma^0(e^+e^-\to \text{hadrons} (+\gamma))}{\sigma_{\rm
    pt}}\,,\quad 
\sigma_{\rm pt}= \frac{4\pi\alpha^2}{3s}\,.
\ee
Due to the factor $K(s)/s$, contributions from the lowest energies
are weighted most strongly in \cref{eq:amuhvsdisplo}. Note that the
superscript in $\sigma^0$ indicates that the total hadronic cross
section in the dispersion integral must be the {\em bare} cross section,
excluding effects from vacuum polarization (VP) (which lead to the running
QED coupling). If these effects are included as part of the measured hadronic cross
section, this data must be ``undressed,'' i.e., VP effects must be
subtracted, see the more detailed discussion below. Otherwise, there
would be a double counting and, as such, iterated VP
insertions are taken into account as part of the higher-order HVP
contributions. 

Conversely, the hadronic cross section used in the
dispersion integral is normally taken to be inclusive with respect to final-state radiation (FSR) of additional photons. While this is in
contradiction to the formal power counting in $\alpha$, it would
basically be impossible to subtract the real and virtual photonic FSR effects in hadron production, especially for higher-multiplicity states for which these QED effects are difficult to
model. As these FSR effects are not included explicitly in the higher-order VP contributions, this procedure is fully consistent. Note that,
in line with these arguments, the threshold for hadron production is
provided by the $\pi^0\gamma$ cross section and hence the 
lower limit of the dispersion integral is $M_{\pi^0}^2$. 

Similar dispersion integrals have been derived for the HVP
contributions at next-to-leading order (NLO)~\cite{Krause:1996rf} and next-to-next-to-leading order (NNLO)~\cite{Kurz:2014wya}. 
They are more complicated and require double and triple
integrations, respectively, and will not be given explicitly
here. As will be discussed in more detail below, the NLO contributions
are numerically of the order of the HLbL
contributions, but negative in sign. The NNLO contributions turn out
to be somewhat larger than naively expected and, therefore, should 
be evaluated as a nonnegligible component of $\amuHVP$.

\paragraph{Hadronic cross section at low energies} 
At low energies, the total hadronic cross section must be obtained
by summing all possible different final states. Numerous measurements for
more than 35 exclusive channels from different experiments have been
published over many years. Due to the size of the cross section and
its dominance at low energies, the most important channel is the two-pion 
channel, which contributes more than 70\% of $\amuHVPLO$.  
This final state stems mainly from decays of the $\rho$ meson, with an 
admixture of the $\omega$. Sub-leading contributions arise from decays
of the $\omega$ and $\phi$ in the three-pion and two-kaon channels,
and from four-pion final states with more complicated production
mechanisms. Note that by taking the incoherent sum over distinct final
states, interferences between different production mechanisms are
taken into account without the need to model their strong dynamics or
to fit them. Even-higher-multiplicity final states (up to six pions)
and final states containing pions and kaons or the $\eta$ have become
important to achieve an accurate description of the total hadronic
cross section. Contributions for which no reliable data exists, but which are
expected not to be negligible, have to be estimated. This is, e.g., the case
for multi-pion channels consisting mostly or entirely of neutral
pions. Such final states can be approximated by assuming isospin
symmetry, which can be used to model relations between measured and
unknown channels. The reliability of such relations is difficult to
quantify and is usually mitigated by assigning a large fractional
error to these final states. However, with more and more channels
having been measured in recent years, the role of these isospin-based
estimates has been largely diminished. 
For leading contributions very close to threshold, where data can be
sparse, the hadronic cross section can be estimated based on
additional constraints, e.g., from chiral perturbation theory (ChPT).
The data for the most relevant channels and recent
developments from the different experiments is reviewed below in \cref{ee-data} in more detail.

\paragraph{Hadronic cross section at higher energies}
For energies beyond about $\sqrt{s}\sim 2\GeV$, summing exclusive
channels becomes unfeasible, as many exclusive measurements do not
extend to higher energies and because more unmeasured higher-multiplicity channels would have to be taken into account. One
therefore relies on measurements of the inclusive cross
section. Alternatively, for energies above the $\tau$ mass and away from
flavor thresholds, perturbative QCD (pQCD) is expected to provide a good
approximation of the total hadronic cross section and is used
widely. Contributions to $R$ from massless quarks are known to order $\alpha_s^4$ in pQCD, whereas the cross section for heavy quarks is
available at order $\alpha_s^2$. QED corrections to the inclusive cross
section are small and can be added easily. A popular routine to
calculate the hadronic $R$-ratio in pQCD is {\tt rhad}
\cite{Harlander:2002ur}, to which we also refer for formulae and a
detailed discussion.

In which energy regions pQCD can be used to replace data is a matter of
debate. Between different groups there is consensus that above the 
open $b\bar b$ threshold, at about $\sqrt{s}\sim 11\GeV$, pQCD can
certainly be trusted and is much more accurate than the available
quite old data for the inclusive hadronic cross section. However, for
energies between the charm and bottom thresholds and above the
exclusive region (i.e., from $1.8\hyph2\GeV$), different analyses either rely 
on the inclusive data or the use of pQCD. 
For a detailed discussion of the resulting differences, see \cref{HVPdisp_evaluations}. 
At higher energies the theoretical uncertainty of the pQCD
predictions can, in a straightforward way, be estimated by varying the
input parameters, the strong coupling $\alpha_s$ and the quark masses,
together with a variation of the renormalization scale. Alternatively one can
consider the size of the highest-order contribution as an indication
of the error induced by the truncation of the perturbative
series. While these procedures have no strict foundation and no clear
statistical interpretation, they are commonly
accepted.\footnote{Another possible estimate of the error would be the
  variation of the renormalization scheme used in the pQCD
  calculations. As the availability of results in schemes different
  from the usually used ${\overline{\rm MS}}$ scheme is limited, such
  an approach has not been adopted commonly, but see, e.g., Ref.~\cite{Gracey:2014pba} for a discussion of results for $R$ in
  different classes of so-called MOM schemes.}
The error estimates of the perturbative cross section obtained in this
way are typically significantly smaller than those obtained when
relying on the available data.
At lower energies, $\sqrt{s}\sim 2\GeV$, residual duality violations are likely to represent a more important correction to the pQCD prediction. These have been estimated in Ref.~\cite{Davier:2019can} and are discussed in more detail in \cref{sec:ConservativeMerging}.

While the density and quality of the available data allows one to resolve
and integrate the contributions from the $\rho$, $\omega$, and $\phi$ resonances
directly and without modeling, the very narrow charm resonances
$J/\psi$ and $\psi'$, as well as the $\Upsilon(1\hyph3\,{\rm S})$ states
have to be added with suitable parameterizations to the continuum
contributions. However, these heavier resonances provide only
subleading contributions to $\amuHVP$ and its error.

\paragraph{Data treatment}
In the hadronic cross section as measured in $e^+e^-$ annihilation,
$e^+e^-\to \gamma^* \to \text{hadrons}$, the physical, ``full'' photon
propagator contains any number of insertions of the VP operator $\Pi(q^2)$. Unless the hadronic cross section is
normalized with respect to the measured muon pair production cross
section, which contains the same VP insertions so that they cancel
exactly, these ``running coupling'' corrections have to be subtracted,
as has been explained above. For many recent data sets, this is
already done and {\em undressed} cross section values $\sigma^0$ are
published. If not, it must be done prior to use in the dispersion
integrals. While the leptonic VP contributions to $\Pi(q^2)$ have been
calculated in QED to three-~\cite{Steinhauser:1998rq} and four-loop accuracy~\cite{Sturm:2013uka}, 
the hadronic VP contributions cannot
be reliably calculated in pQCD.  Instead they are obtained via a
dispersion integral that relates the leading real part of the
hadronic VP operator to its imaginary part, which is provided by the
hadronic cross section (or $R$-ratio), 
\be
\label{alpha_had}
{\rm Re}\, \Pi_{\rm had}(q^2)\, =\, -\frac{\alpha q^2}{3\pi}\, {\rm P}
\int_{\mpi^2}^\infty \frac{R(s)}{s(s-q^2)}\, ds\,. 
\ee
Here P indicates the principal-value prescription and the hadronic $R$-ratio 
is the same as in \cref{eq:amuhvsdisplo}.\footnote{A possible
  experiment designed to measure VP directly is discussed in
  \cref{Sec:MUonE}.} 
The subleading imaginary part is provided by the cross section data
and should, for the best possible accuracy, be included. One therefore
relies on the hadronic data one wants to undress, which is not a
problem as in practice an iterative process converges rapidly.
The main experiments and groups involved in data compilations, such as 
CMD-2 and SND at Novosibirsk, DHMZ, Jegerlehner, and KNT, have
developed their own different routines and parameterizations.

As remarked above, (real and virtual) FSR
must be included in the hadronic cross section. However, it is not
easy to determine to which extent real FSR
may have been excluded in the experimental analyses. Clearly, real
soft and virtual contributions are inevitably part of the measured
cross section, but hard real radiation has been omitted, to some
extent, by selection cuts. For (charged) pion and kaon production,
FSR is typically modeled by scalar QED, which has been shown to be a good
approximation for small photon energies, corresponding to large wavelengths, 
where the composite structure of the mesons is not
resolved~\cite{Hoefer:2001mx,Gluza:2002ui,Lees:2015qna}. 
If subtracted in an experimental analysis,
it can hence be added back in these cases. However, it is difficult to
model FSR in multi-hadronic systems with high precision, which
contributes to the uncertainty of this data (though it should be
noted that for most exclusive channels used for $\amuHVP$
there is limited phase space for hard radiation, which makes this
issue less important). If measurements are based on the method of
radiative return, which in itself is an $\order(\alpha)$ process, the
understanding of FSR and its interplay with the initial-state radiation (ISR) is of paramount
importance and an integral part of these analyses.

It is clear that the accuracy in the treatment of the data with respect to
radiative corrections is limited. Therefore, typically additional
radiative-correction errors are assigned, which aim to take into account
these uncertainties. 

\paragraph{Data combination}
There are different ways in which the hadronic data can be used to
obtain $\amuHVP$ in a combined analysis. In principle, if
cross sections are measured finely enough by a single experiment, one
can first integrate individual data sets, then average. However, this
may prevent the use of sparse data and mask possible tensions in the
spectral function between different experiments (or data sets of the
same experiment), which may be invisible after integrating. Therefore,
most of the recent analyses rely on first combining data, then taking
the $g -  2$ integral.
In this case, the combination (in each exclusive channel or for the
inclusive data) must take into account the different energy
ranges, the different binning, and possible correlations within and
between data sets. To achieve this, different methods are used by
different groups, as will be discussed in \cref{HVPdisp_evaluations}.
For the direct data integration, the $g -  2$ integral can then be
performed using a simple trapezoidal rule or after first applying more
sophisticated methods to smooth the cross section behavior locally. 
Alternatively, additional constraints on the hadronic cross sections
can be imposed. Such constraints can be due to analyticity and
unitarity, see \cref{sec:dispersive}, or from global fits of 
hadronic cross sections based on models like the Hidden Local Symmetry 
models discussed in \cref{sec:other_approaches}. In the latter 
case, the derived model cross sections are used in the dispersion integrals.

The remainder of this section is organized as follows. In \cref{ee-data},
the different experiments and methods, direct scan and radiative
return, are discussed. The hadronic cross section data is reviewed,
with emphasis on the most important channels and comparisons of
data from different experiments for the same channel. This section
also includes a short discussion of radiative corrections and
Monte Carlo generators, and of the possible use of spectral-function
data from hadronic $\tau$ decays. \Cref{HVPdisp_evaluations} contains 
short reviews of the most popular global analyses for the HVP contributions 
to $a_\mu$. It also includes a discussion of additional constraints that
can be used to further improve the two-pion channel, a comparison of the 
different evaluations, and a conservative merging of the main data-driven 
results. \Cref{HVPdisp_prospects} discusses 
prospects for further improvements of the data-driven determination of 
$\amuHVP$ and \cref{HVPdisp_conclusions} contains a short 
summary and the conclusions for this part.

\subsection{Hadronic data}
\label{ee-data}

The dispersive approach for computing HVP contributions to the muon 
anomalous magnetic moment is based on the availability of \epem annihilation
measurements of hadronic cross sections at energies below a few GeV. 
In this section, we present a review of this data,
where a wealth of precision results has been obtained in recent years.

\subsubsection{Experimental approaches}

\paragraph{The scan method}

Until recently, measurements of annihilation cross sections were done by
taking data at fixed CM energies, taking advantage of the 
good beam energy resolution of $e^+e^-$ colliders. Then the full accessible range was scanned at
discrete energy points. At each point the cross section for the process
$\epem \rightarrow X$ is directly obtained
through
\be
  \sigma_X = \frac {N_X}{\epsilon_X (1+\delta) L_{ee}}\,,
\ee
where $N_X$ is the observed number of $X$ events, $\epsilon_X$ is the 
efficiency depending on the detector acceptance and the event selection cuts, 
$(1+\delta)$ the radiative correction, and $L_{ee}$ the integrated \epem 
luminosity 
obtained from registered leptonic events with known QED cross sections 
($\epem \rightarrow e^+e^-$, $\mu^+\mu^-$, or $\gamma\gamma$). All quantities 
depend on the CM energy $\sqrt{s}$ of the scan point. The radiative 
correction takes into account the loss of events by ISR causing them to be rejected by
the selection, which usually imposes constraints on energy-momentum balance.

\begin{figure}[t] \centering
\includegraphics[width=5cm]{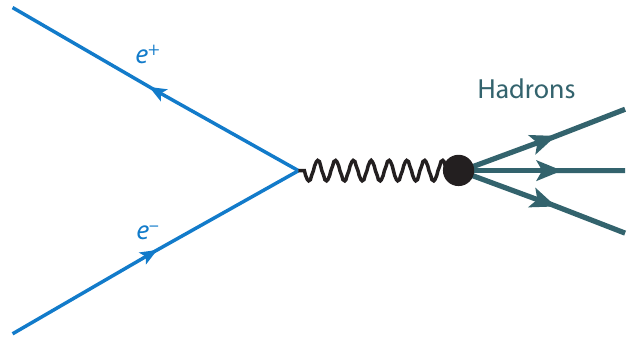}\quad
\includegraphics[width=5cm]{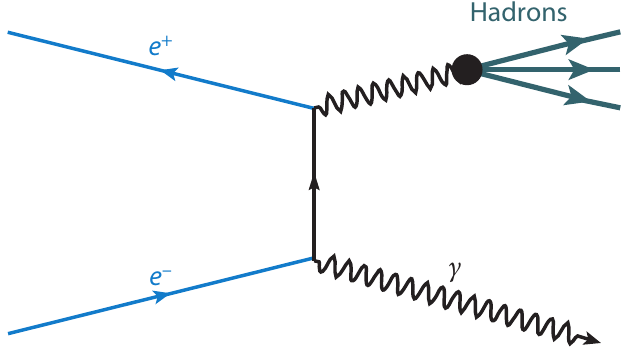} 
\caption{The LO Feynman diagrams for the annihilation processes 
$\epem  \rightarrow {\rm hadrons}$ (left) and 
$\epem \rightarrow \gamma + {\rm hadrons}$ with ISR (right). Reprinted from Ref.~\cite{Davier:2013vna}.}
\label{feynman-ee}
\end{figure}

At LO the process is described by the Feynman diagram shown 
in \cref{feynman-ee}. The beauty of \epem annihilation is its simplicity 
due to the purely leptonic initial state governed by QED and the exchange 
of a highly virtual photon coupled to any charged particles (leptons or 
quarks). Thus strong interaction dynamics can be studied in a very clean 
way as quark pairs are created initially out of the QCD vacuum. 

The advantages of the scan approach are (i) the well-defined CM energy 
(mass of the hadronic system), which applies for both the process being 
investigated and background, thus limiting the number of sources for the 
latter, and (ii) the very good energy resolution, typically 
$\sim 10^{-3}\sqrt{s}$, allowing for the study of the line shape of narrow 
resonances such as the $\omega$ and the $\phi$.
These good points have some negative counterparts, as data taking has to 
be distributed at discrete values, leaving gaps without information, 
while being limited by the operating range of the collider as luminosity 
usually drops steeply at lower energies. The consequence of this fact is 
that the wide range of energies necessary for the dispersion integral has 
to be covered by a number of experiments at different colliders of 
increasing energies. Thus, only for the region from threshold to $2\GeV$, 
three generations of colliders have been used. An additional complication of this 
situation is a lack of continuity in detector performance and therefore 
some difficulties for evaluating systematic uncertainties in a coherent way.

Precise results below $1.4\GeV$ from the CMD-2 and SND detectors at BINP
(Novosibirsk) have been obtained in the scan mode at VEPP-2M, and more 
recently from CMD-3 and SND at VEPP-2000 up to $2\GeV$.
Inclusive measurements with BES-II at the BEPC collider at IHEP (Beijing),
BESIII with the improved BEPCII, and KEDR at BINP are also available above $1.9\GeV$.  Finally, results exist from older experiments at Orsay, Novosibirsk, and Frascati, but they are much less accurate.

\paragraph{The ISR approach}
\label{sec:ISR}

ISR is unavoidable, but it can be turned 
into an advantage by using the
NLO process shown in \cref{feynman-ee} in order
to access the LO cross section. In practice this approach could only be
implemented with the advent of high-luminosity colliders, such as the $\phi$ 
factory DA$\Phi$NE and the $B$ factories KEK-B and PEP-II (all designed for the 
study of $CP$ violation), in order to compensate for the $O(\alpha)$ 
reduction in rate for ISR.

Of course ISR occurs all the time, but the difference between the two 
approaches resides in the fact that in the scan method one selects events with a
radiative photon energy very small compared to the CM energy, whereas in 
the ISR approach one tries to cover the full range of photon energies. 
Keeping a fixed CM energy $\sqrt{s}$ enables the collection of events 
over a wide spectrum of energies $\sqrt{s'}$ controlled by the ISR photon 
energy fraction $x=2E^*_\gamma/\sqrt{s}$ such that $s'=(1-x)s$. 

The cross section for $\epem \rightarrow X$ can be obtained from the measured 
spectrum of $\epem \rightarrow \gamma X$ events through
\be
\label{Eq:def-lumi}
  \frac {dN_X(\gamma)\gamma_{\rm ISR}}{d\sqrt{s'}}=
   \frac {dL_{\rm ISR}^{\rm eff}}{d\sqrt{s'}}~
    \epsilon_{X\gamma}(\sqrt{s'})~\sigma_{X(\gamma)}^0(\sqrt{s'})\,,
\ee
where $dL_{\rm ISR}^{\rm eff}/d\sqrt{s'}$ is the effective ISR luminosity,
$\epsilon_{X\gamma}$ is the full acceptance for the event sample, and
$\sigma_{X(\gamma)}^0$ is the ``bare'' cross section for the process $e^+e^-
\to X(\gamma)$ (including FSR effects, but with leptonic 
and hadronic VP contributions excluded). The latter use 
of the bare cross section, rather than the dressed cross section, is a 
matter of choice, as to where one includes the dressing factor. With the choice made in \cref{Eq:def-lumi}, 
the differential ISR luminosity reads
\be     
\label{lumi-eff}
 \frac {dL_{\rm ISR}^{\rm eff}}{d\sqrt{s'}}=L_{ee}~\frac {dW}{d\sqrt{s'}}
   \left(\frac {\alpha(s')}{\alpha(0)}\right)^2\,. 
\ee
\Cref{lumi-eff} relies on the \epem luminosity measurement 
($L_{ee}$) and on the theoretical radiator function $dW/d\sqrt{s'}$. 
The latter describes the probability to radiate an ISR photon (with possibly
additional ISR photons) so that the produced final state (excluding ISR 
photons) has a mass $\sqrt{s'}$. This probability depends on $s$, $s'$, and 
on the angular range for the ISR photon in the \epem CM system. 

The ISR approach for low-energy cross sections has been followed by
the KLOE experiment at DA$\Phi$NE for the $\pi^+\pi^-$ channel and by BABAR
operating at PEP-II, where an extensive program of multi-channel measurements was conducted
in the few-GeV range. More recently, results have also been obtained with 
BESIII and CLEO-c. 
Different variants have been used, depending on whether or not the ISR 
photon is detected and how the ISR luminosity is determined:
\ben
\item  photon at small angle and undetected, radiator function 
from NLO QED: KLOE-2005~\cite{Aloisio:2004bu};\footnote{The data from this measurement should not be used because of a trigger problem and the need for a reevaluation of the Bhabha cross section in the new version of the Babayaga generator, as explained in Ref.~\cite{Venanzoni:2017ggn}.}  
KLOE-2008 $\pi^+\pi^-$ at $\sqrt{s}=1.02\GeV$~\cite{Ambrosino:2008aa}; BABAR 
$p \bar{p}$ at $\sqrt{s}=10.58\GeV$~\cite{Lees:2013uta};
\item \label{item:2} photon at large angle and detected, radiator function from 
NLO QED: KLOE-2010 $\pi^+\pi^-$ at $\sqrt{s}=1.02\GeV$\,\cite{Ambrosino:2010bv}; 
BABAR multihadronic channels at 
$\sqrt{s}=10.58\GeV$~\cite{Aubert:2004kj,Aubert:2005eg,Lees:2012cr,Aubert:2007ef,Aubert:2006jq,TheBaBar:2017vzo,Aubert:2007ym,Aubert:2007ur,Lees:2011zi,Aubert:2005cb,Lees:2013ebn,Aubert:2007uf};
\item photon at large angle and detected, radiator function from measured
$\mu^+\mu^-(\gamma)$ events: BABAR $\pi^+\pi^-$~\cite{Aubert:2009ad,Lees:2012cj} and $K^+K^-$~\cite{Aubert:2009ad,Lees:2012cj} at $\sqrt{s}=10.58\GeV$; 
BESIII $\pi^+\pi^-$~\cite{Ablikim:2015orh} and CLEO-c $\pi^+\pi^-$~\cite{Xiao:2017dqv} at $\sqrt{s}\sim 4\GeV$;
\item  photon at small angle and undetected, radiator function from measured
$\mu^+\mu^-(\gamma)$ events: 
KLOE-2012 $\pi^+\pi^-$ at $\sqrt{s}=1.02\GeV$~\cite{Babusci:2012rp}).
\een

Specific choices obviously depend on experimental opportunities and are 
optimized as such. At DA$\Phi$NE, small-angle ISR is advantageous in order to
reduce background and LO FSR events. At high-energy colliders, the best approach for precision 
measurements is the large-angle ISR photon detection, which provides a 
kinematic handle against backgrounds and the simultaneous analysis of hadronic
and $\mu^+\mu^-$ final states in order not to depend on 
a Monte Carlo generator
for determining the ISR luminosity. It even allows considering an extra photon 
in the kinematic fit~\cite{Aubert:2009ad,Lees:2012cj,Lees:2013gzt}, ensuring that the ISR process is directly measured at 
NLO, thus reducing the radiative corrections. Also this configuration defines 
a topology where the ISR photon is back-to-back to the produced hadrons, thus 
providing high acceptance and better particle identification due to 
larger momenta.  
High acceptance is important for multi-hadronic final states because it means 
less dependence on internal dynamics for computing the selection efficiency, 
hence a smaller systematic uncertainty.
Particle identification is also easier, particularly with method \ref{item:2} at 
$B$ factories because the final state is strongly boosted. 
ISR luminosity determination with detected muon pairs is 
equivalent to measuring a ratio of events hadrons/$\mu\mu$
in which several effects cancel (particularly extra ISR), thus allowing for
a reduction of systematic uncertainties.

Apart from the points just mentioned the big advantage of the ISR approach is
to yield in one fell swoop a continuous cross section measurement over 
a broad range of energies. 
The practical range extends from threshold (for large-angle ISR) to
energies close to $\sqrt{s}$. At low CM energy (KLOE) the limitation for the 
upper range is the decreasing photon energy and the rapid rise of the LO FSR 
contribution, which has to be subtracted out. At large energy (BABAR) it is
statistics and background that limit the range, but still values of a few GeV
are obtained, depending on the process. The main experimental disadvantage of the ISR 
approach is that many background processes can contribute and some effort is
needed to control them. They range from higher-multiplicity ISR processes to
nonradiative annihilation to hadrons at the beam energies. In the latter case 
the photon from a high-energy $\pi^0$ can mimic an ISR photon and this 
contribution must be estimated directly on data in order not to rely on 
models used in Monte Carlo generators.

\paragraph{Radiative corrections  and Monte Carlo simulation}

A correction of the annihilation event yield because of
extra radiation is mandatory as it can be quite large ($\sim 10\%$ or more). As an overall
precision of less than 1\% is now the state of the art for cross section 
measurements, radiative corrections have to be controlled accordingly. 
Calculations are made at NLO with higher orders resummed for the radiation 
of soft photons in the initial state. The full radiative corrections 
involve ISR and FSR soft and hard photon emission, virtual contributions, 
and VP. As detector acceptance and analysis cuts must 
be taken into account, radiative corrections are implemented using Monte 
Carlo event generators. Dedicated accurate generators have been recently
developed for the determination of the \epem luminosity, such as 
{\small BHWIDE}~\cite{Jadach:1995nk} and
{\small BABAYAGA}~\cite{Balossini:2006wc} at two-loop level (NNLO), and for 
annihilation processes through ISR, such as {\small EVA}~\cite{Binner:1999bt,Czyz:2000wh} and its 
successor {\small PHOKHARA}~\cite{Czyz:2004ua,Czyz:2004rj} with almost complete NLO 
contributions. These generators and their performance are discussed in \cref{MC-radcor}. 

Unlike for the QED $\mu^+\mu^-$ process the simulation of FSR from hadrons
is model-dependent. It is true for the LO part because the measured range
of $s'$ is very close to $s$ for KLOE, thus enhancing the importance of LO FSR,
except for the case of small-angle ISR.
Although the interference between LO ISR and FSR amplitudes vanishes for a 
charge-symmetric detector when integrating over all configurations, some 
control over the $|\text{FSR}|^2$ can be obtained from the measurement of a charge 
asymmetry~\cite{Binner:1999bt,Rodrigo:2001kf}. The additional FSR (NLO) also suffers from 
model-dependence: here the pions are assumed to radiate as pointlike particles
(scalar QED), which is implemented in an approximate way using the 
{\small PHOTOS}~\cite{Barberio:1990ms} package when
a full NLO matrix element is not available, as it is the case for multi-hadronic
processes.

For the BABAR ISR program signal and background ISR processes are simulated 
with a Monte Carlo event generator based on {\small EVA}. Additional ISR 
photons are generated with the structure function method~\cite{Caffo:1997yy}
collinear to the beams, and additional FSR photons with 
{\small PHOTOS}. To study the effects of this 
approximation on the acceptance detailed studies have been performed using
the {\small PHOKHARA} generator. It should be emphasized that for the precision
measurements of the $\pi^+\pi^-$ and $K^+K^-$ processes done with the ratio
method to $\mu^+\mu^-$ the results are essentially independent of the description of higher-order effects in the generator. This independence is exact for the dominant ISR contributions. For different NLO FSR effects, where it is no longer exact, the measurement of events with one additional photon allows corrections to be applied.

Finally the event generation has to be followed in
practice by a full simulation of the detector performance and of the
analysis procedure. Unavoidable differences between real data 
and its simulated counterparts have to be 
thoroughly studied and corrected for within limits that are then translated 
into systematic uncertainties. Modern experiments quote experimental 
uncertainties around 1\% or even below. The smallest values are obtained in the
$\pi^+\pi^-$ and $K\bar{K}$ channels, where the angular distribution is 
known from first principles. In the dominant $\rho$ region, 
the best quoted systematic uncertainties
are  $0.6\hyph0.8\%$ for CMD-2, 1.3\% for SND, $1.0\hyph2.1\hyph0.7\%$ for KLOE, and $0.5\%$ for 
BABAR (multiple values correspond to different data sets and analyses).

\paragraph{Luminosity measurements}

An independent measurement of the \epem luminosity is necessary in most cases,
except for in the ISR approach using the ratio of measured hadrons to muon pairs
in the same data sample. For this purpose Bhabha scattering 
$\epem \rightarrow e^+e^-$ is generally used as the cross section is 
large and electrons are easy to identify. 
The major source of systematic uncertainty comes from differences between the
detector performance and the simulation, mainly regarding the effect of 
angular resolution near acceptance edges. In some cases (for example with
BABAR), several QED processes are used and combined, providing some
cross-checks. Then one should include the uncertainty from the calculation 
of the reference cross section and its implementation in event generators. 
Typical values for the total luminosity uncertainty are $0.4\hyph0.5\%$ for CMD-2 
and SND, $0.5\%$ for BESIII, $0.3\%$ for KLOE, and $0.5\hyph1.0\%$ for BABAR.

\subsubsection{Input data}

\paragraph{Exclusive measurements}

\bi
\item {\it The $\pi^+\pi^-$ channel}

The numerical importance of the $\pi^+\pi^-$ channel for $\amuHVPLO$ has 
triggered a large experimental effort to obtain reliable and precise
data. Thus, although there is no strong reason to ignore them, most older 
measurements are now essentially obsolete. Therefore, we concentrate here on
the results obtained in the last decade or so.

\begin{figure}[t] \centering
\includegraphics[width=7.7cm]{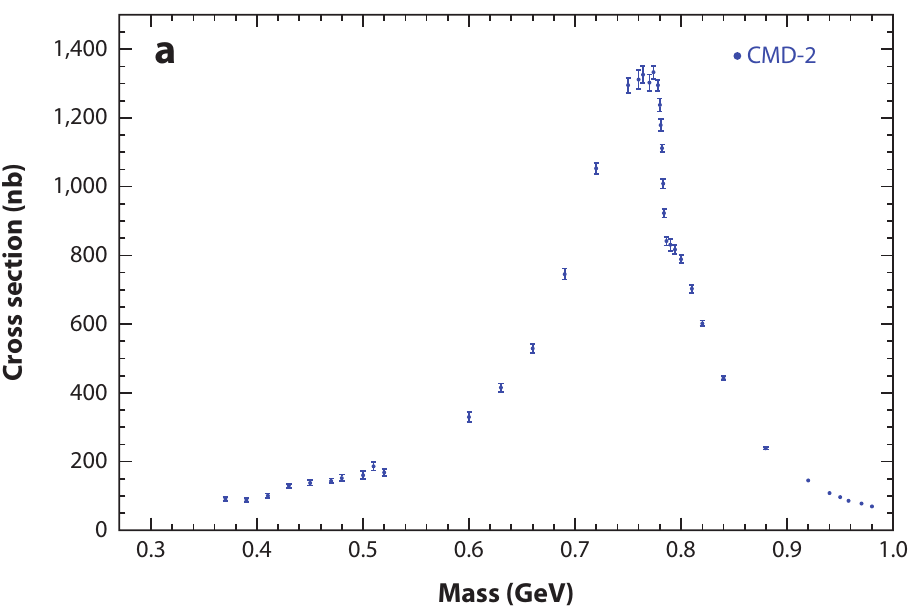} 
\includegraphics[width=7.7cm]{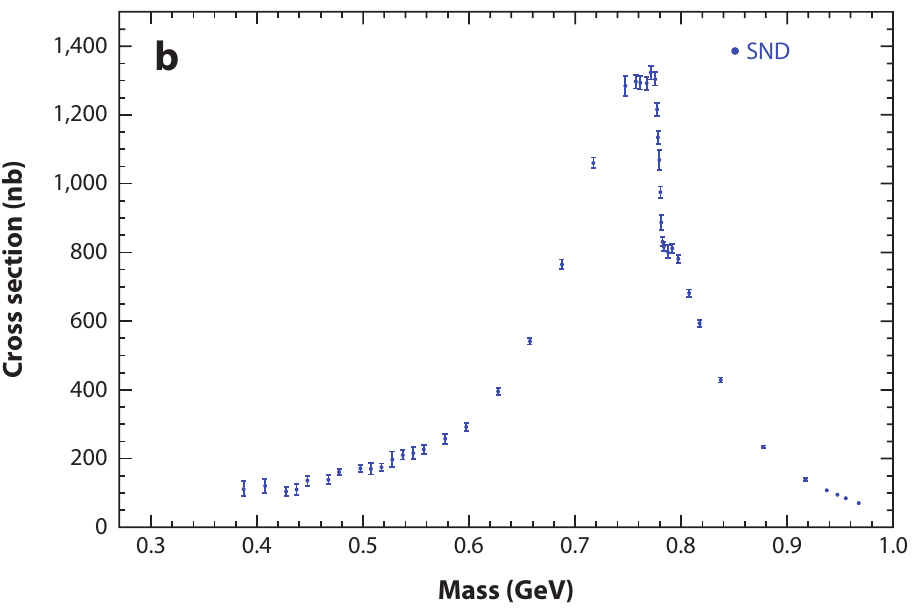}
\caption{The data from CMD-2~\cite{Akhmetshin:2006wh,Akhmetshin:2006bx} (left) and 
SND~\cite{Achasov:2006vp} (right) on 
$\epem \rightarrow \pi^+\pi^-$ in the $\rho$ region. Reprinted from Ref.~\cite{Davier:2013vna}.}
\label{cmd2-pipi}
\end{figure}

Precise measurements in the $\rho$ region came from 
Novosibirsk with CMD-2~\cite{Akhmetshin:2003zn} and SND~\cite{Achasov:2006vp}, revising older results.\footnote{There were problems with the large radiative corrections in previous analyses of CMD-2~\cite{Akhmetshin:2001ig} and SND~\cite{Achasov:2005rg}.}
In addition, CMD-2 has obtained results above the $\rho$ 
region~\cite{Aulchenko:2006na}, as well as a second set of data across
the $\rho$ resonance~\cite{Akhmetshin:2006bx}. Neither experiment can separate pions and muons, except for near threshold using momentum
measurement and kinematics for CMD-2, so that the measured quantity 
is the ratio
$(N_{\pi\pi}+N_{\mu\mu})/N_{ee}$. The pion-pair cross section is obtained
after subtracting the muon-pair contribution and normalizing to the Bhabha
events, using computed QED cross sections for both, including their respective
radiative corrections. 
The results, shown in \cref{cmd2-pipi}, are corrected for leptonic 
and hadronic VP, and 
for photon radiation by the pions, so that the deduced cross section 
corresponds to $\pi^+\pi^-$ including pion-radiated photons 
and virtual final-state QED effects. The uncertainties in the radiative 
corrections (0.4\%) should be considered fully correlated in the two 
experiments as they now use the same programs. 

\begin{figure}[t] \centering
\includegraphics[width=7.5cm]{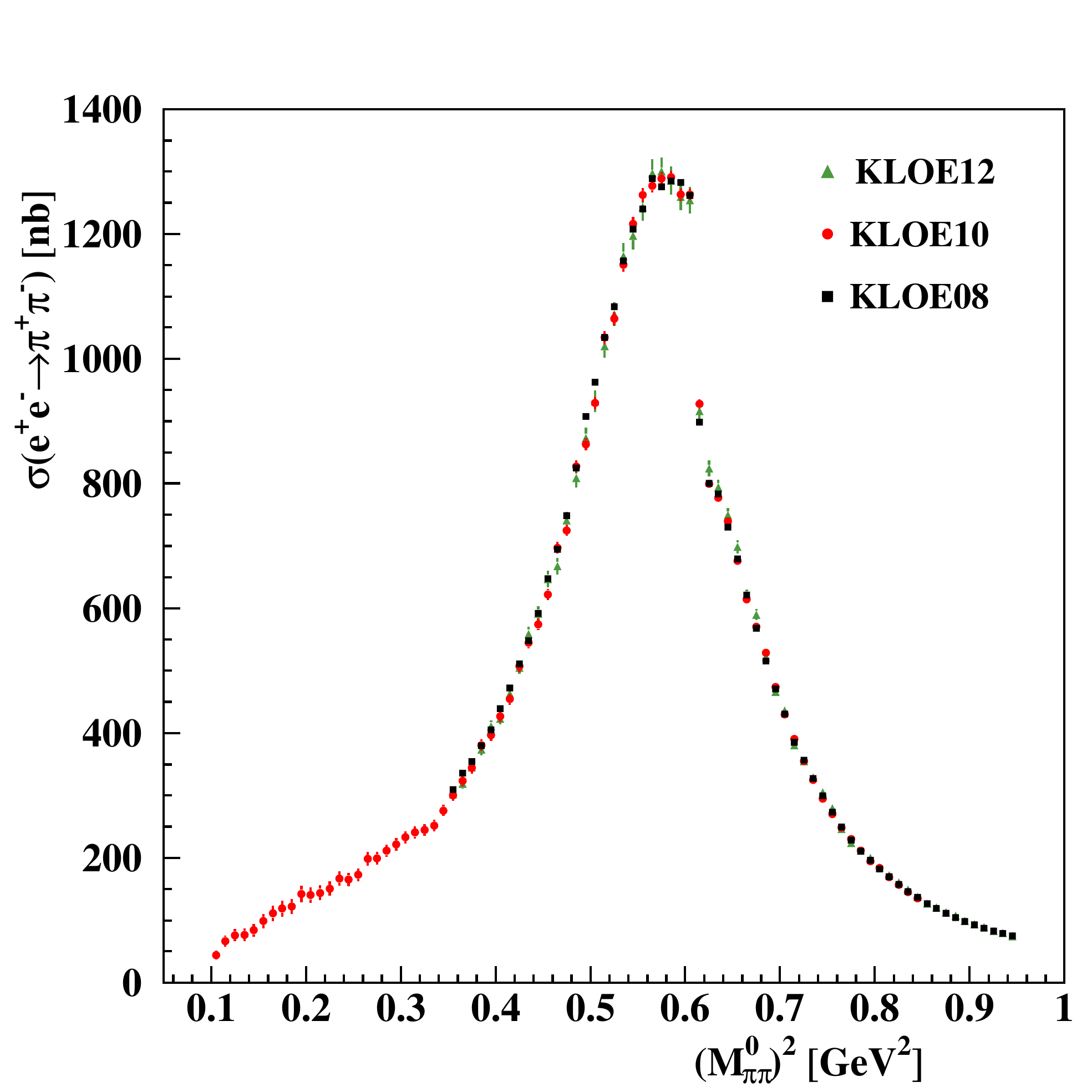} 
\caption{The KLOE data sets on $\epem \rightarrow \pi^+\pi^-$ in the $\rho$ region obtained in the three experimental configurations described in the text~\cite{Ambrosino:2008aa,Ambrosino:2010bv,Babusci:2012rp}. Adapted from Ref.~\cite{Babusci:2012rp}.}
\label{kloe-pipi}
\end{figure}

The KLOE~\cite{Ambrosino:2008aa} and BABAR~\cite{Aubert:2009ad,Lees:2012cj} ISR analyses are 
initially very different. First, the 
CM energy is close to the studied energy in the case of KLOE 
(soft ISR photons), while it is very far in the BABAR case (hard ISR 
photons). In KLOE-2008 and KLOE-2012 the ISR photon is not detected and reconstructed 
kinematically, assuming no extra photon. Since the cross section strongly 
peaks along the beams, a large statistics of ISR events is obtained. Pion pairs
are separated from muon pairs with kinematical constraints. 
In BABAR,
the ISR photon is detected at large angle (about 10\% efficiency) so that the
full event is observed, and an additional photon can be incorporated in the
kinematical fit (undetected forward additional ISR or detected ISR/FSR photon).
Another big difference concerns the ISR luminosity: in the KLOE-2008 and KLOE-2010 analyses it is 
computed using the NLO PHOKHARA generator~\cite{Czyz:2004rj}, 
while in BABAR both pion and muon pairs are measured and the ratio 
$\pi\pi(\gamma)$/$\mu\mu(\gamma)$ directly provides the $\pi\pi(\gamma)$ 
cross section. The small-angle ISR photon provides a suppression of the
sizeable LO $|\text{FSR}|^2$ contribution in KLOE, and the remaining part is computed
from PHOKHARA. In BABAR, the $|\text{FSR}|^2$ contribution is negligible because of
the large value of $s$. The KLOE method with small-angle undetected ISR photons also reduces the range of
$\pi\pi$ masses on the low side because of the limited angular acceptance of
the detector. To overcome this problem, the analysis of KLOE-2010 was performed 
with large-angle ISR~\cite{Ambrosino:2010bv}. Finally, the KLOE-2012 
measurement~\cite{Babusci:2012rp} was obtained using the same ratio method as BABAR 
(\cref{kloe-pipi}), but with undetected small-angle ISR photons.
This ratio is taken in small mass bins (typically $6\MeV$) for KLOE, while for BABAR larger intervals ($50\MeV$) are used in order to reduce statistical fluctuations on the individual cross section values,
taking into account the expected variation of the $\mu \mu (\gamma)$
cross section within each interval and the bin-to-bin correlations in the covariance matrix.

The three KLOE measurements have been recently combined taking into account the
correlations between the different data sets~\cite{Anastasi:2017eio}. The combination method was aimed at providing a coherent KLOE data
set with a fully consistent treatment of uncertainties between the three
analyses.

\begin{figure}[t] \centering
\includegraphics[width=12cm]{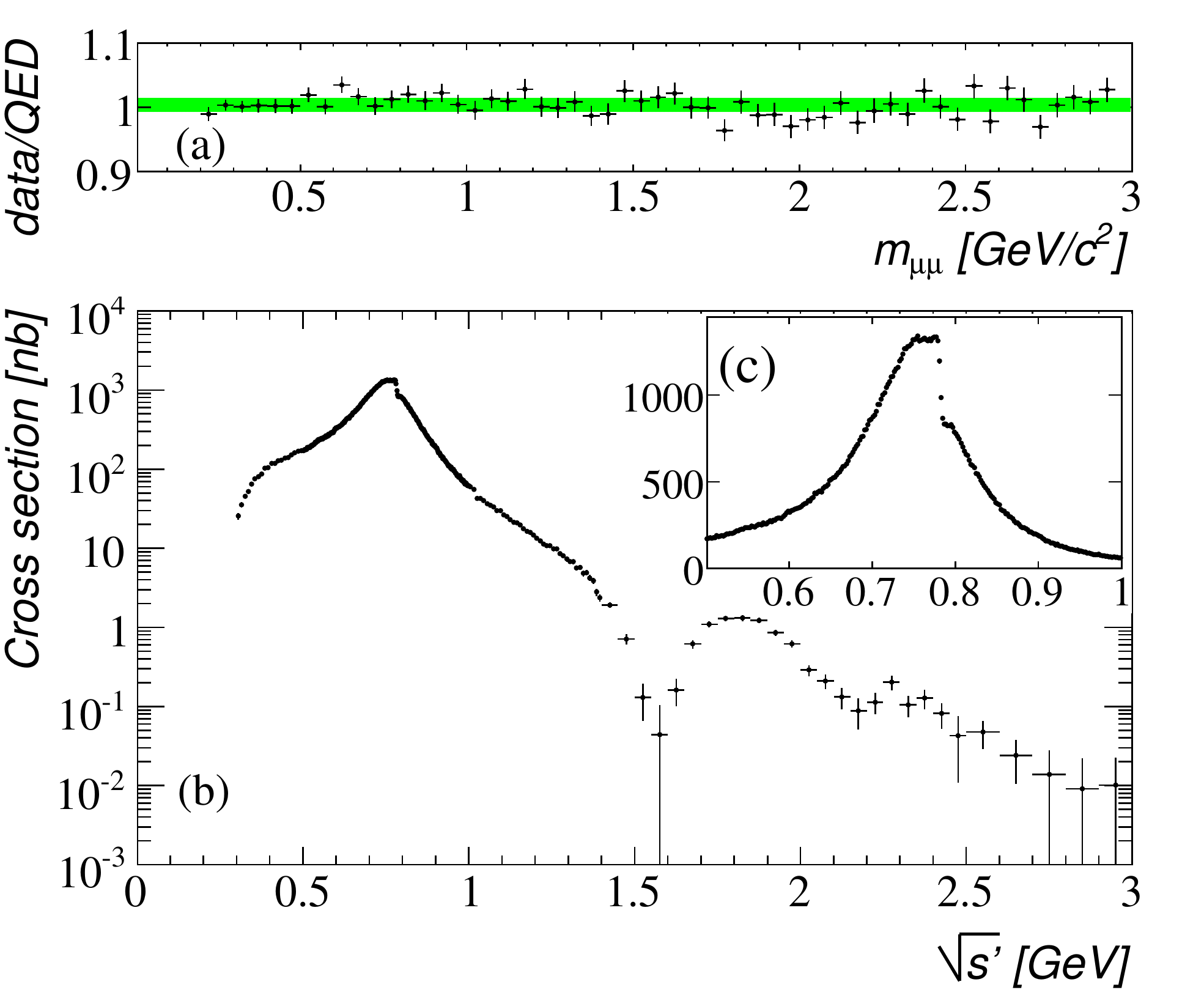} 
\caption{Results from BABAR~\cite{Aubert:2009ad,Lees:2012cj} using the large-angle ISR method: 
$\epem \rightarrow \mu^+\mu^-$ compared to NLO QED (top) and  $
\epem \rightarrow \pi^+\pi^-$ from threshold to $3\GeV$ using the $\pi\pi/\mu\mu$
ratio (bottom). The inset shows the $\rho$ region. Reprinted from Ref.~\cite{Aubert:2009ad}.}
\label{BABAR-pipi}
\end{figure}

In order to reduce  systematic uncertainties, the BABAR method
involves the simultaneous measurement of the process 
$\epem \rightarrow \mu^+\mu^-$, which by itself can be checked against the 
QED prediction using the \epem luminosity. The comparison of the BABAR data 
with NLO QED shows a good agreement from threshold to $3\GeV$ within a total 
uncertainty of 1.1\%, dominated by the luminosity uncertainty 
(\cref{BABAR-pipi}).

\begin{figure}[ht!] \centering
\includegraphics[width=7.9cm]{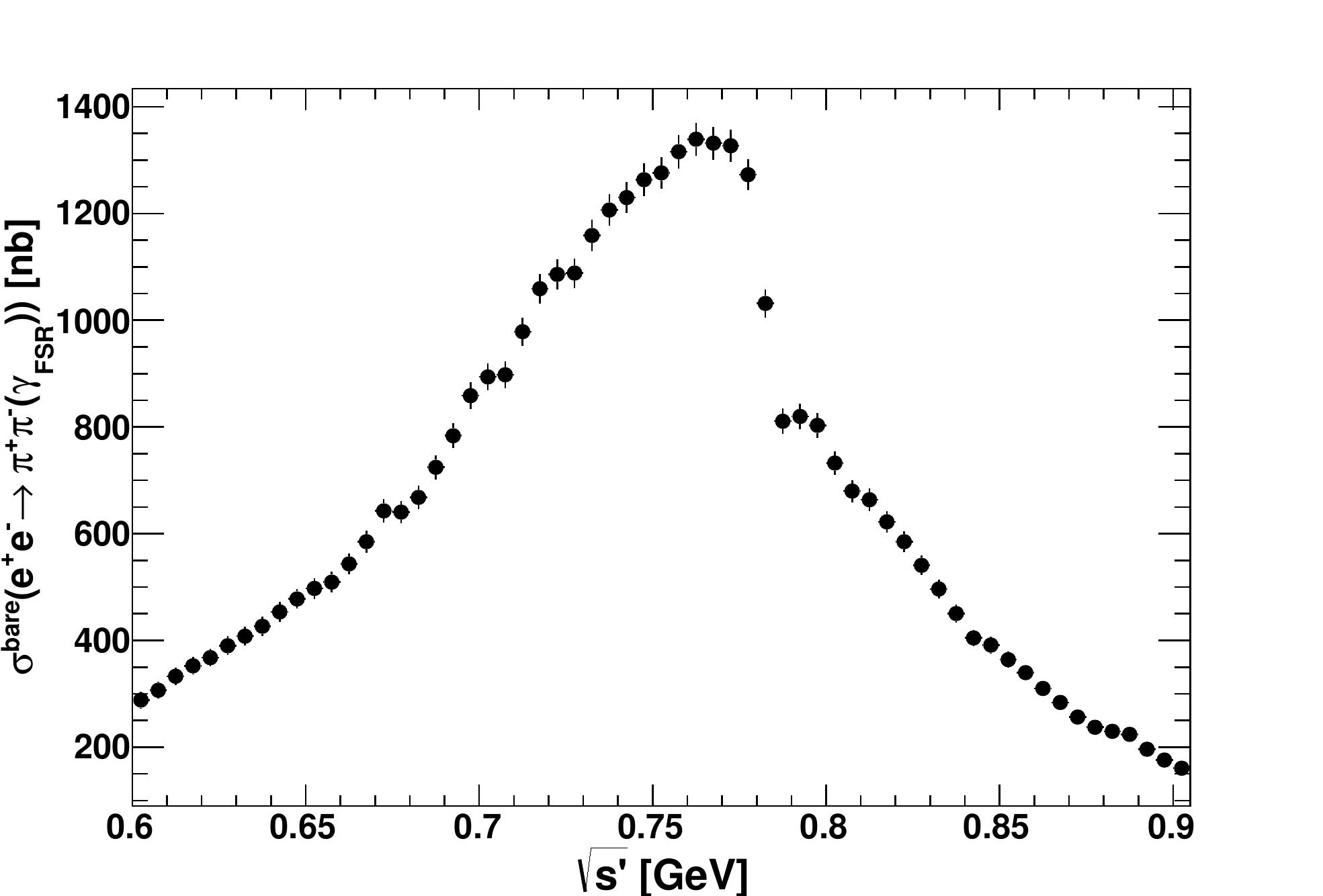} 
\includegraphics[width=7cm]{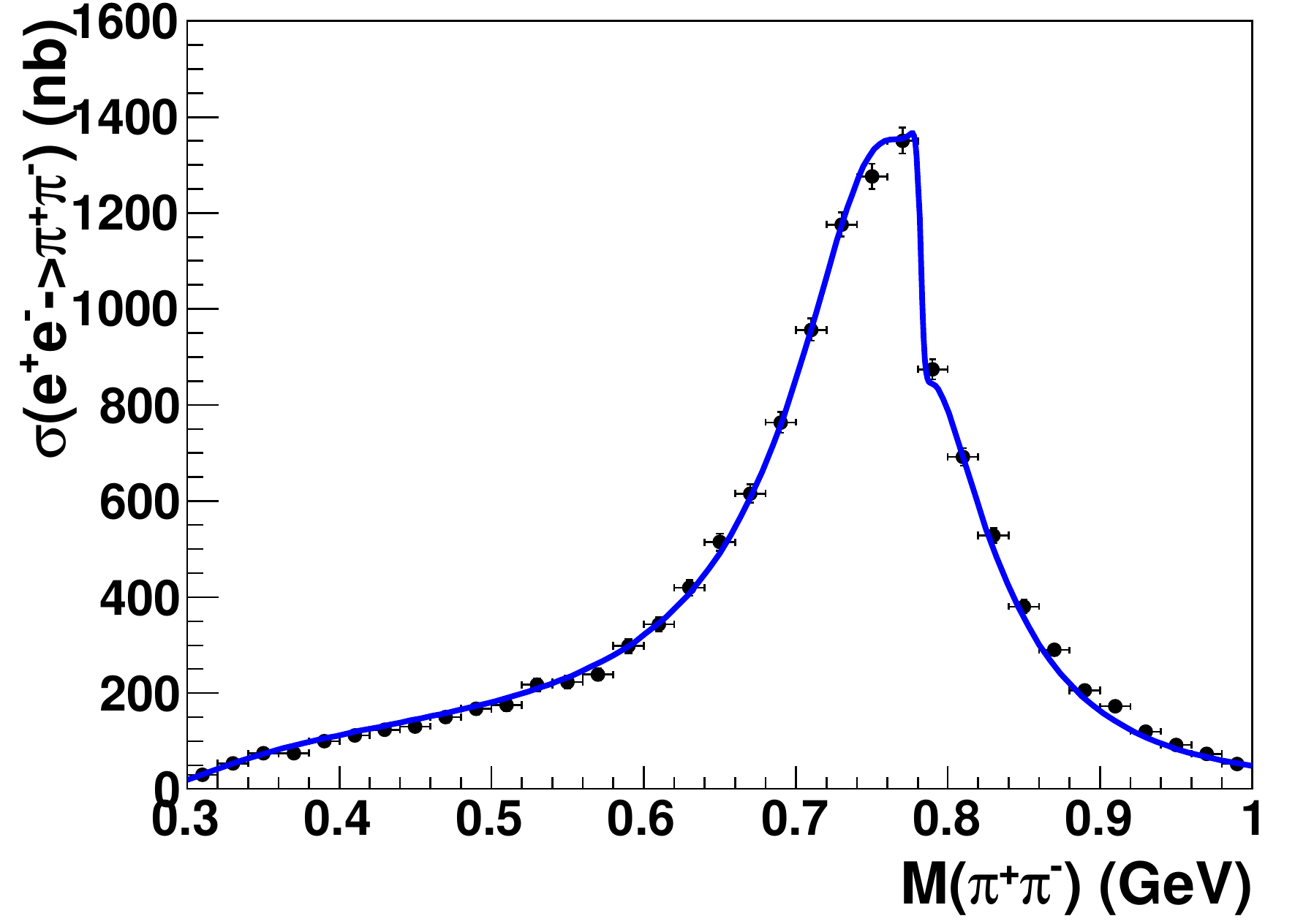}
\caption{The BESIII (left, reprinted from Ref.~\cite{Ablikim:2015orh}) and CLEO-c (right,  reprinted from Ref.~\cite{Xiao:2017dqv}) data on 
$\epem \rightarrow \pi^+\pi^-$ in the $\rho$ region
using large-angle detected ISR photons.}
\label{bes3-cleoc}
\end{figure}

More recently results with the ISR method in the charm region and large-angle 
ISR tagging have been obtained by BESIII~\cite{Ablikim:2015orh} and a group using 
the data from CLEO-c~\cite{Xiao:2017dqv}. Both have a larger statistical 
uncertainty. Their data is shown in \cref{bes3-cleoc}.


Although lots of data have been recorded in the $\pi^+\pi^-$ channel with
precision increasing over time, their consistency is far from excellent. The 
problems encountered will be discussed in \cref{ee-problems}.

\item {\it The other two-body channels}

\begin{figure}[t] \centering
\includegraphics[width=5cm]{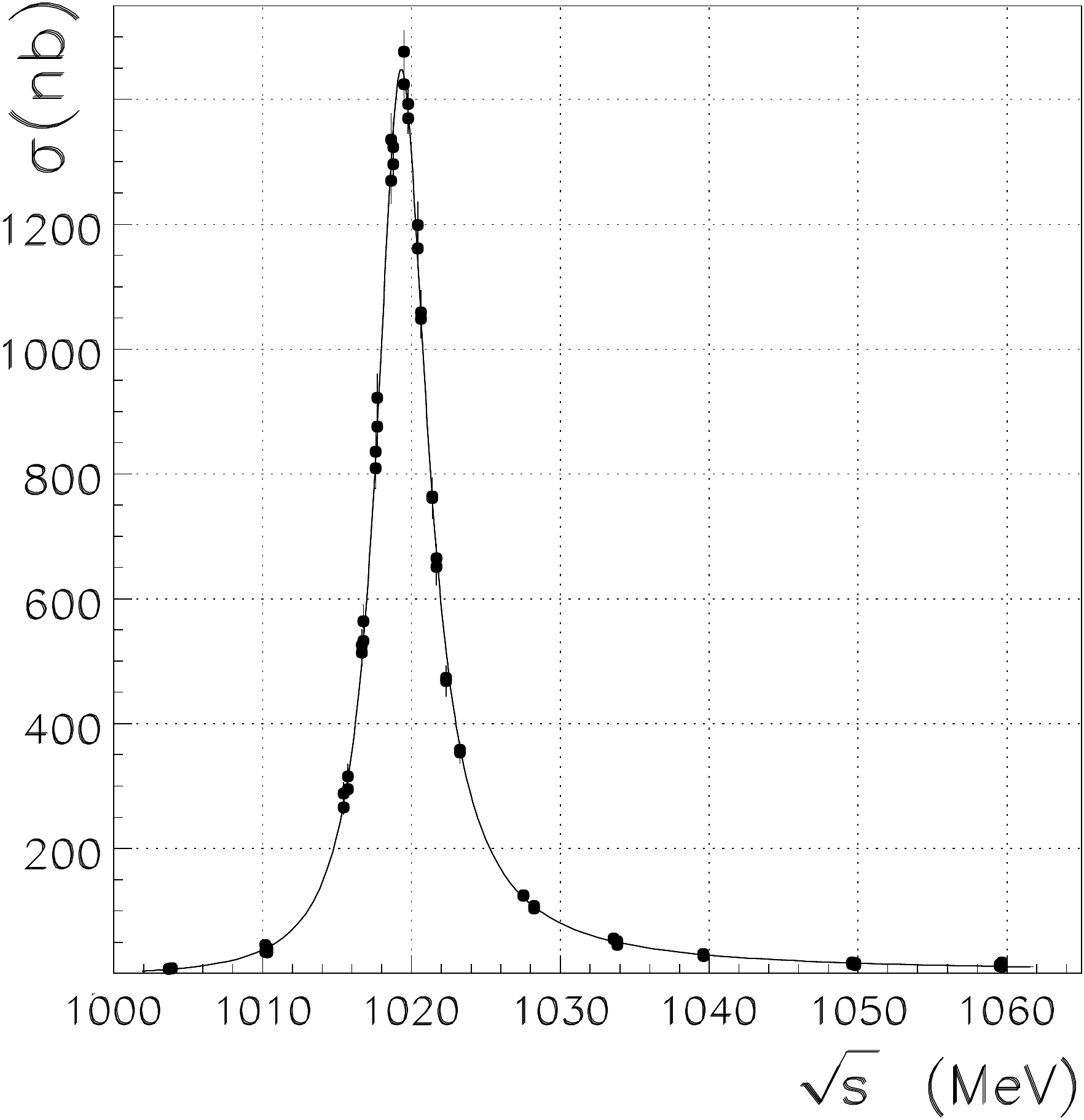}
\includegraphics[width=7.5cm]{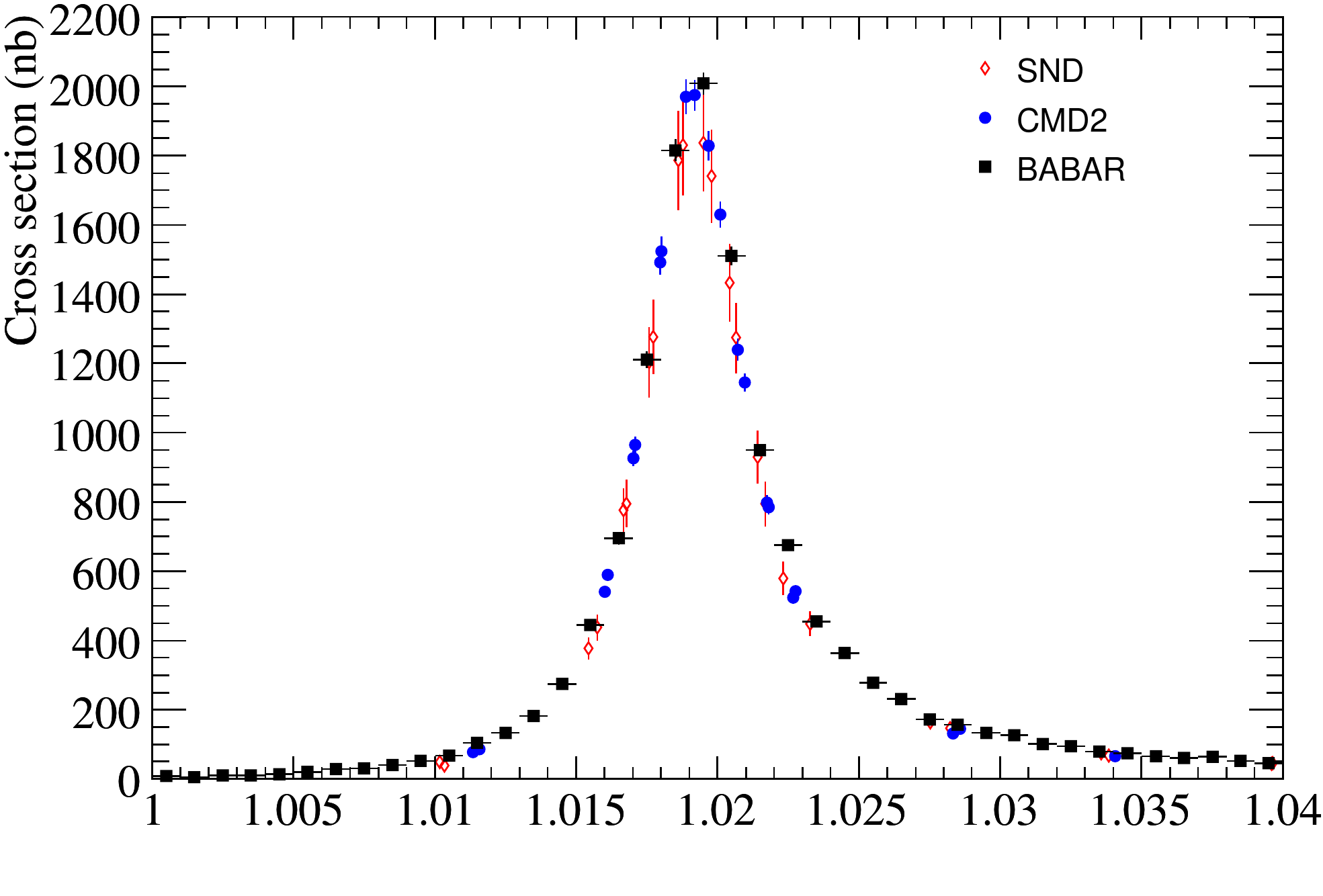}
    \put(-60,4){$\sqrt{s'}$ (GeV)}\\ 
\includegraphics[width=9.5cm]{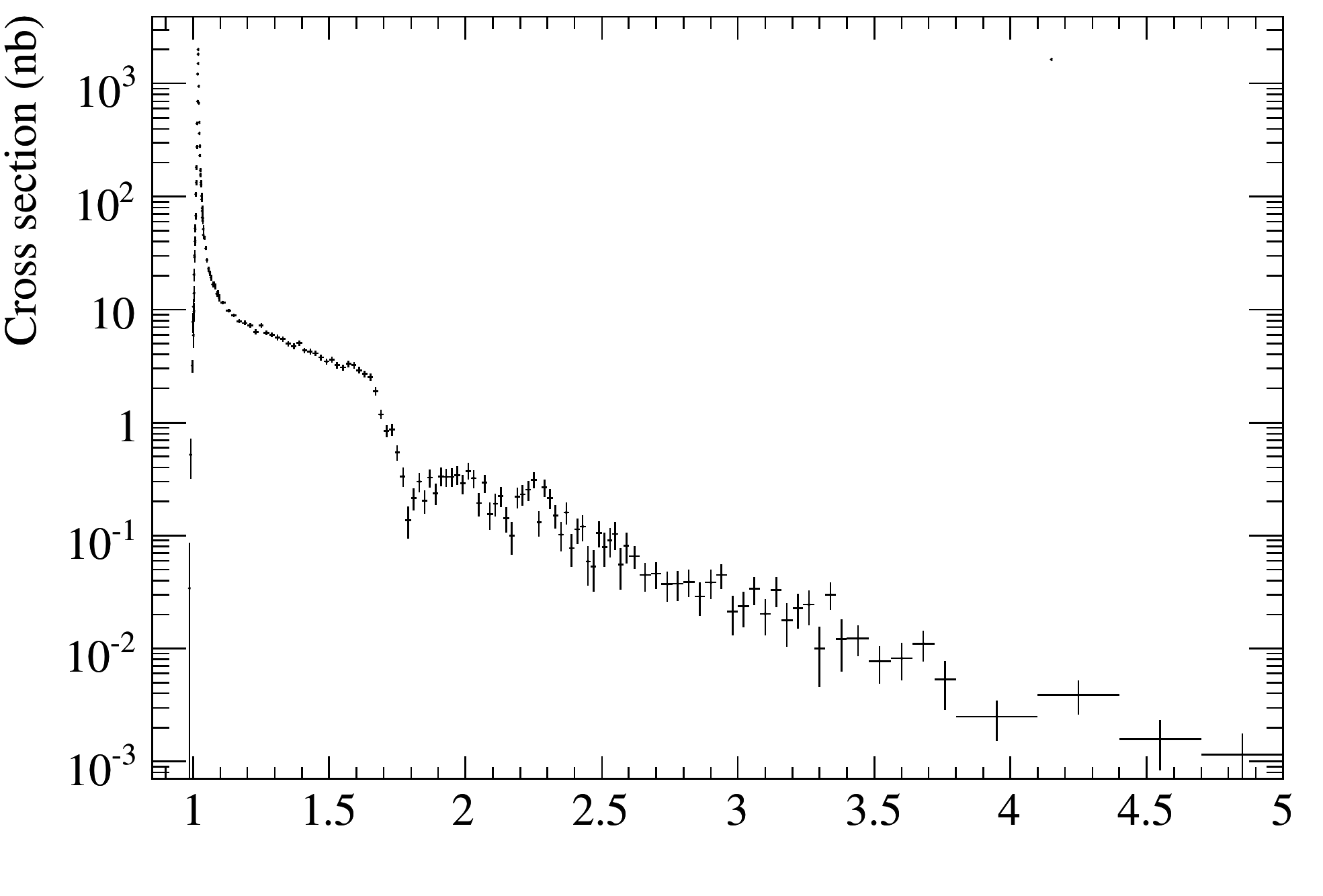}
    \put(-60,4){$\sqrt{s'}$ (GeV)}
\caption{Cross sections for $\epem \rightarrow K_SK_L$ measured by 
SND~\cite{Achasov:2000am} (upper left), and $\epem \rightarrow K^+K^-$ by 
CMD-2~\cite{Akhmetshin:2008gz}, SND~\cite{Achasov:2000am}, and BABAR~\cite{Lees:2013gzt}
(upper right), and BABAR over a wider energy range (bottom). Reprinted from Refs.~\cite{Achasov:2000am,Lees:2013gzt}.}
\label{kkbar}
\end{figure}

\begin{figure}[ht!] \centering
\includegraphics[width=10cm]{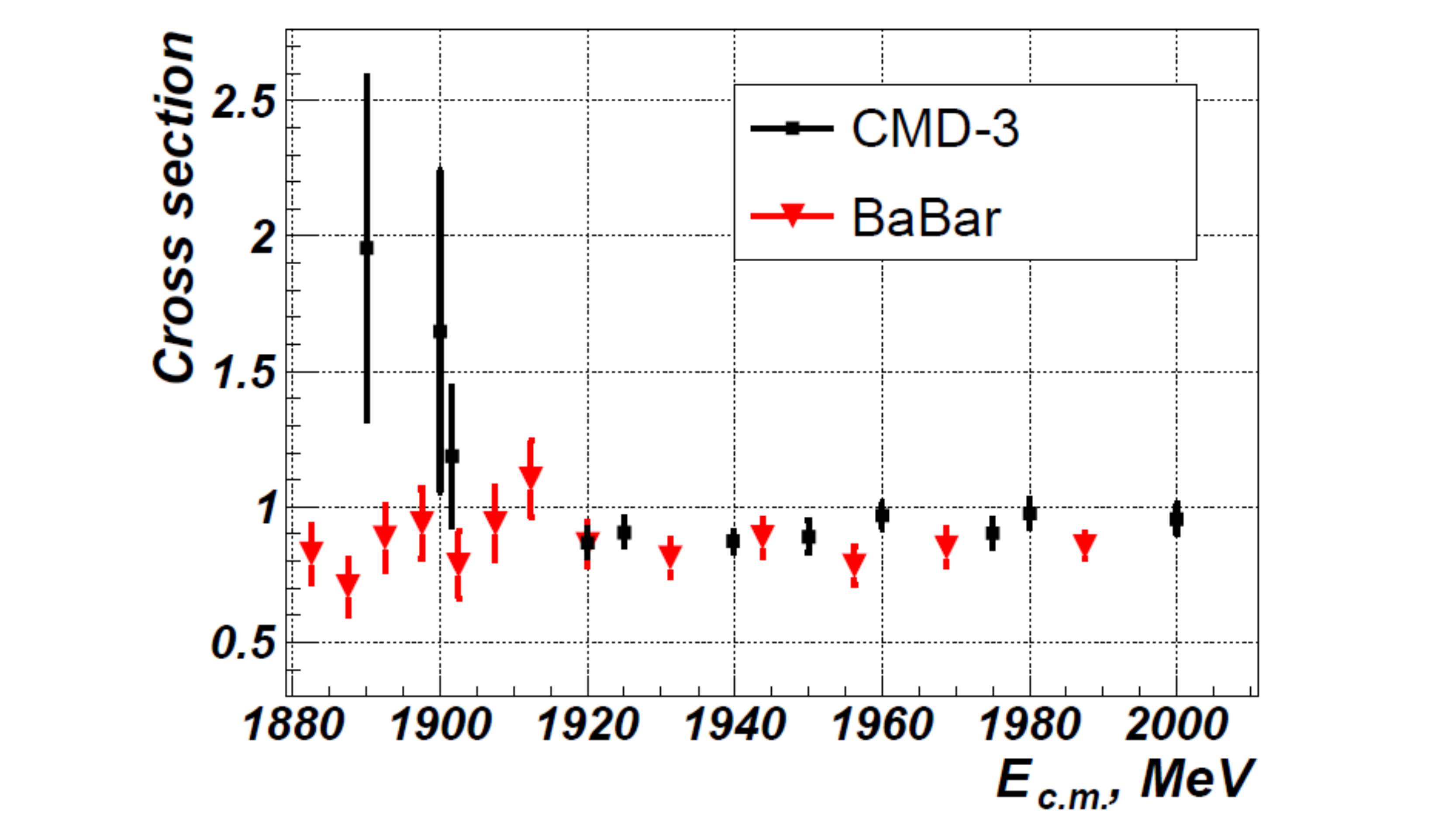} 
\caption{Results from BABAR~\cite{Aubert:2005cb,Lees:2013ebn,Lees:2013uta} and CMD-3~\cite{Akhmetshin:2015ifg} on the cross section (in nb) for $\epem \rightarrow p\bar{p}$. Reprinted from Ref.~\cite{Akhmetshin:2015ifg}.} 
\label{cmd3-babar-pp}
\end{figure}

The $\pi^0\gamma$ final state is the first open hadronic channel and defines the lower limit of integration of the dispersion integral. In addition to older results, recent measurements from SND over the full spectrum up to $2\GeV$ are now available with increased precision~\cite{Achasov:2016bfr,Achasov:2018ujw}.

Cross sections for the $K^+K^-$ and $K_SK_L$ final states are given in
\cref{kkbar} for CMD-2~\cite{Akhmetshin:2008gz,Akhmetshin:1999ym}, 
SND~\cite{Achasov:2000am,Achasov:2007kg}, and BABAR~\cite{Lees:2013gzt}. They are dominated by
 the prominent $\phi$ resonance. Again here the broad mass range available 
through the ISR approach is noteworthy. Very recently results were obtained
at VEPP-2000 by CMD-3~\cite{Kozyrev:2017agm} and SND~\cite{Achasov:2016lbc}, differing
markedly from the earlier CMD-2 and SND measurements at VEPP-2M. While experiments are in good agreement 
for $K_SK_L$, the situation is more problematic for $K^+K^-$, as discussed in
\cref{ee-problems}.  

Precise measurements of the proton--antiproton final state have been achieved
by BABAR~\cite{Aubert:2005cb,Lees:2013ebn,Lees:2013uta} and CMD-3~\cite{Akhmetshin:2015ifg}. 
The cross section 
for $\epem\rightarrow p \bar{p}$ given in \cref{cmd3-babar-pp} shows 
little energy dependence from threshold to $2\GeV$. Again
the ISR method allows the measurement to be performed over a large energy 
range up to $6\GeV$. A measurement of $\epem\rightarrow n \bar{n}$ is
available from SND~\cite{Achasov:2014ncd}, showing a cross section comparable to that
of $p \bar{p}$ from threshold to $2\GeV$ within the modest achieved 
precision. The nucleon-pair production at $2\GeV$ accounts for about 4\% of the 
total hadronic cross section.

\item {\it The multi-hadronic channels}

\begin{figure}[t] \centering
\vspace{-0.4cm}
\includegraphics[width=7cm]{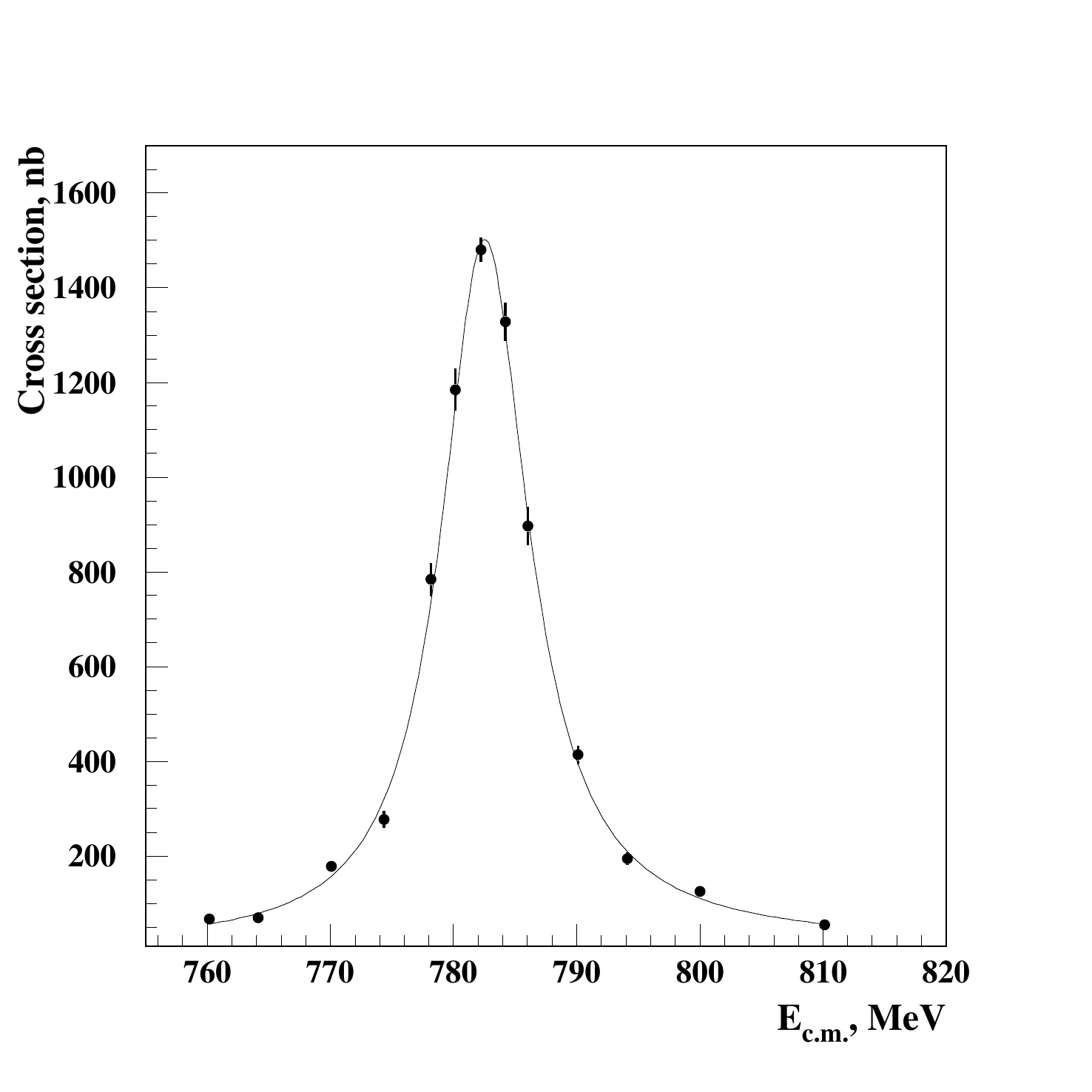}
\caption{The $\omega$ resonance in the $\pi^+ \pi^- \pi^0$ channel from CMD-2. Reprinted from Ref.~\cite{Akhmetshin:2000ca}.} 
\label{cmd2-omega-3pi}
\end{figure}

\begin{figure}[t] \centering
\hspace{5mm}\includegraphics[width=5.9cm]{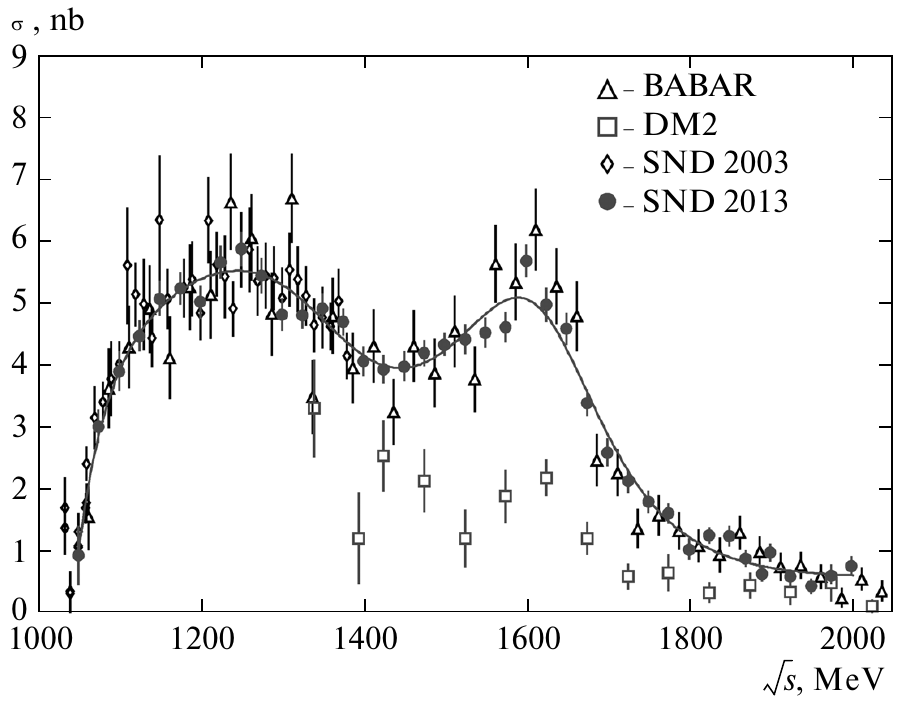}
\includegraphics[width=9.9cm]{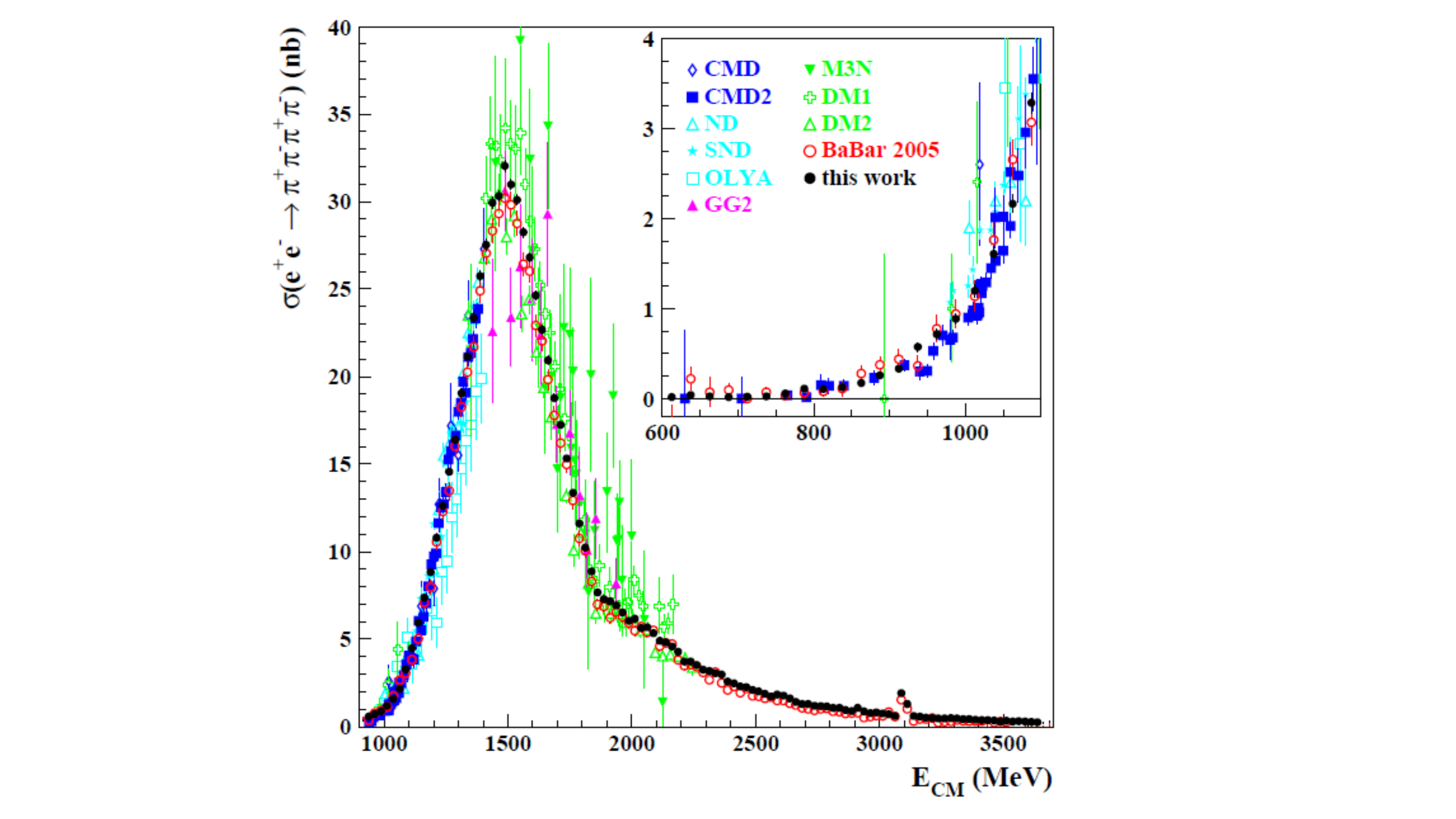}\\
\vspace{0.4cm}
\hspace{-32mm}\includegraphics[width=10.25cm]{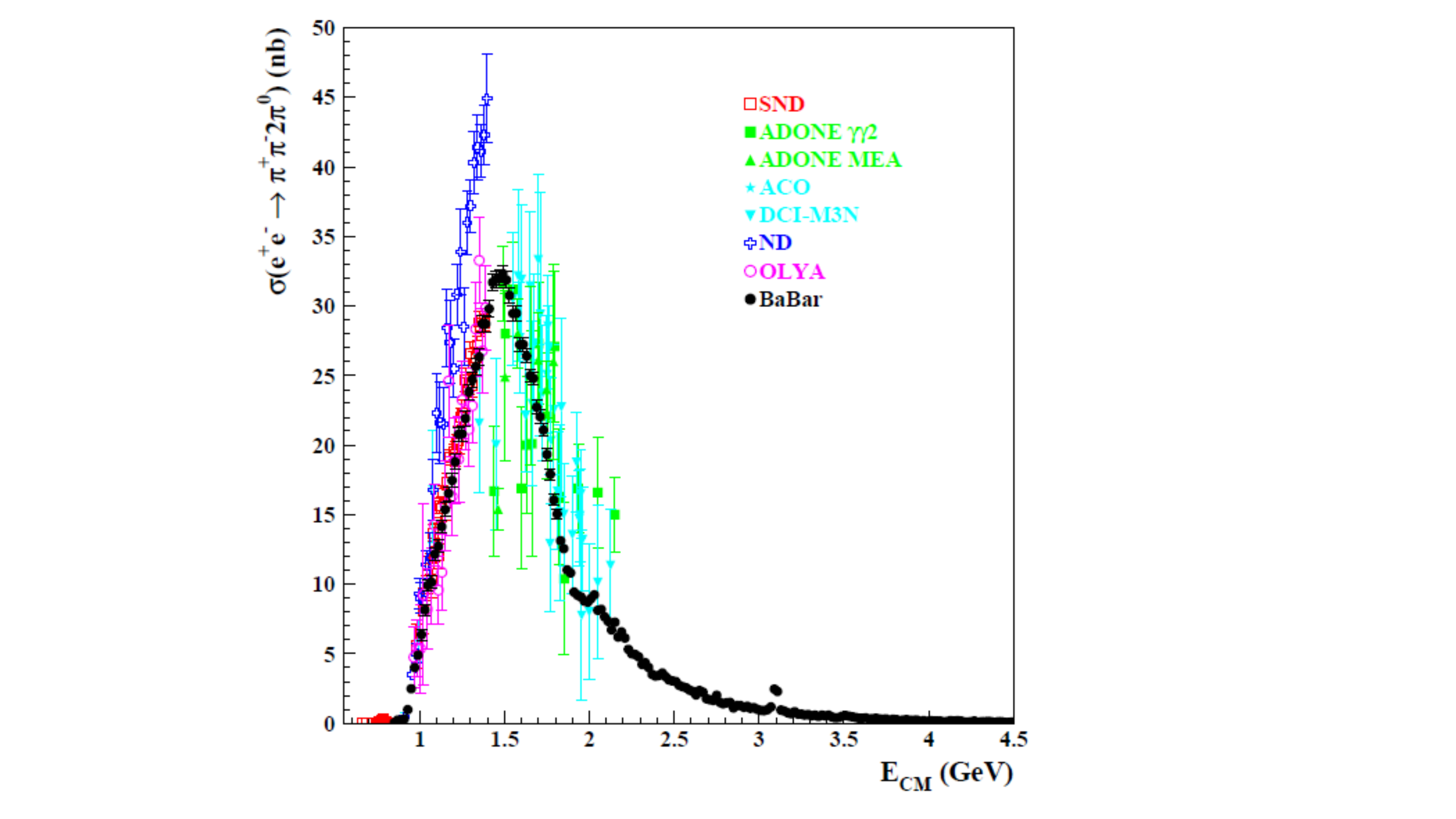}\hspace{-16mm}
\includegraphics[width=5.75cm]{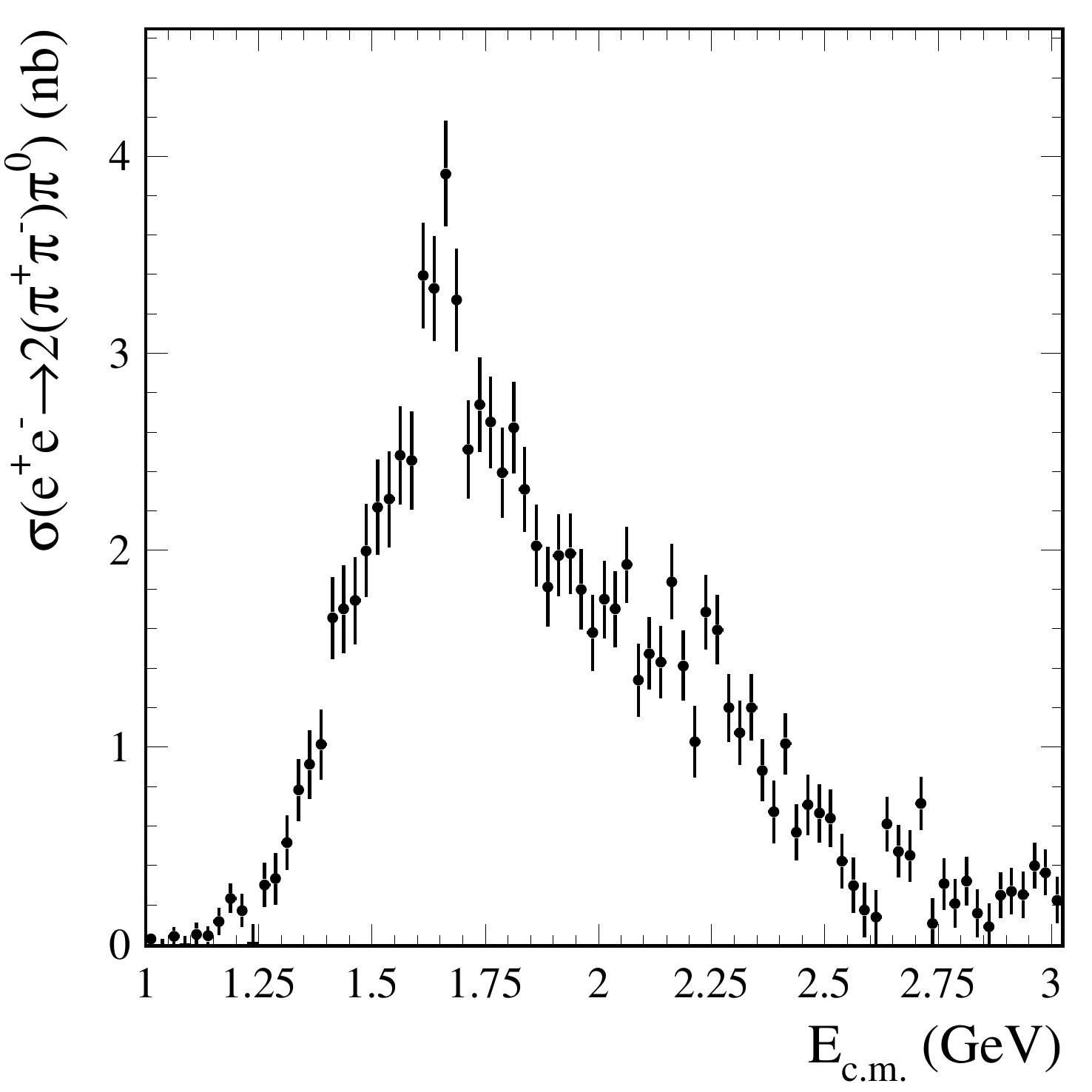}
\caption{Results from some multi-pion cross sections. Top left: $\pi^+ \pi^- \pi^0$
 from BABAR~\cite{Aubert:2004kj}, SND~\cite{Achasov:2002ud,Aulchenko:2015mwt}, and older experiments. Top right:
$2\pi^+ 2\pi^-$BABAR~\cite{Aubert:2005eg,Lees:2012cr} and older 
experiments. Bottom: $\pi^+ \pi^- 2\pi^0$ from BABAR~\cite{TheBaBar:2017vzo} and
older experiments (left), and $2\pi^+ 2\pi^- \pi^0$ from BABAR~\cite{Aubert:2007ef} 
(right). Reprinted from Refs.~\cite{Aulchenko:2015mwt,Lees:2012cr,TheBaBar:2017vzo,Aubert:2007ef}.} 
\label{3-4-5-pion}
\end{figure}

The cross sections for several exclusive channels have been measured with the 
scan method up to $1.4\GeV$ by CMD-2 and SND and extended more recently up to 
$2\GeV$ using the VEPP-2000 collider and the upgraded CMD-3 and SND detectors. 
The ISR approach has been used extensively by 
BABAR, covering the range from threshold to typically $3\hyph5\GeV$. The BABAR
results represent an almost complete set of exclusive measurements up to 
about $2\GeV$. Above $2\GeV$, many channels with higher 
multiplicity open up that in practice cannot be studied individually, so that the 
method using the sum of exclusive cross sections is no longer practicable.
For essentially all the final states measured by the scan method at VEPP-2000, the 
agreement with BABAR is excellent, providing a nice consistency check.

Previous measurements were limited to the maximum energy of $1.4\GeV$ of
VEPP-2M and uncertainties in the efficiency due to limited angular acceptance
and imperfect knowledge of the final-state dynamics. Other experiments were 
plagued by low statistics and large systematic uncertainties. Here the ISR
method with large-angle detected photons profits from the boost of the hadronic
system in the detector acceptance, permitting the detailed study of the 
final states and, therefore, considerably reducing the uncertainty on the
overall efficiency arising from the imperfect knowledge of the hadronic dynamics.

The largest multi-hadronic cross sections below $2\GeV$ are for the 3-pion and 4-pion 
final states. The 3-pion cross section is dominated at low energy by the $\omega$ (\cref{cmd2-omega-3pi}) 
and $\phi$ resonances as measured by the CMD-2~\cite{Akhmetshin:2000ca,Akhmetshin:2006sc} and
SND~\cite{Achasov:2002ud} experiments. Above the $\phi$, results are available from 
BABAR~\cite{Aubert:2004kj} and SND~\cite{Achasov:2014xsa}, which agree with each other 
as seen in \cref{3-4-5-pion}, while both disagree strongly with the earlier results from DM2~\cite{Antonelli:1992jx}. 
For the $2\pi^+2\pi^-$~\cite{Aubert:2005eg,Lees:2012cr} and $\pi^+ \pi^- 2\pi^0$~\cite{TheBaBar:2017vzo}
final states, the improvement provided by the ISR BABAR results is spectacular 
both in terms of precision and mass coverage, as displayed in \cref{3-4-5-pion}. 
Previous results from VEPP-2M~\cite{Akhmetshin:2004dy,Achasov:2003bv,Akhmetshin:1998df} and 
VEPP-2000~\cite{Akhmetshin:2016dtr} only extended to $1.4\GeV$. 
Results on exclusive final states containing up to 6 quasi-stable hadrons are 
available~\cite{Aubert:2006jq,Akhmetshin:2013xc}. The limitation on hadron multiplicity, 
set largely by the difficulty to select and identify multi-$\pi^0$ final states, 
does not permit a reliable reconstruction of the full hadronic rate above $2\GeV$ as 
a sum over individually measured exclusive cross sections.

\begin{figure}[t] \centering
\includegraphics[width=7.cm]{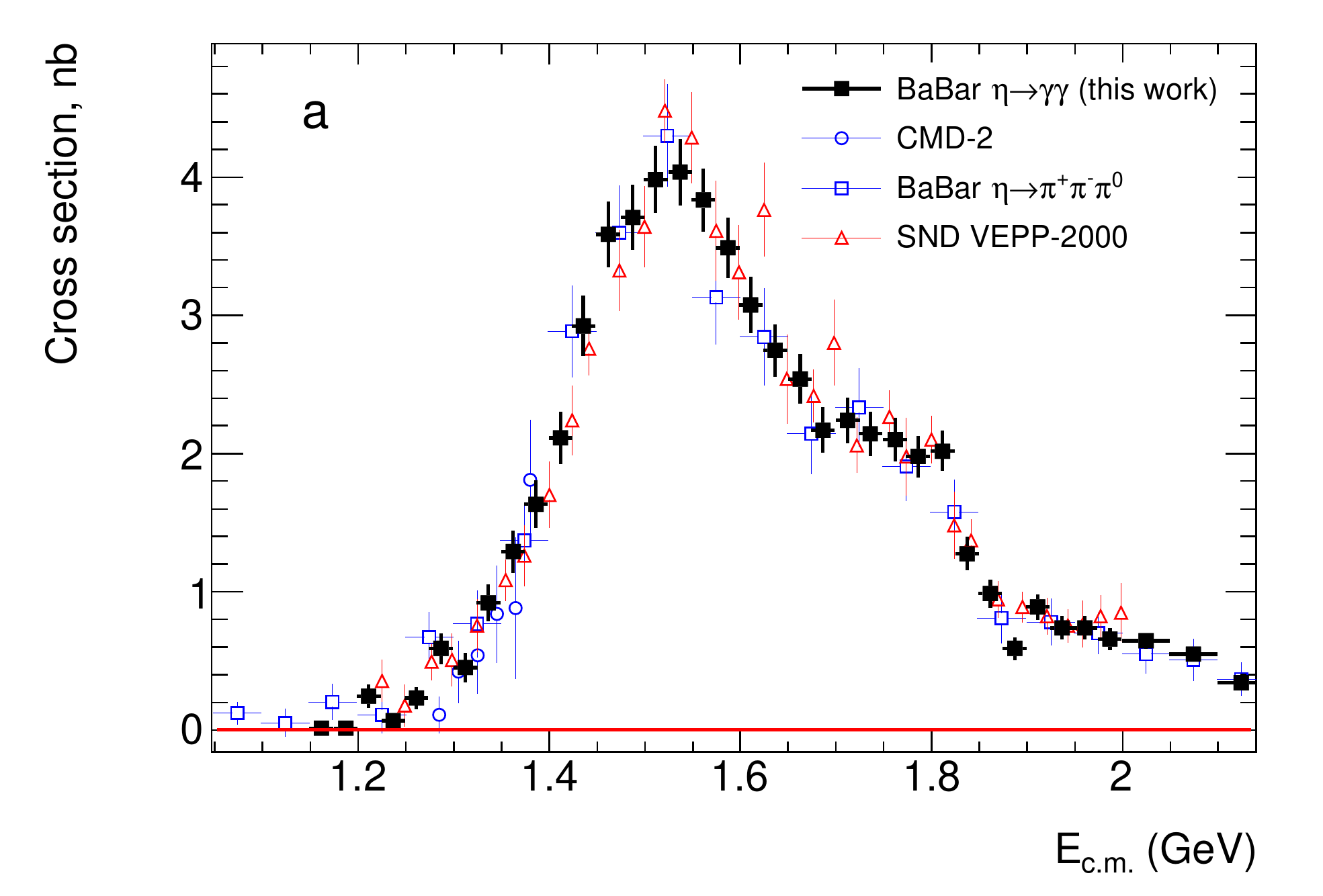}
\includegraphics[width=9.25cm]{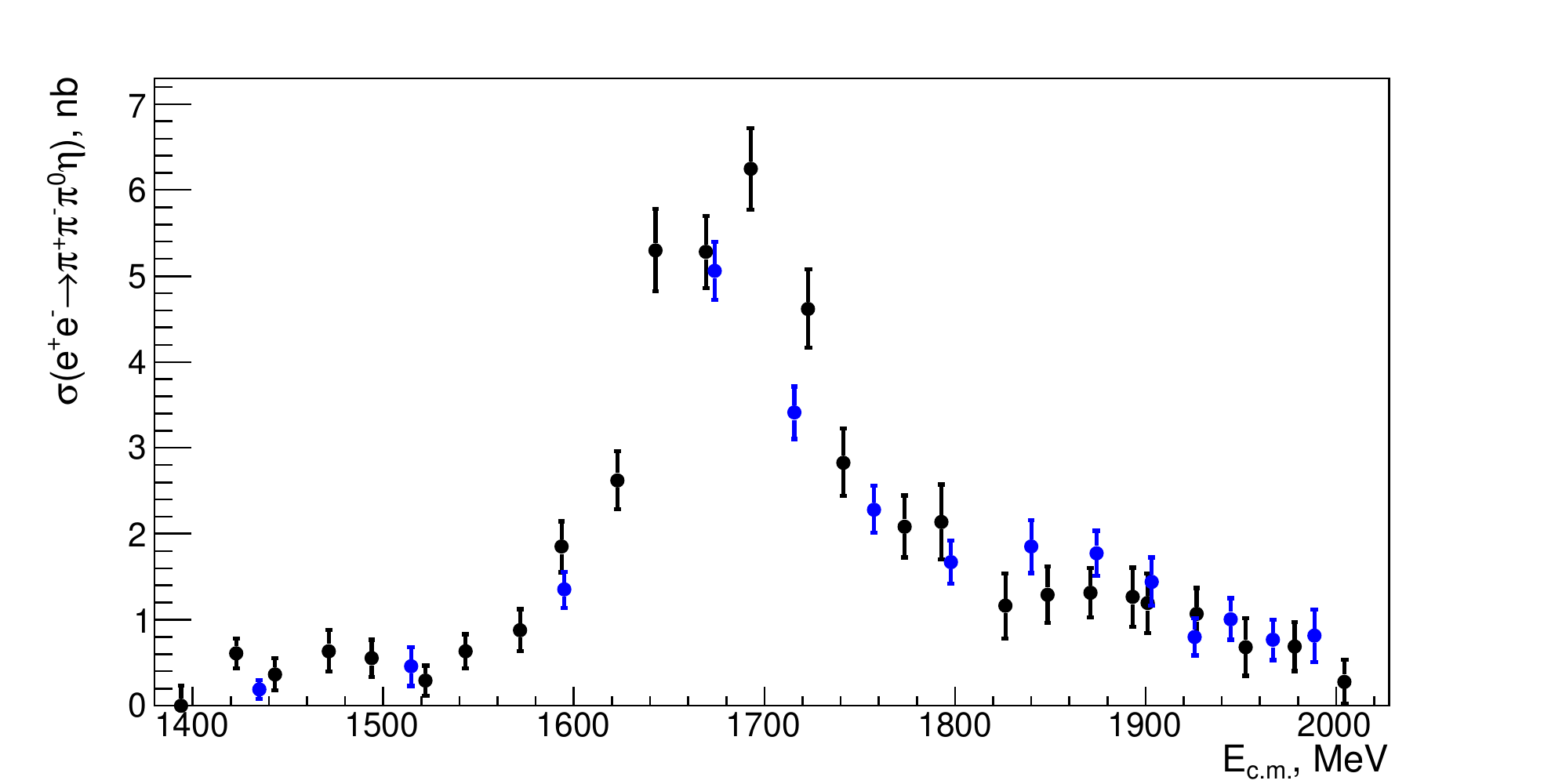}
\caption{Results from BABAR~\cite{Aubert:2007ef,TheBaBar:2018vvb}, 
CMD-2~\cite{Akhmetshin:2000wv}, and SND~\cite{Aulchenko:2014vkn} 
on the cross sections for $\epem \rightarrow \eta \pi^+ \pi^-$ (left), and 
from CMD-3~\cite{CMD-3:2017tgb} for $\epem \rightarrow \eta \pi^+ \pi^- \pi^0$ (right). Reprinted from Refs.~\cite{TheBaBar:2018vvb,CMD-3:2017tgb}.} 
\label{eta-npi}
\end{figure}

\begin{figure}[ht!] \centering
\includegraphics[width=13.5cm]{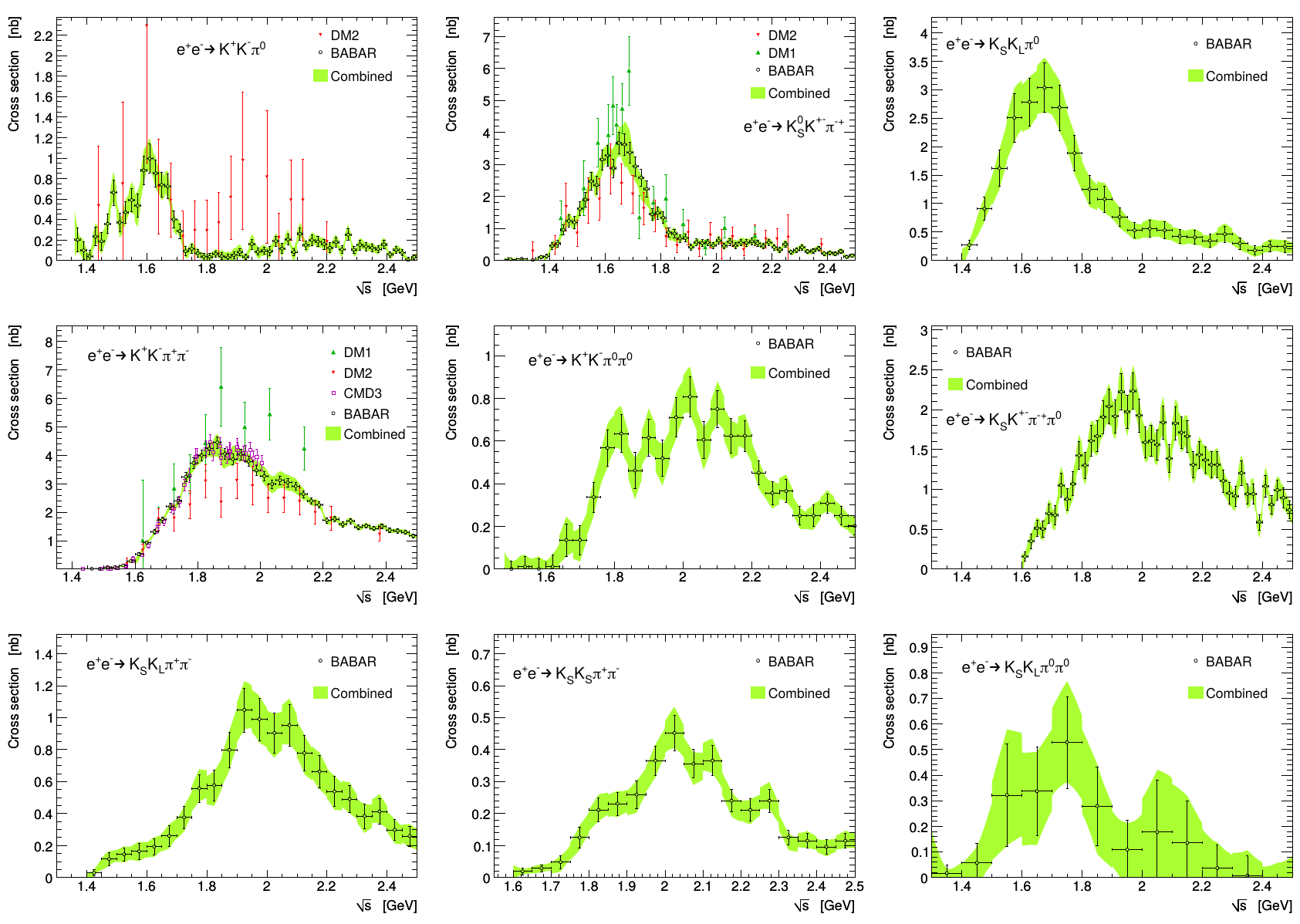}
\caption{Results from BABAR~\cite{Aubert:2007ym,Lees:2014xsh,TheBaBar:2017aph,Aubert:2007ur,Lees:2011zi} 
on the cross sections for $\epem \rightarrow K \bar{K}\pi$ (top row) and
$\epem \rightarrow K \bar{K}\pi\pi$ (second and third rows). Reprinted from Ref.~\cite{Davier:2017zfy}.
} 
\label{BABAR-kkpi-kkpipi}
\end{figure}

Numerous processes with smaller cross sections have to be considered to saturate the
total hadronic rate. \Cref{eta-npi} shows some results on final states including 
$\eta$ mesons, namely $\eta \pi^+ \pi^-$ from 
BABAR~\cite{Aubert:2007ef,TheBaBar:2018vvb},
CMD-2~\cite{Akhmetshin:2000wv}, and SND~\cite{Aulchenko:2014vkn}, and $\eta \pi^+ \pi^- \pi^0$ from
CMD-3~\cite{CMD-3:2017tgb}. 
Further, more recent, data sets for $\eta \pi^+ \pi^-$ exist from
SND~\cite{Achasov:2017kqm} and CMD-3~\cite{Gribanov:2019qgw}.
For the $\eta 4\pi$ final states  only results from BABAR 
are available, both for $\eta 2\pi^+ 2\pi^-$~\cite{Aubert:2007ef} and
$\eta \pi^+ \pi^- 2\pi^0$~\cite{Lees:2018dnv}.
A lot of progress was recently achieved by BABAR on
$K \bar{K} n~\rm{pions}$ final states with the complete set of measurements for all charge configurations with 
$n=1,2$~\cite{Aubert:2007ym,Lees:2014xsh,TheBaBar:2017aph,Aubert:2007ur,Lees:2011zi}, thanks to the detection of $K_S$, $K_L$, charged pions and kaons, and
multiple $\pi^0$. These results are shown in \cref{BABAR-kkpi-kkpipi}.

There are also additional measurements for some specific channels, 
$K^+K^- \pi^+ \pi^-$~\cite{Shemyakin:2015cba} and $K_S K_L \pi^0$~\cite{Achasov:2017vaq}.
Finally, cross sections for $K^+K^- \eta$~\cite{Aubert:2007ym} and
$K_SK_L \eta$~\cite{Lees:2014xsh} are available from BABAR. 

\ei

\paragraph{Narrow resonances}

The contributions of the very narrow resonances $J/ \psi$ and $\psi(2S)$ are obtained 
by numerically integrating their undressed Breit--Wigner line shapes. 
The uncertainties in the integrals are dominated by the knowledge of their bare 
electronic widths available from experiment~\cite{Ablikim:2016xbg,Patrignani:2016xqp}.

\paragraph{Inclusive $R$ measurements}

\begin{figure}[t] \centering
\includegraphics[width=12cm]{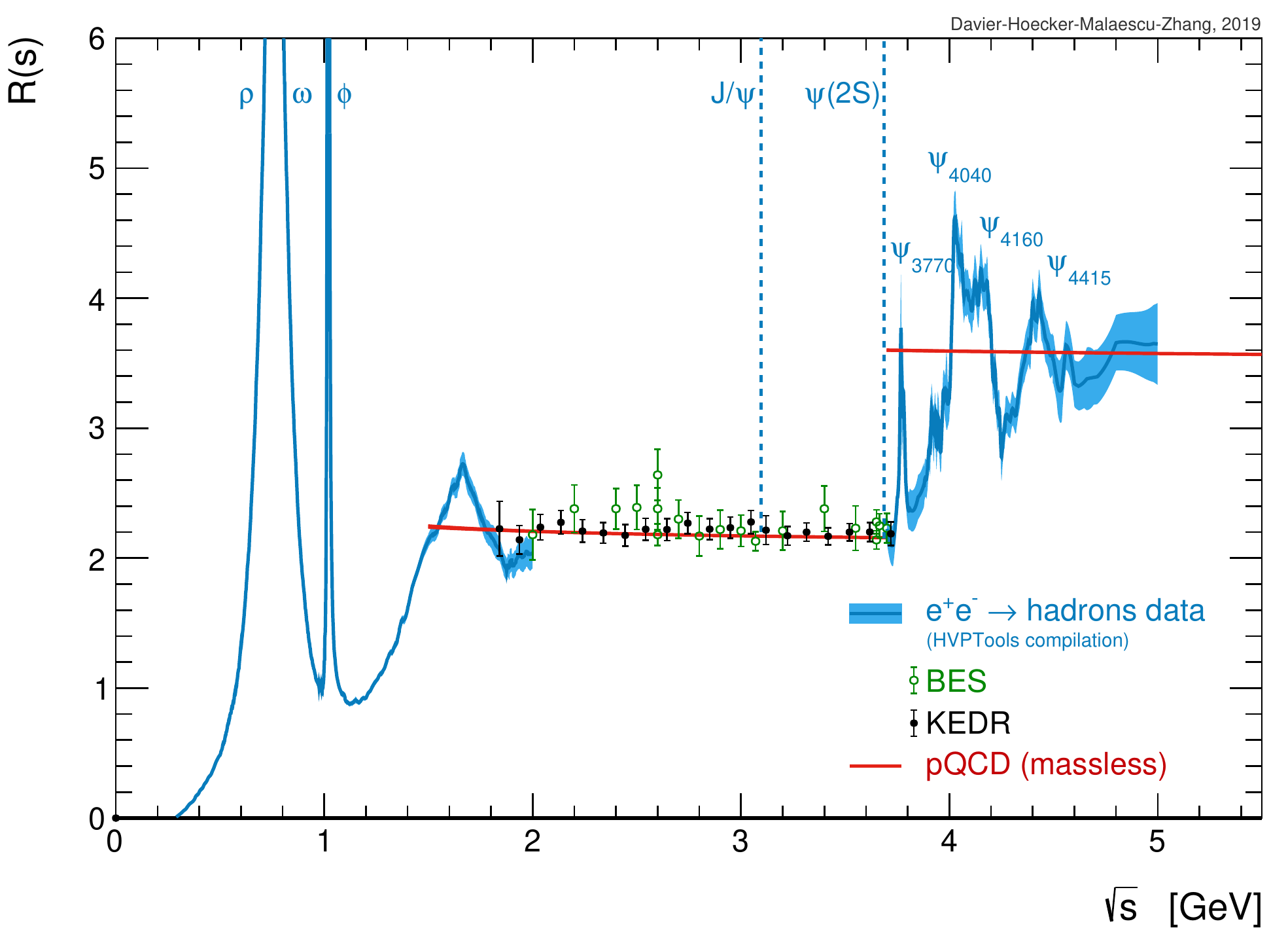}
\caption{The total hadronic \epem annihilation cross section ratio $R$ as a function of $\sqrt{s}$~\cite{Davier:2017zfy}. 
Inclusive measurements from BES~\cite{Bai:1999pk,Bai:2001ct,Ablikim:2009ad} (and references therein) and 
KEDR~\cite{Anashin:2015woa,Anashin:2016hmv,Anashin:2018vdo} are shown as data points, while the sum of exclusive 
channels from this analysis is given by the narrow blue bands. Also shown is the 
prediction from massless pQCD (solid red line). Reprinted from Ref.~\cite{Davier:2019can}.} 
\label{R-2019}
\end{figure}

Above $2\GeV$ the annihilation cross section has to be measured inclusively
because of the large number of open exclusive channels. Precise
results in the $2\hyph4.5\GeV$ range are from BESII~\cite{Bai:1999pk,Bai:2001ct,Ablikim:2009ad}.
The KEDR collaboration has recently published results from an inclusive 
$R$ scan from $\sqrt{s}=1.84$ to $3.05\GeV$~\cite{Anashin:2015woa,Anashin:2018vdo}, complementing their 
previous measurements obtained between 3.12 and $3.72\GeV$~\cite{Anashin:2015woa}. 
This data is the most precise and complete in this energy range with a typical 
systematic uncertainty of 3\%. It constitutes a very valuable input to test the 
validity of the pQCD estimate (cf.\ \cref{R-2019}). Between $2\GeV$ 
and the charm threshold, the $R$ value (hadronic cross section scaled to the 
$s$-channel pointlike fermion-pair lowest-order cross section) behaves
smoothly with a weak energy dependence, and it agrees with the pQCD prediction
within experimental uncertainties. The results on $R$, based on the sum of exclusive 
channels below $2\GeV$~\cite{Davier:2017zfy} and the inclusive measurements above, are given in 
\cref{R-2019}. The matching between the measurements in the two regions is
satisfactory and consistent with the quoted uncertainties.

\subsubsection{The missing channels}

Thanks to the BABAR systematic program of measurements of exclusive cross sections, very few
channels are now missing below $2\GeV$. These involve states with either $K_L$'s, or with high multiplicity, especially with multiple $\pi^0$'s. While single-$K_L$ processes have recently been measured by BABAR~\cite{Lees:2014xsh}, the cross sections for 
$K_L K_L \pi \pi$ can be safely estimated from the corresponding $K_S K_S$ states, 
assuming $CP$ invariance. Also BABAR results on the $\pi^+ \pi^- 3\pi^0$ channel have just 
been released~\cite{Lees:2018dnv}, so that the only relevant final state up to 
6 pions left unmeasured is $\pi^+ \pi^- 4\pi^0$. Its contribution can be estimated 
from the other measured 6-pion final states using isospin constraints obtained 
by  projecting the cross section on Pais isospin classes~\cite{Pais:1960zz,Davier:2010nc}.
When applying these isospin relations, it is important to consider the production of $\eta$ mesons separately because of their isospin-violating decays. At the 
present time the estimated contribution of missing channels contributes a fraction 
of less than 0.05\% of the $\amuHVPLO$ value when integrating the cross sections
up to $1.8\GeV$, which is not an issue anymore (before 2017 this fraction amounted to 
0.7\%). The situation is more problematic between 1.8 and $2\GeV$, since the lack of
measurements of higher-multiplicity final states could introduce
some small systematic effect due to the resulting under-evaluation of $R$. In this respect, the recent measurement by CMD-3~\cite{CMD-3:2019ufp} of the $3\pi^+3\pi^-\pi^0$ final state brings valuable information on this issue.

\subsubsection{Major tensions in hadronic data}
\label{ee-problems}

We present here the status of, and a discussion of the most important discrepancies between,
data from different experiments that affect significantly the precision of the 
combined cross sections used for the evaluation of the dispersion integrals.

\paragraph{Tensions in the $\pi^+\pi^-$ channel}
\label{tensions-pipi}

The $\pi^+\pi^-$ channel accounts for approximately 3/4 of the full hadronic 
contribution to the muon $g-2$. Thus, there is a need for the highest precision. Many experimental measurements have been performed in the last four decades, but it is only in the last 15 years that sufficient statistics and small systematic uncertainties have been achieved. 

\begin{figure}[t] \centering
\includegraphics[width=7.7cm]{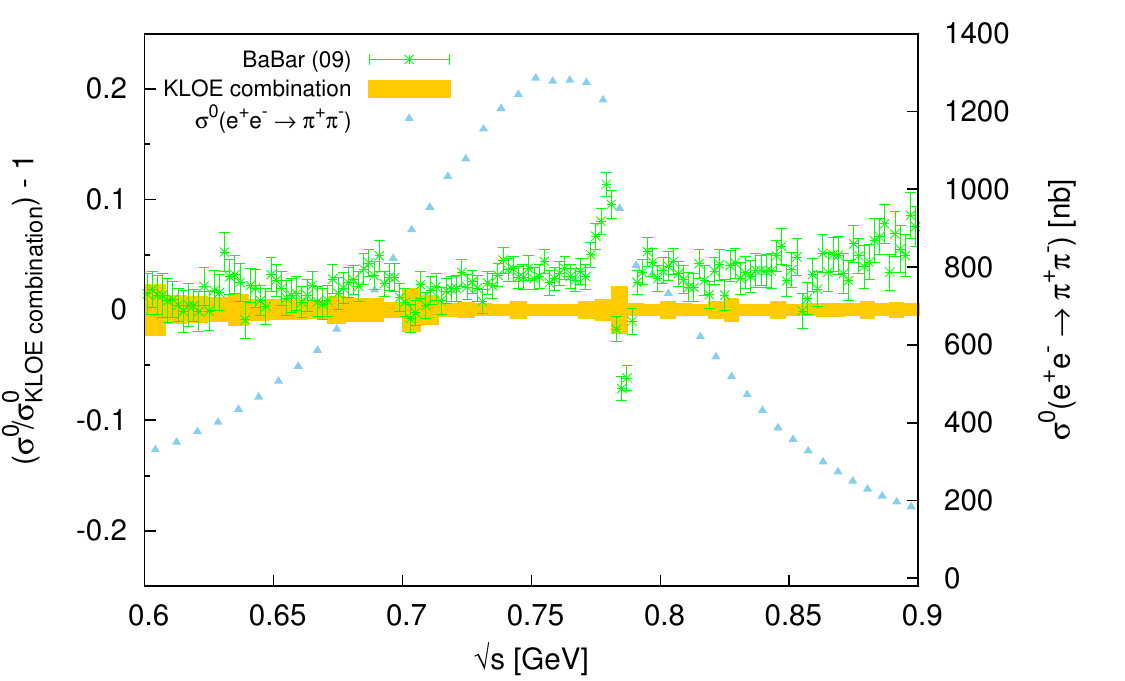}
\includegraphics[width=7.7cm]{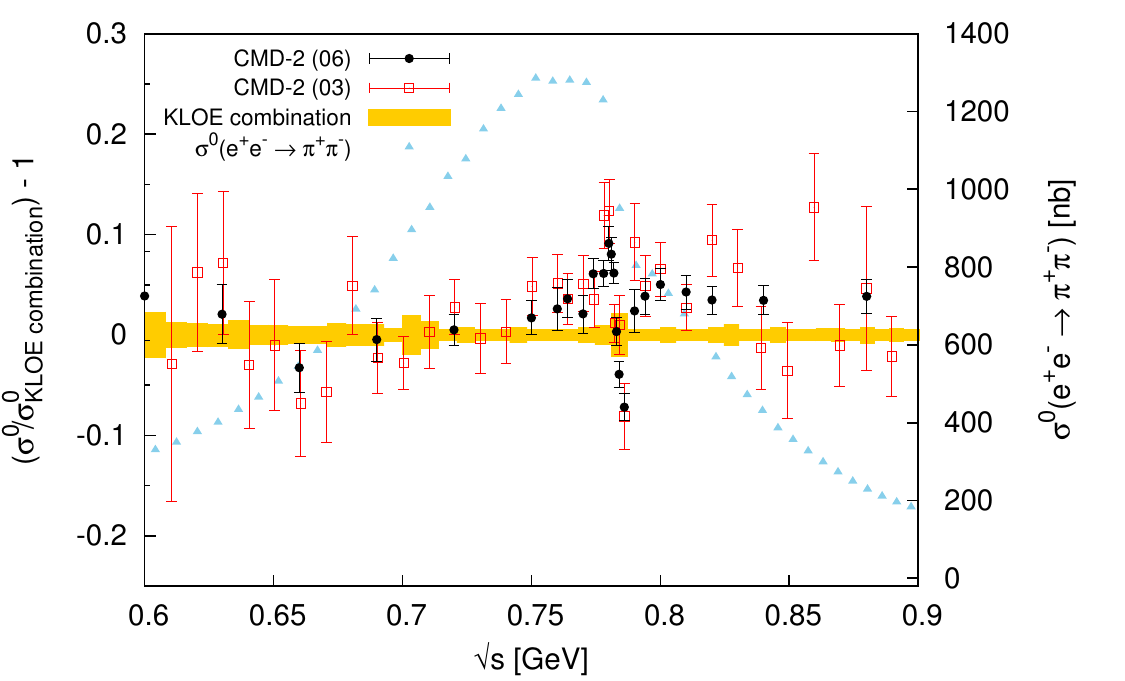}
\includegraphics[width=7.7cm]{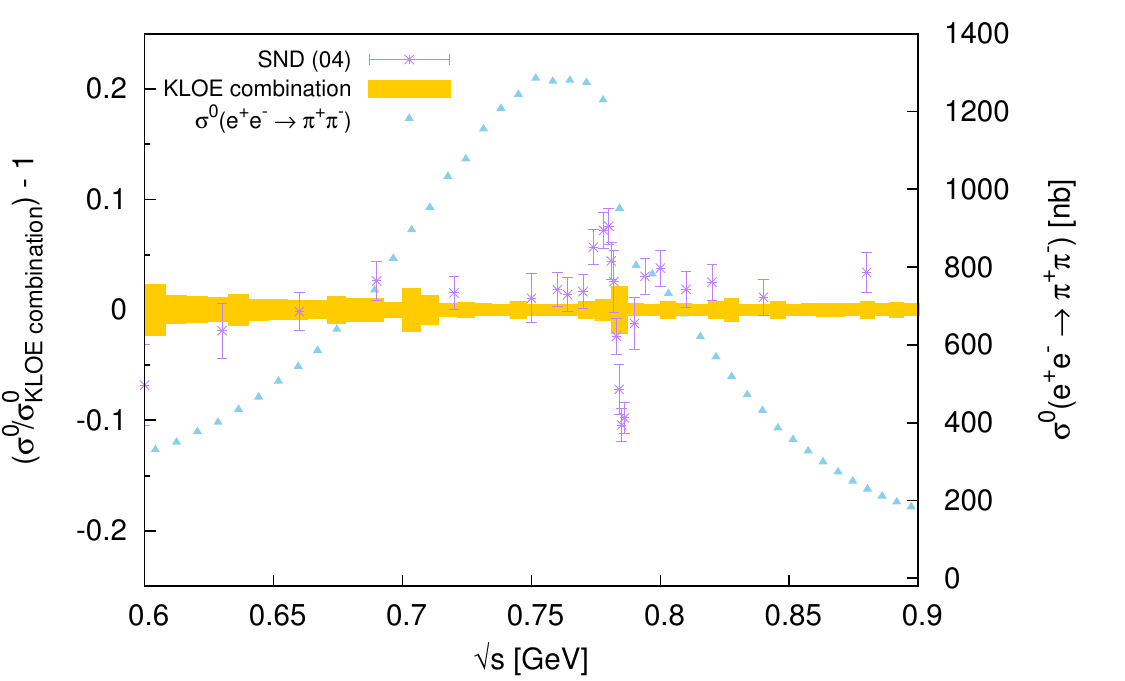}
\includegraphics[width=7.7cm]{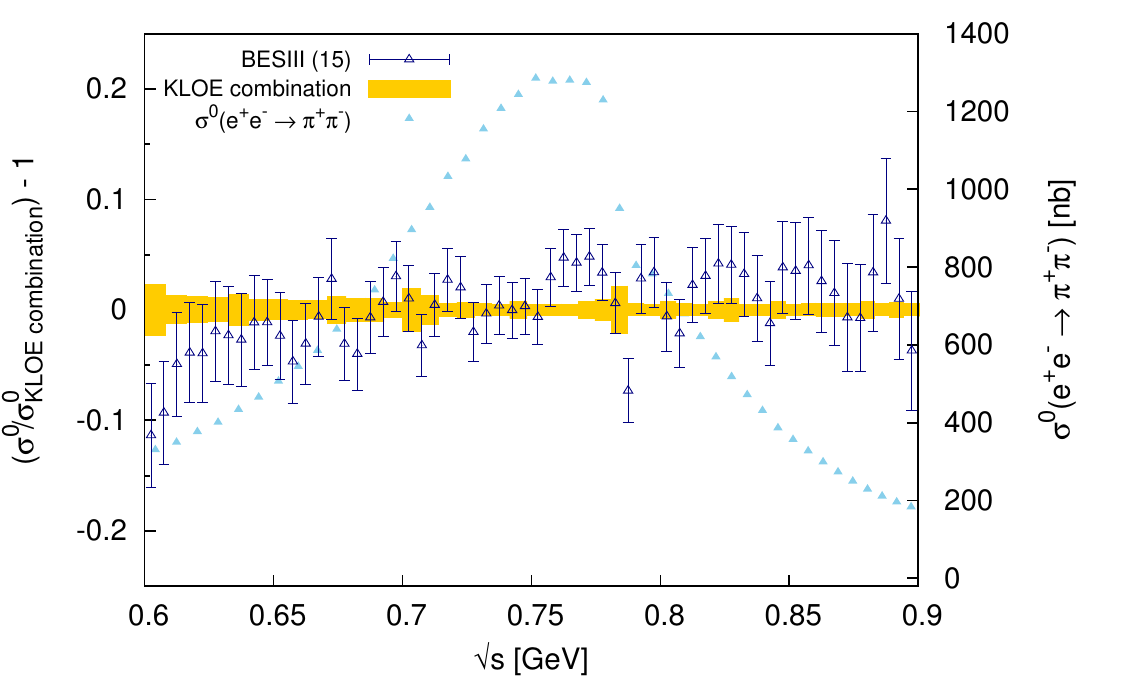}
\caption{The $\pi^+\pi^-$ cross section from the KLOE combination compared to the BABAR, CMD-2, SND, and BESIII data points in the $0.6\hyph0.9\GeV$ range~\cite{Anastasi:2017eio}. The KLOE combination is represented by the yellow band. The uncertainties shown are the diagonal statistical and systematic uncertainties summed in quadrature. Reprinted from Ref.~\cite{Anastasi:2017eio}.} 
\label{allexp-KLOE}
\end{figure}

However, the situation is far from ideal as the two most precise measurements by KLOE and BABAR do not agree well within their quoted uncertainties. 
After the combination~\cite{Anastasi:2017eio} of the three KLOE measurements based on different ISR methods, the reduced uncertainty makes the situation worse. \Cref{allexp-KLOE} taken from Ref.~\cite{Anastasi:2017eio} shows the ratios of the recent measurements by CMD-2, SND, BABAR, and BESIII to the combined KLOE cross section in the $0.6\hyph0.9\GeV$ mass region, where the KLOE band and the data points include the full diagonal error. Several features are apparent: (1) the normalization at the peak is generally higher than KLOE, (2) there is a trend for a linear increase of the ratio with mass, and (3) a clear disagreement is seen in the narrow $\rho$--$\omega$ interference region. Due to the higher precision of the BABAR data, these features are most clearly visible there, but they are also present for the other experiments. While there is reasonable agreement below $0.70\hyph0.75\GeV$, the KLOE data appears noticeably lower on the $\rho$ peak and above by a factor rising to a few percent. 

\begin{figure}[t] \centering
\includegraphics[width=8cm]{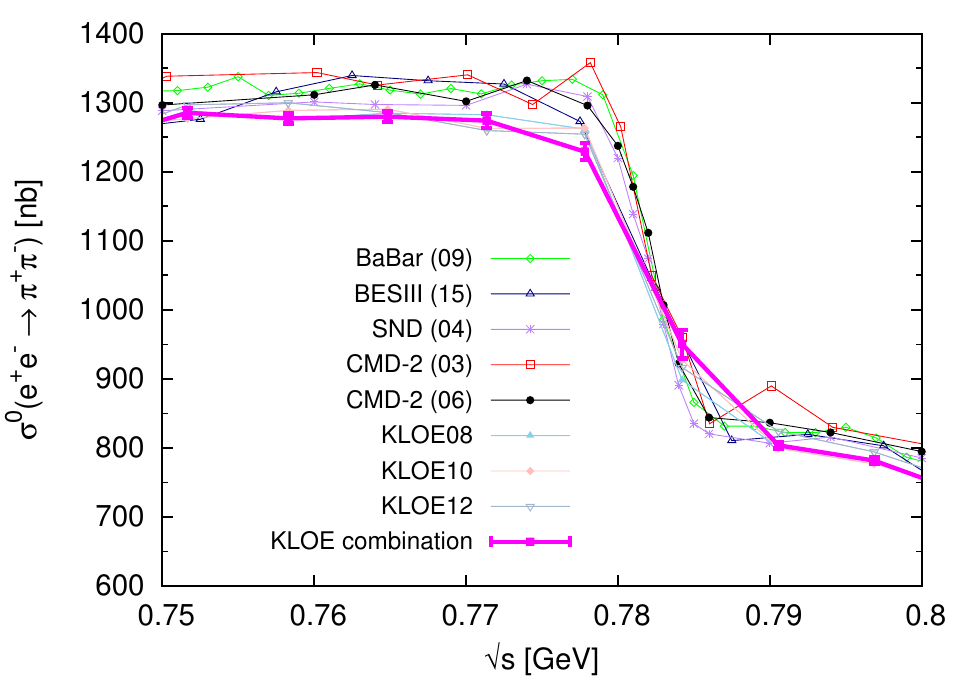}
\caption{The $\pi^+\pi^-$ cross section from KLOE combination, BABAR, CMD-2, SND, and BESIII in the $\rho$--$\omega$ interference region~\cite{Anastasi:2017eio}. Reprinted from Ref.~\cite{Anastasi:2017eio}.}
\label{comp-rho-omega-2pi}
\end{figure}

The ratios in the $\rho$--$\omega$ interference region display a common 
oscillatory pattern. Since in Ref.~\cite{Anastasi:2017eio} the ratio of a given experiment is computed with respect to the linearly interpolated value between adjacent KLOE points, one could expect some bias, especially in the interference region with its fast-varying cross section. Indeed, such oscillation is not present for the ratio KLOE to BABAR~\cite{Lees:2012cj}, where a fit to the BABAR data is used as reference in order to avoid such effects.
As seen in \cref{comp-rho-omega-2pi}, the interference pattern is more
washed out in KLOE, most probably due to the choice of wide mass bins.
A vertical offset is clearly seen in the plot on the $\rho$ peak. It
should be noted that the effect of the $\rho$--$\omega$ interference pattern is
largely canceled when integrating over the mass spectrum. Thus differences in
this region between the experiments are not expected to produce large biases for
the integral values.

\begin{figure}[t] \centering
\includegraphics[width=8cm]{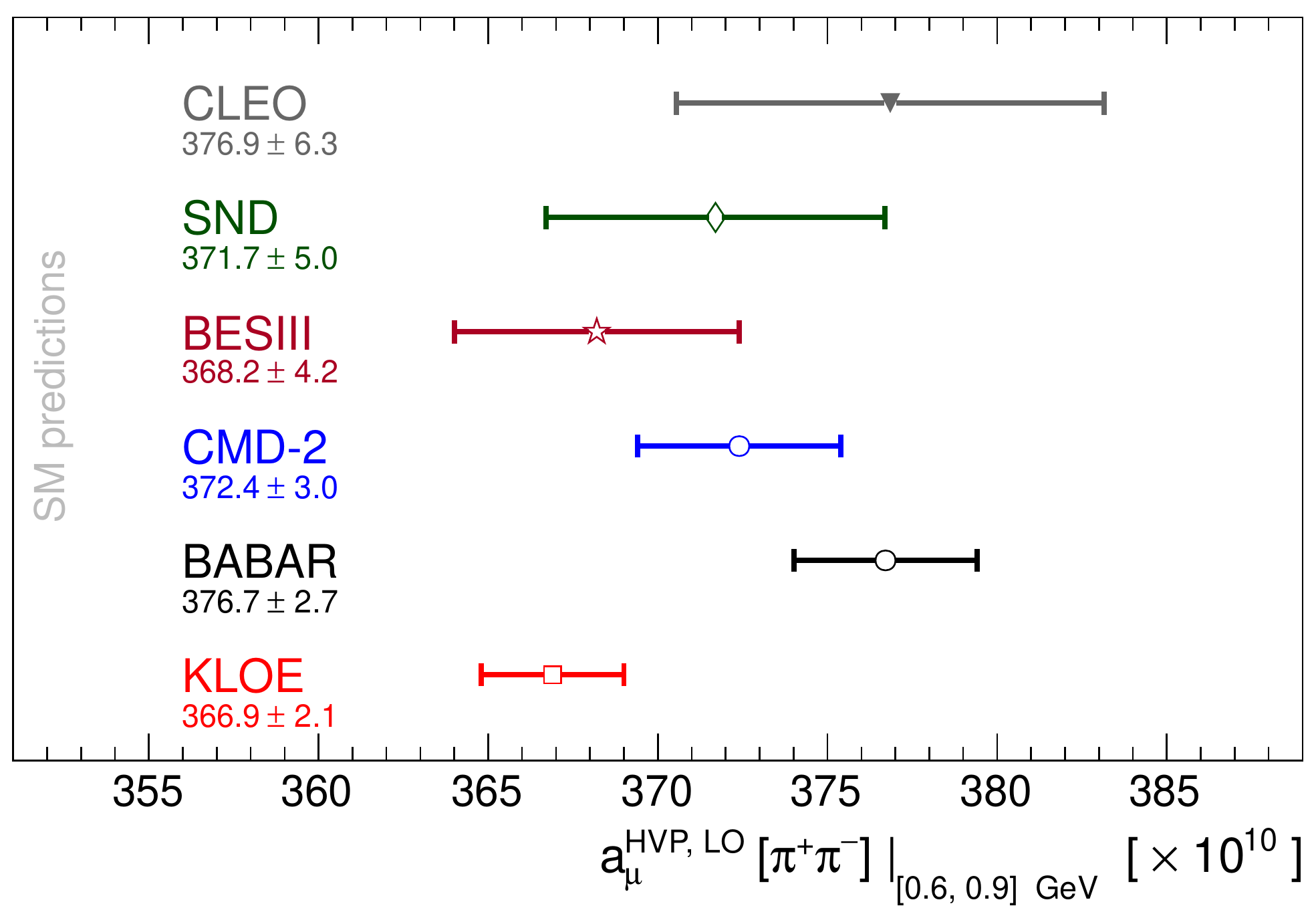}
\caption{Comparison of results for $\amuHVPLO[\pi\pi]$, evaluated between $0.6\GeV$ and $0.9\GeV$ for the various experiments.} 
\label{amu2pi-0.6-0.9}
\end{figure}

The most significant discrepancy between the KLOE and BABAR data points to one or several systematic effects not properly covered by the estimated systematic uncertainties. Here one might hope to appeal to other experiments to resolve this 
discrepancy. Unfortunately, their results are insufficiently precise at present, lying between those of KLOE and BABAR, and overlapping reasonably with both. This can been seen in \cref{amu2pi-0.6-0.9} which shows the contributions to the dispersion integral from the region between 0.6 and 0.9 GeV for each of the experimental data sets. One-parameter fits yield $\chi^2/dof$ values of 4.5/4 
and 3.6/4 for fits including all experiments but BABAR and all experiments but KLOE, respectively.  Thus CMD-2/SND/BESIII/CLEO are compatible with either KLOE or BABAR.

In the combination procedures used by DHMZ (see \cref{sec:DHMZ}) and KNT (see \cref{sec:KNT}), local tensions are dealt with by introducing scaling factors for the uncertainties. Global tension is also accounted for in the DHMZ analysis.

\begin{figure}[t] \centering
\includegraphics[width=7.5cm]{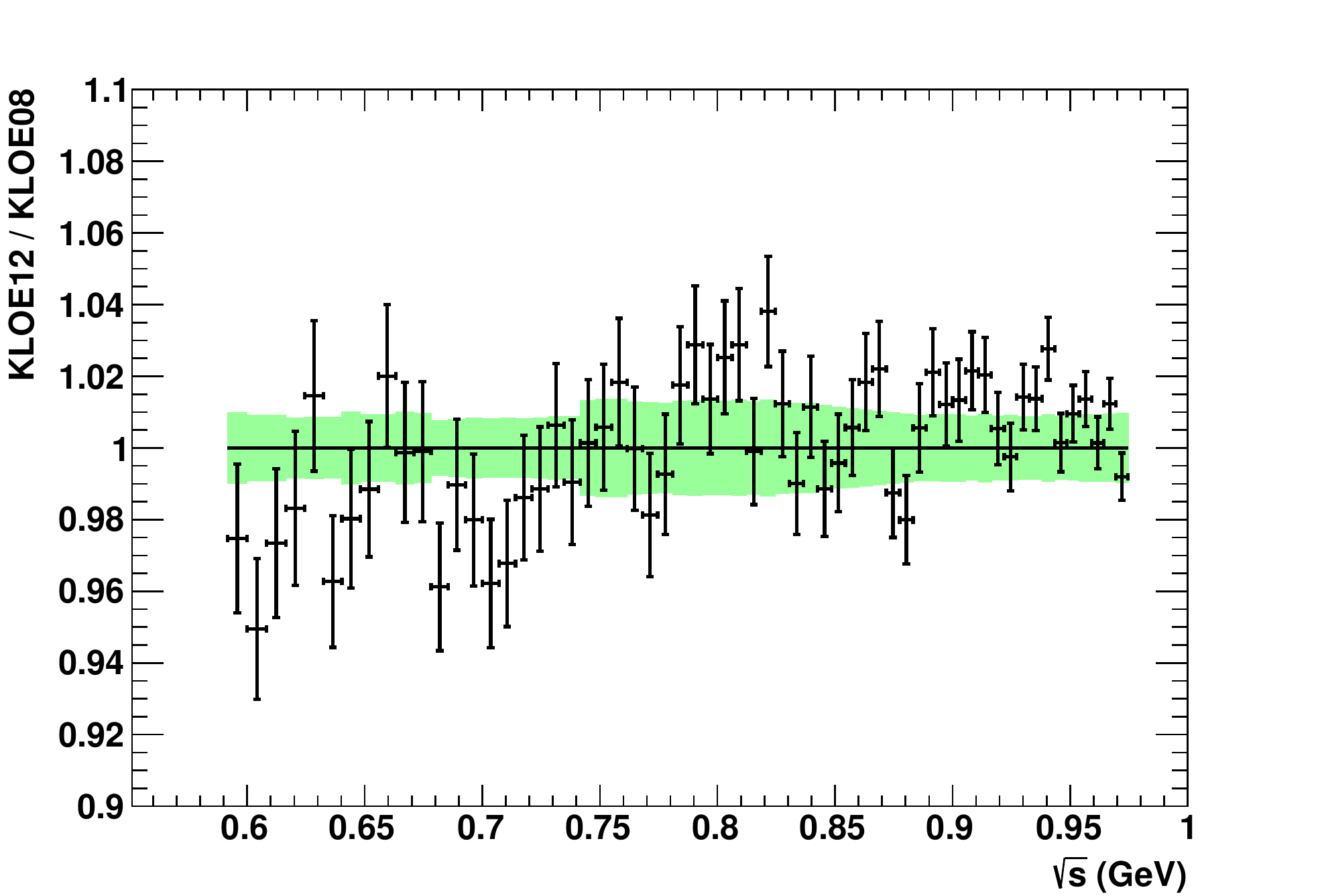}
\includegraphics[width=7.5cm]{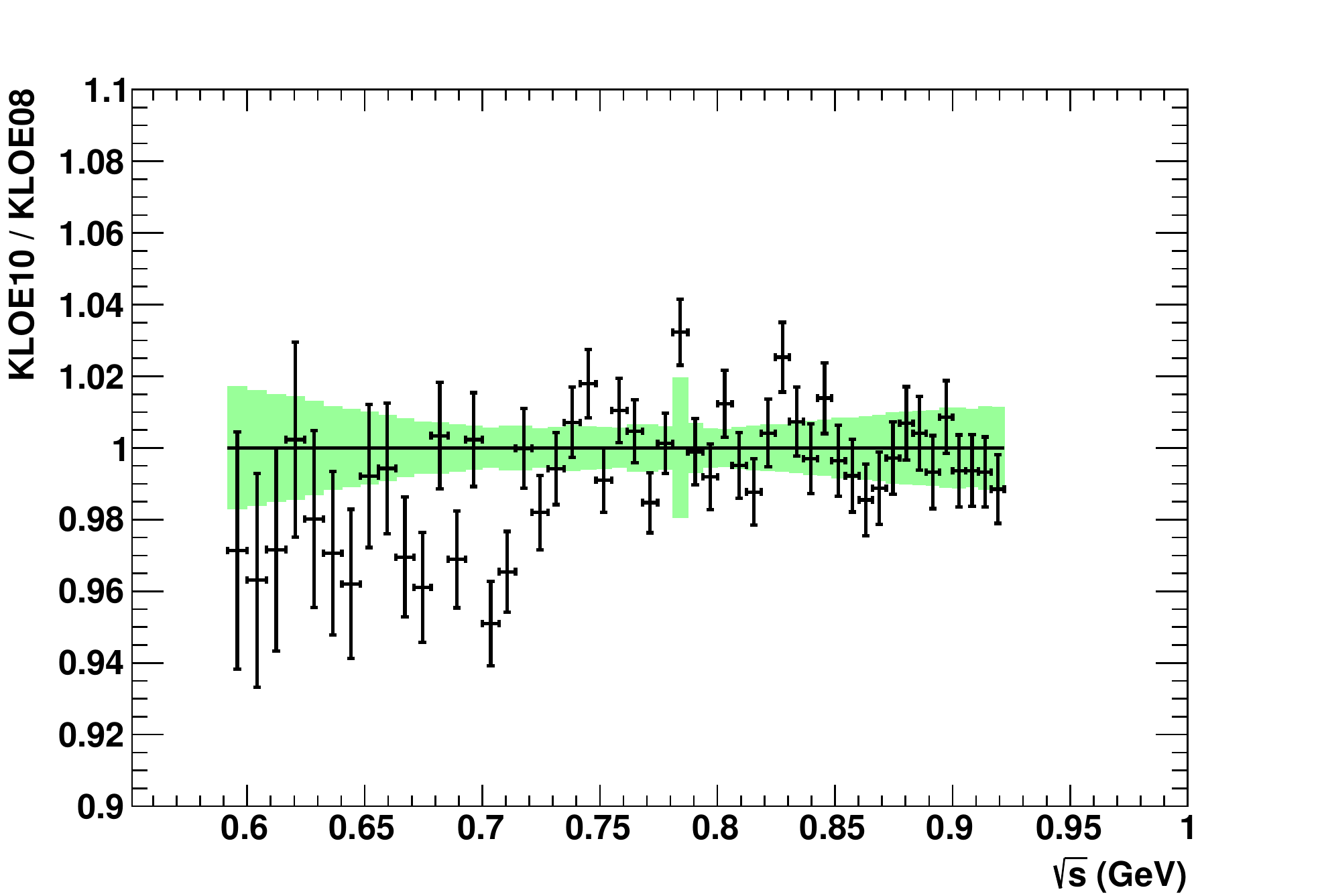} \\
\includegraphics[width=7.5cm]{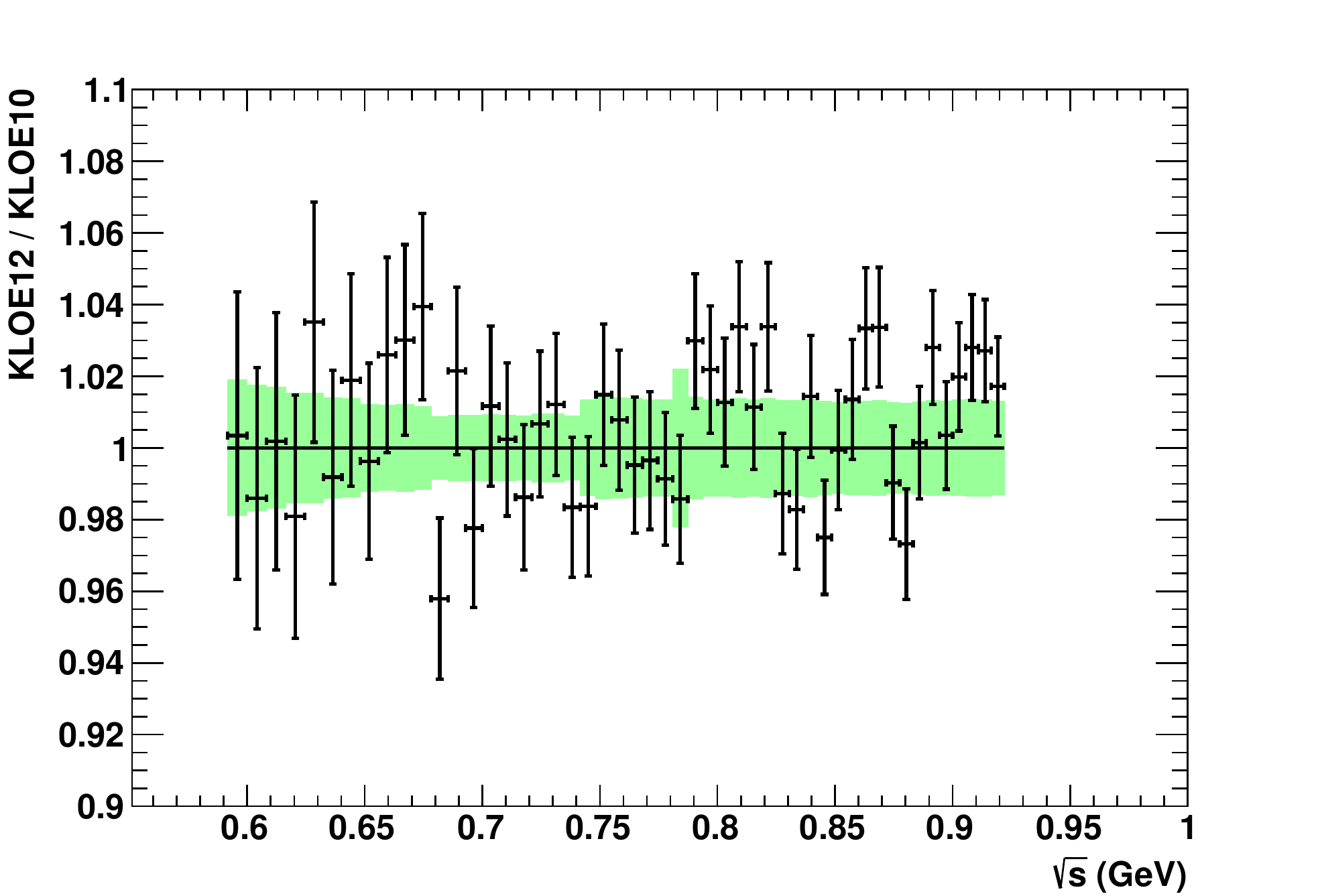}
\caption{Ratios of cross sections~\cite{Anastasi:2017eio} from KLOE-2012 to KLOE-2008~(top left), KLOE-2010 to KLOE-2008~(top right), and KLOE-2012 to KLOE-2010~(bottom). The green bands indicate the uncommon systematic uncertainty in the respective ratios.}
\label{kloe-ratios}
\end{figure}

Some tension also occurs in the combination of the results from the three KLOE measurements~\cite{Anastasi:2017eio}. The ratios of the cross section values between KLOE-2012 and KLOE-2008, as well as KLOE-2010 and KLOE-2008, were computed taking into account all the correlations between the measurements, for both the statistical and systematic uncertainties. They show some systematic deviations from unity (\cref{kloe-ratios}) that are statistically significant and not fully taken into account by the local scaling procedure~\cite{bogdan-kek-2018}, leading to what is likely an underestimated systematic uncertainty in the combined result.
Since these deviations largely cancel when integrating the spectrum, the integral values are consistent~\cite{Anastasi:2017eio}.
These discrepancies are not present in the ratio between the KLOE-2012 and KLOE-2010 measurements, which is consistent with unity in the whole energy range (see \cref{kloe-ratios}).

\begin{figure}[t] \centering
\includegraphics[width=8cm]{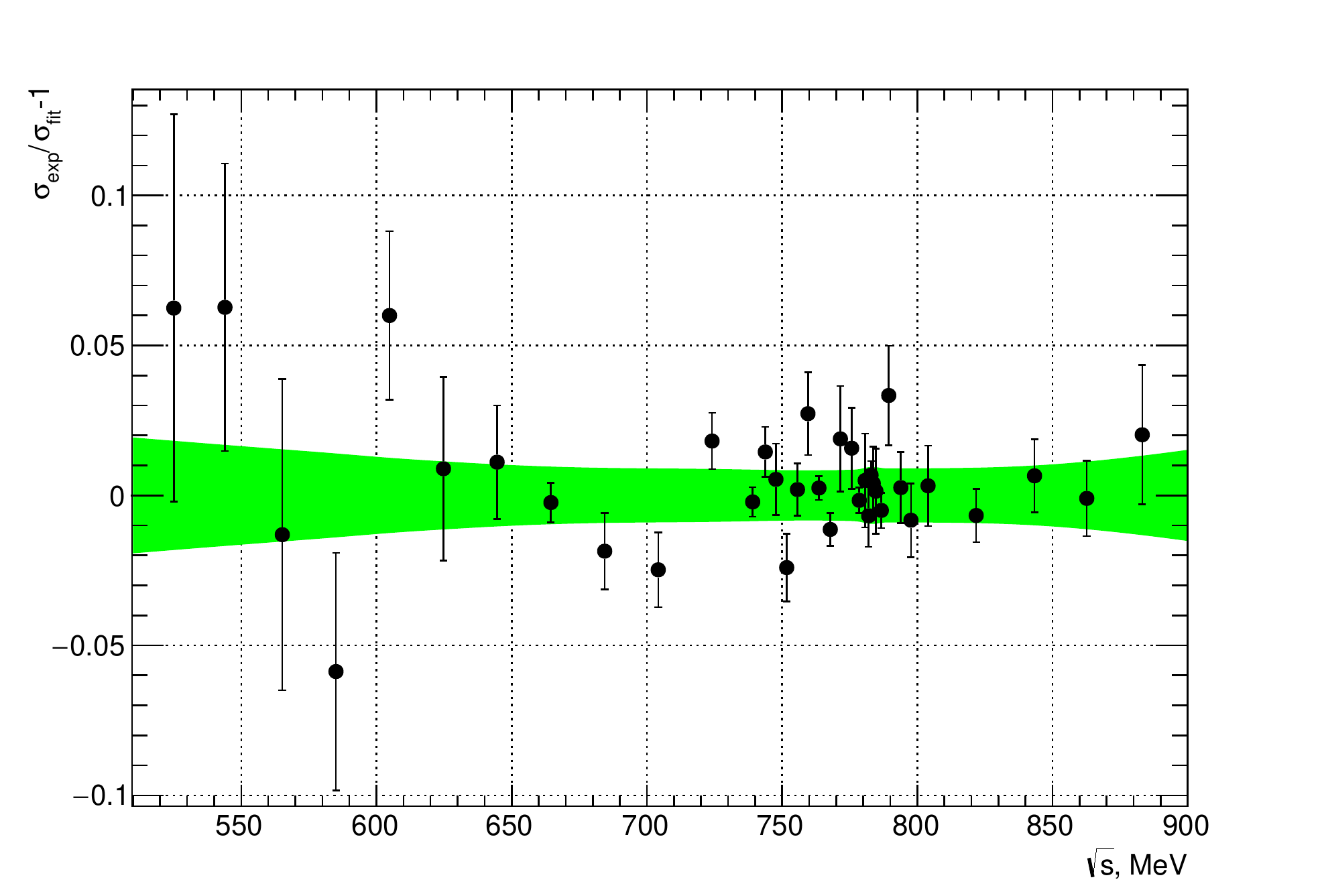}
\includegraphics[width=8cm]{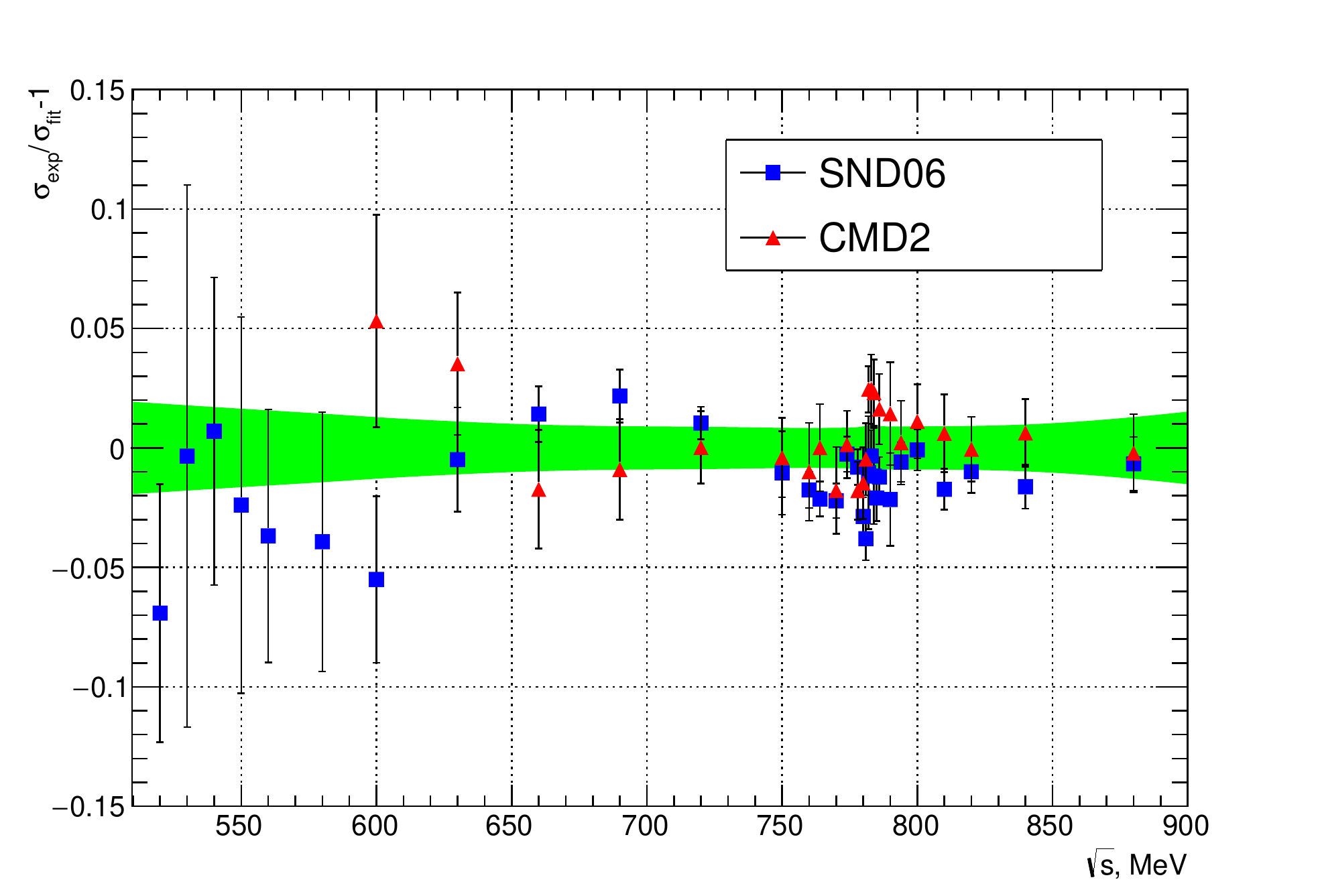}\\
\includegraphics[width=8cm]{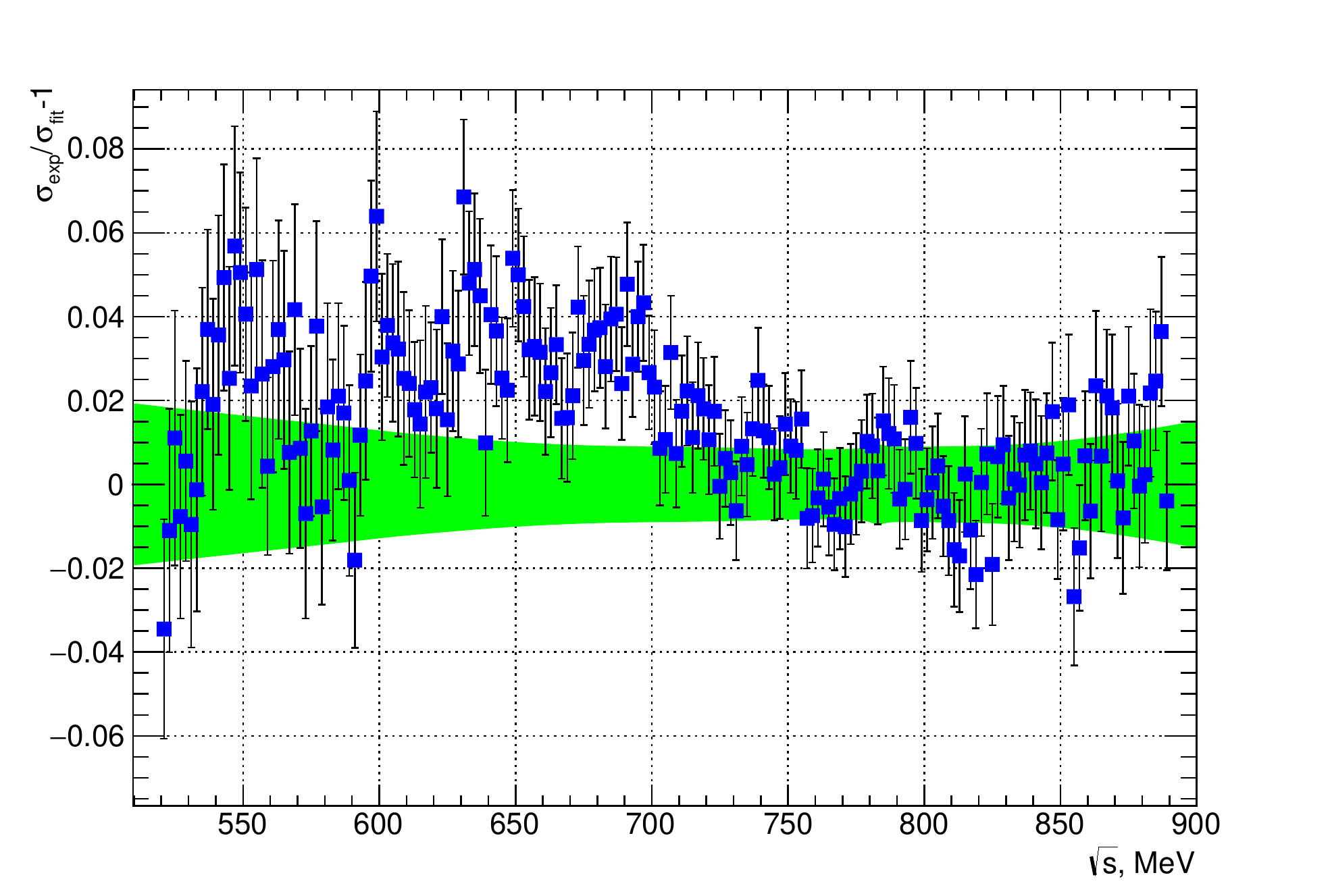}
\includegraphics[width=8cm]{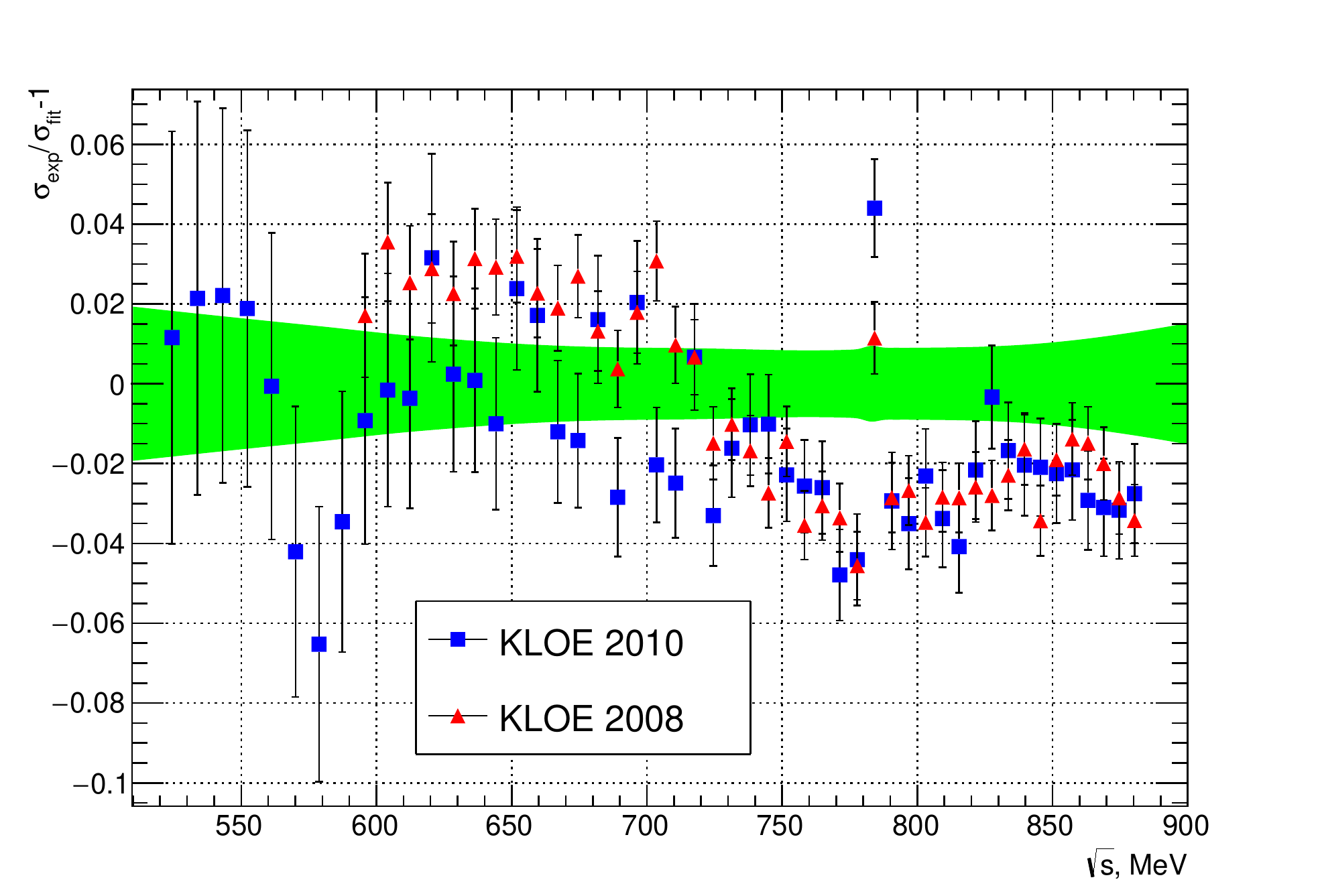}
\caption{Top left: ratio of SND-2020 $\pi^+\pi^-$ cross section values to form factor fit. Top right: ratio of previous SND and CMD-2 cross sections to new SND fit. Bottom: similar ratios for BABAR (left) and KLOE8 and KLOE10 (right). Reprinted from Ref.~\cite{Achasov:2020iys}.} 
\label{SND-2020}
\end{figure}

Very recently the SND collaboration has presented their results at VEPP-2000 on the
$\pi^+\pi^-$ channel~\cite{Achasov:2020iys} with increased statistics and reduced systematic uncertainties (0.8\%) compared to their analysis at VEPP-2M discussed above. They perform a fit of the pion form factor using a vector-meson dominance (VMD) ansatz  for the $\rho$ resonance together with $\omega$ and $\rho'$ contributions. This description of their data is used to compare with existing data in a convenient way. The resulting comparison ratios are shown in \cref{SND-2020} separately for BABAR, KLOE-2008, and KLOE-2010, and VEPP2M results from SND and CMD-2. While there are some small deviations from the latter two results, more severe discrepancies are found with KLOE and BABAR. On the one hand, below $0.7\GeV$  both KLOE-2008 and BABAR are higher than SND by 2\hyph4\%, while KLOE-2010 is more in agreement. On the other hand, above $0.7\GeV$  SND agrees well with BABAR, while both KLOE measurements are below by 2\hyph3\%. If these observations could provide some hints for understanding the KLOE--BABAR discrepancy, it is clear that still more experimental investigations with high precision are needed for further progress in this crucial $\pi^+\pi^-$ contribution. The new SND results are not yet included in the data combinations discussed in this WP version, but will be added later after they are carefully examined and accepted for publication.

\paragraph{Tensions in the $K^+K^-$ channel}

\begin{figure}[t]\centering
\includegraphics[width=7.5cm]{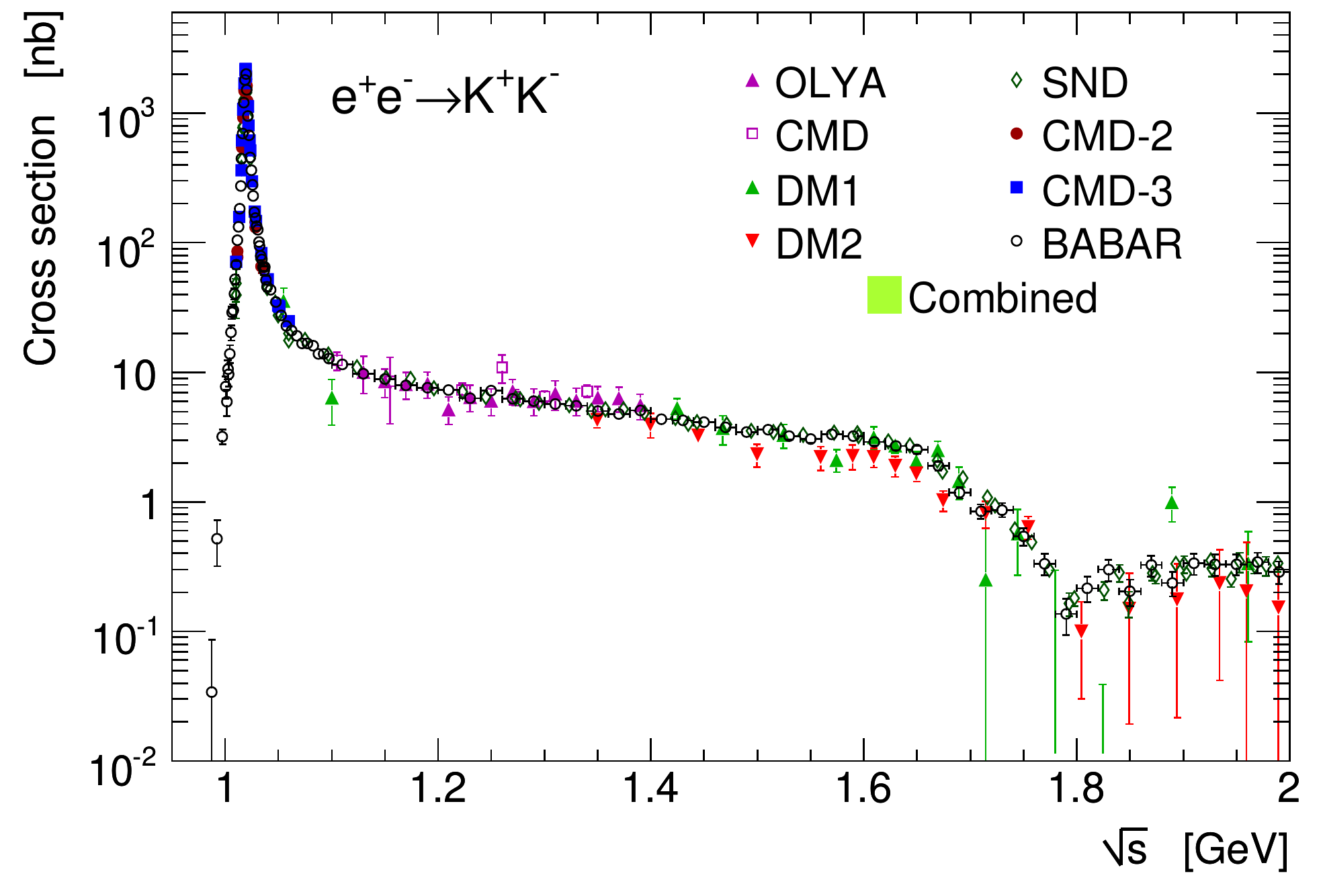}
\includegraphics[width=7.5cm]{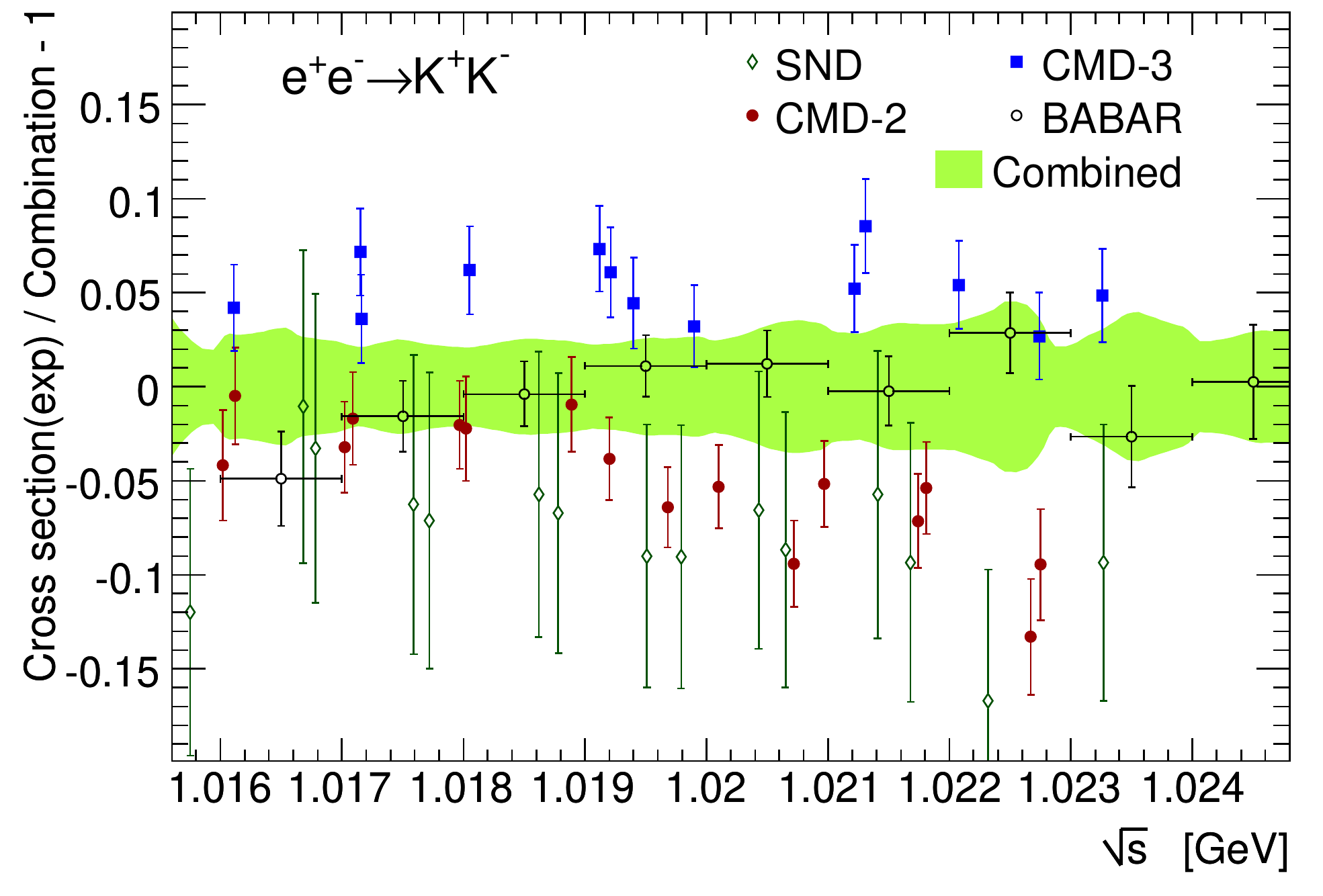}
\includegraphics[width=7.5cm]{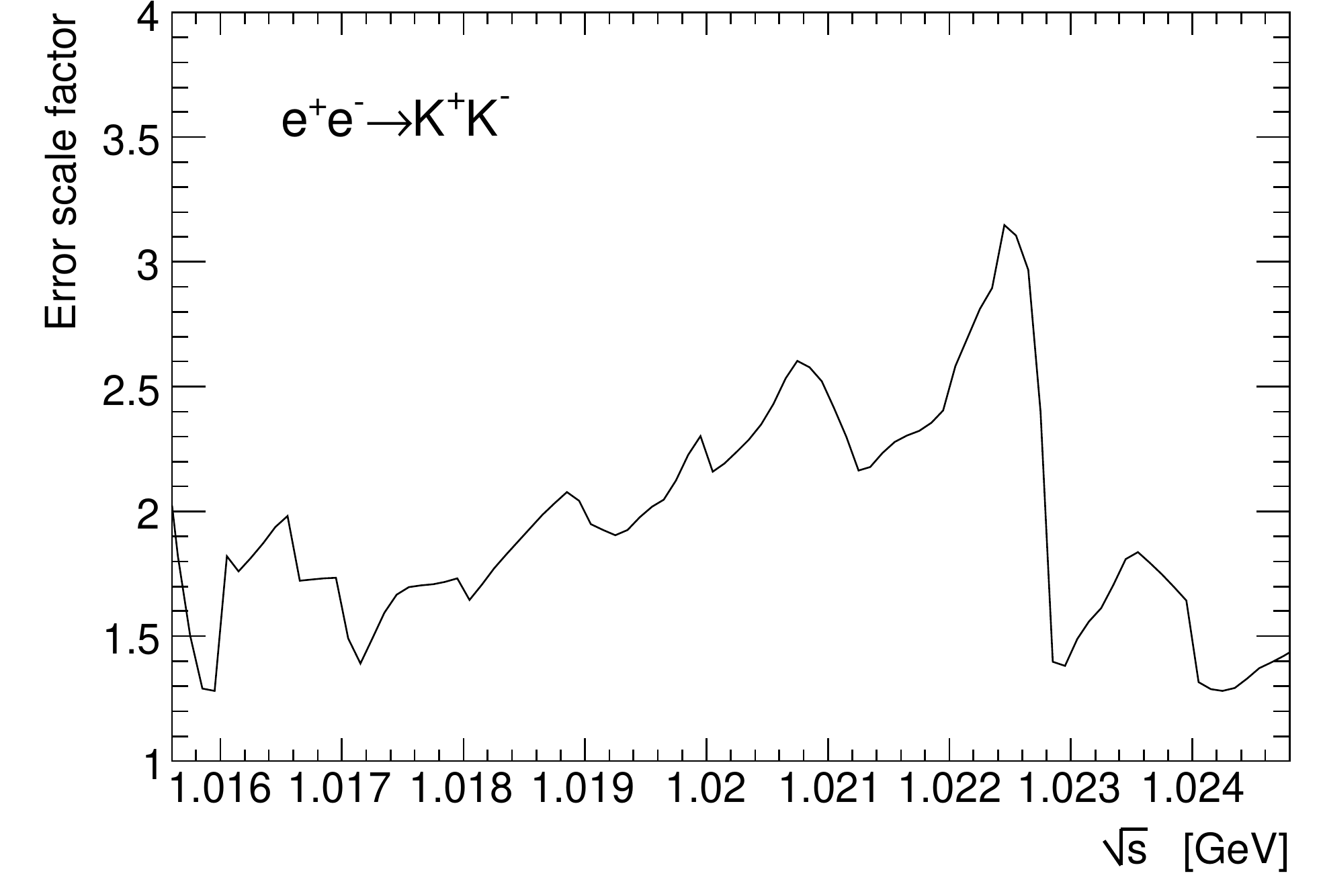}
\caption[.]{ 
            Top left: bare cross sections for  $e^+e^-\rightarrow K^+ K^-  $. See text for a description of the data used. 
            Top right: comparison between individual $e^+e^-\rightarrow K^+K^-$ cross-section measurements from BABAR~\cite{Lees:2013gzt}, CMD-2~\cite{Akhmetshin:2008gz}, CMD-3~\cite{Kozyrev:2017agm}, SND~\cite{Achasov:2000am}, and the HVPTools combination.  
            Bottom: local scale factor vs.\ CM energy applied 
            to the combined $K^+K^-$ cross-section uncertainty to account for inconsistency 
            in the individual measurements. Reprinted from Ref.~\cite{Davier:2019can}.}
\label{fig:kkweights}
\end{figure}

Tensions among data sets are also present in the $K^+K^-$ channel (see top panel of \cref{fig:kkweights} for a display of the available measurements). A discrepancy up to 20\% between BABAR~\cite{Lees:2013gzt} and SND~\cite{Achasov:2007kg} was observed for masses between 1.05 and $1.4\GeV$. Fortunately the problem has been resolved with the most recent SND result~\cite{Achasov:2016lbc}, although the origin of the previous systematic shift is not discussed. It looks like the older SND results should be discarded. 

Concerns also arise regarding data on the $\phi(1020)$ resonance. Previously, a 5.1\% difference between CMD-2~\cite{Akhmetshin:2008gz} at VEPP-2M and BABAR~\cite{Lees:2013gzt}, with the CMD-2 data being lower, was observed. SND~\cite{Achasov:2000am} results are also low compared to BABAR, but the discrepancy is not significant in view of the larger SND systematic uncertainty (6.8\%). Surprisingly, new results from CMD-3 at VEPP-2000~\cite{Kozyrev:2017agm} exhibit the opposite effect: they are 5.5\% higher than BABAR (cf.\ \cref{fig:kkweights} (middle)). The  discrepancy of almost 11\% between the two CMD-2/3 data sets greatly exceeds the  quoted systematic uncertainty of 2.2\%, of which only 1.2\% is assigned to the detection efficiency. The upward cross section shift is claimed to originate from a better understanding of the detection efficiency of kaons with very low energy in the CMD-3 data, since the $\phi(1020)$ lies very close to the $K^+K^-$ threshold. It should be remarked that, in comparison with the CMD-2/3 and SND measurements, the ISR method of BABAR benefits from higher-momentum kaons with better detection efficiency owing to the boost of the final state.

Given the  yet unresolved situation, both CMD-2 and CMD-3 data sets should be kept, which, owing to the uncertainty rescaling procedure, leads to a deterioration of the precision (by about a factor of 2) of the combined data (\cref{fig:kkweights} (bottom)). A better understanding of the data from CMD-2/3 and SND is necessary in order to improve the situation.

\subsubsection{Short-term perspectives}
\label{ee-short-term-perspectives}

Given the progress achieved in the last decade the situation on the contributions 
from multi-hadronic final states appears well under control and new results to come 
from VEPP-2000 will provide additional checks. Thus, the attention should be focused
on the contributions from the low-lying vector mesons, where discrepancies between
experiments remain unresolved. The largest component from the $\rho$ meson is still
the first priority for improvement. In this respect, results are expected from the 
on-going analysis of three experiments: BABAR~\cite{davier-kek-2018} using the full data sample and a new 
method independent of particle ID, which contributed the largest single systematic uncertainty in the 2009 analysis,
and CMD-3~\cite{cmd3-eps} and SND~\cite{snd-eps} taking advantage of their upgraded 
detectors and the larger luminosity delivered by VEPP-2000. First results and comparisons in the $\pi^+\pi^-$ channel have been presented recently by SND~\cite{Achasov:2020iys}, as discussed in \cref{tensions-pipi}.

\subsubsection{Use of hadronic data from $\tau$ decay}
\label{HVP-tau}

The use of data on semileptonic $\tau$ decays in the evaluation of $\amuHVPLO$ and $\Delta\alpha_{\rm had}^{(5)}$ was originally proposed in Ref.~\cite{Alemany:1997tn}. It is based on the fact that in the limit of isospin invariance, the spectral function of the vector current decay $\tau^-\to X^-\nu_\tau$ is related to the $e^+e^-\to X^0$ cross section of the corresponding isovector final state $X^0$ (the so-called conserved vector current (CVC) relation),
\begin{equation}
\sigma^{l=1}_{X^0}(s)=\frac{4\pi\alpha^2}{s}v_{1, X^-}(s)\,,
\end{equation}
where $s$ is the CM energy-squared or equivalently the invariant mass-squared of the $\tau$ final state $X$, $\alpha$ is the fine-structure constant, and $v_{1,X^-}$ is the nonstrange, isospin-one vector spectral function given by
\begin{equation}
v_{1,X^-}(s) = \frac{m^2_\tau}{6|V_{ud}|^2}\frac{{\cal B}_{X^-}}{{\cal B}_e}\frac{1}{N_X}\frac{dN_X}{ds}
\times \left[\left(1-\frac{s}{m^2_\tau}\right)^2\left(1+\frac{2s}{m^2_\tau}\right)\right]^{-1}\frac{R_{\rm IB}(s)}{S_{\rm EW}}\,.\label{eq:sf}
\end{equation}
Here, $m_\tau$ is the $\tau$ mass, $|V_{ud}|$ the CKM matrix element, ${\cal B}_{X^-}$ and ${\cal B}_e$ are the branching fractions of $\tau^-\to X^-(\gamma)\nu_\tau$ (final-state photon radiation is implied for $\tau$ branching fractions) and of $\tau^-\to e^-\bar{\nu}_e\nu_\tau$, $(1/N_x)dN_x/ds$ is the normalized $\tau$ spectral function (invariant mass spectrum) of the hadronic final state, $R_{\rm IB}$ represents $s$-dependent isospin-breaking (IB) corrections, and $S_{\rm EW}$ is the short-distance electroweak radiative correction~\cite{Davier:2009ag}.

\begin{figure}[ht!]
\centering
\includegraphics[width=0.45\columnwidth]{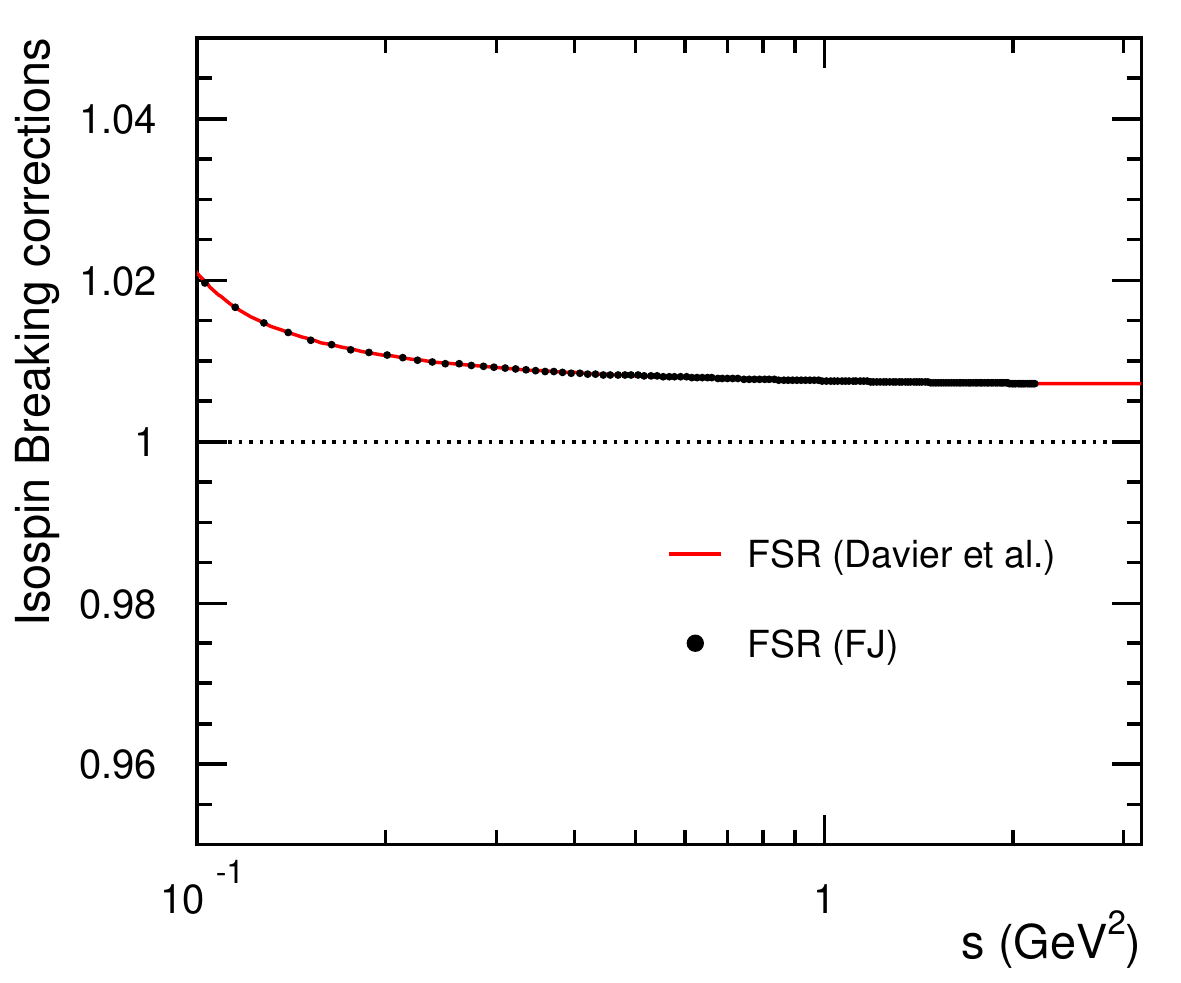}
\includegraphics[width=0.45\columnwidth]{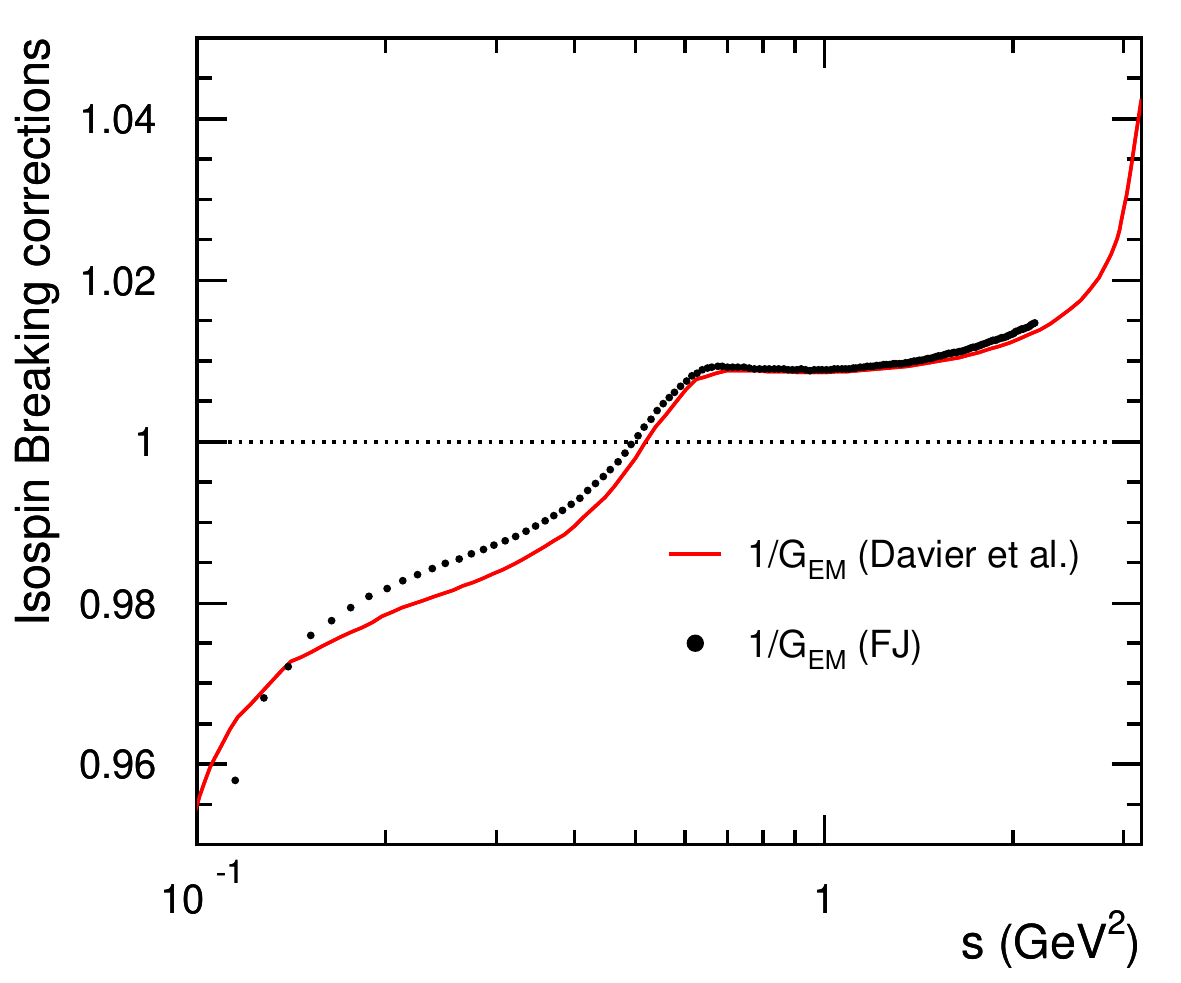}
\includegraphics[width=0.45\columnwidth]{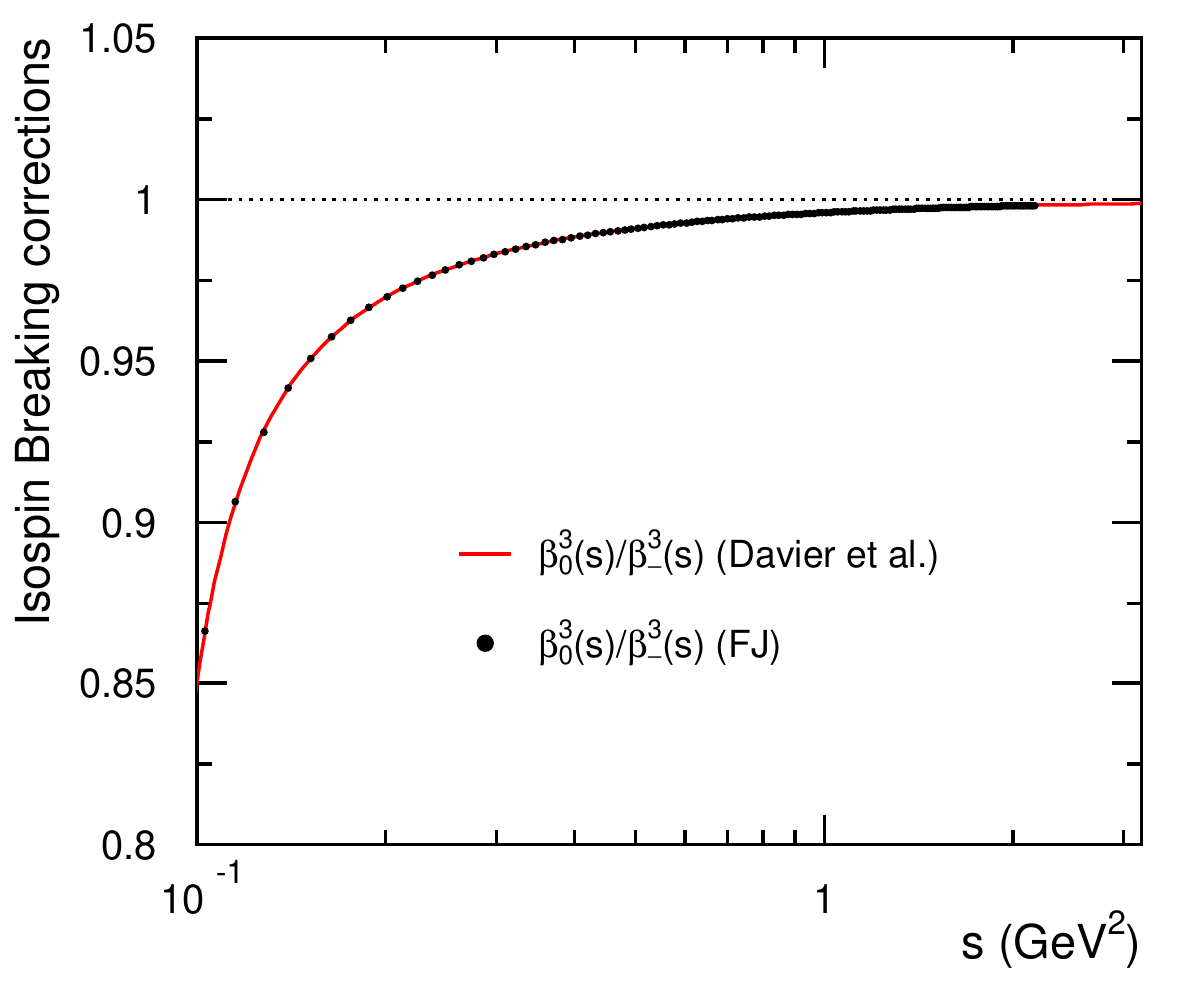}
\includegraphics[width=0.45\columnwidth]{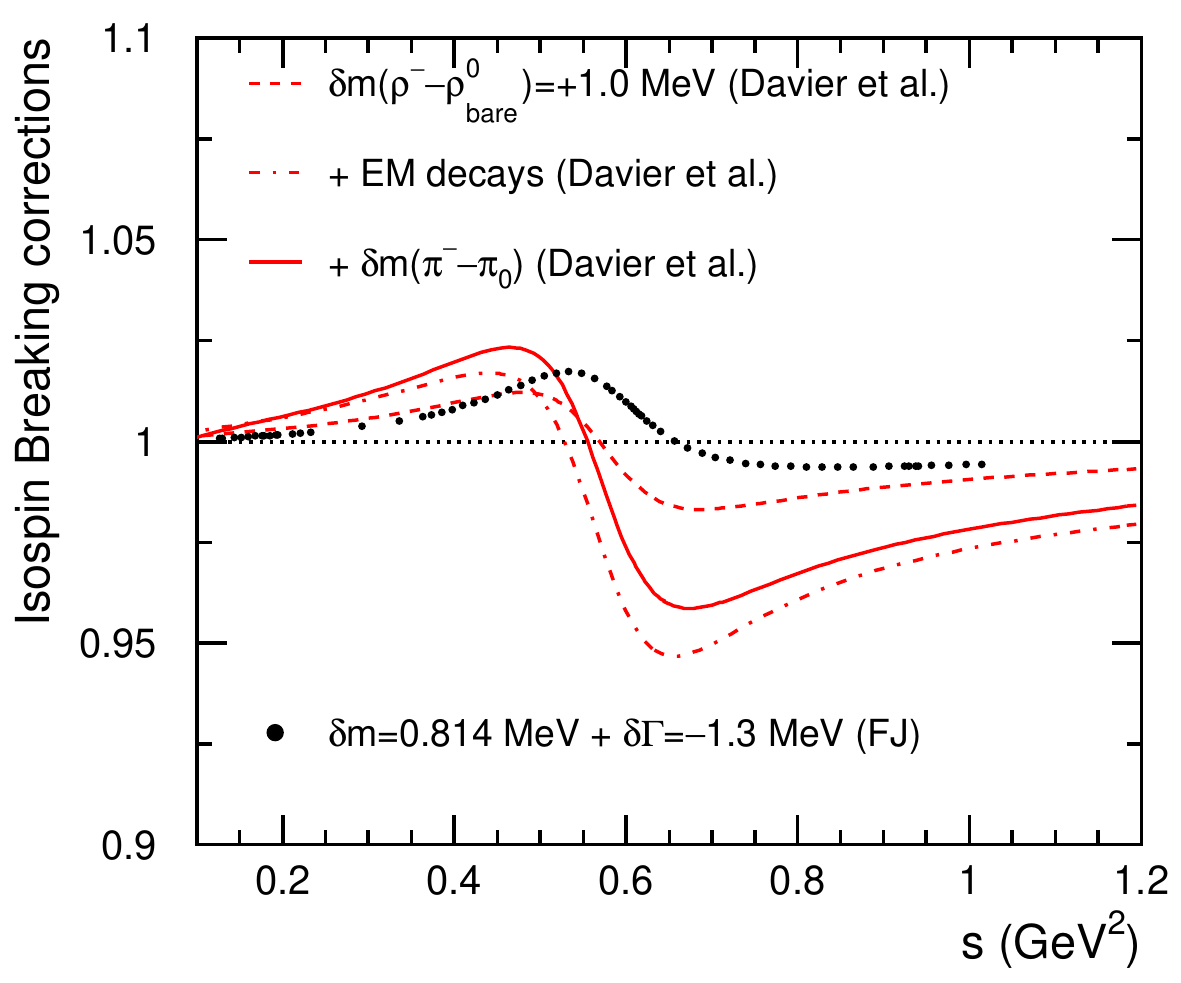}
\includegraphics[width=0.45\columnwidth]{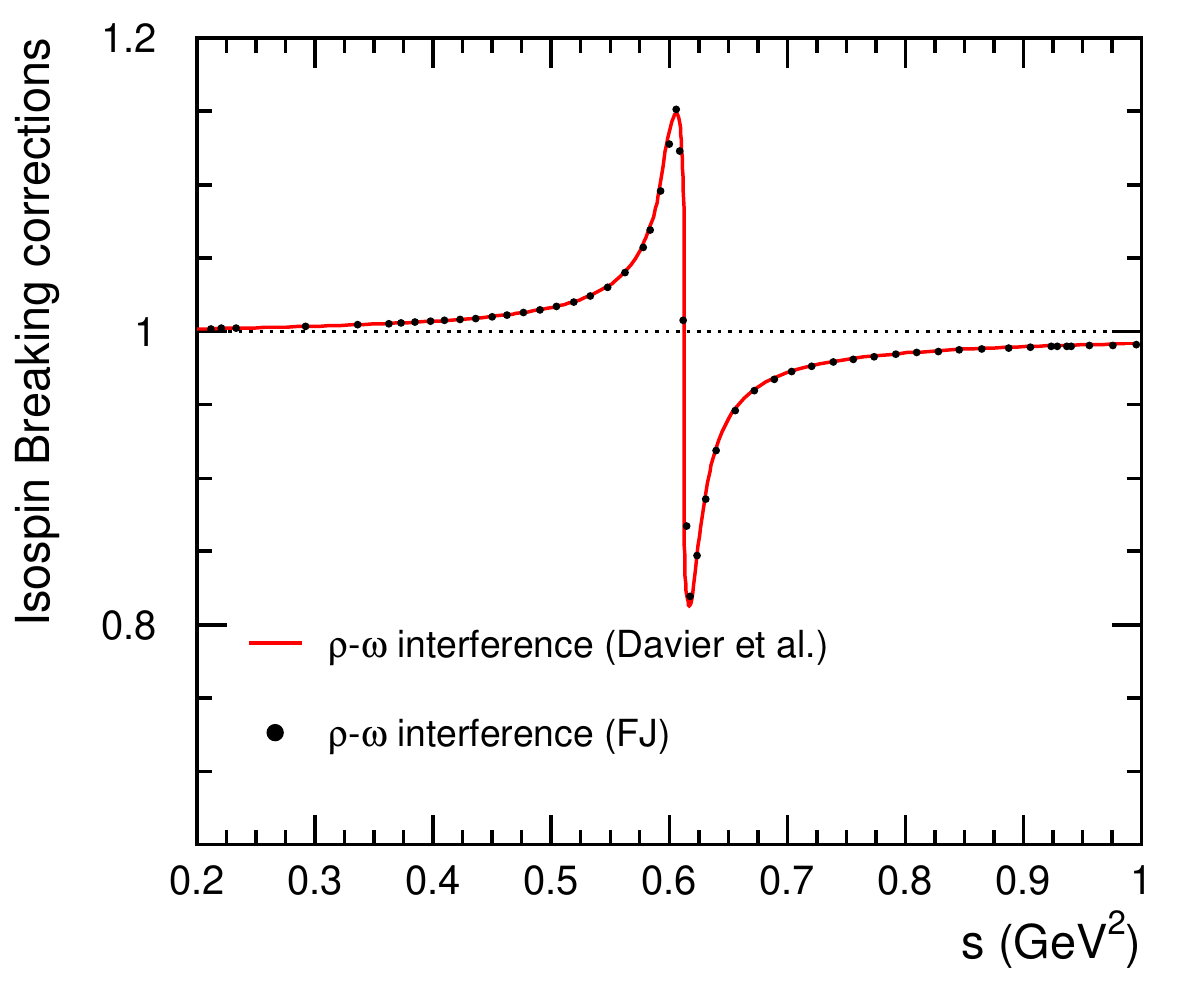}
\includegraphics[width=0.45\columnwidth]{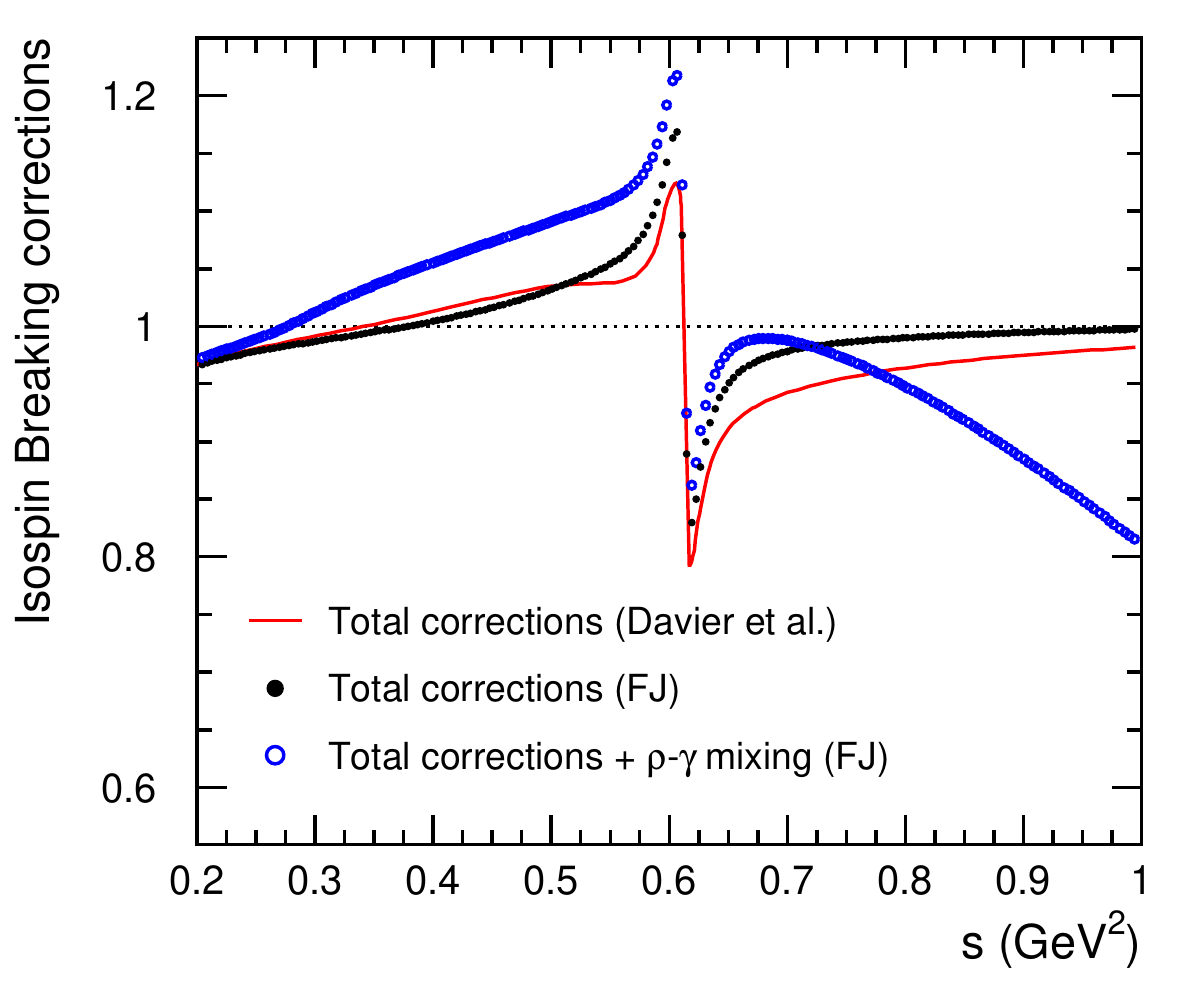}
\caption{Comparison of IB corrections versus $s$ used by Davier et al.~\cite{Davier:2009ag} and by FJ~\cite{fred}. The different plots correspond to FSR (top left),  $1/G_{\rm EM}$ (top right), $\beta^3_0/\beta^3_-$ ratio term (middle left), the effect of the $\rho$ mass and width difference in the $|F_0/F_-|^2$ term (middle right), the effect of the $\rho$--$\omega$ interference in the $|F_0/F_-|^2$ term (bottom left), and the total corrections (bottom right). The difference between the open blue points and the solid black one in the last plot stems from the $\rho$--$\gamma$ mixing corrections proposed in Ref.~\cite{Jegerlehner:2011ti}. Reprinted from Ref.~\cite{Zhang:2015yfi}.}
\label{fig:ib}
\end{figure}

Spectral functions and branching fractions for the $\tau$ have been precisely measured at LEP and at the $B$ factories under very different conditions. Much larger statistics are available at $B$ factories, but overwhelming QCD backgrounds must be reduced at the cost of small efficiencies with corresponding irreducible systematic uncertainty. The reverse occurred at LEP with $Z$ decays into two boosted $\tau$s and small well-understood backgrounds inducing small systematic uncertainties, but with moderately high statistics. As a consequence, branching fractions are well measured at LEP, while the determination of normalized spectral functions profit from the high statistics available at $B$ factories. For the dominant $2\pi$ channel the branching ratio has been best measured by ALEPH~\cite{Schael:2005am} in agreement with the other experiments~\cite{Acciarri:1994vr,Ackerstaff:1998yj,Anderson:1999ui,Abdallah:2003cw,Fujikawa:2008ma} and
the most accurate spectral function has been obtained by Belle~\cite{Fujikawa:2008ma}. Combined spectral functions from all experiments are available~\cite{Davier:2009ag}.

If one focuses on this dominant $2\pi$ channel, the IB correction term $R_{\textrm{IB}}(s)$ has the form
\begin{equation}
R_\textrm{IB}(s)=\frac{\textrm{FSR}(s)}{G_\textrm{EM}(s)}\frac{\beta^3_0(s)}{\beta^3_-(s)}\left|\frac{F_0(s)}{F_-(s)}\right|^2\,,\label{eq:ib}
\end{equation}
where $\textrm{FSR}(s)$ refers to the final-state radiative corrections~\cite{jss89} (see also Ref.~\cite{Drees:1990te}), $G_\textrm{EM}(s)$ denotes the long-distance radiative corrections of order $\alpha$ to the photon-inclusive $\tau^-\to \pi^-\pi^0\nu_\tau (\gamma)$ spectrum~\cite{Davier:2009ag}, $\beta_0^3(s)/\beta_-^3(s)$ accounts for the impact on the ratio of phase space factors of the $\pi^\pm-\pi^0$ mass splitting and is important only close to threshold (see \cref{fig:ib}), and $F_0(s)$ and $F_-(s)$ are the timelike $\pi^+\pi^-$ and $\pi^-\pi^0$ pion form factors, respectively.

The ratio $|F_0(s)/F_-(s)|^2$ is the most difficult to estimate reliably, since a number of different IB effects may contribute. Some, such as the contribution to the numerator of this ratio from the IB part of the $\rho$--$\omega$ interference shoulder, and the impact of IB differences in the masses and widths of the charged and neutral $\rho$ mesons can be estimated from data, albeit with some residual model dependence. In contrast, contributions produced by, for example, an IB difference in the charged and neutral $\rho$ isovector current decay constants and/or a purely IB $\rho^0$ isoscalar current decay constant, both of which are expected to exist on general grounds, would manifest themselves as small IB differences in the broad $\rho$ distributions for which no obvious phenomenological method of estimating their size exists. One could, of course, assume that such contributions are numerically small, estimate the contributions one is able to constrain phenomenologically,
and then see if the sum of that partial set of corrections, when applied
to the $\tau$ $\pi\pi$ distribution, brings the result into agreement with
$e^+e^-\to\pi\pi$ distribution results. If this turned out to be the case, one 
could argue that this provides post facto evidence for the smallness of the 
IB contributions one is unable to estimate phenomenologically. Considerable
effort, described in more detail below, has been expended on investigating 
this possibility. As we will see, the sum of the partial set of IB corrections
that result does not yet provide an understanding of the IB difference
between the $\tau$ and $e^+e^-\to\pi\pi$ distributions, so at present we
are unable to take advantage of the $\tau$ data. It is, however, worth
detailing the work that has been done in this direction to date. The alternate
possibility of using lattice simulations to include all sources of IB 
simultaneously and evaluate the IB inclusive $\tau$\hyph$e^+e^-$ $\amuHVPLO$ 
difference is discussed in \cref{subsec:tau}.

\begin{table}[t]
\centering
\small
\begin{tabular}{lcc}
\toprule
Source      & $\Delta \amuHVPLO[\pi\pi, \tau]$ &  $\Delta{\cal B}^{\rm CVC}_{\pi^-\pi^0}$ \\
\midrule
$S_{\rm EW}$ &  $-12.21(15)$ & $+0.57(1)$  \\
$G_{\rm EM}$  &  $ -1.92(90)$ & $-0.07(17)$  \\
FSR                 &  $+4.67(47)$ & $-0.19(2)$   \\
$\rho$--$\omega$ interference
                    & $+2.80(19)$ & $-0.01(1)$  \\
$M_{\pi^\pm}-M_{\pi^0}$ effect on $\sigma$
                    & $ -7.88$ & $+0.19$        \\
$M_{\pi^\pm}-M_{\pi^0}$ effect on $\Gamma_{\rho}$
                    & $+4.09$ & $-0.22$   \\
$M_{\rho^\pm}-M_{\rho^0_{\rm bare}}$ 
                    & $0.20^{+0.27}_{-0.19}$ & $+0.08(8)$    \\
$\pi\pi\gamma$, electromagnetic decays
                    & $ -5.91(59)$ & $+0.34(3)$ \\
                    $\delta({\rm GS}-{\rm KS})$ & $-0.67$ & $-0.03$ \\
\midrule
Total               & $-16.07(1.85)$ & $+0.69(22)$ \\
\bottomrule
\end{tabular}
\caption{
    Contributions to $\amuHVPLO[\pi\pi](\tau)$ ($\times10^{10}$) and ${\cal B}^{\rm CVC}_{\pi^-\pi^0}$ ($\times 10^{2}$)
    from the IB corrections.
    Corrections shown correspond to the 
    GS parameterization~\cite{Davier:2009ag}. The total uncertainty includes the difference with the KS parameterization quoted as $\delta({\rm GS}-{\rm KS})$.}
    \label{tab:ib}
\end{table}

Below $1\GeV$, the pion form factors are dominated by the $\rho$ meson resonance. Important IB effects are thus expected from the mass and width differences between the $\rho^\pm$ and $\rho^0$ mesons, and $\rho$\hyph$\omega$ mixing. The difference between the corrections used in Ref.~\cite{Davier:2009ag} and those of Refs.~\cite{fred,Jegerlehner:2011ti} is mainly due to different width differences considered. The width difference $\delta\Gamma_\rho=\Gamma_{\rho^0}-\Gamma_{\rho^-}$ used in Ref.~\cite{Davier:2009ag} was based on~\cite{FloresBaez:2007es}
\begin{equation}
\delta\Gamma_\rho(s)=\frac{g^2_{\rho\pi\pi}\sqrt{s}}{48\pi}\left[\beta^3_0(s)(1+\delta_0)-\beta^3_-(s)(1+\delta_-)\right]\,,\label{eq:drho1}
\end{equation}
where $g_{\rho\pi\pi}$ is the strong coupling of the isospin-invariant $\rho\pi\pi$ vertex and $\delta_{0,-}$ denotes radiative corrections for photon-inclusive $\rho\to \pi\pi$ decays and other electromagnetic decays, contrary to
\begin{equation}
\delta\Gamma_\rho=\frac{g^2_{\rho\pi\pi}}{48\pi}\left(\beta^3_0M_{\rho^0}-\beta^3_-M_{\rho^-}\right)\,,\label{eq:drho2}
\end{equation}
used in Ref.~\cite{Jegerlehner:2011ti}. The numerical values of \cref{eq:drho1} and \cref{eq:drho2} at $M_\rho=775$\,MeV are $+0.76\MeV$ and $-1.3\MeV$, respectively. Another small difference that contributes to the IB difference originates from the mass difference $\delta M_\rho=M_{\rho^-}-M_{\rho^0}$ of $1.0(9)\MeV$~\cite{Davier:2009ag} and $0.814\MeV$~\cite{Jegerlehner:2011ti}.
This illustrates the systematic uncertainties when estimating the IB corrections related to phenomenological form factor parameterizations. To avoid a circularity problem, the $\rho$ parameters need to be determined from other reactions than $e^+e^-\to \pi^+\pi^-$ and $\tau^-\to \pi^-\pi^0 \nu_\tau$, but since, e.g., the Breit--Wigner parameters are reaction dependent, this induces a systematic uncertainty that is difficult to control, one aspect of which is the need to define a $\rho^0$ in the presence of electromagnetic interactions and thus a convention for $\rho^0$--$\gamma$ mixing.

The effects of the IB corrections applied to $\amuHVPLO$ using $\tau$ data in the dominant $\pi\pi$ channel are shown in \cref{tab:ib}~\cite{Davier:2009ag} for the energy range between the $2\pi$ mass threshold and $1.8\GeV$. The short-distance correction, $S_{\rm EW}=1.0235(3)$~\cite{Davier:2009ag}, is dominant. The uncertainty of $G_{\rm EM}$ corresponds to the difference of the two $G_{\rm EM}$ corrections shown in \cref{fig:ib}. The quoted 10\% uncertainty on the FSR and $\pi\pi\gamma$ electromagnetic corrections is an estimate of the structure-dependent effects (pion form factor) in virtual corrections and of intermediate resonance contributions to real photon emission~\cite{Davier:2009ag}. The systematic uncertainty assigned to the $\rho$--$\omega$ interference contribution accounts for the difference in $\amuHVPLO$ between two phenomenological fits, where the mass and width of the $\omega$ resonance are either left free to vary or fixed to their world-average values. Some of the IB corrections depend on the form factor parameterization used, and the values quoted in \cref{tab:ib}  correspond to those of Gounaris--Sakurai (GS) parameterizations~\cite{Gounaris:1968mw}, but the total uncertainty includes the full difference between the GS and the K\"uhn--Santamaria (KS) parameterizations~\cite{Davier:2009ag}. 

\begin{figure}[t]
\centering
\includegraphics[width=0.6\columnwidth]{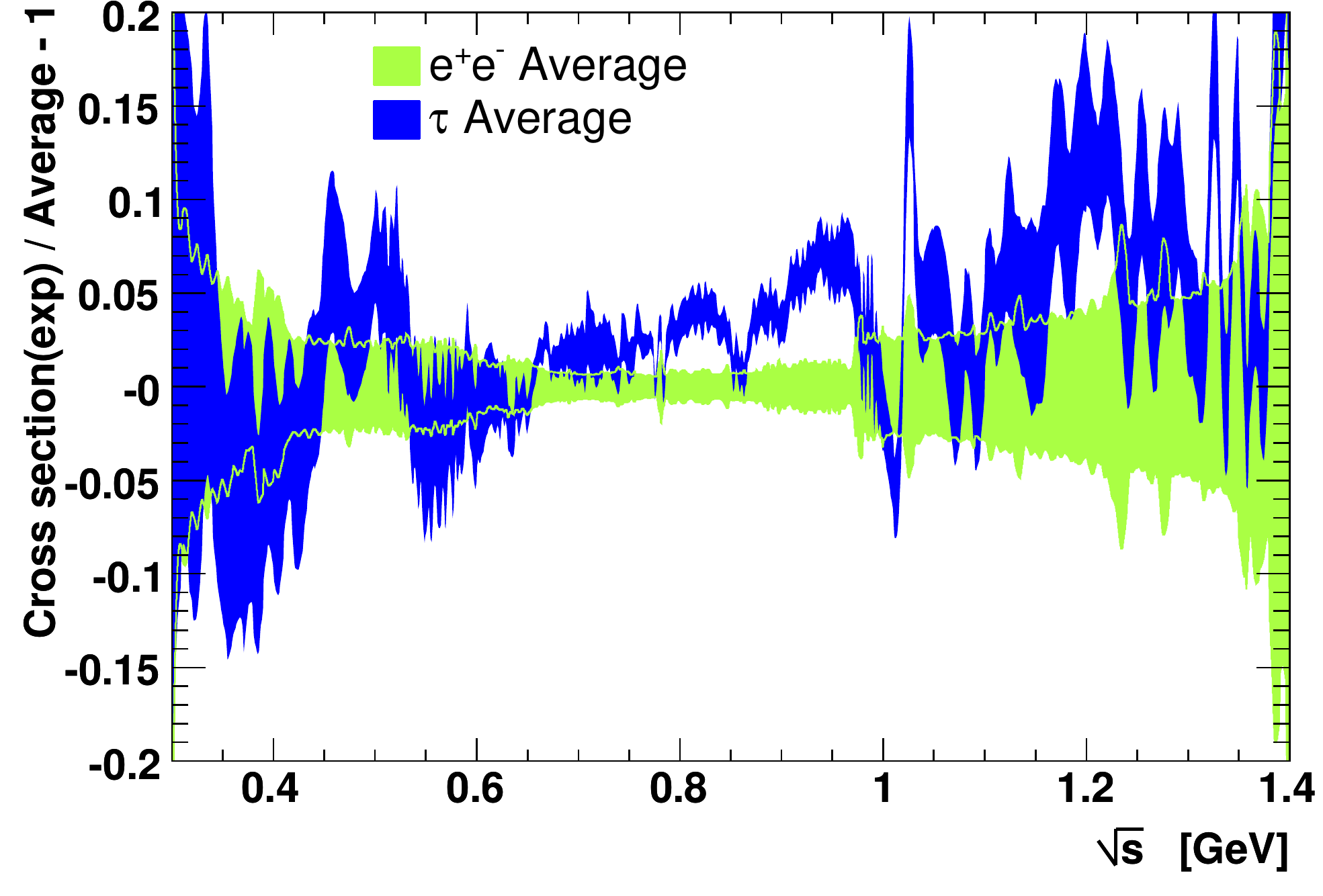}
\caption{Relative comparison between the combined $\tau$ (after the IB corrections) and $e^+e^-\to \pi^+\pi^-$ spectral function contributions. Reprinted from Ref.~\cite{Davier:2009zi}.}
\label{fig:chkib1}
\end{figure}

Another way to compare $e^+e^-$ and $\tau$ spectral functions is to predict
the branching ratio of $\tau$ decays into the $\pi\pi^0(\gamma)\nu_\tau$ final state, ${\cal B}_{\pi\pi^0}$, using $e^+e^-$ data.
In \cref{tab:ib}, the effects of the IB correction to the prediction are also shown. Using CVC, the branching fraction of $\tau$ decaying into a $G$-parity even hadronic final state $X^-$ is given by
\begin{equation}
{\cal B}_X^{\rm CVC}=\frac{3}{2}\frac{{\cal B}_e|V_{ud}|^2}{\pi\alpha^2m^2_\tau}\int^{m^2_\tau}_{s_{\rm min}}ds s \sigma^I_{X^0}(s)
\times \left(1-\frac{s}{m^2_\tau}\right)^2\left(1+\frac{2s}{m^2_\tau}\right)\frac{S_{\rm EW}}{R_{\rm IB}(s)}\,,
\end{equation}
where $s_{\rm min}$ is the threshold of the invariant mass-squared of the final state $X^0$ in $e^+e^-$ annihilation. CVC comparisons of $\tau$ branching fractions are of special interest because they are essentially insensitive to the shape of the $\tau$ spectral function, hence avoiding biases in the unfolding of the raw mass distributions from acceptance and resolution effects.

\begin{figure}[t]
\centering
\includegraphics[width=0.6\columnwidth]{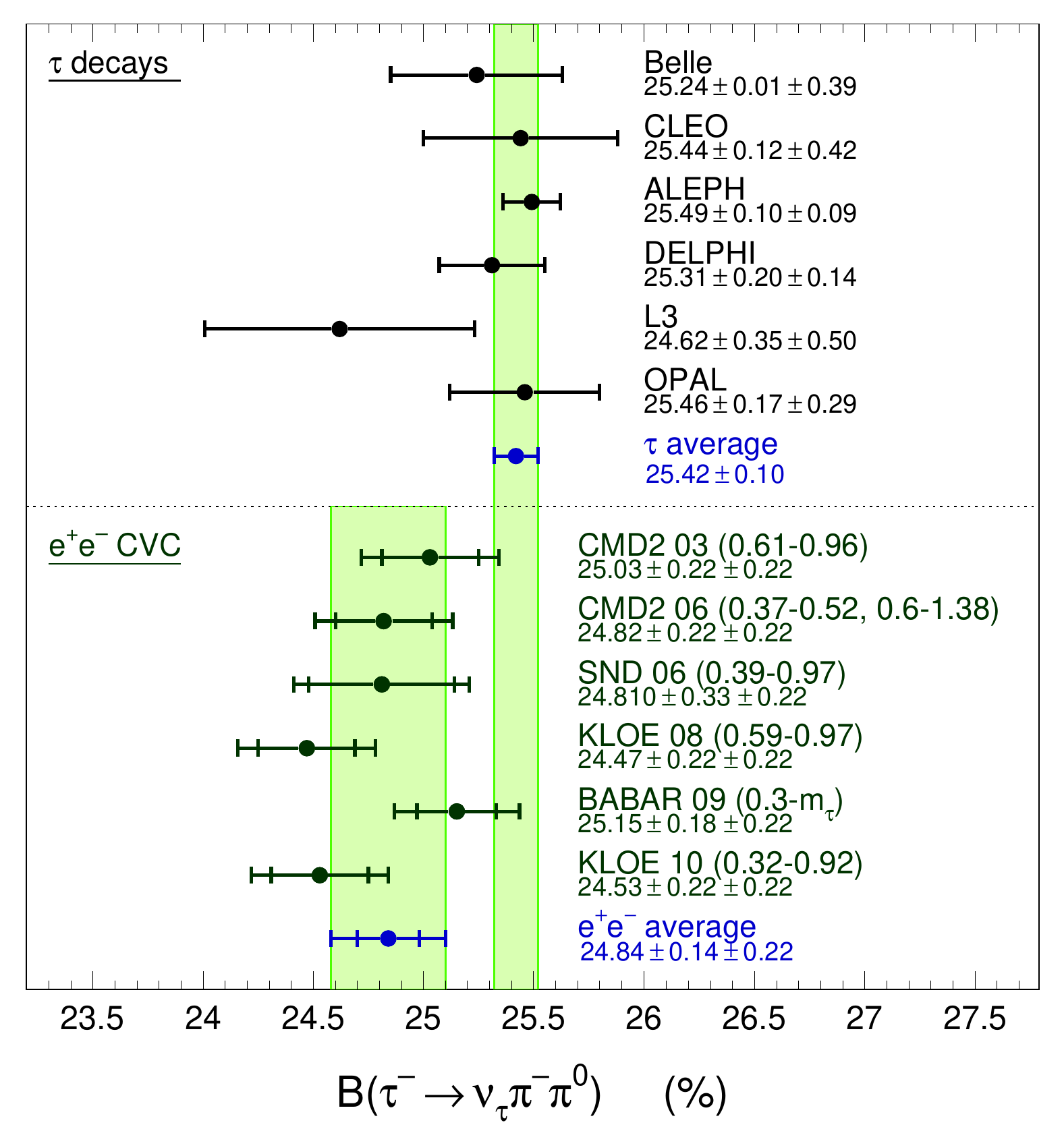}
\caption{The measured branching fractions for $\tau^-\to \pi^-\pi^0\nu_\tau$ compared to the predictions from the $e^+e^-\to \pi^+\pi^-$ spectral functions, applying the IB corrections. The long and short vertical error bands correspond to the $\tau$ and $e^+e^-$ averages, respectively. Reprinted from Ref.~\cite{Davier:2009ag}.}
\label{fig:chkib2}
\end{figure}

Despite the improved IB corrections, there is still a sizable difference between the $e^+e^-$ based prediction of $692.3(4.2)\times 10^{-10}$ and the $\tau$ based one of $703.0(4.4)\times 10^{-10}$~\cite{Davier:2013sfa}. The difference amounts to $10.7(4.9)\times 10^{-10}$, corresponding to a deviation of $2.2\sigma$. The shape of the combined $\tau$ spectral function after the IB corrections in the two-pion channel is also different from the one from $e^+e^-$ data (\cref{fig:chkib1}). The discrepancy is further reflected in the $\tau$ branching fractions (\cref{fig:chkib2}).

\begin{figure}[t]
\centering
\includegraphics[width=0.6\columnwidth]{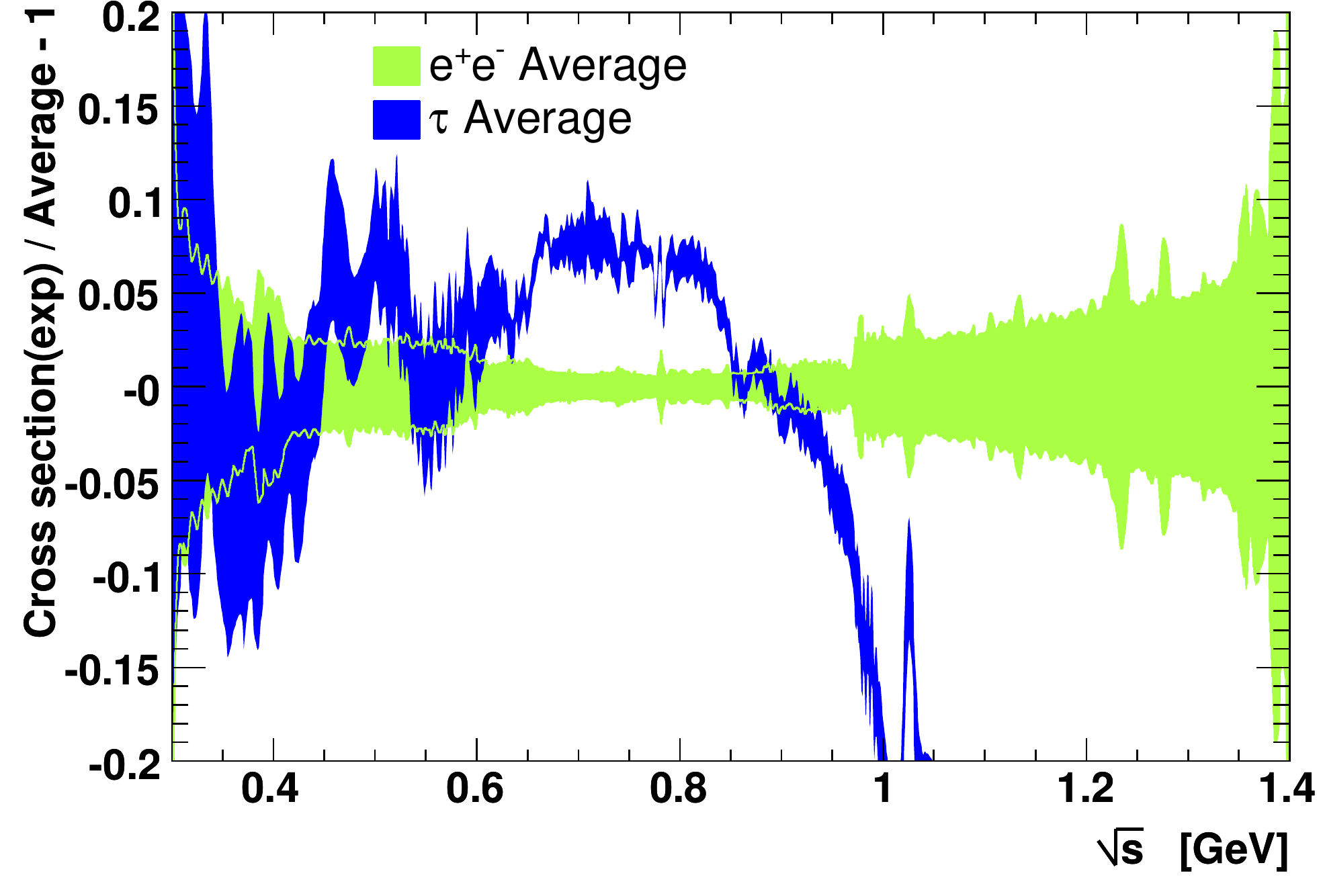}
\caption{Same as \cref{fig:chkib1}, except that the $\rho$--$\gamma$ mixing correction proposed in Ref.~\cite{Jegerlehner:2011ti} has been applied to the $\tau$ data.}
\label{fig:chkrgmixing} 
\end{figure}

A model-dependent $\rho$--$\gamma$ mixing, occurring only in the $e^+e^-$ data, was proposed in Ref.~\cite{Jegerlehner:2011ti} to explain the $e^+e^-$--$\tau$ discrepancy. The proposed correction corresponds to the difference between the open blue points and the solid black points in \cref{fig:ib} (bottom right), showing an increasing effect above the $\rho$ peak that appears uncomfortably large. Unlike $\gamma$--$Z$ mixing on the $Z$ resonance, well established theoretically and experimentally, the description of photon mixing with a strongly interacting $\rho$ may be affected by significant difficult-to-assess uncertainties.   
The correction~\cite{Jegerlehner:2011ti}, shown in \cref{fig:chkrgmixing}, seems to overestimate the observed difference. 

Concluding this part, it appears that, at the required precision to match the $e^+e^-$ data, the present understanding of the IB corrections to $\tau$ data is unfortunately not yet at a level allowing their use for the HVP dispersion integrals.
It remains a possibility, however, that the alternate lattice approach, discussed in \cref{subsec:tau}, may provide a solution to this problem.

\subsubsection{Radiative corrections and Monte Carlo generators}
\label{MC-radcor}

For the scan experiments at Novosibirsk, an event generator
MCGPJ~\cite{Arbuzov:2005pt} is used. Its theoretical precision was estimated
to be better than 0.2\%. It simulates $e^+e^-$, $\mu^+\mu^-$, $\pi^+\pi^-$,
$K^+K^-$, and $K_LK_S$ final states.
The code was cross-checked against BHWIDE~\cite{Jadach:1995nk} for the $e^+e^-$
final state
  and against KKMC~\cite{Jadach:1999vf} for the $\mu^+\mu^-$
  final state. Agreement at a level of 0.2\% was found in both cases.

  For the luminosity measurements, MCGPJ~\cite{Arbuzov:2005pt}
  or BABAYAGA@NLO~\cite{Balossini:2006wc} event generators are used.
  The accuracy of the BABAYAGA@NLO is 0.1\%, which was cross-checked
  by comparisons with the  BHWIDE event generator.

In most of the experiments that used the radiative-return method~\cite{Aubert:2009ad,Anastasi:2017eio,Ablikim:2015orh,Ambrosino:2008aa,Ambrosino:2010bv,Babusci:2012rp},
 the PHOKHARA event generator was used in the experimental analyses.
The generator evolved from its first versions~\cite{Rodrigo:2001kf,Czyz:2002np}, 
where only ISR corrections were included
at NLO.
 The missing ISR NNLO corrections were also estimated there.
 They can be at most 0.3\%, which was confirmed later in  Ref.~\cite{Jadach:2005gx}.
 They will be included soon at the leading logarithmic approximation~\cite{3ph}.
The other NLO corrections were added
gradually for final states with two hadrons and $\mu^+\mu^-$ in the final state:
for $\pi^+\pi^-$ in Refs.~\cite{Czyz:2003ue,Czyz:2004rj,Czyz:2004nq},
for $K^+K^-$ in Ref.~\cite{Czyz:2010hj}, for $p\bar p$ in Ref.~\cite{Czyz:2014sha}, and 
for $\mu^+\mu^-$ in Ref.~\cite{Czyz:2004rj}.

In all these papers, NLO radiative corrections involving the exchange of two virtual
photons between the initial electron--positron line and the final hadron (muon) line
were neglected. The complete NLO radiative corrections for the reaction
$e^+e^-\to \mu^+\mu^-\gamma$ were calculated and included into the generator
PHOKHARA in Ref.~\cite{Campanario:2013uea}. It was shown there that for that process,
used for luminosity monitoring, the missing radiative corrections for the KLOE
experimental event selections were at most 0.1\%, while for the BABAR event selection,
they were not bigger than 0.2\%.
 The generator with the complete NLO radiative corrections
was available during the BESIII analysis. The same type of corrections
for the reaction $e^+e^-\to \pi^+\pi^-\gamma$ were calculated recently~\cite{Campanario:2019mjh}. Again, the previously missing part of the NLO  radiative corrections
is small for all experimental event selections. Around the $\rho$ peak
the corrections do not exceed 0.05\% for all the experiments and are equally small
for other invariant masses for BABAR and BESIII  event selections.
For KLOE it is useful to discuss separately the
most relevant region of the pion pair invariant masses ($0.6\hyph0.9\GeV$)
 for the evaluation of the muon anomalous magnetic moment.
 For the event selections used by  KLOE in 2008
  and 2012~\cite{Ambrosino:2008aa,Babusci:2012rp}, 
 the missing corrections
reach as much as 0.18\% at low pion-pair invariant masses. At the $\rho$
peak, they are well below 0.05\%.  For the KLOE 2010 event selection
 (detected photon)~\cite{Ambrosino:2010bv}, the
radiative corrections can be larger, up to 0.5\% in the relevant region. Yet around
the $\rho$ peak, they are smaller and amount up to only 0.2\%.
Above $0.9\GeV$, which is outside the relevant region, the corrections can 
reach up to 2.4\%. 
 Summarizing, with
the results of Ref.~\cite{Campanario:2019mjh}, it was finally excluded that the differences between
the experimental extractions of the pion form factor can originate from
the missing radiative corrections.

The radiative corrections involving photon--hadron interactions are model-dependent. They were studied carefully for the $\pi^+\pi^-$ final state
(see Ref.~\cite{Actis:2010gg} and references therein), so the model used
in PHOKHARA is reliable with the accuracy of the current experiments.
With a further experimental error reduction it would be good to repeat
the tests with a better accuracy. For other final states, the
systematic experimental tests were not performed, and moreover the
FSR corrections are implemented in PHOKHARA only for a small number
of final states and are modeled in experiments using PHOTOS~\cite{Nanava:2006vv}.
With the increasing experimental accuracy, such tests and a better modeling are
 necessary.

\subsection{Evaluations of HVP}
\label{HVPdisp_evaluations}

\subsubsection{The DHMZ approach}
\label{sec:DHMZ}

The software package HVPTools,\footnote
{HVPTools is written in object-oriented C++ and relies on ROOT functionality~\cite{Brun:1997pa}.
   The cross-section data base is provided in XML format. The systematic errors are introduced component 
   by component as an algebraic function of mass or as a numerical value for each 
   data point (or bin). Systematic errors belonging to the same identifier (name) are
   taken to be fully correlated throughout all measurements affected. 
} developed for the DHMZ approach, features an accurate data interpolation, averaging, and integration method, 
systematic tests, and a statistical analysis based on the generation of large samples of pseudo-experiments.
It has been deployed in Ref.~\cite{Davier:2009ag} for hadronic $\tau$ decay spectra and in Ref.~\cite{Davier:2009zi} for the most important channel, $e^+e^-\rightarrow\pi^+\pi^-$, as well as for $e^+e^-\rightarrow\pi^+\pi^-2\pi^0$.
In Refs.~\cite{Davier:2010nc,Davier:2017zfy,Davier:2019can} it has then been used for performing combinations in all the $\epem\to{\rm hadrons}$ channels under consideration.
HVPTools allows for a comprehensive treatment of the correlations between the measurements of one experiment, as well as inter-experiment and inter-channel correlations.

\paragraph{Cross-section data}

In the dispersion integral for the lowest-order hadronic contribution (see \cref{eq:amuhvsdisplo})
the contribution from the light $u,d,s$ quark states 
is evaluated using exclusive experimental cross-section data up to an energy of 1.8\GeV, 
where resonances dominate, and pQCD to predict the quark continuum beyond that
energy (except for the charmonium region, where the evaluation of the dispersion integral is based on inclusive experimental cross-section data).

A large number of $\epem\to{\rm hadrons}$ cross-section measurements are available (see \cref{ee-data}).
\Cref{eq:amuhvsdisplo} and the treatment of higher-order hadronic contributions
require ISR as well as leptonic and hadronic VP
contributions to be subtracted from the measured cross-section data, while FSR should be included.

Older measurements are affected by an incomplete or undocumented application of radiative corrections.
Because of lack of documentation, the latter contribution of 
approximately $0.9\%$ in the \pp channel has been added to the data, accompanied by a 
100\% systematic error~\cite{Davier:2002dy}. Initial-state radiation and leptonic VP effects are 
corrected by all experiments, however hadronic VP effects are not. They are strongly 
energy dependent, and in average amount to approximately $0.6\%$. In the DHMZ approach this correction is applied,
accompanied by a $50\%$ systematic error~\cite{Davier:2002dy}. These FSR and hadronic VP systematic 
uncertainties are treated as fully correlated between all measurements of one experiment, and also
among different experiments and different channels.

\paragraph{Combining cross-section data}
\label{sec:hvptools}

The requirements for averaging and integrating cross-section data are: 
(i) properly propagate all the uncertainties in the data to the final integral 
error, (ii) minimize biases, \ie, reproduce the true integral as closely as 
possible in average and measure the remaining systematic error, and 
(iii) optimize the integral error after averaging while respecting the two 
previous requirements. The first item practically requires the use of pseudo-data or Monte Carlo
(MC) simulation, which needs to be a faithful representation of the measurement ensemble
and to contain the full data treatment chain (interpolation, averaging, integration). 
The second item requires a flexible data interpolation method and a realistic truth model used to test the 
accuracy of the integral computation with pseudo-data experiments.
Finally, the third item requires optimal data averaging, taking into account all known correlations
to minimize the spread in the integral measured from the pseudo-data sample.
Furthermore, this optimization has to be done without overestimating the precision
with which the uncertainties of the measurements and their correlations are known.

The combination and integration of the $\epem\to{\rm hadrons}$ cross-section data are performed using 
the software package HVPTools.
It transforms the bare cross-section data and associated 
statistical and systematic covariance matrices into fine-grained energy bins, taking 
into account the correlations within each experiment 
as well as between the experiments (such as uncertainties in radiative corrections) to the best available knowledge. 
The covariance matrices are obtained by assuming common systematic error sources to 
be fully correlated. To these matrices are added statistical covariances, present for 
example in binned measurements as provided by KLOE, BABAR, BES, or the $\tau$ data, which 
are subject to bin-to-bin migration that has been unfolded by the experiments, thus 
introducing correlations.

The interpolation between adjacent measurements of a given experiment uses second-order polynomials. This is an improvement with respect to the previously 
applied trapezoidal rule, corresponding to a linear interpolation, which leads
to systematic biases in the integral (see below, and also the discussion in 
Sec.~8.2 and Fig.~12 of Ref.~\cite{Davier:2002dy}). In the case of binned data, the 
interpolation function within a bin is renormalized to keep the integral in that
bin invariant after the interpolation. This may lead to small discontinuities
in the interpolation function across bin boundaries. The final interpolation 
function per experiment within its applicable energy domain is discretized into 
small bins (of 1\MeV, or narrower for the $\omega$ and $\phi$ resonances) for the purpose of averaging and numerical integration. 

The averaging of the interpolated measurements from different experiments contributing 
to a given energy bin is the most delicate step in the analysis chain.
Correlations between measurements and experiments must be taken into account.
Moreover, the experiments have different measurement densities or bin widths within a given energy 
interval, and one must avoid having extrapolated information (through the polynomial interpolation)
substitute for missing information in a region of lower measurement density.
To derive proper {\em averaging weights} given to each experiment, wider 
{\em averaging regions}\footnote
{For example, when averaging two binned measurements with unequal bin widths, a
   useful averaging region would be defined by the experiment with the larger bin width, 
   and the bins of the other experiments would be statistically merged before computing
   the averaging weights.
}  
are defined to ensure that all locally available experiments contribute to the averaging
region, and that in case of binned measurements at least one 
full bin is contained in it.

\begin{figure}[tb]
\includegraphics[width=8cm]{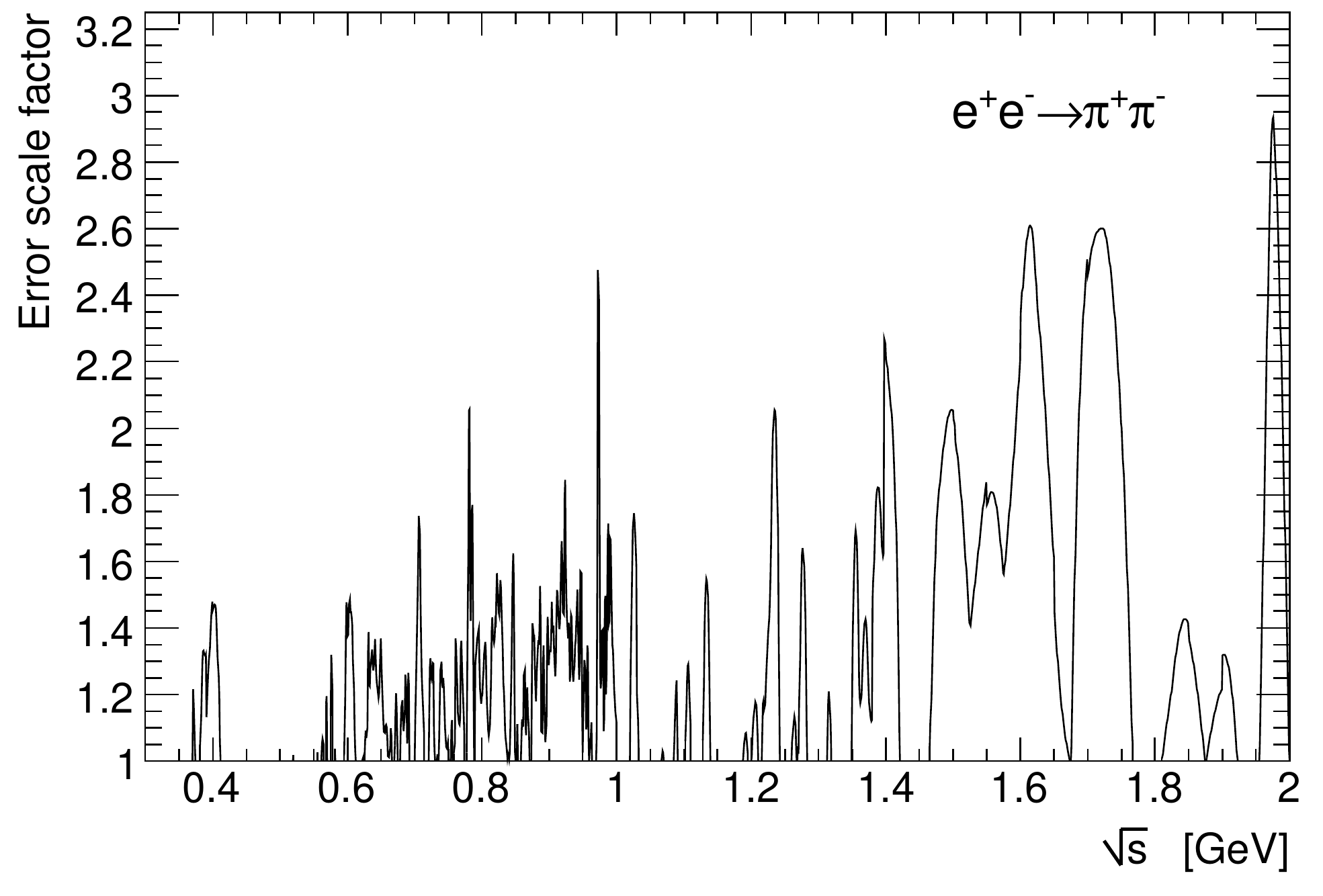}\hspace{0.1cm}
\includegraphics[width=8cm]{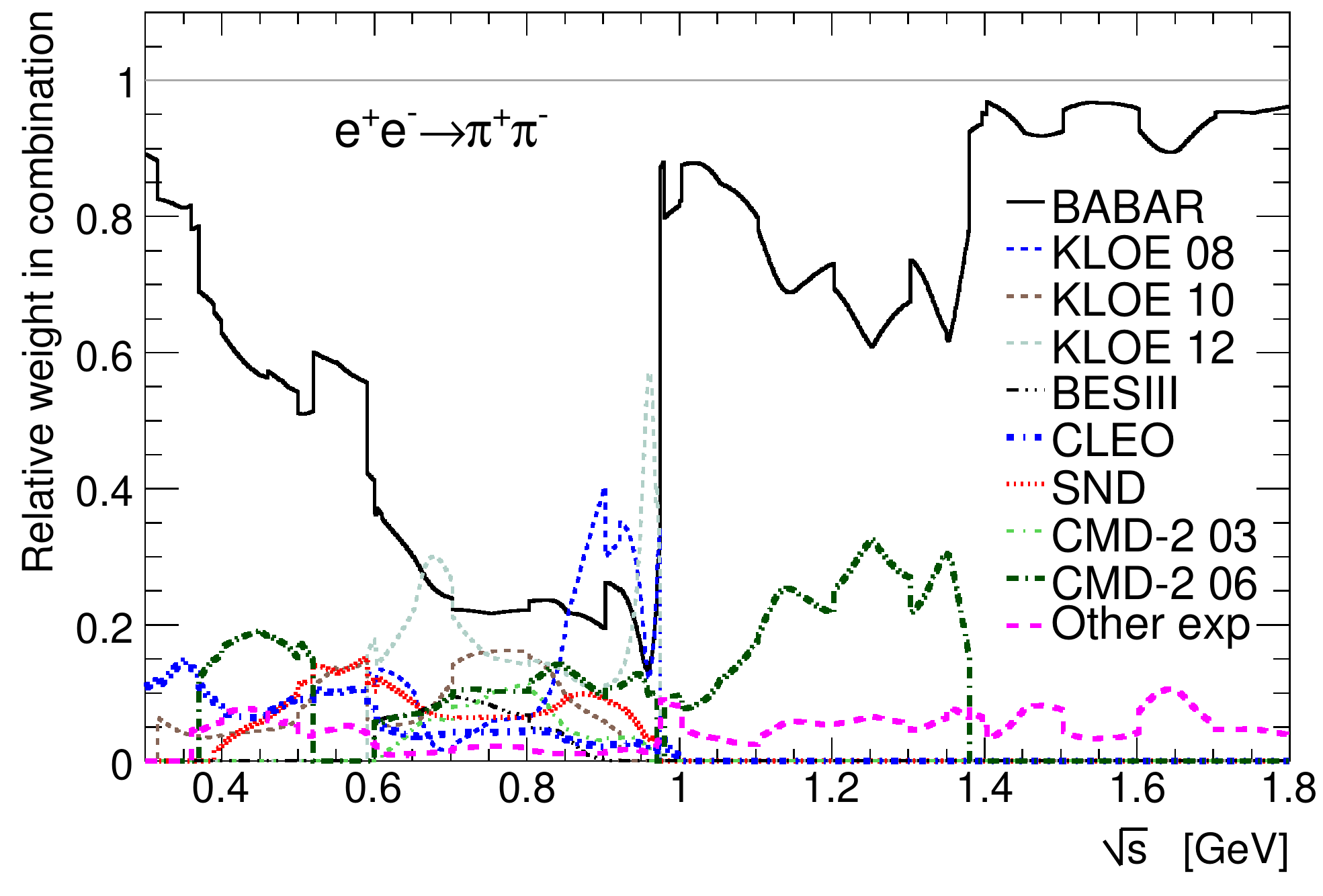}
\vspace{-0.3cm}
\caption{Left: Rescaling factor accounting for inconsistencies among experiments versus $\sqrt{s}$. Right: Relative averaging weights per experiment versus $\sqrt{s}$ (see text). Reprinted from Ref.~\cite{Davier:2019can}.}
\label{figDHMZ:chi2weights}
\end{figure}

The bin-wise average between experiments is computed as follows:
a) Pseudo-data generation fluctuates the data cross sections taking into account all known correlations and
         the polynomial interpolation is redone for each generated pseudo-data;
         for the purpose of determining the averaging weights, the averaging regions are filled and interpolated for each experiment.
b) For each generated pseudo-data, small bins
         are filled for each experiment, 
         in the energy intervals covered by it, using the polynomial interpolation.
c) In each small bin a correlation matrix between the experiments is computed
         and a $\chi^2$ minimization yields the averaging weights.
         This approach avoids relying too much on long-range correlations of uncertainties
         in the determination of the average weights (see \cref{Sec:UncertaintiesOnUncertainties}).
d) The average and its uncertainty are computed in each small bin.
e) If the $\chi^2$ value of a bin-wise average
exceeds the number of degrees of freedom (dof), the uncertainty in this averaged bin
is rescaled by $\sqrt{\chi^2/\text{dof}}$ to account for inconsistencies (cf. the left panel of  \cref{figDHMZ:chi2weights}).\footnote{Such inconsistencies frequently occur because most experiments 
are dominated by systematic uncertainties, which are difficult to estimate. 
In particular, the sharp peak at 0.78\,GeV is due to local discrepancies across the $\rho$--$\omega$ interference and
the bump between 0.85 and 0.95\,GeV is due to a discrepancy between KLOE and BABAR.}

The consistent propagation of all errors into the evaluation of $\amuHVPLO$ is ensured 
by generating large samples of pseudo-experiments, representing the 
full list of available measurements and taking into account all known correlations. 
An eigenvector decomposition technique is used for generating pseudo-experiments
for measurements provided with covariance matrices.
In particular, this technique is used to propagate the statistical and systematic correlations
between the three KLOE measurements~\cite{bogdan-kek-2018-DHMZ,bogdan-Mainz-2018-DHMZ}, for which global covariance matrices have been made available.
For each generated set of pseudo-measurements, the identical interpolation and 
averaging treatment leading to the computation of \cref{eq:amuhvsdisplo} 
as for real data is performed, hence resulting in a probability density distribution
for $\amuHVPLO$, the mean and RMS of which define the $1\sigma$ allowed interval
(and which---by construction---has a proper pull behavior). 
The same pseudo-experiments are also used to derive (statistical and systematic)
covariance matrices of combined cross sections, used then for the integral evaluation.
Uncertainties are also propagated through shifts of one standard deviation of each uncertainty,
allowing one to account for correlations between different channels for integrals and combined spectra.
The consistency between the results obtained with these three different approaches has been checked.
The procedure yielding the weights of the experiments can be optimized with respect 
to the resulting error on $\amuHVPLO$. 

The fidelity of the full analysis chain (polynomial interpolation, 
averaging, integration) has been tested by using as truth representation a Gounaris--Sakurai~\cite{Gounaris:1968mw}
vector-meson resonance model faithfully describing the \pp data. The central values
for each of the available measurements are shifted to agree with the Breit--Wigner model,
leaving their statistical and systematic errors unchanged. The so-created set of 
measurements is then analyzed akin to the original data sets. The difference between 
true and estimated $\amuHVPLO$ values is a measure for the systematic uncertainty due 
to the data treatment. The bias is found to be negligible, below $0.1\times 10^{-10}$, when using second order polynomials. Interpolation using the trapezoidal rule, 
in contrast, leads to a much larger, $\sim 1\times 10^{-10}$, bias.

\begin{figure}[tb]
\begin{center}
\includegraphics[width=10cm]{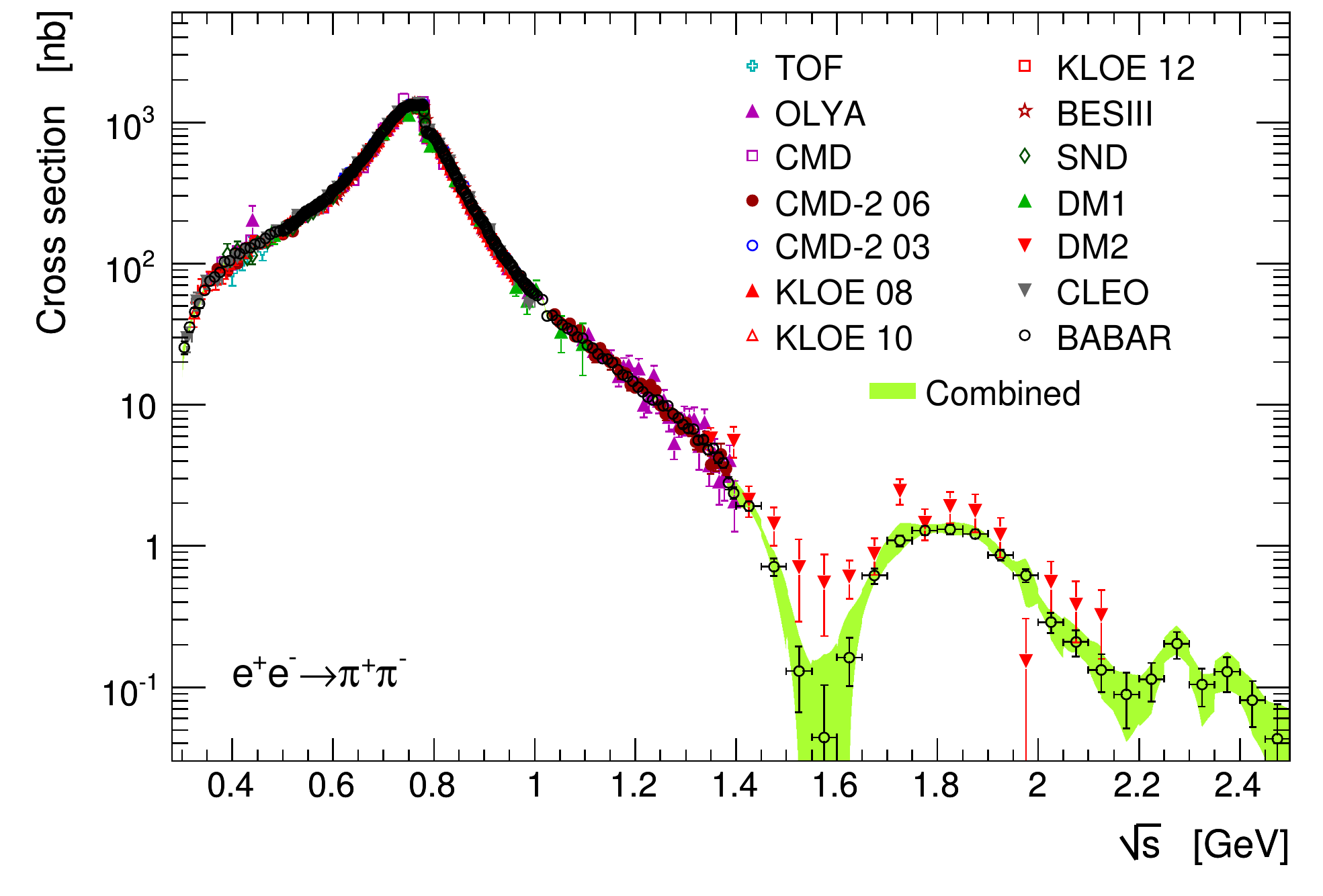}
\end{center}
\vspace{-1.0cm}
\caption{Cross section for $\epem\to\pp$ annihilation measured by the different experiments
            for the entire energy range.
            The error bars contain both statistical and systematic errors, added in 
            quadrature. The shaded (green) band represents the average of all the 
            measurements obtained by HVPTools, which is used for the numerical integration 
            following the procedure discussed in \cref{sec:hvptools}. Reprinted from Ref.~\cite{Davier:2019can}.}
\label{figDHMZ:xsec}
\end{figure}

The individual $\epem\to\pp$ cross-section measurements (dots) and their average (shaded/green 
band) are plotted in \cref{figDHMZ:xsec}. The error bars contain statistical and 
systematic errors.

The right panel of \Cref{figDHMZ:chi2weights} shows the weights vs.\ $\sqrt{s}$ the different 
experiments carry in the average. BABAR and KLOE dominate over the entire energy range. 
The experiments labeled ``other exp'' in the figure correspond to older data with incomplete radiative corrections.
Their weights are small throughout the entire energy domain.

The evaluation of the complete $\amuHVPLOpp$ integral for the \pp contribution from threshold to $1.8\GeV$, using a fit up to $0.6\GeV$ (see \cref{sec:dispersive}) and the HVPTools data combination above, gives $507.0(1.9)\times 10^{-10}$.
The correlation among the uncertainties of these two contributions of 62\% evaluated using pseudo-experiments is taken into account.
Removing BABAR or KLOE from the data set gives $505.1(2.1)\times 10^{-10}$ and $510.6(2.2)\times 10^{-10}$, respectively, with an absolute difference of $5.5\times 10^{-10}$ that is significantly larger than the individual uncertainties.
In light of this discrepancy, which is not fully captured by the local uncertainty rescaling procedure,
half of the full difference between the complete integrals without BABAR and KLOE, respectively, is included as additional systematic uncertainty in the DHMZ study.
The central value of the $\amuHVPLOpp$ contribution is placed half-way between the two results.
To avoid double counting, the local uncertainty rescaling between BABAR and KLOE is not applied, but that between these and the other \pp data sets is kept.
This procedure results in a total \pp contribution of  $\amuHVPLOpp=507.9(0.8)(3.2)\times 10^{-10}$, where the first uncertainty is statistical and the second systematic (dominated by the new uncertainty of $2.8\times 10^{-10}$)~\cite{Davier:2019can}.

Common sources of systematic uncertainties also occur between measurements of different final-state channels and
must be taken into account when summing up the exclusive contributions.
Such correlations mostly arise from luminosity uncertainties, if the data stem from the same
experimental facility, and from radiative corrections. 
In total, 15 categories of correlated systematic uncertainties are distinguished.
Among those, the most significant belong to radiative corrections, which are the same for
CMD2, CMD3, and SND, as well as to luminosity determinations by BABAR, CMD2, and SND (correlated per experiment
for different channels, but independent between different experiments).
Propagating each of these categories of systematic uncertainties separately in the combinations allows for their coherent
treatment in the sum of the contributions of the different channels.
This yields a larger, but more realistic, uncertainty on $\amuHVPLO$.

Adding all lowest-order hadronic contributions together gives
\be
\amuHVPLO = 694.0(4.0)\times 10^{-10},
\ee
which is dominated by experimental systematic uncertainties~\cite{Davier:2019can}.


\subsubsection{The KNT approach}
\label{sec:KNT}

KNT~\cite{Keshavarzi:2018mgv, Keshavarzi:2019abf} provide a predominantly
data-driven compilation for the hadronic $R$-ratio, which is then used
to predict the HVP contributions to precision observables such as the
anomalous magnetic moment of the electron ($a_e$), muon ($a_\mu$), and
$\tau$ lepton ($a_\tau$), to the ground-state hyperfine splitting of
muonium, and also the hadronic contributions to the running of the QED
coupling $\alpha(q^2)$. The obtained $R$-ratio has also been used to
determine the strong coupling $\alpha_s$ at low scales through
finite-energy sum rules~\cite{Boito:2018yvl}.

\paragraph{Data selection and application of radiative
  corrections}
\label{sec:knt_data}
The KNT analyses~\cite{Keshavarzi:2018mgv,Keshavarzi:2019abf} are
based on the earlier
works of Refs.~\cite{Hagiwara:2002ma,Hagiwara:2003da,Hagiwara:2006jt,Teubner:2010ah,Hagiwara:2011af}. With
very few exceptions
where data sets are known to be unreliable, superseded by newer
analyses, or not adding any useful information, all available data sets
for $e^+ e^- \to \text{hadrons}$ in the relevant energy range are
used. However, KNT do not use information from hadronic $\tau$
decays. Predictions from pQCD are only used from
$\sqrt{s}>11.199\GeV$, above all flavor thresholds (apart from the
top quark, which can safely be treated perturbatively). 

As discussed in \cref{sec:hvp_introduction}, in general, the
data must be undressed with respect to VP corrections, but should contain
FSR effects. For many recent data sets, the
cross sections are provided in the required form already, and no
further corrections are applied. However, for cases where ``undressing''
is required, the KNT VP routine is used. This is, in a self-consistent
iterative procedure, obtained from the very same hadronic cross
section data.\footnote{This routine, currently version {\tt
    vp\_knt\_v3.0}, can be obtained by contacting the authors
  directly.} For sets where
VP corrections have been applied only partially (some older data
sets), or with routines known to be unreliable, these corrections are 
undone and the KNT routine is applied. For FSR, one requires that all
photon radiation, real and virtual, be included in the data. Clearly,
unless a subtraction based on theoretical predictions is made, any
measurement includes the virtual and soft real radiation. However,
hard real photons above a resolution or analysis-based cut may be lost
and lead to a smaller measured cross section. In the
$\pi^+\pi^-$ channel, for the recent, precise data sets obtained via
the method of radiative return, photons in the final state are an
integral part of the experimental analyses and their error estimates,
and, hence, additional FSR corrections are not applied. For data obtained
through the energy scan method, if analyses indicate that FSR had been
subtracted or not fully included, inclusive FSR
corrections based on scalar QED are applied, which have been shown to be of
sufficient accuracy at the low energies where data are used. The
situation is different in the $K^+K^-$ and $K_S^0K_L^0$ channels. 
In the numerically
most important region close to the $\phi$ peak, there is very little phase
space for hard photon radiation above a resolution or analysis-imposed 
selection cut, and estimates of the corresponding real photon FSR
correction have shown that their size is
negligible~\cite{Keshavarzi:2018mgv}. FSR corrections are, therefore, not applied in the $K^+K^-$ or $K_S^0K_L^0$ channels. 
For more complicated higher-multiplicity final states, only limited
theoretical predictions and tools (including the Monte Carlo
generators used in the experimental analyses) to calculate FSR with
precision are available.

To account for any under- or over-correction resulting from limited
information on radiative corrections and the rather crude
approximations made when applying them, additional VP and
FSR errors are assigned in the different hadronic channels for the
derived quantities, like $\amuHVPLO$ and
$\Delta\alpha(q^2)$. For details of the applied procedures and
individual numbers
see Refs.~\cite{Hagiwara:2011af,Keshavarzi:2018mgv,Keshavarzi:2019abf}.

\paragraph{Data combination}
\label{sec:knt_combination}
To determine the contributions of a specific hadronic final state to
the hadronic $R$-ratio, data combinations are performed in all channels
for which data are available. For this, the data in each channel are
subjected to a {\em clustering} procedure, which, based on the
available local data density, determines the
optimal binning of the data into a set of  clusters. This is done by
an algorithm that performs a scan over a large range of possible
binnings and the many resulting fits for the data combination. The
optimal clustering is then determined by a combination of criteria;
the best overall fit quality, the goodness of fit locally, i.e., in
the individual clusters and the achieved uncertainty of the
contributions to $\amuHVPLO$, or other derived quantities in
the given channel. While a too narrow clustering would defeat the aim
of a local data combination and hence not lead to the best possible
determination of $R$ and its error, a too wide clustering would
effectively lead to a significant re-binning of individual data sets
and ultimately to a too coarse and wrong representation of the data.

The fit methodology that achieves the actual combination has significantly
evolved from the original approach used up to Ref.~\cite{Hagiwara:2011af}, where
systematic errors due to simple, overall multiplicative uncertainties were
effectively integrated out by fitting related renormalization
(nuisance) factors through a nonlinear $\chi^2$ minimization. To
consistently use the information from full covariance
matrices for correlated (statistical and systematic) uncertainties
given in recent experimental analyses for the leading hadronic
channels, this approach proved not to be sufficient. Hence,
since Ref.~\cite{Keshavarzi:2018mgv}, the KNT compilation uses instead the method of an
iterated $\chi^2$ fit, which fully takes into account all available
covariance matrices and, hence, accounts for all correlations arising
from correlated uncertainties. This method has been shown to be free
from the bias that would arise when naively using correlations in a
$\chi^2$ minimization~\cite{Ball:2009qv,DAgostini:1993arp}, but
accounts for nontrivial, nonlocal correlations that 
not only enter in the error estimate, but also influence the
determination of the mean values of $R$ over the full energy
range. This is important, as measurements from one experiment, with
locally small uncertainties, can help to constrain the normalization
of data points from another experiment at different energies, through
and within its systematic uncertainties. Therefore, the completely
data-driven combination in local clusters, through the correlated fit,
becomes a global fit in each channel.

\begin{figure}[t]
\begin{center}
\includegraphics[width=9cm]{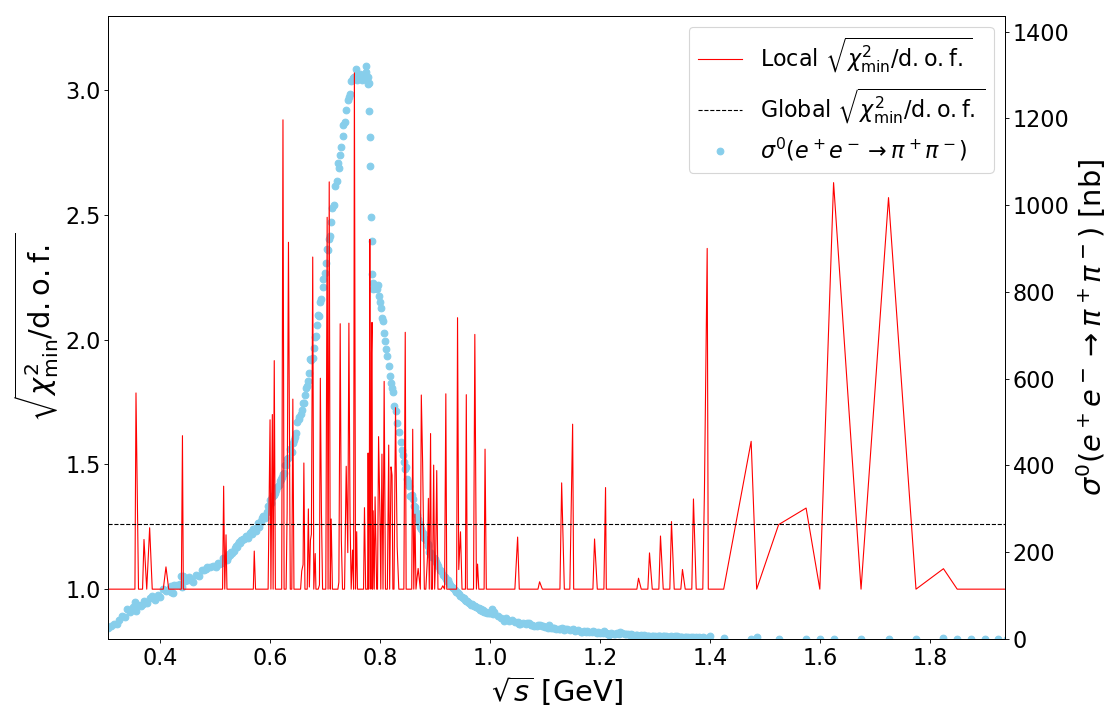}
\end{center}
\vspace{-0.7cm}
\caption{Distribution of  the local $\chi^2_{\rm min}/\text{dof}$ 
used for the error inflation in the two-pion channel.}
\label{figKNT2pichisq}
\end{figure}

The fit is performed by minimizing, iteratively, the $\chi^2$-function 
\begin{equation}
\label{IChi^2}
\chi^2_{I} =
\sum^{N_{\mathrm{tot}}}_{i=1}\sum^{N_{\mathrm{tot}}}_{j=1}\big(R_{i}^{(m)}
- \mathcal{R}_{m}^{i,I}\big) {\bf
  C}_{I}^{-1}\big(i^{(m)},j^{(n)}\big)\big(R_{j}^{(n)} -
\mathcal{R}_{n}^{j,I} \big) \,, 
\end{equation}
where the double sum is over a total of $N_{\mathrm{tot}}$ contributing
data points in the channel under consideration and the labels $m$ and
$n$ refer to the clusters. The quantities
$\mathcal{R}_{m}^{i, I}$ and $\mathcal{R}_{n}^{j,I}$ are the
linear-model interpolant cross section values of clusters $m$ and $n$
against which data points $R_i^{(m)}$ and $R_j^{(n)}$ are compared,
and $I$ denotes the iteration number of the fit. 
The matrix ${\bf C}_{I}^{-1}\big(i^{(m)},j^{(n)}\big)$ is the inverse of the
covariance matrix defined by 
\begin{equation}
\label{ICk}
{\bf C}_I\big(i^{(m)},j^{(n)}\big) = {\text
  C}^{\text{stat}}\big(i^{(m)},j^{(n)}\big) + \frac{{\text
    C}^{\text{sys}}\big(i^{(m)},j^{(n)}\big)}{R_{i}^{(m)}R_{j}^{(n)}}
\mathcal{R}_{m}^{i,I-1}\mathcal{R}_{n}^{j,I-1}\,,
\end{equation}
given as the sum of the statistical covariance matrix ${\rm C}^{\rm
  stat}\big(i^{(m)},j^{(n)}\big)$ and the systematic covariance matrix
${\rm C}^{\rm sys}\big(i^{(m)},j^{(n)}\big)$, redefined at
each step of the iteration. For the first step of the iteration, the
matrix is initialized simply using the experimental $R$ measurements
instead of fit results, hence the first step corresponds to a naive
correlated fit, which would, however, in general lead to a biased
result. In practice, only a few iterations are needed for the fit to
converge.

The fit returns, apart from the mean $R$ values for each cluster, a
full covariance matrix, which contains all correlated and uncorrelated
uncertainties. These are then used to determine derived quantities 
(e.g., $\amuHVPLO$) channel-by-channel and using 
trapezoidal integration.\footnote{It has been checked that this direct
  data integration does not lead to any numerically relevant bias, with the exception of
  the $\omega$ resonance in the $\pi^+\pi^-\pi^0$ channel. There, a quintic
  polynomial is applied, as the linear interpolation would
  lead to an overestimate, while lower polynomial fits would lead to
  unphysical fluctuations~\cite{Keshavarzi:2019abf}.} 
The fit also
returns local $\chi^2_{\rm min}/\text{dof}$ values for each cluster, which
are used for a local $\chi^2_{\rm min}$ error inflation according to
the standard procedure recommended by the PDG. For the two-pion
channel, the distribution of the local  $\chi^2_{\rm min}/\text{dof}$ is shown in
\cref{figKNT2pichisq}. Note that, for all channels, the
local error inflation exceeds the error inflation based on the global
$\chi^2_{\rm min}/\text{dof}$.

Overall, the full use of correlation information in the fit has benefited the KNT analysis, with this approach having further constrained the mean values. Consequently, for the $\amuHVPLO$ integral, this has yielded results that are not expected from a simple local data combination (which does not fully take into account the given correlations) and has achieved a considerable error reduction. It should be stressed that, if correlated uncertainties were neglected in the fit of the mean values and only propagated for the error analysis, this would neglect important information. Especially for derived quantities with a nonflat kernel like $\amuHVPLO$, this could lead to a result that is significantly different ($> 1\sigma$) from the result obtained from a correlated fit (see the discussion below for this effect in the KNT analysis of the $\pi^+\pi^-$ channel).

\begin{figure}[t]
\begin{center}
\includegraphics[width=9cm]{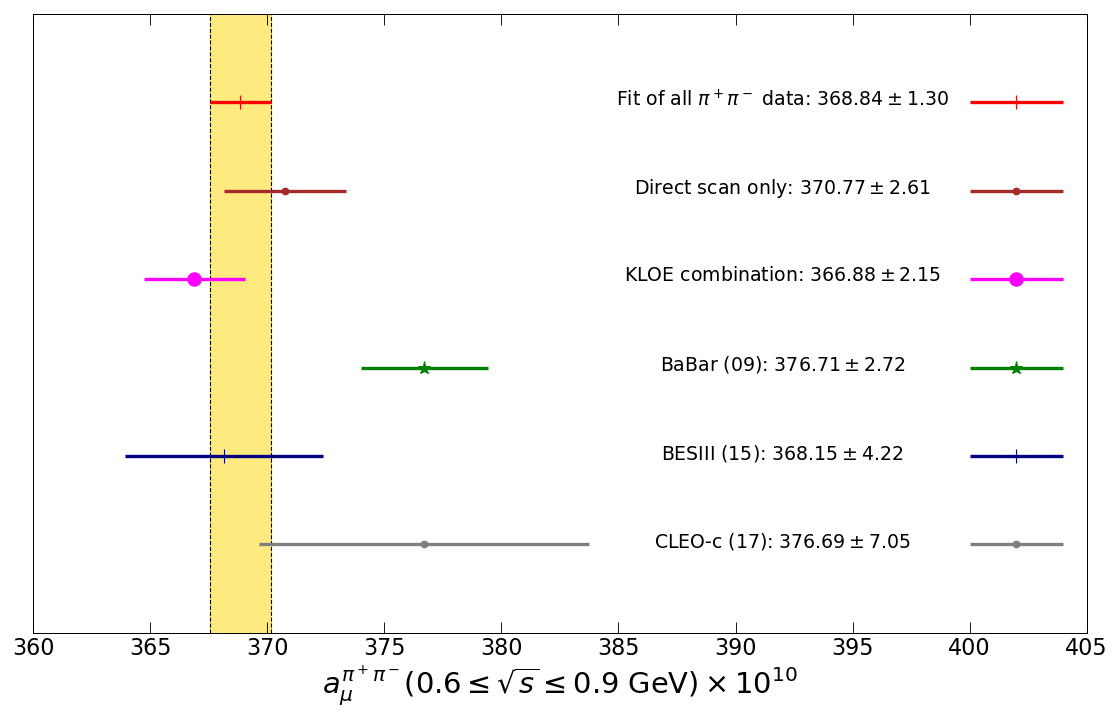}
\end{center}
\vspace{-0.7cm}
\caption{Comparison of $\amuHVPLOpp$ obtained when using
only subsets of the available data to the full KNT result  shown by the yellow band. To allow for the comparison, the energy
range is restricted to $0.6 \le \sqrt{s} \le 0.9\GeV$.
See Refs.~\cite{Keshavarzi:2018mgv,Keshavarzi:2019abf} for details. Reprinted from Ref.~\cite{Keshavarzi:2019abf}.} 
\label{figKNT2picomp}
\end{figure}

\paragraph{Results}
\label{sec:knt_results}

For the most important two-pion channel, \cref{figKNT2picomp} 
demonstrates these nontrivial effects of the correlated data
combination. The result obtained for $\amuHVPLOpp$ from the 
combination of all data differs from what one would
expect from a weighted average of predictions where only individual
data sets (or subsets of the data) are used. Instead, the high
accuracy of the KLOE data, with their strong correlations and good
agreement with the direct scan and the BESIII data, leads to a
rigidity of the combination that limits the influence of the BABAR
data. These BABAR data would, in a local data combination that does not fully take into account the correlations, dominate the fit.\footnote{In the range $\sqrt{s} < 1.937\GeV$, the combination of all $\pi^+\pi^-$ data yields $\amuHVPLOpp(\text{KNT19}) = 503.46(1.91)\times 10^{-10}$. Propagating only local correlations to the mean value (for the energy-binning defined by the KNT clustering algorithm) results in $\amuHVPLOpp (\text{Local average}) =  509.4(2.9)\times 10^{-10}$, compared to the BABAR only result of $\amuHVPLOpp(\text{BABAR data only})  =  513.2(3.8)\times 10^{-10}$.} This is corroborated by the findings from studying combinations where all data
are used but either the KLOE or the BABAR data are excluded. While,
when neglecting BABAR, the mean value in the KNT analysis hardly changes, when neglecting
KLOE, the data from BABAR are dominating the fit and the mean value
moves up significantly, though still not as much as it would in a
simple local average, which would become very much dominated by
BABAR. The impact of different choices for the data combination will
be further discussed in \cref{sec:comparison}.

\begin{figure}[t]
\begin{center}
\includegraphics[width=9cm]{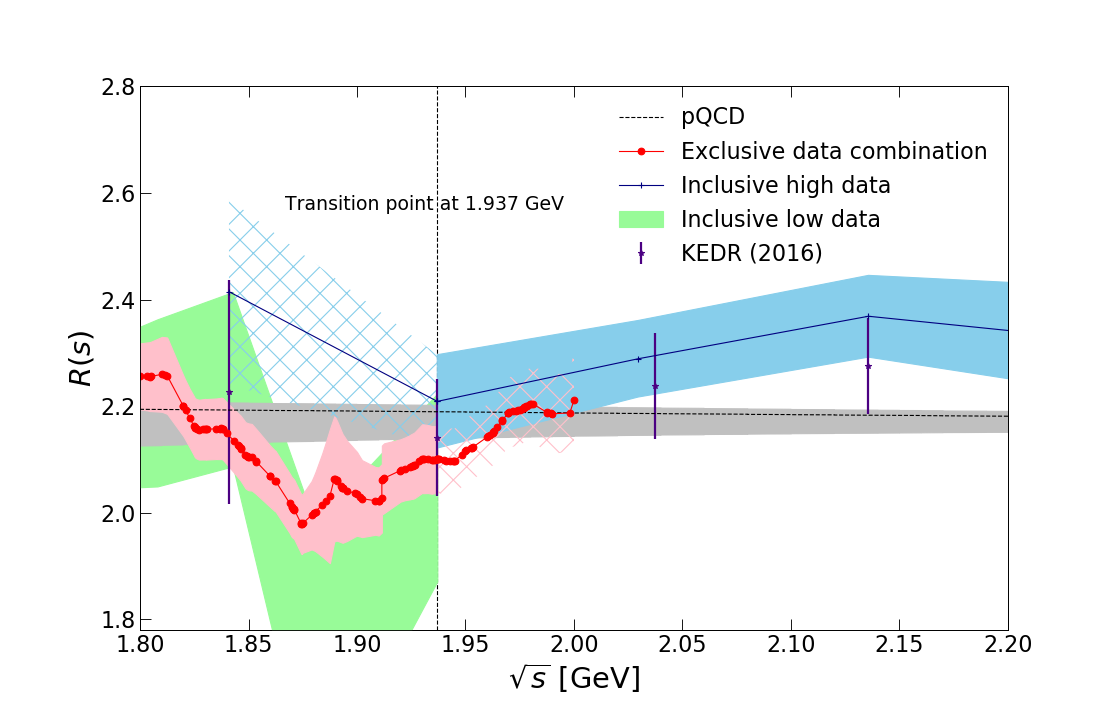}
\end{center}
\vspace{-0.7cm}
\caption{Comparison of the sum over all exclusive channels
  to the compilation of inclusive data and the prediction
  from pQCD in the range from 1.8 to $2.2\GeV$. The blue
  square markers show the data from KEDR~\cite{Anashin:2016hmv}.}
\label{figKNTexclincl}
\end{figure}
As one cannot exclude duality-violating effects to be still effective
at CM energies of about $2\GeV$, KNT prefer to use data as much as
possible and not to rely on pQCD. \Cref{figKNTexclincl}
shows the KNT data combination in the energy region from 1.8 to $2.2\GeV$,
where the transition from using the sum of exclusive channels (red
band) to a compilation of the inclusive data (blue band) is made. The gray
band shows, for comparison, the prediction from pQCD.

In total, the updated KNT19 analysis~\cite{Keshavarzi:2019abf} leads to 
\begin{align}
 \amuHVPLO &= 692.78(1.21)_{\rm stat}(1.97)_{\rm sys}(0.21)_{\rm vp}(0.70)_{\rm fsr} \times 10^{-10}\notag\\
&= 692.78(2.42)_{\rm tot} \times 10^{-10}\,,
\end{align}
where the different errors (statistical, systematic, and the additional errors
due to radiative corrections, VP, and FSR) are added in quadrature.

\subsubsection{Other approaches}
\label{sec:other_approaches}

\paragraph{Approach by F.~Jegerlehner}
\label{sec:FJ}

\begin{figure}[t]
\centering
\includegraphics[height=6cm]{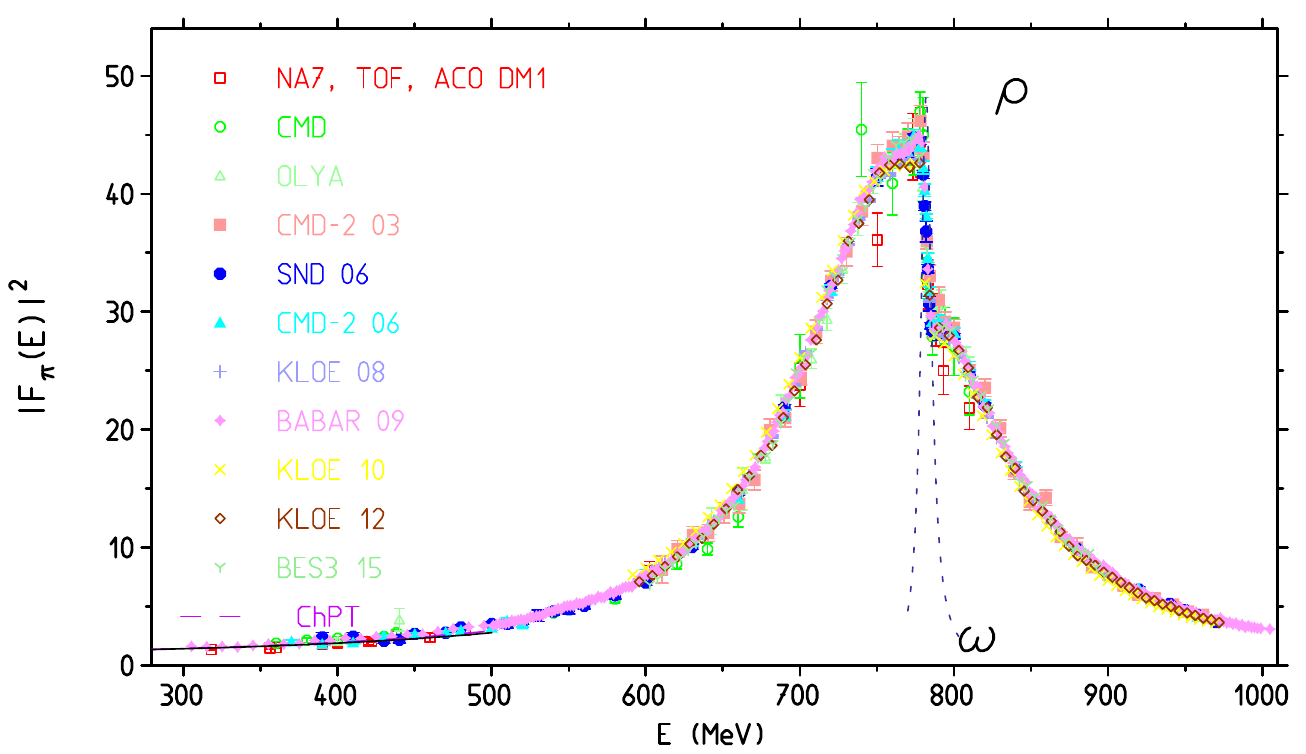}
\caption{A compilation of the modulus squared of the pion form factor in the $\rho$ meson region,
which yields about 75\% of $\amuHVPLO$. Data from CMD-2, SND, KLOE, BABAR, BESIII, and
CLEOc~\cite{Akhmetshin:2003zn,Aulchenko:2006na,Akhmetshin:2006wh,Akhmetshin:2006bx,Achasov:2006vp,Aloisio:2004bu,Ambrosino:2008aa,Ambrosino:2010bv,Babusci:2012rp,Anastasi:2017eio,Venanzoni:2017ggn,Aubert:2009ad,Lees:2012cj,Ablikim:2015orh,Xiao:2017dqv}, besides some older sets. Reprinted from Ref.~\cite{Jegerlehner:2017gek}.}
\label{fig:VPdiadata}
\end{figure}

\begin{figure}[t]
\centering
\includegraphics[]{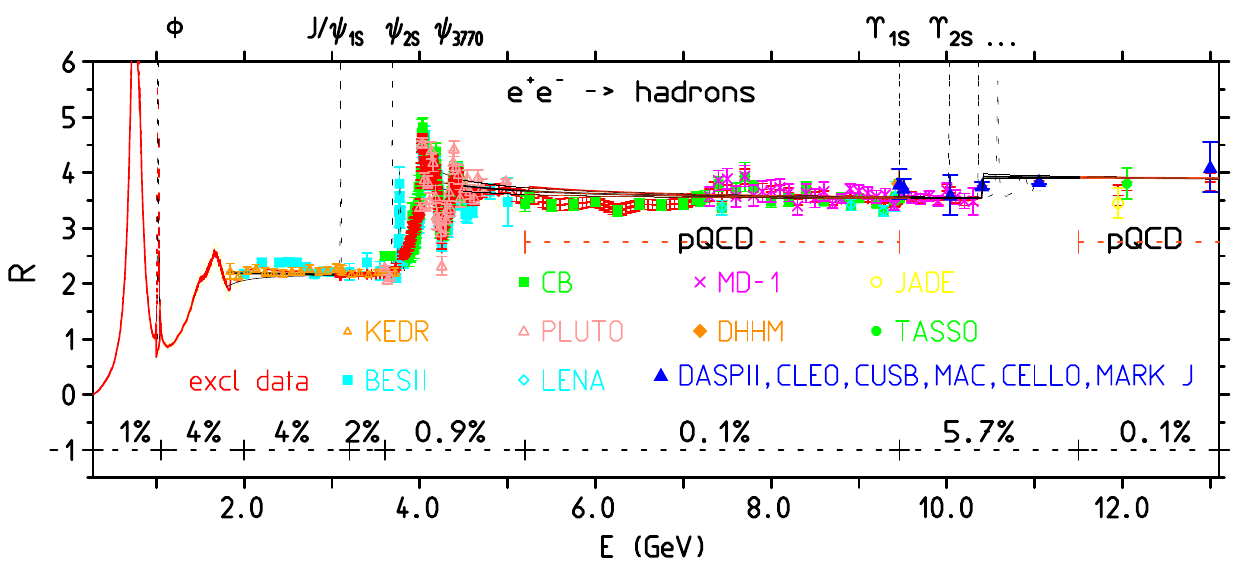}
\caption{The compilation of $R(s)$-data utilized in the 
analyses of Refs.~\cite{Jegerlehner:2015stw,Jegerlehner:2017gek,Jegerlehner:2017lbd,Jegerlehner:2017zsb,Jegerlehner:2018gjd}. The
bottom line shows the relative systematic errors within the split
regions. Different regions are assumed to have uncorrelated
systematics. Data
from Refs.~\cite{Aubert:2004kj,Aubert:2005eg,Aubert:2005cb,Aubert:2006jq,Aubert:2007ur,Aubert:2007ef,Aubert:2007uf,Aubert:2007ym,Lees:2012cr,Lees:2011zi,Lees:2013ebn,Lees:2013gzt,Lees:2014xsh,Davier:2015bka,Davier:2016udg,Akhmetshin:2013xc,Akhmetshin:2015ifg,Kozyrev:2016raz,Achasov:2014ncd,Aulchenko:2014vkn,Achasov:2016bfr,Achasov:2016eyg,Achasov:2016qvd,Achasov:2016lbc,Achasov:2016zvn,Bai:1999pk,Bai:2001ct,Ablikim:2009ad,Anashin:2015woa,Anashin:2016hmv}
and others. Adapted from Ref.~\cite{Jegerlehner:2017gek}.}
\label{fig:rofs} 
\end{figure}

In this section, we review the approach underlying the results recently presented in Refs.~\cite{Jegerlehner:2015stw,Jegerlehner:2017gek,Jegerlehner:2017lbd,Jegerlehner:2017zsb,Jegerlehner:2018gjd}, 
which is also based on the direct integration of cross section data, see \cref{fig:VPdiadata,fig:rofs}.  
The main features of the approach regarding the data integration can be summarized as follows~\cite{Eidelman:1995ny,Jegerlehner:2017lbd,Jegerlehner:2017gek,Jegerlehner:2018gjd}:

\begin{enumerate}
\item
Take undressed data as they are and apply the trapezoidal rule (connecting data
points by straight lines) for integration. Integrating data smoothed
by Chebyshev polynomial fits reproduces the results within insignificant
deviations.
\item
To combine results from different experiments: (i) integrate data for
individual experiments and combine the results for each overlap
region; (ii) combine data from different experiments before integration
and integrate the combined ``integrand.'' Check consistency of the two
possible procedures to estimate the reliability of the results.
\item
Error analysis: (i) statistical errors are added in quadrature; (ii)
systematic errors are added linearly for different experiments; (iii) combined results are obtained by taking weighted averages; (iv) all errors are added in quadrature for ``independent'' data ranges as specified in \cref{fig:rofs}, assuming this to be allowed in
particular for different energy regions and/or different accelerators
and/or detectors; (v) best: apply the true covariance matrix if
available, this is the case for the ISR measurements from meson
factories.
\item
The $\rho$-resonance region is integrated using the 
GS parameterization of the pion form
factor, see \cref{fig:VPdiadata}. Other pronounced resonances have been parameterized by
Breit--Wigner (BW)
shapes with parameters taken from the PDG. For the $\omega$ and $\phi$ one can apply a BW+PDG
evaluation or use the corresponding decay spectra into $3\pi$, $\pi^0\gamma$, $K^+K^-$, $K_L K_S$, and $\eta\gamma$. 
\end{enumerate}

\begin{figure}
\centering
\includegraphics[height=4.5cm]{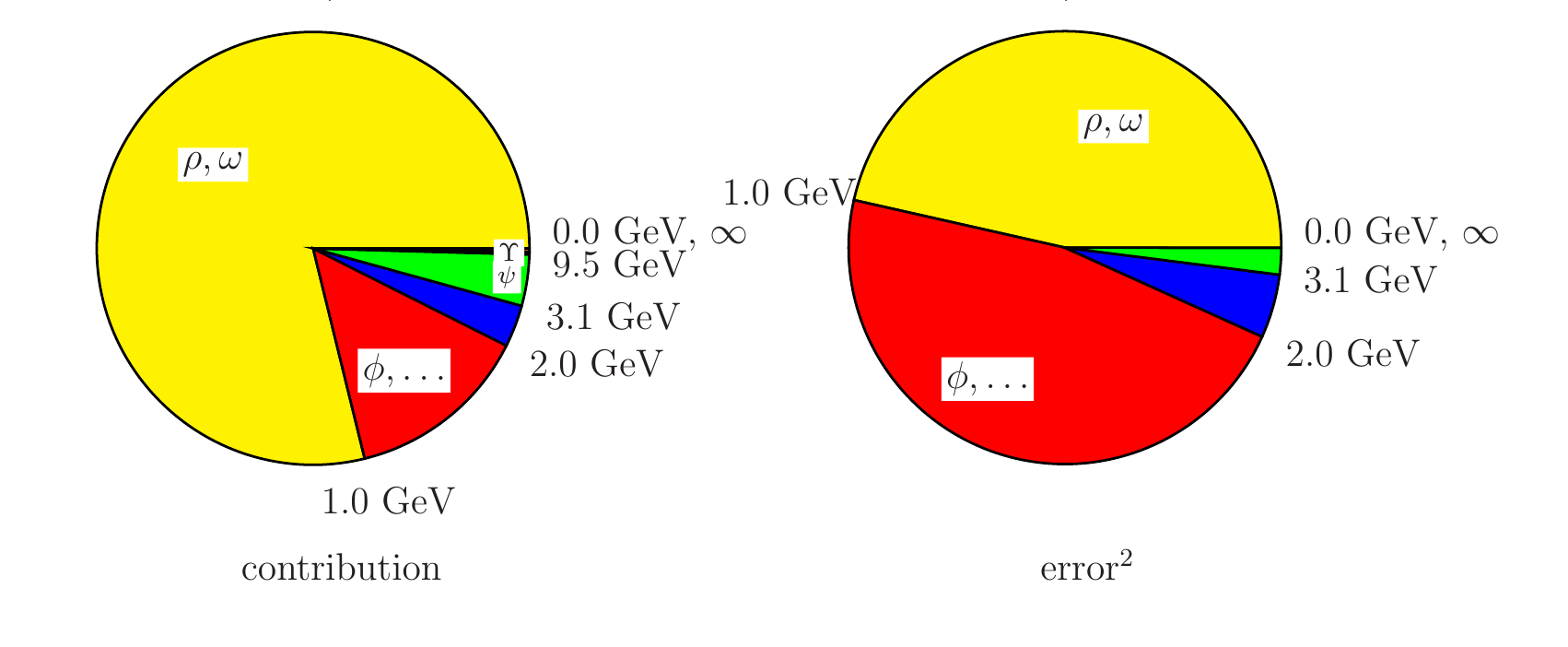}
\caption{Distribution of contributions and error squares from different energy regions. Reprinted from Ref.~\cite{Jegerlehner:2017gek}.}
\label{fig:gm2dist}
\end{figure}

In addition to the data shown in the figures, pQCD is applied from $5.2\GeV$ to $9.46\GeV$ as well as above $11.5\GeV$, see \cref{fig:rofs}, using the
code of Ref.~\cite{Harlander:2002ur}. The central result based on $e^+e^-$ data alone is\footnote{This number, which relies on GS and BW parameterizations as described above, is quoted below in \cref{sec:comparison} as the main result from this approach.}
\begin{equation}
\amuHVPLO=688.1(4.1)\times10^{-10}\,,
\label{LOHVP}
\end{equation}
where the central values and uncertainties are distributed on different energy ranges as shown in \cref{fig:gm2dist}.
In view of the observed discrepancies in the $e^+e^- \to \pi\pi$ data
from BABAR and KLOE, also a combined analysis with the 
$\tau \to \pi\pi \nu_\tau$ data from ALEPH~\cite{Barate:1997hv,Barate:1998uf,Schael:2005am,Davier:2013sfa}, OPAL~\cite{Ackerstaff:1998yj}, CLEO~\cite{Anderson:1999ui}, 
and Belle~\cite{Fujikawa:2008ma} has been considered~\cite{Jegerlehner:2017gek}
\bea
\amuHVPLO=688.8(3.4)\times 10^{-10}\,,
\label{amuhadLOee+tau}
\eea
which is based on the isospin-breaking corrections from Ref.~\cite{Jegerlehner:2011ti} (see \cref{HVP-tau}).
Finally, the combination with analyticity constraints (see \cref{sec:dispersive}) in the 
implementation of Ref.~\cite{Ananthanarayan:2013zua} leads to~\cite{Jegerlehner:2017gek}
\bea
\amuHVPLO=689.5(3.3)\times 10^{-10}\,.
\label{amuhadLOee+tau+pipiphase}
\eea

\paragraph{Hidden-local-symmetry approach}
\label{sec:HLS}

ChPT provides a rigorous 
access to the low-energy part of the nonperturbative sector of QCD, 
but needs to be extended by vector mesons to  go deeper inside the resonance region, leading to
Resonance Chiral Perturbation Theory (R$\chi$PT). As shown in Ref.~\cite{Ecker:1989yg},
R$\chi$PT is, in principle, equivalent to the Hidden Local Symmetry 
(HLS) model~\cite{Bando:1987br,Harada:2003jx}, motivating the use of the HLS model 
for the analysis of annihilation cross sections as input for HVP.
Such effective Lagrangians  share the important feature that they predict physics correlations among the different 
annihilation processes $ {\cal H}=\{ {\cal H}_i, i=1, \ldots,  p \}$  they encompass.  
This means that any given $e^+e^- \to  {\cal H}_i$ cross section
is numerically constrained,  not only by the experimental data collected in the
${\cal H}_i$  channel, but also by those collected in any other final state 
${\cal H}_j$ embodied inside the Lagrangian framework. Stated otherwise,   
the data  collected in any  annihilation channel 
${\cal H}_{i \ne j}$ behave as an increased statistics for any given channel 
${\cal H}_j$.

\begin{figure}[t]
\centering
\includegraphics[width=0.45\textwidth]{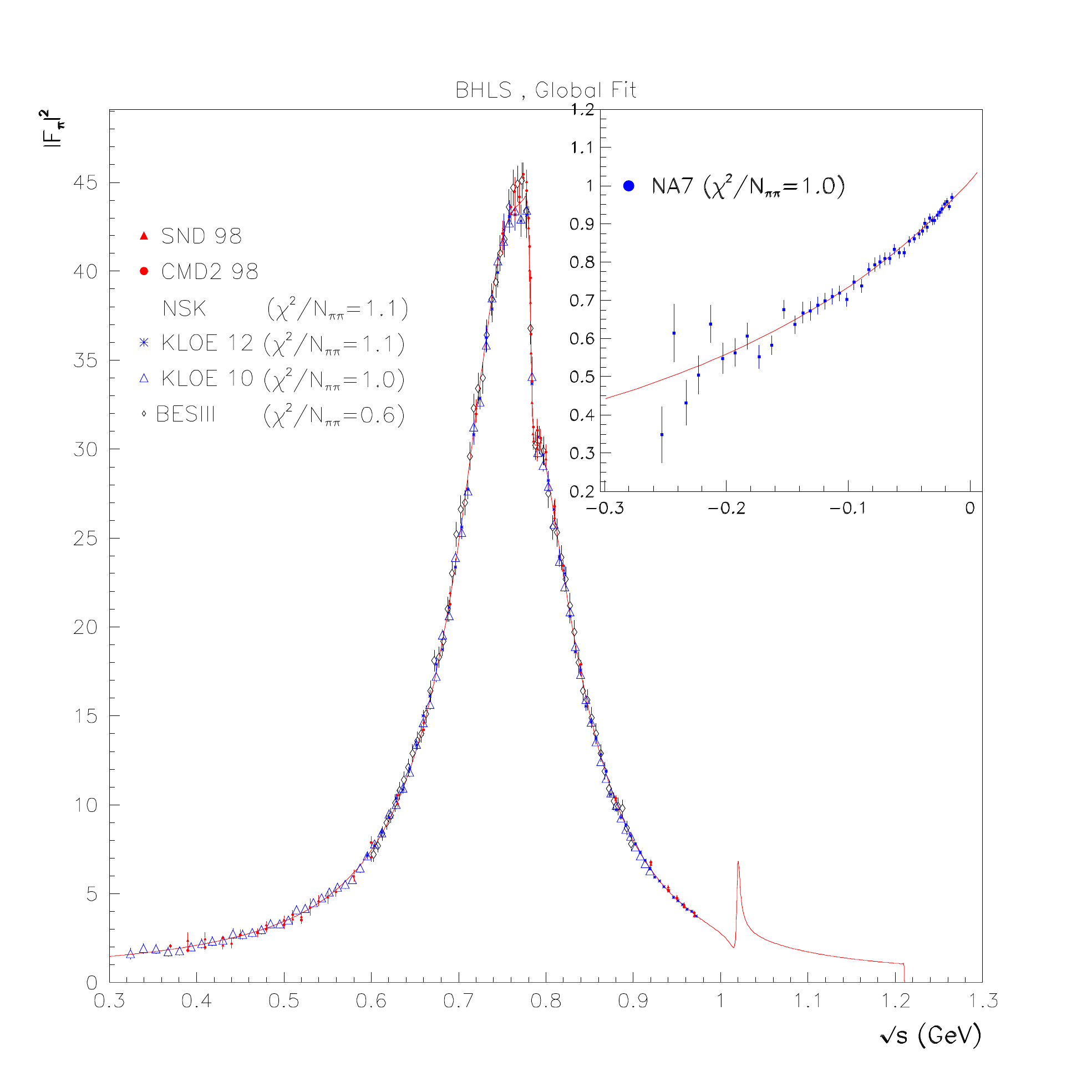}
\includegraphics[width=0.45\textwidth]{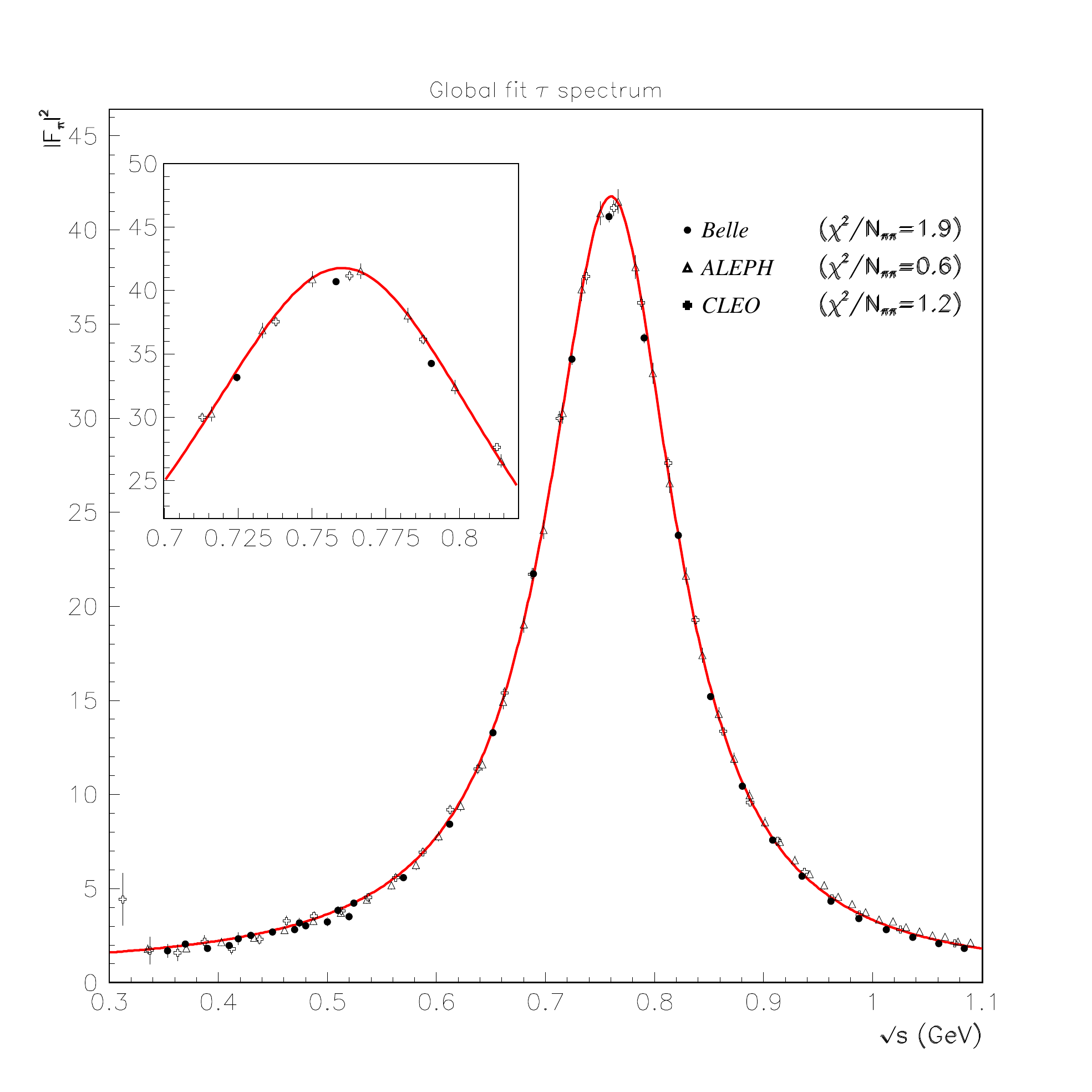}
\caption{The $\pi \pi $ data in the global fit approach: the left panel shows the $e^+e^- \to \pi^+ \pi^-$
data with an inset to show the spacelike region; the right panel displays the 
dipion  spectrum
in the $\tau$ decay. The fit region covers the interval 
$s=[-0.3,1.0]\GeV^2$. Adapted from Ref.~\cite{Benayoun:2019zwh}.}
\label{Fig:PIPI}
\end{figure}

The HLS model is a framework easier to handle than R$\chi$PT. 
However, in order to successfully account for the large amount of precise 
experimental data currently available,  one should go beyond the basic model 
and extend it, consistently  with its framework, using appropriate 
symmetry-breaking mechanisms;  basically, HLS models define interacting 
frameworks for the fundamental pseudoscalar ($P$) and vector ($V$) 
meson nonets together with photons ($\gamma$). Historically, the vector-meson mixing ($V$--$V^\prime$) and the vector-meson--photon mixing ($V$--$\gamma$) 
induced by pseudoscalar meson loops  allowed for a primitive  version of a 
broken HLS model~\cite{Benayoun:2007cu}. This raw broken HLS model 
has evolved towards 
an operating version (BHLS)~\cite{Benayoun:2011mm,Benayoun:2012wc,Benayoun:2015gxa},  which  covers 
a large realm of physics processes involving  the fundamental vector and 
pseudoscalar meson nonets. An enriched version of BHLS, named BHLS$_2$, 
is now available, which has been shown to sharply improve the
description of the very low-energy region~\cite{Benayoun:2019zwh}. Therefore, 
working Lagrangian frameworks now exist that are able to provide a 
consistent  picture of all the data samples covering the $s$ interval ranging
from the nearby spacelike region~\cite{Amendolia:1986wj,Amendolia:1986ui} to slightly above the 
$\phi$ meson mass.
\Cref{Fig:PIPI} illustrates the fit quality on the various $\pi \pi$ 
data (spacelike and timelike form factors and the dipion spectrum in the 
$\tau$ decay).

In their present form, the BHLS frameworks encompass a substantial number of processes of different 
kinds. Beside the 
$VP\gamma$ and $P \gamma \gamma$ partial width decays---where $V$  and $P$ are shorthands to denote any meson pair taken in  the
fundamental $V$ and $P$ nonets---the BHLS$_2$  framework  
describes  the $e^+e^-$ annihilation into the six crucial
channels\footnote{The contribution of these six
hadronic final  states up the the $\phi$ mass ($\simeq 1.05\GeV$) 
represents  about 83\% 
of the total muon HVP and about half of its (squared) uncertainty
in standard approaches.}
 $ \pi^+\pi^-$, $\pi^+\pi^- \pi^0$,  $K^+ K^-$, $K^0 \bar{K}^0$,
$\pi^0 \gamma$, and $\eta\gamma $ (all covered by several data samples),
and, finally, the hadronic spectrum in the $\tau^\pm \to \pi^\pm \pi^0 \nu_\tau$ 
decay. 
Additionally,  the new BHLS$_2$ release~\cite{Benayoun:2019zwh} allows, for 
the first time,  the use of the accurate  spacelike data~\cite{Amendolia:1986wj}  in the 
derivation of $\amuHVPLOpp$.  
Finally, it should be valuable to include within the BHLS$_2$ framework the dipion spectra 
in the $\eta,\eta^\prime \to \pi^+\pi^- \gamma$ decays, as 
they provide additional high-statistics data samples able to further 
constrain the estimate for $\amuHVPLOpp$.

As most BHLS parameters are not known ab initio, 
they should be extracted from data. This is performed by means 
of  global fit techniques permitted by the BHLS frameworks and involves 
the annihilation channels listed before together with the 
($\simeq 50$) available experimental data samples. 
Satisfactory global fits---performed with the {\sc minuit} minimizer---return
the vector of the parameter central values ($\vec{a}_0$) and 
their (full) error covariance matrix  $V(\vec{a})$. The reported statistical 
and systematic errors being closely implemented  in the construction of  
the minimized $\chi^2$~\cite{Benayoun:2015gxa}, $V(\vec{a})$ reflects accurately the 
full error budget of 
the fit data samples. In particular, it
should be noted that the most rigorous way to account for  the
normalization uncertainties  affecting most of the precise data samples is
via a global fit, where the different samples covering the same channel
with different normalization uncertainties 
are allowed to compete and optimally  lead to  their common absolute normalization~\cite{Benayoun:2015gxa,Blobel_2006,Blobel:2003wa}.
Then, the  cross section  derived from the fit intrinsically carries
the most probable normalization for the channel considered; this has been recently confirmed
within the BHLS$_2$ scheme by showing that  the normalization of the pion form factor derived from a global fit to only
experimental data is in close accord~\cite{Benayoun:2019zwh}  with pion form factor spectra produced by lattice QCD groups.
When the fit is  successful, one is motivated to assume that $\vec{a}_0$ and 
$V(\vec{a})$ reliably summarize the full information carried by the data samples involved in the fit---assuming the uncertainties provided by the various experiments
are  reasonably well under control. 
Because of the role of the ``physics correlations,'' one may also infer that, 
for any given annihilation channel within the BHLS scope, the model 
parameter information derived from a global fit 
should be more precise than any treatment of each annihilation channel
separately. This statement is substantiated below in the case of the $\pi^+ \pi^-$
channel.

For a given channel ${\cal H}_i$, its contribution to $a_\mu$ is then obtained by 
replacing  the 
experimental cross sections by the model cross sections
and using the fit outcome to estimate $a_\mu({\cal H}_i)$.  The reference value for each
  $a_\mu({\cal H}_i)$ is derived from a set of estimates  $Q \equiv \{a_\mu({\cal H}_i ,k), k=1, \ldots, p \}$ 
obtained by making a large enough number ($p$)  of samplings of the model 
parameter vector $\vec{a}$ using its central values $\vec{a}_0$ and the  
error covariance matrix  $V(\vec{a})$ returned by the {\sc  minuit} fit. The 
reference value  for $a_\mu({\cal H}_i)$ is then
 defined as the average and its uncertainty as the RMS of the set $Q$
 of evaluations.
 The remaining $\simeq 17 \%$ of the HVP contribution not covered by the BHLS framework should be determined using other methods 
based on direct integration of data samples over 
 wider ranges. This full method has been used in order to derive the 
information for $a_\mu$ presented in Refs.~\cite{Benayoun:2011mm,Benayoun:2015gxa,Benayoun:2019zwh} and 
reported below.

To illustrate the outcome of the method, \cref{Table:T1} contains 
some results derived for the $\pi^+\pi^-$ channel, using the example of the KLOE data sets.  
The results therein are based on almost all $e^+e^-$ data except for other data sets for the $\pi^+\pi^-$ channel,
the spacelike data from Refs.~\cite{Amendolia:1986wj,Amendolia:1986ui},
and the dipion spectra in $\tau$ decays collected by ALEPH, CLEO, and Belle.
The reduction in uncertainty of the BHLS fit result compared to 
the direct integration is precisely driven by the correlations to other channels. 

\begin{table}[t] 
\centering
\small
\begin{tabular}{l c c c c}
\toprule
$\pi^+ \pi^- $ Data set  &  Direct integration~\cite{Anastasi:2017eio} & BHLS$_2$  &  $\chi^2/N_{\pi^+ \pi^-}$ & $p$-value\\\colrule
 KLOE08 	   & $378.9(3.2)$    &   $373.78(1.84)$	& 130/60 &14\%  \\
 KLOE10      & $376.0(3.4)$    & $375.04(2.35)$  	& 69/75 & 78\%	\\
 KLOE12      & $377.4(2.6)$    & $376.74(1.59)$ 	& 59/60 &80\%   \\
 KLOE combination     & $377.5(2.2)$    & $377.17(0.89)$  	& 95/85 & 65\%	 \\
\botrule
\end{tabular}
\caption {\label{Table:T1}
The $\pi^+ \pi^- $ contribution to $\amuHVPLO$ in the range 
$[0.35,0.85]\GeV^2$ in units of $10^{-10}$.
The numbers given in the first data column are read off from Ref.~\cite{Anastasi:2017eio}. The numbers in the second data column are the 
corresponding reconstructed values from  BHLS$_2$ fits; the last two columns display the reduced $\chi^2$ and the $p$-value of the fit, respectively.}
\end{table}
 
The most important outcome of the broken BHLS$_2$ model is the contribution 
of the six channels involved in its framework from the various thresholds up 
to $1.05\GeV$, which results in~\cite{Benayoun:2019zwh}
\begin{equation}
 \amuHVPLO({\rm HLS}, \sqrt{s} < 1.05\GeV) = 572.44(1.20)_\text{fit}\times10^{-10}\,.
\end{equation}
This includes all data sets with the notable exception of the BABAR $\pi^+\pi^-$ data.\footnote{A high quality of the global fits requires the absence of significant tensions 
between the global model and the various candidate data samples 
to be considered. Within the BHLS/BHLS$_2$ global framework, about
90\% of the reported data samples possess quite good fit properties;
the most noticeable exceptions are the KLOE08 and BABAR dipion data sets. 
As discussed in Ref.~\cite{Benayoun:2015gxa}, they severely degrade the fit quality
when included  in the set of fit samples; one also
observes significant discrepancies when using each of them as single 
representative for the $\pi^+\pi^-$ channel within the (full) sample
collecting all the available data sets for the other HLS channels.
This is the reason why in Refs.~\cite{Benayoun:2015gxa,Benayoun:2019zwh} it is found  to be more (statistically) secure to discard
the KLOE08 and BABAR dipion spectra from the set of samples
included in the BHLS global fits.} 
Adding $114.67(2.76)\times 10^{-10}$ for $\sqrt{s} > 1.05\GeV$
and $1.28(17)\times 10^{-10}$ from the tiny  non-HLS channels below $1.05\GeV$,
the BHLS$_2$ final result becomes~\cite{Benayoun:2019zwh}
\begin{equation}
 \amuHVPLO =687.1(3.0)\times 10^{-10}\,.
\end{equation}

\subsubsection{Constraints from analyticity, unitarity, and crossing symmetry}
\label{sec:dispersive}

Since the HVP integral only relies on input for $e^+e^-\to\text{hadrons}$, the corresponding contribution to the anomalous magnetic moment of the muon
can, in principle, be evaluated with minimal model dependence as long as data are available everywhere at the required level of precision. 
For the lowest-multiplicity channels, however, strong constraints on the functional form of the cross section arise from general principles that 
the underlying hadronic amplitudes have to fulfill, such as analyticity, unitarity, and crossing symmetry. 
These additional constraints can allow for an improved evaluation in energy regions where data are scarce, and, more importantly, provide valuable 
cross-checks on the data sets because an incompatibility with the resulting global fit function may point towards inconsistencies in the data. 
Such analyses have recently been performed for the $2\pi$~\cite{Colangelo:2018mtw,Ananthanarayan:2018nyx,Davier:2019can} and $3\pi$~\cite{Hoferichter:2019gzf} channels. 

In the absence of radiative corrections, the hadronic cross section for the $2\pi$ channel is directly related to the pion vector form factor by means of
\be
\sigma(e^+e^-\to\pi^+\pi^-) = \frac{\pi \alpha^2}{3s} \sigma_\pi^3(s) \big| F_\pi^V(s) \big|^2\,,
\ee
where $\sigma_\pi(s) = \sqrt{1-4M_\pi^2/s}$ and $F_\pi^V(s)$ describes the pion matrix element of the electromagnetic current
\be
	\langle \pi^\pm(p') | j^\mu(0) | \pi^\pm(p) \rangle =\pm (p'+p)^\mu F_\pi^V((p'-p)^2)\,.
\ee
This form factor is strongly constrained by analyticity and unitarity, allowing one to derive a global fit function based on fundamental properties of QCD.  
Such representations have been used before to evaluate $\amuHVPLOpp$~\cite{DeTroconiz:2001rip,Leutwyler:2002hm,Colangelo:2003yw,deTroconiz:2004yzs}. 
Here, we concentrate on recent applications~\cite{Colangelo:2018mtw,Ananthanarayan:2018nyx,Davier:2019can} that impose the constraints from analyticity and unitarity on the modern high-statistics data sets, thereby providing a powerful cross-check and a complementary perspective to the direct integration of the data.

We start with the basic structure of a dispersive form factor parameterization from Ref.~\cite{Colangelo:2018mtw} (see Refs.~\cite{Hanhart:2012wi,Hoferichter:2016duk,Hanhart:2016pcd} for similar representations)
\be
\label{eq:VFF}
	F_\pi^V(s) = \Omega_1^1(s) G_\omega(s) G_\mathrm{in}^N(s)\,,
\ee
where the respective factors incorporate $2\pi$, $3\pi$, and higher intermediate states in the unitarity relation. The elastic, $2\pi$, contribution is described via the Omn\`es function~\cite{Omnes:1958hv}
\be
\label{Omnes}
\Omega_1^1(s) = \exp\left\{ \frac{s}{\pi} \int_{4M_\pi^2}^\infty ds' \frac{\delta_1^1(s')}{s'(s'-s)} \right\}\,,
\ee
with $\delta_1^1(s)$ the isospin $I=1$ elastic $\pi\pi$ phase shift. Inelastic corrections in the $3\pi$ channel almost exclusively arise from $\rho$--$\omega$ mixing, whose effect is concentrated 
in a narrow region around the $\omega$ mass.
Inelasticities from higher intermediate states are dominated by the $4\pi$ channel, they are strongly 
suppressed below the $\pi\omega$ threshold~\cite{Lukaszuk:1973jd,Eidelman:2003uh}. 

For the practical application to the modern $e^+e^-\to\pi^+\pi^-$ data sets a number of subtleties need to be taken into account~\cite{Colangelo:2018mtw}. 
First, the representation \cref{eq:VFF} applies to the form factor in pure QCD, while experimentally it is the photon-inclusive cross section that is inserted in the HVP integral. 
Accordingly, one first needs to remove FSR effects from the bare cross sections provided by experiment, then perform the fit, and add the FSR effects back in the end~\cite{Gluza:2002ui,Czyz:2004rj,Bystritskiy:2005ib}. 
Second, to obtain a representation fully based on general QCD properties also the input for the phase shift $\delta_1^1(s)$ needs to be consistent 
with analyticity and unitarity, as well as crossing symmetry, and thus fulfill the Roy equations for $\pi\pi$ scattering~\cite{Roy:1971tc} (see Refs.~\cite{Ananthanarayan:2000ht,GarciaMartin:2011cn,Caprini:2011ky}).
While $\rho$--$\omega$ mixing is adequately described by a narrow resonance approximation
\be
	G_\omega(s) = 1 + \frac{s}{\pi} \int_{9M_\pi^2}^\infty ds' \frac{\Im g_\omega(s')}{s'(s'-s)} \left( \frac{1 - \frac{9M_\pi^2}{s'}}{1 - \frac{9M_\pi^2}{M_\omega^2}} \right)^4\,,\qquad
	g_\omega(s) = 1 + \epsilon_\omega \frac{s}{(M_\omega - \frac{i}{2} \Gamma_\omega)^2 - s}\,,
\ee
or even $g_\omega(s)$ directly, 
an efficient parameterization of higher inelasticities is crucial to obtain 
a representation valid throughout the whole region $\sqrt{s}\lesssim 1\,\text{GeV}$, e.g., in terms of a conformal polynomial whose phase is constrained by the 
Eidelman--\L{}ukaszuk bound~\cite{Lukaszuk:1973jd,Eidelman:2003uh}. Finally, systematic effects related to the high-energy continuation of the phase shift, the solution of the Roy equations,
and the order of the conformal polynomial $N$ need to be taken into account.

In Ref.~\cite{Colangelo:2018mtw}, it is shown that all modern timelike data sets allow for a statistically acceptable description based on \cref{eq:VFF}, as long as the $\omega$ mass is admitted as a free 
parameter.\footnote{The only exception concerns the BESIII data set, for which issues with the statistical covariance matrix were uncovered (the collaboration is aware of the problem and an erratum is currently in preparation). For KLOE08, the global fit function reveals 
two bins with disproportionately large contributions to the $\chi^2$.}
In the fit, an iterative strategy~\cite{Ball:2009qv} is required to account for the systematic covariance matrices while avoiding the D'Agostini bias~\cite{DAgostini:1993arp}, 
the data points from the ISR data sets are to be interpreted as weighted averages over the bins, and the Eidelman--\L{}ukaszuk bound can be implemented as an additional contribution to the $\chi^2$ function.
Combined fits to several data sets should account for the uncertainties in the energy calibration, which is implemented in Ref.~\cite{Colangelo:2018mtw} using separate energy rescalings for each data set, constrained by the respective estimate of the experimental calibration uncertainty. In this way, the central result, including also the spacelike data from NA7~\cite{Amendolia:1986wj}, becomes
\be
\label{eq:amu_disp}
\amuHVPLOpp\big|_{\leq 1\,\text{GeV}}=495.0(1.5)(2.1)\times 10^{-10}=495.0(2.6)\times 10^{-10}\,,
\ee
where the first, fit, uncertainty has been increased by $\sqrt{\chi^2/\text{dof}}\sim 1.1$, mainly driven by the known tension between the KLOE and BABAR data sets, and the second, systematic, uncertainty
is dominated by the sensitivity to $N$. This uncertainty is estimated by the variation seen among fits with different $N$ and comparable $\chi^2/\text{dof}$,
which could be improved by including explicit input on the inelastic channels.
In addition, experimental and systematic uncertainties could be better disentangled 
by an analysis using pseudo-experiments~\cite{Davier:2019can} (to ensure that the experimental uncertainties are not counted multiple times). 
A surprising outcome of the fit is that all data sets unanimously favor an $\omega$ mass significantly below the PDG average~\cite{Tanabashi:2018oca} and the extraction from the $3\pi$ channel (see below), the reason for which is currently not fully understood.
A free complex phase in the $3\pi$ contribution $G_\omega(s)$, as often employed in experimental analyses, violates analyticity, but resonance-enhanced isospin-breaking corrections 
due to the radiative channels $\pi^0\gamma, \pi\pi\gamma, \eta\gamma, \ldots$ can effectively produce a small imaginary part in $\epsilon_\omega$ that removes part of the tension in $M_\omega$~\cite{Kubis}. 
We also quote the result for the low-energy region
\be
\label{eq:amu_disp_low}
\amuHVPLOpp\big|_{\leq 0.63\,\text{GeV}}=132.8(0.4)(1.0)\times 10^{-10}=132.8(1.1)\times 10^{-10}\,.
\ee

An extension to the $3\pi$ channel very close in spirit to Ref.~\cite{Colangelo:2018mtw} has recently been put forward in Ref.~\cite{Hoferichter:2019gzf}. In this case, the 
constraints on the underlying amplitude $\gamma^*\to3\pi$ from analyticity, unitarity, and crossing symmetry are analyzed in the framework of Khuri--Treiman equations~\cite{Khuri:1960zz},
see \cref{sec:pi0DR}, which allow one to resum the final-state rescattering of the pions in terms of the $\pi\pi$ phase shift $\delta_1^1(s)$.  
In addition, the normalization is constrained by the Wess--Zumino--Witten anomaly~\cite{Wess:1971yu,Witten:1983tw} 
in terms of the pion decay constant $F_\pi$~\cite{Adler:1971nq,Terentev:1971cso,Aviv:1971hq}.
The resulting representation leads to
\be
\label{eq:disp_3pi}
\amuHVPLOppp\big|_{\leq 1.8\,\text{GeV}}=46.2(6)(6)\times 10^{-10}=46.2(8)\times 10^{-10}\,,
\ee
where the first, fit, uncertainty includes a scale factor $\sqrt{\chi^2/\text{dof}}\sim 1.2$, and the systematic error is again dominated by inelastic effects 
parameterized by a conformal expansion. Even though the scale factor is larger, the tension among the data sets should be considered much less severe than in the $2\pi$ channel,  
given that the $\chi^2$ inflation mainly reflects tensions among a large number of low-statistics data sets, with far less pronounced 
consequences for the central value than in $2\pi$ if a single experiment is omitted. 
Finally, the masses of $\omega$ and $\phi$ come out in agreement with the PDG parameters, but only once radiative effects in the definition of the mass parameters are 
taken into account consistently.

The main systematic uncertainty in \cref{eq:amu_disp} and \cref{eq:amu_disp_low} arises from the treatment of inelastic effects, as parameterized in terms of the conformal polynomial. 
For the low-energy region there is an alternative strategy that is less sensitive to the inelastic region~\cite{Ananthanarayan:2018nyx}. (See Refs.~\cite{Ananthanarayan:2013zua,Ananthanarayan:2016mns} for earlier work.) Removing $\rho$--$\omega$ mixing via $G_\omega(s)$, this formalism concentrates on the elastic region where Watson's theorem~\cite{Watson:1954uc}, equating the phase of the form factor to $\delta_1^1(s)$,
is exact. In short, input for the spacelike form factor~\cite{Horn:2006tm,Huber:2008id}, the modulus of the timelike form factor in the range $[0.65,0.71]\GeV$, the elastic $\pi\pi$ phase shift, 
and a weighted integral over the modulus squared of the form factor above the inelastic threshold, defines a functional extremal problem whose solution determines optimal upper and lower bounds 
on $|F_\pi^V(s)|$ below $0.63\,\text{GeV}$. The final result
\be
\label{Anant}
\amuHVPLOpp\big|_{\leq 0.63\,\text{GeV}}=132.9(8)\times 10^{-10}
\ee
agrees perfectly with \cref{eq:amu_disp_low}, slightly improving the uncertainty thanks to the reduced sensitivity to the high-energy region.
Note that in this approach the $\omega$ mass and width, the
$\rho$--$\omega$ mixing parameter $\epsilon_\omega$, and the $\pi\pi$
phase shift below the inelastic threshold enter as input into the analysis
(for details see Ref.~\cite{Ananthanarayan:2018nyx}).

The latest DHMZ update~\cite{Davier:2019can} (see \cref{sec:DHMZ}), also incorporates analyticity constraints in the energy range below $0.6\,\text{GeV}$. 
As in Refs.~\cite{Colangelo:2018mtw,Ananthanarayan:2018nyx}, the representation is based on an Omn\`es factor \cref{Omnes}, 
but the details of the implementation as well as data treatment and fit strategy are quite different. 
The phase shift $\delta^1_1$ is not constrained by Roy equations, but fit to data using the parameterization from Ref.~\cite{GarciaMartin:2011cn}, 
and no inelastic phases are included in the representation. 
All the measurements in the timelike region up to $1.0\,\text{GeV}$ are used in the fit. 
When the measurements from different experiments are locally inconsistent, the corresponding uncertainties are scaled according to the local $\chi^2$ value and the number of degrees of freedom 
following the PDG prescription~\cite{Tanabashi:2018oca}. The fit is performed using as test statistic a diagonal $\chi^2$ function ignoring correlations among different measurements, 
avoiding the D'Agostini bias without making too strong assumptions on the knowledge of correlations (see \cref{Sec:UncertaintiesOnUncertainties}).  
Pseudo-experiments that account for the correlations are used to propagate the experimental uncertainties through the fit.
This is a conservative procedure regarding the uncertainty on $\amuHVPLO$~(as exploiting the correlations in the test statistic would reduce its amplitude), 
but comes at the expense of using a test statistics that exploits less information on correlations and is, therefore, less powerful (in the statistical sense) when trying to check for 
possible tensions between a theory and a data set.  
Another set of pseudo-experiments that account for the correlations is used to assess the goodness-of-fit yielding a $p$-value of $0.27$. 
The reliability of the procedure is checked by generating a set of pseudo-experiments and the resulting $p$-values are found to be uniformly distributed between 0 and 1 as expected. 
While the fit is performed between the threshold and $1.0\,\text{GeV}$,
the result is used below $0.6\,\text{GeV}$ only. The choice of the ranges is motivated by the gain of precision of the fit in the low-energy region compared to the combined data integration.
The fit result below $0.63\,\text{GeV}$,
\be
\label{DHMZ}
\amuHVPLOpp\big|_{\leq 0.63\,\text{GeV}}=133.2(5)(4)\times 10^{-10}=133.2(6)\times 10^{-10}\,,
\ee
where the first error estimates experimental and the second model uncertainty (checked to be significant with respect to fluctuations of the experimental uncertainties), 
agrees well with \cref{eq:amu_disp_low} and \cref{Anant}. While the slightly larger
central value could also be due to the differences in the data treatment, 
the smaller systematic uncertainty likely arises when no inelastic effects need to be constrained in the fit.

\subsubsection{Comparison of dispersive HVP evaluations}
\label{sec:comparison}

The different evaluations described in the previous sections all rely on data for $e^+e^-\to\text{hadrons}$, but differ in the treatment of the data as well as the assumptions 
made on the functional form of the cross section.
In short, the evaluations from~\cref{sec:DHMZ} (DHMZ19) and~\cref{sec:KNT} (KNT19) directly use the bare cross section, the one from~\cref{sec:FJ} (FJ17) assumes 
in addition a Breit--Wigner form for some of the resonances, and the evaluation from~\cref{sec:HLS} (BDJ19) relies on a hidden-local-symmetry (HLS) model. 
For certain channels, most notably $2\pi$ and $3\pi$, constraints from analyticity and unitarity define a global fit function or optimal bounds that can be used in the dispersion
integral to integrate the data, see~\cref{sec:dispersive} (ACD18 and CHS18 for $2\pi$). 
In this section, we compare the different evaluations and comment on possible origins of the most notable differences in the numerical results. 

\begin{table}[t]
\small
	\centering
	\begin{tabular}{c r r r r}
	\toprule
	  & BDJ19 & DHMZ19 & FJ17 & KNT19\\
	 $\amuHVPLO\times 10^{10}$ & $687.1(3.0)$  & $694.0(4.0)$ & $688.1(4.1)$ & $692.8(2.4)$\\
	\botrule
	\end{tabular}
	\caption{Full evaluations of $\amuHVPLO$ from FJ17~\cite{Jegerlehner:2017gek}, DHMZ19~\cite{Davier:2019can}, KNT19~\cite{Keshavarzi:2019abf}, and BDJ19~\cite{Benayoun:2019zwh}. The uncertainty in DHMZ19 includes an additional systematic uncertainty to account for the tension between KLOE and BABAR.}
\label{tab:HVP_overview}
\end{table}

\begin{table}[t]
\small
	\centering
	\begin{tabular}{c r r r}
	\toprule
	  & DHMZ19 & KNT19 & Difference\\\colrule
	 $\pi^+\pi^-$ & $507.85(0.83)(3.23)(0.55)$ & $504.23(1.90)$ & $3.62$\\
	 $\pi^+\pi^-\pi^0$ & $46.21(0.40)(1.10)(0.86)$ & $46.63(94)$ & $-0.42$\\
	 $\pi^+\pi^-\pi^+\pi^-$ & $13.68(0.03)(0.27)(0.14)$ & $13.99(19)$ & $-0.31$\\
	 $\pi^+\pi^-\pi^0\pi^0$ & $18.03(0.06)(0.48)(0.26)$ & $18.15(74)$ & $-0.12$\\
	 $K^+K^-$ & $23.08(0.20)(0.33)(0.21)$ & $23.00(22)$ & $0.08$\\
	 $K_SK_L$ & $12.82(0.06)(0.18)(0.15)$ & $13.04(19)$ & $-0.22$\\
	 $\pi^0\gamma$ & $4.41(0.06)(0.04)(0.07)$ & $4.58(10)$ & $-0.17$\\\colrule
	 Sum of the above  & $626.08(0.95)(3.48)(1.47)$ & $623.62(2.27)$ & $2.46$\\\colrule
	 $[1.8,3.7]\GeV$ (without $c \bar c$) & $33.45(71)$ & $34.45(56)$ & $-1.00$\\
	 $J/\psi$, $\psi(2S)$ & $7.76(12)$ & $7.84(19)$ & $-0.08$ \\
	 $[3.7,\infty)\GeV$ & $17.15(31)$ & $16.95(19)$ & $0.20$\\\colrule
	 Total $\amuHVPLO$ & $694.0(1.0)(3.5)(1.6)(0.1)_\psi(0.7)_\textrm{DV+QCD}$ & $692.8(2.4)$ & $1.2$\\ 
	\botrule
	\end{tabular}
	\caption{Selected exclusive-mode contributions to $\amuHVPLO$ from DHMZ19 and KNT19, for the energy range $\leq 1.8\GeV$, in units of $10^{-10}$. Where three (or more) uncertainties are given for DHMZ19, the first is statistical, the second channel-specific systematic, and the third common systematic, which is correlated with at least one other channel. For the $\pi^+\pi^-$ channel, the uncertainty accounting for the tension between BABAR and KLOE (amounting to $2.76\times 10^{-10}$) is included in the channel-specific systematic.}
\label{tab:KNT_DHMZ}
\end{table}

\Cref{tab:HVP_overview} shows the results of recent global evaluations. We start with a more detailed comparison of DHMZ19 and KNT19. At first sight, both evaluation appear in very good agreement, but the 
comparison in the individual channels, see \cref{tab:KNT_DHMZ}, shows significant differences, most notably in the $2\pi$ channel, which differs at the level of the final uncertainty. 
For the $3\pi$ channel, both analyses are now in good agreement, between each other as well as with a fit using analyticity and unitarity constraints~\cite{Hoferichter:2019gzf},
which produces $46.2(8)\times 10^{-10}$, see \cref{eq:disp_3pi}.
Previous tensions could be traced back to different interpolating functions~\cite{Hoferichter:2019gzf,KNT:private,DHMZ:private}:
since the data is relatively scarce off-peak in the $\omega$ region (and similarly, to a lesser extent, for the $\phi$), while the cross section is still sizable, a linear interpolation 
overestimates the integral. 
Both DHMZ19 and KNT19 analyses include evaluations of the threshold region of the $2\pi$ channel, either using ChPT or dispersive fits, as well as, going back to Ref.~\cite{Hagiwara:2003da}, estimates for the threshold regions of $\pi^0\gamma$ and $3\pi$ below the lowest data points, based on the chiral anomaly for the normalization and $\omega$ dominance
for the energy dependence (following Ref.~\cite{Achasov:2002bh} for $\pi^0\gamma$ and Refs.~\cite{Kuraev:1995hc,Ahmedov:2002tg} for $3\pi$). The corresponding estimates, $0.12(1)\times 10^{-10}$ for $\pi^0\gamma$
and $0.01\times 10^{-10}$ for $3\pi$, agree well with recent dispersive analyses, which lead to $0.13\times 10^{-10}$~\cite{Hoid:private} and $0.02\times 10^{-10}$~\cite{Hoferichter:2019gzf}, respectively.\footnote{Since the $3\pi$ threshold contribution is very small, it does not matter for $a_\mu$ that in this case $\omega$ dominance from Refs.~\cite{Kuraev:1995hc,Ahmedov:2002tg}  noticeably underestimates the cross section.}  
Finally, a difference of about $1.0\times 10^{-10}$ arises from the energy region $[1.8,3.7]\GeV$ depending on whether data (KNT19) or pQCD (DHMZ19) is used. 
Summing up these three individual channels already leads to a significant cancellation among the differences, which in combination with the smaller channels, see \cref{tab:KNT_DHMZ}, 
produces the agreement of the central value at the level of $1\times 10^{-10}$.  
The difference in the $[1.8,3.7]\GeV$ interval may reflect the level of agreement between data and pQCD, while for the exclusive channels most differences are well within the uncertainties,
apart from the $2\pi$ channel. Since both evaluations for the latter are based on the same data, this channel deserves further attention.

To this end we consider the detailed breakdown in energy intervals as given in \cref{tab:2pi}, in comparison to the result of a global fit function derived from analyticity and unitarity, see \cref{sec:dispersive}. In addition, the DHMZ19 number includes a systematic error defined as half the difference between evaluations performed either without KLOE or without BaBar, with the central value 
defined as the mean between the two. Without this adjustment (DHMZ19$'$), the total number for the $2\pi$ channel, $507.0(1.9)$, becomes a little closer to KNT19. \begin{table}[t]
	\centering
	\small
	\begin{tabular}{c r r r r r}
	\toprule
	Energy range & ACD18 & CHS18 & DHMZ19 & DHMZ19$'$ & KNT19\\\colrule
	$\leq 0.6\GeV$ & & $110.1(9)$ & $110.4(4)(5)$ & $110.3(4)$ & $108.7(9)$\\
	$\leq 0.7\GeV$ & & $214.8(1.7)$ & $214.7(0.8)(1.1)$ & $214.8(8)$ & $213.1(1.2)$\\
	$\leq 0.8\GeV$ & & $413.2(2.3)$ & $414.4(1.5)(2.3)$ & $414.2(1.5)$ & $412.0(1.7)$\\
	$\leq 0.9\GeV$ & & $479.8(2.6)$ & $481.9(1.8)(2.9)$ & $481.4(1.8)$ & $478.5(1.8)$\\
	$\leq 1.0\GeV$ & & $495.0(2.6)$ & $497.4(1.8)(3.1)$ & $496.8(1.9)$ & $493.8(1.9)$\\
	\colrule
	$[0.6,0.7]\GeV$ & & $104.7(7)$ & $104.2(5)(5)$ & $104.5(5)$ & $104.4(5)$\\
	$[0.7,0.8]\GeV$ & & $198.3(9)$ & $199.8(0.9)(1.2)$ & $199.3(9)$ & $198.9(7)$\\
	$[0.8,0.9]\GeV$ & & $66.6(4)$ & $67.5(4)(6)$ & $67.2(4)$ & $66.6(3)$\\
	$[0.9,1.0]\GeV$ & & $15.3(1)$ & $15.5(1)(2)$ & $15.5(1)$ & $15.3(1)$\\
	\colrule
	$\leq 0.63\GeV$ & $132.9(8)$ & $132.8(1.1)$ & $132.9(5)(6)$ & $132.9(5)$ & $131.2(1.0)$\\
	$[0.6,0.9]\GeV$ & & $369.6(1.7)$ & $371.5(1.5)(2.3)$ & $371.0(1.6)$ & $369.8(1.3)$\\
	$\big[\sqrt{0.1},\sqrt{0.95}\,\big]\GeV$& & $490.7(2.6)$ & $493.1(1.8)(3.1)$ & $492.5(1.9)$ & $489.5(1.9)$\\
	\botrule
	\end{tabular}
	\caption{Comparison of $\amuHVPLOpp$ from CHS18~\cite{Colangelo:2018mtw}, KNT19~\cite{Keshavarzi:2019abf,KNT:private}, DHMZ19 with the BABAR/KLOE adjustment~\cite{Davier:2019can,DHMZ:private} (the second error gives the additional systematic uncertainty included for the BABAR/KLOE tension), and the variant without (DHMZ19$'$). 
	All numbers in units of $10^{-10}$. For the low-energy region $\leq 0.63\GeV$ the comparison is also shown to ACD18~\cite{Ananthanarayan:2018nyx}.}
	\label{tab:2pi}
\end{table} 

The first observation is that the dispersive result, CHS18, lies halfway between DHMZ19 and KNT19 when considered for the full energy range $\leq 1\GeV$ (the contribution above $1\GeV$ is small and differences there negligible). Next, the low-energy region, say below the $\rho$ peak, agrees well with DHMZ19, while the KNT19 results lie about $1.5\times 10^{-10}$ lower. Vice versa, above the $\rho$ peak the global fit agrees with KNT19, while DHMZ19 lies about $1.5\times 10^{-10}$ higher. 
These observations suggest the following interpretation: at low energies data are relatively scarce, so that in the direct integration of the data the treatment of correlations around the $\rho$ peak may well influence the low-energy result. Here, evaluations imposing analyticity and unitarity constraints~\cite{Colangelo:2018mtw,Ananthanarayan:2018nyx,Davier:2019can} favor the higher value.\footnote{Note that the DHMZ19 number in \cref{tab:2pi} for $\leq 0.63\GeV$ 
corresponds to the combined result of the fit ($\leq 0.6\GeV$) and data integration ($[0.6,0.63]\GeV$), while the number in \cref{DHMZ} refers to the fit result only. As indicated by the small change and the uncertainty on it~\cite{Davier:2019can}, good agreement is observed between the direct integration and the fit-based combination also in the low-energy region.} 
For the energy region above the $\rho$, the data display the well-known tension between BABAR and KLOE, so any combination will effectively reflect the relative weight assigned 
to each experiment in the fit. This is also the reason why the difference becomes larger in DHMZ19 than in DHMZ19$'$, because with the central value chosen as the mean between 
evaluations without KLOE and without BABAR, the weight of BABAR in defining the central value slightly increases. 
Finally, in addition to the covariance matrices provided by experiment, in the direct integration also the algorithm to combine the data into bins plays a role. 

 \begin{figure}[t]
\centering
    \includegraphics[width= 0.48\textwidth]{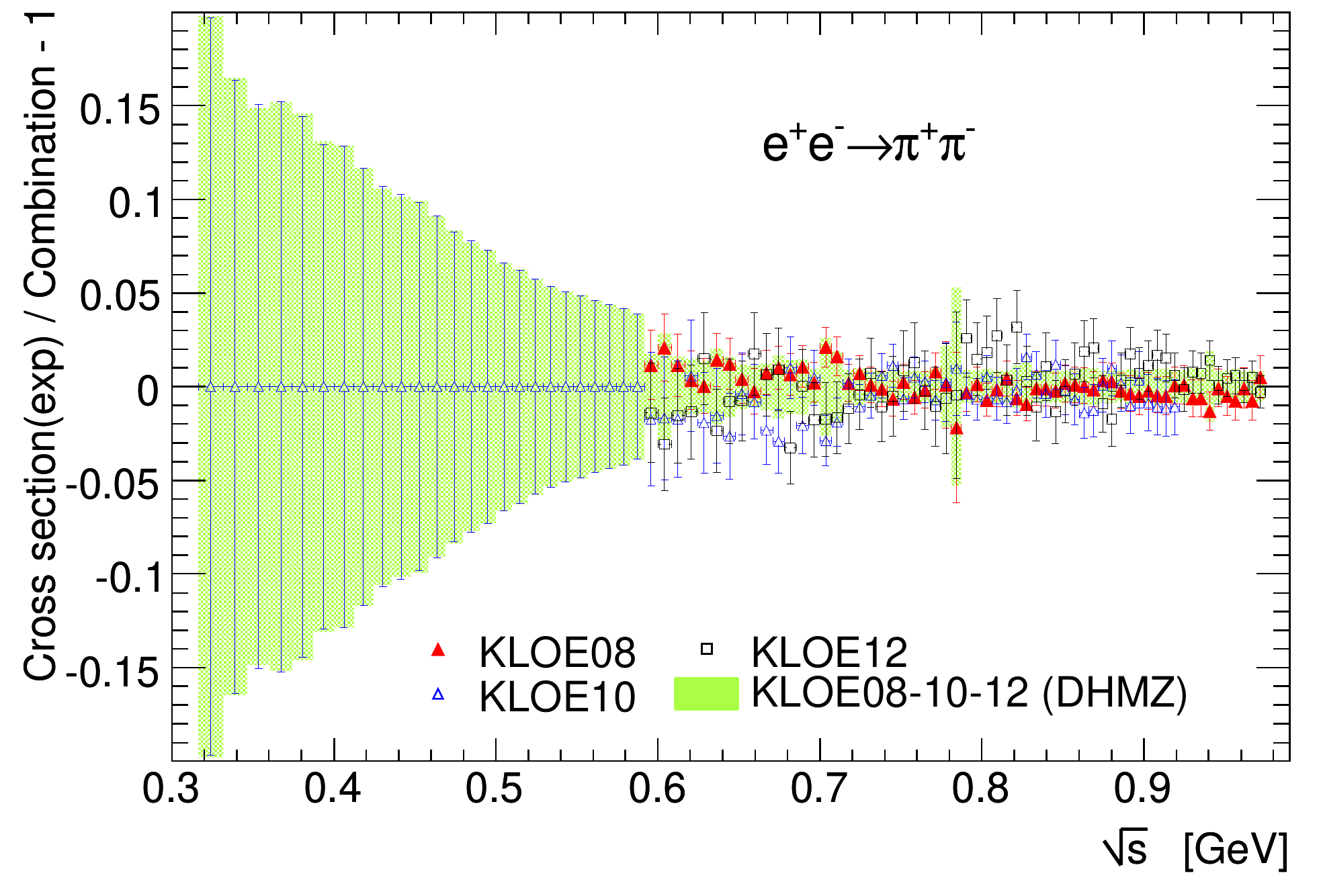}
    \includegraphics[width= 0.5\textwidth]{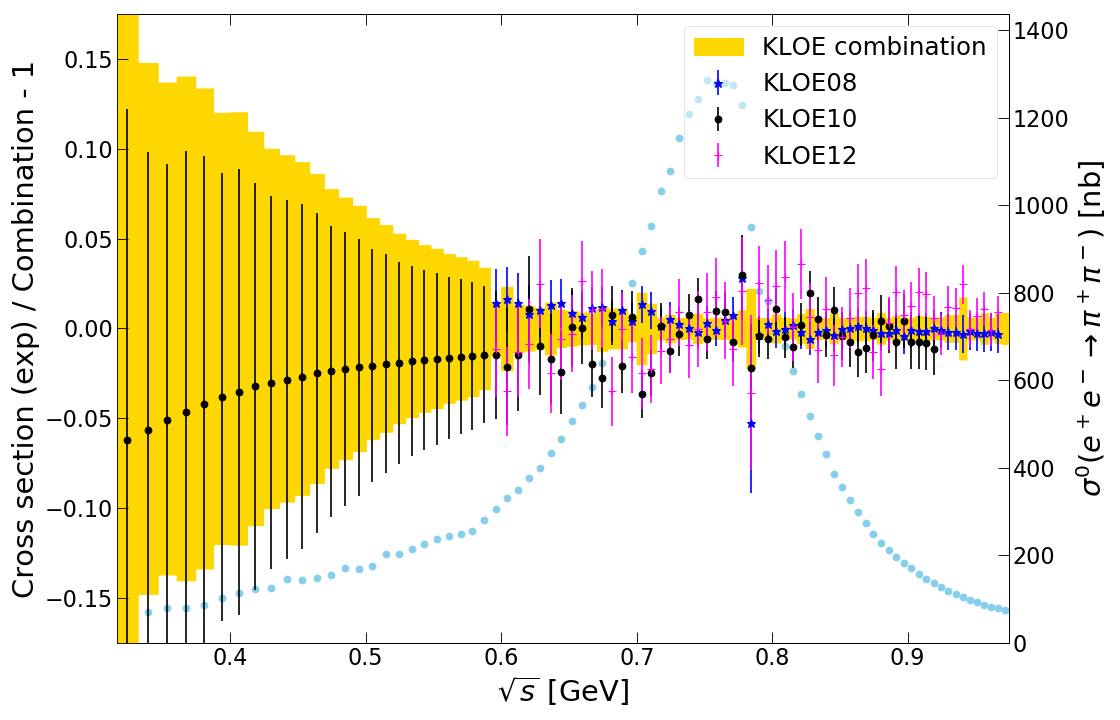}
  \caption{\small The normalized difference of the three KLOE measurements (KLOE-2008, KLOE-2010, and KLOE-2012) of the $\pi^+\pi^-$ cross section with the combination of the three from DHMZ~(left) and KNT~(right, adapted from Ref.~\cite{Anastasi:2017eio}).}\label{fig:KLOE-DHMZvsKNT}
\end{figure} 

In this regard, the use of correlated uncertainties in the DHMZ and KNT approaches deserves detailed consideration. As mentioned previously, in the DHMZ data combination, uncertainties are propagated through large samples from pseudo-experiments produced via Monte Carlo (MC) simulation (see \cref{sec:DHMZ}). This approach results in correlations being propagated to the mean values in local regions in an effort to avoid making too-strong assumptions about the given covariances. In the KNT approach, as described in \cref{sec:KNT}, a correlated fit is implemented, where the available covariances are used in their entirety to constrain the fit and influence the mean values fully. A comparison of these methods is given in \cref{fig:KLOE-DHMZvsKNT}, which shows the normalized difference of the three KLOE measurements of the $\pi^+\pi^-$ cross section with each combination. For DHMZ, the higher-energy data points do not influence the lower-energy data region only covered by KLOE10 and the fit mean values in this region are described only by those KLOE10 data. In the KNT case, the covariances from the energy-independent normalization uncertainties mean that the precision of the higher-energy KLOE08/KLOE12 data is propagated to the lower-energy region through the window allowed by those correlations. For the complete $\pi^+\pi^-$ combination, the KNT analysis is therefore restricted by the correlations of these three precise and highly correlated measurements, consequently favoring a lower resulting $\pi^+\pi^-$ cross section than in the DHMZ analysis. Overall, this results in a smaller value for $\amuHVPLOpp$ in KNT than in DHMZ  (specifically DHMZ19'). It should be noted that in Ref.~\cite{Anastasi:2017eio}, the KNT data combination was compared with the Best Unbiased Linear Estimator (BLUE) approach~\cite{Lyons:1988rp}, where all uncertainties and covariances were propagated via MC pseudo-experiments to the BLUE values. This yielded results that were consistent with those from KNT.
It is known that the BLUE method is equivalent to the minimization of the uncertainty of the output of a weighted average, cf.\ Gauss--Markov theorem (see, for example, Ref.~\cite{Nakamura:2010zzi}).
It is also interesting to note that the central values of the integrals of the KLOE combination from DHMZ and KNT in the dominant $\rho$-region are similar at $\amuHVPLOpp\big|_{[0.6,0.9]\GeV}= 366.5(2.8)\times 10^{-10}$ (DHMZ) and $\amuHVPLOpp\big|_{[0.6,0.9]\GeV} = 366.9(2.1) \times 10^{-10}$ (KNT), although the KNT combination yields a smaller uncertainty.

Next, we turn to the comparison to FJ17. The number quoted in \cref{tab:HVP_overview} refers to the result from~\cref{sec:FJ} using $e^+e^-$ data alone, with input from $\tau$ data
increasing the value by $0.8\times 10^{-10}$. 
In comparison to KNT19 and DHMZ19 two main effects can be identified: first, the contributions from $\omega$ and $\phi$ are fit using Breit--Wigner functions 
with parameters from the PDG~\cite{Tanabashi:2018oca}, the $\rho$ using a Gounaris--Sakurai parameterization,
instead of a direct integration of the data. Second, the data from different experiments are combined by taking weighted averages of integrals in overlapping regions
instead of a locally weighted average. In combination, these effects increase the HVP integral by $2.2\times 10^{-10}$~\cite{Jegerlehner:2018gjd}.
Including $\tau$ data and adapting the low-energy result from Ref.~\cite{Ananthanarayan:2018nyx} below $0.63\GeV$, the best value given in~\cref{sec:FJ} and Ref.~\cite{Jegerlehner:2018gjd}
becomes $689.5(3.3)\times 10^{-10}$, so accounting, in addition, for these two effects the central value would move closer to DHMZ19 and KNT19.    
 
Finally, in Ref.~\cite{Benayoun:2019zwh} the low-energy channels $\pi^+\pi^-$, $\pi^0\gamma$, $\eta\gamma$, $\pi^+\pi^-\pi^0$, $K^+K^-$, $K_LK_S$ are fit in an HLS model below $1.05\GeV$, while for the energy region above as well as the non-HLS channels below the results from Refs.~\cite{Jegerlehner:2017gek,Jegerlehner:2018gjd} are applied. 
By far the biggest numerical effect compared to DHMZ19 and KNT19 arises because the BABAR data for the $\pi^+\pi^-$ channel are not included in the fit, which amounts to about $3.5\times 10^{-10}$~\cite{Benayoun:private}.
The remainder of the difference originates largely from the non-HLS channels, e.g., the difference between KNT19 and BDJ19 in the energy region $[1.05,2]\GeV$ 
is $2.0\times 10^{-10}$~\cite{Benayoun:private,KNT:private}.

\subsubsection{Uncertainties on uncertainties and on their correlations}
\label{Sec:UncertaintiesOnUncertainties}

Modern $\epem$ annihilation data are published as a set of cross section values, their corresponding uncertainties, and
a prescription for evaluating the point-to-point or bin-to-bin correlations of the uncertainties~(see \cref{ee-data}).
It has been pointed out by the ATLAS collaboration, in the context of the jet performance~\cite{Aad:2014bia} and jet cross section studies~\cite{Aaboud:2017dvo,Aaboud:2017wsi},
that the amplitudes of systematic uncertainties and their correlations are generally impacted by uncertainties themselves.
In order to account for uncertainties on uncertainties and on their correlations, several different ``configurations"/``scenarios" of uncertainties were published there.
More recently, similar remarks were made in the context of the $\epem$ annihilation data used for the evaluation of $\amuHVPLO$~\cite{Davier:2019can,bogdan-Mainz-2018-DHMZ-UncOnUnc}, pointing to the fact that their combination
has to be done without overestimating the precision with which
the uncertainties of the measurements and their correlations are known (see also \cref{sec:hvptools}).

The statistical uncertainties of the measured cross sections and their correlations are generally known rather precisely,\footnote{Note, however, that in the case of measurements performed with relatively low statistics, the statistical fluctuations of the number of event counts can itself impact the evaluation of the statistical uncertainty of the measurements~\cite{DAgostini:1993arp}.} both for measurements performed at fixed $\sqrt{s}$ values and for ISR measurements involving unfolding techniques.
However, even if they are published together with the nominal values of the measurement, the systematic uncertainties and their correlations are never really measured, but rather {\it estimated}.
For this reason, the size of the systematic uncertainties of the various measurements, the correlations of a systematic uncertainty component impacting several measurements, as well as the correlations between the systematic uncertainties impacting a given measurement, are not perfectly known.
Furthermore, the uncertainties impacting all these quantities are not available in the current publications of hadronic spectra.

Looking closely at the systematic uncertainties of the hadronic spectra, one can notice that they are evaluated in relatively wide mass ranges, with sharp transitions between them (see, e.g., Refs.~\cite{Ambrosino:2008aa,Lees:2012cj}).
Actually, if one were trying to evaluate them in narrower ranges, fluctuations caused by the limited available statistics would impact the values of the extracted systematics, also yielding an uncertainty on their evaluation.
Furthermore, the change of the event topologies between the measurements performed at different $\sqrt{s}$ values triggers questions about the treatment of a given systematic uncertainty component as being fully correlated across the phase space.
Typical examples are the uncertainties related to acceptance or to tracking.
Similarly, the amount and type of background contributions differ significantly depending on the $\sqrt{s}$ value, but often a single uncertainty component is provided  for their subtraction.
One can also question whether all the systematic uncertainties, e.g., the ones related to the trigger and the tracking~(when the latter is involved in the trigger decision), are really independent.
Finally, one standard deviation is statistically not well defined for systematic uncertainties.
For all these reasons the systematic uncertainties and their correlations can never be considered as being perfectly known.

As discussed earlier, tensions between the input measurements are observed in several channels, in particular \pp, where there are tensions between BABAR and KLOE, and \KK, where measurements differ on the $\phi\rightarrow\KK$ peak.
These tensions are not only local, but manifest as systematic trends having a coherent impact on the dispersion integrals.
They are a direct indication of underestimated uncertainties for the measurements and point to the need for a conservative uncertainty treatment in the data merging and fitting procedure.

The existence of uncertainties on uncertainties and on their correlations is relevant for precision studies and should be taken into account in any data combination exercise.
From this perspective, for the data merging exercise, one can distinguish two main types of approaches.

First, the method used in Refs.~\cite{Davier:2009ag,Davier:2009zi,Davier:2010nc,Davier:2017zfy,Davier:2019can} is based on a local \chiS minimization, performed separately in each of the fine bins used for determining the weights of the measurements in the combination.
Doing so, the \chiS minimization only uses locally the information on the size of the uncertainties and on the correlations between the measurements.\footnote{This information is used on slightly wider ranges, of typically up to a couple of $100\MeV$, when {\it averaging regions} are defined in order to account for the difference between the point-spacing and bin-sizes for the various experiments~\cite{Davier:2009zi}. In this procedure the systematic uncertainties are not constrained (profiled), but rather directly propagated from each input measurement to the {\it averaging regions} and then to the fine bins.}
Afterwards, the full information on the correlations across the phase space for a given experiment, the correlations of the uncertainties between the experiments, as well as between the various channels, is propagated from the input measurements to the final combination result and the corresponding dispersion integrals.
This is done using a series of pseudo-experiments, as well as by varying the experimental inputs by one standard deviation of a given uncertainty and repeating the combination procedure.

A second category of methods uses a \chiS computed globally, for the full mass range of interest, including correlations across all the experimental data points and bins~\cite{Keshavarzi:2018mgv,Anastasi:2017eio}.
Such methods rely on the description of the mass-dependence of the amplitude of each uncertainty component, as well as on the assumption of each uncertainty component being fully correlated across the phase space, which also induces a reduced uncertainty of the combination result compared to local merging procedures.

Similarly, when fitting the experimental spectra, one can either use only the diagonal uncertainties in the \chiS followed by a propagation of the full set of uncertainties and their correlations, as done in Ref.~\cite{Davier:2019can}, or one can use a global \chiS definition~\cite{Colangelo:2018mtw,Hoferichter:2019gzf, Benayoun:2015gxa, Benayoun:2019zwh}.
In the latter case one makes the same assumptions on the perfect knowledge of the uncertainties and the correlations as discussed earlier,
which also reduces the uncertainty on the corresponding dispersion integrals.

The question of the knowledge of correlations between different measurements of an experiment and between experiments was also brought up in Refs.~\cite{Ananthanarayan:2016mns,Ananthanarayan:2018nyx,Ananthanarayan:2019zic}.
Therein, a weighted average is conservatively used for combinations of results from several measurements of a given experiment, in the $\sqrt{s}$ range between $0.65$ and $0.71\GeV$, and then of single (combined) values from each experiment.
However, a global \chiS definition with correlations is used to determine the uncertainty for the results of the combinations  and an attempt is made to estimate the correlations used therein.
This follows an approximate method for dealing with the case when the exact correlations are not known~\cite{Schmelling:1994pz}.
Doing so, the level of correlations between all the pairs of measurements of a given experiment, in the considered $\sqrt{s}$ range, is assumed to be the same.
This is indeed a reasonable approximation for the systematic uncertainties of a given experiment in this rather limited energy interval.
Still, it is well known that the level of statistical correlations between the bins of measurements involving unfolding (e.g., BABAR and KLOE) strongly depends of their $\sqrt{s}$ separation.
Similarly, the level of correlations between all the pairs of results based on different experiments in the timelike region is assumed to be the same.
However, this assumption is at variance with a direct study of the timelike experimental inputs.
Indeed, e.g., the correlations of the systematic uncertainties between the results based on two analyses of the KLOE experiment are much larger than the ones between a measurement from KLOE and a measurement from a different experiment.
In the context of the method applied in Refs.~\cite{Ananthanarayan:2016mns,Ananthanarayan:2018nyx,Ananthanarayan:2019zic}, the use  of the same uncertainties on the phase in the elastic region and on the spacelike data, for all the results, may dampen the effect of different correlations of the timelike data and can justify to a certain extent the assumption.

In summary, there are clear indications for the existence of uncertainties on the uncertainties and on their correlations for the hadronic data and they have a direct impact on the combination procedures.
At the same time, the assumptions that combination or fitting methods make about the knowledge of uncertainties and of their correlations have direct consequences on the central values and the uncertainties of the resulting dispersive integrals.
In the long term, it is desirable to have measurements provided together with the information of uncertainties on uncertainties and on their correlations.

\subsubsection{Conservative merging of model-independent HVP results}
\label{sec:ConservativeMerging}

\paragraph{The methodology}

In this section we describe a procedure for {\it merging} the HVP combination results discussed in the \cref{sec:DHMZ,sec:KNT,sec:dispersive}, labeled DHMZ, KNT, and CHHKS respectively.\footnote{We emphasize that this does not represent an attempt to {\it combine} these results, which would involve evaluating the correlations between them etc.}
The merging procedure is therefore based on the DHMZ and KNT results in exclusive hadronic channels in the mass range below $1.8\GeV$ and on the inclusive evaluations in the various~(complementary) higher-mass ranges.
The CHHKS results are included in the merging for the \pp and $\pi^+\pi^-\piz$ channels, where they are available.

The first requirement for this merging procedure is to be {\it conservative}.
This is motivated by the tensions observed between experimental data (see \cref{ee-data}), as well as by the differences between the results of combinations based on the same data inputs but using different methodologies (see \cref{sec:comparison}), with their various assumptions on the knowledge of uncertainties and of their correlations (see \cref{Sec:UncertaintiesOnUncertainties}).
The second requirement is to {\it account for the correlations of the systematic uncertainties between different channels}, yielding an unavoidable increase of the total uncertainty~\cite{Davier:2010nc,Davier:2017zfy,Davier:2019can}.
Indeed, a detailed study of the uncertainties in each channel allowed one to identify $15$ such correlated uncertainty components in the latest DHMZ analysis~\cite{Davier:2019can}.

In the merging procedure implemented here, the central value of $\amuHVPLO$ is computed as the sum of simple averages (i.e., the arithmetic means) of the DHMZ and KNT results in the relevant hadronic channels and mass ranges.
The CHHKS central values are included in the simple averages for the \pp channel below $1.0\GeV$, as well as for the $\pi^+\pi^-\piz$ channel.

For the experimental and theoretical uncertainties, in each channel and each relevant mass range, the maximum of the DHMZ and KNT results is taken.\footnote{The CHHKS uncertainties are not included here, which is motivated as follows: for the $3\pi$ channel, both the experimental and systematic uncertainties are below the uncertainty from the direct integration. For the $2\pi$ channel, this still holds true for the uncertainty derived from experiment, but the systematic effects as estimated in Ref.~\cite{Colangelo:2018mtw} are slightly larger, indicating that contrary to $3\pi$ the precision that can be obtained with a global fit is slightly worse than from the direct integration. Checks of the significance of the variations used for the systematic uncertainties as well as using explicit input for the inelastic channels are also foreseen (see \cref{sec:dispersive}). Given the conservative treatment of the $2\pi$ channel in view of the BABAR and KLOE tension (see \cref{sec:DHMZ}), we therefore do not see a reason to increase the uncertainty further.}
Correlations between channels are taken into account as in the DHMZ analysis. To account for the cases in which the KNT uncertainties are larger, a quadratic difference is then used to evaluate by which amount the corresponding DHMZ uncertainty would need to be enhanced in order to reach the same value, so that the final uncertainty can be constructed by adding these amounts to
the experimental uncertainty of the DHMZ sum of channels, using a quadratic sum.

In each channel and each relevant mass range, half of the difference between the central values of the DHMZ and KNT combinations is taken as an extra systematic uncertainty.
An exception occurs in the \pp channel, where the maximum between this difference and the uncertainty related to the tension between the BABAR and KLOE measurements, as evaluated by DHMZ~\cite{Davier:2019can}, is taken.
This allows us to stay conservative while at the same time avoiding any double-counting of the effect of the BABAR/KLOE tension, which has a direct impact on the difference between the central values of the DHMZ and KNT results.
This systematic uncertainty is treated as independent between channels and mass ranges, which is motivated in part by the fact that the sign of the corresponding algebraic differences between the DHMZ and KNT results tends to fluctuate randomly.

\paragraph{Numerical results}

The numbers required  to implement this merging procedure have been collected in \cref{sec:comparison}, in particular, \cref{tab:KNT_DHMZ}. We obtain
\begin{align}
 \amuHVPLO&=  693.1(2.8)_\text{exp}(2.8)_\text{sys}(0.7)_\text{DV+QCD}\times 10^{-10}\notag\\
  &=693.1(4.0)\times 10^{-10}\,. 
  \label{merging}
\end{align}
This value is based on  Refs.~\cite{\HVPref}, which should be cited in any work that uses or quotes \cref{merging}, with main experimental input from Refs.~\cite{\HVPexpref}. 
The first error in \cref{merging} refers to the experimental uncertainties, the largest of which is that 
       of the $2\pi$ channel ($1.9\times 10^{-10}$), followed by that associated 
       with the $3\pi$ channel ($1.5\times 10^{-10}$).
The contribution of the experimental systematic uncertainties correlated between at least two channels amounts to $1.6\times 10^{-10}$.
The second, systematic error is nearly completely saturated by the $2\pi$ channel, where the $2.8\times 10^{-10}$ uncertainty assigned in Ref.~\cite{Davier:2019can} as half the difference between evaluations without BABAR and KLOE, respectively, exceeds the systematic error defined as half the difference between the DHMZ and KNT evaluations ($1.8\times 10^{-10}$), and is therefore adopted as the systematic uncertainty in the $2\pi$ channel. With the CHS18 number about half-way in between the DHMZ and KNT evaluations, it is clear that including this dispersive evaluation 
will not further increase the uncertainty.
The second-largest difference between DHMZ and KNT occurs in the energy region $[1.8,3.7]\GeV$ ($0.5\times 10^{-10}$), but is covered by the duality violation uncertainty, as evaluated by DHMZ~\cite{Davier:2019can}.
The corresponding error, computed as the difference of the evaluations of the HVP contribution based on either inclusive data or pQCD input in the region $[1.8,2.0]\GeV$, could either be considered of experimental or systematic origin.
We therefore display this uncertainty, completed by a small contribution from the pQCD calculation itself, separately in \cref{merging} labeled ``DV+QCD.''
The final error is evaluated as the quadratic sum.  

We consider the merged number quoted in \cref{merging} a conservative but realistic 
assessment of the current situation regarding LO HVP. It merges different methodologies for the direct integration of the data, while including analyticity/unitarity constraints where available (covering about $80\%$ of the final value) and accounting for the tension between the BABAR/KLOE data sets beyond 
the standard (local) $\chi^2$ inflation.  
In particular, in this procedure, it is clear how new high-statistics data on the $2\pi$ channel would impact the various components of the uncertainty: by reducing the experimental and the systematic uncertainties when included in the average; because evaluations without BABAR/KLOE would no longer be dominated by KLOE/BABAR; and because differences in methodology will become less important when tensions between the data are less severe.

\subsubsection{Higher-order insertions of HVP}
\label{sec:higher_orders}

 \begin{figure}[t]
\centering
    \includegraphics[width= 0.8\textwidth]{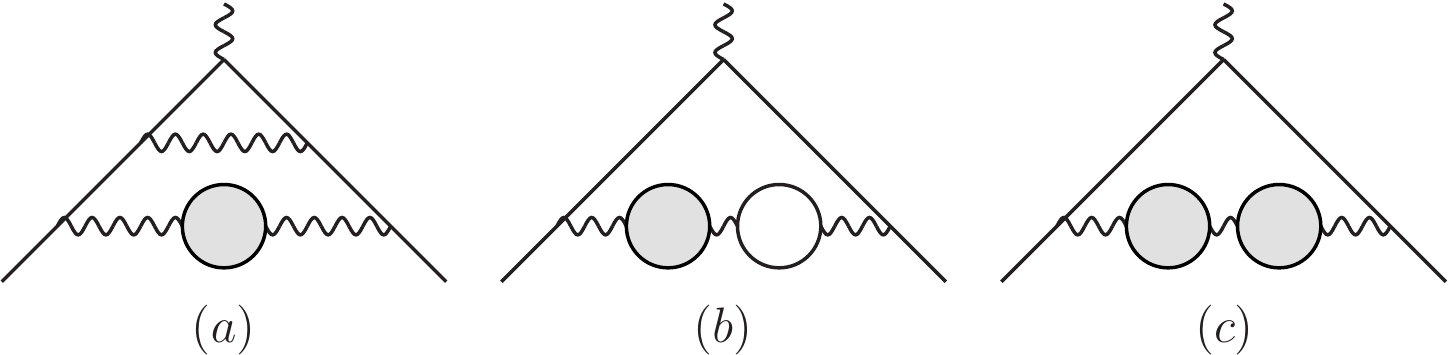}
  \caption{Higher-order insertions of HVP at NLO. The gray blobs refer to HVP, the white one in diagram $(b)$ to leptonic VP.}
  \label{fig:HVPHO}
\end{figure} 

\begin{table}[t]
	\centering
	\small
	\begin{tabular}{c r r r r}
	\toprule
	 & FJ17 & HLMNT11 & KLMS14 & KNT19  \\
	 $\amuHVPNLO\times 10^{10}$ & $-9.93(7)$ & $-9.84(7)$ & $-9.87(9)$ & $-9.83(4)$\\
	\botrule
	\end{tabular}
	\caption{Evaluations of $\amuHVPNLO$ from FJ17~\cite{Jegerlehner:2017gek}, HLMNT11~\cite{Hagiwara:2011af}, KLMS14~\cite{Kurz:2014wya} (using the LO HVP routine from Ref.~\cite{Hagiwara:2011af} as input), and KNT19~\cite{Keshavarzi:2019abf}. The evaluation from Ref.~\cite{Kurz:2014wya} was also adopted in Ref.~\cite{Davier:2019can}.}
\label{tab:HVP_NLO_overview}
\end{table}

Higher-order iterations of HVP have been considered as early as Ref.~\cite{Calmet:1976kd}, leading to the consideration of the topologies shown in \cref{fig:HVPHO}. Explicit kernel functions for the corresponding integrals have been derived in Refs.~\cite{Barbieri:1974nc,Krause:1996rf}. Numerically, 
diagrams $(a)$ and $(b)$ yield the dominant contribution (with a sizable cancellation between them), while diagram $(c)$ is very small. Recent evaluations are shown in \cref{tab:HVP_NLO_overview}.
Within uncertainties there is good agreement among the different evaluations, especially in view of the final accuracy required for $(g-2)_\mu$. To be consistent with the merged value for LO HVP given in~\cref{merging}, we adopt the central value from Ref.~\cite{Keshavarzi:2019abf} but increase the uncertainty accordingly, leading to
\begin{equation}
 \amuHVPNLO= -9.83(7)\times 10^{-10},
 \label{HVPNLO}
\end{equation}
which, in particular, ensures that the uncertainty can still be considered anticorrelated with the one assigned to the LO HVP contribution.

Beyond NLO, it was pointed out in Ref.~\cite{Kurz:2014wya} that even NNLO insertions are not negligible, as their combined effect 
\begin{equation}
 \amuHVPNNLO= 1.24(1)\times 10^{-10}
 \label{HVPNNLO}
\end{equation}
is of a similar size as the final accuracy goal of the Fermilab $g-2$ experiment. We will adopt this value for the NNLO contribution, which agrees well with the subsequent evaluation 
from Ref.~\cite{Jegerlehner:2017gek},
$\amuHVPNNLO= 1.22(1)\times 10^{-10}$.

\subsection{Prospects to improve HVP further}
\label{HVPdisp_prospects}

\subsubsection{The MUonE Project}

\label{Sec:MUonE}

A novel approach has been proposed to determine $\amuHVPLO$, measuring the effective electromagnetic coupling in the spacelike region via scattering data~\cite{Calame:2015fva}. The elastic scattering of high-energy muons on atomic electrons of a low-$Z$ target has been identified as an ideal process for this measurement and a new experiment, MUonE, has been proposed at CERN to measure the shape of the differential cross section of $\mu e$ elastic scattering as a
function of the spacelike squared momentum transfer~\cite{Abbiendi:2016xup}.

Assuming a 150\,GeV muon beam with an average intensity of about $1.3\times10^7$ muons/s, presently available at CERN's muon M2
beamline, incident on a target consisting of 40 beryllium layers, each 1.5\,cm thick, and three years of data taking, one can reach an
integrated luminosity of about $1.5\times10^7\,\text{nb}^{-1}$, which would correspond to a statistical error of $0.3\%$ on the value of
$\amuHVPLO$. The direct measurement of the effective electromagnetic coupling via $\mu e$ scattering would therefore
provide an independent and competitive determination of $\amuHVPLO$. It would consolidate the muon $g-2$ prediction and allow a
firmer interpretation of the upcoming measurements at Fermilab and J-PARC.

In the kinematic configuration described above, MUonE can cover the squared four-momentum transfer region $-0.143\GeV^2<q^2<0$, corresponding to approximately $87\%$ of the $\amuHVPLO$ integral. The rest can be obtained either using timelike data and pQCD, or via lattice-QCD evaluations as described in \cref{subsec:muone}. First lattice-QCD computations of this remaining part are presented in Refs.~\cite{Marinkovic:2019zoi,Giusti:2019hkz}.

\paragraph{The experiment}
The detector is comprised of 40 identical modules, each consisting of a 1.5\,cm-thick layer of Be coupled to three Si tracking layers separated from each other by a distance of 
       $\sim 1$ m (to be optimized) with intermediate air gaps (see \cref{Be-detector})~\cite{MUonE:LoI}. Thin targets are required to minimize the impact of multiple scattering and background on the measurement, and 
       multiple copies of such targets to obtain the necessary statistics. The Si detectors provide the necessary resolution ($\sim 20$\,$\mu$m) with a limited material budget ($<0.07\, X_0$ per unit). This arrangement provides both a distributed low-$Z$ target as well as the tracking system. Downstream of the apparatus, a calorimeter and a muon system (a filter plus active planes) will be used for $e$/$\mu$ particle identification. 

Significant contributions of HVP to the $\mu e \to \mu e$ differential cross section are essentially restricted to electron scattering angles below 10\,mrad, corresponding to electron energies above 10\,GeV. The net effect of these contributions is to increase the cross section by a few permil: a precise determination of $\amuHVPLO$ requires not only high statistics, but also a high systematic accuracy, as the final goal of the experiment is equivalent to a determination of the signal to normalization ratio with an $\order(10\,\text{ppm})$ systematic uncertainty at the peak of the integrand function. Although this determination does not require the knowledge of the absolute cross section (signal and normalization regions will be obtained by $\mu e$ data), it poses severe requirements on the knowledge of the following quantities:
\begin{figure}[t]
\includegraphics[width=.49\textwidth]{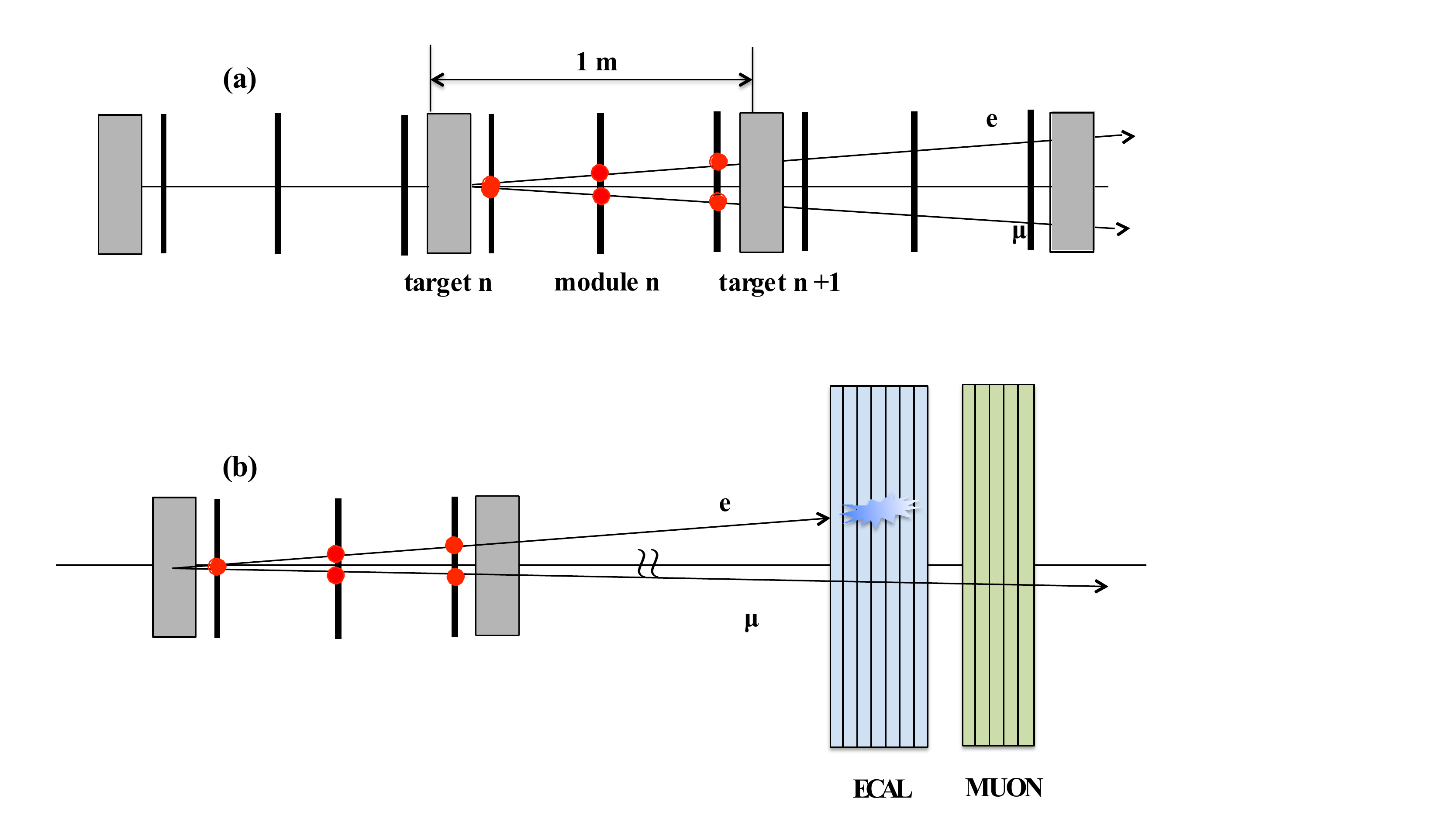}~\includegraphics[width=.49\textwidth]{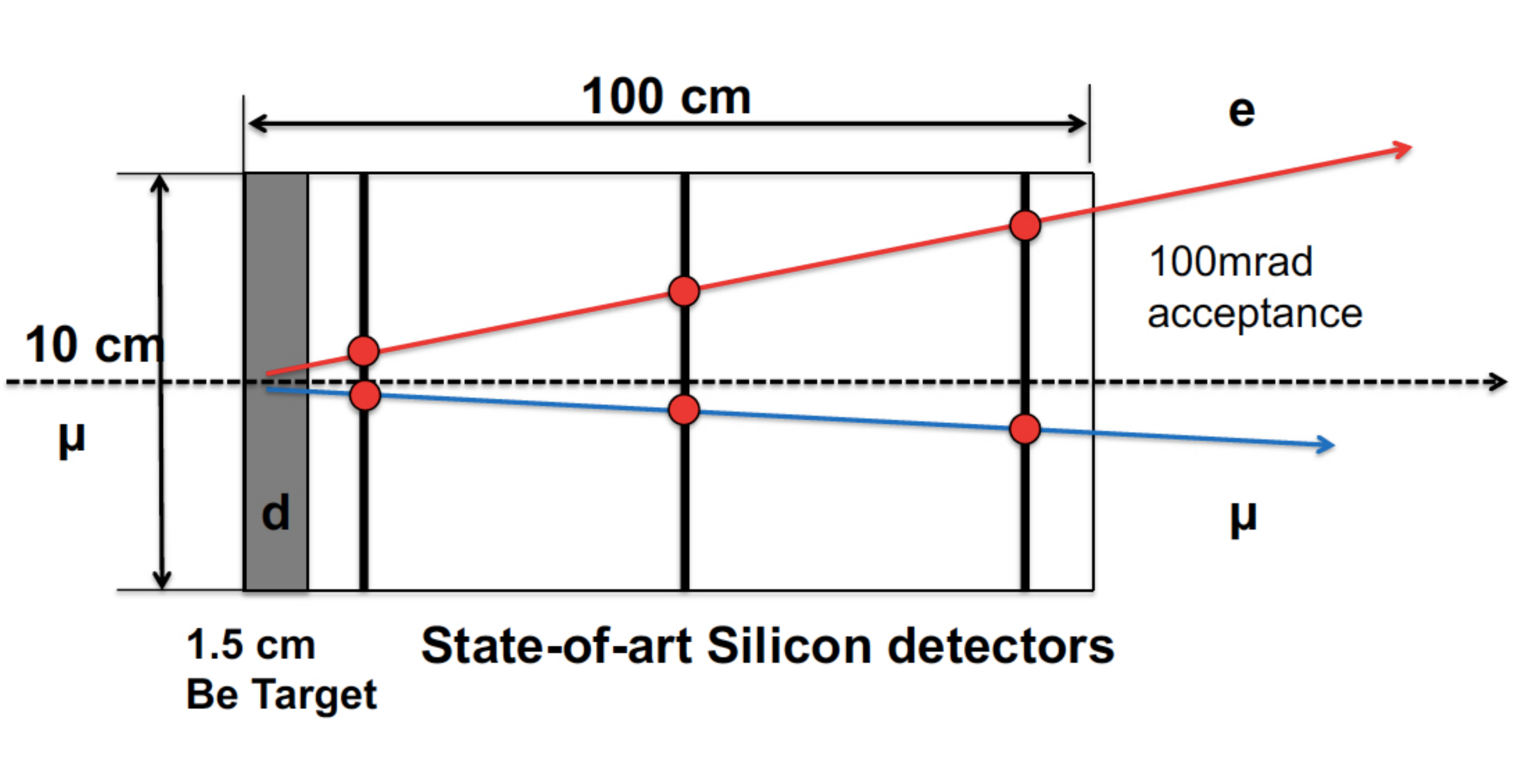}
\caption{Left: design of the baseline detector concept. Right: single station.}
  \label{Be-detector}
\end{figure}

\begin{itemize}

\item Multiple scattering: studies of the systematics indicate that an accuracy of the order of 1\% is required on the knowledge of the multiple-scattering effects in the core region. Results from a test beam at CERN with 12 and $20\GeV$ electrons on 8 and 20\,mm C targets show good agreement between data and GEANT4 simulations, see \cref{ms}~\cite{Abbiendi:2019qtw}.
\begin{figure}[t]
\includegraphics[width=.39\textwidth]{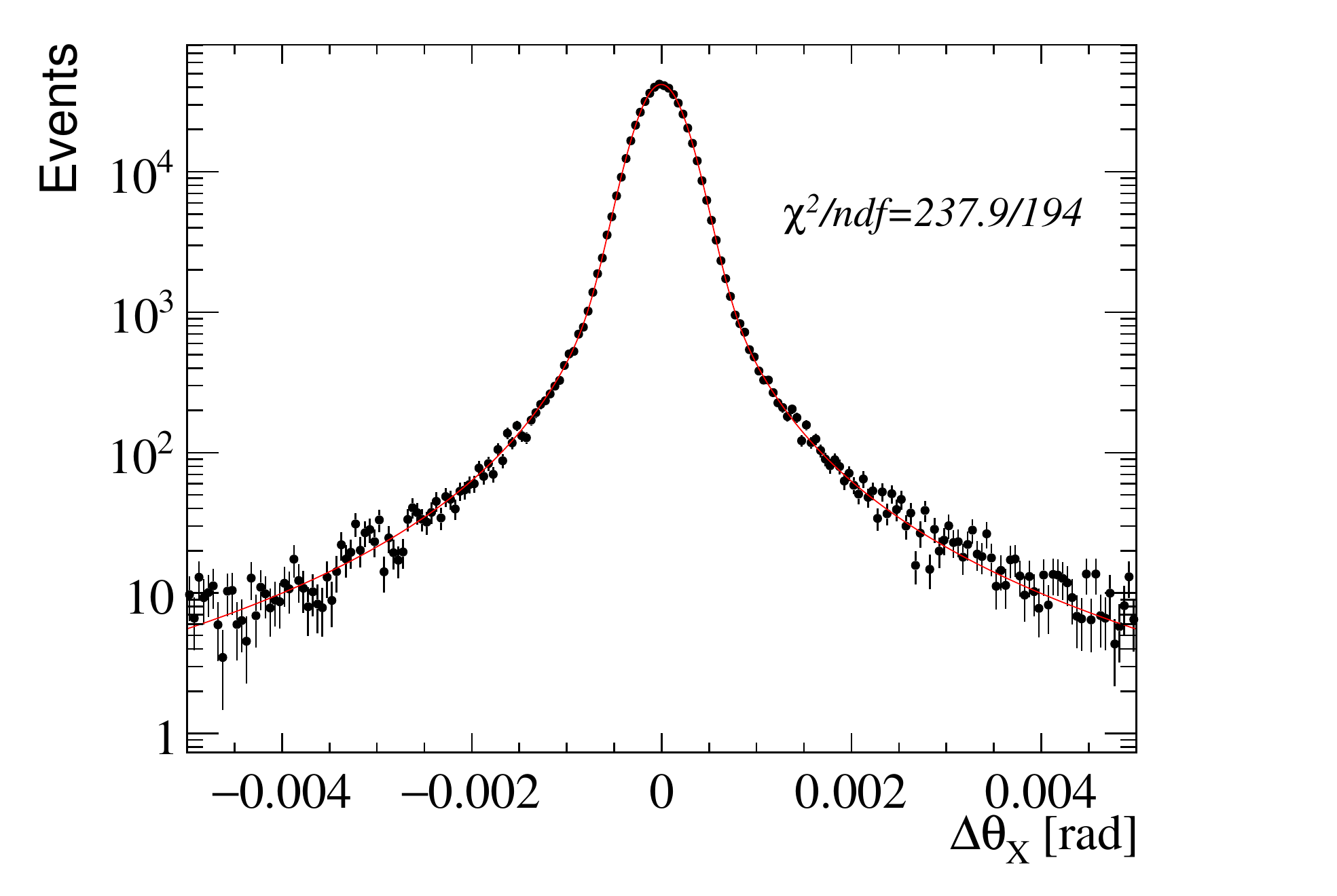}~\includegraphics[width=.59\textwidth]{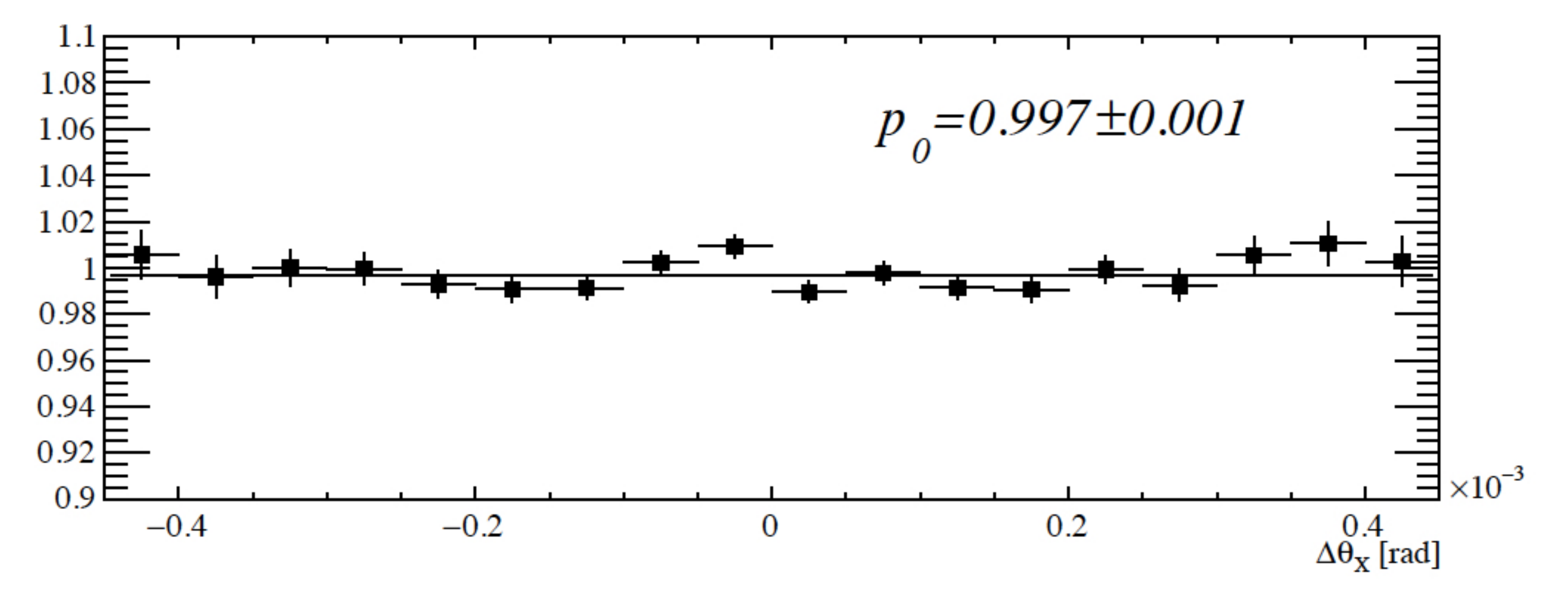}
\caption{Left: $x$-projection of the scattering angle from $12\GeV$ $e^-$ with 8\,mm target compared with the results of the fit based on the sum of a Gaussian and a Student's $t$ distribution. Right: Data/Monte Carlo ratio. Reprinted from Ref.~\cite{Abbiendi:2019qtw}. }
\label{ms}
\end{figure}

\item Tracking uniformity, alignment, and angle reconstruction: it is important to keep the systematic error arising from the nonuniformity of the tracking efficiency and angle reconstruction at the 10\,ppm level. The use of state-of-the-art Si detectors should ensure the required uniformity. Among the considered alternatives, the Si strip sensors being developed for the CMS tracker upgrade represent a good solution. In particular, the Si sensors that are foreseen for the CMS HL-LHC outer tracker in the so-called 2S configuration have been chosen~\cite{cmsu}. They are 320\,$\mu$m-thick sensors with n-in-p polarity produced by Hamamatsu Photonics. They have an area of 10\,cm$\times$10\,cm (sufficient to cover the MUonE acceptance) and a pitch p = 90\,$\mu$m, which means having a single hit
precision $\sim p/\sqrt{12} \sim 26\,\mu$m. The strips are capacitively-coupled and segmented in two approximately 5\,cm-long strips. In the 2S configuration, two closely-spaced Si sensors reading the same coordinate are mounted together and read out by common front-end ASIC.  With their accompanying front-end electronics, they can sustain high readout rate (40\,MHz) and are well suited to track triggering (see \cref{cms}).
\begin{figure}[t]
\begin{center}
\includegraphics[width=0.9\textwidth]{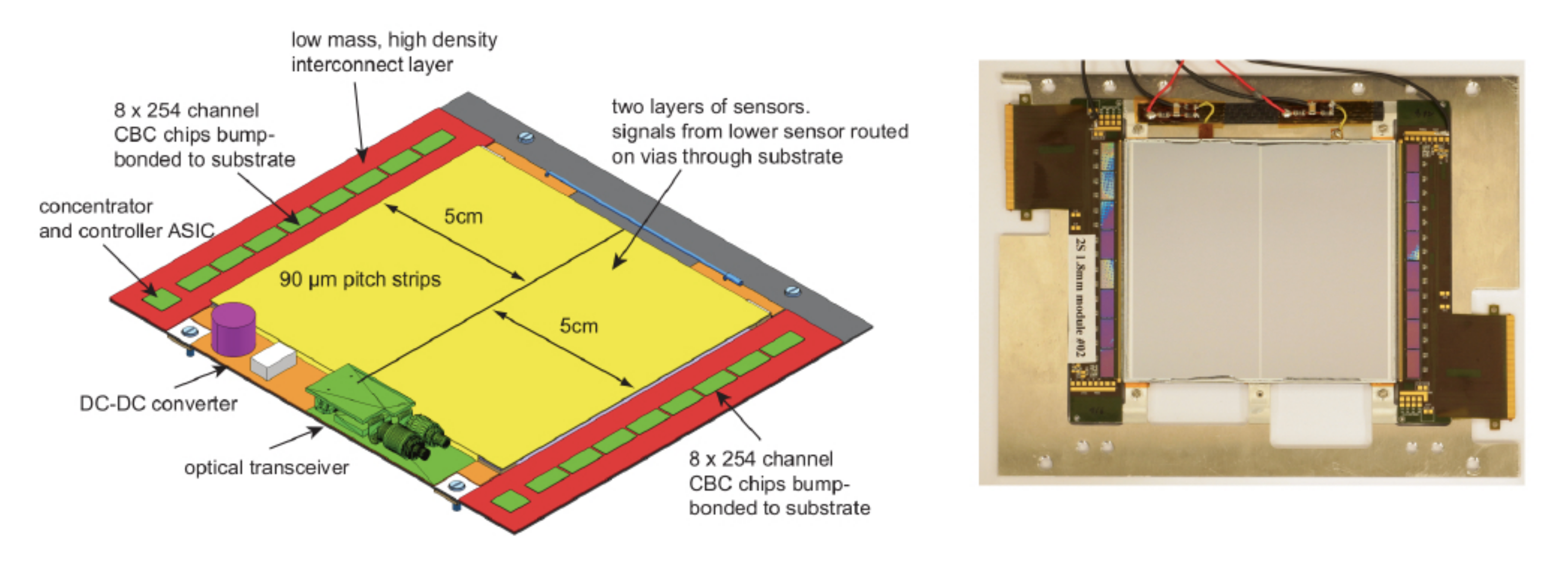}
\caption{Left: a schematic view of the CMS 2S module. Right: a picture of the CMS 2S module.}
\label{cms}
\end{center}
\end{figure}
The relative alignment of the Si detectors will be monitored with the high statistics provided by the muon beam. The mechanics of
the structure is still under optimization.
The relative distance between the Si tracker elements will be monitored by a laser-interferometry system. 

\item Knowledge of the beam: a 0.8\% accuracy on the knowledge of the beam momentum, as obtained by the BMS spectrometer used by COMPASS, is sufficient to control the systematic effects arising from beam spread. The beam scale must be known at the level of about $5\MeV$. This can be obtained by $\mu e$ elastic scattering events exploring the $\mu e$ kinematics~\cite{MUonE:LoI}.

\item Extraction of $\Delta\alpha_\text{had}(q^2)$ in the presence of radiative corrections: the signal extraction is carried out by a template fit method. $\Delta\alpha_\text{had}(q^2)$, the leading hadronic contribution to the effective electromagnetic coupling, is modeled by a parametric analytic function with logarithmic dependence at large $|q^2|$ and linear behavior at small $|q^2|$, as expected from general principles~\cite{MUonE:LoI}. Template distributions for the muon and electron scattering angles $\theta_{\mu}$ and $\theta_e$, both 1D and 2D, have been calculated from NLO Monte Carlo events on a grid of points in the parameter space sampling the region around the expected reference values. The template fit is then carried out by a $\chi^2$ minimization comparing the angular distribution of pseudo-data with the predictions $\amuHVPLO$ obtained for the scanned grid points. The extraction is consistent with the expected value within half a standard deviation.

\end{itemize}

\paragraph{Theory}

The present status of the theory prediction for $\mu e$ scattering is summarized in a recent report of the MUonE Theory Initiative~\cite{Banerjee:2020tdt}. In this section we briefly review some of the main results, recent activities, and future plans.

NLO QED radiative corrections to the differential $\mu e$ scattering cross section were computed a long time ago by applying some approximations and revisited more recently in Ref.~\cite{Kaiser:2010zz}. The complete calculation of the full set of NLO QED corrections and of NLO electroweak corrections with the development of a fully exclusive Monte Carlo event generator for MUonE was completed in Ref.~\cite{Alacevich:2018vez}. The generator is currently used for simulation of MUonE events in the presence of QED radiation.

The QED corrections at NNLO, crucial to interpret MUonE high-precision data, are not yet known, although
some of the two-loop corrections that were computed for Bhabha scattering in QED~\cite{Bern:2000ie,Bonciani:2003te,Bonciani:2003cj}
and for $t{\bar t}$ production in QCD~\cite{Bonciani:2008az, Bonciani:2013ywa} can be applied to $\mu e$ scattering as well. Several steps towards the calculation of the full NNLO QED corrections to $\mu e$ scattering were taken in Refs.~\cite{Mastrolia:2017pfy, DiVita:2018nnh, Mastrolia:2018sso, DiVita:2019lpl}, where all the master integrals for the two-loop planar and nonplanar four-point Feynman diagrams were computed. These results, obtained setting the electron mass to zero while retaining full dependence on the muon one, paved the way for the evaluation of the two-loop QED amplitude for massless electrons, which is now close to completion~\cite{Ronca:2019kcw}. The extraction of the leading electron mass effects from the massless $\mu e$ scattering amplitudes has been recently addressed in Ref.~\cite{Engel:2018fsb} (see also Refs.~\cite{Penin:2005kf,Mitov:2006xs,Becher:2007cu}). In addition to the two-loop amplitude, real--virtual and double-real contributions must be computed. In order to combine these contributions using dimensional regularization, a suitable NNLO subtraction scheme has to be implemented. One example of such a scheme is the $\text{FKS}^2$ scheme~\cite{Engel:2019nfw} developed very recently, which extends the NLO FKS subtraction scheme~\cite{Frixione:1995ms,Frederix:2009yq} to NNLO in the case of massive QED where only soft singularities are present. A large theoretical effort is under way to complete the full NNLO QED calculation~\cite{Banerjee:2020tdt}.

Besides pure QED, NNLO corrections also involve QCD contributions that cannot be computed perturbatively. They have been computed recently in Ref.~\cite{Fael:2019nsf}, using the dispersive approach with hadronic $e^+e^-$ annihilation (timelike) data. This approach, originally based on Ref.~\cite{Cabibbo:1961sz}, has also been employed to calculate the hadronic corrections to muon decay~\cite{vanRitbergen:1998hn, Davydychev:2000ee} and Bhabha scattering~\cite{Actis:2007fs,Kuhn:2008zs,CarloniCalame:2011zq}. The results of Ref.~\cite{Fael:2019nsf} show that these hadronic corrections will play a crucial role in the analysis of MUonE's data. Recently, taking advantage of the hyperspherical integration method, the NNLO hadronic corrections have also been calculated employing HVP in the spacelike region, without using timelike data~\cite{Fael:2018dmz}. This elegant alternative approach can in principle be included in the template fitting procedure.

To NNLO accuracy, electron pair production effects should also be considered and taken into account. Diagrams at one-loop and two-loops with VP insertions in the photon propagator were considered some time ago~\cite{Arbuzov:1995cn,Arbuzov:1995vi} for Bhabha scattering in the massless limit up to 0.1\% accuracy. Pair production shows in fact potentially large corrections of the form $(\alpha/\pi)^2\big[\log (-q^2/m_e^2 )\big]^3$, which are only canceled by the real pair production diagrams at the same order. The resummation of lepton pair production contributions has also been shown to take place to all orders of the perturbative expansion~\cite{Catani:1989et,Skrzypek:1992vk,Arbuzov:2010zzb}.

The extreme accuracy of MUonE demands the resummation of classes of higher-order corrections that are potentially enhanced by large logarithms. They can be organized in a power series of $\alpha/(2\pi)$ times powers of $L = \log\left(-q^2/m_e^2\right)$ (we refer here only to the case of radiation from the electron leg, which is numerically the most relevant one) and $\ell = -2\log\left(2\Delta\omega/\sqrt{s}\right)$, where $L$ and $\ell$ are the so-called {collinear log} and {infrared log} (or {soft log}), respectively. In the definition of $\ell$, $\Delta\omega$ is related to the maximum energy allowed for the radiation, which is in general a function of the applied cuts and the observable under consideration. Thanks to factorization theorems of soft and collinear radiation, the resummation techniques exponentiate the leading-log corrections up to all orders in $\alpha$ (terms of the form $\alpha^n (L-1)^n \ell^n$). A general framework for implementing the leading logarithmic resummation numerically is provided either by the QED Parton Shower (PS) approach or the Yennie\hyph Frautschi\hyph Suura (YFS) formalism (for a review of the approaches see for example Ref.~\cite{Actis:2010gg} and references therein). These methods can be improved to consistently include NLO corrections~\cite{Montagna:1996gw,Balossini:2006wc,Jadach:1995nk}. Going one step further, when the complete NNLO corrections will be available and a NNLO matched PS (or $\order(\alpha^2)$ YFS) will be implemented, we expect that the error due to missing corrections will start at order $\alpha^3L^2$, not enhanced by any infrared log $\ell$.

The precision expected at the MUonE experiment also raises the question whether possible new physics (NP) could affect its measurements. This issue was addressed in Ref.~\cite{Masiero:2020vxk} studying possible NP signals in muon--electron collisions at MUonE due to heavy or light mediators, depending on whether their mass is higher or lower than $\order(1\GeV)$. The former were analyzed in a model-independent way via an effective field theory approach, whereas for the latter the authors discussed scenarios with light spin-0 and spin-1 bosons. Using existing experimental bounds, it was shown that possible NP effects in muon--electron collisions are expected to lie below MUonE's sensitivity, therefore concluding that it is very unlikely that NP contributions will contaminate MUonE's extraction of $\Delta\alpha_\text{ had}(q^2)$. Reference~\cite{Dev:2020drf} addressed the sensitivity of MUonE to new light scalar or vector mediators able to explain the muon $g-2$ discrepancy, concluding that the measurement of $\Delta\alpha_\text{had}(q^2)$ at MUonE is not vulnerable to these NP scenarios. Therefore, the analyses of Refs.~\cite{Masiero:2020vxk,Dev:2020drf} reach similar conclusions where they overlap.\footnote{Some specific NP examples with heavy mediators were also discussed in Ref.~\cite{Schubert:2019nwm}, reaching broadly the same conclusions of the general analyses of Refs.~\cite{Masiero:2020vxk,Dev:2020drf}.} These results confirm and reinforce the physics case of the MUonE proposal.

\paragraph{Status and future plans}

The MUonE collaboration presently consists of groups from CERN, China, Germany, Greece, Italy, Poland, Russia, Switzerland, UK, and USA. These groups have strong expertise in the field of precision physics. A Letter of Intent has been submitted in June 2019 to CERN
SPSC~\cite{MUonE:LoI}, and a test run of a few weeks in 2021, with one two-station detector, has been recently approved. This test run will hopefully be followed by a full-statistics run in 2022\hyph24.

\subsubsection{Impact of future measurements on dispersive HVP}

Although BABAR has already published results of its ISR analyses for most of the final 
states below $2\GeV$, some processes with higher multiplicities and lower cross sections
are still under study: $\pi^+\pi^-3\pi^0$, $\pi^+\pi^-4\pi^0$, $2\pi^+2\pi^-3\pi^0$, 
$\pi^+\pi^-3\pi^0\eta$, $2\pi^+2\pi^-2\pi^0\eta$. Of special importance is the new ongoing 
analysis of the $\pi^+\pi^-$ channel, aimed at reducing systematic uncertainties of the 
published results (0.5\% in total on the $\rho$ resonance). To achieve this goal, a new 
method~\cite{michel17} has been developed to separate the $\pi^+\pi^-$, $K^+K^-$, and $\mu^+\mu^-$ 
processes without using particle identification (ID), which contributed the largest systematic 
uncertainty in the previous analysis. The separation is now performed by fitting the respective 
contributions in the angular distribution in the particle-pair CM system using their 
well-known expected shapes. This approach takes full advantage of the complete coverage of these 
distributions thanks to the large boost of the final state opposite to the high-energy ISR photon 
detected at large angles. The new analysis will also use full BABAR statistics, double that of the previous analysis. This doubling, together with the increase in efficiency from no longer using particle ID with strong cuts, will produce a final data sample for analysis seven times that of the previous BABAR study. 
The new and previous results will have a 
negligible statistical correlation and largely different systematic uncertainties.

Work on collecting larger samples of $e^+e^-$ annihilation data
aimed at obtaining much more precise hadronic cross sections is 
going on in two directions---scan (CMD-3 and SND at VEPP-2000) and ISR 
(BESIII at BEPCII and Belle II at SuperKEKB) measurements.

Two detectors at the only currently operating low-energy $e^+e^-$ collider 
VEPP-2000, CMD-3 and SND, continue data taking by scanning the CM energy 
range from the threshold of hadron production ($\approx 300 \MeV$) up to $2007\MeV$. Currently the integrated luminosity collected by each of the 
detectors is $63\,\text{pb}^{-1}$ in the low-energy range below $1030\MeV$
(the CM energy range with the $\rho$, $\omega$, and $\phi$ mesons
dominated by the low-multiplicity final states and giving the
largest contribution to HVP) and $185\,\text{pb}^{-1}$   
from 1030 to $2007\MeV$ dominated by the multi-particle final states.
The current plan is to continue data taking for another five years
with a goal of collecting about $1\,\text{fb}^{-1}$ in the whole energy range.

\begin{figure}
\centering
\includegraphics[width=0.51\textwidth]{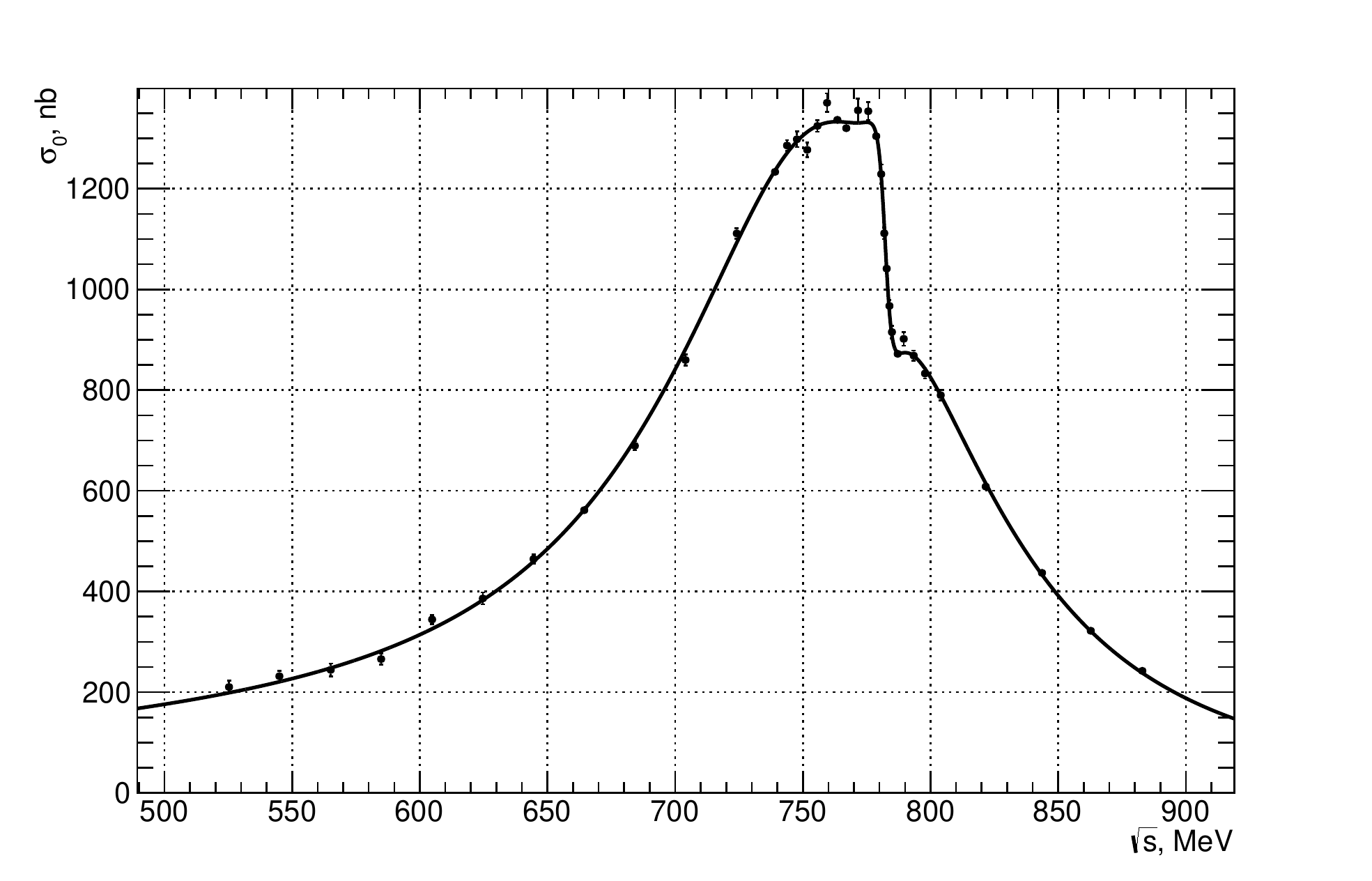}\hfill%
\includegraphics[width=0.49\textwidth]{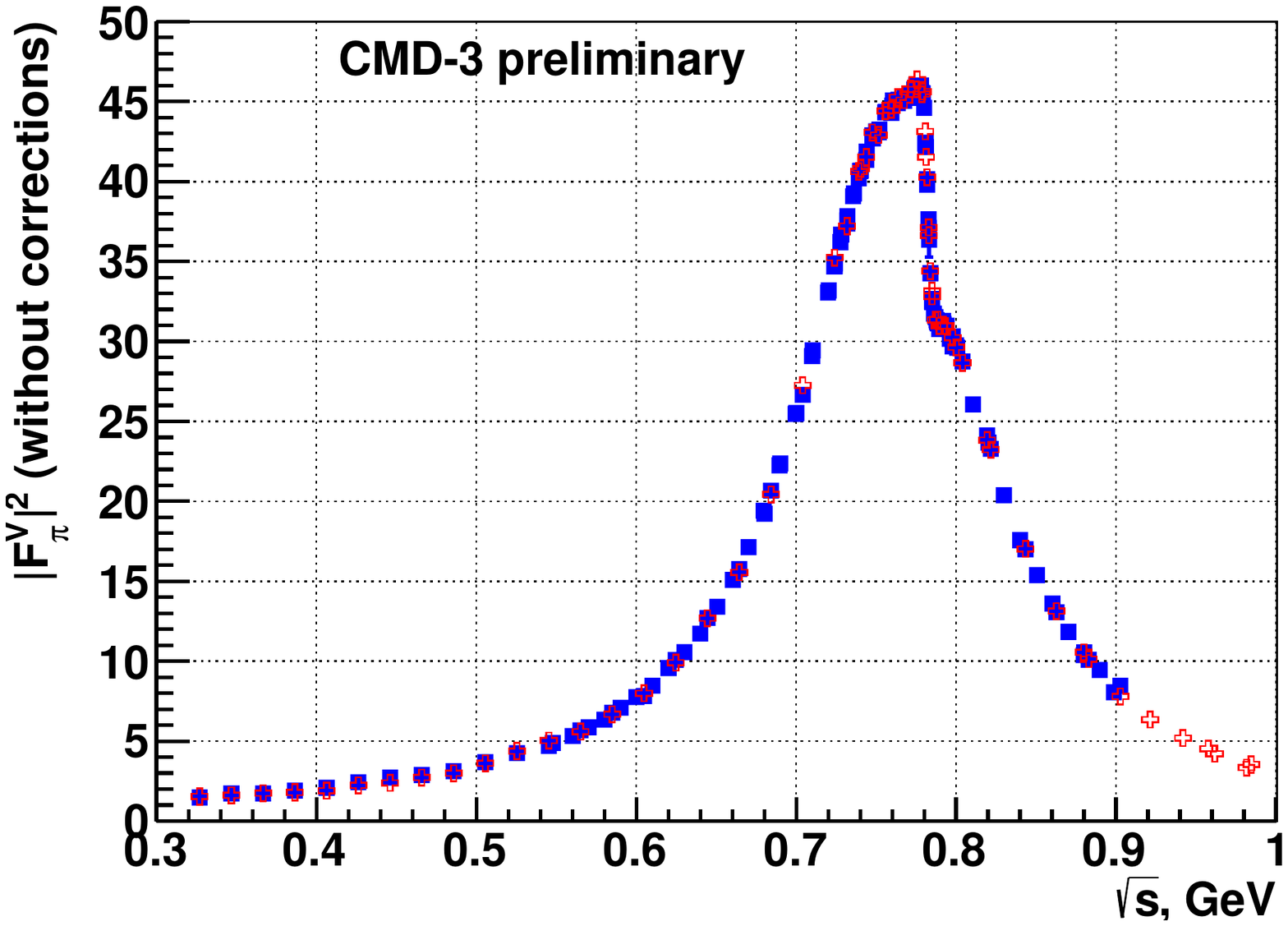}
\caption{\label{fig:twopion}Results on the $e^+e^-\to\pi^+\pi^-$ cross section 
at SND (left, reprinted from Ref.~\cite{Achasov:2020iys}) and on the squared pion
form factor at CMD-3 (right, preliminary, adapted from Ref.~\cite{Ignatov:2019omb}) measured in energy scans at VEPP-2000.}
\end{figure}

At both detectors, the available data sample of $e^+e^- \to \pi^+\pi^-$ events
around the peak of the $\rho$ resonance is already larger than in any other
experiment. The results for the cross section obtained at
SND~\cite{Achasov:2020iys} (left) and the pion form factor at CMD-3~\cite{Ignatov:2019omb} (right, preliminary) are 
shown in \cref{fig:twopion}.
The achieved systematic uncertainty for the squared form factor
is currently 0.8\% for SND and about 0.6\% for CMD-3. Work is in
progress and CMD-3 has a goal of reaching (0.4\hyph0.5)\%~\cite{Ignatov:2019omb}. 

As to the higher energy range, the goal is to reach a systematic uncertainty 
of (2\hyph3)\% on the final states with the largest higher-multiplicity cross section, 
$\pi^+\pi^-\pi^0$, $2\pi^+2\pi^-$, $\pi^+\pi^-2\pi^0$, etc.
Here the challenge is to disentangle various intermediate states leading
to the same final state, e.g., $a^\pm_1(1260)\pi^\mp$, $\rho^0 f_0(980)$,
etc., in the 4$\pi$ case. This task can be accomplished by performing 
       a more complicated multidimensional amplitude analysis in place of 
       the much simpler  one-dimensional analysis, already tried in the processes
$e^+e^- \to 3\pi^+3\pi^-$~\cite{Akhmetshin:2013xc} and 
$e^+e^- \to K^+K^-\pi^+ \pi^-$~\cite{Shemyakin:2015cba},
where the achieved systematic uncertainty is limited by the knowledge
of the production dynamics and reaches 6\% and 10\%, respectively.   
Finally, a scan of the CM energy range from $4.7\GeV$ to $7.0\GeV$
has been recently performed with the KEDR detector at the
VEPP-4M $e^+e^-$ collider in Novosibirsk~\cite{kedr2020}.
Data were taken at 17 energy points with an approximate
step of $150\MeV$. The goal is to measure $R$ with an accuracy of about 3\%. 

The ISR program of the BESIII collaboration focuses so far on the three most 
important channels $e^+e^-\to\pi^+\pi^-$, $e^+e^-\to\pi^+\pi^-\pi^0$, and 
$e^+e^-\to\pi^+\pi^-\pi^0\pi^0$. All investigations are performed on 
$2.93\,\textrm{fb}^{-1}$ of data taken at 
$\sqrt{s}=3.773\GeV$~\cite{Ablikim:2014gna}.

The analysis of the pion form factor is considered as the flagship project 
of the ISR program at BESIII. The first measurement, in which sub-percent 
accuracy has been achieved, is published~\cite{Ablikim:2015orh}. The total 
uncertainty is dominated by the contributions of the theoretical uncertainty 
of the radiator function and the uncertainty of the luminosity measurement, 
each of which is 0.5\%. Progress in the theoretical description of radiative 
corrections and further improvement of the understanding of the BESIII 
detector performance can help to reduce the total uncertainty of the pion 
form factor measurement. Alternatively, a different normalization scheme 
can be used, in which the two dominating uncertainties cancel. The 
normalization to the muon yield is also used in the most precise ISR 
measurements of the $\pi^+\pi^-$ cross section by the 
KLOE~\cite{Anastasi:2017eio} and BABAR~\cite{Aubert:2009ad} collaborations. 
However, based on the data evaluated at BESIII, the final result in this 
approach is limited by the statistics of the muon yield. The 
BESIII collaboration is considering acquiring additional $20\,\textrm{fb}^{-1}$ 
at $\sqrt{s}=3.773\GeV$~\cite{Ablikim:2019hff}. A combination of
these data and already recorded data sets should provide statistics 
sufficient to achieve a final accuracy of 0.5\% at BESIII. Additional 
efforts aim at extending the investigated mass ranges of the two-pion system. 
The published result is limited 
to masses between 600 and 900\,MeV, which  covers the peak of the 
       $\rho$ resonance, and determines approximately 50\% of the LO 
HVP contribution to $a_\mu$. The applied technique of exclusively 
reconstructing the ISR final state also allows one to measure the hadronic 
cross section at smaller masses, down to the $\pi^+\pi^-$ threshold. At 
higher masses, in addition, an inclusive measurement of the ISR production 
of the $\pi^+\pi^-$ channel is performed, where the ISR photon is 
reconstructed from the missing four-momentum of the pions, and events 
where the photon is not emitted along the beam axis are rejected. This 
strategy allows one to measure the cross section at hadronic masses 
above $1\,\textrm{GeV}$ with high statistics.

\begin{figure}
 \centering
 \includegraphics[width=0.49\textwidth]{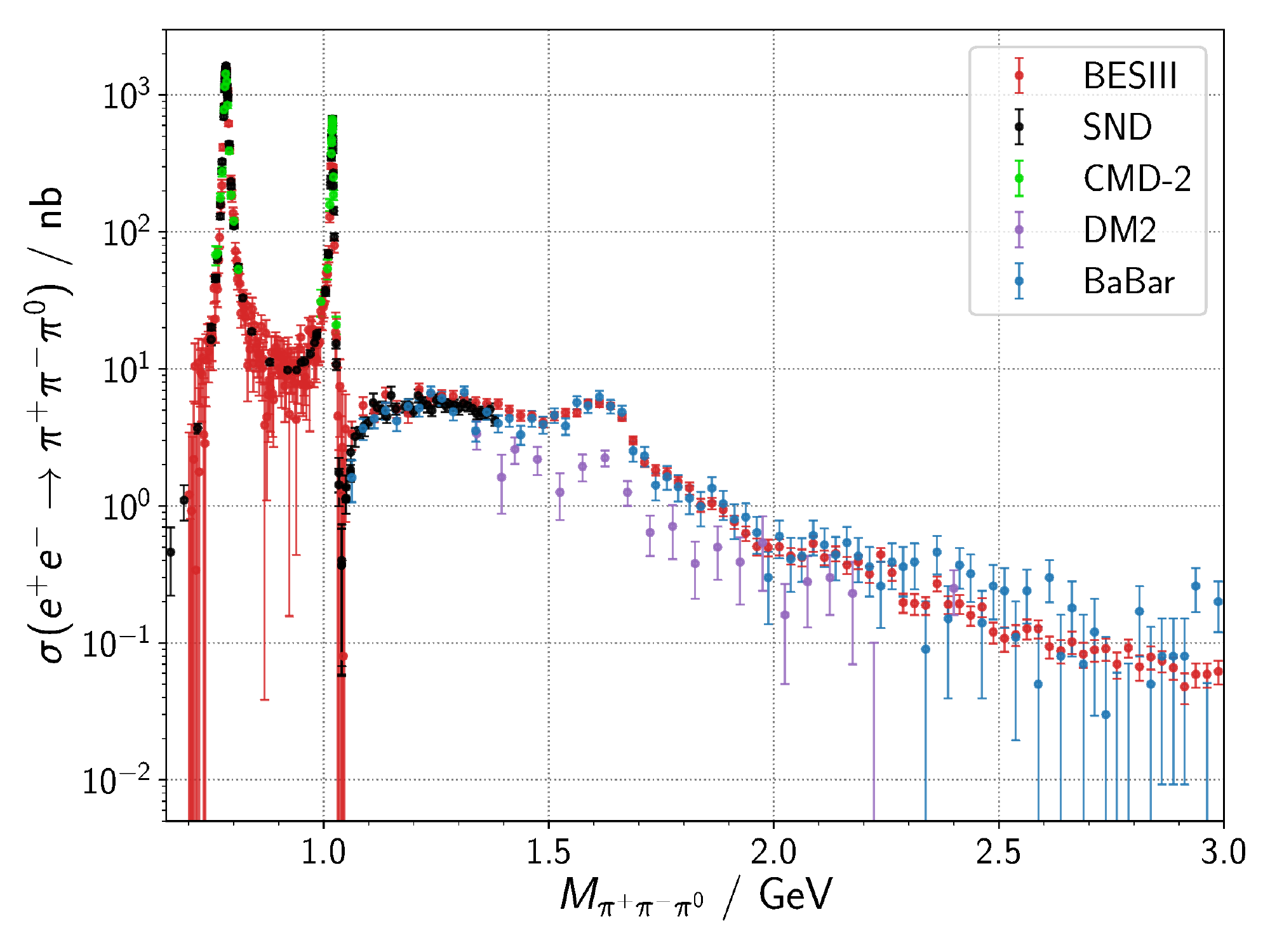}\hfill%
 \includegraphics[width=0.49\textwidth]{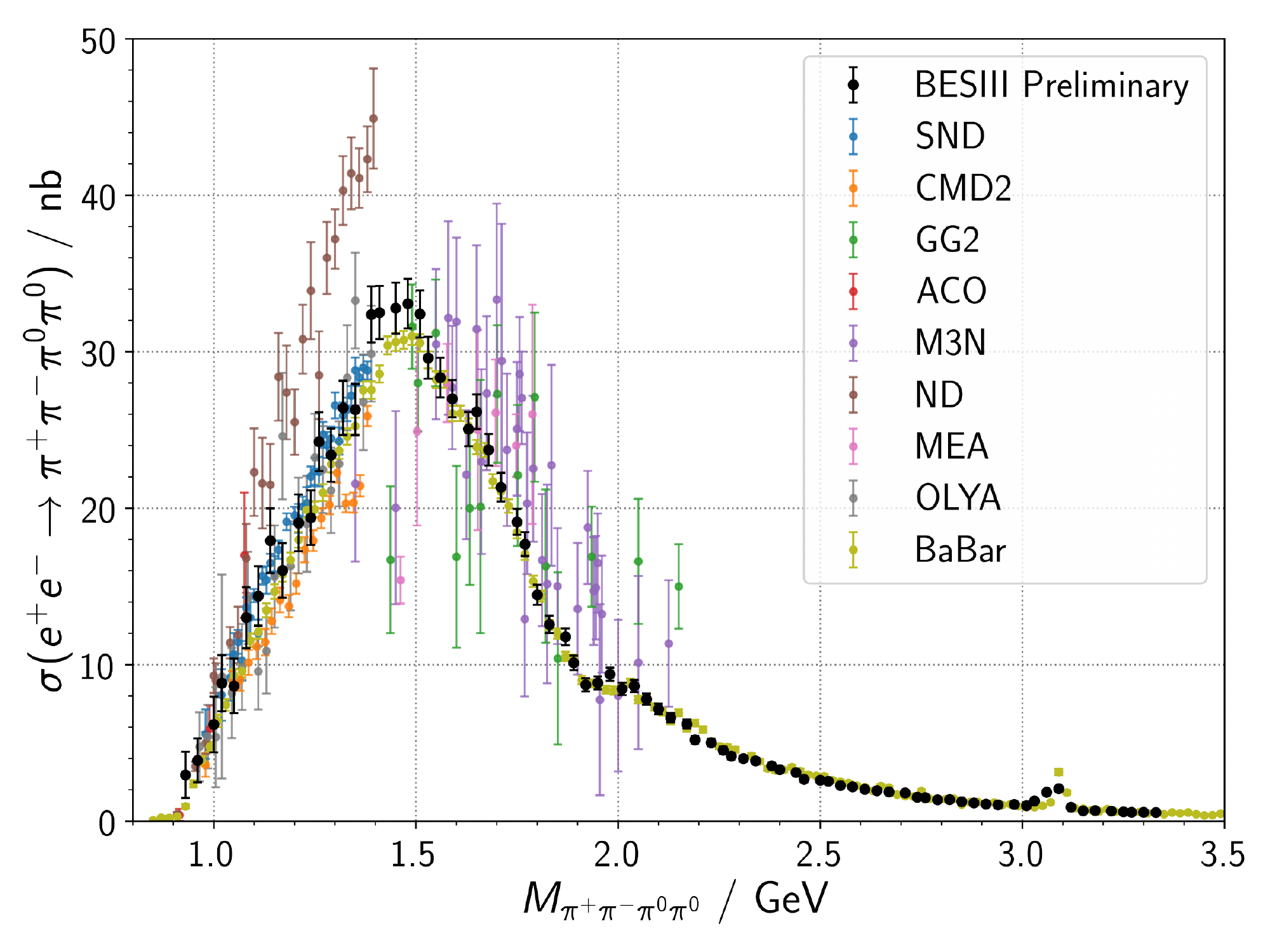}
 \caption{\label{fig:multipion}BESIII results on the 
$e^+e^-\to\pi^+\pi^-\pi^0$~\cite{Ablikim:2019sjw} (left, red) and 
$e^+e^-\to\pi^+\pi^-\pi^0\pi^0$~\cite{Redmer:2018nxj} 
(right, black) cross sections measured with ISR events in 
$2.93\,\textrm{fb}^{-1}$ of data at $\sqrt{s}=3.773\,\textrm{GeV}$. Previous 
results obtained in energy scans and by BABAR using the ISR method are 
shown for comparison~\cite{Antonelli:1992jx,Akhmetshin:1998se,Akhmetshin:2000ca,Achasov:2002ud,Aubert:2004kj,Cosme:1976tf,Cosme:1978qe,Esposito:1981dv,Bacci:1980zs,Kurdadze:1986tc,Dolinsky:1991vq,Akhmetshin:1998df,Achasov:2001gp,TheBaBar:2017vzo}. }
\end{figure}

The investigation of $e^+e^-\to\pi^+\pi^-\pi^0$~\cite{Ablikim:2019sjw} combines 
two analysis strategies, involving 
       either explicitly detecting the ISR photon or 
reconstructing it from the missing four-momentum. In this way, masses of the 
three-pion system covering the range from the peak of the $\omega$ resonance 
up to the $J/\psi$ resonance peak can be studied from the 
$2.93\,\textrm{fb}^{-1}$ of data at $\sqrt{s}=3.773\,\textrm{GeV}$. 
As a spin-off, the branching fractions of the narrow resonances are 
determined. The left panel of \cref{fig:multipion} illustrates the 
BESIII result in comparison with existing ISR and energy scan 
measurements. The systematic uncertainty of the cross section is found 
to be 2\% at the narrow resonances and 3\% in the remaining regions. 
The dominating background contribution to ISR production in the three-pion channel comes from the four-pion production with two neutral 
pions, which is studied separately.

The cross section of $e^+e^-\to\pi^+\pi^-\pi^0\pi^0$ is also measured as 
the error-weighted mean of the results of the 
       two ISR analysis strategies~\cite{Redmer:2019zzr}. The main 
background contribution comes again from the next higher neutral-pion 
multiplicity, which has been measured to tune MC distributions for 
background subtraction from data. The resulting preliminary cross section 
is shown in the right panel of \cref{fig:multipion}. The systematic 
uncertainty of the four-pion cross section is determined as 4\%, where 
the $\pi^0$ reconstruction efficiency becomes one of the important 
contributions, and, consequently, has been studied in detail.

The above results on both three- and four-pion cross sections 
illustrate that from the individual data set used in this analysis, 
BESIII can provide world-class accuracy results at hadronic masses 
above approximately $1.5\,\textrm{GeV}$. With the additional future 
data mentioned above, this accuracy will also be achieved 
at smaller masses.

In the Belle II experiment the Bhabha tagging at the level-one (L1) trigger 
was significantly improved by using accurate determination of  trigger cell 
clusters in the trigger electronics. In the L1 trigger there are several 
independent trigger modes, charged and neutral, to provide a careful measurement and monitoring of the trigger 
       efficiency from the experimental information 
during data taking.

\begin{figure}[t]
	\centering
	\includegraphics[width=0.5\textwidth]{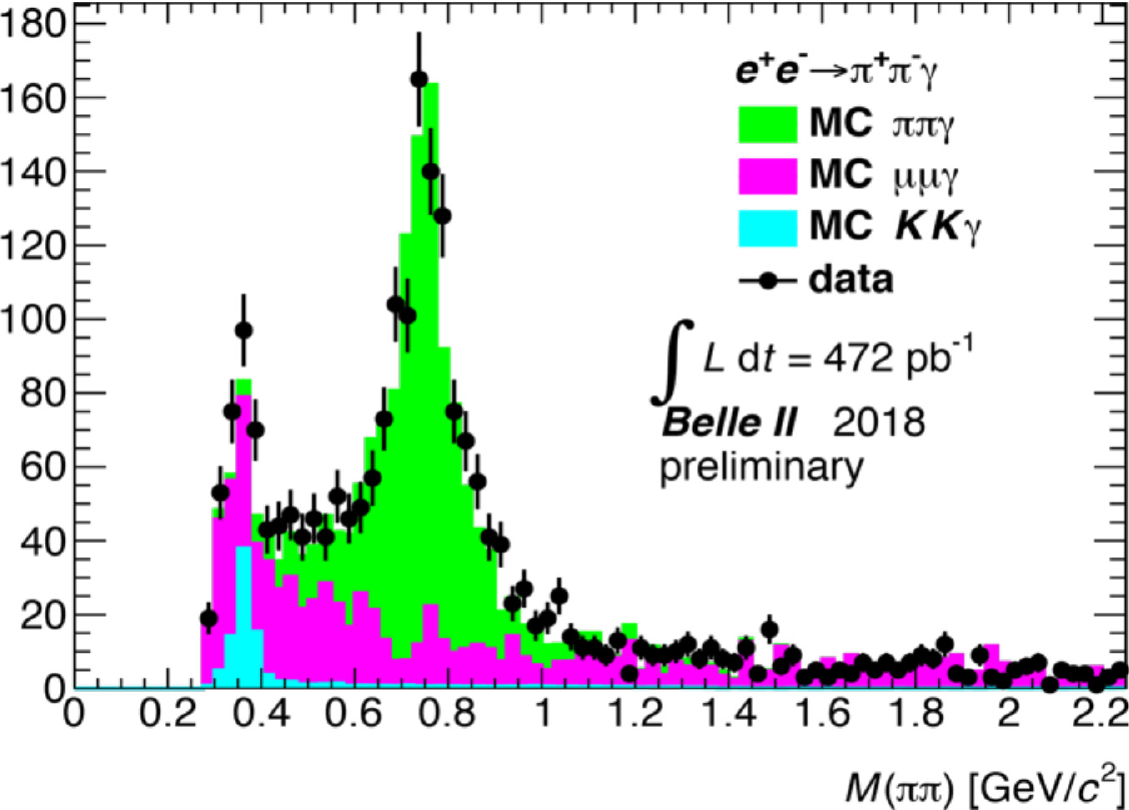}
	\caption{Preliminary results of the ISR studies at Belle II~\cite{Maeda:2018}.
	}
	\label{fig:belle2}
\end{figure}

In  spring\hyph summer of 2018 the first physics run was performed in the 
Belle II experiment. Among other processes, a first look at ISR events was 
carried out. The invariant mass of two charged particles assuming each to 
       have the pion mass is shown in \cref{fig:belle2} in comparison with MC 
simulations performed with the PHOKHARA code~\cite{Maeda:2018}.
It was demonstrated that the trigger efficiency for ISR events was
better than 90\%.

In the second Belle II physics run performed in March\hyph July of 2019,
about $5\,\text{fb}^{-1}$ of integrated luminosity was collected.
The data analysis is ongoing. It focuses on the  $\pi^+\pi^-$ and
 $\pi^+\pi^-\pi^+\pi^-$ channels with the goal of providing a careful check of the data  and their comparison with simulation. Plans for the future
suggest extending these studies to other final states, aiming
at a determination of the total cross sections for most of the hadronic 
channels when the collected luminosity achieves a few hundred $\text{fb}^{-1}$.

With $50\,\text{ab}^{-1}$ of data planned at Belle II, one hopes
to collect an order of magnitude higher data samples of various hadronic
final states. Together with new independent measurements at BESIII,
two detectors at VEPP-2000, as well as with those already existing at BABAR 
and KLOE, this should significantly improve the input, and hence also the 
       precision, of the data-driven HVP determination. Let us consider first 
the $\pi^+\pi^-$ channel. We will assume that in a few years there will
be five new independent data samples (BABAR, Belle II, BESIII, CMD-3, and SND), 
all giving close results. If each measurement has an uncertainty of 0.5\% 
dominated by systematic effects, an optimistic average of these data sets  
results in a $\pi^+\pi^-$-channel uncertainty of the LO hadronic contribution
of the order of $1\times 10^{-10}$ (0.2\%). One should also take into account
additional uncertainties due to higher-order radiative corrections~\cite{Campanario:2019mjh},
at the level of 0.2\% fully correlated between the data. Thus the total uncertainty
due to the $\pi^+\pi^-$ channel is estimated to be $1.5 \times 10^{-10}$. 
If for other exclusive channels we assume that each of the measurements achieves a 2\% 
systematic uncertainty, the total uncertainty of their contribution is $1.2 \times 10^{-10}$. 
Finally, taking from Ref.~\cite{Davier:2019can} $0.7 \times 10^{-10}$ as an estimate for the 
systematic uncertainty of the $R_{\rm QCD}$ contribution, we arrive at $\sim 2 \times 10^{-10}$ 
as a total uncertainty of the LO hadronic contribution or about 2 times better 
than today. Possible hidden correlations between various data and corresponding estimates 
can obviously affect the numbers above.  

\subsection{Summary and conclusions}
\label{HVPdisp_conclusions}

In recent years, data-driven evaluations of the HVP contribution to the muon $g - 2$ have been consolidated and improved to an accuracy below the percent level. In this section, the status of the hadronic data and of the evaluation of $\amuHVP$ has been reviewed, and a comparison and critical assessment of the most popular compilations has been presented. This has led to a detailed understanding of their differences and of the current limitations of the dispersive HVP approach. The limitations are not only due to the published (statistical and systematic) uncertainties of the hadronic data sets used as input, but arise from currently unresolved and significant discrepancies between data sets from different experiments in leading channels and, importantly, also from the limited knowledge of the correlated (systematic) uncertainties, see the discussion in \cref{Sec:UncertaintiesOnUncertainties}.

A main result of this section of the white paper is the prediction of the leading-order hadronic contributions, \cref{merging}, obtained through a merging procedure designed to take into account the differences between recent data-driven analyses and to provide a conservative estimate of the uncertainty. It reads
\begin{equation}
\amuHVPLO = 693.1(4.0) \times 10^{-10}\,.
\end{equation}
For the complete HVP contribution, this result has to be complemented by NLO and NNLO HVP contributions, obtained within the same framework, see \cref{HVPNLO} and \cref{HVPNNLO},
\begin{equation}
 \amuHVPNLO = -9.83(7) \times 10^{-10}\,,\qquad 
\amuHVPNNLO = 1.24(1) \times 10^{-10}\,.
\end{equation}
When adding the contributions, their errors should be taken as fully (anti-)correlated. These evaluations are based on Refs.~\cite{\HVPref} (LO), Ref.~\cite{Keshavarzi:2019abf} (NLO), and Ref.~\cite{Kurz:2014wya} (NNLO), with main experimental input from Refs.~\cite{\HVPexpref}.

To further improve these predictions, it will be crucial to resolve the major tensions between existing data sets, which have been discussed in \cref{ee-problems}, in particular in the leading $\pi^+\pi^-$ and important $K^+K^-$ channels. For the former, several new analyses are underway, see the brief discussion in \cref{ee-short-term-perspectives}. The expected new, high-accuracy data sets have the potential to resolve the puzzle in the $\pi^+\pi^-$ channel and hence to significantly improve the predictions. For the latter, further scrutiny of the existing data analyses should lead to a better understanding of their current discrepancies and in turn allow one to improve the accuracy of the $K^+K^-$ contribution. At present it is not possible to reliably predict how soon and to which extent such further improvements will be achieved. However, as discussed in \cref{HVPdisp_prospects}, with the prospect of many more data from several experiments, including Belle II, and the possibility of a direct and completely independent measurement of HVP in electron--muon scattering, there is a realistic chance to significantly improve the data-driven prediction of $\amuHVP$ within the next few years.

\FloatBarrier

\clearpage

\section{Lattice QCD calculations of HVP}
\label{sec:latticeHVP}

\noindent
 \begin{flushleft}
 \emph{T.~Blum, M.~Bruno, M.~C\`e, C.~T.~H.~Davies, M.~Della Morte, A.~X.~El-Khadra, D.~Giusti, Steven Gottlieb, V.~G\"ulpers, G.~Herdo\'iza, T.~Izubuchi, C.~Lehner, L.~Lellouch, M.~K.~Marinkovi\'c, A.~S.~Meyer, K.~Miura, A.~Portelli, S.~Simula, R.~Van de Water, G.~von Hippel, H.~Wittig}
 \end{flushleft}

\subsection{Introduction}
\label{sec:intro_latHVP}

In this section we review the status of lattice QCD calculations of the HVP contribution to the muon's anomalous magnetic moment. Our discussion is organized as follows: \Cref{sec:intro_latHVP} provides a general introduction followed by \cref{sec:strat}, which details the strategies employed in the various calculations.  In \cref{sec:comp} we compare recent lattice results and in \cref{sec:connect}  we discuss connections of HVP calculations with the MUonE experiment, with $\tau$ decays, and with the running of the electroweak coupling constants $\alpha$ and $\sin^2\theta_{\rm W}$.  Finally, in \cref{sec:summary_latHVP} we conclude with a summary of the current status and prospects for the future.

Within this subsection, we first discuss some of the basic ideas and formulae in \cref{subsec:intro_hvp}.  Then in \cref{subsec:inthvp}, we discuss the calculation of HVP as a function of momentum, its integration over momenta, and which techniques for calculating VP are most useful in different momentum ranges.  In \cref{subsec:moments}, we discuss the time moments method, which is followed by a discussion of the coordinate space representation in \cref{subsec:tmr}.  Finally, in \cref{subsec:common}, we provide a brief discussion of some of the issues common to all the methods.

\subsubsection{Hadronic vacuum polarization}
\label{subsec:intro_hvp}

Any lattice approach aiming to determine the leading hadronic contribution to the anomalous magnetic moment of
the muon starts from the correlator of the electromagnetic current
\begin{equation}
C^{(N_f)}_{\mu\nu}(x) = \left\langle j^{(N_f)}_\mu(x) j^{(N_f)}_\nu(0)\right\rangle\; ,
\label{eq:thecorrelator}
\end{equation}
where $j_\mu^{(N_f)}(x)= \sum_{f=1}^{N_f}Q_f \bar{\psi}_f(x)\gamma_\mu\psi_f(x)$, the index $f$ labeling quark flavors, and
$Q_f$ being the corresponding electric charge in units of the electron charge.
Traditionally, one performs a Fourier transform and introduces the VP tensor, 
\begin{equation}
  \Pi_{\mu\nu}^{(N_f)}(Q)=\int d^4x \,\mathrm{e}^{iQ\cdot x}\, C_{\mu\nu}^{(N_f)}(x) \,.
\label{eq:Four}  
\end{equation}  
In the continuum and in infinite volume, Euclidean invariance and current conservation allow one to rewrite the tensor as
\begin{equation}
  \Pi_{\mu\nu}^{(N_f)}(Q)=(\delta_{\mu\nu}Q^2-Q_\mu Q_\nu)\Pi^{(N_f)}(Q^2)\,.
 \label{eq:tenstoscal} 
\end{equation}
In finite volume and at finite lattice spacing,  the tensor decomposition of HVP is more complicated, because $SO(4)$ symmetry is explicitly broken to the finite hypercubic group through space-time discretization and boundary conditions~\cite{Bernecker:2011gh,Aubin:2015rzx}. The relation above is, however, recovered in the continuum and infinite-volume limits.

In order to obtain the leading  hadronic contribution to the anomalous magnetic moment of the muon ($\amuHVPLO$),
one performs an integration over $Q^2$. Specifically, (and suppressing the index $N_f$)
\begin{equation}
\amuHVPLO= \left( \frac{\alpha}{\pi} \right)^2 \int_0^\infty dQ^2 \, f(Q^2)
\hat{\Pi}(Q^2) 
\,,
\label{eq:thething}
\end{equation}
where  $\hat{\Pi}(Q^2)\equiv 4\pi^2\left[\Pi(0)-\Pi(Q^2)\right]$ and
\begin{equation}
f(Q^2)=\frac{m_\mu^2Q^2Z^3(1-Q^2Z)}{1+m_\mu^2Q^2Z^2}\,, \quad \quad
Z=-\frac{Q^2-\sqrt{Q^4+4m_\mu^2Q^2 }}{2m_\mu^2Q^2}\,,
\end{equation}
as derived in Refs.~\cite{Lautrup:1971jf,deRafael:1993za,Blum:2002ii,Gockeler:2003cw} for spacelike momenta. 

We see that in going from \cref{eq:thecorrelator} to \cref{eq:thething} 
one needs to perform
a Fourier transform (which implies a volume integral in coordinate space) and a weighted integral over momenta,
with a weight function (or kernel) $f(Q^2)$. One has the flexibility of performing these operations in different
orders, which produces the different approaches described in the following. While the final quantity is always
$\amuHVPLO$, intermediate expressions (e.g., concerning kernels) differ substantially and in practical implementations
each approach has its own virtues and drawbacks.

\subsubsection{\texorpdfstring{Calculating and integrating $\Pi(Q^2)$ to obtain $\amuHVPLO$}{Calculating and integrating Pi to obtain (g-2)}}
\label{subsec:inthvp}

Let us first consider the case where the  VP tensor, $\Pi_{\mu\nu}(Q)$, has been computed for a number of lattice momenta, perhaps including the use of {\it{twisted boundary conditions}}~\cite{DellaMorte:2011aa,Aubin:2013daa} in order to obtain a finer momentum resolution. 
What is usually computed is the zero-mode-subtracted VP tensor (obtained by replacing $e^{iQ\cdot x}$ with $e^{iQ\cdot x}-1$ in \cref{eq:Four}), as proposed in Ref.~\cite{Bernecker:2011gh}. 
This reduces contamination  from finite-volume effects~\cite{Aubin:2015rzx,Malak:2015sla}, and also removes contact terms, which would otherwise be present in the case $\mu=\nu$ (see, for example, Ref.~\cite{Gockeler:2003cw}). 
One can then obtain the scalar VP by using \cref{eq:tenstoscal}. In the often used setup, $\mu=\nu$, that equation takes the simple form
\begin{equation}
  \Pi(Q^2)=\frac{\sum_\mu \Pi_{\mu\mu}(Q)}{3Q^2}\,,
\end{equation}
from which one clearly sees that any imprecision in $\sum_\mu \Pi_{\mu\mu}(Q)$ at low $Q^2$ is enhanced in $\Pi(Q^2)$.  This has important consequences on the achievable accuracy for $\amuHVPLO$. The reason is that the kernel $f(Q^2)$ in \cref{eq:thething} diverges as $Q^2 \to 0$ so that the integrand $f(Q^2)\hat{\Pi}(Q^2)$ is peaked at values of $Q^2$
around a quarter of the lepton mass squared. These very low momenta cannot be directly accessed on current lattices since extensions of about 10\,fm would be required. Therefore, the lattice results in the accessible but most noisy region (the region of lowest precision) have to be extrapolated towards $Q^2=0$. 
The value of the scalar VP at vanishing momentum is in addition needed in order to compute the corresponding ultraviolet finite quantity $\hat{\Pi}(Q^2)$, which eventually enters the weighted integral discussed above. This subtraction constant can be obtained by fitting the data, as we describe in the following, but it can also be directly computed either through operator insertions, as proposed in Ref.~\cite{deDivitiis:2012vs}, or by calculating the low time moments of the vector correlator~\cite{Chakraborty:2014mwa}. The latter approach can be generalized to determine the HVP and Adler functions, as well as higher derivatives, at finite values of $Q^2$~\cite{Malak:2015sla}. 
An alternative approach to compute $\Pi(Q^2)$ is to utilize the derivatives of the free-energy density with respect to the background magnetic field~\cite{Bali:2015msa}.

Initial lattice determinations of $\amuHVPLO$ relied on fit ans\"atze to describe $\Pi(Q^2)$ over the entire range of available momenta (possibly extending it to infinity by matching to perturbation theory)~\cite{Blum:2002ii,Gockeler:2003cw,Feng:2011zk,Aubin:2006xv,Boyle:2011hu,DellaMorte:2011aa}. Those ans\"atze were inspired either by vector-meson dominance or more generally based on the use of rational functions.
In such approaches, however, the derivatives of $\Pi$ with respect to $Q^2$ of all orders are monotonic and decrease in magnitude with increasing $Q^2$, leaving such fits dominated by the high-precision, higher-$Q^2$ region where all curvatures are smaller than they are in the crucial low-$Q^2$ region, see for instance Ref.~\cite{Golterman:2014ksa}.
The so-called hybrid method~\cite{Golterman:2014ksa} was proposed to circumvent this problem.
\begin{figure}
\centering
\includegraphics[width=0.68\textwidth]{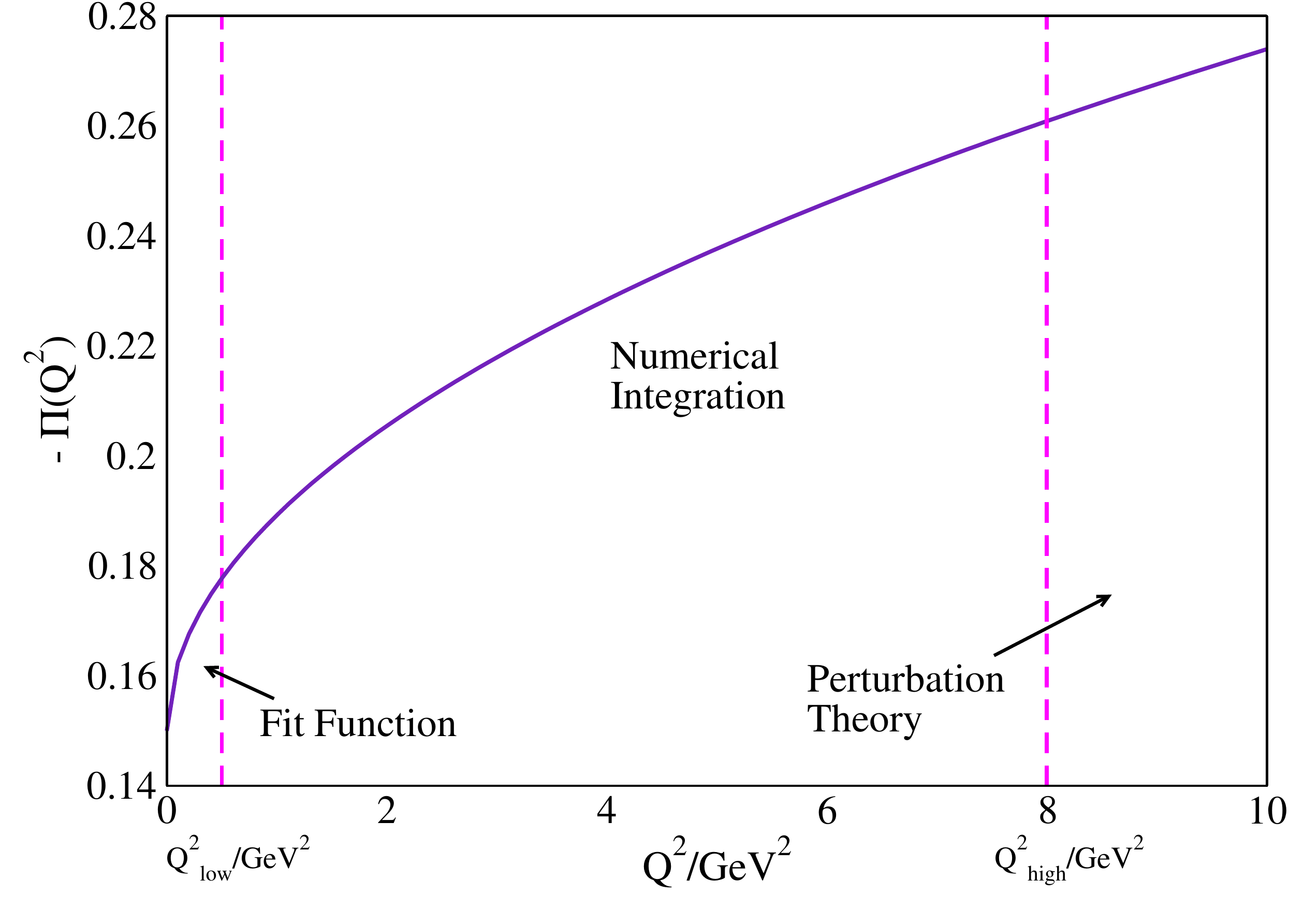}
\caption{Pictorial representation of the hybrid method.
  Different procedures are used in the three regions to obtain partial contributions to $\amuHVPLO$.}
  \label{fig:hybrid}
\end{figure}
The approach is depicted in \cref{fig:hybrid}. The entire $Q^2$ region is divided into three parts by introducing two cuts, $\Qlow$ and $\Qhigh$. The integral in \cref{eq:thething} is split accordingly, 
\begin{align}
\amuHVPLO&= I_0 + I_1 + I_2\,, \label{eq:intsum}\\
I_0&=\left(\frac{\alpha}{\pi}\right)^2 \int_{0}^{\Qlow}dQ^{2} f(Q^2) \times \hat{\Pi}(Q^2)\,,  \label{eq:subint0} \\ 
I_1&=\left(\frac{\alpha}{\pi}\right)^2 \int_{\Qlow}^{\Qhigh}dQ^{2} f(Q^2) \times \hat{\Pi}(Q^2)\,, \label{eq:subint1}\\
I_2&=\left(\frac{\alpha}{\pi}\right)^2 \int_{\Qhigh}^{\infty}dQ^{2} f(Q^2) \times \hat{\Pi}_\text{pert}(Q^2)\,,
\label{eq:subint2}
\end{align}
and different procedures are used in the three different regions. 
The integration from $\Qhigh$ (a few GeV$^2$) to infinity in \cref{eq:subint2} is done using perturbation theory. This is typically a very tiny contribution, at the permil level for the muon.
In the region between $\Qlow$ and $\Qhigh$ (the former being a hadronic scale $\approx 0.1 \GeV^2$), lattice results are very accurate and one can perform a simple numerical integration, for example by using the trapezoidal rule. This region contributes 10--20\% to $\amuHVPLO$, mostly depending on the actual choice for $\Qlow$.

The interval 0--$\Qlow$ in \cref{eq:subint0} is obviously the most important one and obtaining a reliable description of $\hat{\Pi}(Q^2)$ in this region is therefore crucial. The authors of Refs.~\cite{Aubin:2012me,Golterman:2014ksa} start from the observation that $\hat{\Pi}(Q^2)$ is related to a Stieltjes function, whose mathematical properties are well studied~\cite{Baker:1969sw}, in particular concerning the convergence of representations via Pad\'e approximants. They proposed, in fact, to use Pad\'e functions of the form
\begin{equation}
 \Pi_{[N,M]}(Q^2)=\Pi(0) + \frac{\sum_{i=1}^N a_i \, Q^{2i}}{ 1+\sum_{i=1}^M b_i \,Q^{2i} } \,, 
  \end{equation}
to  model the VP in the low-$Q^2$ regime. Mathematical theorems guarantee that asymptotically in $M$ and $N$ one obtains a model-independent description of the data
and the convergence of the Pad{\'e} sequence provides rigorous lower and upper bounds to the exact function $\hat{\Pi}(Q^2)$ of the form
\be
\Pi_{[N-1,N]}(Q^2) \leq \Pi_{[N,N+1]}(Q^2) \leq \hat{\Pi}(Q^2) \leq \Pi_{[N,N]}(Q^2)
 \leq \Pi_{[N-1,N-1]}(Q^2) \;.
\ee
However, in practical applications $M$ and $N$ are chosen to be equal to 2 or 3 at most. 
Another approach put forward in Ref.~\cite{Golterman:2014ksa} relies on a conformal change of variables in order to improve the radius of convergence of a simple Taylor expansion. In detail, the proposal consists of adopting the following fit model
\begin{equation}
  \Pi_N(Q^2)=\Pi(0)+ \sum_{i=1}^N p_i w^i\,, \quad w=\frac{1-\sqrt{1+z}}{1+\sqrt{1+z}}\,, \quad z=Q^2/4M_\pi^2\,.
\end{equation}  
In both cases (Pad\'e functions or conformal polynomials) the stability of the fits can be improved by supplementing them with estimates of the derivatives of $\Pi(Q^2)$ at zero momentum either through numerical differentiation or from the time moments, as we discuss in the following subsection.

Finally, we note that the hybrid method can naturally be adapted to include information on $\hat{\Pi}(Q^2)$ from experimental data at low $Q^2$. Indeed, the proposed MUonE experiment~\cite{MUonE:LoI} aims to provide a measurement of the VP function at spacelike $Q^2$ in exactly the low-$Q^2$ region that is problematic for lattice calculations. Here the split into three $Q^2$ regions as in \cref{eq:intsum,eq:subint0,eq:subint1,eq:subint2} is an integral part of the MUonE experiment's strategy, see \cref{subsec:muone} for more details. 

\subsubsection{Time moments}
\label{subsec:moments}

The method of time moments was introduced in Ref.~\cite{Chakraborty:2014mwa} as a way to calculate
the VP for small $Q^2$.  Starting from \cref{eq:tenstoscal}, we can look at the VP tensor with two identical spatial indices with $Q$ having only a time-component, \ie, $Q^\mu = (\omega, 0, 0, 0)$. Again dropping the superscript $(N_f)$, we have
\begin{equation}
\Pi_{kk}(Q) = Q^2 \Pi(Q^2) = \omega^2 \Pi(\omega^2)\,.
\label{eq:Pikkofq}
\end{equation}
Using \cref{eq:Four}, we can reexpress the right-hand side (RHS) in terms of the Fourier transform of the vector-current correlator:
\begin{equation}
 \omega^2 \Pi(\omega^2) = \int d^4 x \, e^{iQ\cdot x} C_{kk}(x)\,.
 \label{eq:omegasqPiOfOmega}
\end{equation}
Recognizing that we can pick any spatial index $k$, we can increase statistics by averaging over all three spatial directions. Further, since $Q\cdot x = \omega x_0$, we can define
\begin{equation}
C(x_0) = -\frac{1}{3}\sum_{k=1}^3 \int d^3 x \, C_{kk}(x)\ ,\label{eq:C_t}
\end{equation}
and write the RHS of \cref{eq:omegasqPiOfOmega} as
\begin{equation}
-\int d x_0 \, e^{i \omega x_0} C(x_0)\,.
\end{equation}
At this point, we may either consider the coefficients resulting from an expansion of the exponential in a power series or successively differentiate the RHS with respect to $\omega$ to Taylor expand around $\omega=0$.  In either case, the integrals with odd powers of $x_0$ vanish because $C(x_0)$ is an even function. The time moments are given by
\begin{equation}
G_{2n} \equiv \int_{-\infty}^{\infty} d x_0\, x_0^{2n} C(x_0) =
 (-1)^{n+1} \frac{\partial^{2n}}{\partial\omega^{2n}} \left(\omega^2 {\Pi}(\omega^2)\right)_{\omega=0}\,.
\end{equation}
Reverting to $Q^2$ rather than $\omega$ as the kinematic variable, we may write a power series for $\Pi(Q^2)$,
\begin{equation}
\Pi(Q^2) = \Pi_0 + \sum_{n=1}^\infty \Pi_n Q^{2n}\,,
\end{equation}
with
\begin{equation}
\Pi_0 = \Pi(0) = -\frac{1}{2}G_2\,, \qquad 
\Pi_{n} = \frac{(-1)^{n+1}}{(2n+2)!} G_{2n+2}\,.\label{eq:moments}
\end{equation}
A  Pad\'e approximation to $\Pi$ (and $\hat{\Pi}$) can be constructed from the lowest few time moments of the correlator~\cite{Chakraborty:2014mwa,Chakraborty:2015ugp}.

The time moments extend to infinitely large $x_0$; however, the lattices are of finite extent, so one usually models the long-time behavior of the current correlator $C(x_0)$ to extend the moment
integral (or sum) to infinity.  The same issue also arises in the related time-momentum method (see  \cref{subsec:tmr}), where it is discussed in more detail.  For the time moments, this issue clearly becomes more important for higher moments.  The achievable precision was discussed in Refs.~\cite{Chakraborty:2014mwa,Golterman:2014ksa}, for example.  

Time moments can also be used as input to a collection of approximants put forward in Refs.~\cite{deRafael:2014gxa,Charles:2017snx}. These arise from the use of Mellin--Barnes techniques and, in the cases analyzed in Ref.~\cite{deRafael:2014gxa}, are shown to converge to the full result very rapidly with the number of moments used. They have the advantage, over Pad\'e approximants, of allowing for a systematic matching to perturbation theory at short distance, though this advantage is of more formal than practical relevance, given the very small size of perturbative contributions to $\amuHVPLO$ (see \cref{subsec:inthvp}). 

The Taylor coefficients themselves are also useful as intermediate quantities enabling detailed comparisons between independent lattice calculations. In particular, the $\Pi_{n}$ for different $n$ have different sensitivities to the short- and long-distance systematic effects in lattice calculations (see \cref{sec:comp}). Finally, since the Taylor coefficients can also be evaluated using the data-driven methods discussed in \cref{sec:dataHVP}, they can be used to provide valuable tests of the lattice methods. 

\subsubsection{Coordinate-space representation}
\label{subsec:tmr}

An alternative way to write the subtracted VP in terms of the current correlator is given by~\cite{Bernecker:2011gh}
\begin{equation}
\hat\Pi(Q^2) = 4\pi^2 \int_0^\infty d x_0\, C(x_0) \left[x_0^2-\frac{4}{Q^2}\sin^2\left(\frac{Qx_0}{2}\right)\right] \,.
\label{eq:tmrpihat}
\end{equation}
Inserting this formulation of $\hat\Pi(Q^2)$ into \cref{eq:thething} for $\amuHVPLO$, one finds that
\begin{equation}
\amuHVPLO = \left(\frac{\alpha}{\pi}\right)^2 \int_0^\infty d x_0\, C(x_0) \widetilde{f}(x_0) \, ,
\label{eq:tmramu}
\end{equation}
where the kernel function
\begin{equation}
\widetilde{f}(x_0) = 8\pi^2\int_0^\infty \frac{d \omega}{\omega} f(\omega^2)\left[\omega^2x_0^2-4\sin^2\left(\frac{\omega x_0}{2}\right)\right]
\label{eq:tmrkernel}
\end{equation}
can be written explicitly in terms of a modified Bessel function of the second kind and Meijer's $G$ function~\cite{DellaMorte:2017dyu} as
\begin{equation}
\widetilde{f}(x_0) = \frac{2\pi^2}{m_\mu^2}\left[-2+8\gamma_{\rm E} + \frac{4}{\hat{t}^2} + \hat{t}^2-\frac{8}{\hat{t}}K_1(2\hat{t})+8\log\hat{t}+G^{2,1}_{1,3}\left(\left.\begin{array}{c}\frac{3}{2}\\0,1,\frac{1}{2}\end{array}\right|\hat{t}^2\right)\right]\, ,
\label{eq:tmrexplicitkernel}
\end{equation}
where $\hat{t}=m_\mu x_0$; numerically convenient series expansions for $\widetilde{f}$ are given in Appendix~B of Ref.~\cite{DellaMorte:2017dyu}. Alternatively, $\widetilde{f}(x_0)$
is evaluated numerically (e.g., as in Ref.~\cite{Borsanyi:2017zdw}).

While the main difficulty in the determination of $\amuHVPLO$ via $\hat{\Pi}(Q^2)$ lies in getting an accurate estimate of $\Pi(Q^2)$ in the low-$Q^2$ region, the main difficulty in determining $\amuHVPLO$ via \cref{eq:tmramu} lies in controlling the large-$x_0$ behavior of the integrand. The main issues are the exponential growth of the relative statistical error of $C(x_0)$ at large time separations, the presence of finite-volume (and potentially finite-temperature) effects in this regime, and the need to extend the $x_0$ integration beyond the region where lattice data are available.

To address the latter issue, it becomes necessary to split the integration range at some point $x_0^{\rm cut}$, where for $x_0\le x_0^{\rm cut}$ the correlator $C(x_0)$ is estimated by a local interpolation of the lattice data (with cubic splines working well in practice), while for $x_0> x_0^{\rm cut}$ a suitable extension derived from the lattice data supplemented with additional information is used instead.

The value chosen for $x_0^{\rm cut}$ impacts the overall error on $\amuHVPLO$ in two ways: if $x_0^{\rm cut}$ is chosen too large, the statistical accuracy deteriorates quickly due to the rapidly decaying signal-to-noise (StN\footnote{StN problems in lattice QCD have been studied since the pioneering works of Parisi~\cite{Parisi:1983ae} and Lepage~\cite{Lepage:1989hd} and arise when there are states contributing to a variance correlation function with less than twice the energy of the ground state of the correlation function. A possible solution to this problem can be found in the framework of multi-level Monte Carlo integration techniques for fermionic systems~\cite{Ce:2016idq,Ce:2016ajy,Giusti:2017ksp}.}) ratio of the correlator data; if $x_0^{\rm cut}$ is chosen smaller, the systematic error due to the model dependence of the extension of the correlator grows. In practice, at least for pion masses above the physical one, the effect is found to be negligible for the strange and charm-quark~\cite{Chakraborty:2014mwa,Giusti:2017jof} contributions as long as $x_0^{\rm cut}\ge 1.2$\,fm, whereas for the light-quark contribution~\cite{Chakraborty:2016mwy,Giusti:2018mdh,Giusti:2019xct} a window can be found within which the value of $\amuHVPLO$ is not significantly impacted by the precise choice of $x_0^{\rm cut}$ at least for pion masses larger than $200\MeV$.

\begin{figure}
\centering
\includegraphics[width=0.5\textwidth,keepaspectratio=]{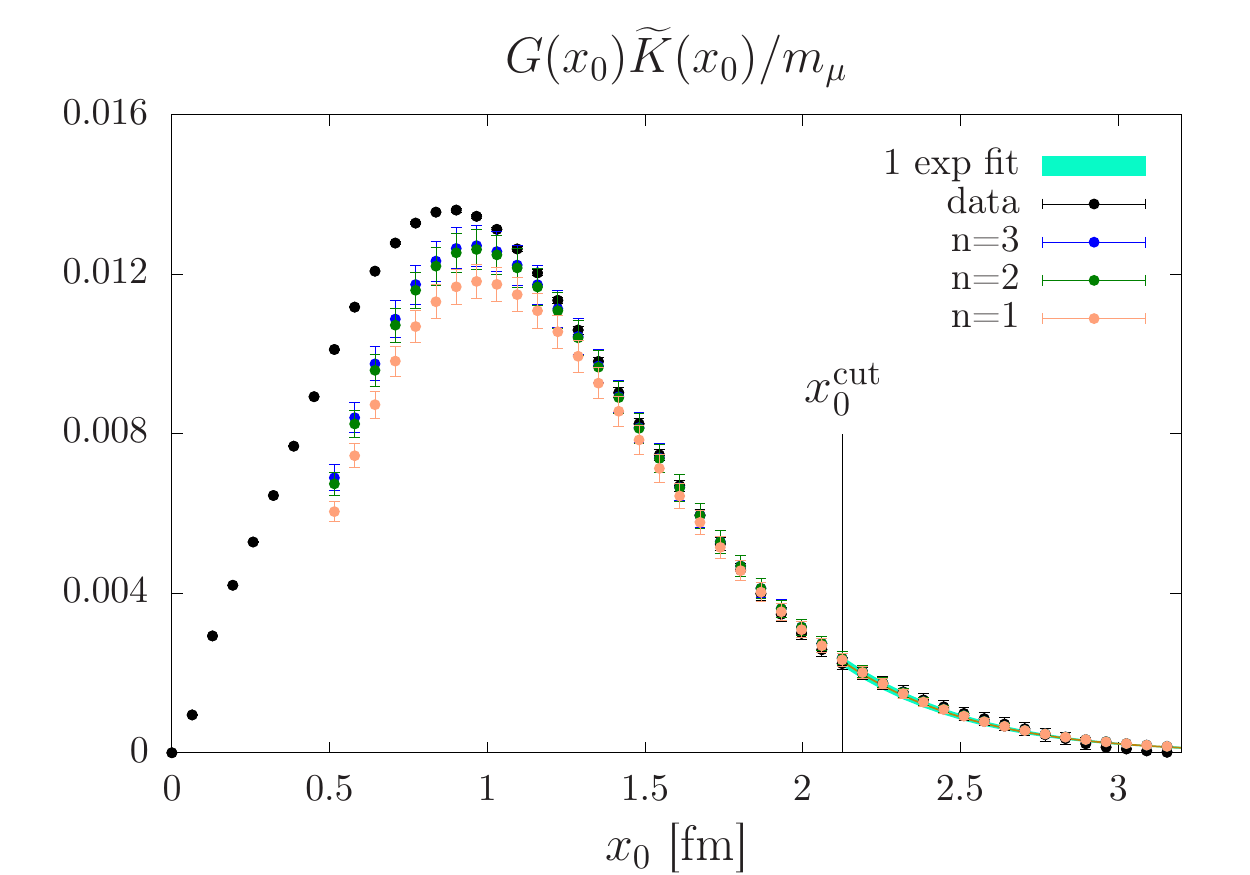}%
\includegraphics[width=0.5\textwidth,keepaspectratio=]{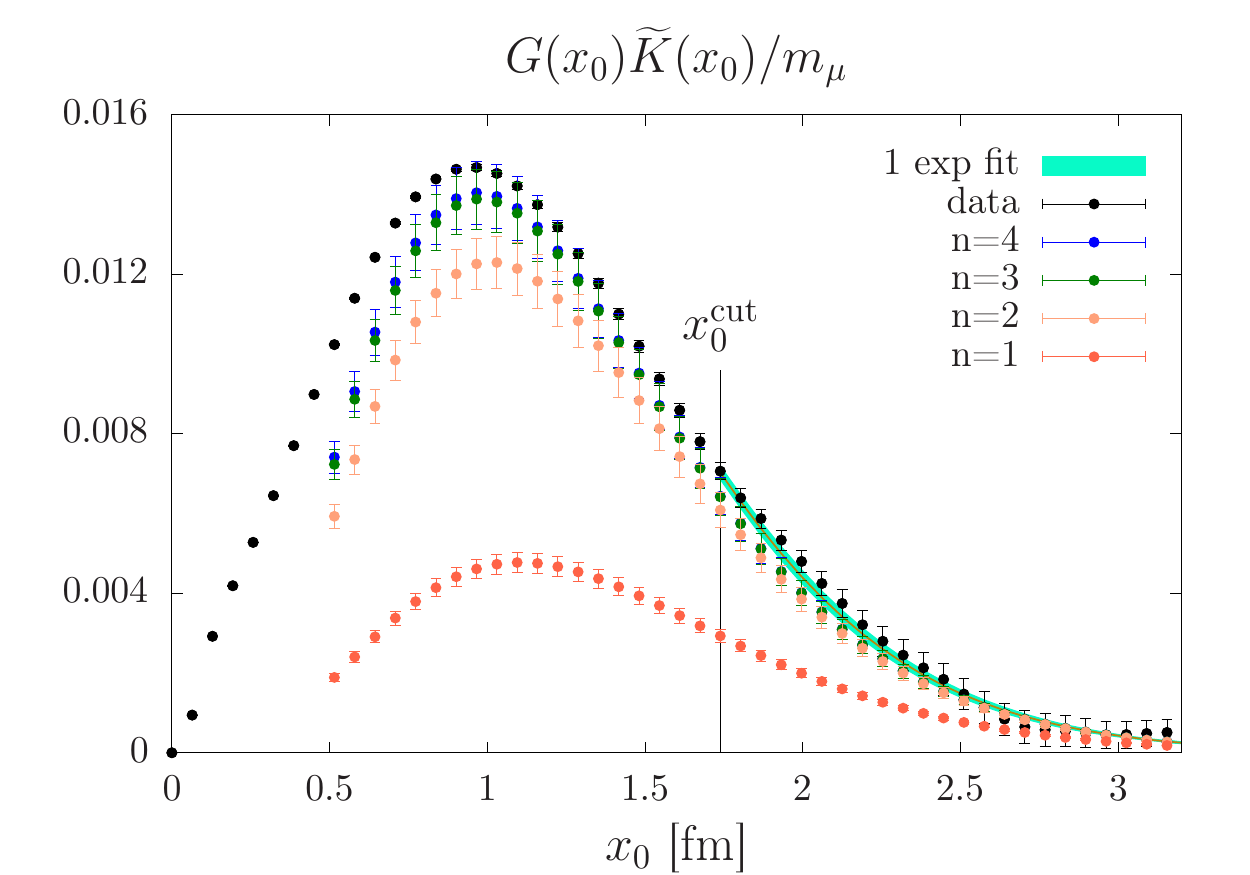}
\caption{The integrand of \cref{eq:tmramu} for the evaluation of the light-quark contribution to $\amuHVPLO$ in the time-momentum representation on $N_f=2+1$ lattice ensembles with pion masses of $M_\pi=280\MeV$ (left panel) and $M_\pi=200\MeV$ (right panel). Also shown are the results from reconstructing the correlator using $n_{\rm max}=1,\ldots,4$ states in \cref{eq:tmrspecrep} and the reconstruction of the long-time tail using a single-exponential extension. Left panel from Ref.~\cite{vonHippel:2018talk}, right panel adapted from Ref.~\cite{Gerardin:2018sin}.}
\label{fig:tmrlongtimetail}
\end{figure}

The basic form of the extension of the correlator is given by the spectral representation in a finite volume,
\begin{equation}
C(x_0) = \sum_{n=1}^\infty A_n {\rm e}^{-E_n x_0}\,,
\label{eq:tmrspecrep}
\end{equation}
where $E_n$ is the energy of an energy eigenstate $\left|n\right\rangle$ belonging to the representation $T_1$ of the cubic group, and $A_n$ is the associated matrix element of the electromagnetic current. Ideally, the low-lying finite-volume spectrum is known explicitly from a dedicated spectroscopic study, permitting the use of a truncated spectral sum for $C(x_0)$ beyond $x_0^{\rm cut}$~\cite{DellaMorte:2017khn}. 
Alternatively, the large-time behavior of the correlator can be modeled in various ways. The simplest model is a single-exponential extension, i.e., taking only one term in the series of \cref{eq:tmrspecrep} and fixing $E_1$ and $A_1$ from a fit to data at shorter time separations (using a smeared version of the vector correlator, where available, to extract $E_1$ with better precision)~\cite{DellaMorte:2017dyu,Giusti:2018mdh}. This model (which is essentially vector-meson dominance) is of course overly simplistic, and while it tends to describe the data well at heavy pion masses, it becomes a poor description of the very-long-time tail at light pion mass, where the two-pion channel opens (cf. \cref{fig:tmrlongtimetail}). 
A more sophisticated approach in the absence of detailed spectroscopic information is to model the finite-volume spectrum via the L\"uscher formalism~\cite{Luscher:1991cf,Meyer:2011um} applied to the Gounaris--Sakurai parameterization~\cite{Gounaris:1968mw} of the timelike pion form factor with parameters $\Gamma_\rho$, $M_\rho$ fixed via a fit to the lattice data~\cite{DellaMorte:2017dyu,Giusti:2018mdh}. 
The latter procedure also allows for correcting the leading finite-size effects by calculating the vector correlator in infinite volume from the timelike pion form factor and calculating $\amuHVPLO$ from there~\cite{Meyer:2011um,Francis:2013qna,Giusti:2018mdh}. Future studies, however, should perform a dedicated spectroscopic companion study. 

A third possibility is to implement rigorous upper and lower bounds on the correlation function~\cite{Borsanyi:2017zdw,Blum:2018mom}. These can then be used to replace the correlation function, at large $x_0$ where noise takes over, by a statistically more precise representation in terms of these bounds (see below).

We note that the coordinate space representation described in this section is related to the method of time moments (cf.\ \cref{subsec:moments}) in that the Taylor expansion of $\tilde f(x_0)$ in the integrand of \cref{eq:tmramu} yields the sum over time moments that gives $\amuHVPLO$ in that method. For a discussion of other related methods see Ref.~\cite{Meyer:2018til}.

\subsubsection{Windows in Euclidean time}

In the $a_\mu$ integral in \cref{eq:tmramu}, it is useful to consider different time regions in order to separate the short- and long-distance systematic lattice effects (discretization, finite volume, etc.). To this end, the RBC/UKQCD collaboration has proposed the window method~\cite{Blum:2018mom}, which breaks the time integral into three parts:
\begin{align}
\amuHVPLO &= \amuSD + \amuW + \amuLD \,, \notag\\
\amuSD &=  \left(\frac{\alpha}{\pi}\right)^2 \int_0^\infty d x_0\, C(x_0) \widetilde{f}(x_0) [1-\Theta(x_0,t_0,\Delta)]\,, \notag\\
\amuW &= \left(\frac{\alpha}{\pi}\right)^2 \int_0^\infty d x_0\, C(x_0) \widetilde{f}(x_0) [\Theta(x_0,t_0,\Delta)-\Theta(x_0,t_1,\Delta)]\,, \notag\\
\amuLD &= \left(\frac{\alpha}{\pi}\right)^2 \int_0^\infty d x_0\, C(x_0) \widetilde{f}(x_0) \Theta(x_0,t_1,\Delta)\,, \label{eq:win}
\end{align}
where $\Theta(t, t', \Delta) = [1 + \tanh [(t-t')/\Delta]] /2$ defines the window as a smooth function in Euclidean time with width parameter $\Delta$ that smoothes the window on both sides of the interval. A convenient choice is $\Delta = 0.15$\,fm. The parameters $t_0$ and $t_1$ are chosen to separate short- and long-distance effects, typically $t_0 \sim 0.4$\,fm and $t_1 \sim1.0$\,fm. Corrections due to systematic effects or subleading contributions can be applied to each window separately, and each window can also be separately extrapolated to the continuum and infinite-volume limits. In particular, the intermediate window $\amuW$ is, by design, expected to be less sensitive to discretization errors than the short-distance window $\amuSD$. It is also designed to be less sensitive to long-distance effects than $\amuLD$ and much less affected by the StN problems at large $x_0$ (see \cref{sec:comp}). In summary, we expect that $\amuW$ can, in principle, be calculated with better precision and better control over systematics than the complete $\amuHVPLO$. 

The windows, same as the Taylor coefficients of \cref{subsec:moments}, are useful intermediate quantities enabling detailed comparisons between independent lattice calculations. 
The window method also enables a comparison of intermediate quantities with results from data-driven methods. Here one first translates  the $R$-ratio data for $e^+e^-\to$ hadrons, see \cref{Rratio}, to Euclidean time via 
\begin{align}
    C(x_0) &= \frac{1}{12\pi^2}\int_{0}^{\infty} d(\sqrt{s})R(s) s e^{-\sqrt{s}x_0}\,,
    \label{eq:winwR}
\end{align}
which can then be used to construct the windows. In particular, as advocated by the RBC/UKQCD collaboration, the window method can be used to combine evaluations of $\amuSD$ and $\amuLD$ using $R$-ratio data with lattice determinations of $\amuW$ to yield a hybrid determination of $\amuHVPLO$ that is more precise than each alone~\cite{Blum:2018mom}. 

\subsubsection{Common issues}
\label{subsec:common}

One of the main issues affecting the calculation of $\amuHVPLO$ on the lattice is the difficulty of determining the long-distance behavior of the vector correlator with good statistical accuracy. This manifests itself in the direct approach to $\hat{\Pi}(Q^2)$ as a lack of precise information on the low-momentum regime even when using twisted boundary conditions to probe momenta below the lowest Fourier momentum on a given lattice size. 
In the method of time moments, the problem manifests itself as an increase of the statistical error for the higher moments. In the time-momentum approach, it manifests itself in the perhaps most direct way as a loss of statistical precision on the correlation function at large Euclidean time distances.

While the problem is the same in terms of its origin, the available remedies differ between the different methods. In the direct approach, the hybrid strategy uses analyticity to describe the low-momentum region with a Pad\'e approximant, whereas in the time-momentum approach, the spectral representation can be used to model the long-time behavior of the correlation function. In either case, additional information (such as from a dedicated spectroscopic study~\cite{Gerardin:2018sin,Feng:2014gba,Andersen:2018mau,Erben:2019nmx}) is needed in order to make the best use of the lattice data.

Control of finite-volume effects and the determination of the lattice scale are other important issues common to all lattice QCD calculations.  
The scale-setting uncertainty is of particular importance because, while $\amuHVPLO$ is a dimensionless quantity, it depends on the scale through the value $am_\mu$ of the muon mass in lattice units. Standard error propagation then shows that the uncertainty in $\amuHVPLO$ is related to the relative uncertainty in the quantity $\Lambda$ used to set the scale via~\cite{DellaMorte:2017dyu}
\begin{equation}
\Delta \amuHVPLO = \left|\Lambda \frac{d \amuHVPLO}{d\Lambda}\right| \times \frac{\Delta\Lambda}{\Lambda} = \left|M_\mu \frac{d \amuHVPLO}{dM_\mu}\right| \times \frac{\Delta\Lambda}{\Lambda}\,,
\end{equation}
where $M_\mu=m_\mu/\Lambda$ is the muon mass in units of $\Lambda$. By working in the coordinate-space representation, it can then be shown~\cite{DellaMorte:2017dyu} that $M_\mu\frac{d \amuHVPLO}{dM_\mu}=1.22\times 10^{-7}$, corresponding to an amplification factor of 1.8 from the relative scale setting error $\Delta\Lambda/\Lambda$ to the relative error $\Delta \amuHVPLO/\amuHVPLO$. This implies that reaching an accuracy of better than $1\%$ for $\amuHVPLO$ requires the lattice scale to be known at the few-permil level.

\subsection{Strategies}
\label{sec:strat}

As described in the previous section, at order ${\cal{O}}(\alpha^2)$  the hadronic contribution $\amuHVPLO (\alpha^2)$ is related to the average of the $T$-product of two electromagnetic currents over gluon and fermion fields, see \cref{eq:thecorrelator}.
Since all quark flavors contribute to the current, $\amuHVPLO (\alpha^2)$ can be split into two main contributions
\begin{equation}
     \amuHVPLO (\alpha^2)= \amuHVPLOconn  +  \amuHVPLOdisc\,,
     \label{eq:HLO}
\end{equation}
where the subscripts ``conn'' and ``disc'' indicate quark-connected and -disconnected contractions, respectively, for all flavors.
The various flavor-connected components have different statistical and systematic uncertainties and, therefore, they are usually calculated separately.
Thus, the quark-connected contribution $\amuHVPLOconn (\alpha^2)$ can be written as
\begin{equation}
    \amuHVPLOconn = \amuHVPLOud + \amuHVPLOs + \amuHVPLOc + \amuHVPLOb\,,
    \label{eq:connectedHLO}
\end{equation}
where the terms on the RHS correspond to the contributions of the light $u$- and $d$-quarks 
(treated in the isosymmetric limit $m_u = m_d$), 
of the strange, charm, and bottom quarks, respectively.

By definition, $\amuHVPLO (\alpha^2)$ and the terms in \cref{eq:HLO,eq:connectedHLO} do not include effects due to the electric charges of the (valence and sea) quarks, which would add corrections of ${\cal{O}}(\alpha)$ to $\amuHVPLO (\alpha^2)$. Effects due to the up--down mass difference yield corrections that are similar in size, since both $\delta m = (m_d - m_u)$ and $\alpha$ are parameters of order $1 \%$, which should be interpreted relatively to $\Lambda_{\text{QCD}}$ for $\delta m$. We note that almost all lattice QCD ensembles employed in current studies contain light sea quarks with degenerate masses, i.e., $\delta m = 0$. Hence, $\amuHVPLO (\alpha^2)$ and the terms in \cref{eq:HLO,eq:connectedHLO} are further defined to be evaluated in the isosymmetric limit ($m_u=m_d$).  Specifically, in this review the isosymmetric point is defined as the isospin-corrected pion mass suggested by FLAG~\cite{Aoki:2016frl}, namely $M_\pi = 134.8(3)\MeV$.  

Thanks to the recent progress in lattice determinations of $\amuHVPLO(\alpha^2)$ it becomes necessary to include the strong and electromagnetic isospin-breaking (IB) corrections discussed above.
A simultaneous expansion in $\delta m$ and $\alpha$ leads to contributions to $\amuHVPLO$ of order ${\cal{O}}(\alpha^2\delta m)$ and ${\cal{O}}(\alpha^3)$. Thus, the total LO HVP contribution $\amuHVPLO$ is given by
\begin{equation}
 \amuHVPLO  = \amuHVPLO (\alpha^2) + \delta \amuHVPLO\,, 
    \label{eq:HVP}
\end{equation}
with
\begin{equation}
    \delta \amuHVPLO = \delta \amuHVPLOud + \delta \amuHVPLOs + \delta \amuHVPLOc + \delta \amuHVPLOdisc\,,
    \label{eq:deltaHVP}
\end{equation}
where $\delta \amuHVPLOud$ includes both the strong and the QED IB corrections to the connected light-quark contribution, while $\delta \amuHVPLOs$ and $\delta \amuHVPLOc$ contain only QED effects of order ${\cal{O}}(\alpha^3)$: strong IB (SIB) corrections in these contributions only appear at subleading, $(m_d-m_u)^2$ order, which can be safely neglected even for a permil-precision calculation of $\amuHVPLO$.

It should be stressed that the separation in \cref{eq:HVP} and \cref{eq:deltaHVP}, into the isospin-symmetric flavor terms ($\alpha=0$ and $m_u=m_d$) $\amuHVPLO (\alpha^2)$ and the IB corrections $\delta \amuHVPLO$ is prescription and scheme dependent. Since the individual terms in \cref{eq:connectedHLO,eq:deltaHVP}  are typically calculated separately, it is important that the prescription used is fully specified (see \cref{sec:prescript}). To enable detailed comparisons between results from independent lattice calculations for these quantities, it is desirable to understand their prescription dependence. 
However, the total LO HVP contribution, i.e., $\amuHVPLO (\alpha^2) + \delta \amuHVPLO$, evaluated in the full QCD+QED theory is, of course, unambiguous. 

The following subsections describe the strategies adopted by lattice collaborations to compute the separate flavor terms appearing in \cref{eq:HVP,eq:deltaHVP}. This also includes descriptions of how the uncertainties coming from statistics, discretization effects, scale setting, finite-volume effects, long-distance effects due to $\pi\pi$ intermediate states, chiral extrapolation/interpolation, and quark mass tuning are estimated. The main results with their error budgets are collected and compared in \cref{sec:comp}. 

\subsubsection{Separation prescriptions}
\label{sec:prescript}

As discussed above, the impressive precision improvements of modern lattice QCD simulations now require calculations in the full QCD+QED theory, which includes explicit QED and strong IB effects, in order to produce physical results. An interesting question, which is of practical importance to the organization of lattice simulations, is to address the impact of SIB effects and QED corrections separately. This requires a careful definition of  ``QCD without electromagnetism'' and, generally speaking, it corresponds to stating the conditions that are used to determine the quark masses and the lattice spacing. Therefore, the separation of the QCD+QED theory into QCD plus corrections is unavoidably prescription dependent.

Presently the various lattice collaborations have adopted a diversity of separation schemes~\cite{deDivitiis:2013xla,Borsanyi:2013lga,Borsanyi:2014jba,Fodor:2016bgu,Giusti:2017dmp,Basak:2018yzz,DiCarlo:2019thl,Borsanyi:2020mff}. These schemes are based on defining a complete set of conditions fixing the physics when $\alpha$ becomes unphysical. These conditions are typically defined in term of hadronic masses or renormalized Lagrangian parameters.
A review of the different methods and a discussion of the subtleties concerning a precise definition of ``QCD without electromagnetism'' is given in Ref.~\cite{Aoki:2019cca}. 

For the reader's reference, here we provide details of the specific renormalization prescriptions adopted by the three groups that performed the first lattice calculations of the IB corrections $\delta \amuHVPLO$, namely ETM, Fermilab--HPQCD--MILC, and RBC/UKQCD (see \cref{sec:HVP_IB} for detailed descriptions of the calculations).
The renormalization scheme used by the ETM collaboration~\cite{Giusti:2019xct,Giusti:2017jof} is to impose that $m_{ud}$, $m_s$, $m_c$, and the strong coupling constant $\alpha_s$ match at $\overline{{\rm MS}}(2\,\text{GeV})$ in QCD$+$QED and in pure QCD~\cite{deDivitiis:2011eh,deDivitiis:2013xla}. This prescription was introduced by Gasser, Rusetsky, and Scimemi (GRS) in Ref.~\cite{Gasser:2003hk} and is known as the GRS renormalization prescription.
Many lattice groups, including the ETM collaboration, use the pion decay constant\footnote{We use the notation $f_\pi=\sqrt{2}F_\pi$ throughout this review.\label{pion_decay constant}} $f_{\pi^\pm} = 130.41 (20)\MeV$~\cite{Tanabashi:2018oca} to determine the lattice scale in physical units (see the discussion on scale setting in \cref{sec:HLO_ud}). This value of the pion decay constant is obtained from experimental measurements of the leptonic $\pi_{\mu 2}$ decay rate by subtracting the radiative corrections to the decay rate using a ChPT based estimate from Ref.~\cite{Cirigliano:2011tm}.\footnote{The $f_{\pi^\pm}$ result also uses $|V_{ud}|$ taken from nuclear beta decay as input.}  
In Refs.~\cite{Giusti:2017dwk,DiCarlo:2019thl} it has been shown that the IB corrections to the leptonic $\pi_{\mu 2}$ decay rate computed on the lattice in the GRS scheme agree with the corresponding ChPT estimate and the resulting change of $f_{\pi^\pm}$ turns out to be less than $\sim 0.2\%$. Such variation is found to be well within the level of the statistical precision of the lattice spacing determination by ETM.
The Fermilab Lattice, HPQCD, and MILC collaboration~\cite{Chakraborty:2017tqp} calculates SIB corrections to HVP using up- and down-quark masses determined by the MILC collaboration~\cite{Basak:2016jnn,Basak:2018yzz} to reproduce experimental hadron masses with QED corrections removed. A similar scheme was used by the RBC/UKQCD collaboration~\cite{Blum:2018mom,Gulpers:2018mim} to calculate IB corrections to HVP as a perturbative expansion around the isospin-symmetric result with light and strange quark masses determined in Ref.~\cite{Blum:2014tka}. IB corrections were determined by tuning the up-, down-, and strange-quark mass such that the experimental $\pi^+$, $K^+$, and $K^0$ masses are reproduced. In addition the RBC/UKQCD collaboration has determined the lattice spacing including QED by fixing the $\Omega^-$ mass to its experimental value. The shift of the $\Omega^-$ mass due to QED corrections is found to be significantly smaller than the lattice spacing uncertainty.

Ambiguities in choosing different prescription schemes are of ${\cal O}(\alpha m_f)$, where $m_f$ is the mass of a quark with flavor $f$. For light quarks these ambiguities are numerically of the same order as higher-order IB effects at the physical point~\cite{Borsanyi:2013lga,Aoki:2019cca} since $m_{ud}=\frac{1}{2}(m_u+m_d)\simeq m_d-m_u$. Hence, for the purposes of the present review the impact of the prescription dependence is expected to be small compared to the current uncertainties, particularly in the case of the light-quark contribution $\amuHVPLO(ud)$, which provides almost 90\% of the total $\amuHVPLO$. We stress again that $\amuHVPLO$ is free from ambiguities. However, in order to enable detailed comparisons between lattice QCD results for the individual contributions in \cref{eq:HVP} and \cref{eq:deltaHVP}, it will be important that future higher-precision calculations explicitly describe the prescription employed. Ideally, each lattice group would perform and discuss comparisons between different schemes.

\subsubsection{Connected light-quark contribution}
\label{sec:HLO_ud}

The connected contribution of light $u$- and $d$-quarks, $\amuHVPLOud$, is the most important hadronic contribution to $\amuHVPLO(\alpha^2)$. It represents $\approx 90 \%$ of $\amuHVPLO(\alpha^2)$. As a reminder, this intermediate quantity is defined in the isosymmetric limit ($\delta m=0$), where we take the physical point as the isospin-corrected neutral pion mass, $M_\pi = 134.8(3)\MeV$. 

In this subsection we consider the latest determinations of $\amuHVPLOud$ obtained by various lattice collaborations, namely HPQCD~\cite{Chakraborty:2016mwy},\footnote{The intermediate light-quark quantity reported by HPQCD as $\amuHVPLO (u/d)$ is evaluated at the physical charged pion mass, and so includes some SIB and QED corrections. It should not be confused with $\amuHVPLOud$ as defined here.} BMW~\cite{Borsanyi:2017zdw}, RBC/UKQCD~\cite{Blum:2018mom}, ETM~\cite{Giusti:2018mdh,Giusti:2019hkz}, Fermilab--HPQCD--MILC~\cite{Davies:2019efs}, PACS~\cite{Shintani:2019wai},  Mainz/CLS~\cite{Gerardin:2019rua}, and Aubin {\em et al.}~\cite{Aubin:2019usy}.
Details of the various lattice formulations adopted for the QCD action can be found in the corresponding references. 
However, note that the HPQCD, BWM, ETM, Fermilab--HPQCD--MILC, and Aubin {\em et al.} calculations are based on $N_f=2+1+1$ ensembles, while those of RBC/UKQCD, PACS, and Mainz/CLS  are performed on $N_f=2+1$ ensembles.  The former allow short-distance matching to the full SM at renormalization scales equal to $m_b$, up to $1/m_b^2$ corrections, while that scale is $m_c$ and corrections are of order $1/m_c^2$ for calculations based on $N_f=2+1$ ensembles. 
The BMW~\cite{Borsanyi:2017zdw}, RBC/UKQCD~\cite{Blum:2018mom}, ETM~\cite{Giusti:2018mdh}, Mainz/CLS~\cite{Gerardin:2019rua}, and Aubin {\em et al.}~\cite{Aubin:2019usy} evaluations of $\amuHVPLOud$ are obtained adopting the discretized version of the time-momentum representation \cref{eq:tmramu}, expressed directly in terms of the vector correlator $C(x_0)$ determined on the lattice at zero spatial momentum. 
Instead, the HPQCD~\cite{Chakraborty:2016mwy} results are based on  \cref{eq:thething},  
where $\hat{\Pi}(q^2)$ is constructed from the time-moments of the vector correlator $C(x_0)$ using the Pad\'e approximation (see \cref{subsec:moments}).
The Fermilab--HPQCD--MILC determination~\cite{Davies:2019efs}  is obtained using both the time-momentum representation and the HPQCD time-moment procedure~\cite{Chakraborty:2016mwy}. Finally, the PACS calculation~\cite{Shintani:2019wai} employs both the time-momentum representation and momentum-space integration methods. 

\paragraph{Statistical errors}

Both the signal and the noise in the vector correlator $C(x_0)$ decrease exponentially with the time separation $x_0$ between the source and the sink. 
The statistical precision of the vector correlator $C(x_0)$ is governed by the number of gauge configurations and by the number of sources (either point or stochastic) adopted in the calculation.
However, in spite of the high statistics generally used, in the case of the light $u$ and $d$ quarks the StN ratio becomes a severe issue, since in this case the signal deteriorates very quickly with increasing $x_0$.
Indeed, the exponential decay of the noise is controlled by the pion mass, while the one of the signal is dominated by vector-meson states at intermediate times and two-pion states at large times. 

A possible strategy for controlling the StN ratio is the replacement of the correlator $C(x_0)$ at large values of $x_0$ with its value determined by a (multi-)exponential fit to the data performed at smaller values of $x_0$.
This strategy has been adopted by Mainz/CLS~\cite{DellaMorte:2017dyu} (for $N_f=2$), HPQCD~\cite{Chakraborty:2016mwy}, ETM~\cite{Giusti:2018mdh,Giusti:2019hkz}, and Fermilab--HPQCD--MILC~\cite{Davies:2019efs}.
Typical values of the replacement point are between $\sim 2$ and $\sim 2.5$\,fm.

Instead, BMW~\cite{Borsanyi:2017zdw}, RBC/UKQCD~\cite{Blum:2018mom}, Mainz/CLS~\cite{Gerardin:2019rua} (for $N_f=2+1$), and Aubin {\em et al.}~\cite{Aubin:2019usy} introduce a cut $x_c$ beyond which the correlator $C(x_0)$ is replaced by the average of a lower and an upper bound.
The value of $x_c$ is chosen such that the upper and lower bounds agree within the statistical errors when the latter are not too large. 
All three adopt a value of $x_c$ close to $\sim 3$\,fm.

An improved version of the bounding method was investigated by RBC/UKQCD in Refs.~\cite{Lehner:2018LAT,Bruno:2019nzm}. It uses a set of correlation functions built from other interpolating operators in conjunction with the standard vector current operator to probe the long-distance region  to better precision.
The other correlation functions in the basis are designed to have strong overlap with the difficult-to-measure states in the spectrum, namely the lowest-lying $\pi\pi$ scattering state with back-to-back momentum.
From these correlation functions, the spectrum and matrix elements needed to reconstruct the long-distance region are computed precisely. These parameters are used to reconstruct the contribution to $C(x_0)$ from the lowest-lying states in the spectrum.
These states are subtracted from $C(x_0)$ and the bounding method is applied to the subtracted correlation function. Doing so improves the convergence of the upper and lower bounds and permits smaller values of $x_c$ to be taken, where the correlation functions have better statistical precision.
The remaining contribution from the subtracted states is added back after bounding.

\paragraph{Finite-volume effects and long-distance two-pion contributions}
\label{subsec:FVEs}

Finite-volume effects (FVEs) and long-distance two-pion contributions represent two very delicate, interconnected issues in the lattice evaluation of $\amuHVPLOud$.

The RBC/UKQCD~\cite{Blum:2018mom} and BMW~\cite{Borsanyi:2017zdw} determinations of $\amuHVPLOud$ include an estimate of FVEs based on ChPT at NLO. For the current--current correlation function NLO ChPT simply corresponds to scalar QED, and its prediction coincides with the estimate corresponding to the case of two noninteracting pions~\cite{Aubin:2015rzx}.
The HPQCD~\cite{Chakraborty:2016mwy} and Fermilab--HPQCD--MILC~\cite{Davies:2019efs} results include an extended version of ChPT, which accounts for $\gamma$\hyph$\rho$ mixing. 
However, the changes with respect to the standard NLO ChPT are found to be numerically small~\cite{Chakraborty:2016mwy}. 

Aubin {\em et al.} computed FVEs to NNLO in ChPT~\cite{Aubin:2019usy} in coordinate space. NNLO corrections were first computed by Bijnens and Relefors~\cite{Bijnens:2017esv} in momentum space.  The NNLO corrections are significant, almost 50\% of the NLO contribution with the same sign, for physical masses and lattices of size $L\sim5\hyph6$\,fm.

A study of FVEs on the lattice has been carried out by Mainz/CLS~\cite{DellaMorte:2017khn} and, more systematically, by ETM~\cite{Giusti:2018mdh}. They compared the results obtained for $\amuHVPLOud$ using different gauge ensembles sharing approximately the same pion mass  and lattice spacing, but differing in the lattice volume. Then, they adopt a representation of the vector correlator at intermediate and large time distances based on the L\"uscher formalism~\cite{Luscher:1991cf,Meyer:2011um} adopting the Gounaris--Sakurai parameterization~\cite{Gounaris:1968mw} for the timelike pion form factor with parameters $\Gamma_\rho$ and $M_\rho$ fixed via a fit to the lattice data.
Such a procedure enables an estimate of the vector correlator in the infinite-volume limit and, consequently, to estimate FVEs coming from intermediate and long-distance two-pion contributions.
The ETM collaboration has extended the representation of the vector correlator by including, at short times, a contribution based on quark--hadron duality, obtaining a nice reproduction of all their lattice data for time distances from $\sim 0.2$\,fm up to $\sim 2 \hyph2.5$\,fm. 
This procedure allows them to extrapolate the vector correlator to the physical pion point both at finite volumes and in the infinite-volume limit. 
The corresponding estimate of FVEs at the physical pion point turns out to be larger than the NLO ChPT prediction by a factor of $\sim 1.5 \hyph 2$~\cite{Giusti:2018mdh,Giusti:2019dmu}. This is in accord with the NNLO FVEs in Ref.~\cite{Aubin:2019usy}. 

Two groups, RBC/UKQCD~\cite{Lehner:2018LAT} and PACS~\cite{Izubuchi:2018tdd,Shintani:2019wai}, have studied FVEs by comparing  results at a few lattice volumes directly at or close to the physical pion point. Both obtain FV corrections that are larger than the NLO ChPT predictions by a factor of $\approx 1.7$. Using the bounding method, it is possible to calculate $\amuHVPLOud$ directly
 on two ensembles at the same lattice spacing with enough precision
 to resolve the FVEs, showing good agreement with the prediction
 from combining the Gounaris--Sakurai and L\"uscher formalisms~\cite{Lehner:2018LAT}. A novel proposal to use the input of the pion electromagnetic form factor to estimate finite-volume effects to $\amuHVPLO$ is discussed in Ref.~\cite {Hansen:2019rbh}. It should be mentioned that in the case of the HPQCD~\cite{Chakraborty:2016mwy}, BMW~\cite{Borsanyi:2017zdw}, and Fermilab--HPQCD--MILC~\cite{Davies:2019efs} calculations, which are performed with staggered fermions, the FVEs are expected to be significantly mitigated by the contributions of pions with different tastes, which are more massive than the pseudo-Goldstone pion. However, as discussed in the next paragraph, these effects are a source of additional discretization errors, which can also be estimated and corrected using ChPT. Indeed, as shown in Refs.~\cite{Chakraborty:2016mwy,Davies:2019efs,Aubin:2019usy}, it is advantageous to consider FV and taste-breaking effects together and apply them to $\amuHVPLO$ prior to the continuum extrapolation. 

\paragraph{Discretization errors and scale setting}

Discretization effects due to finite lattice spacing depend on the specific lattice formulation adopted for describing both the fermion and the gluon fields.
The QCD action on the lattice is usually on-shell ${\cal{O}}(a)$-improved, i.e., hadron masses are affected by discretization effects starting at second order in the lattice spacing, while the calculation of the HVP tensor in \cref{eq:thecorrelator} requires in general a dedicated improvement, since it receives off-shell contributions.
The vector correlator employed in Refs.~\cite{Chakraborty:2016mwy,Blum:2018mom,Borsanyi:2017zdw,Giusti:2018mdh,Davies:2019efs,Gerardin:2019rua} is ${\cal{O}}(a)$-improved.

In the case of the staggered formulations adopted in Refs.~\cite{Chakraborty:2016mwy,Borsanyi:2017zdw,Davies:2019efs,Aubin:2019usy}, generic discretization effects start at ${\cal{O}}(a^2)$~\cite{Borsanyi:2017zdw} and  ${\cal{O}}(\alpha_s a^2)$~\cite{Chakraborty:2016mwy,Davies:2019efs,Aubin:2019usy}, respectively, but additional ${\cal{O}}(\alpha_s^2 a^2)$~\cite{Borsanyi:2017zdw} and ${\cal{O}}(\alpha_s^2 a^2)$~\cite{Chakraborty:2016mwy,Davies:2019efs,Aubin:2019usy} effects that generate the taste splittings among the pions result in a sizable lattice spacing dependence of $a_\mu$. 
However, when the LO taste-breaking effects are taken into account using the predictions of one-loop staggered ChPT, the lattice spacing dependence is significantly reduced~\cite{Chakraborty:2016mwy,Davies:2019efs,Aubin:2019usy}. 
The NNLO ChPT results of Refs.~\cite{Aubin:2019usy,Bijnens:2017esv} can, in principle, be modified to also include high-order taste-breaking effects~\cite{Chakraborty:2016mwy,Davies:2019efs}. 

As discussed in \cref{subsec:common}, $\amuHVPLOud$ is sensitive to the scale setting despite being a dimensionless quantity, since the lepton kernel $f(Q^2)$ involves the physical lepton mass.
The scale is set using the PDG~\cite{Tanabashi:2018oca} value of the pion decay constant $f_{\pi^\pm}$ in Refs.~\cite{Chakraborty:2016mwy,Borsanyi:2017zdw,Giusti:2018mdh,Davies:2019efs,Aubin:2019usy}, a linear combination of pion and kaon decay constants in Ref.~\cite{Gerardin:2019rua}, the $\Omega^-$ mass in Ref.~\cite{Blum:2018mom}, and the $\Xi$-mass in Ref.~\cite{Shintani:2019wai}.

A strategy to reduce such a sensitivity significantly was first proposed by ETM~\cite{Burger:2013jya} and consists in replacing the physical muon mass in lattice units, $m_\mu^\text{phys}$, by an effective mass given by $m_\mu^\text{eff} = m_\mu^\text{phys} M_V / M_V^\text{phys}$, where $M_V$ is a hadronic quantity (also in lattice units) that can be calculated precisely in lattice QCD. In Ref.~\cite{Giusti:2018mdh}, $M_V$ is taken as the lightest vector-meson mass, determined from a finite-volume spectrum decomposition of the vector lattice correlator using the L\"uscher formalism. In Ref.~\cite{Gerardin:2019rua}, instead, the pion decay constant $f_\pi$ is used for this purpose.

The continuum limit is based on simulations at different values of the lattice spacing, where it is desirable to include at least three lattice spacings in the analysis. The extrapolation to the continuum is typically performed in a combined fit  together with the chiral extrapolation/interpolation to the physical pion point discussed next.

\paragraph{Chiral extrapolation/interpolation}

The QCD simulations are carried out around the isospin-symmetric physical pion point ($M_\pi = 134.8(3)\MeV$) in Refs.~\cite{Borsanyi:2017zdw,Blum:2018mom,Davies:2019efs,Shintani:2019wai,Aubin:2019usy}. In these cases only a smooth interpolation of the lattice data is required and the systematic uncertainties associated with this procedure are negligible. 
Instead, for the Mainz/CLS-17~\cite{DellaMorte:2017dyu} and ETM~\cite{Giusti:2018mdh} results, the QCD simulations are performed at heavier pion masses and the extrapolation to the physical pion point is carried out using ChPT-inspired fit functions~\cite{Golowich:1995kd,Amoros:1999dp,Bijnens:2016ndo,Golterman:2017njs}.
These chiral extrapolations are a significant source of error, but acceptable at the current few percent level precision. Finally,  Refs.~\cite{Chakraborty:2016mwy,Gerardin:2019rua} include  ensembles close to the physical point, as well as ensembles at heavier-than-physical pion masses. These calculations also employ ChPT-inspired fit functions to include the data at the heavy pion masses in the analysis, albeit to interpolate to the physical point. 
However, for subpercent precision, lattice calculations should be based entirely on ensembles at the physical point, which, as we discussed above, is already the case for some lattice efforts, and will soon be universal.

\subsubsection{Connected strange, charm, and bottom contributions}
\label{sec:HLO_sc}

The connected contribution of strange and charm quarks, $\amuHVPLOs$ and $\amuHVPLOc$, are important hadronic contributions to HVP at order ${\cal{O}}(\alpha^2)$.
They represent $\approx 8 \%$ and $\approx 2 \%$ of $\amuHVPLO(\alpha^2)$, respectively .
The contribution of the heavier $b$-quark was calculated on the lattice in Ref.~\cite{Colquhoun:2014ica}. The value obtained, $0.27(4) \times 10^{-10}$, is negligible at the current level of overall uncertainty on $\amuHVPLO$.  

The evaluation of $\amuHVPLOb$ and $\amuHVPLOc$ do not suffer from the difficulties discussed for the light-quark contribution $\amuHVPLOud$ in \cref{sec:HLO_ud}. The two-point vector heavyonium correlation function has a good StN ratio, but also falls very rapidly with time, $x_0$, on the lattice, because of the large mass of the heavyonium states. This means that, for example, time-moments can be calculated very accurately without the need for high precision at large times~\cite{Donald:2012ga}. Because heavyonium states are relatively small in spatial extent, finite-volume effects are also negligible on the lattice sizes typically being used. Discretization effects are more problematic for heavy quarks than light ones, and they will typically be the main source of uncertainty.

The time-moments of charmonium and bottomonium vector current--current correlators calculated in lattice QCD can be directly compared, in the continuum limit, to values of $q^2$-derivative moments extracted from experimental data for the $R$-ratio \cref{Rratio}. This is because the charm- and bottom-quark contributions to the experimental data can be isolated and the quark-line disconnected contribution, not included in the lattice QCD results, can be neglected. The lattice QCD calculations of the moments for the charm case in Refs.~\cite{Donald:2012ga,Nakayama:2016atf} and for the bottom case in Ref.~\cite{Colquhoun:2014ica} agree well with the values extracted from experiment. They provide a test against $R(s)$ on a flavor-by-flavor basis, which is equivalent to testing part of HVP. 

The latest determinations of $\amuHVPLOc$ come from HPQCD~\cite{Chakraborty:2014mwa} (using Ref.~\cite{Donald:2012ga}), ETM~\cite{Giusti:2017jof}, BMW~\cite{Borsanyi:2017zdw}, RBC/UKQCD~\cite{Blum:2018mom}, and Mainz/CLS~\cite{Gerardin:2019rua}. They agree well, as is seen by the comparison in \cref{subsec:compare_fbf}.

The evaluation of $\amuHVPLOs$ is similarly straightforward in lattice QCD~\cite{Chakraborty:2014mwa}. The vector strangeonium states have relatively heavy masses ($\gtrsim 1\GeV$) so that, although the StN is worse than for the charm and bottom cases and the numerical cost is larger for high statistics, accurate correlators can be obtained for the full range of $x_0$ values needed to determine the contribution to HVP by any of the methods discussed in \cref{sec:intro_latHVP}. 
Consequently, the choice of specific strategies made by the various collaborations is not important. Finite-volume effects might be expected here from the fact that the ground-state of the physical strangeonium vector correlator is a $K\bar{K}$ configuration, which will be distorted on the lattice at finite volume. In practice this has very little impact because the $\phi$ is so close to threshold and finite-volume effects are seen to be very small~\cite{Chakraborty:2014mwa, Giusti:2017jof}. $\amuHVPLOs$ is sensitive to the tuning of the valence strange quark mass and to light and strange sea-quark masses~\cite{Blum:2016xpd}. The impact of this in different calculations depends on how these are handled. Those uncertainties are at present generally smaller than the uncertainty coming from the determination of the lattice spacing. 

Again lattice QCD calculations for $\amuHVPLOs$ agree well between different lattice formalisms and methods, see \cref{subsec:compare_fbf}. This is a good test of the different approaches to calculating  HVP in lattice QCD, independent of the specific issues that arise in the $\amuHVPLOud$ case.

\subsubsection{Disconnected contributions}
\label{sec:HLO_disconn}

Results for the quark-disconnected contribution $\amuHVPLOdisc$ have been presented in
Refs.~\cite{Chakraborty:2015ugp, Blum:2015you, Borsanyi:2016lpl,
Borsanyi:2017zdw, DellaMorte:2017dyu, Gerardin:2019rua}. The estimate of $\amuHVPLOdisc$ in Ref.~~\cite{Davies:2019efs} is based on a ChPT-inspired model. In addition, the size of the disconnected
contribution to the VP function $\hat\Pi$ has been
analyzed in ChPT at NLO~\cite{DellaMorte:2010aq} and NNLO
\cite{Bijnens:2016ndo}. Direct calculations~\cite{Blum:2015you,
Borsanyi:2017zdw, Gerardin:2019rua} show that the absolute magnitude
of disconnected diagrams amounts to around $-2$\% of the total value
of $\amuHVPLO$. Hence, while the disconnected contribution is not
dominant, its accurate determination is crucial regarding the
precision target.

Calculations of $\amuHVPLOdisc$
typically employ the time-momentum representation and the spatially
summed vector--vector correlator $C(x_0)\equiv C_{\rm conn}(x_0)+C_{\rm disc}(x_0)$,
where $C_{\rm conn}(x_0)$ denotes the connected part of
the correlator with individual contributions from the light ($ud$),
strange ($s$), charm ($c$), and bottom ($b$) quarks (see \cref{eq:HLO} and \cref{eq:connectedHLO} and surrounding discussion). 
The disconnected contribution $\amuHVPLOdisc$ is then
defined via the integral representation of \cref{eq:tmramu}, i.e., 
\begin{equation}
\amuHVPLOdisc
= \left(\frac{\alpha}{\pi}\right)^2 \int_0^\infty dx_0\, 
C_{\rm disc}(x_0)\, \widetilde{f}(x_0) \, .
\end{equation}
It is instructive to analyze the long-distance behavior of the
correlator $C(x_0)$ and its isospin decomposition
$C(x_0)=C^{I=1}(x_0)+C^{I=0}(x_0)$, where $I=1, 0$ denote the
isovector and isoscalar contributions. First, one realizes that the
isovector component $C^{I=1}(x_0)$ is entirely given in terms of the
connected light-quark contribution. Therefore, in the isospin limit
one has
\begin{equation}
C^{I=1}(x_0) = \frac{9}{10}C^{ud}_{\rm conn}(x_0)\, ,
\end{equation}
where the prefactor is equal to $1/2$ times the inverse of the sum of the squared electric
charges. In the long-distance regime, the correlator $C(x_0)$ is
saturated by the two-pion contribution, which implies
\begin{equation}
  C(x_0) \stackrel{x_0\to\infty}{\longrightarrow}
  C^{I=1}(x_0) \left(1+ \order(e^{-M_\pi x_0}) \right)\,.
\end{equation}
In addition, from the observation that the isoscalar spectral function
vanishes below the three-pion threshold, one can derive the asymptotic
behavior of the disconnected/total correlator
ratio~\cite{Francis:2013qna,DellaMorte:2017dyu} as
\begin{equation}
  \frac{C_{\rm disc}(x_0)}{C(x_0)}
  \stackrel{x_0\to\infty}{\longrightarrow} -\frac{1}{9}\,.
\label{eq:asympratio}
\end{equation}
The relative size of the disconnected and connected contributions to
the VP function has also been studied in ChPT. At
one-loop order one finds $\hat\Pi_{\rm disc}/\hat\Pi_{\rm
conn}=-1/10$~\cite{DellaMorte:2010aq}, while the two-loop correction
computed in Ref.~\cite{Bijnens:2016ndo} reduces the magnitude
significantly, 
to $\hat\Pi_{\rm disc}/\hat\Pi_{\rm conn} \simeq -0.04$.\footnote{The big change between the one- and two-loop contributions is due to the $\rho$ resonance, which appears in ChPT only at two-loop order.}

The quark-disconnected contribution $G_{\rm disc}(x_0)$ is obtained from 
\begin{equation}
  G_{\rm disc}(x_0) = -\frac{1}{3}\sum_{k=1}^3 \sum_{f,f^\prime}q^f
  q^{f^\prime} \left\langle \Delta^f_k(x_0)\Delta^{f,f^\prime}_k(0)
  \right\rangle\, , 
\end{equation}
where $q^f, q^{f^\prime}$ are the electric charges of quarks flavors
$f$ and $f^\prime$, respectively, and $\Delta^f_\mu$ is given by
\begin{equation}
  \Delta^f_\mu(x_0) = a^3\sum_{\mathbf{x}}\,{\rm
  Tr}\,\left[\gamma_\mu S^f(x,x) \right]\,. 
\label{eq:loops}
\end{equation}
The quark propagator $S^f(x,y) \equiv [D^{f}]^{-1}(x,y)$ is obtained
by solving the Dirac equation $\sum_z D^f(x, z)\phi(z) = \delta_{xy}$.
For a given timeslice $x_0$, the loop $\Delta_{\mu}^f(x_0)$ includes
the sum over spatial coordinates $\mathbf{x}$, and one must solve the
Dirac equation for every spatial coordinate $\mathbf{x}$. This must be
repeated for various timeslices $x_0$ to evaluate $G_{\rm
disc}(x_0)$. Therefore, the computational cost for the required {\em
all-to-all} propagator is increased by a factor proportional to the
four-volume of the lattice, which is typically of order $10^7$, and
prohibitively costly. In practice, the quantity $\Delta^f_\mu$ in
\cref{eq:loops} is computed using stochastic techniques, by
solving the linear system $\sum_z D^f(x,z)\phi^{(r)}(z)
= \eta^{(r)}(y)$ for a set of random noise vectors $\eta^{(r)}(y)$,
$r=1,\ldots,N_r$, that satisfy
\begin{equation}
  \lim_{N_r\to \infty}\frac{1}{N_r}\sum_{r=1}^{N_r}
  \eta^{(r)}(x)\eta^{(r)}(y)^{\dagger} = \delta_{xy}\,.\label{eq:eta_prop} 
\end{equation}
After inserting the solution $\phi^{(r)}$ into \cref{eq:loops},
one recovers $\Delta^f_\mu$ from the stochastic average. The
statistical error in $G_{\rm disc}(x_0)$ is therefore a combination of
the statistical uncertainty arising from the gauge average and the
additional stochastic error due to summing over noise vectors. The
number $N_r$ of source vectors $\eta^{(r)}$ is optimized in
order to minimize the associated stochastic noise for given numerical
cost. Compared to the exact calculation of all-to-all propagators, the
numerical cost is typically reduced by $3\hyph4$ orders of magnitude.

Restricting the discussion to QCD with only light ($ud$) and strange
($s$) quarks, the statistical accuracy of $G_{\rm disc}(x_0)$ can be
further enhanced. In Ref.~\cite{Francis:2014hoa} it was shown that
$G_{\rm disc}(x_0)$ factorizes according to
\begin{equation}
G_{\rm disc}(x_0) = -\frac{1}{9}\frac{1}{3}\sum_{k=1}^3
\Big\langle \Big(\Delta^{ud}_k(x_0) - \Delta^{s}_k(x_0)\Big)\Big(\Delta^{ud}_k(0)
- \Delta^{s}_k(0)\Big)\Big\rangle\,.\label{eq:disc_iso}
\end{equation}
Hence, the stochastic noise largely cancels in the difference
$\Delta^{ud}_k - \Delta^{s}_k$, provided that the same source vectors
are used to compute the individual light and strange contributions. In
Ref.~\cite{Francis:2014hoa} it was demonstrated that the gain in
statistical accuracy amounts to two orders of magnitude.

There are a number of additional variance reduction techniques
designed to improve the statistical precision in the determination of
$G_{\rm disc}(x_0)$. These include low-mode
deflation~\cite{Foley:2005ac, Blum:2015you}, which can be combined either with
hierarchical probing and Hadamard vectors~\cite{Stathopoulos:2013aci,Gerardin:2019rua}
or with a hopping parameter expansion that suppresses stochastic noise
by inverse powers of the quark
mass~\cite{Bali:2009hu,Gulpers:2013uca}, as well as all-mode-averaging
(AMA)~\cite{Blum:2012uh}, which is based on a combination of multiple
low-precision calculations of the quark propagator with a subsequent
bias correction.

We now discuss recent evaluations of $\amuHVPLOdisc$. A collection of results is shown and
compared in \cref{subsec:compare_fbf} (the lower-right panel of \cref{fig:amu_compare_fbyf}).

The calculation by the BMW
collaboration~\cite{Borsanyi:2016lpl,Borsanyi:2017zdw} uses staggered
quarks at the physical pion mass and five different lattice
spacings. The StN ratio was enhanced via a combination of
several variance reduction techniques, including AMA and low-mode
deflation and also exploited the stochastic noise cancellation between
light and strange quark loops. For a given quark flavor $f$, the loop
$\Delta_k^f(x_0)$ is computed via a combination of the exact treatment
based on the spectral representation for a finite number of the lowest
eigenmodes and the stochastic estimate of the quark propagator in the
orthogonal complement of the subspace spanned by the lowest
eigenmodes. For the latter, BMW used a combination of high-precision
 and low-precision  solutions of the projected Dirac equation,
such that the bias correction involving the high-precision solves is
only computed for a subset of noise vectors, thereby reducing the
numerical effort.
Clear signals are obtained for five lattice spacings / twelve ensembles
at almost physical mass point,
and the continuum extrapolation is taken reliably.
%
%

In their calculations using domain wall fermions, the RBC/UKQCD
collaboration~\cite{Blum:2015you,Blum:2018mom} has determined the
disconnected contribution from $ud$ and $s$ quarks at the physical pion
mass and at a single lattice spacing $a=0.114$\,fm. The calculation is
based on low-mode deflation and the spectral representation of the
quark propagator for the lowest eigenmodes, combined with the
stochastic evaluation of the contributions corresponding to the high
modes. In order to control the stochastic noise associated with the
latter, RBC/UKQCD employed a particular spatial distribution of
stochastic source vectors, designed to minimize unwanted long-distance
contributions~\cite{Blum:2015you}.
%
%

The recent calculation by the Mainz group~\cite{Gerardin:2019rua} is
based on six ensembles generated using $\order(a)$ improved Wilson
fermions, which cover four different lattice spacings with pion masses
down to $200\MeV$. The calculation of the quark loops $\Delta_k^f$ was
performed using hierarchical probing~\cite{Stathopoulos:2013aci}, and
a significant reduction of the stochastic noise was observed when
using two noise vectors with 512 Hadamard vectors each. The
quark-disconnected contribution was determined by first applying the
``bounding method'' to the full isoscalar correlator before
subtracting the connected light- and strange-quark contributions. Since
the ensembles satisfy ${\rm Tr}\,M_q=\rm const$, where $M_q$ is the
bare quark mass matrix, the results were extrapolated to the physical
point in the variable $M_K^2-M_\pi^2$.
%
%

\begin{sloppypar}
The Fermilab--HPQCD--MILC collaboration~\cite{Davies:2019efs} has
estimated the quark-disconnected contribution using a ChPT-inspired model, which includes  $\rho$ mesons and photons~\cite{Chakraborty:2016mwy}. The contribution from $\pi\pi$
states to $\amuHVPLOdisc$ was
determined by multiplying the full $\pi\pi$ contribution to
$\amuHVPLO$ by $-1/9$, in accordance with
\cref{eq:asympratio}. The additional contributions from
non-$\pi\pi$ states were estimated by considering the difference
between the isoscalar and isovector correlators, which are assumed to
be saturated by the $\omega$ and $\rho$ mesons,
respectively~\cite{Chakraborty:2015ugp}. A direct lattice calculation of the quark-line disconnected contribution by this collaboration is in progress~\cite{Yamamoto:2018cqm,Davies:2019acq}.
\end{sloppypar}
%
%

In \cref{tab:amu_fbf} and \cref{fig:amu_compare_fbyf} (lower-right panel),
we show a compilation of results for
$\amuHVPLOdisc$. While the lattice results from
BMW~\cite{Borsanyi:2017zdw} and RBC/UKQCD~\cite{Blum:2018mom} are in good agreement, the
estimate from the Mainz group~\cite{Gerardin:2019rua} is somewhat lower than the rest.

\subsubsection{Strong and QED isospin-breaking contributions}
\label{sec:HVP_IB}

In the previous sections we have discussed the determination of HVP in the isospin-symmetric limit, where the up- and down-quark are treated as being equal in lattice calculations. However, in nature isospin symmetry is broken by the mass difference between the up and down quarks as well as by their electric charges. Effects due to the up--down mass difference ($\delta m = m_d-m_u$) and corrections from QED, i.e., due to the interactions of the charged quarks with photons, are, in general, of ${\cal{O}}(\delta m/\Lambda_{\text{QCD}})$ and ${\cal{O}}(\alpha)$, respectively, and thus, both of the order of $1\%$. The current efforts to determine HVP from lattice QCD at a precision level of $1\%$ or better make it now necessary to include these effects in the calculations.
Recent results are summarized in \cref{tab:ib_summary}.

To include QED effects in lattice calculations one needs to determine the Euclidean path integral in QCD$+$QED for a given observable $O$
\begin{equation}
  \left<O\right>_{\text{QCD}+\text{QED}} = \frac{1}{Z} \int\mathcal{D}[U,A,\psi,\bar{\psi}]\,O[\psi,\bar{\psi} ,A, U]\,e^ { -S[\psi,\bar{\psi} ,A, U]}\,,
\label{eq:QCDplusQEDpathintegral}
\end{equation}
with the quark fields $\psi$ and $\bar{\psi}$, the $SU(3)$ gluon fields $U$, and the photon fields $A$. The action $S[\psi,\bar{\psi} ,A, U]$ now also contains couplings of quarks to photons as well as a kinetic term for the photon fields. There are mainly two methods that are used to include QED in the calculation of the path integral of \cref{eq:QCDplusQEDpathintegral}. It was first proposed in Ref.~\cite{Duncan:1996xy} to treat QED nonperturbatively by using stochastic photon fields and to calculate quantities with QCD$+$QED gauge configurations. On the other hand, since the fine-structure constant is a small parameter, one can treat QED in a perturbative fashion. The RM123 collaboration proposed in Ref.~\cite{deDivitiis:2013xla} to expand the QCD$+$QED path integral of \cref{eq:QCDplusQEDpathintegral} in the fine-structure constant, 
\begin{equation}
 \left<O\right>_{\text{QCD}+\text{QED}} = \left<O\right>_{\text{QCD}} + 
\frac{1}{2}\,e^2\left.\frac{\partial^2}{\partial 
e^2}\left<O\right>\right|_{e=0} + {\cal{O}}(\alpha^2) \,,
\label{eq:pathintegraleexpansion}
\end{equation}
and to explicitly calculate the contributions to ${\cal{O}}(\alpha)$. In general, this will amount to calculating contributions from diagrams with one photon propagator (see below for a discussion on the diagrams required for HVP). A detailed comparison of the stochastic and the perturbative treatment of QED can be found in Ref.~\cite{Boyle:2017gzv}. The perturbative expansion was used to calculate QED corrections to HVP by the ETM collaboration~\cite{Giusti:2017jof,Giusti:2017ier,Giusti:2018vrc,Giusti:2019xct,Giusti:2019dmu,Giusti:2019hkz} and the RBC/UKQCD collaboration~\cite{Blum:2018mom,Gulpers:2018mim}.

One important issue when including QED in lattice calculations is the treatment of the zero-mode of the photon field, which cannot be constrained by a gauge-fixing procedure (see Ref.~\cite{Patella:2017fgk} for a detailed discussion). Both the ETM and RBC/UKQCD collaborations have used the QED$_L$ prescription, where all spatial zero-modes of the photon fields are removed~\cite{Hayakawa:2008an,Borsanyi:2014jba}, i.e., $\sum_{\mathbf{x}} A_\mu(x_0,\mathbf{x}) = 0$ for all $\mu, x_0$.

In addition to including QED one has to account for the different masses of the up- and down-quark. One possibility is to use different input masses for up- and down-quarks as done by the Fermilab--HPQCD--MILC collaboration~\cite{Chakraborty:2017tqp} to calculate SIB corrections to $a_\mu$. Another approach to treating the difference in the light-quark masses was proposed by the RM123 collaboration in Ref.~\cite{deDivitiis:2011eh}. Here, the path integral is expanded with respect to the differences $\Delta m_f = m_{f} - \hat{m}_f$,
\begin{equation}
 \left<O\right>_{m_f\neq \hat{m}_f} = \left<O\right>_{m_f=\hat{m}} + \Delta m_f \left.\frac{\partial}{\partial m_f}\left<O\right>\right|_{m_f=\hat{m}} + \mathcal{O}\left(\Delta m_f^2\right)\,,
\label{eq:patintegralmexpansion}
\end{equation}
where $\hat{m}_f$ is the mass of the quark with flavor $f$ in the isospin-symmetric theory. To ${\cal{O}}\big(\Delta m_f\big)$ one finds diagrams with an insertion of a scalar current (see below for a discussion on the specific diagrams required for HVP). The differences $\Delta m_f$ are free parameters that can be tuned \textit{a posteriori}. A perturbative expansion in $\Delta m_f$ was used by the ETM collaboration~\cite{Giusti:2019xct,Giusti:2017jof} and the RBC/UKQCD collaboration~\cite{Blum:2018mom,Gulpers:2018mim} to calculate IB corrections to HVP.

As discussed in \cref{sec:prescript}, it is important to stress that the physical point is only unambiguously defined in full QCD$+$QED with $m_u\neq m_d$.
On the contrary, defining the physical point in a pure QCD setup is ambiguous and relies on imposing a renormalization prescription to separate strong and QED IB effects. 
The renormalization schemes adopted by the ETM, Fermilab--HPQCD--MILC, and RBC/UKQCD collaborations when calculating IB corrections to HVP are reviewed in \cref{sec:prescript}.

The calculation of the strong and QED IB corrections $\delta \amuHVPLO$ of \cref{eq:deltaHVP} requires the evaluation of both quark-connected and -disconnected contractions for all flavors.
Since the various quark-connected components are diagonal in flavor and show different statistical and systematic uncertainties, they are usually computed separately on the lattice.
As in the case of the ${\cal O} (\alpha^2)$ term $\amuHVPLOconn$, the main contribution to $\delta \amuHVPLOconn$ is given by the light $u$- and $d$-quark, $\delta \amuHVPLOud$.

Only a few lattice calculations of $\delta \amuHVPLO$ are present in the literature so far, namely those performed by the ETM~\cite{Giusti:2017jof,Giusti:2019xct}, RBC/UKQCD~\cite{Blum:2018mom}, and Fermilab--HPQCD--MILC~\cite{Chakraborty:2017tqp} collaborations. While none of them include all of the contributions to $\delta \amuHVPLO$, they already provide useful results, as we discuss below.  
Details of the various lattice formulations adopted for the QCD action can be found in the corresponding references.
In Ref.~\cite{Borsanyi:2017zdw} the BMW collaboration provides a phenomenological estimate of IB corrections to $\amuHVPLO (\alpha^2)$ based on dispersive methods and ChPT.
Furthermore, a new lattice calculation of the HVP tensor with dynamical QCD and QED fields by the CSSM/QCDSF/UKQCD collaboration is ongoing~\cite{Westin:2019tgc}.

\begin{figure}[t!]
\centering{\includegraphics[width=\linewidth]{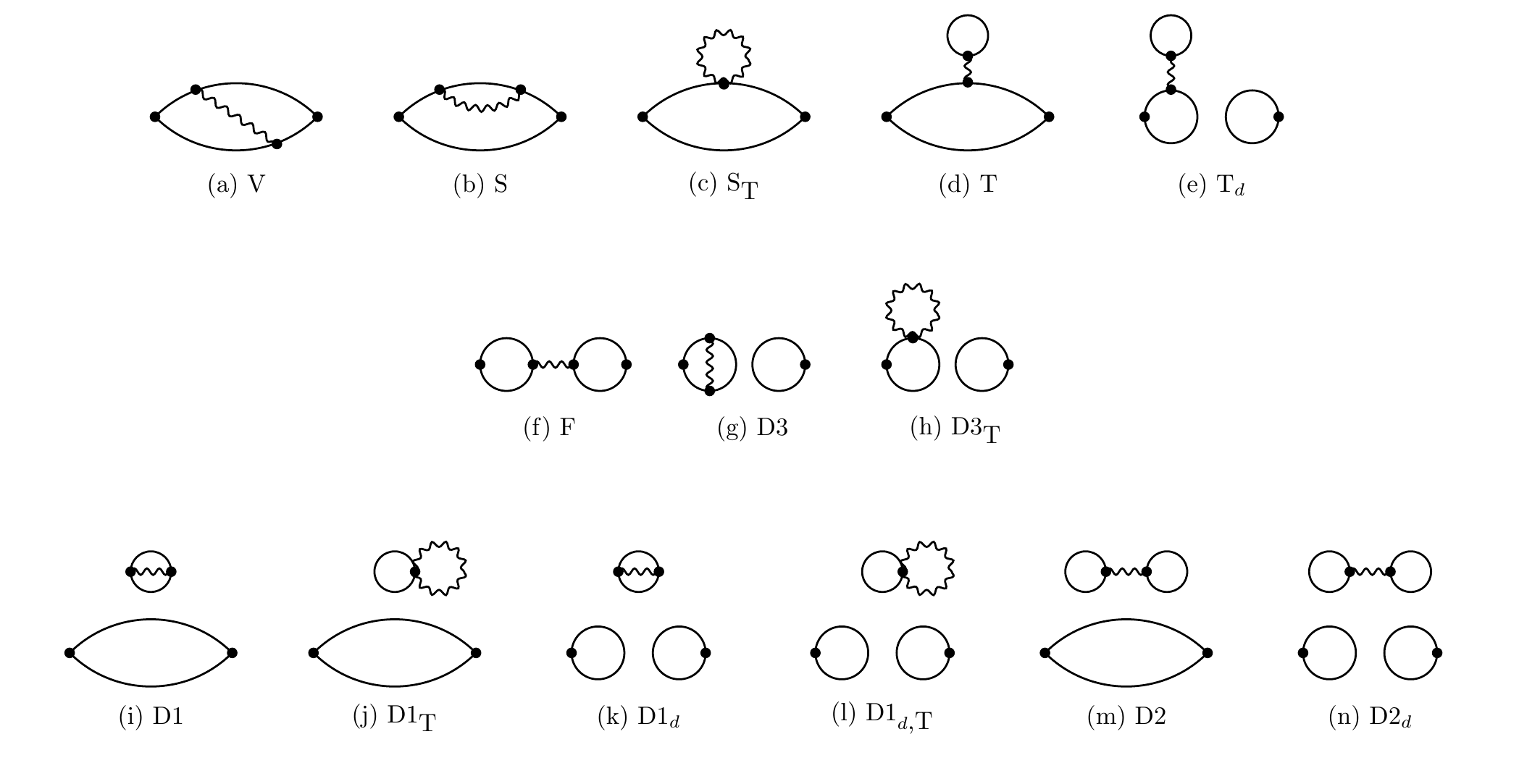}}
\caption{Photonic-correction diagrams to order ${\cal O}(\alpha)$ with external vector operators. Adapted from Ref.~\cite{Gulpers:2018mim}.}
\label{fig:ph_ins}
\end{figure}

\begin{figure}[t!]
\centering{\includegraphics[width=\linewidth]{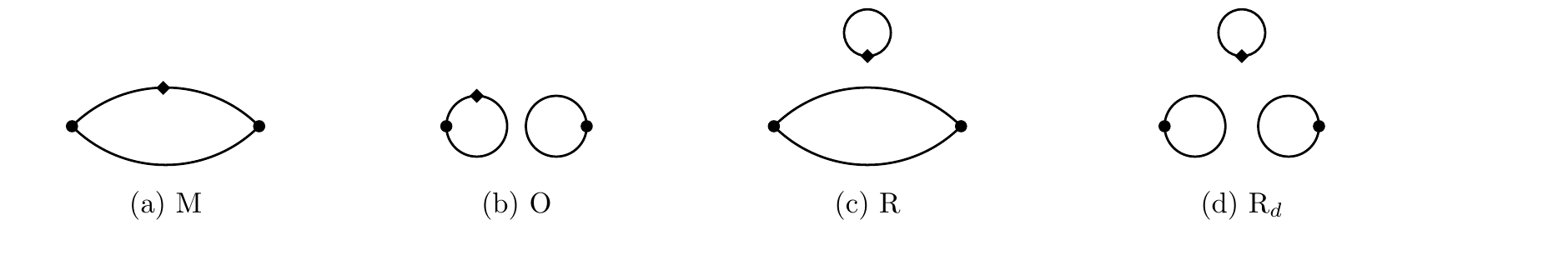}}
\caption{SIB-correction diagrams to order ${\cal O}(\Delta m)$. The diamonds denote the insertion of a scalar operator. Adapted from Ref.~\cite{Gulpers:2018mim}.}
\label{fig:s_ins}
\end{figure}

When using the perturbative method of Refs.~\cite{deDivitiis:2011eh,deDivitiis:2013xla}, as done by the ETM~\cite{Giusti:2017jof,Giusti:2019xct} and RBC/UKQCD~\cite{Blum:2018mom} collaborations, $\amuHVPLO$ is expanded into a lowest-order contribution $\amuHVPLO$, evaluated in isospin-symmetric QCD (i.e., $m_u = m_d$ and $\alpha = 0$), and a correction $\delta \amuHVPLO$ computed at LO in the {\emph small} parameters $(m_d - m_u)/\Lambda_{\text{QCD}}$ and $\alpha$.
The IB correction $\delta \amuHVPLO$ encodes both strong and QED corrections contributing to $\amuHVPLO$ to ${\cal O}(\alpha^2 (m_d - m_u) / \Lambda_{\text{QCD}})$ and ${\cal O}(\alpha^3)$, respectively, and its evaluation requires the computation of the diagrams depicted in 
\cref{fig:ph_ins,fig:s_ins}.
The set of diagrams coming from the expansion to ${\cal O}(\alpha)$ in the QED coupling is shown in \cref{fig:ph_ins}.\footnote{The use of lattice conserved vector currents at the source and at the sink requires the evaluation of additional diagrams~\cite{Boyle:2017gzv}.}
These diagrams can be divided into three different classes: QED corrections to the quark-connected contributions are given by diagrams $V$, $S$, and $S_T$; diagrams $F$, $D3$, and $D3_T$ represent corrections to the quark-disconnected contributions; while diagrams $T$, $D1$, $D1_T$, $D2$ and $T_d$, $D1_d$, $D1_{d, T}$, $D2_d$ account for QED IB effects coming from dynamical sea quarks for the quark-connected and quark-disconnected contributions (the latter denoted by a subscript $d$), respectively.
Diagrams $S_T$, $D1_T$, $D3_T$, $D1_{d, T}$ of \cref{fig:ph_ins}, denoted by subscripts $T$, correspond to the insertions of the tadpole operator.
The latter is a feature of lattice discretization and plays a crucial role in order to preserve gauge invariance to ${\cal O}(\alpha)$ in the expansion of the quark action.
In the calculation performed in Ref.~\cite{Blum:2018mom} tadpole contributions are absent since insertions of local vector currents are used.
Regularization-specific IB effects associated with the tuning of the quark critical masses in the presence of QED interactions~\cite{deDivitiis:2013xla,Giusti:2017dmp} arise when the lattice fermionic action does not preserve chiral symmetry, as in the case of Wilson and twisted-mass fermions (see, e.g., Refs.~\cite{Giusti:2017jof,Giusti:2019xct}).
This requires in addition the evaluation of diagrams with insertions of pseudoscalar densities (i.e., replacing the insertions of a scalar operator in \cref{fig:s_ins} by pseudoscalar currents).
IB corrections due to the expansion in the quark masses are given by the diagrams in \cref{fig:s_ins}.

The ETM~\cite{Giusti:2017jof,Giusti:2019xct} and RBC/UKQCD~\cite{Blum:2018mom} calculations are performed within the so-called electroquenched approximation, i.e., by treating sea quarks as neutral particles with respect to electromagnetism.
Thus, diagrams where photonic corrections or insertions of scalar/pseudoscalar operators are applied to sea-quark loops (i.e., diagrams $T$, $D1$, $D1_T$, $D2$, $T_d$, $D1_d$, $D1_{d, T}$, $D2_d$ of \cref{fig:ph_ins} and $R$, $R_d$ of \cref{fig:s_ins}) have not been included so far.
The ETM collaboration~\cite{Giusti:2017jof,Giusti:2019xct} evaluates the LO SIB and QED quark-connected corrections\footnote{In the GRS renormalization prescription~\cite{Gasser:2003hk} adopted in Refs.~\cite{Giusti:2017jof,Giusti:2019xct} the light-quark contribution includes both strong and QED IB corrections, while the strange- and charm-quark ones are induced by purely QED effects of order ${\cal O} (\alpha^3)$. SIB corrections to the latter contributions only appear at higher orders in the quark-mass difference $(m_d-m_u)$.} (due to diagrams $V$, $S$, $S_T$, $M$, and an analogous one with one insertion of the pseudoscalar density) with dynamical up, down, strange, and charm quarks at three values of the lattice spacing, at several lattice volumes and with pion masses between $\simeq 210$ and $\simeq 450\MeV$.
The RBC/UKQCD~\cite{Blum:2018mom} calculation (including diagrams $V$, $S$,\footnote{RBC/UKQCD uses local currents for the QED vertices and, therefore, diagram $S_T$ is absent.} $F$, $M$) is performed with three dynamical quarks, on a single ensemble at a nearly-physical pion mass.
In addition to calculating IB corrections to the connected HVP contribution, RBC/UKQCD calculates the leading QED correction to the disconnected contribution due to diagram $F$ in \cref{fig:ph_ins}.
Here one is only interested in contributions where the quark lines are connected by gluons in addition to the photon.
If no additional gluons connect the two quark lines, these contributions are conventionally counted as higher-order HVP terms.
Estimates of the uncertainties related to the neglect of the remaining SIB and QED diagrams are provided by both collaborations.

Since the ETM and RBC/UKQCD collaborations use local versions of the external vector currents, QED corrections to the corresponding renormalization constants are included in the calculations of $\delta \amuHVPLO$.
ETM~\cite{Giusti:2019xct} makes use of the nonperturbative determinations of the quark-bilinear-operator renormalization constants obtained in Refs.~\cite{DiCarlo:2019thl,DiCarlo:2019knp} within the RI$^\prime$-MOM scheme~\cite{Martinelli:1994ty} to first order in the electromagnetic coupling and to all orders in the strong one, while RBC/UKQCD~\cite{Boyle:2017gzv} determines the multiplicative renormalization constant of the vector current from the ratio of local--conserved to local--local vector two-point functions.

The Fermilab--HPQCD--MILC collaboration~\cite{Chakraborty:2017tqp} calculates the strong IB corrections using two QCD gauge ensembles generated with four flavors of dynamical quarks, at the same lattice spacing and volume.
The two ensembles share the same strange- and charm-quark masses, which are both fixed close to their physical values, and differ in the values of the light sea-quark masses.
One is isospin-symmetric, while the other has fully nondegenerate quark masses with nearly-physical values of $m_u$ and $m_d$.
By comparing the results on the two ensembles the authors of Ref.~\cite{Chakraborty:2017tqp} provide an estimate of the tiny, leading sea-IB contributions to $\amuHVPLO$ (see diagram $R$ of \cref{fig:s_ins}), which are expected to be quadratic in the quark mass difference $(m_d - m_u)$.

We close the present subsection by describing the main sources of uncertainty current lattice calculations of $\delta \amuHVPLO$ have to deal with, and we refer the reader to \cref{sec:comp} for results and comparisons.
As in the case of the leading hadronic contribution $\amuHVPLO (\alpha^2)$, typical uncertainties are related to statistics; long-distance effects due to light intermediate states; finite-volume and discretization effects; scale setting and chiral extrapolation/interpolation.
The light $u$- and $d$-quark contributions suffer from StN problems, which are controlled by adopting strategies similar to those described in \cref{sec:HLO_ud}.
ETM applies an exponential analytic representation at large time distances directly to the strong and QED IB corrections, by expanding the functional form of the total correlator in $\Delta m$ and $\alpha$ (\`a la RM123~\cite{deDivitiis:2011eh,deDivitiis:2013xla}), while Fermilab--HPQCD--MILC adopts this strategy for the full vector correlator $C(x_0)$ including only strong IB corrections due to different up- and down-quark input masses.
The RBC/UKQCD~\cite{Blum:2018mom} collaboration evaluates the leading strong and QED IB corrections $\delta \amuHVPLO$ replacing the data in the integration over the whole Euclidean time region by an exponential fit ansatz.

Due to the long-ranged nature of QED interactions, the inclusion of the latter in lattice calculations is known to generate potentially large finite-volume effects (FVEs) suppressed only by powers of the inverse spatial extent $L$.
The infinite-volume limit of the QED contributions $\delta \amuHVPLO(f)$, with $f = u, d, s, c$, is performed by ETM~\cite{Giusti:2017jof,Giusti:2019xct} using several gauge ensembles of different lattice extent and it is combined together with the continuum limit and the chiral extrapolation to the physical pion mass.
Four ensembles sharing approximately the same pion mass and lattice spacing, but differing in the lattice volume allow for a direct estimate of FVEs and hence control over the infinite-volume extrapolation.
As mentioned in Refs.~\cite{Giusti:2017jof,Giusti:2019xct}, it is expected that the leading FVEs in the QED$_L$ prescription start at order ${\cal O}(1 / L^3)$ because of the overall neutrality of the system, instead of ${\cal O}(1 / L^2)$ obtained with a naive power-counting. This behavior was formally established in Ref.~\cite{Bijnens:2019ejw}, by constraining the analytic structure of the forward Compton scattering amplitude for a vector two-point function using Ward--Takahashi identities. This object is identical to the LbL function described in Ref.~\cite{Colangelo:2017urn}, and this work was used as an input in one of the two derivations in Ref.~\cite{Bijnens:2019ejw}. Moreover, an analytic two-loop expression for the QED$_L$ finite-volume corrections to the HVP two-pion contribution is derived in Ref.~\cite{Bijnens:2019ejw} using scalar QED as an effective theory of elementary pions. It is confirmed in this framework that the ${\cal O}(1 / L^2)$ contribution cancels for neutral currents, and is nonzero for charged current. As a consequence, it is argued in Ref.~\cite{Bijnens:2019ejw} that for typical physical simulations with $M_{\pi}L\simeq4$, one has $\exp(-M_{\pi}L)\simeq 1/(M_{\pi}L)^3\simeq 2\%$. Therefore in principle the power-like FVEs introduced by QED are expected to have a size similar to the exponential-like QCD FVEs. In both cases the magnitude of these effects on the IB corrections to the HVP is roughly a factor of $5$ smaller than a target relative uncertainty of $0.1\%$ on the total HVP contribution.
These theoretical expectations are consistent with the lattice results $\delta \amuHVPLO$ of the four ETM gauge ensembles~\cite{Giusti:2019xct} and FVEs do not affect significantly both $\delta \amuHVPLOs$ and $\delta \amuHVPLOc$.
The estimate of FVEs provided by RBC/UKQCD~\cite{Blum:2018mom} arises from the difference between the computation using an infinite-volume photon propagator and the QED$_L$ result.
For the SIB corrections FVEs differ qualitatively and quantitatively.
In the GRS renormalization prescription adopted by ETM~\cite{Giusti:2017jof,Giusti:2019xct} the SIB correlator receives nonvanishing contributions only in the light-quark sector and, since a fixed value of the light quark mass difference $(m_d - m_u)$ is used for all gauge ensembles, an exponential dependence in terms of the quantity $M_\pi L$ is expected.
By fitting the results of numerous twisted-mass ensembles with several ans{\"a}tze, an estimate of the systematic uncertainty due to FVEs is provided.
The Fermilab--HPQCD--MILC~\cite{Chakraborty:2017tqp} and RBC/UKQCD~\cite{Blum:2018mom} collaborations estimate FVEs for the SIB contributions using ChPT.

As in the case of the LO contribution, the quantity $\delta \amuHVPLO$ is sensitive to the scale setting through the lepton kernel $f(q^2)$.
Scale settings adopted by the various collaborations (see \cref{sec:HLO_ud}) are part of the ingredients used to calibrate the lattices nonperturbatively and, thus, contribute to defining what one means by QCD in the \emph{full} (QCD+QED) theory.
The separation of the full theory into isospin-symmetric QCD and radiative corrections requires a convention (see Refs.~\cite{DiCarlo:2019thl,Giusti:2018guw} for a detailed discussion); only the complete quantity $\amuHVPLO$ of \cref{eq:HVP} evaluated in the full theory is unambiguous and prescription free (see \cref{sec:prescript}).
In order to reduce the uncertainty related to the scale setting a possible strategy proposed by ETM~\cite{Giusti:2017jof,Giusti:2019xct} is to consider the ratio of the IB corrections $\delta \amuHVPLO$ over the LO terms $\amuHVPLO$ for each flavor channel.
The attractive feature of this ratio is to be less sensitive to some of the systematic effects.
The continuum limit is then performed via a combined fit by taking advantage of several gauge ensembles at different values of the lattice spacing.
As in the case of the isospin-symmetric calculation, discretization effects play a minor role in the light-quark sector, while in the evaluation of $\delta \amuHVPLOs$ and $\delta \amuHVPLOc$ they are larger.
In the case of the staggered formulation adopted in Ref.~\cite{Chakraborty:2017tqp} lattice data are corrected for the additional discretization effects due to the taste splitting among pions by using the predictions of staggered ChPT.
RBC/UKQCD~\cite{Blum:2018mom} calculates IB corrections at a single lattice spacing and takes a simple $(a \Lambda)^2$ estimate, with $\Lambda = 400\MeV$ as discretization error.

Finally, for simulations carried out close to the physical pion mass, such as the ones performed by the Fermilab--HPQCD--MILC~\cite{Chakraborty:2017tqp} and RBC/UKQCD~\cite{Blum:2018mom} collaborations, only a smooth interpolation of the lattice data is required.
Instead, the ETM gauge ensembles used in Refs.~\cite{Giusti:2017jof,Giusti:2019xct} have been generated at heavier pion masses and the extrapolation to the physical point, namely $M_\pi = 134.8~(3)\MeV$, is performed using a phenomenological fit ansatz.
The pion mass dependence of $\delta \amuHVPLOs$ is quite mild and even weaker in $\delta \amuHVPLOc$, being driven only by the sea quarks.

\subsection{Comparisons}
\label{sec:comp}

This section is devoted to presenting and comparing lattice results for $\amuHVPLO$ and the various intermediate quantities, such as the flavor specific contributions or subleading IB corrections that can be calculated separately. 
All the results presented here are extrapolated to the continuum and infinite-volume limits and interpolated or extrapolated to the physical point. The quoted errors in all lattice results include statistical and systematic uncertainties, where the latter estimates effects from scale setting, input parameters, continuum extrapolation, infinite-volume extrapolation, and chiral interpolations/extrapolations. Typically, these systematic errors are estimated by varying the chiral, continuum, or finite-volume fit functions, including adding higher-order terms in the corresponding EFT expansions, or varying which  lattice data are included, among other things.

\subsubsection{Total leading-order HVP contribution}\label{subsec:compare_lat}

\begin{table}[t]
\begin{center}
\small
\begin{tabular}{lclcc}
\toprule
Collaboration&
$N_f$&
$\amuHVPLO\times 10^{10}$&
Fermion&
$\hat{\Pi}(Q^2)$\\
\midrule
ETM-18/19~\cite{Giusti:2018mdh,Giusti:2019hkz}&
2+1+1&
692.1\,(16.3)&
tmQCD&
TMR\\
FHM-19~\cite{Davies:2019efs}&
2+1+1&
699\,(15)&
HISQ&
Pad\'{e} w. Moments/TMR \\
BMW-17~\cite{Borsanyi:2017zdw}&
2+1+1&
711.1\,(7.5)(17.5)&
Stout4S&
TMR\\
HPQCD-16~\cite{Chakraborty:2016mwy}&
2+1+1&
667\,(6)(12)&
HISQ&
Pad\'{e} w. Moments \\
ETM-13~\cite{Burger:2013jya}&
2+1+1&
674\,(21)(18)${{}^{\ast}}$&
tmQCD&
VMD\\
\midrule
Mainz/CLS-19~\cite{Gerardin:2019rua}&
2+1&
720.0\,(12.4)(9.9)&
Clover&
TMR\\
PACS-19~\cite{Shintani:2019wai}&
2+1&
$737\,(9)({}^{+13}_{-18})$&
StoutW&
TMR/Pad\'{e}\\
RBC/UKQCD-18~\cite{Blum:2018mom}&
2+1&
717.4\,(16.3)(9.2)&
DWF&
TMR\\
\midrule
Mainz-17~\cite{DellaMorte:2017dyu}&
2&
$654\,(32)({}^{+21}_{-23}){{}^{\ast}}$&
Clover&
TMR\\
\midrule
KNT-19~\cite{Keshavarzi:2019abf}&
pheno.&
692.8\,(2.4)&
$-$&
dispersion\\
DHMZ-19~\cite{Davier:2019can}&
pheno.&
694.0\,(4.0)&
$-$&
dispersion\\
BDJ-19~\cite{Benayoun:2019zwh}&
pheno.&
687.1\,(3.0)&
$-$&
dispersion\\
FJ-17~\cite{Jegerlehner:2017gek}&
pheno.&
688.1\,(4.1)&
$-$&
dispersion\\
\midrule
RBC/UKQCD-18~\cite{Blum:2018mom}&
lat.+pheno.&
692.5\,(1.4)(2.3)&
DWF&
TMR + disp.\\
\bottomrule
\end{tabular}
\caption{Summary of results for $\amuHVPLO$; see also \cref{fig:amu_compare}. All lattice results fully take into account the corrections and systematic errors, except for those marked with $\ast$, which are older results that did not include SIB and QED corrections  in the quoted values and errors. In some cases, the lattice results include phenomenological estimates of the SIB/QED corrections instead of direct lattice calculations.
Results for which the second column states $N_f = 2+1$  include charm contributions in the valence sector, but not in the sea. 
Results with $N_f = 2$ also omit strange sea-quark effects.
When results are displayed with two errors, the first is the statistical uncertainty and the second the systematic one. With only one quoted error, the statistical and systematic uncertainties are combined. 
HISQ = highly improved staggered quarks,
Stout4S = 4 steps stout-smeared staggered quarks,
tmQCD = twisted mass QCD,
DWF = domain wall fermions,
Clover = $\mathcal{O}(a)$ improved Wilson quarks,
StoutW = stout-smeared $\mathcal{O}(a)$ improved Wilson quarks.
Simulations with staggered quarks employ ``rooted'' determinants, to remove the extra doublers from the sea.
TMR = time-momentum representation,
VMD = vector-meson dominance.}
\label{tab:amu_cmp}
\end{center}
\end{table}

\begin{figure}[ht]
\begin{center}
\includegraphics[width=0.6\textwidth]{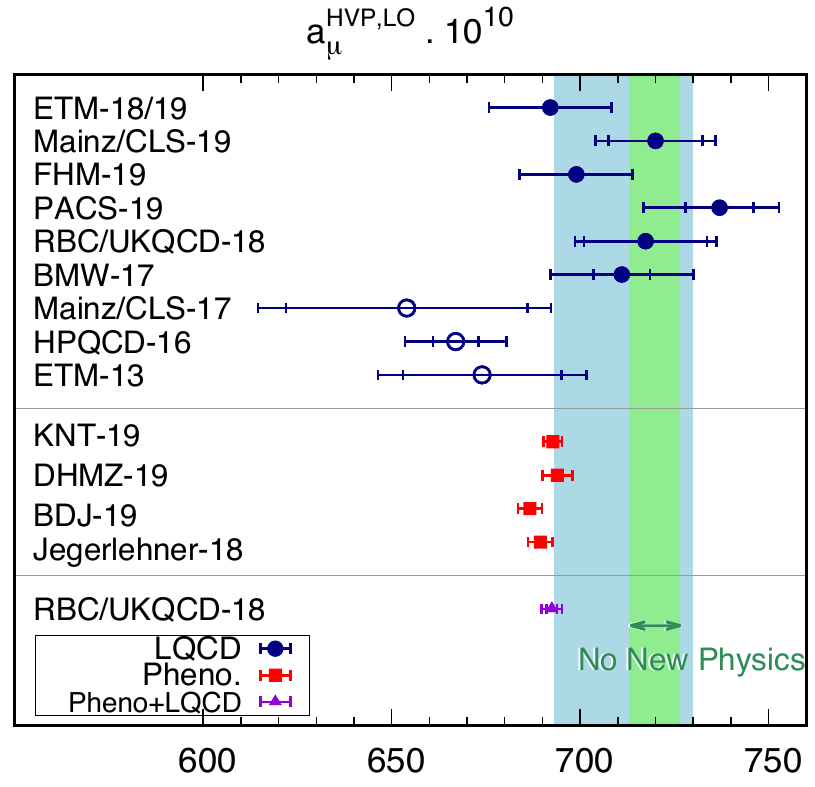}
\caption{Compilation of recent results for $\amuHVPLO$ in units of $10^{-10}$. The filled dark blue circles are lattice results that are included in the ``lattice world average''. The average, which is obtained from a conservative averaging procedure in \cref{subsec:status}, is indicated by a light blue band, while the light-green band indicates the ``no new physics'' scenario, where $\amuHVPLO$ results are large enough to bring the SM prediction of $a_\mu$ into agreement with experiment. The unfilled dark blue circles are lattice results that are older or superseded by more recent calculations. The red squares indicate results obtained from the data-driven methods reviewed in \cref{sec:dataHVP}.
See \cref{tab:amu_cmp} for more information on the results included in the plot. Adapted from Ref.~\cite{Miura:2019xtd}.}
\label{fig:amu_compare}
\end{center}
\end{figure}

In \cref{fig:amu_compare} and \cref{tab:amu_cmp}, we compare the results for $\amuHVPLO$ reported by the various lattice QCD groups as well as those obtained from the data-driven methods described in \cref{sec:dataHVP}. Note that lattice results based on gauge ensembles with $N_f=2$ sea quarks are not included in our averages.  
The results from the BMW collaboration (BMW-17~\cite{Borsanyi:2017zdw}), the RBC/UKQCD collaboration (RBC/UKQCD-18~\cite{Blum:2018mom}), and the Mainz/CLS collaboration (Mainz/CLS-19~\cite{Gerardin:2019rua}) agree well within their errors. The results reported by the Fermilab Lattice, HPQCD, and MILC collaborations (FHM-19~\cite{Davies:2019efs}) and by the ETM collaboration (ETM-18/19~\cite{Giusti:2018mdh,Giusti:2019hkz}), while in good agreement with each other, are slightly lower than BMW-17, RBC/UKQCD-18, and Mainz/CLS-19, but still consistent, at about 1$\sigma$. 
The BMW-17, RBC/UKQCD-18, and Mainz/CLS-19 results are in slight tension with the data-driven evaluations, and consistent with the ``no new physics'' scenario (black vertical lines), i.e., the value that $\amuHVPLO$ would need to have in order to bring the SM prediction of $a_\mu$ into agreement with the experimental measurement~\cite{Bennett:2006fi}, while keeping all other SM contributions unchanged. The situation is reversed for the FHM-19 and ETM-18/19 results, which are consistent with the data-driven results but in slight tension with the ``no new physics scenario''. 
The PACS collaboration (PACS-19~\cite{Shintani:2019wai}) has reported a value of $\amuHVPLO$ which is slightly larger than the ``no new physics'' scenario and in more than $2\sigma$ tension with the dispersive predictions. This result is also in mild tension with BMW-17, RBC/UKQCD-18, and Mainz/CLS-19 and has a $2\sigma$ tension with FHM-19 and ETM-18/19. Excluding older lattice calculations, which have been superseded~\cite{Burger:2013jya,Chakraborty:2016mwy,DellaMorte:2017dyu}, the BMW-17, RBC/UKQCD-18, FHM-19, Mainz/CLS-19, PACS-19, and ETM-18/19 results are combined in \cref{subsec:status} into a ``lattice world average'' using a conservative procedure, which is shown as a blue band in \cref{fig:amu_compare}. Given the relatively large spread between the lattice results, it is not surprising that the  ``lattice world average'' is consistent with both the data-driven results and the ``no new physics'' scenario. 

\subsubsection{Flavor-specific and subleading contributions}\label{subsec:compare_fbf}

\begin{table}[t]
\begin{center}
\small
\begin{tabular}{lcllll}
\toprule
Collaboration&\hspace{-0mm}
$N_f$&\hspace{-0mm}
$\amuHVPLOud\times 10^{10}$&\hspace{-0mm}
$\amuHVPLOs\times 10^{10}$&\hspace{-0mm}
$\amuHVPLOc\times 10^{10}$&\hspace{-0mm}
$\amuHVPLOdisc\times 10^{10}$\\
\midrule
ETM-18/19~\cite{Giusti:2018mdh,Giusti:2019hkz}&\hspace{-0mm}
2+1+1&\hspace{-0mm}
629.1(13.7)&\hspace{-0mm}
53.1(1.6)(2.0)&\hspace{-0mm}
14.75(42)(37)&\hspace{-0mm}
$-$\\
Aubin {\it et al.}-19~\cite{Aubin:2019usy}&\hspace{-0mm}
2+1+1&\hspace{-0mm}
659 (22) &\hspace{-0mm}
$-$&\hspace{-0mm}
$-$&\hspace{-0mm}
$-$\\
FHM-19~\cite{Davies:2019efs}&\hspace{-0mm}
2+1+1&\hspace{-0mm}
637.8(8.8) &\hspace{-0mm}
$-$&\hspace{-0mm}
$-$&\hspace{-0mm}
$-13(5)$\\
BMW-17~\cite{Borsanyi:2017zdw}&\hspace{-0mm}
2+1+1&\hspace{-0mm}
647.6(7.5)(17.7) &\hspace{-0mm}
53.73(04)(49)&\hspace{-0mm}
14.74(04)(16)&\hspace{-0mm}
$-12.8(1.1)(1.6)$\\
HPQCD-16~\cite{Chakraborty:2016mwy}&\hspace{-0mm}
2+1+1&\hspace{-0mm}
$599.0(6.0)(11.0)^{\dagger}$&\hspace{-0mm}
$-$&\hspace{-0mm}
$-$&\hspace{-0mm}
$0(9)(-)$\\
HPQCD-14~\cite{Chakraborty:2014mwa}&\hspace{-0mm}
2+1+1/2+1&\hspace{-0mm}
$-$&\hspace{-0mm}
53.41(00)(59)&\hspace{-0mm}
14.42(00)(39)&\hspace{-0mm}
$-$\\
\midrule
Mainz/CLS-19~\cite{Gerardin:2019rua}&\hspace{-0mm}
2+1&\hspace{-0mm}
674(12)(5) &\hspace{-0mm}
54.5(2.4)(0.6)&\hspace{-0mm}
14.66(45)(6)&\hspace{-0mm}
$-23.2(2.2)(4.5)$\\
PACS-19~\cite{Shintani:2019wai}&\hspace{-0mm}
2+1&\hspace{-0mm}
673(9)(11) &\hspace{-0mm}
52.1(2)(5)&\hspace{-0mm}
11.7(0.2)(1.6)&\hspace{-0mm}
$-$\\
RBC/UKQCD-18~\cite{Blum:2018mom}&\hspace{-0mm}
2+1&\hspace{-0mm}
649.7(14.2)(4.9) &\hspace{-0mm}
53.2(4)(3)&\hspace{-0mm}
14.3(0)(7)&\hspace{-0mm}
$-11.2(3.3)(2.3)$\\
\midrule
Mainz/CLS-17~\cite{DellaMorte:2017dyu}&\hspace{-0mm}
2&\hspace{-0mm}
588.2(31.7)(16.6) &\hspace{-0mm}
51.1(1.7)(0.4)&\hspace{-0mm}
14.3(2)(1)&\hspace{-0mm}
$-$\\
\bottomrule
\end{tabular}
\caption{Flavor-specific contributions to $\amuHVPLO$, see also \cref{fig:amu_compare_fbyf}. The HPQCD-16 result  for the light-quark connected contribution (marked by a $\dagger$) is evaluated at the physical charged pion mass, and so includes some SIB and QED corrections. It is therefore not directly comparable with the other lattice results for $\amuHVPLOud$. The ETM-19 result \cite{Giusti:2019hkz} is an update of ETM-18 \cite{Giusti:2018mdh} and uses the same gauge ensembles and analysis methods. The FHM-19~\cite{Davies:2019efs} entry for $\amuHVPLOdisc$ is a phenomenological (non-lattice) estimate obtained from a ChPT-inspired model. When results are displayed with two errors, the first is the statistical uncertainty and the second the systematic one. With only one quoted error, the statistical and systematic uncertainties are combined. }
\label{tab:amu_fbf}
\end{center}
\end{table}

\begin{figure}[ht]
\begin{center}
\includegraphics[width=0.48\textwidth]{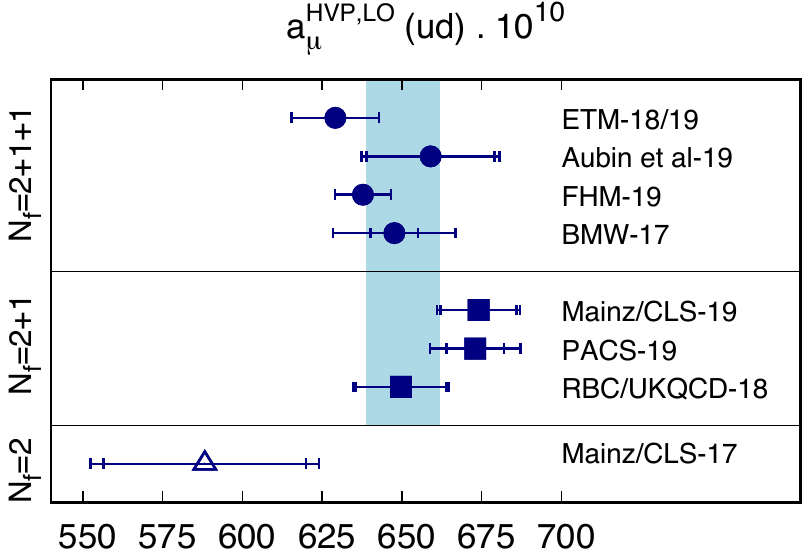}
\includegraphics[width=0.48\textwidth]{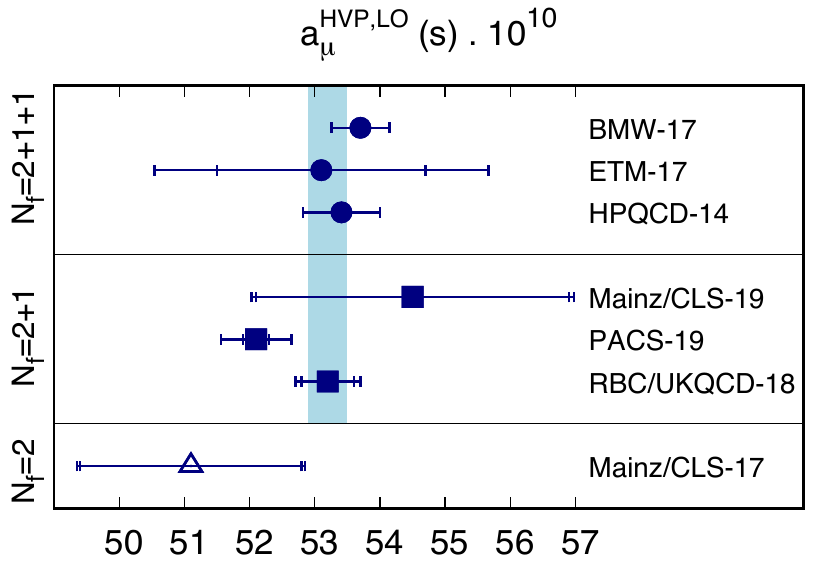}
\includegraphics[width=0.48\textwidth]{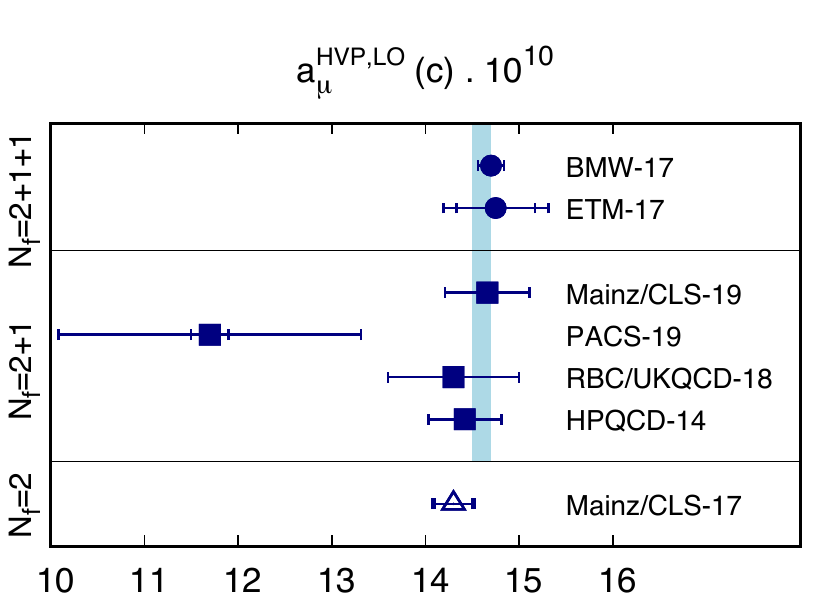}
\includegraphics[width=0.48\textwidth]{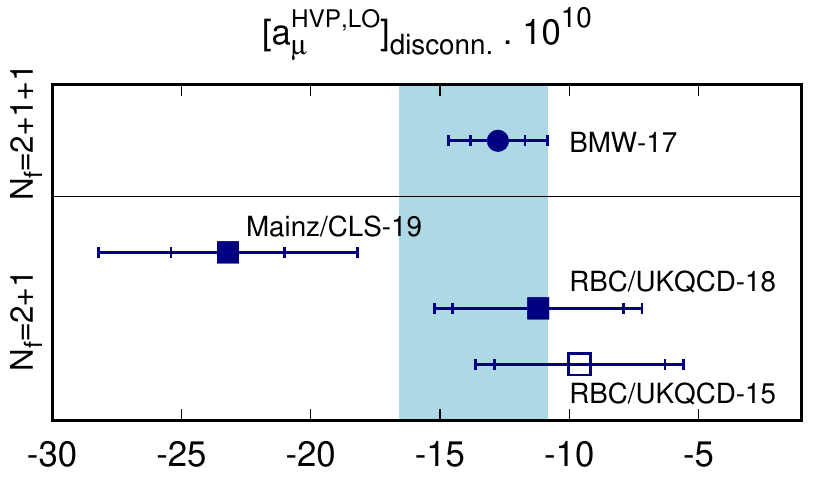}
\caption{
Comparisons of lattice results for flavor-specific contributions to $\amuHVPLO(\alpha^2)$. (Upper-Left) Light-quark connected contribution $\amuHVPLOud$. (Upper-Right) Strange-quark connected contribution $\amuHVPLOs$. (Lower-Left) Charm-quark connected contribution $\amuHVPLOs$. (Lower-Right) Quark-disconnected contribution $\amuHVPLOdisc$. The lattice results in each panel are grouped by the number of sea quarks in the gauge ensembles employed in the underlying calculations, where ``$N_f = 2 + 1+1$'' (circles) labels ensembles with up, down, strange, and charm quarks in the sea, for ``$N_f = 2 + 1$'' (squares) charm quarks are not included in the sea, while for ``$N_f = 2$,'' (up triangles) strange quarks are also omitted in the sea. Filled symbols indicate results included in the lattice averages of \cref{subsec:status}, which are shown here as light blue bands. Open symbols indicate results that have been updated or superseded, see \cref{tab:amu_fbf} for further details. Adapted from Ref.~\cite{Miura:2019xtd}.}
\label{fig:amu_compare_fbyf}
\end{center}
\end{figure}

In \cref{fig:amu_compare_fbyf} and \cref{tab:amu_fbf,tab:ib_summary} we consider the flavor-specific contributions, namely $\amuHVPLOud$, $\amuHVPLOs$, $\amuHVPLOc$, and the subleading corrections $\amuHVPLOdisc$ and $\delta \amuHVPLO$, which allow for more detailed comparisons of the various lattice results. 
The plots in the four panels of \cref{fig:amu_compare_fbyf} indicate the number of sea quarks in the gauge ensembles on which the lattice results are based, where $N_f = 2 + 1+1$ labels ensembles with up, down, strange, and charm quarks in the sea, $N_f = 2 + 1$ means that charm-quark effects are taken into account in the valence sector, but not in the sea, while for $N_f = 2$, strange-quark effects in the sea are also missing. Filled symbols indicate results included in the lattice averages of \cref{subsec:status}, which are shown in each panel as a light blue band. Open symbols indicate results that are not included in the averages.  

The lattice results for the dominant light-quark connected contribution $\amuHVPLOud$, shown in the upper-right panel of \cref{fig:amu_compare_fbyf}, exhibit a similar spread in central values as those for $\amuHVPLO$. There is a $2.4\sigma$ tension between the results with lowest (ETM-18/19~\cite{Giusti:2018mdh,Giusti:2019hkz}) and highest (Mainz/CLS-19~\cite{Gerardin:2019rua}, PACS-19~\cite{Shintani:2019wai}) central values, while BMW-17~\cite{Borsanyi:2017zdw}, RBC/UKQCD-18~\cite{Blum:2018mom}, FHM-19~\cite{Davies:2019efs}, and Aubin {\it et al.}-19~\cite{Aubin:2019usy} lie in between. 
In HPQCD-16~\cite{Chakraborty:2016mwy}, the light-quark connected contribution is not defined in the same way as in this review, as it is evaluated there at the physical charged pion mass. As a result it cannot be directly compared to lattice results for $\amuHVPLOud$ and is therefore omitted from \cref{fig:amu_compare_fbyf}.
As discussed in \cref{sec:strat}, $\amuHVPLOud$ is sensitive to the long-distance (large Euclidean time) behavior of the vector-current correlator,  which is the region where the correlator suffers from a StN problem. It is possible that the above tension is related to the different strategies employed to control and model this important region. Further investigations, including comparisons of other intermediate quantities with different levels of sensitivity to the short- and long-distance contributions, would be useful. The fact that the tension between different results is larger than the individual errors may be an  indication that some systematic effects are underestimated. We expect that this situation will improve in future high-precision studies, which will enable more refined analyses of the underlying systematic errors. 

The strange- and charm-quark connected contributions $\amuHVPLOs$ and $\amuHVPLOc$ are shown in the upper-right and lower-left panels of \cref{fig:amu_compare_fbyf} respectively. These quantities are already calculated at close to target precision.
The results for $\amuHVPLOs$ and for $\amuHVPLOc$ from Refs.~\cite{Chakraborty:2014mwa,Borsanyi:2017zdw,Blum:2018mom,Giusti:2018mdh,Giusti:2019hkz} are nicely consistent. However, the PACS-19~\cite{Shintani:2019wai} result for $\amuHVPLOs$ is in $1\sigma$ tension with the other lattice results while for $\amuHVPLOc$ it is in almost $2\sigma$ tension with the rest. The strange- and charm-quark connected contributions, while insensitive to FVEs and StN problems from large Euclidean times, suffer from larger discretization effects. This is especially true for $\amuHVPLOc$, and we note that the PACS-19 calculation has ${\mathcal O}(a)$ artifacts, which are not present in the other lattice results.  

As explained in \cref{sec:HLO_disconn}, the calculation of the quark-disconnected contribution $\amuHVPLOdisc$
is an especially challenging part of the lattice-QCD calculation of $\amuHVPLO$. In fact, as shown in \cref{fig:amu_compare_fbyf} (lower-right panel) the results for $\amuHVPLOdisc$ exhibit the second-largest tension among the individual contributions to $\amuHVPLO$. 
While the BMW-17~\cite{Borsanyi:2017zdw} and RBCC/UKQCD-18~\cite{Blum:2018mom} results are nicely consistent with each other, they disagree with the Mainz/CLS-19~\cite{Gerardin:2019rua} result. Unlike  BMW-17 and RBCC/UKQCD-18, the lattice calculation in Mainz/CLS-19 employs ensembles at unphysically large pion masses and therefore requires a chiral extrapolation to the physical point. One of the fit ans\"atze employed in the chiral extrapolation takes the $1/M^2_{\pi}$ singularity into account, which leads to a significantly lower value for $\amuHVPLOdisc$ at the physical point. 

\begin{table}[t]
\begin{center}
\small
\begin{tabular}{lcl}
\toprule
Collaboration &
\quad $\delta \amuHVPLO\times 10^{10}$&
\qquad Comments\\
\midrule
ETM-19~\cite{Giusti:2019xct}&
7.1\,(2.9)&
SIB+QED, perturbative method:\\
&
&
QED = $(V, S, S_{T})$ in \cref{fig:ph_ins}, SIB = $M$ in \cref{fig:s_ins}.\\
&
&
$M\ni$ scalar/pseudoscalar(PS) masses,\\
&
&
where PS is for keeping maximal twist.\\
\midrule
RBC/UKQCD-18~\cite{Blum:2018mom,Gulpers:2018mim}&
9.5\,(10.2)&
SIB+QED, perturbative method:\\
&
&
QED = $(V, S, F)$ in \cref{fig:ph_ins}, SIB = $M$ in \cref{fig:s_ins}.\\
&
&
$F$ with no gluon between two quark-loops\\
&
&
belongs to NNLO and is excluded.\\
\midrule
FHM-17~\cite{Chakraborty:2017tqp}&
9.5\,(4.5)&
Simulations with full-SIB for $ud$-conn:\\
&
&
$m_d - m_u \neq 0$ while $\alpha = 0$.\\
\midrule
BMW-17~\cite{Borsanyi:2017zdw}&
7.8\,(5.1)&
(SIB + QED) using ChPT and dispersion:\\
&
&
$\rho$\hyph$\omega$ mix., FSR, $M_{\pi}^{\rm ISLim}\to M_{\pi^\pm}$, $\pi^0\gamma$, $\eta\gamma$.\\
\midrule
CSSM/QCDSF/UKQCD &
$\lesssim 1\% \times \amuHVPLO$\quad &
Simulations with Full-QED for $ud$-conn:\\
Preliminary~\cite{Westin:2019tgc}&
&
$\alpha \neq 0$ while $m_d - m_u = 0$. $M_{\pi}\sim 400$ MeV.\\
\bottomrule
\end{tabular}
\caption{Summary of SIB and/or QED corrections: $\delta \amuHVPLO$. See \cref{sec:HVP_IB} for further details.}
\label{tab:ib_summary}
\end{center}
\end{table}

Finally, the challenging nonperturbative calculation of the subleading IB contributions $\delta \amuHVPLO$ has been performed by only a few collaborations so far, as can be seen in \cref{tab:ib_summary} where we have collected the current lattice evaluations (see \cref{sec:HVP_IB} for a detailed discussion of calculations).
Of the five results listed in \cref{tab:ib_summary} only FHM-17~\cite{Chakraborty:2017tqp}, RBC/UKQCD-18~\cite{Blum:2018mom,Gulpers:2018mim}, and ETM-19~\cite{Giusti:2019xct} are based on actual lattice calculations that are precise enough to quote results. While none of the three collaborations provide a complete lattice computation of all the contributions to $\delta \amuHVPLO$, the omitted contributions are estimated phenomenologically in all cases. In Ref.~\cite{Chakraborty:2017tqp} (FHM-17) a result for the connected SIB correction is presented, while Refs.~\cite{Blum:2018mom,Gulpers:2018mim,Giusti:2019xct} (RBC/UKQCD-18 and ETM-19) present a calculation of the connected SIB and QED corrections. No disconnected contributions are included in the lattice calculations of  Refs.~\cite{Chakraborty:2017tqp,Giusti:2019xct}, while Refs.~\cite{Blum:2018mom,Gulpers:2018mim} (RBC/UKQCD-18) 
include only the leading QED diagram. Nevertheless, within the quoted uncertainties an overall agreement among the present estimates is evident.

\subsubsection{Taylor coefficients}

\begin{table}[t]
\begin{center}
\small
\begin{tabular}{lcllll}
\toprule
Collaboration &\hspace{-0mm}
$N_f$&\hspace{-0mm}
$\Pi_1^{ud}$&\hspace{-0mm}
$-\Pi_2^{ud}$&\hspace{-0mm}
$\Pi_1^{\rm tot}$&\hspace{-0mm}
$-\Pi_2^{\rm tot}$\\
\midrule
ETM-18/19~\cite{Giusti:2018mdh,Giusti:2019hkz}&\hspace{-0mm}
2+1+1&\hspace{-0mm}
0.1642(33)&\hspace{-0mm}
0.383(16)&\hspace{-0mm}
0.1002(23)&\hspace{-0mm}
$-$\\
Aubin {\it et al.}-19~\cite{Aubin:2019usy}&\hspace{-0mm}
2+1+1&\hspace{-0mm}
${0.159(15)}^{\star}$&\hspace{-0mm}
$-$&\hspace{-0mm}
$-$&\hspace{-0mm}
$-$\\
FHM-19~\cite{Davies:2019efs}&\hspace{-0mm}
2+1+1&\hspace{-0mm}
${0.16776(25)}^{\star}$&\hspace{-0mm}
${0.3760(115)}^{\star}$&\hspace{-0mm}
$0.1011(24)$&\hspace{-0mm}
$0.2089(95)$\\
BMW-16~\cite{Borsanyi:2016lpl}&\hspace{-0mm}
2+1+1&\hspace{-0mm}
0.1660(17)(30)&\hspace{-0mm}
0.313(10)(13)&\hspace{-0mm}
0.1000(10)(28)&\hspace{-0mm}
0.181(6)(11)\\
HPQCD-16~\cite{Chakraborty:2016mwy}&\hspace{-0mm}
2+1+1&\hspace{-0mm}
${0.1606(22)(14)}^{\star}$&\hspace{-0mm}
${0.362(7)(14)}^{\star}$&\hspace{-0mm}
0.0984(14)&\hspace{-0mm}
0.2070(89)\\
\midrule
RBC/UKQCD-18~\cite{Blum:2018mom}&\hspace{-0mm}
2+1&\hspace{-0mm}
0.1713(46)(14)&\hspace{-0mm}
0.352(37)(10)&\hspace{-0mm}
$-$&\hspace{-0mm}
$-$\\%
\midrule
Benayoun-16~\cite{Benayoun:2016krn}&\hspace{-0mm}
pheno.&\hspace{-0mm}
$-$&\hspace{-0mm}
$-$&\hspace{-0mm}
0.09896(73)&\hspace{-0mm}
0.20569(162)\\
Charles-18~\cite{Charles:2017snx}&\hspace{-0mm}
pheno.&\hspace{-0mm}
$-$&\hspace{-0mm}
$-$&\hspace{-0mm}
0.10043(36)&\hspace{-0mm}
0.20914(113)\\
\bottomrule
\end{tabular}
\caption{Up/down-contribution and total for $\Pi_{1,2}$.
The former correspond to the navy circles in \cref{fig:Pi_l_compare}
and include FV corrections.
The light components, $\Pi_{n}^{\rm ud}$, do not include SIB/QED corrections,
but the total $\Pi_{n}^{\rm tot}$ does.
The results for HPQCD-16~\cite{Chakraborty:2016mwy}, FHM-19~\cite{Davies:2019efs}, and Aubin {\em et al.}-19~\cite{Aubin:2019usy} (annotated with $\star$) have been multiplied by a charge factor of $(9/5)$
to convert them to the convention used by the other groups.}
\label{tab:tm_fbf}
\end{center}
\end{table}

\begin{figure}[ht]
\begin{center}
\includegraphics[width=0.48\textwidth]{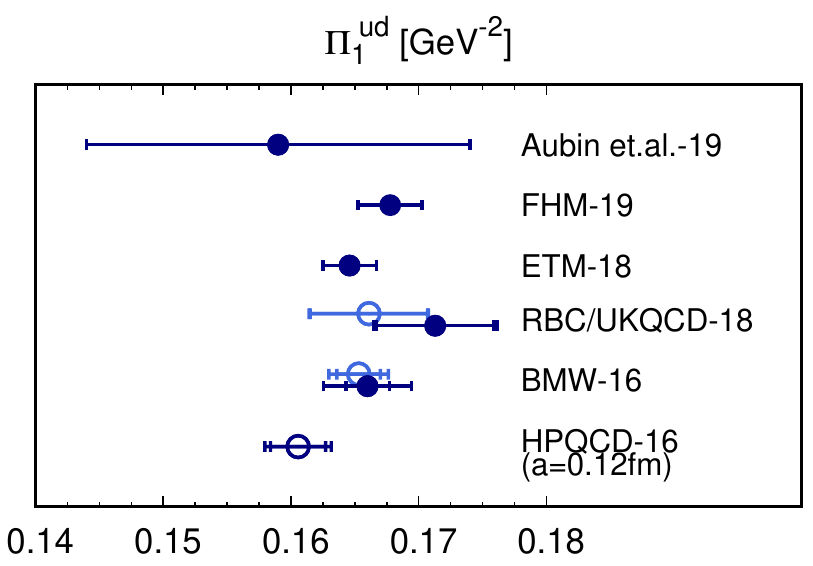}
\includegraphics[width=0.48\textwidth]{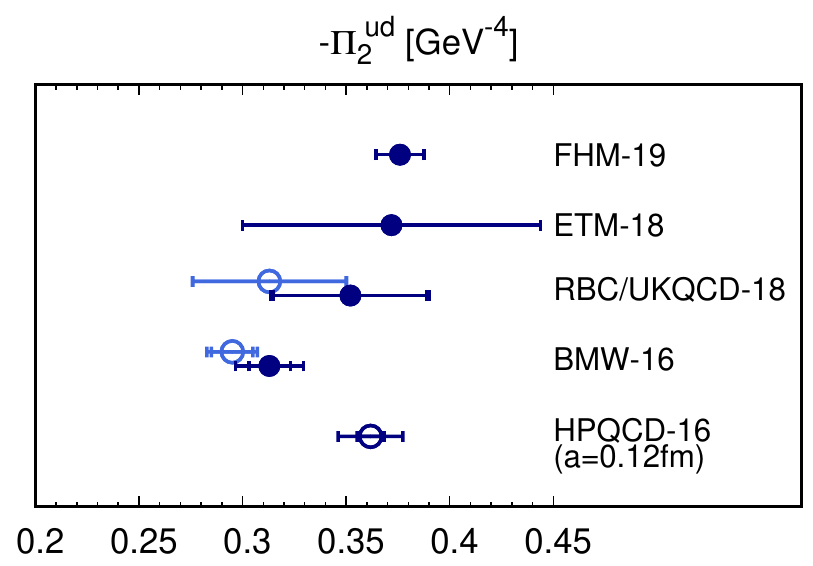}
\caption{
The first (left) and second (right) time moments of the $ud$ contribution to the current--current correlator. The filled blue circles indicate continuum-limit and FV corrected lattice results, while the open light blue circles correspond to results with FV corrections removed. The open dark blue circle shows the HPQCD-16 result at a finite lattice spacing of $a\approx 0.12\,\text{fm}$ with leading discretization and FV corrections added. 
See \cref{tab:tm_fbf} for detailed numbers.
}\label{fig:Pi_l_compare}
\end{center}
\end{figure}

We now consider the moments of the vector current correlator, $\Pi_{n}$, defined in \cref{eq:moments}.
As discussed in \cref{subsec:moments} the  Taylor coefficients $\Pi_{n}$ become sensitive to larger distance scales as $n$ increases. For example, the dominant contributions to $\Pi_{1}$ and $\Pi_{2}$ are from the correlator at $t \sim 1$\,fm and $1.5$\,fm, respectively~\cite{Borsanyi:2016lpl}. Thus, the moments allow a distance-scale-dependent comparison of HVP.
\Cref{fig:Pi_l_compare} shows the light-quark connected contribution to the Taylor coefficients 
($\Pi_{1,2}^{ud}$) reported by various lattice QCD groups.
The open circles are lattice results obtained after the continuum and physical point extrapolation, but without FV corrections, while FV corrections are added in the filled blue circles, making the latter suitable for comparison.\footnote{By definition, $\Pi_{n}^{ud}$ does not include SIB/QED corrections. However, the results for $\Pi_{n}^{\rm tot}$ listed in \cref{tab:tm_fbf} do take SIB/QED effects into account.} The HPQCD-16 ~\cite{Chakraborty:2016mwy} results (open circles) are obtained at a finite lattice spacing ($a \approx 0.12$\,fm), because Ref.~\cite{Chakraborty:2016mwy} did not report continuum limit results for $\Pi_{n}^{ud}$.  
As shown in the left panel of \cref{fig:Pi_l_compare}, all lattice results~\cite{Chakraborty:2016mwy,Borsanyi:2016lpl,Blum:2018mom,Giusti:2018mdh,Davies:2019efs} for $\Pi_1^{ud}$ are nicely consistent, while there is a $2\sigma$ tension between BMW-16~\cite{Borsanyi:2016lpl} and FHM-19~\cite{Davies:2019efs} for $\Pi_2^{ud}$ (right panel).  
Considering the $[1,1]$ Pad\'{e} approximant for the HVP scalar function, $\hat{\Pi}(Q^2)$, the light-quark connected anomaly is evaluated as
\begin{align}
\amuHVPLOud = \Bigl(\frac{\alpha}{\pi}\Bigr)^2
\int_{0}^{\infty} dQ^2 \omega(Q^2/m_{\mu}^2)
\frac{Q^2\Pi_1^{ud}}{ 1 - Q^2\big(\Pi_2^{ud} / \Pi_1^{ud}\big)}\, .
\end{align}
The Pad\'{e} approximants thus tell us that a larger $\Pi_1^{ud}$ and smaller $(-\Pi_2^{ud})$ result in a larger $\amuHVPLOud$~\cite{Borsanyi:2016lpl}. This may explain why BMW-17 (with BMW-16) and RBC/UKQCD-18 obtain a somewhat larger $\amuHVPLOud$ than ETM-18/19 and FHM-19.

\subsubsection{Intermediate window}

\begin{figure}[htb]
    \centering
    \includegraphics[height=0.225\textheight]{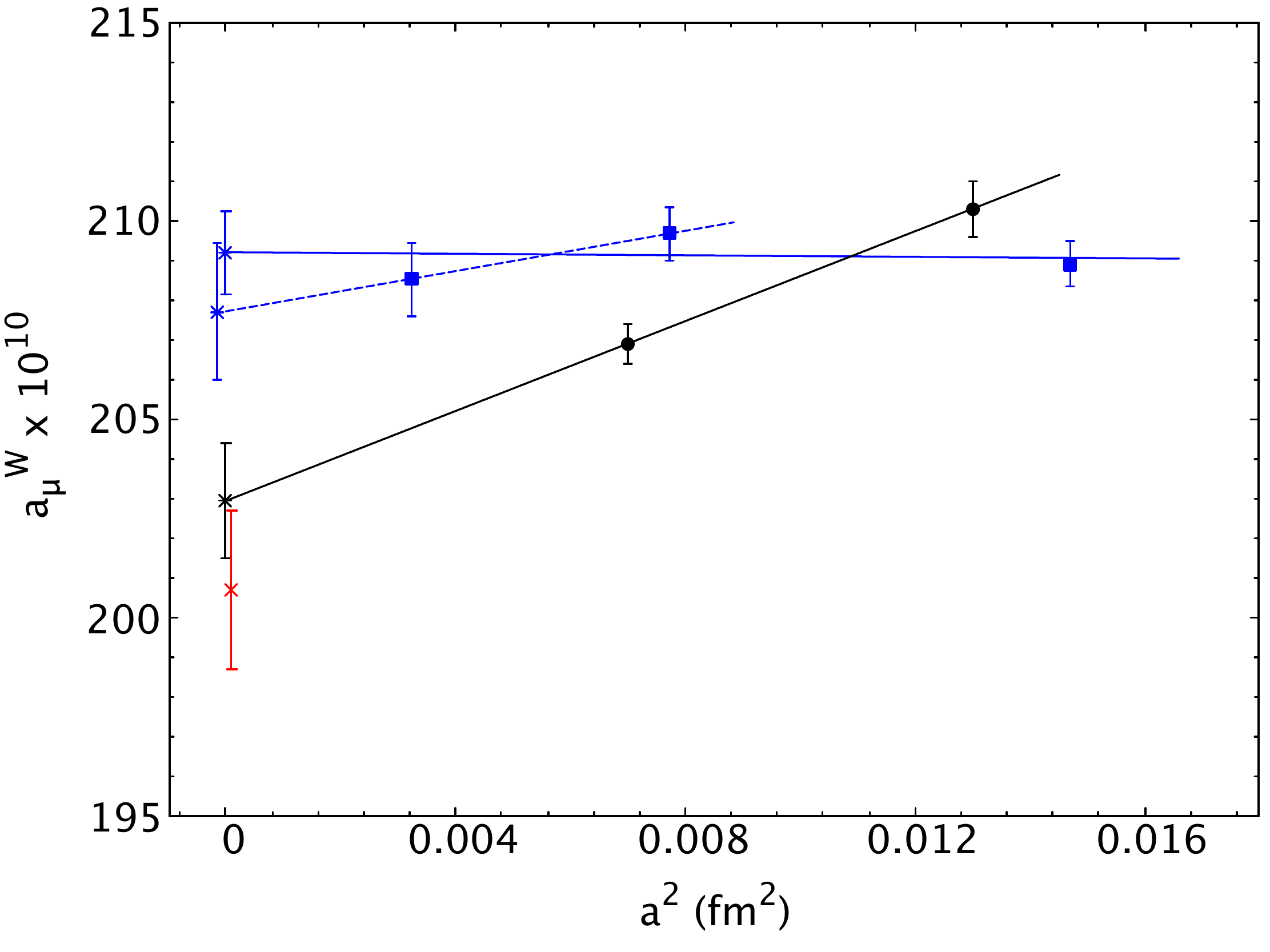}
    \includegraphics[width=0.45\textwidth]{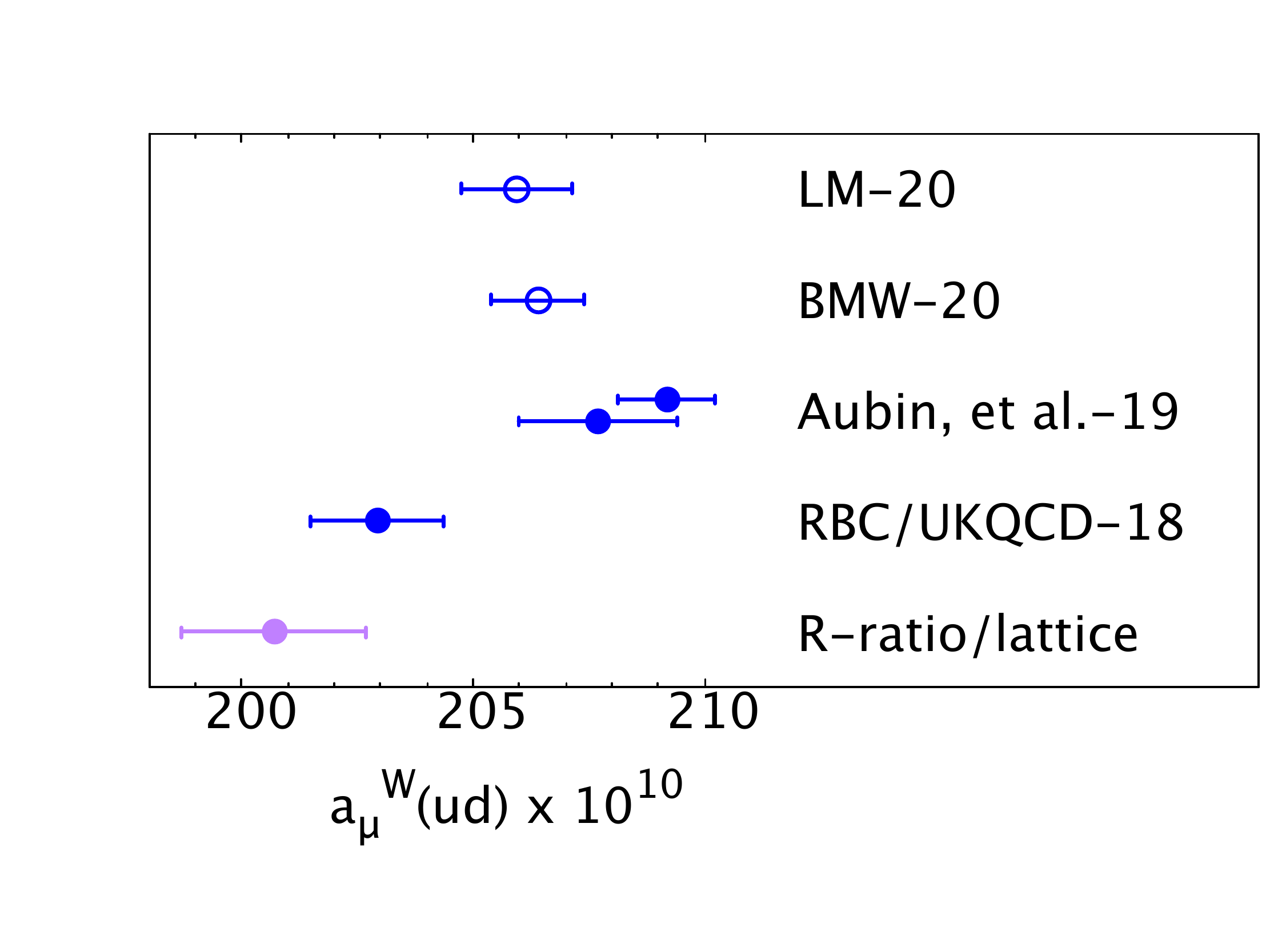}
    \caption{(Left) Continuum limit of the intermediate window $\amuW (ud)$ ($\Delta=0.15$\,fm, $t_0= 0.4$\,fm, $t_1= 1$\,fm) for  DW (circles) and  HISQ (squares) fermions. Lines denote fits linear in $a^2$ for the $a\to0$ limit (bursts).   The $R$-ratio result (cross, Ref.~\cite{CL:private}) is also shown. It corresponds to the difference of the total $R$-ratio window result based on Ref.~\cite{Keshavarzi:2018mgv} and the sum of all but the isospin-symmetric light-quark lattice results of Ref.~\cite{Blum:2018mom}.  Finite-volume (DWF and HISQ) and taste-breaking (HISQ) corrections are included to NLO in ChPT. Lattice spacing uncertainties, added in quadrature with statistical errors, are also included. Plot reproduced from Ref.~\cite{Aubin:2019usy}. (Right) Comparison of $\amuW (ud)$ results after continuum and infinite-volume extrapolation (blue circles). The RBC/UKQCD-18 result~\cite{Blum:2018mom} is based on 2+1 flavors of DW fermions. All other lattice results are based on 2+1+1 flavors of HISQ fermions~\cite{Aubin:2019usy,Borsanyi:2020mff,Lehner:2020crt}. Both extrapolations for HISQ shown in the left panel are reproduced in the right panel as well for comparison.  The DW and $R$-ratio results are the same as shown in the left panel. }
    \label{fig:wincomp}
\end{figure}

The window method developed in Ref.~\cite{Blum:2018mom} can be used to compare lattice results where they are most precise, for example when $0.4 \lesssim x_0\lesssim 1.5$\,fm, which defines the intermediate window $\amuW$. By design, the upper end of the $x_0$ range is chosen to remove contributions from the large Euclidean time region, which is sensitive to FV and two-pion effects and suffers from severe StN problems. In addition, the cut on the lower end of the $x_0$ range is expected to result in reduced discretization errors. This quantity can therefore be calculated with much better (statistical and systematic) precision than the total $\amuHVPLO$, and hence is a powerful diagnostic tool for comparing different lattice methods. Given the precision goals, careful studies of the remaining systematic effects, in particular discretization errors are certainly needed. Finally, as proposed in Ref.~\cite{Blum:2018mom} and discussed in \cref{subsec:tmr}, one can also evaluate $\amuW$ using experimental $R$-ratio data for a more detailed comparison between lattice and data-driven results. However, until very recently, only two groups, RBC/UKQCD-18~\cite{Blum:2018mom} and Aubin {\it et al.}-19~\cite{Aubin:2019usy}, had used their lattice data to evaluate the intermediate window $\amuW$ in the continuum and infinite-volume limits (see the left panel of \cref{fig:wincomp}). The two panels in \cref{fig:wincomp} show lattice results for $\amuW(ud)$, which is defined in isosymmetric QCD (without SIB and QED corrections) specifically for the light-quark contributions. In order to compare the lattice results with an $R$-ratio derived evaluation, the ``$R$-ratio/lattice'' point in \cref{fig:wincomp} is constructed in Ref.~\cite{CL:private} by first using the analysis in Ref.~\cite{Keshavarzi:2018mgv} to evaluate the $R$-ratio window and then subtracting from it the contributions from the heavier flavors, the disconnected, and the IB terms using the lattice results of Ref.~\cite{Blum:2018mom}.
The right panel of \cref{fig:wincomp} shows, in addition to the published RBC/UKQCD-18 and Aubin {\it et al.}-19 results, two new lattice results for $\amuW$, BMW-20~\cite{Borsanyi:2020mff} and LM-20~\cite{Lehner:2020crt} (open blue circles). Both appeared only very recently, and have therefore not yet been reviewed in depth in this paper. The three staggered results, Aubin {\it et al.}-19, BMW-20, and LM-20, lie above the RBC/UKQCD-18 and ``$R$-ratio/lattice''  values. The quoted uncertainties on the new BMW-20 and LM-20 lattice results  and on Aubin {\it et al.}-19 for the three lattice spacing fit are significantly smaller than the ``$R$-ratio/lattice'' one, with which only the RBC/UKQCD-18 value is clearly compatible. If one considers the spread of the two and three lattice spacing fit results from Aubin {\it et al.}-19 as a systematic uncertainty for this calculation,
the RBC/UKQCD-18 evaluation is also compatible with the Aubin {\it et al.}-19, BMW-20, and LM-20 results at the 2$\sigma$ level.  It will be important to see what happens to the spread of the lattice results as more high-precision calculations of this quantity become available. Needless to say, such calculations should include detailed analyses of the associated systematics, in particular, discretization effects. 

\subsection{Connections}
\label{sec:connect}

In this section, we highlight some of the connections between lattice calculations of HVP and other efforts related to the anomalous magnetic moment of the muon and other electroweak precision quantities. In \cref{subsec:muone}, we explain the impact of lattice calculations of HVP on the proposed
MUonE experiment. In \cref{subsec:tau}, we show how lattice calculations of IB contributions to
HVP can be used to assess the IB corrections to $\tau$ spectral functions in the evaluation of HVP
from $\tau$ decays. Finally, in \cref{subsec:alpha}, we demonstrate the impact of lattice
calculations of HVP on the determination of the running of the electromagnetic coupling $\alpha$
and the weak mixing angle $\sin^2{\theta_W}$.

\subsubsection{HVP from lattice QCD and the MUonE experiment}
\label{subsec:muone}

A novel effort proposing a muon--electron scattering experiment (MUonE) at CERN is gaining a lot of attention as a clean, direct measurement of 
the HVP function, $\hat\Pi(Q^2)$, in a {\it spacelike} regime~\cite{Calame:2015fva,Abbiendi:2016xup} that will help improve the precision of the determination of $\amuHVPLO$.
Initially a part of the {\it Physics Beyond Colliders} program for a dedicated fixed-target facility in CERN's North Area, the request for a test run of the MUonE experiment~\cite{MUonE:LoI} has recently been approved by the SPS Committee at CERN. Pending a successful feasibility study in 2021, 
MUonE plans to measure HVP up to $Q^2=0.14\,\gev^2$
with statistical precision of roughly 0.3\% after two years of data taking (cf.\ \cref{Sec:MUonE}). 

Lattice QCD and the MUonE experiment will determine HVP to high accuracy in complementary momentum ranges. Thus, a hybrid strategy including both experimental and lattice data sets~\cite{Marinkovic:2019zoi} will give an estimate of the HVP contribution to $a_\mu$ that is independent of $e^+e^-\to\text{hadrons}$ data. This strategy is similar to the approach described in \cref{subsec:inthvp}, which allows for a better control of the systematics in $\amuHVPLO$ by dividing the Euclidean $Q^2$-range into three sub-intervals and probing different integration boundaries of these domains~\cite{Golterman:2014ksa}. 
MUonE will yield a measurement of the effective electromagnetic coupling $\Delta \alpha (Q^2(x))$ in the kinematic range with maximal momentum corresponding to $x_\text{max}=0.932$, $Q^2 = \frac{x^2m_{\mu}^2}{1-x}$. 
 Hence, writing $\amuHVPLO = I_0 + I_1 + I_2$ as in  \cref{eq:intsum},  this suggests that $I_0$ can be evaluated using MUonE data. Rephrasing \cref{eq:subint0} in terms of $x$ gives
\begin{equation}
I_0 =\frac{\alpha}{\pi}\int_{0}^{0.932}dx (1-x) \Delta \alpha_\text{had}[Q^2(x)]\,,  
\label{eq:subintegral0}
\end{equation}
where the upper limit, $x_\text{max}=0.932$, corresponds to $\Qlow =0.14\GeV^2$  and the integrand will be determined by  the data collected by the MUonE experiment. The strategy to reach the aimed precision is discussed in \cref{Sec:MUonE}.  
As noted in \cref{subsec:inthvp}, the data in the integration range of $I_0$ (cf.\ \cref{fig:hybrid} and text below) is crucial for the overall precision of the HVP spacelike integral. The expected high accuracy of the MUonE experiment in this $Q^2$-region, combined with lattice input for $I_1$, defined in~\cref{eq:subint1}, for the intermediate-$Q^2$ range, will lead to a precise estimate of HVP entirely independent of the dispersive approach. 
Precise determinations of $I_1$ from the lattice will be used in two ways: 
\begin{enumerate} 
\item A lattice estimate of $I_1$  will be combined with the MUonE experimental measurements in the low-$Q^2$ region and the perturbative contribution $I_2$ to obtain a precise estimate of the total HVP contribution.  A preliminary result for $I_1$ with $\order(a$) improved Wilson fermions is discussed in Ref.~\cite{Marinkovic:2019zoi}. 
Recently~\cite{Giusti:2018mdh,Giusti:2019hoy,Giusti:2019hkz}, the sum $I_1+I_2$ (including disconnected contributions and IB effects) has been estimated to be equal to $91.6(2.0) \times 10^{-10}$, using the time momentum representation and the twisted-mass discretization of the fermion QCD action. It is encouraging that the uncertainty in the above result is close to the statistical error expected for the integral $I_0$ after two years of data taking.
\item The infinite-volume, continuum values of the integrals $I_1$ and $I_2$ for a predefined set of integration boundaries can serve as benchmarks between different lattice collaborations, in addition to the comparisons discussed in \cref{sec:comp}.
\end{enumerate}
Furthermore, the MUonE experimental data at fixed-$Q^2$ in the measured range could be used to validate the independent lattice determinations (after the infinite-volume and continuum limits are taken) at fixed $Q^2$-values. 

\subsubsection{HVP from $\tau$ decays}
\label{subsec:tau}

Hadronic decays of $\tau$ leptons have been used in the 
past as an alternative source of experimental data
in the evaluation of the HVP contribution to $(g-2)_\mu$ with 
the dispersive methods.
As discussed in \cref{HVP-tau},  
the sum of contributions from the subset of IB effects 
that have been identified and estimated phenomenologically to date 
have so far failed to explain the observed experimental 
$\tau$--$e^+e^-$ $\pi\pi$ difference. 
The possibility, moreover, exists that contributions from sources of 
IB for which phenomenological estimates are not
possible may not be numerically negligible. 
In this situation, it is worth exploring the alternate 
approach based on lattice QCD+QED simulations, 
which would automatically take into
account all sources of IB.

The calculation of the QED and  strong IB corrections 
from lattice simulations to HVP has been presented in detail in \cref{sec:HVP_IB}. Here we start our discussion from the 
isospin decomposition of the two-flavor current, 
$j_\mu = j_\mu^{I=0,I_3=0} + j_\mu^{I=1,I_3=0}$. 
The contributions to HVP can be split into three classes of two-point correlation functions $C(x_0)$ 
(see Ref.~\cite{Bruno:2018ono} for more details): 
isoscalar $C_{00}$, isovector $C_{11}$, and 
mixed contributions $C_{01}$. 
Similarly, from the charged (vector current) operators $j_\mu^{I=1,I_3=\pm1}$
we define the two-point correlation function $C_{+-}$: 
if isospin was a preserved
symmetry, the mixed contribution would vanish identically and 
$C_{+-} = C_{11}$.
However, the presence of QED interactions together with the 
up/down mass difference breaks isospin symmetry, leading to
two corrections that must be applied to the charged spectral functions: 
the first, given by the mixed contribution 
$C_{01}$, is completely absent in the $\tau$ data, 
which carries most of the $\rho\hyph\omega$ mixing physics; 
the second is given by the difference $C_{11} - C_{+-}$.
The additional channels not present in $\tau$ decays, 
such as $\pi^0 \gamma$ or $\pi^+ \pi^- \pi^0$, can be supplemented
directly from experimental data or via $C_{00}$.

If we imagine a perturbative expansion in $\alpha$ and 
$(m_u - m_d)$ around the isosymmetric world,
the difference $C_{11} - C_{+-}$ only depends on two specific 
QED diagrams, used by the RBC/UKQCD collaborations 
in Ref.~\cite{Bruno:2018ono}
to provide a preliminary estimate.
$C_{01}$ instead involves many more diagrams and is dominated by
the SIB corrections, which in turn are noisier and are currently 
under investigation.\footnote{See \cref{sec:HVP_IB} for a discussion of recent progress in lattice calculations of 
the connected SIB contribution.}
Note, however, that the overall precision required
for the total correction is approximately 10\hyph15\%, a goal that is 
well within reach of state-of-the-art lattice simulations.

In conclusion, the calculation of the IB correction
to inclusive $\tau$ spectral densities involves essentially a manipulation of
the QED and SIB diagrams of quark-connected and disconnected HVP, 
described in \cref{sec:HVP_IB}. We refer the reader to that section for a proper discussion and 
understanding of the systematic errors, 
such as finite-volume and discretization effects. 
Before concluding this subsection, we focus our attention on two important
systematic effects, peculiar to this specific 
IB correction factor.
First, we must remember that lattice correlation functions are fully
inclusive, and as such, the correction factor obtained from lattice simulations
should be applied to the sum of the $\pi^-\pi^0$ spectral
density with other channels, like the four-pion one.
Higher excited states are expected to be suppressed by the $(g-2)_\mu$ kernel, 
but a study has to be performed to properly understand this subtlety.
Second, when comparing 
isospin-corrected $\tau$ spectral
functions (not $a_\mu$) between a lattice and a phenomenological 
prediction, or an experimental measurement,
particular care is required to make sure that the
treatment of the radiative effects is the same.
As discussed in \cref{HVP-tau}, in $\tau$ decays to $\pi^- \pi^0$ final states, radiative events are removed from the experimental data via the function $G_\mathrm{EM}$ introduced in \cref{eq:ib}. In contrast, only FSR is present in the lattice calculation of IB effects discussed in Ref.~\cite{Bruno:2018ono}. Therefore a redefinition of $G_\mathrm{EM}$ to remove the remaining radiative corrections from the experimental $\tau$ data is required, before one applies the weighted integral over the spectral function difference evaluated on the lattice to the weighted integral over the experimental $\tau$ data.

As new exciting results are expected in the upcoming years 
on the experimental determination of the charged spectral functions,
especially from Belle II, 
a controlled theoretical input of the IB correction is crucial 
to provide an alternative determination of the HVP 
contribution to $a_\mu$, and lattice QCD+QED calculations will play
an essential role in the understanding of the final uncertainties, given
the difficulties so far encountered in the phenomenological approach.

\subsubsection{Hadronic corrections to the running of the electromagnetic coupling and the weak mixing angle}
\label{subsec:alpha}

The lattice QCD computation of HVP can also provide a first-principles
determination of the hadronic contributions to the running of the
electromagnetic coupling $\alpha$ and of the weak mixing angle
$\sin^2\theta_W$.

A usual parameterization of the momentum dependence of the QED coupling
reads $\alpha(q^2)\ =\ \alpha/\big(1- \Delta \alpha(q^2)\big)$, where
$\alpha$ refers to its value at $q^2=0$, corresponding to the fine-structure constant, and where $\Delta \alpha(q^2)$ collects the sum of
quark and lepton contributions. The largest fraction of the overall
uncertainty on the running of $\alpha(q^2)$ from $q^2=0$ up to the
$Z$-pole mass is due to low-energy hadronic effects, denoted by
$\Delta \alpha_{\rm had}(q^2)$. The prospects for a future
International Linear Collider indicate that the uncertainty on the
running of $\alpha$ could become a limiting factor in the global fit
of the electroweak sector of the SM~\cite{Jegerlehner:2011mw,Baak:2014ora}.

The dispersive approach can be used to determine $\Delta \alpha_{\rm
  had}(q^2)$ from the lightest five quark
flavors~\cite{Davier:2017zfy,Keshavarzi:2018mgv,Jegerlehner:2017zsb},
where the dispersive integral is dominated by higher-energy scales 
compared to the case of $\amuHVPLO$.
On the lattice, the subtracted HVP function $\hat{\Pi}(Q^2) = 4\pi^2[\Pi(Q^2)-\Pi(0)]$ at spacelike momenta $Q^2=-q^2$ is used to estimate the hadronic contribution to the running
\begin{equation}
  \Delta\alpha_\textrm{had}(-Q^2) = \frac{\alpha}{\pi} \hat{\Pi}(Q^2)\,.
\end{equation}
The connection to the running to the $Z$~pole in the timelike region is given by the so-called Euclidean split technique (or Adler function approach)~\cite{Eidelman:1998vc,Jegerlehner:2008rs}
\begin{equation}
\label{eq:Eu_split}
  \Delta\alpha_\textrm{had}(M_Z^2) = \Delta\alpha_\textrm{had}(-Q_0^2) + \left[ \Delta\alpha_\textrm{had}(-M_Z^2)-\Delta\alpha_\textrm{had}(-Q_0^2) \right] + \left[ \Delta\alpha_\textrm{had}(M_Z^2)-\Delta\alpha_\textrm{had}(-M_Z^2) \right]\,.
\end{equation}
The second and third contributions in the RHS are computed using the perturbative expansion of the Adler function $D(Q^2)$, which works well starting from $Q_0^2\approx 4\,\textrm{GeV}^2$, while the first term on the RHS of \cref{eq:Eu_split} accounts for the nonperturbative contribution.
Its lattice determination provides an alternative to the dispersive approach, in which the nonperturbative running is derived from $e^+e^-\to\text{hadrons}$ data by means of \cref{alpha_had}.
While the relative uncertainty
at the $Z$~pole is below $0.5\%$, at intermediate values of $Q^2
\approx 4\,{\rm GeV}^2$, the uncertainty from experimental data in the
dispersive approach amounts to a $\sim 1\%$ error on $\Delta
\alpha_{\rm had}(-Q^2)$~\cite{Jegerlehner:2017zsb}. A lattice QCD
determination at this level of precision would therefore already
provide important information about the hadronic corrections to the
running of $\alpha$.

In the range of $Q^2$-values between $0.5$ and $4\,{\rm GeV}^2$, the
subtracted HVP function $\hat{\Pi}(Q^2)$ can be computed on the
lattice with high statistical precision. Contrary to the case of
$\amuHVPLO$, where smaller $Q^2$-values are responsible for the
largest share of the total error, in the case of $\Delta \alpha_{\rm
  had}(-Q^2)$ short-distance contributions play a more significant
role. Particular attention is thus required to the assessment of
systematic effects from lattice artifacts and scale setting. A first
comparison of lattice determinations of $\Delta \alpha_{\rm had}(-Q^2)$
to the results from the dispersive approach has been reported
in Refs.~\cite{Burger:2015lqa,Francis:2014yga,Francis:2015grz,Ce:2019imp}. We remark
that detailed comparisons among lattice QCD and dispersive 
results at definite values of $Q^2$ could provide stringent
cross-checks of both approaches.
One way to achieve this is using the window method described in \cref{sec:intro_latHVP}.

In the ``no new physics'' scenario the theoretical estimate of $\amuHVPLO$ would shift the SM prediction for the total $a_\mu$ towards the experimental value. While the data-driven determinations of $\amuHVPLO$ are incompatible with this scenario, the current lattice results cover a large enough range to be compatible with both, ``no new physics'' and the data-driven result for $\amuHVPLO$ (see \cref{sec:summary_latHVP}). It is in principle possible that future, precise lattice results will consolidate around a high enough value to bring the corresponding SM prediction into agreement with experiment. Indeed, very recently, the BMW collaboration reported a first lattice QCD+QED calculation of $\amuHVPLO$ with subpercent precision~\cite{Borsanyi:2020mff}, which is compatible with the ``no new physics'' scenario and discrepant with the $R$-ratio data. 
However, in addition to disagreeing with the $R$-ratio data, which would need to be understood, such a shift in $\amuHVPLO$ would also affect $\Delta\alpha_\textrm{had}$ and would therefore have consequences for global SM fits~\cite{Passera:2008jk,Crivellin:2020zul}.
Indeed, excluding input from $R$-ratio data, global SM fits are sufficiently overconstrained that it is possible to predict $\Delta\alpha^{(5)}_\mathrm{had}(M_Z^2)$ using only electroweak precision probes. For instance, the 2018 Gfitter result, $\Delta\alpha^{(5)}_\mathrm{had}(M_Z^2)|_\text{EW}=271.6(3.9)\times 10^{-4}$~\cite{Haller:2018nnx},\footnote{The precise value depends on the choice of input parameters and details of the fit, e.g., $\Delta\alpha^{(5)}_\mathrm{had}(M_Z^2)|_\text{EW}=270.3(3.0)\times 10^{-4}$~\cite{Crivellin:2020zul} using the PDG values~\cite{Tanabashi:2018oca} of $m_t$, $M_H$ and the Bayesian implementation \texttt{HEPfit}~\cite{deBlas:2019okz}.} is compatible with, albeit slightly lower than, the data-driven evaluation   $\Delta\alpha^{(5)}_\mathrm{had}(M_Z^2)|_{e^+e^-}=276.1(1.1)\times 10^{-4}$~\cite{Davier:2019can,Keshavarzi:2019abf} based on the $R$-ratio data.
In summary, a ``no new physics'' result for $\amuHVPLO$ is likely to also yield a higher value of $\Delta\alpha_\mathrm{had}$. However, because the two quantities have different, positive weights in each energy bin of the $e^+e^-$ cross section, for lattice calculations like Ref.~\cite{Borsanyi:2020mff}, information about the $Q^2$ dependence of the HVP function $\hat{\Pi}(Q^2)$ is needed to quantify the effect on $\Delta\alpha^{(5)}_\mathrm{had}(M_Z^2)$. Indeed, high-precision lattice calculations for $\Delta\alpha_\textrm{had}(Q^2)$ up to some energy scale $Q^2$ (where connection with the perturbative running up to the $Z$~pole can be made) would exactly provide the needed information.

The weak mixing angle $\sin^2\theta_W$ governs the $\gamma$--$Z$ mixing and induces a constraint on the coupling constants of the electroweak theory. The value of $\sin^2\theta_W$ at the $Z$~pole has been determined with high precision by the LEP and Tevatron experiments and is heavily constrained by the global fit of the SM in the electroweak sector~\cite{Haller:2018nnx}.

Various ongoing and future low-energy experiments~\cite{Erler:2019hds}
target a high-precision measurement of $\sin^2\theta_W$ at energy
scales below $4\,{\rm GeV}^2$, where nonperturbative QCD effects play
a prominent role. In contrast to the case of $\amuHVPLO$ and
$\alpha$, a straightforward application of a dispersive approach
suffers from the difficulty to isolate the contributions from up- and
down-type quarks. Such a flavor separation is, however, naturally
provided by a lattice QCD calculation.
As an example, flavor separation based on neglecting OZI-violating effects has been shown to be incompatible with lattice data, while assuming isovector $\rho$-meson dominance works better~\cite{Jegerlehner:2017zsb}.
The hadronic contribution to
the running of the weak mixing angle, $\Delta^{\rm
  had}\sin^2\theta_W(-Q^2)$, can be expressed in terms of the HVP
function once an independent input for the value of the $SU(2)_L$
coupling $\alpha_2$ at $Q^2=0$ is
employed~\cite{Erler:2017knj}. Lattice QCD computations of
$\Delta^{\rm had}\sin^2\theta_W(-Q^2)$ have been reported
in Refs.~\cite{Burger:2015lqa,Francis:2015grz,Guelpers:2015nfb,Ce:2018ziv,Ce:2019imp}. These
results suggest that subpercent precision could be achieved
in upcoming lattice determinations of $\Delta^{\rm
  had}\sin^2\theta_W(-Q^2)$ at intermediate $Q^2$-values, thus
providing an essential theoretical counterpart to the ongoing
experimental efforts.

\subsection{Summary and conclusions}
\label{sec:summary_latHVP}

\subsubsection{Current status}\label{subsec:status}

As discussed in \cref{sec:comp} and shown in \cref{tab:amu_cmp,tab:amu_fbf}, there are up to seven high-quality calculations of the different flavor contributions to $\amuHVPLO$. Thus, it makes sense to consider some form of world average of lattice results for this quantity. However, not all groups compute all contributions and, as seen above, the various contributions suffer from significantly different systematic errors, some of which are correlated from calculation to calculation. Therefore, we have chosen to average  the individual contributions separately, taking into account correlations so as not to underestimate errors. These flavor-by-flavor averages will also be useful to the lattice community as benchmarks.

All contributions have a systematic uncertainty associated with the determination of the lattice scale in common. This uncertainty is significant, but largely uncorrelated 
among the various calculations, because it is dominated by the underlying statistical errors. Small correlations due to common, subleading systematics may be present, in particular for SIB and QED effects. For example, for all calculations except for those of RBC/UKQCD-18~\cite{Blum:2018mom} and ETM-19~\cite{Giusti:2019hkz}, the lattice spacing is determined from lattice calculations based on ensembles with $m_u=m_d$ and $\alpha = 0$, with QED effects estimated phenomenologically. While in some cases SIB effects are calculated directly, in others they are also estimated phenomenologically. The resulting systematic errors on the overall scale are in the range of a few permil to one percent.
In summary, we treat the scale errors as uncorrelated between independent lattice results, except for Aubin {\em et al.}-19~\cite{Aubin:2019usy} and FHM-19~\cite{Davies:2019efs}, since the Aubin {\em et al.}-19 result uses the scale uncertainty obtained in Ref.~\cite{Davies:2019efs}. 

The $ud$ and disconnected contributions to $\amuHVPLO$ also suffer from significant FV uncertainties, even on lattices with $L\sim 6\,\fm$, as discussed in \cref{sec:HLO_ud,sec:HLO_disconn}. Indeed, FVEs have been mostly estimated using ChPT~\cite{Aubin:2015rzx,Bijnens:2017esv,Aubin:2019usy}, ChPT-inspired models~\cite{Chakraborty:2016mwy}, or direct estimates in still imprecise lattice calculations at the physical pion mass~\cite{Shintani:2019wai}. In the case of Mainz/CLS-19~\cite{Gerardin:2019rua} and ETM-18/19~\cite{Giusti:2018mdh,Giusti:2019hkz}, models of the $m_{ud}$ and $L$ dependencies of FVEs have been studied more carefully using ensembles with smaller volumes and with a number of average, light, $u$ and $d$ quark masses that are larger than the physical value. These studies have then been used to extrapolate their results for $\amuHVPLOud$ to infinite volume and physical quark masses. All of these estimates point to FV corrections around 3\% for $L\sim 6\,\fm$ and physical $m_{ud}$. However, the uncertainties on these results are still quite large and the methods used by different teams to estimate them are quite similar. Thus, we consider the uncertainties associated with these FV corrections, as well as those associated with extrapolations in $m_{ud}$, as 100\% correlated across calculations.

In addition, we treat the statistical errors in FHM-19~\cite{Davies:2019efs} and Aubin {\em et al.}-19~\cite{Aubin:2019usy} as 100\% correlated, because these two calculations are based on overlapping MILC configurations. 

To summarize, we go through the uncertainties, listed in each calculation, for each of the contributions to $\amuHVPLO$, and isolate the FV and chiral uncertainties and, for FHM-19~\cite{Davies:2019efs} and Aubin {\em et al.}-19~\cite{Aubin:2019usy}, the statistical and scale errors. We then average the results for each of the contributions to $\amuHVPLO$, assuming 100\% correlation between these uncertainties. If these cannot be isolated, but rather are subsumed in a more encompassing systematic error, we correlate this larger uncertainty with the uncertainties of the other calculations. All other errors are treated as uncorrelated. To perform the averages, we use the FLAG strategy described in Ref.~\cite{Aoki:2019cca}. 

Sea-charm-quark effects on $\amuHVPLO$ can be estimated in QCD perturbation theory (see, e.g., Ref.~\cite{Blum:2018mom}) and are found to be below the permil level. Thus, we can combine $N_f=2+1$ and $N_f=2+1+1$ results. However, when permil-level calculations of $\amuHVPLO$ start being considered, it will be important to make sure that sea-charm effects do not alter the scale setting at that level of precision. Note that $b$-quark effects can also be estimated in perturbation theory (see, for example,  Ref.~\cite{Davies:2019efs}) and are found to be below 1 permil. These will be neglected here. 

To determine the average $\amuHVPLOud$, we use the calculations of BMW-17~\cite{Borsanyi:2017zdw}, RBC/UKQCD-18~\cite{Blum:2018mom}, PACS-19~\cite{Shintani:2019wai}, FHM-19~\cite{Davies:2019efs}, Mainz/CLS-19~\cite{Gerardin:2019rua}, Aubin {\em et al.}-19~\cite{Aubin:2019usy}, and ETM-18/19~\cite{Giusti:2018mdh,Giusti:2019hkz}. As can be seen in \cref{fig:amu_compare_fbyf}, agreement between the different collaborations is not that good. The average carries a $\chi^2/\text{dof}=10.3/6=1.7$ and we follow the PDG recipe of increasing the final error on the average by a factor equal to the square root of that number.

The averages for $\amuHVPLOs$ and $\amuHVPLOc$ are obtained from the same calculations, except for Aubin {\em et al.}-19 that does not include results for those quantities. Here the $\chi^2/\text{dof}$ are $5.8/5=1.2$ and $4.3/5=0.9$, so that there is only a small inflation of the final error for $\amuHVPLOs$. Note that the inclusion of PACS-19 for $\amuHVPLOs$ increases the total $\chi^2$ from 0.7 to 5.8, indicating that its agreement with the average of the other results is poor.

For $\amuHVPLOdisc$, there are only three independent lattice results with a signal: BMW-17, RBC/UKQCD-18, and Mainz/CLS-19, which we average. While the first two agree very well, the third is significantly lower, due to a large chiral extrapolation. The average carries a $\chi^2/\text{dof}=4.2/2=2.1$ and the final error is appropriately enlarged.

All of the contributions discussed up to now have been obtained in the isospin limit defined in \cref{sec:strat}, with $m_u=m_d$ and $\alpha=0$. To compare lattice results to those obtained in phenomenology, SIB and QED corrections must be added. Of the results given \cref{tab:ib_summary}, only RBC/UKQCD-18, FHM-17, and ETM-19 present results for the SIB corrections from direct lattice calculations. Further, only RBC/UKQCD-18 and ETM-19 also include connected QED contributions in their IB calculation, where RBC/UKQCD-18 alone includes the leading disconnected QED diagram. As these are first-generation calculations of some of those effects and those effects are small, we choose the conservative approach of averaging RBC/UKQCD-18 and ETM-19 under the assumption that the uncertainties of the two calculations are 100\% correlated.

\begin{table}[t]
\begin{center}
\small
\begin{tabular}{c@{\hskip 24pt}c@{\hskip 24pt}c@{\hskip 24pt}c@{\hskip 24pt}c}
\toprule
$\amuHVPLOud$ & $\amuHVPLOs$ & $\amuHVPLOc$ & $\amuHVPLOdisc$
& $\delta a_\mu^\text{HVP}$\\
\midrule
$650.2 (11.6)$ & $53.2 (0.3)$ & $14.6 (0.1)$ & $-13.7  (2.9)$ & $7.2 (3.4)$\\
\bottomrule
\end{tabular}
\caption{\label{tab:amuflavsummary}Summary results for the individual flavor contributions to $\amuHVPLO (\alpha^2)$ (in the isospin limit, $\delta m = 0$ and $\alpha=0$) and for the SIB and QED corrections in units of $10^{-10}$. }
\end{center}
\end{table}

Our averages for the individual flavor contributions to $\amuHVPLO$, in the isospin limit and for the leading SIB and QED corrections, are summarized in \cref{tab:amuflavsummary}. These contributions can now be combined to obtain average lattice numbers for the LO HVP contribution in the isospin limit, $\amuHVPLO (\alpha^2)$. Adding to this result the SIB and QED corrections yields the total LO HVP correction $\amuHVPLO$, which can be compared with the phenomenological determinations of $\amuHVPLO$.
A final choice has to be made on how to combine uncertainties in adding up the individual flavor and SIB+QED contributions. The uncertainty associated with the determination of the overall scale is 100\% correlated among these contributions in a same calculation. Those associated with FV corrections are anticorrelated in the $ud$ and disconnected contributions, but those in the latter are only $-1/10$ of the former. Also, in a calculation of the various contributions performed on the same set of ensembles, statistical errors will be correlated. Thus, we chose the conservative approach of adding the errors of the individual contributions linearly in our final results. This procedure leads the following world average for the ${\cal O}(\alpha^2)$ contribution to the total LO HVP correction:
\begin{equation}
\label{eq:amuhlo_iso_lat}
    \amuHVPLO (\alpha^2) = 704.3(15.0)\times 10^{-10}\ .
\end{equation}
Adding to this result the average SIB+QED correction, we obtain the following lattice world average for the total LO HVP contribution to the muon $(g-2)$:

\begin{equation}
\label{eq:amuhlo_IB_lat}
    \amuHVPLO = 711.6(18.4) \times 10^{-10}\ ,
\end{equation}
where we include subleading digits at intermediate stages before rounding to the digits shown. This result has an uncertainty of 2.6\%, which is a factor of 4.5 larger than the data-driven one quoted in  \cref{HVPdisp_conclusions}. Unfortunately this means that it is not precise enough to distinguish between the data-driven evaluation and the ``no new physics'' scenario, where $\amuHVPLO$ would be large enough to bring the SM prediction into agreement with the current experimental measurement of $a_\mu$. Thus, it is imperative that lattice calculations of $\amuHVPLO$ significantly improve in the coming years. The prospects for this are discussed in the following subsection. Finally, we note that the lattice average for $\amuHVPLO$ is based on Refs.~\cite{\HVPlatticeref}, which should be cited in any work that uses or quotes \cref{eq:amuhlo_IB_lat}. 

\subsubsection{Towards permil-level precision} 

Thanks to a dedicated and sustained effort by the lattice community in recent years, remarkable progress has been made in lattice calculations of the hadronic contributions to $a_\mu$ and related quantities, as we have outlined in the preceding subsections (see also \cref{sec:latticeHLbL}). This work continues, as evidenced by the several new lattice results for $\amuHVPLO$~\cite{Borsanyi:2020mff,Lehner:2020crt,Giusti:2020efo} (and related quantities) that have appeared so recently that they could not be reviewed in depth in this paper.\footnote{See \cref{sec:introWP} for more information about the deadline and rules adopted in this review concerning the inclusion of papers as essential inputs.} 

We note that lattice QCD calculations relevant for quark flavor physics have also improved significantly, where a growing number of quantities have now been calculated with permil-level precision (see, for example, Refs.~\cite{Aoki:2019cca,Lehner:2019wvv} for an overview). In that case, however, the relevant quantities, hadron masses, decay constants, form factors, and other hadronic matrix elements of local operators involving stable, single-particle states do not suffer from the same complications as the HVP function discussed in this review. As discussed in \cref{sec:intro_latHVP,sec:strat}, the complications are due to HVP (which is computed from the vector-current two-point function, see \cref{subsec:intro_hvp}) receiving contributions from vector-meson intermediate states, chiefly the $\rho$ meson, and hence also from multi-pion states, which cause significant finite-volume effects and exponentially growing signal-to-noise problems at large Euclidean times. A further complication is the presence of disconnected contributions, which are notoriously difficult to calculate with precision. Finally, a common challenge is that permil-level precision requires the inclusion of corrections due to strong IB ($m_u \neq m_d$) and QED effects. 

Nevertheless, despite the added complications in lattice QCD+QED calculations of HVP, the existence of reliable permil-level lattice results for other quantities is encouraging. Also encouraging is that a significant effort by the international lattice community to add HVP to this list is currently underway. Indeed, the first lattice QCD+QED result for $\amuHVPLO$ with subpercent precision was very recently presented by the BMW collaboration~\cite{Borsanyi:2020mff}, an impressive achievement.  The quoted result, $\amuHVPLO = 712.4(4.5) \times 10^{-10}$, differs from the data-driven evaluation obtained in \cref{HVPdisp_conclusions} by 3.2$\sigma$ and is consistent with the ``no new physics'' scenario. This surprising disagreement with the data-driven $R$-ratio results must be checked by other lattice calculations with commensurate precision and, if confirmed, must be understood. Moreover, as pointed out in \cref{subsec:alpha} and in Refs.~\cite{Passera:2008jk,Crivellin:2020zul}, while bringing the SM into agreement with experiment for $a_\mu$, this result would impact existing tensions within the global electroweak fit, since a change in HVP also affects $\Delta\alpha_\text{had}$. 

The results presented in Ref.~\cite{Borsanyi:2020mff} will, of course, be scrutinized in detail in the ongoing efforts by other lattice groups. In particular, the extent to which Ref.~\cite{Borsanyi:2020mff} has successfully leveraged the methods described in this review to overcome the above-mentioned challenges must be checked by other groups. Some results may indeed appear later this year, but it is much too early to speculate on the outcome of these efforts. Here we emphasize that any upcoming study should include detailed results not just for the individual flavor contributions to $\amuHVPLO$, but also for the other intermediate quantities that are defined in \cref{sec:intro_latHVP} and discussed in \cref{sec:comp}, including the Taylor coefficients ($\Pi_n$), hybrid integral ($I_1$), and the window quantities ($\amuSD$, $\amuW$, and $\amuLD$).  These intermediate quantities are sensitive to different distance scales, and hence differ in how they are affected by statistical and systematic uncertainties. They can therefore yield valuable insights into any tensions between different lattice results and also between lattice and data-driven methods. In particular, the intermediate window $\amuW$, being relatively insensitive to long-distance effects, may be a powerful diagnostic tool to assess compatibility among all HVP calculations.  To summarize the lessons learned from the current lattice efforts, the following key features will pave the road towards permil-level precision: 
\begin{enumerate}[label=(\alph*)]
\item To reduce the statistical uncertainty on the dominant light-quark contribution to $\amuHVPLO$, noise reduction methods such as all-mode averaging~\cite{Blum:2012uh} and low-mode averaging~\cite{Giusti:2004yp,DeGrand:2004qw,Neff:2001zr} that have been used in Refs.~\cite{Blum:2018mom,Aubin:2019usy,Borsanyi:2020mff}, together with high statistics, are crucial. A promising approach for further significant improvements is the analysis of additional correlation functions with multi-pion states, which enables a precise and model-independent reconstruction of the long-distance tail~\cite{Erben:2017hvr,Erben:2019nmx,Bruno:2019nzm}. Additionally, such an analysis can yield valuable information on finite-volume effects.  

\item It is understood that the numerical simulations must be performed at the physical point, employ improved actions and operators, so that discretization errors start at ${\cal O}(a^2)$ (or higher), and include multiple lattice spacings.  Several of the lattice results for $\amuHVPLO$ presented in this review are already based on such ensembles. For permil-level precision, it will also be important to resolve higher-order discretization effects, which requires the inclusions of ensembles at more than three lattice spacings in the analysis. The inclusion of dedicated finite-volume studies and ensembles with large spatial volumes ($L \ge 6$\,fm) or the use of other model-independent methods to estimate finite-volume effects are also desirable. 

\item  The corrections due to strong isospin breaking and QED effects must be calculated with sufficient precision and must include the complete set of ${\cal O}(\alpha)$ and ${\cal O}(\delta m)$ terms described in \cref{sec:HVP_IB}, including disconnected (sea-quark) effects. 

\item Given the sensitivity of $\amuHVPLO$ to the scale setting uncertainty (see \cref{subsec:common}), a dedicated effort based on a similarly comprehensive study is necessary. In particular, strong isospin breaking and QED corrections  must also be included in  such a study. 
\end{enumerate}

Fortunately, the technology and methods needed to meet these challenges are already in place. Indeed, if  the pace of innovation seen in the past few years continues, there may be further improvements to the methods along the way. In conclusion, with sustained, dedicated effort, there are no roadblocks that would prevent the lattice community from reaching the permil-level precision goal for $\amuHVPLO$.

\FloatBarrier

\clearpage

\section{Data-driven and dispersive approach to HLbL}
\label{sec:dispHLbL}

\noindent
\begin{flushleft}
\emph{J.~Bijnens, G.~Colangelo, F.~Curciarello, H.~Czy\.z, I.~Danilkin, F.~Hagelstein, M.~Hoferichter, B.~Kubis, A.~Kup\'s\'c, A.~Nyffeler, V.~Pascalutsa, E.~Perez del Rio, M.~Procura, C.~F.~Redmer, P.~S\'anchez-Puertas, P.~Stoffer, M.~Vanderhaeghen}
\end{flushleft}

\subsection{Introduction}
\label{sec:intro_HLbL_DR}

\subsubsection{The HLbL contribution to the muon $g-2$} 

\begin{figure}[t]
\centering
\includegraphics[width=\linewidth]{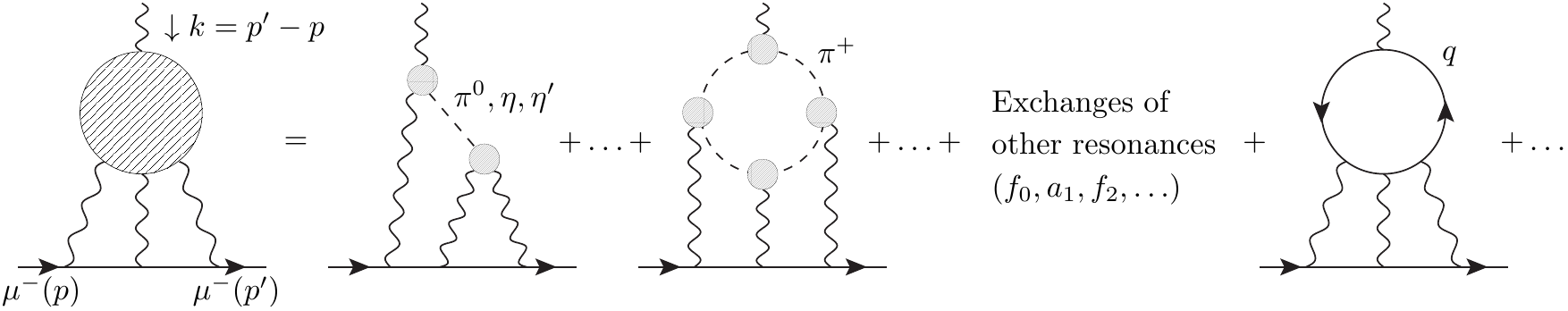}
\caption{HLbL in the muon $g-2$ in model calculations. The blobs on the
  right-hand side of the equal sign are form factors that describe the
  interaction of photons with hadrons. Adapted from Ref.~\cite{Nyffeler:2017ohp}.} 
\label{genericHLbL}
\end{figure}

One of the largest uncertainties of the SM evaluation of
$(g-2)_\mu$ at present comes from the HLbL
scattering contribution depicted on the left-hand side of
\cref{genericHLbL}. Unlike its QED counterpart, this contribution
cannot be calculated in perturbation theory, and thus one should rely
on either lattice QCD or data-driven evaluations, similarly to how it
is done for the HVP contribution.

HLbL scattering is, however, more complicated than HVP, because 
the LbL contributions enter through a
four-point function---the LbL scattering amplitude---rather
than a two-point function as in the case of VP.  To
the right-hand side of the equal sign in \cref{genericHLbL} various
contributions to the HLbL tensor are shown. This picture was used in
early model calculations, but it is to a large extent still valid,
though defined more precisely in modern, data-driven approaches as
will be explained here. At low energies, there are exchanges of single
mesons, like the light pseudoscalars~$\pi^0,\eta,\eta^\prime$, heavier
scalars like $f_0(980), a_0(980)$, or axial-vector mesons $a_1, f_1$,
and tensor mesons $f_2, a_2$ above 1\,GeV. Furthermore, there are loops
with charged pions and kaons. Finally, when all momenta are large,
HLbL can be described by a perturbative quark loop. Since the HLbL
contribution to the $(g-2)_\mu$ is obtained through integration of the
HLbL tensor over all momenta, it is a priori not clear whether any momentum
expansion of the tensor could be usefully applied.  In the integral
there is a weight function (arising from muon and photon propagators)
in which the only scale is the muon mass. One could therefore expect
that low momenta should dominate the integral, but translating this
expectation into an algebraic expansion scheme has not been possible
so far. A detailed analysis of the respective merits of the chiral and
the large-$N_c$ expansions in QCD have been discussed in a key paper by
Eduardo de Rafael~\cite{deRafael:1993za} (see, however, Appendix C of Ref.~\cite{Hoferichter:2018kwz} for a discussion of the large-$N_c$ expansion in the SM). According to this analysis,
the leading contribution in ChPT is the charged-pion loop at order $p^4$, while it is subleading in large $N_c$. On
the other hand, the exchanges of single mesons are leading in large
$N_c$ (as is the quark loop), with the light pseudoscalars starting at
order $p^6$ and the heavier mesons at order $p^8$. Since in general
the interaction of photons with mesons is described by form factors
the situation becomes very involved and in general rather
model-dependent. Note that the charged-pion loop is actually finite
without a form factor (scalar QED), whereas the neutral pion exchange
with pointlike vertices diverges like $\log^2\Lambda$ with some UV
cutoff
$\Lambda$~\cite{Melnikov:2001uw,Knecht:2001qf,Knecht:2001qg}. Some
model calculations will be summarized at the end of
\cref{sec:tensor}. They show that the exchanges of the
light pseudoscalars~$\pi^0,\eta,\eta^\prime$ are numerically dominant
compared to the charged-pion loop with form factors. Actually, the
identification of the different contributions on the right-hand side
of \cref{genericHLbL} is not unique in model calculations. Only the
dispersive framework discussed below allows for an unambiguous
definition of individual contributions to HLbL. Finally, in many model
calculations, the high-energy part of the HLbL amplitude is modeled by
a constituent quark loop, denoted by $q$ in the figure, again with
form factors describing the interaction with the photons. This leads
to the problem of a possible double-counting of such a dressed
constituent quark loop with the mesonic contributions discussed
above. Note, however, that in the HLbL contribution to the muon $g-2$
there are always mixed regions in the loop integrals where not all
momenta are large, but some are small. Within the dispersive framework
this amounts to a proper matching of the dispersive contribution to
HLbL with short-distance constraints from pQCD, as will be
discussed in \cref{sec:asymptotic}.

At the same time, the HLbL contribution to the muon $g-2$ is
suppressed by an extra power of the fine-structure constant $\alpha$
compared to HVP and indeed is about two orders of magnitude smaller:
hence it does not need to be determined to the same level of
accuracy. To meet the precision goal of the ongoing Fermilab
experiment it suffices to know the HLbL contribution with about 10\%
relative accuracy. Past evaluations of HLbL before 2009 were
summarized and combined with some new estimates in
Ref.~\cite{Prades:2009tw} (``Glasgow consensus'' by Prades, de Rafael, and
Vainshtein) and in the review by Jegerlehner and
Nyffeler~\cite{Jegerlehner:2009ry}:
\begin{align}
\amuHLbL (\mbox{Glasgow consensus}) & = 105(26)
\times 10^{-11}\,,\\ 
\amuHLbL (\mbox{Jegerlehner--Nyffeler}) & = 116(39)\times 10^{-11}\,. 
\end{align}
They are consistent with each other and are actually largely based on
the same model calculations, but the estimates fall short of the
present accuracy goal. Moreover, error estimates of individual
contributions of similar size were combined in quadrature in
Ref.~\cite{Prades:2009tw} and linearly in
Ref.~\cite{Jegerlehner:2009ry}. Therefore, of even greater concern is
the question of ``model dependence'' of these values and their
uncertainties, as they are not based on lattice QCD or data-driven
dispersive evaluations.  The aim of this section is to summarize
recent progress in the evaluation of the HLbL contribution based on
analytic, data-driven approaches that were proposed in the year 2014
and thereafter. As we will show, improvements with respect to the
above evaluations will concern both accuracy and rigor. Some more
recent estimates for HLbL that partly use results from the
dispersive, data-driven approaches will be summarized in
\cref{sec:result_HLbL_DR}.

\subsubsection{Dispersive approach to the HLbL amplitude} \label{sec:DRLbL}

As most of the earlier work based on analytic approaches to HLbL relied on
models, not only was it difficult to assess uncertainties, but also
to compare different calculations in detail. A common language was
nonetheless used and was mainly based on the identification of the
different contributions by the particles that are exchanged inside
the hadronic blob. This identification was however not free from
ambiguities as soon as one tried to compare the same contributions between
different models. These were mainly related to the off-shell contributions
of the different particles. In order to clearly and uniquely identify the
various contributions on the basis of the particles exchanged, it was
necessary to adopt a dispersive language. The underlying logic is that
if an amplitude can be reconstructed from its singularities, and these
are related by unitarity to physical subamplitudes obtained by cutting the hadronic blobs in all possible ways and taking into account all
possible (on-shell) intermediate states, then the whole amplitude can be
split into a number of contributions clearly identified by the (on-shell)
intermediate states.

This approach has long been used for HVP, and indeed once it is implemented
for a certain amplitude there can hardly be any reason not to use it. For
HLbL it was not known how and whether at all it could be implemented. The
problem is that some preliminary steps are necessary if one wants to
translate the operation of cutting the hadronic blob and identifying the
relevant intermediate states into uniquely defined mathematical
expressions. In the case of $n$-point functions of currents, which are rank
$n$ Lorentz tensors, one first needs to find a basis of Lorentz tensors of
that rank, written in terms of the available independent momenta, and make
sure that this basis also satisfies the relevant Ward identities. This step
is almost trivial for HVP: the basis of Lorentz-invariant rank-2 tensors
built out of one momentum consists of two elements and if one imposes gauge
invariance one ends up with one single tensor. In addition one has to make
sure that the basis tensors do not have kinematic singularities or zeros,
because only for such scalar functions one can write down a dispersion
relation in a straightforward manner. This step too is trivially performed for HVP. Solving these problems
for HLbL is a much more formidable task, which has been achieved for the
first time only a few years ago~\cite{Colangelo:2015ama}. A summary and
some details of the steps that lead to the derivation of such a basis are
presented below in \cref{sec:tensor}. There one can also find a derivation
of a master formula that provides a compact representation of the
contribution of the relevant basis functions of the HLbL tensor to the
$(g-2)_\mu$ in terms of a three-dimensional integral. The latter has to be
evaluated numerically.

This is only the preparatory work, however, and writing down a dispersion
relation still requires to list all possible cuts and intermediate states
that can appear in the hadronic blob. Here again the case of HVP is much
simpler than that of HLbL. For the former, there is only one possible cut
and the form of the dispersion relation is the same for any of the possible
intermediate states. This implies that one can write a general formula for
all possible intermediate states. Indeed, as can be seen in the section
dedicated to HVP, this contribution to the $(g-2)_\mu$ can be written as a
single integral over the cross section $\sigma(e^+ e^- \to
\mbox{hadrons})$, because the kernel function in the integral does not
depend on the intermediate state. For HLbL things are significantly more
complicated, and the approach that has been followed so
far~\cite{Colangelo:2014dfa,Colangelo:2015ama} was to write down a
dispersion relation separately for each intermediate state. Those that
have been dealt with so far are: {(i)} pion pole; {(ii)} pion box;
{(iii)} two-pion cut. In the following, one section is dedicated to each of these. The first two contributions are currently known with very
good accuracy, and in particular the second one, which can be expressed in
terms of the vector form factor only, known to remarkable accuracy thanks
to the $e^+ e^- \to \pi^+ \pi^-$ measurements, has been calculated with
negligible uncertainties. The first one also, now known to better than 5\%,
can be considered a closed chapter. The third one has been estimated to
about 10\% accuracy for the contribution of the $S$-waves (a precision which
is more than sufficient when considering the fact that this contribution
is itself less than 10\% of the total). Going beyond the $S$-waves is still
challenging, however, also in view of conceptual developments which have
not been completed yet and mostly concern the application of the formalism
to $D$-waves and higher.

This approach can be trivially extended to heavier states with similar
quantum numbers: the formula for the single-pion-pole contribution applies as well
to the $\eta$ and $\eta'$ contributions, that for the pion box to the
kaon box, and that for two-pion exchanges for any heavier pairs of pseudoscalars (two kaons, $\pi\eta$, $\eta\eta$, etc.). Indeed all these
contributions can be estimated, as discussed below. What still needs to be done is an
estimate of contributions that would require an extension of the formalism
just described: in particular the contribution of three-pion intermediate
states and higher. Calculating these contributions at the
same level of rigor as the ones discussed here does not seem feasible at
the moment. An approach based on resonance saturation (for example
representing this contribution via an axial-meson exchange) will be pursued
instead. These estimates confirm the initial hope that their uncertainty is already under good control at the present level of precision.

The final point that needs to be addressed in this introduction concerns
the behavior of the HLbL at asymptotically large values of the momenta, or
short-distance constraints (SDCs) in short. The relevance of these and an
estimate of the numerical impact for the calculation of the HLbL
contribution to the $(g-2)_\mu$ have been pointed out in
Ref.~\cite{Melnikov:2003xd}. The contributions of the various
intermediate states that have been estimated and put together here do not satisfy the SDCs. As pointed out by Melnikov and Vainshtein, the most important
contribution, that of a single pion, cannot satisfy these constraints. This 
implies that not even by adding any finite number of such intermediate
states the problem can be solved. Two alternative proposals have been recently put forward for satisfying these constraints, one based on an infinite tower of excited pseudoscalars and one based on a tower of axial-vector mesons. Despite the completely different implementations the estimates of the impact of the SDCs on the HLbL contribution to the $(g-2)_\mu$ agree on a value significantly lower than what the original suggestion in
Ref.~\cite{Melnikov:2003xd} would give. There are still open questions, however, and a more satisfactory understanding of the role of the SDCs, and in particular of the axial-vector mesons in satisfying them, will require additional work.

\subsubsection{Dispersion relation for the Pauli form factor $F_2$} \label{sec:DRF2}

Another dispersive data-driven approach for the calculation of
$\amuHLbL$ was proposed in Ref.~\cite{Pauk:2014rfa}. It
is based on the analytic properties of the electromagnetic vertex
function of the muon and expresses $\amuHLbL$ through a
dispersive integral over the discontinuity of the vertex function,
which in turn can be related to observables. Experimentally, the
amplitudes involved in the unitarity equations for the required
discontinuities can be measured in two-photon and $e^+e^-$ production
processes, see \cref{sec:exp_MC}.

The HLbL contribution from \cref{genericHLbL} to the Pauli form
factor $F_2(k^2)$ is obtained from the following two-loop integral,
using well-known projection techniques, see, e.g.,
Refs.~\cite{Aldins:1970id,Barbieri:1974nc,Jegerlehner:2008zza} 
\begin{align}
F_2(k^2) &= e^6 \sum\limits_{\lambda_1,\lambda_2,\lambda_3,\lambda}
(-1)^{\lambda+\lambda_1+\lambda_2+\lambda_3}  
\int\frac{d^4 q_1}{(2\pi)^4}\int\frac{d^4 q_2}{(2\pi)^4}
\nonumber \\  
&\qquad \qquad \times
\frac{L_{\lambda_1\lambda_2\lambda_3\lambda}(p,q_1,k-q_1-q_2,q_2) \, 
\Pi_{\lambda_1\lambda_2\lambda_3\lambda}(q_1,k-q_1-q_2,q_2,k)}{q_1^2q_2^2(k-q_1-q_2)^2
\left[(p+q_1)^2-m_\mu^2\right]\left[(p+k-q_2)^2-m_\mu^2\right]}\,, 
\label{eq:F2loopexplicit}   
\end{align}
where $\Pi_{\lambda_1\lambda_2\lambda_3\lambda}$ and
$L_{\lambda_1\lambda_2\lambda_3\lambda}$ are the projections on the
helicity basis of the fourth-rank HLbL tensor (see \cref{eq:defPI} for its definition, even though with only three arguments, the fourth being fixed by momentum conservation) and the corresponding lepton tensor from the vertices along the muon line and the projection on $F_2(k^2)$, see Ref.~\cite{Pauk:2014rfa} for details. The anomalous magnetic moment is then defined as the static limit of the form factor, $k^2=0$, with $k$ being the photon momentum.

When analytically continued to complex values of the virtuality $k^2$
of the external photon, the electromagnetic vertex function possesses
branch point singularities joining the physical production thresholds,
as is dictated by unitarity. Assuming that the Pauli form factor falls
off sufficiently fast at infinity, one obtains the following
unsubtracted dispersion relation for the form factor in the static
limit, i.e., the HLbL contribution to the anomalous magnetic moment 
\be
\amuHLbL = F_2(0) = \frac1{2\pi i}
\int\limits_0^\infty\frac{dk^2}{k^2}\mathrm{Disc}_{k^2}\,F_2(k^2)\,,  
\label{eq:F2disp}
\ee
where the integration involves the discontinuity of the form factor
$\mathrm{Disc}_{k^2}F_2(k^2)$ along the cut in the $k^2$-plane starting from
the lowest branch point. 

As can be seen from the structure of the two-loop integral in
\cref{eq:F2loopexplicit}, the branch cuts of the form factor
$F_2(k^2)$ are related to the propagators of the virtual muons and
photons and nonanalyticities of the HLbL tensor. The latter possesses
two types of discontinuities, the corner (one-photon) and cross
(two-photon) cuts. The corner cuts are related to a conversion of a
virtual photon to a hadronic state with negative $C$-parity, while the
cross cuts are related to the two-photon production of a $C$-even
hadronic state. As the dominant contributions originate from the
lowest thresholds, they are mainly governed by intermediate states
including pions. In particular, the lowest threshold in the $C$-odd
channel is related to $\pi^+\pi^-$-pair production (pion loop in HLbL)
and in the $C$-even channel to a $\pi^0$ intermediate state (pion pole
in HLbL). By virtue of unitarity, these discontinuities are related to
amplitudes of physical hadron production processes.

\begin{figure}
\centering
  \includegraphics[width=7.8cm]{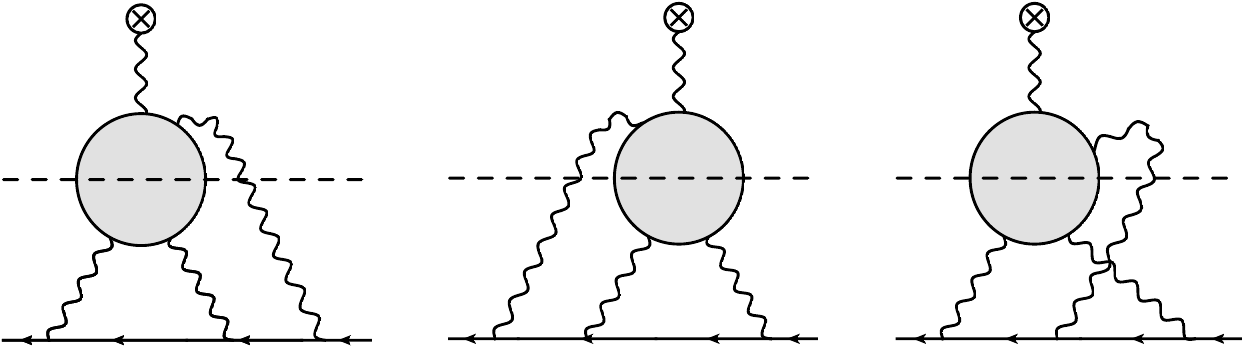}
  \vspace{.3cm}\\
  \includegraphics[width=7.8cm]{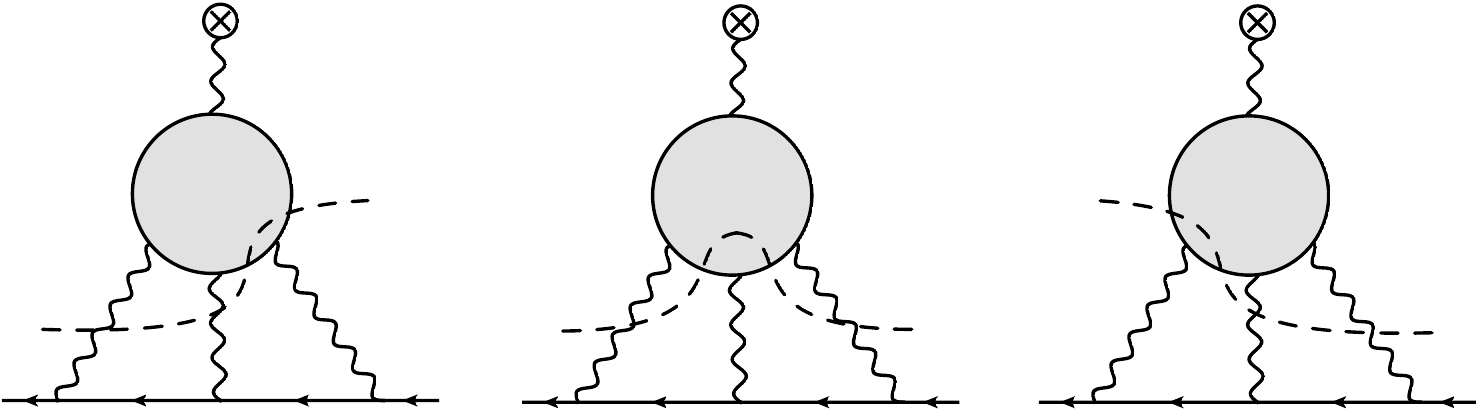}
  \vspace{.3cm}\\
  \includegraphics[width=2.5cm]{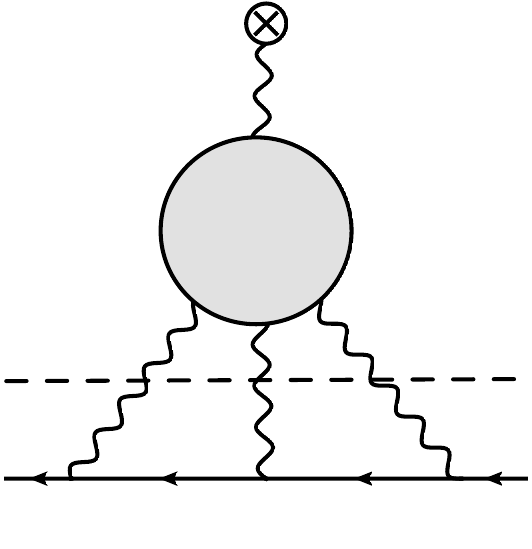}
  \hspace{.2cm}
    \includegraphics[width=2.5cm]{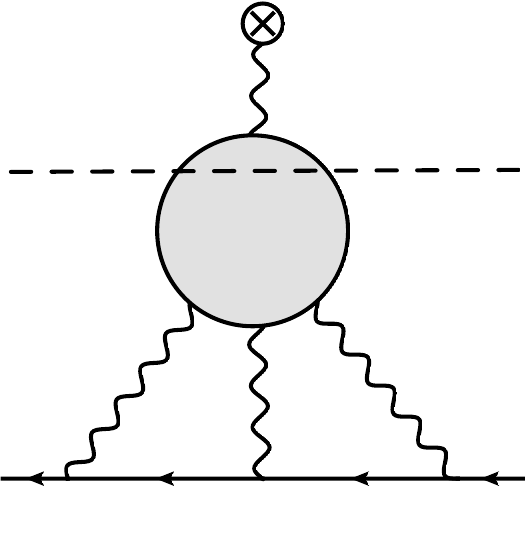}
  \hspace{.2cm}
  \includegraphics[width=2.5cm]{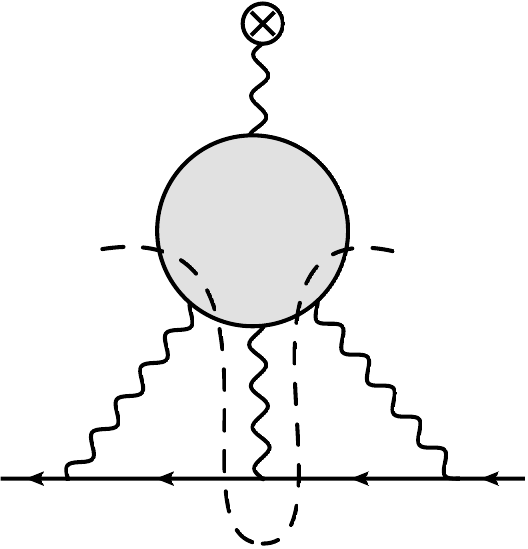}
  \caption{Unitarity diagrams contributing to the imaginary part of
    the vertex function. The cut indicates the on-shell intermediate
    state. Reprinted from Ref.~\cite{Pauk:2014rfa}.}  
  \label{fig:undiagr} 
\end{figure}

The discontinuity in \cref{eq:F2disp} is obtained as a sum of nine
topologically different contributions, which are graphically
represented by unitarity diagrams in \cref{fig:undiagr}. As an
example, for the first diagram it implies that we obtain
$\mathrm{Disc} F_2(k^2)$ from \cref{eq:F2loopexplicit}, if we replace the cut photon propagator $1/q_2^2$ by $(2\pi i)
\delta(q_2^2)$ and the cut function
$\Pi_{\lambda_1\lambda_2\lambda_3\lambda}(q_1,k-q_1-q_2,q_2,k)$ by
$\mathrm{Disc}_{(k-q_2)^2}\Pi_{\lambda_1\lambda_2\lambda_3\lambda}(q_1,k-q_1-q_2,q_2,k)$ in the integral. The
latter nonperturbative discontinuity function is directly related to
amplitudes of processes $\gamma^\ast\gamma^\ast\to X$ and
$\gamma^\ast\to\gamma X$, with $X$ denoting a $C$-even hadronic state,
which are accessible experimentally.

The approach was tested by considering the pion-pole contribution to
HLbL using a simple VMD model for the transition form factor (TFF), using
the methods described in Ref.~\cite{Pauk:2014jza}. Taking into account
the discontinuities from all relevant two-particle and
three-particle cuts through the propagators of the muons, photons,
vector mesons, and neutral pions, the result from
Ref.~\cite{Knecht:2001qf} was reproduced exactly. 
The method allows one to quantify the relative contribution of $e^+e^- \to \pi^0 \gamma$ and $e^+e^- \to \gamma \gamma \gamma$ processes to the $a_\mu$ integral, and involves a partial compensation between those contributions. 
The pion-pole
prescription in this dispersive approach involves a
singly-virtual TFF at the external vertex, as in
the dispersive approach from Refs.~\cite{Colangelo:2014dfa,
  Colangelo:2015ama, Colangelo:2017fiz}, but differs from the
prescription given in Ref.~\cite{Melnikov:2003xd}. So far, no explicit
expressions have been given in this approach to HLbL to obtain the
pion-pole contribution for a generic pion TFF or
the precise input that is needed from the scattering process
$\gamma^\ast\gamma^\ast \to \pi\pi$ to obtain the pion-loop
contribution to HLbL from data.

\subsubsection{Schwinger sum rule}

The Schwinger sum rule~\cite{Schwinger:1975ti,Schwinger:1975uq} is an exact dispersive formula that expresses
the anomalous magnetic moment in terms of a single integral over an observable cross section.
As such, the Schwinger sum rule is not only a conceptual analog of the dispersive approach
used to evaluate the HVP contribution, 
it generalizes the HVP formula to arbitrary topologies (see Ref.~\cite{Hagelstein:2017obr} for more details). 

The Schwinger sum rule for the muon reads:
\begin{equation}
\label{eq:SchwingerSR}
a_\mu  = \frac{m_\mu^2}{\pi^2\alpha}
\int_{\nu_0}^\infty \!d \nu \, \left[\frac{\sigma_{LT}(\nu,Q^2)}{Q}\right]_{Q^2=0}\,,
\end{equation}
where $\sigma_{LT}(\nu,Q^2)$---the longitudinal-transverse photoabsorption cross section---is the response
function corresponding  to an absorption of a polarized virtual
photon with energy $\nu$ and spacelike 
virtuality $Q^2$ on the muon with mass $m_\mu$. This response function is rather common
in the studies of nucleon spin structure via electron scattering.
Equivalently, it can be written in terms of the sum of the muon spin structure
functions $g_1$ and $g_2$:
\begin{equation}
a_\mu = \lim_{Q^2\to 0} \frac{8m_\mu^2}{Q^2}
 \int_0^{x_0} d x \;  
 \big[ g_1(x,Q^2)+ g_2(x,Q^2)\big]\,,
\end{equation}
where $x=Q^2/(2m_\mu \nu)$ is the Bjorken variable; $\nu_0$ and $x_0$ denote the photoabsorption thresholds. 

\begin{figure}[tb]
\centering
\includegraphics[height=2.5cm]{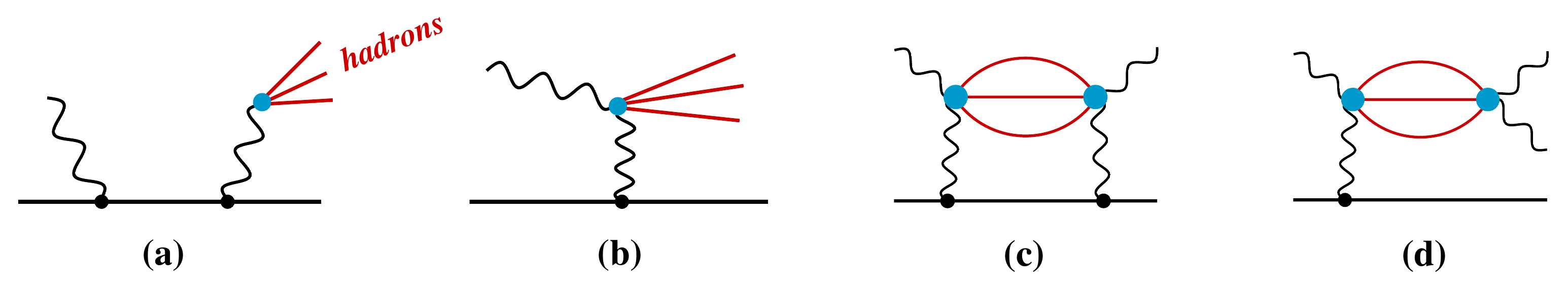}
\caption{Different channels contributing to the photoabsorption process: hadron photoproduction through timelike Compton scattering (a) and the Primakoff mechanism (b), and HLbL contributions to the electromagnetic channels (c, d)  (crossed diagrams omitted). Adapted from Ref.~\cite{Hagelstein:2017obr}.}
\label{Fig:Hphotoprod}
\end{figure}

 To evaluate the hadronic contributions to the photoabsorption on the muon, it is useful to distinguish 
 the following two ways hadrons can contribute:
\begin{itemize}
\item[(i)]
explicit hadron photoproduction, see, e.g., \cref{Fig:Hphotoprod} (a, b), 
\item[(ii)] virtual contributions via the electromagnetic channels, see \cref{Fig:Hphotoprod} (c, d). 
\end{itemize}
The standard HVP formula is reproduced when using the timelike Compton scattering channel \cref{Fig:Hphotoprod} (a) as input, see Ref.~\cite{Hagelstein:2017obr} for a detailed proof.
The rest of the contributions in \cref{Fig:Hphotoprod} are of the HLbL type.
The prospects for an experimental determination of the latter contributions are unclear at the moment.
The hadron-production-channel contributions to the muon spin structure functions
can in principle be done at a high-energy muon beam facility, such as SPS at CERN,
where the polarized muon beam is scattered off atomic electrons with hadrons in the final state.
The virtual HLbL contributions into the electromagnetic channels are not measurable directly and will need to be evaluated using the dispersive approach to the HLbL scattering amplitude discussed above.  
It therefore remains to be seen whether the virtues of the Schwinger sum rule
can be useful in practical evaluations of the HLbL contribution, but it is certainly a new avenue for approaching this difficult calculation, which offers reasons of interest and is worthwhile pursuing.

\subsection{Hadronic light-by-light tensor}
\label{sec:tensor}

\subsubsection{Definitions, kinematics, notation}

The HLbL contribution to the muon anomalous magnetic moment is governed by the polarization tensor for fully off-shell photon--photon scattering in pure QCD,
\begin{align}
\label{eq:defPI}
	\Pi^{\mu\nu\lambda\sigma}(q_1,q_2,q_3)= -i \int d^4x \, d^4y \, d^4z \, e^{-i(q_1 \cdot x + q_2 \cdot y + q_3 \cdot z)} 
	\langle 0 | T \{ j^\mu(x) j^\nu(y) j^\lambda(z) j^\sigma(0) \} | 0 \rangle\,.
\end{align}
This involves four electromagnetic quark currents, incoming momenta $\{q_1, q_2\}$ and outgoing momenta $\{-q_3, q_4\}$, with Mandelstam invariants 
\begin{equation}
s=(q_1+q_2)^2\,,\quad t=(q_1+q_3)^2\,,\quad u=(q_2+q_3)^2\;\;\;\;\;{\mathrm{fulfilling}}\;\;\;\;\; s+t+u = \sum_i q_i^2\,.
\end{equation}

\subsubsection{Lorentz and gauge invariant representation}
\label{sec:BTT}

Based on Lorentz covariance, this tensor can be decomposed into 138 structures~\cite{Karplus:1950zza,Leo:1975fb,Bijnens:1995xf},
\begin{equation}
\Pi^{\mu\nu\lambda\sigma} = \sum_{i=1}^{138} L_i^{\mu\nu\lambda\sigma}\; \Xi_i \,,
\end{equation}
where the scalar functions $\Xi_i$ depend on 6 kinematic variables. However, this set is largely redundant since gauge invariance implies 95 linear relations between the $\Xi_i$. Furthermore, two additional constraints arise in four space-time dimensions due to a degeneracy in the Lorentz structures that follows from the Schouten identity~\cite{Eichmann:2014ooa}. 

By generalizing the procedure introduced for doubly-virtual Compton scattering by Bardeen, Tung~\cite{Bardeen:1969aw}, and Tarrach~\cite{Tarrach:1975tu} (BTT), it is possible to derive a generating set of 54 structures
\begin{equation} \label{eq:BTT}
\Pi^{\mu\nu\lambda\sigma} = \sum_{i=1}^{54} T_i^{\mu\nu\lambda\sigma}\; \Pi_i
\end{equation}
that is manifestly gauge invariant and closed with respect to crossing relations~\cite{Colangelo:2015ama}. Only 7 of the $T_i^{\mu\nu\lambda\sigma}$ are genuinely different, the remaining 47 being determined by crossing (see Ref.~\cite{Colangelo:2017fiz} for their explicit form). The connection between the two sets of $\Xi_i$ and $\Pi_i$ is presented in detail in Ref.~\cite{Colangelo:2015ama}. Crucially, only the BTT scalar functions $\Pi_i$ are free of kinematic singularities. This makes them suitable for a dispersive representation. The BTT set is still redundant since a real tensor basis for $\Pi^{\mu\nu\lambda\sigma}$ contains 41 elements, which matches the number of helicity amplitudes for fully off-shell photon--photon scattering. However, the resulting ambiguities in the definition of the scalar functions cancel out in the calculation of the observable $a_\mu$. 

The HLbL contribution to $a_\mu$ can be derived from $\Pi^{\mu\nu\lambda\sigma}$ using standard projection operator techniques~\cite{Aldins:1970id,Barbieri:1974nc,Jegerlehner:2008zza}. 
By expanding the photon--muon vertex function around $q_4=0$ and taking into account that $\Pi^{\mu\nu\lambda\sigma}$ vanishes linearly with $q_4$, one obtains~\cite{Colangelo:2015ama}
\begin{align} \label{eq:mf}
		\amuHLbL &= - \frac{e^6}{48 m_\mu}  \int \frac{d^4q_1}{(2\pi)^4} \frac{d^4q_2}{(2\pi)^4} \frac{1}{q_1^2 q_2^2 (q_1+q_2)^2} \frac{1}{(p+q_1)^2 - m_\mu^2} \frac{1}{(p-q_2)^2 - m_\mu^2} \notag \\
			& \quad \times \mathrm{Tr}\left( (\slashed p + m_\mu) [\gamma^\rho,\gamma^\sigma] (\slashed p + m_\mu) \gamma^\mu (\slashed p + \slashed q_1 + m_\mu) \gamma^\lambda (\slashed p - \slashed q_2 + m_\mu) \gamma^\nu \right)  \notag \\
			& \quad \times  \sum_{i=1}^{54} \left( \frac{\partial}{\partial q_4^\rho} T^i_{\mu\nu\lambda\sigma}(q_1,q_2,q_4-q_1-q_2) \right) \bigg|_{q_4=0} \Pi_i(q_1,q_2,-q_1-q_2)\,,
\end{align}
where $p$ is the muon momentum.
In this expression, the limit $q_4 \to 0$ can be taken straightforwardly since the BTT scalar functions are free of kinematic singularities and zeros. 
Since the $\Pi_i$ depend
only on $q_1^2$, $q_2^2$, and $q_1 \cdot q_2$, five out of the eight integrals in \cref{eq:mf} can be performed analytically without any knowledge of the scalar functions, through angular averages after a Wick rotation to Euclidean momenta, using the technique of Gegenbauer polynomials (hyperspherical approach)~\cite{Rosner:1967zz,Levine:1974xh}. Furthermore, the symmetry properties in \cref{eq:mf} under $q_1 \leftrightarrow -q_2$ allow one to derive
a master formula involving  the sum of only 12 terms~\cite{Colangelo:2015ama}
\begin{align}
	\label{eq:MasterFormula3Dim}
	\amuHLbL &= \frac{2 \alpha^3}{3 \pi^2} \int_0^\infty dQ_1 \int_0^\infty dQ_2 \int_{-1}^1 d\tau \sqrt{1-\tau^2} \,Q_1^3 Q_2^3 \,\sum_{i=1}^{12} T_i(Q_1,Q_2,\tau)\, \bar \Pi_i(Q_1,Q_2,\tau)\,.
\end{align}
Here $Q_1 := |Q_1|$ and $Q_2 := |Q_2|$ denote the norm of the Euclidean four-momentum vectors, and $\tau$ is the cosine of the angle between these vectors. The integral kernels $T_i(Q_1,Q_2,\tau)$, which involve traces and derivatives at $q_4=0$ according to \cref{eq:mf} for the structures in \cref{eq:BTT}, are fully general and explicitly given in Appendix~B of Ref.~\cite{Colangelo:2017fiz}. The functions $\bar \Pi_i$ are linear combinations of the $\Pi_i$ and have to be evaluated for the reduced spacelike kinematics
\begin{align}
		s &= - Q_3^2 = -Q_1^2 - 2 Q_1 Q_2 \tau - Q_2^2 , \quad t = -Q_2^2 , \quad u = -Q_1^2 , \notag \\
		q_1^2 &= -Q_1^2, \quad q_2^2 = -Q_2^2, \quad q_3^2 = - Q_3^2 = - Q_1^2 - 2 Q_1 Q_2 \tau - Q_2^2 , \quad q_4^2 = 0\,.
\end{align}
There are only $6$ distinct $\bar\Pi_i$, the remaining ones being related to these by crossing symmetry.  
Eq.~\eqref{eq:MasterFormula3Dim} generalizes the three-dimensional integral formula for the pion-pole contribution in Ref.~\cite{Jegerlehner:2009ry}. It is a consequence of gauge invariance and crossing and does not depend upon the method used to compute $\bar\Pi_i$, which fully parameterize the hadronic content in $a_\mu^{\rm {HLbL}}$. This master formula is well-suited for a direct numerical implementation. In particular, the energy regions generating the bulk of the contribution can be identified by numerical integration.

\subsubsection{Dispersive representation and definition of individual contributions}
\label{sec:DefinitionOfIndividualContributionsToHLbL}

\begin{figure}[t]
 \centering
 \includegraphics[width=0.75\linewidth]{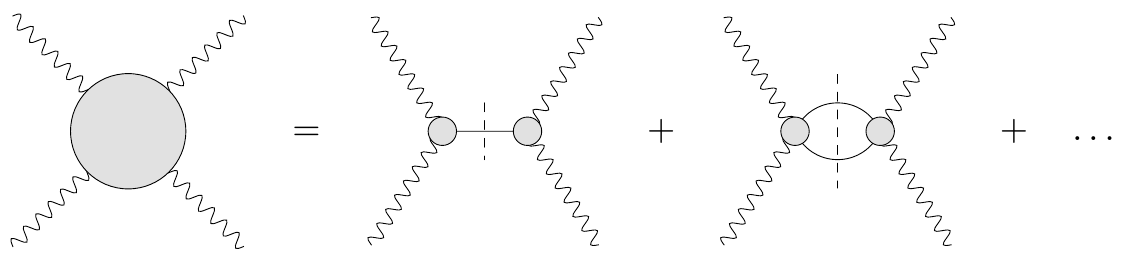}
	\caption{Intermediate states in the direct channel for HLbL scattering: one- and two-particle cut. The gray blobs denote hadronic amplitudes involving $\gamma^*\gamma^{(*)}$. Reprinted from Ref.~\cite{Colangelo:2017fiz}.}
	\label{HLbLIntermediateStates}
\end{figure}

\begin{figure}[t]
 \centering
 \includegraphics[width=0.75\linewidth]{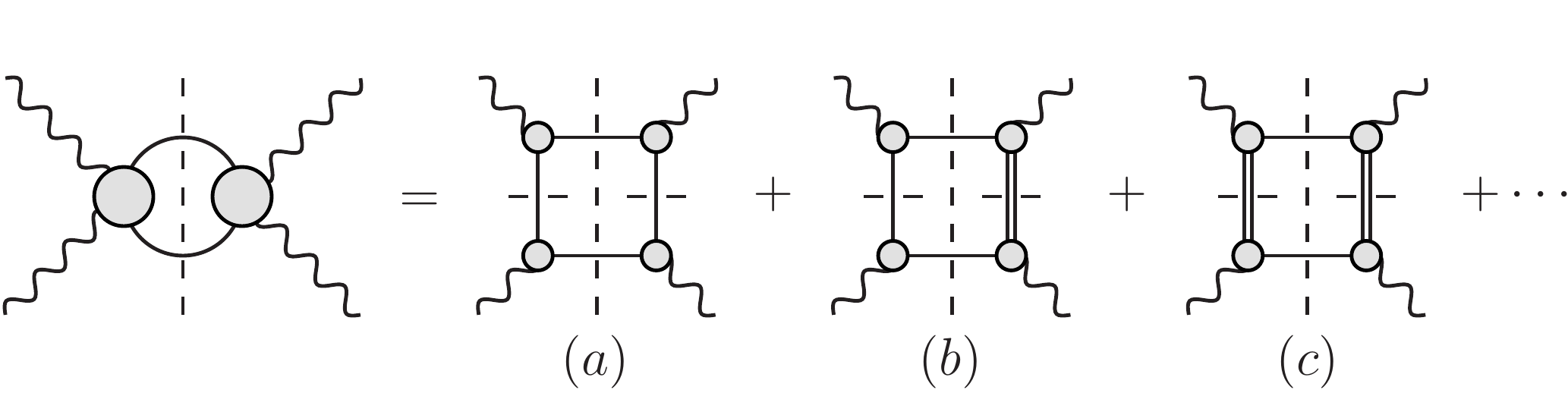}
 \caption{Two-pion-cut contributions to HLbL scattering. Solid/wiggly lines refer to pions/photons, respectively. Double lines generically denote heavier intermediate states (multi-pion states, heavier resonances etc.). Crossed diagrams are omitted. Reprinted from Ref.~\cite{Colangelo:2017qdm}.}
	\label{HLbLTwoPionContributions}
\end{figure}

The complexity of the analytic structure of the four-point function $\Pi^{\mu\nu\lambda\sigma}$ prevents us from summing over all possible intermediate states at once as done in the case of HVP. A dispersive representation of the BTT scalar functions at fixed photon virtualities involves two independent kinematic variables instead of only one, and the double spectral functions that fully determine $\bar\Pi_i$ cannot be directly measured in an inclusive manner. Therefore, the only way to make use of a Mandelstam representation of the scalar functions~\cite{Colangelo:2015ama} is to consider {\it individual} intermediate states and for each of these construct a relation between the double spectral function and physical observables like form factors and cross sections involving on-shell hadrons. In this framework, the unitarity relation makes it possible to rigorously define the contributions to $\bar\Pi_i$ and thus to $\amuHLbL$ stemming from one-, two-hadron etc.\ intermediate states in direct and crossed channels. Unitarity diagrams for one- and two-particle cuts in the direct channel are shown in \cref{HLbLIntermediateStates}. 
By making the assumption---corroborated by all model calculations---that the importance of the contribution of an intermediate state decreases as the corresponding threshold increases, one can focus on the most important contributions and treat them explicitly. This is the approach pursued in Refs.~\cite{Hoferichter:2013ama,Colangelo:2014dfa,Colangelo:2014pva,Colangelo:2015ama,Colangelo:2017qdm,Colangelo:2017fiz,Hoferichter:2018dmo,Hoferichter:2018kwz}. The bulk of $\amuHLbL$ is expected to originate from states at energies up to about $1.5\,{\rm{GeV}}$, with the numerically dominant role played by the $\pi^0$-pole. Compared to this contribution, other one-particle states, like $\eta$ and $\eta'$, are suppressed. Two-pion effects are further suppressed but still more important than two kaons etc. Within this framework, one can unambiguously split $\amuHLbL$ into the following sum:
\begin{align}
	\amuHLbL = a_\mu^{\pi^0\text{-pole}} + a_\mu^{\pi\text{-box}} +  a_\mu^{\pi\pi} + \ldots\,
\end{align}
where $a_\mu^{\pi^0\text{-pole}}$ is generated by the exchange of a $\pi^0$ in one of the channels, $a_\mu^{\pi\text{-box}}$ has two-pion discontinuities simultaneously in two channels  (as shown in diagram (a) of \cref{HLbLTwoPionContributions}), while $a_\mu^{\pi\pi}$ is characterized by a two-pion cut only in one of the three channels and denotes the two-pion contribution beyond the pion box shown in diagrams (b) and (c). The ellipsis refers to higher intermediate states. 
The extensions to $\eta$-, $\eta'$-exchange and two-kaon states are straightforward. The dispersive representations of all these contributions will be discussed in \cref{sec:pion-pole,sec:two-pion}.

\subsubsection{Summary of earlier calculations}

An early attempt at calculating the HLbL contribution
was performed in Ref.~\cite{Kinoshita:1984it} with essentially either the pion-pole
contribution or a constituent quark loop as input. Both approaches gave numbers in the same ballpark
but it was not clear how to combine them. A solution to this was
given in Ref.~\cite{deRafael:1993za} where chiral and large-$N_c$ counting were used
to distinguish a number of contributions and show that they are separate.
Soon after, two groups used this underlying method to do a full
evaluation of the HLbL contribution, namely Hayakawa, Kinoshita, and Sanda (HKS)~\cite{Hayakawa:1995ps,Hayakawa:1996ki,Hayakawa:1997rq} and
Bijnens, Pallante, and Prades (BPP)~\cite{Bijnens:1995cc,Bijnens:1995xf,Bijnens:2001cq}. 
HKS used pseudo-scalar exchange with vector-meson dominance (VMD) suppression in all photon legs, a quark loop also with VMD suppression in all photon legs
and the pion loop evaluated within a hidden local symmetry (HLS) model.
BPP used the framework of the extended Nambu--Jona-Lasinio (ENJL) model
plus a number
of phenomenological adjustments and a short-distance quark loop. This model
gave a decent description of low-energy results. The modeling was necessary
since very few experimental results with one off-shell photon existed at the
time and none with two or more off-shell photons.

Both groups made, albeit for different reasons, a sign mistake
in the pion ``pole'' contribution\footnote{We use quotation marks
since in the model calculations this contribution has some ambiguities
concerning exactly what is included in it.} discovered by Ref.~\cite{Knecht:2001qf}. Both groups obtained similar total numbers with the main contribution
from pion exchange, sizable contributions from $\eta$ and $\eta^\prime$ exchange
as well as from the pion and quark loops. The pion pole has since been
recalculated by many groups (see the reviews Refs.~\cite{Jegerlehner:2009ry,Jegerlehner:2017gek} with numbers that are in reasonable agreement with each other).
The leading logarithmic contributions are in fact fixed from chiral perturbation theory~\cite{Knecht:2001qg,RamseyMusolf:2002cy}. The present best value
for the pion-pole term is discussed in \cref{sec:pion-pole} and is in good agreement with the
old estimates but much more precise.

The next largest contribution was the pion loop but the two models
gave very different results. The HLS model was later realized to not
live up to a short-distance constraint with two photons off-shell~\cite{Bijnens:2007pz,Bijnens:2016hgx}. The ENJL approach of BPP is very similar
to a pion box contribution with a model description for the form-factor.
There was some confusion due to the effect of pion polarizabilities~\cite{Engel:2012xb,Engel:2013kda}, which was finally confirmed to be small~\cite{Bijnens:2016hgx,Colangelo:2017qdm}. This is discussed in \cref{sec:two-pion} on the two-pion
contributions.

The final contributions included in the early work were scalar and axial-vector
exchange. Both were found to be small but not fully negligible. The former
is partly included in the two-pion contribution within the dispersive approach and the latter is discussed in \cref{sec:higher_states}.

The next big improvement was the inclusion of short-distance constraints by
Ref.~\cite{Melnikov:2003xd}. This work gave a larger value for HLbL due
to the way the short-distance constraint was included, which relied on a particular way to treat pseuodoscalar and axial-vector exchange. Short-distance constraints are discussed in \cref{sec:asymptotic} on asymptotics.
It was later agreed that the short-distance quark loop
also satisfied the new short-distance constraint leading to numbers in the
ballpark of $10 \times 10^{-11}$ for HLbL~\cite{Bijnens:2007pz,Prades:2009tw,Jegerlehner:2009ry}.

Alternative approaches include the constituent chiral quark model where only
the pion exchange and the quark loop is taken into
account~\cite{Greynat:2012ww}, variations thereof~\cite{Dorokhov:2012qa,Dorokhov:2015psa},
a number of discussion with only the quark loop~\cite{Boughezal:2011vw,Masjuan:2012qn},
and an in principle more complete
model using the Dyson--Schwinger framework~\cite{Goecke:2010if,Eichmann:2014ooa}.
The former include only a limited number of contributions and the latter
has not yet reached final status. In \cref{tab:earlier} we collect earlier results in various approaches. The dispersive approach is in good agreement with the better model estimates and improves on the precision as well as on the quality of the error estimates. This table is essentially the same as Table~13 in Ref.~\cite{Jegerlehner:2009ry}.

\begin{table}
\centering
\small
\begin{tabular}{crrrrrrr}
\toprule
   Contribution  &  BPP(96) & HKS(96)  &  KnN(02)  &  MV(04)  &  BP(07)  &  PdRV(09)  &  N/JN(09)  \tabularnewline
\midrule 
 $\pi^0,\eta,\eta'$ & $85(13)$ & $82.7(6.4)$   & $83(12)$   & $114(10)$ &  &$114(13)$ & $99(16)$  \tabularnewline

$\pi,K$ loops & $-19(13)$ &  $-4.5(8.1)$    &  &  &  & $-19(19)$ & $-19(13)$  \tabularnewline

$\pi,K$ loops + subleading in $N_c$\; & &   &      & $0(10)$ &  &  &    \tabularnewline

axial vectors & $2.5(1.0)$ & $1.7(1.7)$   &      & $22(5)$ &  & $15(10)$ & $22(5)$   \tabularnewline

scalars & $-6.8(2.0)$ &  &      &  &  & $-7(7)$ & $-7(2)$   \tabularnewline

quark loops & $21(3)$ & $9.7(11.1)$ &      &  &  & $2.3$ & $21(3)$   \tabularnewline

\midrule

total & $83(32)$ & $89.6(15.4)$ & $80(40)$     & $136(25)$ & $110(40)$  & $105(26)$ & $116(39)$   \tabularnewline

\bottomrule
\end{tabular}
\caption{Selection of earlier estimates for HLbL in units of $10^{-11}$. Legend: BPP = Bijnens, Pallante, Prades; HKS = Hayakawa, Kinoshita, Sanda; KnN = Knecht, Nyffeler; MV = Melnikhov, Vainshtein; BP = Bijnens, Prades; PdRV = Prades, de Rafael, Vainshtein; N/JN = Nyffeler /  Jegerlehner, Nyffeler.}
\label{tab:earlier}
\end{table}

\subsection{Experimental inputs and related Monte Carlo studies}
\label{sec:exp_MC}

The numerically most important contributions to HLbL come from the neutral-pion-pole diagram, poles
of other light pseudoscalar mesons ($P=\eta,\eta'$), and two-pion contributions~\cite{Colangelo:2017fiz}. The necessary input to
calculate the pseudo\-scalar-meson-pole contribution to $(g-2)_\mu$ are
TFFs $F_{P\gamma^*\gamma^*}(q^2_1,q^2_2)$ 
of the pseudoscalar
mesons. The TFFs can be
measured experimentally and the status of recent experiments is
presented in \cref{sec:PS_TFF_exp}. In particular 
the normalization of the TFFs, $F_{P\gamma^*\gamma^*}(0,0)$, is given by the radiative decay width 
$\Gamma(P\to\gamma\gamma)$. The most important input is 
the radiative decay width of the $\pi^0$ and the recent results from the PrimEx experiment are discussed in \cref{ssec:pionG}.  In principle the $\pi^0$, $\eta$, and
$\eta'$ pole contributions to HLbL can be obtained from experimental data. However, the doubly off-shell TFFs $F_{P\gamma^*\gamma^*}(q^2_1<0,q^2_2<0)$  are suppressed and
the statistical accuracy is limited.  An alternative
method uses large-statistics data sets on 
hadronic and radiative processes to determine the TFFs using 
dispersion relations.  For the $\pi^0$ TFF the most 
important pieces of input are $\pi\pi$ $P$-wave phase shifts 
and the cross section for $e^+e^-\to\pi^+\pi^-\pi^0$, 
as discussed in \cref{sec:pi0DR}. The next-important contribution is 
from two-photon processes involving a pseudoscalar meson pair,
where the contributions of $\pi^+\pi^-$ and $\pi^0\pi^0$ pairs dominate. 

\subsubsection{Pseudoscalar transition form factors}
\label{sec:PS_TFF_exp}
Pseudoscalar TFFs are experimentally accessible in three different processes.\footnote{In addition, the doubly-virtual TFFs are also probed by dilepton decays~\cite{Abegg:1994wx,Abouzaid:2006kk,Akhmetshin:2014hxv,Achasov:2015mek,Achasov:2018idb,Achasov:2019wtd}.} While the spacelike regime can be studied in two-photon collisions, investigations of the timelike regime are possible in Dalitz decays of pseudoscalar mesons and the radiative production  of pseudoscalar mesons in $e^+e^-$ annihilations. Furthermore, the TFFs' normalizations can be accessed in the decays into two real photons. The theoretical description of TFFs is discussed in \cref{sec:pion-pole}.

\paragraph{Radiative decay widths}
\label{ssec:pionG}
In the case of real photons, the TFF $F_{P\gamma^*\gamma^*}(0,0)$
of a pseudoscalar meson $P$ with mass $M_P$ is related to the radiative decay width according to
\begin{equation}
\Gamma(P \rightarrow \gamma\gamma) = \frac{\pi\alpha^2 M_{P}^3}{4} \left | F_{P\gamma^*\gamma^*}(0,0) \right |^2\,.
\end{equation}
Thus, the precise knowledge of the decay width is important, as it provides the normalization of the TFF measurements.

The radiative decay width of the pion can be determined precisely with the chiral axial anomaly in the limit of vanishing quark masses~\cite{Adler:1969gk,Bell:1969ts}:\footnote{See \cref{pion_decay constant} for our conventions regarding $F_\pi$.}
\begin{equation}
\Gamma(\pi^0 \rightarrow \gamma\gamma) = \frac{\alpha^2 M_{\pi^0}^3}{64\pi^3F_{\pi}^2} = 7.750(16) \,\textrm{eV}\,.
\end{equation}
Higher-order corrections have been calculated in the framework of chiral perturbation theory~\cite{Moussallam:1994xp,Goity:2002nn,Ananthanarayan:2002kj,Kampf:2009tk} up to order $p^8$. Also, corrections to the chiral anomaly have been estimated using dispersion relations and sum rules~\cite{Ioffe:2007pl}. The estimated uncertainty in the ChPT prediction is 1.4\%~\cite{Kampf:2009tk}.
The PrimEx collaboration at Jefferson Lab has performed two Primakoff-type experiments 
to measure the $\pi^0 \to \gamma\gamma$ decay width with the matching accuracy.
In the Primakoff method the  $\pi^0$ photoproduction cross section is measured at forward angles in the Coulomb field of a heavy nucleus. The two experiments, PrimEx-I and PrimEx-II with an upgraded setup, have measured the cross sections on carbon and lead (PrimEx-I) and on carbon and silicon (PrimEx-II).

The PrimEx-I result~\cite{Larin:2010kq}, with an extracted value of the decay width of $\Gamma(\pi^0 \to \gamma\gamma) = 7.82(14)_{\rm stat}(17)_{\rm syst}\,\textrm{eV}$, with a total uncertainty of 2.8\%, is the most precise published measurement. The preliminary improved value from PrimEx-II is  $\Gamma(\pi^0 \to \gamma\gamma) = 7.790(56)_{\rm stat}(109)_{\rm syst} \,\textrm{eV}$~\cite{Larin:2020}. Therefore the combined result is:
\begin{equation}
 \Gamma(\pi^0 \to \gamma\gamma) = 7.802(52)_{\rm stat}(105)_{\rm syst}\,\textrm{eV} = 7.802(117) \,\textrm{eV} \,,
\end{equation}
with a 1.5\% accuracy matching the ChPT predictions.

The best result for
$\Gamma(\eta\to\gamma\gamma)$ is determined from the cross section of
the $e^+ e^- \to e^+ e^- \eta$ process at CM\ energy
of $1\,\text{GeV}$ measured by the KLOE-2 experiment, where the
final electrons escape the detector at low scattering angles. This
cross section is dominated by the contributions from
photons with virtualities close to
zero, leading to a 4.6\% accuracy for the $\eta$ radiative width:
$520(20)_{\rm stat}(13)_{\rm stat}\,\text{eV}$~\cite{Babusci:2012ik}.

Based on the same experimental approach, the best individual result for a measurement of  $\Gamma(\eta^\prime\to\gamma\gamma)$ is reported by the L3 collaboration. The reaction $e^+ e^- \to e^+ e^- \eta^\prime$ is studied at $\sqrt{s}=91\,\textrm{GeV}$ and the radiative width of the $\eta^\prime$ is determined as $4.17(10)_{\rm stat}(27)_{\rm stat}\,\textrm{keV}$~\cite{Acciarri:1997yx}, hence with a relative accuracy of 6.9\%.

\paragraph{Spacelike transition form factors}
Two-photon collisions are studied at $e^+e^-$ colliders. The cross section of pseudoscalar meson production in two-photon collisions is proportional to the square of the respective spacelike TFF, which is a function of the momentum transfers $Q_{1,2}^2=-q_{1,2}^2\geq0$ of the two photons. Due to the dependence of the cross section on the momentum transfers, published data on pseudoscalar TFFs almost exclusively correspond to the case where one of the photons is quasi-real, i.e., $Q^2\approx0$.

\begin{figure}[t]
    \centering
    \includegraphics[width=0.6\textwidth]{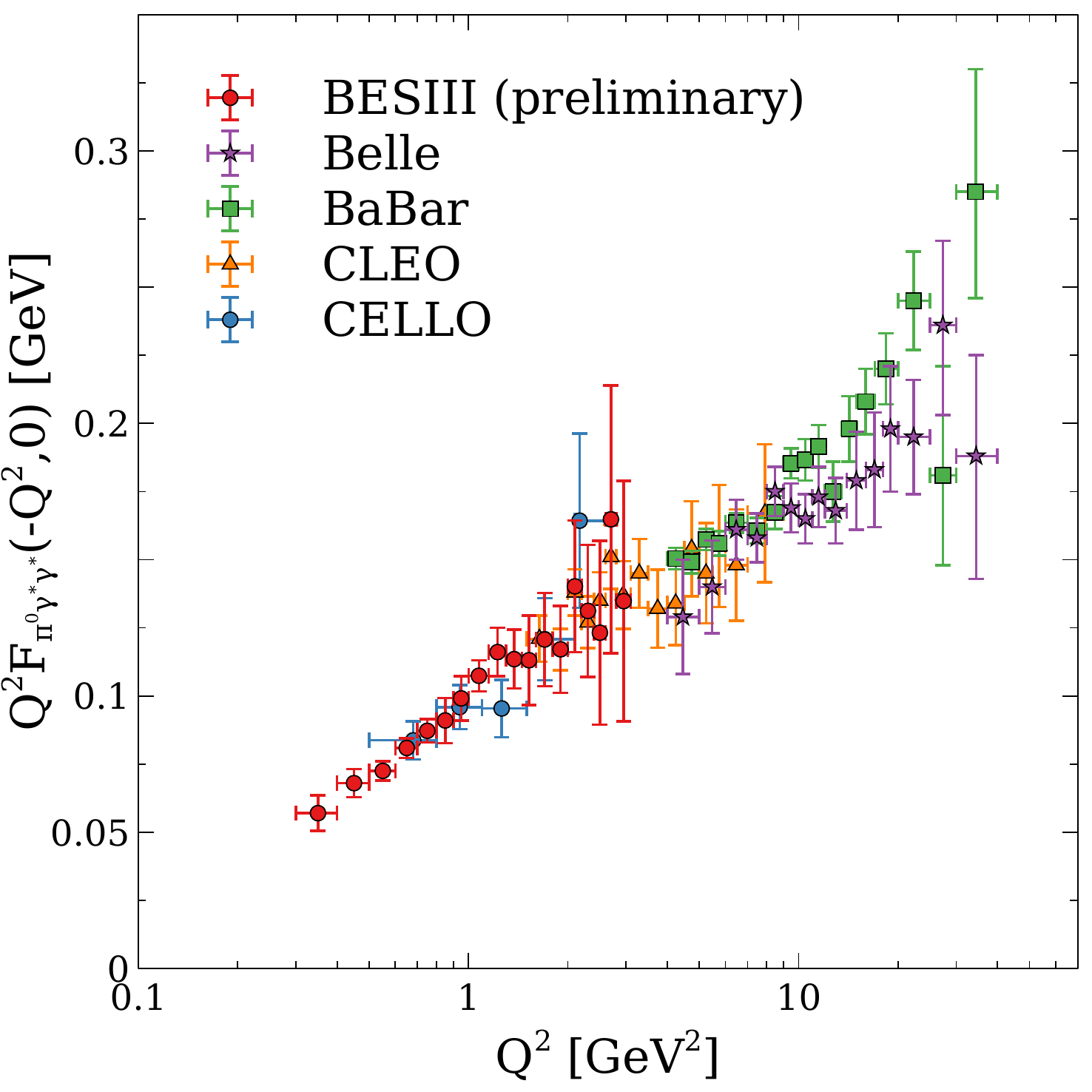}
    \caption{\label{fig:expslpion}Available information on the spacelike TFF of the $\pi^0$. CELLO~\cite{Behrend:1990sr} (blue circles), CLEO~\cite{Gronberg:1997fj} (yellow triangles), BABAR~\cite{Aubert:2009mc} (green boxes), Belle~\cite{Uehara:2012ag} (purple stars), and preliminary BESIII data (red) are displayed. The error bars indicate total uncertainties.}
\end{figure}

Published results on the $\pi^0$ TFF from the CELLO experiment at DESY~\cite{Behrend:1990sr} are illustrated with blue circles in \cref{fig:expslpion}, they cover momentum transfers $0.5\,\textrm{GeV}^2\leq Q^2\leq 2.7\,\textrm{GeV}^2$. The CLEO results~\cite{Gronberg:1997fj} have better accuracy, albeit at higher momentum transfers of $1.5\,\textrm{GeV}^2 \leq Q^2 \leq 9.0\,\textrm{GeV}^2$, as shown with yellow triangles in \cref{fig:expslpion}. The most recent measurements are from BABAR and Belle~\cite{Aubert:2009mc,Uehara:2012ag}, and they are illustrated with green boxes and purple stars, respectively, in \cref{fig:expslpion}. Due to kinematic limitations only momentum transfers $Q^2>4\,\textrm{GeV}^2$ are measured. On the other hand, the results cover the full range up to $40\,\textrm{GeV}^2$, allowing for tests of pQCD applicability. Here, a peculiar feature of the BABAR results is the excess of the TFF beyond the predicted asymptotic limit of pQCD, which also shows a tension with the Belle result. It should be noted that the BABAR result is the first one to explicitly take into account radiative corrections. Potential issues in the treatment are addressed in \cref{sec:expmc_radcor}. More recently, the BESIII collaboration released preliminary data on the $\pi^0$ TFF in the range $0.3\,\textrm{GeV}^2 \leq Q^2 \leq 3.1\,\textrm{GeV}^2$~\cite{Redmer:2018uew}. As can be seen from the red circles in \cref{fig:expslpion}, the statistical accuracy is unprecedented at $Q^2<1.5\,\textrm{GeV}^2$ and compatible with the accuracy achieved by CLEO at larger values of $Q^2$. The final BESIII result will include the latest developments on radiative corrections discussed in \cref{sec:expmc_radcor}.

\begin{figure}[t]
    \centering
    \includegraphics[width=0.49\textwidth]{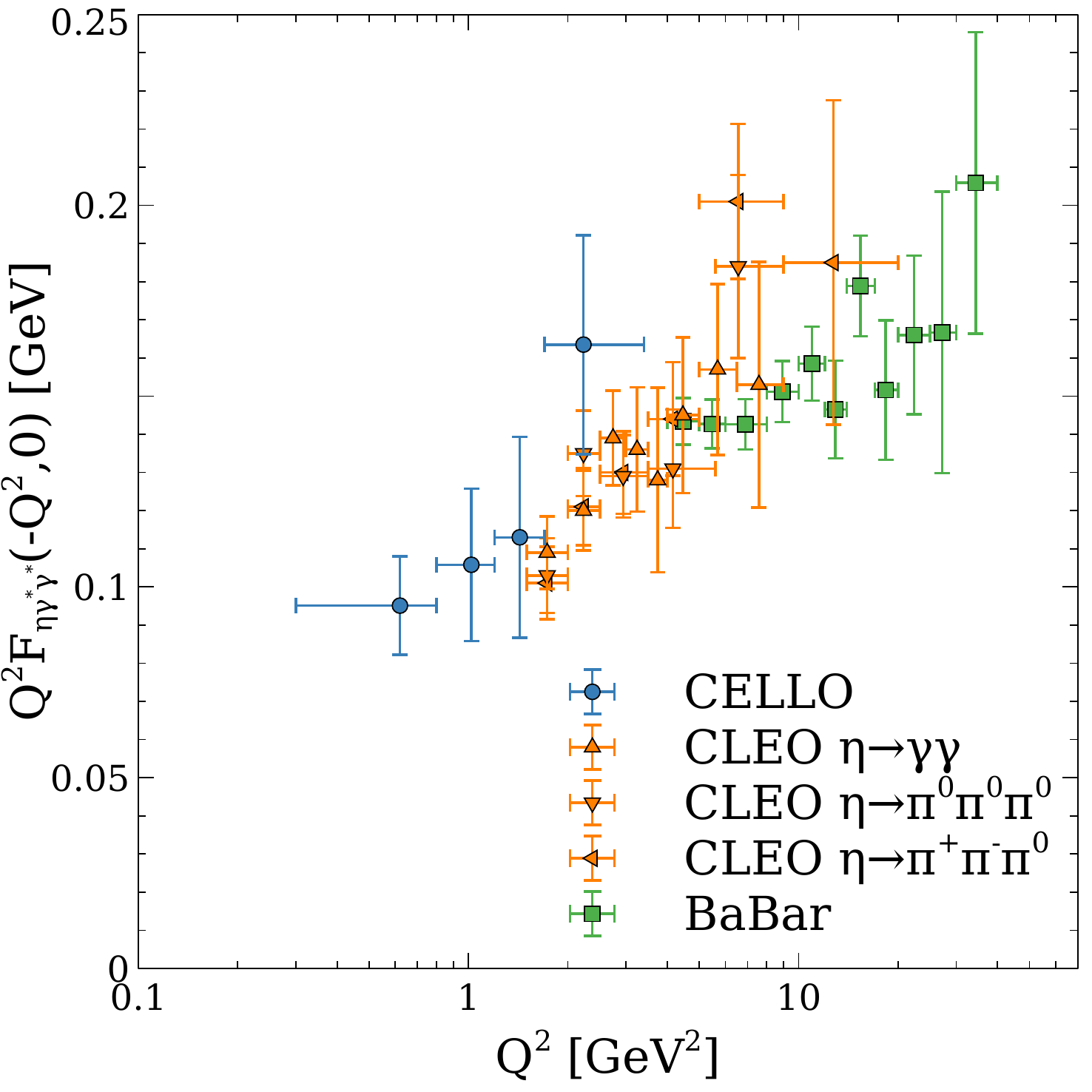}\hfill%
    \includegraphics[width=0.49\textwidth]{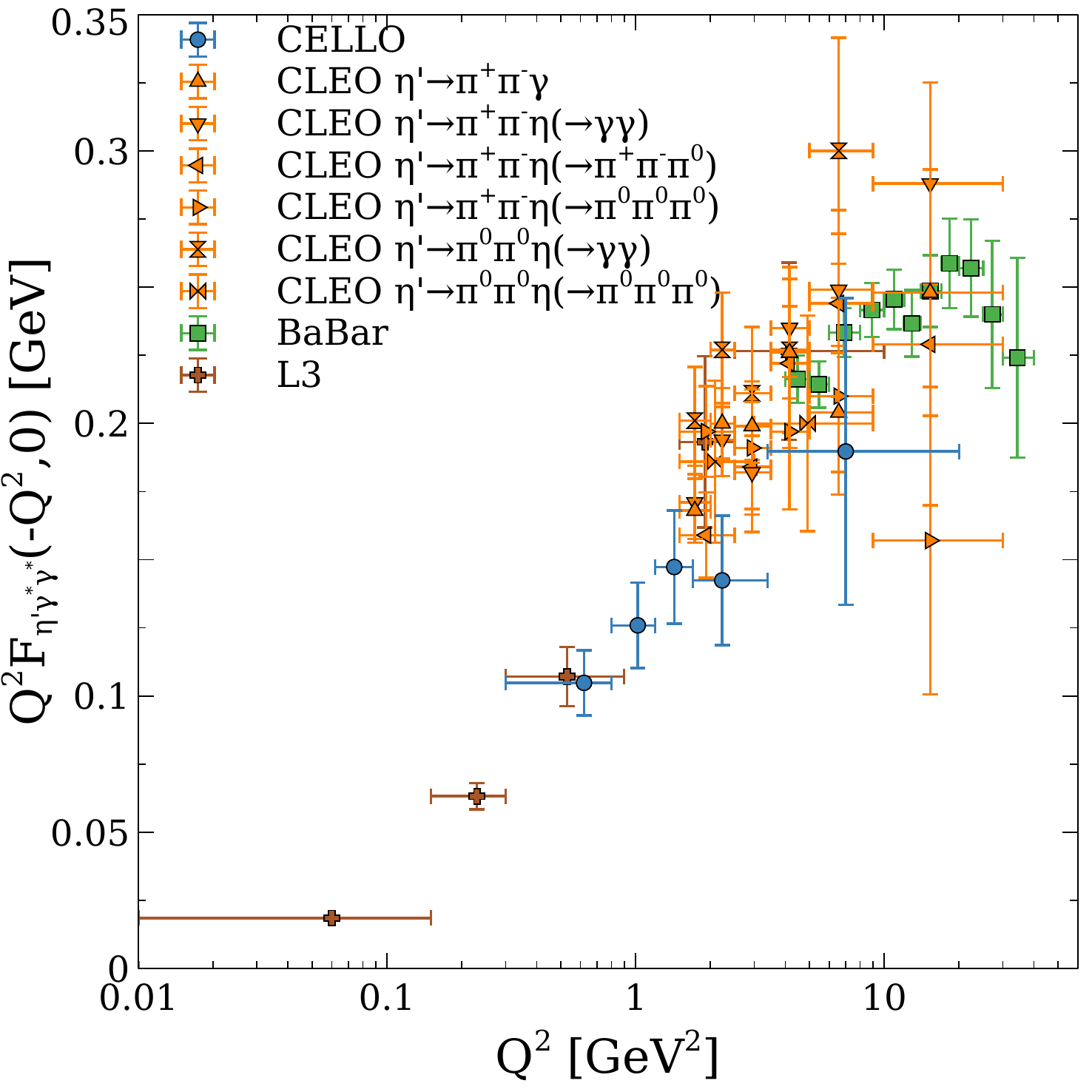}
    \caption{\label{fig:expsletap}Available information on the spacelike TFFs of $\eta$ (left) and $\eta^\prime$ (right). CELLO~\cite{Behrend:1990sr} (blue circles), CLEO~\cite{Gronberg:1997fj} (yellow triangles), BABAR~\cite{BABAR:2011ad} (green boxes), and L3 data~\cite{Acciarri:1997yx} (brown crosses) are shown. The error bars indicate total uncertainties.}
\end{figure}

Data on the spacelike $\eta$ TFF have been published by the CELLO, CLEO, and BABAR~\cite{BABAR:2011ad} collaborations and are illustrated in the left panel of \cref{fig:expsletap}. The CELLO result (blue circles) combines the analysis of the decay modes $\eta\to\gamma\gamma, \eta\to\pi^+\pi^-\pi^0$, and $\eta\to\pi^+\pi^-\gamma$~\cite{Behrend:1990sr}. Momentum transfers $0.3\,\textrm{GeV}^2 \leq Q^2\leq3.4\,\textrm{GeV}^2$ are covered. The CLEO result (yellow triangles), in contrast, treats every investigated decay mode independently~\cite{Gronberg:1997fj}. For $\eta\to\gamma\gamma$ and $\eta\to3\pi^0$ momentum transfers in the range $1.5\,\textrm{GeV}^2 \leq Q^2 \leq 9.0\,\textrm{GeV}^2$ are studied, while information on $\eta\to\pi^+\pi^-\pi^0$ is available for $Q^2\leq20\,\textrm{GeV}^2$. The size of the  $Q^2$ intervals differs for the individual decay modes.
At BABAR (green boxes) only the decay mode $\eta\to\pi^+\pi^-\pi^0$ is investigated, providing information in the momentum range $4.0\,\textrm{GeV}^2 \leq Q^2 \leq 40.0\,\textrm{GeV}^2$~\cite{BABAR:2011ad}.

The same experiments also provided data on the spacelike $\eta^\prime$ TFF, which are shown in the right panel of \cref{fig:expsletap}. The CELLO result (blue circles) is a combination of the investigated decay modes $\eta^\prime\to\pi^+\pi^-\gamma$ and $\eta^\prime\to\pi^+\pi^-\eta$~\cite{Behrend:1990sr}. For the latter, the subsequent decay modes $\eta\to\gamma\gamma$, $\eta\to\pi^+\pi^-\pi^0$, and $\eta\to\pi^+\pi^-\gamma$ are considered. Information on the momentum transfer dependence is available for $0.3\,\textrm{GeV}^2 \leq Q^2\leq20.0\,\textrm{GeV}^2$. The CLEO result (yellow triangles) covers momentum transfers $1.5\,\textrm{GeV}^2 \leq Q^2\leq30.0\,\textrm{GeV}^2$~\cite{Gronberg:1997fj}. The decay modes $\eta^\prime\to\pi^+\pi^-\gamma$, $\eta^\prime\to\pi^+\pi^-\eta$, and $\eta^\prime\to\pi^0\pi^0\eta$ are considered along with several subsequent decay modes of the $\eta$ meson. The momentum transfer dependence of the TFF is provided separately for the six final states. At BABAR (green boxes), only the decay $\eta^\prime\to\pi^+\pi^-\eta$ with the subsequent two-photon decay of the $\eta$ meson has been used, providing information in the momentum range $4.0\,\textrm{GeV}^2 \leq Q^2 \leq 40.0\,\textrm{GeV}^2$~\cite{BABAR:2011ad}.
The L3 experiment (brown crosses) at LEP, CERN investigated $\eta^\prime\to\pi^+\pi^-\gamma$ for momentum transfers up to $Q^2\leq10.0\,\textrm{GeV}^2$~\cite{Acciarri:1997yx}. In addition to the single-tag measurement, the proportionality of the transverse momentum of the $\eta^\prime$ to the total momentum transfer $Q^2=Q_1^2+Q_2^2$ is exploited to provide information down to $Q^2\geq0.1\,\textrm{GeV}^2$.

Recently, the BABAR collaboration published the first measurement of the doubly-virtual TFF of the $\eta^\prime$ based on $\eta^\prime\to\pi^+\pi^-\eta$ and the subsequent $\eta\to\gamma\gamma$~\cite{BaBar:2018zpn}. Based on $46.2^{+8.3}_{-7.0}$ signal events, the TFF is determined in seven intervals of $(Q_1^2,Q_2^2)$ in the range $2.0\,\textrm{GeV}^2 \leq Q_{1,2}^2\leq60.0\,\textrm{GeV}^2$. Three of the intervals are along the diagonal of $Q_1^2=Q_2^2$, four are off-diagonal. A one-dimensional representation of the result is shown in \cref{fig:etapDV}.

\paragraph{Timelike transition form factors}
Dalitz decays of pseudoscalar mesons allow one to determine the respective TFFs in the momentum range of $m_{ll}^2\leq q^2\leq M_P^2$, with $m_{ll}^2$ the CM energy of the dilepton. Thus, they are ideally suited to determine the slope of the TFF at $q^2=0$, defined according to $F_{P\gamma^*\gamma^*}(q^2,0) \approx F_{P\gamma^*\gamma^*}(0,0)[1+q^2/\Lambda_P^2 + \mathcal{O}(q^4)]$.

Until recently the PDG value of the slope of the $\pi^0$ TFF was dominated by a model-dependent extrapolation of the spacelike results of the CELLO collaboration~\cite{Behrend:1990sr}, which provided a better accuracy than the existing data~\cite{Farzanpay:1992pz,MeijerDrees:1992qb} on the $\pi^0$ Dalitz decay. Recently, two high-statistics measurements of the latter were published by the A2~\cite{Adlarson:2016ykr} and NA62~\cite{TheNA62:2016fhr} collaborations. The former studied the Dalitz decay in $\pi^0$ photoproduction with a nonmagnetic spectrometer, obtaining a slope parameter of $\Lambda_\pi^2 = 0.61(20)\,\textrm{GeV}^2$. The latter exploited the $K^\pm\to\pi^\pm\pi^0$ decay using a secondary particle beam at CERN to obtain a slope parameter value of $\Lambda_\pi^2=0.495(76)\,\textrm{GeV}^2$. Both results take into account the recent radiative corrections of Ref.~\cite{Husek:2015sma}.

Due to its larger mass, the Dalitz decay of the $\eta$ and $\eta^\prime$ mesons can proceed via electron--positron as well as muon pairs. The latest results on $\eta\to\mu^+\mu^-\gamma$ have been obtained by the NA60 collaboration in $p$--$A$ collisions~\cite{Arnaldi:2016pzu}, where the inclusively measured dimuon mass spectrum is fit with the expected contributions after background subtraction to determine the slope parameter as $\Lambda^2_\eta = 0.517(22)\,\textrm{GeV}^{2}$. In contrast, the A2 collaboration studied $\eta\to e^+ e^-\gamma$ exclusively in photo-induced production with a nonmagnetic spectrometer yielding $\Lambda^2_\eta = 0.507(28)\,\textrm{GeV}^{2}$ in their latest result~\cite{Adlarson:2016hpp}. In contrast to the $\pi^0$ case, radiative corrections~\cite{Husek:2017vmo} have not been applied here yet.

For the Dalitz decays of the $\eta^\prime$ meson, the available experimental information is rather scarce. A first measurement based on muon pairs in the final state was reported from a pion-induced experiment in Serpukhov~\cite{Landsberg:1986fd}. Only recently, the  $\eta^\prime\to e^+ e^-\gamma$ decay was observed by BESIII and  the slope parameter of  $\Lambda_{\eta^\prime}^2=0.625(73)\,\textrm{GeV}^{2}$~\cite{Ablikim:2015wnx} was determined. 

The Dalitz decays of pseudoscalar mesons can also proceed via two virtual photons, resulting in a four-lepton final state. The most recent result for the $\pi^0$ double Dalitz decay comes from the KTeV experiment at Fermilab~\cite{Abouzaid:2008cd}. Besides the momentum dependence of the TFF, the branching ratio and the parity of the pion are determined with high accuracy. For the heavier pseudoscalar mesons only the double Dalitz decay of the $\eta$ meson was observed at KLOE-2~\cite{KLOE2:2011aa} and the branching fraction determined.

At timelike momentum transfers larger than the squared rest mass of the pseudoscalar meson, information on the TFF can be obtained from the radiative production of the mesons in $e^+e^-$ annihilations. The production cross section is related to the TFF according to \begin{equation}
    \sigma_{M\gamma}(q^2) = \frac{2\pi^2\alpha^3}{3}\frac{(q^2-M_P^2)^3}{q^6} |F_{P\gamma^*\gamma^*}(q^2,0)|^2 \,,
\end{equation} 
where $q^2 = s$ is equivalent to the squared CM energy $\sqrt{s}$ defined by the accelerator.

\begin{figure}[t]
    \centering
    \includegraphics[width=0.49\textwidth]{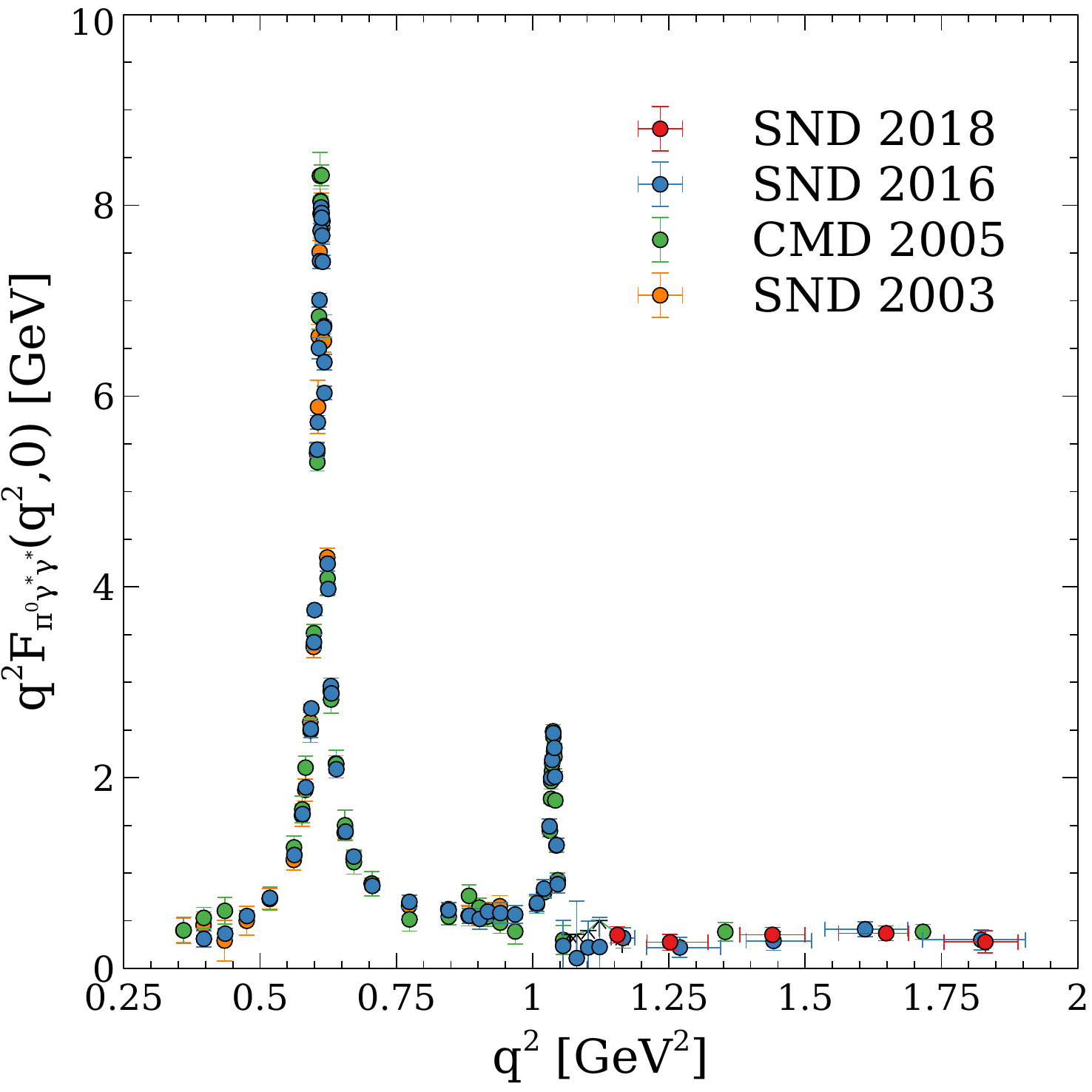}\hfill%
    \includegraphics[width=0.49\textwidth]{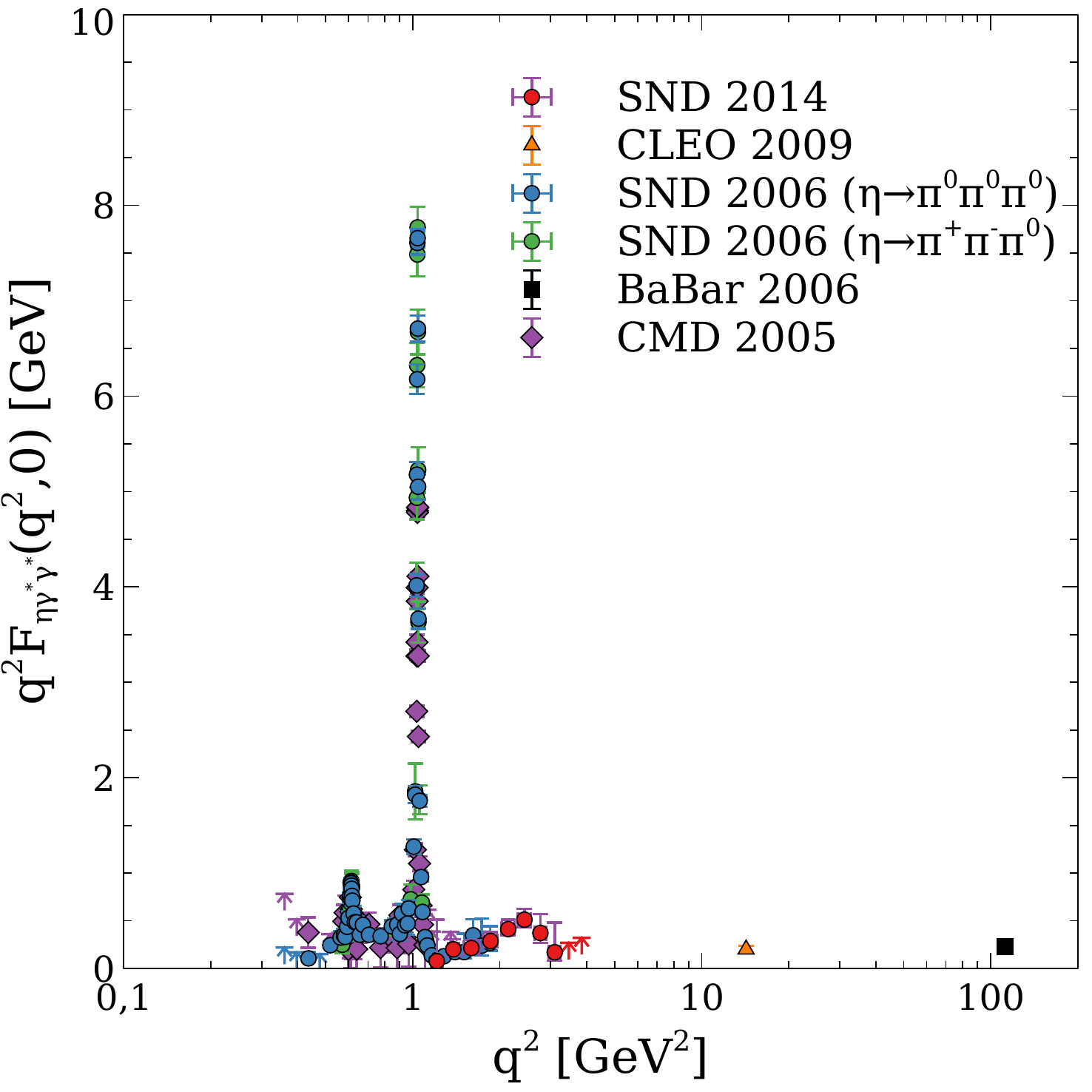}
    \caption{\label{fig:exptlpieta}Available information on the timelike TFF of $\pi^0$ (left) and $\eta$ (right), from CMD-2~\cite{Akhmetshin:2004gw}, SND~\cite{Achasov:2006dv,Achasov:2013eli,Achasov:2016bfr,Achasov:2018ujw}, CLEO~\cite{Pedlar:2009aa}, and BABAR~\cite{Aubert:2006cy}. The error bars indicate total uncertainties.}
\end{figure}

The CMD and SND experiments in Novosibirsk have a long tradition in studying $e^+e^-\to\gamma\pi^0$ and $e^+e^-\to\gamma\eta$ in energy scan experiments up to $\sqrt{s}=2\,\textrm{GeV}$, exploiting the potential of the detectors and the accelerator complex~\cite{Akhmetshin:2001hm,Achasov:2002ud,Akhmetshin:2004gw,Achasov:2006dv,Achasov:2013eli,Achasov:2016bfr,Achasov:2018ujw} upgrades. The most recent cross section results are shown in \cref{fig:exptlpieta}. The data on the $\eta^\prime$ production are available, at significantly larger $\sqrt{s}$, from the CLEO and BABAR collaborations, where the $\eta$ and $\eta^\prime$ TFFs were 
measured at $\sqrt{s}=3.773\,\textrm{GeV}$ and $\sqrt{s}=10.6\,\textrm{GeV}$~\cite{Pedlar:2009aa,Aubert:2006cy}. In addition, there is information on $\phi\to\eta'\gamma$ from KLOE~\cite{Ambrosino:2006gk}, SND~\cite{Aulchenko:2003wv}, and CMD-2~\cite{Akhmetshin:2000hp}.

Dalitz decays of vector mesons into pseudoscalars and lepton pairs also provide information of pseudoscalar TFFs.
Special interest is in $\omega\to\pi^0\mu^+\mu^-$, where the NA60 collaboration reported a strong deviation of the $q^2$ dependence from the expected VMD behavior~\cite{Arnaldi:2009aa}. The result, obtained by fitting known contributions to the inclusive dimuon spectrum measured in peripheral In--In collisions was later confirmed in $p$--$A$ collisions with an order of magnitude better statistics~\cite{Arnaldi:2016pzu}. An exclusive measurement of $\omega\to\pi^0 e^+ e^-$ is provided by the A2 collaboration~\cite{Adlarson:2016hpp}. However, due to limited statistics especially at large values of $q^2$, an unambiguous confrontation of experimental results and calculations is not yet possible.

The KLOE-2 collaboration reported results on the Dalitz decays of the $\phi$ into $e^+e^-$ pairs and $\pi^0$ and $\eta$ mesons, respectively, with world-leading accuracy~\cite{Anastasi:2016qga,Babusci:2014ldz}. While the $q^2$ dependence in case of the $\eta$ meson can still be reasonably described by a VMD model, a better description is achieved for the case of the $\pi^0$ by a calculation with parameters fit to the $\omega\to\pi^0\mu^+\mu^-$ result obtained by the NA60 collaboration.

\paragraph{Data available in the near future} 

The experimental input to the dispersive calculations of the hadronic light-by-light contributions for the pseudoscalar pole contributions are the TFFs $F_{P\gamma^*\gamma^*}(-Q_1^2,-Q_2^2)$ at arbitrary, yet small values of momentum transfer $Q_i^2\approx1\,\textrm{GeV}^2$. The preliminary result for $\pi^0$ presented by the BESIII collaboration (see \cref{sec:PS_TFF_exp}) covering $0.3\,\textrm{GeV}^2 \leq Q^2 \leq 3.1\,\textrm{GeV}^2$ thus provides singly-virtual information in the relevant region. The final publication of the result is expected within the next months. Additionally, the analyses are being extended to the TFFs of $\eta$ and $\eta^\prime$, combining different decay modes for better statistical accuracy. First tests indicate the feasibility of the measurements covering the same range of momentum transfer as in the  $\pi^0$ case.

\begin{figure}
    \centering
    \includegraphics[width=0.49\textwidth]{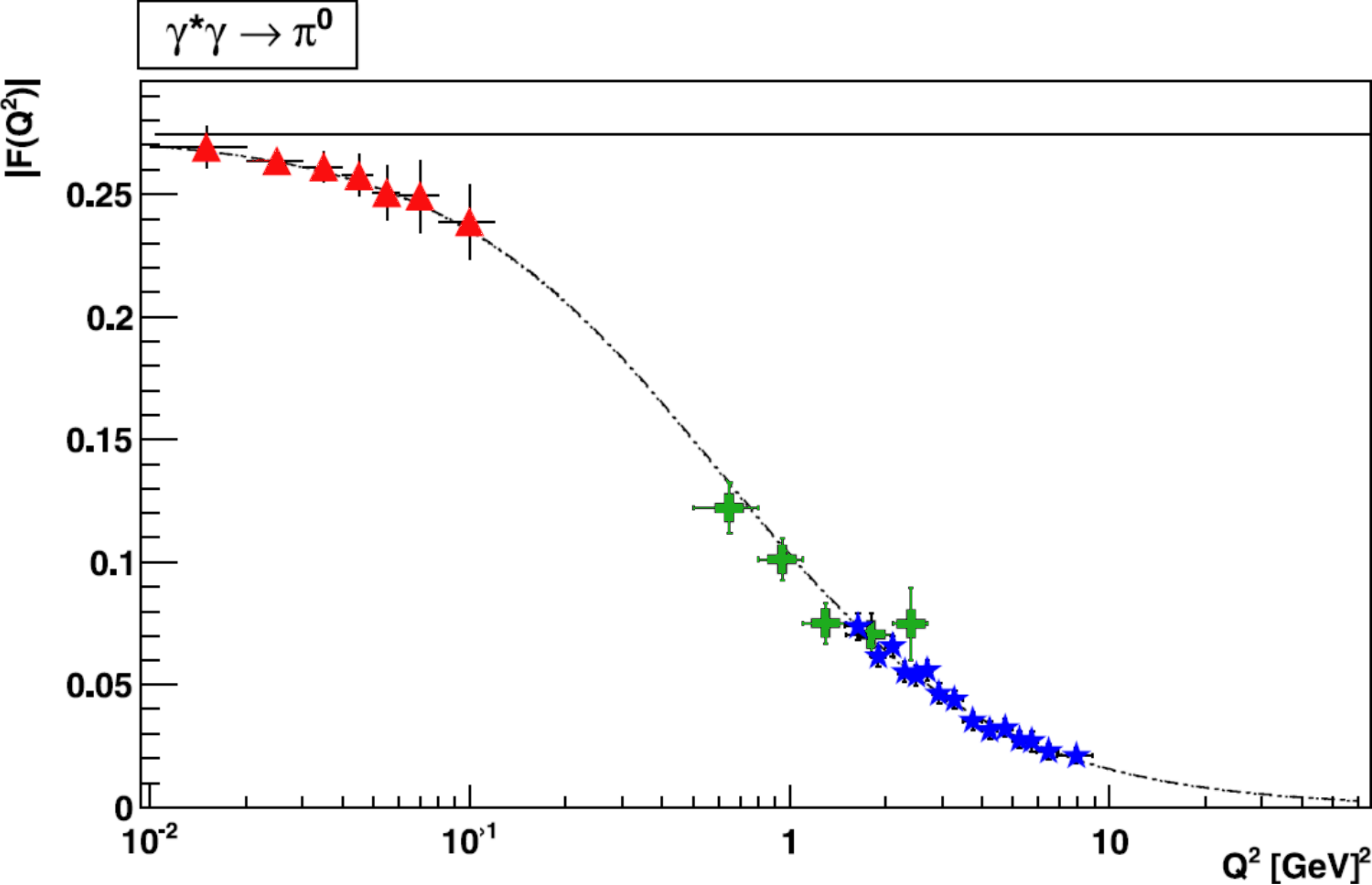}
    \caption{\label{fig:kloesimu} Simulation of KLOE-2 measurement of the $\gamma^* \gamma \to \pi^0$ TFF $F_{\pi^0\gamma^*\gamma^*}(-Q^2,0)$ (red triangles) with statistical errors for $5\,\text{fb}^{-1}$~\cite{Babusci:2011bg}. Dashed line is $F_{\pi^0\gamma^*\gamma^*}(-Q^2,0)$ according to the LMD+V model~\cite{Knecht:2001xc}, solid line is $F_{\pi^0\gamma^*\gamma^*}(0,0)$ as given by the chiral anomaly~\cite{Adler:1969gk,Bell:1969ts}. CELLO~\cite{Behrend:1990sr} (green crosses) and CLEO~\cite{Gronberg:1997fj} (blue stars) data at high $Q^2$ are also shown for illustration. Reprinted from Ref.~\cite{Babusci:2011bg}.}
\end{figure}

Further relevant results are expected from the KLOE-2 experiment, whose data taking campaign has ended in March 2018 after collecting more than 5\,fb$^{-1}$ at the CM\ energy corresponding to the $\phi$ meson mass. The goal of the ongoing analysis is to measure the $\pi^0 \to \gamma\gamma$ decay width with 1\% statistical accuracy, and the $\gamma^* \gamma \to \pi^0$ TFF with 6\% accuracy for each bin of the spacelike momentum transfer. Covering the range of $0.01\,\textrm{GeV}^2\leq Q^2 \leq0.1\,\textrm{GeV}^2$~\cite{Babusci:2011bg}, for the first time data at smallest values of $Q^2$ will be provided, as illustrated in \cref{fig:kloesimu}. The result will serve both as an input to the data-driven approaches to HLbL discussed in \cref{sec:pion-pole}
and as an intermediate cross-check of the respective calculations including the results obtained by lattice QCD~\cite{Gerardin:2019vio}.

Recently, the KLOE-2 collaboration for the first time~\cite{Curciarello:2019} presented a preliminary statistical evidence of correlated coincidence events between the tagger stations~\cite{Archilli:2010zza} and the KLOE calorimeter, obtained from the analysis of a data sample of about $500\,\text{pb}^{-1}$. Efforts are still needed to reduce the big amount of uncorrelated coincidence events affecting the feasibility and the precision of the measurement.

New measurements at similarly small virtualities are also proposed at JLab~\cite{JLab12pCDR:2014,Gan:2015nyc}. After the $12\,\text{GeV}$ upgrade the Primakoff program can on the one hand be used to measure the radiative widths of $\eta$ and $\eta^\prime$ with high precision and in a systematically independent approach compared to the $e^+e^-$ experiments. On the other hand, the Primakoff program can be extended to measure the momentum dependence of the meson TFFs at smallest values of $Q^2$. A momentum range of $0.001\,\textrm{GeV}^2$ up to $0.5\,\textrm{GeV}^2$ can be covered, which is not only complementary to the efforts at KLOE-2, but also bridges the gap between the expected KLOE-2 data and the BESIII results, leading to a fully mapped out momentum dependence of the singly-virtual meson TFFs.

To actually provide data at arbitrary values of both virtualities, double-tagged measurements are required. So far only the BABAR result on $\eta^\prime$ is available at rather large virtualities. The BESIII collaboration started feasibility studies of double-tagged measurements for the three lightest pseudoscalar mesons~\cite{Redmer:2018gah}. Within the next years results can be expected on $F_{P\gamma^*\gamma^*}(-Q_1^2,-Q_2^2)$ with $0.5\,\textrm{GeV}^2\leq Q_1^2,Q_2^2 \leq 2.0\,\textrm{GeV}^2$. The impact of this measurement for the $\pi^0$ pole contribution has already been estimated based on an optimistic Monte Carlo study~\cite{Nyffeler:2016gnb}.

\subsubsection{$\gamma^{(*)} \gamma \to \pi \pi$ and other pseudoscalar meson pairs}
\label{ssec:ggPP}
Most of the investigations on two-photon production of pairs of pions or other pseudoscalar mesons have been performed studying collisions of quasi-real photons. In terms of accuracy, the studies compiled in Refs.~\cite{Morgan:1994ip,Whalley:2001mk}, which were performed mostly at SLAC and DESY, have been superseded by the results obtained at Belle for $\pi^0\pi^0$~\cite{Uehara:2008ep,Uehara:2009cka} $\pi^+\pi^-$~\cite{Nakazawa:2004gu,Mori:2007bu}, $\pi^0\eta$~\cite{Uehara:2009cf}, $\eta\eta$~\cite{Uehara:2010mq}, $K^+K^-$~\cite{Abe:2003vn,Nakazawa:2004gu}, and $K^0_SK^0_S$~\cite{Uehara:2013mbo}. The data allows one to determine properties of scalar and tensor resonances, especially their radiative widths.

Recently, the Belle collaboration published the first measurements of $\pi^0$ and $K^0_S$ pairs production in a singly off-shell photon process~\cite{Masuda:2015yoh,Masuda:2017rhm}. The $\pi^0\pi^0$ system is studied for masses between $0.5\,\textrm{GeV}\leq m_{\pi\pi}\leq2.1\,\textrm{GeV}$ and helicity angles of the pions with $|\cos\theta|<1.0$. Momentum transfers between $Q^2\geq3.0\,\textrm{GeV}^2$ and $Q^2\leq30\,\textrm{GeV}^2$ are covered. The TFF of $f_0(980)$ and of the helicity-$0$, -$1$, and -$2$ components of $f_2(1270)$ are determined separately. The analogous measurement of the $K^0_sK^0_s$ system allowed for the first time for equivalent investigations of the TFF of the $f^\prime_2(1525)$. Finally, further constraints arise from timelike processes, which have been measured for singly-virtual kinematics~\cite{Achasov:2000zr,Achasov:2002jv,Akhmetshin:2003ag,Akhmetshin:2003rg,Ambrosino:2006hb,Achasov:2016zvn}.

\subsubsection{Other relevant measurements}

Anticipating the combined estimate in \cref{sec:result_HLbL_DR}, we discuss here which other, future, measurements will be particularly useful to improve on the data-driven determination of the HLbL contribution.

\begin{table}
     \centering
     \small
    \begin{tabular}{ll}
    \toprule
issue   & experimental input [I] or cross-checks [C] \\
    \midrule
     axials, tensors, higher pseudoscalars     & $\gamma^{(*)}\gamma^*\to 3\pi,\,  4\pi, \, K\bar{K}\pi,\,\eta\pi\pi,\,\eta'\pi\pi$ [I]\\
 missing states                     & inclusive $\gamma^{(*)}\gamma^*\to\textrm{hadrons}$ at 1--3\,GeV [I]\\ 
  dispersive analysis of $\eta^{(\prime)}$ TFFs 
                                    & $e^+e^-\to\eta\pi^+\pi^-$ [I] \\
                                    & $\eta^\prime\to\pi^+\pi^-\pi^+\pi^-$ [I] \\
                                    & $\eta^\prime\to\pi^+\pi^-e^+e^-$ [I] \\
                                    & $\gamma\pi^-\to\pi^-\eta$ [C]\\
 dispersive analysis of $\pi^0$ TFF & $\gamma\pi\to\pi\pi$ [I] \\
 & high accuracy Dalitz plot $\omega\to\pi^+\pi^-\pi^0$ [C]\\
                                    & $e^+e^-\to\pi^+\pi^-\pi^0$ [C]\\
                                    & $\omega, \phi\to\pi^0l^+l^-$ [C]\\
    pseudoscalar TFF                & $\gamma^{(*)}\gamma^*\to \pi^0,\eta,\eta^\prime$ at arbitrary virtualities [I,C]\\
    pion, kaon, $\pi\eta$ loops  & $\gamma^{(*)}\gamma^*\to \pi\pi,\,K\bar{K},\,\pi\eta$ at arbitrary virtualities,\\ 
     \quad (including scalars and tensors) & \quad partial waves [I,C]\\
 \bottomrule
 \end{tabular}
 \caption{Priorities for new experimental input and cross-checks.}
    \label{tab:wishlist}
\end{table}

Apart from the uncertainty originating from the short-distance regime, the largest individual error is currently attributed to the axial-vector contributions; beyond that, also scalars and tensors above $1\,\text{GeV}$ come with a very large relative uncertainty and the role of 
excited pseudoscalar states  has been stressed recently in the context of short-distance constraints~\cite{Colangelo:2019lpu,Colangelo:2019uex}.
For the estimate of such contributions, data on three- or four-pion as well as other multi-hadron final states ($K\bar K\pi$, $\eta\pi\pi$, $\eta'\pi\pi$) are needed. In the past, mostly measurements of the two-photon production using quasi-real photons were performed. In view of the Landau--Yang theorem~\cite{Landau:1948kw,Yang:1950rg} that forbids the coupling of axial vectors to two real photons, new measurements should go beyond that restriction. Studies on the four-pion final states involving a single virtuality focused on double vector-meson production~\cite{Achard:2003qa,Achard:2004ux,Achard:2004us,Achard:2005pb}. 

An experimentally challenging task will be a measurement of the inclusive hadron production cross section in two-photon collisions at masses between 1 and 3\,GeV (see Refs.~\cite{Berger:1984uk,Bintinger:1984as,Aihara:1989kr,Baru:1991bt} for the on-shell case). The inclusive mass spectra with one or both of the photons off-shell will settle the issue of missing states in the calculations of $a_\mu^{\rm HLbL}$, and may lead to an improved matching of this intermediate region to quark-loop estimates that interpolate towards the short-distance limits.

Beyond these altogether rather poorly known contributions, there is a strong incentive to further improve upon the dominant, large pieces.
For a dispersive analysis of the singly- and doubly-virtual pseudoscalar TFFs, as discussed in \cref{sec:pion-pole}, additional, independent experimental information is needed. 
The data can be divided into necessary input to the calculations that, together with theory 
uncertainties, will determine the accuracy of the predictions; and experimental cross-checks.

For the dispersive description of the TFFs of $\eta$ and $\eta^\prime$ (that has not been completed yet)~\cite{Kubis:2018bej}, experimental input to constrain the doubly-virtual behavior are of utmost importance.  To this end, detailed differential data on $e^+e^-\to\eta\pi^+\pi^-$ will contribute to an improved understanding of the deviations of the doubly-virtual TFF from the factorization hypothesis at intermediate energies.  Similarly, differential decay data on $\eta^\prime\to \pi^+\pi^-\pi^+\pi^-$ will allow one to develop a double spectral function, and corresponding measurement of $\eta^\prime\to \pi^+\pi^-e^+e^-$ will help complete the dispersive framework for the $\eta^{(\prime)}$ TFFs, although in either case the kinematic range is limited by the decay kinematics.
Finally, data on the Primakoff-type reaction $\gamma\pi^-\to\pi^-\eta$ would be very helpful to better constrain the high-energy continuation of the dispersive input. 

For the $\pi^0$ TFF~\cite{Leupold:2018mgr}, precision data on the $e^+e^-\to\pi^+\pi^-\pi^0$ cross section would be desirable to settle tensions between the existing data. 
 In addition to the cross section
studies in the context of HVP, the analysis of
$\omega,\phi\to\pi^+\pi^-\pi^0$ decay dynamics provides 
a valuable cross-check of the dispersive formalism. For $\phi\to\pi^+\pi^-\pi^0$
precision data from KLOE and CMD-2 can be used~\cite{Aloisio:2003ur,Akhmetshin:2006sc} but
until recently, surprisingly little
information had been available on the $\omega\to\pi^+\pi^-\pi^0$
Dalitz plot. First observation of a deviation from $P$-wave phase
space consistent with $\rho$ meson contribution was reported by
WASA-at-COSY~\cite{Adlarson:2016wkw}. Recently high-statistics results from BESIII were released~\cite{Ablikim:2018yen},
with accuracy allowing one to test assumptions in the dispersive calculations~\cite{Niecknig:2012sj,Danilkin:2014cra}. 
New data on the Primakoff-type reaction $\gamma\pi\to \pi\pi$ will further help to improve the predictions of the dispersive frameworks. Additional data on the Dalitz decays $\omega\to\pi^0l^+l^-$ and $\phi\to\pi^0l^+l^-$ can be used to cross-check the calculations.

As experimental input to the calculations of the pion loop contribution to HLbL in $(g-2)_\mu$, information on the partial waves in $\gamma^*\gamma^* \to \pi\pi$ are required at arbitrary values of momenta transfer. The only result with a single photon off-shell is for the $\pi^0\pi^0$ system from the Belle collaboration~\cite{Masuda:2015yoh} with $Q^2>3.0\,\textrm{GeV}^2$. The BESIII collaboration announced a measurement of the $\pi^+\pi^-$ system at $0.3\,\textrm{GeV}^2\leq Q^2 \leq 3.0\,\textrm{GeV}^2$, covering invariant masses from the two pion threshold up to 2.0\,GeV and the full range of the pion helicity angle~\cite{Guo:2019gjf}. Similar studies are considered for the neutral two-meson final states $\pi^0\pi^0, \pi^0\eta$, and $\eta\eta$.

We summarize our recommendations for intensified experimental activities to improve on HLbL in \cref{tab:wishlist}.

\subsubsection{Radiative corrections and Monte Carlo event generators}\label{sec:expmc_radcor}
With the error on $(g-2)_\mu$ coming from the HVP and the HLbL contributions almost at the same level~\cite{Davier:2017zfy,Keshavarzi:2018mgv,Davier:2019can,Keshavarzi:2019abf}, the relevance of the error reduction in the HLbL is as important as the error reduction in HVP. As a consequence, all possible sources of the error have to be scrutinized once more. The most important information on these contributions comes from experimental data on  $\gamma^{(*)}\gamma^{(*)}\to{\rm hadrons}$ amplitudes, where one of the sources of the systematic error is the accuracy of the Monte Carlo event generators. The most important one for the evaluation of the $(g-2)_\mu$  are the $\gamma^{(*)}\gamma^{(*)}\to{\rm pseudoscalar(s)}$ amplitudes. In the spacelike region they were reported in Refs.~\cite{Behrend:1990sr,Gronberg:1997fj,BABAR:2011ad,Aubert:2009mc,Uehara:2012ag}. A strong tension was found between the BABAR data~\cite{Aubert:2009mc} and other measurements~\cite{Behrend:1990sr,Gronberg:1997fj,Uehara:2012ag} of the pion TFF.

For the recent experimental measurements two event generators were used: TREPSPST~\cite{Uehara:2013dna,Uehara:2012ag} in the BELLE analysis and  GGRESRC~\cite{Druzhinin:2014sba} in the BABAR analysis. Both of them include radiative corrections by means of a structure function method. 
Recently the event generator EKHARA~\cite{Czyz:2006dm,Czyz:2010sp} was upgraded~\cite{Czyz:2018jpp} and the QED corrections were included exactly into the code. The EKHARA code predicts much bigger, up to 20\%, radiative corrections for the single tag event selection of BABAR experiment as compared to the GGRESRC event generator prediction, which is of order of 1\%.
If the predictions by the EKHARA generator are correct, reanalyses of the experimental data are necessary. It is not straightforward to get information on how this large change in the radiative corrections influences the measurements, as it affects not only the radiative correction factor, but also the experimental efficiencies; and to obtain the experimental efficiencies the detector simulations are necessary.
Ignoring the effect on the efficiencies and taking into account only the radiative correction factor, the pion TFF extracted from the BABAR data is about 20\% higher. 
However, as can be seen from \cref{eq:amuPpole} and \cref{fig:pi0-etap-weights}, the weighting factors $w_{1,2}$ dampen the influence of the BABAR data to $\amu{\pi^0}$, as they are only available at $Q^2>4\,\textrm{GeV}^2$. Nevertheless, these considerations strongly suggest that reanalyses of the experimental data and a confirmation of the results of Ref.~\cite{Czyz:2018jpp} are necessary, as the radiative corrections might affect the measured form factors at a scale relevant for the $(g-2)_\mu$ predictions.

\subsection{Contribution of the pion pole and other pseudoscalar poles}
\label{sec:pion-pole}

\begin{figure}[t]
    \centering
    \includegraphics[width=0.8\textwidth]{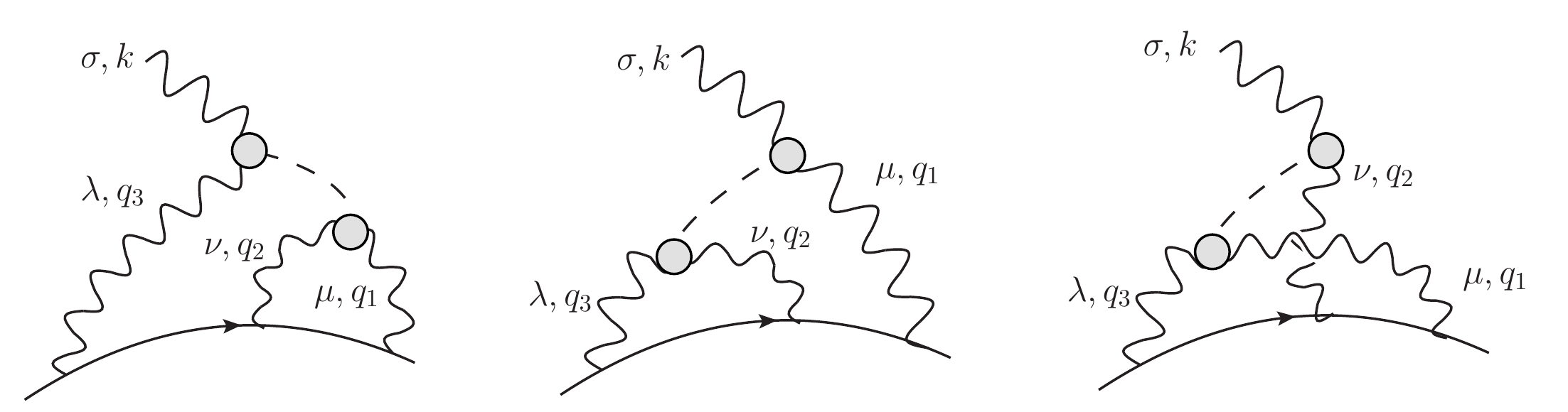}
    \caption{The pseudoscalar-pole contribution: the dashed lines stand for the pseudoscalar meson, while the blobs can be unambiguously related to the TFFs. Reprinted from Ref.~\cite{Masjuan:2017tvw}.}
    \label{fig:PSpoles}
\end{figure}

The pseudoscalar-pole contributions (see \cref{fig:PSpoles}) require the on-shell pseudoscalar TFFs in the spacelike region as the only input and represent the largest individual contributions to HLbL: typically, $a_{\mu}^{P\textrm{-pole}}\sim 100\times 10^{-11}$, which is similar in size to the whole HLbL. As a consequence, and accounting for the expected experimental uncertainty $(15\times 10^{-11})$, these should be understood below the $10\%$ level. Still, given the relevant scales, the $\pi^0$ represents the major contribution, yet the currently sought precision requires also a careful evaluation of the $\eta$ and $\eta'$. 

Given their prominent role, these contributions have been widely explored in the past, starting from the pioneering work in Refs.~\cite{Bijnens:1995cc,Bijnens:1995xf,Bijnens:2001cq,Hayakawa:1995ps,Hayakawa:1996ki,Hayakawa:1997rq} based on hadronic effective Lagrangians at a time when experimental data were scarce, implying potentially large uncertainties. 
As a result, it was necessary to use more phenomenological descriptions for the pseudoscalar TFFs, based on vector-meson-dominance (VMD) ideas and guided by the few existing experimental data~\cite{Bijnens:1995xf,Hayakawa:1997rq}. Later, building on large-$N_c$ ideas and new data, the inclusion of additional resonances allowed the authors to satisfy (certain) known low- and high-energy QCD constraints and to better fit and interpolate the data~\cite{Knecht:2001qf}.
We do not discuss the nonpole (``$\pi^0$-exchange'') contributions or even variants in which one vertex contains a constant form factor~\cite{Melnikov:2003xd,Nyffeler:2009tw,Jegerlehner:2009ry} for the reasons outlined in \cref{sec:DefinitionOfIndividualContributionsToHLbL}.

With the advent of the new generation of $(g-2)_\mu$ experiments, systematic uncertainties of such approaches (related to the finite number of resonances and the large-$N_c$ limit), previously irrelevant, must be improved upon far beyond the typical $30\%$ estimates. Consequently, the phenomenological determinations must be model-independent and data-driven to as large an extent as possible, making use of all experimental data on the corresponding TFFs in order to achieve a new standard of precision, and also to provide a competitive cross-check on the lattice calculation in \cref{sec:pion pole}.
In the following, we review what we believe are the most up-to-date evaluations of the pseudoscalar-pole contributions in the literature, with a special emphasis on the $\pi^0$.  
In particular, we demand that three criteria be met:
\begin{enumerate}
\item in addition to the TFF normalization given by the real-photon decay widths, also high-energy constraints must be fulfilled; 
\item at least the spacelike experimental data for the singly-virtual TFF must be reproduced;
\item systematic uncertainties must be assessed with a reasonable procedure.
\end{enumerate}
We distinguish two different strategies fulfilling these criteria: the dispersive one, which could in principle reconstruct the TFF from completely unrelated data based on analyticity constraints; and the one based in the mathematical framework of Pad\'e approximants along with experimental data in the spacelike (and low-energy timelike) region. As both are based on very different approaches, the numerical agreement that is found between the two, and also with the lattice determination in \cref{sec:pion pole}, gives us further confidence in the reliability of the $\pi^0$-pole contribution thus determined. In addition, we also comment on recent progress in other approaches. Finally, we summarize the status of the $\eta$ and $\eta'$ contributions.

\subsubsection{Definitions, asymptotic constraints}
\label{sec:asymptotic_constraint_PS}

\begin{figure}
    \centering
    \includegraphics[width=0.45\textwidth]{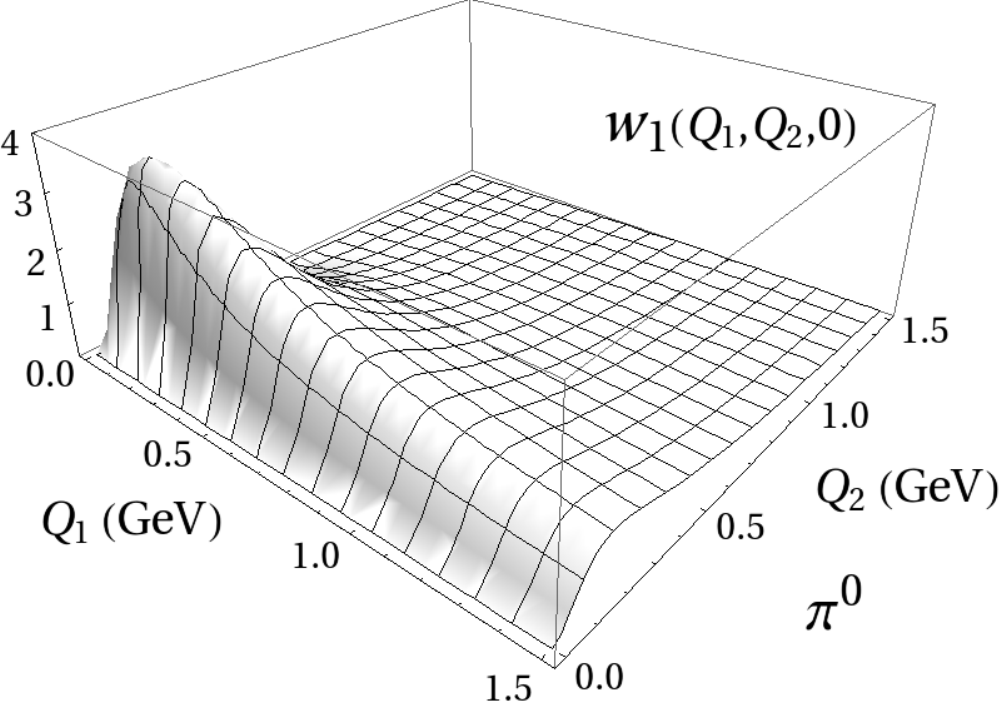}
    \includegraphics[width=0.45\textwidth]{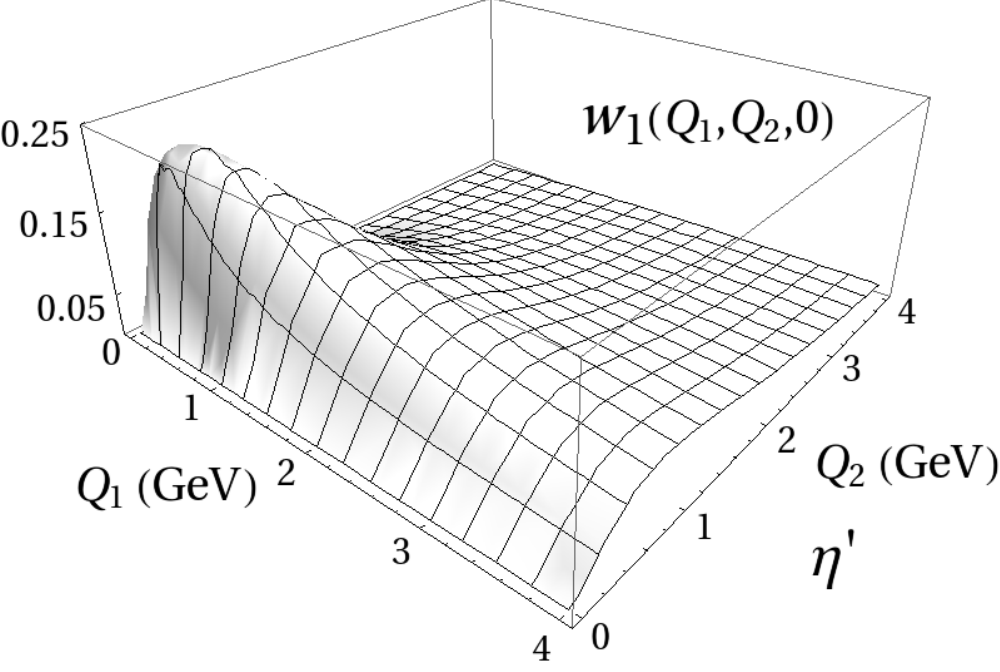}
    \caption{Weight function $w_1(Q_1,Q_2,0)$ for $\pi^0$ (left) and $\eta'$ (right); cf.\ \cref{eq:amuPpole}. Reprinted from Ref.~\cite{Masjuan:2017tvw}.}
    \label{fig:pi0-etap-weights}
\end{figure}

The pseudoscalar-pole contributions are given according to
\begin{align}\label{eq:amuPpole}
  a_{\mu}^{P\textrm{-pole}} = \left(\frac{\alpha}{\pi}\right)^3\int dQ_1 dQ_2 d\tau\Big[&w_1(Q_1,Q_2,\tau) F_{P\gamma^*\gamma^*}(-Q_1^2,-Q_3^2)F_{P\gamma^*\gamma^*}(-Q_2^2,0)  \notag\\
     &+w_2(Q_1,Q_2,\tau) F_{P\gamma^*\gamma^*}(-Q_1^2,-Q_2^2)F_{P\gamma^*\gamma^*}(-Q_3^2,0)\Big]\,,
\end{align}
where $Q_3^2\equiv Q_1^2+Q_2^2+2\tau Q_1Q_2$. 
The explicit form of the weight functions $w_{1/2}(Q_1,Q_2,\tau)$ can be found in the literature~\cite{Jegerlehner:2009ry,Nyffeler:2016gnb,Masjuan:2017tvw,Hoferichter:2018kwz}. A numerical evaluation of $w_1(Q_1,Q_2,0)$ is shown in \cref{fig:pi0-etap-weights}: their most important property is the fact that they are peaked at low energies, for the $\pi^0$ in the range $Q_i < 1\,\text{GeV}$.
The TFF appearing above is defined as
\begin{equation}
    i \int  d^4x\, e^{iq_1 \cdot x} \langle 0 | T\{ j_{\mu}(x)j_{\nu}(0) \}  | P(q_1+q_2) \rangle =  \epsilon_{\mu\nu\rho\sigma}q_1^{\rho}q_2^{\sigma} F_{P\gamma^*\gamma^*}(q_1^2,q_2^2)\,,
\end{equation}
where $j_{\mu}(x)=\bar{q}(x)\mathcal{Q}\gamma_{\mu}q(x)$, with $\mathcal{Q}=\textrm{diag}(2,-1,-1)/3$, and $\epsilon^{0123}=+1$.
For real photons and in the chiral limit $m_{u,d,s}\to0$ (supplemented by the large-$N_c$ limit for the $\eta'$, so that the latter remains massless), it is related to the anomaly~\cite{Adler:1969gk,Bell:1969ts,Bardeen:1969md}, 
\begin{equation}
    \sum_P F_P^a F_{P\gamma^*\gamma^*}(0,0) =\frac{3}{4\pi^2 }\Tr(\mathcal{Q}^2\lambda^a)\,,
\end{equation}
where $a=0,\ldots,8$ is the corresponding flavor index  associated to the Gell-Mann matrices $\lambda^a$, extended to include $\lambda^0\equiv\sqrt{2/3}\operatorname{diag}(1,1,1)$, and $\langle 0| j_{5\mu}^a |P\rangle \equiv ip_{\mu} F_P^a$ with $j_{5\mu}^a = \bar{q}\gamma_{\mu}\gamma_5\frac{\lambda^a}{2}q$. Away from the chiral limit, corrections arise and $\eta$--$\eta'$ mixing must be accounted for, see Refs.~\cite{Feldmann:1999uf,Escribano:2015yup} and references therein.
The high-energy behavior can be obtained by expanding the product of electromagnetic currents on the light-cone, obtaining at leading order in pQCD and at leading-twist~\cite{Lepage:1979zb,Lepage:1980fj}
\begin{equation}\label{eq:TFF-OPE}
    F_{P\gamma^*\gamma^*}(-Q_1^2,-Q_2^2) = \sum_a 2\Tr(\mathcal{Q}^2\lambda^a)F_P^a \int_0^1 dx \frac{\phi_P^a(x)}{x Q_1^2 +(1-x)Q_2^2}\,.
\end{equation}
Higher-order corrections in pQCD have been derived as well~\cite{delAguila:1981nk,Braaten:1982yp}. Since  for large momenta $\phi_P^a(x)\to 6x(1-x)$~\cite{Lepage:1980fj,Braun:2003rp}, the following limits can be inferred
\begin{align}
    \lim_{Q^2\to\infty}Q^2F_{P\gamma^*\gamma^*}(-Q^2,0) &{}= \sum_a6\Tr(\mathcal{Q}^2\lambda^a)F_P^a\left[1-\delta^{a0}\frac{2N_f}{\pi\beta_0}\alpha_s(\mu_0)\right]\,,\\
    \lim_{Q^2\to\infty}Q^2F_{P\gamma^*\gamma^*}(-Q^2,-Q^2) &{}= \sum_a2\Tr(\mathcal{Q}^2\lambda^a)F_P^a\left[1-\delta^{a0}\frac{2N_f}{\pi\beta_0}\alpha_s(\mu_0)\right]\,,
\label{eq:TFFOPE}
\end{align}
where we include $\beta_0\equiv 11N_c/3 -2N_f/3$, with $N_f$ the number of effective active flavors.
The first limit is commonly known as the Brodsky--Lepage (BL) limit~\cite{Lepage:1979zb,Lepage:1980fj}, while the latter can be rigorously obtained from the operator product expansion (OPE)~\cite{Nesterenko:1982dn,Novikov:1983jt,Gorsky:1987mu,Manohar:1990hu}. The $\eta$ and $\eta'$ cases receive important $\alpha_s$ corrections due to the anomalous dimension of the singlet axial current~\cite{Leutwyler:1997yr}, which have been accounted for by the last factor~\cite{Agaev:2014wna,Escribano:2015yup,Alte:2015dpo}. Finally, higher-order corrections have been calculated using the OPE, which, for the $\pi^0$, multiply \cref{eq:TFFOPE} by $(1-\frac{8}{9}\frac{\delta^2}{Q^2})$, with the estimate $\delta^2 = 0.20(2)\,\text{GeV}^2$ determined from sum rules~\cite{Novikov:1983jt} already used in Refs.~\cite{Melnikov:2003xd,Nyffeler:2009tw,Jegerlehner:2009ry} and also supported by lattice results~\cite{Bali:2018spj,Gerardin:2019vio}.

\subsubsection{The pion pole in a dispersive approach}\label{sec:pi0DR}

The central idea behind the dispersive analysis of the $\pi^0$ TFF~\cite{Hoferichter:2014vra,Hoferichter:2018dmo,Hoferichter:2018kwz} is to reconstruct this object from its dominant low-energy singularities.  As \cref{fig:pi0-etap-weights} (left) demonstrates, the main weight for the HLbL integration in \cref{eq:amuPpole} lies in the region of $Q_i < 1\,\text{GeV}$; in this range, where a precise and reliable theoretical description is therefore of prime importance, the intermediate states dominating the discontinuities in the two form factor virtualities are given by two- and three-pion intermediate states.  In particular, these discontinuities can be reconstructed from data on $e^+e^-\to 2\pi,\,3\pi$ and automatically contain the effects of the lowest-lying resonances in these channels, the $\rho(770)$, $\omega(782)$, and $\phi(1020)$, in a model-independent way.  Beyond this dominant part constructed rigorously from dispersion theory, two further pieces are added in order to fulfill all asymptotic constraints described in the previous section: an effective pole that parameterizes heavier intermediate states; and an asymptotic contribution constructed on the basis of a pion distribution amplitude.  Altogether, the TFF is therefore written as~\cite{Hoferichter:2018dmo,Hoferichter:2018kwz}
\begin{equation}\label{eq:disp-decomp}
F_{\pi^0\gamma^*\gamma^*}=F_{\pi^0\gamma^*\gamma^*}^\text{disp}+F_{\pi^0\gamma^*\gamma^*}^\text{eff}+F_{\pi^0\gamma^*\gamma^*}^\text{asym} \,.
\end{equation}

For the dispersive part, it is useful to decompose the TFF according to the photons' isovector ($v$) and isoscalar ($s$) components as
\begin{equation}
F^\text{disp}_{\pi^0\gamma^*\gamma^*}(q_1^2,q_2^2)=F^\text{disp}_{vs}(q_1^2,q_2^2)+F^\text{disp}_{vs}(q_2^2,q_1^2)\,. 
\end{equation}
This function obeys a double-spectral representation
\begin{align}
\label{eq:low-energy}
F_{vs}^\text{disp}(-Q_1^2,-Q_2^2)&=
\frac{1}{\pi^2} \int_{4M_\pi^2}^{s_\text{iv}}  dx \int_{s_\text{thr}}^{s_\text{is}}  \frac{d y  \, \rho(x,y)}{\big(x+Q_1^2\big)\big(y+Q_2^2\big)},\notag\\
\rho(x,y)&=\frac{q_\pi^3(x)}{12\pi\sqrt{x}}\Im \Big[\big(F_\pi^{V}(x)\big)^*f_1(x,y)\Big]\,,
\end{align}
where $q_\pi(s)=\sqrt{s/4-M_\pi^2}$, $F_\pi^{V}(s)$ is the electromagnetic form factor of the pion, and $f_1(s,q^2)$ the partial-wave amplitude for $\gamma^*(q)\pi\to\pi\pi$. 
The onset of the isoscalar discontinuity is $s_\text{thr} = 9M_\pi^2$ in the absence of electromagnetic effects, while taking into account the significant decay $\omega\to\pi^0\gamma$ lowers it to $M_{\pi^0}^2$. 
$s_{\text{iv}/\text{is}}$ represent isovector and isoscalar cutoffs.  
$F_\pi^{V}(s)$ is described in terms of an Omn\`es representation~\cite{Omnes:1958hv} based on a variety of inputs for the $\pi\pi$ $P$-wave phase shift and fit to data on $\tau^- \to \pi^-\pi^0\nu_\tau$~\cite{Fujikawa:2008ma}.  The amplitude $f_1(s,q^2)$ is constructed based on solutions of Khuri--Treiman equations~\cite{Khuri:1960zz}, with a normalization function $a(q^2)$ that needs to be adjusted to $e^+e^-\to3\pi$ data.  At the real-photon point, $f_1(s,q^2=0)$ can be tested experimentally in the reaction $\gamma\pi\to\pi\pi$~\cite{Hoferichter:2012pm,Hoferichter:2017ftn}, while Dalitz plot distributions on $\omega\to3\pi$~\cite{Adlarson:2016wkw,Ablikim:2018yen} and $\phi\to3\pi$~\cite{Aloisio:2003ur,Akhmetshin:2006sc} probe it on the narrow isoscalar vector resonances~\cite{Niecknig:2012sj}.  For general $q^2$, a representation of $a(q^2)$ with good analytic properties~\cite{Hoferichter:2018kwz} is fit to $e^+e^-\to3\pi$ cross section data by SND~\cite{Achasov:2002ud,Achasov:2003ir} and BABAR~\cite{Aubert:2004kj} up to $q^2 = (1.8\,\text{GeV})^2$.  
In particular, a single-variable dispersion relation yields a prediction for the timelike singly-virtual $\pi^0$ TFF~\cite{Schneider:2012ez,Hoferichter:2014vra,Hoferichter:2018kwz} that is in very good agreement with precise data~\cite{Achasov:2000zd,Achasov:2003ed,Akhmetshin:2004gw,Achasov:2016bfr}: with its correct analytic properties, the dispersive TFF representation links the timelike and the spacelike form factor seamlessly, such that timelike data helps constrain the spacelike low-energy region where data is still relatively scarce.  Even more importantly, the dispersive formulation as in \cref{eq:low-energy} fixes the doubly-virtual TFF, for which no data at all is available as yet, from singly-virtual input.
 
To account for the asymptotic behavior of the TFF in doubly-virtual kinematics, we realize that \cref{eq:TFF-OPE} can be written in a double-spectral form akin to \cref{eq:low-energy}, with a formal asymptotic double-spectral function
\begin{equation}\label{eq:rhoasym}
\rho^\text{asym}(x,y)=-2\pi^2F_\pi x y\delta''(x-y)\,,  
\end{equation}
where $\delta''(.)$ denotes the second derivative of a delta function.
Inserting \cref{eq:rhoasym} into the double-spectral representation \cref{eq:low-energy} and integrating over all $x$ and $y$ reproduces the pQCD form of \cref{eq:TFF-OPE} for the asymptotic distribution amplitude.
Restraining however the support for this contribution to energies above a lower matching point $s_m$, the asymptotic TFF contribution becomes 
\begin{equation}
\label{eq:asym}
F_{\pi^0\gamma^*\gamma^*}^\text{asym}(q_1^2,q_2^2)
 = 2F_\pi\int_{s_m}^\infty d x \frac{q_1^2q_2^2}{(x-q_1^2)^2(x-q_2^2)^2}\,,
\end{equation}
which does not contribute for singly-virtual kinematics, but ensures the asymptotic behavior \cref{eq:TFF-OPE} otherwise.  Finally, the addition of an effective pole 
\begin{equation}
F_{\pi^0\gamma^*\gamma^*}^\text{eff}(q_1^2,q_2^2)=\frac{g_{\text{eff}}}{4\pi^2F_\pi}\frac{M_{\text{eff}}^4}{(M_{\text{eff}}^2-q_1^2)(M_{\text{eff}}^2-q_2^2)}
\end{equation}
allows the combined representation to fulfill the TFF normalization exactly via an effective coupling $g_{\text{eff}}$ around 10\%, and an effective mass $M_{\text{eff}}$ as the only free parameter that is fit to the high-energy singly-virtual spacelike data~\cite{Behrend:1990sr,Gronberg:1997fj,Aubert:2009mc,Uehara:2012ag}.  $M_{\text{eff}}$ turns out to be of the order of $1.5$--$2\,\text{GeV}$, consistent with the interpretation as an effective pole that summarizes the corrections due to higher energies and higher intermediate states.

Altogether, the TFF representation \cref{eq:disp-decomp} leads to the $\pi^0$-pole contribution 
\begin{equation}
\label{eq:result_final}
 a_\mu^{\pi^0\text{-pole}}=63.0(0.9)_{F_{\pi\gamma\gamma}}(1.1)_\text{disp}(^{2.2}_{1.4})_\text{BL}(0.6)_\text{asym}\times 10^{-11}
 =63.0^{+2.7}_{-2.1}\times 10^{-11}\,.
\end{equation}
The individual errors reflect the experimental uncertainty in the TFF normalization, in the dispersive representation (due to phase-shift input and the variation of cutoffs), the fit to high-energy singly-virtual data approaching the Brodsky--Lepage limit, and the onset of the asymptotic contribution in \cref{eq:asym}.  
This has been updated compared to the published result, 
$a_\mu^{\pi^0\text{-pole}}=62.6(1.7)_{F_{\pi\gamma\gamma}}(1.1)_\text{disp}(^{2.2}_{1.4})_\text{BL}(0.5)_\text{asym}\times 10^{-11}
 =62.6^{+3.0}_{-2.5}\times 10^{-11}$~\cite{Hoferichter:2018dmo,Hoferichter:2018kwz}\,,
by adapting both central value and uncertainty of the TFF normalization to the final PrimEx result~\cite{Larin:2020}.
The decomposition of the uncertainty in \cref{eq:result_final} therefore paves the way to further scrutinizing by means of dedicated experimental measurements, requiring (at most) singly-virtual input only. Obviously, however, doubly-virtual input can be of help as well.

\subsubsection{Pion pole: Pad\'e and Canterbury approximants}\label{sec:pi0PA}

Both success and limitations of resonance saturation approaches, despite expected $30\%$ large-$N_c$ corrections, were understood and reinterpreted from a purely mathematical point of view in the context of Pad\'e approximants (PAs)~\cite{BakerMorris,Baker} in Refs.~\cite{Peris:2006ds,Golterman:2006gv,Masjuan:2007ay,Masjuan:2008fr,Masjuan:2009wy}, and employed to describe the pseudoscalar TFFs in Refs.~\cite{Masjuan:2012wy,Escribano:2013kba,Escribano:2015nra,Escribano:2015yup}. 
PAs are rational functions that, under certain conditions, can be rigorously applied to describe QCD Green's functions---in contrast to VMD models, without the restriction to the large-$N_c$ limit---provided one abandons the idea to interpret the poles therein in terms of physical states, and reinterprets the parameters in terms of the underlying function's series expansion. Specifically, to have an improved convergence around the region of interest $Q^2\approx 0$, one imposes the matching 
\begin{equation}\label{eq:PAdef}
  P^N_M(Q^2) = \frac{\sum_i^N a_i Q^{2i}}{1+\sum_j^M b_j Q^{2j}} = 
  F_{P\gamma^*\gamma^*}(0,0)\left[1 + b_P \frac{Q^2}{M_P^2} + c_P\frac{Q^4}{M_P^4} + \ldots  + \mathcal{O}\!\left(Q^{2(N+M+1)}\right) \right]\,,    
\end{equation}
where $b_P,\, c_P,\,\ldots$ stand for the coefficients of the (singly-virtual) TFF series expansion at low energies. With such a construction, convergence theorems exist for meromorphic functions~\cite{BakerMorris,Masjuan:2007ay} 
and, more remarkably, for Stieltjes functions (this essentially means that the spectral function is positive---a textbook example is the HVP~\cite{Peris:2006ds,Masjuan:2009wy}) 
 that apply to $P^{N+J}_N$ ($J \geq-1$) sequences in the whole complex plane except on the branch cuts. Further, $P^N_{N+1}$ and $P^N_N$ sequences provide, respectively, lower and upper bounds to the approximated function. In addition, the high-energy behavior can also be incorporated~\cite{Queralt:2010sv}. As an important point, then, the convergence of a sequence can help to estimate the sought systematic uncertainty as outlined in Refs.~\cite{Masjuan:2008fr,Masjuan:2017tvw}.

For the TFFs at hand, their analytic structure is a priori unknown, which impedes using any theorem in a straightforward way. However, beyond the large-$N_c$ limit and in line with the previous section, the most relevant aspects at low energies are elastic $\pi\pi$ rescattering along with the narrow $\omega,\phi$ resonances. For such features, the former being Stieltjes once subleading left-hand cuts are disregarded (see the discussion in Ref.~\cite{Kubis:2015sga}), and the $\omega,\phi$ resonances essentially meromorphic (in their zero-width approximation), one recovers a so-called meromorphic function of Stieltjes type, for which convergence with Pad\'e approximants is guaranteed~\cite{BakerMorris,Masjuan:2008cp,Queralt:2010sv}; in particular $P^{N+J}_{N}$ ($J\geq-1$) sequences converge. As such, Pad\'e theory can be straightforwardly applied to the $\pi^0$ TFF with anticipated success---with systematics expected of the order of the subleading effects, well beyond the precision obtained in this framework with the current data, and completely different from other approaches.

The generalization to the doubly-virtual case involves an extension of PAs, Canterbury approximants (CAs)~\cite{Baker,Chisholm:1973,Chisholm:1974,Jones:1976,Masjuan:2015lca,Sanchez-Puertas:2017sih}, whose definition parallels that of PAs and extends the notion of convergence to meromorphic and Stieltjes functions~\cite{Alabiso:1974vk}---previous comments about convergence to the TFF also apply here~\cite{Masjuan:2017tvw}. For simplicity, only the lowest-lying elements within the $C^N_{N+1}$ sequence of approximants employed in determining the pseudoscalar pole are given here~\cite{Masjuan:2017tvw}, 
\begin{align}
 C^0_1(Q_1^2,Q_2^2) &= \frac{F_{P\gamma^*\gamma^*}(0,0)}{1+\frac{b_P}{M_P^2}(Q_1^2+Q_2^2)}, \notag\\
 C^1_2(Q_1^2,Q_2^2) &= \frac{F_{P\gamma^*\gamma^*}(0,0)(1+\alpha_{1}(Q_1^2+Q_2^2) + \alpha_{1,1}Q_1^2Q_2^2)}{1+\beta_{1}(Q_1^2+Q_2^2)+\beta_{2}(Q_1^{4}+Q_2^{4})+\beta_{1,1}Q_1^2Q_2^2+\beta_{2,1}Q_1^2Q_2^2(Q_1^2+Q_2^2)}\,. \label{eq:TFF-CA}
\end{align}
For the $C^0_1$ approximant, all parameters are fixed from the $P\to\gamma\gamma$ decays and the slope at $Q^2=0$ that, together with additional terms in the series expansion, were determined from a fit to data~\cite{Tanabashi:2018oca,Behrend:1990sr,Gronberg:1997fj,Aubert:2009mc,Uehara:2012ag,Adlarson:2016ykr,TheNA62:2016fhr} using PAs~\cite{Masjuan:2009wy,Masjuan:2017tvw}. This leads to a first estimate $\amu{\pi^0}=65.3(2.8)\times10^{-11}$~\cite{Masjuan:2017tvw}. 
For the second case, the singly-virtual parameters can be fixed from the curvature and the BL limit. Regarding doubly-virtual parameters, the OPE limit (including the $\delta^2$ term) is imposed to supply the necessary constraints in the absence of data. This fixes all but one single parameter, which can be related to a term $a_{P;1,1}Q_1^2 Q_2^2$ in the low-energy series expansion and is left free within a range $(a_{P;1,1}^{\textrm{min}}\leq a_{P;1,1} \leq a_{P;1,1}^{\textrm{max}})$ that avoids spacelike poles and results in the band $63.0(1.2)\times10^{-11}\leq \amu{\pi^0} \leq64.1(1.3)\times10^{-11}$, that can be easily improved with the advent of doubly-virtual data. The systematic uncertainties are obtained from the difference with respect to the $C^0_1$ estimate, which proves to be a conservative estimate for a set of toy models~\cite{Masjuan:2017tvw}. The final result reads
\begin{equation}\label{eq:pi0-PA}
  \amu{\pi^0}=63.6(1.3)_{\textrm{stat}}(0.6)_{a_{P;1,1}}(2.3)_{\textrm{sys}}\times10^{-11} \to  63.6(2.7)\times10^{-11}\,.    
\end{equation}
The first error arises from singly-virtual parameters alone, including the normalization and the slope; the second is connected to doubly-virtual uncertainties, connected to $a_{P;1,1}$; the third is the systematic uncertainty inherent to the sequence truncation. The error reduction with respect to the $C^0_1$ estimate is related to the $C^1_2$ approximant complying exactly with the pQCD predictions in contrast to the $C^0_1$ one, which is also partly responsible for the difference in the central values.
This approach will benefit from forthcoming BESIII results (see \cref{sec:PS_TFF_exp}) that should allow us to improve and increase the number of parameters obtained for the series expansion \cref{eq:PAdef}. 
Further, the prospects for measuring doubly-virtual data at BESIII, see \cref{sec:PS_TFF_exp}, will help in constraining doubly-virtual parameters, such as $a_{P;1,1}$. Also, it will benefit from new $\pi^0\to\gamma\gamma$ measurements. In particular, preliminary studies show that latest PrimEx result~\cite{Larin:2020} shifts the central value up to $64.8\times10^{-11}$, with similar uncertainties, yet this has to be taken with caution until a full reanalysis is carried out.

\subsubsection{Pion pole: other approaches}

Several other approaches to the pion-pole contribution have been pursued, which however do not fulfill all criteria laid out above.  

The $\pi^0$ TFF has been computed using Dyson--Schwinger 
equations~\cite{Goecke:2010if,Raya:2015gva,Eichmann:2017wil,Weil:2017knt,Eichmann:2019tjk,Raya:2019dnh}, i.e., in a microscopic scheme that attempts to solve QCD in terms of quark and gluon degrees of freedom, introducing certain approximations on the way. Both Refs.~\cite{Eichmann:2017wil,Raya:2019dnh} reproduce the spacelike TFF data and the OPE-like behavior becomes obvious already at moderate energies. 
In contrast, Ref.~\cite{Eichmann:2017wil} shows significant deviations from the BL limit at higher energies---irrelevant for $(g-2)_\mu$---, while Refs.~\cite{Raya:2015gva,Raya:2019dnh} find only small corrections at moderate energies due to a nonasymptotic nature of the distribution amplitude also advocated in other approaches~\cite{Chang:2013pq,Raya:2015gva,Bali:2019dqc}, with the BL limit actually restored at higher energies. 
The authors obtain $\amu{\pi^0}=62.6(1.3)\times10^{-11}$~\cite{Eichmann:2019tjk} and $\amu{\pi^0}=61.4(2.1)\times10^{-11}$~\cite{Raya:2019dnh}, in perfect agreement with \cref{eq:result_final} as well as \cref{eq:pi0-PA}, but systematic uncertainties due to truncations in the system of coupled integral equations are hard to gauge.

In addition, holographic models of (large-$N_c$) QCD have been used to
determine
$a_{\mu}^{\pi^0\textrm{-pole}}$~\cite{Hong:2009zw,Cappiello:2010uy,Leutgeb:2019zpq}.
 In particular, Ref.~\cite{Leutgeb:2019zpq} is the first one to use the
full holographic result without any further approximations, obtaining
the range $a_{\mu}^{\pi^0\textrm{-pole}} = 59(2)\times10^{-11}$ based on
different models.  Among them, the only model exactly reproducing the
pQCD constraints leads to $a_{\mu}^{\pi^0\textrm{-pole}} =
61\times10^{-11}$, in perfect agreement with \cref{eq:result_final} and \cref{eq:pi0-PA}, but with systematic errors
due to model dependence and large-$N_c$ corrections that are hard to
estimate.

Furthermore, there are various effective-Lagrangian models (sometimes dubbed resonance chiral theory)~\cite{Czyz:2017veo,Guevara:2018rhj} that reproduce a large variety of space- and timelike data~\cite{Czyz:2017veo}, yet commonly fail to incorporate all the high-energy constraints, in particular the OPE limit~\cite{Czyz:2017veo,Guevara:2018rhj} and the BL one for a second small but nonzero virtuality. 
Reference~\cite{Guevara:2018rhj} also investigates the inclusion of loop, $1/N_c$, and heavier vector resonance corrections. 
Recently, a modification of the BL interpolation formula~\cite{Brodsky:1981rp} to the doubly-virtual region has been proposed~\cite{Danilkin:2019mhd}, motivated by pQCD models accounting for higher-twist effects. The model incorporates the high-energy pQCD scaling and depends on a single parameter $\Lambda$ that is fit to the available spacelike data. Such a one-parameter ansatz then fixes the doubly-virtual behavior relying on the pQCD-based interpolation. Compared to Refs.~\cite{Hoferichter:2018dmo,Hoferichter:2018kwz,Masjuan:2017tvw,Gerardin:2019vio,Eichmann:2019tjk}, it leads to a smaller doubly-virtual form factor in the low $Q^2$ range (by around $20\%$ around $Q^2 = 0.5\,\textrm{GeV}^2$), resulting in a value for $\amu{\pi^0}$ that is around $10\%$ smaller. Systematic uncertainties inherent in VMD-like models and the pQCD-motivated interpolation are hard to estimate.
We note that all modern evaluations to $\amu{\pi^0}$ agree with the early calculations~\cite{Bijnens:1995xf,Bijnens:2001cq,Hayakawa:1997rq,Knecht:2001qf}, given the larger uncertainty estimates of the latter. Additional approaches calculating a nonpole contribution~\cite{Hong:2009zw,Cappiello:2010uy,Bartos:2001pg,Nyffeler:2009tw,Jegerlehner:2009ry,Dorokhov:2008pw,Dorokhov:2011zf,Dorokhov:2012qa,Dorokhov:2015psa,Goecke:2010if,Kampf:2011ty,Masjuan:2012qn,Roig:2014uja} are omitted from the discussion for the reasons discussed in \cref{sec:DefinitionOfIndividualContributionsToHLbL}.

\subsubsection{$\eta$- and $\eta'$-pole contributions}

The $\eta$ and $\eta'$ contributions closely resemble that of the $\pi^0$, with the additional difficulties inherent to the $\eta$--$\eta'$ mixing, their higher masses, and singularities related to their singlet component. These imply theoretical difficulties for reproducing their $\gamma\gamma$ decays and asymptotic behavior, \cref{eq:TFFOPE}, producing marked differences in the literature. \Cref{fig:pi0-etap-weights} (right) also demonstrates that the weight functions in the integral \cref{eq:amuPpole} are such that for the heavier pseudoscalars, larger momentum ranges play a relevant role.

A full dispersive analysis of the $\eta$- and $\eta'$-pole contributions to $a_\mu$ along the lines of the $\pi^0$ analysis described in \cref{sec:pi0DR} has not been completed yet.  This is partly due to the different isospin decomposition, 
\begin{equation}
F_{\eta^{(\prime)}\gamma^*\gamma^*}(q_1^2,q_2^2)=F_{vv^{(\prime)}}(q_1^2,q_2^2)+F_{ss^{(\prime)}}(q_1^2,q_2^2)\,,
\end{equation}
which depends on two different functions and makes the full form factor less amenable to a complete reconstruction from singly-virtual input only.  The dispersive formalism for the singly-virtual $\eta$/$\eta'$ TFF has been established~\cite{Hanhart:2013vba}: while the isoscalar part at low energies can be described in a VMD-type approximation due to the narrowness of the $\omega(782)$ and $\phi(1020)$ resonances, the isovector contribution relies, next to the pion vector form factor, heavily on data for the decays $\eta^{(\prime)}\to\pi^+\pi^-\gamma$~\cite{Adlarson:2011xb,Babusci:2012ft,Ablikim:2017fll}, which show strong deviations from a simple-minded $\rho$-dominance picture~\cite{Stollenwerk:2011zz,Kubis:2015sga,Hanhart:2016pcd}.  First steps towards an investigation of the doubly-virtual isovector contribution have been taken~\cite{Xiao:2015uva}, which analyze the dipion invariant mass distribution in data on $e^+e^-\to\pi^+\pi^-\eta$~\cite{Aubert:2007ef,TheBaBar:2018vvb,Gribanov:2019qgw}.  Ultimately, a construction of the double-spectral representation for $F_{vv^{(\prime)}}(q_1^2,q_2^2)$ will need to proceed based on an amplitude for $\eta^{(\prime)}\to 2(\pi^+\pi^-)$~\cite{Guo:2011ir,Ablikim:2014eoc}, which is still work in progress.  The completion of this low-energy dispersive representation by an effective pole and a pQCD-inspired asymptotic
contribution in analogy to \cref{eq:disp-decomp} should then be straightforward.

\begin{figure}[t]
    \centering
    \includegraphics*[width=0.49\textwidth]{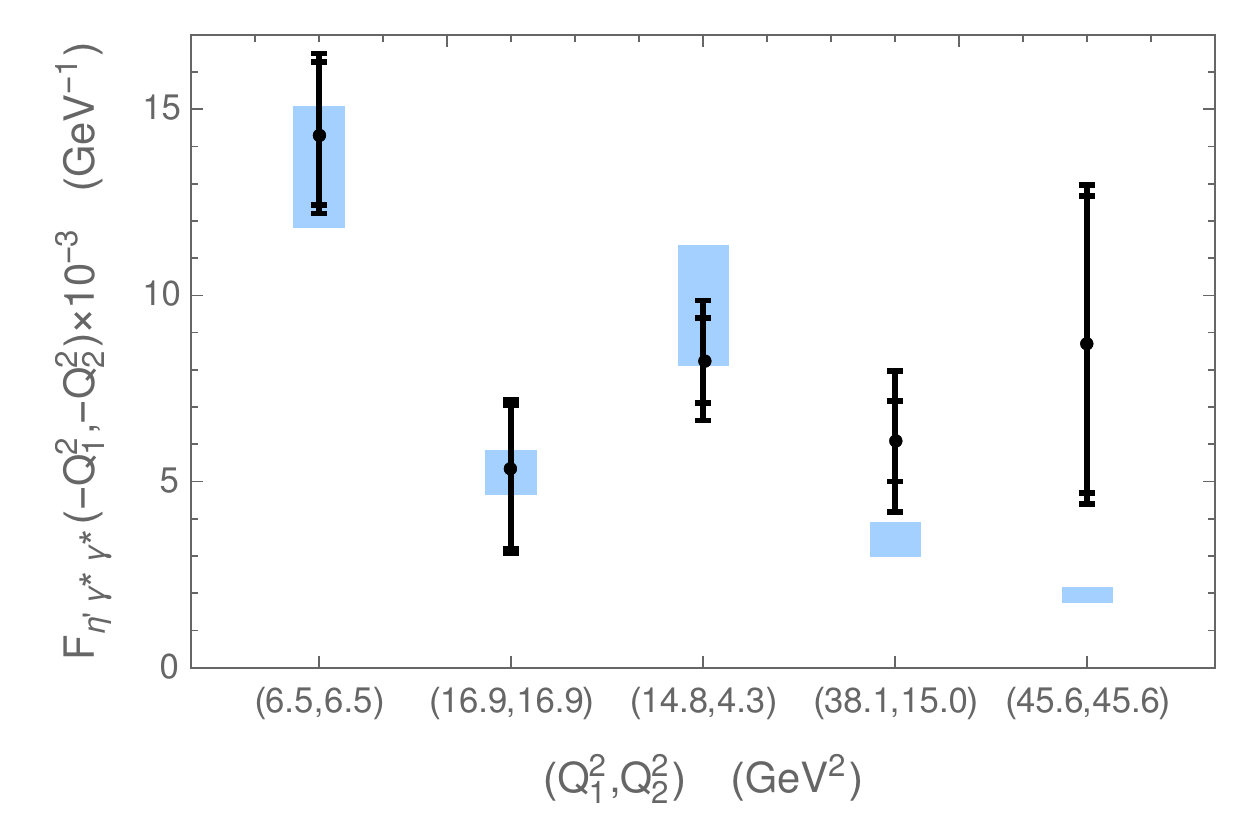} \hfill
    \includegraphics*[width=0.49\textwidth]{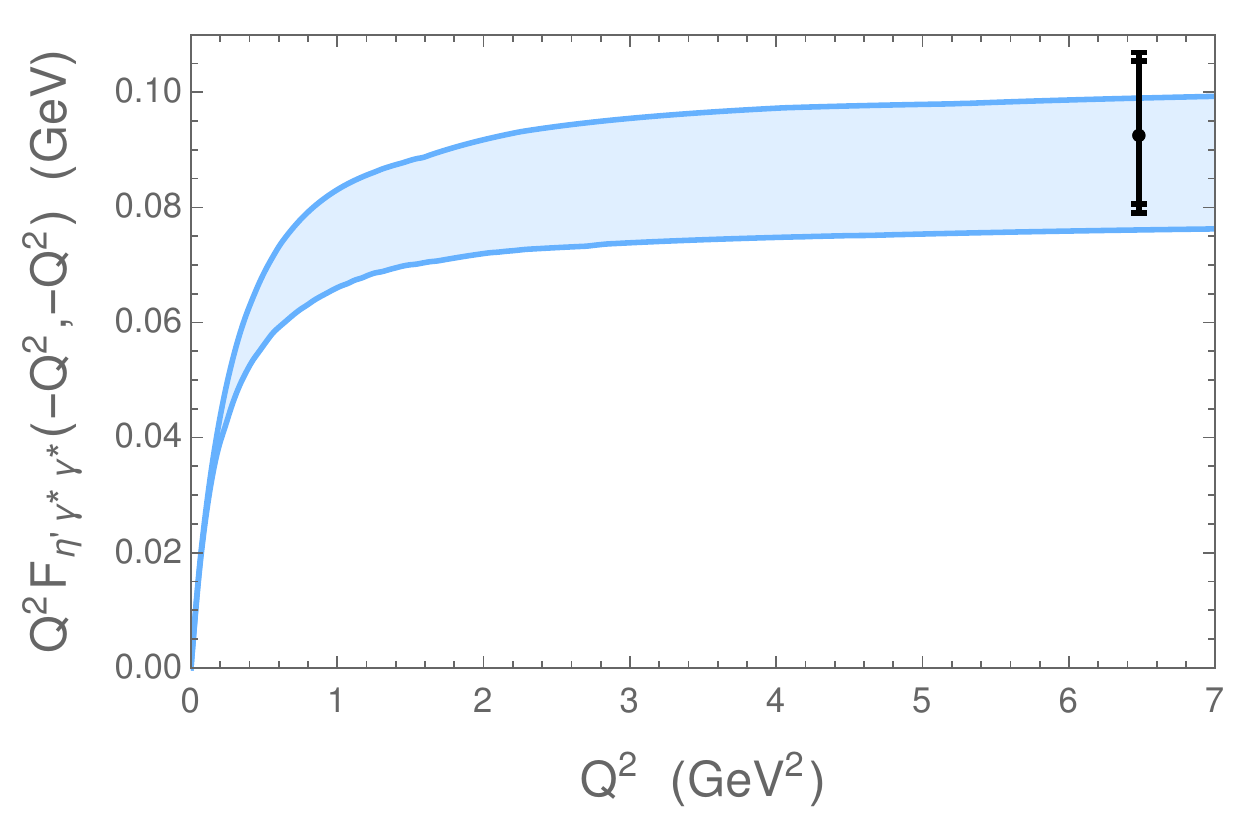}
    \caption{Left: BABAR data points~\cite{BaBar:2018zpn} with statistical errors (inner bars) and statistical and systematic combined (outer bars) in black, together with the CA prediction  including errors (blue bands). Right: The analogous plot for the diagonal $Q^2F_{\eta'\gamma^*\gamma^*}(-Q^2,-Q^2)$ TFF.}
    \label{fig:etapDV}
\end{figure}

The approach of CAs to the $\eta$ and $\eta'$ contributions bypasses several of the complications above arising in most theoretical approaches. The methodology is analogous to the $\pi^0$ with two differences. First, the available data~\cite{Arnaldi:2009aa,Berghauser:2011zz,Aguar-Bartolome:2013vpw,Arnaldi:2016pzu,Adlarson:2016hpp,Acciarri:1997yx,Behrend:1990sr,Gronberg:1997fj,BABAR:2011ad,Ablikim:2015wnx} at lower energies allows one to extract additional terms in the series expansion and avoids the use of the BL limit---the latter can be nevertheless obtained from data and is employed to predict the OPE limit. Second, the OPE parameter $\delta^2$ has not been determined for the $\eta$ or $\eta'$. Instead, an additional $30\%$ uncertainty for $SU(3)_F$-breaking has been assumed~\cite{Masjuan:2017tvw}. The prediction thus obtained can be checked against the recently published BABAR data~\cite{BaBar:2018zpn} for the $\eta'$ doubly-virtual TFF, the first measurement of its kind ever, albeit out of the $\{Q_{1,2}^2|~ Q_{1,2}^2 < 6.5\,\text{GeV}^2\}$ region that represents $95\%$ of the total $\amu{\eta'}$ contribution. The results are shown in \cref{fig:etapDV} and are in good agreement, while the statistics are not sufficient yet to further constrain the doubly-virtual parameters.
This results in 
\begin{align}
    \amu{\eta} = 16.3(1.0)_{\textrm{stat}}(0.5)_{a_{P;1,1}}(0.9)_{\textrm{sys}}\times10^{-11} \to  16.3(1.4)\times10^{-11}\,, \notag \\
    \amu{\eta'} = 14.5(0.7)_{\textrm{stat}}(0.4)_{a_{P;1,1}}(1.7)_{\textrm{sys}}\times10^{-11} \to  14.5(1.9)\times10^{-11}\,. \label{eq:eta-PA}
\end{align}
Given the excellent numerical agreement between dispersive analysis and the CA one for the $\pi^0$-pole contribution and the very good performance for low-energy timelike data~\cite{Escribano:2015nra,Escribano:2015yup}, we expect a similar scenario for $\eta/\eta'$, while an independent dispersive calculation ought to be completed as soon as possible.

Similar to the $\pi^0$ case, very recently two different groups~\cite{Eichmann:2019tjk,Raya:2019dnh} have computed these contributions using Dyson--Schwinger equations. The first group accounts for quark-mass effects when computing the---unphysical---light- and strange-quark TFFs, that are later employed to compute the---physical---$\eta/\eta'$ TFF using the mixing parameters determined from phenomenology in Ref.~\cite{Feldmann:1999uf}, obtaining $\amu{\eta/\eta'}=[15.8(1.2)/13.3(0.9)]\times 10^{-11}$~\cite{Eichmann:2019tjk}. In contrast, the second group~\cite{Raya:2019dnh} includes the $U(1)_A$ anomaly effects explicitly, while incorporating additional parameters subsequently constrained from phenomenology. This allows them to work directly with the physical form factors~\cite{Ding:2018xwy}. Interestingly enough, the results, $\amu{\eta/\eta'}=[14.7(1.9)/13.6(0.8)]\times 10^{-11}$~\cite{Raya:2019dnh}, are consistent with Ref.~\cite{Eichmann:2019tjk} and with \cref{eq:eta-PA}.

In addition, holographic estimates for the $\eta$ and $\eta'$
exist~\cite{Hong:2009zw,Leutgeb:2019zpq}.
Furthermore, there are estimates from effective Lagrangian models~\cite{Czyz:2017veo,Guevara:2018rhj}, which again fail to reproduce the high-energy constraints, as well as from the interpolation formula in Ref.~\cite{Danilkin:2019mhd}. 
Once more, we omit calculations of nonpole contributions~\cite{Bartos:2001pg,Melnikov:2003xd,Hong:2009zw,Goecke:2010if,Dorokhov:2011zf,Dorokhov:2012qa,Dorokhov:2015psa,Roig:2014uja}.

\subsubsection{Conclusion}

\begin{figure}[t]
    \centering
    \includegraphics*[width=0.49\textwidth]{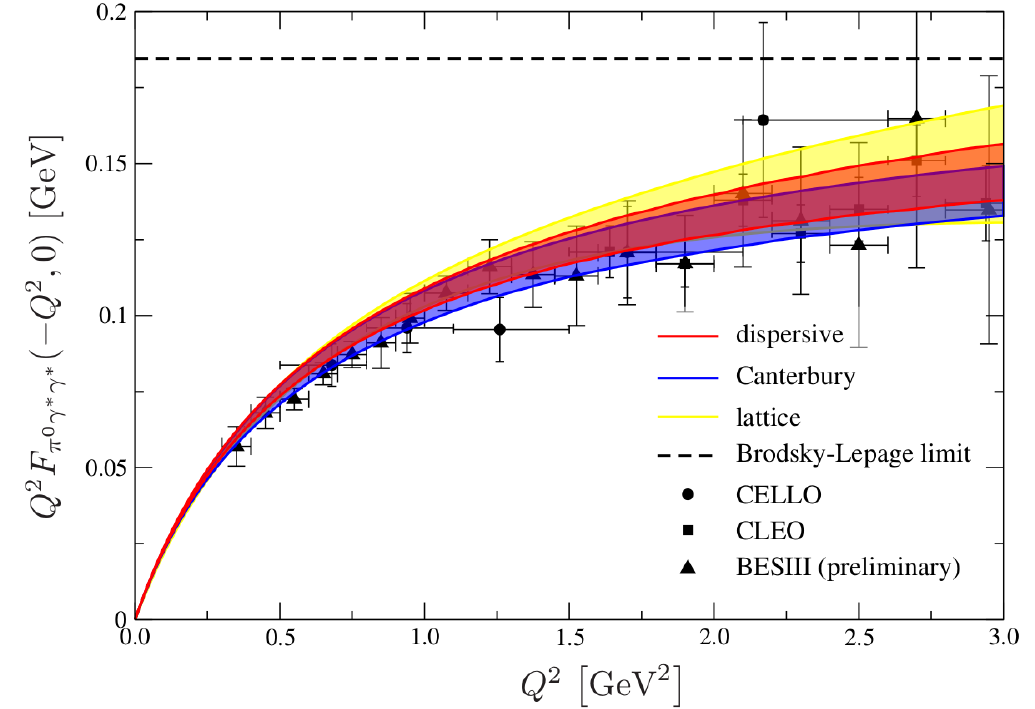} \hfill
    \includegraphics*[width=0.49\textwidth]{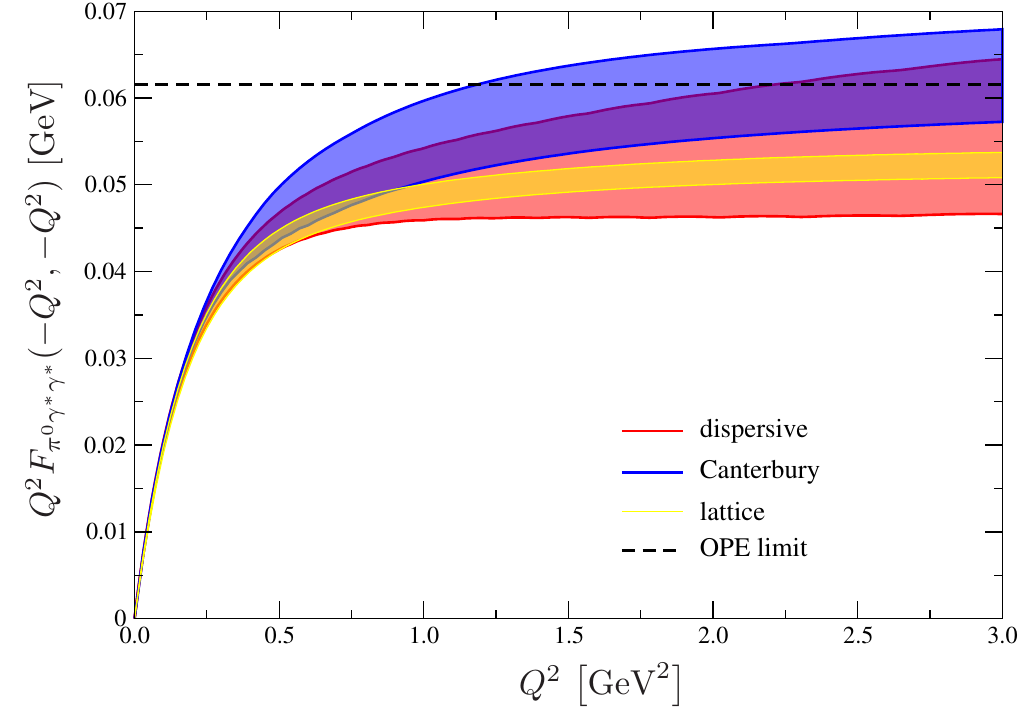}
    \caption{Comparison of the $\pi^0$ TFF from dispersion theory~\cite{Hoferichter:2018dmo,Hoferichter:2018kwz} (red), CA~\cite{Masjuan:2017tvw} (blue), and lattice QCD~\cite{Gerardin:2019vio} (yellow). We show both the singly- (left) and the doubly-virtual (right) form factors.} 
    \label{fig:pi0-comparison}
\end{figure}
The most compelling determinations of $\amu{\pi^0}$ fulfilling the outlined quality criteria are in good agreement with each other and with the lattice determination~\cite{Gerardin:2019vio} (see \cref{sec:pion pole} for details) that is obtained from a $z$-expansion fit, adjusting the form factor normalization for the experimental PrimEx result~\cite{Larin:2010kq}:\footnote{Note that these determinations use slightly different input for the TFF normalization,  e.g., with the CA result updated to PrimEx-II the spread among the central values would become slightly larger---see previous comments.}
\begin{align}
\label{pion_pole}
    \amu{\pi^0}(\text{disp}) &= 63.0^{+2.7}_{-2.1}\times10^{-11} \,, \notag\\
    \amu{\pi^0}(\text{CA}) &= 63.6(2.7)\times10^{-11} \,, \notag\\
    \amu{\pi^0}(\text{lattice}) &= 62.3(2.3)\times10^{-11} \,.
\end{align}
Moreover, all three underlying $\pi^0$ TFFs are consistent with each other, as can be observed from \cref{fig:pi0-comparison}.
Combining the dispersive evaluation of the $\pi^0$ with the CA for $\eta$ and $\eta'$, which are in good agreement with Dyson--Schwinger equations determinations, we arrive at the following current estimate for the pseudoscalar-pole contributions: 
\begin{equation}
\amu{\pi^0+\eta+\eta'} = 93.8^{+4.0}_{-3.6}\times 10^{-11}\,,
\end{equation}
where individual systematic uncertainties of CA have been combined linearly. Similarly, combining the result obtained from CA for the $\pi^0$, $\eta$, and $\eta'$ contributions, we obtain
\begin{equation}
\amu{\pi^0+\eta+\eta'} = 94.3(5.3)\times 10^{-11}\,,
\end{equation}
with differences in uncertainties mainly due to the different treatment in systematic errors.

\subsection{Contribution of two-pion intermediate states}
\label{sec:two-pion}

As discussed in \cref{sec:DefinitionOfIndividualContributionsToHLbL}, Mandelstam's double-dispersion relation~\cite{Mandelstam:1958xc} can be used to define the contributions of different hadronic intermediate states to the HLbL tensor and $a_\mu$~\cite{Colangelo:2015ama} in a model-independent way. In the unitarity relation, the lightest intermediate state is a single neutral pion, giving rise to the pion-pole contribution $a_\mu^{\pi^0\text{-pole}}$ discussed in \cref{sec:pion-pole}. The next lightest intermediate state is given by two pions, either $\pi^0\pi^0$ or $\pi^+\pi^-$. The unitarity relation for the HLbL tensor expresses the discontinuity due to two-pion intermediate states in terms of the sub-process $\gamma^*\gamma^*\to\pi\pi$. If one applies unitarity a second time and considers intermediate states in the crossed sub-process $\gamma^*\pi\to\gamma^*\pi$, one can split the full two-pion contribution to HLbL into a sum of different box topologies, as illustrated in \cref{HLbLTwoPionContributions}. In the following, we discuss the different two-pion contributions and their numerical evaluation based on dispersion theory.

\subsubsection{Pion box}

\begin{figure}[t]
	\centering
	\begin{align*}
 		\minidiagSize{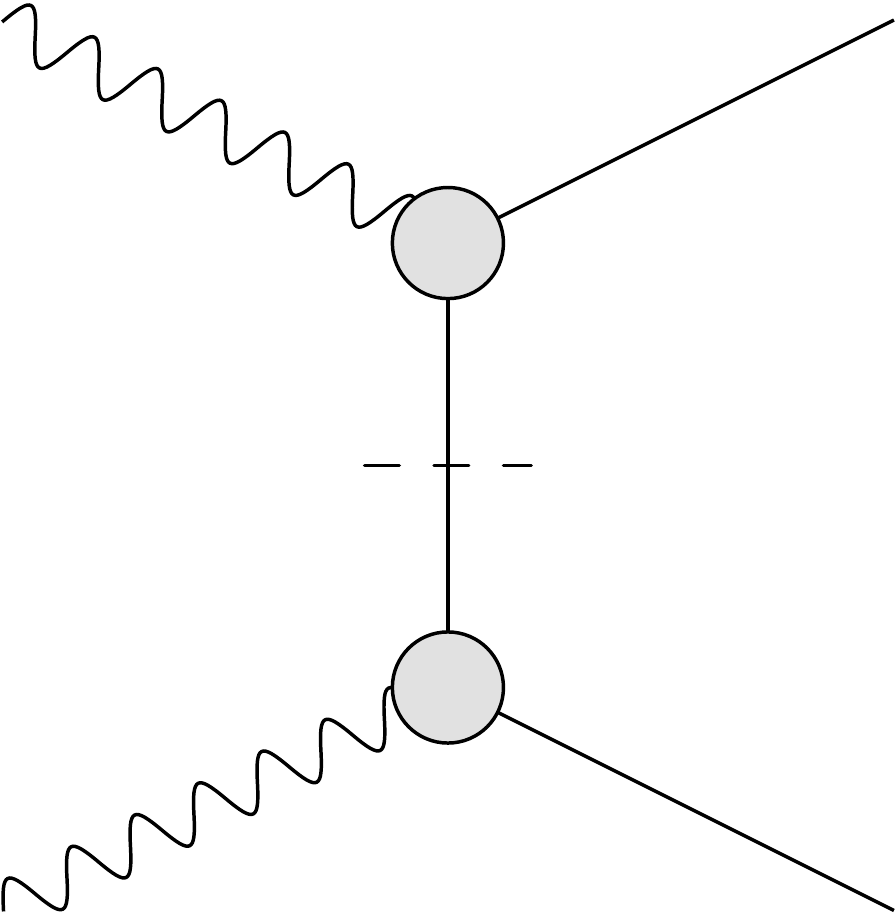}{2cm} + \minidiagSize{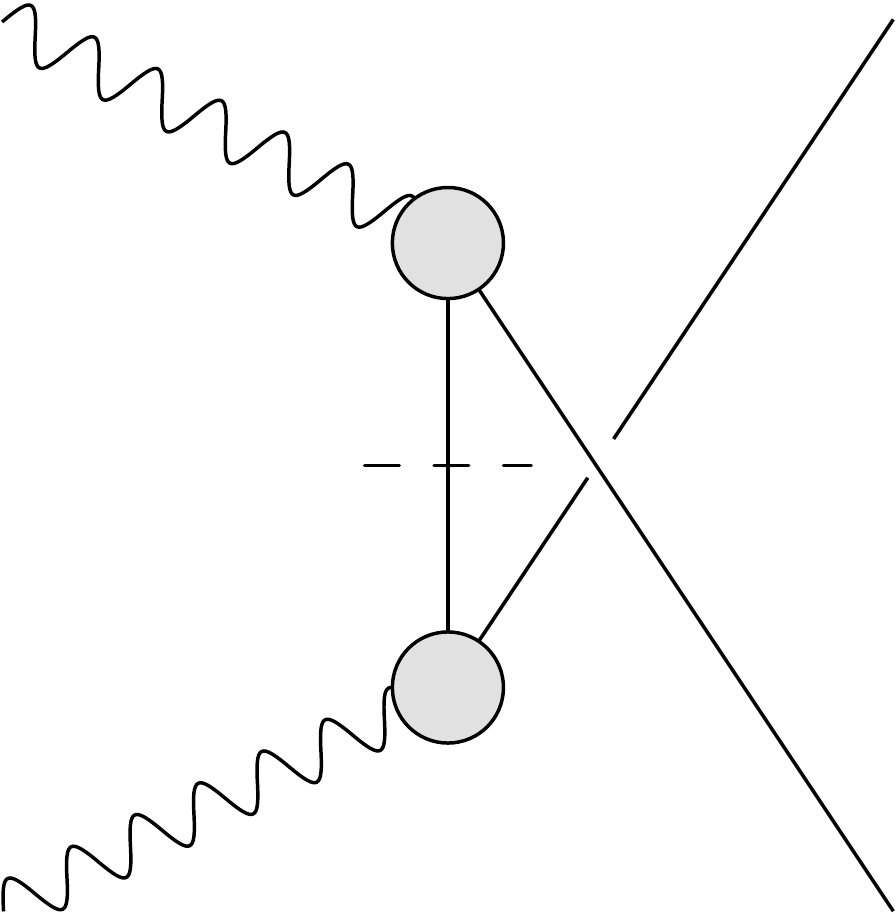}{2cm} = F_\pi^V(q_1^2) F_\pi^V(q_2^2) \times \left[ \; \minidiagSize{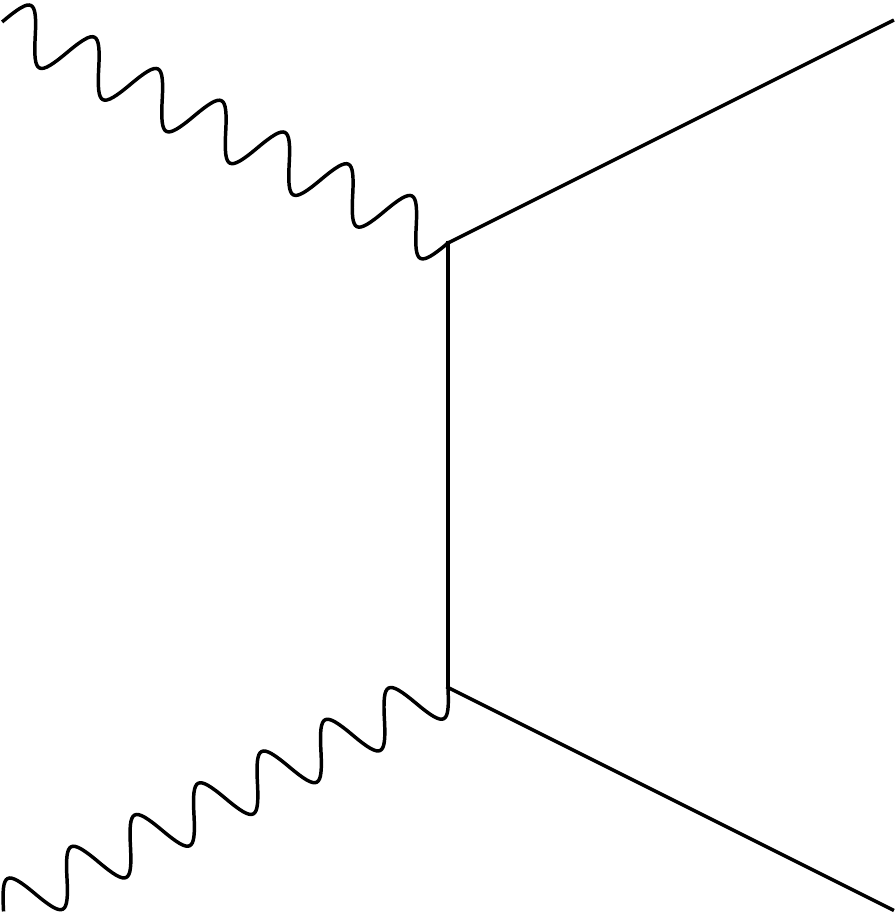}{2cm} + \minidiagSize{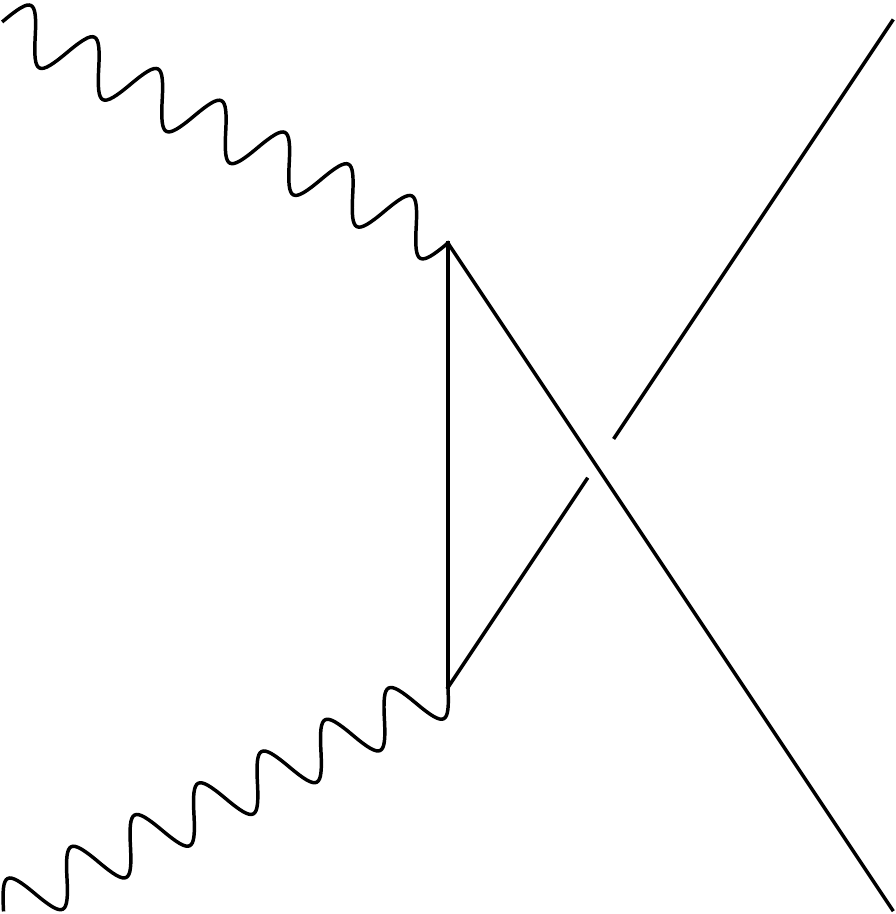}{2cm} + \minidiagSize{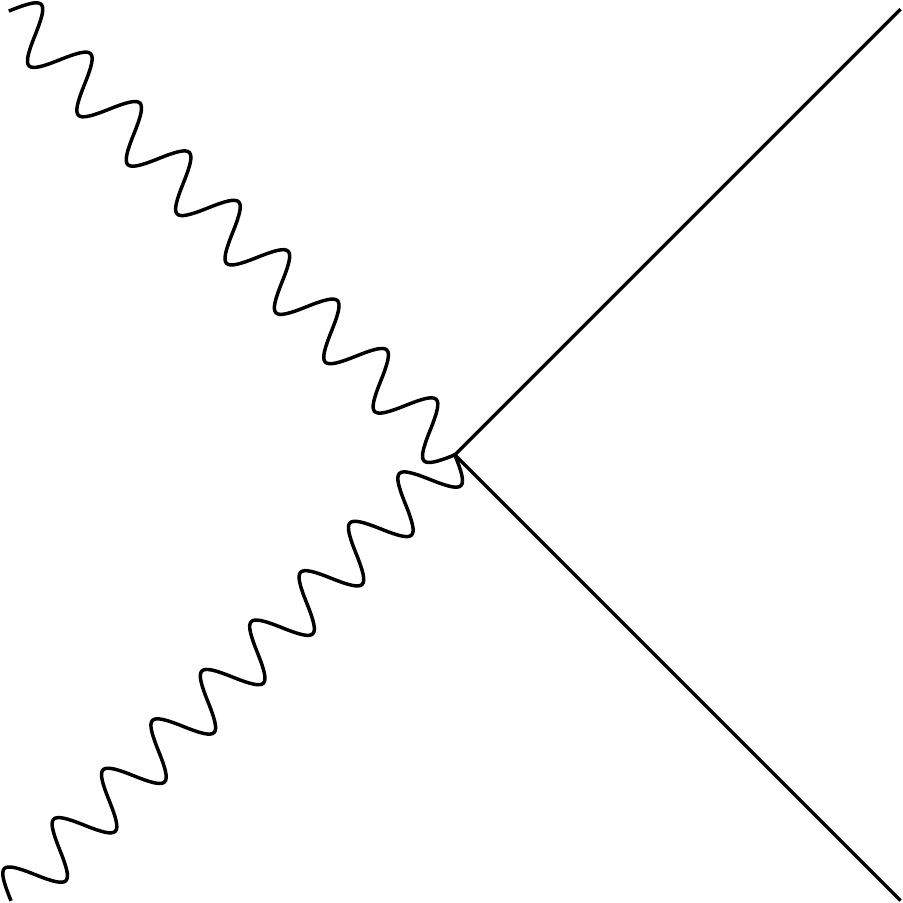}{2cm} \; \right]
	\end{align*}
	\caption{Contribution to $\gamma^*\gamma^*\to\pi^+\pi^-$ of a charged-pion pole in the crossed channels. The dispersively defined pole contribution is equivalent to the scalar QED Born terms, multiplied by pion vector form factors for the off-shell photons. Adapted from Ref.~\cite{Colangelo:2015ama}.}
	\label{fig:ggpipiPole}
\end{figure}

In the unitarity relation for $\gamma^*\pi^+\to\gamma^*\pi^+$, the lightest intermediate state is the charged pion. The starting-point to define this contribution is a fixed-$s$ dispersion relation for $\gamma^*\gamma^*\to\pi^+\pi^-$. In Ref.~\cite{Colangelo:2015ama}, it was shown that the pure pole contribution in this dispersion relation exactly coincides with the Born term in a scalar QED (sQED) calculation of $\gamma^*\gamma^*\to\pi^+\pi^-$, multiplied by a pion vector form factor (VFF) for each of the two off-shell photons as illustrated in \cref{fig:ggpipiPole}. The seagull term is required in a sQED Feynman-diagram calculation to obtain a gauge-invariant expression. In the dispersive approach, one first defines a BTT tensor decomposition for $\gamma^*\gamma^*\to\pi\pi$~\cite{Bardeen:1969aw,Tarrach:1975tu}, which fully takes care of gauge invariance, in analogy to the case of the HLbL tensor itself as described in \cref{sec:BTT}. In this representation, gauge invariance is manifest and there are two pion-pole unitarity diagrams present. Note that cutting the propagator puts the pion on shell: only on-shell states enter the sum of intermediate states in the unitarity relation. The two hadronic blobs in \cref{fig:ggpipiPole} therefore correspond to the pion VFF and only depend on the photon virtualities.

\begin{figure}[t]
	\centering
	\begin{align*}
 		& \minidiagSize{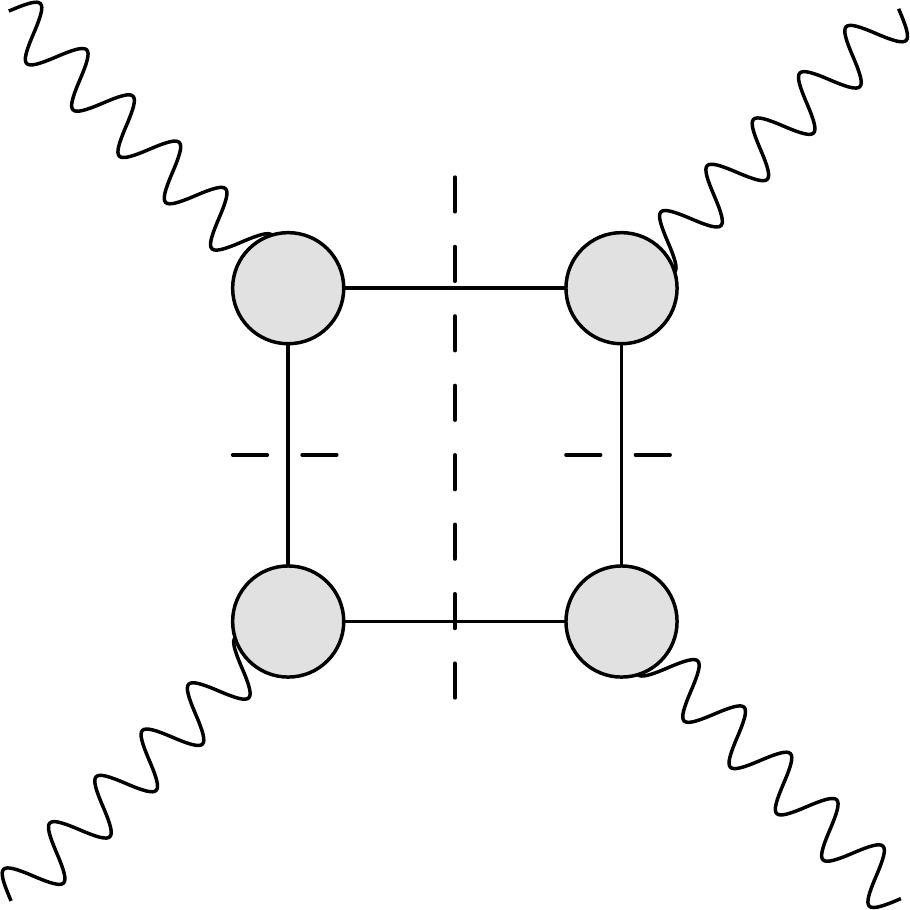}{2cm} + \minidiagSize{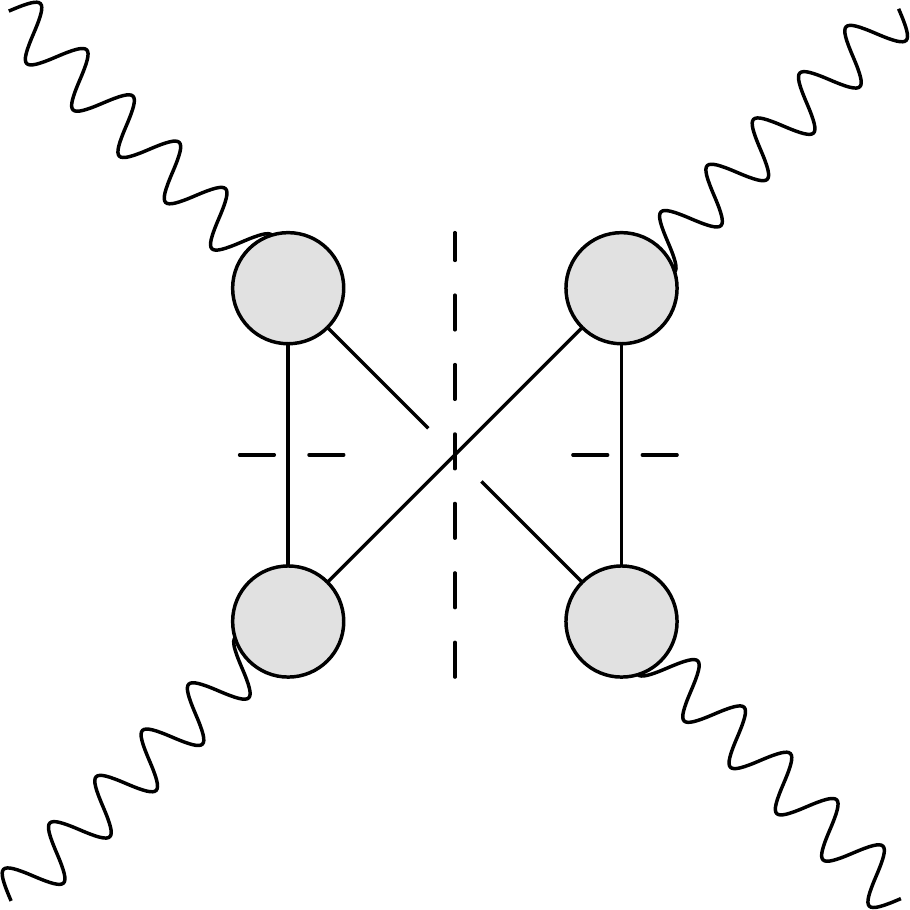}{2cm}  + \minidiagSize{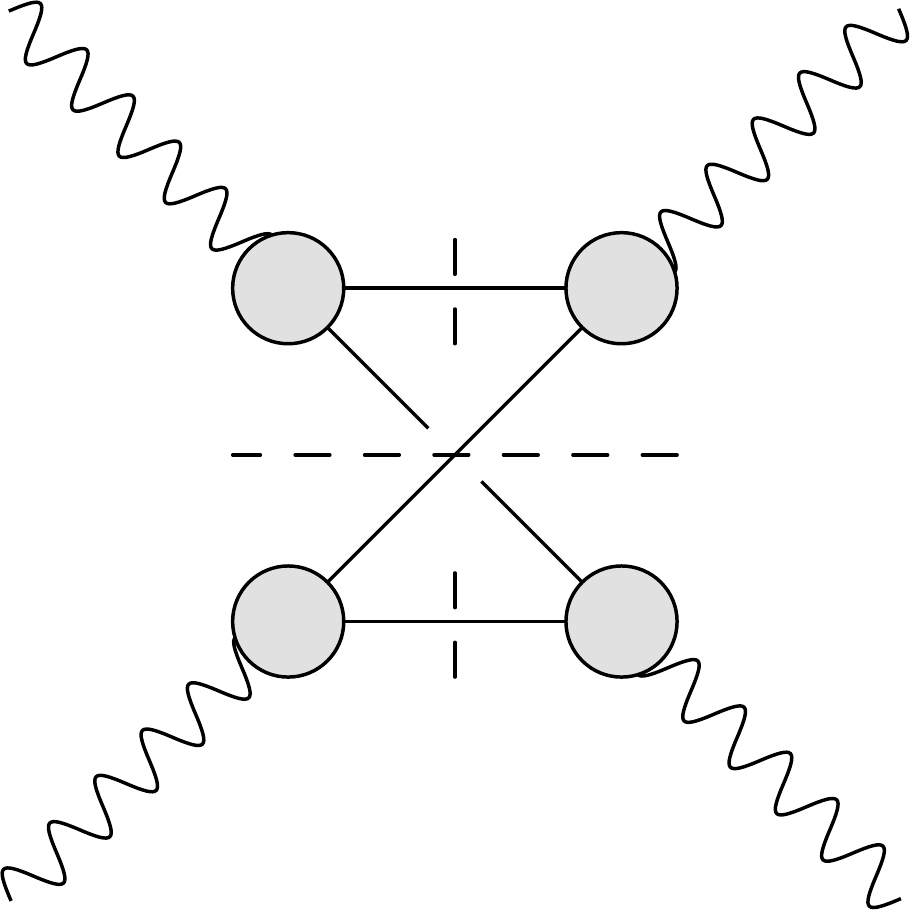}{2cm} \\
		&= F_\pi^V(q_1^2) F_\pi^V(q_2^2) F_\pi^V(q_3^2) F_\pi^V(q_4^2) \times \left[ \; \minidiagSize{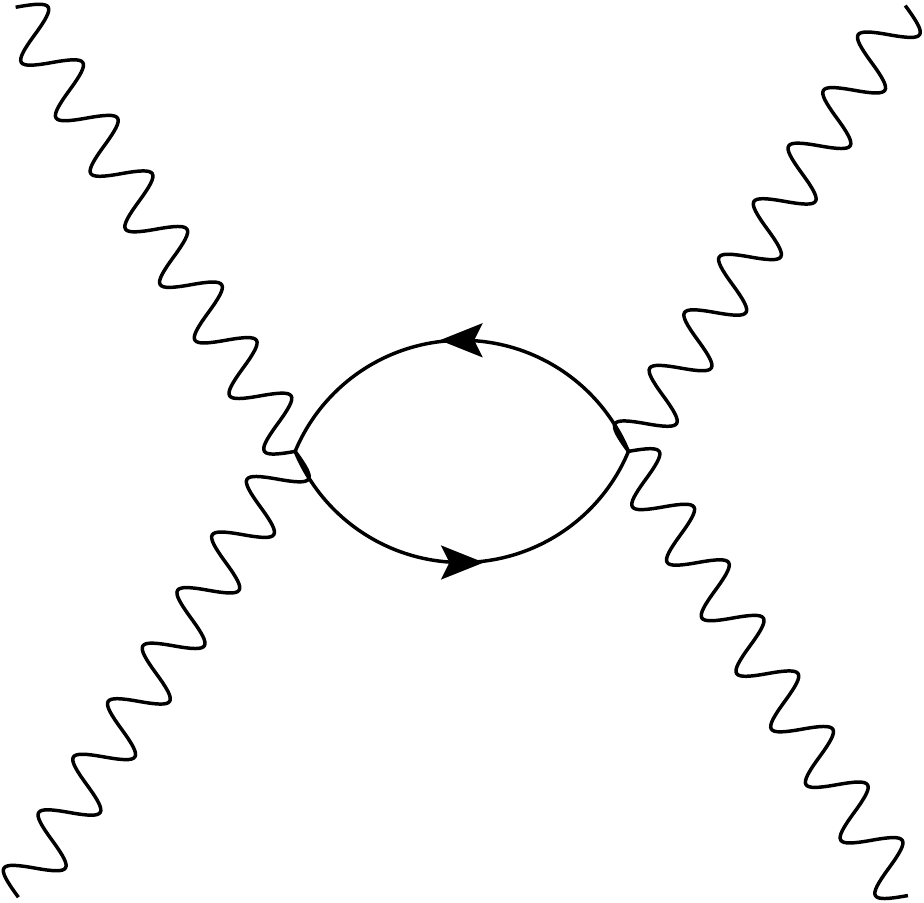}{2cm} + \minidiagSize{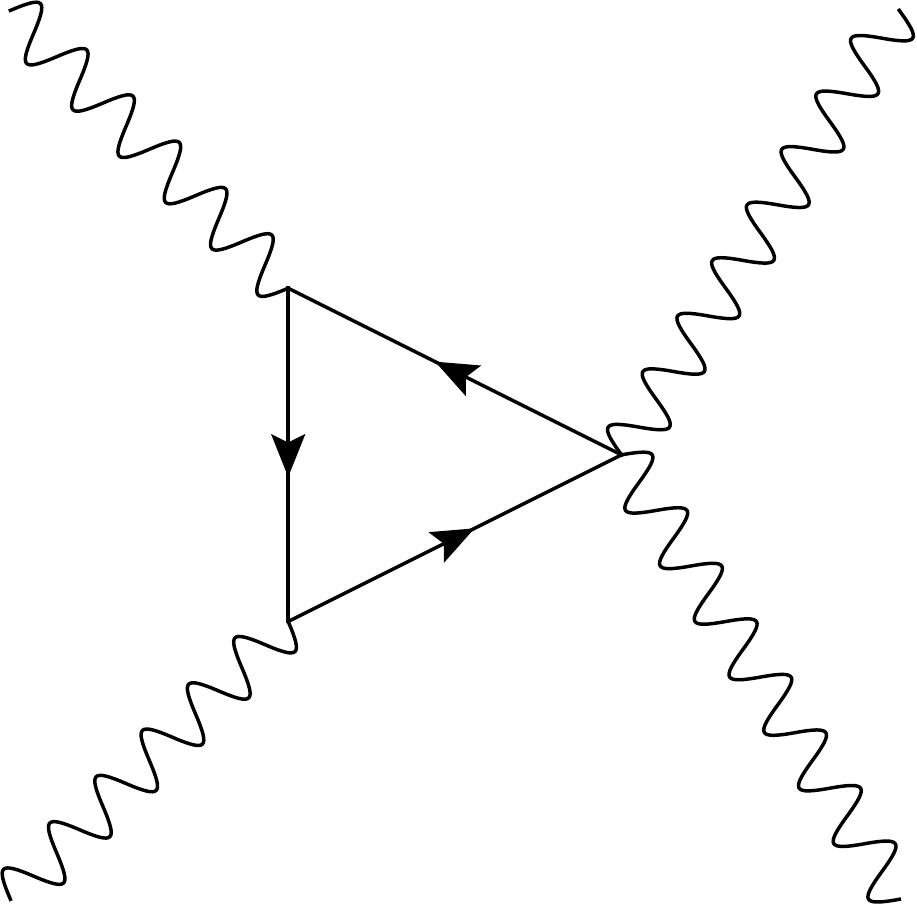}{2cm} + \minidiagSize{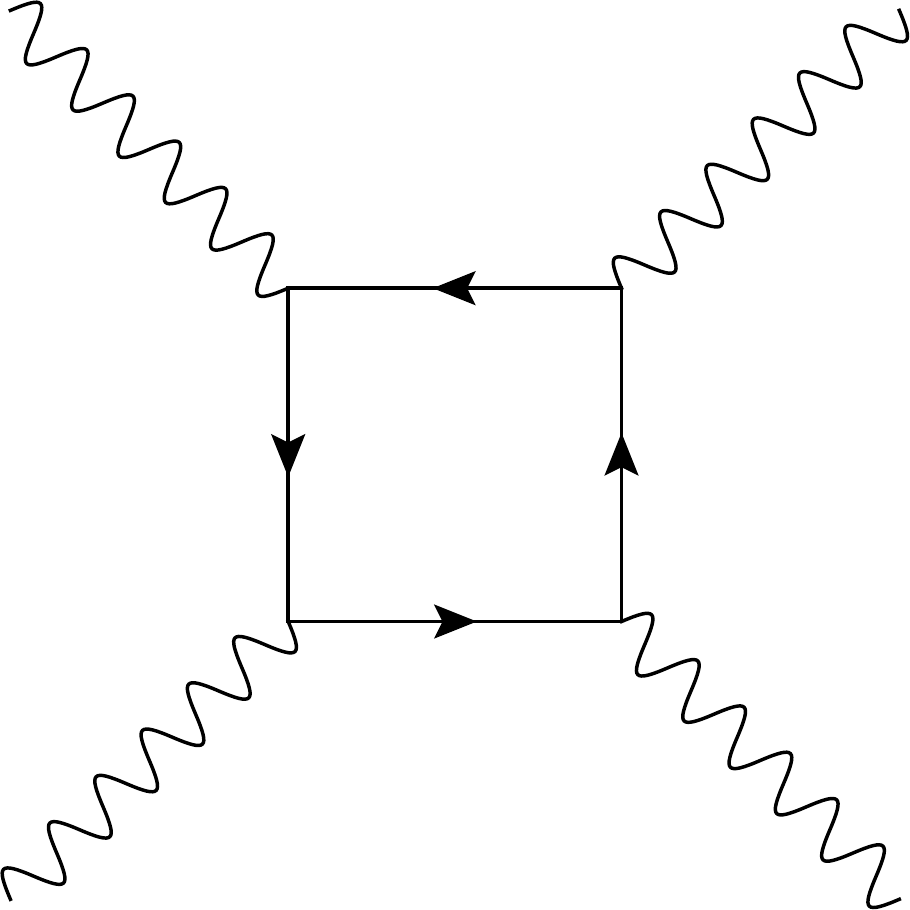}{2cm} + \text{crossed} \; \right]
	\end{align*}
	\caption{The dispersively defined box topologies are identical to the sQED one-loop expression, multiplied by pion vector form factors for the off-shell photons.}
	\label{fig:HLbLPionBox}
\end{figure}

If this pure pole term is singled out in both of the sub-processes in the unitarity relation for HLbL, one obtains the pion-box topology shown in \cref{HLbLTwoPionContributions}(a). On both sides of the unitarity cut, the hadronic blobs refer to the pion VFF, which is a function of the respective squared momentum $q_i^2$ of the off-shell photon. In particular, these factors do not depend on the Mandelstam variables, hence in a double-dispersion relation that treats the $q_i^2$ as fixed external quantities, they can be taken out of the dispersion integrals and multiply a double-dispersion relation for a pion box with pointlike vertices:
\begin{align}
	\label{eq:PionBoxMandelstamRepresentation}
	\Pi_i^{\pi\text{-box}}(s,t,u; q_i^2) &= F_\pi^V(q_1^2) F_\pi^V(q_2^2) F_\pi^V(q_3^2) F_\pi^V(q_4^2) \\
		&\times \bigg(  \frac{1}{\pi^2} \int ds' dt' \frac{\rho_{i;st}(s',t')}{(s'-s)(t'-t)} + \frac{1}{\pi^2} \int ds' du' \frac{\rho_{i;su}(s',u')}{(s'-s)(u'-u)}  + \frac{1}{\pi^2} \int dt' du' \frac{\rho_{i;tu}(t',u')}{(t'-t)(u'-u)} \bigg)\, .\notag
\end{align}
Explicit calculation of the double-spectral densities $\rho_i$ shows that \cref{eq:PionBoxMandelstamRepresentation} without the form factors is mathematically equivalent to the one-loop light-by-light expression in sQED~\cite{Colangelo:2015ama}. Therefore, the situation is analogous to the pion pole in the sub-process $\gamma^*\gamma^*\to\pi^+\pi^-$: the dispersively defined box topologies are identical to the sQED one-loop expression, multiplied by one VFF for each of the off-shell photons. Note that the sQED loop contribution in terms of Feynman diagrams consists of
boxes, triangles, and bulbs, but that the corresponding unitarity diagrams are just the three box topologies shown in \cref{fig:HLbLPionBox}. The bulb and triangle diagrams in sQED are required for a gauge-invariant result. However, upon projection on gauge-invariant BTT structures, the kinematic and dynamic singularities are separated and the sQED contribution to the coefficient functions only have the dynamic singularities of pure box topologies, which can be expressed in terms of double-spectral representations.

This relation is useful because it allows one to express the dispersively defined boxes in terms of Feynman-parameter integrals and to obtain very compact expressions for the scalar functions. In the master formula for $a_\mu$ \cref{eq:MasterFormula3Dim}, the scalar functions enter in the limit $q_4\to0$. For this reduced kinematics, one can explicitly perform the integral over one Feynman parameter, which leaves a two-dimensional integral representation
\begin{align}
	\bar \Pi_i^{\pi\text{-box}}(q_1^2,q_2^2,q_3^2) = F_\pi^V(q_1^2) F_\pi^V(q_2^2) F_\pi^V(q_3^2) \frac{1}{16\pi^2} \int_0^1 dx \int_0^{1-x} dy \, I_i(x,y)\, ,
\end{align}
where, e.g.,
\begin{align}
	I_1(x,y) &= \frac{8 xy (1-2x) (1-2y)}{\Delta_{123} \Delta_{23}} \, , \nn
	\Delta_{ijk} &= M_\pi^2 - x y q_i^2 - x (1-x-y) q_j^2 - y(1-x-y) q_k^2 \,, \nn
	\Delta_{ij} &= M_\pi^2 - x (1-x) q_i^2 - y (1-y) q_j^2 \,.
	\label{Ipibox}
\end{align}
The expressions for the other scalar functions contributing to $a_\mu$ can be found in Ref.~\cite{Colangelo:2017fiz}.

Therefore, the only input quantity in the pion-box contribution is the pion VFF. In the master formula \cref{eq:MasterFormula3Dim}, one has to integrate over spacelike $q_i^2$, i.e., it is also the spacelike VFF that is required for the calculation of the pion-box contribution. The VFF has been studied in much detail in the context of the HVP contribution to $a_\mu$ and can be precisely determined in fits to $e^+e^-\to\pi^+\pi^-$ data~\cite{Leutwyler:2002hm,Colangelo:2003yw,Colangelo:2018mtw}. For the application of the pion box in HLbL scattering, the available precision of the VFF is beyond what is needed and leads to the result~\cite{Colangelo:2017fiz}
\be
	\label{eq:PionBox}
	a_\mu^{\pi\text{-box}}=-15.9(2)\times 10^{-11}\,.
\ee
The dispersively defined pion box can be understood as a model-independent notion of a pion loop. The clean definition of the pion box practically eliminates the uncertainty on this particular contribution.

\subsubsection{Pion rescattering, $S$-waves}

The unitarity relation for the HLbL tensor in general provides a connection between the two-pion contribution to HLbL and the sub-process $\gamma^*\gamma^*\to\pi\pi$. Unitarity is diagonal in the space of helicity partial waves, i.e., simply given by
\begin{align}
	\Imspipi h^J_{\lambda_1\lambda_2,\lambda_3\lambda_4}(s) = \eta_i \eta_f \frac{\sigma_\pi(s)}{16\pi S} h_{J,\lambda_1\lambda_2}(s) h_{J,\lambda_3\lambda_4}^*(s)\, ,
\end{align}
where $\eta_{i,f}=\pm1$ take care of the sign conventions for the partial waves (see Ref.~\cite{Colangelo:2017fiz} for details), $S$ is the symmetry factor for the two pions, $\sigma_\pi(s) = \sqrt{1-4M_\pi^2/s}$ is the phase space. The photon helicities are denoted by $\lambda_i$ and only partial waves with even $J$ are present. A given input for the helicity partial waves for $\gamma^*\gamma^*\to\pi\pi$ defines the imaginary part of the HLbL partial waves. The difficulty lies in the reconstruction of the real part. As is well known, helicity amplitudes contain kinematic singularities, hence it is mandatory to perform a basis change to a set of scalar functions that are free of kinematic singularities in order to enable a dispersive reconstruction of the real part. The basis change from the BTT scalar functions to helicity amplitudes can be easily performed by contraction with polarization vectors. The inversion of this basis change is nontrivial and has been worked out in Ref.~\cite{Colangelo:2017fiz}. In particular, it is necessary to first resolve the redundancy in the BTT set, at least for the simplified kinematics of fixed-$s$, -$t$, or -$u$ dispersion relations. This turns out to be possible provided that one makes use of sum rules for the HLbL scalar functions for fixed-$s$/-$t$/-$u$ kinematics~\cite{Colangelo:2017fiz}. These sum rules are related to the ones for the special case of forward scattering, derived in Ref.~\cite{Pascalutsa:2012pr}.

\begin{figure}[t]
	\includegraphics[width=8cm,page=1]{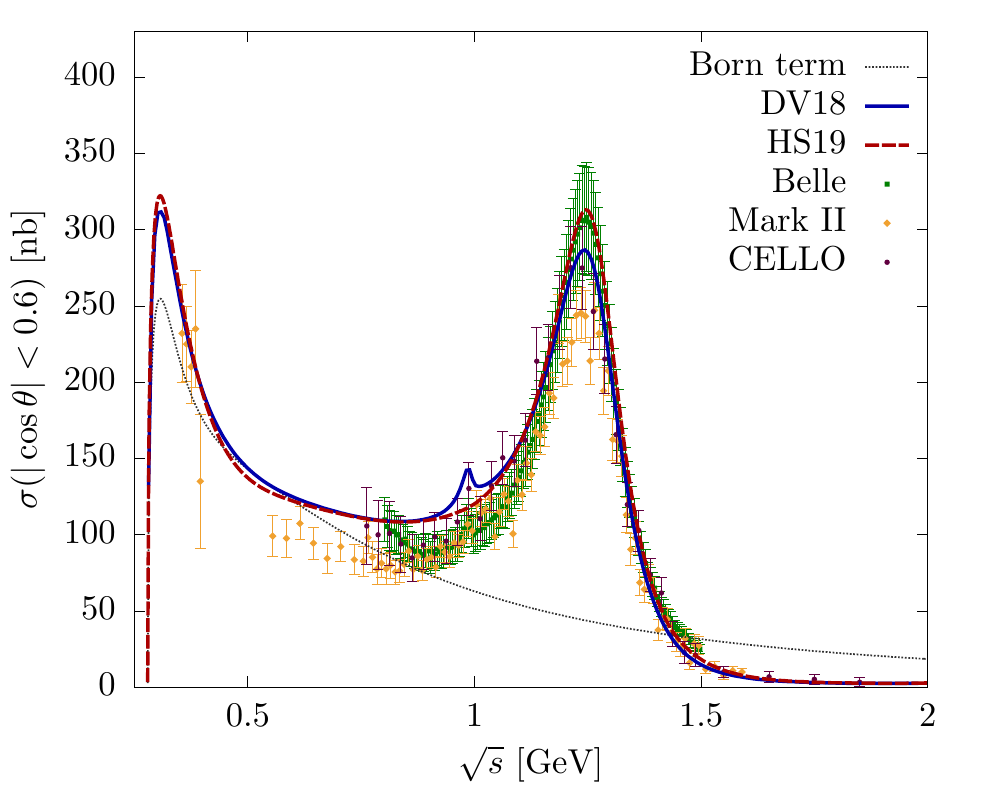}
	\includegraphics[width=8cm,page=2]{dispHLbL/figures/ggpipi-xs}
	\caption{Cross section for $\gamma\gamma\to\pi^+\pi^-$ (left) and $\gamma\gamma\to\pi^0\pi^0$ (right) from DV18~\cite{Danilkin:2018qfn} and HS19~\cite{Hoferichter:2019nlq}, in comparison to the data from Belle~\cite{Mori:2007bu,Uehara:2009cka}, Mark II~\cite{Boyer:1990vu}, CELLO~\cite{Behrend:1992hy}, and Crystal Ball~\cite{Marsiske:1990hx}.}
	\label{fig:ggpipiCrossSection}
\end{figure}

In Ref.~\cite{Colangelo:2017fiz}, a first evaluation of the $S$-wave low-energy contribution has been performed. The input for the doubly-virtual $\gamma^*\gamma^*\to\pi\pi$ partial waves $h_{0,00}$ and $h_{0,++}$ was constructed as follows. Although the general tensor decomposition for $\gamma^*\gamma^*\to\pi\pi$ involves five structures~\cite{Tarrach:1975tu,Drechsel:1997xv,Colangelo:2015ama}, only two of them get contributions from $S$-waves. One therefore has to solve a simplified $2\times2$ system, where the dispersion relation is written in the basis of scalar functions that are free of kinematic singularities and the elastic unitarity relation is diagonal in the basis of helicity amplitudes. The solution of the dispersion relation is given by an inhomogeneous Omn\`es representation~\cite{Colangelo:2017fiz}:
\begin{align}
	\label{eq:MOSolutionSWaves}
	h_{0,++}(s)&=\Delta_{0,++}(s) \nn*
		 &+\frac{\Omega_0(s)}{\pi}\int_{4\mpi^2}^\infty ds'\frac{\sin\delta_0(s')}{|\Omega_0(s')|}\bigg[\bigg(\frac{1}{s'-s}-\frac{s'-q_1^2-q_2^2}{\lambda_{12}(s')}\bigg)\Delta_{0,++}(s')+\frac{2q_1^2q_2^2}{\lambda_{12}(s')}\Delta_{0,00}(s')\bigg]\,, \nn
	h_{0,00}(s)&=\Delta_{0,00}(s) \nn
		&+\frac{\Omega_0(s)}{\pi}\int_{4\mpi^2}^\infty ds'\frac{\sin\delta_0(s')}{|\Omega_0(s')|}\bigg[\bigg(\frac{1}{s'-s}-\frac{s'-q_1^2-q_2^2}{\lambda_{12}(s')}\bigg)\Delta_{0,00}(s')+\frac{2}{\lambda_{12}(s')}\Delta_{0,++}(s')\bigg] \,,
\end{align}
where $\lambda_{12}(s) = \lambda(q_1^2,q_2^2,s)$, $\lambda(a,b,c) = a^2 + b^2 + c^2 - 2(ab+bc+ca)$ is the K\"all\'en function,
\be
	\Omega_0(s)=\exp\Bigg\{\frac{s}{\pi}\int_{4\mpi^2}^\infty ds'\frac{\delta_0(s')}{s'(s'-s)}\Bigg\}
\ee
is the Omn\`es function, and $\delta_0$ denotes the elastic $S$-wave $\pi\pi$-scattering phase shift (isospin indices are suppressed). $\Delta_{J,\lambda_1\lambda_2}$ denotes the inhomogeneity due to the left-hand cut (given by the singularities in the $t$- and $u$-channel $\gamma^*\pi\to\gamma^*\pi$). In Ref.~\cite{Colangelo:2017fiz}, the left-hand cut has been approximated by a pion pole. In this case, $\Delta_{J,\lambda_1\lambda_2}$ is the partial-wave-projected Born term, multiplied by VFFs describing the dependence on the photon virtualities. The $\pi\pi$ phase shift is an input in the dispersion relation. In Ref.~\cite{Colangelo:2017fiz}, a simple phase-shift representation based on a modified inverse-amplitude method~\cite{GomezNicola:2007qj} was used, which at low energies agrees well with the full phenomenological phase shift from Roy-equation analyses~\cite{Colangelo:2001df,GarciaMartin:2011cn,Caprini:2011ky} and accurately reproduces the parameters of the $f_0(500)$ resonance, but cuts off the $f_0(980)$ resonance that would require a coupled-channel treatment together with two-kaon intermediate states.

Properly subtracting the contribution of two Born terms on both sides of the unitarity relation to avoid double-counting with the pion box, the $S$-wave low-energy contribution with a pion-pole left-hand cut is given by~\cite{Colangelo:2017qdm,Colangelo:2017fiz}
\begin{align}
	\label{eq:pipiSwave}
	a_{\mu,J=0}^{\pi\pi,\pi\text{-pole LHC}}=-8(1)\times 10^{-11} \,.
\end{align}
The isospin-0 component thereof can be understood as a dispersive description of the $f_0(500)$ contribution in terms of two-pion intermediate states.

\begin{figure}[t]
	\includegraphics[width=7.5cm,page=3]{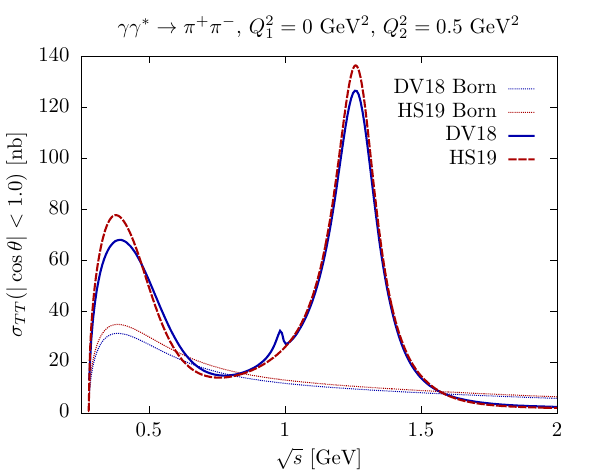}
	\includegraphics[width=7.5cm,page=5]{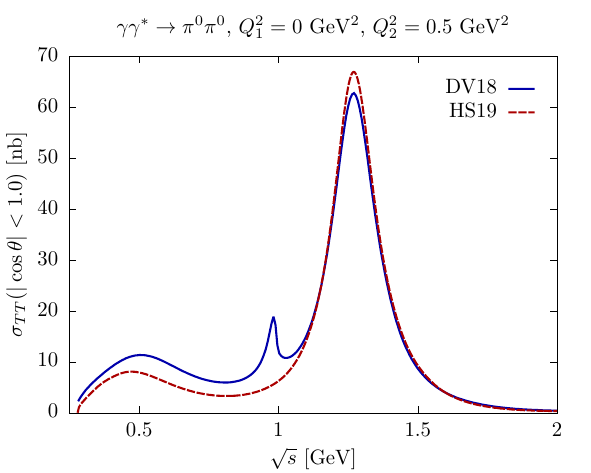}
	\\[0.5cm]
	\includegraphics[width=7.5cm,page=4]{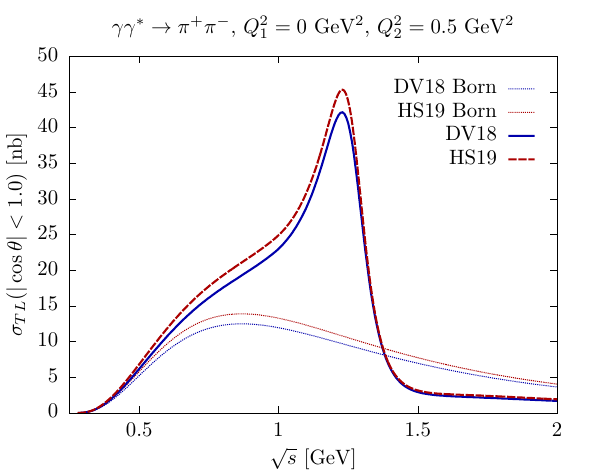}
	\includegraphics[width=7.5cm,page=6]{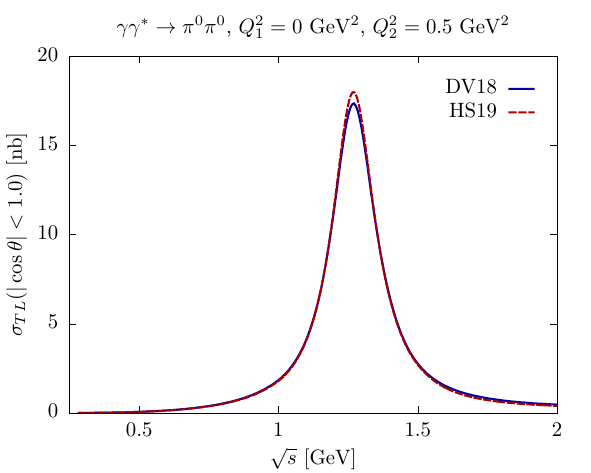}
	\caption{Predictions for the cross section for $\gamma\gamma^*\to\pi^+\pi^-$ (left) and $\gamma\gamma^*\to\pi^0\pi^0$ (right) for $Q^2=0.5\GeV^2$ from DV18~\cite{Danilkin:2018qfn} and HS19~\cite{Hoferichter:2019nlq} compared to the Born results (dotted curves).}
	\label{fig:ggpipiCrossSection2}
\end{figure}

\subsubsection{Pion rescattering, $D$- and higher waves}

The partial-wave framework developed in Ref.~\cite{Colangelo:2017fiz} is in principle valid for arbitrary partial waves. This has been tested by expanding the Born term for $\gamma^*\gamma^*\to\pi^+\pi^-$ into partial waves and resumming the contribution to $a_\mu$ up to $J=20$. A nice convergence pattern to the full pion-box contribution is observed, providing a strong cross-check of the partial-wave framework.

\begin{figure}[ht!]
	\includegraphics[width=7.5cm,page=7]{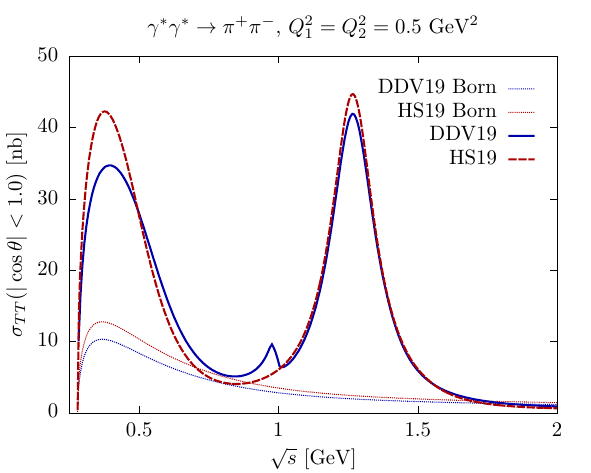}
	\includegraphics[width=7.5cm,page=10]{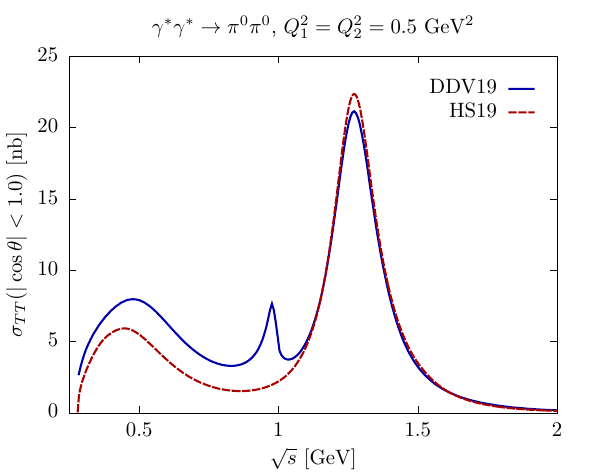}
	\\[0.5cm]
	\includegraphics[width=7.5cm,page=8]{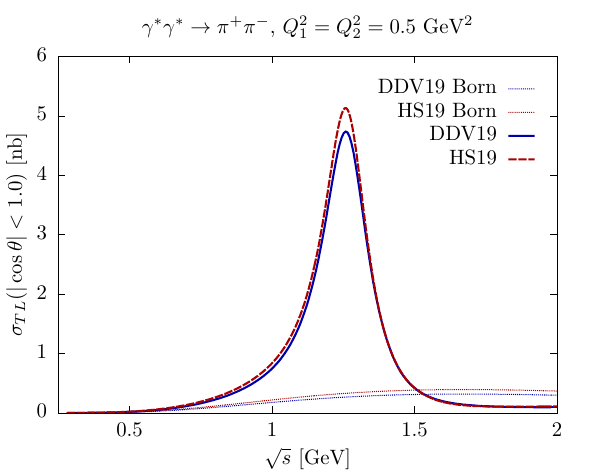}
	\includegraphics[width=7.5cm,page=11]{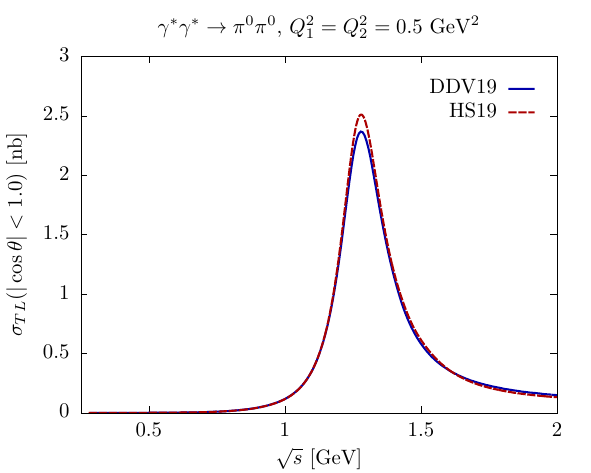}
	\\[0.5cm]
	\includegraphics[width=7.5cm,page=9]{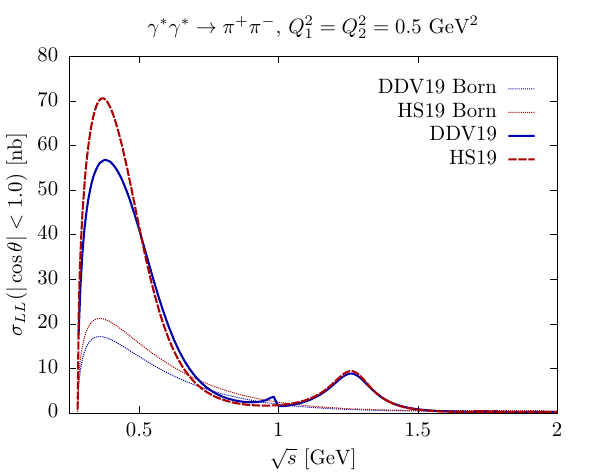}
	\includegraphics[width=7.5cm,page=12]{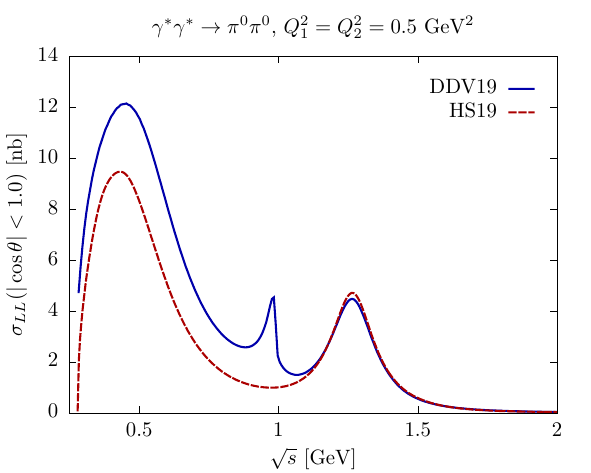}
	\caption{Predictions for the cross section for $\gamma^*\gamma^*\to\pi^+\pi^-$ (left) and $\gamma^*\gamma^*\to\pi^0\pi^0$ (right) for $Q_1^2=Q_2^2=0.5\GeV^2$ from HS19~\cite{Hoferichter:2019nlq} and DDV19~\cite{Danilkin:2019opj} compared to the Born results (dotted curves).}
	\label{fig:ggpipiCrossSection3}
\end{figure}

\begin{figure}[t]
	\centering
	\includegraphics[width=7.5cm,page=1]{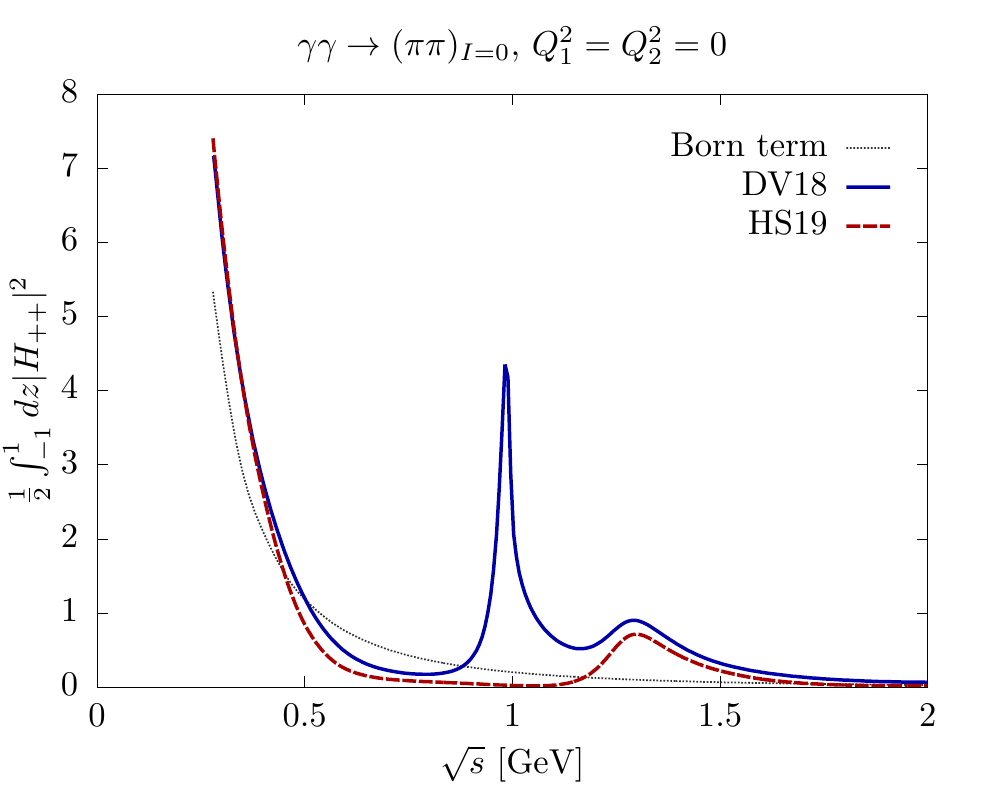}
	\includegraphics[width=7.5cm,page=2]{dispHLbL/figures/ggpipi-helamps}
	\caption{Helicity amplitudes for $\gamma\gamma\to\pi\pi$ in the isospin $I=0$ channel including $S$- and $D$-waves from DV18~\cite{Danilkin:2018qfn} and HS19~\cite{Hoferichter:2019nlq}, integrated over the scattering angle $z=\cos\theta$, in comparison to the Born result (dotted curves).}
	\label{fig:ggpipiHelAmps1}
\end{figure}

The complication of a generalization to $D$- and higher partial waves then mainly lies in the construction of the appropriate input for the $\gamma^*\gamma^*\to\pi\pi$ helicity partial waves. In the case of on-shell photons $\gamma\gamma\to\pi\pi$, there is enough data available~\cite{Marsiske:1990hx,Boyer:1990vu,Behrend:1992hy,Mori:2007bu,Uehara:2008ep,Uehara:2009cka} to enable a partial-wave analysis~\cite{Dai:2014zta}. However, the continuation in the photon virtualities requires the solution of dispersion relations for $\gamma^*\gamma^*\to\pi\pi$. The on-shell case has already been studied in detail in Refs.~\cite{GarciaMartin:2010cw,Hoferichter:2011wk,Moussallam:2013una}. The generalization to the singly-virtual case $\gamma^*\gamma\to\pi\pi$ has first been studied for $S$-waves in Ref.~\cite{Moussallam:2013una}. In Ref.~\cite{Danilkin:2018qfn}, this work was extended to a coupled-channel treatment (with $ K \bar{K}$) of the $S$-waves as well as a single-channel treatment for $D$-waves. For the $S$-wave isoscalar amplitude the coupled-channel Omn\`es function was used from a dispersive summation scheme~\cite{Gasparyan:2010xz, Danilkin:2010xd, Danilkin:2011fz} that is based on the $N/D$ ansatz~\cite{Chew:1960iv}. The set of coupled-channel integral equations was solved with the input from the left-hand cuts, which were present in a model-independent form as an expansion in a suitably constructed conformal mapping variable. These coefficients were determined from fitting to Roy analyses for $\pi\pi \to \pi\pi$~\cite{Ananthanarayan:2000ht,GarciaMartin:2011cn}, $\pi\pi \to K\bar{K}$ (for $I=0$)~\cite{Buettiker:2003pp,Pelaez:2018qny}, and existing experimental data for these channels. The solution of the unsubtracted dispersion relation with the left-hand cut given by the pion (kaon) pole allowed one to cover the region of the scalar $f_0(980)$ resonance, which has a dynamical $\{\pi\pi, K\bar{K}\}$ origin. In order to reproduce the $f_2(1270)$ resonance in the on-shell $D$-wave data, higher left-hand cuts beyond the pion pole are required~\cite{GarciaMartin:2010cw}. In Ref.~\cite{Danilkin:2018qfn} they have been approximated with light vector-pole contributions, i.e., $\omega$ and $\rho$ exchanges. To describe the real-photon data, the radiative decay coupling $\omega/\rho\to \pi \gamma$ (in the $SU(3)$ limit) was fixed at the $f_2(1270)$ resonance position from the $\gamma\gamma\to \pi^0\pi^0$ cross section. Using the phenomenological couplings extracted from the radiative widths~\cite{Tanabashi:2018oca} does not exactly reproduce the on-shell $\gamma\gamma\to\pi\pi$ cross sections, pointing to a small correction from even heavier intermediate states in the left-hand cut and potentially inelastic effects in the $\pi\pi$ $D$-wave~\cite{Hoferichter:2019nlq}. The photon virtuality dependence in the vertex was included by the corresponding electromagnetic TFFs.  For the $\omega$ the dispersive analysis from Ref.~\cite{Danilkin:2014cra} was used (see also Ref.~\cite{Schneider:2012ez}), while for the sub-dominant $\rho$ contribution the VMD model was adopted. As a result, the first dispersive prediction of the cross sections for the finite spacelike $q^2$ including the $f_0(980)$ and $f_2(1270)$ regions was obtained~\cite{Danilkin:2018qfn}.

For $a_\mu$ the doubly-virtual $\gamma^*\gamma^*  \to \pi\pi$ process is needed, where compared to the singly-virtual case $\gamma\gamma^*  \to \pi\pi$ the number of helicity amplitudes increases from three to five and there is an additional complication related to the behavior of the
left-hand cuts at large spacelike virtualities. For $q_1^2 q_2^2 > (M_V^2-M_\pi^2)^2$ a narrow-width resonance of mass $M_V$ produces a singularity that moves from the unphysical sheet through the left-hand branch cut onto the physical sheet, a situation similar to anomalous thresholds in triangle diagrams~\cite{Hoferichter:2019nlq}. For virtualities below the anomalous point, preliminary plots of the $\gamma^*\gamma^*  \to \pi\pi$ cross sections were shown in Ref.~\cite{Danilkin:2019mhd}.\footnote{The preliminary curves shown in Ref.~\cite{Danilkin:2019mhd} suffered from a numerical instability in the calculation of one of the five dispersive integrals, which led to an overestimation of $\sigma_{LL}$ in the $f_2(1270)$ region by a factor $\sim 2.4$, leaving the predictions for $\sigma_{TT}$ and $\sigma_{TL}$ mainly unchanged~\cite{Danilkin:2019opj}.} 
In~\cite{Hoferichter:2019nlq}, the full doubly-virtual $D$-wave system was derived from the Roy--Steiner equations~\cite{Colangelo:2015ama} and subsequently solved using the modified Muskhelishvili--Omn\`es method introduced in Ref.~\cite{GarciaMartin:2010cw}. In particular, the solution of the complication of spacelike anomalous thresholds was given in terms of a path deformation, which enabled the calculation of all $\gamma^*\gamma^*\to\pi\pi$ $D$-waves for arbitrary virtualities. In Ref.~\cite{Danilkin:2019opj}, the singly-virtual analysis of Ref.~\cite{Danilkin:2018qfn} based on partial-wave dispersion relations was extended to the doubly-virtual case $\gamma^*\gamma^*\to\pi\pi$ where an alternative strategy for the treatment of the anomalous thresholds was shown. Below we provide a comparison of the two independent dispersive analyses Refs.~\cite{Danilkin:2018qfn,Danilkin:2019opj} and \cite{Colangelo:2017fiz,Hoferichter:2019nlq}.
The inclusion of these results into the dispersion relation for HLbL and the numerical evaluation of the contribution of $\pi\pi$ $D$-waves to $a_\mu$ is the focus of ongoing work.

\begin{figure}[t]
	\centering
	\includegraphics[width=7.5cm,page=3]{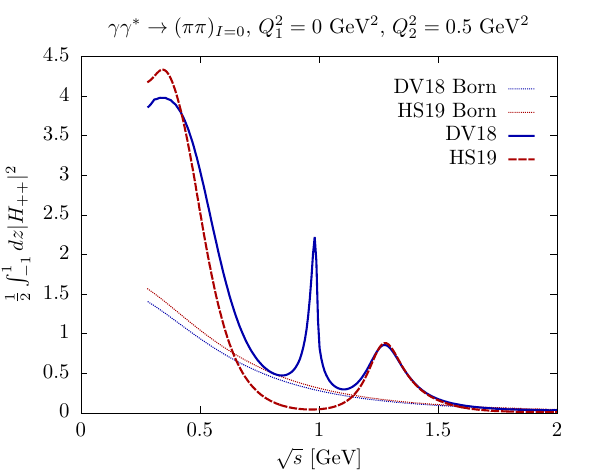}
	\includegraphics[width=7.5cm,page=4]{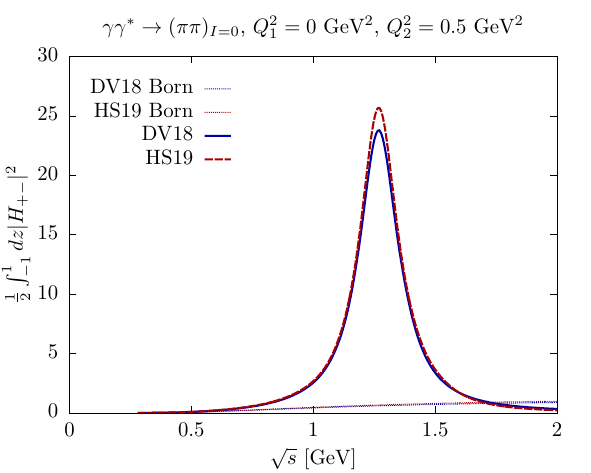}
	\includegraphics[width=7.5cm,page=5]{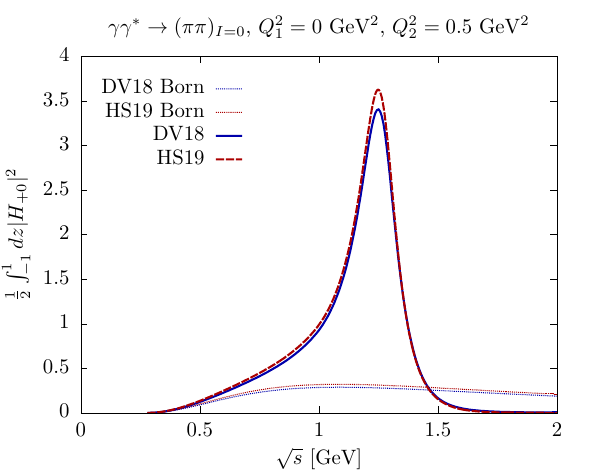}
	\caption{Helicity amplitudes for $\gamma\gamma^*\to\pi\pi$ in the isospin $I=0$ channel including $S$- and $D$-waves from DV18~\cite{Danilkin:2018qfn} and HS19~\cite{Hoferichter:2019nlq}, integrated over the scattering angle $z=\cos\theta$, in comparison to the Born results (dotted curves). The photon virtuality is $Q^2=0.5\GeV^2$.}
	\label{fig:ggpipiHelAmps2}
\end{figure}

\begin{figure}[t]
	\centering
	\includegraphics[width=7.5cm,page=6]{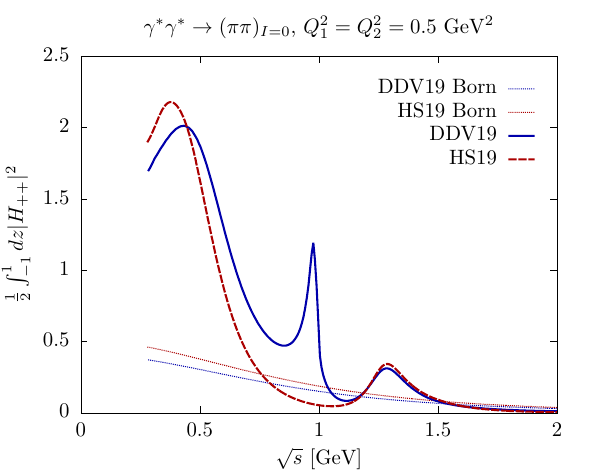}
	\includegraphics[width=7.5cm,page=7]{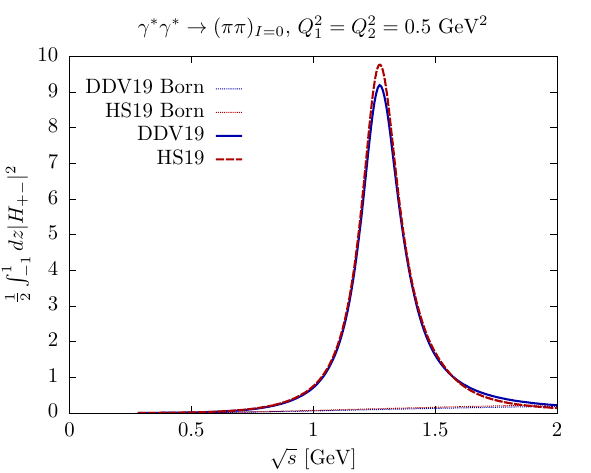}
	\includegraphics[width=7.5cm,page=8]{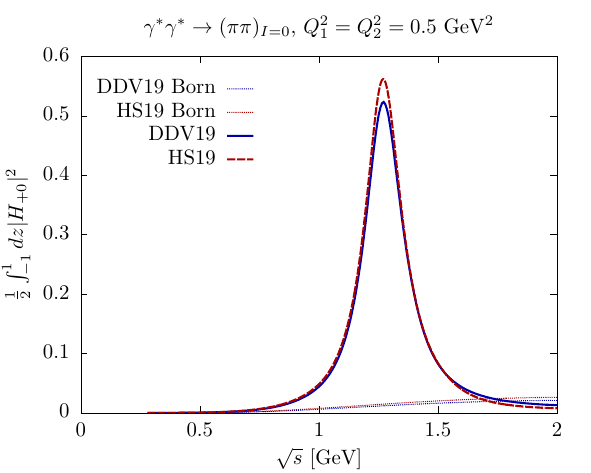}
	\includegraphics[width=7.5cm,page=9]{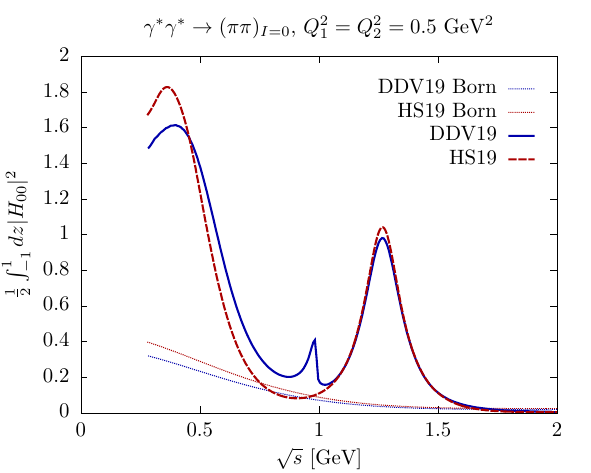}
	\caption{Helicity amplitudes for $\gamma^*\gamma^*\to\pi\pi$ in the isospin $I=0$ channel including $S$- and $D$-waves from HS19~\cite{Hoferichter:2019nlq} and DDV19~\cite{Danilkin:2019opj}, integrated over the scattering angle $z=\cos\theta$, in comparison to the Born results (dotted curves). The photon virtualities are $Q_1^2 = Q_2^2 =0.5\GeV^2$.}
	\label{fig:ggpipiHelAmps3}
\end{figure}

In \cref{fig:ggpipiCrossSection,fig:ggpipiCrossSection2,fig:ggpipiCrossSection3}, we compare the results for the $\gamma^{(*)}\gamma^{(*)}\to\pi\pi$ cross sections (for different virtualities) defined by
\begin{align}\label{Eq:Cross_section1}
&\frac{d \sigma_{TT}}{d\,\Omega}(\gamma^*\gamma^*\to \pi^+ \pi^-)=\frac{\alpha^2\,\sigma_{\pi}(s)}{8\,\lambda^{1/2}(s,-Q_1^2,-Q_2^2)}\left(|H_{++}|^2+|H_{+-}|^2\right)\,,\nonumber\\
&\frac{d \sigma_{TL}}{d \Omega}(\gamma^*\gamma^*\to \pi^+ \pi^-)=\frac{\alpha^2\,\sigma_{\pi}(s)}{4\,\lambda^{1/2}(s,-Q_1^2,-Q_2^2)}\, |H_{+0}|^2\,,\quad
\nonumber\\
&\frac{d \sigma_{LL}}{d \Omega}(\gamma^*\gamma^*\to \pi^+ \pi^-)=\frac{\alpha^2\,\sigma_{\pi}(s)}{4\,\lambda^{1/2}(s,-Q_1^2,-Q_2^2)}\, |H_{00}|^2\,,
\end{align}
where for the neutral pions one has to include an additional factor of $1/2$ and $\lambda$ is the K\"all\'en triangle function. The quantities $\sigma_{TT}$, $\sigma_{TL}$, $\sigma_{LT}$, and $\sigma_{LL}$
enter the cross section for the process $e^+e^{-}$ $\to$ $e^{+}e^{+}\pi\pi$ given in Refs.~\cite{Budnev:1974de,Pascalutsa:2012pr}. The latter sets the convention for the flux factor, while for the longitudinal polarization vectors the convention from Ref.~\cite{Danilkin:2019opj} is adopted. In \cref{fig:ggpipiHelAmps1,fig:ggpipiHelAmps2,fig:ggpipiHelAmps3}, we show the helicity amplitudes for $\gamma^{(*)}\gamma^{(*)}\to\pi\pi$ for the isospin $I=0$ channel, where the $f_0(500)$, $f_0(980)$ and $f_2(1270)$ resonances occur. We display the sum of $S$- and $D$-waves including unitarization, integrated over the full scattering angle. The Born terms are shown for comparison.
In \cref{fig:ggpipiCrossSection,fig:ggpipiCrossSection2,fig:ggpipiCrossSection3,fig:ggpipiHelAmps1,fig:ggpipiHelAmps2,fig:ggpipiHelAmps3} the theoretical curves from the dispersive analyses of Refs.~\cite{Danilkin:2018qfn,Danilkin:2019opj} and \cite{Colangelo:2017fiz,Hoferichter:2019nlq} are shown, with some additional adjustments as explained below.
Overall, we observe relatively good agreement between the two analyses. The deviations can be attributed to a different treatment of the vector-meson couplings~\cite{Danilkin:2018qfn,Hoferichter:2019nlq}, the external input for the form factors, as well as the inclusion of the coupled-channel $S$-wave in Refs.~\cite{Danilkin:2018qfn,Danilkin:2019opj}, which is responsible for the $f_0(980)$ peak. The curves in Ref.~\cite{Hoferichter:2019nlq} used pure VMD form factors with $M_V = 0.77\GeV$. For the curves shown in \cref{fig:ggpipiCrossSection2,fig:ggpipiCrossSection3}, the $\omega$ TFF was adjusted to reproduce the form-factor slope determined in the dispersive analysis~\cite{Schneider:2012ez}. The differences in the Born terms between the analyses Refs.~\cite{Danilkin:2018qfn,Danilkin:2019opj} and \cite{Hoferichter:2019nlq} reflect a slightly different mass in the VMD form factor. For more details on the dispersive predictions for the off-shell processes, we refer to Refs.~\cite{Danilkin:2018qfn,Hoferichter:2019nlq,Danilkin:2019opj}. The solution of the technical difficulties in the dispersion relations for $\gamma^*\gamma^*\to\pi\pi$~\cite{Hoferichter:2019nlq,Danilkin:2019opj} is independent of the form-factor input. The evaluation of the contribution to $a_\mu$ will rely on a refined analysis with dispersive input for the form factors, which can be validated by singly-virtual measurements. The BESIII collaboration is currently analyzing both $\pi^+ \pi^-$ and $\pi^0 \pi^0$ production in single-tag measurements in the $0.2\GeV^2 \lesssim Q^2 \lesssim 2\GeV^2$ range~\cite{Redmer:2017fhg}. 

\subsubsection{Comparison with earlier work}
\label{sec:TwoPionEarlierWork}

The available dispersive evaluations of two-pion contributions can be compared to ingredients of more model-dependent evaluations of the HLbL contribution. However, such a direct comparison is of very limited significance: in principle, one should only compare pole residues and discontinuities to avoid the model-dependent notion of off-shell hadronic intermediate states. In addition, even residues suffer from a basis dependence due to the presence of HLbL sum rules, as will be discussed in \cref{sec:higher_states}. This makes the separation into contributions from different intermediate states only unique for those contributions that fulfill the sum rules exactly.

The pion-box contribution \cref{eq:PionBox} can be contrasted with the pion-loop contribution, which had been rather controversial in the past. Estimates of pion and kaon loops were given as
\begin{align}
	a_\mu^{\pi,K\text{-loop}} &= -4.5(8.1) \times 10^{-11} \; \text{\cite{Hayakawa:1997rq}} \,, \quad -19(13) \times 10^{-11} \; \text{\cite{Bijnens:1995xf}} \,, \quad 0(10) \times 10^{-11} \; \text{\cite{Melnikov:2003xd}}\, ,
\end{align}
where the charged-pion loop dominates and the kaon-loop contribution is responsible for approximately 5\% of the above estimates. The estimate of Ref.~\cite{Bijnens:1995xf} was later refined to
\begin{align}
	a_\mu^{\pi\text{-loop}} = -20(5) \times 10^{-11} \quad \text{\cite{Bijnens:2016hgx}} \,,
\end{align}
which includes the effect of pion polarizabilities.

The low-energy $\pi\pi$ $S$-wave contribution \cref{eq:pipiSwave} might be compared to a model-dependent scalar-exchange contribution:
\begin{align}
	a_\mu^{\text{scalar}} &= -6.8(2.0) \times 10^{-11} \quad \text{\cite{Bijnens:1995xf}}\,.
\end{align}
According to Ref.~\cite{Bijnens-private}, this contribution in the ENJL model~\cite{Bijnens:1995xf} can be interpreted to be related to a large part to the broad $f_0(500)$ resonance and therefore to $\pi\pi$-rescattering effects.
The model estimate of Ref.~\cite{Jegerlehner:2017gek} for scalar contributions
\begin{align}
    \label{eq:ScalarsJegerlehner}
    a_\mu^{\text{scalars}}[a_0,f_0,f_0'] &= -6.0(1.2) \times 10^{-11} \quad \text{\cite{Jegerlehner:2017gek}}
\end{align}
includes the $f_0(500)$ as well as the narrow scalars $f_0(980)$ and $a_0(980)$. The $f_0(500)$ scalar is responsible for $\sim50\%\hyph85\%$ of \cref{eq:ScalarsJegerlehner}. In the narrow-width model of Ref.~\cite{Knecht:2018sci}, the contribution of the broad $f_0(500)$ was estimated to lie in the range
\begin{align}
    a_\mu^{\text{scalar}}[f_0(500)] &= [ -(3.1^{+0.8}_{-0.7}) , -(0.3^{+0.4}_{-0.8})] \times 10^{-11} \quad \text{\cite{Knecht:2018sci}}\,.
\end{align}

The scalar contributions beyond the $f_0(500)$ region have been estimated in Refs.~\cite{Pauk:2014rta,Knecht:2018sci}, see \cref{sec:NarrowResonances}.
Finally, the $\pi\pi$ $D$-wave contribution could also be compared to a narrow-width approximation for the $f_2(1270)$ tensor meson (which has a width of $186.7_{-2.5}^{+2.2}\MeV$~\cite{Tanabashi:2018oca}). Existing estimates~\cite{Pauk:2014rta,Danilkin:2016hnh} are based on propagator expressions for narrow tensor resonances. The required input for the resonance TFFs is constrained by using HLbL forward-scattering sum rules and new data by the Belle collaboration~\cite{Masuda:2015yoh}, see \cref{sec:NarrowResonances}.

\subsubsection{Conclusion}
\label{sec:TwoPionConclusion}

The dispersive formalism for the HLbL tensor of Refs.~\cite{Colangelo:2014dfa,Colangelo:2015ama} allowed for the first time a model-independent definition of the contribution of one- and two-particle intermediate states. The clean definition of separate contributions allows one to significantly reduce uncertainties, as in the case of the pion box. To complete this program, one has to make sure that the summation of the hadronic intermediate states at low energies leads to a description that reproduces or at least can be matched to the asymptotic behavior demanded by QCD, see \cref{sec:asymptotic}.

The dispersive framework for HLbL relates the pion-box contribution to the pion VFF in the spacelike region. With the solution for the pion VFF from an extended Omn\`es solution~\cite{Leutwyler:2002hm,Colangelo:2003yw,Colangelo:2018mtw}, the pion-box contribution is determined as~\cite{Colangelo:2017qdm,Colangelo:2017fiz}
\be
	\label{eq:PionBoxSummary}
	a_\mu^{\pi\text{-box}}=-15.9(2)\times 10^{-11}\,.
\ee
Since the relation to the pion VFF is exact, the only uncertainty comes from the dispersive determination of the pion VFF itself, which is negligible in the present context.

Two-pion contributions beyond the pion box are due to $\pi\pi$-rescattering effects in the direct channel as well as heavier left-hand cuts in the crossed channel $\gamma^*\pi\to\gamma^*\pi$. So far, the $S$-wave low-energy rescattering contribution has been calculated, in an approximation that only takes the pion pole as a left-hand cut into account and cuts off the $f_0(980)$ resonance (and higher scalars)~\cite{Colangelo:2017qdm,Colangelo:2017fiz}:
\be
	\label{eq:pipiSwaveSummary}
	a_{\mu,J=0}^{\pi\pi,\pi\text{-pole LHC}}=-8(1)\times 10^{-11} \,.
\ee
At the moment, no estimate of the impact of heavier states in the left-hand cut on the $S$-wave contribution is available, but it is assumed to be moderate~\cite{Colangelo:2017fiz}. We abstain from guessing its value and assume this contribution to be covered by the uncertainty estimates assigned to heavier states and the asymptotic contribution. For the contribution of $D$- and higher partial waves, these heavier left-hand cuts need to be taken into account, but a model-independent dispersive evaluation of the contribution to $a_\mu$ is not yet available. Therefore, we consider all two-pion contributions beyond \cref{eq:PionBoxSummary,eq:pipiSwaveSummary} to be covered by the discussion in \cref{sec:higher_states}.

\subsection{Contribution of higher hadronic intermediate states}
\label{sec:higher_states} 

The framework for two-pion intermediate states discussed in \cref{sec:two-pion} can in principle be generalized directly to other two-particle intermediate states, such as $K \bar K$, $\pi\eta$, $\eta\eta$. Probably, a generalization to intermediate states of three and more particles cannot be done with the same level of rigor, necessitating further approximations.
A systematic numerical analysis of intermediate states beyond two pions within the framework defined in \cref{sec:tensor} is not yet available. In this section, we discuss possible generalizations and existing estimates that could be combined with the dispersive framework, and we provide some observations on the suppression of higher intermediate states.

\subsubsection{Kaon box, two-kaon, $\pi \eta$, and $\eta \eta$ intermediate states}

When considering two-meson intermediate states beyond two pions, we can be guided by the strength of the measured $\gamma\gamma \to M M$ cross sections in comparison with the $\gamma\gamma \to \pi \pi$ cross section to get an idea of the most important channels contributing to $a_\mu$, although the weighting by kernel functions in the integrals for $a_\mu$ might change the naive picture. 
In the region around 1\,GeV, corresponding to the scalar resonances $f_0(980)$ and $a_0(980)$, the largest measured cross sections come from the $\pi^+ \pi^-$ channel (around 120\,nb), followed by the $\pi^0 \eta$ channel (around 40\,nb). In the region of the tensor resonances $f_2(1270)$ and $a_2(1320)$, the two-photon cross sections for the $\pi^+ \pi^-$, $\pi^0 \eta$, and $K^+ K^-$ channels amount to around 300, 45, and 40\,nb, respectively. Other two-meson channels, such as $K^0 \bar K^0$ or $\eta \eta$ have much smaller cross sections in the same energy range.  
For the contributions to $a_\mu$ we can therefore expect, beyond the $\pi \pi$ channels, the $\pi^0 \eta$ channel to be the next important channel, followed by $K^+ K^-$. 

The exact same formalism of two-particle intermediate states discussed in \cref{sec:two-pion} for two pions can in principle be applied directly to the case of two kaons. Selecting in the light-by-light unitarity relation the two-kaon intermediate state leads to a relation between the light-by-light two-kaon discontinuity and the sub-process $\gamma^*\gamma^*\to K\bar K$ (in the charge basis, the two kaons are either $K^+K^-$ or $K^0 \bar K^0$). Selecting in the crossed-channel unitarity relation for $\gamma^*K\to\gamma^*K$ the one-kaon intermediate state defines the double-spectral contribution of the kaon box. In complete analogy to the pion box, the kaon box can be written as a one-loop sQED Feynman integral, multiplied by kaon VFFs for the off-shell photons. In the limit $q_4\to0$, e.g., one easily finds
\begin{align}
    \label{eq:KaonBoxFeynmanParametrization}
	\bar \Pi_i^{K\text{-box}}(q_1^2,q_2^2,q_3^2) = F_K^V(q_1^2) F_K^V(q_2^2) F_K^V(q_3^2) \frac{1}{16\pi^2} \int_0^1 dx \int_0^{1-x} dy \, I_i^K(x,y) \,,
\end{align}
where $I_i^K$ is given by analogous expressions as for the pion case, \cref{Ipibox}.

In the case of the pion VFF, a dispersive parameterization that is fit to $e^+e^-$ data~\cite{Colangelo:2017fiz,Colangelo:2018mtw} turns out to be rather close to a simple VMD parameterization of the form factor. The pion-box result with a VMD form factor
\begin{align}
	F_{\pi,\mathrm{VMD}}^V(s) = \frac{M_\rho^2}{M_\rho^2-s} 
\end{align}
is given by
\be
	\label{eq:PionBoxVMD}
	a_\mu^{\pi\text{-box,VMD}}=-16.4\times 10^{-11} \,,
\ee
deviating only by $3\%$ from the pion-box result \cref{eq:PionBox} with a dispersive VFF. Assuming that also in the case of the kaon a VMD parameterization of the VFF
\be
	F_{K^+,\mathrm{VMD}}^V(s) = 1 + \frac{s}{2} \left( \frac{1}{M_\rho^2-s} + \frac{1}{3} \frac{1}{M_\omega^2-s} + \frac{2}{3} \frac{1}{M_\phi^2-s} \right)
\ee
works reasonably well, we find the VMD estimate
\be
	a_\mu^{K^+\text{-box,VMD}} = -0.50 \times 10^{-11} \,.
\ee
A recent calculation based on Dyson--Schwinger equations for the kaon VFF~\cite{Eichmann:2019bqf} obtains a result very close to the VMD estimate,
\be
    a_\mu^{K^+\text{-box,DSE}} = -0.48(2)(4) \times 10^{-11} \quad \text{\cite{Eichmann:2019bqf}}\,.
\ee
The first error is due to the variation of the model parameters. The second error is a numerical uncertainty due to the employed nine-dimensional Monte Carlo integration. It can be eliminated by using the master formula and the Feynman parameterization \cref{eq:KaonBoxFeynmanParametrization} from~\cite{Colangelo:2015ama,Colangelo:2017fiz} together with the kaon VFF input from~\cite{Eichmann:2019bqf}. This leads to
\be
	a_\mu^{K^+\text{-box,DSE}} = -0.46(2) \times 10^{-11} \,,
\ee
where the error is only due to the variation of the model parameters and does not include possible further intrinsic model uncertainties.
These estimates could be improved by implementing a model-independent dispersive analysis of the kaon vector form factor~\cite{Blatnik:1978wj}. 

In principle, there is also the contribution of the neutral kaon box. 
However, a VMD representation~\cite{Ecker:1988te} of the neutral kaon vector form factor (the $SU(3)$ singlet is included assuming ideal mixing)
\be
	F_{K^0,\mathrm{VMD}}^V(s) = \frac{s}{2} \left( - \frac{1}{M_\rho^2-s} + \frac{1}{3} \frac{1}{M_\omega^2-s} + \frac{2}{3} \frac{1}{M_\phi^2-s} \right)
\ee
leads to a completely negligible contribution of
\be
	a_\mu^{K^0\text{-box,VMD}} = -1.2 \times 10^{-15} \,.
\ee

A coupled-channel solution for the helicity partial waves for $\gamma^*\gamma^*\to\pi\pi$ and $\gamma^*\gamma^*\to K\bar K$ would allow one to calculate the kaon-rescattering contribution using the same partial-wave formalism as for the pion-rescattering contribution. Such an analysis could be used to extend the calculation of the low-energy $S$-wave $\pi\pi$ contribution~\cite{Colangelo:2017qdm,Colangelo:2017fiz} to higher energies and include the $f_0(980)$ resonance, which strongly couples to both two-pion and two-kaon states.

Turning to the second most important two-meson channel $\gamma^*\gamma^*\to \pi\eta$, a description of the 
$a_0(980)$ resonance requires a coupled-channel treatment of the $\{\pi\eta, K\bar{K}\}$ system. It was found that the inclusion of $K\bar{K}$ intermediate states appears to be necessary in order to describe $\gamma\gamma \to \pi\eta$ data~\cite{Oller:1997yg,Danilkin:2012ua,Danilkin:2017lyn}.  
In a recent dispersive analysis~\cite{Danilkin:2017lyn} the left-hand cuts coming from the $t$- and $u$ - channel vector-meson exchanges were tested against the data in the crossed process, the $\eta \to \pi^0 \gamma \gamma$ decay. The $a_2(1320)$ resonance was taken into account explicitly within the assumption that it is predominantly produced by the helicity-2 state (similar to the $f_2(1270)$). Together with the proposed dispersive method for the $a_0(980)$ this yields a parameter-free description of the $\gamma\gamma\to\pi^0\eta$ cross section~\cite{Danilkin:2017lyn}, which is in  reasonable agreement with the data from the Belle collaboration~\cite{Uehara:2009cf}. An extension to the singly- and doubly-virtual processes of $\gamma^* \gamma^* \to \pi\eta$ is also under development~\cite{Deineka:2018nuh}, and can be tested with future data from BESIII~\cite{Redmer:2018gah}.

\subsubsection{Estimates of higher scalar and tensor resonances}
\label{sec:NarrowResonances}

The unitarity relation involves a sum over all possible asymptotic intermediate states. In practice, for the HLbL tensor, we are not able to perform this complete sum: higher multi-particle intermediate states lead to more and more complicated topologies for the hadronic four-point function. Hence the goal of the dispersive approach is to determine the contribution of the lightest intermediate states as precisely as possible. The contribution of hadronic states in the energy regime around $1\ldots2\GeV$ must be estimated using additional approximations. This should be done in a way that is compatible with the dispersive framework explained in \cref{sec:tensor}, in particular a double-counting of contributions calculated dispersively must be avoided. For the contribution of even heavier states, the matching to asymptotic constraints has to be considered, see \cref{sec:asymptotic}.

Some basic relations of the partial-wave framework of Ref.~\cite{Colangelo:2017fiz} are valid beyond two-particle intermediate states. E.g., fixed-$s$/-$t$/-$u$ dispersion relations for the scalar basis functions of the light-by-light tensor can be written without yet referring to the unitarity relation. The basis change to light-by-light helicity amplitudes is generic as well. The expansion into helicity partial waves can be generalized to arbitrary intermediate states by including also odd partial waves. The contribution of multi-particle intermediate states beyond two pions is expected to be suppressed by phase space and the higher thresholds. The expectation is that these contributions are negligible unless the light-by-light helicity partial wave is resonantly enhanced.

The contribution of resonant partial waves is tractable if the resonance is narrow. In the narrow-width approximation, the spectral function of the partial wave becomes a $\delta$ function and the dispersive integral collapses to a resonance-pole contribution. In analogy to the pole contribution of the light pseudoscalars, the contribution of resonance poles can then be expressed in terms of (on-shell) TFFs $F_{R\gamma^*\gamma^*}(q_1^2,q_2^2)$.

Such narrow-resonance calculations may provide an estimate for the size of higher resonance contributions, and consequently the resulting uncertainty on $a_\mu$. 
As an example, recent estimates for the narrow scalars with masses $\gtrsim$~1\,GeV yield values in the range 
\begin{align}
	\label{eq:narrowScalars}
	a_\mu^\mathrm{scalars} &= [-3.1(8), -0.9(2)] \times 10^{-11} \quad \text{\cite{Pauk:2014rta}}\,, \nn
	a_\mu^\mathrm{scalars} &= [- (2.2^{+3.2}_{-0.7}),  -(1.0^{+2.0}_{-0.4})] \times 10^{-11} \quad \text{\cite{Knecht:2018sci}}\,.
\end{align}
The numbers from Ref.~\cite{Pauk:2014rta} give the contribution of the scalars $f_0(980)$, $a_0(980)$, and $f_0(1370)$, whereas the numbers from Ref.~\cite{Knecht:2018sci} in addition include the narrow scalars $a_0(1450)$ and $f_0(1500)$. The numbers in \cref{eq:narrowScalars} exclude the broad $f_0(500)$ resonance, which we discussed in \cref{sec:two-pion}.

A first estimate of the contributions of the tensor mesons $f_2(1270)$, $f_2(1565)$, $a_2(1320)$, and $a_2(1700)$ to $a_\mu$ in the narrow-resonance approximation was performed in Ref.~\cite{Pauk:2014rta}. The dominant helicity-2 TFFs of the tensor mesons were constrained from forward light-by-light sum rules. This estimate was updated in Ref.~\cite{Danilkin:2016hnh} in light of new data on the $f_2(1270)$ TFFs from the Belle collaboration, by allowing the production of tensor states in helicity-0, -1, and -2 states. 
The resulting contribution to $a_\mu$ was found to be
\be
	\label{eq:tensors}
	a_\mu^\mathrm{tensors} = 0.9(1) \times 10^{-11} \quad \text{\cite{Danilkin:2016hnh}}\,. 
\ee

The narrow-width approximation can be tested in the case of the resonances $f_0(980)$ and $f_2(1270)$ by comparing with a dispersive determination in terms of two-pion and two-kaon intermediate states~\cite{CHPSinProgress}. In such a comparison, the residue of the resonance pole should be extracted for the same set of scalar functions that is used in the dispersive determination of the two-pion/-kaon contribution. Unfortunately, the contribution of resonance poles depends on the choice of the tensor basis. In Ref.~\cite{Colangelo:2017fiz} it was shown that $a_\mu$ is not affected by a basis change provided that the HLbL tensor fulfills a set of sum rules for general fixed-$s$/-$t$/-$u$ kinematics, which generalize the sum rules for forward scattering~\cite{Pascalutsa:2012pr}. Since a single resonance does not saturate the light-by-light sum rules, its contribution to $a_\mu$ is basis dependent. Only the contribution of a tower of resonances that saturates the sum rules will be basis independent. This complication will have to be taken into account in a comparison with the existing narrow-width estimates~\cite{Pauk:2014rta,Danilkin:2016hnh}.

The present estimates of scalars and tensors could be confronted with future evaluations of two-pion/-kaon contributions including the effects of the resonances $f_0(980)$ and $f_2(1270)$ in $S$- and $D$-waves, see \cref{sec:TwoPionEarlierWork}.

\subsubsection{Axial-vector-meson contributions}
\label{sec:axials}

A potentially important contribution to $a_\mu$ beyond the two-meson channel input  comes from axial-vector mesons. 
Although the production of an axial-vector meson by two real photons is forbidden by the Landau--Yang theorem~\cite{Landau:1948kw,Yang:1950rg}, an axial-vector meson can be produced when one or both photons are virtual, and thus must be taken into account in the HLbL contribution to $a_\mu$. Early estimates in the ENJL~\cite{Bijnens:1995xf} and HLS models~\cite{Hayakawa:1995ps,Hayakawa:1996ki} found contributions to $a_\mu$ in the range 
$(2 - 3) \times 10^{-11}$.
In Ref.~\cite{Melnikov:2003xd}, OPE constraints on the light-by-light tensor were derived, see \cref{sec:asymptotic}, and a model was constructed where the saturation of the asymptotic constraints was achieved by dropping the momentum dependence of the singly-virtual TFFs, thereby introducing substantial model dependence at low energies. The axial-vector contributions in this model due to $a_1(1260)$, $f_1(1285)$, and $f_1(1420)$ were ten times larger than the previous estimates. Such a value corresponds to about 20\% of the total HLbL contribution. 

A general discussion of the $A \gamma^\ast \gamma^\ast$ vertex leads to three independent Lorentz structures and form factors that are subject to symmetry constraints, so that the Landau--Yang result is satisfied. The axial contribution was reevaluated in Refs.~\cite{Pauk:2014rta,Jegerlehner:2017gek}, which incorrectly claimed that the calculation of Ref.~\cite{Melnikov:2003xd} was in contradiction with the Landau--Yang theorem.\footnote{A closer inspection in fact shows that the model of Ref.~\cite{Melnikov:2003xd} respects the Landau--Yang theorem~\cite{Hoferichter-Seattle,Hoferichter:2020lap}, as recently confirmed in Ref.~\cite{Roig:2019reh}.} The L3 collaboration measured the two-photon fusion to $f_1(1285)$ and $f_1(1420)$ at LEP using the decays to $\pi^+\pi^-\eta$~\cite{Achard:2001uu} and $K^0_S K^\pm \pi^\mp$~\cite{Achard:2007hm}. The $Q^2$ dependence of one linear combination of TFFs was studied assuming the squared transverse momentum of the reconstructed final state to be equivalent to $Q^2$. In Ref.~\cite{Pauk:2014rta}, this empirical information was used to model the TFFs of the $f_1(1285)$ and $f_1(1420)$, leading to a contribution to $a_\mu$ of
\be
	\label{eq:axialsPV}
	a_\mu^\mathrm{axials}[f_1, f_1'] = 6.4(2.0) \times 10^{-11} \quad  \text{\cite{Pauk:2014rta}} \,.
\ee
Ref.~\cite{Jegerlehner:2017gek} has also estimated the $a_1(1260)$ contribution (responsible for $25\%$ of the estimated axial contribution) and imposed short-distance constraints with the result
\be
	\label{eq:axialsJ}
	a_\mu^\mathrm{axials}[a_1, f_1, f_1'] = 7.6(2.7) \times 10^{-11} \quad  \text{\cite{Jegerlehner:2017gek}} \,.
\ee
Both results are around 3 times smaller than the estimate of Ref.~\cite{Melnikov:2003xd}.
Recently, a new calculation of the axial-vector contribution was performed using resonance chiral theory~\cite{Roig:2019reh}, obtaining 
\be
	\label{eq:axialsRS}
	a_\mu^\mathrm{axials}[a_1, f_1, f_1'] = (0.8_{-0.8}^{+3.5}) \times 10^{-11} \quad  \text{\cite{Roig:2019reh}}\, .
\ee
The even smaller value obtained in this model is explained by the fact that resonance chiral theory at the considered order only contributes to antisymmetric form factors, while in Refs.~\cite{Achard:2007hm,Pauk:2014rta} only the symmetric part of one form factor was retained and parameterized by a dipole form. At low energies the antisymmetric part of the form factors only gives a suppressed contribution to $a_\mu$, leading to the substantially smaller value of \cref{eq:axialsRS}. In resonance chiral theory, the symmetric form factor of Refs.~\cite{Achard:2007hm,Pauk:2014rta} appears at higher chiral/large-$N_c$ order~\cite{Roig:2019reh}. The uncertainty estimate in \cref{eq:axialsRS} is dominated by the poorly known asymptotic behavior of the form factors, and also covers uncertainties due to the neglected symmetric form factor.

In Refs.~\cite{Leutgeb:2019gbz,Cappiello:2019hwh}, the contribution of a tower of axial-vector resonances was considered within holographic-QCD models in order to satisfy short-distance constraints. The contribution of the lightest axial multiplet was given in Ref.~\cite{Leutgeb:2019gbz} as
\be
    a_\mu^\mathrm{axials}[a_1,f_1,f_1'] = 17.4(4.0) \times 10^{-11} \quad \text{\cite{Leutgeb:2019gbz}} \,.
\ee
However, this result is entangled with the contribution that ensures the fulfillment of short-distance constraints discussed in \cref{sec:asymptotic}, in particular its longitudinal component is part of $\Delta a_\mu^\mathrm{LSDC}$ in \cref{HoloLong1}. Hence, we do not include this result in our estimate of the lowest axial-vector contribution. In Ref.~\cite{Leutgeb:2019gbz}, $58\%$ of the contribution of the whole tower of axial-vector resonances are due to the longitudinal component. The difficulty in the bookkeeping between axial-vector contributions and short-distance constraints is related to the necessary matching procedure. This entanglement will be reflected in a linear combination of the uncertainties, see the detailed discussion in \cref{sec:result_HLbL_DR}.

\subsubsection{Conclusion}
\label{sec:HigherStatesConclusion}

At present, contributions beyond the pion box and the low-energy $S$-wave $\pi\pi$-rescattering discussed in \cref{sec:two-pion} are not calculated in a model-independent dispersive framework. The most important contributions of heavier states concern scalars beyond the low-energy region (the $f_0(500)$ is covered by the $S$-wave $\pi\pi$-rescattering contribution), axial vectors, and tensor mesons. These contributions have been estimated mainly within a narrow-width approximation and based on Lagrangian formulations for the contribution to HLbL, which may differ from dispersively defined residues depending on the choice of basis, as discussed in \cref{sec:NarrowResonances}. The narrow-width approximation relates the contribution to $a_\mu$ to TFFs for the resonances. Experimental information on these TFFs is available only for singly-off-shell kinematics and in the case of axials and tensors only for a subset of the independent TFFs. Current estimates therefore model the full off-shell dependence of the TFFs as a factorizing monopole or dipole.

The impact of the mentioned approximations and model dependences is difficult to quantify. However, at least in the case of scalar and tensor resonances in the two-pion channel, the situation is expected to improve with future extensions of the dispersive framework. At present, the existing published estimates \cref{eq:narrowScalars,eq:tensors,eq:axialsPV,eq:axialsJ,eq:axialsRS} suggest
\begin{align}
	a_\mu^\mathrm{scalars+tensors} &= -1(3) \times 10^{-11}\, , \nn
	a_\mu^\mathrm{axials} &= 6(6) \times 10^{-11} \,.
\end{align}
The central value for scalars and tensors is given by adding up the average of \cref{eq:narrowScalars} with \cref{eq:tensors}, the central value for the axials is given by an unweighted average of \cref{eq:axialsPV} (increased by 33\% to account for the contribution of the $a_1$), \cref{eq:axialsJ}, and \cref{eq:axialsRS}. The errors are chosen to cover the range of these model estimates but do not include unquantified uncertainties due to approximations and model dependences. In order to take them into account, one might inflate the uncertainties from the existing estimates by an arbitrary amount, e.g., by a factor of $2$ as in Ref.~\cite{Jegerlehner:2017gek}, but the situation remains unsatisfactory and calls for a dedicated effort to obtain a model-independent determination of these contributions.

\subsection{Asymptotic region and short-distance constraints}
\label{sec:asymptotic}

The asymptotic regime is not expected to give a large contribution
to HLbL but it plays an important role in the estimate of the error. 
The main object underlying HLbL is the Green function of four vector currents
defined in \cref{eq:defPI}. HLbL itself is then given by \cref{eq:MasterFormula3Dim}. 
The previous sections discuss in detail how
intermediate states of increasing mass contribute. For higher
virtualities $q_i^2$, more and more intermediate states can
contribute and these become progressively more difficult to estimate using
a dispersive approach. Here one needs to start using QCD directly as much as
possible.

The approach with summing over intermediate states as exchange
of resonances cannot reproduce all
QCD constraints unless an infinite number of states is included. The reason is that the resonance couplings to off-shell
photons satisfy QCD constraints, usually called Brodsky--Lepage (BL) constraints~\cite{Lepage:1980fj,Brodsky:1981rp},
and that these require the full four-point Green function to have too strong
a falloff at high virtualities for a finite number of tree level exchanges. 
This was shown in Ref.~\cite{Bijnens:2003rc} for a three-point function, but a
similar argument applies here.

SDCs are used in several different ways in HLbL.
First are the constraints from the operator product expansion (OPE) and BL arguments
on the form factors involved in coupling hadrons to photons. These are useful
in obtaining the high-virtuality contributions from the individual
intermediate states as discussed in earlier sections. Second, we have SDCs on the full four-point function with all four legs highly virtual.
These can be used to see if a particular set of intermediate states is
sufficient. This regime is less relevant for HLbL since we require one of the
legs at zero momentum. Third is the regime where one of the $q_i^2$,
say $q_3^2$, is much smaller than the other two, implying from momentum
conservations that $q_1^2\sim q_2^2$ (for $q_4=0$). The SDC for this
case has been derived in Ref.~\cite{Melnikov:2003xd}. 
Finally, there are contributions from the regime where all three virtualities
are large, here the main contribution is the quark loop and possible
higher-order corrections as derived in Ref.~\cite{Bijnens:2019ghy}. A further important observation is how fast short-distance contributions are expected to fall with a lower cutoff $Q_{\mathrm{min}}$. For sizable $q_i^2$ quark masses play a minor role. The scale of contributions must then be a function of $m_\mu^2/q_i^2$. We can expand the functions $T_i(Q_1,Q_2,\tau)$ of \cref{eq:MasterFormula3Dim} in this case and obtain $T_1\sim m_\mu^4/q_i^4$ and all others $T_i\sim m_\mu^2/q_i^2$. With no other scale involved we thus expect contributions to fall as $1/Q_{\mathrm{min}}^2$.

\subsubsection{Derivation of the short-distance constraints}

The derivation of HLbL SDCs is based on the OPE for configurations in which some of the virtualities become
large. Since always $q_4^2=0$, this leaves two distinct kinematic configurations that are relevant for HLbL scattering: either all remaining virtualities are large,
$q_1^2\sim q_2^2\sim q_3^2$,
or one is much smaller than the others, $q_3^2\ll q_1^2\sim q_2^2$. 
In the first case, when we take the standard first term in the OPE with all Wick contractions the free propagator (i.e., the free quark loop), the limit $q_1+q_2+q_3\to0$ can be taken. The next term involving $m_q\langle\bar qq\rangle$
however does not allow this limit~\cite{Bijnens:2019ghy}. An alternative OPE where the $q_4$ is treated as a background external field is needed here. The first two nonzero orders have been worked out in Ref.~\cite{Bijnens:2019ghy}. The leading term is indeed analytically the same as the quark loop. The quark loop is thus the first contribution in a systematic expansion. The next term is suppressed by two powers of the large scale and is proportional to $m_q X_q$ with $\left\langle \bar q \sigma_{\alpha\beta}q\right\rangle = e_q F_{\alpha\beta} X_q$ the external field induced condensate. $X_q$ has been determined for the light quarks in lattice QCD~\cite{Bali:2012jv}. The numerical contribution of the next term is negligible since $m_q$ and $X_q$ are numerically small~\cite{Bijnens:2019ghy}. The numerical results for the quark loop are discussed below in \cref{sec:quark_loop} and in Ref.~\cite{Bijnens:2019ghy}.

The SDCs for the second configuration, $q_3^2\ll q_1^2\sim q_2^2$, were studied in detail in Ref.~\cite{Melnikov:2003xd}. A key role is played by the $VVA$ correlator, based on the observation that the OPE of the two vector currents corresponding to $q_1$ and $q_2$ produces an axial current\footnote{This expansion assumes $q_1+q_2\ne0$, otherwise there are additional terms.}
\begin{align}
\label{asymptoticjj}
\Pi_{\mu\nu}(q_1,q_2)&=i \int d^4x \, d^4y \, e^{-i(q_1 \cdot x + q_2 \cdot y)}T \{ j_\mu(x) j_\nu(y)\}\notag\\
&=\int d^4z \, e^{-i(q_1 + q_2)\cdot z}\bigg[-\frac{2i}{\hat q^2}\epsilon_{\mu\nu\alpha\beta}\hat q^\alpha j_5^\beta(z)
+\order\big(\hat q^{-2}\big)\bigg]\,,
\end{align}
with currents defined as
\be
j^\mu(x) = \bar \psi(x) \mathcal{Q} \gamma^\mu \psi(x)\,, \qquad
j_5^\mu(x)=\bar \psi(x)\mathcal{Q}^2\gamma^\mu\gamma_5\psi(x)\,, \qquad
\psi = ( u , d, s )^T\,,
\ee
as well as $\mathcal{Q} = \text{diag}\big(2,-1,-1\big)/3$, $\hat q=(q_1-q_2)/2$, and $\epsilon^{0123}=+1$. Following similar work in the context of the electroweak contribution to $(g-2)_\mu$~\cite{Knecht:2002hr,Czarnecki:2002nt}, 
the analysis of Ref.~\cite{Melnikov:2003xd} is then based on nonrenormalization theorems for the remaining $VVA$ correlator $W_{\mu\nu\rho}(q_1,q_2)$~\cite{Vainshtein:2002nv,Knecht:2003xy}. The HLbL tensor becomes
\begin{align}
\Pi_{\mu\nu\lambda\sigma}(q_1,q_2,q_3)
&=\frac{2i}{\hat q^2}\epsilon_{\mu\nu\alpha\beta}\hat q^\alpha\int d^4 x\, d^4 y\, 
e^{-iq_3\cdot x}e^{iq_4\cdot y}\langle 0 | T \{j_\lambda(x)j_\sigma(y)j_5^\beta(0) \} | 0 \rangle\notag\\
&=\frac{8}{\hat q^2}\epsilon_{\mu\nu\alpha\beta}\hat q^\alpha W_{\lambda\sigma}^\beta(-q_3,q_4)\sum_{a=0,3,8} C_a^2\,,
\end{align}
with flavor
\be
C_a=\frac{1}{2}\Tr(\mathcal{Q}^2\lambda_a)\,,\qquad C_3=\frac{1}{6}\,,\qquad C_8=\frac{1}{6\sqrt{3}}\,,\qquad C_0=\frac{2}{3\sqrt{6}}\,,
\ee
and Lorentz decomposition
\begin{align}
\label{VVA_decomposition}
 W_{\mu\nu\rho}(q_1,q_2)&=\frac{1}{8\pi^2}\bigg[w_L\big(q_1^2,q_2^2,(q_1+q_2)^2\big)\epsilon_{\mu\nu\alpha\beta}q_1^\alpha q_2^\beta(q_1+q_2)_\rho
 -w_T^+\big(q_1^2,q_2^2,(q_1+q_2)^2\big)t_{\mu\nu\rho}^+\notag\\
 &-w_T^-\big(q_1^2,q_2^2,(q_1+q_2)^2\big)t_{\mu\nu\rho}^-
 -\tilde w_T^-\big(q_1^2,q_2^2,(q_1+q_2)^2\big)\tilde t_{\mu\nu\rho}^-\bigg]\,,
\end{align}
in terms of longitudinal and transverse structure functions $w_L$, $w_T^\pm$, $\tilde w_T^-$ (see Ref.~\cite{Knecht:2003xy} for the definition of $t_{\mu\nu\rho}^\pm$, $\tilde t_{\mu\nu\rho}^-$).
The axial anomaly provides the constraint
\be
\label{w_L}
w_L(q_1^2,q_2^2,(q_1+q_2)^2)=\frac{6}{(q_1+q_2)^2}\,,
\ee
and the full set of transverse nonrenormalization theorems reads~\cite{Knecht:2003xy}
\begin{align}
\label{non_renormalization}
 0&=(w_T^+ + w_T^-)\big(q_1^2,q_2^2,(q_1+q_2)^2\big)-(w_T^+ + w_T^-)\big((q_1+q_2)^2,q_2^2,q_1^2\big)\,,\notag\\
 0&=(\tilde w_T^- + w_T^-)\big(q_1^2,q_2^2,(q_1+q_2)^2\big)+(\tilde w_T^- + w_T^-)\big((q_1+q_2)^2,q_2^2,q_1^2\big)\,,\notag\\
 w_L\big((q_1+q_2)^2,q_2^2,q_1^2\big)&=(w_T^+ + \tilde w_T^-)\big(q_1^2,q_2^2,(q_1+q_2)^2\big)+(w_T^+ + \tilde w_T^-)\big((q_1+q_2)^2,q_2^2,q_1^2\big)\notag\\
 &+\frac{2q_2\cdot(q_1+ q_2)}{q_1^2}w_T^+\big((q_1+q_2)^2,q_2^2,q_1^2\big) - \frac{2q_1\cdot q_2}{q_1^2}w_T^-\big((q_1+q_2)^2,q_2^2,q_1^2\big)\,.
\end{align}
While the nonrenormalization theorem for \cref{w_L} applies to both perturbative and nonperturbative corrections, see Ref.~\cite{Vainshtein:2002nv}, the relations \cref{non_renormalization} 
do receive nonperturbative corrections.
The singlet part receives both perturbative and nonperturbative corrections. All constraints are derived in the chiral limit. 

Similarly, SDCs can be derived for the hadronic quantities that define 
the contribution from individual intermediate states, see \cref{sec:asymptotic_constraint_PS} for the pseudoscalar poles and SDCs on the corresponding TFFs.
In particular couplings to two off-shell photons with large and nearly equal virtualities
can use the same expansion as in \cref{asymptoticjj} to obtain constraints. For $\pi^0$
exchange including this is standard. For two-pion intermediate states, the matrix element of 
an axial current with two pions vanishes, implying that the amplitude has to drop as $1/\hat q^2$ in the asymptotic limit~\cite{Bijnens:2016hgx}. Similar constraints can be derived
for all intermediate one-particle exchanges on the two off-shell photon couplings using the
matrix elements of the particle and the axial current in \cref{asymptoticjj}.

\subsubsection{Hadronic approaches to satisfy short-distance constraints}
\label{sec:hadronic_SDC}

In the limit $q_3^2\ll q_1^2\sim q_2^2$, the longitudinal contribution from \cref{VVA_decomposition} only appears in $\Pi_1$,
\be
\label{Pi1L}
\Pi_1^L(q_1^2,q_2^2,q_3^2)=-\frac{4w_L\big(q_3^2,0,q_3^2\big)}{\pi^2(2q_1^2+2q_2^2-q_3^2)}\sum_{a=0,3,8} C_a^2
=-\frac{6}{\pi^2 \hat q^2 q_3^2}\sum_{a=0,3,8} C_a^2\,.
\ee
This section will be mainly concerned with this part.
The constraint \cref{Pi1L} should be compared with the contribution from the pseudoscalar poles
\be
\Pi_1^{\pi^0\text{-pole}}(q_1^2,q_2^2,q_3^2)=\frac{F_{\pi^0\gamma^*\gamma^*}(q_1^2,q_2^2)F_{\pi^0\gamma^*\gamma^*}(q_3^2,0)}{q_3^2-M_{\pi^0}^2}\,,
\ee
and similarly for $\eta$, $\eta'$. The low-energy theorem
\be
F_{\pi^0\gamma^*\gamma^*}(0,0)=\frac{3 C_3}{2\pi^2F_\pi}
\ee
and the asymptotic behavior
\be
F_{\pi^0\gamma^*\gamma^*}(\hat q^2,\hat q^2)=-\frac{4 C_3F_\pi}{\hat q^2}+\order\big(\hat q^{-4}\big)\,,
\ee
together with the corresponding relations for $\eta$, $\eta'$, ensure that, formally, in the chiral limit and for $q_3^2\to 0$ the sum of the pseudoscalar
poles indeed saturates \cref{Pi1L}. 

However, as pointed out in Ref.~\cite{Melnikov:2003xd}, this SDC still applies for $|q_3^2|>\Lambda_\text{QCD}^2$ as long as the hierarchy $q_3^2\ll q_1^2\sim q_2^2$ is fulfilled. 
Since the TFFs drop with $1/q_3^2$, $F_{\pi^0\gamma^*\gamma^*}(q^2,0)\sim -12 C_3F_\pi/q^2$ asymptotically, the pseudoscalar poles no longer saturate the SDC in this regime.
As a remedy, it was suggested in Ref.~\cite{Melnikov:2003xd} to neglect the momentum dependence of the singly-virtual TFF altogether, leading to an increase by $13.5\times 10^{-11}$ for the $\pi^0$ and $5\times 10^{-11}$ each for $\eta$, $\eta'$ (using the LMD+V model from Ref.~\cite{Knecht:2001qf} for the $\pi^0$ and VMD for $\eta$, $\eta'$) for a total increase of about $23.5\times 10^{-11}$. Further arguments for this procedure were recently given in Ref.~\cite{Melnikov:2019xkq}.

In Ref.~\cite{Melnikov:2003xd} it was also argued that the effect of replacing the full singly-virtual TFF by its normalization could be interpreted as the result of summing an infinite tower of excited pseudoscalar poles,
similarly to the Regge model for the pion TFF itself~\cite{RuizArriola:2006jge,Arriola:2010aq}. Another alternative is the use of holographic models that automatically incorporate an infinite tower of resonances as recently discussed in Refs.~\cite{Leutgeb:2019gbz,Cappiello:2019hwh}. Two facts make implementing the constraint difficult, \cref{Pi1L} is exact in the chiral limit as the first term in the expansion in $q_3^2/\hat q^2$ and the decay constants $f_P$ for the massive pseudoscalars must vanish in the chiral limit.

In the holographic approach~\cite{Leutgeb:2019gbz,Cappiello:2019hwh} models are employed that consist of an infinite tower of vector and axial-vector resonances but not of the pseudoscalars. The SDCs are satisfied in the longitudinal sector as well due to axial-vector exchanges. The parameters in the models are chosen such that
as much as possible of low-energy phenomenology is reproduced while keeping the SDCs satisfied. The SDC can only be satisfied when the infinite tower of axial vectors is included, they are violated for any finite number as expected from the arguments of Ref.~\cite{Bijnens:2003rc}. For the longitudinal part Ref.~\cite{Leutgeb:2019gbz} (including ground-state axial vectors) quotes
\begin{equation}
\label{HoloLong1}
    \Delta a_\mu^\text{LSDC} = \left(17(\text{HW2})\text{~or~}23(\text{HW1})\right)\times10^{-11}\,.
\end{equation}
Both HW1 and HW2 fit low-energy phenomenology, but HW1 has additional freedom to fit the SDC, while HW2 does so only partially.
Reference~\cite{Cappiello:2019hwh} instead obtains
\begin{equation}
\label{HoloLong2}
\Delta a_\mu^\text{LSDC} = 14\times10^{-11} (\text{both Set 1 and Set 2})\,.
\end{equation}
Set 1 and Set 2 refer to different sets of input parameters, 
Set 1 has the $\rho$ mass at its
physical value, but satisfies the SDC only qualitatively,
while Set 2 enforces the correct SDC at the expense of a
$\rho$ mass which is then too large.
In both papers effects of quark masses are only taken into account in the pseudoscalar-pole contribution.

Finally, the longitudinal SDCs can also be saturated explicitly by a large-$N_c$-inspired Regge model for the corresponding scalar functions~\cite{Colangelo:2019lpu,Colangelo:2019uex}, albeit away from the chiral limit. Using phenomenological input for 
excited pion and $\eta$ states as well as their two-photon couplings~\cite{Acciarri:1997rb,Acciarri:2000ev,Ahohe:2005ug,Achard:2007hm,Zhang:2012tj,Ablikim:2018ajr}, 
the increase compared to the leading pseudoscalar poles alone is found to be~\cite{Colangelo:2019lpu,Colangelo:2019uex}
\begin{align}
\label{CHHLS_PS}
 \Delta a_\mu^{\text{PS-poles}}&=\Delta a_\mu^{\pi\text{-poles}}+\Delta a_\mu^{\eta\text{-poles}}+\Delta a_\mu^{\eta'\text{-poles}}\notag\\
 &=12.6^{\,+1.6}_{\,-1.5}\big\vert_\text{Model}\,(3.8)_\text{syst}\times 10^{-11}=12.6(4.1)\times 10^{-11}\,,
\end{align}
where the first error propagates the uncertainty from the input model parameters and the second estimates the systematic uncertainty due to the modeling of the TFFs. This is much smaller 
than obtained when neglecting the momentum dependence of the second form factor of $23.5\times 10^{-11}$~\cite{Melnikov:2003xd}.\footnote{Note that Ref.~\cite{Colangelo:2019uex} found that with modern input for the TFFs the result with the prescription from Ref.~\cite{Melnikov:2003xd} increases to $38\times 10^{-11}$, mainly because TFFs including the BL short-distance behavior increase the result over the VMD form factors used for $\eta$, $\eta'$ in Ref.~\cite{Melnikov:2003xd}. Taking the singly-virtual TFF pointlike as in Ref.~\cite{Melnikov:2003xd} implies neglecting $2\pi$ and $3\pi$ cuts in the dispersive point of view of Ref.~\cite{Colangelo:2019uex}.} 
The estimated model dependence is
further reduced in Refs.~\cite{Colangelo:2019lpu,Colangelo:2019uex} by matching to pQCD, see \cref{sec:quark_loop}, 
with the following result for the increase due to longitudinal SDCs~\cite{Colangelo:2019lpu,Colangelo:2019uex}
\begin{equation}
 \Delta a_\mu^\text{LSDC}=\left[8.7(5.5)_{\text{PS-poles}}+4.6(9)_{\text{pQCD}}\right]\times 10^{-11}
 = 13(6)\times 10^{-11}\,,
\label{CHHLS}
\end{equation}
where the second contribution, evaluated from the pQCD quark loop with $m_q=0$, implements the asymptotic region and the first is dominated by the lowest pseudoscalar excitations. The latter contribution can be interpreted as an estimate for the mixed regions in the $(g-2)_\mu$ integral and carries large uncertainties 
due to limited phenomenological information on the two-photon couplings of these states. In fact, in the chiral limit the pseudoscalar excitations would decouple, but $\eta(1475)$~\cite{Achard:2007hm,Ablikim:2018ajr} and $\eta(1760)$~\cite{Zhang:2012tj} have been seen in $\gamma\gamma$ collision, with two-photon couplings in line with the expectation from the Regge models 
constrained by the HLbL SDCs~\cite{Colangelo:2019lpu,Colangelo:2019uex}. The matching scale in \cref{CHHLS} is chosen at $Q_\text{match}=Q_\text{min}=1.7(5)\GeV$ (with $Q_\text{min}$ as defined in \cref{sec:quark_loop}), and the contribution from the ground-state pseudoscalar states is taken into account in the matching. 

\begin{figure}[t]
 \centering
 \includegraphics[width=0.65\linewidth]{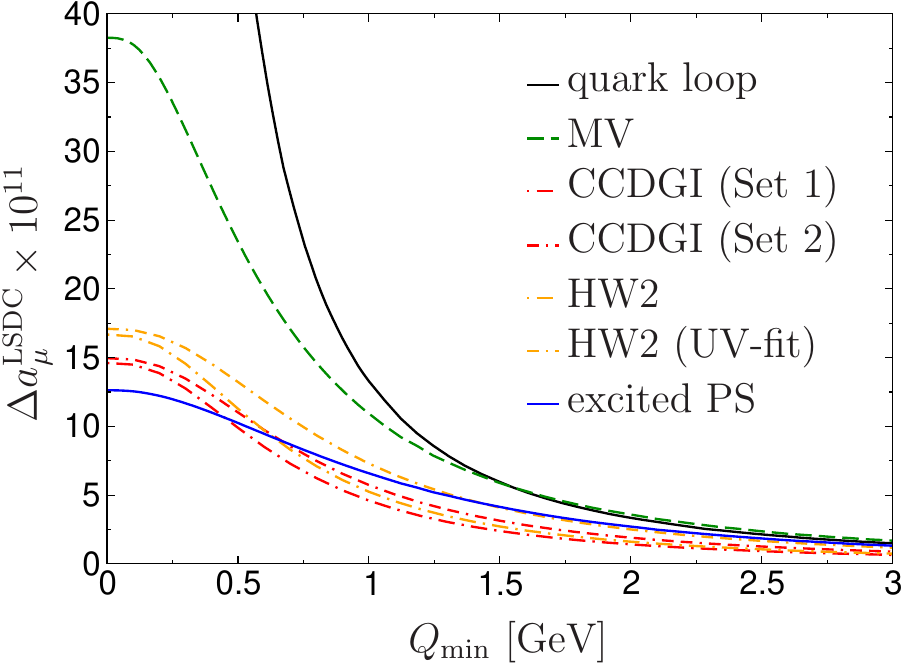}
	\caption{$\Delta a_\mu^\text{LSDC}$ as a function of a cut $Q_\text{min}\leq Q_i$ in the $(g-2)_\mu$ integral. The quark loop diverges for $Q_\text{min}\to 0$ (and $m_q=0$) and is included for its asymptotic behavior, the other curves illustrate the implementations of the longitudinal SDCs discussed in the main text: MV~\cite{Melnikov:2003xd,Melnikov:2019xkq}, CCDGI~\cite{Cappiello:2019hwh} (two parameter sets), HW2~\cite{Leutgeb:2019gbz} (two variants of the HW2 model), and excited pseudoscalars~\cite{Colangelo:2019lpu,Colangelo:2019uex}.}
	\label{SD_comparison}
\end{figure}

The different implementations discussed above are illustrated in \cref{SD_comparison} as a function of a cutoff $Q_\text{min}$ in the $(g-2)_\mu$ integral. Asymptotically, all curves approach the pQCD quark loop (except for HW2, which fits the SDC only partially). The figure shows that given all the caveats in the comparison, the $Q_\text{min}$-dependence of the various implementations in terms of a summation of axial-vector and pseudoscalar resonances agree reasonably well, while the significant numerical difference to the MV model originates predominantly from the low-energy region.

Work on the transversal constraints related to $w_T$ also exists. In Ref.~\cite{Melnikov:2003xd}
it was suggested to saturate these SDCs with axial-vector contributions, but the proposed model 
is not matched to the two-photon couplings~\cite{Achard:2001uu,Achard:2007hm} and again 
neglects the momentum dependence of the singly-virtual form factor in the same spirit as done for the longitudinal part. The holographic model estimates are similar in size but a bit smaller than the longitudinal contributions of \cref{HoloLong1,HoloLong2}, again including the lowest axial vectors.
With the resummation of excited axial-vector states also the transversal SDCs are satisfied.  
The role of axial-vector resonances in HLbL is discussed in more detail in \cref{sec:axials} concentrating on the lowest multiplet.

\subsubsection{Quark loop}
\label{sec:quark_loop}

The pQCD quark loop is the first term in an OPE expansion with the soft photon treated via a background external field~\cite{Bijnens:2019ghy}.
In addition, it does, in the chiral limit, 
reproduce both the longitudinal OPE constraint \cref{Pi1L} and the transversal constraints~\cite{Bijnens:2007pz,Colangelo:2019uex}. Accordingly, it is instructive to study its numerical
contribution when various cuts on the integration momenta are imposed.

\begin{figure}[t]
 \centering
 \includegraphics[width=0.65\linewidth]{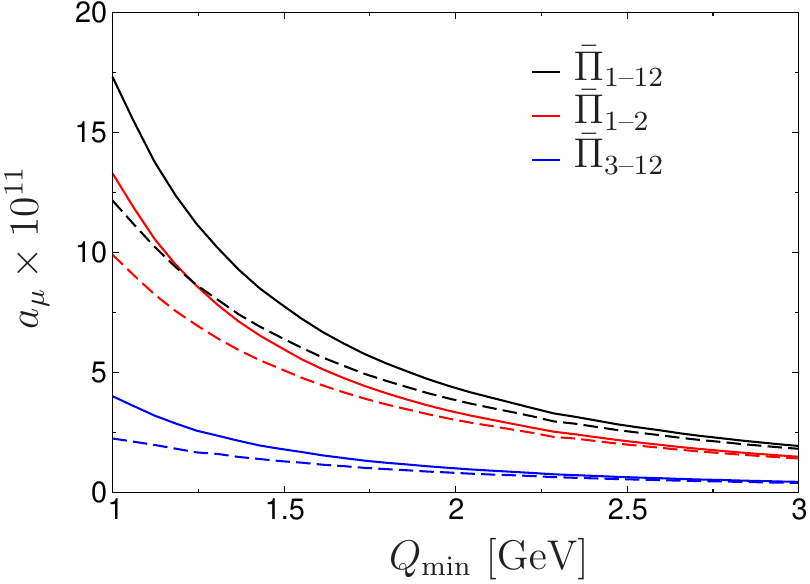}
	\caption{Contribution of the pQCD quark loop to $a_\mu$ for $Q_i\geq Q_\text{min}$. Solid lines for vanishing quark masses, dashed lines for $m_q=0.3\GeV$. The total contribution from $\bar\Pi_{1\text{--}12}$ is shown in black, together with the partial ones from $\bar\Pi_{1\text{--}2}$ (red) and $\bar\Pi_{3\text{--}12}$ (blue).}
	\label{quark_loop_pQCD}
\end{figure}

The numerical result for the pQCD quark loop with a cut on each $Q_i\geq Q_\text{min}$ is shown in \cref{quark_loop_pQCD}, both in the chiral limit and for a constituent quark mass $m_q=0.3\GeV$. We further separate the total contribution into the partial ones from $\bar\Pi_{1\text{--}2}$ and $\bar\Pi_{3\text{--}12}$, motivated by the fact that in the configuration $q_3^2\ll q_1^2\sim q_2^2$ the longitudinal SDCs are expressed in terms of $\bar\Pi_{1\text{--}2}$ and the transversal ones in terms of the others. The figure shows that the numerical result for the quark loop is dominated by the ``longitudinal'' scalar functions. The difference between solid ($m_q=0$) and dashed ($m_q=0.3\GeV$) lines indicates the potential impact of chiral corrections.

\begin{figure}[t]
 \centering
 \includegraphics[width=0.49\linewidth]{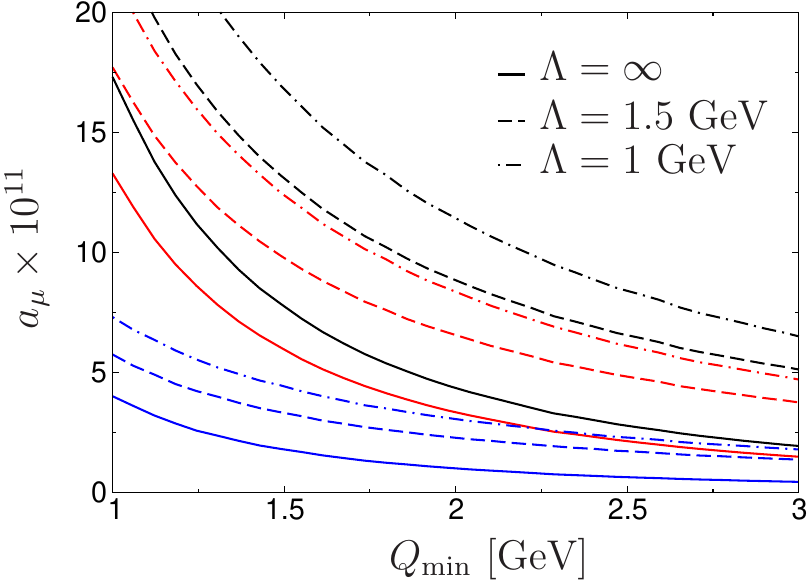}
  \includegraphics[width=0.49\linewidth]{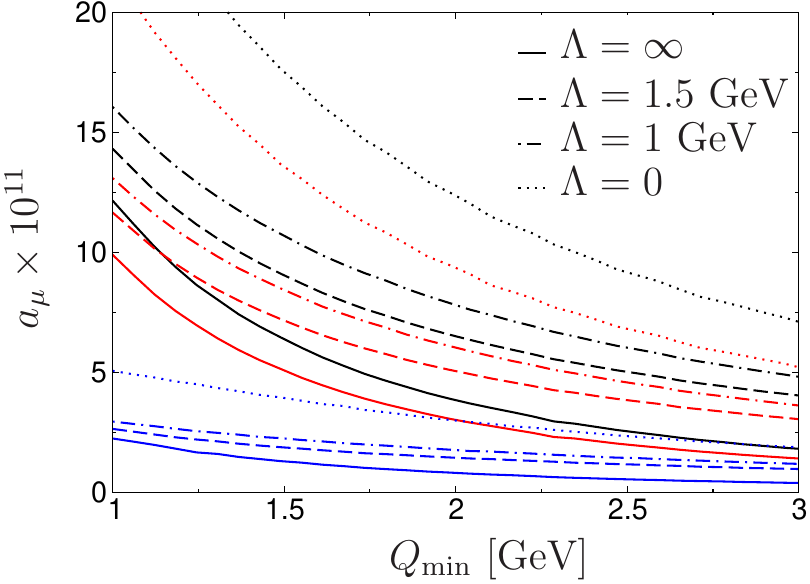}
	\caption{Contribution of the pQCD quark loop to $a_\mu$ for $Q_{1,2}\geq Q_\text{min}$ and $Q_3^2$ damped by $Q_3^2/(Q_3^2+\Lambda^2)$ below $Q_\text{min}$ (plus crossed), see main text, for vanishing quark mass (left) and $m_q=0.3\GeV$ (right).
	Color coding as in \cref{quark_loop_pQCD}, which is reproduced in the limit $\Lambda\to\infty$. The limit $\Lambda\to0$ does not exist for $m_q=0$. Left diagram reprinted from Ref.~\cite{Colangelo:2019lpu}.}
	\label{quark_loop_MV}
\end{figure}

Fig.~\ref{quark_loop_MV} extends the integration region by including configurations in which two momenta are above $Q_\text{min}$, but the third may take arbitrary values, to mimic the case $q_3^2\ll q_1^2\sim q_2^2$. Since the result diverges in the chiral limit, we dampen the additional integration regions by $Q^2/(Q^2+\Lambda^2)$ whenever $Q<Q_\text{min}$, so that for $\Lambda\to\infty$ \cref{quark_loop_pQCD} is recovered. Again, the result is dominated by the ``longitudinal'' functions, while the sensitivity to $\Lambda$ gives some indication as for the impact of the mixed regions to the HLbL integral, and should be matched onto hadronic realizations of the SDCs as described in \cref{sec:hadronic_SDC}.
In fact, for scales $Q_\text{min}\sim\Lambda\sim 1.4\GeV$, which reproduce \cref{CHHLS} for the longitudinal SDCs, one would expect a transversal component 
$\Delta a_\mu^\text{TSDC}\sim 4\times 10^{-11}$~\cite{Colangelo:2019lpu,Colangelo:2019uex}.

\subsubsection{Estimate of the high-energy contribution to HLbL}

The final goal is to obtain an error of about 10\% on hadronic HLbL. As mentioned above, we do not expect the pure short-distance contribution to be a large one, but it does become important in
the error estimates. We face here two questions: how much of the short-distance region has already been included
in the other parts and how well does the quark loop describe this contribution?

First, let us consider the case for all $Q_i\ge Q_\text{min}$. For $Q_\text{min}=2\GeV$, the quark loop should be a good approximation, we also see in \cref{quark_loop_pQCD} that
taking massless quarks or a constituent quark mass does not make much of a difference.
Lowering $Q_\text{min}$ to 1\,GeV increases the uncertainty on the quark-loop dominance
but again we notice that the difference with and without a constituent quark mass is
about $3\times 10^{-11}$, giving an indication of the sensitivity to low-energy effects.
The main question at the moment is how much of this is already included via the
dispersive contributions and the other resonance exchanges. The expectation is that most
will be included and we estimate the missing parts as $5(5)\times10^{-11}$. 

The more difficult part is the case where two $Q_i$ are large and the third is small. The effects are estimated in \cref{quark_loop_MV} by the difference between the full lines and the various dashed lines.
Alternatively, one may consider the implementation of the SDCs in terms of hadronic states as discussed
in \cref{sec:hadronic_SDC}. We expect a pointlike TFF at the second vertex as in Ref.~\cite{Melnikov:2003xd} to be an overestimate due to the modifications that this model implies in the low-energy region. In Ref.~\cite{Bijnens:2007pz} it was determined that a large part of the enhancement is from regions other than where the SDC is valid, which agrees with the conclusion from Refs.~\cite{Colangelo:2019lpu,Colangelo:2019uex} that the main contribution beyond the asymptotic pQCD part
is dominated by the lowest pseudoscalar excitations in the Regge sum.
From the curves in \cref{quark_loop_MV} it is also clear that a total contribution significantly larger then $20\times10^{-11}$ seems unlikely.

The estimates from the quark-loop and the Regge model with a small enhancement expected from the transversal contributions lead us to estimate
\begin{equation}
\Delta a_\mu^\text{SDC}=15(10)\times10^{-11}\,.
\label{eq:SDC}
\end{equation}
This does not include the axial-vector estimate of \cref{sec:axials}
nor the short-distance part of the lowest pseudoscalar exchange as discussed in \cref{sec:pion-pole}.
The uncertainty is meant to cover the impact of the transversal SDCs as well as the interplay with states whose masses lie in the matching region between $1\text{--}2\GeV$. In particular, 
this needs to be taken into account when combining the uncertainty in \cref{eq:SDC} with the ones quoted for these states in \cref{sec:higher_states}. In contrast, double counting with the ground-state pseudoscalar poles is not an issue, since their (small) contribution has been taken into account 
in the matching leading to \cref{CHHLS}, they are also not included in the holographic estimates we quoted in \cref{HoloLong1,HoloLong2}. 
A more complete estimate beyond \cref{eq:SDC} will require a detailed study of the interplay and matching between different contributions, in particular for the transversal SDCs and axial-vector resonances as well as the matching to the quark loop.

For completeness, we also comment on the contribution from the charm quark.
Assuming that this contribution is fully perturbative, with mass $m_c=1.27(2)\GeV$~\cite{Tanabashi:2018oca}, the quark loop evaluates to
$a_\mu^{c\text{-loop}}=3.1(1)\times 10^{-11}$. In analogy to the light quarks, one would expect 
the most important nonperturbative effect to be related to the pole contribution from
the lowest-lying $c \bar c$ resonance, the $\eta_c(1S)$ with mass $m_{\eta_c(1S)}=2.9839(5)\GeV$
and two-photon width $\Gamma(\eta_c(1S)\to\gamma\gamma)=5.0(4)\keV$~\cite{Tanabashi:2018oca}.
Using a VMD-type form factor with scale set by the $J/\psi$ (as suggested by a significant branching fraction $\BR(J/\psi\to\eta_c(1S)\gamma)=1.7(4)\%$~\cite{Tanabashi:2018oca}), this leads to the estimate
$a_\mu^{\eta_c(1S)}=0.8\times 10^{-11}$~\cite{Colangelo:2019uex} (this estimate agrees with the Dyson--Schwinger-equation result $a_\mu^{\eta_c(1S)}=0.9(1)\times 10^{-11}$ from Refs.~\cite{Raya:2019dnh,Raya:2016yuj}). Given the relatively low scale set by $m_c$ one may also expect
$\alpha_s$ corrections in a similar ballpark. To avoid a potential double counting issue, we do  not add the $\eta_c$ to the quark loop, but consider its contribution as an indication for the uncertainty from nonperturbative effects.
Altogether, we estimate 
\begin{equation}
a_\mu^{c\text{-quark}}=3(1)\times 10^{-11}\,,    
\end{equation}
while the $b$-quark contribution is already suppressed to the level of $10^{-13}$.

\subsection{Hadronic light-by-light scattering at NLO}
\label{sec:HLbL_NLO}

\begin{figure}[t]
\centering
\includegraphics[width=0.25\linewidth]{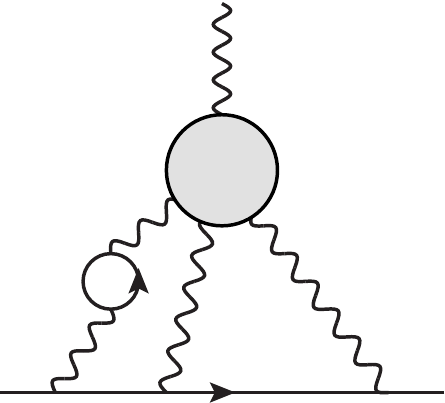}
\caption{HLbL scattering combined with lepton vacuum
  polarization. Diagrams where the lepton loop is inserted into the 
  other photon propagators are not shown. Reprinted from Ref.~\cite{Colangelo:2014qya}.}
\label{fig:HLbL_NLO}
\end{figure}

In Ref.~\cite{Kurz:2014wya} the contribution of HVP at ${\cal O}(\alpha^4)$ to the muon $g-2$ was
calculated as $\amuHVPNNLO=12.4(1) \times
10^{-11}$. This result is only suppressed by a factor $1/8$ compared
to the absolute value of the HVP at NLO. This raises the question 
whether other contributions at order ${\cal O}(\alpha^4)$, i.e., HLbL
combined with lepton VP (HLbL at NLO), see
\cref{fig:HLbL_NLO}, could become relevant in case of a large
prefactor in combination with the expected enhancement factor $\log (m_\mu/m_e)$. If HLbL at NLO were $1/8$ of the result for HLbL as estimated in
\cref{eq:final-estimate2} in the next section, it would be almost of the same size as the
final precision goal of the Fermilab muon $g-2$ experiment.

In Ref.~\cite{Colangelo:2014qya} the size of HLbL at NLO was estimated
by evaluating the presumably numerically dominating pion-pole
contribution with the additional inclusion of the VP
of an electron loop. To allow for a quick numerical evaluation of the
loop integrals, a simple VMD model for the pion TFF
was used, which yields a result for HLbL close to the dispersive
approach. One thus obtains the estimate
\be
\label{HLbL_NLO_e}
a_\mu^{\pi^0\text{-pole, NLO}}=1.5 \times 10^{-11}\,,
\ee
a correction of about $2.6\%$ to the pion-pole contribution to HLbL
with VMD. In fact, from renormalization-group
arguments~\cite{Lautrup:1974ic} one would have expected a suppression
of 
\be
\label{suppression_factor}
3\times\frac{\alpha}{\pi}\times\frac{2}{3}\log\frac{m_\mu}{m_e}\approx 2.5\%\,,
\ee
in remarkable agreement with the explicit calculation. In this
estimate, the factor $3$ originates from the fact that each of the
photon propagators can be renormalized, and the prefactor of the
logarithm can be derived from the expression of the VP in the limit $m_\ell\to 0$.

Other contributions of order ${\cal O}(\alpha^4)$ were estimated in
Ref.~\cite{Colangelo:2014qya} to be of the order of $1/4$ of the
logarithmically-enhanced diagrams with electron VP,
based on the corresponding results for radiative corrections to LbL in QED
where HLbL itself is replaced by a muon loop, arguing that the muon mass is close
to a typical hadronic scale. The factor of $4$ enhancement for the diagrams with 
electron VP is close to the RG estimate $(2/3) \log (m_\mu/m_e) \sim 3.6$.
Also radiative corrections to the HLbL blob itself should be negligible, at least for the dominant
pseudoscalar-pole contribution, where electrically neutral particles are exchanged.

In this way, the conservative estimate $\amuHLbLNLO=2.3(1.1) \times 10^{-11}$ 
is obtained, using the updated value for the total HLbL from \cref{eq:final-estimate2}, which should 
be compared to the total $116(39)\times 10^{-11}$ used in Ref.~\cite{Colangelo:2014qya},
and the suppression factor from \cref{suppression_factor}. Accordingly, we find a modest change in the 
central value, but a noticeable reduction in the uncertainty when compared to $\amuHLbLNLO=2.9(1.7) \times 10^{-11}$ 
from Ref.~\cite{Colangelo:2014qya} (rounded therein to $3(2)\times 10^{-11}$). The errors propagated from the total HLbL estimate, 
$0.5\times 10^{-11}$, as well as the possible effect of nonlogarithmically enhanced contributions, $0.6\times 10^{-11}$, 
i.e., a quarter of the central value as observed for QED, have been added linearly to be conservative. In view of the precision 
obtained for the total HLbL in \cref{eq:final-estimate2}, we therefore get the following updated value for HLbL at NLO:  
\be \label{HLbL_NLO}
\amuHLbLNLO=2(1) \times 10^{-11}\,. 
\ee

\subsection{Final result}
\label{sec:result_HLbL_DR}

\subsubsection{Combining all contributions and estimating missing ones}
Analytic approaches to the calculation of the HLbL contribution to the muon $(g-2)_\mu$ have a long history, but in very recent years have been extended and improved significantly, even though more work is still needed to achieve a complete analysis and a satisfactory estimate of uncertainties. The most significant progress, as described in this section, has been made in the calculation of the pseudoscalar-pole contributions and in the pion box and related rescattering effects. Both have been calculated with reliable uncertainty estimates at the level of 5\% or below. Recent work on other contributions (resonances in particular) has also been made but not with the same level of rigor, which is a consequence of the very nontrivial related difficulties described in detail in the corresponding sections. Finally the issue of matching to the pQCD behavior at short distances is being addressed, but final results are not yet available.

The summary of the present status for all these contributions is as follows (expressed in units of $10^{-11}$): 
\begin{itemize}
\item pseudoscalar poles
\[ a_\mu^\text{PS-poles} = 93.8^{+4.0}_{-3.6} \,; \quad \]
\item pion box
\[ a_\mu^{\pi\text{-box}} = -15.9(2) \,;\]
\item $S$-wave $\pi \pi$ rescattering
\[ a_{\mu,J=0}^{\pi\pi,\pi\text{-pole LHC}} =-8(1) \,; \qquad\qquad \]
\item kaon box
\[  a_\mu^{K\text{-box}} = -0.5(1) \,;\]
\item scalars and tensors with $M_R \gtrsim 1$\,GeV
\[ a_\mu^\text{scalars+tensors} \sim -1(3) \,; \qquad\quad \]
\item axial vectors
\[ a_\mu^\text{axials} \sim  6(6) \,; \quad \]
\item short-distance contribution
\[ \Delta a_\mu^\text{SDC} \sim 15(10) \,; \quad \]
\item charm and other heavy-quark contribution
\[ a_\mu^\text{c} \sim 3(1) \,. \]
\end{itemize}
Summing all these contributions we obtain a central value of $92 \times 10^{-11}$. As for the uncertainty, summing all of them in quadrature gives $13 \times 10^{-11}$, which is somewhat higher than the goal of the 10\% accuracy for this contribution, but already significantly lower than previous estimates. A thorough discussion of the uncertainty and whether adding errors of individual contributions in quadrature is appropriate at the present stage is important and will be provided in the next subsection. Note that the three different estimates of the pion pole discussed in
\cref{sec:pion-pole}, see \cref{pion_pole},  would lead to virtually the same final result even
once updated to the exact same TFF normalizations.

\subsubsection{Uncertainty estimate}
Given the current rather unsatisfactory theoretical status of the calculation of some contributions and what is at stake (a possible discovery of physics beyond the SM), it is worthwhile discussing the uncertainty estimate in some detail and whether it should be made more conservative. Clearly, the theoretically more difficult estimates concern the following four contributions: scalars, tensors, axial vectors, and short distance.
As explained above, single-resonance-exchange contributions beyond the pseudoscalar ones are not unambiguously defined: whenever one calculates a particular contribution to HLbL, one has to choose a ``basis'' (BTT set, to be more precise) for the HLbL Lorentz tensor. The contribution to the muon $g-2$ will depend on this choice unless a set of sum rules is satisfied, but as explained above if one considers a single resonance at a time, this is the case only for pseudoscalars. This poses the problem of how to quantify the uncertainty coming from an ambiguity, which has no unique answer. Progress in addressing this issue will require a calculation of these single-resonance-exchange contributions that does not suffer from such an ambiguity. A possible way to argue is that in the presence of an ambiguity one would expect a specific calculation within a given framework (and basis) to provide the correct order of magnitude, but not more. This suggests to assign roughly a 100\% uncertainty to any contribution affected by this ambiguity, which is indeed the case for the sum of scalars and tensors and for the axial-vector contribution. 

A rather extreme approach would be to combine all errors linearly instead of in quadrature, including those not affected by this ambiguity, but we do not think that this is justified and would be overly conservative. One should also keep in mind that the errors assigned to the short-distance, axial-vector, heavy-scalar, and tensor contributions are not independent, see the discussion in \cref{sec:asymptotic}, to the extent that currently some of these uncertainties may actually be double counted, providing another rationale for not simply adding all uncertainties linearly.  

We opted for the following procedure, which we consider more sensible. We first add the errors from the independent data-driven, dispersive estimates for the pseudoscalar poles, the pion box, and $\pi\pi$ rescattering in quadrature, yielding $\pm 4.1 \times 10^{-11}$, then we add the errors for the model-dependent estimates for the sum of scalars and tensors, the axial-vector contribution, and the short-distance contribution linearly, yielding $\pm 19 \times 10^{-11}$, and finally we combine these two errors and the one from the charm quark loop in quadrature. This leads to our final estimate  $\amuHLbL= 92(19) \times 10^{-11}$.

\subsubsection{Comparison to the Glasgow consensus and other compilations}

The intense activity on the HLbL contribution of the last five years
based on the dispersive approach has been reported in this section and
summarized above. It is useful to discuss here in some detail what are
the reasons behind the changes in the numbers compared to the
estimates used in 2009, even though on the surface they do not seem to be
so large. We will also comment on a few recent estimates.
In \cref{tab:compilations} we have collected the frequently used
compilations for HLbL from 2009 by Prades, de Rafael, and Vainshtein
(``Glasgow consensus,'' PdRV(09))~\cite{Prades:2009tw,Glasgow2007} and Jegerlehner
and Nyffeler~(N/JN(09))~\cite{Nyffeler:2009tw, Jegerlehner:2009ry}, and
a recent update of the latter that has appeared in the book by Jegerlehner
(2nd edition, J(17))~\cite{Jegerlehner:2017gek}. Our estimate is also shown for comparison.

\begin{table}
\centering
\small 
\begin{tabular}{crrrr}
\toprule
  Contribution & \quad PdRV(09)~\cite{Prades:2009tw} \quad & 
  \quad N/JN(09)~\cite{Nyffeler:2009tw, Jegerlehner:2009ry} \quad & 
  \quad J(17)~\cite{Jegerlehner:2017gek} \quad & 
  \quad Our estimate \quad
  \tabularnewline  
\midrule
 $\pi^0,\eta,\eta'$-poles & $ 114(13) $ & $ 99(16) $ & $ 95.45
 (12.40) $ & $ 93.8(4.0) $ 
\tabularnewline 

$\pi,K$-loops/boxes & $ -19(19) $ & $ -19(13) $ & $ -20(5) $ & 
 $ -16.4(2) $
\tabularnewline 
$S$-wave $\pi\pi$ rescattering & $ -7(7) $ & $ -7(2) $ & $-5.98(1.20) $ &  
 $ -8(1) $
\tabularnewline 
\midrule
subtotal & $88(24)$ & $73(21)$ & $69.5(13.4)$ & $69.4(4.1)$
\tabularnewline 
\midrule
scalars & $-$ & $-$ & $-$ &    
  \multirow{2}{*}{$\bigg\}\qquad -1(3)$} 
 \tabularnewline 

tensors & $-$ & $-$ & $ 1.1(1) $ & 
 \tabularnewline 

axial vectors & $ 15(10) $ & $ 22(5) $ & $ 7.55(2.71) $ &  
 $ 6(6) $ 
\tabularnewline 

~$u,d,s$-loops / short-distance~ & $-$ & $ 21(3) $ & $ 20(4) $
&   $ 15(10) $   
\tabularnewline \midrule
$c$-loop & $2.3$ & $-$ & $2.3(2)$
&   $ 3(1)$   
\tabularnewline 
\midrule
total & $ 105(26) $ & $ 116(39) $ & $ 100.4(28.2) $ &   
  $ 92(19) $ 
\tabularnewline 
\bottomrule
\end{tabular}
\caption{Comparison of two frequently used compilations for HLbL in units of
  $10^{-11}$ from 2009 and a recent update with our estimate. Legend: PdRV = Prades, de Rafael, Vainshtein (``Glasgow
  consensus''); N/JN = Nyffeler / Jegerlehner, Nyffeler; J = Jegerlehner.} 
  \label{tab:compilations}
\end{table}

The main difference of the first three estimates by
PdRV~\cite{Prades:2009tw}, N/JN~\cite{Nyffeler:2009tw,
  Jegerlehner:2009ry}, and J~\cite{Jegerlehner:2017gek} to our result
is that they are based purely on model calculations, see also
\cref{tab:earlier} in \cref{sec:tensor} for details of the original
works for some of the individual contributions. Some constraints from
theory, e.g., from ChPT at low energies or from short distances in
pQCD, and from experiment are taken into account in those
models, e.g., on the singly-virtual pseudoscalar TFFs. But this model dependence makes it very difficult to estimate
the uncertainty in a reliable way. On the other hand, our estimates
for the numerically dominant contributions from the light
pseudoscalar poles $\pi^0,\eta,\eta^\prime$ and for a substantial part
of the two-pion intermediate state in HLbL (pion-box and $S$-wave
$\pi\pi$ rescattering) are now based on model-independent dispersion
relations or Canterbury approximants and the error estimates are
largely driven by the precision of the input data. To emphasize this significant progress we have evaluated the sum of these contributions and compared the different evaluations for the corresponding subtotal in the line labeled as ``subtotal'' in \cref{tab:compilations}.\footnote{To make a meaningful comparison, since the largest contribution among the scalars is due to the $\sigma/f_0(500)$, which is treated as a $\pi\pi$ rescattering effect here, we have considered the contribution of the scalars of earlier evaluations in the line labeled ``$S$-wave $\pi \pi$ rescattering.'' This is indeed justified for the scalar contribution $-6.8(2.0) \times 10^{-11}$ in the ENJL model from Ref.~\cite{Bijnens:1995xf}, as confirmed in Ref.~\cite{Bijnens-private}. The $\sigma/f_0(500)$ is also responsible for $50\hyph80\%$ of the value $-6.0(1.2) \times 10^{-11}$ from Ref.~\cite{Jegerlehner:2017gek}, depending on the mixing.} While the central values are all quite close to each other (the largest discrepancy is with the Glasgow consensus, which, however, includes a large part of the short-distance contribution in the pseudoscalar poles) and all compatible within errors, the largest improvement is in the uncertainty, which has been reduced by a factor 6 to 3.

The lower part of the table contains the remaining contributions, which still suffer from significant uncertainties, further separated into the contribution from light quarks as well as the $c$-loop. For these a comparison among different evaluations is more difficult, because model dependence is still affecting all contributions (with the exception of the short-distance contribution evaluated here). It is in this second part of the table that future progress will have to happen.

We have described above how we obtained our final error estimate. Just for comparison, in 
PdRV~\cite{Prades:2009tw} all errors have been added in quadrature, in  N/JN~\cite{Nyffeler:2009tw,
  Jegerlehner:2009ry} all errors have been added linearly, and in J~\cite{Jegerlehner:2017gek} the errors have been added in quadrature and then multiplied by a factor 2 to account for possible model uncertainties so far unaccounted for.

We also briefly comment on the numbers in the recent review by Danilkin, Redmer, and Vanderhaeghen~\cite{Danilkin:2019mhd}. The main difference is their estimate of the pseudoscalar-pole contribution, $84(4) \times 10^{-11}$, lower than our value by about $2.5\sigma$, which is incompatible with what we know about this contribution as explained in \cref{sec:pion-pole}.
The smaller value for the PS-poles is compensated by the quark-loop contribution, $20(4) \times 10^{-11}$, which is a bit larger than our estimate of the short-distance contribution, leading to a central value, $87(13) \times 10^{-11}$, very close to ours.  The errors in Ref.~\cite{Danilkin:2019mhd} are added linearly, but in particular the uncertainties for the axial-vectors and the short-distance contribution are much smaller than ours, which is the main reason for their rather small total uncertainty.

The comparison discussed here clearly shows that there has been
significant progress since the time of the Glasgow consensus. The
development of a more systematic approach to the calculation of the HLbL
contribution has led to improved estimates of several of the underlying
contributions. The shifts in the central values are relatively moderate,
never larger than two sigmas with respect to older estimates, but the
overall shift is quite significant and in the negative direction, thus
increasing the discrepancy with the measured value. Even more important
than the shift in the central value is our ability to make better
uncertainty estimates. In some cases these have been drastically reduced
with respect to the time of the Glasgow consensus, but in some others a
better theoretical understanding of the formalism has led to a more
cautious attitude. The upshot is that even taking a conservative approach
we could bring the total uncertainty down to about $20\%$ of the central
value and the prospects for an even further reduction in the coming years,
towards the $10\%$ goal, are very good as will be sketched in the next subsection.

\subsubsection{Final estimate and outlook \label{sec:HLbL}}

Our final estimates for HLbL from \cref{tab:compilations} and HLbL at NLO~\cite{Colangelo:2014qya} from \cref{HLbL_NLO} read as follows: 
\begin{align}
   \amuHLbL&= (69.3(4.1) + 20(19) + 3(1)) \times 10^{-11} \nonumber \\ 
   &= 92(19) \times 10^{-11} \, , \label{eq:final-estimate2} \\
   \amuHLbLNLO & = 2(1) \times 10^{-11} \, , 
\end{align}
where we want to stress again that now the numerically dominant contributions from the single-pseudoscalar poles and large parts of the two-pion contribution rely on a data-driven dispersive approach and are under good control, as shown in the first line. The errors for the model-dependent estimates for the sum of scalars and tensors, the axial-vector contribution, and the short-distance contribution have been added linearly. Finally, the errors have been combined in quadrature to obtain the second line in \cref{eq:final-estimate2}. The total error is about 20\% and is completely dominated by the model estimates of a numerically subdominant part of the total. 
The final number in \cref{eq:final-estimate2} is thus mainly based on Refs.~\cite{Masjuan:2017tvw,Hoferichter:2018kwz,Gerardin:2019vio} for the pseudoscalar poles, Ref.~\cite{Colangelo:2017fiz} for $2\pi$ intermediate states, and Refs.~\cite{Melnikov:2003xd,Bijnens:2019ghy,Colangelo:2019uex} for OPE constraints and the charm loop. In addition, Refs.~\cite{Pauk:2014rta, Danilkin:2016hnh,Jegerlehner:2017gek,Knecht:2018sci,Eichmann:2019bqf,Roig:2019reh} entered the estimates of heavy intermediate states. We recommend that these papers be cited as well when using \cref{eq:final-estimate2}, because they are important to determine the ranges, but as an absolute minimum those that directly affect the central values and short-distance behavior. The main experimental input that enters the phenomenological determination of HLbL scattering, besides $e^+e^-\to\text{hadrons}$ cross sections, is from Refs.~\cite{\HLbLexpref}.  

What needs to be done to improve the estimate for HLbL and what is the final uncertainty that one can expect to achieve? The most urgent next steps are the following:
\begin{enumerate}
\item
complete the calculation of the $\eta$ and $\eta'$ contributions with a dispersive approach. We do not expect significant shifts neither in the central value nor in the error, but it will be very important to confirm our current estimates for these two  contributions with a second and theoretically better founded method;
\item
carry out a calculation of single-resonance contributions that is unambiguous (basis independent) and that satisfies all phenomenological constraints on the resonance properties;
\item
understand how this can be implemented for the axial vectors and how these contributions can be matched to the short-distance constraints in the transverse channel;
\item
further study the matching between the short-distance constraints and the dispersive representation in the longitudinal channel analytically and numerically, and in particular understand the role of axial-vector resonances and their possible interplay with the pseudoscalar ones in this matching.
\end{enumerate}

At the present status of our knowledge none of the steps above seem to be unfeasible. Unless there are surprises, we expect that the final uncertainties of the four contributions listed above will be slightly or even significantly reduced. With this, it will be possible to achieve the goal of a final and conservatively evaluated $10\%$ uncertainty on the HLbL contribution to the muon $(g-2)_\mu$.

\FloatBarrier

\clearpage

\section{Lattice approaches to HLbL}
\label{sec:latticeHLbL}

\noindent
\begin{flushleft}
\emph{N.~Asmussen, T.~Blum, A.~G{\'e}rardin, M.~Hayakawa, R.~J.~Hudspith, T.~Izubuchi, L.~Jin, C.~Lehner, H.~B.~Meyer, A.~Nyffeler}
\end{flushleft}

\subsection{Introduction}
\begin{figure}[tb]
\centerline{\includegraphics[width=0.4\textwidth]{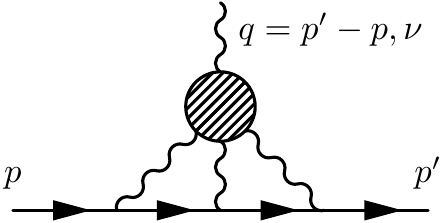}}
\caption{HLbL contribution to the muon
  $g-2$. The shaded blob represents all possible intermediate hadronic
  states. Reprinted from Ref.~\cite{Blum:2015gfa}.
  }
\label{fig:hlbl_general}
\end{figure}

The HLbL scattering contribution to the
anomalous magnetic moment of the muon is depicted in
\cref{fig:hlbl_general}, where the external soft and on-shell photon
interacts through a hadronic blob with three off-shell photons that
then couple to the muon.  In this section, we discuss lattice QCD
approaches to calculating this contribution, which allow for a first-principles
calculation with systematically improvable uncertainty.

\begin{figure}[bt]
    \centering
    \includegraphics[width=0.4\textwidth]{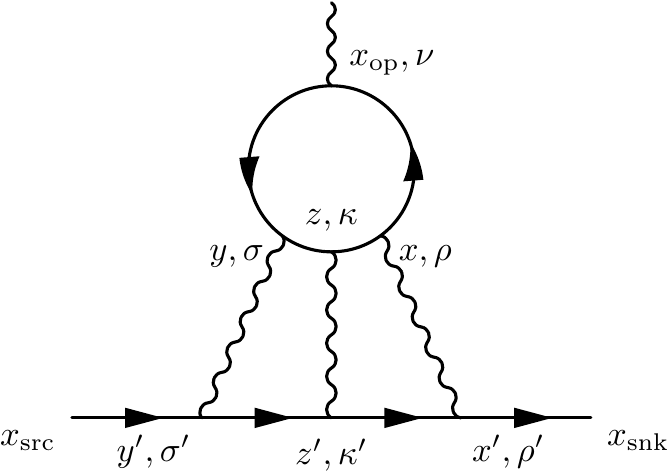}
    \includegraphics[width=0.4\textwidth]{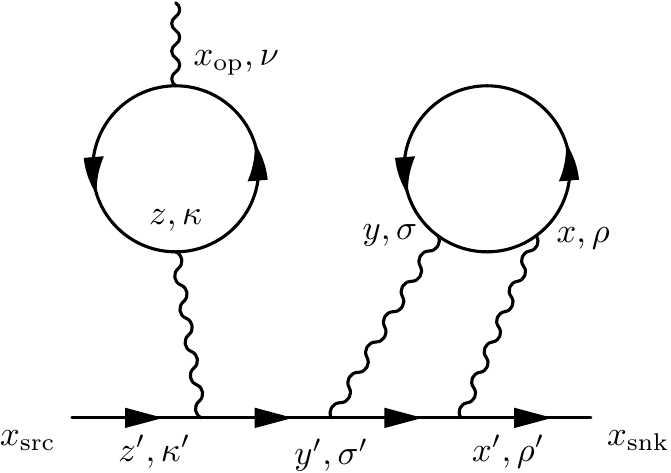}
    \caption{Diagrams contributing to HLbL
      scattering at order $\order(\alpha^3)$. Quark connected (left) and
      leading disconnected (right) diagrams are shown. Reprinted from Ref.~\cite{Blum:2016lnc}.} 
    \label{fig:diagrams_intro}
\end{figure}

\begin{figure}[bt]
    \centering
    \includegraphics[width=0.3\textwidth]{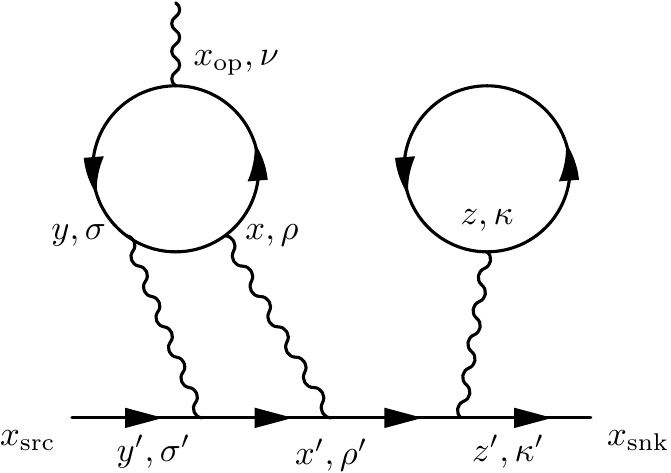}
    \includegraphics[width=0.3\textwidth]{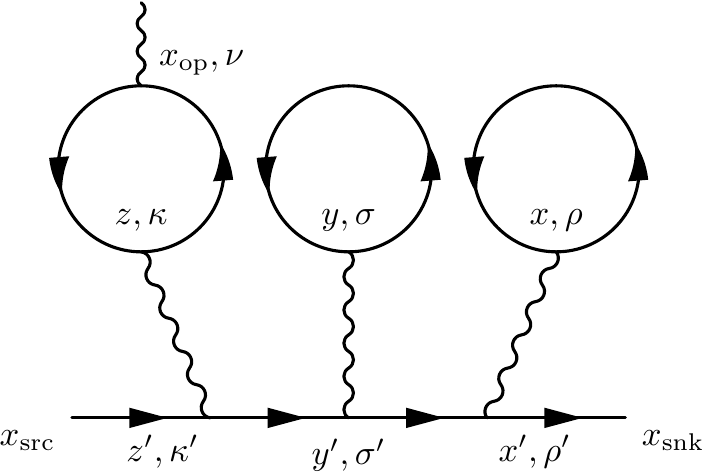}
    \includegraphics[width=0.3\textwidth]{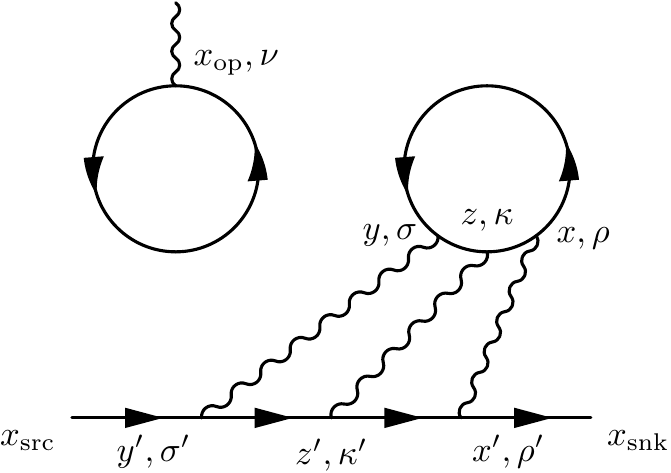}
    \includegraphics[width=0.3\textwidth]{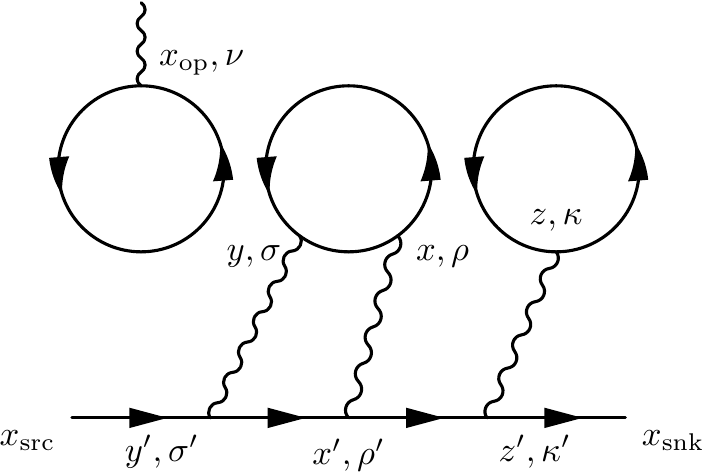}
    \includegraphics[width=0.4\textwidth]{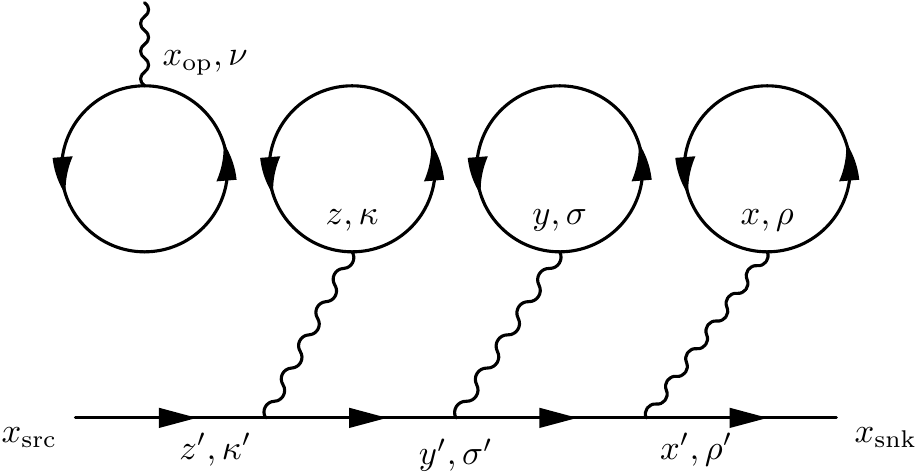}
    \caption{Sub-leading disconnected diagrams contributing to
      HLbL at order $\order(\alpha^3)$. Reprinted from Ref.~\cite{Blum:2016lnc}.}  
    \label{fig:disco_diagrams_intro}
\end{figure}

In a perturbative framework for QED, the HLbL scattering contribution
to the muon $g-2$ in a lattice QCD calculation arises at order
$\alpha^3$ from the diagrams shown in
\cref{fig:diagrams_intro,fig:disco_diagrams_intro}. They
are classified as connected or disconnected depending on whether the
quark lines are (dis-)connected.  Though they are not shown explicitly,
it is understood that for a given
diagram quark--gluon interactions to all orders are included.  In a lattice
QCD calculation, these additional diagrams are generated by statistical sampling
from a gauge-configuration ensemble.  The diagrams in
\cref{fig:diagrams_intro} dominate the HLbL contribution.  Because at
least one loop has a single photon attached, each of the sub-leading
disconnected diagrams shown in \cref{fig:disco_diagrams_intro} vanish
in the $SU(3)$ flavor limit since $Q_u+Q_d+Q_s=0$, and each are quark-mass and color suppressed.  The total $\amuHLbL$ comes from
summing the contributions from the diagrams shown in
~\cref{fig:diagrams_intro,fig:disco_diagrams_intro} and
permutations generated by attaching the three photons to the muon line
in six possible ways. 

\subsection{HLbL on the lattice}
The HLbL contribution to the muon $g-2$ has been calculated from first
principles using the lattice regularization by RBC and later by the
Mainz group. The first successful such calculation treated QED
nonperturbatively~\cite{Blum:2014oka}. Subsequent calculations that
treat QED
perturbatively~\cite{Blum:2015gfa,Green:2015sra,Blum:2016lnc,Asmussen:2018oip} have
proven more efficient. Furthermore, the calculations are now performed in coordinate space
by both groups, which is most natural on the
lattice, and a moment method is applied to calculate
$F_2(q^2=0)$ directly, which avoids the extrapolation from finite to vanishing
$q^2$. The QED parts of the amplitudes have been computed in
finite~\cite{Blum:2014oka,Blum:2015gfa,Blum:2016lnc} and infinite
volume~\cite{Asmussen:2016lse,Asmussen:2017bup,Blum:2017cer}. The former exhibits large
$1/L^2$ corrections from infinite volume~\cite{Blum:2015gfa}, where
$L$ is the linear size of the lattice, due to the massless photons
while the latter exhibits the usual exponentially suppressed errors in
QCD due to its mass gap. Nonzero lattice spacing errors are also
large~\cite{Blum:2015gfa,Blum:2016lnc} though they can be
significantly controlled using clever subtractions that vanish in the
continuum and infinite-volume limit~\cite{Blum:2016lnc,Asmussen:2019act}.

Generally in lattice gauge theory physical observables are computed from Euclidean time correlation functions on a hypercubic lattice with spacing $a$ and four-volume $VT=L^3T$. Physical results are obtained from extrapolations to the continuum ($a\to0$) and infinite-volume ($L,T\to\infty$) limits.

 The anomalous magnetic moment
 is computed from the correlation of muon fields and the quark electromagnetic current (see \cref{fig:hlbl_general}),
\begin{equation}
    C(t) = \langle \psi_\mu^\dagger(t^\prime,{\bf p^\prime}) j_\nu({\bf q},t)\psi_\mu(0,{\bf p})\rangle\,, \qquad
    j_\nu({\bf q},t) = \sum_{f}Q_f \bar q_f\gamma_\nu q_f({\bf q},t)\,,
\end{equation}
where $q=p^\prime-p$ is the momentum transferred to the muon by the external photon.
In the limit of large time separations, $t^\prime\gg t\gg 0$, the ground state of the muon dominates the correlation function, and it is proportional to the matrix element of the electromagnetic current between on-shell muon states.
\begin{equation}
    \lim_{t^\prime\gg t\gg 0}C({\bf q},t) \propto \langle p^\prime | j_\nu |p\rangle
    =\bar u({\bf p^\prime})\left(F_1(q^2)\gamma_\nu+i\frac{[\gamma_\nu,\gamma_\mu]q_\mu}{2 m_\mu}F_2(q^2)\right)u({\bf p})\,,
    \label{eq:me}
\end{equation}
where the form factors $F_1$ and $F_2$ depend on $q^2$.
The form factors evaluated in the static limit, $q^2=0$, contain all information on the muon's intrinsic couplings to the photon. In particular, it is a straightforward exercise to show $F_2(0)=(g-2)/2$. 

The form of \cref{eq:me} is dictated by the Ward identity (or charge conservation). In Refs.~\cite{Blum:2015gfa,Green:2015mva}, a nontrivial ``moment'' method for computing the correlation function at precisely $q^2=0$ is given, which is a crucial ingredient in both the finite-volume and infinite-volume QED approaches.  This method is explained in detail below.

The total $\amuHLbL$ comes from summing the contributions from diagrams shown in \cref{fig:diagrams_intro} and \cref{fig:disco_diagrams_intro} and permutations generated by attaching the three photons to the muon line in six possible ways.
In practice the correlation functions corresponding to these diagrams
are computed in coordinate space, and the integrals (sums) over the
positions of the internal vertices conserve momentum flowing from the
external vertex to the muon. Exact sums are prohibitively expensive,
so they are done stochastically. The RBC group chooses random point
pairs $(x,y)$ to emphasize the important region $|x-y|\simle 1$\,fm
since the contribution is exponentially suppressed by the QCD mass gap
whenever any pair of vertices becomes separated by a long
distance~\cite{Blum:2015gfa}. Using $x$ and $y$ as point source
locations to compute quark propagators, the remaining two vertices are
summed exactly. 

In the aforementioned moment method, the momentum projection of the external vertex becomes a coordinate-space moment. This adds a complication. Since translational invariance is used to take the moment with respect to $(x+y)/2$, it can no longer be used to ensure the Ward identity, i.e., that the photon attaches at all possible points on the loop. To overcome this, the external photon is explicitly attached in all possible ways, requiring extra (sequential) propagator calculations. The benefits of the moment method far outweigh the extra cost~\cite{Blum:2015gfa}.  If one were to perform this moment method with respect to
$(x+y+z)/3$, the use of sequential propagators could be avoided~\cite{Blum:2015gfa}.

The finite-volume QED method is currently only being pursued by RBC, while the infinite-volume QED method is used by both Mainz and RBC. Since the two groups do not use the same notation, we first present the different
approaches separately, using RBC's notation for the finite-volume QED method and Mainz's notation for the infinite-volume QED method.
In \cref{sec:cross-check_QCD} we then perform a dedicated comparison study
between both groups for QCD ensembles with similar quark masses and volumes. \Cref{sec:phys mass results} describes results computed directly with physical masses.

\subsubsection{The HLbL calculation using finite-volume QED (RBC)}

The RBC group has performed calculations in finite-volume QED~\cite{Blum:2014oka,Blum:2015gfa,Blum:2016lnc}. Due to gauge invariance, nonzero charge cannot exist in a finite volume with periodic boundaries. This obstacle can be overcome by omitting all photons with zero spatial momentum~\cite{Hayakawa:2008an}. This finite-volume version of QED is known as QED$_L$, and QED is recovered in the infinite-volume limit. 

The photons and muons are implemented on a hypercubic lattice of size $L$ with spacing $a$. For the muons, the group uses domain wall fermions (DWF) with the size of the extra fifth
dimension, $L_s$, set to infinity. DWF define a discrete
version of the Dirac operator that has exact chiral symmetry at nonzero lattice spacing.
Due to chiral symmetry, the leading discretization error for DWF is $\order(a^2)$. For the QCD part of the amplitude a similar lattice Dirac operator is used, but with finite $L_s$, in which case a small explicit breaking of chiral symmetry results in an additive shift of the quark masses.
The photon propagators are evaluated in the Feynman gauge using the noncompact formalism.

\begin{figure}
    \centering
    \includegraphics[width=0.75\textwidth]{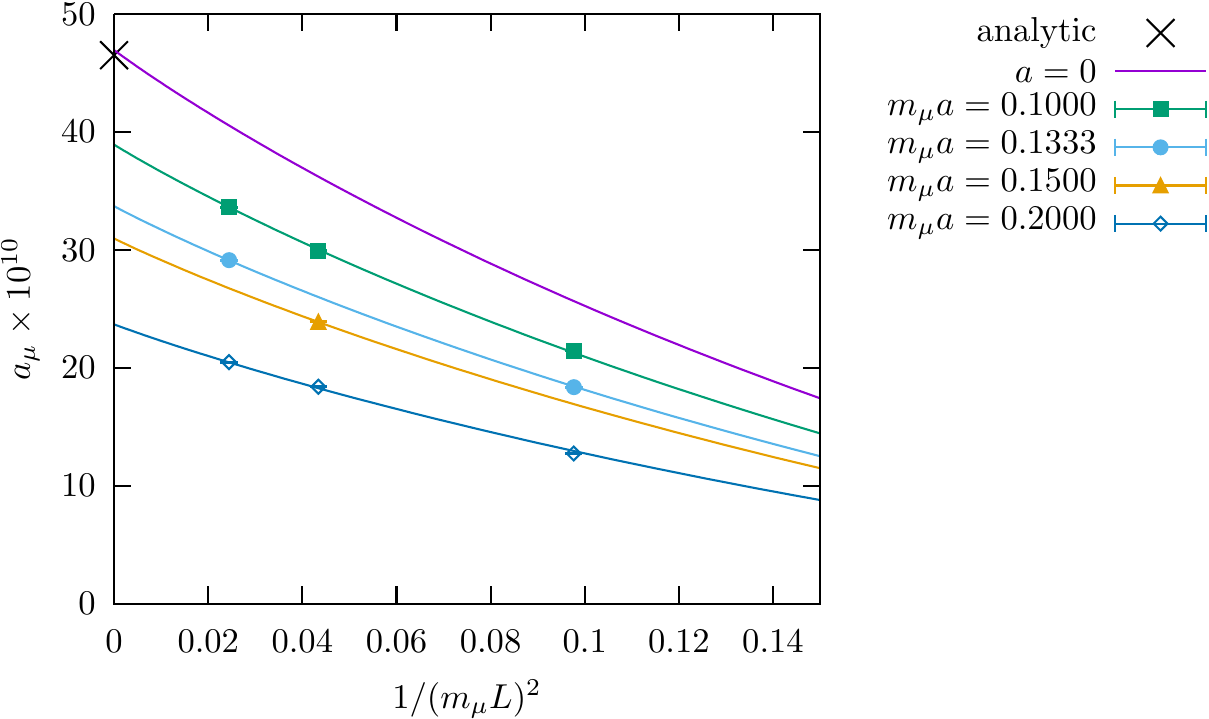}
    \caption{QED$_L$ LbL scattering contribution to the
      muon anomaly. Curves are evaluated from a fit using
      \cref{eq:QED_L fit} at fixed lattice spacing. The uppermost
      curve is the continuum limit and the cross is the value
      evaluated in perturbative QED~\cite{Laporta:1991zw}. Reprinted from Ref.~\cite{Blum:2019ugy}.} 
    \label{fig:QED_L}
\end{figure}

As a check on the method, including systematic errors, the calculation was first performed in the case of pure QED$_L$. In \cref{fig:QED_L}, the LbL scattering contribution in pure QED$_L$ is shown for several volumes and lattice spacings~\cite{Blum:2015gfa}. A simple ansatz is used to extrapolate to the infinite-volume and continuum limits,
\begin{equation}
\label{eq:QED_L fit}
    a_\mu^{\rm LbL}(a,L) = a_\mu^{\rm LbL} \left(1-b_1 a^2+b_2 a^4\right)
    \left(1-c_1 \frac{1}{m_\mu^2L^2}+c_2 \frac{1}{m_\mu^3L^3}\right)\,.
\end{equation}

From \cref{fig:QED_L} one sees the extrapolated value agrees with the well-known analytic result~\cite{Laporta:1991zw}. However, the size of the discretization and finite-volume errors is large. The typical lattice scale of QCD simulations is $a^{-1}\sim 1\hyph2$\,GeV or higher, so corresponding curves compared to those shown in 
\cref{fig:QED_L} would lie closer to the continuum limit. The typical lattice size for QCD simulations is 5\hyph6\,fm, which corresponds to $1/(m_\mu L)^2\approx 0.1$. Thus large finite-volume effects for the hadronic contributions computed in QED$_L$ are expected.  

\subsubsection{The HLbL calculation using infinite-volume QED (Mainz and RBC)} 

The Mainz group has pioneered the infinite-volume QED method and we
follow the Mainz approach unless otherwise noted.  RBC has also
contributed to the development of this method and is actively using it
as well, as explained in the following section.  The infinite-volume QED
method provides an explicit formula for \(F_2(q^2)\) at \(q^2=0\)
that is manifestly Lorentz covariant.  The formula separates the task
into a QED part and a QCD part.  For the QED part a semi-analytical
kernel function has been derived using infinite-volume, continuum perturbative
QED. This avoids $1/L^2$ finite-volume errors that would arise in a
finite box due to propagating massless photons.  The Mainz group has
pre-computed and stored the kernel function for multiple use. The QCD
part is an integral over a four-point position-space correlation
function that can be computed using lattice QCD.

\paragraph{Semi-analytical calculation of QED kernel and master formula}

The derivation in the Mainz approach starts from the well-known
projection~\cite{Brodsky:1966mv,Aldins:1970id} that allows one to obtain  
the value of \(F_2(q^2)\) at \(q^2=0\),
\begin{align}
   F_2(0)
   =\frac{-i}{48m_\mu}
   \trace\{[\gamma_\rho,\gamma_\tau](-i\slashed{p}+m_\mu)\Gamma_{\rho\tau}(p,p)(-i\slashed{p}+m_\mu)\} 
   \label{eq:KinoshitaTrace}
   \,,
\end{align}
with the vertex function $\Gamma$ defined as
\begin{align}
\begin{split}
& \Gamma_{\rho\sigma}(p',p)=
   -e^6\int_{q_1,q_2} 
   \frac{1}{q_1^2q_2^2(q_1+q_2-k)^2}
   \frac{1}{(p'-q_1)^2+m_\mu^2}\frac{1}{(p'-q_1-q_2)^2+m_\mu^2}
   \\& \qquad \quad 
   \times \Big(\gamma_\mu(i\slashed{p}'-i\slashed{q}_1-m_\mu)
   \gamma_\nu(i\slashed{p}'-i\slashed{q}_1-i\slashed{q}_2-m_\mu)
   \gamma_\lambda\Big)
   \frac{\partial}{\partial k_\rho}\Pi_{\mu\nu\lambda\sigma}(q_1,q_2,k-q_1-q_2)\,,
\end{split}
\notag\\
& \Pi_{\mu\nu\lambda\sigma}(q_1,q_2,q_3)=
\int_{x_1,x_2,x_3}e^{-i(q_1 \cdot x_1+q_2 \cdot x_2+q_3 \cdot x_3)}
\Big\langle j_\mu(x_1)j_\nu(x_2)j_\lambda(x_3)j_\sigma(0)\Big\rangle\,,
\label{eq:Pimomentumsp}
\end{align}
with photon momenta \(q_i\) and the coordinates \(x_i\). The integrals are 
defined as \(\int_{x_i}=\int d^4x_i\) and 
\(\int_{q_i}=\int\frac{d^4q_i}{(2\pi)^4}\).
The momentum-space
\cref{eq:KinoshitaTrace,eq:Pimomentumsp}  
already show the trace over the gamma structure and the propagators
that will be contained in the QED kernel function, and the QCD
correlation function \(\Pi_{\mu\nu\lambda\sigma}\).

Parameterizing the on-shell muon momentum with an arbitrary four-dimensional 
direction~\(\hat\epsilon\),
\begin{align}
   p=im_\mu\hat\epsilon\quad (p^2=-m_\mu^2)
   \,,
\end{align}
and performing the Fourier transformation of  
\cref{eq:KinoshitaTrace,eq:Pimomentumsp} into position space 
yields an expression involving a kernel \(\mathcal 
L_{[\rho,\sigma];\mu\nu\lambda}(\hat\epsilon,x,y)\):
\begin{align}
   \hat F_2(0)&=\frac{m_\mu e^6}{3}\int_y
   \int_x\mcL_{[\rho,\sigma];\mu\nu\lambda}(\hat\epsilon,x,y)\;
   i\,\widehat\Pi_{\rho;\mu\nu\lambda\sigma}(x,y)\,,
\end{align}
with
\begin{align}
   i\widehat \Pi_{\rho;\mu\nu\lambda\sigma}( x, y)
   &= \int d^4z\;
   (-z_\rho)\widetilde\Pi_{\mu\nu\sigma\lambda}(x,y,z),\notag\\ 
 \widetilde\Pi_{\mu\nu\sigma\lambda}(x,y,z) &=
 \Big\langle j_\mu(x)\,j_\nu(y)\,j_\sigma(z)\,
 j_\lambda(0)\Big\rangle\,, \label{eq:Pitilde} 
\end{align}
i.e., \(i\widehat \Pi \) is a spatial moment of the QCD correlation
function \( \widetilde\Pi \). This expression is valid for any
\(\hat\epsilon\). The average over \(\hat\epsilon\), 
\begin{align}
\bmcL_{[\rho,\sigma];\mu\nu\lambda}(x,y)=
\frac{1}{2\pi^2}\int d\Omega_\epsilon\;
\mcL_{[\rho,\sigma];\mu\nu\lambda}(\hat\epsilon,x,y)
\equiv \Big\langle \mcL_{[\rho,\sigma];\mu\nu\lambda}(\hat\epsilon,x,y)\Big\rangle_{\hat\epsilon}
\,,
\end{align}
removes the dependence on the direction, and leads to the master formula,
\begin{align}
   \amuHLbL=F_2(0)&=\frac{m_\mu e^6}{3}\int_y
   \int_x\bmcL_{[\rho,\sigma];\mu\nu\lambda}(x,y)\,i\widehat\Pi_{\rho;\mu\nu\lambda\sigma}(x,y)
   \, 
   \notag \\
   &= \frac{m_\mu e^6}{3}2\pi^2 \int d|y| |y|^3
   \int_x\bmcL_{[\rho,\sigma];\mu\nu\lambda}(x,y)i\,\widehat\Pi_{\rho;\mu\nu\lambda\sigma}(x,y) 
   \notag \\
   &= \int d|y| f(|y|)\,.
   \label{eq:masterMainz}
\end{align}
The kernel is defined as
\begin{align}
	\bar {\cal L}_{[\rho,\sigma];\mu\nu\lambda}(x,y) 
	= \sum_{A={\rm I,II,III}} {\cal G}^A_{\delta[\rho\sigma]\mu\alpha\nu\beta\lambda}
	T^{(A)}_{\alpha\beta\delta}(x,y)\,,
\end{align}
with, e.g.,
\begin{align}
	{\cal G}^{\rm I}_{\delta[\rho\sigma]\mu\alpha\nu\beta\lambda} &\equiv 
	{\frac{1}{8}} {\rm Tr}\Big\{\Big( \gamma_\delta [\gamma_\rho,\gamma_\sigma] +
	2 (\delta_{\delta\sigma}\gamma_\rho - \delta_{\delta\rho}\gamma_\sigma)\Big)
	\gamma_\mu \gamma_\alpha \gamma_\nu \gamma_\beta \gamma_\lambda \Big\}\,,
\end{align}
and the tensors are given by
\begin{align}
	T^{({\rm I})}_{\alpha\beta\delta}(x,y) &=   \partial^{(x)}_\alpha (\partial^{(x)}_\beta + \partial^{(y)}_\beta) 
	V_\delta(x,y)\,, \notag
	\\
	T^{({\rm II})}_{\alpha\beta\delta}(x,y) &= 
	m_\mu \partial^{(x)}_\alpha 
	\bigg( T_{\beta\delta}(x,y) + \frac{1}{4}\delta_{\beta\delta} S(x,y)\bigg)\,, \notag
	\\
	T^{({\rm III})}_{\alpha\beta\delta}(x,y) &=  m_\mu (\partial^{(x)}_\beta + \partial^{(y)}_\beta)
	\bigg( T_{\alpha\delta}(x,y) + \frac{1}{4}\delta_{\alpha\delta} S(x,y)\bigg)\,.
\end{align}
 \(S\), \(V\), and \(T\) are parameterized by the six weight functions $\mathfrak{\bar g}^{(1,2,3)}$ and  $\mathfrak{\bar l}^{(1,2,3)}$,
\begin{align}
   S(x,y) &= \mathfrak{\bar g}^{(0)}\big(|x|,\hat x\cdot \hat y, |y|\big)\,,  
   \phantom{\frac{1}{1}} \notag
   \\
   V_\delta(x,y)
   &= x_\delta  \bar{\mathfrak{g}}^{(1)}\big(|x|,\hat x\cdot\hat y,|y|\big)
   + y_\delta  \bar{\mathfrak{g}}^{(2)}\big(|x|,\hat x\cdot\hat y,|y|\big)\,, \notag
   \\
       T_{\alpha\beta}(x,y) &= \bigg(x_\alpha x_\beta - 
       \frac{x^2}{4}\delta_{\alpha\beta}\bigg)\; \bar{\mathfrak{l}}^{(1)}\big(|x|,\hat 
       x\cdot\hat y,|y|\big)
      + \bigg(y_\alpha y_\beta - \frac{y^2}{4}\delta_{\alpha\beta}\bigg)\; 
      \bar{\mathfrak{l}}^{(2)}\big(|x|,\hat x\cdot\hat y,|y|\big)
      \notag\\ &
      \quad + \bigg(x_\alpha y_\beta + y_\alpha x_\beta - \frac{x\cdot
        y}{2}\delta_{\alpha\beta}\bigg)\; 
      \bar{\mathfrak{l}}^{(3)}\big(|x|,\hat x\cdot\hat y,|y|\big)\,.
\end{align}
These functions depend on the three variables \(\lvert x \rvert \),
\(\lvert y \rvert \), and \(\cos\beta=\hat x\cdot \hat y\), where
\(\hat x = x / \lvert x\rvert \) and \(\hat y = y / \lvert y\rvert \).

If all weight functions are known, it is very cheap to compute the kernel.  
Therefore the Mainz group has stored pre-computed representations of the six 
weight functions. Reference~\cite{Asmussen:2017bup} shows the corresponding 
plots.  These representations allow one to compute the kernel during the lattice 
computation on the fly.

\begin{figure}[hbt]
   \centerline{\includegraphics[width=0.4\textwidth]{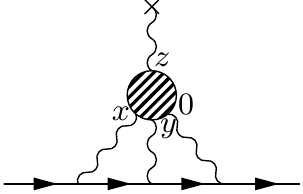}}
   \caption{Mainz definition of the coordinates \(x,y,z,0\) in 
     HLbL in the $(g-2)_\mu$. The muon
     and the photon propagators are contained in the QED kernel
     function~\(\bar{\mathcal L}\).  The blob denotes the QCD
     correlation function~\(i\hat\Pi\) to be evaluated on the
     lattice. Adapted from Ref.~\cite{Asmussen:2017bup}.}
   \label{fig:MainzCoord}
\end{figure}

Note that the points \(x\), \(y\), \(z\), and the origin $0$ used in the
Mainz approach differ from the corresponding coordinates as defined in
the RBC method.  \Cref{fig:MainzCoord} shows a graphical
representation of the Mainz definitions. See Refs.~\cite{Asmussen:2016lse}
and~\cite{MainzPrep} for more details of the derivation and an
explicit formula for the weight function \(\bar{\mathfrak g}^{(2)}\).
The Mainz group uses $O(4)$ rotational invariance to choose points
along a diagonal of the lattice~\cite{Asmussen:2018oip}.

Note that the QED kernel is not unique. RBC introduced a subtracted
kernel that leaves the integral unchanged, but changes the integrand,
in particular it smoothens its short-distance behavior, i.e., when the vertices approach each
other~\cite{Blum:2017cer}. Only in this way was it possible to
reproduce the known result for a lepton loop in QED. The Mainz group
has seen this behavior as well in both the continuum and in infinite
volume, when studying the case of the pion-pole contribution to HLbL
with a vector-meson dominance (VMD) model and for a lepton loop in
QED~\cite{Asmussen:2019act}. In both of these test cases, the
integrand $f(|y|)$ for $\amuHLbL$ from
\cref{eq:masterMainz} as function of $|y|$, after integrating
over $|x|$ and $\cos\beta$, rises steeply at small $|y|$ and has a
long negative tail at large $|y|$. The first behavior poses a
challenge for use in lattice computations due to the finite lattice
spacing, while the second property of the integrand demands a large
lattice volume in physical units to cover the tail.

The derivation of the subtracted QED kernels starts from the following
observation: in the continuum and in infinite volume one can write the
conserved current $j_\mu(x)$ as a total derivative,
\begin{align}
   j_\mu(x) = \partial_\nu^x \big(x_\mu j_\nu(x)\big)\,,
\end{align}
and therefore
\begin{align}
   \int_x j_\mu(x) = 0\,.
\end{align}
The correlation function inherits this property,
\begin{align}
   \int_x i\hat\Pi(x,y)=\int_y i\hat\Pi(x,y)=0\,.
   \label{eq:corrintzero}
\end{align}

One can define subtracted kernels \(\mcL^{(1,2,3)}\), such that for all kernels 
\(\mcL^{(i)}\) (omitting the Lorentz indices),
\begin{align}
   \mcL^{(0)}(x,y)&=\bar{\mcL}(x,y)\,,\quad \text{(standard
     kernel)}\notag
   \\
   \mcL^{(1)}(x,y)&=\bar{\mcL}(x,y)-\frac{1}{2}\bar{\mcL}(x,x)-\frac{1}{2}\bar{\mcL}(y,y)\,,
   \notag \\
   \mcL^{(2)}(x,y)&=\bar{\mcL}(x,y)-\bar{\mcL}(0,y)-\bar{\mcL}(x,0)\,,
   \notag \\
   \mcL^{(3)}(x,y)&=\bar{\mcL}(x,y)-\bar{\mcL}(0,y)-\bar{\mcL}(x,x)+\bar{\mcL}(0,x)\,,
   \label{eq:Lkernels}
\end{align}
the master formula retains its form due to \cref{eq:corrintzero}:
\begin{align}\begin{split}
   F_2(0)
   &=\frac{m_\mu e^6}{3}\int_y
   \int_x\mcL^{(i)}_{[\rho,\sigma];\mu\nu\lambda}(x,y)\,
   i\widehat\Pi_{\rho;\mu\nu\lambda\sigma}(x,y)\,.
\end{split}\end{align}

The standard kernel obeys
\begin{align}
   \mcL^{(0)}(0,0)=0\,.
\end{align}
The subtracted kernels have the following additional properties:
\begin{align}
   \mcL^{(1)}(x,x)&=0\,,\notag\\
   \mcL^{(2)}(0,y)&=\mcL^{(2)}(x,0)=0\,,\notag\\
   \mcL^{(3)}(x,x)&=\mcL^{(3)}(0,y)=0\,.
\end{align}

The quark-connected part of the four-point function \(i\widehat\Pi\)
involves three different contractions. Computing all three of them and
applying \cref{eq:masterMainz} amounts to what we call ``method
1.'' In a lattice implementation of this method, for \(N\) evaluations
of the \(y\) integrand, \(1+N\) propagators and \(6(1+N)\) sequential
propagators are needed.  If we define
\begin{align} 
\Pi^{(1)}_{\mu\nu\sigma\lambda}(x,y,z)\equiv -2{\rm ReTr}\{ S(0,x)
\gamma_\mu S(x,y)  \gamma_\nu S(y,z)  \gamma_\sigma S(z,0)
\gamma_\lambda \}\,, 
\end{align} 
where the \(S(x,y)\) are quark propagators, which represents one of the Wick
contractions of the quark-connected part
$\widetilde\Pi^c_{\mu\nu\sigma\lambda}(x,y,z)$, we can write (for any
given background gauge field)
\begin{align}
\widetilde\Pi^c_{\mu\nu\sigma\lambda}(x,y,z) = 
   \Pi^{(1)}_{\mu\nu\sigma\lambda}(x,y,z)+\Pi^{(1)}_{\nu\mu\sigma\lambda}(y,x,z)+\Pi^{(1)}_{\nu\sigma\mu\lambda}(y,z,x)\,.
\end{align}
Note that $\partial_\mu^x \widetilde
\Pi^c_{\mu\nu\sigma\lambda}(x,y,z) = 0$ for all $x$.  The computation
can be arranged in a different way, such that only the contraction
$\Pi^{(1)}$ is computed and the others are implemented by permuting
the way that the photons are attached to the vertices of the
four-point function. We call this ``method~2,'' which
reads~\cite{Asmussen:2019act} 
\begin{align}
   a_\mu^{\text{HLbL,c}} = \frac{m_\mu e^6}{3}
   \int_{y,x,z}\Big(&
      \left[\bmcL_{[\rho,\sigma];\mu\nu\lambda}(x,y)
      +\bmcL_{[\rho,\sigma];\nu\mu\lambda}(y,x)
      -\bmcL_{[\rho,\sigma];\lambda\nu\mu}(x,x-y)\right]
      (-z_\rho)\,\Pi^{(1)}_{\mu\nu\sigma\lambda}(x,y,z)
\nonumber\\ &
      +\bmcL_{[\rho,\sigma];\lambda\nu\mu}(x,x-y)\,(-x_\rho) \Pi^{(1)}_{\mu\nu\sigma\lambda}(x,y,z)
   \Big)\,.
\end{align}

\begin{figure}[t]
   \centerline{\includegraphics[width=0.6\textwidth]{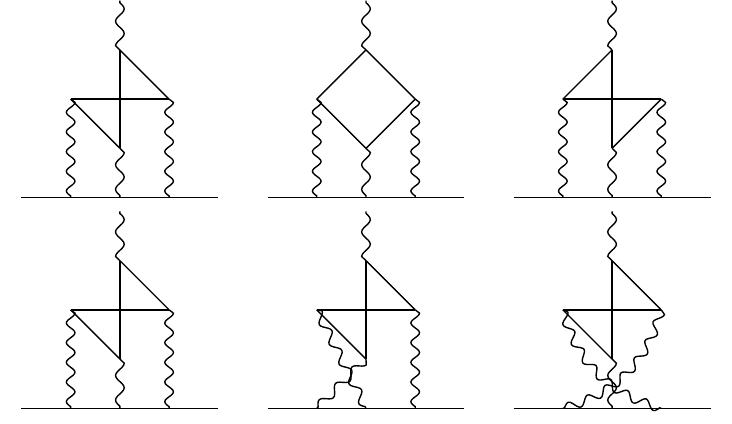}}
   \caption{%
     Contractions needed to compute \(g-2\). Upper row: The three
     connected Wick contractions needed in method~1.  Bottom row: In
     method~2 the different contraction types are implemented in the
     QED part of the diagram. Reprinted from Ref.~\cite{Asmussen:2019act}.}
   \label{fig:methods}
\end{figure}

A diagrammatic representation of the contractions for methods~1 and 2
is shown in~\cref{fig:methods}. While method~2 requires the
calculation of far fewer propagators, its integrand receives
contributions from the exchange of resonances odd under charge
conjugation, which cancel out upon fully integrating over \(x,y,\) and \(z\).

The freedom one has in choosing the QED kernel without affecting
$\amuHLbL$ allows for the suppression of certain
discretization effects via subtractions; see \cref{eq:Lkernels}.  However, control over the finite-size
effects turns out to be challenging as these subtractions tend to make the integrand broader. Therefore the Mainz group is
currently investigating the benefit of a new class of
kernels~\cite{Asmussen:2019act}
\be {\cal
  L}^{(2,\bar\lambda)}_{\rho\sigma;\mu\nu\lambda} = {\cal
  L}^{(0)}_{\rho\sigma;\mu\nu\lambda}(x,y) - \partial_\mu^{(x)}\Big(
x_\alpha \,e^{-\bar\lambda m_\mu^2 x^2/2}\Big)\; {\cal
  L}^{(0)}_{\rho\sigma;\alpha\nu\lambda}(0,y)
- \partial_\nu^{(y)}\Big( y_\alpha \,e^{-\bar\lambda m_\mu^2
  y^2/2}\Big)\; {\cal L}^{(0)}_{\rho\sigma;\mu\alpha\lambda}(x,0)\,, 
\ee
which reduces to ${\cal L}^{(2)}$ for $\bar\lambda=0$, and ${\cal L}^{(0)}$ for $\bar\lambda=\infty$. Such a kernel has the property of vanishing whenever $x$ or $y$ does, but does not
qualitatively alter the long-distance behavior of the original kernel
${\cal L}^{(0)}$ for positive values of $\lambda$.

\paragraph{Numerical tests of the QED kernel}

This section provides numerical studies, with the aim of verifying the
correctness of the Mainz approach and learning about its numerical
properties. The literature provides results for the pion-pole
contribution to HLbL with the VMD model and for the lepton-loop
contribution to LbL scattering in QED. The Mainz group has
implemented the correlation function \(i\widehat\Pi\) for both
contributions in position space. 

\begin{figure}[t]
   \centering 
   \includegraphics[width=0.49\textwidth]{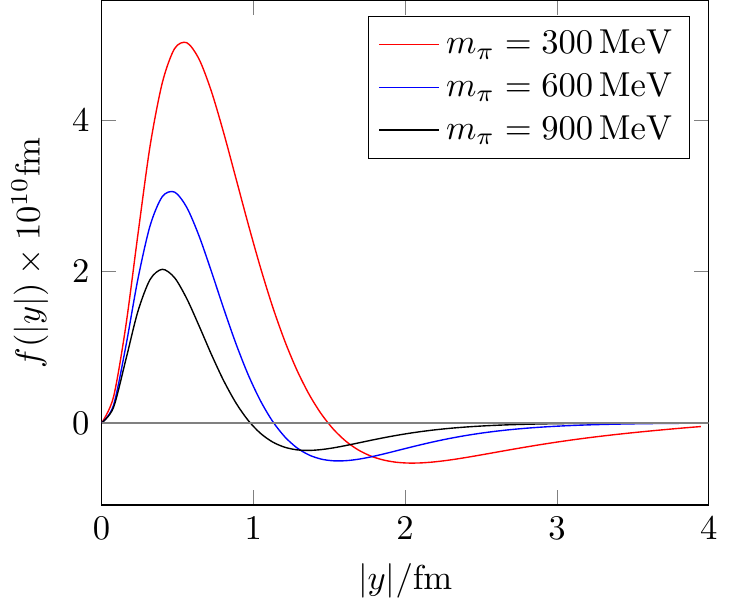}
   \includegraphics[width=0.49\textwidth]{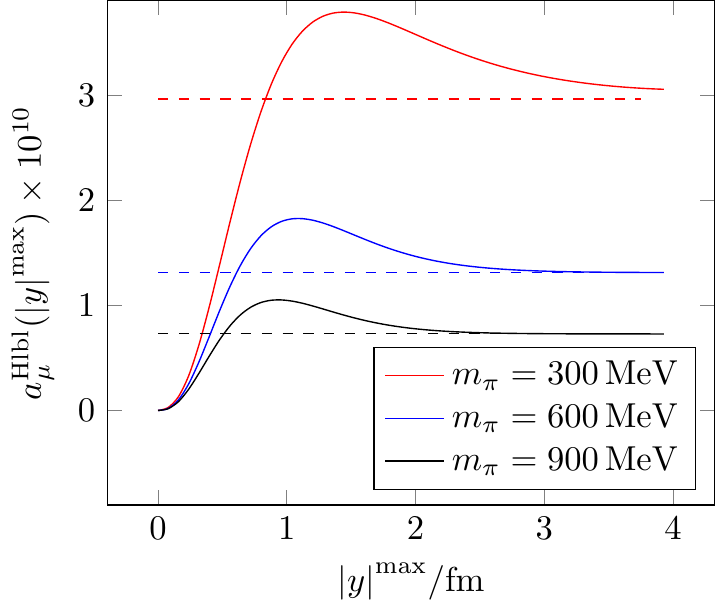}
   \caption{Numerical study of the pion-pole contribution
     with the VMD model. Left: integrand for \(\amuHLbL\)
     after the integration over \(\lvert x\rvert\) and
     \(\cos\beta\). Right: \(\amuHLbL\) integrated in  
      \(\lvert y\rvert\) up to the cutoff \(\lvert y\rvert^{\text{max}}\).
      Solid curves: result of the Mainz approach. Dashed curves: results obtained in 
      momentum space. Adapted from Ref.~\cite{Asmussen:2017bup}.}
   \label{fig:pi0vmd}
\end{figure}

Assuming a VMD model transition form factor (TFF),\footnote{See \cref{pion_decay constant} for our conventions regarding $F_\pi$.}
\begin{align}
   \FF(-q_1^2,-q_2^2) = \frac{c_\pi}{(q_1^2+M_V^2)(q_2^2+M_V^2)}, \qquad 
   c_\pi = \frac{M_V^4}{4\pi^2 F_\pi}\,,
\end{align}
allows one to derive the pion-pole correlation function
\begin{align}
    i \widehat\Pi_{\rho;\mu\nu\lambda\sigma}(x,y)
   &= \frac{c_\pi^2}{M_V^2(M_V^2-M_\pi^2)} \frac{\partial}{\partial 
   x_\alpha}\frac{\partial}{\partial y_\beta}
   \bigg\{ 
    \epsilon_{\mu\nu\alpha\beta}\epsilon_{\sigma\lambda\rho\gamma}
   \bigg(\frac{\partial}{\partial x_\gamma}+\frac{\partial}{\partial y_\gamma}\bigg) 
   K_\pi(x,y)
   \nonumber\\ & \quad + \epsilon_{\mu\lambda\alpha\beta} 
   \epsilon_{\nu\sigma\gamma\rho}
    \frac{\partial}{\partial y_\gamma} 
   K_\pi(y-x,y)
   + \epsilon_{\mu\sigma\alpha\rho}\epsilon_{\nu\lambda\beta\gamma} 
   \frac{\partial}{\partial x_\gamma}
   K_\pi(x,x-y)
    \bigg\}\,.
\end{align}
The function \(K_\pi\) is given by
\begin{align}
   K_\pi(x,y) \equiv \int d^4u \Big(G_{M_\pi}(u) - G_{M_V}(u)\Big)
   G_{M_V}(x-u) G_{M_V}(y-u) = K_\pi(y,x)\,,
\end{align}
where \(G_M(x)\) is the massive scalar propagator in position space
\begin{align}
   G_M(x)=\int_p\frac{e^{ip\cdot x}}{p^2+M^2}
   =\frac{M}{4\pi^2\lvert x\rvert}K_1(M\lvert x\rvert)\,,
\end{align}
with the modified Bessel function \(K_1\).  \Cref{fig:pi0vmd}
shows the result of the computation using this \(i\widehat\Pi\) for
different pion masses. The Mainz approach reproduces the known result
for the pion-pole contribution at the percent level and therefore
demonstrates that the approach is correct. Note that the integrand
gets a longer negative tail for pion masses as they approach the physical value.

\begin{figure}[hbt]
   \centerline{\includegraphics{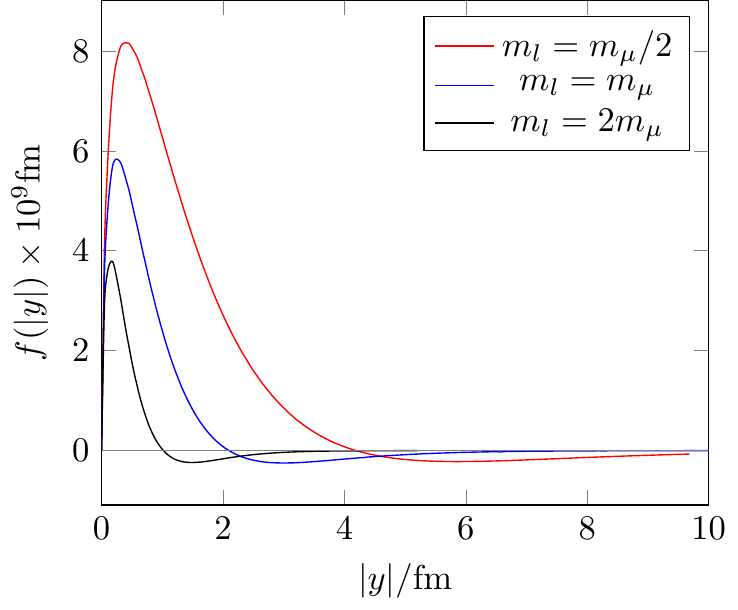}}
   \caption{Integrand of the lepton loop for different loop-particle 
   masses~\(m_l\). Reprinted from Ref.~\cite{Asmussen:2018lcw}.}
   \label{fig:leploop}
\end{figure}

Similarly, an explicit expression for the lepton-loop
correlation function \(i\widehat\Pi\) in terms of 
Bessel functions and traces of products of gamma matrices can be found in 
Ref.~\cite{Asmussen:2017bup}.  \Cref{fig:leploop} and 
\cref{tab:leploop} show the corresponding results.

\begin{table}[hbt]
   \begin{center}
   \small
      \begin{tabular}{ccr@{}lcc}
        \toprule
           $m_l / m_\mu$
         & {$a_\mu^{\text{LbL}} \times 10^{11}$ (exact)}
         & \multicolumn{2}{c}{$a_\mu^{\text{LbL}} \times 10^{11}$}
         & {Precision}
         & {Deviation}
         \\ \midrule
         1/2 & 1229.07 & 1257.5&(6.2)(2.4) & 0.5\% & 2.3\% \\
         1   &  464.97 &  470.6&(2.3)(2.1) & 0.7\% & 1.2\% \\
         2   &  150.31 &  150.4&(0.7)(1.7) & 1.2\% & 0.06\%\\
         \bottomrule
      \end{tabular}
   \end{center}
   \caption{Lepton-loop contribution in QED. Comparison of the exact 
      result~\cite{Laporta:1992pa,Passera:PC} and the Mainz approach.
      The first uncertainty stems from the three-dimensional integration, the 
   second from the extrapolation to small \(\lvert y\rvert\).}
   \label{tab:leploop}
\end{table}

\begin{figure}[t]
   \centering 
   \includegraphics[width=0.33\textwidth]{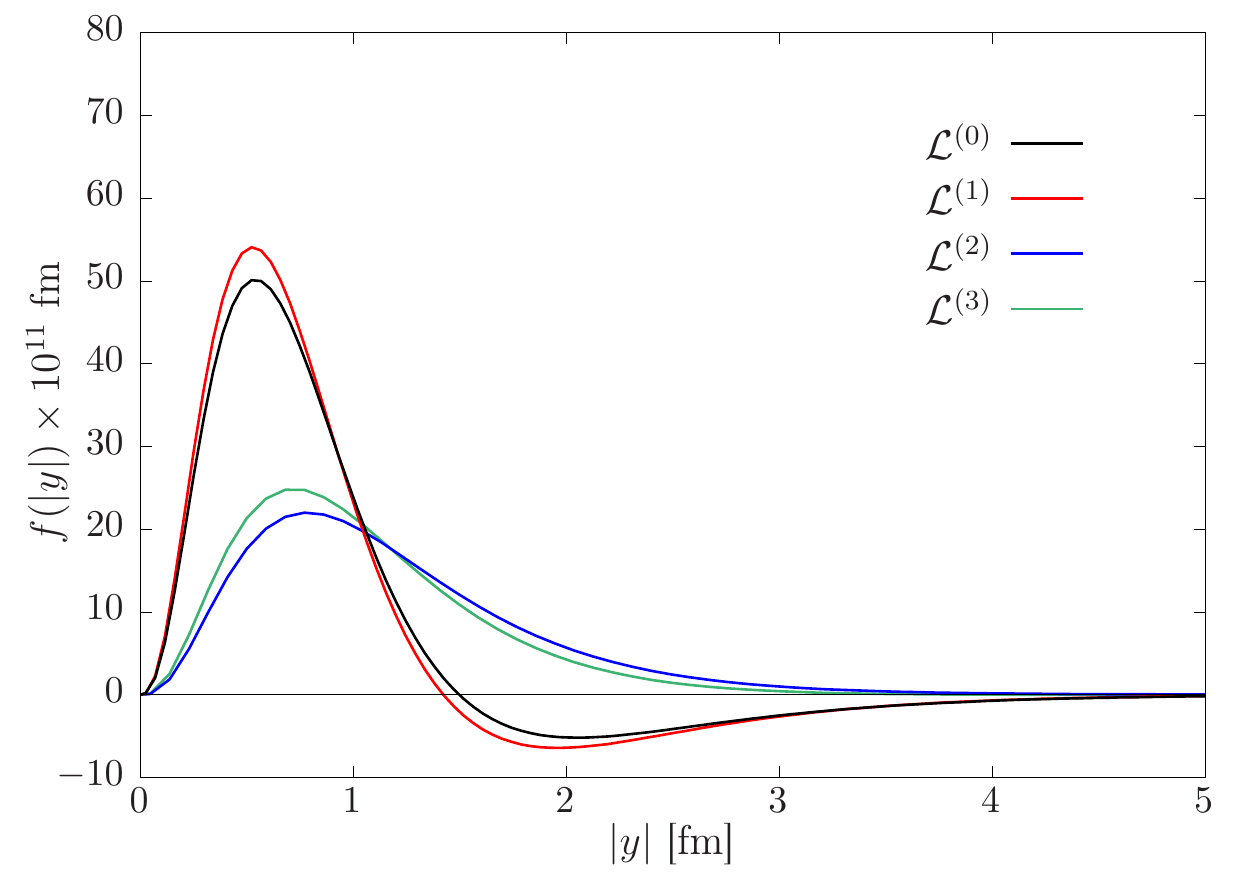}\hfill
   \includegraphics[width=0.33\textwidth]{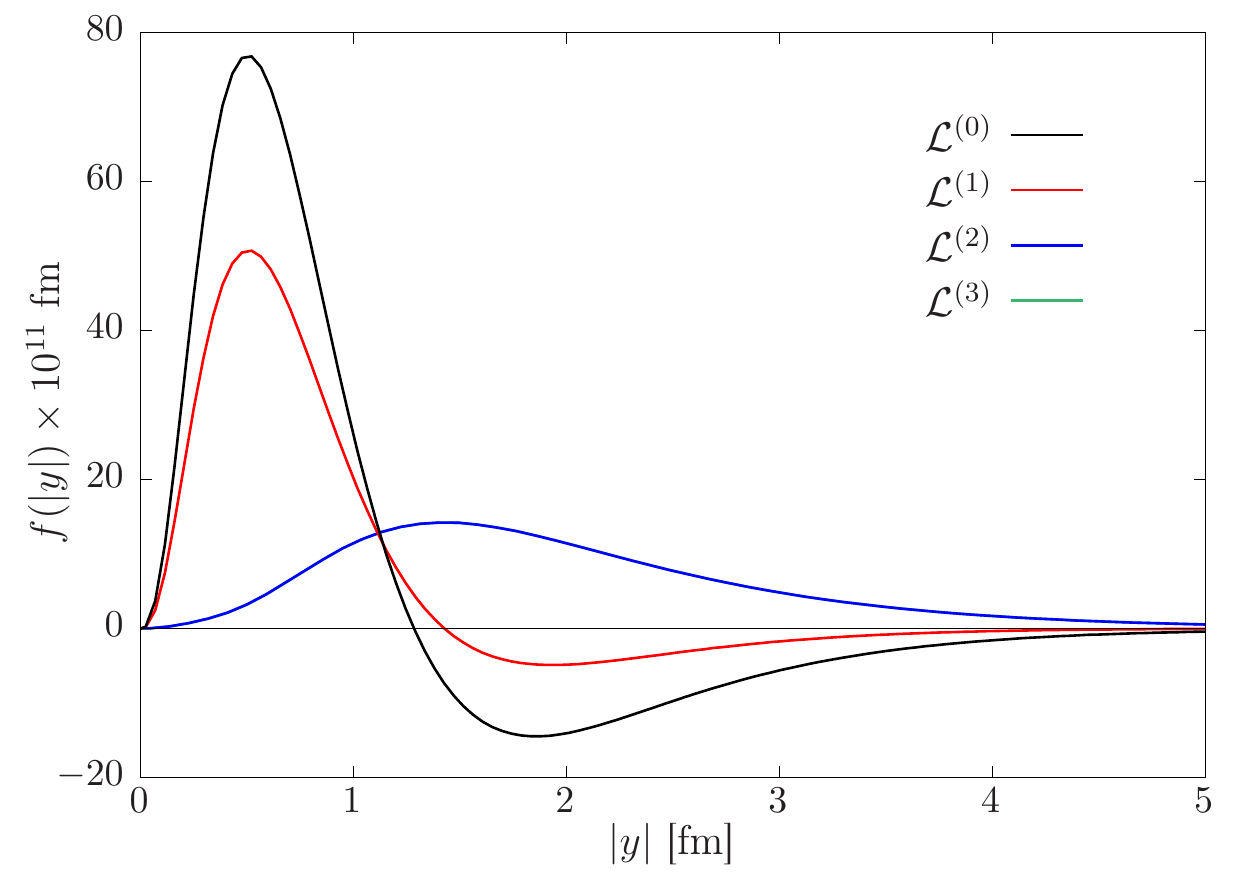}\hfill 
   \includegraphics[width=0.33\textwidth]{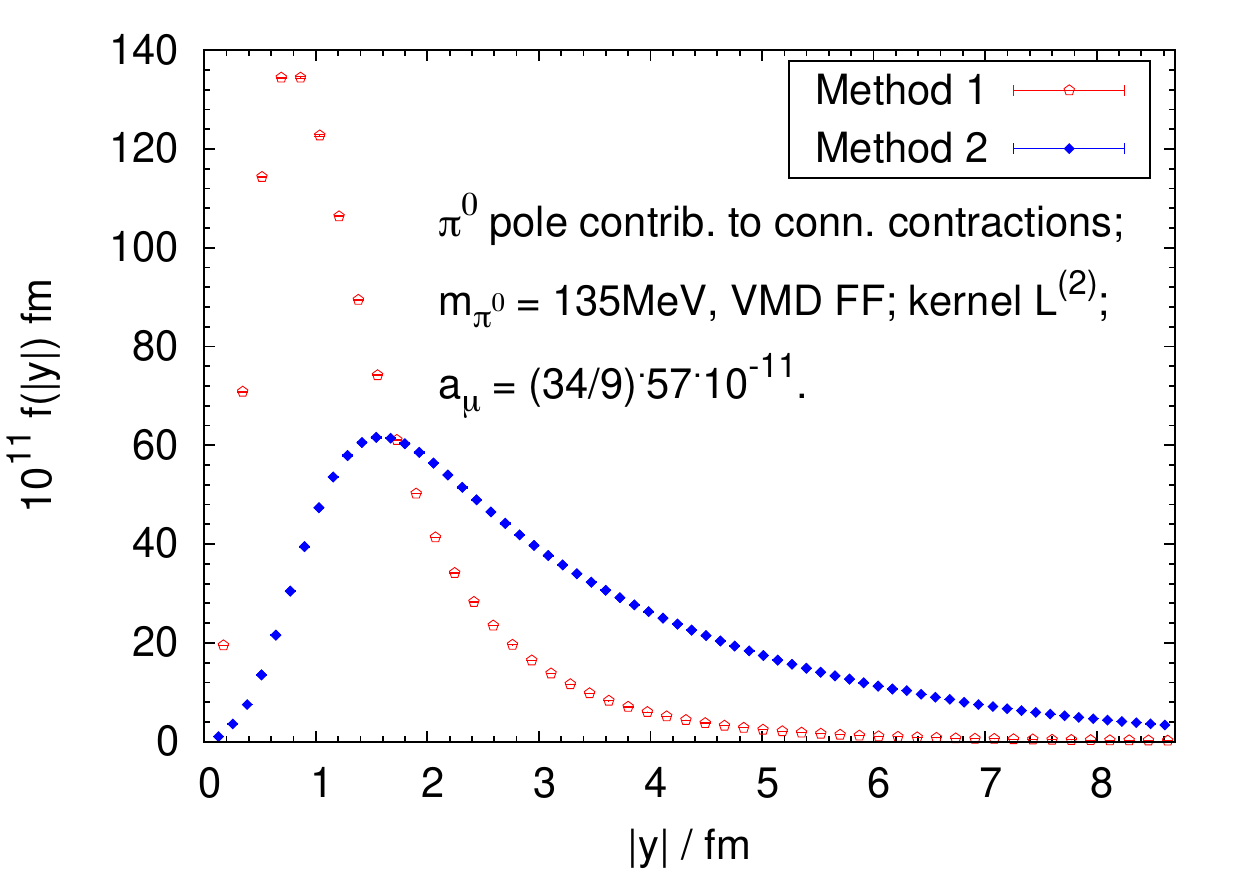}
   \caption{%
     The integrand for the $\pi^0$-pole contribution using the VMD
     form factor based on the standard kernel \(\mcL^{(0)}\) and the
     subtracted kernels \(\mcL^{(1,2,3)}\) at \(M_\pi=300\MeV\),
     using method~1 (left panel) and method~2 (middle panel) for the
     contractions. In the plot in the middle, the \(\mcL^{(3)}\) curve
     is hidden behind the \(\mcL^{(2)}\) curve. The right panel
     compares the integrands with kernel \(\mcL^{(2)}\) for methods~1
     and 2 for a pion with physical mass. Reprinted from Ref.~\cite{Asmussen:2019act}.
   }
   \label{fig:subtractions_methods}
\end{figure}

As explained above, the integral does not change with the choice of
the QED kernel; however, the integrand does. The left panel
of~\cref{fig:subtractions_methods} displays the integrands $f(|y|)$
from \cref{eq:masterMainz} of the final integration over~$|y|$
corresponding to the different kernels, for the neutral pion-pole
contribution for $M_\pi = 300$~MeV with the VMD model. Here method~1
for the contractions is used. Compared to the standard kernel,
\(\mcL^{(2)}\) and \(\mcL^{(3)}\) have less pronounced peaks at short
distances, there is no negative tail, and they approach zero faster at
long distances.  We expect these subtracted kernels to have smaller
lattice artifacts and therefore to be favorable in lattice
computations.  For comparison, the middle panel shows the effect of
using the method~2 for the contractions. As explained above, method~2
is computationally much cheaper than method~1, but now the integrand, e.g.,
for kernel \(\mcL^{(2)}\) becomes again more long-ranged. This is even
more pronounced in the right panel, which shows the integrand for the
VMD form factor with the physical pion mass, again using kernel
\(\mcL^{(2)}\).

\subsubsection{Differences between the RBC and Mainz infinite-volume QED methods and QED loop tests}
\label{sec:LbL_QED}

The idea to compute the QED part analytically in the continuum was first proposed by the Mainz group~\cite{Green:2015mva}. Subsequently, the RBC group developed a similar method~\cite{Blum:2017cer}.  
\begin{figure}
    \centering
    \includegraphics[width=0.35\textwidth]{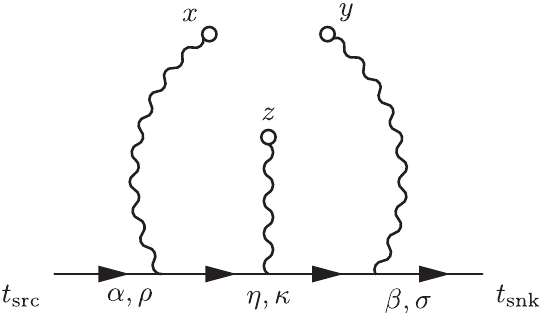}
    \caption{The QED weighting function for the HLbL scattering
      amplitude with the conventions by RBC. Reprinted from Ref.~\cite{Blum:2017cer}.}
        \label{fig:qed inf}
\end{figure}

The amplitude, or weighting function, to be computed is shown in
\cref{fig:qed inf}. The Mainz group employs a fully Lorentz-covariant formulation. Using translational and $O(4)$
rotational invariance, the space-time dependence is reduced to three
parametric variables for the points $x$, $y$, and
$z$~\cite{Asmussen:2016lse}. The RBC group instead uses a setup
consistent with their QED$_L$ simulations where the direction of the
muon line is fixed along the time axis, which results in five
parameters. In either case, the function is computed once and saved,
and then used repeatedly for each set of points ($x,y,z$) to be
evaluated for the hadronic amplitude. In practice, the QED weighting
functions are computed on a regular grid in parameter space that can
be interpolated for arbitrary values of ($x,y,z$). 

Both the Mainz and RBC infinite-volume QED method can be tested by replacing
the quark loop with a lepton loop, while keeping the rest of the lattice calculation
the same.  In practice this means that the quark propagators are solved on a unit gauge background instead of a QCD gauge configuration.  This therefore is a strong test of the diagrammatic sampling methods that are used.

\begin{figure}[t]
    \centering
    \includegraphics[width=0.49\textwidth]{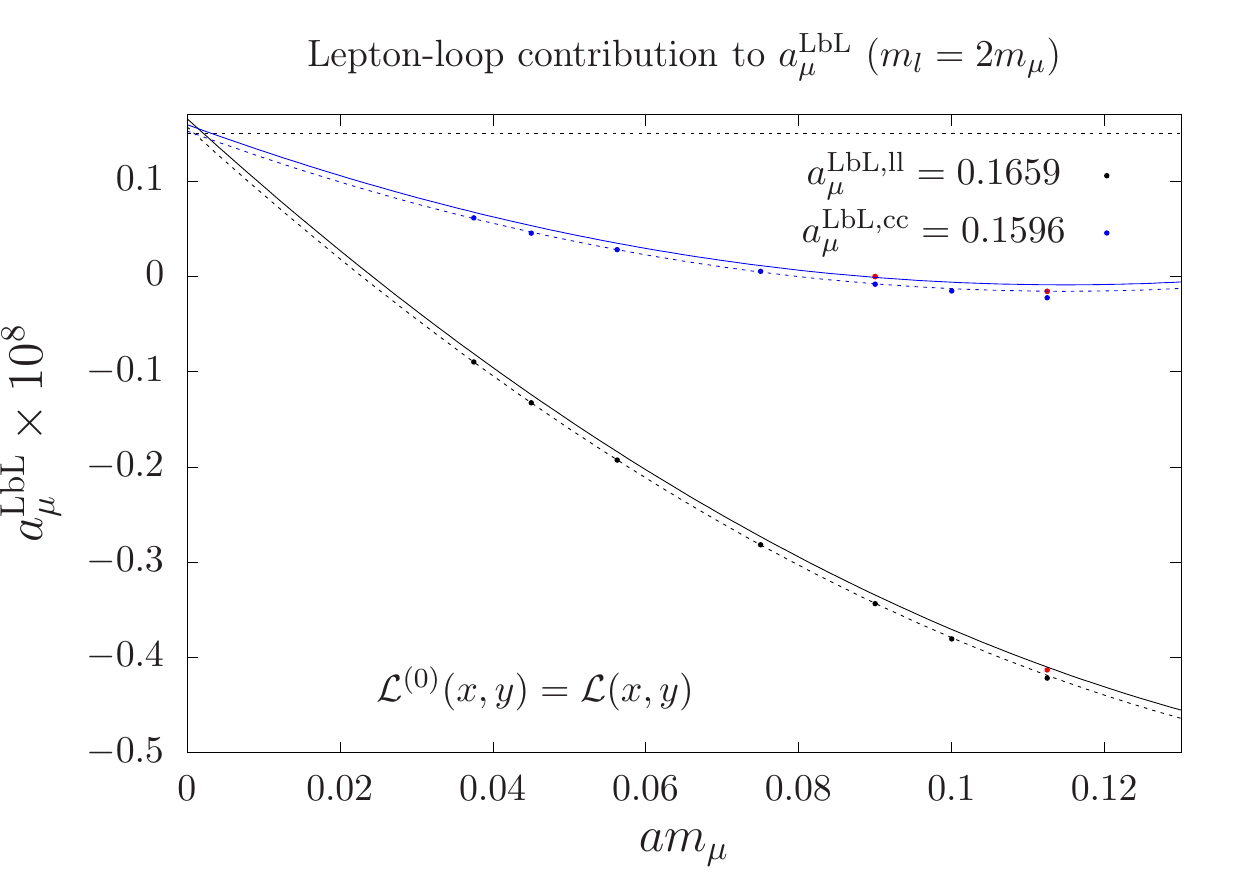}
    \includegraphics[width=0.49\textwidth]{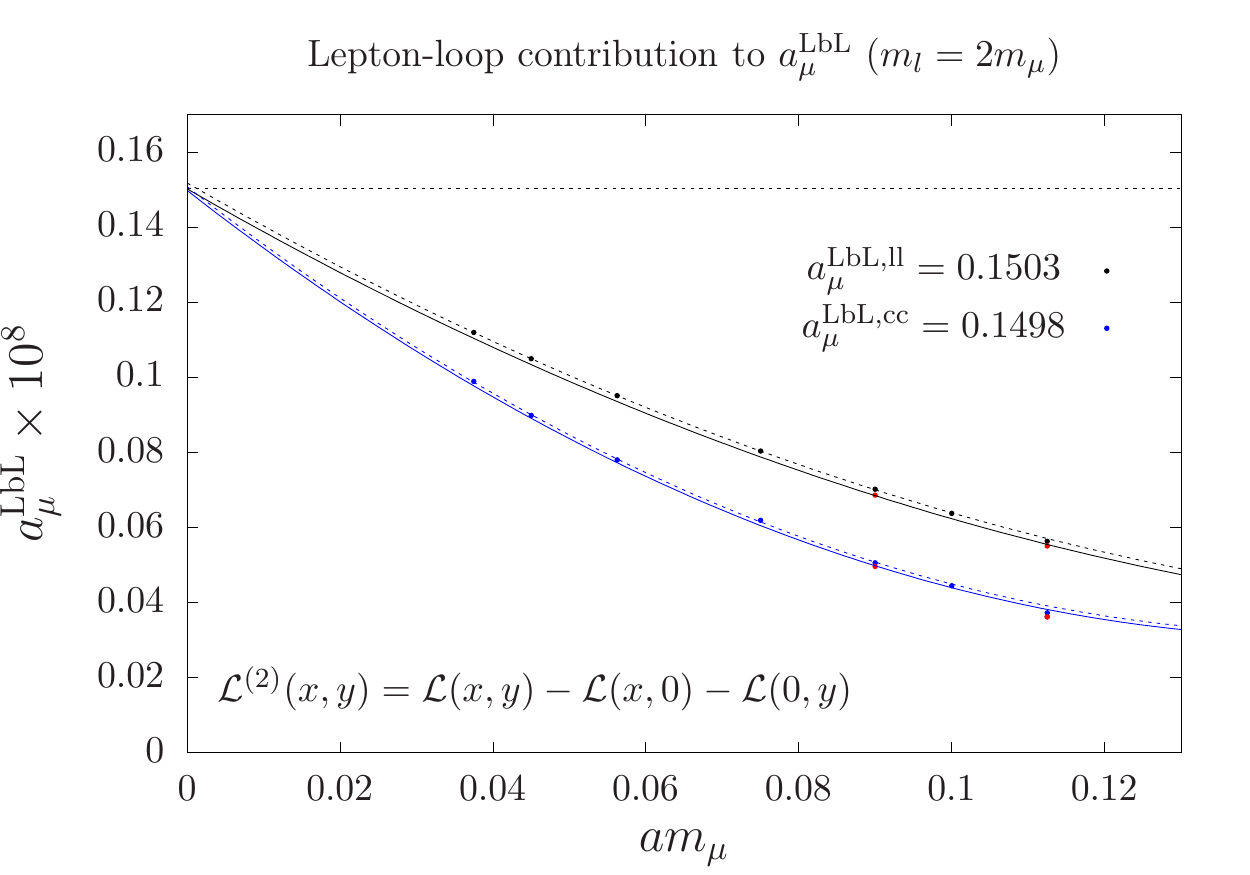}
    \caption{LbL scattering contribution to the muon
      anomaly for QED$_\infty$ using the Mainz
      approach~\cite{Asmussen:2018oip} (loop mass $m_l$ $=$ 2$\times$ muon
      mass $m_\mu$) with 4 local (ll), or 2  conserved and 2 local currents (cc). The left panel shows results for the original QED
      weighting function. The right panel is for the subtracted case
      (see text). The lepton loop is evaluated for various lattice
      spacings, and $m_\mu L=7.2$ (dashed lines). Larger volume (14.4)
      curves (solid lines) are obtained by shifting the $m_\mu L=7.2$ curve
      by the difference between 7.2 and 14.4 simulations at two
      lattice spacings ($am_\mu\approx0.09$ and 0.11).  The horizontal lines denote the known analytic result.}
        \label{fig:qedinfmainz}
\end{figure}

The left panel of \cref{fig:qedinfmainz} shows infinite-volume QED results from the Mainz group. The lepton loop is evaluated on a discrete lattice ($m_\mu L=7.2$) using improved Wilson fermions
(leading $\order(a)$ error). 
Again, typically $a^{-1}\sim 1\hyph2$\,GeV for lattice QCD simulations, which corresponds to $am_\mu\sim 0.1\hyph0.05$ in the figure. Even though the QED weighting function is computed entirely in the continuum and infinite volume, the residual lattice spacing [$\order(a^2)$] error is significant. Results for local and point-split (conserved) currents are shown, the former having much larger discretization effects---even the wrong sign for the entire range of the simulations. We come back to this below. The finite-volume errors are barely detectable, a consequence of the large volume used ($m_\mu L=7.2$). Typical sizes for QCD simulations satisfy $m_\mu L \approx 4$.

\begin{figure}[ht!]
    \centering
    \includegraphics[width=0.46\textwidth]{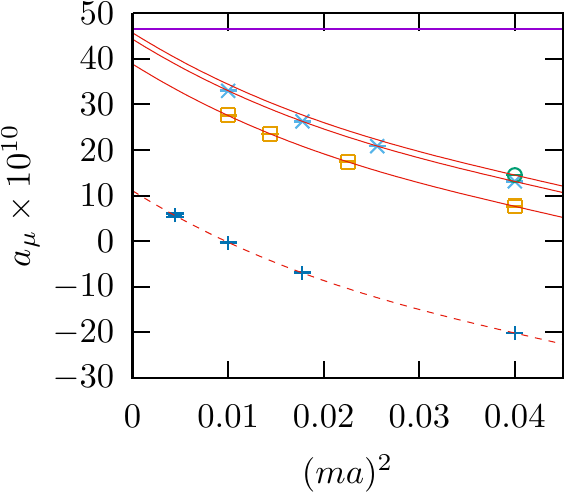} \hfill
    \includegraphics[width=0.485\textwidth]{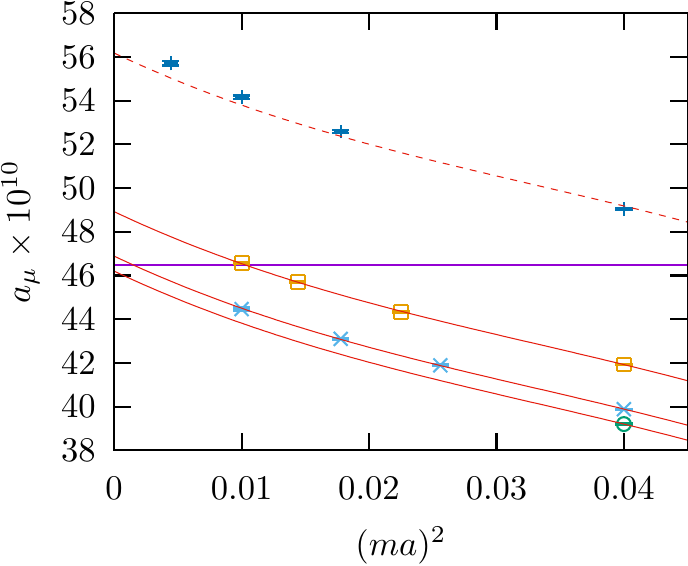}
    \caption{LbL scattering contribution to the muon
      anomaly for QED$_\infty$ using the RBC
      approach~\cite{Blum:2017cer} (loop mass $m_l$ $=$ muon mass $m_\mu$ $=$ axis label $m$).The left
      panel shows results for the original QED weighting function. The
      right panel is for the subtracted case (see text). The lepton
      loop is evaluated for various lattice spacings and sizes
      [$m_\mu L=3.2$, 4.8, 6.4, 9.6, bottom to top (left) and top to bottom
      (right)]. In both panels the results are fit up to order $a^6$
      while assuming that $F_2(L=\infty)=F_2(m_\mu L=9.6)$.  The horizontal lines denote the known analytic result. Adapted from Ref.~\cite{Blum:2017cer}.}
        \label{fig:qedinfrbc}
\end{figure}

The left panel of \cref{fig:qedinfrbc} shows infinite-volume QED results from the RBC group~\cite{Blum:2017cer}. The lepton loop is again evaluated on a discrete lattice for various spacings and sizes, and only local currents. Similarly to the Mainz results, residual lattice spacing ($\order(a^2)$) effects are large. We also see large finite-volume corrections since here sizes start at $m_\mu L=3.2$.

In Ref.~\cite{Blum:2017cer} an improved weighting function is constructed by subtracting the same weighting function, but evaluated for one point coincident with another. Using the notation in Ref.~\cite{Blum:2017cer} to avoid confusion, the RBC subtracted weighting function is 
\begin{equation}
  \mathfrak{G}^{(2)}_{\sigma, \kappa, \rho} (y, z, x)  = 
  \mathfrak{G}^{(1)}_{\sigma, \kappa, \rho} (y, z, x)
  -\mathfrak{G}^{(1)}_{\sigma, \kappa, \rho} (z, z, x)
  -\mathfrak{G}^{(1)}_{\sigma, \kappa, \rho} (y, z, z)\,.
  \label{eq:muon-line-v2}
\end{equation}
In this case the subtraction term vanishes in the continuum and infinite-volume limits when combined with the QCD part as a consequence of the Ward identity, or charge conservation for the quark loop~\cite{Blum:2017cer}. Since the subtraction removes contributions where two of the three points are the same, it is expected that lattice spacing errors will be significantly reduced.
In the Mainz notation the equivalent subtraction reads 
\begin{equation}
  \bar{\mcL}(x,y) - \bar{\mcL}(x-y,0) - \bar{\mcL}(y,y)\,.
\end{equation}
In the right-hand panels of \cref{fig:qedinfmainz,fig:qedinfrbc} results are shown for the subtracted case. One sees the residual lattice spacing and finite-volume effects are substantially reduced (comparing RBC results, discretization errors are reduced by a factor of four, and finite-volume by two~\cite{Blum:2017cer}). For the Mainz results the local-current values are now closer to the continuum one. This suggests the subtractions are working as intended, removing large short-distance artifacts that arise when currents collide.

The known analytic values for loop masses one and two times the muon mass are
listed in \cref{tab:leploop}.
The extrapolated values agree well with these, but it should be noted that subtraction terms are important for reliable extrapolations.

\subsection{Cross-checks between RBC and Mainz}
\label{sec:cross-check_QCD}

\begin{figure}[t]
    \centering
    \includegraphics[width=0.49\textwidth]{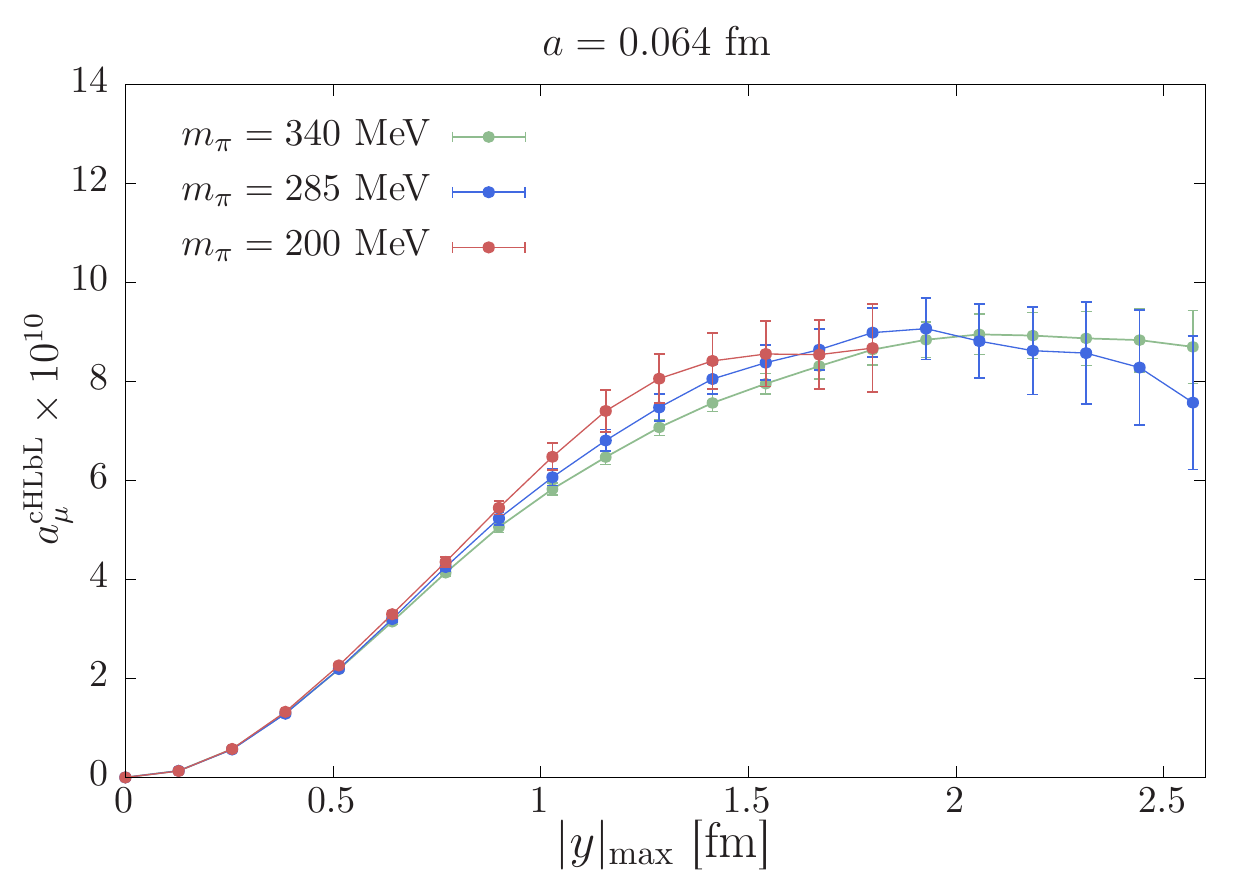} \hfill
    \includegraphics[width=0.48\textwidth]{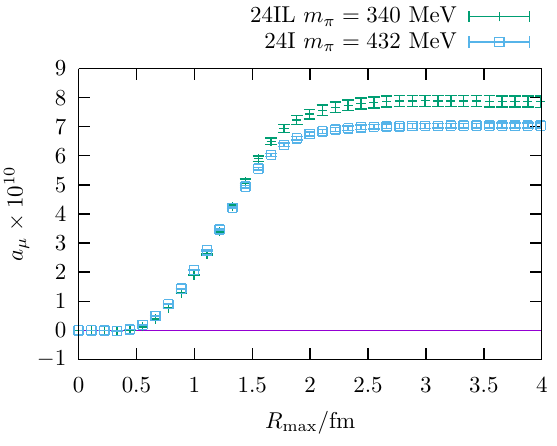}
    \caption{Pion mass dependence of partial sums for the HLbL contribution to the muon anomaly. Mainz (left) and RBC (right). QED$_\infty$ with subtraction is used in both [${\cal L}^{(2)}$ (left) and $\mathfrak{G}^{(2)}$ (right)]. }
        \label{fig:mpi compare}
\end{figure}

\begin{figure}[ht!]
    \centering
    \includegraphics[width=0.49\textwidth]{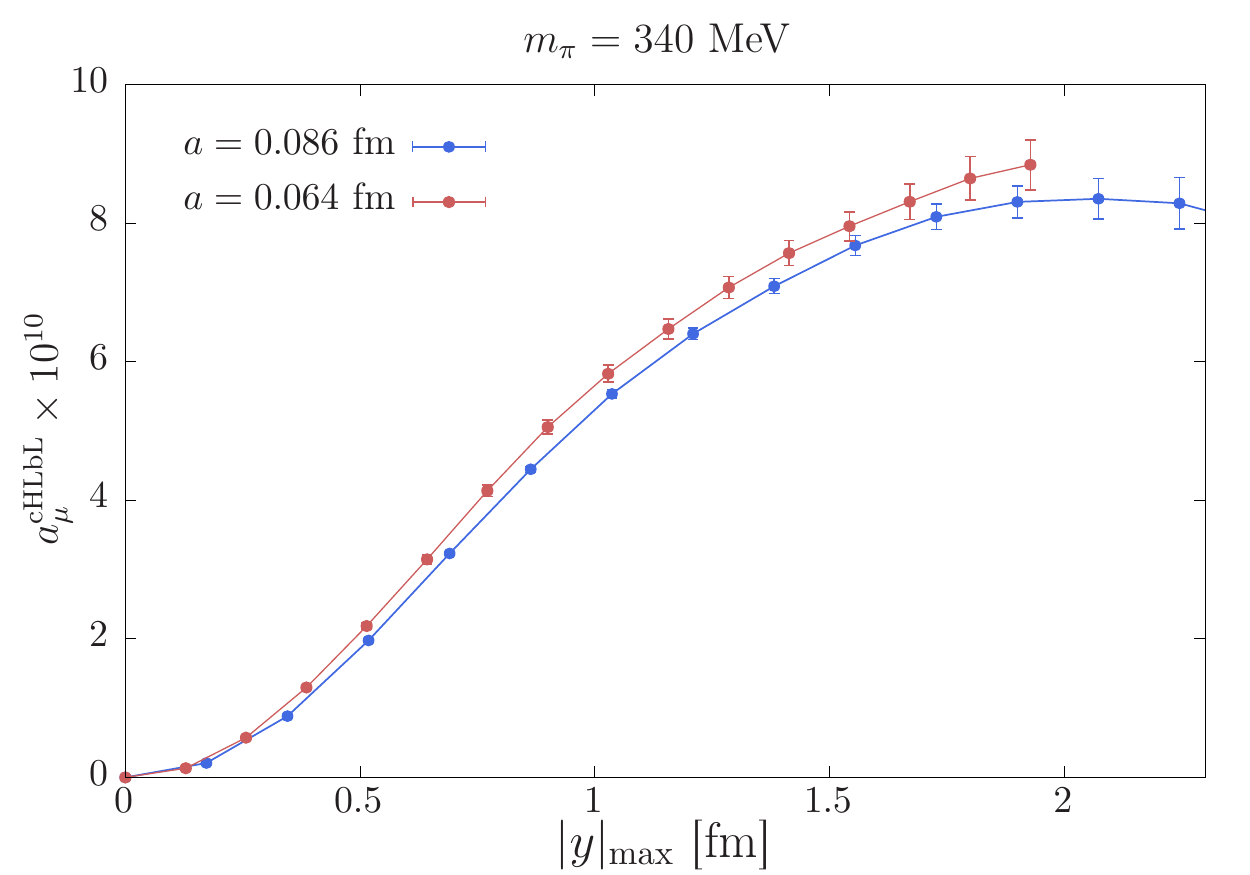}
    \hfill 
    \includegraphics[width=0.48\textwidth]{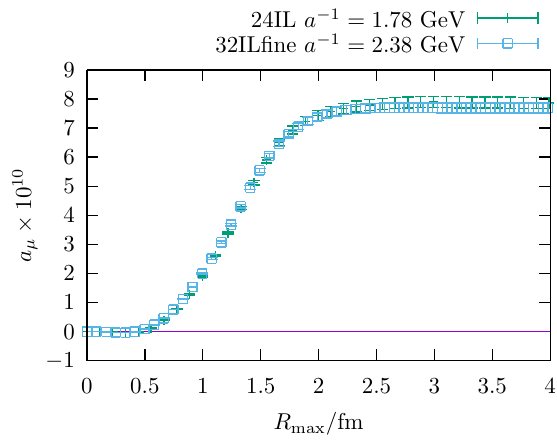}
    \caption{Lattice spacing dependence of partial sums for the connected HLbL contribution to the muon anomaly. Mainz, $M_\pi=340$\,MeV (left) and RBC, $M_\pi=340$ and 357\,MeV (right). QED$_\infty$ with subtraction is used in both [${\cal L}^{(2)}$ (left) and $\mathfrak{G}^{(2)}$ (right)]. }
        \label{fig:a compare}
\end{figure}

A cross-check of results between the Mainz and RBC groups has been carried out for heavier than physical masses. Strictly speaking, one should only compare values in the infinite-volume and continuum limits since both groups use different QED$_\infty$ weighting functions and lattice discretizations. Such extrapolations, at a fiducial pion mass, have not been done by the Mainz group, and only for the connected diagram by the RBC group. However, the parameters of the calculations are reasonably close and so a comparison is still worthwhile. 

Mainz uses the Wilson-clover fermion and
L\"uscher--Weisz gauge action with tree-level coefficients.
The parameters for the ensembles are $a^{-1}=2.29$ and 3.08\,GeV, $L=2.7\hyph4.2$\,fm, and $M_\pi=200$, 285, and 340\,MeV. RBC uses the Domain Wall fermion and both the Iwasaki and Iwasaki--DSDR gauge actions with $a^{-1}=1.015$, 1.73, and 2.38\,GeV, $L=2.7\hyph4.7$\,fm, and $M_\pi=340$, 357, and 432\,MeV.  The Iwasaki--DSDR gauge action includes an additional dislocation-suppressing determinant ratio (DSDR) compared to the Iwasaki gauge action that reduces lattice artifacts for coarser gauge ensembles.

In \cref{fig:mpi compare} the pion mass dependence is shown for $\amuHLbL$. In the left panel (Mainz) a weak pion mass dependence is observed (note the signal for the 200\,MeV case disappears beyond $|y|=1.5$\,fm). In the right panel (RBC), the mass dependence is clearly resolved as $R_{\rm max}$ increases
($R_{\rm max}$ is the upper bound on distances between positions $x$, $y$, $z$ of three internal QED vertices). For the common 340\,MeV pion ensembles, the curves plateau to consistent values, around $8\times10^{-10}$. However this good agreement might be coincidental, due to the differing approaches, and could fail in the limits $a\to0$, $L\to\infty$.

To further investigate the consistency between the calculations, we show the lattice spacing dependence in \cref{fig:a compare}. Mainz (left) uses 340\,MeV pions, and RBC 340 and 357\,MeV pions (the small shift in the pion mass should not have a large effect here, cf.\ \cref{fig:mpi compare}). The signal degrades above 2\,fm in the left panel, before a sharp conclusion can be drawn. However, at shorter distance where the lattice artifacts should be more pronounced, there is no discernible difference. The right panel shows no difference within relatively small errors, and from this we conclude that for these lattice spacings and pion masses the discretization errors are small, at most.

\begin{figure}[t]
    \centering
    \includegraphics[width=0.49\textwidth]{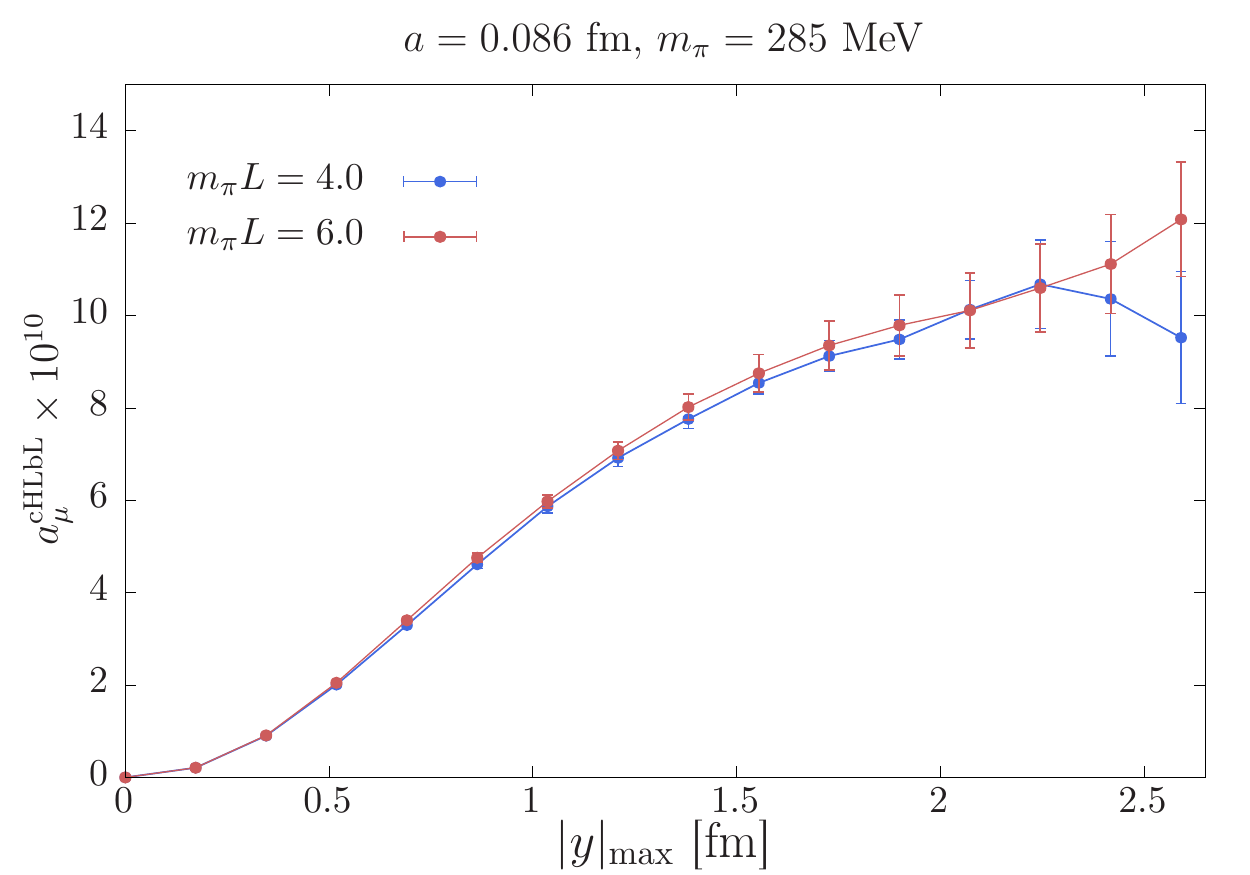} \hfill
    \includegraphics[width=0.45\textwidth]{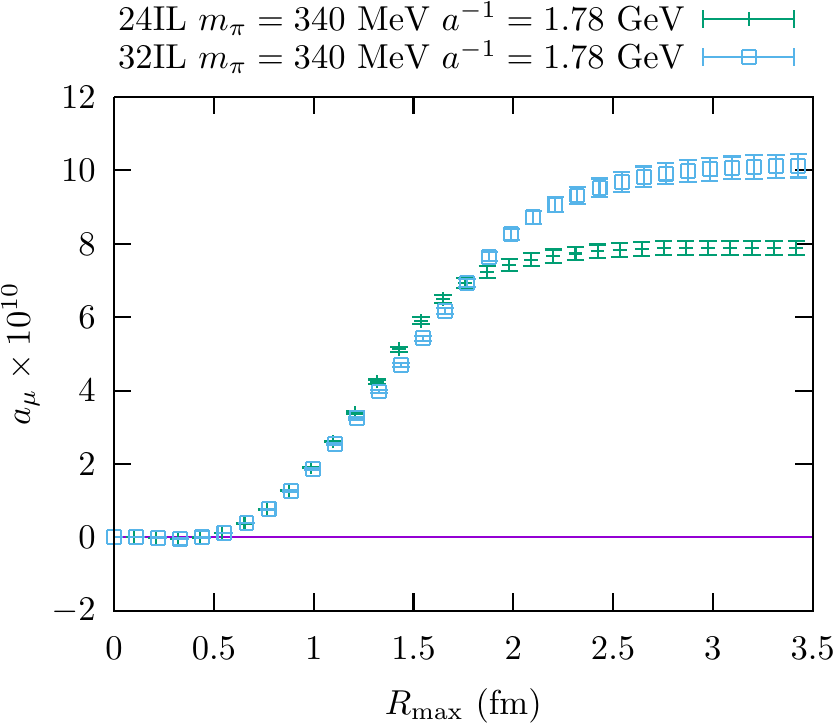}
    \caption{Volume dependence of partial sums for the HLbL contribution to the muon anomaly. Mainz, $M_\pi=285$\,MeV (left) and RBC, $M_\pi=340$\,MeV (right). QED$_\infty$ with subtraction is used in both [${\cal L}^{(2)}$ (left) and $\mathfrak{G}^{(2)}$ (right)]. }
        \label{fig:fv compare}
\end{figure}

Finally, we discuss the finite-size effects in the calculations.
In \cref{fig:fv compare} results are shown for fixed pion mass, with $M_\pi=285$ and 340\,MeV for the Mainz and RBC groups, respectively. In the left panel (Mainz) there is no effect outside of statistical errors, which grow appreciably at large distance. In the right panel (RBC) there is a significant volume dependence, which is clear after about 2\,fm. In the left panel, the curves diverge at about the same distance, though not outside of statistical errors. The ratio of lattice sizes is $4.13/2.75=1.5$ and $3.55/2.66=1.33$ for Mainz and RBC, respectively, but the Mainz pion mass is lighter, so the effects should be roughly the same.

\begin{figure}[t]
    \centering
    \includegraphics[width=0.49\textwidth]{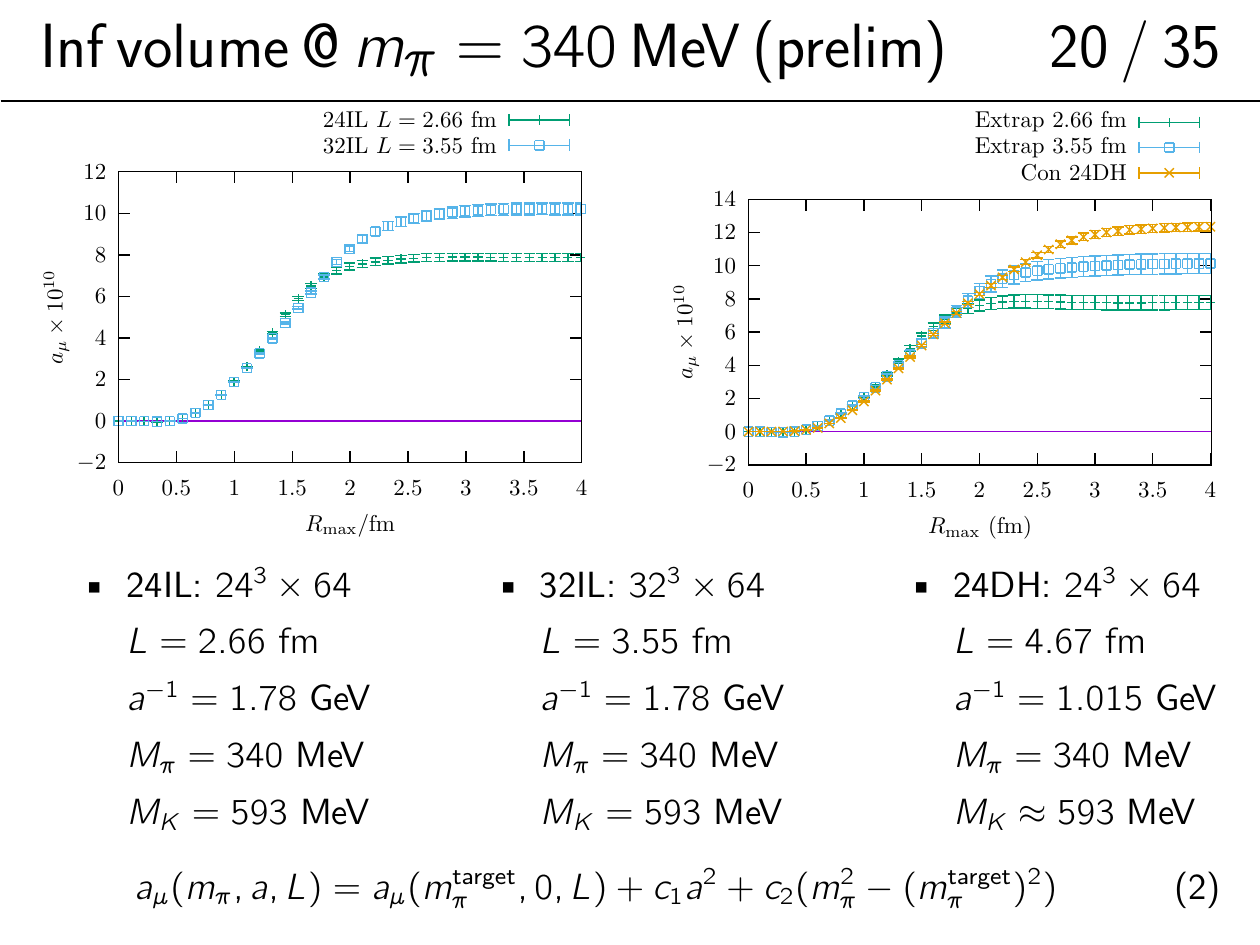}
    \includegraphics[width=0.49\textwidth]{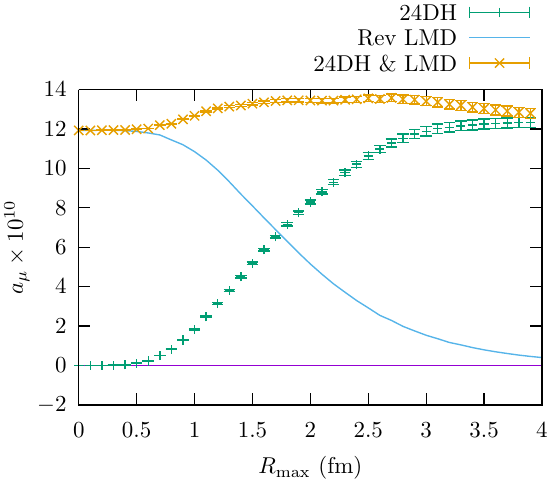}
    \caption{Connected HLbL contribution to the muon anomaly (RBC). In the left panel the two Iwasaki ensembles, $L=2.66$ and 3.55\,fm, have been extrapolated to the continuum limit, and the Iwasaki--DSDR, 1\,GeV, $L=4.7$\,fm ensemble is shown for comparison. The right panel shows the combination of the Iwasaki--DSDR ensemble with the LMD pion-pole model result for long distance (reverse partial sum).}
        \label{fig:qed inf vol}
\end{figure}

Both the RBC and the Mainz~\cite{Asmussen:2019act} groups have gone further to investigate the finite-volume effects. In \cref{fig:qed inf vol} the results in \cref{fig:fv compare} have been extrapolated to the continuum limit, and another ensemble with $a^{-1}=1$\,GeV and $L=4.7$\,fm is shown for comparison. Neglecting nonzero $a$ effects, which are expected to be small for QED$_\infty$, there is still a potentially sizable finite-volume effect. Since the finite-volume effects arise at long distance from the pion contribution, the lattice results are combined with the pion-pole contribution in infinite volume (see \cref{sec:pion pole}). The combination is shown in the right panel of \cref{fig:qed inf vol}. If the switch over from the full lattice result to the pion-pole contribution is done at large enough distance (but not too large), a plateau should develop that gives the infinite-volume result for the entire contribution. Such a plateau exists for $R_{\rm max}$ between 1.5 and 3\,fm. The value is about $13.5\times10^{-10}$, which indicates the 1\,GeV curve in the left panel is already close to the infinite-volume limit.

From the above we conclude the Mainz and RBC connected HLbL results for heavy pion masses are compatible. It would be worthwhile to carry out a similar comparison for the disconnected diagram when those results become available. The results for the connected diagram should also be improved to make a stringent comparison.

\subsection{Results for physical pion mass}
\label{sec:phys mass results}

\begin{figure}[t]
    \centering
    \includegraphics[width=0.45\textwidth]{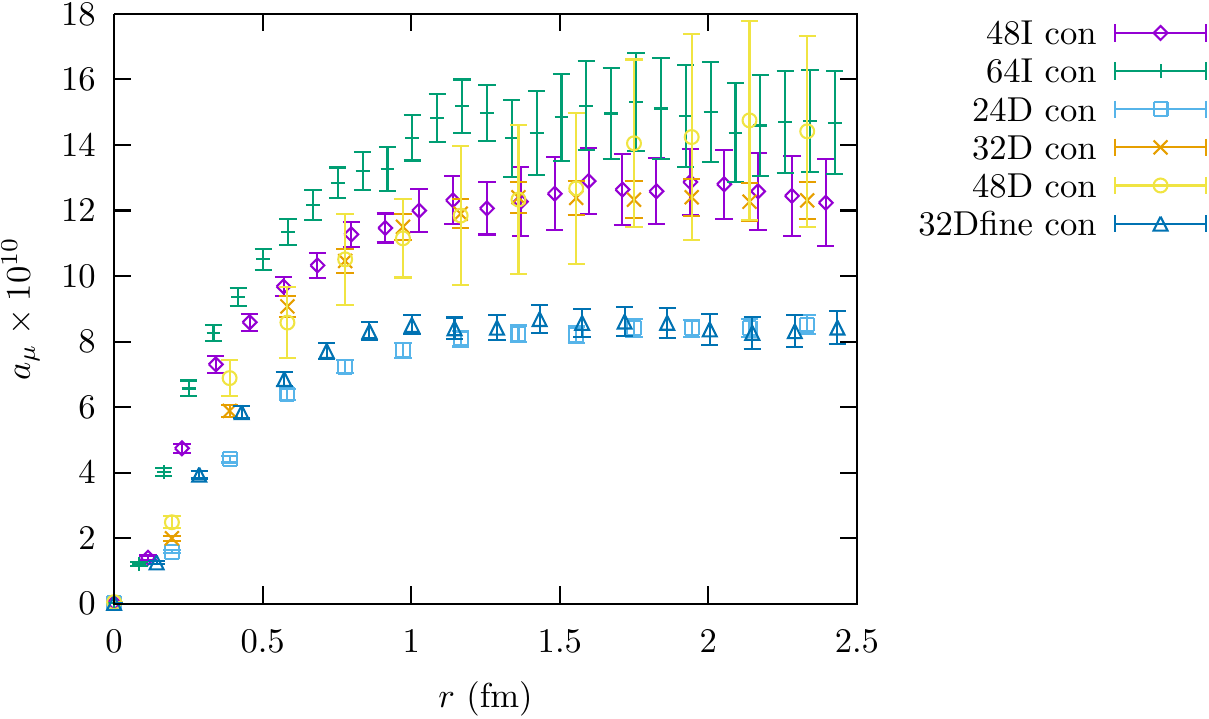}
    \hskip 0.05\textwidth
    \includegraphics[width=0.45\textwidth]{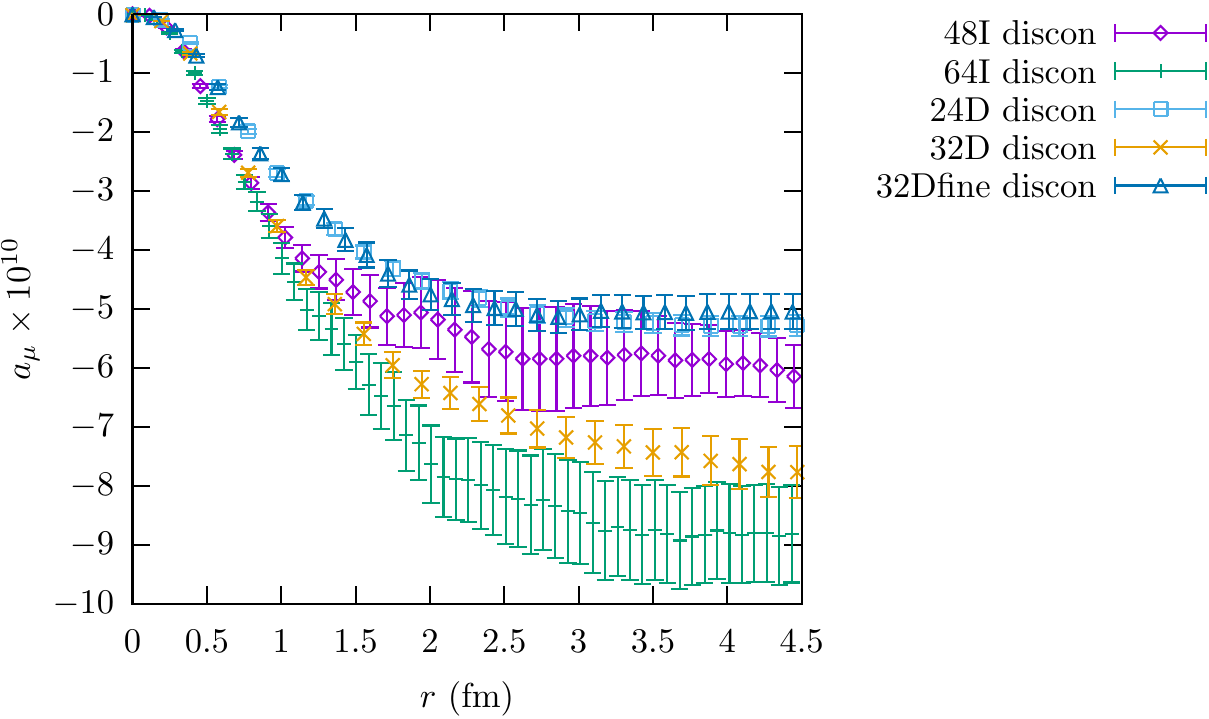}
    \caption{Partial sums for connected (left) and leading disconnected (right) HLbL contributions to the muon anomaly at the physical mass point (RBC). ``$r$'' denotes the distance between the two sampled vertices on the hadronic loop(s). Most of the contributions come from the region $r \simle 1\hyph2$\,fm. Reprinted from Ref.~\cite{Blum:2019ugy}.}
    \label{fig:qedl phys mass results}
\end{figure}

So far only the RBC group has obtained results at the physical mass point (see \cref{tab:rbc ensembles} for a summary of ensembles and lattice parameters). In Ref.~\cite{Blum:2016lnc} using QED$_L$, they obtained 
\begin{equation}
    \amuHLbL=5.35(1.35)\times 10^{-10}
\end{equation}
for $a=0.114$\,fm, in finite volume ($L\approx 5.5$\,fm), using the first ensemble listed in \cref{tab:rbc ensembles}. To obtain this result only connected and leading disconnected diagrams were computed (\cref{fig:qedl phys mass results} displays results for all ensembles). While this value is about a factor of two smaller than model and data-driven ones, as observed in the pure QED$_L$ case, there may be severe power-law finite-volume effects (and somewhat less severe nonzero lattice spacing effects). Further inspection of \cref{fig:qedl phys mass results} reveals a large cancellation between the leading contributions, rendering the statistical precision on the final result significantly worse than the connected or disconnected ones alone. Such a cancellation is expected from simple charge counting arguments together with pion dominance~\cite{Bijnens:2016hgx,Jin:2016rmu,Gerardin:2017ryf}. 

\begin{table}[t]
    \centering
    \small
    \begin{tabular}{cccccccc}
    \toprule
    fermion & gluon & $M_\pi$ (MeV) & size & $a^{-1}$ (GeV) & $a$ (fm) & $L$ (fm)& meas\\
    \midrule
       MDWF  & I & 139 & $48^3$ & 1.730 & 0.114 & 5.476 & 65, 124 \\
       MDWF  & I & 135 & $64^3$ & 2.359 & 0.0837 & 5.354 & 43, 105 \\
       MDWF  & ID & 142 & $24^3$ & 1.015 & 0.194 & 4.656 & 157, 156\\
       MDWF  & ID & 142 & $32^3$ & 1.015 & 0.194 & 6.21 & 70, 69\\
       MDWF  & ID & 142 & $48^3$ & 1.015 & 0.194 & 9.32 & 8, 0\\
       MDWF  & ID & 144 & $32^3$ & 1.378 & 0.141 & 4.51 & 75, 69\\
    \bottomrule
    \end{tabular}
    \caption{Lattice parameters for RBC ensembles. M\"obius domain wall fermions (MDWF) and Iwasaki (I), or Iwasaki--DSDR (ID) gauge fields. The QCD amplitudes are combined with QED$_L$ and QED$_\infty$ weighting functions. For the former, free DWFs and photons are used. The muon mass is always taken at the physical point~\cite{Tanabashi:2018oca}. In the ``meas'' column the entries refer to the number of configurations used for connected and disconnected diagram measurements, respectively.}
    \label{tab:rbc ensembles}
\end{table}

The calculation has been repeated~\cite{Blum:2019ugy} on the ensembles listed in \cref{tab:rbc ensembles}, which include smaller lattice spacing at the same volume and larger spacing with larger volumes.

\begin{figure}[t]
    \centering
    \includegraphics[width=0.45\textwidth]{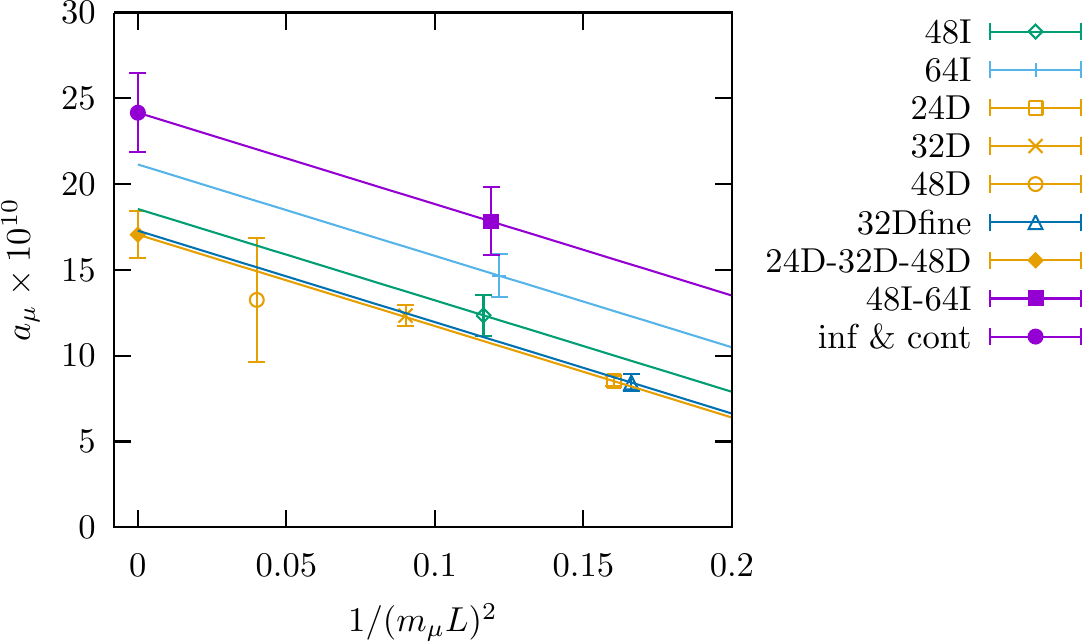}
    \includegraphics[width=0.45\textwidth]{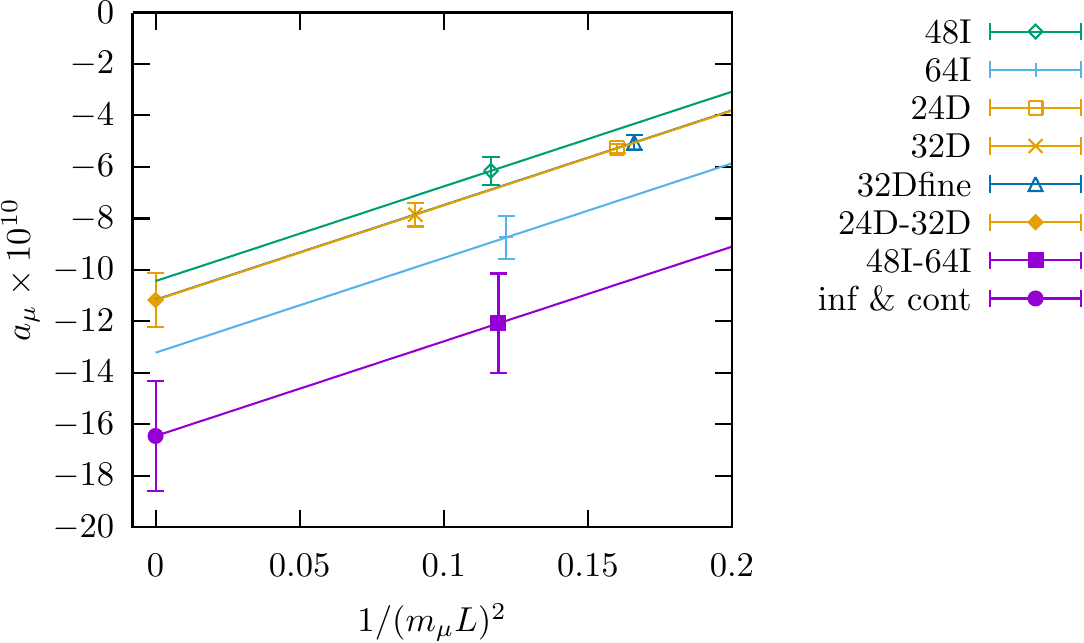}
    \includegraphics[width=0.45\textwidth]{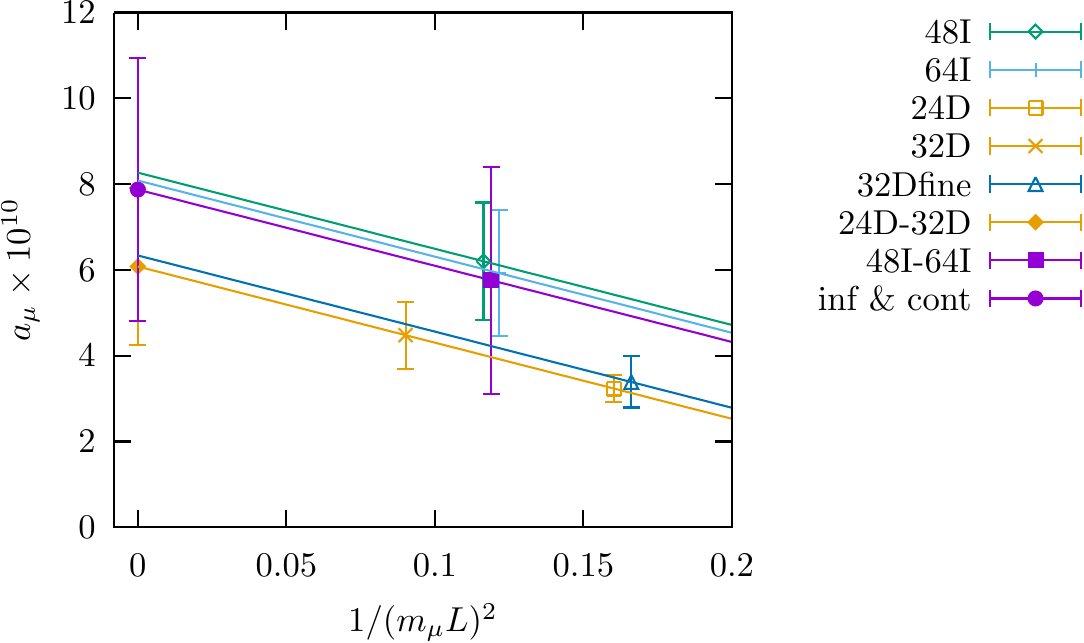}
    \caption{Continuum and infinite-volume extrapolation in QED$_L$ at the physical mass point (RBC). Connected (top-left), disconnected (top-right), and total (bottom) hadronic contributions to the muon anomaly from HLbL scattering. The solid lines are evaluated from a fit to \cref{eq:QED_L+QCD fit}. Upper (connected, total) and lower (disconnected) lines correspond to $a=0$. Reprinted from Ref.~\cite{Blum:2019ugy}.}
    \label{fig:qedl extrap}
\end{figure}

The data points are combined using a simple ansatz,
\begin{equation}
\label{eq:QED_L+QCD fit}
    a_\mu(L,a^{\rm I},a^{\rm D}) =  a_\mu\left(1-\frac{b_1}{(m_\mu L)^2}-c_1 \big(a^{\rm I}\big)^2 -c_1 \big(a^{\rm D}\big)^2 +c_2 \big(a^{\rm D}\big)^4 \right),
\end{equation}
where ``I'' and ``D'' refer to Iwasaki and Iwasaki--DSDR gauge actions, respectively. 
The extrapolation to the limits $L\to\infty$ and $a\to0$ is shown in \cref{fig:qedl extrap} and yields~\cite{Blum:2019ugy}
\begin{align}
a^{\rm cHLbL} &= ~~24.16(2.30)(5.11)\times10^{-10}\,,\notag\\
a^{\rm dHLbL} &= -16.45(2.13)(3.99)\times10^{-10}\,,\notag\\
\amuHLbL &= ~~~~7.87(3.06)(1.77)\times10^{-10}\,,
\end{align}
for the connected, leading disconnected, and total hadronic contribution to the muon anomaly from HLbL scattering. The first quoted error is statistical, the second systematic. The latter is estimated by taking the difference of central values from \cref{eq:QED_L+QCD fit} and fit forms with various higher-order corrections, including cross terms, i.e., terms that depend on both $a$ and $L$. 

\begin{figure}[t]
    \centering
    \includegraphics[width=0.45\textwidth]{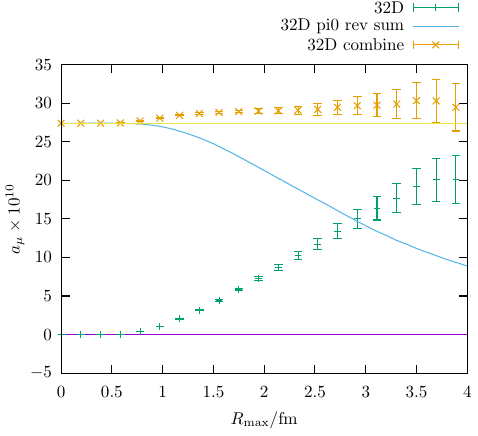}
    \includegraphics[width=0.45\textwidth]{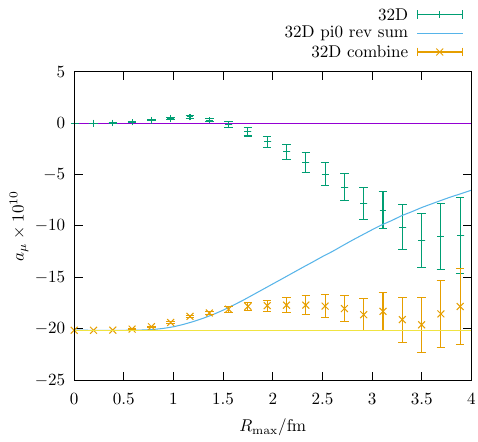}
    \includegraphics[width=0.45\textwidth]{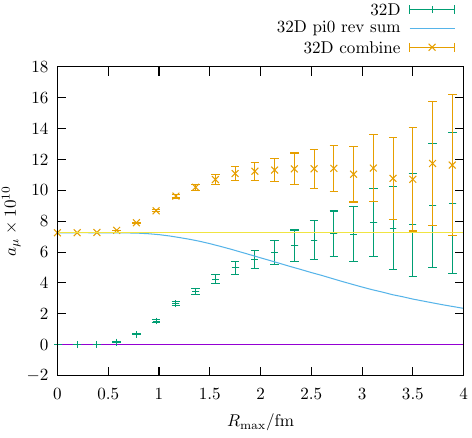}
    \caption{Combined lattice and pion-pole contributions to the HLbL scattering part of the muon anomaly. Partial sums for the hadronic contributions, connected (top-left), leading disconnected (top-right), and total (bottom), computed with QED$_\infty$. $a^{-1}=1$\,GeV, $L=6.4$\,fm, and $M_\pi=142$\,MeV. Lines denote the $\pi^0$-pole contribution computed from the LMD model and are summed right-to-left.}
    \label{fig:qed inf phys mass}
\end{figure}

\begin{figure}[t]
    \centering
    \includegraphics[width=0.5\textwidth]{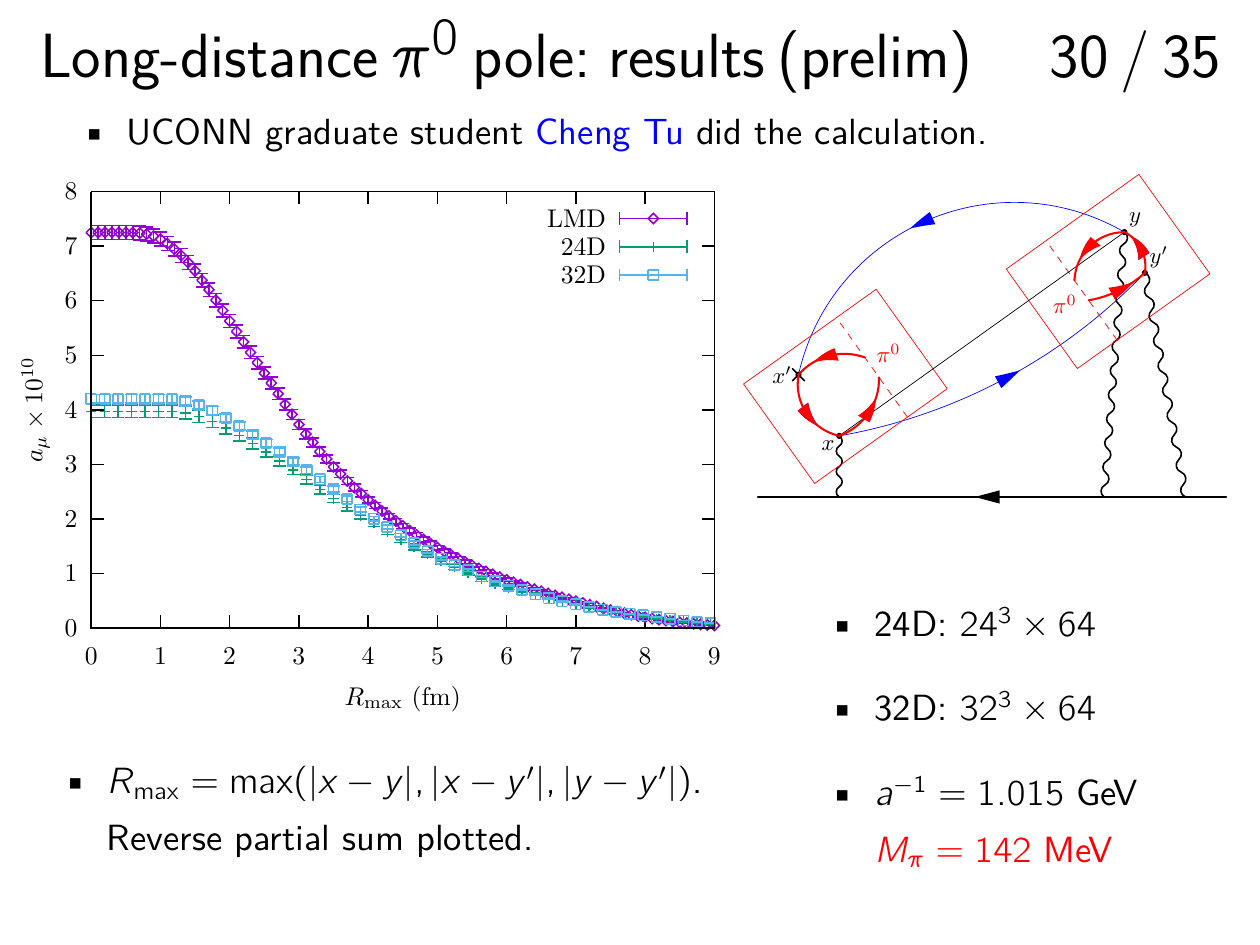}
    \caption{Contribution of the $\pi^0$ pole contribution summed right-to-left.
    The LMD curve uses the LMD model converted to coordinate space and combined with
    the RBC's subtracted QED kernel.
    The other two curves, 24D and 32D, use the $\pi^0 \to \gamma\gamma$
    amplitude calculated with these two ensembles.
    However, different from the LMD curve, the calculation of these two curves
    only captures the leading $1/(M_\pi R_\text{max})$ $\pi^0$ term
    in the pole contribution.
    Therefore the difference is due to
    (i) the pion TFF an  d
    (ii) the long distance ($M_\pi R_\text{max} \gg 1 $)
    approximation used in lattice calculation.
    }
    \label{fig:rbc pion pole}
\end{figure}

The RBC group has also computed the hadronic contributions within the QED$_\infty$ framework. \Cref{fig:qed inf phys mass} shows values computed on the $L=6.4$\,fm, $a^{-1}=1$\,GeV ensemble. In this demonstration the lattice result has been combined with a model calculation of the dominant pion-pole piece, which is free of statistical fluctuations, discretization errors, and finite-volume effects. At long distance the pion pole dominates, while for shorter distances other contributions are important as well. Therefore one should search for a distance to switch between the two where the statistical and finite-volume errors from the full calculation are under control, but is still large enough that the pole term dominates. From \cref{fig:qed inf phys mass} one sees a range, $1.5\le R_{\rm max}\le 3.0$\,fm, where the total contribution remains constant, while the statistical error stays under control. The value of this plateau is a bit higher, but consistent within errors, to the QED$_L$ result quoted above. While a continuum limit of the QCD part is yet to be taken, because the subtracted weighting function has been used, the residual lattice spacing error is expected to be small. Likewise, the systematic error due to the model is likely to be small (see \cref{sec:pion pole}), and eventually the model will be replaced by the lattice calculation as is shown in \cref{fig:rbc pion pole}.

\subsection{Pion-pole contribution}
\label{sec:pion pole}

In Minkowski space-time, the TFF describing the interaction between a neutral pion and two off-shell photons with momenta $q_1$ and $q_2$ is defined via the following matrix element,
\begin{equation} 
 i \int d^4 x \, e^{i q_1 \cdot x} \, \langle \Omega | T \{ j_{\mu}(x) j_{\nu}(0) \} | \pi^0(p) \rangle =
\epsilon_{\mu\nu\alpha\beta} \, q_1^{\alpha} \, q_2^{\beta} \, \FF(q_1^2, q_2^2) \,,
\label{eq:M}
\end{equation}
where $j_{\mu}$ is the hadronic component of the electromagnetic current, $p=q_1+q_2$, and $\epsilon_{\mu\nu\alpha\beta}$ is the fully antisymmetric tensor with  $\epsilon^{0123} = +1$.

In lattice QCD, the starting point is the three-point correlation function defined as 
\begin{equation}
C^{(3)}_{\mu\nu}(\tau,t_{\pi}) \equiv a^6\sum_{{\bf x}, {\bf z}} \, \big\langle   j_{\mu}({\bf z}, t_i) j_{\nu}({\bf 0}, t_f)  P^{\dag}({\bf x},t_0) \big\rangle \, e^{i {\bf p}\cdot{\bf x}} \, e^{-i {\bf q}_1\cdot{\bf z}} \,.
\label{eq:C3} 
\end{equation}
Here $\tau=t_i-t_f$ is the time separation between the two vector currents, 
and  $t_{\pi}={\rm min}(t_f-t_0,t_i-t_0)$ is the minimal time separation between the pion interpolating operator $P^\dagger$ 
and the two vector currents.
From here, the amplitude $\widetilde A_{\mu\nu}$ is extracted,
\begin{gather}
\widetilde{A}_{\mu\nu}(\tau) \equiv \lim_{t_{\pi} \rightarrow + \infty} e^{E_\pi (t_f-t_0)} C^{(3)}_{\mu\nu}(\tau,t_{\pi}) \,.
\label{eq:Amunu}
\end{gather}
With $i{Z_{\pi}}/{\sqrt{2E_\pi}}=i\langle0|P(0)|\pi\rangle/\sqrt{\langle \pi|\pi\rangle}>0$ 
parameterizing the overlap of the pseudoscalar operator with the pion
state, $\widetilde A_{\mu\nu}$ is related to the TFF via
\begin{align}
\widetilde{A}_{\mu\nu}(\tau)  &= -i Q^{E}_{\mu\nu} \ \widetilde{A}^{(1)}(\tau) + P^{E}_{\mu\nu} \ \frac{ d \widetilde{A}^{(1)} }{d\tau}(\tau) \,,
\label{eq:decomp}
\\
\widetilde{A}^{(1)}(\tau) &= \frac{iZ_{\pi} }{ 4 \pi E_{\pi} }  \int_{-\infty}^{\infty} \, d \widetilde{\omega} \, \FF(q_1^2,q_2^2) e^{-i \widetilde{\omega} \tau} \,.
\label{eq:A1}
\end{align}
The arguments of the TFF in \cref{eq:A1} are given by 
\begin{equation}
q_1^2 = \omega_1^2 - {\bf q}_1^{\, 2}\,,\qquad 
q_2^2 = (E_{\pi} - \omega_1)^2 - ({\bf p}-{\bf q}_1)^2\,,
\label{eq:kin}
\end{equation}
with $\omega_1$ set to $i \widetilde{\omega}$.
The ($\omega_1$-independent) tensors appearing on the right-hand side of \cref{eq:decomp} are defined by
$ \epsilon_{\mu\nu\alpha\beta} q_1^{\alpha} q_2^{\beta}  = -i P^{E}_{\mu\nu} \omega_1 + i^{n_0} Q^{E}_{\mu\nu}$,
where $n_0$ denotes the number of temporal indices carried by the two vector currents.

\Cref{eq:decomp,eq:A1} allow one to predict the amplitude $\widetilde A_{\mu\nu}(\tau)$,
given the TFF. The inversion of the Fourier transform in \cref{eq:A1}, followed by the analytic continuation 
to imaginary $\widetilde\omega$, 
i.e., real $\omega_1$, allows one to obtain the TFF from $\widetilde{A}^{(1)}(\tau)$
for all spacelike and lightlike virtualities, $q_i^2\leq 0$~\cite{Ji:2001wha}.

\begin{figure}[hbt]

\begin{center}
\includegraphics[width=0.49\textwidth]{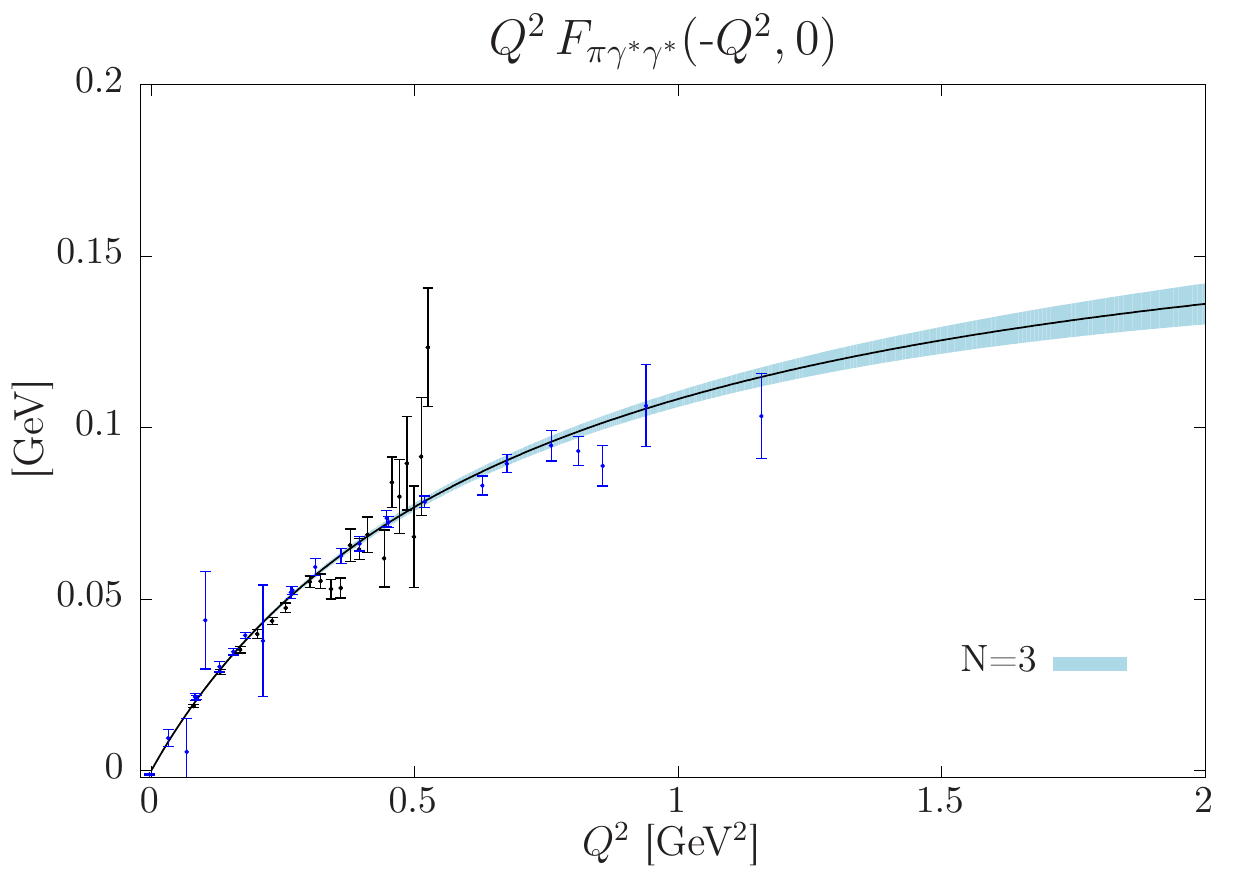}
\includegraphics[width=0.49\textwidth]{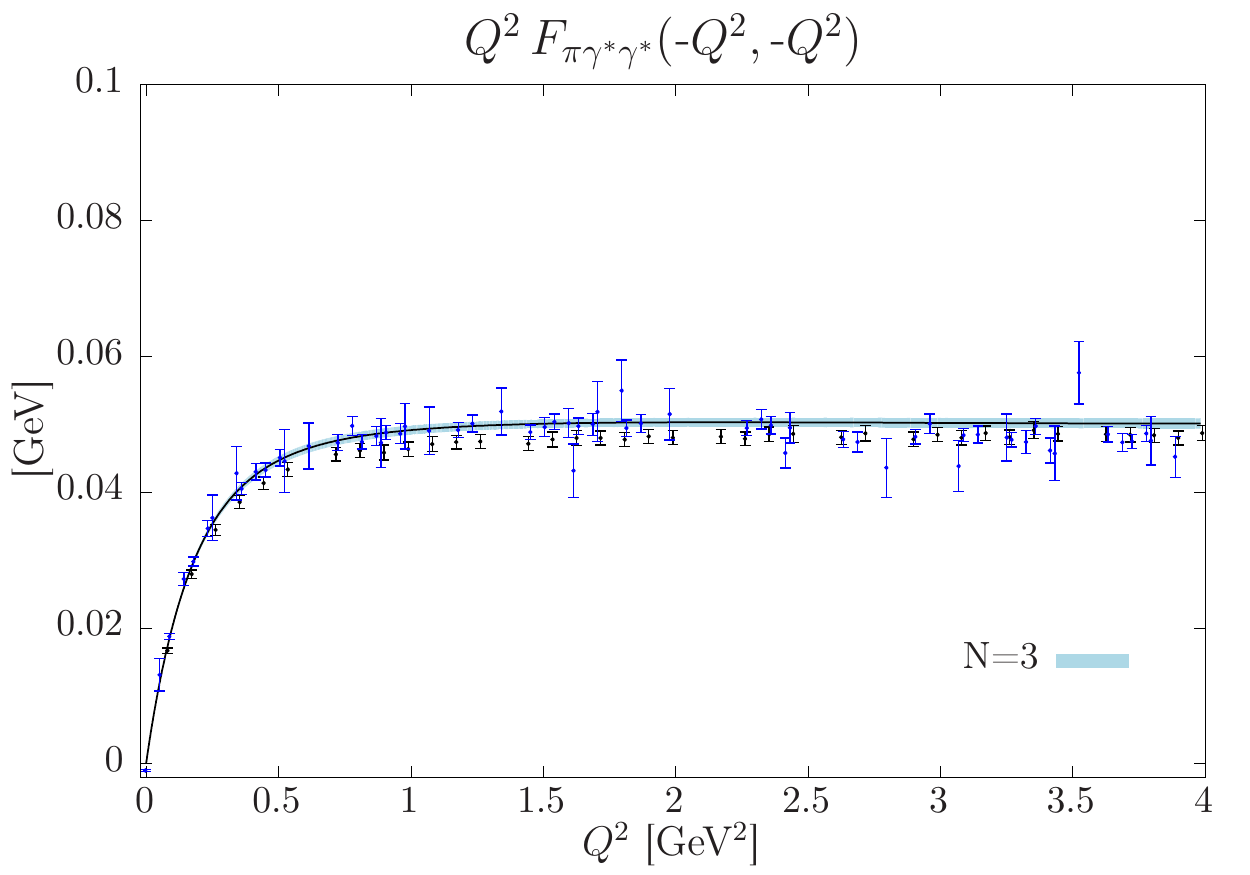}
\end{center}
\caption{Lattice results for the pion TFF~\cite{Gerardin:2019vio} using local vector currents
on a $64^3\times128$ ensemble (labeled D200) with a pion mass of 200\,MeV and a lattice spacing of 0.064\,fm.
The left panel shows the TFF at singly-virtual kinematics, the right panel at equal-virtuality kinematics. 
Black points correspond to the results obtained in the pion rest frame, while the blue points are obtained in the pion moving frame. 
Error bands correspond to the global $z$-expansion fitting procedure
described in the main text. Reprinted from Ref.~\cite{Gerardin:2019vio}.}
\label{fig:TFF_D200}
\end{figure}

Two calculations have been performed to date with the goal of covering the kinematics relevant to $\amuHLbL$.
The first one was carried out with $N_{\rm f}=2$ flavors of dynamical quarks~\cite{Gerardin:2016cqj},
the second~\cite{Gerardin:2019vio} with $N_{\rm f}=2+1$ flavors (i.e., dynamical up, down, and strange quarks).
Both used an $\order(a)$ improved Wilson quark action, however only the second employed $\order(a)$ improvement 
for the vector currents, with improvement coefficients computed in Ref.~\cite{Gerardin:2018kpy}.
Moreover, two lattice discretizations of the vector current have been used to constrain the continuum extrapolation.
An example of the lattice data obtained at $M_\pi=200\,$MeV is given in \cref{fig:TFF_D200}.
Two reference frames were used, thus providing a check for the Lorentz covariance of the results,
and allowing for a better kinematic coverage than with the pion rest-frame alone.

As an important development since Ref.~\cite{Gerardin:2016cqj}, 
a systematically improvable parameterization of the TFF was used in Ref.~\cite{Gerardin:2019vio},
inspired by the analysis of other hadronic form factors,
\begin{equation}\label{eq:z_exp_mod}
\Bigg(1 + \frac{Q_1^2 + Q_2^2}{M_V^2}\Bigg) \, \FF(-Q_1^2, -Q_2^2)  =   \sum_{n,m=0}^{N} c_{nm} \, \left( z_1^n - (-1)^{N+n+1} \frac{n}{N+1} \, z_1^{N+1} \right) \, \left( z_2^m -  (-1)^{N+m+1} \frac{m}{N+1} \, z_2^{N+1} \right) \,,
\end{equation}
with $c_{nm}=c_{mn}$, $M_V=775\,{\rm MeV}$,
\be
 z_k = \frac{ \sqrt{t_c+Q_k^2} - \sqrt{t_c - t_0} }{ \sqrt{t_c+Q_k^2} + \sqrt{t_c - t_0} }\,,
\ee
the threshold $t_c$ set to $4M_\pi^2$, and $t_0$ chosen so as to
minimize the maximum value of $z_k$ for $Q_k^2\in [0 ,4\,{\rm GeV}^2]$.
The subtractions to the powers $z_1^n$ and $z_2^m$ in \cref{eq:z_exp_mod} are designed to enforce the property that the
imaginary part of the TFF open proportionally to
$(q^2-t_c)^{\ell+1/2}$ with $\ell=1$.  The $z$-expansion at order
$N=3$, fit to virtualities up to $4\,{\rm GeV}^2$, is used.  The
parameters of the expansion are themselves expanded to linear order in
$M_\pi^2$ and $a^2$ and fit globally to the data from thirteen
lattice ensembles covering four lattice spacings and pion masses
ranging from 200\,MeV to 420\,MeV. Plots of the singly- and
doubly-virtual form factors extrapolated to the physical point can be
found in \cref{fig:pi0-comparison} in \cref{sec:pion-pole}. The
agreement with experimental data in the singly-virtual case from
BESIII, CELLO,
and CLEO, see \cref{sec:PS_TFF_exp}, and with theoretical calculations for
the singly- and doubly-virtual form factors using dispersion
relations~\cite{Hoferichter:2018dmo, Hoferichter:2018kwz} and
Canterbury approximants~\cite{Masjuan:2017tvw} is very good.

Using the values of the parameters in the continuum (i.e., at $a=0$)
and at the physical pion mass, the pion-pole contribution to
$\amuHLbL$ is calculated using the relevant weight functions, see
\cref{eq:amuPpole}, and the
final result is~\cite{Gerardin:2019vio},
\be
a_\mu^{\pi^0\text{-pole}} = 59.7(3.4)(0.9)(0.5) \times
10^{-11} =  59.7(3.6)\times 10^{-11}. 
\label{eq:amu_zexp}
\ee
The first error is statistical, the second is the systematic error
associated with the parameterization of the TFF, and the third comes
from the disconnected contribution; in the last equality, the
individual error estimates have been added in quadrature. The total
absolute error is reduced by a factor of $2.3$ with respect to the
earlier calculation~\cite{Gerardin:2016cqj} and corresponds to a
relative precision of $6\%$.

An important benchmark of the lattice calculation is the normalization
of the TFF, which is known to 0.8\% from the PrimEx-II
experiment~\cite{Larin:2020}, $\FF(0,0)=0.275(2)\,{\rm GeV}^{-1}$.
The lattice calculation~\cite{Gerardin:2019vio} finds
\be
\FF(0,0)=0.264(8)(4)\,{\rm GeV}^{-1}
\ee
by fitting the virtuality, pion-mass, and lattice-spacing dependence of $f_\pi \FF(Q_1^2,Q_2^2)$
for $Q_i^2 \in [0,1{\rm GeV}^2]$ with a $z$-expansion truncated at
$N=1$. Thus the lattice result is statistically consistent with the
experimental measurement at the 1.2$\sigma$ level, and lies on the lower side. ChPT predicts the absence of a chiral logarithm in the
leading pion-mass dependence of $[f_\pi
\FF(0,0)]$~\cite{Donoghue:1986wv,Bijnens:1988kx}, and phenomenological
studies estimate the slope in $M_\pi^2$ to be small compared to the
typical size $0.5\times 10^{-3}{\rm GeV}^{-2}$ of low-energy constants
(see in particular Ref.~\cite{Kampf:2009tk}). The fit performed to lattice
data~\cite{Gerardin:2019vio} is consistent with these predictions.

It is also interesting to ask what value would be obtained for
$a_\mu^{\pi^0\text{-pole}}$ by combining the lattice data with the
precise experimental information on $\FF(0,0)$.  The result is then
$62.3(2.3)\times 10^{-11}$~\cite{Gerardin:2019vio}, where the
central value, higher by about one standard deviation than
\cref{eq:amu_zexp}, reflects the higher value of the
experimental determination of $\FF(0,0)$ as compared to the lattice
one.

Finally, we note that a further application of the lattice TFF
calculation is the treatment of long-distance effects in the lattice
calculation of the full $\amuHLbL$. The $\pi^0$-exchange
contribution dominates the tail of the coordinate-space integral that
yields $\amuHLbL$ as well as its leading finite-size effect.
In addition, the dependence of $\amuHLbL$ on $M_\pi$ receives
a major contribution from the $\pi^0$ exchange.  These effects can be
better controlled if the TFF is known at the parameters of the gauge
ensembles used to compute $\amuHLbL$.

\subsection{LbL forward scattering amplitudes}

The HLbL amplitudes can be studied for their own sake in lattice QCD,
at kinematics such that no hadrons could be produced in the $\gamma^*\gamma^*$ collision~\cite{Green:2015sra,Gerardin:2017ryf}.
Dispersive sum rules have been derived for the forward amplitudes~\cite{Pascalutsa:2010sj,Pascalutsa:2012pr}
in terms of the cross sections for $\gamma^*\gamma^*\to{\rm hadrons}$, allowing for comparisons with
phenomenology.
The Euclidean momentum-space four-point function of the electromagnetic current
$j_\mu$ at spacelike virtualities 
\be \label{eq:masterl}
 \Pi_{\mu_1\mu_2\mu_3\mu_4}(P_4;P_1,P_2) \equiv 
\int_{X_1,X_2,X_4} 
e^{-i \sum_{a} P_a\cdot X_a} \Big\langle j_{\mu_1}(X_1)j_{\mu_2}(X_2)j_{\mu_3}(0)j_{\mu_4}(X_4)\Big\rangle
\ee
can be computed in lattice QCD and projected to one of the eight forward $\gamma^*\gamma^*\to\gamma^*\gamma^*$ scattering amplitudes, 
in particular
\be
 {\cal M}_{\rm TT}(-Q_1^{\,2},-Q_2^{\,2},-Q_1\cdot Q_2)
= \frac{ e^4}{4} {R_{\mu_1\mu_3}R_{\mu_2\mu_4}}  \Pi_{\mu_1\mu_3\mu_4\mu_2}(-Q_2;-Q_1,Q_1)\,.
\ee
The projectors $R_{\mu\nu}$ project onto the plane orthogonal to the vectors $Q_1$ and $Q_2$;
thus ${\cal M}_{\rm TT}$  corresponds to the amplitude involving transversely polarized photons.
With 
 $\nu = \frac{1}{2}(s+Q_1^2+Q_2^2)$, a crossing-symmetric variable parameterizing the CM energy $\sqrt{s}$, 
the subtracted dispersion relation
\be
 {\cal M}_{\rm TT}(q_1^2,q_2^2,\nu) - {\cal M}_{\rm TT}(q_1^2,q_2^2,0) 
  = \frac{2\nu^2}{\pi}
\int_{\nu_0}^\infty d\nu'\frac{ \sqrt{ \nu'{}^2 - q_1^2 q_2^2  }}{\nu'(\nu'{}^2-\nu^2-i\epsilon)} (\sigma_0+\sigma_2)(\nu')
\ee
can be written, 
where $\sigma_{J}$ corresponds to the total cross section for the photon--photon fusion reaction $\gamma^*\gamma^*\to{\rm hadrons}$ 
with total helicity $J$.

\begin{figure}[t]
\centerline{\includegraphics[width=0.49\textwidth]{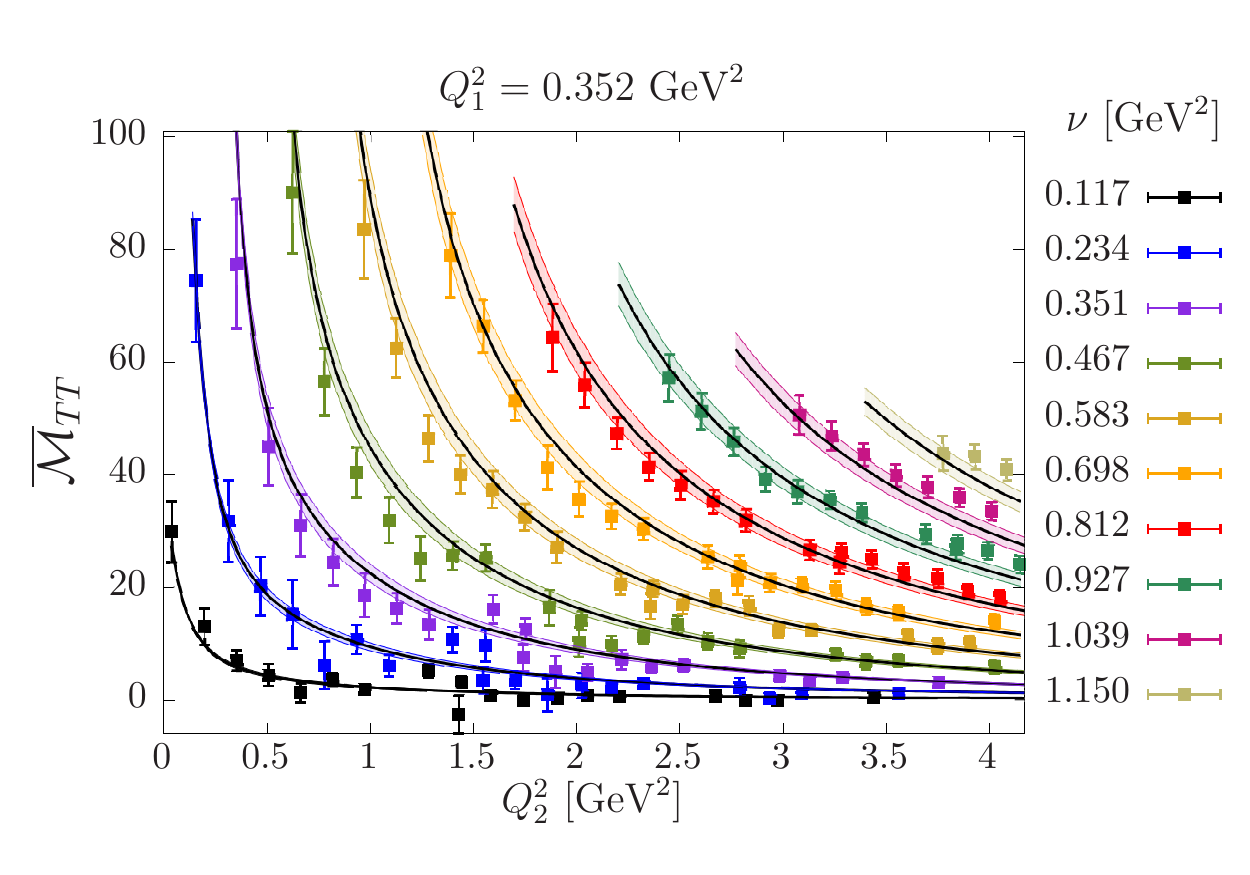}
\includegraphics[width=0.49\textwidth]{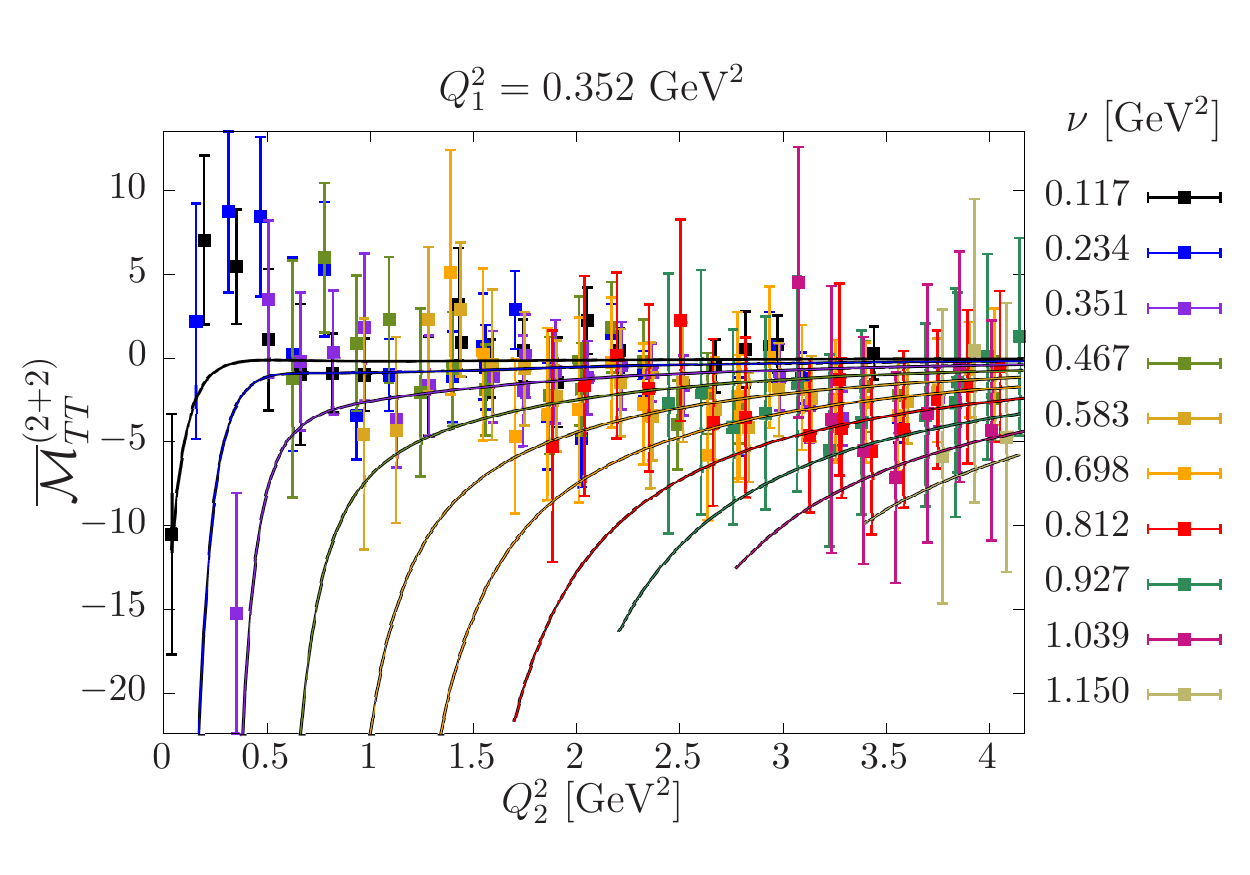}}
\vspace{-0.3cm}
\caption{\label{fig:MTT} 
The subtracted forward HLbL amplitude ${\cal M}_{\rm TT}(-Q_1^2,-Q_2^2,\nu)-{\cal M}_{\rm TT}(-Q_1^2,-Q_2^2,0)$,
multiplied by $10^6$, computed on a $48^3\times96$ lattice ensemble with $M_\pi=314\,$MeV and lattice spacing $a=0.065\,$fm. Left: contribution of the fully connected 
class of quark contractions. Right: contribution of the (2+2) quark-contraction class. Reprinted from Ref.~\cite{Gerardin:2017ryf}.}
\end{figure}

While experimental data exists for the fusion of real photons into hadrons, no such data is available for spacelike photons. 
In order to model the corresponding cross section, we note that the contribution of a narrow meson resonance is 
\be
\sigma_{\gamma^*\gamma^*\to{\rm resonance}} 
\propto ~\delta(s-M^2)\times \Gamma_{\gamma\gamma}\times \Bigg[\frac{F_{M\gamma^*\gamma^*}(Q_1^2,Q_2^2)}{F_{M\gamma^*\gamma^*}(0,0)} \Bigg]^2\,.
\ee
To what extent all eight forward LbL amplitudes obtained from lattice computations can be described by such
a sum of resonances via the dispersive sum rule is an interesting question. Essential ingredients
in this parameterization of $\sigma_{\gamma^*\gamma^*\to{\rm hadrons}}$
are the TFFs $F_{M\gamma^*\gamma^*}(Q_1^2,Q_2^2)$,
describing the coupling of the resonance to two virtual photons.  In
the case of the neutral pion, dedicated lattice QCD calculations of
$F_{\pi^0\gamma^*\gamma^*}$ have been performed (see \cref{sec:pion pole}),
thus allowing for a definite prediction for this contribution. For the other
included hadronic resonances, which have quantum numbers $J^{PC}=0^{\pm+},~1^{++},~2^{++}$, 
a monopole or dipole parameterization of the virtuality-dependence of the TFFs was chosen and
fit to the lattice data for the forward LbL amplitudes. In addition to the resonances, the Born expression for 
$\sigma_{\gamma^*\gamma^*\to\pi\pi}$ was included in the cross section.
A satisfactory description of the data was obtained in this way; see \cref{fig:MTT}.

Five classes of Wick contractions contribute to the full four-point
correlation function.  While the fully connected class of diagrams can
be computed cost-effectively using ``sequential'' propagators, the other
classes require the use of stochastic methods.  In
Ref.~\cite{Gerardin:2017ryf}, only the first two classes, denoted by the
symbols (4) and (2+2), were computed, because the other three classes
(3+1), (2+1+1), and (1+1+1+1) are expected to yield significantly
smaller contributions. If this expectation is correct, and if the LbL
amplitude is dominated by resonance exchanges, one can infer with what
weight factors the isovector and the isoscalar resonances contribute
to the leading contraction topologies (4) and (2+2).  The isoscalar
resonances contribute with unit weight to the class (2+2); the
isovector resonances over-contribute with a weight factor $34/9$ to
class (4), while the (2+2) contractions compensate with a weight
factor of $-25/9$~\cite{Bijnens:2016hgx}. These counting rules have
been used in describing the lattice data in \cref{fig:MTT}. In
particular, the large-$N_c$-inspired counting rules suggest that there
is a large cancellation between the isovector resonances and the
isoscalar resonances in the (2+2) class of diagrams, with the
exception of the pseudoscalar mesons, due to the large mass difference
between the $\pi^0$ and the $\eta'$ meson. Therefore,
in Ref.~\cite{Gerardin:2017ryf} the contribution of the (2+2) diagrams to
the LbL amplitudes was modeled as the $\eta'$
contribution, minus $\frac{25}{9}$ times the $\pi^0$ contribution.
Within the $\sim30\%$ uncertainties, the lattice data was successfully
reproduced.

Thus the exploratory study~\cite{Gerardin:2017ryf} found that the LbL
tensor \cref{eq:masterl} at moderate spacelike virtualities can be
described by a set of resonance poles, much in the same way that
$\amuHLbL$ is obtained in model calculations.  In the future, a 
lattice/phenomenology comparison at a higher degree of precision would
be worthwhile to perform. 

\subsection{Summary of current knowledge from the lattice \label{sec:LHLBL}}

The HLbL contribution to the muon $g-2$ has been
calculated by RBC with both QED and QCD gauge fields simulated on
the finite-volume lattice using the QED$_L$ scheme.
The calculation was performed for several lattice ensembles, with different lattice spacing and volume and all particles at their physical masses.
After the infinite-volume and continuum extrapolations, RBC
obtained the value~\cite{Blum:2019ugy}
\begin{align}\label{eqn:hlbllatrbcres}
\amuHLbL = 7.87(3.06)_\text{stat}(1.77)_\text{sys}\times 10^{-10} \,,
\end{align}
which includes contributions from both connected diagrams and disconnected diagrams.
Large discretization and finite-volume corrections are apparent but under control, and the value in the continuum and infinite-volume limits is compatible with previous model and dispersive treatments, albeit with a large statistical error. This error comes from a large cancellation between the connected and the disconnected diagrams, which are each determined relatively precisely.  The result of \cref{eqn:hlbllatrbcres} currently represents the best ab-initio knowledge from lattice QCD for the complete $\amuHLbL$ contribution.  It builds on the crucial methodological developments of Refs.~\cite{\HLbLlatticemethods}.

Using infinite-volume QED, there are currently two similar approaches (RBC and Mainz) to directly computing $\amuHLbL$ from the lattice.
Both reproduce the literature result for the analogous lepton-loop case, and both seem consistent at comparable lattice QCD simulation parameters for the hadronic contribution. It remains unclear whether one methodology has more desirable features than the other but it could be that their systematics are somewhat different, making a comparison between the two in the continuum limit at the physical point even more valuable.

Both approaches have illustrated the practical necessity of performing a subtraction in the QED kernel, in order to remove contributions coming from the region where two vertices coincide. Although such a subtraction contributes nothing to the final value of $\amuHLbL$, purely from the lattice calculation perspective it helps remove unwanted discretization effects by altering the shape of the integrand. Several choices of subtracted kernels have now been proposed, but this is not an exhaustive list and there is the possibility of further improvements by a good choice of subtracted kernel.

The comparison between the two approaches focused on the connected diagram as its statistical uncertainty is small in both approaches. Unfortunately, the leading disconnected contribution is known to be of comparable size to the connected one but with opposite sign, thus creating a significant cancellation in the full determination of $\amuHLbL$.  While disconnected diagrams are notoriously difficult to calculate accurately in lattice QCD, improved sampling strategies can be helpful.
RBC's results using the so-called M$^2$ trick, e.g., exhibit similar statistical precision as the connected diagram.  It is expected that the higher-order disconnected diagrams are negligible, and preliminary results for one of the additional diagrams, from both groups, indicate this is true. However, all the diagrams need to be calculated to verify this expectation.

The RBC group has carried out preliminary calculations of both connected and leading disconnected diagrams in QED$_\infty$ with physical masses. When combined with model or lattice calculations of the long-distance part of the pion-pole contribution, precise results are obtained that are consistent with QED$_L$. While the results are obtained on a coarse lattice, in finite volume, the subtraction kernel combined with the smaller lattice artifacts of QED$_\infty$ suggest the agreement may hold even after residual QCD artifacts are extrapolated away.

Currently the lattice determination of the HLbL contribution is not as precise as the analytic methods. However, it is a first-principles determination that is statistically and systematically improvable. The best determination from lattice QCD so far is approximately at the current experimental precision and therefore serves as a very important consistency check. Now that the methodology has been formulated and tested, improvements in precision are expected with additional statistics and possible
further methodological improvements.

\subsection{Expected progress in the next few years}

RBC is now improving its physical point QED$_L$ calculation by quadrupling the statistics on the leading disconnected diagram on the smallest-lattice-spacing ensemble, which is expected to reduce the total uncertainty in the current continuum extrapolation by approximately 50\%.  At the same time results for QED$_\infty$ will be computed to investigate residual QCD discretization errors. Together, these new computations should produce a significant reduction of the theory uncertainty in the near term.

In the next few years it is expected that there will be a first-principles, continuum result for $\amuHLbL$ from both the Mainz and RBC approaches using infinite-volume QED and lattice QCD. After this the target will then be that of reducing statistical uncertainty, which can be done by additional lattice QCD measurements. It is possible that new methodologies and techniques will be invented to aid this calculation, but even without them an uncertainty of the order of $10^{-10}$ appears feasible by the end of the Fermilab experiment.

Although thought to be small, it is still worthwhile to attempt to compute the contributions from the remaining disconnected diagrams beyond leading order. This poses a significant challenge as they may
be statistically noisy and computationally expensive to compute. It is expected that an estimate of the size of these contributions directly measured from the lattice will be made soon.

Over the next few years it is possible that more lattice collaborations may also choose to compute $\amuHLbL$, and this would be most welcome, as different groups tend to have different systematics in their approaches allowing for an extended global comparison. It is expected that the Mainz code for computing the QED kernel will be made public, which should help facilitate such endeavors.

\FloatBarrier

\clearpage

\section{The QED contributions to $\boldsymbol{a_\mu}$}
\label{sec:QED}

\noindent
\emph{T.~Aoyama, T.~Kinoshita, M.~Nio}

\subsection{Introduction}

 In the SM of elementary particles, the anomalous magnetic moment of the muon, $a_\mu\equiv (g-2)_\mu/2$, 
 can be divided into
electromagnetic, hadronic, and electroweak contributions
\begin{equation}
a_\mu = \amuQED + \amuhad + \amuEW\,.
\label{eq:anomaly}
\end{equation}
The QED contribution is further divided according to its lepton-mass dependence.
Since the anomaly $a_\mu$ is dimensionless, the lepton-mass dependence appears in the form of the ratio between
lepton masses. Thus, we may rewrite
\begin{equation}
\amuQED = A_1 + A_2(m_\mu/m_e) + A_2(m_\mu/m_\tau) + A_3(m_\mu/m_e, m_\mu/m_\tau)\,,
\label{eq:amu_QED}
\end{equation}
where $m_e$, $m_\mu$, and $m_\tau$ are masses of the electron, muon, and $\tau$-lepton, respectively.
The term $A_1$ is independent of the lepton-mass ratios and universal for all lepton species.

The smallness of the QED coupling constant, the fine-structure constant  $\alpha =1/137.035\dots$, allows us to calculate
each $A_i$ by using perturbation theory for QED:
\begin{equation}
A_i = \left( \frac{\alpha}{\pi} \right ) A_i^{(2)} + \left( \frac{\alpha}{\pi} \right )^2 A_i^{(4)} + \left( \frac{\alpha}{\pi} \right )^3 A_i^{(6)} + \cdots \,,
   ~~~~~~~~\text{for}~~i=1, 2, 3\,.
\end{equation}
Because of renormalizability of QED, every $A_i^{(n)}$ is finite and
calculable by using the Feynman-diagram techniques.  Since Schwinger's result of 1947, $A_1^{(2)} =1/2$, many researchers have
been involved in the calculation of  higher-order terms.  By 2018, all terms up to the eighth order have been obtained
and cross-checked by different  groups using different methods. 
On the other hand, the entire tenth-order contribution has been calculated only by one group with numerical means.
Only small portions of  the tenth-order contribution have  been independently double-checked.
In the following sections, we summarize all perturbative coefficients $A_i^{(2n)}$ up to the tenth order.  

\subsection{Mass-independent contributions}

\begin{figure}[t]
\centering
\includegraphics[width=5cm]{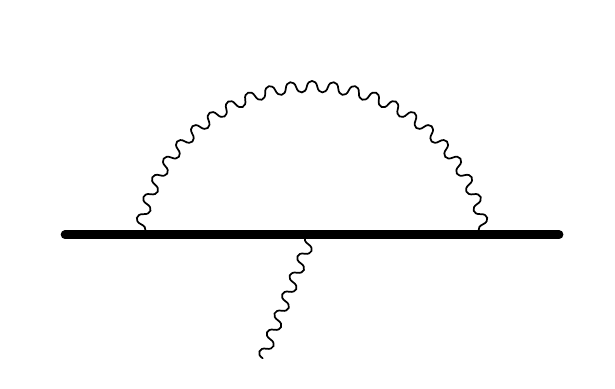}
\caption{Second-order vertex diagram. There is only one diagram. The straight and wavy lines represent
lepton and photon propagators, respectively. Reprinted from Ref.~\cite{Aoyama:2012qma}.}
\label{fig:2nd-order}
\end{figure}

\begin{figure}[t]
\centering
\includegraphics[width=15cm]{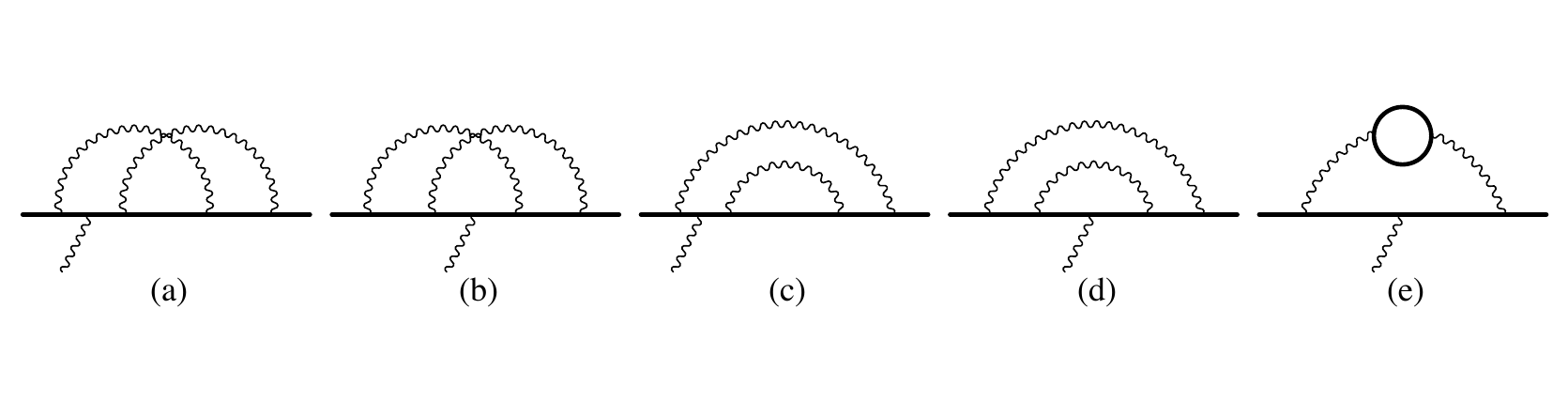}
\caption{Fourth-order
vertex diagrams. There are seven diagrams in total.
The time-reversed diagrams of ({a},{c}) are not shown.
The solid and wavy lines represent
lepton and photon propagators, respectively. Reprinted from Ref.~\cite{Aoyama:2012qma}.}
\label{fig:4th-order}
\end{figure}

\begin{figure}[t]
\centering
\includegraphics[width=15cm]{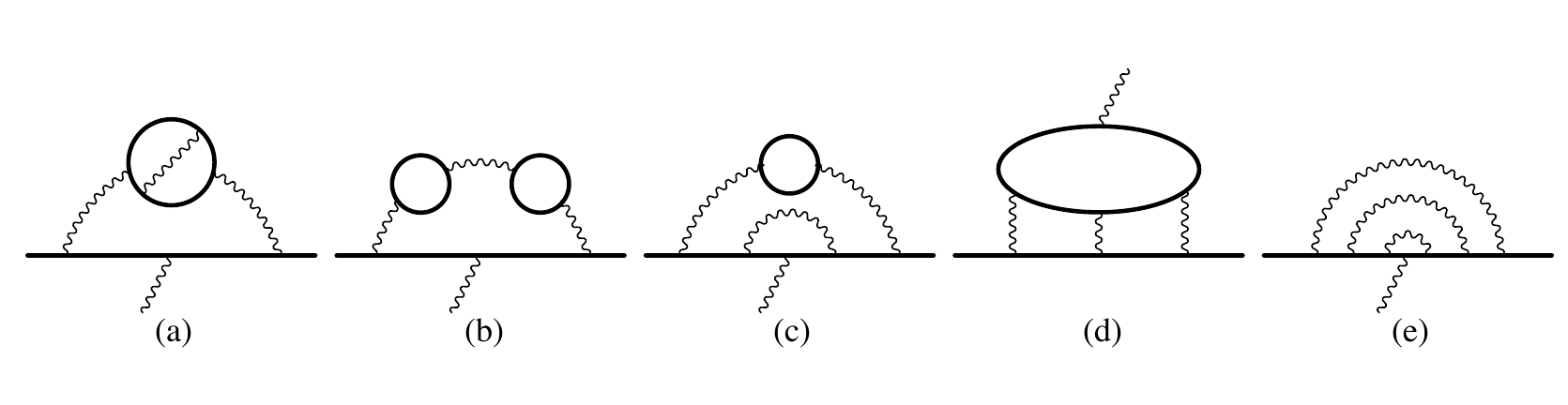}
\caption{Sixth-order vertex diagrams. There are 72 diagrams in total, and
they are divided into five gauge-invariant sets.
Typical diagrams from each set are shown as ({a})--({e}).
There are 3 diagrams in set (a), 1 diagram in set (b), 12 diagrams in set (c), 6 diagrams in set (d), and 50 diagrams in set (e).
The solid and wavy lines represent
lepton and photon propagators, respectively. Reprinted from Ref.~\cite{Aoyama:2012qma}.}
\label{fig:6th-order}
\end{figure}

The mass-independent contribution $A_1$  is known up to the tenth-order of perturbation theory.
The number of vertex Feynman diagrams contributing to the second-, fourth-, and sixth-order terms are
1, 7, and 72, respectively, see \cref{fig:2nd-order,fig:4th-order,fig:6th-order}.
Their contributions are known in closed analytic forms 
\cite{Schwinger:1948iu,Petermann:1957hs,Sommerfield:1958,Laporta:1996mq}:
\begin{align}
A_1^{(2)} & =\frac{1}{2}
\label{eq:A1_2}\,,
\\
A_1^{(4)} &=  \frac{197}{144} + \left ( \frac{1}{2} - 3 \log 2 \right ) \zeta(2) + \frac{3}{4} \zeta (3) \nonumber \\
               & =  -0.328~478~965~579~193~784~582\dots \,,
\label{eq:A1_4}               
\\
A_1^{(6)}& =   \frac{83}{72} \pi^2 \zeta(3) -\frac{215}{24} \zeta(5)
         + \frac{100}{3}a_4  + \frac{25}{18} \log^4 2 
         - \frac{25}{18} \pi^2 \log^2 2   - \frac{239}{2160} \pi^4
         + \frac{139}{18} \zeta(3)   \nonumber \\
                    &  - \frac{298}{9} \pi^2 \log 2
         + \frac{17101}{810} \pi^2 + \frac{28259}{5184}  \nonumber \\
                &=    1.181~241~456~587~200\dots \,,
\label{eq:A1_6}
\end{align}
where $a_4$ is the polylogarithm  
\begin{equation}
a_4 \equiv \text{Li}_4 (1/2)= \sum_{n=1}^{\infty} \frac{1}{2^n n^4}=0.517~479~061~673~899~386\dots~.
\end{equation}
An earlier attempt at the fourth-order
calculation was also made in Ref.~\cite{Karplus:1950zzb}, but gave the wrong answer.
The sixth-order terms were numerically calculated in Refs.~\cite{Kinoshita:1973ph,Cvitanovic:1974um,Levine:1974cb,Carroll:1974bu,Carroll:1975jf,Kinoshita:1995ym}.

\begin{figure}[t]
\centering
\includegraphics[width=16cm]{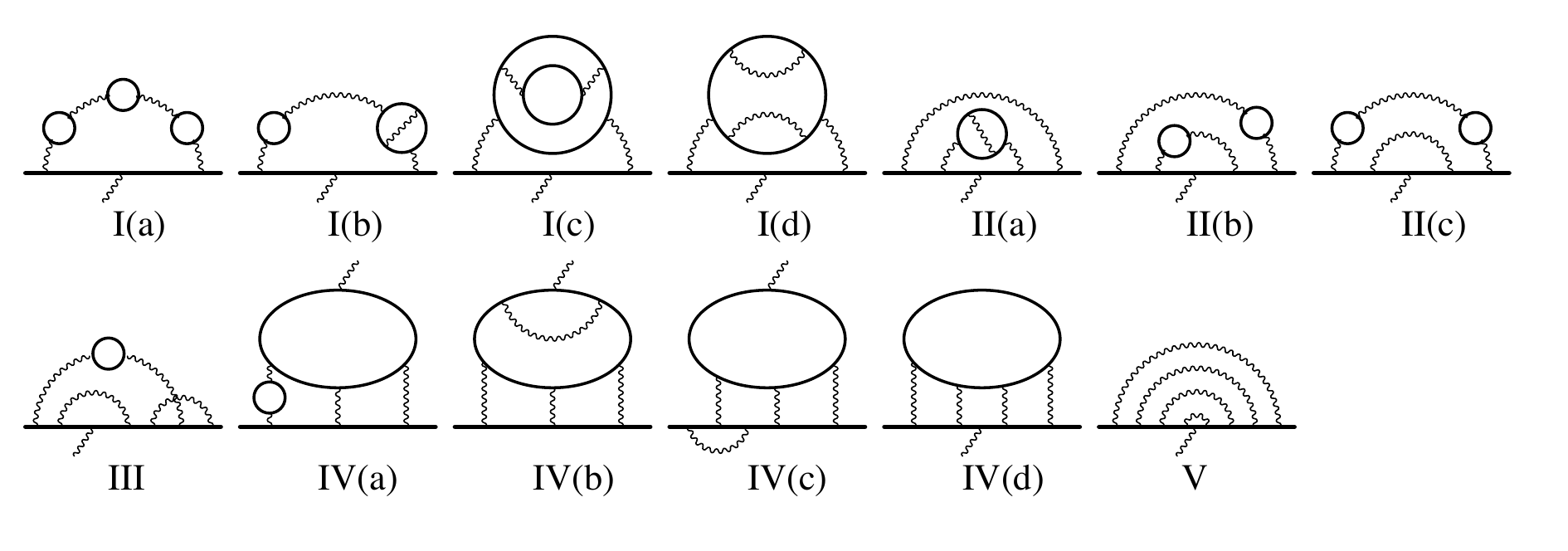}
\caption{Eighth-order vertex diagrams. There are 891 diagrams in total, and
they are divided into 13~gauge-invariant subsets of five super sets.
A typical diagram from each subset is shown as {I}({a})--({d}), {II}({a})--({c}), {III}, {IV}({a})--({d}), and {V}.
Set I(a) consists of a single diagram, while there are 6 diagrams in set I(b), 3 in set I(c), 15 in set I(d), 36 in set II(a), 6 in set {II}({b}),  12 in set {II}({c}), 150 in set {III}, 18 in set {IV}({a}), 60 in set {IV}({b}), 48 in set {IV}({c}),  18 in set IV(d), and 518 in set V.
The~straight and wavy lines represent
lepton and photon propagators, respectively. Reprinted from Ref.~\cite{Aoyama:2012wj}.}
\label{fig:8th-order}
\end{figure}

Recently, the eighth-order mass-independent contribution $A_1^{(8)}$ 
has been  
calculated in near-analytical form~\cite{Laporta:2017okg}.  The analytic forms of only a small number of integrals remain undetermined, but their numerical
values are precisely known.  Thus, the contribution from 891 vertex diagrams, $A_1^{(8)}$,
is obtained up to  1100 digits:
\begin{equation}
A_1^{(8)} =  -1.912~245~764~926~445~574 \dots \,,
\label{eq:A1_8}
\end{equation}
which is consistent with the  latest fully numerical calculation~\cite{Aoyama:2014sxa}.
Earlier numerical calculations are found in Refs.~\cite{Kinoshita:1981vs,Kinoshita:1979dy, Kinoshita:1979ej, Kinoshita:1979ei,  Kinoshita:1981wx,Kinoshita:1981ww,
Kinoshita:1981wm,Kinoshita:2002ns,Kinoshita:2005zr,Aoyama:2007dv, Aoyama:2007mn}.
A less accurate but consistent result was also obtained in Ref.~\cite{Marquard:2017iib} by using analytical means.
The contribution from the gauge-invariant Set~V of \cref{fig:8th-order} is also independently and numerically calculated
in Refs.~\cite{Volkov:2017xaq,Volkov:2018jhy}, showing a consistent result.

\begin{figure}[th!]
\centering
\includegraphics[width=14cm]{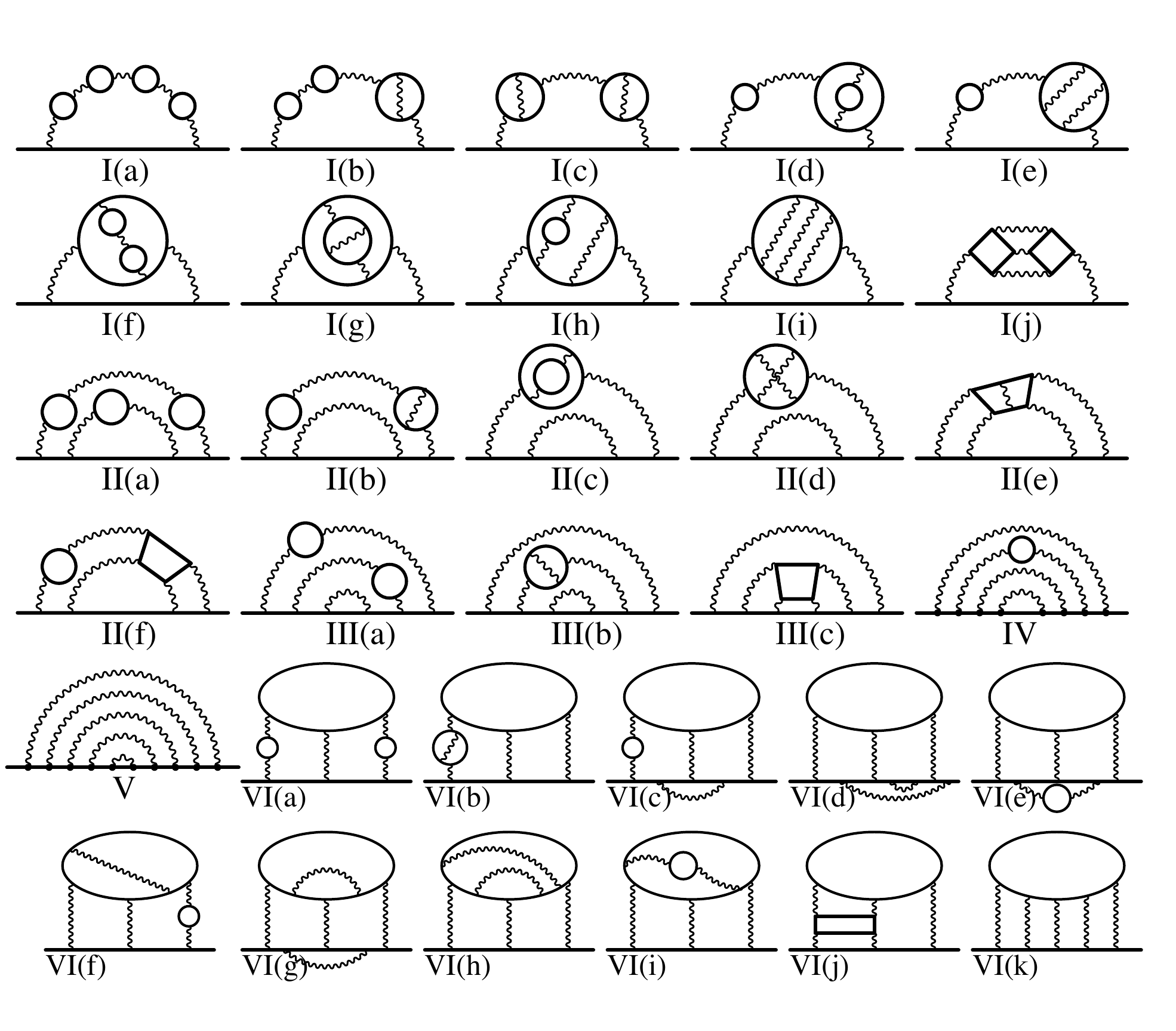}
\caption{Tenth-order vertex diagrams. There are 12~672 diagrams in total, and
they are divided into 32~gauge-invariant subsets over six super sets.
Typical diagrams of each subsets are shown as {I}({a})--({j}), {II}({a})--({f}), {III}({a})--({c}), {IV}, {V},
and {VI}({a})--({k}).
There are
208 Set~I diagrams
(1 for {I}({a}), 9 for {I}({b}), 9 for {I}({c}), 6 for {I}({d}), 30 for \mbox{{I}({e})}, 3 for  {I}({f}), 9 for {I}({g}),  30 for {I}({h}), 105 for {I}({i}), and 6 for {I}({j})),
600 Set~{II} diagrams
(24 for {II}({a}), 108 for {II}({b}), 36 for {II}({c}), 180 for {II}({d}), 180 for {II}({e}), and 72 for {II}({f})),
1140 Set~{III }diagrams
(300 for {III}({a}), 450 for {III}({b}), and 390 for {III}({c})),
2072 Set~{IV} diagrams,
6354 Set~{V} diagrams, and
2298 Set~{VI} diagrams
(36 for {VI}({a}),  54 for {VI}({b}), 144 for {VI}({c}), 492 for {VI}({d}), 48 for {VI}({e}), 180 for {VI}({f}), 480 for \mbox{{VI}({g})}, 630 for {VI}({h}),
60 for {VI}({i}), 54 for {VI}({j}), and 120 for {VI}({k})).
The straight and wavy lines represent
lepton and photon propagators, respectively.
The external photon vertex is omitted for simplicity and
can be attached to one of the lepton propagators of the bottom straight line in super sets {I}--{V}
or the large ellipse in super set {VI}.
Reprinted from Ref.~\cite{Aoyama:2012wj}. }
\label{fig:10th-order}
\end{figure}

The tenth-order mass-independent term $A_1^{(10)}$ receives contributions from $12~672$ vertex diagrams 
and has been calculated only by numerical means.
The latest value is found in Ref.~\cite{Aoyama:2019ryr}:
\begin{equation}
A_1^{(10)}  = 6.737(159)\,.
\label{eq:A1_10}
\end{equation}
The improvement of this result over the previous value $A_1^{10}=6.675(192)$ in Ref.~\cite{Aoyama:2017uqe} is a consequence of the continued accumulation of more statistics in the numerical evaluation of the relevant integrals.
Each gauge-invariant set of \cref{fig:10th-order}  is described in detail in Refs.~\cite{Aoyama:2008gy,Aoyama:2008hz,Aoyama:2010yt,
Aoyama:2010pk,Aoyama:2010zp,Aoyama:2011rm,Aoyama:2011zy,Aoyama:2011dy,Aoyama:2012fc,Aoyama:2012wj,Aoyama:2014sxa}.
A part of the tenth-order contribution coming from the diagrams without
a fermion loop shown as Set~V of \cref{fig:10th-order} has been
independently calculated by numerical means
in Ref.~\cite{Volkov:2019phy}.
The two available numerical results, however, differ from each other
by $4.8 \sigma$.
This difference is still negligible for the current precision
of the electron $g-2$,
but may be crucial for future measurements.
The difference is not relevant for the muon $g-2$,
since the electron-loop contribution is nearly a hundred times greater
than the mass-independent contribution at the tenth order.

\subsection{Mass-dependent contributions}
 
Lepton-mass dependence of the QED contribution to $a_\mu$  comes from vertex diagrams with  at least one  closed fermion loop and thus starts to appear at fourth
order in the perturbative expansion.
As the muon mass is located in the middle of the lepton mass hierarchy,  
one has two types of asymptotic expansion parameter, either 
$m_\mu/m_e > 1 $ or $m_\mu/m_\tau <  1$.  Therefore,  for the analytic calculation, 
the evaluation of one Feynman diagram requires different asymptotic expansions for
the closed-electron and closed-$\tau$ loop
cases.  
For the numerical calculation, on the other hand,  all that is needed is to change the input value for the mass of the lepton in the loop. 

The mass ratios used for the evaluation are 
the muon-to-electron mass ratio 
$m_\mu/m_e=206.768~2827(47)$~\cite{Mohr:2015ccw} 
and  the muon-to-$\tau$ mass ratio
$m_\mu/m_\tau = 5.946~35(40) \times 10^{-2}$~\cite{Tanabashi:2018oca}.
It is the mass-dependent terms that distinguish between
the electron and the muon anomalous magnetic moments in QED.
Particularly in  higher-order terms, the electron loop contribution is logarithmically enhanced and 
gives rise to much larger contributions than  the mass-independent term of the same order. 

\subsubsection{Fourth-order}
The fourth-order mass-dependent term was obtained in the asymptotic expansion with $x=m_e/m_\mu < 1 $ as 
an expansion parameter in Refs.~\cite{Suura:1957zz,Petermann:1957ir}.  
The closed analytic expression valid for any mass ratio $x$  of a closed fermion loop was then obtained in Refs.~\cite{Elend:1966a,Li:1992xf} 
and the following more convenient form was given in Ref.~\cite{Passera:2006gc}:
\begin{align}
A_2^{(4)}(1/x) &= -\frac{25}{36} - \frac{ \log x }{3} + x^2 ( 4+ 3 \log x )  \nonumber \\
&+ \frac{x}{2} (1- 5 x^2) \left [ \frac{\pi^2}{2}  
- \log x \log \left ( \frac{ 1-x}{1+x} \right ) - \text{Li}_2 (x) + \text{Li}_2(-x) \right ] \nonumber \\
&+ x^4 \left [  \frac{ \pi^2}{3} - 2 \log x \log \bigg( \frac{1}{x} - x \bigg) - \text{Li}_2 \big(x^2\big) \right ]  \,,
\end{align}
where 
$\text{Li}_2(z) $ is the dilogarithm
and for $|z| < 1$, 
\begin{equation}
\text{Li}_2(z) = - \int_0^z \frac{dt}{t} \log (1-t) \,.
\end{equation}
For  $|z| >1$,  the logarithm $\log(1-z)$ is analytically continued and its principal
value $\text{Log}(1-z)$ is instead  used:
\begin{align}
\text{Log}  (1-z) = \log | 1-z|  +  i \text{Arg} (1-z) \,.  
\end{align}

For $x=m_e/m_\mu < 1$, the expansion is found to be
\begin{align}
A_2^{(4)}(m_\mu/m_e)  &=  -\frac{25}{36}-\frac{\log x}{3} + \frac{ \pi^2 x}{4}
           +(3+4 \log x)x^2-\frac{5\pi^2 x^3}{4}  \nonumber \\
            & +\left ( \frac{\pi^2}{3}+\frac{44}{9}-\frac{14 \log x}{3}
              +2  \log^2 x  \right ) x^4 
            +\left ( \frac{8  \log x}{15}
            -\frac{109 }{225} \right ) x^6 + \dots
                       \nonumber \\
                       & = 1.094~258~3093(76)\,,
\end{align} 
where the uncertainty comes from the muon mass  $m_\mu$.
For $x= m_\tau/m_\mu >1$,  the expansion is
\begin{align}
A_2^{(4)}(m_\mu/m_\tau) &= \frac{x^{-2}}{45} + x^{-4} \left ( -\frac{\log x}{70} 
                        + \frac{9}{19600} \right )  + x^{-6} \left (  - \frac{4 \log x}{315} - \frac{131}{99225 }\right )   + \dots
                        \nonumber \\ 
                                                &= 0.000~078~076(11) \,,
\end{align}       
where the uncertainty comes from the $\tau$-lepton mass  $m_\tau$.                 

\subsubsection{Sixth-order}
The sixth-order mass-dependent term is  
known  in closed analytic form for arbitrary values of the mass ratios~\cite{Laporta:1992pa,Laporta:1993ju}.
Since the form is too lengthy to list in papers, only the expansions  are given in Refs.~\cite{Laporta:1992pa,Laporta:1993ju}.
With $x=m_e/m_\mu$, the electron contribution   $A_2^{(6)}(m_\mu/m_e)$ 
is given by 
\begin{align}
A_2^{(6)}(m_\mu/m_e ) & =         
\frac{2}{9} \log^2 x  - \left ( \zeta(3) - \frac{2}{3} \pi^2 \log 2 + \frac{7\pi^2}{9} + \frac{31}{27} \right ) \log x + \frac{97 \pi^4}{360} \nonumber \\
&    - \frac{2}{9} \pi^2 \log ^2 2 - \frac{8}{3} a_4 - \frac{ \log^4 2 }{9} - 6 \zeta (3) + \frac{5}{3} \pi^2 \log 2 
- \frac{85 \pi^2}{18} + \frac{1219}{216}   
\nonumber \\
& + x \left ( - \frac{4}{3} \pi^2 \log x - \frac{604}{9} \pi^2 \log 2 + \frac{54079 \pi^2 }{1080} - \frac{13 \pi^3}{18} \right )
\nonumber \\
&  + x^2 \left [  \frac{2}{3} \log^3 x  + \left ( \frac{\pi^2}{9} - \frac{10}{3} \right ) \log^2 x  \right .
                        + \left ( \frac{16\pi^4 }{135} + 4 \zeta(3) - \frac{32 \pi^2}{9} + \frac{194}{9} \right ) \log x 
\nonumber  \\
&    + \frac{4}{3} \zeta(3) \pi^2 - \frac{61 \pi^4}{270} + \zeta(3) +\frac{197 \pi^2}{36} - \frac{2809}{108} 
- \left .   \frac{14}{3} \pi^2 \log 2 \right ] + \mathcal{O}(x^3)
\nonumber \\
& =    22.868~379~98(20)\,,                    
\end{align}         
where the dominant contribution $20.947~924~85(15)$ comes from the  LbL scattering diagrams shown in 
\cref{fig:6th-order}(d)  and the uncertainty comes from the muon mass  $m_\mu$.
Before arriving at this analytic expression, many analytic and/or numerical calculations had been carried out
\cite{Kinoshita:1967,Lautrup:1969fr,Lautrup:1969uk,Aldins:1970id,Lautrup:1971yp,Lautrup:1977tc,Samuel:1976yt,Kinoshita:1989mf,Kinoshita:1990wp,Samuel:1990qf}.

The $\tau$-lepton contribution is as follows, with $x=m_\tau/m_\mu$~\cite{Laporta:1992pa,Laporta:1993ju}:
\begin{align}
A_2^{(6)}(m_\mu/m_\tau) & =      
x^{-2} \left ( - \frac{23}{135} \log x - \frac{74957}{97200} - \frac{ 2 \pi^2}{45} + \frac{3}{2} \zeta(3) \right )
\nonumber \\
& +x^{-4} \left ( - \frac{4337}{22680} \log^2 x  - \frac{209891}{476280} \log x - \frac{451205689}{533433600} 
          -\frac{1919\pi^2}{68040} + \frac{1811}{2304} \zeta(3) \right )
 + \mathcal{O}(x^{-6})
 \nonumber \\
 & =0.000~360~671(94)\,,
\end{align}
where the uncertainty comes from the $\tau$-lepton mass  $m_\tau$.            

The three-mass term $A_3^{(6)}(m_\mu/m_e, m_\mu/m_\tau)$ appears for the first time at sixth order
from the diagram with two VP insertions shown in \cref{fig:6th-order}(b), see Ref.~\cite{Czarnecki:1998rc,Friot:2005cu,Ananthanarayan:2020acj}:
\begin{align}
A_3^{(6)}(m_\mu/m_e, m_\mu/m_\tau)& =
   y^{-2}  \left ( -\frac{4}{135} \log x - \frac{1}{135}  + \frac{2}{15} x^2  -\frac{4 \pi^2 }{45} x^3 \right )
   \nonumber \\ 
& + y^{-4}   \left ( -  \frac{229213}{12348000}  + \frac{\pi^2}{630}-\frac{37}{11025} \log y 
                                                      -\frac{1}{105} \log y  \log \left (\frac{y}{x^2} \right ) 
                                                     - \frac{3}{4900} \log x \right )
\nonumber \\
& + \mathcal{O}(y^{-6})
\nonumber \\
&= 0.000~527~738(75)\,,
\end{align}
where $x=m_e/m_\mu$ and $y=m_\tau/m_\mu$ and the uncertainty comes from the $\tau$-lepton mass  $m_\tau$.

\subsubsection{Eighth-order}

The electron contribution to the eighth-order mass-dependent term, $A_2^{(8)}(m_\mu/m_e)$, comes from the
12 gauge-invariant sets I(a)--(d), II(a)--(c), III,  and IV(a)--(d) in \cref{fig:8th-order}.
The dominant contribution comes from diagram IV(a), which has a double-logarithmic enhancement:
one enhancement comes from the LbL scattering loop and the other from the second-order VP
insertion~\cite{Lautrup:1972iw}.  The complete result is 
\begin{equation}
A_2^{(8)}(m_\mu/m_e) = 123.785~51(44)  + 8.8997(59) = 132.6852(60)\,,
\label{eq:A2_8me}
\end{equation}
where the first term is the contribution from the LbL diagram IV(a) and the second term is the sum of
the other diagrams. 
The result~\eqref{eq:A2_8me}  is obtained mostly by  numerical evaluation of
the Feynman integrals~\cite{Laporta:1993ds,Kinoshita:2005zr,Aoyama:2012wk}.
The uncertainties are due to numerical integration by VEGAS~\cite{Lepage:1977sw}.
An independent check has been done  by means of an asymptotic expansion. 
The expansion itself is analytic, but the coefficients of the small expansion parameter $x=m_e/m_\mu$ 
are calculated numerically~\cite{Kurz:2016bau}.   
For some simple integrals such as the diagrams I(a--c) of \cref{fig:8th-order}, the asymptotic expansion 
provides much better precision, while
for other complicated diagrams, the entirely numerical integration provides more precise results.
Both results are in good agreement on a diagram-by-diagram basis.
 
The eighth-order $\tau$-lepton contributions, $A_2^{(8)}(m_\mu/m_\tau)$ and $A_3^{(8)}(m_\mu/m_e, m_\mu/m_\tau)$,
 are also independently checked. This is done in two ways, first by numerical calculation~\cite{Aoyama:2012wk} and second by use of an asymptotic expansion~\cite{Kurz:2013exa}.
The latter produces the higher-precision results,
\begin{align}
&A_2^{(8)}(m_\mu/m_\tau ) = 0.042~4941(53)\,,  \\
&A_3^{(8)}(m_\mu/m_e, m_\mu/m_\tau)  = 0.062~722(10) \,.
\end{align}
The contribution from the LbL diagram  IV(b) has also been independently checked~\cite{Kataev:2012kn}.

\subsubsection{Tenth-order}
Although the asymptotic expansions  of some diagrams  are available~\cite{Laporta:1994md},  most of the diagrams shown in
\cref{fig:10th-order} have been calculated only by numerical means~\cite{Kinoshita:2005sm,Aoyama:2008gy,Aoyama:2010yt,Aoyama:2008hz,Aoyama:2010pk,Aoyama:2011rm,Aoyama:2010zp,Aoyama:2011zy,Aoyama:2011dy,Aoyama:2012fc}. 
However, using confirmed lower-order results, 
the application of the renormalization group provides reliable estimates of the electron contributions of some sets of \cref{fig:10th-order}, see Refs.~\cite{Kataev:1991cp,Kataev:1994rw,Kataev:2006yh}. The nonrelativistic calculation is also useful to investigate the LbL scattering diagram VI(k) of \cref{fig:10th-order}, see Ref.~\cite{Karshenboim:1993rt}.
Another approach to the diagrams  I(a)--(j) of \cref{fig:10th-order} is  to construct  the Pad\'{e} approximant of the
four-loop VP function~\cite{Baikov:2013ula}  by applying the method developed for the three-loop VP function~\cite{Baikov:1995ui}.   This method  is powerful enough to improve the asymptotic expansion and yields very good
agreement with the numerical results of each subset I(a)--(j).

The numerical results for the sum of all diagrams with one or more fermion loops are given by
\begin{align}
&A_2^{(10)} (m_\mu/m_e)     =  742. 32(86)   \label{eq:A2_10me} \,,\\
&A_2^{(10)} (m_\mu/m_\tau) =     -0. 0656(45)  \label{eq:A2_10mt}\, , \\
&A_3^{(10)} (m_\mu/m_e, m_\mu/m_\tau ) =  2.011(10)  \label{eq:A3_10met} \,,
\end{align}
where all uncertainties are statistical ones from the numerical integration by VEGAS.
The values given in \cref{eq:A2_10me,eq:A2_10mt} are slightly different from
those given in Ref.~\cite{Aoyama:2012wk}.  This is because the numerical precision of
the contributions from the diagrams I(i) and  III(c)  of \cref{fig:10th-order} 
has been  improved since the publication of Ref.~\cite{Aoyama:2012wk}.

\subsubsection{Twelfth-order}
In view of the rather large value of 
$A_2^{(10)} (m_\mu/m_e)$,
one might wonder about the size of the twelfth-order perturbative QED contributions
For this purpose, note that the dominant contributions to
$A_2^{(8)} (m_\mu/m_e)$ and
$A_2^{(10)}(m_\mu/m_e)$ both 
come from the LbL diagram with  insertion of second-order VP functions.
Analogously, the leading contribution to the twelfth-order term
would  come from the insertion of three VPs into  the sixth-order LbL
diagram of \cref{fig:6th-order}(d), see Ref.~\cite{Aoyama:2012wk}.
One thus expects
\begin{equation}
A_2^{(12)}(m_\mu/m_e) 
 \sim A_2^{(6)} (m_\mu/m_e; \text{LbL}\,) 
\times \left \{ \frac{2}{3} \log \left (\frac{m_\mu}{m_e} \right ) 
    - \frac{5}{9} \right\}^3 \times 10 \sim 5400\,,
\end{equation}
and thus 
\begin{equation}
A_2^{(12)}(m_\mu/m_e) \times \left(\frac{\alpha}{\pi}\right)^6
                      \sim 0.8 \times 10^{-12}\,,
\end{equation}
where  $A_2^{(6)} (m_\mu/m_e; \text{LbL}\,) \sim 20$
and the factor 10 accounts for the possible ways of VP insertions.
Including the contribution of the other diagrams, the size of the twelfth-order
term might be as large as $10^{-12}$. 
This is smaller than the target uncertainty $15 \times 10^{-11}$ of the new muon $g-2$ experiment at Fermilab,
but larger than that of the tenth-order QED term.

\subsection{Fine-structure constant}
\label{sec:fine_structure_constant}
In order to obtain the theoretical prediction of the QED contribution to $a_\mu$, 
we need a precise value for the fine-structure constant $\alpha$.
Currently, the best value of $\alpha$ comes from the Cs atom-interferometry experiment~\cite{Parker:2018vye}:
\begin{equation}
\alpha^{-1}(\text{Cs} ) = 137.035~999~046(27)\,.
\label{eq:alpha_Cs}
\end{equation}
The value is consistent with the one previously obtained by the Rb-atom interferometry~\cite{Bouchendira:2010es}
and the uncertainty  of \cref{eq:alpha_Cs} represents  a 3.1 fold improvement over $\alpha(\text{Rb})$.
A quotient of the Planck constant and
the mass of Cs, $h/m_\text{Cs}$,  is measured
and $\alpha (\text{Cs})$ determined via
\begin{equation}
\alpha (\text{Cs}) = \left [ \frac{2 R_\infty}{c} \frac{ A_r(\text{Cs})}{A_r(e)}\frac{h}{m_\text{Cs}} \right ]^{1/2}\,,
 \end{equation}
where $R_\infty $ is the Rydberg constant
and $A_r(\text{Cs})$ and $A_r(e)$ are the relative atomic masses of a Cs atom and an electron, respectively,
which are defined by $m_\text{Cs}/u$ and $m_e/u$, $u$ being the unified atomic mass unit.
As all three are precisely known and found in the CODATA
 2014 adjustment~\cite{Mohr:2015ccw},  the uncertainty of the quotient $h/m_\text{Cs}$ governs the
 uncertainty of $\alpha(\text{Cs})$.   To determine the values of the Rydberg constant
 $R_\infty$ and the relative atomic mass $A_r(e)$, a number of QED loop calculations are needed, but the uncertainties
 due to QED corrections are small.  Thus, the value of $\alpha(\text{Cs})$ can be regarded as independent from QED.
 We also note that there exists a $2.7\sigma$ tension between two values of $R_\infty$ reported in
 Refs.~\cite{Beyer:2017,Fleurbaey:2018fih}.
But both are sufficiently accurate and the difference does not affect the determination of the value of $\alpha(\text{Cs})$.
 
 \begin{table}[t]
\centering
\small
\begin{tabular}{lD{.}{.}{-1}D{.}{.}{-1}}
\toprule
\multicolumn{1}{l}{Order} &
\multicolumn{1}{c}{with $\alpha(\text{Cs})$} &
\multicolumn{1}{c}{with $\alpha(a_e)$} \\
\midrule
2 &
116~140~973.321(23) &
116~140~973.233(28) \\
4 &
413~217.6258(70) &
413~217.6252(70) \\
6 &
30~141.90233(33) &
30~141.90226(33) \\
8 &
381.004(17) &
381.004(17) \\
10 &
5.0783(59) &
5.0783(59) \\
\hline
$a_\mu(\text{QED})$ &
116~584~718.931(30) &
116~584~718.842(34) \\\bottomrule
\end{tabular}
\caption{Contributions to muon $g-2$ from QED perturbation terms
in units of $10^{-11}$. 
They are evaluated with two values of the fine-structure constant 
$\alpha$ determined by the Cs experiment and the electron $g-2$.
The uncertainties of the second-order term is due to $\alpha$.
Those of the  fourth- and sixth-order terms are due to the $\tau$-lepton mass.
Those of the eighth-  and tenth-order terms are due to
numerical integrations.
The twelfth-order contribution is unlisted in this table and might
be $0.0(1)$ from the estimate of the leading-order contribution.  
}
\label{tbl:amu}
\end{table} 
 
Another determination of the value of $\alpha$ relies on the electron anomalous magnetic moment $a_e$.
The best measurement of $a_e$ at Harvard~\cite{Hanneke:2008tm}
\begin{equation}
a_e^\text{exp}= 1~159~652~180.73(28) \times 10^{-12}
\label{eq:ae_HV}
\end{equation}
is equated to the theoretical formula of $a_e$ similar to  \cref{eq:anomaly}.
The value of $\alpha(a_e)$ thus obtained is  
\cite{Aoyama:2019ryr}
\begin{equation}
\alpha^{-1}(a_e) = 137.035~999~1496(13)(14)(330)\,,
\label{eq:alpha_ae}
\end{equation}
where the uncertainties are from the numerical evaluation of the tenth-order QED term, the hadronic contribution,
and the measurement \cref{eq:ae_HV}.
The QED mass-independent terms $A_1^{(2n)}$ are common for $a_e$ and $a_\mu$ and given
in  \crefrange{eq:A1_2}{eq:A1_10}.
The QED mass-dependent terms for $a_e$ are found in Refs.~\cite{Aoyama:2012wj, Aoyama:2014sxa}
and the contribution is $2.747~5720(14) \times 10^{-12} $ in total.
The hadronic and electroweak corrections are  $a_e^\text{had} = 1.693(12)\times 10^{-12} $ and $a_e^\text{EW} = 0.03053(23)\times 10^{-12}$,
respectively, both quoted from Refs.~\cite{Jegerlehner:2017zsb,Jegerlehner:2017gek}.\footnote{See also the recent evaluation in Ref.~\cite{Keshavarzi:2019abf}, $a_e^\text{had}=1.7030(77)\times 10^{-12}$, which is fully compatible but more precise. Both evaluations use $a_e^\text{HLbL}=0.037(5)\times 10^{-12}$~\cite{Jegerlehner:2017zsb,Jegerlehner:2017gek}, whose central value is close to previous estimates, $a_e^\text{HLbL}=0.035(10)\times 10^{-12}$~\cite{Prades:2009tw} and $a_e^\text{HLbL}=0.039(13)\times 10^{-12}$~\cite{Jegerlehner:2009ry}, but in view of the relative accuracy that we quote in \cref{eq:final-estimate2}, its uncertainty may be underestimated.}
The obtained  $\alpha^{-1}(a_e)$ is   $ 0.104(43)\times 10^{-6} $ smaller than \cref{eq:alpha_Cs}
and the discrepancy is $2.4 \sigma$.
When  $\alpha(a_e)$ is used to evaluate $\amuQED$, 
one must keep in mind that
$\alpha(a_e)$ and the theoretical formula \cref{eq:amu_QED} are strongly
correlated with each other.  The mass-independent terms $A_1^{(2n)}$ are common
to the QED formulae for both $a_e$ and  $a_\mu$.
Even for the mass-dependent terms, 
the same computer programs are used for numerical calculation
just by changing loop-fermion masses.

\subsection{QED contribution to $a_\mu$ \label{sec:QED_sum}}
Summing the terms in the perturbative QED expansion up to tenth order,
we obtain the QED contribution to the muon anomalous magnetic moment, 
as summarized in \cref{tbl:amu}.
The two possible choices for the fine-structure constant, $\alpha(\text{Cs})$ of \cref{eq:alpha_Cs}  and $\alpha(a_e)$ of \cref{eq:alpha_ae},  lead to
\begin{align}
\amuQED(\alpha(\text{Cs})) &= 116~584~718.931(7) (17) (6)(100) (23) [104]  \times 10^{-11}\,,
\label{eq:amuQED_Cs}
\\
\amuQED(\alpha(a_e) ) &=  116~584~718.842(7) (17)(6)(100)(28)[106] \times 10^{-11}\,,
\label{eq:amuQED_ae}
\end{align}
where the uncertainties are due to the $\tau$-lepton mass $m_\tau$,
the eighth-order QED,  the tenth-order QED, the estimate of the twelfth-order QED, the fine-structure constant $\alpha$, and the sum in quadrature of all of these.
Apart from the respective input for $\alpha$ and the lepton masses, these final values are based on the latest QED calculations from Refs.~\cite{\QEDref}, which should be cited in any work that uses or quotes \cref{eq:amuQED_Cs,eq:amuQED_ae}.
The difference
between \cref{eq:amuQED_Cs} and \cref{eq:amuQED_ae}
is  $0.09 \times 10^{-11}$,  so that
we may use either one as far as comparison with the
on-going experiments is concerned.

\FloatBarrier

\clearpage

\section{The electroweak contributions to $\boldsymbol{a_\mu}$}
\label{sec:EW}

\noindent
 \emph{D.~St\"ockinger, H.~St\"ockinger-Kim}

\subsection{Introduction}

\begin{figure}[t]
\centerline{  \includegraphics[scale=.45]{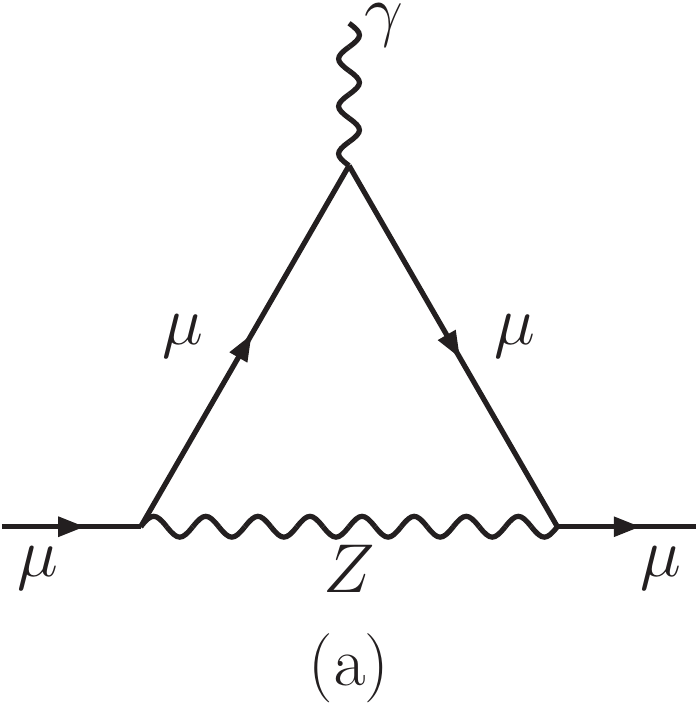}
  \includegraphics[scale=.45]{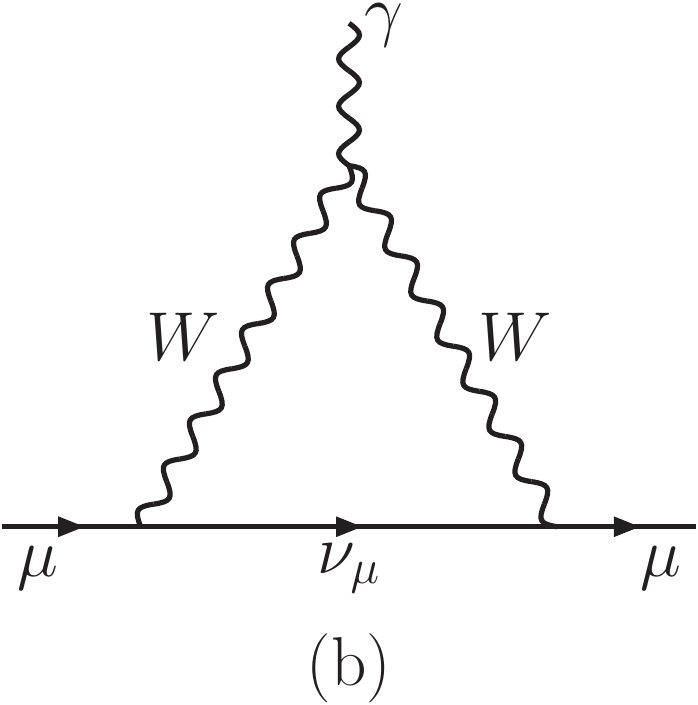}
  \includegraphics[scale=.45]{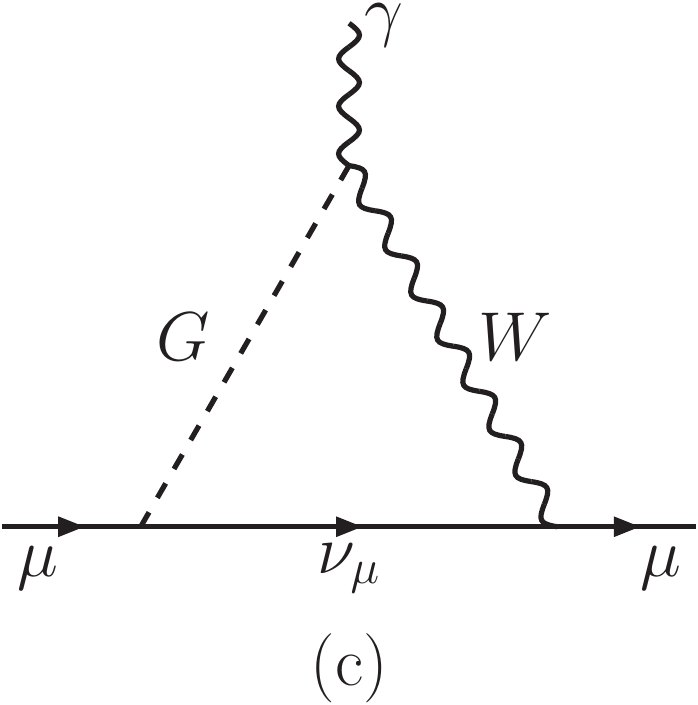}
  \includegraphics[scale=.45]{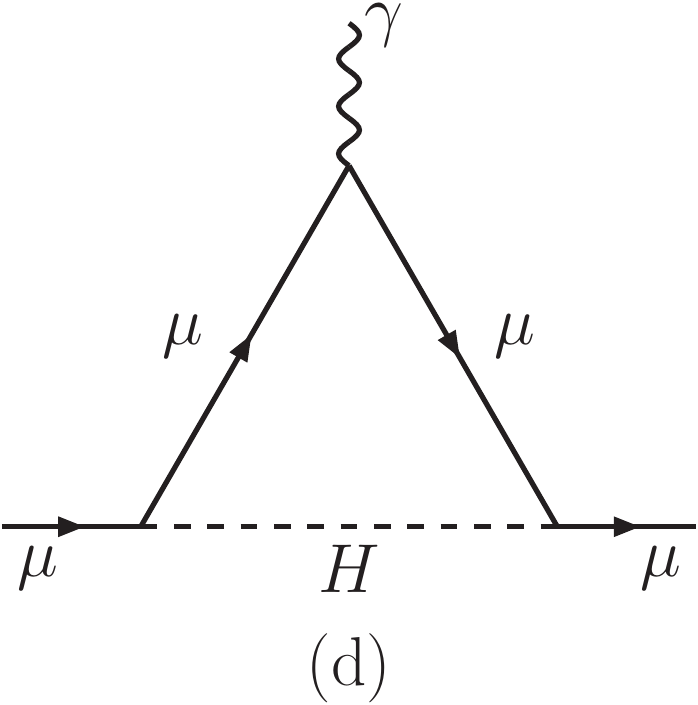}}
  \caption{One-loop Feynman diagrams contributing to $\amuEW$.}
  \label{fig:SMEW1}
\end{figure}

In this section we describe the electroweak (EW) SM contributions to $a_\mu$.
These contributions are defined as all SM contributions that are not
contained in the pure QED, the HVP, or the HLbL contributions. 
Equivalently, the EW SM contributions are given by Feynman diagrams
that contain at least one of the EW bosons $W$, $Z$, or the Higgs.

\begin{figure}[t]
\centerline{  \includegraphics[scale=.45]{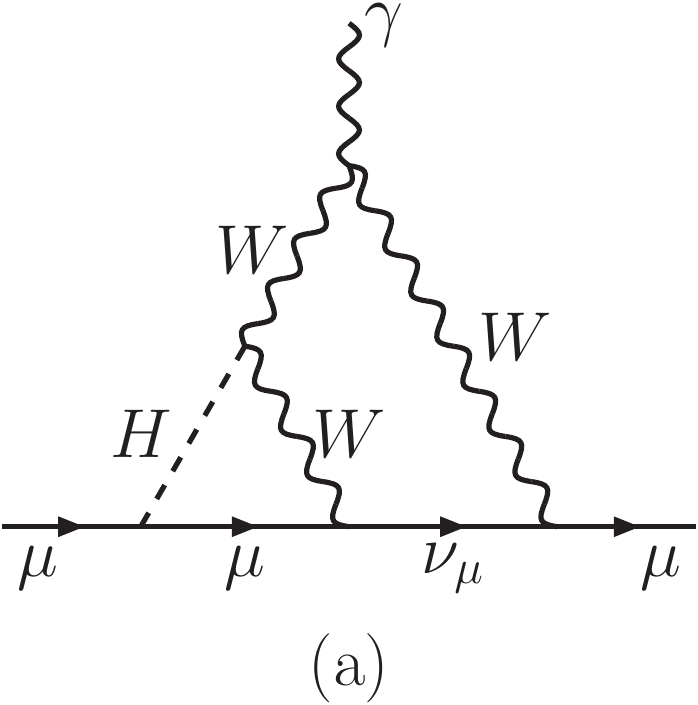}
\quad  \includegraphics[scale=.45]{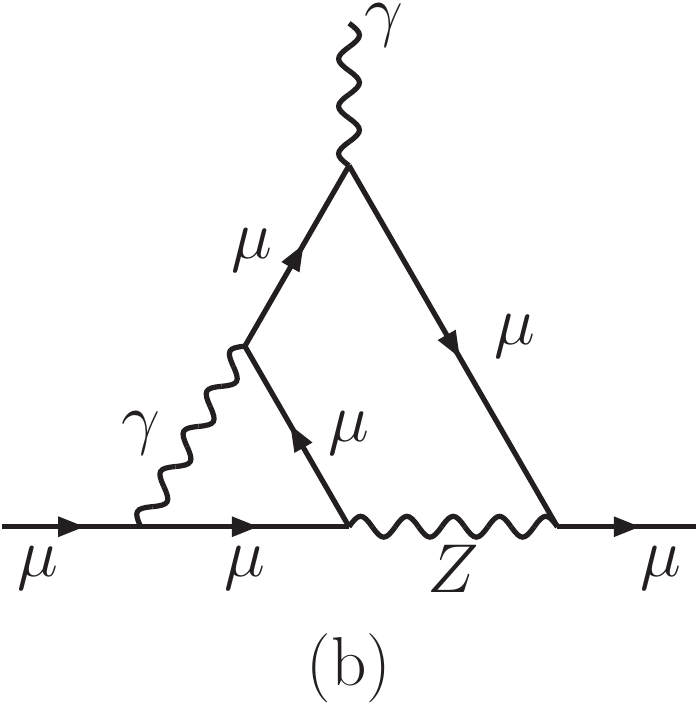}
\quad  \includegraphics[scale=.45]{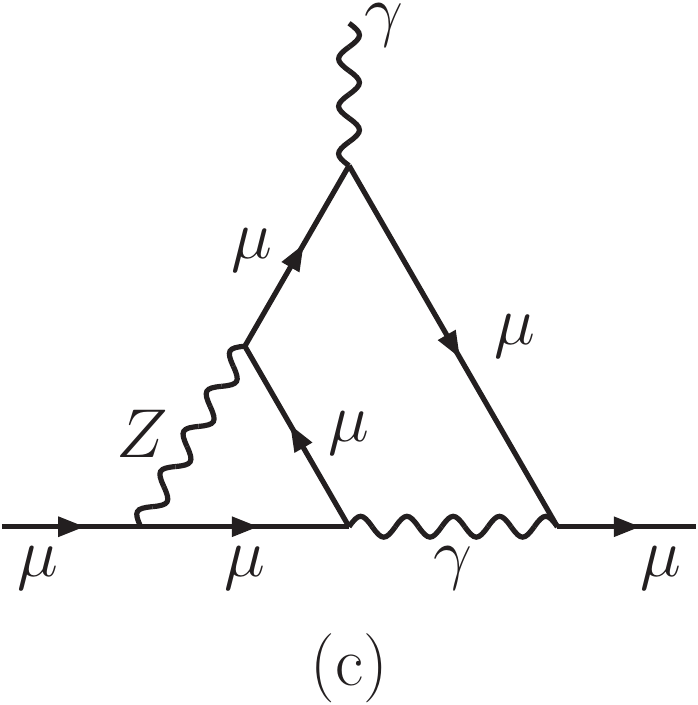}}
  \caption{Sample bosonic two-loop Feynman diagrams contributing to $\amuEW$.}
  \label{fig:SMEW2b}
\end{figure}

\begin{figure}[t]
\centerline{  \includegraphics[scale=.45]{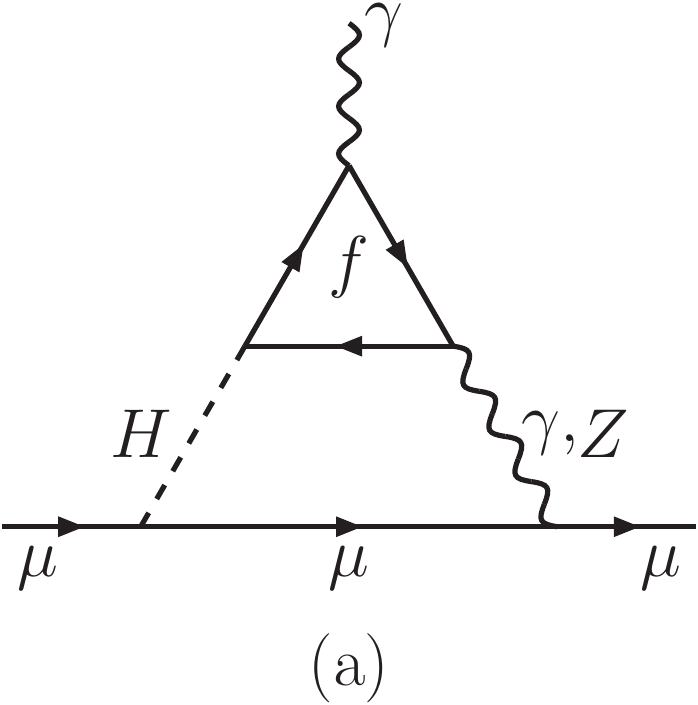}
\quad  \includegraphics[scale=.45]{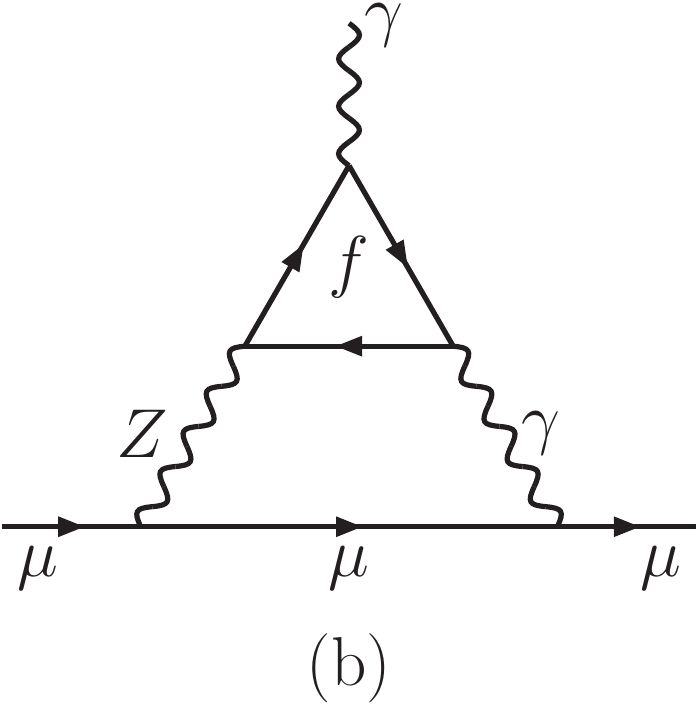}
\quad  \includegraphics[scale=.45]{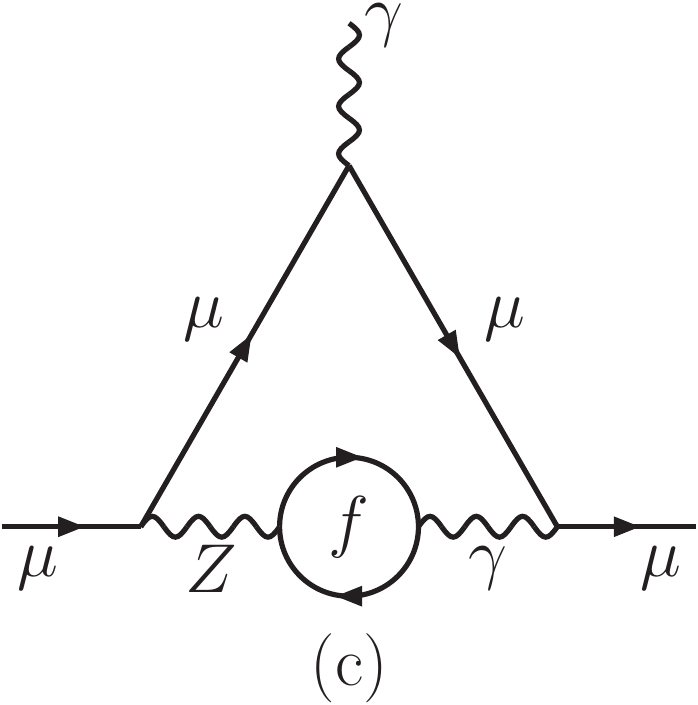}}
  \caption{Sample fermionic two-loop Feynman diagrams contributing to $\amuEW$.}
  \label{fig:SMEW2f}
\end{figure}

\Crefrange{fig:SMEW1}{fig:SMEW2f} show sample
one-loop and two-loop diagrams. 
The EW contributions are strongly suppressed by the heavy masses of
the EW bosons; numerically they contribute at the same order as the
HLbL correction. They involve diverse and
interesting physical effects. The heaviest SM particles including the
top quark and Higgs boson enter, EW gauge and Yukawa interactions and
EW parameters are relevant. At higher orders large logarithmic
corrections and nonperturbative hadronic corrections need to be
included.

In the following we first provide an overview of the EW contributions
and their most interesting qualitative
features. \Cref{sec:EWLeadinglogsAndEWhad} gives details on the
logarithmically enhanced and the nonperturbative hadronic
higher-order corrections. \Cref{sec:fullEWresult} presents
full up-to-date numerical results. Our presentation and the updated
numerical evaluation is based on Refs.~\cite{Czarnecki:2002nt,Gnendiger:2013pva}. For an extensive review we also refer to
Ref.~\cite{Jegerlehner:2009ry}. 

\subsection{Brief overview}
The EW one-loop contributions can be written as
\begin{align}
  \label{eq:amuew1}
  \amuewl &=
  \frac{G_{\text{F}}}{\sqrt{2}}
  \frac{m _\mu ^{2}}{8 \pi ^2}
  \bigg[
    \frac{5}{3}
    +
    \frac{1}{3}(1-4 s_{\text{W}} ^2)^2
    \bigg]=194.79(1)\times 10^{-11}\,,
\end{align}
with the Fermi constant $G_{\text{F}}$ and the on-shell weak mixing
angle  $s_{\text{W}}^2=(1-M_W^2/M_Z^2)$ defined via the $W$- and
$Z$-boson pole masses.  In the numerical evaluation the current values
of the input parameters~\cite{Tanabashi:2018oca} have been used, including the
SM prediction for $M_W$; the given uncertainty is the parametric uncertainty.
The one-loop contributions are essentially given by diagrams with $W$
or $Z$ exchange. 
Although the full one-loop contribution result is gauge independent, the
individual diagrams depend on the choice of gauge, and
in most common gauges also diagrams with unphysical Goldstone bosons $G$
enter as in \cref{fig:SMEW1}c.
The one-loop contribution of the physical Higgs
boson \cref{fig:SMEW1}d is negligible: 
it is suppressed by the additional factor $m _\mu ^2 / M_H ^2$ because of
the two muon--Higgs Yukawa couplings.

The most prominent feature of \cref{eq:amuew1} is its smallness.
Written in a different way, the one-loop contribution is proportional to
\begin{align}
  \label{eq:amuew1approx}
  \amuewl \propto
  \frac{\alpha}{4 \pi s_{\text{W}}^2}
  \frac{m _\mu ^2}{M _W ^2}\,,
\end{align}
i.e., it is not only  suppressed by the loop factor but also by the
small mass ratio
$m_\mu ^2 / M _W ^2 \approx 10 ^{-6}$.
Such a mass suppression is actually a general property of
contributions of heavy particles to $a_\mu$.

The two powers of the heavy mass and one power of $m_\mu$ in \cref{eq:amuew1approx} arise from
the dimensionality of the appropriate operator and its relation to
$a_\mu$;
the other power of $m_ \mu$ arises from the need for a chirality flip
between a left- and right-handed muon, which in the SM is generated by
the muon mass.
It is noteworthy that a
similar suppression happens in new physics models with new heavy
particles 
of mass $M _{\text{NP}}$.
Typically such particles contribute terms $\sim\alpha m _\mu ^2 / M
_{\text{NP}} ^2$,
but there might be important additional factors if the new couplings
are different from $\alpha$ 
and if there are new potentially enhanced mechanisms for left--right
chirality flips. 
For an extensive discussion of the role of chirality flips in the context of $a_\mu$ we refer to Refs.~\cite{Czarnecki:2001pv,Stockinger:1900zz}; surveys of possible deviations from the naive scaling due to chiral enhancement, relevant in particular in the context of $a_e$ (see \cref{sec:fine_structure_constant}), are given in Refs.~\cite{Giudice:2012ms,Crivellin:2018qmi}.

Three interesting physical effects appear at higher orders.
\begin{enumerate}
  \item
There are very large logarithmically enhanced
corrections to the one-loop result \cref{eq:amuew1approx}.
These arise from diagrams like the ones in \cref{fig:SMEW2b}b,c,
\cref{fig:SMEW2f}b,c, generally from 
two-loop diagrams 
that contain heavy particles and a photon. 
The resulting large logarithms $\log M_Z ^2 / m _f ^2$, where $m _f$ is one
of the light fermions
partially compensate the two-loop suppression.
Numerically, these logarithmic two-loop effects reduce the one-loop result by
approximately $20\%$. 
\item
The top quark and Higgs boson appear, most significantly from
\cref{fig:SMEW2b}a, \cref{fig:SMEW2f}a,b.  
The top quark contributes, e.g., via diagram \cref{fig:SMEW2f}b. 
Because of anomaly cancellation, this diagram is only well defined for
the combination of all fermions of one generation, but the
contribution from the third generation $(t,\,b,\,\tau)$ amounts to
approximately $ -8\times10^{-11}$.
The Higgs boson contributes, e.g., via diagrams
\cref{fig:SMEW2b}a, \cref{fig:SMEW2f}a. As an example, \cref{fig:SMEW2f}a is gauge independent and 
finite by itself and amounts to $- 1.5 \times 10^{-11}$. 
\item
Nonperturbative hadronic corrections are important. In particular in the diagrams in
\cref{fig:SMEW2f}b,c the computation using perturbative quark loops
and, e.g., constituent quark masses is not satisfactory. In
a better treatment the $\gamma$--$Z$ or $\gamma$--$\gamma$--$Z$ Green 
functions are evaluated without  perturbation theory using,
e.g., dispersion relations or hadronic models similarly to the
evaluation of the HVP and HLbL contributions. The shift obtained by replacing the
perturbative quark loop by the improved treatment amounts to
approximately $+2\times10^{-11}$.
\end{enumerate}

Taking into account all up-to-date
corrections given in detail in \cref{sec:fullEWresult} one 
obtains the final result 
\begin{align}
\amuEW &=153.6(1.0)\times10^{-11}\,.
\label{amuEWtotal}
\end{align}

\subsection{Leading two-loop logarithms and hadronic electroweak corrections}
\label{sec:EWLeadinglogsAndEWhad}

A special class of higher-order diagrams are those that contain  heavy
EW particles and also a photon. Examples are diagrams in 
\cref{fig:SMEW2b}b,c or the diagrams in
\cref{fig:SMEW2f}b,c.
These can lead to large logarithms of the form $\log M _Z ^2 / m _\mu
^2 \approx 13.6$ or  $\log M _Z ^2 / m _f ^2$, where  $m _f ^2$ is a
light SM fermion. 

These logarithmic two-loop contributions were first extensively
discussed and computed in Ref.~\cite{Kukhto:1992qv}, however omitting certain diagrams. 
Later, as a part of the first full two-loop computation, Refs.~\cite{Czarnecki:1995wq,Czarnecki:1995sz}
obtained a full expression of the large two-loop logarithms in the
approximation $(1 - 4 \sin ^2 \theta _{\text{W}}) \to 0$.
Using an elegant approach based on effective field theory (EFT) and
renormalization group methods, 
Refs.~\cite{Degrassi:1998es,Czarnecki:2002nt} obtained the large two-loop logarithms without
approximation and also extended the computation to the leading
three-loop logarithms.
The final result for the two-loop logarithms using a perturbative
treatment of quarks can be written in the
form~\cite{Czarnecki:2002nt}
\begin{align}
  a_\mu^{\text{EW(2), logs}} &=
  -4\frac{\alpha}{\pi}
  \log\frac{M_Z}{m_\mu} \amuewl
  \nonumber\\
&  +\frac{G_{\text{F}}m_\mu^2}{8\pi^2\sqrt{2}} \, \frac{\alpha}{\pi} \,
  \log\frac{M_Z}{m_\mu}
  \left[-\frac{47}{9} -\frac{11}{9}(1-4s_{\text{W}}^2)^2\right]
\label{EW2Llogs}
  \\
  &+\frac{G_{\text{F}}m_\mu^2}{8\pi^2\sqrt{2}}
  \,\frac{\alpha}{\pi}
  \sum_{f}\log\frac{M_Z}{\text{max}(m_f,m_\mu)}
  \left[-6g_A^\mu g_A^f N_fQ_f^2
    +\frac{4}{9}g_V^\mu g_V^f N_f Q_f\right]\,,
\nonumber
\end{align}
where in the last line the sum runs over all SM quarks and leptons and
$g_V^f=2I_3^f-4s_{\text{W}}^2Q_f$ and $g_A^f=2I_3^f$ in terms of 
the fermion weak isospin and charge.
The result reflects the contributions in the EFT approach.
The first line  originates from diagrams like
\cref{fig:SMEW2b}c, which contain an insertion of an $\amuEW$
one-loop diagram. In the EFT it corresponds to the running of the dimension-5 dipole
operator contribution to $a_\mu$; the coefficient $4\alpha/\pi$
is the anomalous dimension of this operator.
In fact, the same logarithmic correction also applies to a large class
of new physics models~\cite{Degrassi:1998es,vonWeitershausen:2010zr}.
The other terms correspond to 4-fermion operators, which arise in the
EFT upon integrating out tree-level $Z$-exchange diagrams. Via
renormalization group evolution, the 4-fermion operators mix with the
dipole operator and thus contribute to $a_\mu$.
The terms in the second line of \cref{EW2Llogs} correspond to
bosonic two-loop diagrams 
such as \cref{fig:SMEW2b}b, in which integrating out a  $Z$-boson
generates an effective 4-muon 
vertex.
The terms in the third line  correspond to the fermionic two-loop
diagrams in \cref{fig:SMEW2f}b,c; here the masses of the fermions
enter, except in case of the 
electron loop, where the muon mass sets the relevant light scale.  
Interestingly, the qualitative behavior of the fermion loop
contributions is affected by anomaly cancellation in the SM, which
requires in particular that the sum
\begin{align}
  \label{EWanomalycancellation}
  \sum_f N_f I_3^f
  Q_f^2 &=0
\end{align}
  for each fermion generation. The
prefactor of $\log M_Z$ in the first term of the last line is given by this sum; hence
$\log M_Z$ actually cancels and the true enhancement of the diagram
\cref{fig:SMEW2f}b comes from the mass ratios of the different fermions of
each generation, not from the smallness of the fermion masses compared
to $M_Z$~\cite{Czarnecki:1995wq}.

The two diagrams in \cref{fig:SMEW2f}b,c contain loops of
the light quarks $u,d,s$. As shown in the third line of \cref{EW2Llogs},
a perturbative evaluation produces large logarithms of light quark
masses. Because of confinement, these quark masses are not well
defined and significant nonperturbative corrections to these Feynman
diagrams are expected. A nonperturbative evaluation of these
diagrams, which can replace the perturbative result,
has been pioneered in Ref.~\cite{Peris:1995bb} and
improved in Refs.~\cite{Knecht:2002hr,Czarnecki:2002nt}.

The simpler diagram in \cref{fig:SMEW2f}c contains a two-point
$\gamma$--$Z$ subdiagram. Lorentz invariance implies that only the
vectorial coupling of the $Z$ boson can contribute, as is reflected by the
appearance of the factors $g_V^\mu g_V^f$ in the 
corresponding terms in \cref{EW2Llogs}. This
allows one to obtain the value of the $\gamma$--$Z$ subdiagram and then of the
contribution to $a_\mu$ from the diagram in \cref{fig:SMEW2f}c
nonperturbatively via a dispersion relation from $e^+e^-\to \text{hadrons}$
data. The nonperturbative correction can be taken into account by
replacing 
\begin{align}
2/3\sum_{q=u,d,s,c,b} N_c\left(I^3_q Q_q-2Q_q^2
s_{\text{W}}^2\right)\text{log} M_Z/m_q\to6.88
\end{align} 
in \cref{EW2Llogs}, see Ref.~\cite{Czarnecki:2002nt}.

The diagram in \cref{fig:SMEW2f}b contains a three-point
$\gamma^*$--$\gamma$--$Z^*$ subdiagram with quark or lepton loops. Because
of the Furry theorem, here only the axial $Z$ boson coupling can
contribute, and as a consequence gauge anomaly cancellation is
important.
The basic quantity of interest for the nonperturbative evaluation is
the Green function for the $VVA$ 
currents $\langle0|T j^\mu(x)j^\nu(y)j_5^\rho(z)|0\rangle$. After
contraction with one photon polarization vector and in the limit of
small external photon momentum it can be expressed in terms of two
scalar functions $w_{L,T}(Q^2)$, which only depend on the $Z^*$
momentum scale $Q^2$. The contribution to $a_\mu$ can be expressed in terms
of integrals of $w_{L,T}$ over $Q^2$, and using a one-loop evaluation of
$w_{L,T}$ one would recover the perturbative result for $a_\mu$. 
Gauge anomaly cancellation is reflected by the fact that
\begin{align}
  w_{L,T}^{\text{1-loop}}(Q^2)&\propto
  \sum_f  N_f I_3^f Q_f^2
  \left(
  \frac{1}{Q^2}
  - \frac{2 m_f ^2}{Q^4}\log\frac{Q^2}{m_f ^2}
  +{\cal O}\left(\frac{1}{Q^6}\right)
  \right)\,.
\end{align}
For each individual
fermion, there is only a $1/Q^2$ suppression and the contribution to
$a_\mu$ is UV divergent, but because of
\cref{EWanomalycancellation} the sum over each fermion generation is
well-behaved.

References~\cite{Peris:1995bb,Knecht:2002hr,Czarnecki:2002nt} have investigated
constraints on these functions from nonrenormalization theorems and
from operator product expansions. Specifically, in the chiral limit
$m_{u,d,s}\to0$, Ref.~\cite{Czarnecki:2002nt} obtains for the longitudinal
function $w_L^{[u,d]}(Q^2)=-3w_L^{[s]}(Q^2)=2/Q^2$; $w_T$ is similar
with specific ${\cal O}(1/Q^6)$ corrections. For the actual calculation outside the chiral limit,
Ref.~\cite{Czarnecki:2002nt} employs a model ansatz
involving pion and other meson masses
for all four functions
$w_{L,T}^{[u,d],[s]}$ that is compatible with all constraints.
Combining this with the perturbative results for the electron and muon
loop contributions, Ref.~\cite{Czarnecki:2002nt} obtains
\begin{align}
\label{EWVVA12}
a_\mu^{\rm EW(2)}(e,\mu,u,c,d,s)=-6.91(20)(30)\times10^{-11}\,,
\end{align}
for the nonperturbative evaluation of the first- and second-generation
contributions of diagram \cref{fig:SMEW2f}b,
where the uncertainties refer to the first and second generation, respectively. The perturbative evaluation for the third
generation yields
\begin{align}
\label{EWVVA3}
a_\mu^{\rm EW(2)}(\tau,t,b)&=-8.21(10)\times10^{-11}\,.
\end{align}
The uncertainties given in \cref{EWVVA12,{EWVVA3}} are obtained in Ref.~\cite{Czarnecki:2002nt} by varying respective input parameters of the hadronic models and perturbative calculations in generous intervals. The total theory uncertainty associated with these contributions is then conservatively inflated by Ref.~\cite{Czarnecki:2002nt} to $\pm1.0\times10^{-11}$.

\subsection{Full result including all known higher-order corrections}
\label{sec:fullEWresult}
Here we present the full numerical results of the EW contributions to
$a_\mu$, up to leading three-loop order, following the presentation in
Ref.~\cite{Gnendiger:2013pva}.  
The one-loop result \cref{eq:amuew1} has been expressed in terms of
the muon decay constant $G_{\text{F}}$. It is important that
higher-order contributions are evaluated in a renormalization scheme
compatible with this. The following results all correspond to a
parameterization of higher-order $n$-loop results in terms of
$G_{\text{F}}\alpha^{n-1}$.

In order to avoid double counting between leading logarithmic and
nonlogarithmic and between nonperturbative and perturbative
contributions, it is useful to split up the EW contributions
as
\begin{align}
\label{EWsplitup}
a_\mu^{\rm EW} &=
a_\mu^{\rm EW(1)} + \amub + \amuf + a_\mu^{\rm EW(\ge3)},
\end{align}
and to subdivide the fermionic two-loop contributions further into
\begin{align}
  \label{EWfermsplit}
\amuf &= 
a_\mu^{\rm EW(2)}(e,\mu,u,c,d,s)
+a_\mu^{\rm EW(2)}(\tau,t,b)
+\amufrestH
 + \amufrestnoH \, .
\end{align}
The first two terms on the RHS of \cref{EWfermsplit}
denote contributions from the
diagrams in \cref{fig:SMEW2f}b with a
$\gamma^*$--$\gamma$--$Z^*$-subdiagram, as already discussed in
\cref{sec:EWLeadinglogsAndEWhad}. 
The third term denotes the Higgs-dependent fermion-loop diagrams 
like in \cref{fig:SMEW2f}a; the fourth collects all remaining
fermionic two-loop contributions, e.g., from $W$-boson exchange or from diagram in
\cref{fig:SMEW2f}c.

The bosonic two-loop contributions $\amub$ are defined by two-loop and
associated counterterm diagrams without closed fermion loops, as
in \cref{fig:SMEW2b}. 
These contributions contain the large logarithms in the first and second lines
of \cref{EW2Llogs} and nonlogarithmic terms, which depend in
particular on the Higgs boson mass.
Their first full computation in Ref.~\cite{Czarnecki:1995sz} was a milestone---the first full computation of a SM observable at the
two-loop level. Reference~\cite{Czarnecki:1995sz} employed an approximation assuming $M_H\gg M_W$. Reference~\cite{Heinemeyer:2004yq} later confirmed the result but provided the full
$M_H$-dependence; Ref.~\cite{Gribouk:2005ee} then published the result in semianalytical
form. Recently a fully numerical computation in
Ref.~\cite{Ishikawa:2018rlv} again confirmed the result. 
Using the PDG value  $M_H=125.18(16)$\,GeV~\cite{Tanabashi:2018oca} now fixes the
value of these contributions 
and we obtain 
\begin{align}
\label{Higgsbosonicres}
\amub&=-19.96(1)\times 10^{-11}\,.
\end{align}
The given theory uncertainty is the parametric uncertainty resulting
from the experimental uncertainty of the Higgs boson and $W$-boson
masses.

The most critical types of fermion-loop contributions $a_\mu^{\rm
  EW(2)}(e,\mu,u,c,d,s;\tau,t,b)$ from the
diagrams of \cref{fig:SMEW2f}b have already 
been discussed in \cref{sec:EWLeadinglogsAndEWhad}, and the numerical
results have been given in \cref{EWVVA12,EWVVA3}.

Next we
focus on the Higgs-dependent fermion-loop corrections. They are given
by the diagrams of  \cref{fig:SMEW2f}a with 
Higgs--$\gamma$--$\gamma$ or Higgs--$\gamma$--$Z$ subdiagram. These
diagrams were computed in various limits of the ratio $M_H/m_t$ in
Ref.~\cite{Czarnecki:1995wq}; an exact expression can be found, e.g., in
Ref.~\cite{Gnendiger:2013pva}, and further discussion of the validity of
large-mass expansions for such contributions has been given recently
in Ref.~\cite{Czarnecki:2017rlm}.
Inserting the measured value of the Higgs boson mass, and taking into
account all fermions (only the heavy fermions top, bottom, charm, and $\tau$
are relevant) and diagrams with Higgs and $Z$-boson exchange, we obtain
\begin{align}
\amufrestH=-1.51(1)\times10^{-11}\,,
\label{Higgsfermionicres}
\end{align}
where the indicated uncertainty arises essentially from the uncertainty of
the input parameters $m_t$ and $M_H$.

The non-Higgs dependent contributions $\amufrestnoH$ have also been
computed in Ref.~\cite{Czarnecki:1995wq} in the approximation
$(1-4s_{\text{W}}^2)\to0$; the neglected terms have then been added in
Ref.~\cite{Czarnecki:2002nt} and include the hadronic corrections to diagrams
with $\gamma$--$Z$ interaction
\cref{fig:SMEW2f}c. Reference~\cite{Gnendiger:2013pva} provides the
full analytic result. Numerically, it obtains
\begin{align}
\amufrestnoH=-4.64(10)\times10^{-11}\,.
\label{fermrestnoHiggs}
\end{align}
The uncertainty corresponds to an estimate of still neglected terms
 suppressed by a factor $(1-4s_{\text{W}}^2)$ or $M_Z^2/m_t^2$ and not 
enhanced by anything. 

\begin{figure}[t]
\centering
\includegraphics[width=0.5\textwidth]{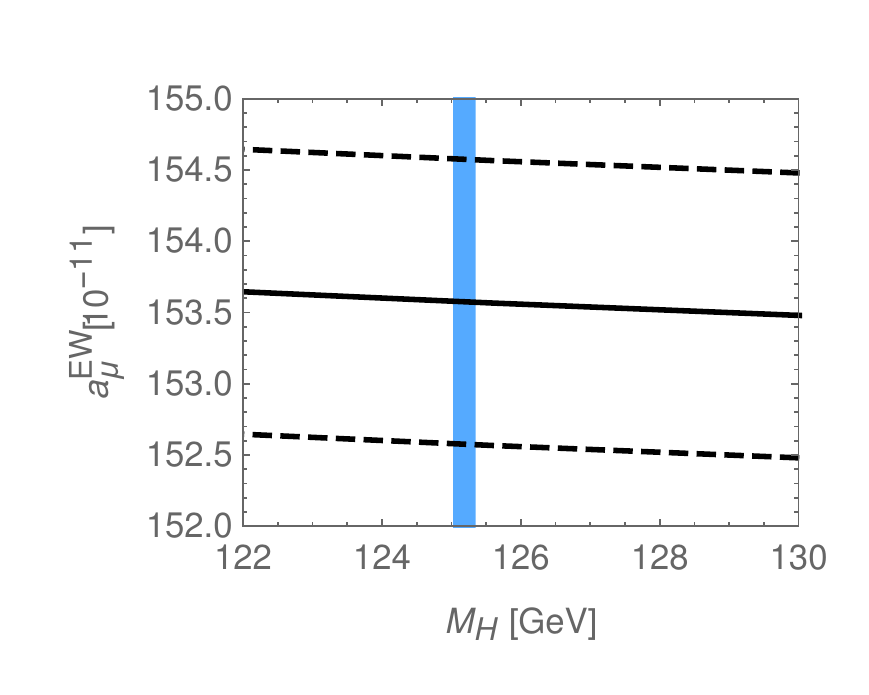}
\caption{
Update of Fig.\ 5 of Ref.~\cite{Gnendiger:2013pva} for the numerical
result for $a_\mu^{\rm EW}$ as a function of the Higgs boson mass. 
The vertical band indicates the measured value of $M_H$~\cite{Tanabashi:2018oca}.
The dashed lines correspond to the uncertainty of the final result, 
quoted in \cref{amuEWNew}.}
\label{fig:amuEWfull}
\end{figure}

Contributions from beyond the two-loop level have been considered in
Refs.~\cite{Degrassi:1998es,Czarnecki:2002nt}, using the EFT and
renormalization group methods mentioned in
\cref{sec:EWLeadinglogsAndEWhad}. The value of these logarithmically
enhanced three-loop corrections depends on the parameterization of the
one- and two-loop results. Specifically, an accidental cancellation
among the three-loop corrections
was observed in Ref.~\cite{Czarnecki:2002nt} if the
two-loop result is parameterized in terms of $G_F\,\alpha$. In this
case the three-loop
logarithms are numerically negligible.  Hence, 
\begin{align}
a_\mu^{\rm EW(\ge3)}=0(0.20)\times10^{-11}\,,
\label{EWthreeloop}
\end{align}
where the uncertainty estimate is from Ref.~\cite{Czarnecki:2002nt}. It corresponds to estimating the
nonleading logarithmic three-loop contributions to be below a percent
of the two-loop contributions.

Summing up the previous numerical results of
the one-loop contributions \cref{eq:amuew1}, the bosonic two-loop contributions \cref{Higgsbosonicres},
the four fermionic two-loop contributions
\cref{Higgsfermionicres,fermrestnoHiggs,EWVVA3,EWVVA12}, and the leading three-loop
logarithms \cref{EWthreeloop},
we obtain
\begin{align}
a_\mu^{\rm EW}&= 153.6(1.0)\times10^{-11}\,,
\label{amuEWNew}
\end{align}
as already given in \cref{amuEWtotal}. This value is mainly based on Refs.~\cite{\EWref}, which should be cited in any work that uses or quotes \cref{amuEWNew}.
The result is illustrated in \cref{fig:amuEWfull}, which
is an update of Fig.\ 5 from
Ref.~\cite{Gnendiger:2013pva}.
We assess the final theory uncertainty of these contributions to be
$\pm1.0\times10^{-11}$, the estimate of Ref.~\cite{Czarnecki:2002nt} for the
overall hadronic uncertainty from the 
diagrams of \cref{fig:SMEW2f}b, which is 
now by far the dominant source of uncertainty of the EW contributions.
The uncertainty from unknown three-loop contributions and neglected two-loop terms
suppressed by $M_Z^2/m_t^2$ and $(1-4s_{\text{W}}^2)$ is significantly smaller
and the uncertainty due to the 
experimental uncertainty of the Higgs-boson, $W$-boson, and top-quark
masses is well below $10^{-12}$ and thus negligible.

\FloatBarrier

\clearpage

\section{Conclusions and outlook}
\label{sec:conclusionsWP}
In this paper we provide a detailed analysis and review of the
SM calculation of the muon anomalous magnetic moment
$a_\mu$. The emphasis is on the hadronic contributions, since
they dominate the final uncertainty, but the QED and electroweak
contributions are also discussed in detail and up-to-date numbers are
provided. 

The QED contribution, which has been calculated up to tenth order in the 
perturbative expansion, i.e., $\mathcal{O}(\alpha^5)$, is reviewed in
\cref{sec:QED}. The final number depends on the input used for the
fine-structure constant $\alpha$ and at present there are two independent
determinations that differ by about 2.4 standard deviations. The impact of
this discrepancy on the final number for $a_\mu$ is however well below the
uncertainty of the QED contribution itself, which is dominated by the
estimated effect of the $\mathcal{O}(\alpha^6)$ contribution.
As final number we take the one based on the value of $\alpha$ obtained
from atom-interferometry measurements of the Cs atom~\cite{Parker:2018vye}, see \cref{eq:amuQED_Cs}, and the latest QED calculations from Refs.~\cite{\QEDref}:
\begin{align}
\amuQED(\alpha(\text{Cs})) &= 116~584~718.931(104)  \times 10^{-11} \, .
\end{align}

Electroweak contributions are reviewed in \cref{sec:EW}: they
have been calculated up to two loops and an estimate of the leading
logarithmic contribution beyond two-loop level is also included in
the final estimate. The hadronic loops, which appear at two-loop level, are also included and dominate the uncertainty of the EW contribution. 
The final result~\cref{amuEWNew} (mainly based on Refs.~\cite{\EWref}) reads
\begin{align}
a_\mu^{\rm EW}&= 153.6(1.0)\times10^{-11}\, ,
\end{align}
with an uncertainty ten times larger than the QED one, but still negligible
with respect to the hadronic uncertainties.

In the section on data-driven evaluations of HVP we reviewed both the available data sets for the $e^+e^-\to\text{hadrons}$ cross section and the techniques applied for the evaluation of the HVP dispersive integral. In particular, we provide a detailed discussion of the differences between these approaches and the current limitations of the dispersive HVP evaluation, as they arise from the published experimental uncertainties as well as, crucially, from unresolved tensions among the data sets, especially in the dominant $\pi\pi$ channel. As the main result, \cref{merging}, we devised a merging procedure that adequately takes into account these tensions, which also drive the differences between the available HVP evaluations. The resulting estimate, based on Refs.~\cite{\HVPref} as well as the main experimental input from Refs.~\cite{\HVPexpref},
\begin{equation}
\label{amuHVPLO}
\amuHVPLO = 6931(40) \times 10^{-11}
\end{equation}
should provide a conservative but realistic assessment of the current precision of data-driven HVP evaluations. In the same framework, the LO result is complemented by NLO~\cite{Keshavarzi:2019abf} and NNLO~\cite{Kurz:2014wya} HVP iterations, see \cref{HVPNLO} and \cref{HVPNNLO}, 
\begin{equation}
\amuHVPNLO = -98.3(7) \times 10^{-11}\,,\qquad 
\amuHVPNNLO = 12.4(1) \times 10^{-11}\,, 
\end{equation}
leading to the sum
\begin{equation}
\label{HVPtotal}
  \amuHVPLO+\amuHVPNLO+\amuHVPNNLO=6845(40)\times 10^{-11}\,.  
\end{equation}
Finally, we discussed the prospects for future improvements, including new data from several $e^+e^-$ experiments as well as the possibility to measure HVP independently in electron--muon scattering.

The status of lattice QCD+QED calculations of HVP is reviewed in \cref{sec:latticeHVP}.
While lattice calculations can, in principle, provide an alternate, {\em ab initio} determination of the HVP contribution, they are, at present, not precise enough to confront the data-driven evaluations. The current ``lattice world average,'' obtained in \cref{subsec:status} from a conservative combination of current, published lattice QCD+QED results, is consistent with the data-driven result of \cref{amuHVPLO} but with a large enough uncertainty to also cover the  ``no new physics'' scenario: 
\begin{equation}
\label{eq:latHVPfinal}
    \amuHVPLO = 7116(184) \times 10^{-11}\,,
\end{equation}
based on Refs.~\cite{\HVPlatticeref}.

The phenomenological estimate of HLbL scattering as reviewed in
\cref{sec:dispHLbL} is essentially based on a dispersive approach, in
analogy to HVP. The various contributions to HLbL can be collected into
three main pieces depending on how they have been estimated: (1) the
numerically dominant contributions from the single-pseudoscalar poles and
large parts of the two-pion intermediate states, both of which rely on
data-driven approaches and are under good control; (2) the model-dependent
estimates for the sum of scalar, tensor, and axial-vector contributions,
as well as the impact of short-distance constraints; all of these still
suffer from significant uncertainties, which in the total have been added
linearly; (3) the $c$-quark contribution, which can be estimated using
perturbative QCD, with a conservative uncertainty estimate in view of the
low scale and potential nonperturbative effects.  The final estimates for
HLbL from \cref{tab:compilations} (mainly based on Refs.~\cite{\HLbLref} and, in addition to $e^+e^-\to\text{hadrons}$ cross sections, the experimental input from Refs.~\cite{\HLbLexpref}) and HLbL at NLO~\cite{Colangelo:2014qya} from \cref{HLbL_NLO} read
as follows:
\begin{align}
   \amuHLbL &= (69.3(4.1) + 20(19) + 3(1)) \times 10^{-11} \nonumber \\ 
   &= 92(19) \times 10^{-11} \, ,\\
   \amuHLbLNLO & = 2(1) \times 10^{-11} \,,\label{amuHLbLNLO_concl}
\end{align}
where the first line gives the three pieces in the same order as discussed
above and the total in the second line is obtained by adding the central
values of the three contributions and combining the errors in quadrature.
The final error is about 20\% and is completely dominated by the model
estimates of a numerically subdominant part of the total.

The lattice determination of HLbL scattering is reviewed in \cref{sec:latticeHLbL}.  The lattice methodology for this quantity has advanced significantly in the last years~\cite{\HLbLlatticemethods} and has now reached a mature stage, resulting in a calculation~\cite{\HLbLlatticeref} with reliable estimates of both statistical and systematic uncertainties (\cref{eqn:hlbllatrbcres}):
\begin{align}
\amuHLbL = 78.7(30.6)_\text{stat}(17.7)_\text{sys}\times 10^{-11} \,.
\end{align}
There have been extensive checks between different groups working on the lattice HLbL as well as internal checks of the calculations such as the regression against the leptonic loop or pion-pole contributions.  These checks are explained in detail in \cref{sec:latticeHLbL}.

To obtain a recommendation for the full SM prediction we proceed as follows: for HLbL scattering, there is excellent agreement between phenomenology and lattice QCD, to the extent that it is justified to consider a weighted average. Taking into account that the lattice-QCD value does not include the $c$-quark loop, we first average the light-quark contribution and add the $c$ quark as estimated phenomenologically in the end. This produces
\be
\amuHLbL (\text{phenomenology + lattice QCD})=90(17)\times 10^{-11}\,, 
\label{eq:HLbL_comb}\ee
and, using \cref{amuHLbLNLO_concl},
\be
\label{HLbLtotal}
\amuHLbL (\text{phenomenology + lattice QCD}) + \amuHLbLNLO =92(18)\times 10^{-11}\,.
\ee
For HVP, the current uncertainties in lattice calculations are too large to perform a similar average 
and the future confrontation of phenomenology and lattice QCD  crucially depends on the outcome of forthcoming lattice studies. For this reason, we adopt \cref{amuHVPLO} as our final estimate, emphasizing that the uncertainty estimate already accounts for the tensions in the $e^+e^-$ data base. Combined with the QED and EW contributions, we obtain
\begin{align}
\label{SM_prediction}
 \amuSM&=\amuQED+\amuEW+\amuHVPLO+\amuHVPNLO+\amuHVPNNLO+\amuHLbL+\amuHLbLNLO\notag\\
 &=116\,591\,810(43)\times 10^{-11}\,.
\end{align}
This value is mainly based on Refs.~\cite{\SMref}, which should be cited in any work that uses or quotes \cref{SM_prediction}. 
It differs from the Brookhaven measurement~\cite{Bennett:2006fi} 
\be
\label{amuexp}
\amuexp=116\,592\,089(63)\times 10^{-11}\,,
\ee
where the central value is adjusted to the latest value of $\lambda=\mu_\mu/\mu_p=3.183345142(71)$~\cite{Mohr:2015ccw}, by
\be
\label{amudiff}
\Delta a_\mu:= \amuexp-\amuSM=279(76)\times 10^{-11}\,,
\ee
corresponding to a $3.7 \sigma$ discrepancy.
In constructing \cref{HVPtotal,HLbLtotal,SM_prediction}, we have taken into account the correlations between the uncertainties in the leading and subleading HVP contributions as well as the partial correlation in the case of HLbL, with numbers rounded including subleading digits from the individual contributions. 

The prospects for near-term and long-term improvements of the uncertainties in the SM prediction are excellent. As discussed in \cref{sec:dataHVP}, a new measurement of the crucial $2\pi$ channel by SND is currently under review,  
and more measurements of the $2\pi$ channel and others are forthcoming, 
leading to the realistic prospects of 
reducing the dispersive HVP error by a factor of $2$. In addition, independent data-driven input could be provided by the MUonE project. 
The past five years have seen great progress in the development of methods to address the challenges associated with lattice determinations of $\amuHVPLO$ at the target precision, as discussed in detail in \cref{sec:latticeHVP}.
This is also evident in the recent high-precision lattice result for $\amuHVPLO$~\cite{Borsanyi:2020mff}, which, however, still needs to be scrutinized in detail. With these methods now in place, and with sustained, dedicated effort, lattice results with permil-level precision will be forthcoming.
 The phenomenological determination of HLbL scattering has been consolidated at a level well below the Glasgow consensus, see \cref{sec:dispHLbL}, with the dominant contributions derived using data-driven methods in analogy to the dispersive HVP approach. With expected progress on the subleading contributions, a $10\%$ calculation of HLbL scattering now appears feasible.  Finally, we expect more independent lattice calculations of the HLbL to appear in the next years.  Building on the newly developed methodologies, a $10\%$ lattice calculation of the HLbL also appears feasible by the end of the Fermilab experiment.

\clearpage

\section*{Acknowledgements}

We are very grateful to the Fermilab Directorate and the Fermilab Theoretical Physics Department for their financial and logistical support of the first workshop of the Muon $g-2$ Theory Initiative (held near Fermilab in June 2017)~\cite{FNAL2017}, which was crucial for its success, and indeed for the successful start of the Initiative. Financial support for this workshop was also provided by the Fermilab Distinguished Scholars program, the Universities Research Association through a URA Visiting Scholar award, the Riken Brookhaven Research Center, and the Japan Society for the Promotion of Science under Grant Number KAKEHNHI-17H02906.
We thank Shoji Hashimoto, Toru Iijima, Takashi Kaneko, and Shohei Nishida for hosting the the HVP workshop at KEK \cite{KEK2018} and the KEK Theory Center and the U.S.--Japan Science and Technology Cooperation Program in High Energy Physics for providing logistical and financial support. 
The HLbL workshop at the University of Connecticut \cite{UConn2018} was hosted by the University of Connecticut Physics Department. 
We also gratefully acknowledge support for the second plenary workshop in Mainz \cite{Mainz2018} from the Deutsche Forschungsgemeinschaft via the Cluster of Excellence ``Precision Physics, Fundamental Interactions and Structure of Matter'' (PRISMA), the Collaborative Research Centre ``The low-energy frontier of the Standard Model'' (SFB 1044), as well as the Helmholtz Institute Mainz.
And finally, we thank the Institute for Nuclear Theory at the University of Washington for hosting the third plenary workshop \cite{INT2019} and for its kind hospitality and stimulating research environment. This workshop was supported in part by the U.S.\ Department of Energy, Office of Science, under Award Numbers DE-FG02-00ER41132, DE-SC0020106, and by the U.S.--Japan Science and Technology Cooperation Program in High Energy Physics. 
This review benefited from discussions with 
O.~Cat{\`a},
N.~Christ,
L.~Y.~Dai,
H.~Davoudiasl,
S.~Fayer,
S.~Ganguly,
A.~Gasparian,
S.~Hashimoto,
T.~Iijima,
K.~Kampf,
D.~Kawall,
I.~Larin,
Z.~Pagel,
M.~Petschlies,
A.~Rebhan,
K.~Schilcher,
K.~Shimomura,
E.~Shintani,
D.~Steffen,
S.~Tracz,
C.~Tu,
and T.~Yamazaki. 

The work in this paper was supported by CNRS, by Conacyt (Ciencia B{\'a}sica 2015) under Grant Number 250628, by CONACyT-Mexico under Grant Number CB2014-22117, by Coordinaci{\'o}n de la Investigaci{\'o}n Cient{\'i}fica (CIC-UMSNH) under Grant Number 4.10, by Danmarks Frie Forskningsfond under Grant Number 8021-00122B, by Deutsche Forschungsgemeinschaft Collaborative Research Centers CRC~1044, CRC~1044 -204404729, CRC~110, and under Grant Numbers HI~2048/1-1, Prisma Cluster for Excellence PRISMA$^+$ EXC2118/1, STO/876/6-1, by European Research Council under the European Union's Horizon 2020 research and innovation programme under Grant Agreement Numbers 668679, 757646, 771971-SIMDAMA, 813942, by the European Union H2020-MSCA-COFUND2016 under Grant Number 754510, by the European Union's Horizon 2020 research and innovation programme under the Marie Sk\l{}odowska-Curie Grant Agreement number 843134, by European Union  EuroPLEx Grant H2020-MSCA-ITN-2018-813942, by European Union STRONG 2020 project under Grant Agreement Number 824093, by the Excellence Initiative of Aix-Marseille University - A*MIDEX, a French ``Investissements d'Avenir'' program, through the Chaire d'Excellence program and the OCEVU Laboratoire d'Excellence (ANR-11-LABX-0060), by the Fermilab Distinguished Scholars program, by Fondo SEP-Cinvestav under Grant Number 142, Funda\c{c}{\~a}o para a Ci{\^e}ncia e a Tecnologia under Grant Number SFRH/BPD/109443/2015, by Generalitat de Catalunya under Grant Number 2017SGR1069, by the Helmholtz Association (German Federal Ministry of Education and Research), by the  Helmholtz-Institut Mainz, by the Istituto Nazionale di Fisica Nucleare (INFN), by the Isaac Newton Trust, by the Japan Society for the Promotion of Science under Grant Numbers KAKENHI-15H05742, 16K05317, 16K05323, 16K05338, 17H01133, 17H02906, 18H05226, 19K21872, 20K03926, 20K03960, by Junta de Andaluc\'{\i}a under Grant Number A-FQM-467-UGR18, by KEK, by Ministerio de Ciencia e Innovacion under Grant Number  CICYTFEDER-FPA2017-86989-P, by Ministerio de Industria, Econom\'{\i}a y Competitividad under Grant Numbers FPA2016-78220-C3-3-P, FPA2017-86989-P, PGC2018-094857-B-I00, SEV-2016-0588, SEV-2016-0597, by Laboratoires d'Excellence FIRST-TF grants, a French ``Investissements d'Avenir'' program, by the Ministry of Science and Higher Education of the Russian Federation under  Grant Agreement Number 14.W03.31.0026, by the National Research Foundation of South Africa, by the Natural Sciences and Engineering Research Council of Canada, by the Portuguese Science Foundation (FCT) Investigator Grant IF/00898/2015, by the Romanian Ministry of Education and Research under Grant Number PN19060101, by the Russian Science Foundation under Grant Number RSF 18-12-00128, by the Secretaria d'Universitats i Recerca del Departament d'Economia i Coneixement de la Generalitat de Catalunya under Grant Number 2017 SGR 1069, by the Swedish Research Council under Grant Numbers 2016-05996, 2019-03779, by the Swiss National Science Foundation under Grant Numbers PP00P2\_176884, PCEFP2\_181117, by The Leverhulme Trust under Grant Number ECF-2019-223, by the UK Science and Technology Facilities Council (STFC) under Grant Numbers ST/N504130/1, ST/P000290/1, ST/P00055X/1,  ST/P000630/1, ST/P000711/1, ST/P000746/1, ST/S000879/1, ST/S000925/1, by the U.S.\ Department of Energy, Office of Science, Office of High Energy Physics under Award Numbers DE-SC0009919, DE-SC0009998, DE-SC0010005, DE-SC0010120, DE-SC0010339, DE-SC0012391, DE-SC0012704, DE-SC0013682, DE-SC0013895, DE-SC0015655, by the U.S.\ Department of Energy, Office of Science, Office of Nuclear Physics under Award Numbers DE-AC02-05CH11231, DE-FG02-00ER41132, DE-FG02-97ER41020,  by the U.S.--Japan Science and Technology Cooperation Program in High Energy Physics, ``Incubation Platform for Intensity Frontier,'' by the U.S.\ National Science Foundation under Grant Numbers NSF-PHY-1316222, PHY14-14614, PHY17-19626, and PHY19-23131, and by the U.S.\ National Institute of Standards and Technlology (NIST) Precision Measurement Grant Program under Award Number 60NANB16D271.  
This manuscript has been authored by Fermi Research Alliance, LLC under Contract No.~DE-AC02-07CH11359 with the U.S.\ Department of Energy, Office of Science, Office of High Energy Physics.

\clearpage

\bibliographystyle{apsrev4-1_mod}
\bibliography{SM}

\begin{thebibliography}{824}%
\makeatletter
\providecommand \@ifxundefined [1]{%
 \@ifx{#1\undefined}
}%
\providecommand \@ifnum [1]{%
 \ifnum #1\expandafter \@firstoftwo
 \else \expandafter \@secondoftwo
 \fi
}%
\providecommand \@ifx [1]{%
 \ifx #1\expandafter \@firstoftwo
 \else \expandafter \@secondoftwo
 \fi
}%
\providecommand \natexlab [1]{#1}%
\providecommand \enquote  [1]{``#1''}%
\providecommand \bibnamefont  [1]{#1}%
\providecommand \bibfnamefont [1]{#1}%
\providecommand \citenamefont [1]{#1}%
\providecommand \href@noop [0]{\@secondoftwo}%
\providecommand \href [0]{\begingroup \@sanitize@url \@href}%
\providecommand \@href[1]{\@@startlink{#1}\@@href}%
\providecommand \@@href[1]{\endgroup#1\@@endlink}%
\providecommand \@sanitize@url [0]{\catcode `\\12\catcode `\$12\catcode
  `\&12\catcode `\#12\catcode `\^12\catcode `\_12\catcode `\%12\relax}%
\providecommand \@@startlink[1]{}%
\providecommand \@@endlink[0]{}%
\providecommand \url  [0]{\begingroup\@sanitize@url \@url }%
\providecommand \@url [1]{\endgroup\@href {#1}{\urlprefix }}%
\providecommand \urlprefix  [0]{URL }%
\providecommand \Eprint [0]{\href }%
\providecommand \doibase [0]{http://dx.doi.org/}%
\providecommand \selectlanguage [0]{\@gobble}%
\providecommand \bibinfo  [0]{\@secondoftwo}%
\providecommand \bibfield  [0]{\@secondoftwo}%
\providecommand \translation [1]{[#1]}%
\providecommand \BibitemOpen [0]{}%
\providecommand \bibitemStop [0]{}%
\providecommand \bibitemNoStop [0]{.\EOS\space}%
\providecommand \EOS [0]{\spacefactor3000\relax}%
\providecommand \BibitemShut  [1]{\csname bibitem#1\endcsname}%
\let\auto@bib@innerbib\@empty
\bibitem [{\citenamefont {Bennett}\ \emph {et~al.}(2006)\citenamefont {Bennett}
  \emph {et~al.}}]{Bennett:2006fi}%
  \BibitemOpen
  \bibfield  {author} {\bibinfo {author} {\bibfnamefont {G.~W.}\ \bibnamefont
  {Bennett}} \emph {et~al.} (\bibinfo {collaboration} {Muon $g-2$}),\ }\href
  {\doibase 10.1103/PhysRevD.73.072003} {\bibfield  {journal} {\bibinfo
  {journal} {Phys. Rev.}\ }\textbf {\bibinfo {volume} {D73}},\ \bibinfo {pages}
  {072003} (\bibinfo {year} {2006})},\ \Eprint
  {http://arxiv.org/abs/hep-ex/0602035} {arXiv:hep-ex/0602035
  [hep-ex]}\BibitemShut {NoStop}%
\bibitem [{\citenamefont {Davier}\ \emph {et~al.}(2017)\citenamefont {Davier},
  \citenamefont {Hoecker}, \citenamefont {Malaescu},\ and\ \citenamefont
  {Zhang}}]{Davier:2017zfy}%
  \BibitemOpen
  \bibfield  {author} {\bibinfo {author} {\bibfnamefont {M.}~\bibnamefont
  {Davier}}, \bibinfo {author} {\bibfnamefont {A.}~\bibnamefont {Hoecker}},
  \bibinfo {author} {\bibfnamefont {B.}~\bibnamefont {Malaescu}}, \ and\
  \bibinfo {author} {\bibfnamefont {Z.}~\bibnamefont {Zhang}},\ }\href
  {\doibase 10.1140/epjc/s10052-017-5161-6} {\bibfield  {journal} {\bibinfo
  {journal} {Eur. Phys. J.}\ }\textbf {\bibinfo {volume} {C77}},\ \bibinfo
  {pages} {827} (\bibinfo {year} {2017})},\ \Eprint
  {http://arxiv.org/abs/1706.09436} {arXiv:1706.09436 [hep-ph]}\BibitemShut
  {NoStop}%
\bibitem [{\citenamefont {Keshavarzi}\ \emph {et~al.}(2018)\citenamefont
  {Keshavarzi}, \citenamefont {Nomura},\ and\ \citenamefont
  {Teubner}}]{Keshavarzi:2018mgv}%
  \BibitemOpen
  \bibfield  {author} {\bibinfo {author} {\bibfnamefont {A.}~\bibnamefont
  {Keshavarzi}}, \bibinfo {author} {\bibfnamefont {D.}~\bibnamefont {Nomura}},
  \ and\ \bibinfo {author} {\bibfnamefont {T.}~\bibnamefont {Teubner}},\ }\href
  {\doibase 10.1103/PhysRevD.97.114025} {\bibfield  {journal} {\bibinfo
  {journal} {Phys. Rev.}\ }\textbf {\bibinfo {volume} {D97}},\ \bibinfo {pages}
  {114025} (\bibinfo {year} {2018})},\ \Eprint
  {http://arxiv.org/abs/1802.02995} {arXiv:1802.02995 [hep-ph]}\BibitemShut
  {NoStop}%
\bibitem [{\citenamefont {Colangelo}\ \emph {et~al.}(2019)\citenamefont
  {Colangelo}, \citenamefont {Hoferichter},\ and\ \citenamefont
  {Stoffer}}]{Colangelo:2018mtw}%
  \BibitemOpen
  \bibfield  {author} {\bibinfo {author} {\bibfnamefont {G.}~\bibnamefont
  {Colangelo}}, \bibinfo {author} {\bibfnamefont {M.}~\bibnamefont
  {Hoferichter}}, \ and\ \bibinfo {author} {\bibfnamefont {P.}~\bibnamefont
  {Stoffer}},\ }\href {\doibase 10.1007/JHEP02(2019)006} {\bibfield  {journal}
  {\bibinfo  {journal} {JHEP}\ }\textbf {\bibinfo {volume} {02}},\ \bibinfo
  {pages} {006} (\bibinfo {year} {2019})},\ \Eprint
  {http://arxiv.org/abs/1810.00007} {arXiv:1810.00007 [hep-ph]}\BibitemShut
  {NoStop}%
\bibitem [{\citenamefont {Hoferichter}\ \emph {et~al.}(2019)\citenamefont
  {Hoferichter}, \citenamefont {Hoid},\ and\ \citenamefont
  {Kubis}}]{Hoferichter:2019gzf}%
  \BibitemOpen
  \bibfield  {author} {\bibinfo {author} {\bibfnamefont {M.}~\bibnamefont
  {Hoferichter}}, \bibinfo {author} {\bibfnamefont {B.-L.}\ \bibnamefont
  {Hoid}}, \ and\ \bibinfo {author} {\bibfnamefont {B.}~\bibnamefont {Kubis}},\
  }\href {\doibase 10.1007/JHEP08(2019)137} {\bibfield  {journal} {\bibinfo
  {journal} {JHEP}\ }\textbf {\bibinfo {volume} {08}},\ \bibinfo {pages} {137}
  (\bibinfo {year} {2019})},\ \Eprint {http://arxiv.org/abs/1907.01556}
  {arXiv:1907.01556 [hep-ph]}\BibitemShut {NoStop}%
\bibitem [{\citenamefont {Davier}\ \emph {et~al.}(2020)\citenamefont {Davier},
  \citenamefont {Hoecker}, \citenamefont {Malaescu},\ and\ \citenamefont
  {Zhang}}]{Davier:2019can}%
  \BibitemOpen
  \bibfield  {author} {\bibinfo {author} {\bibfnamefont {M.}~\bibnamefont
  {Davier}}, \bibinfo {author} {\bibfnamefont {A.}~\bibnamefont {Hoecker}},
  \bibinfo {author} {\bibfnamefont {B.}~\bibnamefont {Malaescu}}, \ and\
  \bibinfo {author} {\bibfnamefont {Z.}~\bibnamefont {Zhang}},\ }\href
  {\doibase 10.1140/epjc/s10052-020-7792-2} {\bibfield  {journal} {\bibinfo
  {journal} {Eur. Phys. J.}\ }\textbf {\bibinfo {volume} {C80}},\ \bibinfo
  {pages} {241} (\bibinfo {year} {2020})},\ \bibinfo {note} {[Erratum: Eur.
  Phys. J. {\bf C80}, 410 (2020)]},\ \Eprint {http://arxiv.org/abs/1908.00921}
  {arXiv:1908.00921 [hep-ph]}\BibitemShut {NoStop}%
\bibitem [{\citenamefont {Keshavarzi}\ \emph {et~al.}(2020)\citenamefont
  {Keshavarzi}, \citenamefont {Nomura},\ and\ \citenamefont
  {Teubner}}]{Keshavarzi:2019abf}%
  \BibitemOpen
  \bibfield  {author} {\bibinfo {author} {\bibfnamefont {A.}~\bibnamefont
  {Keshavarzi}}, \bibinfo {author} {\bibfnamefont {D.}~\bibnamefont {Nomura}},
  \ and\ \bibinfo {author} {\bibfnamefont {T.}~\bibnamefont {Teubner}},\ }\href
  {\doibase 10.1103/PhysRevD.101.014029} {\bibfield  {journal} {\bibinfo
  {journal} {Phys. Rev.}\ }\textbf {\bibinfo {volume} {D101}},\ \bibinfo
  {pages} {014029} (\bibinfo {year} {2020})},\ \Eprint
  {http://arxiv.org/abs/1911.00367} {arXiv:1911.00367 [hep-ph]}\BibitemShut
  {NoStop}%
\bibitem [{\citenamefont {Kurz}\ \emph
  {et~al.}(2014{\natexlab{a}})\citenamefont {Kurz}, \citenamefont {Liu},
  \citenamefont {Marquard},\ and\ \citenamefont {Steinhauser}}]{Kurz:2014wya}%
  \BibitemOpen
  \bibfield  {author} {\bibinfo {author} {\bibfnamefont {A.}~\bibnamefont
  {Kurz}}, \bibinfo {author} {\bibfnamefont {T.}~\bibnamefont {Liu}}, \bibinfo
  {author} {\bibfnamefont {P.}~\bibnamefont {Marquard}}, \ and\ \bibinfo
  {author} {\bibfnamefont {M.}~\bibnamefont {Steinhauser}},\ }\href {\doibase
  10.1016/j.physletb.2014.05.043} {\bibfield  {journal} {\bibinfo  {journal}
  {Phys. Lett.}\ }\textbf {\bibinfo {volume} {B734}},\ \bibinfo {pages} {144}
  (\bibinfo {year} {2014}{\natexlab{a}})},\ \Eprint
  {http://arxiv.org/abs/1403.6400} {arXiv:1403.6400 [hep-ph]}\BibitemShut
  {NoStop}%
\bibitem [{\citenamefont {Chakraborty}\ \emph {et~al.}(2018)\citenamefont
  {Chakraborty} \emph {et~al.}}]{Chakraborty:2017tqp}%
  \BibitemOpen
  \bibfield  {author} {\bibinfo {author} {\bibfnamefont {B.}~\bibnamefont
  {Chakraborty}} \emph {et~al.} (\bibinfo {collaboration} {Fermilab Lattice,
  LATTICE-HPQCD, MILC}),\ }\href {\doibase 10.1103/PhysRevLett.120.152001}
  {\bibfield  {journal} {\bibinfo  {journal} {Phys. Rev. Lett.}\ }\textbf
  {\bibinfo {volume} {120}},\ \bibinfo {pages} {152001} (\bibinfo {year}
  {2018})},\ \Eprint {http://arxiv.org/abs/1710.11212} {arXiv:1710.11212
  [hep-lat]}\BibitemShut {NoStop}%
\bibitem [{\citenamefont {Borsanyi}\ \emph {et~al.}(2018)\citenamefont
  {Borsanyi} \emph {et~al.}}]{Borsanyi:2017zdw}%
  \BibitemOpen
  \bibfield  {author} {\bibinfo {author} {\bibfnamefont {S.}~\bibnamefont
  {Borsanyi}} \emph {et~al.} (\bibinfo {collaboration}
  {Budapest-Marseille-Wuppertal}),\ }\href {\doibase
  10.1103/PhysRevLett.121.022002} {\bibfield  {journal} {\bibinfo  {journal}
  {Phys. Rev. Lett.}\ }\textbf {\bibinfo {volume} {121}},\ \bibinfo {pages}
  {022002} (\bibinfo {year} {2018})},\ \Eprint
  {http://arxiv.org/abs/1711.04980} {arXiv:1711.04980 [hep-lat]}\BibitemShut
  {NoStop}%
\bibitem [{\citenamefont {Blum}\ \emph {et~al.}(2018)\citenamefont {Blum},
  \citenamefont {Boyle}, \citenamefont {G{\"u}lpers}, \citenamefont {Izubuchi},
  \citenamefont {Jin}, \citenamefont {Jung}, \citenamefont {J{\"u}ttner},
  \citenamefont {Lehner}, \citenamefont {Portelli},\ and\ \citenamefont
  {Tsang}}]{Blum:2018mom}%
  \BibitemOpen
  \bibfield  {author} {\bibinfo {author} {\bibfnamefont {T.}~\bibnamefont
  {Blum}}, \bibinfo {author} {\bibfnamefont {P.~A.}\ \bibnamefont {Boyle}},
  \bibinfo {author} {\bibfnamefont {V.}~\bibnamefont {G{\"u}lpers}}, \bibinfo
  {author} {\bibfnamefont {T.}~\bibnamefont {Izubuchi}}, \bibinfo {author}
  {\bibfnamefont {L.}~\bibnamefont {Jin}}, \bibinfo {author} {\bibfnamefont
  {C.}~\bibnamefont {Jung}}, \bibinfo {author} {\bibfnamefont {A.}~\bibnamefont
  {J{\"u}ttner}}, \bibinfo {author} {\bibfnamefont {C.}~\bibnamefont {Lehner}},
  \bibinfo {author} {\bibfnamefont {A.}~\bibnamefont {Portelli}}, \ and\
  \bibinfo {author} {\bibfnamefont {J.~T.}\ \bibnamefont {Tsang}} (\bibinfo
  {collaboration} {RBC, UKQCD}),\ }\href {\doibase
  10.1103/PhysRevLett.121.022003} {\bibfield  {journal} {\bibinfo  {journal}
  {Phys. Rev. Lett.}\ }\textbf {\bibinfo {volume} {121}},\ \bibinfo {pages}
  {022003} (\bibinfo {year} {2018})},\ \Eprint
  {http://arxiv.org/abs/1801.07224} {arXiv:1801.07224 [hep-lat]}\BibitemShut
  {NoStop}%
\bibitem [{\citenamefont {Giusti}\ \emph
  {et~al.}(2019{\natexlab{a}})\citenamefont {Giusti}, \citenamefont {Lubicz},
  \citenamefont {Martinelli}, \citenamefont {Sanfilippo},\ and\ \citenamefont
  {Simula}}]{Giusti:2019xct}%
  \BibitemOpen
  \bibfield  {author} {\bibinfo {author} {\bibfnamefont {D.}~\bibnamefont
  {Giusti}}, \bibinfo {author} {\bibfnamefont {V.}~\bibnamefont {Lubicz}},
  \bibinfo {author} {\bibfnamefont {G.}~\bibnamefont {Martinelli}}, \bibinfo
  {author} {\bibfnamefont {F.}~\bibnamefont {Sanfilippo}}, \ and\ \bibinfo
  {author} {\bibfnamefont {S.}~\bibnamefont {Simula}} (\bibinfo {collaboration}
  {ETM}),\ }\href {\doibase 10.1103/PhysRevD.99.114502} {\bibfield  {journal}
  {\bibinfo  {journal} {Phys. Rev.}\ }\textbf {\bibinfo {volume} {D99}},\
  \bibinfo {pages} {114502} (\bibinfo {year} {2019}{\natexlab{a}})},\ \Eprint
  {http://arxiv.org/abs/1901.10462} {arXiv:1901.10462 [hep-lat]}\BibitemShut
  {NoStop}%
\bibitem [{\citenamefont {Shintani}\ and\ \citenamefont
  {Kuramashi}(2019)}]{Shintani:2019wai}%
  \BibitemOpen
  \bibfield  {author} {\bibinfo {author} {\bibfnamefont {E.}~\bibnamefont
  {Shintani}}\ and\ \bibinfo {author} {\bibfnamefont {Y.}~\bibnamefont
  {Kuramashi}},\ }\href {\doibase 10.1103/PhysRevD.100.034517} {\bibfield
  {journal} {\bibinfo  {journal} {Phys. Rev.}\ }\textbf {\bibinfo {volume}
  {D100}},\ \bibinfo {pages} {034517} (\bibinfo {year} {2019})},\ \Eprint
  {http://arxiv.org/abs/1902.00885} {arXiv:1902.00885 [hep-lat]}\BibitemShut
  {NoStop}%
\bibitem [{\citenamefont {Davies}\ \emph {et~al.}(2020)\citenamefont {Davies}
  \emph {et~al.}}]{Davies:2019efs}%
  \BibitemOpen
  \bibfield  {author} {\bibinfo {author} {\bibfnamefont {C.~T.~H.}\
  \bibnamefont {Davies}} \emph {et~al.} (\bibinfo {collaboration} {Fermilab
  Lattice, LATTICE-HPQCD, MILC}),\ }\href {\doibase
  10.1103/PhysRevD.101.034512} {\bibfield  {journal} {\bibinfo  {journal}
  {Phys. Rev.}\ }\textbf {\bibinfo {volume} {D101}},\ \bibinfo {pages} {034512}
  (\bibinfo {year} {2020})},\ \Eprint {http://arxiv.org/abs/1902.04223}
  {arXiv:1902.04223 [hep-lat]}\BibitemShut {NoStop}%
\bibitem [{\citenamefont {G\'erardin}\ \emph {et~al.}(2019)\citenamefont
  {G\'erardin}, \citenamefont {C\`e}, \citenamefont {von Hippel}, \citenamefont
  {H{\"o}rz}, \citenamefont {Meyer}, \citenamefont {Mohler}, \citenamefont
  {Ottnad}, \citenamefont {Wilhelm},\ and\ \citenamefont
  {Wittig}}]{Gerardin:2019rua}%
  \BibitemOpen
  \bibfield  {author} {\bibinfo {author} {\bibfnamefont {A.}~\bibnamefont
  {G\'erardin}}, \bibinfo {author} {\bibfnamefont {M.}~\bibnamefont {C\`e}},
  \bibinfo {author} {\bibfnamefont {G.}~\bibnamefont {von Hippel}}, \bibinfo
  {author} {\bibfnamefont {B.}~\bibnamefont {H{\"o}rz}}, \bibinfo {author}
  {\bibfnamefont {H.~B.}\ \bibnamefont {Meyer}}, \bibinfo {author}
  {\bibfnamefont {D.}~\bibnamefont {Mohler}}, \bibinfo {author} {\bibfnamefont
  {K.}~\bibnamefont {Ottnad}}, \bibinfo {author} {\bibfnamefont
  {J.}~\bibnamefont {Wilhelm}}, \ and\ \bibinfo {author} {\bibfnamefont
  {H.}~\bibnamefont {Wittig}},\ }\href {\doibase 10.1103/PhysRevD.100.014510}
  {\bibfield  {journal} {\bibinfo  {journal} {Phys. Rev.}\ }\textbf {\bibinfo
  {volume} {D100}},\ \bibinfo {pages} {014510} (\bibinfo {year} {2019})},\
  \Eprint {http://arxiv.org/abs/1904.03120} {arXiv:1904.03120
  [hep-lat]}\BibitemShut {NoStop}%
\bibitem [{\citenamefont {Aubin}\ \emph {et~al.}(2020)\citenamefont {Aubin},
  \citenamefont {Blum}, \citenamefont {Tu}, \citenamefont {Golterman},
  \citenamefont {Jung},\ and\ \citenamefont {Peris}}]{Aubin:2019usy}%
  \BibitemOpen
  \bibfield  {author} {\bibinfo {author} {\bibfnamefont {C.}~\bibnamefont
  {Aubin}}, \bibinfo {author} {\bibfnamefont {T.}~\bibnamefont {Blum}},
  \bibinfo {author} {\bibfnamefont {C.}~\bibnamefont {Tu}}, \bibinfo {author}
  {\bibfnamefont {M.}~\bibnamefont {Golterman}}, \bibinfo {author}
  {\bibfnamefont {C.}~\bibnamefont {Jung}}, \ and\ \bibinfo {author}
  {\bibfnamefont {S.}~\bibnamefont {Peris}},\ }\href {\doibase
  10.1103/PhysRevD.101.014503} {\bibfield  {journal} {\bibinfo  {journal}
  {Phys. Rev.}\ }\textbf {\bibinfo {volume} {D101}},\ \bibinfo {pages} {014503}
  (\bibinfo {year} {2020})},\ \Eprint {http://arxiv.org/abs/1905.09307}
  {arXiv:1905.09307 [hep-lat]}\BibitemShut {NoStop}%
\bibitem [{\citenamefont {Giusti}\ and\ \citenamefont
  {Simula}(2019{\natexlab{a}})}]{Giusti:2019hkz}%
  \BibitemOpen
  \bibfield  {author} {\bibinfo {author} {\bibfnamefont {D.}~\bibnamefont
  {Giusti}}\ and\ \bibinfo {author} {\bibfnamefont {S.}~\bibnamefont
  {Simula}},\ }\href {\doibase 10.22323/1.363.0104} {\bibfield  {journal}
  {\bibinfo  {journal} {PoS}\ }\textbf {\bibinfo {volume} {LATTICE2019}},\
  \bibinfo {pages} {104} (\bibinfo {year} {2019}{\natexlab{a}})},\ \Eprint
  {http://arxiv.org/abs/1910.03874} {arXiv:1910.03874 [hep-lat]}\BibitemShut
  {NoStop}%
\bibitem [{\citenamefont {Melnikov}\ and\ \citenamefont
  {Vainshtein}(2004)}]{Melnikov:2003xd}%
  \BibitemOpen
  \bibfield  {author} {\bibinfo {author} {\bibfnamefont {K.}~\bibnamefont
  {Melnikov}}\ and\ \bibinfo {author} {\bibfnamefont {A.}~\bibnamefont
  {Vainshtein}},\ }\href {\doibase 10.1103/PhysRevD.70.113006} {\bibfield
  {journal} {\bibinfo  {journal} {Phys. Rev.}\ }\textbf {\bibinfo {volume}
  {D70}},\ \bibinfo {pages} {113006} (\bibinfo {year} {2004})},\ \Eprint
  {http://arxiv.org/abs/hep-ph/0312226} {arXiv:hep-ph/0312226
  [hep-ph]}\BibitemShut {NoStop}%
\bibitem [{\citenamefont {Masjuan}\ and\ \citenamefont
  {S{\'a}nchez-Puertas}(2017)}]{Masjuan:2017tvw}%
  \BibitemOpen
  \bibfield  {author} {\bibinfo {author} {\bibfnamefont {P.}~\bibnamefont
  {Masjuan}}\ and\ \bibinfo {author} {\bibfnamefont {P.}~\bibnamefont
  {S{\'a}nchez-Puertas}},\ }\href {\doibase 10.1103/PhysRevD.95.054026}
  {\bibfield  {journal} {\bibinfo  {journal} {Phys. Rev.}\ }\textbf {\bibinfo
  {volume} {D95}},\ \bibinfo {pages} {054026} (\bibinfo {year} {2017})},\
  \Eprint {http://arxiv.org/abs/1701.05829} {arXiv:1701.05829
  [hep-ph]}\BibitemShut {NoStop}%
\bibitem [{\citenamefont {Colangelo}\ \emph
  {et~al.}(2017{\natexlab{a}})\citenamefont {Colangelo}, \citenamefont
  {Hoferichter}, \citenamefont {Procura},\ and\ \citenamefont
  {Stoffer}}]{Colangelo:2017fiz}%
  \BibitemOpen
  \bibfield  {author} {\bibinfo {author} {\bibfnamefont {G.}~\bibnamefont
  {Colangelo}}, \bibinfo {author} {\bibfnamefont {M.}~\bibnamefont
  {Hoferichter}}, \bibinfo {author} {\bibfnamefont {M.}~\bibnamefont
  {Procura}}, \ and\ \bibinfo {author} {\bibfnamefont {P.}~\bibnamefont
  {Stoffer}},\ }\href {\doibase 10.1007/JHEP04(2017)161} {\bibfield  {journal}
  {\bibinfo  {journal} {JHEP}\ }\textbf {\bibinfo {volume} {04}},\ \bibinfo
  {pages} {161} (\bibinfo {year} {2017}{\natexlab{a}})},\ \Eprint
  {http://arxiv.org/abs/1702.07347} {arXiv:1702.07347 [hep-ph]}\BibitemShut
  {NoStop}%
\bibitem [{\citenamefont {Hoferichter}\ \emph
  {et~al.}(2018{\natexlab{a}})\citenamefont {Hoferichter}, \citenamefont
  {Hoid}, \citenamefont {Kubis}, \citenamefont {Leupold},\ and\ \citenamefont
  {Schneider}}]{Hoferichter:2018kwz}%
  \BibitemOpen
  \bibfield  {author} {\bibinfo {author} {\bibfnamefont {M.}~\bibnamefont
  {Hoferichter}}, \bibinfo {author} {\bibfnamefont {B.-L.}\ \bibnamefont
  {Hoid}}, \bibinfo {author} {\bibfnamefont {B.}~\bibnamefont {Kubis}},
  \bibinfo {author} {\bibfnamefont {S.}~\bibnamefont {Leupold}}, \ and\
  \bibinfo {author} {\bibfnamefont {S.~P.}\ \bibnamefont {Schneider}},\ }\href
  {\doibase 10.1007/JHEP10(2018)141} {\bibfield  {journal} {\bibinfo  {journal}
  {JHEP}\ }\textbf {\bibinfo {volume} {10}},\ \bibinfo {pages} {141} (\bibinfo
  {year} {2018}{\natexlab{a}})},\ \Eprint {http://arxiv.org/abs/1808.04823}
  {arXiv:1808.04823 [hep-ph]}\BibitemShut {NoStop}%
\bibitem [{\citenamefont {G{\'e}rardin}\ \emph
  {et~al.}(2019{\natexlab{a}})\citenamefont {G{\'e}rardin}, \citenamefont
  {Meyer},\ and\ \citenamefont {Nyffeler}}]{Gerardin:2019vio}%
  \BibitemOpen
  \bibfield  {author} {\bibinfo {author} {\bibfnamefont {A.}~\bibnamefont
  {G{\'e}rardin}}, \bibinfo {author} {\bibfnamefont {H.~B.}\ \bibnamefont
  {Meyer}}, \ and\ \bibinfo {author} {\bibfnamefont {A.}~\bibnamefont
  {Nyffeler}},\ }\href {\doibase 10.1103/PhysRevD.100.034520} {\bibfield
  {journal} {\bibinfo  {journal} {Phys. Rev.}\ }\textbf {\bibinfo {volume}
  {D100}},\ \bibinfo {pages} {034520} (\bibinfo {year} {2019}{\natexlab{a}})},\
  \Eprint {http://arxiv.org/abs/1903.09471} {arXiv:1903.09471
  [hep-lat]}\BibitemShut {NoStop}%
\bibitem [{\citenamefont {Bijnens}\ \emph
  {et~al.}(2019{\natexlab{a}})\citenamefont {Bijnens}, \citenamefont
  {Hermansson-Truedsson},\ and\ \citenamefont
  {Rodr{\'i}guez-S{\'a}nchez}}]{Bijnens:2019ghy}%
  \BibitemOpen
  \bibfield  {author} {\bibinfo {author} {\bibfnamefont {J.}~\bibnamefont
  {Bijnens}}, \bibinfo {author} {\bibfnamefont {N.}~\bibnamefont
  {Hermansson-Truedsson}}, \ and\ \bibinfo {author} {\bibfnamefont
  {A.}~\bibnamefont {Rodr{\'i}guez-S{\'a}nchez}},\ }\href {\doibase
  10.1016/j.physletb.2019.134994} {\bibfield  {journal} {\bibinfo  {journal}
  {Phys. Lett.}\ }\textbf {\bibinfo {volume} {B798}},\ \bibinfo {pages}
  {134994} (\bibinfo {year} {2019}{\natexlab{a}})},\ \Eprint
  {http://arxiv.org/abs/1908.03331} {arXiv:1908.03331 [hep-ph]}\BibitemShut
  {NoStop}%
\bibitem [{\citenamefont {Colangelo}\ \emph
  {et~al.}(2020{\natexlab{a}})\citenamefont {Colangelo}, \citenamefont
  {Hagelstein}, \citenamefont {Hoferichter}, \citenamefont {Laub},\ and\
  \citenamefont {Stoffer}}]{Colangelo:2019uex}%
  \BibitemOpen
  \bibfield  {author} {\bibinfo {author} {\bibfnamefont {G.}~\bibnamefont
  {Colangelo}}, \bibinfo {author} {\bibfnamefont {F.}~\bibnamefont
  {Hagelstein}}, \bibinfo {author} {\bibfnamefont {M.}~\bibnamefont
  {Hoferichter}}, \bibinfo {author} {\bibfnamefont {L.}~\bibnamefont {Laub}}, \
  and\ \bibinfo {author} {\bibfnamefont {P.}~\bibnamefont {Stoffer}},\ }\href
  {\doibase 10.1007/JHEP03(2020)101} {\bibfield  {journal} {\bibinfo  {journal}
  {JHEP}\ }\textbf {\bibinfo {volume} {03}},\ \bibinfo {pages} {101} (\bibinfo
  {year} {2020}{\natexlab{a}})},\ \Eprint {http://arxiv.org/abs/1910.13432}
  {arXiv:1910.13432 [hep-ph]}\BibitemShut {NoStop}%
\bibitem [{\citenamefont {Pauk}\ and\ \citenamefont
  {Vanderhaeghen}(2014{\natexlab{a}})}]{Pauk:2014rta}%
  \BibitemOpen
  \bibfield  {author} {\bibinfo {author} {\bibfnamefont {V.}~\bibnamefont
  {Pauk}}\ and\ \bibinfo {author} {\bibfnamefont {M.}~\bibnamefont
  {Vanderhaeghen}},\ }\href {\doibase 10.1140/epjc/s10052-014-3008-y}
  {\bibfield  {journal} {\bibinfo  {journal} {Eur. Phys. J.}\ }\textbf
  {\bibinfo {volume} {C74}},\ \bibinfo {pages} {3008} (\bibinfo {year}
  {2014}{\natexlab{a}})},\ \Eprint {http://arxiv.org/abs/1401.0832}
  {arXiv:1401.0832 [hep-ph]}\BibitemShut {NoStop}%
\bibitem [{\citenamefont {Danilkin}\ and\ \citenamefont
  {Vanderhaeghen}(2017)}]{Danilkin:2016hnh}%
  \BibitemOpen
  \bibfield  {author} {\bibinfo {author} {\bibfnamefont {I.}~\bibnamefont
  {Danilkin}}\ and\ \bibinfo {author} {\bibfnamefont {M.}~\bibnamefont
  {Vanderhaeghen}},\ }\href {\doibase 10.1103/PhysRevD.95.014019} {\bibfield
  {journal} {\bibinfo  {journal} {Phys. Rev.}\ }\textbf {\bibinfo {volume}
  {D95}},\ \bibinfo {pages} {014019} (\bibinfo {year} {2017})},\ \Eprint
  {http://arxiv.org/abs/1611.04646} {arXiv:1611.04646 [hep-ph]}\BibitemShut
  {NoStop}%
\bibitem [{\citenamefont {Jegerlehner}(2017)}]{Jegerlehner:2017gek}%
  \BibitemOpen
  \bibfield  {author} {\bibinfo {author} {\bibfnamefont {F.}~\bibnamefont
  {Jegerlehner}},\ }\href {\doibase 10.1007/978-3-319-63577-4} {\bibfield
  {journal} {\bibinfo  {journal} {Springer Tracts Mod. Phys.}\ }\textbf
  {\bibinfo {volume} {274}},\ \bibinfo {pages} {1} (\bibinfo {year}
  {2017})}\BibitemShut {NoStop}%
\bibitem [{\citenamefont {Knecht}\ \emph {et~al.}(2018)\citenamefont {Knecht},
  \citenamefont {Narison}, \citenamefont {Rabemananjara},\ and\ \citenamefont
  {Rabetiarivony}}]{Knecht:2018sci}%
  \BibitemOpen
  \bibfield  {author} {\bibinfo {author} {\bibfnamefont {M.}~\bibnamefont
  {Knecht}}, \bibinfo {author} {\bibfnamefont {S.}~\bibnamefont {Narison}},
  \bibinfo {author} {\bibfnamefont {A.}~\bibnamefont {Rabemananjara}}, \ and\
  \bibinfo {author} {\bibfnamefont {D.}~\bibnamefont {Rabetiarivony}},\ }\href
  {\doibase 10.1016/j.physletb.2018.10.048} {\bibfield  {journal} {\bibinfo
  {journal} {Phys. Lett.}\ }\textbf {\bibinfo {volume} {B787}},\ \bibinfo
  {pages} {111} (\bibinfo {year} {2018})},\ \Eprint
  {http://arxiv.org/abs/1808.03848} {arXiv:1808.03848 [hep-ph]}\BibitemShut
  {NoStop}%
\bibitem [{\citenamefont {Eichmann}\ \emph {et~al.}(2020)\citenamefont
  {Eichmann}, \citenamefont {Fischer},\ and\ \citenamefont
  {Williams}}]{Eichmann:2019bqf}%
  \BibitemOpen
  \bibfield  {author} {\bibinfo {author} {\bibfnamefont {G.}~\bibnamefont
  {Eichmann}}, \bibinfo {author} {\bibfnamefont {C.~S.}\ \bibnamefont
  {Fischer}}, \ and\ \bibinfo {author} {\bibfnamefont {R.}~\bibnamefont
  {Williams}},\ }\href {\doibase 10.1103/PhysRevD.101.054015} {\bibfield
  {journal} {\bibinfo  {journal} {Phys. Rev.}\ }\textbf {\bibinfo {volume}
  {D101}},\ \bibinfo {pages} {054015} (\bibinfo {year} {2020})},\ \Eprint
  {http://arxiv.org/abs/1910.06795} {arXiv:1910.06795 [hep-ph]}\BibitemShut
  {NoStop}%
\bibitem [{\citenamefont {Roig}\ and\ \citenamefont
  {S{\'a}nchez-Puertas}(2020)}]{Roig:2019reh}%
  \BibitemOpen
  \bibfield  {author} {\bibinfo {author} {\bibfnamefont {P.}~\bibnamefont
  {Roig}}\ and\ \bibinfo {author} {\bibfnamefont {P.}~\bibnamefont
  {S{\'a}nchez-Puertas}},\ }\href {\doibase 10.1103/PhysRevD.101.074019}
  {\bibfield  {journal} {\bibinfo  {journal} {Phys. Rev.}\ }\textbf {\bibinfo
  {volume} {D101}},\ \bibinfo {pages} {074019} (\bibinfo {year} {2020})},\
  \Eprint {http://arxiv.org/abs/1910.02881} {arXiv:1910.02881
  [hep-ph]}\BibitemShut {NoStop}%
\bibitem [{\citenamefont {Colangelo}\ \emph
  {et~al.}(2014{\natexlab{a}})\citenamefont {Colangelo}, \citenamefont
  {Hoferichter}, \citenamefont {Nyffeler}, \citenamefont {Passera},\ and\
  \citenamefont {Stoffer}}]{Colangelo:2014qya}%
  \BibitemOpen
  \bibfield  {author} {\bibinfo {author} {\bibfnamefont {G.}~\bibnamefont
  {Colangelo}}, \bibinfo {author} {\bibfnamefont {M.}~\bibnamefont
  {Hoferichter}}, \bibinfo {author} {\bibfnamefont {A.}~\bibnamefont
  {Nyffeler}}, \bibinfo {author} {\bibfnamefont {M.}~\bibnamefont {Passera}}, \
  and\ \bibinfo {author} {\bibfnamefont {P.}~\bibnamefont {Stoffer}},\ }\href
  {\doibase 10.1016/j.physletb.2014.06.012} {\bibfield  {journal} {\bibinfo
  {journal} {Phys. Lett.}\ }\textbf {\bibinfo {volume} {B735}},\ \bibinfo
  {pages} {90} (\bibinfo {year} {2014}{\natexlab{a}})},\ \Eprint
  {http://arxiv.org/abs/1403.7512} {arXiv:1403.7512 [hep-ph]}\BibitemShut
  {NoStop}%
\bibitem [{\citenamefont {Blum}\ \emph {et~al.}(2020)\citenamefont {Blum},
  \citenamefont {Christ}, \citenamefont {Hayakawa}, \citenamefont {Izubuchi},
  \citenamefont {Jin}, \citenamefont {Jung},\ and\ \citenamefont
  {Lehner}}]{Blum:2019ugy}%
  \BibitemOpen
  \bibfield  {author} {\bibinfo {author} {\bibfnamefont {T.}~\bibnamefont
  {Blum}}, \bibinfo {author} {\bibfnamefont {N.}~\bibnamefont {Christ}},
  \bibinfo {author} {\bibfnamefont {M.}~\bibnamefont {Hayakawa}}, \bibinfo
  {author} {\bibfnamefont {T.}~\bibnamefont {Izubuchi}}, \bibinfo {author}
  {\bibfnamefont {L.}~\bibnamefont {Jin}}, \bibinfo {author} {\bibfnamefont
  {C.}~\bibnamefont {Jung}}, \ and\ \bibinfo {author} {\bibfnamefont
  {C.}~\bibnamefont {Lehner}},\ }\href {\doibase
  10.1103/PhysRevLett.124.132002} {\bibfield  {journal} {\bibinfo  {journal}
  {Phys. Rev. Lett.}\ }\textbf {\bibinfo {volume} {124}},\ \bibinfo {pages}
  {132002} (\bibinfo {year} {2020})},\ \Eprint
  {http://arxiv.org/abs/1911.08123} {arXiv:1911.08123 [hep-lat]}\BibitemShut
  {NoStop}%
\bibitem [{\citenamefont {Aoyama}\ \emph
  {et~al.}(2012{\natexlab{a}})\citenamefont {Aoyama}, \citenamefont {Hayakawa},
  \citenamefont {Kinoshita},\ and\ \citenamefont {Nio}}]{Aoyama:2012wk}%
  \BibitemOpen
  \bibfield  {author} {\bibinfo {author} {\bibfnamefont {T.}~\bibnamefont
  {Aoyama}}, \bibinfo {author} {\bibfnamefont {M.}~\bibnamefont {Hayakawa}},
  \bibinfo {author} {\bibfnamefont {T.}~\bibnamefont {Kinoshita}}, \ and\
  \bibinfo {author} {\bibfnamefont {M.}~\bibnamefont {Nio}},\ }\href {\doibase
  10.1103/PhysRevLett.109.111808} {\bibfield  {journal} {\bibinfo  {journal}
  {Phys. Rev. Lett.}\ }\textbf {\bibinfo {volume} {109}},\ \bibinfo {pages}
  {111808} (\bibinfo {year} {2012}{\natexlab{a}})},\ \Eprint
  {http://arxiv.org/abs/1205.5370} {arXiv:1205.5370 [hep-ph]}\BibitemShut
  {NoStop}%
\bibitem [{\citenamefont {Aoyama}\ \emph {et~al.}(2019)\citenamefont {Aoyama},
  \citenamefont {Kinoshita},\ and\ \citenamefont {Nio}}]{Aoyama:2019ryr}%
  \BibitemOpen
  \bibfield  {author} {\bibinfo {author} {\bibfnamefont {T.}~\bibnamefont
  {Aoyama}}, \bibinfo {author} {\bibfnamefont {T.}~\bibnamefont {Kinoshita}}, \
  and\ \bibinfo {author} {\bibfnamefont {M.}~\bibnamefont {Nio}},\ }\href
  {\doibase 10.3390/atoms7010028} {\bibfield  {journal} {\bibinfo  {journal}
  {Atoms}\ }\textbf {\bibinfo {volume} {7}},\ \bibinfo {pages} {28} (\bibinfo
  {year} {2019})}\BibitemShut {NoStop}%
\bibitem [{\citenamefont {Czarnecki}\ \emph {et~al.}(2003)\citenamefont
  {Czarnecki}, \citenamefont {Marciano},\ and\ \citenamefont
  {Vainshtein}}]{Czarnecki:2002nt}%
  \BibitemOpen
  \bibfield  {author} {\bibinfo {author} {\bibfnamefont {A.}~\bibnamefont
  {Czarnecki}}, \bibinfo {author} {\bibfnamefont {W.~J.}\ \bibnamefont
  {Marciano}}, \ and\ \bibinfo {author} {\bibfnamefont {A.}~\bibnamefont
  {Vainshtein}},\ }\href {\doibase 10.1103/PhysRevD.67.073006} {\bibfield
  {journal} {\bibinfo  {journal} {Phys. Rev.}\ }\textbf {\bibinfo {volume}
  {D67}},\ \bibinfo {pages} {073006} (\bibinfo {year} {2003})},\ \bibinfo
  {note} {[Erratum: Phys. Rev. {\bf D73}, 119901 (2006)]},\ \Eprint
  {http://arxiv.org/abs/hep-ph/0212229} {arXiv:hep-ph/0212229
  [hep-ph]}\BibitemShut {NoStop}%
\bibitem [{\citenamefont {Gnendiger}\ \emph {et~al.}(2013)\citenamefont
  {Gnendiger}, \citenamefont {St{\"o}ckinger},\ and\ \citenamefont
  {St{\"o}ckinger-Kim}}]{Gnendiger:2013pva}%
  \BibitemOpen
  \bibfield  {author} {\bibinfo {author} {\bibfnamefont {C.}~\bibnamefont
  {Gnendiger}}, \bibinfo {author} {\bibfnamefont {D.}~\bibnamefont
  {St{\"o}ckinger}}, \ and\ \bibinfo {author} {\bibfnamefont {H.}~\bibnamefont
  {St{\"o}ckinger-Kim}},\ }\href {\doibase 10.1103/PhysRevD.88.053005}
  {\bibfield  {journal} {\bibinfo  {journal} {Phys. Rev.}\ }\textbf {\bibinfo
  {volume} {D88}},\ \bibinfo {pages} {053005} (\bibinfo {year} {2013})},\
  \Eprint {http://arxiv.org/abs/1306.5546} {arXiv:1306.5546
  [hep-ph]}\BibitemShut {NoStop}%
\bibitem [{\citenamefont {Bai}\ \emph {et~al.}(2000)\citenamefont {Bai} \emph
  {et~al.}}]{Bai:1999pk}%
  \BibitemOpen
  \bibfield  {author} {\bibinfo {author} {\bibfnamefont {J.~Z.}\ \bibnamefont
  {Bai}} \emph {et~al.} (\bibinfo {collaboration} {BES}),\ }\href {\doibase
  10.1103/PhysRevLett.84.594} {\bibfield  {journal} {\bibinfo  {journal} {Phys.
  Rev. Lett.}\ }\textbf {\bibinfo {volume} {84}},\ \bibinfo {pages} {594}
  (\bibinfo {year} {2000})},\ \Eprint {http://arxiv.org/abs/hep-ex/9908046}
  {arXiv:hep-ex/9908046 [hep-ex]}\BibitemShut {NoStop}%
\bibitem [{\citenamefont {Akhmetshin}\ \emph
  {et~al.}(2000{\natexlab{a}})\citenamefont {Akhmetshin} \emph
  {et~al.}}]{Akhmetshin:2000ca}%
  \BibitemOpen
  \bibfield  {author} {\bibinfo {author} {\bibfnamefont {R.~R.}\ \bibnamefont
  {Akhmetshin}} \emph {et~al.} (\bibinfo {collaboration} {CMD-2}),\ }\href
  {\doibase 10.1016/S0370-2693(00)00123-4} {\bibfield  {journal} {\bibinfo
  {journal} {Phys. Lett.}\ }\textbf {\bibinfo {volume} {B476}},\ \bibinfo
  {pages} {33} (\bibinfo {year} {2000}{\natexlab{a}})},\ \Eprint
  {http://arxiv.org/abs/hep-ex/0002017} {arXiv:hep-ex/0002017
  [hep-ex]}\BibitemShut {NoStop}%
\bibitem [{\citenamefont {Akhmetshin}\ \emph
  {et~al.}(2000{\natexlab{b}})\citenamefont {Akhmetshin} \emph
  {et~al.}}]{Akhmetshin:2000wv}%
  \BibitemOpen
  \bibfield  {author} {\bibinfo {author} {\bibfnamefont {R.~R.}\ \bibnamefont
  {Akhmetshin}} \emph {et~al.} (\bibinfo {collaboration} {CMD-2}),\ }\href
  {\doibase 10.1016/S0370-2693(00)00937-0} {\bibfield  {journal} {\bibinfo
  {journal} {Phys. Lett.}\ }\textbf {\bibinfo {volume} {B489}},\ \bibinfo
  {pages} {125} (\bibinfo {year} {2000}{\natexlab{b}})},\ \Eprint
  {http://arxiv.org/abs/hep-ex/0009013} {arXiv:hep-ex/0009013
  [hep-ex]}\BibitemShut {NoStop}%
\bibitem [{\citenamefont {Achasov}\ \emph
  {et~al.}(2001{\natexlab{a}})\citenamefont {Achasov} \emph
  {et~al.}}]{Achasov:2000am}%
  \BibitemOpen
  \bibfield  {author} {\bibinfo {author} {\bibfnamefont {M.~N.}\ \bibnamefont
  {Achasov}} \emph {et~al.} (\bibinfo {collaboration} {SND}),\ }\href {\doibase
  10.1103/PhysRevD.63.072002} {\bibfield  {journal} {\bibinfo  {journal} {Phys.
  Rev.}\ }\textbf {\bibinfo {volume} {D63}},\ \bibinfo {pages} {072002}
  (\bibinfo {year} {2001}{\natexlab{a}})},\ \Eprint
  {http://arxiv.org/abs/hep-ex/0009036} {arXiv:hep-ex/0009036
  [hep-ex]}\BibitemShut {NoStop}%
\bibitem [{\citenamefont {Bai}\ \emph {et~al.}(2002)\citenamefont {Bai} \emph
  {et~al.}}]{Bai:2001ct}%
  \BibitemOpen
  \bibfield  {author} {\bibinfo {author} {\bibfnamefont {J.~Z.}\ \bibnamefont
  {Bai}} \emph {et~al.} (\bibinfo {collaboration} {BES}),\ }\href {\doibase
  10.1103/PhysRevLett.88.101802} {\bibfield  {journal} {\bibinfo  {journal}
  {Phys. Rev. Lett.}\ }\textbf {\bibinfo {volume} {88}},\ \bibinfo {pages}
  {101802} (\bibinfo {year} {2002})},\ \Eprint
  {http://arxiv.org/abs/hep-ex/0102003} {arXiv:hep-ex/0102003
  [hep-ex]}\BibitemShut {NoStop}%
\bibitem [{\citenamefont {Achasov}\ \emph
  {et~al.}(2002{\natexlab{a}})\citenamefont {Achasov} \emph
  {et~al.}}]{Achasov:2002ud}%
  \BibitemOpen
  \bibfield  {author} {\bibinfo {author} {\bibfnamefont {M.~N.}\ \bibnamefont
  {Achasov}} \emph {et~al.} (\bibinfo {collaboration} {SND}),\ }\href {\doibase
  10.1103/PhysRevD.66.032001} {\bibfield  {journal} {\bibinfo  {journal} {Phys.
  Rev.}\ }\textbf {\bibinfo {volume} {D66}},\ \bibinfo {pages} {032001}
  (\bibinfo {year} {2002}{\natexlab{a}})},\ \Eprint
  {http://arxiv.org/abs/hep-ex/0201040} {arXiv:hep-ex/0201040
  [hep-ex]}\BibitemShut {NoStop}%
\bibitem [{\citenamefont {Akhmetshin}\ \emph
  {et~al.}(2004{\natexlab{a}})\citenamefont {Akhmetshin} \emph
  {et~al.}}]{Akhmetshin:2003zn}%
  \BibitemOpen
  \bibfield  {author} {\bibinfo {author} {\bibfnamefont {R.~R.}\ \bibnamefont
  {Akhmetshin}} \emph {et~al.} (\bibinfo {collaboration} {CMD-2}),\ }\href
  {\doibase 10.1016/j.physletb.2003.10.108} {\bibfield  {journal} {\bibinfo
  {journal} {Phys. Lett.}\ }\textbf {\bibinfo {volume} {B578}},\ \bibinfo
  {pages} {285} (\bibinfo {year} {2004}{\natexlab{a}})},\ \Eprint
  {http://arxiv.org/abs/hep-ex/0308008} {arXiv:hep-ex/0308008
  [hep-ex]}\BibitemShut {NoStop}%
\bibitem [{\citenamefont {Aubert}\ \emph {et~al.}(2004)\citenamefont {Aubert}
  \emph {et~al.}}]{Aubert:2004kj}%
  \BibitemOpen
  \bibfield  {author} {\bibinfo {author} {\bibfnamefont {B.}~\bibnamefont
  {Aubert}} \emph {et~al.} (\bibinfo {collaboration} {BABAR}),\ }\href
  {\doibase 10.1103/PhysRevD.70.072004} {\bibfield  {journal} {\bibinfo
  {journal} {Phys. Rev.}\ }\textbf {\bibinfo {volume} {D70}},\ \bibinfo {pages}
  {072004} (\bibinfo {year} {2004})},\ \Eprint
  {http://arxiv.org/abs/hep-ex/0408078} {arXiv:hep-ex/0408078
  [hep-ex]}\BibitemShut {NoStop}%
\bibitem [{\citenamefont {Aubert}\ \emph {et~al.}(2005)\citenamefont {Aubert}
  \emph {et~al.}}]{Aubert:2005eg}%
  \BibitemOpen
  \bibfield  {author} {\bibinfo {author} {\bibfnamefont {B.}~\bibnamefont
  {Aubert}} \emph {et~al.} (\bibinfo {collaboration} {BABAR}),\ }\href
  {\doibase 10.1103/PhysRevD.71.052001} {\bibfield  {journal} {\bibinfo
  {journal} {Phys. Rev.}\ }\textbf {\bibinfo {volume} {D71}},\ \bibinfo {pages}
  {052001} (\bibinfo {year} {2005})},\ \Eprint
  {http://arxiv.org/abs/hep-ex/0502025} {arXiv:hep-ex/0502025
  [hep-ex]}\BibitemShut {NoStop}%
\bibitem [{\citenamefont {Aubert}\ \emph
  {et~al.}(2006{\natexlab{a}})\citenamefont {Aubert} \emph
  {et~al.}}]{Aubert:2005cb}%
  \BibitemOpen
  \bibfield  {author} {\bibinfo {author} {\bibfnamefont {B.}~\bibnamefont
  {Aubert}} \emph {et~al.} (\bibinfo {collaboration} {BABAR}),\ }\href
  {\doibase 10.1103/PhysRevD.73.012005} {\bibfield  {journal} {\bibinfo
  {journal} {Phys. Rev.}\ }\textbf {\bibinfo {volume} {D73}},\ \bibinfo {pages}
  {012005} (\bibinfo {year} {2006}{\natexlab{a}})},\ \Eprint
  {http://arxiv.org/abs/hep-ex/0512023} {arXiv:hep-ex/0512023
  [hep-ex]}\BibitemShut {NoStop}%
\bibitem [{\citenamefont {Aubert}\ \emph
  {et~al.}(2006{\natexlab{b}})\citenamefont {Aubert} \emph
  {et~al.}}]{Aubert:2006jq}%
  \BibitemOpen
  \bibfield  {author} {\bibinfo {author} {\bibfnamefont {B.}~\bibnamefont
  {Aubert}} \emph {et~al.} (\bibinfo {collaboration} {BABAR}),\ }\href
  {\doibase 10.1103/PhysRevD.73.052003} {\bibfield  {journal} {\bibinfo
  {journal} {Phys. Rev.}\ }\textbf {\bibinfo {volume} {D73}},\ \bibinfo {pages}
  {052003} (\bibinfo {year} {2006}{\natexlab{b}})},\ \Eprint
  {http://arxiv.org/abs/hep-ex/0602006} {arXiv:hep-ex/0602006
  [hep-ex]}\BibitemShut {NoStop}%
\bibitem [{\citenamefont {Aul'chenko}\ \emph {et~al.}(2005)\citenamefont
  {Aul'chenko} \emph {et~al.}}]{Aulchenko:2006na}%
  \BibitemOpen
  \bibfield  {author} {\bibinfo {author} {\bibfnamefont {V.~M.}\ \bibnamefont
  {Aul'chenko}} \emph {et~al.} (\bibinfo {collaboration} {CMD-2}),\ }\href
  {\doibase 10.1134/1.2175241} {\bibfield  {journal} {\bibinfo  {journal} {JETP
  Lett.}\ }\textbf {\bibinfo {volume} {82}},\ \bibinfo {pages} {743} (\bibinfo
  {year} {2005})},\ \bibinfo {note} {[Pisma Zh. Eksp. Teor. Fiz. {\bf 82}, 841
  (2005)]},\ \Eprint {http://arxiv.org/abs/hep-ex/0603021}
  {arXiv:hep-ex/0603021 [hep-ex]}\BibitemShut {NoStop}%
\bibitem [{\citenamefont {Achasov}\ \emph
  {et~al.}(2006{\natexlab{a}})\citenamefont {Achasov} \emph
  {et~al.}}]{Achasov:2006vp}%
  \BibitemOpen
  \bibfield  {author} {\bibinfo {author} {\bibfnamefont {M.~N.}\ \bibnamefont
  {Achasov}} \emph {et~al.} (\bibinfo {collaboration} {SND}),\ }\href {\doibase
  10.1134/S106377610609007X} {\bibfield  {journal} {\bibinfo  {journal} {J.
  Exp. Theor. Phys.}\ }\textbf {\bibinfo {volume} {103}},\ \bibinfo {pages}
  {380} (\bibinfo {year} {2006}{\natexlab{a}})},\ \bibinfo {note} {[Zh. Eksp.
  Teor. Fiz. {\bf 130}, 437 (2006)]},\ \Eprint
  {http://arxiv.org/abs/hep-ex/0605013} {arXiv:hep-ex/0605013
  [hep-ex]}\BibitemShut {NoStop}%
\bibitem [{\citenamefont {Aul'chenko}\ \emph {et~al.}(2006)\citenamefont
  {Aul'chenko} \emph {et~al.}}]{Akhmetshin:2006wh}%
  \BibitemOpen
  \bibfield  {author} {\bibinfo {author} {\bibfnamefont {V.~M.}\ \bibnamefont
  {Aul'chenko}} \emph {et~al.} (\bibinfo {collaboration} {CMD-2}),\ }\href
  {\doibase 10.1134/S0021364006200021} {\bibfield  {journal} {\bibinfo
  {journal} {JETP Lett.}\ }\textbf {\bibinfo {volume} {84}},\ \bibinfo {pages}
  {413} (\bibinfo {year} {2006})},\ \bibinfo {note} {[Pisma Zh. Eksp. Teor.
  Fiz. {\bf 84}, 491 (2006)]},\ \Eprint {http://arxiv.org/abs/hep-ex/0610016}
  {arXiv:hep-ex/0610016 [hep-ex]}\BibitemShut {NoStop}%
\bibitem [{\citenamefont {Akhmetshin}\ \emph {et~al.}(2007)\citenamefont
  {Akhmetshin} \emph {et~al.}}]{Akhmetshin:2006bx}%
  \BibitemOpen
  \bibfield  {author} {\bibinfo {author} {\bibfnamefont {R.~R.}\ \bibnamefont
  {Akhmetshin}} \emph {et~al.} (\bibinfo {collaboration} {CMD-2}),\ }\href
  {\doibase 10.1016/j.physletb.2007.01.073} {\bibfield  {journal} {\bibinfo
  {journal} {Phys. Lett.}\ }\textbf {\bibinfo {volume} {B648}},\ \bibinfo
  {pages} {28} (\bibinfo {year} {2007})},\ \Eprint
  {http://arxiv.org/abs/hep-ex/0610021} {arXiv:hep-ex/0610021
  [hep-ex]}\BibitemShut {NoStop}%
\bibitem [{\citenamefont {Akhmetshin}\ \emph {et~al.}(2006)\citenamefont
  {Akhmetshin} \emph {et~al.}}]{Akhmetshin:2006sc}%
  \BibitemOpen
  \bibfield  {author} {\bibinfo {author} {\bibfnamefont {R.~R.}\ \bibnamefont
  {Akhmetshin}} \emph {et~al.} (\bibinfo {collaboration} {CMD-2}),\ }\href
  {\doibase 10.1016/j.physletb.2006.09.041} {\bibfield  {journal} {\bibinfo
  {journal} {Phys. Lett.}\ }\textbf {\bibinfo {volume} {B642}},\ \bibinfo
  {pages} {203} (\bibinfo {year} {2006})}\BibitemShut {NoStop}%
\bibitem [{\citenamefont {Aubert}\ \emph
  {et~al.}(2007{\natexlab{a}})\citenamefont {Aubert} \emph
  {et~al.}}]{Aubert:2007ur}%
  \BibitemOpen
  \bibfield  {author} {\bibinfo {author} {\bibfnamefont {B.}~\bibnamefont
  {Aubert}} \emph {et~al.} (\bibinfo {collaboration} {BABAR}),\ }\href
  {\doibase 10.1103/PhysRevD.76.012008} {\bibfield  {journal} {\bibinfo
  {journal} {Phys. Rev.}\ }\textbf {\bibinfo {volume} {D76}},\ \bibinfo {pages}
  {012008} (\bibinfo {year} {2007}{\natexlab{a}})},\ \Eprint
  {http://arxiv.org/abs/0704.0630} {arXiv:0704.0630 [hep-ex]}\BibitemShut
  {NoStop}%
\bibitem [{\citenamefont {Aubert}\ \emph
  {et~al.}(2007{\natexlab{b}})\citenamefont {Aubert} \emph
  {et~al.}}]{Aubert:2007ef}%
  \BibitemOpen
  \bibfield  {author} {\bibinfo {author} {\bibfnamefont {B.}~\bibnamefont
  {Aubert}} \emph {et~al.} (\bibinfo {collaboration} {BABAR}),\ }\href
  {\doibase 10.1103/PhysRevD.76.092005} {\bibfield  {journal} {\bibinfo
  {journal} {Phys. Rev.}\ }\textbf {\bibinfo {volume} {D76}},\ \bibinfo {pages}
  {092005} (\bibinfo {year} {2007}{\natexlab{b}})},\ \bibinfo {note} {[Erratum:
  Phys. Rev. {\bf D77}, 119902 (2008)]},\ \Eprint
  {http://arxiv.org/abs/0708.2461} {arXiv:0708.2461 [hep-ex]}\BibitemShut
  {NoStop}%
\bibitem [{\citenamefont {Aubert}\ \emph
  {et~al.}(2007{\natexlab{c}})\citenamefont {Aubert} \emph
  {et~al.}}]{Aubert:2007uf}%
  \BibitemOpen
  \bibfield  {author} {\bibinfo {author} {\bibfnamefont {B.}~\bibnamefont
  {Aubert}} \emph {et~al.} (\bibinfo {collaboration} {BABAR}),\ }\href
  {\doibase 10.1103/PhysRevD.76.092006} {\bibfield  {journal} {\bibinfo
  {journal} {Phys. Rev.}\ }\textbf {\bibinfo {volume} {D76}},\ \bibinfo {pages}
  {092006} (\bibinfo {year} {2007}{\natexlab{c}})},\ \Eprint
  {http://arxiv.org/abs/0709.1988} {arXiv:0709.1988 [hep-ex]}\BibitemShut
  {NoStop}%
\bibitem [{\citenamefont {Aubert}\ \emph {et~al.}(2008)\citenamefont {Aubert}
  \emph {et~al.}}]{Aubert:2007ym}%
  \BibitemOpen
  \bibfield  {author} {\bibinfo {author} {\bibfnamefont {B.}~\bibnamefont
  {Aubert}} \emph {et~al.} (\bibinfo {collaboration} {BABAR}),\ }\href
  {\doibase 10.1103/PhysRevD.77.092002} {\bibfield  {journal} {\bibinfo
  {journal} {Phys. Rev.}\ }\textbf {\bibinfo {volume} {D77}},\ \bibinfo {pages}
  {092002} (\bibinfo {year} {2008})},\ \Eprint {http://arxiv.org/abs/0710.4451}
  {arXiv:0710.4451 [hep-ex]}\BibitemShut {NoStop}%
\bibitem [{\citenamefont {Akhmetshin}\ \emph {et~al.}(2008)\citenamefont
  {Akhmetshin} \emph {et~al.}}]{Akhmetshin:2008gz}%
  \BibitemOpen
  \bibfield  {author} {\bibinfo {author} {\bibfnamefont {R.~R.}\ \bibnamefont
  {Akhmetshin}} \emph {et~al.} (\bibinfo {collaboration} {CMD-2}),\ }\href
  {\doibase 10.1016/j.physletb.2008.09.053} {\bibfield  {journal} {\bibinfo
  {journal} {Phys. Lett.}\ }\textbf {\bibinfo {volume} {B669}},\ \bibinfo
  {pages} {217} (\bibinfo {year} {2008})},\ \Eprint
  {http://arxiv.org/abs/0804.0178} {arXiv:0804.0178 [hep-ex]}\BibitemShut
  {NoStop}%
\bibitem [{\citenamefont {Ambrosino}\ \emph {et~al.}(2009)\citenamefont
  {Ambrosino} \emph {et~al.}}]{Ambrosino:2008aa}%
  \BibitemOpen
  \bibfield  {author} {\bibinfo {author} {\bibfnamefont {F.}~\bibnamefont
  {Ambrosino}} \emph {et~al.} (\bibinfo {collaboration} {KLOE}),\ }\href
  {\doibase 10.1016/j.physletb.2008.10.060} {\bibfield  {journal} {\bibinfo
  {journal} {Phys. Lett.}\ }\textbf {\bibinfo {volume} {B670}},\ \bibinfo
  {pages} {285} (\bibinfo {year} {2009})},\ \Eprint
  {http://arxiv.org/abs/0809.3950} {arXiv:0809.3950 [hep-ex]}\BibitemShut
  {NoStop}%
\bibitem [{\citenamefont {Ablikim}\ \emph {et~al.}(2009)\citenamefont {Ablikim}
  \emph {et~al.}}]{Ablikim:2009ad}%
  \BibitemOpen
  \bibfield  {author} {\bibinfo {author} {\bibfnamefont {M.}~\bibnamefont
  {Ablikim}} \emph {et~al.} (\bibinfo {collaboration} {BES}),\ }\href {\doibase
  10.1016/j.physletb.2009.05.055} {\bibfield  {journal} {\bibinfo  {journal}
  {Phys. Lett.}\ }\textbf {\bibinfo {volume} {B677}},\ \bibinfo {pages} {239}
  (\bibinfo {year} {2009})},\ \Eprint {http://arxiv.org/abs/0903.0900}
  {arXiv:0903.0900 [hep-ex]}\BibitemShut {NoStop}%
\bibitem [{\citenamefont {Aubert}\ \emph
  {et~al.}(2009{\natexlab{a}})\citenamefont {Aubert} \emph
  {et~al.}}]{Aubert:2009ad}%
  \BibitemOpen
  \bibfield  {author} {\bibinfo {author} {\bibfnamefont {B.}~\bibnamefont
  {Aubert}} \emph {et~al.} (\bibinfo {collaboration} {BABAR}),\ }\href
  {\doibase 10.1103/PhysRevLett.103.231801} {\bibfield  {journal} {\bibinfo
  {journal} {Phys. Rev. Lett.}\ }\textbf {\bibinfo {volume} {103}},\ \bibinfo
  {pages} {231801} (\bibinfo {year} {2009}{\natexlab{a}})},\ \Eprint
  {http://arxiv.org/abs/0908.3589} {arXiv:0908.3589 [hep-ex]}\BibitemShut
  {NoStop}%
\bibitem [{\citenamefont {Ambrosino}\ \emph
  {et~al.}(2011{\natexlab{a}})\citenamefont {Ambrosino} \emph
  {et~al.}}]{Ambrosino:2010bv}%
  \BibitemOpen
  \bibfield  {author} {\bibinfo {author} {\bibfnamefont {F.}~\bibnamefont
  {Ambrosino}} \emph {et~al.} (\bibinfo {collaboration} {KLOE}),\ }\href
  {\doibase 10.1016/j.physletb.2011.04.055} {\bibfield  {journal} {\bibinfo
  {journal} {Phys. Lett.}\ }\textbf {\bibinfo {volume} {B700}},\ \bibinfo
  {pages} {102} (\bibinfo {year} {2011}{\natexlab{a}})},\ \Eprint
  {http://arxiv.org/abs/1006.5313} {arXiv:1006.5313 [hep-ex]}\BibitemShut
  {NoStop}%
\bibitem [{\citenamefont {Lees}\ \emph
  {et~al.}(2012{\natexlab{a}})\citenamefont {Lees} \emph
  {et~al.}}]{Lees:2011zi}%
  \BibitemOpen
  \bibfield  {author} {\bibinfo {author} {\bibfnamefont {J.~P.}\ \bibnamefont
  {Lees}} \emph {et~al.} (\bibinfo {collaboration} {BABAR}),\ }\href {\doibase
  10.1103/PhysRevD.86.012008} {\bibfield  {journal} {\bibinfo  {journal} {Phys.
  Rev.}\ }\textbf {\bibinfo {volume} {D86}},\ \bibinfo {pages} {012008}
  (\bibinfo {year} {2012}{\natexlab{a}})},\ \Eprint
  {http://arxiv.org/abs/1103.3001} {arXiv:1103.3001 [hep-ex]}\BibitemShut
  {NoStop}%
\bibitem [{\citenamefont {Lees}\ \emph
  {et~al.}(2012{\natexlab{b}})\citenamefont {Lees} \emph
  {et~al.}}]{Lees:2012cr}%
  \BibitemOpen
  \bibfield  {author} {\bibinfo {author} {\bibfnamefont {J.~P.}\ \bibnamefont
  {Lees}} \emph {et~al.} (\bibinfo {collaboration} {BABAR}),\ }\href {\doibase
  10.1103/PhysRevD.85.112009} {\bibfield  {journal} {\bibinfo  {journal} {Phys.
  Rev.}\ }\textbf {\bibinfo {volume} {D85}},\ \bibinfo {pages} {112009}
  (\bibinfo {year} {2012}{\natexlab{b}})},\ \Eprint
  {http://arxiv.org/abs/1201.5677} {arXiv:1201.5677 [hep-ex]}\BibitemShut
  {NoStop}%
\bibitem [{\citenamefont {Lees}\ \emph
  {et~al.}(2012{\natexlab{c}})\citenamefont {Lees} \emph
  {et~al.}}]{Lees:2012cj}%
  \BibitemOpen
  \bibfield  {author} {\bibinfo {author} {\bibfnamefont {J.~P.}\ \bibnamefont
  {Lees}} \emph {et~al.} (\bibinfo {collaboration} {BABAR}),\ }\href {\doibase
  10.1103/PhysRevD.86.032013} {\bibfield  {journal} {\bibinfo  {journal} {Phys.
  Rev.}\ }\textbf {\bibinfo {volume} {D86}},\ \bibinfo {pages} {032013}
  (\bibinfo {year} {2012}{\natexlab{c}})},\ \Eprint
  {http://arxiv.org/abs/1205.2228} {arXiv:1205.2228 [hep-ex]}\BibitemShut
  {NoStop}%
\bibitem [{\citenamefont {Babusci}\ \emph
  {et~al.}(2013{\natexlab{a}})\citenamefont {Babusci} \emph
  {et~al.}}]{Babusci:2012rp}%
  \BibitemOpen
  \bibfield  {author} {\bibinfo {author} {\bibfnamefont {D.}~\bibnamefont
  {Babusci}} \emph {et~al.} (\bibinfo {collaboration} {KLOE}),\ }\href
  {\doibase 10.1016/j.physletb.2013.02.029} {\bibfield  {journal} {\bibinfo
  {journal} {Phys. Lett.}\ }\textbf {\bibinfo {volume} {B720}},\ \bibinfo
  {pages} {336} (\bibinfo {year} {2013}{\natexlab{a}})},\ \Eprint
  {http://arxiv.org/abs/1212.4524} {arXiv:1212.4524 [hep-ex]}\BibitemShut
  {NoStop}%
\bibitem [{\citenamefont {Akhmetshin}\ \emph {et~al.}(2013)\citenamefont
  {Akhmetshin} \emph {et~al.}}]{Akhmetshin:2013xc}%
  \BibitemOpen
  \bibfield  {author} {\bibinfo {author} {\bibfnamefont {R.~R.}\ \bibnamefont
  {Akhmetshin}} \emph {et~al.} (\bibinfo {collaboration} {CMD-3}),\ }\href
  {\doibase 10.1016/j.physletb.2013.04.065} {\bibfield  {journal} {\bibinfo
  {journal} {Phys. Lett.}\ }\textbf {\bibinfo {volume} {B723}},\ \bibinfo
  {pages} {82} (\bibinfo {year} {2013})},\ \Eprint
  {http://arxiv.org/abs/1302.0053} {arXiv:1302.0053 [hep-ex]}\BibitemShut
  {NoStop}%
\bibitem [{\citenamefont {Lees}\ \emph
  {et~al.}(2013{\natexlab{a}})\citenamefont {Lees} \emph
  {et~al.}}]{Lees:2013ebn}%
  \BibitemOpen
  \bibfield  {author} {\bibinfo {author} {\bibfnamefont {J.~P.}\ \bibnamefont
  {Lees}} \emph {et~al.} (\bibinfo {collaboration} {BABAR}),\ }\href {\doibase
  10.1103/PhysRevD.87.092005} {\bibfield  {journal} {\bibinfo  {journal} {Phys.
  Rev.}\ }\textbf {\bibinfo {volume} {D87}},\ \bibinfo {pages} {092005}
  (\bibinfo {year} {2013}{\natexlab{a}})},\ \Eprint
  {http://arxiv.org/abs/1302.0055} {arXiv:1302.0055 [hep-ex]}\BibitemShut
  {NoStop}%
\bibitem [{\citenamefont {Lees}\ \emph
  {et~al.}(2013{\natexlab{b}})\citenamefont {Lees} \emph
  {et~al.}}]{Lees:2013uta}%
  \BibitemOpen
  \bibfield  {author} {\bibinfo {author} {\bibfnamefont {J.~P.}\ \bibnamefont
  {Lees}} \emph {et~al.} (\bibinfo {collaboration} {BABAR}),\ }\href {\doibase
  10.1103/PhysRevD.88.072009} {\bibfield  {journal} {\bibinfo  {journal} {Phys.
  Rev.}\ }\textbf {\bibinfo {volume} {D88}},\ \bibinfo {pages} {072009}
  (\bibinfo {year} {2013}{\natexlab{b}})},\ \Eprint
  {http://arxiv.org/abs/1308.1795} {arXiv:1308.1795 [hep-ex]}\BibitemShut
  {NoStop}%
\bibitem [{\citenamefont {Lees}\ \emph {et~al.}(2014)\citenamefont {Lees} \emph
  {et~al.}}]{Lees:2014xsh}%
  \BibitemOpen
  \bibfield  {author} {\bibinfo {author} {\bibfnamefont {J.~P.}\ \bibnamefont
  {Lees}} \emph {et~al.} (\bibinfo {collaboration} {BABAR}),\ }\href {\doibase
  10.1103/PhysRevD.89.092002} {\bibfield  {journal} {\bibinfo  {journal} {Phys.
  Rev.}\ }\textbf {\bibinfo {volume} {D89}},\ \bibinfo {pages} {092002}
  (\bibinfo {year} {2014})},\ \Eprint {http://arxiv.org/abs/1403.7593}
  {arXiv:1403.7593 [hep-ex]}\BibitemShut {NoStop}%
\bibitem [{\citenamefont {Achasov}\ \emph
  {et~al.}(2014{\natexlab{a}})\citenamefont {Achasov} \emph
  {et~al.}}]{Achasov:2014ncd}%
  \BibitemOpen
  \bibfield  {author} {\bibinfo {author} {\bibfnamefont {M.~N.}\ \bibnamefont
  {Achasov}} \emph {et~al.} (\bibinfo {collaboration} {SND}),\ }\href {\doibase
  10.1103/PhysRevD.90.112007} {\bibfield  {journal} {\bibinfo  {journal} {Phys.
  Rev.}\ }\textbf {\bibinfo {volume} {D90}},\ \bibinfo {pages} {112007}
  (\bibinfo {year} {2014}{\natexlab{a}})},\ \Eprint
  {http://arxiv.org/abs/1410.3188} {arXiv:1410.3188 [hep-ex]}\BibitemShut
  {NoStop}%
\bibitem [{\citenamefont {Aulchenko}\ \emph {et~al.}(2015)\citenamefont
  {Aulchenko} \emph {et~al.}}]{Aulchenko:2014vkn}%
  \BibitemOpen
  \bibfield  {author} {\bibinfo {author} {\bibfnamefont {V.~M.}\ \bibnamefont
  {Aulchenko}} \emph {et~al.} (\bibinfo {collaboration} {SND}),\ }\href
  {\doibase 10.1103/PhysRevD.91.052013} {\bibfield  {journal} {\bibinfo
  {journal} {Phys. Rev.}\ }\textbf {\bibinfo {volume} {D91}},\ \bibinfo {pages}
  {052013} (\bibinfo {year} {2015})},\ \Eprint {http://arxiv.org/abs/1412.1971}
  {arXiv:1412.1971 [hep-ex]}\BibitemShut {NoStop}%
\bibitem [{\citenamefont {Akhmetshin}\ \emph {et~al.}(2016)\citenamefont
  {Akhmetshin} \emph {et~al.}}]{Akhmetshin:2015ifg}%
  \BibitemOpen
  \bibfield  {author} {\bibinfo {author} {\bibfnamefont {R.~R.}\ \bibnamefont
  {Akhmetshin}} \emph {et~al.} (\bibinfo {collaboration} {CMD-3}),\ }\href
  {\doibase 10.1016/j.physletb.2016.04.048} {\bibfield  {journal} {\bibinfo
  {journal} {Phys. Lett.}\ }\textbf {\bibinfo {volume} {B759}},\ \bibinfo
  {pages} {634} (\bibinfo {year} {2016})},\ \Eprint
  {http://arxiv.org/abs/1507.08013} {arXiv:1507.08013 [hep-ex]}\BibitemShut
  {NoStop}%
\bibitem [{\citenamefont {Ablikim}\ \emph
  {et~al.}(2016{\natexlab{a}})\citenamefont {Ablikim} \emph
  {et~al.}}]{Ablikim:2015orh}%
  \BibitemOpen
  \bibfield  {author} {\bibinfo {author} {\bibfnamefont {M.}~\bibnamefont
  {Ablikim}} \emph {et~al.} (\bibinfo {collaboration} {BESIII}),\ }\href
  {\doibase 10.1016/j.physletb.2015.11.043} {\bibfield  {journal} {\bibinfo
  {journal} {Phys. Lett.}\ }\textbf {\bibinfo {volume} {B753}},\ \bibinfo
  {pages} {629} (\bibinfo {year} {2016}{\natexlab{a}})},\ \Eprint
  {http://arxiv.org/abs/1507.08188} {arXiv:1507.08188 [hep-ex]}\BibitemShut
  {NoStop}%
\bibitem [{\citenamefont {Shemyakin}\ \emph {et~al.}(2016)\citenamefont
  {Shemyakin} \emph {et~al.}}]{Shemyakin:2015cba}%
  \BibitemOpen
  \bibfield  {author} {\bibinfo {author} {\bibfnamefont {D.~N.}\ \bibnamefont
  {Shemyakin}} \emph {et~al.} (\bibinfo {collaboration} {CMD-3}),\ }\href
  {\doibase 10.1016/j.physletb.2016.02.072} {\bibfield  {journal} {\bibinfo
  {journal} {Phys. Lett.}\ }\textbf {\bibinfo {volume} {B756}},\ \bibinfo
  {pages} {153} (\bibinfo {year} {2016})},\ \Eprint
  {http://arxiv.org/abs/1510.00654} {arXiv:1510.00654 [hep-ex]}\BibitemShut
  {NoStop}%
\bibitem [{\citenamefont {Anashin}\ \emph {et~al.}(2016)\citenamefont {Anashin}
  \emph {et~al.}}]{Anashin:2015woa}%
  \BibitemOpen
  \bibfield  {author} {\bibinfo {author} {\bibfnamefont {V.~V.}\ \bibnamefont
  {Anashin}} \emph {et~al.} (\bibinfo {collaboration} {KEDR}),\ }\href
  {\doibase 10.1016/j.physletb.2015.12.059} {\bibfield  {journal} {\bibinfo
  {journal} {Phys. Lett.}\ }\textbf {\bibinfo {volume} {B753}},\ \bibinfo
  {pages} {533} (\bibinfo {year} {2016})},\ \Eprint
  {http://arxiv.org/abs/1510.02667} {arXiv:1510.02667 [hep-ex]}\BibitemShut
  {NoStop}%
\bibitem [{\citenamefont {Achasov}\ \emph
  {et~al.}(2016{\natexlab{a}})\citenamefont {Achasov} \emph
  {et~al.}}]{Achasov:2016bfr}%
  \BibitemOpen
  \bibfield  {author} {\bibinfo {author} {\bibfnamefont {M.~N.}\ \bibnamefont
  {Achasov}} \emph {et~al.} (\bibinfo {collaboration} {SND}),\ }\href {\doibase
  10.1103/PhysRevD.93.092001} {\bibfield  {journal} {\bibinfo  {journal} {Phys.
  Rev.}\ }\textbf {\bibinfo {volume} {D93}},\ \bibinfo {pages} {092001}
  (\bibinfo {year} {2016}{\natexlab{a}})},\ \Eprint
  {http://arxiv.org/abs/1601.08061} {arXiv:1601.08061 [hep-ex]}\BibitemShut
  {NoStop}%
\bibitem [{\citenamefont {Achasov}\ \emph
  {et~al.}(2016{\natexlab{b}})\citenamefont {Achasov} \emph
  {et~al.}}]{Achasov:2016lbc}%
  \BibitemOpen
  \bibfield  {author} {\bibinfo {author} {\bibfnamefont {M.~N.}\ \bibnamefont
  {Achasov}} \emph {et~al.} (\bibinfo {collaboration} {SND}),\ }\href {\doibase
  10.1103/PhysRevD.94.112006} {\bibfield  {journal} {\bibinfo  {journal} {Phys.
  Rev.}\ }\textbf {\bibinfo {volume} {D94}},\ \bibinfo {pages} {112006}
  (\bibinfo {year} {2016}{\natexlab{b}})},\ \Eprint
  {http://arxiv.org/abs/1608.08757} {arXiv:1608.08757 [hep-ex]}\BibitemShut
  {NoStop}%
\bibitem [{\citenamefont {Lees}\ \emph
  {et~al.}(2017{\natexlab{a}})\citenamefont {Lees} \emph
  {et~al.}}]{TheBaBar:2017aph}%
  \BibitemOpen
  \bibfield  {author} {\bibinfo {author} {\bibfnamefont {J.~P.}\ \bibnamefont
  {Lees}} \emph {et~al.} (\bibinfo {collaboration} {BABAR}),\ }\href {\doibase
  10.1103/PhysRevD.95.092005} {\bibfield  {journal} {\bibinfo  {journal} {Phys.
  Rev.}\ }\textbf {\bibinfo {volume} {D95}},\ \bibinfo {pages} {092005}
  (\bibinfo {year} {2017}{\natexlab{a}})},\ \Eprint
  {http://arxiv.org/abs/1704.05009} {arXiv:1704.05009 [hep-ex]}\BibitemShut
  {NoStop}%
\bibitem [{\citenamefont {Akhmetshin}\ \emph
  {et~al.}(2017{\natexlab{a}})\citenamefont {Akhmetshin} \emph
  {et~al.}}]{CMD-3:2017tgb}%
  \BibitemOpen
  \bibfield  {author} {\bibinfo {author} {\bibfnamefont {R.~R.}\ \bibnamefont
  {Akhmetshin}} \emph {et~al.} (\bibinfo {collaboration} {CMD-3}),\ }\href
  {\doibase 10.1016/j.physletb.2017.08.019} {\bibfield  {journal} {\bibinfo
  {journal} {Phys. Lett.}\ }\textbf {\bibinfo {volume} {B773}},\ \bibinfo
  {pages} {150} (\bibinfo {year} {2017}{\natexlab{a}})},\ \Eprint
  {http://arxiv.org/abs/1706.06267} {arXiv:1706.06267 [hep-ex]}\BibitemShut
  {NoStop}%
\bibitem [{\citenamefont {Lees}\ \emph
  {et~al.}(2017{\natexlab{b}})\citenamefont {Lees} \emph
  {et~al.}}]{TheBaBar:2017vzo}%
  \BibitemOpen
  \bibfield  {author} {\bibinfo {author} {\bibfnamefont {J.~P.}\ \bibnamefont
  {Lees}} \emph {et~al.} (\bibinfo {collaboration} {BABAR}),\ }\href {\doibase
  10.1103/PhysRevD.96.092009} {\bibfield  {journal} {\bibinfo  {journal} {Phys.
  Rev.}\ }\textbf {\bibinfo {volume} {D96}},\ \bibinfo {pages} {092009}
  (\bibinfo {year} {2017}{\natexlab{b}})},\ \Eprint
  {http://arxiv.org/abs/1709.01171} {arXiv:1709.01171 [hep-ex]}\BibitemShut
  {NoStop}%
\bibitem [{\citenamefont {Kozyrev}\ \emph {et~al.}(2018)\citenamefont {Kozyrev}
  \emph {et~al.}}]{Kozyrev:2017agm}%
  \BibitemOpen
  \bibfield  {author} {\bibinfo {author} {\bibfnamefont {E.~A.}\ \bibnamefont
  {Kozyrev}} \emph {et~al.} (\bibinfo {collaboration} {CMD-3}),\ }\href
  {\doibase 10.1016/j.physletb.2018.01.079} {\bibfield  {journal} {\bibinfo
  {journal} {Phys. Lett.}\ }\textbf {\bibinfo {volume} {B779}},\ \bibinfo
  {pages} {64} (\bibinfo {year} {2018})},\ \Eprint
  {http://arxiv.org/abs/1710.02989} {arXiv:1710.02989 [hep-ex]}\BibitemShut
  {NoStop}%
\bibitem [{\citenamefont {Anastasi}\ \emph {et~al.}(2018)\citenamefont
  {Anastasi} \emph {et~al.}}]{Anastasi:2017eio}%
  \BibitemOpen
  \bibfield  {author} {\bibinfo {author} {\bibfnamefont {A.}~\bibnamefont
  {Anastasi}} \emph {et~al.} (\bibinfo {collaboration} {KLOE-2}),\ }\href
  {\doibase 10.1007/JHEP03(2018)173} {\bibfield  {journal} {\bibinfo  {journal}
  {JHEP}\ }\textbf {\bibinfo {volume} {03}},\ \bibinfo {pages} {173} (\bibinfo
  {year} {2018})},\ \Eprint {http://arxiv.org/abs/1711.03085} {arXiv:1711.03085
  [hep-ex]}\BibitemShut {NoStop}%
\bibitem [{\citenamefont {Achasov}\ \emph
  {et~al.}(2018{\natexlab{a}})\citenamefont {Achasov} \emph
  {et~al.}}]{Achasov:2017vaq}%
  \BibitemOpen
  \bibfield  {author} {\bibinfo {author} {\bibfnamefont {M.~N.}\ \bibnamefont
  {Achasov}} \emph {et~al.} (\bibinfo {collaboration} {SND}),\ }\href {\doibase
  10.1103/PhysRevD.97.032011} {\bibfield  {journal} {\bibinfo  {journal} {Phys.
  Rev.}\ }\textbf {\bibinfo {volume} {D97}},\ \bibinfo {pages} {032011}
  (\bibinfo {year} {2018}{\natexlab{a}})},\ \Eprint
  {http://arxiv.org/abs/1711.07143} {arXiv:1711.07143 [hep-ex]}\BibitemShut
  {NoStop}%
\bibitem [{\citenamefont {Xiao}\ \emph {et~al.}(2018)\citenamefont {Xiao},
  \citenamefont {Dobbs}, \citenamefont {Tomaradze}, \citenamefont {Seth},\ and\
  \citenamefont {Bonvicini}}]{Xiao:2017dqv}%
  \BibitemOpen
  \bibfield  {author} {\bibinfo {author} {\bibfnamefont {T.}~\bibnamefont
  {Xiao}}, \bibinfo {author} {\bibfnamefont {S.}~\bibnamefont {Dobbs}},
  \bibinfo {author} {\bibfnamefont {A.}~\bibnamefont {Tomaradze}}, \bibinfo
  {author} {\bibfnamefont {K.~K.}\ \bibnamefont {Seth}}, \ and\ \bibinfo
  {author} {\bibfnamefont {G.}~\bibnamefont {Bonvicini}},\ }\href {\doibase
  10.1103/PhysRevD.97.032012} {\bibfield  {journal} {\bibinfo  {journal} {Phys.
  Rev.}\ }\textbf {\bibinfo {volume} {D97}},\ \bibinfo {pages} {032012}
  (\bibinfo {year} {2018})},\ \Eprint {http://arxiv.org/abs/1712.04530}
  {arXiv:1712.04530 [hep-ex]}\BibitemShut {NoStop}%
\bibitem [{\citenamefont {Lees}\ \emph
  {et~al.}(2018{\natexlab{a}})\citenamefont {Lees} \emph
  {et~al.}}]{TheBaBar:2018vvb}%
  \BibitemOpen
  \bibfield  {author} {\bibinfo {author} {\bibfnamefont {J.~P.}\ \bibnamefont
  {Lees}} \emph {et~al.} (\bibinfo {collaboration} {BABAR}),\ }\href {\doibase
  10.1103/PhysRevD.97.052007} {\bibfield  {journal} {\bibinfo  {journal} {Phys.
  Rev.}\ }\textbf {\bibinfo {volume} {D97}},\ \bibinfo {pages} {052007}
  (\bibinfo {year} {2018}{\natexlab{a}})},\ \Eprint
  {http://arxiv.org/abs/1801.02960} {arXiv:1801.02960 [hep-ex]}\BibitemShut
  {NoStop}%
\bibitem [{\citenamefont {Anashin}\ \emph {et~al.}(2019)\citenamefont {Anashin}
  \emph {et~al.}}]{Anashin:2018vdo}%
  \BibitemOpen
  \bibfield  {author} {\bibinfo {author} {\bibfnamefont {V.~V.}\ \bibnamefont
  {Anashin}} \emph {et~al.} (\bibinfo {collaboration} {KEDR}),\ }\href
  {\doibase 10.1016/j.physletb.2018.11.012} {\bibfield  {journal} {\bibinfo
  {journal} {Phys. Lett.}\ }\textbf {\bibinfo {volume} {B788}},\ \bibinfo
  {pages} {42} (\bibinfo {year} {2019})},\ \Eprint
  {http://arxiv.org/abs/1805.06235} {arXiv:1805.06235 [hep-ex]}\BibitemShut
  {NoStop}%
\bibitem [{\citenamefont {Achasov}\ \emph
  {et~al.}(2018{\natexlab{b}})\citenamefont {Achasov} \emph
  {et~al.}}]{Achasov:2018ujw}%
  \BibitemOpen
  \bibfield  {author} {\bibinfo {author} {\bibfnamefont {M.~N.}\ \bibnamefont
  {Achasov}} \emph {et~al.} (\bibinfo {collaboration} {SND}),\ }\href {\doibase
  10.1103/PhysRevD.98.112001} {\bibfield  {journal} {\bibinfo  {journal} {Phys.
  Rev.}\ }\textbf {\bibinfo {volume} {D98}},\ \bibinfo {pages} {112001}
  (\bibinfo {year} {2018}{\natexlab{b}})},\ \Eprint
  {http://arxiv.org/abs/1809.07631} {arXiv:1809.07631 [hep-ex]}\BibitemShut
  {NoStop}%
\bibitem [{\citenamefont {Lees}\ \emph
  {et~al.}(2018{\natexlab{b}})\citenamefont {Lees} \emph
  {et~al.}}]{Lees:2018dnv}%
  \BibitemOpen
  \bibfield  {author} {\bibinfo {author} {\bibfnamefont {J.~P.}\ \bibnamefont
  {Lees}} \emph {et~al.} (\bibinfo {collaboration} {BABAR}),\ }\href {\doibase
  10.1103/PhysRevD.98.112015} {\bibfield  {journal} {\bibinfo  {journal} {Phys.
  Rev.}\ }\textbf {\bibinfo {volume} {D98}},\ \bibinfo {pages} {112015}
  (\bibinfo {year} {2018}{\natexlab{b}})},\ \Eprint
  {http://arxiv.org/abs/1810.11962} {arXiv:1810.11962 [hep-ex]}\BibitemShut
  {NoStop}%
\bibitem [{\citenamefont {Akhmetshin}\ \emph {et~al.}(2019)\citenamefont
  {Akhmetshin} \emph {et~al.}}]{CMD-3:2019ufp}%
  \BibitemOpen
  \bibfield  {author} {\bibinfo {author} {\bibfnamefont {R.~R.}\ \bibnamefont
  {Akhmetshin}} \emph {et~al.} (\bibinfo {collaboration} {CMD-3}),\ }\href
  {\doibase 10.1016/j.physletb.2019.04.007} {\bibfield  {journal} {\bibinfo
  {journal} {Phys. Lett.}\ }\textbf {\bibinfo {volume} {B792}},\ \bibinfo
  {pages} {419} (\bibinfo {year} {2019})},\ \Eprint
  {http://arxiv.org/abs/1902.06449} {arXiv:1902.06449 [hep-ex]}\BibitemShut
  {NoStop}%
\bibitem [{\citenamefont {Behrend}\ \emph {et~al.}(1991)\citenamefont {Behrend}
  \emph {et~al.}}]{Behrend:1990sr}%
  \BibitemOpen
  \bibfield  {author} {\bibinfo {author} {\bibfnamefont {H.~J.}\ \bibnamefont
  {Behrend}} \emph {et~al.} (\bibinfo {collaboration} {CELLO}),\ }\href
  {\doibase 10.1007/BF01549692} {\bibfield  {journal} {\bibinfo  {journal} {Z.
  Phys.}\ }\textbf {\bibinfo {volume} {C49}},\ \bibinfo {pages} {401} (\bibinfo
  {year} {1991})}\BibitemShut {NoStop}%
\bibitem [{\citenamefont {Gronberg}\ \emph {et~al.}(1998)\citenamefont
  {Gronberg} \emph {et~al.}}]{Gronberg:1997fj}%
  \BibitemOpen
  \bibfield  {author} {\bibinfo {author} {\bibfnamefont {J.}~\bibnamefont
  {Gronberg}} \emph {et~al.} (\bibinfo {collaboration} {CLEO}),\ }\href
  {\doibase 10.1103/PhysRevD.57.33} {\bibfield  {journal} {\bibinfo  {journal}
  {Phys. Rev.}\ }\textbf {\bibinfo {volume} {D57}},\ \bibinfo {pages} {33}
  (\bibinfo {year} {1998})},\ \Eprint {http://arxiv.org/abs/hep-ex/9707031}
  {arXiv:hep-ex/9707031 [hep-ex]}\BibitemShut {NoStop}%
\bibitem [{\citenamefont {Acciarri}\ \emph {et~al.}(1998)\citenamefont
  {Acciarri} \emph {et~al.}}]{Acciarri:1997yx}%
  \BibitemOpen
  \bibfield  {author} {\bibinfo {author} {\bibfnamefont {M.}~\bibnamefont
  {Acciarri}} \emph {et~al.} (\bibinfo {collaboration} {L3}),\ }\href {\doibase
  10.1016/S0370-2693(97)01219-7} {\bibfield  {journal} {\bibinfo  {journal}
  {Phys. Lett.}\ }\textbf {\bibinfo {volume} {B418}},\ \bibinfo {pages} {399}
  (\bibinfo {year} {1998})}\BibitemShut {NoStop}%
\bibitem [{\citenamefont {Achard}\ \emph {et~al.}(2002)\citenamefont {Achard}
  \emph {et~al.}}]{Achard:2001uu}%
  \BibitemOpen
  \bibfield  {author} {\bibinfo {author} {\bibfnamefont {P.}~\bibnamefont
  {Achard}} \emph {et~al.} (\bibinfo {collaboration} {L3}),\ }\href {\doibase
  10.1016/S0370-2693(01)01477-0} {\bibfield  {journal} {\bibinfo  {journal}
  {Phys. Lett.}\ }\textbf {\bibinfo {volume} {B526}},\ \bibinfo {pages} {269}
  (\bibinfo {year} {2002})},\ \Eprint {http://arxiv.org/abs/hep-ex/0110073}
  {arXiv:hep-ex/0110073 [hep-ex]}\BibitemShut {NoStop}%
\bibitem [{\citenamefont {Achard}\ \emph {et~al.}(2007)\citenamefont {Achard}
  \emph {et~al.}}]{Achard:2007hm}%
  \BibitemOpen
  \bibfield  {author} {\bibinfo {author} {\bibfnamefont {P.}~\bibnamefont
  {Achard}} \emph {et~al.} (\bibinfo {collaboration} {L3}),\ }\href {\doibase
  10.1088/1126-6708/2007/03/018} {\bibfield  {journal} {\bibinfo  {journal}
  {JHEP}\ }\textbf {\bibinfo {volume} {03}},\ \bibinfo {pages} {018} (\bibinfo
  {year} {2007})}\BibitemShut {NoStop}%
\bibitem [{\citenamefont {Arnaldi}\ \emph {et~al.}(2009)\citenamefont {Arnaldi}
  \emph {et~al.}}]{Arnaldi:2009aa}%
  \BibitemOpen
  \bibfield  {author} {\bibinfo {author} {\bibfnamefont {R.}~\bibnamefont
  {Arnaldi}} \emph {et~al.} (\bibinfo {collaboration} {NA60}),\ }\href
  {\doibase 10.1016/j.physletb.2009.05.029} {\bibfield  {journal} {\bibinfo
  {journal} {Phys. Lett.}\ }\textbf {\bibinfo {volume} {B677}},\ \bibinfo
  {pages} {260} (\bibinfo {year} {2009})},\ \Eprint
  {http://arxiv.org/abs/0902.2547} {arXiv:0902.2547 [hep-ph]}\BibitemShut
  {NoStop}%
\bibitem [{\citenamefont {Aubert}\ \emph
  {et~al.}(2009{\natexlab{b}})\citenamefont {Aubert} \emph
  {et~al.}}]{Aubert:2009mc}%
  \BibitemOpen
  \bibfield  {author} {\bibinfo {author} {\bibfnamefont {B.}~\bibnamefont
  {Aubert}} \emph {et~al.} (\bibinfo {collaboration} {BABAR}),\ }\href
  {\doibase 10.1103/PhysRevD.80.052002} {\bibfield  {journal} {\bibinfo
  {journal} {Phys. Rev.}\ }\textbf {\bibinfo {volume} {D80}},\ \bibinfo {pages}
  {052002} (\bibinfo {year} {2009}{\natexlab{b}})},\ \Eprint
  {http://arxiv.org/abs/0905.4778} {arXiv:0905.4778 [hep-ex]}\BibitemShut
  {NoStop}%
\bibitem [{\citenamefont {del Amo~Sanchez}\ \emph {et~al.}(2011)\citenamefont
  {del Amo~Sanchez} \emph {et~al.}}]{BABAR:2011ad}%
  \BibitemOpen
  \bibfield  {author} {\bibinfo {author} {\bibfnamefont {P.}~\bibnamefont {del
  Amo~Sanchez}} \emph {et~al.} (\bibinfo {collaboration} {BABAR}),\ }\href
  {\doibase 10.1103/PhysRevD.84.052001} {\bibfield  {journal} {\bibinfo
  {journal} {Phys. Rev.}\ }\textbf {\bibinfo {volume} {D84}},\ \bibinfo {pages}
  {052001} (\bibinfo {year} {2011})},\ \Eprint {http://arxiv.org/abs/1101.1142}
  {arXiv:1101.1142 [hep-ex]}\BibitemShut {NoStop}%
\bibitem [{\citenamefont {Bergh\"auser}\ \emph {et~al.}(2011)\citenamefont
  {Bergh\"auser} \emph {et~al.}}]{Berghauser:2011zz}%
  \BibitemOpen
  \bibfield  {author} {\bibinfo {author} {\bibfnamefont {H.}~\bibnamefont
  {Bergh\"auser}} \emph {et~al.},\ }\href {\doibase
  10.1016/j.physletb.2011.06.069} {\bibfield  {journal} {\bibinfo  {journal}
  {Phys. Lett.}\ }\textbf {\bibinfo {volume} {B701}},\ \bibinfo {pages} {562}
  (\bibinfo {year} {2011})}\BibitemShut {NoStop}%
\bibitem [{\citenamefont {Uehara}\ \emph {et~al.}(2012)\citenamefont {Uehara}
  \emph {et~al.}}]{Uehara:2012ag}%
  \BibitemOpen
  \bibfield  {author} {\bibinfo {author} {\bibfnamefont {S.}~\bibnamefont
  {Uehara}} \emph {et~al.} (\bibinfo {collaboration} {Belle}),\ }\href
  {\doibase 10.1103/PhysRevD.86.092007} {\bibfield  {journal} {\bibinfo
  {journal} {Phys. Rev.}\ }\textbf {\bibinfo {volume} {D86}},\ \bibinfo {pages}
  {092007} (\bibinfo {year} {2012})},\ \Eprint {http://arxiv.org/abs/1205.3249}
  {arXiv:1205.3249 [hep-ex]}\BibitemShut {NoStop}%
\bibitem [{\citenamefont {Babusci}\ \emph
  {et~al.}(2013{\natexlab{b}})\citenamefont {Babusci} \emph
  {et~al.}}]{Babusci:2012ik}%
  \BibitemOpen
  \bibfield  {author} {\bibinfo {author} {\bibfnamefont {D.}~\bibnamefont
  {Babusci}} \emph {et~al.} (\bibinfo {collaboration} {KLOE-2}),\ }\href
  {\doibase 10.1007/JHEP01(2013)119} {\bibfield  {journal} {\bibinfo  {journal}
  {JHEP}\ }\textbf {\bibinfo {volume} {01}},\ \bibinfo {pages} {119} (\bibinfo
  {year} {2013}{\natexlab{b}})},\ \Eprint {http://arxiv.org/abs/1211.1845}
  {arXiv:1211.1845 [hep-ex]}\BibitemShut {NoStop}%
\bibitem [{\citenamefont {Aguar-Bartolome}\ \emph {et~al.}(2014)\citenamefont
  {Aguar-Bartolome} \emph {et~al.}}]{Aguar-Bartolome:2013vpw}%
  \BibitemOpen
  \bibfield  {author} {\bibinfo {author} {\bibfnamefont {P.}~\bibnamefont
  {Aguar-Bartolome}} \emph {et~al.} (\bibinfo {collaboration} {A2}),\ }\href
  {\doibase 10.1103/PhysRevC.89.044608} {\bibfield  {journal} {\bibinfo
  {journal} {Phys. Rev.}\ }\textbf {\bibinfo {volume} {C89}},\ \bibinfo {pages}
  {044608} (\bibinfo {year} {2014})},\ \Eprint {http://arxiv.org/abs/1309.5648}
  {arXiv:1309.5648 [hep-ex]}\BibitemShut {NoStop}%
\bibitem [{\citenamefont {Ablikim}\ \emph {et~al.}(2015)\citenamefont {Ablikim}
  \emph {et~al.}}]{Ablikim:2015wnx}%
  \BibitemOpen
  \bibfield  {author} {\bibinfo {author} {\bibfnamefont {M.}~\bibnamefont
  {Ablikim}} \emph {et~al.} (\bibinfo {collaboration} {BESIII}),\ }\href
  {\doibase 10.1103/PhysRevD.92.012001} {\bibfield  {journal} {\bibinfo
  {journal} {Phys. Rev.}\ }\textbf {\bibinfo {volume} {D92}},\ \bibinfo {pages}
  {012001} (\bibinfo {year} {2015})},\ \Eprint
  {http://arxiv.org/abs/1504.06016} {arXiv:1504.06016 [hep-ex]}\BibitemShut
  {NoStop}%
\bibitem [{\citenamefont {Masuda}\ \emph {et~al.}(2016)\citenamefont {Masuda}
  \emph {et~al.}}]{Masuda:2015yoh}%
  \BibitemOpen
  \bibfield  {author} {\bibinfo {author} {\bibfnamefont {M.}~\bibnamefont
  {Masuda}} \emph {et~al.} (\bibinfo {collaboration} {Belle}),\ }\href
  {\doibase 10.1103/PhysRevD.93.032003} {\bibfield  {journal} {\bibinfo
  {journal} {Phys. Rev.}\ }\textbf {\bibinfo {volume} {D93}},\ \bibinfo {pages}
  {032003} (\bibinfo {year} {2016})},\ \Eprint
  {http://arxiv.org/abs/1508.06757} {arXiv:1508.06757 [hep-ex]}\BibitemShut
  {NoStop}%
\bibitem [{\citenamefont {Arnaldi}\ \emph {et~al.}(2016)\citenamefont {Arnaldi}
  \emph {et~al.}}]{Arnaldi:2016pzu}%
  \BibitemOpen
  \bibfield  {author} {\bibinfo {author} {\bibfnamefont {R.}~\bibnamefont
  {Arnaldi}} \emph {et~al.} (\bibinfo {collaboration} {NA60}),\ }\href
  {\doibase 10.1016/j.physletb.2016.04.013} {\bibfield  {journal} {\bibinfo
  {journal} {Phys. Lett.}\ }\textbf {\bibinfo {volume} {B757}},\ \bibinfo
  {pages} {437} (\bibinfo {year} {2016})},\ \Eprint
  {http://arxiv.org/abs/1608.07898} {arXiv:1608.07898 [hep-ex]}\BibitemShut
  {NoStop}%
\bibitem [{\citenamefont {Adlarson}\ \emph
  {et~al.}(2017{\natexlab{a}})\citenamefont {Adlarson} \emph
  {et~al.}}]{Adlarson:2016hpp}%
  \BibitemOpen
  \bibfield  {author} {\bibinfo {author} {\bibfnamefont {P.}~\bibnamefont
  {Adlarson}} \emph {et~al.} (\bibinfo {collaboration} {A2}),\ }\href {\doibase
  10.1103/PhysRevC.95.035208} {\bibfield  {journal} {\bibinfo  {journal} {Phys.
  Rev.}\ }\textbf {\bibinfo {volume} {C95}},\ \bibinfo {pages} {035208}
  (\bibinfo {year} {2017}{\natexlab{a}})},\ \Eprint
  {http://arxiv.org/abs/1609.04503} {arXiv:1609.04503 [hep-ex]}\BibitemShut
  {NoStop}%
\bibitem [{\citenamefont {Adlarson}\ \emph
  {et~al.}(2017{\natexlab{b}})\citenamefont {Adlarson} \emph
  {et~al.}}]{Adlarson:2016ykr}%
  \BibitemOpen
  \bibfield  {author} {\bibinfo {author} {\bibfnamefont {P.}~\bibnamefont
  {Adlarson}} \emph {et~al.} (\bibinfo {collaboration} {A2}),\ }\href {\doibase
  10.1103/PhysRevC.95.025202} {\bibfield  {journal} {\bibinfo  {journal} {Phys.
  Rev.}\ }\textbf {\bibinfo {volume} {C95}},\ \bibinfo {pages} {025202}
  (\bibinfo {year} {2017}{\natexlab{b}})},\ \Eprint
  {http://arxiv.org/abs/1611.04739} {arXiv:1611.04739 [hep-ex]}\BibitemShut
  {NoStop}%
\bibitem [{\citenamefont {Lazzeroni}\ \emph {et~al.}(2017)\citenamefont
  {Lazzeroni} \emph {et~al.}}]{TheNA62:2016fhr}%
  \BibitemOpen
  \bibfield  {author} {\bibinfo {author} {\bibfnamefont {C.}~\bibnamefont
  {Lazzeroni}} \emph {et~al.} (\bibinfo {collaboration} {NA62}),\ }\href
  {\doibase 10.1016/j.physletb.2017.02.042} {\bibfield  {journal} {\bibinfo
  {journal} {Phys. Lett.}\ }\textbf {\bibinfo {volume} {B768}},\ \bibinfo
  {pages} {38} (\bibinfo {year} {2017})},\ \Eprint
  {http://arxiv.org/abs/1612.08162} {arXiv:1612.08162 [hep-ex]}\BibitemShut
  {NoStop}%
\bibitem [{\citenamefont {Lees}\ \emph
  {et~al.}(2018{\natexlab{c}})\citenamefont {Lees} \emph
  {et~al.}}]{BaBar:2018zpn}%
  \BibitemOpen
  \bibfield  {author} {\bibinfo {author} {\bibfnamefont {J.~P.}\ \bibnamefont
  {Lees}} \emph {et~al.} (\bibinfo {collaboration} {BABAR}),\ }\href {\doibase
  10.1103/PhysRevD.98.112002} {\bibfield  {journal} {\bibinfo  {journal} {Phys.
  Rev.}\ }\textbf {\bibinfo {volume} {D98}},\ \bibinfo {pages} {112002}
  (\bibinfo {year} {2018}{\natexlab{c}})},\ \Eprint
  {http://arxiv.org/abs/1808.08038} {arXiv:1808.08038 [hep-ex]}\BibitemShut
  {NoStop}%
\bibitem [{\citenamefont {Larin}\ \emph {et~al.}(2020)\citenamefont {Larin}
  \emph {et~al.}}]{Larin:2020}%
  \BibitemOpen
  \bibfield  {author} {\bibinfo {author} {\bibfnamefont {I.}~\bibnamefont
  {Larin}} \emph {et~al.} (\bibinfo {collaboration} {PrimEx II}),\ }\href
  {\doibase 10.1126/science.aay6641} {\bibfield  {journal} {\bibinfo  {journal}
  {Science}\ }\textbf {\bibinfo {volume} {368}},\ \bibinfo {pages} {506}
  (\bibinfo {year} {2020})}\BibitemShut {NoStop}%
\bibitem [{\citenamefont {Blum}\ \emph {et~al.}(2015)\citenamefont {Blum},
  \citenamefont {Chowdhury}, \citenamefont {Hayakawa},\ and\ \citenamefont
  {Izubuchi}}]{Blum:2014oka}%
  \BibitemOpen
  \bibfield  {author} {\bibinfo {author} {\bibfnamefont {T.}~\bibnamefont
  {Blum}}, \bibinfo {author} {\bibfnamefont {S.}~\bibnamefont {Chowdhury}},
  \bibinfo {author} {\bibfnamefont {M.}~\bibnamefont {Hayakawa}}, \ and\
  \bibinfo {author} {\bibfnamefont {T.}~\bibnamefont {Izubuchi}},\ }\href
  {\doibase 10.1103/PhysRevLett.114.012001} {\bibfield  {journal} {\bibinfo
  {journal} {Phys. Rev. Lett.}\ }\textbf {\bibinfo {volume} {114}},\ \bibinfo
  {pages} {012001} (\bibinfo {year} {2015})},\ \Eprint
  {http://arxiv.org/abs/1407.2923} {arXiv:1407.2923 [hep-lat]}\BibitemShut
  {NoStop}%
\bibitem [{\citenamefont {Green}\ \emph {et~al.}(2016)\citenamefont {Green},
  \citenamefont {Asmussen}, \citenamefont {Gryniuk}, \citenamefont {von
  Hippel}, \citenamefont {Meyer}, \citenamefont {Nyffeler},\ and\ \citenamefont
  {Pascalutsa}}]{Green:2015mva}%
  \BibitemOpen
  \bibfield  {author} {\bibinfo {author} {\bibfnamefont {J.}~\bibnamefont
  {Green}}, \bibinfo {author} {\bibfnamefont {N.}~\bibnamefont {Asmussen}},
  \bibinfo {author} {\bibfnamefont {O.}~\bibnamefont {Gryniuk}}, \bibinfo
  {author} {\bibfnamefont {G.}~\bibnamefont {von Hippel}}, \bibinfo {author}
  {\bibfnamefont {H.~B.}\ \bibnamefont {Meyer}}, \bibinfo {author}
  {\bibfnamefont {A.}~\bibnamefont {Nyffeler}}, \ and\ \bibinfo {author}
  {\bibfnamefont {V.}~\bibnamefont {Pascalutsa}},\ }\href {\doibase
  10.22323/1.251.0109} {\bibfield  {journal} {\bibinfo  {journal} {PoS}\
  }\textbf {\bibinfo {volume} {LATTICE2015}},\ \bibinfo {pages} {109} (\bibinfo
  {year} {2016})},\ \Eprint {http://arxiv.org/abs/1510.08384} {arXiv:1510.08384
  [hep-lat]}\BibitemShut {NoStop}%
\bibitem [{\citenamefont {Blum}\ \emph
  {et~al.}(2016{\natexlab{a}})\citenamefont {Blum}, \citenamefont {Christ},
  \citenamefont {Hayakawa}, \citenamefont {Izubuchi}, \citenamefont {Jin},\
  and\ \citenamefont {Lehner}}]{Blum:2015gfa}%
  \BibitemOpen
  \bibfield  {author} {\bibinfo {author} {\bibfnamefont {T.}~\bibnamefont
  {Blum}}, \bibinfo {author} {\bibfnamefont {N.}~\bibnamefont {Christ}},
  \bibinfo {author} {\bibfnamefont {M.}~\bibnamefont {Hayakawa}}, \bibinfo
  {author} {\bibfnamefont {T.}~\bibnamefont {Izubuchi}}, \bibinfo {author}
  {\bibfnamefont {L.}~\bibnamefont {Jin}}, \ and\ \bibinfo {author}
  {\bibfnamefont {C.}~\bibnamefont {Lehner}},\ }\href {\doibase
  10.1103/PhysRevD.93.014503} {\bibfield  {journal} {\bibinfo  {journal} {Phys.
  Rev.}\ }\textbf {\bibinfo {volume} {D93}},\ \bibinfo {pages} {014503}
  (\bibinfo {year} {2016}{\natexlab{a}})},\ \Eprint
  {http://arxiv.org/abs/1510.07100} {arXiv:1510.07100 [hep-lat]}\BibitemShut
  {NoStop}%
\bibitem [{\citenamefont {Blum}\ \emph
  {et~al.}(2017{\natexlab{a}})\citenamefont {Blum}, \citenamefont {Christ},
  \citenamefont {Hayakawa}, \citenamefont {Izubuchi}, \citenamefont {Jin},
  \citenamefont {Jung},\ and\ \citenamefont {Lehner}}]{Blum:2016lnc}%
  \BibitemOpen
  \bibfield  {author} {\bibinfo {author} {\bibfnamefont {T.}~\bibnamefont
  {Blum}}, \bibinfo {author} {\bibfnamefont {N.}~\bibnamefont {Christ}},
  \bibinfo {author} {\bibfnamefont {M.}~\bibnamefont {Hayakawa}}, \bibinfo
  {author} {\bibfnamefont {T.}~\bibnamefont {Izubuchi}}, \bibinfo {author}
  {\bibfnamefont {L.}~\bibnamefont {Jin}}, \bibinfo {author} {\bibfnamefont
  {C.}~\bibnamefont {Jung}}, \ and\ \bibinfo {author} {\bibfnamefont
  {C.}~\bibnamefont {Lehner}},\ }\href {\doibase
  10.1103/PhysRevLett.118.022005} {\bibfield  {journal} {\bibinfo  {journal}
  {Phys. Rev. Lett.}\ }\textbf {\bibinfo {volume} {118}},\ \bibinfo {pages}
  {022005} (\bibinfo {year} {2017}{\natexlab{a}})},\ \Eprint
  {http://arxiv.org/abs/1610.04603} {arXiv:1610.04603 [hep-lat]}\BibitemShut
  {NoStop}%
\bibitem [{\citenamefont {Asmussen}\ \emph {et~al.}(2016)\citenamefont
  {Asmussen}, \citenamefont {Green}, \citenamefont {Meyer},\ and\ \citenamefont
  {Nyffeler}}]{Asmussen:2016lse}%
  \BibitemOpen
  \bibfield  {author} {\bibinfo {author} {\bibfnamefont {N.}~\bibnamefont
  {Asmussen}}, \bibinfo {author} {\bibfnamefont {J.}~\bibnamefont {Green}},
  \bibinfo {author} {\bibfnamefont {H.~B.}\ \bibnamefont {Meyer}}, \ and\
  \bibinfo {author} {\bibfnamefont {A.}~\bibnamefont {Nyffeler}},\ }\href
  {\doibase 10.22323/1.256.0164} {\bibfield  {journal} {\bibinfo  {journal}
  {PoS}\ }\textbf {\bibinfo {volume} {LATTICE2016}},\ \bibinfo {pages} {164}
  (\bibinfo {year} {2016})},\ \Eprint {http://arxiv.org/abs/1609.08454}
  {arXiv:1609.08454 [hep-lat]}\BibitemShut {NoStop}%
\bibitem [{\citenamefont {Blum}\ \emph
  {et~al.}(2017{\natexlab{b}})\citenamefont {Blum}, \citenamefont {Christ},
  \citenamefont {Hayakawa}, \citenamefont {Izubuchi}, \citenamefont {Jin},
  \citenamefont {Jung},\ and\ \citenamefont {Lehner}}]{Blum:2017cer}%
  \BibitemOpen
  \bibfield  {author} {\bibinfo {author} {\bibfnamefont {T.}~\bibnamefont
  {Blum}}, \bibinfo {author} {\bibfnamefont {N.}~\bibnamefont {Christ}},
  \bibinfo {author} {\bibfnamefont {M.}~\bibnamefont {Hayakawa}}, \bibinfo
  {author} {\bibfnamefont {T.}~\bibnamefont {Izubuchi}}, \bibinfo {author}
  {\bibfnamefont {L.}~\bibnamefont {Jin}}, \bibinfo {author} {\bibfnamefont
  {C.}~\bibnamefont {Jung}}, \ and\ \bibinfo {author} {\bibfnamefont
  {C.}~\bibnamefont {Lehner}},\ }\href {\doibase 10.1103/PhysRevD.96.034515}
  {\bibfield  {journal} {\bibinfo  {journal} {Phys. Rev.}\ }\textbf {\bibinfo
  {volume} {D96}},\ \bibinfo {pages} {034515} (\bibinfo {year}
  {2017}{\natexlab{b}})},\ \Eprint {http://arxiv.org/abs/1705.01067}
  {arXiv:1705.01067 [hep-lat]}\BibitemShut {NoStop}%
\bibitem [{\citenamefont {Asmussen}\ \emph
  {et~al.}(2019{\natexlab{a}})\citenamefont {Asmussen}, \citenamefont {Chao},
  \citenamefont {G{\'e}rardin}, \citenamefont {Green}, \citenamefont
  {Hudspith}, \citenamefont {Meyer},\ and\ \citenamefont
  {Nyffeler}}]{Asmussen:2019act}%
  \BibitemOpen
  \bibfield  {author} {\bibinfo {author} {\bibfnamefont {N.}~\bibnamefont
  {Asmussen}}, \bibinfo {author} {\bibfnamefont {E.-H.}\ \bibnamefont {Chao}},
  \bibinfo {author} {\bibfnamefont {A.}~\bibnamefont {G{\'e}rardin}}, \bibinfo
  {author} {\bibfnamefont {J.~R.}\ \bibnamefont {Green}}, \bibinfo {author}
  {\bibfnamefont {R.~J.}\ \bibnamefont {Hudspith}}, \bibinfo {author}
  {\bibfnamefont {H.~B.}\ \bibnamefont {Meyer}}, \ and\ \bibinfo {author}
  {\bibfnamefont {A.}~\bibnamefont {Nyffeler}},\ }\href {\doibase
  10.22323/1.363.0195} {\bibfield  {journal} {\bibinfo  {journal} {PoS}\
  }\textbf {\bibinfo {volume} {LATTICE2019}},\ \bibinfo {pages} {195} (\bibinfo
  {year} {2019}{\natexlab{a}})},\ \Eprint {http://arxiv.org/abs/1911.05573}
  {arXiv:1911.05573 [hep-lat]}\BibitemShut {NoStop}%
\bibitem [{\citenamefont {Parker}\ \emph {et~al.}(2018)\citenamefont {Parker},
  \citenamefont {Yu}, \citenamefont {Zhong}, \citenamefont {Estey},\ and\
  \citenamefont {M{\"u}ller}}]{Parker:2018vye}%
  \BibitemOpen
  \bibfield  {author} {\bibinfo {author} {\bibfnamefont {R.~H.}\ \bibnamefont
  {Parker}}, \bibinfo {author} {\bibfnamefont {C.}~\bibnamefont {Yu}}, \bibinfo
  {author} {\bibfnamefont {W.}~\bibnamefont {Zhong}}, \bibinfo {author}
  {\bibfnamefont {B.}~\bibnamefont {Estey}}, \ and\ \bibinfo {author}
  {\bibfnamefont {H.}~\bibnamefont {M{\"u}ller}},\ }\href {\doibase
  10.1126/science.aap7706} {\bibfield  {journal} {\bibinfo  {journal}
  {Science}\ }\textbf {\bibinfo {volume} {360}},\ \bibinfo {pages} {191}
  (\bibinfo {year} {2018})},\ \Eprint {http://arxiv.org/abs/1812.04130}
  {arXiv:1812.04130 [physics.atom-ph]}\BibitemShut {NoStop}%
\bibitem [{Muo(2020)}]{MuonInitiative}%
  \BibitemOpen
  \href@noop {} {\enquote {\bibinfo {title} {{Muon $g-2$ Theory Initiative}},}\
  }\bibinfo {howpublished} {\url{https://muon-gm2-theory.illinois.edu/}}
  (\bibinfo {year} {2020})\BibitemShut {NoStop}%
\bibitem [{\citenamefont {Grange}\ \emph {et~al.}(2015)\citenamefont {Grange}
  \emph {et~al.}}]{Grange:2015fou}%
  \BibitemOpen
  \bibfield  {author} {\bibinfo {author} {\bibfnamefont {J.}~\bibnamefont
  {Grange}} \emph {et~al.} (\bibinfo {collaboration} {Muon $g-2$}),\
  }\href@noop {} {\  (\bibinfo {year} {2015})},\ \Eprint
  {http://arxiv.org/abs/1501.06858} {arXiv:1501.06858
  [physics.ins-det]}\BibitemShut {NoStop}%
\bibitem [{\citenamefont {Abe}\ \emph {et~al.}(2019)\citenamefont {Abe} \emph
  {et~al.}}]{Abe:2019thb}%
  \BibitemOpen
  \bibfield  {author} {\bibinfo {author} {\bibfnamefont {M.}~\bibnamefont
  {Abe}} \emph {et~al.},\ }\href {\doibase 10.1093/ptep/ptz030} {\bibfield
  {journal} {\bibinfo  {journal} {PTEP}\ }\textbf {\bibinfo {volume} {2019}},\
  \bibinfo {pages} {053C02} (\bibinfo {year} {2019})},\ \Eprint
  {http://arxiv.org/abs/1901.03047} {arXiv:1901.03047
  [physics.ins-det]}\BibitemShut {NoStop}%
\bibitem [{\citenamefont {Blum}\ \emph
  {et~al.}(2013{\natexlab{a}})\citenamefont {Blum}, \citenamefont {Denig},
  \citenamefont {Logashenko}, \citenamefont {de~Rafael}, \citenamefont
  {Lee~Roberts}, \citenamefont {Teubner},\ and\ \citenamefont
  {Venanzoni}}]{Blum:2013xva}%
  \BibitemOpen
  \bibfield  {author} {\bibinfo {author} {\bibfnamefont {T.}~\bibnamefont
  {Blum}}, \bibinfo {author} {\bibfnamefont {A.}~\bibnamefont {Denig}},
  \bibinfo {author} {\bibfnamefont {I.}~\bibnamefont {Logashenko}}, \bibinfo
  {author} {\bibfnamefont {E.}~\bibnamefont {de~Rafael}}, \bibinfo {author}
  {\bibfnamefont {B.}~\bibnamefont {Lee~Roberts}}, \bibinfo {author}
  {\bibfnamefont {T.}~\bibnamefont {Teubner}}, \ and\ \bibinfo {author}
  {\bibfnamefont {G.}~\bibnamefont {Venanzoni}},\ }\href@noop {} {\  (\bibinfo
  {year} {2013}{\natexlab{a}})},\ \Eprint {http://arxiv.org/abs/1311.2198}
  {arXiv:1311.2198 [hep-ph]}\BibitemShut {NoStop}%
\bibitem [{\citenamefont {Aubin}\ \emph {et~al.}(2014)\citenamefont {Aubin}
  \emph {et~al.}}]{Benayoun:2014tra}%
  \BibitemOpen
  \bibfield  {author} {\bibinfo {author} {\bibfnamefont {C.}~\bibnamefont
  {Aubin}} \emph {et~al.},\ }\href@noop {} {\  (\bibinfo {year} {2014})},\
  \Eprint {http://arxiv.org/abs/1407.4021} {arXiv:1407.4021
  [hep-ph]}\BibitemShut {NoStop}%
\bibitem [{FNA(2017)}]{FNAL2017}%
  \BibitemOpen
  \href@noop {} {\enquote {\bibinfo {title} {{First Workshop of the Muon $g-2$
  Theory Initiative}},}\ }\bibinfo {howpublished}
  {\url{https://indico.fnal.gov/event/13795/}} (\bibinfo {year} {2017}),\
  \bibinfo {note} {{held at Fermilab, St.\ Charles, IL, USA, June
  3--6.}}\BibitemShut {Stop}%
\bibitem [{KEK(2018)}]{KEK2018}%
  \BibitemOpen
  \href@noop {} {\enquote {\bibinfo {title} {{Workshop on Hadronic Vacuum
  Polarization Contributions to Muon $g-2$}},}\ }\bibinfo {howpublished}
  {\url{https://www-conf.kek.jp/muonHVPws/}} (\bibinfo {year} {2018}),\
  \bibinfo {note} {{held at KEK, Tsukuba, Japan, February 12--14.}}\BibitemShut
  {Stop}%
\bibitem [{UCo(2018)}]{UConn2018}%
  \BibitemOpen
  \href@noop {} {\enquote {\bibinfo {title} {{Muon $g-2$ Theory Initiative
  Hadronic Light-by-Light working group workshop}},}\ }\bibinfo {howpublished}
  {\url{https://indico.phys.uconn.edu/event/1/}} (\bibinfo {year} {2018}),\
  \bibinfo {note} {{held at University of Connecticut, Storrs, CT, USA, March
  12--14.}}\BibitemShut {Stop}%
\bibitem [{Mai(2018)}]{Mainz2018}%
  \BibitemOpen
  \href@noop {} {\enquote {\bibinfo {title} {{Second Workshop of the Muon $g-2$
  Theory Initiative}},}\ }\bibinfo {howpublished}
  {\url{https://wwwth.kph.uni-mainz.de/g-2/}} (\bibinfo {year} {2018}),\
  \bibinfo {note} {{held at the Helmholtz Institute Mainz, University of Mainz,
  Mainz, Germany, June 18--22.}}\BibitemShut {Stop}%
\bibitem [{INT(2019)}]{INT2019}%
  \BibitemOpen
  \href@noop {} {\enquote {\bibinfo {title} {{Hadronic contributions to
  $(g-2)_\mu$}},}\ }\bibinfo {howpublished}
  {\url{https://indico.fnal.gov/event/21626/}} (\bibinfo {year} {2019}),\
  \bibinfo {note} {{held at the Institute for Nuclear Theory, University of
  Washington, Seattle, WA, USA, September 9--13.}}\BibitemShut {Stop}%
\bibitem [{KEK(2020)}]{KEK2020}%
  \BibitemOpen
  \href@noop {} {\enquote {\bibinfo {title} {{Fourth Plenary Workshop of the
  Muon $g-2$ Theory Initiative}},}\ }\bibinfo {howpublished}
  {\url{https://www-conf.kek.jp/muong-2theory/}} (\bibinfo {year} {2020}),\
  \bibinfo {note} {{to be held at KEK, Tsukuba, Japan, new dates to be
  announced.}}\BibitemShut {Stop}%
\bibitem [{\citenamefont {Brodsky}\ and\ \citenamefont
  {de~Rafael}(1968)}]{Brodsky:1967sr}%
  \BibitemOpen
  \bibfield  {author} {\bibinfo {author} {\bibfnamefont {S.~J.}\ \bibnamefont
  {Brodsky}}\ and\ \bibinfo {author} {\bibfnamefont {E.}~\bibnamefont
  {de~Rafael}},\ }\href {\doibase 10.1103/PhysRev.168.1620} {\bibfield
  {journal} {\bibinfo  {journal} {Phys. Rev.}\ }\textbf {\bibinfo {volume}
  {168}},\ \bibinfo {pages} {1620} (\bibinfo {year} {1968})}\BibitemShut
  {NoStop}%
\bibitem [{\citenamefont {Lautrup}\ and\ \citenamefont
  {de~Rafael}(1968)}]{Lautrup:1969fr}%
  \BibitemOpen
  \bibfield  {author} {\bibinfo {author} {\bibfnamefont {B.~E.}\ \bibnamefont
  {Lautrup}}\ and\ \bibinfo {author} {\bibfnamefont {E.}~\bibnamefont
  {de~Rafael}},\ }\href {\doibase 10.1103/PhysRev.174.1835} {\bibfield
  {journal} {\bibinfo  {journal} {Phys. Rev.}\ }\textbf {\bibinfo {volume}
  {174}},\ \bibinfo {pages} {1835} (\bibinfo {year} {1968})}\BibitemShut
  {NoStop}%
\bibitem [{\citenamefont {Krause}(1997)}]{Krause:1996rf}%
  \BibitemOpen
  \bibfield  {author} {\bibinfo {author} {\bibfnamefont {B.}~\bibnamefont
  {Krause}},\ }\href {\doibase 10.1016/S0370-2693(96)01346-9} {\bibfield
  {journal} {\bibinfo  {journal} {Phys. Lett.}\ }\textbf {\bibinfo {volume}
  {B390}},\ \bibinfo {pages} {392} (\bibinfo {year} {1997})},\ \Eprint
  {http://arxiv.org/abs/hep-ph/9607259} {arXiv:hep-ph/9607259
  [hep-ph]}\BibitemShut {NoStop}%
\bibitem [{\citenamefont {Harlander}\ and\ \citenamefont
  {Steinhauser}(2003)}]{Harlander:2002ur}%
  \BibitemOpen
  \bibfield  {author} {\bibinfo {author} {\bibfnamefont {R.~V.}\ \bibnamefont
  {Harlander}}\ and\ \bibinfo {author} {\bibfnamefont {M.}~\bibnamefont
  {Steinhauser}},\ }\href {\doibase 10.1016/S0010-4655(03)00204-2} {\bibfield
  {journal} {\bibinfo  {journal} {Comput. Phys. Commun.}\ }\textbf {\bibinfo
  {volume} {153}},\ \bibinfo {pages} {244} (\bibinfo {year} {2003})},\ \Eprint
  {http://arxiv.org/abs/hep-ph/0212294} {arXiv:hep-ph/0212294
  [hep-ph]}\BibitemShut {NoStop}%
\bibitem [{\citenamefont {Gracey}(2014)}]{Gracey:2014pba}%
  \BibitemOpen
  \bibfield  {author} {\bibinfo {author} {\bibfnamefont {J.~A.}\ \bibnamefont
  {Gracey}},\ }\href {\doibase 10.1103/PhysRevD.90.094026} {\bibfield
  {journal} {\bibinfo  {journal} {Phys. Rev.}\ }\textbf {\bibinfo {volume}
  {D90}},\ \bibinfo {pages} {094026} (\bibinfo {year} {2014})},\ \Eprint
  {http://arxiv.org/abs/1410.6715} {arXiv:1410.6715 [hep-ph]}\BibitemShut
  {NoStop}%
\bibitem [{\citenamefont {Steinhauser}(1998)}]{Steinhauser:1998rq}%
  \BibitemOpen
  \bibfield  {author} {\bibinfo {author} {\bibfnamefont {M.}~\bibnamefont
  {Steinhauser}},\ }\href {\doibase 10.1016/S0370-2693(98)00503-6} {\bibfield
  {journal} {\bibinfo  {journal} {Phys. Lett.}\ }\textbf {\bibinfo {volume}
  {B429}},\ \bibinfo {pages} {158} (\bibinfo {year} {1998})},\ \Eprint
  {http://arxiv.org/abs/hep-ph/9803313} {arXiv:hep-ph/9803313
  [hep-ph]}\BibitemShut {NoStop}%
\bibitem [{\citenamefont {Sturm}(2013)}]{Sturm:2013uka}%
  \BibitemOpen
  \bibfield  {author} {\bibinfo {author} {\bibfnamefont {C.}~\bibnamefont
  {Sturm}},\ }\href {\doibase 10.1016/j.nuclphysb.2013.06.009} {\bibfield
  {journal} {\bibinfo  {journal} {Nucl. Phys.}\ }\textbf {\bibinfo {volume}
  {B874}},\ \bibinfo {pages} {698} (\bibinfo {year} {2013})},\ \Eprint
  {http://arxiv.org/abs/1305.0581} {arXiv:1305.0581 [hep-ph]}\BibitemShut
  {NoStop}%
\bibitem [{\citenamefont {Hoefer}\ \emph {et~al.}(2002)\citenamefont {Hoefer},
  \citenamefont {Gluza},\ and\ \citenamefont {Jegerlehner}}]{Hoefer:2001mx}%
  \BibitemOpen
  \bibfield  {author} {\bibinfo {author} {\bibfnamefont {A.}~\bibnamefont
  {Hoefer}}, \bibinfo {author} {\bibfnamefont {J.}~\bibnamefont {Gluza}}, \
  and\ \bibinfo {author} {\bibfnamefont {F.}~\bibnamefont {Jegerlehner}},\
  }\href {\doibase 10.1007/s100520200916} {\bibfield  {journal} {\bibinfo
  {journal} {Eur. Phys. J.}\ }\textbf {\bibinfo {volume} {C24}},\ \bibinfo
  {pages} {51} (\bibinfo {year} {2002})},\ \Eprint
  {http://arxiv.org/abs/hep-ph/0107154} {arXiv:hep-ph/0107154
  [hep-ph]}\BibitemShut {NoStop}%
\bibitem [{\citenamefont {Gluza}\ \emph {et~al.}(2003)\citenamefont {Gluza},
  \citenamefont {Hoefer}, \citenamefont {Jadach},\ and\ \citenamefont
  {Jegerlehner}}]{Gluza:2002ui}%
  \BibitemOpen
  \bibfield  {author} {\bibinfo {author} {\bibfnamefont {J.}~\bibnamefont
  {Gluza}}, \bibinfo {author} {\bibfnamefont {A.}~\bibnamefont {Hoefer}},
  \bibinfo {author} {\bibfnamefont {S.}~\bibnamefont {Jadach}}, \ and\ \bibinfo
  {author} {\bibfnamefont {F.}~\bibnamefont {Jegerlehner}},\ }\href {\doibase
  10.1140/epjc/s2003-01146-0} {\bibfield  {journal} {\bibinfo  {journal} {Eur.
  Phys. J.}\ }\textbf {\bibinfo {volume} {C28}},\ \bibinfo {pages} {261}
  (\bibinfo {year} {2003})},\ \Eprint {http://arxiv.org/abs/hep-ph/0212386}
  {arXiv:hep-ph/0212386 [hep-ph]}\BibitemShut {NoStop}%
\bibitem [{\citenamefont {Lees}\ \emph {et~al.}(2015)\citenamefont {Lees} \emph
  {et~al.}}]{Lees:2015qna}%
  \BibitemOpen
  \bibfield  {author} {\bibinfo {author} {\bibfnamefont {J.~P.}\ \bibnamefont
  {Lees}} \emph {et~al.} (\bibinfo {collaboration} {BABAR}),\ }\href {\doibase
  10.1103/PhysRevD.92.072015} {\bibfield  {journal} {\bibinfo  {journal} {Phys.
  Rev.}\ }\textbf {\bibinfo {volume} {D92}},\ \bibinfo {pages} {072015}
  (\bibinfo {year} {2015})},\ \Eprint {http://arxiv.org/abs/1508.04008}
  {arXiv:1508.04008 [hep-ex]}\BibitemShut {NoStop}%
\bibitem [{\citenamefont {Davier}(2013)}]{Davier:2013vna}%
  \BibitemOpen
  \bibfield  {author} {\bibinfo {author} {\bibfnamefont {M.}~\bibnamefont
  {Davier}},\ }\href {\doibase 10.1146/annurev-nucl-102212-170554} {\bibfield
  {journal} {\bibinfo  {journal} {Ann. Rev. Nucl. Part. Sci.}\ }\textbf
  {\bibinfo {volume} {63}},\ \bibinfo {pages} {407} (\bibinfo {year}
  {2013})}\BibitemShut {NoStop}%
\bibitem [{\citenamefont {Aloisio}\ \emph {et~al.}(2005)\citenamefont {Aloisio}
  \emph {et~al.}}]{Aloisio:2004bu}%
  \BibitemOpen
  \bibfield  {author} {\bibinfo {author} {\bibfnamefont {A.}~\bibnamefont
  {Aloisio}} \emph {et~al.} (\bibinfo {collaboration} {KLOE}),\ }\href
  {\doibase 10.1016/j.physletb.2004.11.068} {\bibfield  {journal} {\bibinfo
  {journal} {Phys. Lett.}\ }\textbf {\bibinfo {volume} {B606}},\ \bibinfo
  {pages} {12} (\bibinfo {year} {2005})},\ \Eprint
  {http://arxiv.org/abs/hep-ex/0407048} {arXiv:hep-ex/0407048
  [hep-ex]}\BibitemShut {NoStop}%
\bibitem [{\citenamefont {Venanzoni}(2018)}]{Venanzoni:2017ggn}%
  \BibitemOpen
  \bibfield  {author} {\bibinfo {author} {\bibfnamefont {G.}~\bibnamefont
  {Venanzoni}} (\bibinfo {collaboration} {KLOE-2}),\ }\href {\doibase
  10.1051/epjconf/201816600021} {\bibfield  {journal} {\bibinfo  {journal} {EPJ
  Web Conf.}\ }\textbf {\bibinfo {volume} {166}},\ \bibinfo {pages} {00021}
  (\bibinfo {year} {2018})},\ \Eprint {http://arxiv.org/abs/1705.10365}
  {arXiv:1705.10365 [hep-ex]}\BibitemShut {NoStop}%
\bibitem [{\citenamefont {Lees}\ \emph
  {et~al.}(2013{\natexlab{c}})\citenamefont {Lees} \emph
  {et~al.}}]{Lees:2013gzt}%
  \BibitemOpen
  \bibfield  {author} {\bibinfo {author} {\bibfnamefont {J.~P.}\ \bibnamefont
  {Lees}} \emph {et~al.} (\bibinfo {collaboration} {BABAR}),\ }\href {\doibase
  10.1103/PhysRevD.88.032013} {\bibfield  {journal} {\bibinfo  {journal} {Phys.
  Rev.}\ }\textbf {\bibinfo {volume} {D88}},\ \bibinfo {pages} {032013}
  (\bibinfo {year} {2013}{\natexlab{c}})},\ \Eprint
  {http://arxiv.org/abs/1306.3600} {arXiv:1306.3600 [hep-ex]}\BibitemShut
  {NoStop}%
\bibitem [{\citenamefont {Jadach}\ \emph {et~al.}(1997)\citenamefont {Jadach},
  \citenamefont {Placzek},\ and\ \citenamefont {Ward}}]{Jadach:1995nk}%
  \BibitemOpen
  \bibfield  {author} {\bibinfo {author} {\bibfnamefont {S.}~\bibnamefont
  {Jadach}}, \bibinfo {author} {\bibfnamefont {W.}~\bibnamefont {Placzek}}, \
  and\ \bibinfo {author} {\bibfnamefont {B.~F.~L.}\ \bibnamefont {Ward}},\
  }\href {\doibase 10.1016/S0370-2693(96)01382-2} {\bibfield  {journal}
  {\bibinfo  {journal} {Phys. Lett.}\ }\textbf {\bibinfo {volume} {B390}},\
  \bibinfo {pages} {298} (\bibinfo {year} {1997})},\ \Eprint
  {http://arxiv.org/abs/hep-ph/9608412} {arXiv:hep-ph/9608412
  [hep-ph]}\BibitemShut {NoStop}%
\bibitem [{\citenamefont {Balossini}\ \emph {et~al.}(2006)\citenamefont
  {Balossini}, \citenamefont {Carloni~Calame}, \citenamefont {Montagna},
  \citenamefont {Nicrosini},\ and\ \citenamefont
  {Piccinini}}]{Balossini:2006wc}%
  \BibitemOpen
  \bibfield  {author} {\bibinfo {author} {\bibfnamefont {G.}~\bibnamefont
  {Balossini}}, \bibinfo {author} {\bibfnamefont {C.~M.}\ \bibnamefont
  {Carloni~Calame}}, \bibinfo {author} {\bibfnamefont {G.}~\bibnamefont
  {Montagna}}, \bibinfo {author} {\bibfnamefont {O.}~\bibnamefont {Nicrosini}},
  \ and\ \bibinfo {author} {\bibfnamefont {F.}~\bibnamefont {Piccinini}},\
  }\href {\doibase 10.1016/j.nuclphysb.2006.09.022} {\bibfield  {journal}
  {\bibinfo  {journal} {Nucl. Phys.}\ }\textbf {\bibinfo {volume} {B758}},\
  \bibinfo {pages} {227} (\bibinfo {year} {2006})},\ \Eprint
  {http://arxiv.org/abs/hep-ph/0607181} {arXiv:hep-ph/0607181
  [hep-ph]}\BibitemShut {NoStop}%
\bibitem [{\citenamefont {Binner}\ \emph {et~al.}(1999)\citenamefont {Binner},
  \citenamefont {K{\"u}hn},\ and\ \citenamefont {Melnikov}}]{Binner:1999bt}%
  \BibitemOpen
  \bibfield  {author} {\bibinfo {author} {\bibfnamefont {S.}~\bibnamefont
  {Binner}}, \bibinfo {author} {\bibfnamefont {J.~H.}\ \bibnamefont
  {K{\"u}hn}}, \ and\ \bibinfo {author} {\bibfnamefont {K.}~\bibnamefont
  {Melnikov}},\ }\href {\doibase 10.1016/S0370-2693(99)00658-9} {\bibfield
  {journal} {\bibinfo  {journal} {Phys. Lett.}\ }\textbf {\bibinfo {volume}
  {B459}},\ \bibinfo {pages} {279} (\bibinfo {year} {1999})},\ \Eprint
  {http://arxiv.org/abs/hep-ph/9902399} {arXiv:hep-ph/9902399
  [hep-ph]}\BibitemShut {NoStop}%
\bibitem [{\citenamefont {Czy\.z}\ and\ \citenamefont
  {K{\"u}hn}(2001)}]{Czyz:2000wh}%
  \BibitemOpen
  \bibfield  {author} {\bibinfo {author} {\bibfnamefont {H.}~\bibnamefont
  {Czy\.z}}\ and\ \bibinfo {author} {\bibfnamefont {J.~H.}\ \bibnamefont
  {K{\"u}hn}},\ }\href {\doibase 10.1007/s100520000553} {\bibfield  {journal}
  {\bibinfo  {journal} {Eur. Phys. J.}\ }\textbf {\bibinfo {volume} {C18}},\
  \bibinfo {pages} {497} (\bibinfo {year} {2001})},\ \Eprint
  {http://arxiv.org/abs/hep-ph/0008262} {arXiv:hep-ph/0008262
  [hep-ph]}\BibitemShut {NoStop}%
\bibitem [{\citenamefont {Czy\.z}\ \emph
  {et~al.}(2004{\natexlab{a}})\citenamefont {Czy\.z}, \citenamefont {K{\"u}hn},
  \citenamefont {Nowak},\ and\ \citenamefont {Rodrigo}}]{Czyz:2004ua}%
  \BibitemOpen
  \bibfield  {author} {\bibinfo {author} {\bibfnamefont {H.}~\bibnamefont
  {Czy\.z}}, \bibinfo {author} {\bibfnamefont {J.~H.}\ \bibnamefont
  {K{\"u}hn}}, \bibinfo {author} {\bibfnamefont {E.}~\bibnamefont {Nowak}}, \
  and\ \bibinfo {author} {\bibfnamefont {G.}~\bibnamefont {Rodrigo}},\ }\href
  {\doibase 10.1140/epjc/s2004-01864-7} {\bibfield  {journal} {\bibinfo
  {journal} {Eur. Phys. J.}\ }\textbf {\bibinfo {volume} {C35}},\ \bibinfo
  {pages} {527} (\bibinfo {year} {2004}{\natexlab{a}})},\ \Eprint
  {http://arxiv.org/abs/hep-ph/0403062} {arXiv:hep-ph/0403062
  [hep-ph]}\BibitemShut {NoStop}%
\bibitem [{\citenamefont {Czy\.z}\ \emph
  {et~al.}(2005{\natexlab{a}})\citenamefont {Czy\.z}, \citenamefont
  {Grzeli\'nska}, \citenamefont {K{\"u}hn},\ and\ \citenamefont
  {Rodrigo}}]{Czyz:2004rj}%
  \BibitemOpen
  \bibfield  {author} {\bibinfo {author} {\bibfnamefont {H.}~\bibnamefont
  {Czy\.z}}, \bibinfo {author} {\bibfnamefont {A.}~\bibnamefont
  {Grzeli\'nska}}, \bibinfo {author} {\bibfnamefont {J.~H.}\ \bibnamefont
  {K{\"u}hn}}, \ and\ \bibinfo {author} {\bibfnamefont {G.}~\bibnamefont
  {Rodrigo}},\ }\href {\doibase 10.1140/epjc/s2004-02103-1} {\bibfield
  {journal} {\bibinfo  {journal} {Eur. Phys. J.}\ }\textbf {\bibinfo {volume}
  {C39}},\ \bibinfo {pages} {411} (\bibinfo {year} {2005}{\natexlab{a}})},\
  \Eprint {http://arxiv.org/abs/hep-ph/0404078} {arXiv:hep-ph/0404078
  [hep-ph]}\BibitemShut {NoStop}%
\bibitem [{\citenamefont {Rodrigo}\ \emph {et~al.}(2002)\citenamefont
  {Rodrigo}, \citenamefont {Czy\.z}, \citenamefont {K{\"u}hn},\ and\
  \citenamefont {Szopa}}]{Rodrigo:2001kf}%
  \BibitemOpen
  \bibfield  {author} {\bibinfo {author} {\bibfnamefont {G.}~\bibnamefont
  {Rodrigo}}, \bibinfo {author} {\bibfnamefont {H.}~\bibnamefont {Czy\.z}},
  \bibinfo {author} {\bibfnamefont {J.~H.}\ \bibnamefont {K{\"u}hn}}, \ and\
  \bibinfo {author} {\bibfnamefont {M.}~\bibnamefont {Szopa}},\ }\href
  {\doibase 10.1007/s100520200912} {\bibfield  {journal} {\bibinfo  {journal}
  {Eur. Phys. J.}\ }\textbf {\bibinfo {volume} {C24}},\ \bibinfo {pages} {71}
  (\bibinfo {year} {2002})},\ \Eprint {http://arxiv.org/abs/hep-ph/0112184}
  {arXiv:hep-ph/0112184 [hep-ph]}\BibitemShut {NoStop}%
\bibitem [{\citenamefont {Barberio}\ \emph {et~al.}(1991)\citenamefont
  {Barberio}, \citenamefont {van Eijk},\ and\ \citenamefont
  {Was}}]{Barberio:1990ms}%
  \BibitemOpen
  \bibfield  {author} {\bibinfo {author} {\bibfnamefont {E.}~\bibnamefont
  {Barberio}}, \bibinfo {author} {\bibfnamefont {B.}~\bibnamefont {van Eijk}},
  \ and\ \bibinfo {author} {\bibfnamefont {Z.}~\bibnamefont {Was}},\ }\href
  {\doibase 10.1016/0010-4655(91)90012-A} {\bibfield  {journal} {\bibinfo
  {journal} {Comput. Phys. Commun.}\ }\textbf {\bibinfo {volume} {66}},\
  \bibinfo {pages} {115} (\bibinfo {year} {1991})}\BibitemShut {NoStop}%
\bibitem [{\citenamefont {Caffo}\ \emph {et~al.}(1997)\citenamefont {Caffo},
  \citenamefont {Czy\.z},\ and\ \citenamefont {Remiddi}}]{Caffo:1997yy}%
  \BibitemOpen
  \bibfield  {author} {\bibinfo {author} {\bibfnamefont {M.}~\bibnamefont
  {Caffo}}, \bibinfo {author} {\bibfnamefont {H.}~\bibnamefont {Czy\.z}}, \
  and\ \bibinfo {author} {\bibfnamefont {E.}~\bibnamefont {Remiddi}},\ }\href
  {\doibase 10.1007/BF03035898} {\bibfield  {journal} {\bibinfo  {journal}
  {Nuovo Cim.}\ }\textbf {\bibinfo {volume} {A110}},\ \bibinfo {pages} {515}
  (\bibinfo {year} {1997})},\ \Eprint {http://arxiv.org/abs/hep-ph/9704443}
  {arXiv:hep-ph/9704443 [hep-ph]}\BibitemShut {NoStop}%
\bibitem [{\citenamefont {Akhmetshin}\ \emph {et~al.}(2002)\citenamefont
  {Akhmetshin} \emph {et~al.}}]{Akhmetshin:2001ig}%
  \BibitemOpen
  \bibfield  {author} {\bibinfo {author} {\bibfnamefont {R.~R.}\ \bibnamefont
  {Akhmetshin}} \emph {et~al.} (\bibinfo {collaboration} {CMD-2}),\ }\href
  {\doibase 10.1016/S0370-2693(02)01168-1} {\bibfield  {journal} {\bibinfo
  {journal} {Phys. Lett.}\ }\textbf {\bibinfo {volume} {B527}},\ \bibinfo
  {pages} {161} (\bibinfo {year} {2002})},\ \Eprint
  {http://arxiv.org/abs/hep-ex/0112031} {arXiv:hep-ex/0112031
  [hep-ex]}\BibitemShut {NoStop}%
\bibitem [{\citenamefont {Achasov}\ \emph {et~al.}(2005)\citenamefont {Achasov}
  \emph {et~al.}}]{Achasov:2005rg}%
  \BibitemOpen
  \bibfield  {author} {\bibinfo {author} {\bibfnamefont {M.~N.}\ \bibnamefont
  {Achasov}} \emph {et~al.} (\bibinfo {collaboration} {SND}),\ }\href {\doibase
  10.1134/1.2163921} {\bibfield  {journal} {\bibinfo  {journal} {J. Exp. Theor.
  Phys.}\ }\textbf {\bibinfo {volume} {101}},\ \bibinfo {pages} {1053}
  (\bibinfo {year} {2005})},\ \bibinfo {note} {[Zh. Eksp. Teor. Fiz. {\bf 128},
  1201 (2005)]},\ \Eprint {http://arxiv.org/abs/hep-ex/0506076}
  {arXiv:hep-ex/0506076 [hep-ex]}\BibitemShut {NoStop}%
\bibitem [{\citenamefont {Akhmetshin}\ \emph
  {et~al.}(1999{\natexlab{a}})\citenamefont {Akhmetshin} \emph
  {et~al.}}]{Akhmetshin:1999ym}%
  \BibitemOpen
  \bibfield  {author} {\bibinfo {author} {\bibfnamefont {R.~R.}\ \bibnamefont
  {Akhmetshin}} \emph {et~al.} (\bibinfo {collaboration} {CMD-2}),\ }\href
  {\doibase 10.1016/S0370-2693(99)00973-9} {\bibfield  {journal} {\bibinfo
  {journal} {Phys. Lett.}\ }\textbf {\bibinfo {volume} {B466}},\ \bibinfo
  {pages} {385} (\bibinfo {year} {1999}{\natexlab{a}})},\ \bibinfo {note}
  {[Erratum: Phys. Lett. {\bf B508}, 217 (2001)]},\ \Eprint
  {http://arxiv.org/abs/hep-ex/9906032} {arXiv:hep-ex/9906032
  [hep-ex]}\BibitemShut {NoStop}%
\bibitem [{\citenamefont {Achasov}\ \emph {et~al.}(2007)\citenamefont {Achasov}
  \emph {et~al.}}]{Achasov:2007kg}%
  \BibitemOpen
  \bibfield  {author} {\bibinfo {author} {\bibfnamefont {M.~N.}\ \bibnamefont
  {Achasov}} \emph {et~al.} (\bibinfo {collaboration} {SND}),\ }\href {\doibase
  10.1103/PhysRevD.76.072012} {\bibfield  {journal} {\bibinfo  {journal} {Phys.
  Rev.}\ }\textbf {\bibinfo {volume} {D76}},\ \bibinfo {pages} {072012}
  (\bibinfo {year} {2007})},\ \Eprint {http://arxiv.org/abs/0707.2279}
  {arXiv:0707.2279 [hep-ex]}\BibitemShut {NoStop}%
\bibitem [{\citenamefont {Aul'chenko}\ \emph {et~al.}(2015)\citenamefont
  {Aul'chenko} \emph {et~al.}}]{Aulchenko:2015mwt}%
  \BibitemOpen
  \bibfield  {author} {\bibinfo {author} {\bibfnamefont {V.~M.}\ \bibnamefont
  {Aul'chenko}} \emph {et~al.} (\bibinfo {collaboration} {SND}),\ }\href
  {\doibase 10.1134/S1063776115060023} {\bibfield  {journal} {\bibinfo
  {journal} {J. Exp. Theor. Phys.}\ }\textbf {\bibinfo {volume} {121}},\
  \bibinfo {pages} {27} (\bibinfo {year} {2015})},\ \bibinfo {note} {[Zh. Eksp.
  Teor. Fiz. {\bf 148}, 34 (2015)]}\BibitemShut {NoStop}%
\bibitem [{\citenamefont {Achasov}\ \emph
  {et~al.}(2014{\natexlab{b}})\citenamefont {Achasov} \emph
  {et~al.}}]{Achasov:2014xsa}%
  \BibitemOpen
  \bibfield  {author} {\bibinfo {author} {\bibfnamefont {M.~N.}\ \bibnamefont
  {Achasov}} \emph {et~al.} (\bibinfo {collaboration} {CMD-3, SND}),\ }\href
  {\doibase 10.1051/epjconf/20147100121} {\bibfield  {journal} {\bibinfo
  {journal} {EPJ Web Conf.}\ }\textbf {\bibinfo {volume} {71}},\ \bibinfo
  {pages} {00121} (\bibinfo {year} {2014}{\natexlab{b}})}\BibitemShut {NoStop}%
\bibitem [{\citenamefont {Antonelli}\ \emph {et~al.}(1992)\citenamefont
  {Antonelli} \emph {et~al.}}]{Antonelli:1992jx}%
  \BibitemOpen
  \bibfield  {author} {\bibinfo {author} {\bibfnamefont {A.}~\bibnamefont
  {Antonelli}} \emph {et~al.} (\bibinfo {collaboration} {DM2}),\ }\href
  {\doibase 10.1007/BF01589702} {\bibfield  {journal} {\bibinfo  {journal} {Z.
  Phys.}\ }\textbf {\bibinfo {volume} {C56}},\ \bibinfo {pages} {15} (\bibinfo
  {year} {1992})}\BibitemShut {NoStop}%
\bibitem [{\citenamefont {Akhmetshin}\ \emph
  {et~al.}(2004{\natexlab{b}})\citenamefont {Akhmetshin} \emph
  {et~al.}}]{Akhmetshin:2004dy}%
  \BibitemOpen
  \bibfield  {author} {\bibinfo {author} {\bibfnamefont {R.~R.}\ \bibnamefont
  {Akhmetshin}} \emph {et~al.} (\bibinfo {collaboration} {CMD-2}),\ }\href
  {\doibase 10.1016/j.physletb.2004.05.056} {\bibfield  {journal} {\bibinfo
  {journal} {Phys. Lett.}\ }\textbf {\bibinfo {volume} {B595}},\ \bibinfo
  {pages} {101} (\bibinfo {year} {2004}{\natexlab{b}})},\ \Eprint
  {http://arxiv.org/abs/hep-ex/0404019} {arXiv:hep-ex/0404019
  [hep-ex]}\BibitemShut {NoStop}%
\bibitem [{\citenamefont {Achasov}\ \emph
  {et~al.}(2003{\natexlab{a}})\citenamefont {Achasov} \emph
  {et~al.}}]{Achasov:2003bv}%
  \BibitemOpen
  \bibfield  {author} {\bibinfo {author} {\bibfnamefont {M.~N.}\ \bibnamefont
  {Achasov}} \emph {et~al.} (\bibinfo {collaboration} {SND}),\ }\href {\doibase
  10.1134/1.1581933} {\bibfield  {journal} {\bibinfo  {journal} {J. Exp. Theor.
  Phys.}\ }\textbf {\bibinfo {volume} {96}},\ \bibinfo {pages} {789} (\bibinfo
  {year} {2003}{\natexlab{a}})},\ \bibinfo {note} {[Zh. Eksp. Teor. Fiz. {\bf
  123}, 899 (2003)]}\BibitemShut {NoStop}%
\bibitem [{\citenamefont {Akhmetshin}\ \emph
  {et~al.}(1999{\natexlab{b}})\citenamefont {Akhmetshin} \emph
  {et~al.}}]{Akhmetshin:1998df}%
  \BibitemOpen
  \bibfield  {author} {\bibinfo {author} {\bibfnamefont {R.~R.}\ \bibnamefont
  {Akhmetshin}} \emph {et~al.} (\bibinfo {collaboration} {CMD-2}),\ }\href
  {\doibase 10.1016/S0370-2693(99)01080-1} {\bibfield  {journal} {\bibinfo
  {journal} {Phys. Lett.}\ }\textbf {\bibinfo {volume} {B466}},\ \bibinfo
  {pages} {392} (\bibinfo {year} {1999}{\natexlab{b}})},\ \Eprint
  {http://arxiv.org/abs/hep-ex/9904024} {arXiv:hep-ex/9904024
  [hep-ex]}\BibitemShut {NoStop}%
\bibitem [{\citenamefont {Akhmetshin}\ \emph
  {et~al.}(2017{\natexlab{b}})\citenamefont {Akhmetshin} \emph
  {et~al.}}]{Akhmetshin:2016dtr}%
  \BibitemOpen
  \bibfield  {author} {\bibinfo {author} {\bibfnamefont {R.~R.}\ \bibnamefont
  {Akhmetshin}} \emph {et~al.} (\bibinfo {collaboration} {CMD-3}),\ }\href
  {\doibase 10.1016/j.physletb.2017.03.022} {\bibfield  {journal} {\bibinfo
  {journal} {Phys. Lett.}\ }\textbf {\bibinfo {volume} {B768}},\ \bibinfo
  {pages} {345} (\bibinfo {year} {2017}{\natexlab{b}})},\ \Eprint
  {http://arxiv.org/abs/1612.04483} {arXiv:1612.04483 [hep-ex]}\BibitemShut
  {NoStop}%
\bibitem [{\citenamefont {Achasov}\ \emph
  {et~al.}(2018{\natexlab{c}})\citenamefont {Achasov} \emph
  {et~al.}}]{Achasov:2017kqm}%
  \BibitemOpen
  \bibfield  {author} {\bibinfo {author} {\bibfnamefont {M.~N.}\ \bibnamefont
  {Achasov}} \emph {et~al.} (\bibinfo {collaboration} {SND}),\ }\href {\doibase
  10.1103/PhysRevD.97.012008} {\bibfield  {journal} {\bibinfo  {journal} {Phys.
  Rev.}\ }\textbf {\bibinfo {volume} {D97}},\ \bibinfo {pages} {012008}
  (\bibinfo {year} {2018}{\natexlab{c}})},\ \Eprint
  {http://arxiv.org/abs/1711.08862} {arXiv:1711.08862 [hep-ex]}\BibitemShut
  {NoStop}%
\bibitem [{\citenamefont {Gribanov}\ \emph {et~al.}(2020)\citenamefont
  {Gribanov} \emph {et~al.}}]{Gribanov:2019qgw}%
  \BibitemOpen
  \bibfield  {author} {\bibinfo {author} {\bibfnamefont {S.~S.}\ \bibnamefont
  {Gribanov}} \emph {et~al.} (\bibinfo {collaboration} {CMD-3}),\ }\href
  {\doibase 10.1007/JHEP01(2020)112} {\bibfield  {journal} {\bibinfo  {journal}
  {JHEP}\ }\textbf {\bibinfo {volume} {01}},\ \bibinfo {pages} {112} (\bibinfo
  {year} {2020})},\ \Eprint {http://arxiv.org/abs/1907.08002} {arXiv:1907.08002
  [hep-ex]}\BibitemShut {NoStop}%
\bibitem [{\citenamefont {Ablikim}\ \emph
  {et~al.}(2016{\natexlab{b}})\citenamefont {Ablikim} \emph
  {et~al.}}]{Ablikim:2016xbg}%
  \BibitemOpen
  \bibfield  {author} {\bibinfo {author} {\bibfnamefont {M.}~\bibnamefont
  {Ablikim}} \emph {et~al.} (\bibinfo {collaboration} {BESIII}),\ }\href
  {\doibase 10.1016/j.physletb.2016.08.011} {\bibfield  {journal} {\bibinfo
  {journal} {Phys. Lett.}\ }\textbf {\bibinfo {volume} {B761}},\ \bibinfo
  {pages} {98} (\bibinfo {year} {2016}{\natexlab{b}})},\ \Eprint
  {http://arxiv.org/abs/1604.01924} {arXiv:1604.01924 [hep-ex]}\BibitemShut
  {NoStop}%
\bibitem [{\citenamefont {Patrignani}\ \emph {et~al.}(2016)\citenamefont
  {Patrignani} \emph {et~al.}}]{Patrignani:2016xqp}%
  \BibitemOpen
  \bibfield  {author} {\bibinfo {author} {\bibfnamefont {C.}~\bibnamefont
  {Patrignani}} \emph {et~al.} (\bibinfo {collaboration} {Particle Data
  Group}),\ }\href {\doibase 10.1088/1674-1137/40/10/100001} {\bibfield
  {journal} {\bibinfo  {journal} {Chin. Phys.}\ }\textbf {\bibinfo {volume}
  {C40}},\ \bibinfo {pages} {100001} (\bibinfo {year} {2016})}\BibitemShut
  {NoStop}%
\bibitem [{\citenamefont {Anashin}\ \emph {et~al.}(2017)\citenamefont {Anashin}
  \emph {et~al.}}]{Anashin:2016hmv}%
  \BibitemOpen
  \bibfield  {author} {\bibinfo {author} {\bibfnamefont {V.~V.}\ \bibnamefont
  {Anashin}} \emph {et~al.} (\bibinfo {collaboration} {KEDR}),\ }\href
  {\doibase 10.1016/j.physletb.2017.04.073} {\bibfield  {journal} {\bibinfo
  {journal} {Phys. Lett.}\ }\textbf {\bibinfo {volume} {B770}},\ \bibinfo
  {pages} {174} (\bibinfo {year} {2017})},\ \Eprint
  {http://arxiv.org/abs/1610.02827} {arXiv:1610.02827 [hep-ex]}\BibitemShut
  {NoStop}%
\bibitem [{\citenamefont {Pais}(1960)}]{Pais:1960zz}%
  \BibitemOpen
  \bibfield  {author} {\bibinfo {author} {\bibfnamefont {A.}~\bibnamefont
  {Pais}},\ }\href {\doibase 10.1016/0003-4916(60)90108-1} {\bibfield
  {journal} {\bibinfo  {journal} {Annals Phys.}\ }\textbf {\bibinfo {volume}
  {9}},\ \bibinfo {pages} {548} (\bibinfo {year} {1960})}\BibitemShut {NoStop}%
\bibitem [{\citenamefont {Davier}\ \emph {et~al.}(2011)\citenamefont {Davier},
  \citenamefont {Hoecker}, \citenamefont {Malaescu},\ and\ \citenamefont
  {Zhang}}]{Davier:2010nc}%
  \BibitemOpen
  \bibfield  {author} {\bibinfo {author} {\bibfnamefont {M.}~\bibnamefont
  {Davier}}, \bibinfo {author} {\bibfnamefont {A.}~\bibnamefont {Hoecker}},
  \bibinfo {author} {\bibfnamefont {B.}~\bibnamefont {Malaescu}}, \ and\
  \bibinfo {author} {\bibfnamefont {Z.}~\bibnamefont {Zhang}},\ }\href
  {\doibase 10.1140/epjc/s10052-010-1515-z} {\bibfield  {journal} {\bibinfo
  {journal} {Eur. Phys. J.}\ }\textbf {\bibinfo {volume} {C71}},\ \bibinfo
  {pages} {1515} (\bibinfo {year} {2011})},\ \bibinfo {note} {[Erratum: Eur.
  Phys. J. {\bf C72}, 1874 (2012)]},\ \Eprint {http://arxiv.org/abs/1010.4180}
  {arXiv:1010.4180 [hep-ph]}\BibitemShut {NoStop}%
\bibitem [{\citenamefont {Malaescu}(2018{\natexlab{a}})}]{bogdan-kek-2018}%
  \BibitemOpen
  \bibfield  {author} {\bibinfo {author} {\bibfnamefont {B.}~\bibnamefont
  {Malaescu}},\ }\href@noop {} {\enquote {\bibinfo {title} {{KLOE} data
  comparisons},}\ }\bibinfo {howpublished}
  {\url{https://kds.kek.jp/indico/event/26780/session/9/material/0/0.pdf}}
  (\bibinfo {year} {2018}{\natexlab{a}}),\ \bibinfo {note} {{Muon} $g-2$
  {Theory Initiative workshop KEK}}\BibitemShut {NoStop}%
\bibitem [{\citenamefont {Achasov}\ \emph
  {et~al.}(2020{\natexlab{a}})\citenamefont {Achasov} \emph
  {et~al.}}]{Achasov:2020iys}%
  \BibitemOpen
  \bibfield  {author} {\bibinfo {author} {\bibfnamefont {M.}~\bibnamefont
  {Achasov}} \emph {et~al.} (\bibinfo {collaboration} {SND}),\ }\href@noop {}
  {\  (\bibinfo {year} {2020}{\natexlab{a}})},\ \Eprint
  {http://arxiv.org/abs/2004.00263} {arXiv:2004.00263 [hep-ex]}\BibitemShut
  {NoStop}%
\bibitem [{\citenamefont {Davier}(2018)}]{davier-kek-2018}%
  \BibitemOpen
  \bibfield  {author} {\bibinfo {author} {\bibfnamefont {M.}~\bibnamefont
  {Davier}},\ }\href@noop {} {\enquote {\bibinfo {title} {{$e^+e^-$ results
  from BABAR and implications for the muon $g-2$}},}\ }\bibinfo {howpublished}
  {\url{https://kds.kek.jp/indico/event/26780/session/5/contribution/10/material/slides/0.pdf}}
  (\bibinfo {year} {2018}),\ \bibinfo {note} {{Muon $g-2$ Theory Initiative
  workshop KEK}}\BibitemShut {NoStop}%
\bibitem [{\citenamefont {Ivanov}(2019)}]{cmd3-eps}%
  \BibitemOpen
  \bibfield  {author} {\bibinfo {author} {\bibfnamefont {V.}~\bibnamefont
  {Ivanov}},\ }\href@noop {} {\enquote {\bibinfo {title} {{Measurement of
  hadronic cross section with CMD-3}},}\ }\bibinfo {howpublished}
  {\url{https://indico.cern.ch/event/577856/contributions/3420236/attachments/1879432/3096048/Ivanov_EPSHEP.pdf}}
  (\bibinfo {year} {2019}),\ \bibinfo {note} {{EPS HEP Ghent}}\BibitemShut
  {NoStop}%
\bibitem [{\citenamefont {Druzhinin}(2019)}]{snd-eps}%
  \BibitemOpen
  \bibfield  {author} {\bibinfo {author} {\bibfnamefont {V.}~\bibnamefont
  {Druzhinin}},\ }\href@noop {} {\enquote {\bibinfo {title} {Study of $e^+e^-$
  annihilation into hadrons with the {SND} detector at the {VEPP}-2000
  collider},}\ }\bibinfo {howpublished}
  {\url{https://indico.cern.ch/event/577856/contributions/3420234/attachments/1879251/3095727/druzhinin_epshep2019.pdf}}
  (\bibinfo {year} {2019}),\ \bibinfo {note} {{EPS HEP Ghent}}\BibitemShut
  {NoStop}%
\bibitem [{\citenamefont {Alemany}\ \emph {et~al.}(1998)\citenamefont
  {Alemany}, \citenamefont {Davier},\ and\ \citenamefont
  {Hoecker}}]{Alemany:1997tn}%
  \BibitemOpen
  \bibfield  {author} {\bibinfo {author} {\bibfnamefont {R.}~\bibnamefont
  {Alemany}}, \bibinfo {author} {\bibfnamefont {M.}~\bibnamefont {Davier}}, \
  and\ \bibinfo {author} {\bibfnamefont {A.}~\bibnamefont {Hoecker}},\ }\href
  {\doibase 10.1007/s100520050127} {\bibfield  {journal} {\bibinfo  {journal}
  {Eur. Phys. J.}\ }\textbf {\bibinfo {volume} {C2}},\ \bibinfo {pages} {123}
  (\bibinfo {year} {1998})},\ \Eprint {http://arxiv.org/abs/hep-ph/9703220}
  {arXiv:hep-ph/9703220 [hep-ph]}\BibitemShut {NoStop}%
\bibitem [{\citenamefont {Davier}\ \emph
  {et~al.}(2010{\natexlab{a}})\citenamefont {Davier}, \citenamefont {Hoecker},
  \citenamefont {L\'opez~Castro}, \citenamefont {Malaescu}, \citenamefont {Mo},
  \citenamefont {Toledo~S\'anchez}, \citenamefont {Wang}, \citenamefont
  {Yuan},\ and\ \citenamefont {Zhang}}]{Davier:2009ag}%
  \BibitemOpen
  \bibfield  {author} {\bibinfo {author} {\bibfnamefont {M.}~\bibnamefont
  {Davier}}, \bibinfo {author} {\bibfnamefont {A.}~\bibnamefont {Hoecker}},
  \bibinfo {author} {\bibfnamefont {G.}~\bibnamefont {L\'opez~Castro}},
  \bibinfo {author} {\bibfnamefont {B.}~\bibnamefont {Malaescu}}, \bibinfo
  {author} {\bibfnamefont {X.~H.}\ \bibnamefont {Mo}}, \bibinfo {author}
  {\bibfnamefont {G.}~\bibnamefont {Toledo~S\'anchez}}, \bibinfo {author}
  {\bibfnamefont {P.}~\bibnamefont {Wang}}, \bibinfo {author} {\bibfnamefont
  {C.~Z.}\ \bibnamefont {Yuan}}, \ and\ \bibinfo {author} {\bibfnamefont
  {Z.}~\bibnamefont {Zhang}},\ }\href {\doibase 10.1140/epjc/s10052-009-1219-4}
  {\bibfield  {journal} {\bibinfo  {journal} {Eur. Phys. J.}\ }\textbf
  {\bibinfo {volume} {C66}},\ \bibinfo {pages} {127} (\bibinfo {year}
  {2010}{\natexlab{a}})},\ \Eprint {http://arxiv.org/abs/0906.5443}
  {arXiv:0906.5443 [hep-ph]}\BibitemShut {NoStop}%
\bibitem [{\citenamefont {Jegerlehner}()}]{fred}%
  \BibitemOpen
  \bibfield  {author} {\bibinfo {author} {\bibfnamefont {F.}~\bibnamefont
  {Jegerlehner}},\ }\href@noop {} {}\bibinfo {howpublished} {private
  communication}\BibitemShut {NoStop}%
\bibitem [{\citenamefont {Jegerlehner}\ and\ \citenamefont
  {Szafron}(2011)}]{Jegerlehner:2011ti}%
  \BibitemOpen
  \bibfield  {author} {\bibinfo {author} {\bibfnamefont {F.}~\bibnamefont
  {Jegerlehner}}\ and\ \bibinfo {author} {\bibfnamefont {R.}~\bibnamefont
  {Szafron}},\ }\href {\doibase 10.1140/epjc/s10052-011-1632-3} {\bibfield
  {journal} {\bibinfo  {journal} {Eur. Phys. J.}\ }\textbf {\bibinfo {volume}
  {C71}},\ \bibinfo {pages} {1632} (\bibinfo {year} {2011})},\ \Eprint
  {http://arxiv.org/abs/1101.2872} {arXiv:1101.2872 [hep-ph]}\BibitemShut
  {NoStop}%
\bibitem [{\citenamefont {Zhang}(2016)}]{Zhang:2015yfi}%
  \BibitemOpen
  \bibfield  {author} {\bibinfo {author} {\bibfnamefont {Z.}~\bibnamefont
  {Zhang}},\ }\href {\doibase 10.1051/epjconf/201611801036} {\bibfield
  {journal} {\bibinfo  {journal} {EPJ Web Conf.}\ }\textbf {\bibinfo {volume}
  {118}},\ \bibinfo {pages} {01036} (\bibinfo {year} {2016})},\ \Eprint
  {http://arxiv.org/abs/1511.05405} {arXiv:1511.05405 [hep-ph]}\BibitemShut
  {NoStop}%
\bibitem [{\citenamefont {Schael}\ \emph {et~al.}(2005)\citenamefont {Schael}
  \emph {et~al.}}]{Schael:2005am}%
  \BibitemOpen
  \bibfield  {author} {\bibinfo {author} {\bibfnamefont {S.}~\bibnamefont
  {Schael}} \emph {et~al.} (\bibinfo {collaboration} {ALEPH}),\ }\href
  {\doibase 10.1016/j.physrep.2005.06.007} {\bibfield  {journal} {\bibinfo
  {journal} {Phys. Rept.}\ }\textbf {\bibinfo {volume} {421}},\ \bibinfo
  {pages} {191} (\bibinfo {year} {2005})},\ \Eprint
  {http://arxiv.org/abs/hep-ex/0506072} {arXiv:hep-ex/0506072
  [hep-ex]}\BibitemShut {NoStop}%
\bibitem [{\citenamefont {Acciarri}\ \emph {et~al.}(1995)\citenamefont
  {Acciarri} \emph {et~al.}}]{Acciarri:1994vr}%
  \BibitemOpen
  \bibfield  {author} {\bibinfo {author} {\bibfnamefont {M.}~\bibnamefont
  {Acciarri}} \emph {et~al.} (\bibinfo {collaboration} {L3}),\ }\href {\doibase
  10.1016/0370-2693(94)01587-3} {\bibfield  {journal} {\bibinfo  {journal}
  {Phys. Lett.}\ }\textbf {\bibinfo {volume} {B345}},\ \bibinfo {pages} {93}
  (\bibinfo {year} {1995})}\BibitemShut {NoStop}%
\bibitem [{\citenamefont {Ackerstaff}\ \emph {et~al.}(1999)\citenamefont
  {Ackerstaff} \emph {et~al.}}]{Ackerstaff:1998yj}%
  \BibitemOpen
  \bibfield  {author} {\bibinfo {author} {\bibfnamefont {K.}~\bibnamefont
  {Ackerstaff}} \emph {et~al.} (\bibinfo {collaboration} {OPAL}),\ }\href
  {\doibase 10.1007/s100520050430, 10.1007/s100529901061} {\bibfield  {journal}
  {\bibinfo  {journal} {Eur. Phys. J.}\ }\textbf {\bibinfo {volume} {C7}},\
  \bibinfo {pages} {571} (\bibinfo {year} {1999})},\ \Eprint
  {http://arxiv.org/abs/hep-ex/9808019} {arXiv:hep-ex/9808019
  [hep-ex]}\BibitemShut {NoStop}%
\bibitem [{\citenamefont {Anderson}\ \emph {et~al.}(2000)\citenamefont
  {Anderson} \emph {et~al.}}]{Anderson:1999ui}%
  \BibitemOpen
  \bibfield  {author} {\bibinfo {author} {\bibfnamefont {S.}~\bibnamefont
  {Anderson}} \emph {et~al.} (\bibinfo {collaboration} {CLEO}),\ }\href
  {\doibase 10.1103/PhysRevD.61.112002} {\bibfield  {journal} {\bibinfo
  {journal} {Phys. Rev.}\ }\textbf {\bibinfo {volume} {D61}},\ \bibinfo {pages}
  {112002} (\bibinfo {year} {2000})},\ \Eprint
  {http://arxiv.org/abs/hep-ex/9910046} {arXiv:hep-ex/9910046
  [hep-ex]}\BibitemShut {NoStop}%
\bibitem [{\citenamefont {Abdallah}\ \emph {et~al.}(2006)\citenamefont
  {Abdallah} \emph {et~al.}}]{Abdallah:2003cw}%
  \BibitemOpen
  \bibfield  {author} {\bibinfo {author} {\bibfnamefont {J.}~\bibnamefont
  {Abdallah}} \emph {et~al.} (\bibinfo {collaboration} {DELPHI}),\ }\href
  {\doibase 10.1140/epjc/s2006-02494-9} {\bibfield  {journal} {\bibinfo
  {journal} {Eur. Phys. J.}\ }\textbf {\bibinfo {volume} {C46}},\ \bibinfo
  {pages} {1} (\bibinfo {year} {2006})},\ \Eprint
  {http://arxiv.org/abs/hep-ex/0603044} {arXiv:hep-ex/0603044
  [hep-ex]}\BibitemShut {NoStop}%
\bibitem [{\citenamefont {Fujikawa}\ \emph {et~al.}(2008)\citenamefont
  {Fujikawa} \emph {et~al.}}]{Fujikawa:2008ma}%
  \BibitemOpen
  \bibfield  {author} {\bibinfo {author} {\bibfnamefont {M.}~\bibnamefont
  {Fujikawa}} \emph {et~al.} (\bibinfo {collaboration} {Belle}),\ }\href
  {\doibase 10.1103/PhysRevD.78.072006} {\bibfield  {journal} {\bibinfo
  {journal} {Phys. Rev.}\ }\textbf {\bibinfo {volume} {D78}},\ \bibinfo {pages}
  {072006} (\bibinfo {year} {2008})},\ \Eprint {http://arxiv.org/abs/0805.3773}
  {arXiv:0805.3773 [hep-ex]}\BibitemShut {NoStop}%
\bibitem [{\citenamefont {Schwinger}(1989)}]{jss89}%
  \BibitemOpen
  \bibfield  {author} {\bibinfo {author} {\bibfnamefont {J.}~\bibnamefont
  {Schwinger}},\ }\href@noop {} {\emph {\bibinfo {title} {Particles, Sources
  and Fields}}},\ Vol.\ \bibinfo {volume} {III}\ (\bibinfo  {publisher}
  {Addison-Wesley, Reading},\ \bibinfo {year} {1989})\BibitemShut {NoStop}%
\bibitem [{\citenamefont {Drees}\ and\ \citenamefont
  {Hikasa}(1990)}]{Drees:1990te}%
  \BibitemOpen
  \bibfield  {author} {\bibinfo {author} {\bibfnamefont {M.}~\bibnamefont
  {Drees}}\ and\ \bibinfo {author} {\bibfnamefont {K.-I.}\ \bibnamefont
  {Hikasa}},\ }\href {\doibase 10.1016/0370-2693(90)91094-R} {\bibfield
  {journal} {\bibinfo  {journal} {Phys. Lett.}\ }\textbf {\bibinfo {volume}
  {B252}},\ \bibinfo {pages} {127} (\bibinfo {year} {1990})}\BibitemShut
  {NoStop}%
\bibitem [{\citenamefont {Flores-Ba\'ez}\ \emph {et~al.}(2007)\citenamefont
  {Flores-Ba\'ez}, \citenamefont {L\'opez~Castro},\ and\ \citenamefont
  {Toledo~S\'anchez}}]{FloresBaez:2007es}%
  \BibitemOpen
  \bibfield  {author} {\bibinfo {author} {\bibfnamefont {F.~V.}\ \bibnamefont
  {Flores-Ba\'ez}}, \bibinfo {author} {\bibfnamefont {G.}~\bibnamefont
  {L\'opez~Castro}}, \ and\ \bibinfo {author} {\bibfnamefont {G.}~\bibnamefont
  {Toledo~S\'anchez}},\ }\href {\doibase 10.1103/PhysRevD.76.096010} {\bibfield
   {journal} {\bibinfo  {journal} {Phys. Rev.}\ }\textbf {\bibinfo {volume}
  {D76}},\ \bibinfo {pages} {096010} (\bibinfo {year} {2007})},\ \Eprint
  {http://arxiv.org/abs/0708.3256} {arXiv:0708.3256 [hep-ph]}\BibitemShut
  {NoStop}%
\bibitem [{\citenamefont {Gounaris}\ and\ \citenamefont
  {Sakurai}(1968)}]{Gounaris:1968mw}%
  \BibitemOpen
  \bibfield  {author} {\bibinfo {author} {\bibfnamefont {G.~J.}\ \bibnamefont
  {Gounaris}}\ and\ \bibinfo {author} {\bibfnamefont {J.~J.}\ \bibnamefont
  {Sakurai}},\ }\href {\doibase 10.1103/PhysRevLett.21.244} {\bibfield
  {journal} {\bibinfo  {journal} {Phys. Rev. Lett.}\ }\textbf {\bibinfo
  {volume} {21}},\ \bibinfo {pages} {244} (\bibinfo {year} {1968})}\BibitemShut
  {NoStop}%
\bibitem [{\citenamefont {Davier}\ \emph
  {et~al.}(2010{\natexlab{b}})\citenamefont {Davier}, \citenamefont {Hoecker},
  \citenamefont {Malaescu}, \citenamefont {Yuan},\ and\ \citenamefont
  {Zhang}}]{Davier:2009zi}%
  \BibitemOpen
  \bibfield  {author} {\bibinfo {author} {\bibfnamefont {M.}~\bibnamefont
  {Davier}}, \bibinfo {author} {\bibfnamefont {A.}~\bibnamefont {Hoecker}},
  \bibinfo {author} {\bibfnamefont {B.}~\bibnamefont {Malaescu}}, \bibinfo
  {author} {\bibfnamefont {C.~Z.}\ \bibnamefont {Yuan}}, \ and\ \bibinfo
  {author} {\bibfnamefont {Z.}~\bibnamefont {Zhang}},\ }\href {\doibase
  10.1140/epjc/s10052-010-1246-1} {\bibfield  {journal} {\bibinfo  {journal}
  {Eur. Phys. J.}\ }\textbf {\bibinfo {volume} {C66}},\ \bibinfo {pages} {1}
  (\bibinfo {year} {2010}{\natexlab{b}})},\ \Eprint
  {http://arxiv.org/abs/0908.4300} {arXiv:0908.4300 [hep-ph]}\BibitemShut
  {NoStop}%
\bibitem [{\citenamefont {Davier}\ \emph {et~al.}(2014)\citenamefont {Davier},
  \citenamefont {Hoecker}, \citenamefont {Malaescu}, \citenamefont {Yuan},\
  and\ \citenamefont {Zhang}}]{Davier:2013sfa}%
  \BibitemOpen
  \bibfield  {author} {\bibinfo {author} {\bibfnamefont {M.}~\bibnamefont
  {Davier}}, \bibinfo {author} {\bibfnamefont {A.}~\bibnamefont {Hoecker}},
  \bibinfo {author} {\bibfnamefont {B.}~\bibnamefont {Malaescu}}, \bibinfo
  {author} {\bibfnamefont {C.-Z.}\ \bibnamefont {Yuan}}, \ and\ \bibinfo
  {author} {\bibfnamefont {Z.}~\bibnamefont {Zhang}},\ }\href {\doibase
  10.1140/epjc/s10052-014-2803-9} {\bibfield  {journal} {\bibinfo  {journal}
  {Eur. Phys. J.}\ }\textbf {\bibinfo {volume} {C74}},\ \bibinfo {pages} {2803}
  (\bibinfo {year} {2014})},\ \Eprint {http://arxiv.org/abs/1312.1501}
  {arXiv:1312.1501 [hep-ex]}\BibitemShut {NoStop}%
\bibitem [{\citenamefont {Arbuzov}\ \emph {et~al.}(2006)\citenamefont
  {Arbuzov}, \citenamefont {Fedotovich}, \citenamefont {Ignatov}, \citenamefont
  {Kuraev},\ and\ \citenamefont {Sibidanov}}]{Arbuzov:2005pt}%
  \BibitemOpen
  \bibfield  {author} {\bibinfo {author} {\bibfnamefont {A.~B.}\ \bibnamefont
  {Arbuzov}}, \bibinfo {author} {\bibfnamefont {G.~V.}\ \bibnamefont
  {Fedotovich}}, \bibinfo {author} {\bibfnamefont {F.~V.}\ \bibnamefont
  {Ignatov}}, \bibinfo {author} {\bibfnamefont {E.~A.}\ \bibnamefont {Kuraev}},
  \ and\ \bibinfo {author} {\bibfnamefont {A.~L.}\ \bibnamefont {Sibidanov}},\
  }\href {\doibase 10.1140/epjc/s2006-02532-8} {\bibfield  {journal} {\bibinfo
  {journal} {Eur. Phys. J.}\ }\textbf {\bibinfo {volume} {C46}},\ \bibinfo
  {pages} {689} (\bibinfo {year} {2006})},\ \Eprint
  {http://arxiv.org/abs/hep-ph/0504233} {arXiv:hep-ph/0504233
  [hep-ph]}\BibitemShut {NoStop}%
\bibitem [{\citenamefont {Jadach}\ \emph {et~al.}(2000)\citenamefont {Jadach},
  \citenamefont {Ward},\ and\ \citenamefont {Was}}]{Jadach:1999vf}%
  \BibitemOpen
  \bibfield  {author} {\bibinfo {author} {\bibfnamefont {S.}~\bibnamefont
  {Jadach}}, \bibinfo {author} {\bibfnamefont {B.~F.~L.}\ \bibnamefont {Ward}},
  \ and\ \bibinfo {author} {\bibfnamefont {Z.}~\bibnamefont {Was}},\ }\href
  {\doibase 10.1016/S0010-4655(00)00048-5} {\bibfield  {journal} {\bibinfo
  {journal} {Comput. Phys. Commun.}\ }\textbf {\bibinfo {volume} {130}},\
  \bibinfo {pages} {260} (\bibinfo {year} {2000})},\ \Eprint
  {http://arxiv.org/abs/hep-ph/9912214} {arXiv:hep-ph/9912214
  [hep-ph]}\BibitemShut {NoStop}%
\bibitem [{\citenamefont {Czy\.z}\ \emph {et~al.}(2003)\citenamefont {Czy\.z},
  \citenamefont {Grzeli\'nska}, \citenamefont {K{\"u}hn},\ and\ \citenamefont
  {Rodrigo}}]{Czyz:2002np}%
  \BibitemOpen
  \bibfield  {author} {\bibinfo {author} {\bibfnamefont {H.}~\bibnamefont
  {Czy\.z}}, \bibinfo {author} {\bibfnamefont {A.}~\bibnamefont
  {Grzeli\'nska}}, \bibinfo {author} {\bibfnamefont {J.~H.}\ \bibnamefont
  {K{\"u}hn}}, \ and\ \bibinfo {author} {\bibfnamefont {G.}~\bibnamefont
  {Rodrigo}},\ }\href {\doibase 10.1140/epjc/s2003-01134-4} {\bibfield
  {journal} {\bibinfo  {journal} {Eur. Phys. J.}\ }\textbf {\bibinfo {volume}
  {C27}},\ \bibinfo {pages} {563} (\bibinfo {year} {2003})},\ \Eprint
  {http://arxiv.org/abs/hep-ph/0212225} {arXiv:hep-ph/0212225
  [hep-ph]}\BibitemShut {NoStop}%
\bibitem [{\citenamefont {Jadach}(2005)}]{Jadach:2005gx}%
  \BibitemOpen
  \bibfield  {author} {\bibinfo {author} {\bibfnamefont {S.}~\bibnamefont
  {Jadach}},\ }\href
  {https://www.actaphys.uj.edu.pl/fulltext?series=Reg&vol=36&page=2387}
  {\bibfield  {journal} {\bibinfo  {journal} {Acta Phys. Polon.}\ }\textbf
  {\bibinfo {volume} {B36}},\ \bibinfo {pages} {2387} (\bibinfo {year}
  {2005})},\ \Eprint {http://arxiv.org/abs/hep-ph/0506180}
  {arXiv:hep-ph/0506180 [hep-ph]}\BibitemShut {NoStop}%
\bibitem [{\citenamefont {Czy\.z}\ and\ \citenamefont {Tracz}()}]{3ph}%
  \BibitemOpen
  \bibfield  {author} {\bibinfo {author} {\bibfnamefont {H.}~\bibnamefont
  {Czy\.z}}\ and\ \bibinfo {author} {\bibfnamefont {S.}~\bibnamefont {Tracz}},\
  }\href@noop {} {}\bibinfo {howpublished} {in preparation}\BibitemShut
  {NoStop}%
\bibitem [{\citenamefont {Czy\.z}\ \emph
  {et~al.}(2004{\natexlab{b}})\citenamefont {Czy\.z}, \citenamefont
  {Grzeli\'nska}, \citenamefont {K{\"u}hn},\ and\ \citenamefont
  {Rodrigo}}]{Czyz:2003ue}%
  \BibitemOpen
  \bibfield  {author} {\bibinfo {author} {\bibfnamefont {H.}~\bibnamefont
  {Czy\.z}}, \bibinfo {author} {\bibfnamefont {A.}~\bibnamefont
  {Grzeli\'nska}}, \bibinfo {author} {\bibfnamefont {J.~H.}\ \bibnamefont
  {K{\"u}hn}}, \ and\ \bibinfo {author} {\bibfnamefont {G.}~\bibnamefont
  {Rodrigo}},\ }\href {\doibase 10.1140/epjc/s2004-01605-0} {\bibfield
  {journal} {\bibinfo  {journal} {Eur. Phys. J.}\ }\textbf {\bibinfo {volume}
  {C33}},\ \bibinfo {pages} {333} (\bibinfo {year} {2004}{\natexlab{b}})},\
  \Eprint {http://arxiv.org/abs/hep-ph/0308312} {arXiv:hep-ph/0308312
  [hep-ph]}\BibitemShut {NoStop}%
\bibitem [{\citenamefont {Czy\.z}\ \emph
  {et~al.}(2005{\natexlab{b}})\citenamefont {Czy\.z}, \citenamefont
  {Grzeli\'nska},\ and\ \citenamefont {K{\"u}hn}}]{Czyz:2004nq}%
  \BibitemOpen
  \bibfield  {author} {\bibinfo {author} {\bibfnamefont {H.}~\bibnamefont
  {Czy\.z}}, \bibinfo {author} {\bibfnamefont {A.}~\bibnamefont
  {Grzeli\'nska}}, \ and\ \bibinfo {author} {\bibfnamefont {J.~H.}\
  \bibnamefont {K{\"u}hn}},\ }\href {\doibase 10.1016/j.physletb.2005.02.027}
  {\bibfield  {journal} {\bibinfo  {journal} {Phys. Lett.}\ }\textbf {\bibinfo
  {volume} {B611}},\ \bibinfo {pages} {116} (\bibinfo {year}
  {2005}{\natexlab{b}})},\ \Eprint {http://arxiv.org/abs/hep-ph/0412239}
  {arXiv:hep-ph/0412239 [hep-ph]}\BibitemShut {NoStop}%
\bibitem [{\citenamefont {Czy\.z}\ \emph {et~al.}(2010)\citenamefont {Czy\.z},
  \citenamefont {Grzeli\'nska},\ and\ \citenamefont {K{\"u}hn}}]{Czyz:2010hj}%
  \BibitemOpen
  \bibfield  {author} {\bibinfo {author} {\bibfnamefont {H.}~\bibnamefont
  {Czy\.z}}, \bibinfo {author} {\bibfnamefont {A.}~\bibnamefont
  {Grzeli\'nska}}, \ and\ \bibinfo {author} {\bibfnamefont {J.~H.}\
  \bibnamefont {K{\"u}hn}},\ }\href {\doibase 10.1103/PhysRevD.81.094014}
  {\bibfield  {journal} {\bibinfo  {journal} {Phys. Rev.}\ }\textbf {\bibinfo
  {volume} {D81}},\ \bibinfo {pages} {094014} (\bibinfo {year} {2010})},\
  \Eprint {http://arxiv.org/abs/1002.0279} {arXiv:1002.0279
  [hep-ph]}\BibitemShut {NoStop}%
\bibitem [{\citenamefont {Czy\.z}\ \emph {et~al.}(2014)\citenamefont {Czy\.z},
  \citenamefont {Kühn},\ and\ \citenamefont {Tracz}}]{Czyz:2014sha}%
  \BibitemOpen
  \bibfield  {author} {\bibinfo {author} {\bibfnamefont {H.}~\bibnamefont
  {Czy\.z}}, \bibinfo {author} {\bibfnamefont {J.~H.}\ \bibnamefont {Kühn}}, \
  and\ \bibinfo {author} {\bibfnamefont {S.}~\bibnamefont {Tracz}},\ }\href
  {\doibase 10.1103/PhysRevD.90.114021} {\bibfield  {journal} {\bibinfo
  {journal} {Phys. Rev.}\ }\textbf {\bibinfo {volume} {D90}},\ \bibinfo {pages}
  {114021} (\bibinfo {year} {2014})},\ \Eprint {http://arxiv.org/abs/1407.7995}
  {arXiv:1407.7995 [hep-ph]}\BibitemShut {NoStop}%
\bibitem [{\citenamefont {Campanario}\ \emph {et~al.}(2014)\citenamefont
  {Campanario}, \citenamefont {Czy\.z}, \citenamefont {Gluza}, \citenamefont
  {Gunia}, \citenamefont {Riemann}, \citenamefont {Rodrigo},\ and\
  \citenamefont {Yundin}}]{Campanario:2013uea}%
  \BibitemOpen
  \bibfield  {author} {\bibinfo {author} {\bibfnamefont {F.}~\bibnamefont
  {Campanario}}, \bibinfo {author} {\bibfnamefont {H.}~\bibnamefont {Czy\.z}},
  \bibinfo {author} {\bibfnamefont {J.}~\bibnamefont {Gluza}}, \bibinfo
  {author} {\bibfnamefont {M.}~\bibnamefont {Gunia}}, \bibinfo {author}
  {\bibfnamefont {T.}~\bibnamefont {Riemann}}, \bibinfo {author} {\bibfnamefont
  {G.}~\bibnamefont {Rodrigo}}, \ and\ \bibinfo {author} {\bibfnamefont
  {V.}~\bibnamefont {Yundin}},\ }\href {\doibase 10.1007/JHEP02(2014)114}
  {\bibfield  {journal} {\bibinfo  {journal} {JHEP}\ }\textbf {\bibinfo
  {volume} {02}},\ \bibinfo {pages} {114} (\bibinfo {year} {2014})},\ \Eprint
  {http://arxiv.org/abs/1312.3610} {arXiv:1312.3610 [hep-ph]}\BibitemShut
  {NoStop}%
\bibitem [{\citenamefont {Campanario}\ \emph {et~al.}(2019)\citenamefont
  {Campanario}, \citenamefont {Czy\.z}, \citenamefont {Gluza}, \citenamefont
  {Jeliński}, \citenamefont {Rodrigo}, \citenamefont {Tracz},\ and\
  \citenamefont {Zhuridov}}]{Campanario:2019mjh}%
  \BibitemOpen
  \bibfield  {author} {\bibinfo {author} {\bibfnamefont {F.}~\bibnamefont
  {Campanario}}, \bibinfo {author} {\bibfnamefont {H.}~\bibnamefont {Czy\.z}},
  \bibinfo {author} {\bibfnamefont {J.}~\bibnamefont {Gluza}}, \bibinfo
  {author} {\bibfnamefont {T.}~\bibnamefont {Jeliński}}, \bibinfo {author}
  {\bibfnamefont {G.}~\bibnamefont {Rodrigo}}, \bibinfo {author} {\bibfnamefont
  {S.}~\bibnamefont {Tracz}}, \ and\ \bibinfo {author} {\bibfnamefont
  {D.}~\bibnamefont {Zhuridov}},\ }\href {\doibase 10.1103/PhysRevD.100.076004}
  {\bibfield  {journal} {\bibinfo  {journal} {Phys. Rev.}\ }\textbf {\bibinfo
  {volume} {D100}},\ \bibinfo {pages} {076004} (\bibinfo {year} {2019})},\
  \Eprint {http://arxiv.org/abs/1903.10197} {arXiv:1903.10197
  [hep-ph]}\BibitemShut {NoStop}%
\bibitem [{\citenamefont {Actis}\ \emph {et~al.}(2010)\citenamefont {Actis}
  \emph {et~al.}}]{Actis:2010gg}%
  \BibitemOpen
  \bibfield  {author} {\bibinfo {author} {\bibfnamefont {S.}~\bibnamefont
  {Actis}} \emph {et~al.} (\bibinfo {collaboration} {Working Group on Radiative
  Corrections and Monte Carlo Generators for Low Energies}),\ }\href {\doibase
  10.1140/epjc/s10052-010-1251-4} {\bibfield  {journal} {\bibinfo  {journal}
  {Eur. Phys. J.}\ }\textbf {\bibinfo {volume} {C66}},\ \bibinfo {pages} {585}
  (\bibinfo {year} {2010})},\ \Eprint {http://arxiv.org/abs/0912.0749}
  {arXiv:0912.0749 [hep-ph]}\BibitemShut {NoStop}%
\bibitem [{\citenamefont {Nanava}\ and\ \citenamefont
  {Was}(2007)}]{Nanava:2006vv}%
  \BibitemOpen
  \bibfield  {author} {\bibinfo {author} {\bibfnamefont {G.}~\bibnamefont
  {Nanava}}\ and\ \bibinfo {author} {\bibfnamefont {Z.}~\bibnamefont {Was}},\
  }\href {\doibase 10.1140/epjc/s10052-007-0316-5} {\bibfield  {journal}
  {\bibinfo  {journal} {Eur. Phys. J.}\ }\textbf {\bibinfo {volume} {C51}},\
  \bibinfo {pages} {569} (\bibinfo {year} {2007})},\ \Eprint
  {http://arxiv.org/abs/hep-ph/0607019} {arXiv:hep-ph/0607019
  [hep-ph]}\BibitemShut {NoStop}%
\bibitem [{\citenamefont {Brun}\ and\ \citenamefont
  {Rademakers}(1997)}]{Brun:1997pa}%
  \BibitemOpen
  \bibfield  {author} {\bibinfo {author} {\bibfnamefont {R.}~\bibnamefont
  {Brun}}\ and\ \bibinfo {author} {\bibfnamefont {F.}~\bibnamefont
  {Rademakers}},\ }\href {\doibase 10.1016/S0168-9002(97)00048-X} {\bibfield
  {journal} {\bibinfo  {journal} {Nucl. Instrum. Meth.}\ }\textbf {\bibinfo
  {volume} {A389}},\ \bibinfo {pages} {81} (\bibinfo {year}
  {1997})}\BibitemShut {NoStop}%
\bibitem [{\citenamefont {Davier}\ \emph {et~al.}(2003)\citenamefont {Davier},
  \citenamefont {Eidelman}, \citenamefont {Hoecker},\ and\ \citenamefont
  {Zhang}}]{Davier:2002dy}%
  \BibitemOpen
  \bibfield  {author} {\bibinfo {author} {\bibfnamefont {M.}~\bibnamefont
  {Davier}}, \bibinfo {author} {\bibfnamefont {S.}~\bibnamefont {Eidelman}},
  \bibinfo {author} {\bibfnamefont {A.}~\bibnamefont {Hoecker}}, \ and\
  \bibinfo {author} {\bibfnamefont {Z.}~\bibnamefont {Zhang}},\ }\href
  {\doibase 10.1140/epjc/s2003-01136-2} {\bibfield  {journal} {\bibinfo
  {journal} {Eur. Phys. J.}\ }\textbf {\bibinfo {volume} {C27}},\ \bibinfo
  {pages} {497} (\bibinfo {year} {2003})},\ \Eprint
  {http://arxiv.org/abs/hep-ph/0208177} {arXiv:hep-ph/0208177
  [hep-ph]}\BibitemShut {NoStop}%
\bibitem [{\citenamefont
  {Malaescu}(2018{\natexlab{b}})}]{bogdan-kek-2018-DHMZ}%
  \BibitemOpen
  \bibfield  {author} {\bibinfo {author} {\bibfnamefont {B.}~\bibnamefont
  {Malaescu}},\ }\href@noop {} {\enquote {\bibinfo {title} {{Hadronic
  contribution to $(g-2)_\mu$ from $e^+e^-$ annihilations}},}\ }\bibinfo
  {howpublished}
  {\url{https://kds.kek.jp/indico/event/26780/session/1/contribution/1/material/slides/0.pdf}}
  (\bibinfo {year} {2018}{\natexlab{b}}),\ \bibinfo {note} {{Muon $g-2$ Theory
  Initiative workshop KEK}}\BibitemShut {NoStop}%
\bibitem [{\citenamefont
  {Malaescu}(2018{\natexlab{c}})}]{bogdan-Mainz-2018-DHMZ}%
  \BibitemOpen
  \bibfield  {author} {\bibinfo {author} {\bibfnamefont {B.}~\bibnamefont
  {Malaescu}},\ }\href@noop {} {\enquote {\bibinfo {title} {{Hadronic
  contribution to $(g-2)_\mu$ from $e^+e^-$ annihilations}},}\ }\bibinfo
  {howpublished}
  {\url{https://indico.him.uni-mainz.de/event/11/session/1/contribution/39/material/slides/0.pdf}}
  (\bibinfo {year} {2018}{\natexlab{c}}),\ \bibinfo {note} {{Muon $g-2$ Theory
  Initiative workshop Mainz}}\BibitemShut {NoStop}%
\bibitem [{\citenamefont {Boito}\ \emph {et~al.}(2018)\citenamefont {Boito},
  \citenamefont {Golterman}, \citenamefont {Keshavarzi}, \citenamefont
  {Maltman}, \citenamefont {Nomura}, \citenamefont {Peris},\ and\ \citenamefont
  {Teubner}}]{Boito:2018yvl}%
  \BibitemOpen
  \bibfield  {author} {\bibinfo {author} {\bibfnamefont {D.}~\bibnamefont
  {Boito}}, \bibinfo {author} {\bibfnamefont {M.}~\bibnamefont {Golterman}},
  \bibinfo {author} {\bibfnamefont {A.}~\bibnamefont {Keshavarzi}}, \bibinfo
  {author} {\bibfnamefont {K.}~\bibnamefont {Maltman}}, \bibinfo {author}
  {\bibfnamefont {D.}~\bibnamefont {Nomura}}, \bibinfo {author} {\bibfnamefont
  {S.}~\bibnamefont {Peris}}, \ and\ \bibinfo {author} {\bibfnamefont
  {T.}~\bibnamefont {Teubner}},\ }\href {\doibase 10.1103/PhysRevD.98.074030}
  {\bibfield  {journal} {\bibinfo  {journal} {Phys. Rev.}\ }\textbf {\bibinfo
  {volume} {D98}},\ \bibinfo {pages} {074030} (\bibinfo {year} {2018})},\
  \Eprint {http://arxiv.org/abs/1805.08176} {arXiv:1805.08176
  [hep-ph]}\BibitemShut {NoStop}%
\bibitem [{\citenamefont {Hagiwara}\ \emph {et~al.}(2003)\citenamefont
  {Hagiwara}, \citenamefont {Martin}, \citenamefont {Nomura},\ and\
  \citenamefont {Teubner}}]{Hagiwara:2002ma}%
  \BibitemOpen
  \bibfield  {author} {\bibinfo {author} {\bibfnamefont {K.}~\bibnamefont
  {Hagiwara}}, \bibinfo {author} {\bibfnamefont {A.~D.}\ \bibnamefont
  {Martin}}, \bibinfo {author} {\bibfnamefont {D.}~\bibnamefont {Nomura}}, \
  and\ \bibinfo {author} {\bibfnamefont {T.}~\bibnamefont {Teubner}},\ }\href
  {\doibase 10.1016/S0370-2693(03)00138-2} {\bibfield  {journal} {\bibinfo
  {journal} {Phys. Lett.}\ }\textbf {\bibinfo {volume} {B557}},\ \bibinfo
  {pages} {69} (\bibinfo {year} {2003})},\ \Eprint
  {http://arxiv.org/abs/hep-ph/0209187} {arXiv:hep-ph/0209187
  [hep-ph]}\BibitemShut {NoStop}%
\bibitem [{\citenamefont {Hagiwara}\ \emph {et~al.}(2004)\citenamefont
  {Hagiwara}, \citenamefont {Martin}, \citenamefont {Nomura},\ and\
  \citenamefont {Teubner}}]{Hagiwara:2003da}%
  \BibitemOpen
  \bibfield  {author} {\bibinfo {author} {\bibfnamefont {K.}~\bibnamefont
  {Hagiwara}}, \bibinfo {author} {\bibfnamefont {A.~D.}\ \bibnamefont
  {Martin}}, \bibinfo {author} {\bibfnamefont {D.}~\bibnamefont {Nomura}}, \
  and\ \bibinfo {author} {\bibfnamefont {T.}~\bibnamefont {Teubner}},\ }\href
  {\doibase 10.1103/PhysRevD.69.093003} {\bibfield  {journal} {\bibinfo
  {journal} {Phys. Rev.}\ }\textbf {\bibinfo {volume} {D69}},\ \bibinfo {pages}
  {093003} (\bibinfo {year} {2004})},\ \Eprint
  {http://arxiv.org/abs/hep-ph/0312250} {arXiv:hep-ph/0312250
  [hep-ph]}\BibitemShut {NoStop}%
\bibitem [{\citenamefont {Hagiwara}\ \emph {et~al.}(2007)\citenamefont
  {Hagiwara}, \citenamefont {Martin}, \citenamefont {Nomura},\ and\
  \citenamefont {Teubner}}]{Hagiwara:2006jt}%
  \BibitemOpen
  \bibfield  {author} {\bibinfo {author} {\bibfnamefont {K.}~\bibnamefont
  {Hagiwara}}, \bibinfo {author} {\bibfnamefont {A.~D.}\ \bibnamefont
  {Martin}}, \bibinfo {author} {\bibfnamefont {D.}~\bibnamefont {Nomura}}, \
  and\ \bibinfo {author} {\bibfnamefont {T.}~\bibnamefont {Teubner}},\ }\href
  {\doibase 10.1016/j.physletb.2007.04.012} {\bibfield  {journal} {\bibinfo
  {journal} {Phys. Lett.}\ }\textbf {\bibinfo {volume} {B649}},\ \bibinfo
  {pages} {173} (\bibinfo {year} {2007})},\ \Eprint
  {http://arxiv.org/abs/hep-ph/0611102} {arXiv:hep-ph/0611102
  [hep-ph]}\BibitemShut {NoStop}%
\bibitem [{\citenamefont {Teubner}\ \emph {et~al.}(2010)\citenamefont
  {Teubner}, \citenamefont {Hagiwara}, \citenamefont {Liao}, \citenamefont
  {Martin},\ and\ \citenamefont {Nomura}}]{Teubner:2010ah}%
  \BibitemOpen
  \bibfield  {author} {\bibinfo {author} {\bibfnamefont {T.}~\bibnamefont
  {Teubner}}, \bibinfo {author} {\bibfnamefont {K.}~\bibnamefont {Hagiwara}},
  \bibinfo {author} {\bibfnamefont {R.}~\bibnamefont {Liao}}, \bibinfo {author}
  {\bibfnamefont {A.~D.}\ \bibnamefont {Martin}}, \ and\ \bibinfo {author}
  {\bibfnamefont {D.}~\bibnamefont {Nomura}},\ }\href {\doibase
  10.1088/1674-1137/34/6/019} {\bibfield  {journal} {\bibinfo  {journal} {Chin.
  Phys.}\ }\textbf {\bibinfo {volume} {C34}},\ \bibinfo {pages} {728} (\bibinfo
  {year} {2010})},\ \Eprint {http://arxiv.org/abs/1001.5401} {arXiv:1001.5401
  [hep-ph]}\BibitemShut {NoStop}%
\bibitem [{\citenamefont {Hagiwara}\ \emph {et~al.}(2011)\citenamefont
  {Hagiwara}, \citenamefont {Liao}, \citenamefont {Martin}, \citenamefont
  {Nomura},\ and\ \citenamefont {Teubner}}]{Hagiwara:2011af}%
  \BibitemOpen
  \bibfield  {author} {\bibinfo {author} {\bibfnamefont {K.}~\bibnamefont
  {Hagiwara}}, \bibinfo {author} {\bibfnamefont {R.}~\bibnamefont {Liao}},
  \bibinfo {author} {\bibfnamefont {A.~D.}\ \bibnamefont {Martin}}, \bibinfo
  {author} {\bibfnamefont {D.}~\bibnamefont {Nomura}}, \ and\ \bibinfo {author}
  {\bibfnamefont {T.}~\bibnamefont {Teubner}},\ }\href {\doibase
  10.1088/0954-3899/38/8/085003} {\bibfield  {journal} {\bibinfo  {journal} {J.
  Phys.}\ }\textbf {\bibinfo {volume} {G38}},\ \bibinfo {pages} {085003}
  (\bibinfo {year} {2011})},\ \Eprint {http://arxiv.org/abs/1105.3149}
  {arXiv:1105.3149 [hep-ph]}\BibitemShut {NoStop}%
\bibitem [{\citenamefont {Ball}\ \emph {et~al.}(2010)\citenamefont {Ball},
  \citenamefont {Del~Debbio}, \citenamefont {Forte}, \citenamefont {Guffanti},
  \citenamefont {Latorre}, \citenamefont {Rojo},\ and\ \citenamefont
  {Ubiali}}]{Ball:2009qv}%
  \BibitemOpen
  \bibfield  {author} {\bibinfo {author} {\bibfnamefont {R.~D.}\ \bibnamefont
  {Ball}}, \bibinfo {author} {\bibfnamefont {L.}~\bibnamefont {Del~Debbio}},
  \bibinfo {author} {\bibfnamefont {S.}~\bibnamefont {Forte}}, \bibinfo
  {author} {\bibfnamefont {A.}~\bibnamefont {Guffanti}}, \bibinfo {author}
  {\bibfnamefont {J.~I.}\ \bibnamefont {Latorre}}, \bibinfo {author}
  {\bibfnamefont {J.}~\bibnamefont {Rojo}}, \ and\ \bibinfo {author}
  {\bibfnamefont {M.}~\bibnamefont {Ubiali}} (\bibinfo {collaboration}
  {NNPDF}),\ }\href {\doibase 10.1007/JHEP05(2010)075} {\bibfield  {journal}
  {\bibinfo  {journal} {JHEP}\ }\textbf {\bibinfo {volume} {05}},\ \bibinfo
  {pages} {075} (\bibinfo {year} {2010})},\ \Eprint
  {http://arxiv.org/abs/0912.2276} {arXiv:0912.2276 [hep-ph]}\BibitemShut
  {NoStop}%
\bibitem [{\citenamefont {D'Agostini}(1994)}]{DAgostini:1993arp}%
  \BibitemOpen
  \bibfield  {author} {\bibinfo {author} {\bibfnamefont {G.}~\bibnamefont
  {D'Agostini}},\ }\href {\doibase 10.1016/0168-9002(94)90719-6} {\bibfield
  {journal} {\bibinfo  {journal} {Nucl. Instrum. Meth.}\ }\textbf {\bibinfo
  {volume} {A346}},\ \bibinfo {pages} {306} (\bibinfo {year}
  {1994})}\BibitemShut {NoStop}%
\bibitem [{\citenamefont {Jegerlehner}(2016)}]{Jegerlehner:2015stw}%
  \BibitemOpen
  \bibfield  {author} {\bibinfo {author} {\bibfnamefont {F.}~\bibnamefont
  {Jegerlehner}},\ }\href {\doibase 10.1051/epjconf/201611801016} {\bibfield
  {journal} {\bibinfo  {journal} {EPJ Web Conf.}\ }\textbf {\bibinfo {volume}
  {118}},\ \bibinfo {pages} {01016} (\bibinfo {year} {2016})},\ \Eprint
  {http://arxiv.org/abs/1511.04473} {arXiv:1511.04473 [hep-ph]}\BibitemShut
  {NoStop}%
\bibitem [{\citenamefont {Jegerlehner}(2018)}]{Jegerlehner:2017lbd}%
  \BibitemOpen
  \bibfield  {author} {\bibinfo {author} {\bibfnamefont {F.}~\bibnamefont
  {Jegerlehner}},\ }\href {\doibase 10.1051/epjconf/201816600022} {\bibfield
  {journal} {\bibinfo  {journal} {EPJ Web Conf.}\ }\textbf {\bibinfo {volume}
  {166}},\ \bibinfo {pages} {00022} (\bibinfo {year} {2018})},\ \Eprint
  {http://arxiv.org/abs/1705.00263} {arXiv:1705.00263 [hep-ph]}\BibitemShut
  {NoStop}%
\bibitem [{\citenamefont
  {Jegerlehner}(2019{\natexlab{a}})}]{Jegerlehner:2017zsb}%
  \BibitemOpen
  \bibfield  {author} {\bibinfo {author} {\bibfnamefont {F.}~\bibnamefont
  {Jegerlehner}},\ }\href {\doibase 10.1051/epjconf/201921801003} {\bibfield
  {journal} {\bibinfo  {journal} {EPJ Web Conf.}\ }\textbf {\bibinfo {volume}
  {218}},\ \bibinfo {pages} {01003} (\bibinfo {year} {2019}{\natexlab{a}})},\
  \Eprint {http://arxiv.org/abs/1711.06089} {arXiv:1711.06089
  [hep-ph]}\BibitemShut {NoStop}%
\bibitem [{\citenamefont
  {Jegerlehner}(2019{\natexlab{b}})}]{Jegerlehner:2018gjd}%
  \BibitemOpen
  \bibfield  {author} {\bibinfo {author} {\bibfnamefont {F.}~\bibnamefont
  {Jegerlehner}},\ }\href {\doibase 10.1051/epjconf/201919901010} {\bibfield
  {journal} {\bibinfo  {journal} {EPJ Web Conf.}\ }\textbf {\bibinfo {volume}
  {199}},\ \bibinfo {pages} {01010} (\bibinfo {year} {2019}{\natexlab{b}})},\
  \Eprint {http://arxiv.org/abs/1809.07413} {arXiv:1809.07413
  [hep-ph]}\BibitemShut {NoStop}%
\bibitem [{\citenamefont {Eidelman}\ and\ \citenamefont
  {Jegerlehner}(1995)}]{Eidelman:1995ny}%
  \BibitemOpen
  \bibfield  {author} {\bibinfo {author} {\bibfnamefont {S.}~\bibnamefont
  {Eidelman}}\ and\ \bibinfo {author} {\bibfnamefont {F.}~\bibnamefont
  {Jegerlehner}},\ }\href {\doibase 10.1007/BF01553984} {\bibfield  {journal}
  {\bibinfo  {journal} {Z. Phys.}\ }\textbf {\bibinfo {volume} {C67}},\
  \bibinfo {pages} {585} (\bibinfo {year} {1995})},\ \Eprint
  {http://arxiv.org/abs/hep-ph/9502298} {arXiv:hep-ph/9502298
  [hep-ph]}\BibitemShut {NoStop}%
\bibitem [{\citenamefont {Davier}(2015)}]{Davier:2015bka}%
  \BibitemOpen
  \bibfield  {author} {\bibinfo {author} {\bibfnamefont {M.}~\bibnamefont
  {Davier}} (\bibinfo {collaboration} {BABAR}),\ }\href {\doibase
  10.1016/j.nuclphysbps.2015.02.021} {\bibfield  {journal} {\bibinfo  {journal}
  {Nucl. Part. Phys. Proc.}\ }\textbf {\bibinfo {volume} {260}},\ \bibinfo
  {pages} {102} (\bibinfo {year} {2015})}\BibitemShut {NoStop}%
\bibitem [{\citenamefont {Davier}\ \emph {et~al.}(2016)\citenamefont {Davier},
  \citenamefont {Hoecker}, \citenamefont {Malaescu},\ and\ \citenamefont
  {Zhang}}]{Davier:2016udg}%
  \BibitemOpen
  \bibfield  {author} {\bibinfo {author} {\bibfnamefont {M.}~\bibnamefont
  {Davier}}, \bibinfo {author} {\bibfnamefont {A.}~\bibnamefont {Hoecker}},
  \bibinfo {author} {\bibfnamefont {B.}~\bibnamefont {Malaescu}}, \ and\
  \bibinfo {author} {\bibfnamefont {Z.}~\bibnamefont {Zhang}},\ }\href
  {\doibase 10.1142/9789814733519_0007} {\bibfield  {journal} {\bibinfo
  {journal} {Adv. Ser. Direct. High Energy Phys.}\ }\textbf {\bibinfo {volume}
  {26}},\ \bibinfo {pages} {129} (\bibinfo {year} {2016})}\BibitemShut
  {NoStop}%
\bibitem [{\citenamefont {Kozyrev}\ \emph {et~al.}(2016)\citenamefont {Kozyrev}
  \emph {et~al.}}]{Kozyrev:2016raz}%
  \BibitemOpen
  \bibfield  {author} {\bibinfo {author} {\bibfnamefont {E.~A.}\ \bibnamefont
  {Kozyrev}} \emph {et~al.} (\bibinfo {collaboration} {CMD-3}),\ }\href
  {\doibase 10.1016/j.physletb.2016.07.003} {\bibfield  {journal} {\bibinfo
  {journal} {Phys. Lett.}\ }\textbf {\bibinfo {volume} {B760}},\ \bibinfo
  {pages} {314} (\bibinfo {year} {2016})},\ \Eprint
  {http://arxiv.org/abs/1604.02981} {arXiv:1604.02981 [hep-ex]}\BibitemShut
  {NoStop}%
\bibitem [{\citenamefont {Achasov}\ \emph
  {et~al.}(2016{\natexlab{c}})\citenamefont {Achasov} \emph
  {et~al.}}]{Achasov:2016eyg}%
  \BibitemOpen
  \bibfield  {author} {\bibinfo {author} {\bibfnamefont {M.~N.}\ \bibnamefont
  {Achasov}} \emph {et~al.} (\bibinfo {collaboration} {SND}),\ }\href {\doibase
  10.1103/PhysRevD.94.032010} {\bibfield  {journal} {\bibinfo  {journal} {Phys.
  Rev.}\ }\textbf {\bibinfo {volume} {D94}},\ \bibinfo {pages} {032010}
  (\bibinfo {year} {2016}{\natexlab{c}})},\ \Eprint
  {http://arxiv.org/abs/1606.06481} {arXiv:1606.06481 [hep-ex]}\BibitemShut
  {NoStop}%
\bibitem [{\citenamefont {Achasov}\ \emph
  {et~al.}(2016{\natexlab{d}})\citenamefont {Achasov} \emph
  {et~al.}}]{Achasov:2016qvd}%
  \BibitemOpen
  \bibfield  {author} {\bibinfo {author} {\bibfnamefont {M.~N.}\ \bibnamefont
  {Achasov}} \emph {et~al.} (\bibinfo {collaboration} {SND}),\ }\href {\doibase
  10.1103/PhysRevD.94.092002} {\bibfield  {journal} {\bibinfo  {journal} {Phys.
  Rev.}\ }\textbf {\bibinfo {volume} {D94}},\ \bibinfo {pages} {092002}
  (\bibinfo {year} {2016}{\natexlab{d}})},\ \Eprint
  {http://arxiv.org/abs/1607.00371} {arXiv:1607.00371 [hep-ex]}\BibitemShut
  {NoStop}%
\bibitem [{\citenamefont {Achasov}\ \emph
  {et~al.}(2016{\natexlab{e}})\citenamefont {Achasov} \emph
  {et~al.}}]{Achasov:2016zvn}%
  \BibitemOpen
  \bibfield  {author} {\bibinfo {author} {\bibfnamefont {M.~N.}\ \bibnamefont
  {Achasov}} \emph {et~al.} (\bibinfo {collaboration} {SND}),\ }\href {\doibase
  10.1103/PhysRevD.94.112001} {\bibfield  {journal} {\bibinfo  {journal} {Phys.
  Rev.}\ }\textbf {\bibinfo {volume} {D94}},\ \bibinfo {pages} {112001}
  (\bibinfo {year} {2016}{\natexlab{e}})},\ \Eprint
  {http://arxiv.org/abs/1610.00235} {arXiv:1610.00235 [hep-ex]}\BibitemShut
  {NoStop}%
\bibitem [{\citenamefont {Barate}\ \emph {et~al.}(1997)\citenamefont {Barate}
  \emph {et~al.}}]{Barate:1997hv}%
  \BibitemOpen
  \bibfield  {author} {\bibinfo {author} {\bibfnamefont {R.}~\bibnamefont
  {Barate}} \emph {et~al.} (\bibinfo {collaboration} {ALEPH}),\ }\href
  {\doibase 10.1007/s002880050523} {\bibfield  {journal} {\bibinfo  {journal}
  {Z. Phys.}\ }\textbf {\bibinfo {volume} {C76}},\ \bibinfo {pages} {15}
  (\bibinfo {year} {1997})}\BibitemShut {NoStop}%
\bibitem [{\citenamefont {Barate}\ \emph {et~al.}(1998)\citenamefont {Barate}
  \emph {et~al.}}]{Barate:1998uf}%
  \BibitemOpen
  \bibfield  {author} {\bibinfo {author} {\bibfnamefont {R.}~\bibnamefont
  {Barate}} \emph {et~al.} (\bibinfo {collaboration} {ALEPH}),\ }\href
  {\doibase 10.1007/s100520050217} {\bibfield  {journal} {\bibinfo  {journal}
  {Eur. Phys. J.}\ }\textbf {\bibinfo {volume} {C4}},\ \bibinfo {pages} {409}
  (\bibinfo {year} {1998})}\BibitemShut {NoStop}%
\bibitem [{\citenamefont {Ananthanarayan}\ \emph {et~al.}(2014)\citenamefont
  {Ananthanarayan}, \citenamefont {Caprini}, \citenamefont {Das},\ and\
  \citenamefont {Sentitemsu~Imsong}}]{Ananthanarayan:2013zua}%
  \BibitemOpen
  \bibfield  {author} {\bibinfo {author} {\bibfnamefont {B.}~\bibnamefont
  {Ananthanarayan}}, \bibinfo {author} {\bibfnamefont {I.}~\bibnamefont
  {Caprini}}, \bibinfo {author} {\bibfnamefont {D.}~\bibnamefont {Das}}, \ and\
  \bibinfo {author} {\bibfnamefont {I.}~\bibnamefont {Sentitemsu~Imsong}},\
  }\href {\doibase 10.1103/PhysRevD.89.036007} {\bibfield  {journal} {\bibinfo
  {journal} {Phys. Rev.}\ }\textbf {\bibinfo {volume} {D89}},\ \bibinfo {pages}
  {036007} (\bibinfo {year} {2014})},\ \Eprint {http://arxiv.org/abs/1312.5849}
  {arXiv:1312.5849 [hep-ph]}\BibitemShut {NoStop}%
\bibitem [{\citenamefont {Ecker}\ \emph
  {et~al.}(1989{\natexlab{a}})\citenamefont {Ecker}, \citenamefont {Gasser},
  \citenamefont {Leutwyler}, \citenamefont {Pich},\ and\ \citenamefont
  {de~Rafael}}]{Ecker:1989yg}%
  \BibitemOpen
  \bibfield  {author} {\bibinfo {author} {\bibfnamefont {G.}~\bibnamefont
  {Ecker}}, \bibinfo {author} {\bibfnamefont {J.}~\bibnamefont {Gasser}},
  \bibinfo {author} {\bibfnamefont {H.}~\bibnamefont {Leutwyler}}, \bibinfo
  {author} {\bibfnamefont {A.}~\bibnamefont {Pich}}, \ and\ \bibinfo {author}
  {\bibfnamefont {E.}~\bibnamefont {de~Rafael}},\ }\href {\doibase
  10.1016/0370-2693(89)91627-4} {\bibfield  {journal} {\bibinfo  {journal}
  {Phys. Lett.}\ }\textbf {\bibinfo {volume} {B223}},\ \bibinfo {pages} {425}
  (\bibinfo {year} {1989}{\natexlab{a}})}\BibitemShut {NoStop}%
\bibitem [{\citenamefont {Bando}\ \emph {et~al.}(1988)\citenamefont {Bando},
  \citenamefont {Kugo},\ and\ \citenamefont {Yamawaki}}]{Bando:1987br}%
  \BibitemOpen
  \bibfield  {author} {\bibinfo {author} {\bibfnamefont {M.}~\bibnamefont
  {Bando}}, \bibinfo {author} {\bibfnamefont {T.}~\bibnamefont {Kugo}}, \ and\
  \bibinfo {author} {\bibfnamefont {K.}~\bibnamefont {Yamawaki}},\ }\href
  {\doibase 10.1016/0370-1573(88)90019-1} {\bibfield  {journal} {\bibinfo
  {journal} {Phys. Rept.}\ }\textbf {\bibinfo {volume} {164}},\ \bibinfo
  {pages} {217} (\bibinfo {year} {1988})}\BibitemShut {NoStop}%
\bibitem [{\citenamefont {Harada}\ and\ \citenamefont
  {Yamawaki}(2003)}]{Harada:2003jx}%
  \BibitemOpen
  \bibfield  {author} {\bibinfo {author} {\bibfnamefont {M.}~\bibnamefont
  {Harada}}\ and\ \bibinfo {author} {\bibfnamefont {K.}~\bibnamefont
  {Yamawaki}},\ }\href {\doibase 10.1016/S0370-1573(03)00139-X} {\bibfield
  {journal} {\bibinfo  {journal} {Phys. Rept.}\ }\textbf {\bibinfo {volume}
  {381}},\ \bibinfo {pages} {1} (\bibinfo {year} {2003})},\ \Eprint
  {http://arxiv.org/abs/hep-ph/0302103} {arXiv:hep-ph/0302103
  [hep-ph]}\BibitemShut {NoStop}%
\bibitem [{\citenamefont {Benayoun}\ \emph {et~al.}(2008)\citenamefont
  {Benayoun}, \citenamefont {David}, \citenamefont {DelBuono}, \citenamefont
  {Leitner},\ and\ \citenamefont {O'Connell}}]{Benayoun:2007cu}%
  \BibitemOpen
  \bibfield  {author} {\bibinfo {author} {\bibfnamefont {M.}~\bibnamefont
  {Benayoun}}, \bibinfo {author} {\bibfnamefont {P.}~\bibnamefont {David}},
  \bibinfo {author} {\bibfnamefont {L.}~\bibnamefont {DelBuono}}, \bibinfo
  {author} {\bibfnamefont {O.}~\bibnamefont {Leitner}}, \ and\ \bibinfo
  {author} {\bibfnamefont {H.~B.}\ \bibnamefont {O'Connell}},\ }\href {\doibase
  10.1140/epjc/s10052-008-0586-6} {\bibfield  {journal} {\bibinfo  {journal}
  {Eur. Phys. J.}\ }\textbf {\bibinfo {volume} {C55}},\ \bibinfo {pages} {199}
  (\bibinfo {year} {2008})},\ \Eprint {http://arxiv.org/abs/0711.4482}
  {arXiv:0711.4482 [hep-ph]}\BibitemShut {NoStop}%
\bibitem [{\citenamefont {Benayoun}\ \emph {et~al.}(2012)\citenamefont
  {Benayoun}, \citenamefont {David}, \citenamefont {DelBuono},\ and\
  \citenamefont {Jegerlehner}}]{Benayoun:2011mm}%
  \BibitemOpen
  \bibfield  {author} {\bibinfo {author} {\bibfnamefont {M.}~\bibnamefont
  {Benayoun}}, \bibinfo {author} {\bibfnamefont {P.}~\bibnamefont {David}},
  \bibinfo {author} {\bibfnamefont {L.}~\bibnamefont {DelBuono}}, \ and\
  \bibinfo {author} {\bibfnamefont {F.}~\bibnamefont {Jegerlehner}},\ }\href
  {\doibase 10.1140/epjc/s10052-011-1848-2} {\bibfield  {journal} {\bibinfo
  {journal} {Eur. Phys. J.}\ }\textbf {\bibinfo {volume} {C72}},\ \bibinfo
  {pages} {1848} (\bibinfo {year} {2012})},\ \Eprint
  {http://arxiv.org/abs/1106.1315} {arXiv:1106.1315 [hep-ph]}\BibitemShut
  {NoStop}%
\bibitem [{\citenamefont {Benayoun}\ \emph {et~al.}(2013)\citenamefont
  {Benayoun}, \citenamefont {David}, \citenamefont {DelBuono},\ and\
  \citenamefont {Jegerlehner}}]{Benayoun:2012wc}%
  \BibitemOpen
  \bibfield  {author} {\bibinfo {author} {\bibfnamefont {M.}~\bibnamefont
  {Benayoun}}, \bibinfo {author} {\bibfnamefont {P.}~\bibnamefont {David}},
  \bibinfo {author} {\bibfnamefont {L.}~\bibnamefont {DelBuono}}, \ and\
  \bibinfo {author} {\bibfnamefont {F.}~\bibnamefont {Jegerlehner}},\ }\href
  {\doibase 10.1140/epjc/s10052-013-2453-3} {\bibfield  {journal} {\bibinfo
  {journal} {Eur. Phys. J.}\ }\textbf {\bibinfo {volume} {C73}},\ \bibinfo
  {pages} {2453} (\bibinfo {year} {2013})},\ \Eprint
  {http://arxiv.org/abs/1210.7184} {arXiv:1210.7184 [hep-ph]}\BibitemShut
  {NoStop}%
\bibitem [{\citenamefont {Benayoun}\ \emph {et~al.}(2015)\citenamefont
  {Benayoun}, \citenamefont {David}, \citenamefont {DelBuono},\ and\
  \citenamefont {Jegerlehner}}]{Benayoun:2015gxa}%
  \BibitemOpen
  \bibfield  {author} {\bibinfo {author} {\bibfnamefont {M.}~\bibnamefont
  {Benayoun}}, \bibinfo {author} {\bibfnamefont {P.}~\bibnamefont {David}},
  \bibinfo {author} {\bibfnamefont {L.}~\bibnamefont {DelBuono}}, \ and\
  \bibinfo {author} {\bibfnamefont {F.}~\bibnamefont {Jegerlehner}},\ }\href
  {\doibase 10.1140/epjc/s10052-015-3830-x} {\bibfield  {journal} {\bibinfo
  {journal} {Eur. Phys. J.}\ }\textbf {\bibinfo {volume} {C75}},\ \bibinfo
  {pages} {613} (\bibinfo {year} {2015})},\ \Eprint
  {http://arxiv.org/abs/1507.02943} {arXiv:1507.02943 [hep-ph]}\BibitemShut
  {NoStop}%
\bibitem [{\citenamefont {Benayoun}\ \emph {et~al.}(2020)\citenamefont
  {Benayoun}, \citenamefont {Delbuono},\ and\ \citenamefont
  {Jegerlehner}}]{Benayoun:2019zwh}%
  \BibitemOpen
  \bibfield  {author} {\bibinfo {author} {\bibfnamefont {M.}~\bibnamefont
  {Benayoun}}, \bibinfo {author} {\bibfnamefont {L.}~\bibnamefont {Delbuono}},
  \ and\ \bibinfo {author} {\bibfnamefont {F.}~\bibnamefont {Jegerlehner}},\
  }\href {\doibase 10.1140/epjc/s10052-020-7611-9} {\bibfield  {journal}
  {\bibinfo  {journal} {Eur. Phys. J.}\ }\textbf {\bibinfo {volume} {C80}},\
  \bibinfo {pages} {81} (\bibinfo {year} {2020})},\ \bibinfo {note} {[Erratum:
  Eur. Phys. J. {\bf C80}, 244 (2020)]},\ \Eprint
  {http://arxiv.org/abs/1903.11034} {arXiv:1903.11034 [hep-ph]}\BibitemShut
  {NoStop}%
\bibitem [{\citenamefont {Amendolia}\ \emph
  {et~al.}(1986{\natexlab{a}})\citenamefont {Amendolia} \emph
  {et~al.}}]{Amendolia:1986wj}%
  \BibitemOpen
  \bibfield  {author} {\bibinfo {author} {\bibfnamefont {S.~R.}\ \bibnamefont
  {Amendolia}} \emph {et~al.} (\bibinfo {collaboration} {NA7}),\ }\href
  {\doibase 10.1016/0550-3213(86)90437-2} {\bibfield  {journal} {\bibinfo
  {journal} {Nucl. Phys.}\ }\textbf {\bibinfo {volume} {B277}},\ \bibinfo
  {pages} {168} (\bibinfo {year} {1986}{\natexlab{a}})}\BibitemShut {NoStop}%
\bibitem [{\citenamefont {Amendolia}\ \emph
  {et~al.}(1986{\natexlab{b}})\citenamefont {Amendolia} \emph
  {et~al.}}]{Amendolia:1986ui}%
  \BibitemOpen
  \bibfield  {author} {\bibinfo {author} {\bibfnamefont {S.~R.}\ \bibnamefont
  {Amendolia}} \emph {et~al.} (\bibinfo {collaboration} {NA7}),\ }\href
  {\doibase 10.1016/0370-2693(86)91407-3} {\bibfield  {journal} {\bibinfo
  {journal} {Phys. Lett.}\ }\textbf {\bibinfo {volume} {B178}},\ \bibinfo
  {pages} {435} (\bibinfo {year} {1986}{\natexlab{b}})}\BibitemShut {NoStop}%
\bibitem [{\citenamefont {Blobel}(2006)}]{Blobel_2006}%
  \BibitemOpen
  \bibfield  {author} {\bibinfo {author} {\bibfnamefont {V.}~\bibnamefont
  {Blobel}},\ }\href@noop {} {\enquote {\bibinfo {title} {{Dealing with
  systematics for chi-square and for log likelihood goodness of fit}},}\
  }\bibinfo {howpublished}
  {\url{https://www.birs.ca/workshops/2006/06w5054/files/volker-blobel.pdf}}
  (\bibinfo {year} {2006}),\ \bibinfo {note} {{Banff International Research
  Station}}\BibitemShut {NoStop}%
\bibitem [{\citenamefont {Blobel}(2003)}]{Blobel:2003wa}%
  \BibitemOpen
  \bibfield  {author} {\bibinfo {author} {\bibfnamefont {V.}~\bibnamefont
  {Blobel}},\ }\href
  {https://www.slac.stanford.edu/econf/C030908/papers/MOET002.pdf} {\bibfield
  {journal} {\bibinfo  {journal} {eConf}\ }\textbf {\bibinfo {volume}
  {C030908}},\ \bibinfo {pages} {MOET002} (\bibinfo {year} {2003})}\BibitemShut
  {NoStop}%
\bibitem [{\citenamefont {Ananthanarayan}\ \emph {et~al.}(2018)\citenamefont
  {Ananthanarayan}, \citenamefont {Caprini},\ and\ \citenamefont
  {Das}}]{Ananthanarayan:2018nyx}%
  \BibitemOpen
  \bibfield  {author} {\bibinfo {author} {\bibfnamefont {B.}~\bibnamefont
  {Ananthanarayan}}, \bibinfo {author} {\bibfnamefont {I.}~\bibnamefont
  {Caprini}}, \ and\ \bibinfo {author} {\bibfnamefont {D.}~\bibnamefont
  {Das}},\ }\href {\doibase 10.1103/PhysRevD.98.114015} {\bibfield  {journal}
  {\bibinfo  {journal} {Phys. Rev.}\ }\textbf {\bibinfo {volume} {D98}},\
  \bibinfo {pages} {114015} (\bibinfo {year} {2018})},\ \Eprint
  {http://arxiv.org/abs/1810.09265} {arXiv:1810.09265 [hep-ph]}\BibitemShut
  {NoStop}%
\bibitem [{\citenamefont {de~Troc{\'o}niz}\ and\ \citenamefont
  {Yndur{\'a}in}(2002)}]{DeTroconiz:2001rip}%
  \BibitemOpen
  \bibfield  {author} {\bibinfo {author} {\bibfnamefont {J.~F.}\ \bibnamefont
  {de~Troc{\'o}niz}}\ and\ \bibinfo {author} {\bibfnamefont {F.~J.}\
  \bibnamefont {Yndur{\'a}in}},\ }\href {\doibase 10.1103/PhysRevD.65.093001}
  {\bibfield  {journal} {\bibinfo  {journal} {Phys. Rev.}\ }\textbf {\bibinfo
  {volume} {D65}},\ \bibinfo {pages} {093001} (\bibinfo {year} {2002})},\
  \Eprint {http://arxiv.org/abs/hep-ph/0106025} {arXiv:hep-ph/0106025
  [hep-ph]}\BibitemShut {NoStop}%
\bibitem [{\citenamefont {Leutwyler}(2002)}]{Leutwyler:2002hm}%
  \BibitemOpen
  \bibfield  {author} {\bibinfo {author} {\bibfnamefont {H.}~\bibnamefont
  {Leutwyler}},\ }\href {\doibase 10.1142/9789812776310_0002} {\bibfield
  {journal} {\bibinfo  {journal} {{Continuous advances in QCD}}\ }\textbf
  {\bibinfo {volume} {2002}},\ \bibinfo {pages} {23} (\bibinfo {year}
  {2002})},\ \Eprint {http://arxiv.org/abs/hep-ph/0212324}
  {arXiv:hep-ph/0212324 [hep-ph]}\BibitemShut {NoStop}%
\bibitem [{\citenamefont {Colangelo}(2004)}]{Colangelo:2003yw}%
  \BibitemOpen
  \bibfield  {author} {\bibinfo {author} {\bibfnamefont {G.}~\bibnamefont
  {Colangelo}},\ }\href {\doibase 10.1016/j.nuclphysbps.2004.02.025} {\bibfield
   {journal} {\bibinfo  {journal} {Nucl. Phys. Proc. Suppl.}\ }\textbf
  {\bibinfo {volume} {131}},\ \bibinfo {pages} {185} (\bibinfo {year}
  {2004})},\ \Eprint {http://arxiv.org/abs/hep-ph/0312017}
  {arXiv:hep-ph/0312017 [hep-ph]}\BibitemShut {NoStop}%
\bibitem [{\citenamefont {de~Troc{\'o}niz}\ and\ \citenamefont
  {Yndur{\'a}in}(2005)}]{deTroconiz:2004yzs}%
  \BibitemOpen
  \bibfield  {author} {\bibinfo {author} {\bibfnamefont {J.~F.}\ \bibnamefont
  {de~Troc{\'o}niz}}\ and\ \bibinfo {author} {\bibfnamefont {F.~J.}\
  \bibnamefont {Yndur{\'a}in}},\ }\href {\doibase 10.1103/PhysRevD.71.073008}
  {\bibfield  {journal} {\bibinfo  {journal} {Phys. Rev.}\ }\textbf {\bibinfo
  {volume} {D71}},\ \bibinfo {pages} {073008} (\bibinfo {year} {2005})},\
  \Eprint {http://arxiv.org/abs/hep-ph/0402285} {arXiv:hep-ph/0402285
  [hep-ph]}\BibitemShut {NoStop}%
\bibitem [{\citenamefont {Hanhart}(2012)}]{Hanhart:2012wi}%
  \BibitemOpen
  \bibfield  {author} {\bibinfo {author} {\bibfnamefont {C.}~\bibnamefont
  {Hanhart}},\ }\href {\doibase 10.1016/j.physletb.2012.07.038} {\bibfield
  {journal} {\bibinfo  {journal} {Phys. Lett.}\ }\textbf {\bibinfo {volume}
  {B715}},\ \bibinfo {pages} {170} (\bibinfo {year} {2012})},\ \Eprint
  {http://arxiv.org/abs/1203.6839} {arXiv:1203.6839 [hep-ph]}\BibitemShut
  {NoStop}%
\bibitem [{\citenamefont {Hoferichter}\ \emph {et~al.}(2016)\citenamefont
  {Hoferichter}, \citenamefont {Kubis}, \citenamefont {Ruiz~de Elvira},
  \citenamefont {Hammer},\ and\ \citenamefont
  {Mei{\ss}ner}}]{Hoferichter:2016duk}%
  \BibitemOpen
  \bibfield  {author} {\bibinfo {author} {\bibfnamefont {M.}~\bibnamefont
  {Hoferichter}}, \bibinfo {author} {\bibfnamefont {B.}~\bibnamefont {Kubis}},
  \bibinfo {author} {\bibfnamefont {J.}~\bibnamefont {Ruiz~de Elvira}},
  \bibinfo {author} {\bibfnamefont {H.-W.}\ \bibnamefont {Hammer}}, \ and\
  \bibinfo {author} {\bibfnamefont {U.-G.}\ \bibnamefont {Mei{\ss}ner}},\
  }\href {\doibase 10.1140/epja/i2016-16331-7} {\bibfield  {journal} {\bibinfo
  {journal} {Eur. Phys. J.}\ }\textbf {\bibinfo {volume} {A52}},\ \bibinfo
  {pages} {331} (\bibinfo {year} {2016})},\ \Eprint
  {http://arxiv.org/abs/1609.06722} {arXiv:1609.06722 [hep-ph]}\BibitemShut
  {NoStop}%
\bibitem [{\citenamefont {Hanhart}\ \emph {et~al.}(2017)\citenamefont
  {Hanhart}, \citenamefont {Holz}, \citenamefont {Kubis}, \citenamefont
  {Kup{\'s\'c}}, \citenamefont {Wirzba},\ and\ \citenamefont
  {Xiao}}]{Hanhart:2016pcd}%
  \BibitemOpen
  \bibfield  {author} {\bibinfo {author} {\bibfnamefont {C.}~\bibnamefont
  {Hanhart}}, \bibinfo {author} {\bibfnamefont {S.}~\bibnamefont {Holz}},
  \bibinfo {author} {\bibfnamefont {B.}~\bibnamefont {Kubis}}, \bibinfo
  {author} {\bibfnamefont {A.}~\bibnamefont {Kup{\'s\'c}}}, \bibinfo {author}
  {\bibfnamefont {A.}~\bibnamefont {Wirzba}}, \ and\ \bibinfo {author}
  {\bibfnamefont {C.-W.}\ \bibnamefont {Xiao}},\ }\href {\doibase
  10.1140/epjc/s10052-017-4651-x} {\bibfield  {journal} {\bibinfo  {journal}
  {Eur. Phys. J.}\ }\textbf {\bibinfo {volume} {C77}},\ \bibinfo {pages} {98}
  (\bibinfo {year} {2017})},\ \bibinfo {note} {[Erratum: Eur. Phys. J. {\bf
  C78}, 450 (2018)]},\ \Eprint {http://arxiv.org/abs/1611.09359}
  {arXiv:1611.09359 [hep-ph]}\BibitemShut {NoStop}%
\bibitem [{\citenamefont {Omn{\`e}s}(1958)}]{Omnes:1958hv}%
  \BibitemOpen
  \bibfield  {author} {\bibinfo {author} {\bibfnamefont {R.}~\bibnamefont
  {Omn{\`e}s}},\ }\href {\doibase 10.1007/BF02747746} {\bibfield  {journal}
  {\bibinfo  {journal} {Nuovo Cim.}\ }\textbf {\bibinfo {volume} {8}},\
  \bibinfo {pages} {316} (\bibinfo {year} {1958})}\BibitemShut {NoStop}%
\bibitem [{\citenamefont {\L{}ukaszuk}(1973)}]{Lukaszuk:1973jd}%
  \BibitemOpen
  \bibfield  {author} {\bibinfo {author} {\bibfnamefont {L.}~\bibnamefont
  {\L{}ukaszuk}},\ }\href {\doibase 10.1016/0370-2693(73)90567-4} {\bibfield
  {journal} {\bibinfo  {journal} {Phys. Lett.}\ }\textbf {\bibinfo {volume}
  {47B}},\ \bibinfo {pages} {51} (\bibinfo {year} {1973})}\BibitemShut
  {NoStop}%
\bibitem [{\citenamefont {Eidelman}\ and\ \citenamefont
  {\L{}ukaszuk}(2004)}]{Eidelman:2003uh}%
  \BibitemOpen
  \bibfield  {author} {\bibinfo {author} {\bibfnamefont {S.}~\bibnamefont
  {Eidelman}}\ and\ \bibinfo {author} {\bibfnamefont {L.}~\bibnamefont
  {\L{}ukaszuk}},\ }\href {\doibase 10.1016/j.physletb.2003.12.030} {\bibfield
  {journal} {\bibinfo  {journal} {Phys. Lett.}\ }\textbf {\bibinfo {volume}
  {B582}},\ \bibinfo {pages} {27} (\bibinfo {year} {2004})},\ \Eprint
  {http://arxiv.org/abs/hep-ph/0311366} {arXiv:hep-ph/0311366
  [hep-ph]}\BibitemShut {NoStop}%
\bibitem [{\citenamefont {Bystritskiy}\ \emph {et~al.}(2005)\citenamefont
  {Bystritskiy}, \citenamefont {Kuraev}, \citenamefont {Fedotovich},\ and\
  \citenamefont {Ignatov}}]{Bystritskiy:2005ib}%
  \BibitemOpen
  \bibfield  {author} {\bibinfo {author} {\bibfnamefont {Y.~M.}\ \bibnamefont
  {Bystritskiy}}, \bibinfo {author} {\bibfnamefont {E.~A.}\ \bibnamefont
  {Kuraev}}, \bibinfo {author} {\bibfnamefont {G.~V.}\ \bibnamefont
  {Fedotovich}}, \ and\ \bibinfo {author} {\bibfnamefont {F.~V.}\ \bibnamefont
  {Ignatov}},\ }\href {\doibase 10.1103/PhysRevD.72.114019} {\bibfield
  {journal} {\bibinfo  {journal} {Phys. Rev.}\ }\textbf {\bibinfo {volume}
  {D72}},\ \bibinfo {pages} {114019} (\bibinfo {year} {2005})},\ \Eprint
  {http://arxiv.org/abs/hep-ph/0505236} {arXiv:hep-ph/0505236
  [hep-ph]}\BibitemShut {NoStop}%
\bibitem [{\citenamefont {Roy}(1971)}]{Roy:1971tc}%
  \BibitemOpen
  \bibfield  {author} {\bibinfo {author} {\bibfnamefont {S.~M.}\ \bibnamefont
  {Roy}},\ }\href {\doibase 10.1016/0370-2693(71)90724-6} {\bibfield  {journal}
  {\bibinfo  {journal} {Phys. Lett.}\ }\textbf {\bibinfo {volume} {36B}},\
  \bibinfo {pages} {353} (\bibinfo {year} {1971})}\BibitemShut {NoStop}%
\bibitem [{\citenamefont {Ananthanarayan}\ \emph {et~al.}(2001)\citenamefont
  {Ananthanarayan}, \citenamefont {Colangelo}, \citenamefont {Gasser},\ and\
  \citenamefont {Leutwyler}}]{Ananthanarayan:2000ht}%
  \BibitemOpen
  \bibfield  {author} {\bibinfo {author} {\bibfnamefont {B.}~\bibnamefont
  {Ananthanarayan}}, \bibinfo {author} {\bibfnamefont {G.}~\bibnamefont
  {Colangelo}}, \bibinfo {author} {\bibfnamefont {J.}~\bibnamefont {Gasser}}, \
  and\ \bibinfo {author} {\bibfnamefont {H.}~\bibnamefont {Leutwyler}},\ }\href
  {\doibase 10.1016/S0370-1573(01)00009-6} {\bibfield  {journal} {\bibinfo
  {journal} {Phys. Rept.}\ }\textbf {\bibinfo {volume} {353}},\ \bibinfo
  {pages} {207} (\bibinfo {year} {2001})},\ \Eprint
  {http://arxiv.org/abs/hep-ph/0005297} {arXiv:hep-ph/0005297
  [hep-ph]}\BibitemShut {NoStop}%
\bibitem [{\citenamefont {Garc{\'i}a-Mart{\'i}n}\ \emph
  {et~al.}(2011)\citenamefont {Garc{\'i}a-Mart{\'i}n}, \citenamefont
  {Kami{\'n}ski}, \citenamefont {Pel{\'a}ez}, \citenamefont {Ruiz~de Elvira},\
  and\ \citenamefont {Yndur{\'a}in}}]{GarciaMartin:2011cn}%
  \BibitemOpen
  \bibfield  {author} {\bibinfo {author} {\bibfnamefont {R.}~\bibnamefont
  {Garc{\'i}a-Mart{\'i}n}}, \bibinfo {author} {\bibfnamefont {R.}~\bibnamefont
  {Kami{\'n}ski}}, \bibinfo {author} {\bibfnamefont {J.~R.}\ \bibnamefont
  {Pel{\'a}ez}}, \bibinfo {author} {\bibfnamefont {J.}~\bibnamefont {Ruiz~de
  Elvira}}, \ and\ \bibinfo {author} {\bibfnamefont {F.~J.}\ \bibnamefont
  {Yndur{\'a}in}},\ }\href {\doibase 10.1103/PhysRevD.83.074004} {\bibfield
  {journal} {\bibinfo  {journal} {Phys. Rev.}\ }\textbf {\bibinfo {volume}
  {D83}},\ \bibinfo {pages} {074004} (\bibinfo {year} {2011})},\ \Eprint
  {http://arxiv.org/abs/1102.2183} {arXiv:1102.2183 [hep-ph]}\BibitemShut
  {NoStop}%
\bibitem [{\citenamefont {Caprini}\ \emph {et~al.}(2012)\citenamefont
  {Caprini}, \citenamefont {Colangelo},\ and\ \citenamefont
  {Leutwyler}}]{Caprini:2011ky}%
  \BibitemOpen
  \bibfield  {author} {\bibinfo {author} {\bibfnamefont {I.}~\bibnamefont
  {Caprini}}, \bibinfo {author} {\bibfnamefont {G.}~\bibnamefont {Colangelo}},
  \ and\ \bibinfo {author} {\bibfnamefont {H.}~\bibnamefont {Leutwyler}},\
  }\href {\doibase 10.1140/epjc/s10052-012-1860-1} {\bibfield  {journal}
  {\bibinfo  {journal} {Eur. Phys. J.}\ }\textbf {\bibinfo {volume} {C72}},\
  \bibinfo {pages} {1860} (\bibinfo {year} {2012})},\ \Eprint
  {http://arxiv.org/abs/1111.7160} {arXiv:1111.7160 [hep-ph]}\BibitemShut
  {NoStop}%
\bibitem [{\citenamefont {Tanabashi}\ \emph {et~al.}(2018)\citenamefont
  {Tanabashi} \emph {et~al.}}]{Tanabashi:2018oca}%
  \BibitemOpen
  \bibfield  {author} {\bibinfo {author} {\bibfnamefont {M.}~\bibnamefont
  {Tanabashi}} \emph {et~al.} (\bibinfo {collaboration} {Particle Data
  Group}),\ }\href {\doibase 10.1103/PhysRevD.98.030001} {\bibfield  {journal}
  {\bibinfo  {journal} {Phys. Rev.}\ }\textbf {\bibinfo {volume} {D98}},\
  \bibinfo {pages} {030001} (\bibinfo {year} {2018})}\BibitemShut {NoStop}%
\bibitem [{\citenamefont {Kubis}\ \emph {et~al.}()\citenamefont {Kubis} \emph
  {et~al.}}]{Kubis}%
  \BibitemOpen
  \bibfield  {author} {\bibinfo {author} {\bibfnamefont {B.}~\bibnamefont
  {Kubis}} \emph {et~al.},\ }\href@noop {} {}\bibinfo {howpublished} {in
  preparation}\BibitemShut {NoStop}%
\bibitem [{\citenamefont {Khuri}\ and\ \citenamefont
  {Treiman}(1960)}]{Khuri:1960zz}%
  \BibitemOpen
  \bibfield  {author} {\bibinfo {author} {\bibfnamefont {N.~N.}\ \bibnamefont
  {Khuri}}\ and\ \bibinfo {author} {\bibfnamefont {S.~B.}\ \bibnamefont
  {Treiman}},\ }\href {\doibase 10.1103/PhysRev.119.1115} {\bibfield  {journal}
  {\bibinfo  {journal} {Phys. Rev.}\ }\textbf {\bibinfo {volume} {119}},\
  \bibinfo {pages} {1115} (\bibinfo {year} {1960})}\BibitemShut {NoStop}%
\bibitem [{\citenamefont {Wess}\ and\ \citenamefont
  {Zumino}(1971)}]{Wess:1971yu}%
  \BibitemOpen
  \bibfield  {author} {\bibinfo {author} {\bibfnamefont {J.}~\bibnamefont
  {Wess}}\ and\ \bibinfo {author} {\bibfnamefont {B.}~\bibnamefont {Zumino}},\
  }\href {\doibase 10.1016/0370-2693(71)90582-X} {\bibfield  {journal}
  {\bibinfo  {journal} {Phys. Lett.}\ }\textbf {\bibinfo {volume} {37B}},\
  \bibinfo {pages} {95} (\bibinfo {year} {1971})}\BibitemShut {NoStop}%
\bibitem [{\citenamefont {Witten}(1983)}]{Witten:1983tw}%
  \BibitemOpen
  \bibfield  {author} {\bibinfo {author} {\bibfnamefont {E.}~\bibnamefont
  {Witten}},\ }\href {\doibase 10.1016/0550-3213(83)90063-9} {\bibfield
  {journal} {\bibinfo  {journal} {Nucl. Phys.}\ }\textbf {\bibinfo {volume}
  {B223}},\ \bibinfo {pages} {422} (\bibinfo {year} {1983})}\BibitemShut
  {NoStop}%
\bibitem [{\citenamefont {Adler}\ \emph {et~al.}(1971)\citenamefont {Adler},
  \citenamefont {Lee}, \citenamefont {Treiman},\ and\ \citenamefont
  {Zee}}]{Adler:1971nq}%
  \BibitemOpen
  \bibfield  {author} {\bibinfo {author} {\bibfnamefont {S.~L.}\ \bibnamefont
  {Adler}}, \bibinfo {author} {\bibfnamefont {B.~W.}\ \bibnamefont {Lee}},
  \bibinfo {author} {\bibfnamefont {S.~B.}\ \bibnamefont {Treiman}}, \ and\
  \bibinfo {author} {\bibfnamefont {A.}~\bibnamefont {Zee}},\ }\href {\doibase
  10.1103/PhysRevD.4.3497} {\bibfield  {journal} {\bibinfo  {journal} {Phys.
  Rev.}\ }\textbf {\bibinfo {volume} {D4}},\ \bibinfo {pages} {3497} (\bibinfo
  {year} {1971})}\BibitemShut {NoStop}%
\bibitem [{\citenamefont {Terent'ev}(1972)}]{Terentev:1971cso}%
  \BibitemOpen
  \bibfield  {author} {\bibinfo {author} {\bibfnamefont {M.~V.}\ \bibnamefont
  {Terent'ev}},\ }\href {\doibase 10.1016/0370-2693(72)90171-2} {\bibfield
  {journal} {\bibinfo  {journal} {Phys. Lett.}\ }\textbf {\bibinfo {volume}
  {38B}},\ \bibinfo {pages} {419} (\bibinfo {year} {1972})}\BibitemShut
  {NoStop}%
\bibitem [{\citenamefont {Aviv}\ and\ \citenamefont {Zee}(1972)}]{Aviv:1971hq}%
  \BibitemOpen
  \bibfield  {author} {\bibinfo {author} {\bibfnamefont {R.}~\bibnamefont
  {Aviv}}\ and\ \bibinfo {author} {\bibfnamefont {A.}~\bibnamefont {Zee}},\
  }\href {\doibase 10.1103/PhysRevD.5.2372} {\bibfield  {journal} {\bibinfo
  {journal} {Phys. Rev.}\ }\textbf {\bibinfo {volume} {D5}},\ \bibinfo {pages}
  {2372} (\bibinfo {year} {1972})}\BibitemShut {NoStop}%
\bibitem [{\citenamefont {Ananthanarayan}\ \emph {et~al.}(2016)\citenamefont
  {Ananthanarayan}, \citenamefont {Caprini}, \citenamefont {Das},\ and\
  \citenamefont {Sentitemsu~Imsong}}]{Ananthanarayan:2016mns}%
  \BibitemOpen
  \bibfield  {author} {\bibinfo {author} {\bibfnamefont {B.}~\bibnamefont
  {Ananthanarayan}}, \bibinfo {author} {\bibfnamefont {I.}~\bibnamefont
  {Caprini}}, \bibinfo {author} {\bibfnamefont {D.}~\bibnamefont {Das}}, \ and\
  \bibinfo {author} {\bibfnamefont {I.}~\bibnamefont {Sentitemsu~Imsong}},\
  }\href {\doibase 10.1103/PhysRevD.93.116007} {\bibfield  {journal} {\bibinfo
  {journal} {Phys. Rev.}\ }\textbf {\bibinfo {volume} {D93}},\ \bibinfo {pages}
  {116007} (\bibinfo {year} {2016})},\ \Eprint
  {http://arxiv.org/abs/1605.00202} {arXiv:1605.00202 [hep-ph]}\BibitemShut
  {NoStop}%
\bibitem [{\citenamefont {Watson}(1954)}]{Watson:1954uc}%
  \BibitemOpen
  \bibfield  {author} {\bibinfo {author} {\bibfnamefont {K.~M.}\ \bibnamefont
  {Watson}},\ }\href {\doibase 10.1103/PhysRev.95.228} {\bibfield  {journal}
  {\bibinfo  {journal} {Phys. Rev.}\ }\textbf {\bibinfo {volume} {95}},\
  \bibinfo {pages} {228} (\bibinfo {year} {1954})}\BibitemShut {NoStop}%
\bibitem [{\citenamefont {Horn}\ \emph {et~al.}(2006)\citenamefont {Horn} \emph
  {et~al.}}]{Horn:2006tm}%
  \BibitemOpen
  \bibfield  {author} {\bibinfo {author} {\bibfnamefont {T.}~\bibnamefont
  {Horn}} \emph {et~al.} (\bibinfo {collaboration} {Jefferson Lab $F_\pi$}),\
  }\href {\doibase 10.1103/PhysRevLett.97.192001} {\bibfield  {journal}
  {\bibinfo  {journal} {Phys. Rev. Lett.}\ }\textbf {\bibinfo {volume} {97}},\
  \bibinfo {pages} {192001} (\bibinfo {year} {2006})},\ \Eprint
  {http://arxiv.org/abs/nucl-ex/0607005} {arXiv:nucl-ex/0607005
  [nucl-ex]}\BibitemShut {NoStop}%
\bibitem [{\citenamefont {Huber}\ \emph {et~al.}(2008)\citenamefont {Huber}
  \emph {et~al.}}]{Huber:2008id}%
  \BibitemOpen
  \bibfield  {author} {\bibinfo {author} {\bibfnamefont {G.~M.}\ \bibnamefont
  {Huber}} \emph {et~al.} (\bibinfo {collaboration} {Jefferson Lab $F_\pi$}),\
  }\href {\doibase 10.1103/PhysRevC.78.045203} {\bibfield  {journal} {\bibinfo
  {journal} {Phys. Rev.}\ }\textbf {\bibinfo {volume} {C78}},\ \bibinfo {pages}
  {045203} (\bibinfo {year} {2008})},\ \Eprint {http://arxiv.org/abs/0809.3052}
  {arXiv:0809.3052 [nucl-ex]}\BibitemShut {NoStop}%
\bibitem [{\citenamefont {Keshavarzi}\ \emph {et~al.}()\citenamefont
  {Keshavarzi}, \citenamefont {Nomura},\ and\ \citenamefont
  {Teubner}}]{KNT:private}%
  \BibitemOpen
  \bibfield  {author} {\bibinfo {author} {\bibfnamefont {A.}~\bibnamefont
  {Keshavarzi}}, \bibinfo {author} {\bibfnamefont {D.}~\bibnamefont {Nomura}},
  \ and\ \bibinfo {author} {\bibfnamefont {T.}~\bibnamefont {Teubner}},\
  }\href@noop {} {}\bibinfo {howpublished} {{private
  communication}}\BibitemShut {NoStop}%
\bibitem [{\citenamefont {Davier}\ \emph {et~al.}()\citenamefont {Davier},
  \citenamefont {Hoecker}, \citenamefont {Malaescu},\ and\ \citenamefont
  {Zhang}}]{DHMZ:private}%
  \BibitemOpen
  \bibfield  {author} {\bibinfo {author} {\bibfnamefont {M.}~\bibnamefont
  {Davier}}, \bibinfo {author} {\bibfnamefont {A.}~\bibnamefont {Hoecker}},
  \bibinfo {author} {\bibfnamefont {B.}~\bibnamefont {Malaescu}}, \ and\
  \bibinfo {author} {\bibfnamefont {Z.}~\bibnamefont {Zhang}},\ }\href@noop {}
  {}\bibinfo {howpublished} {{private communication}}\BibitemShut {NoStop}%
\bibitem [{\citenamefont {Achasov}\ and\ \citenamefont
  {Kiselev}(2002)}]{Achasov:2002bh}%
  \BibitemOpen
  \bibfield  {author} {\bibinfo {author} {\bibfnamefont {N.~N.}\ \bibnamefont
  {Achasov}}\ and\ \bibinfo {author} {\bibfnamefont {A.~V.}\ \bibnamefont
  {Kiselev}},\ }\href {\doibase 10.1103/PhysRevD.65.097302} {\bibfield
  {journal} {\bibinfo  {journal} {Phys. Rev.}\ }\textbf {\bibinfo {volume}
  {D65}},\ \bibinfo {pages} {097302} (\bibinfo {year} {2002})},\ \Eprint
  {http://arxiv.org/abs/hep-ph/0202047} {arXiv:hep-ph/0202047
  [hep-ph]}\BibitemShut {NoStop}%
\bibitem [{\citenamefont {Kuraev}\ and\ \citenamefont
  {Silagadze}(1995)}]{Kuraev:1995hc}%
  \BibitemOpen
  \bibfield  {author} {\bibinfo {author} {\bibfnamefont {E.~A.}\ \bibnamefont
  {Kuraev}}\ and\ \bibinfo {author} {\bibfnamefont {Z.~K.}\ \bibnamefont
  {Silagadze}},\ }\href@noop {} {\bibfield  {journal} {\bibinfo  {journal}
  {Phys. Atom. Nucl.}\ }\textbf {\bibinfo {volume} {58}},\ \bibinfo {pages}
  {1589} (\bibinfo {year} {1995})},\ \bibinfo {note} {[Yad. Fiz. {\bf 58N9},
  1687 (1995)]},\ \Eprint {http://arxiv.org/abs/hep-ph/9502406}
  {arXiv:hep-ph/9502406 [hep-ph]}\BibitemShut {NoStop}%
\bibitem [{\citenamefont {Ahmedov}\ \emph {et~al.}(2002)\citenamefont
  {Ahmedov}, \citenamefont {Fedotovich}, \citenamefont {Kuraev},\ and\
  \citenamefont {Silagadze}}]{Ahmedov:2002tg}%
  \BibitemOpen
  \bibfield  {author} {\bibinfo {author} {\bibfnamefont {A.~I.}\ \bibnamefont
  {Ahmedov}}, \bibinfo {author} {\bibfnamefont {G.~V.}\ \bibnamefont
  {Fedotovich}}, \bibinfo {author} {\bibfnamefont {E.~A.}\ \bibnamefont
  {Kuraev}}, \ and\ \bibinfo {author} {\bibfnamefont {Z.~K.}\ \bibnamefont
  {Silagadze}},\ }\href {\doibase 10.1088/1126-6708/2002/09/008,
  10.1134/1.1755389} {\bibfield  {journal} {\bibinfo  {journal} {JHEP}\
  }\textbf {\bibinfo {volume} {09}},\ \bibinfo {pages} {008} (\bibinfo {year}
  {2002})},\ \bibinfo {note} {[Yad. Fiz. {\bf 67}, 1006 (2004)]},\ \Eprint
  {http://arxiv.org/abs/hep-ph/0201157} {arXiv:hep-ph/0201157
  [hep-ph]}\BibitemShut {NoStop}%
\bibitem [{\citenamefont {Hoid}()}]{Hoid:private}%
  \BibitemOpen
  \bibfield  {author} {\bibinfo {author} {\bibfnamefont {B.-L.}\ \bibnamefont
  {Hoid}},\ }\href@noop {} {}\bibinfo {howpublished} {{private
  communication}}\BibitemShut {NoStop}%
\bibitem [{\citenamefont {Lyons}\ \emph {et~al.}(1988)\citenamefont {Lyons},
  \citenamefont {Gibaut},\ and\ \citenamefont {Clifford}}]{Lyons:1988rp}%
  \BibitemOpen
  \bibfield  {author} {\bibinfo {author} {\bibfnamefont {L.}~\bibnamefont
  {Lyons}}, \bibinfo {author} {\bibfnamefont {D.}~\bibnamefont {Gibaut}}, \
  and\ \bibinfo {author} {\bibfnamefont {P.}~\bibnamefont {Clifford}},\ }\href
  {\doibase 10.1016/0168-9002(88)90018-6} {\bibfield  {journal} {\bibinfo
  {journal} {Nucl. Instrum. Meth.}\ }\textbf {\bibinfo {volume} {A270}},\
  \bibinfo {pages} {110} (\bibinfo {year} {1988})}\BibitemShut {NoStop}%
\bibitem [{\citenamefont {Nakamura}\ \emph {et~al.}(2010)\citenamefont
  {Nakamura} \emph {et~al.}}]{Nakamura:2010zzi}%
  \BibitemOpen
  \bibfield  {author} {\bibinfo {author} {\bibfnamefont {K.}~\bibnamefont
  {Nakamura}} \emph {et~al.} (\bibinfo {collaboration} {Particle Data Group}),\
  }\href {\doibase 10.1088/0954-3899/37/7A/075021} {\bibfield  {journal}
  {\bibinfo  {journal} {J. Phys.}\ }\textbf {\bibinfo {volume} {G37}},\
  \bibinfo {pages} {075021} (\bibinfo {year} {2010})}\BibitemShut {NoStop}%
\bibitem [{\citenamefont {Benayoun}()}]{Benayoun:private}%
  \BibitemOpen
  \bibfield  {author} {\bibinfo {author} {\bibfnamefont {M.}~\bibnamefont
  {Benayoun}},\ }\href@noop {} {}\bibinfo {howpublished} {{private
  communication}}\BibitemShut {NoStop}%
\bibitem [{\citenamefont {Aad}\ \emph {et~al.}(2015)\citenamefont {Aad} \emph
  {et~al.}}]{Aad:2014bia}%
  \BibitemOpen
  \bibfield  {author} {\bibinfo {author} {\bibfnamefont {G.}~\bibnamefont
  {Aad}} \emph {et~al.} (\bibinfo {collaboration} {ATLAS}),\ }\href {\doibase
  10.1140/epjc/s10052-014-3190-y} {\bibfield  {journal} {\bibinfo  {journal}
  {Eur. Phys. J.}\ }\textbf {\bibinfo {volume} {C75}},\ \bibinfo {pages} {17}
  (\bibinfo {year} {2015})},\ \Eprint {http://arxiv.org/abs/1406.0076}
  {arXiv:1406.0076 [hep-ex]}\BibitemShut {NoStop}%
\bibitem [{\citenamefont {Aaboud}\ \emph {et~al.}(2017)\citenamefont {Aaboud}
  \emph {et~al.}}]{Aaboud:2017dvo}%
  \BibitemOpen
  \bibfield  {author} {\bibinfo {author} {\bibfnamefont {M.}~\bibnamefont
  {Aaboud}} \emph {et~al.} (\bibinfo {collaboration} {ATLAS}),\ }\href
  {\doibase 10.1007/JHEP09(2017)020} {\bibfield  {journal} {\bibinfo  {journal}
  {JHEP}\ }\textbf {\bibinfo {volume} {09}},\ \bibinfo {pages} {020} (\bibinfo
  {year} {2017})},\ \Eprint {http://arxiv.org/abs/1706.03192} {arXiv:1706.03192
  [hep-ex]}\BibitemShut {NoStop}%
\bibitem [{\citenamefont {Aaboud}\ \emph {et~al.}(2018)\citenamefont {Aaboud}
  \emph {et~al.}}]{Aaboud:2017wsi}%
  \BibitemOpen
  \bibfield  {author} {\bibinfo {author} {\bibfnamefont {M.}~\bibnamefont
  {Aaboud}} \emph {et~al.} (\bibinfo {collaboration} {ATLAS}),\ }\href
  {\doibase 10.1007/JHEP05(2018)195} {\bibfield  {journal} {\bibinfo  {journal}
  {JHEP}\ }\textbf {\bibinfo {volume} {05}},\ \bibinfo {pages} {195} (\bibinfo
  {year} {2018})},\ \Eprint {http://arxiv.org/abs/1711.02692} {arXiv:1711.02692
  [hep-ex]}\BibitemShut {NoStop}%
\bibitem [{\citenamefont
  {Malaescu}(2018{\natexlab{d}})}]{bogdan-Mainz-2018-DHMZ-UncOnUnc}%
  \BibitemOpen
  \bibfield  {author} {\bibinfo {author} {\bibfnamefont {B.}~\bibnamefont
  {Malaescu}},\ }\href@noop {} {\enquote {\bibinfo {title} {{Treatment of
  uncertainties and correlations in combinations of $e^+e^-$ annihilation
  data}},}\ }\bibinfo {howpublished}
  {\url{https://indico.him.uni-mainz.de/event/11/session/1/contribution/42/material/slides/0.pdf}}
  (\bibinfo {year} {2018}{\natexlab{d}}),\ \bibinfo {note} {{Muon $g-2$ Theory
  Initiative workshop Mainz}}\BibitemShut {NoStop}%
\bibitem [{\citenamefont {Ananthanarayan}\ \emph {et~al.}(2019)\citenamefont
  {Ananthanarayan}, \citenamefont {Caprini},\ and\ \citenamefont
  {Das}}]{Ananthanarayan:2019zic}%
  \BibitemOpen
  \bibfield  {author} {\bibinfo {author} {\bibfnamefont {B.}~\bibnamefont
  {Ananthanarayan}}, \bibinfo {author} {\bibfnamefont {I.}~\bibnamefont
  {Caprini}}, \ and\ \bibinfo {author} {\bibfnamefont {D.}~\bibnamefont
  {Das}},\ }\href {http://www.nipne.ro/rjp/2019_64_7-8/RomJPhys.64.401.pdf}
  {\bibfield  {journal} {\bibinfo  {journal} {Rom. J. Phys.}\ }\textbf
  {\bibinfo {volume} {64}},\ \bibinfo {pages} {401} (\bibinfo {year} {2019})},\
  \Eprint {http://arxiv.org/abs/1907.01767} {arXiv:1907.01767
  [hep-ph]}\BibitemShut {NoStop}%
\bibitem [{\citenamefont {Schmelling}(1995)}]{Schmelling:1994pz}%
  \BibitemOpen
  \bibfield  {author} {\bibinfo {author} {\bibfnamefont {M.}~\bibnamefont
  {Schmelling}},\ }\href {\doibase 10.1088/0031-8949/51/6/002} {\bibfield
  {journal} {\bibinfo  {journal} {Phys. Scripta}\ }\textbf {\bibinfo {volume}
  {51}},\ \bibinfo {pages} {676} (\bibinfo {year} {1995})}\BibitemShut
  {NoStop}%
\bibitem [{\citenamefont {Calmet}\ \emph {et~al.}(1976)\citenamefont {Calmet},
  \citenamefont {Narison}, \citenamefont {Perrottet},\ and\ \citenamefont
  {de~Rafael}}]{Calmet:1976kd}%
  \BibitemOpen
  \bibfield  {author} {\bibinfo {author} {\bibfnamefont {J.}~\bibnamefont
  {Calmet}}, \bibinfo {author} {\bibfnamefont {S.}~\bibnamefont {Narison}},
  \bibinfo {author} {\bibfnamefont {M.}~\bibnamefont {Perrottet}}, \ and\
  \bibinfo {author} {\bibfnamefont {E.}~\bibnamefont {de~Rafael}},\ }\href
  {\doibase 10.1016/0370-2693(76)90150-7} {\bibfield  {journal} {\bibinfo
  {journal} {Phys. Lett.}\ }\textbf {\bibinfo {volume} {61B}},\ \bibinfo
  {pages} {283} (\bibinfo {year} {1976})}\BibitemShut {NoStop}%
\bibitem [{\citenamefont {Barbieri}\ and\ \citenamefont
  {Remiddi}(1975)}]{Barbieri:1974nc}%
  \BibitemOpen
  \bibfield  {author} {\bibinfo {author} {\bibfnamefont {R.}~\bibnamefont
  {Barbieri}}\ and\ \bibinfo {author} {\bibfnamefont {E.}~\bibnamefont
  {Remiddi}},\ }\href {\doibase 10.1016/0550-3213(75)90645-8} {\bibfield
  {journal} {\bibinfo  {journal} {Nucl. Phys.}\ }\textbf {\bibinfo {volume}
  {B90}},\ \bibinfo {pages} {233} (\bibinfo {year} {1975})}\BibitemShut
  {NoStop}%
\bibitem [{\citenamefont {Carloni~Calame}\ \emph {et~al.}(2015)\citenamefont
  {Carloni~Calame}, \citenamefont {Passera}, \citenamefont {Trentadue},\ and\
  \citenamefont {Venanzoni}}]{Calame:2015fva}%
  \BibitemOpen
  \bibfield  {author} {\bibinfo {author} {\bibfnamefont {C.~M.}\ \bibnamefont
  {Carloni~Calame}}, \bibinfo {author} {\bibfnamefont {M.}~\bibnamefont
  {Passera}}, \bibinfo {author} {\bibfnamefont {L.}~\bibnamefont {Trentadue}},
  \ and\ \bibinfo {author} {\bibfnamefont {G.}~\bibnamefont {Venanzoni}},\
  }\href {\doibase 10.1016/j.physletb.2015.05.020} {\bibfield  {journal}
  {\bibinfo  {journal} {Phys. Lett.}\ }\textbf {\bibinfo {volume} {B746}},\
  \bibinfo {pages} {325} (\bibinfo {year} {2015})},\ \Eprint
  {http://arxiv.org/abs/1504.02228} {arXiv:1504.02228 [hep-ph]}\BibitemShut
  {NoStop}%
\bibitem [{\citenamefont {Abbiendi}\ \emph {et~al.}(2017)\citenamefont
  {Abbiendi} \emph {et~al.}}]{Abbiendi:2016xup}%
  \BibitemOpen
  \bibfield  {author} {\bibinfo {author} {\bibfnamefont {G.}~\bibnamefont
  {Abbiendi}} \emph {et~al.},\ }\href {\doibase 10.1140/epjc/s10052-017-4633-z}
  {\bibfield  {journal} {\bibinfo  {journal} {Eur. Phys. J.}\ }\textbf
  {\bibinfo {volume} {C77}},\ \bibinfo {pages} {139} (\bibinfo {year}
  {2017})},\ \Eprint {http://arxiv.org/abs/1609.08987} {arXiv:1609.08987
  [hep-ex]}\BibitemShut {NoStop}%
\bibitem [{\citenamefont {Marinkovi{\'c}}\ and\ \citenamefont
  {Cardoso}(2019)}]{Marinkovic:2019zoi}%
  \BibitemOpen
  \bibfield  {author} {\bibinfo {author} {\bibfnamefont {M.~K.}\ \bibnamefont
  {Marinkovi{\'c}}}\ and\ \bibinfo {author} {\bibfnamefont {N.}~\bibnamefont
  {Cardoso}},\ }\href@noop {} {\  (\bibinfo {year} {2019})},\ \Eprint
  {http://arxiv.org/abs/1910.06467} {arXiv:1910.06467 [hep-lat]}\BibitemShut
  {NoStop}%
\bibitem [{\citenamefont {Abbiendi}\ \emph {et~al.}(2019)\citenamefont
  {Abbiendi} \emph {et~al.}}]{MUonE:LoI}%
  \BibitemOpen
  \bibfield  {author} {\bibinfo {author} {\bibfnamefont {G.}~\bibnamefont
  {Abbiendi}} \emph {et~al.} (\bibinfo {collaboration} {{MUonE}}),\ }\href
  {https://cds.cern.ch/record/2677471} {\emph {\bibinfo {title} {{Letter of
  Intent: the MUonE project}}}},\ \bibinfo {type} {Tech. Rep.}\ \bibinfo
  {number} {CERN-SPSC-2019-026, SPSC-I-252}\ (\bibinfo {year}
  {2019})\BibitemShut {NoStop}%
\bibitem [{\citenamefont {Abbiendi}\ \emph {et~al.}(2020)\citenamefont
  {Abbiendi} \emph {et~al.}}]{Abbiendi:2019qtw}%
  \BibitemOpen
  \bibfield  {author} {\bibinfo {author} {\bibfnamefont {G.}~\bibnamefont
  {Abbiendi}} \emph {et~al.},\ }\href {\doibase 10.1088/1748-0221/15/01/P01017}
  {\bibfield  {journal} {\bibinfo  {journal} {JINST}\ }\textbf {\bibinfo
  {volume} {15}},\ \bibinfo {pages} {01} (\bibinfo {year} {2020})},\ \Eprint
  {http://arxiv.org/abs/1905.11677} {arXiv:1905.11677
  [physics.ins-det]}\BibitemShut {NoStop}%
\bibitem [{\citenamefont {{CMS collaboration}}(2017)}]{cmsu}%
  \BibitemOpen
  \bibfield  {author} {\bibinfo {author} {\bibnamefont {{CMS collaboration}}},\
  }\href {https://cds.cern.ch/record/2272264} {\emph {\bibinfo {title} {{The
  Phase-2 Upgrade of the CMS Tracker Tech. Rep.}}}},\ \bibinfo {type} {Tech.
  Rep.}\ \bibinfo {number} {CERNLHCC-2017-009, CMS-TDR-014}\ (\bibinfo {year}
  {2017})\BibitemShut {NoStop}%
\bibitem [{\citenamefont {Banerjee}\ \emph {et~al.}(2020)\citenamefont
  {Banerjee} \emph {et~al.}}]{Banerjee:2020tdt}%
  \BibitemOpen
  \bibfield  {author} {\bibinfo {author} {\bibfnamefont {P.}~\bibnamefont
  {Banerjee}} \emph {et~al.},\ }\href {\doibase 10.1140/epjc/s10052-020-8138-9}
  {\bibfield  {journal} {\bibinfo  {journal} {Eur. Phys. J. C}\ }\textbf
  {\bibinfo {volume} {80}},\ \bibinfo {pages} {591} (\bibinfo {year} {2020})},\
  \Eprint {http://arxiv.org/abs/2004.13663} {arXiv:2004.13663
  [hep-ph]}\BibitemShut {NoStop}%
\bibitem [{\citenamefont {Kaiser}(2010)}]{Kaiser:2010zz}%
  \BibitemOpen
  \bibfield  {author} {\bibinfo {author} {\bibfnamefont {N.}~\bibnamefont
  {Kaiser}},\ }\href {\doibase 10.1088/0954-3899/37/11/115005} {\bibfield
  {journal} {\bibinfo  {journal} {J. Phys.}\ }\textbf {\bibinfo {volume}
  {G37}},\ \bibinfo {pages} {115005} (\bibinfo {year} {2010})}\BibitemShut
  {NoStop}%
\bibitem [{\citenamefont {Alacevich}\ \emph {et~al.}(2019)\citenamefont
  {Alacevich}, \citenamefont {Carloni~Calame}, \citenamefont {Chiesa},
  \citenamefont {Montagna}, \citenamefont {Nicrosini},\ and\ \citenamefont
  {Piccinini}}]{Alacevich:2018vez}%
  \BibitemOpen
  \bibfield  {author} {\bibinfo {author} {\bibfnamefont {M.}~\bibnamefont
  {Alacevich}}, \bibinfo {author} {\bibfnamefont {C.~M.}\ \bibnamefont
  {Carloni~Calame}}, \bibinfo {author} {\bibfnamefont {M.}~\bibnamefont
  {Chiesa}}, \bibinfo {author} {\bibfnamefont {G.}~\bibnamefont {Montagna}},
  \bibinfo {author} {\bibfnamefont {O.}~\bibnamefont {Nicrosini}}, \ and\
  \bibinfo {author} {\bibfnamefont {F.}~\bibnamefont {Piccinini}},\ }\href
  {\doibase 10.1007/JHEP02(2019)155} {\bibfield  {journal} {\bibinfo  {journal}
  {JHEP}\ }\textbf {\bibinfo {volume} {02}},\ \bibinfo {pages} {155} (\bibinfo
  {year} {2019})},\ \Eprint {http://arxiv.org/abs/1811.06743} {arXiv:1811.06743
  [hep-ph]}\BibitemShut {NoStop}%
\bibitem [{\citenamefont {Bern}\ \emph {et~al.}(2001)\citenamefont {Bern},
  \citenamefont {Dixon},\ and\ \citenamefont {Ghinculov}}]{Bern:2000ie}%
  \BibitemOpen
  \bibfield  {author} {\bibinfo {author} {\bibfnamefont {Z.}~\bibnamefont
  {Bern}}, \bibinfo {author} {\bibfnamefont {L.~J.}\ \bibnamefont {Dixon}}, \
  and\ \bibinfo {author} {\bibfnamefont {A.}~\bibnamefont {Ghinculov}},\ }\href
  {\doibase 10.1103/PhysRevD.63.053007} {\bibfield  {journal} {\bibinfo
  {journal} {Phys. Rev.}\ }\textbf {\bibinfo {volume} {D63}},\ \bibinfo {pages}
  {053007} (\bibinfo {year} {2001})},\ \Eprint
  {http://arxiv.org/abs/hep-ph/0010075} {arXiv:hep-ph/0010075
  [hep-ph]}\BibitemShut {NoStop}%
\bibitem [{\citenamefont {Bonciani}\ \emph {et~al.}(2003)\citenamefont
  {Bonciani}, \citenamefont {Mastrolia},\ and\ \citenamefont
  {Remiddi}}]{Bonciani:2003te}%
  \BibitemOpen
  \bibfield  {author} {\bibinfo {author} {\bibfnamefont {R.}~\bibnamefont
  {Bonciani}}, \bibinfo {author} {\bibfnamefont {P.}~\bibnamefont {Mastrolia}},
  \ and\ \bibinfo {author} {\bibfnamefont {E.}~\bibnamefont {Remiddi}},\ }\href
  {\doibase 10.1016/S0550-3213(03)00299-2} {\bibfield  {journal} {\bibinfo
  {journal} {Nucl. Phys.}\ }\textbf {\bibinfo {volume} {B661}},\ \bibinfo
  {pages} {289} (\bibinfo {year} {2003})},\ \bibinfo {note} {[Erratum: Nucl.
  Phys. {\bf B702}, 359 (2004)]},\ \Eprint
  {http://arxiv.org/abs/hep-ph/0301170} {arXiv:hep-ph/0301170
  [hep-ph]}\BibitemShut {NoStop}%
\bibitem [{\citenamefont {Bonciani}\ \emph {et~al.}(2004)\citenamefont
  {Bonciani}, \citenamefont {Ferroglia}, \citenamefont {Mastrolia},
  \citenamefont {Remiddi},\ and\ \citenamefont {van~der
  Bij}}]{Bonciani:2003cj}%
  \BibitemOpen
  \bibfield  {author} {\bibinfo {author} {\bibfnamefont {R.}~\bibnamefont
  {Bonciani}}, \bibinfo {author} {\bibfnamefont {A.}~\bibnamefont {Ferroglia}},
  \bibinfo {author} {\bibfnamefont {P.}~\bibnamefont {Mastrolia}}, \bibinfo
  {author} {\bibfnamefont {E.}~\bibnamefont {Remiddi}}, \ and\ \bibinfo
  {author} {\bibfnamefont {J.~J.}\ \bibnamefont {van~der Bij}},\ }\href
  {\doibase 10.1016/j.nuclphysb.2004.01.026} {\bibfield  {journal} {\bibinfo
  {journal} {Nucl. Phys.}\ }\textbf {\bibinfo {volume} {B681}},\ \bibinfo
  {pages} {261} (\bibinfo {year} {2004})},\ \bibinfo {note} {[Erratum: Nucl.
  Phys. {\bf B702}, 364 (2004)]},\ \Eprint
  {http://arxiv.org/abs/hep-ph/0310333} {arXiv:hep-ph/0310333
  [hep-ph]}\BibitemShut {NoStop}%
\bibitem [{\citenamefont {Bonciani}\ \emph {et~al.}(2008)\citenamefont
  {Bonciani}, \citenamefont {Ferroglia}, \citenamefont {Gehrmann},
  \citenamefont {Ma{\^i}tre},\ and\ \citenamefont
  {Studerus}}]{Bonciani:2008az}%
  \BibitemOpen
  \bibfield  {author} {\bibinfo {author} {\bibfnamefont {R.}~\bibnamefont
  {Bonciani}}, \bibinfo {author} {\bibfnamefont {A.}~\bibnamefont {Ferroglia}},
  \bibinfo {author} {\bibfnamefont {T.}~\bibnamefont {Gehrmann}}, \bibinfo
  {author} {\bibfnamefont {D.}~\bibnamefont {Ma{\^i}tre}}, \ and\ \bibinfo
  {author} {\bibfnamefont {C.}~\bibnamefont {Studerus}},\ }\href {\doibase
  10.1088/1126-6708/2008/07/129} {\bibfield  {journal} {\bibinfo  {journal}
  {JHEP}\ }\textbf {\bibinfo {volume} {07}},\ \bibinfo {pages} {129} (\bibinfo
  {year} {2008})},\ \Eprint {http://arxiv.org/abs/0806.2301} {arXiv:0806.2301
  [hep-ph]}\BibitemShut {NoStop}%
\bibitem [{\citenamefont {Bonciani}\ \emph {et~al.}(2013)\citenamefont
  {Bonciani}, \citenamefont {Ferroglia}, \citenamefont {Gehrmann},
  \citenamefont {von Manteuffel},\ and\ \citenamefont
  {Studerus}}]{Bonciani:2013ywa}%
  \BibitemOpen
  \bibfield  {author} {\bibinfo {author} {\bibfnamefont {R.}~\bibnamefont
  {Bonciani}}, \bibinfo {author} {\bibfnamefont {A.}~\bibnamefont {Ferroglia}},
  \bibinfo {author} {\bibfnamefont {T.}~\bibnamefont {Gehrmann}}, \bibinfo
  {author} {\bibfnamefont {A.}~\bibnamefont {von Manteuffel}}, \ and\ \bibinfo
  {author} {\bibfnamefont {C.}~\bibnamefont {Studerus}},\ }\href {\doibase
  10.1007/JHEP12(2013)038} {\bibfield  {journal} {\bibinfo  {journal} {JHEP}\
  }\textbf {\bibinfo {volume} {12}},\ \bibinfo {pages} {038} (\bibinfo {year}
  {2013})},\ \Eprint {http://arxiv.org/abs/1309.4450} {arXiv:1309.4450
  [hep-ph]}\BibitemShut {NoStop}%
\bibitem [{\citenamefont {Mastrolia}\ \emph {et~al.}(2017)\citenamefont
  {Mastrolia}, \citenamefont {Passera}, \citenamefont {Primo},\ and\
  \citenamefont {Schubert}}]{Mastrolia:2017pfy}%
  \BibitemOpen
  \bibfield  {author} {\bibinfo {author} {\bibfnamefont {P.}~\bibnamefont
  {Mastrolia}}, \bibinfo {author} {\bibfnamefont {M.}~\bibnamefont {Passera}},
  \bibinfo {author} {\bibfnamefont {A.}~\bibnamefont {Primo}}, \ and\ \bibinfo
  {author} {\bibfnamefont {U.}~\bibnamefont {Schubert}},\ }\href {\doibase
  10.1007/JHEP11(2017)198} {\bibfield  {journal} {\bibinfo  {journal} {JHEP}\
  }\textbf {\bibinfo {volume} {11}},\ \bibinfo {pages} {198} (\bibinfo {year}
  {2017})},\ \Eprint {http://arxiv.org/abs/1709.07435} {arXiv:1709.07435
  [hep-ph]}\BibitemShut {NoStop}%
\bibitem [{\citenamefont {Di~Vita}\ \emph {et~al.}(2018)\citenamefont
  {Di~Vita}, \citenamefont {Laporta}, \citenamefont {Mastrolia}, \citenamefont
  {Primo},\ and\ \citenamefont {Schubert}}]{DiVita:2018nnh}%
  \BibitemOpen
  \bibfield  {author} {\bibinfo {author} {\bibfnamefont {S.}~\bibnamefont
  {Di~Vita}}, \bibinfo {author} {\bibfnamefont {S.}~\bibnamefont {Laporta}},
  \bibinfo {author} {\bibfnamefont {P.}~\bibnamefont {Mastrolia}}, \bibinfo
  {author} {\bibfnamefont {A.}~\bibnamefont {Primo}}, \ and\ \bibinfo {author}
  {\bibfnamefont {U.}~\bibnamefont {Schubert}},\ }\href {\doibase
  10.1007/JHEP09(2018)016} {\bibfield  {journal} {\bibinfo  {journal} {JHEP}\
  }\textbf {\bibinfo {volume} {09}},\ \bibinfo {pages} {016} (\bibinfo {year}
  {2018})},\ \Eprint {http://arxiv.org/abs/1806.08241} {arXiv:1806.08241
  [hep-ph]}\BibitemShut {NoStop}%
\bibitem [{\citenamefont {Mastrolia}\ \emph {et~al.}(2018)\citenamefont
  {Mastrolia}, \citenamefont {Passera}, \citenamefont {Primo}, \citenamefont
  {Schubert},\ and\ \citenamefont {Torres~Bobadilla}}]{Mastrolia:2018sso}%
  \BibitemOpen
  \bibfield  {author} {\bibinfo {author} {\bibfnamefont {P.}~\bibnamefont
  {Mastrolia}}, \bibinfo {author} {\bibfnamefont {M.}~\bibnamefont {Passera}},
  \bibinfo {author} {\bibfnamefont {A.}~\bibnamefont {Primo}}, \bibinfo
  {author} {\bibfnamefont {U.}~\bibnamefont {Schubert}}, \ and\ \bibinfo
  {author} {\bibfnamefont {W.}~\bibnamefont {Torres~Bobadilla}},\ }\href
  {\doibase 10.1051/epjconf/201817901014} {\bibfield  {journal} {\bibinfo
  {journal} {EPJ Web Conf.}\ }\textbf {\bibinfo {volume} {179}},\ \bibinfo
  {pages} {01014} (\bibinfo {year} {2018})}\BibitemShut {NoStop}%
\bibitem [{\citenamefont {Di~Vita}\ \emph {et~al.}(2019)\citenamefont
  {Di~Vita}, \citenamefont {Gehrmann}, \citenamefont {Laporta}, \citenamefont
  {Mastrolia}, \citenamefont {Primo},\ and\ \citenamefont
  {Schubert}}]{DiVita:2019lpl}%
  \BibitemOpen
  \bibfield  {author} {\bibinfo {author} {\bibfnamefont {S.}~\bibnamefont
  {Di~Vita}}, \bibinfo {author} {\bibfnamefont {T.}~\bibnamefont {Gehrmann}},
  \bibinfo {author} {\bibfnamefont {S.}~\bibnamefont {Laporta}}, \bibinfo
  {author} {\bibfnamefont {P.}~\bibnamefont {Mastrolia}}, \bibinfo {author}
  {\bibfnamefont {A.}~\bibnamefont {Primo}}, \ and\ \bibinfo {author}
  {\bibfnamefont {U.}~\bibnamefont {Schubert}},\ }\href {\doibase
  10.1007/JHEP06(2019)117} {\bibfield  {journal} {\bibinfo  {journal} {JHEP}\
  }\textbf {\bibinfo {volume} {06}},\ \bibinfo {pages} {117} (\bibinfo {year}
  {2019})},\ \Eprint {http://arxiv.org/abs/1904.10964} {arXiv:1904.10964
  [hep-ph]}\BibitemShut {NoStop}%
\bibitem [{\citenamefont {Ronca}(2020)}]{Ronca:2019kcw}%
  \BibitemOpen
  \bibfield  {author} {\bibinfo {author} {\bibfnamefont {J.}~\bibnamefont
  {Ronca}},\ }\href {\doibase 10.1051/epjconf/202023401015} {\bibfield
  {journal} {\bibinfo  {journal} {EPJ Web Conf.}\ }\textbf {\bibinfo {volume}
  {234}},\ \bibinfo {pages} {01015} (\bibinfo {year} {2020})},\ \Eprint
  {http://arxiv.org/abs/1912.05397} {arXiv:1912.05397 [hep-ph]}\BibitemShut
  {NoStop}%
\bibitem [{\citenamefont {Engel}\ \emph {et~al.}(2019)\citenamefont {Engel},
  \citenamefont {Gnendiger}, \citenamefont {Signer},\ and\ \citenamefont
  {Ulrich}}]{Engel:2018fsb}%
  \BibitemOpen
  \bibfield  {author} {\bibinfo {author} {\bibfnamefont {T.}~\bibnamefont
  {Engel}}, \bibinfo {author} {\bibfnamefont {C.}~\bibnamefont {Gnendiger}},
  \bibinfo {author} {\bibfnamefont {A.}~\bibnamefont {Signer}}, \ and\ \bibinfo
  {author} {\bibfnamefont {Y.}~\bibnamefont {Ulrich}},\ }\href {\doibase
  10.1007/JHEP02(2019)118} {\bibfield  {journal} {\bibinfo  {journal} {JHEP}\
  }\textbf {\bibinfo {volume} {02}},\ \bibinfo {pages} {118} (\bibinfo {year}
  {2019})},\ \Eprint {http://arxiv.org/abs/1811.06461} {arXiv:1811.06461
  [hep-ph]}\BibitemShut {NoStop}%
\bibitem [{\citenamefont {Penin}(2005)}]{Penin:2005kf}%
  \BibitemOpen
  \bibfield  {author} {\bibinfo {author} {\bibfnamefont {A.~A.}\ \bibnamefont
  {Penin}},\ }\href {\doibase 10.1103/PhysRevLett.95.010408} {\bibfield
  {journal} {\bibinfo  {journal} {Phys. Rev. Lett.}\ }\textbf {\bibinfo
  {volume} {95}},\ \bibinfo {pages} {010408} (\bibinfo {year} {2005})},\
  \Eprint {http://arxiv.org/abs/hep-ph/0501120} {arXiv:hep-ph/0501120
  [hep-ph]}\BibitemShut {NoStop}%
\bibitem [{\citenamefont {Mitov}\ and\ \citenamefont
  {Moch}(2007)}]{Mitov:2006xs}%
  \BibitemOpen
  \bibfield  {author} {\bibinfo {author} {\bibfnamefont {A.}~\bibnamefont
  {Mitov}}\ and\ \bibinfo {author} {\bibfnamefont {S.}~\bibnamefont {Moch}},\
  }\href {\doibase 10.1088/1126-6708/2007/05/001} {\bibfield  {journal}
  {\bibinfo  {journal} {JHEP}\ }\textbf {\bibinfo {volume} {05}},\ \bibinfo
  {pages} {001} (\bibinfo {year} {2007})},\ \Eprint
  {http://arxiv.org/abs/hep-ph/0612149} {arXiv:hep-ph/0612149
  [hep-ph]}\BibitemShut {NoStop}%
\bibitem [{\citenamefont {Becher}\ and\ \citenamefont
  {Melnikov}(2007)}]{Becher:2007cu}%
  \BibitemOpen
  \bibfield  {author} {\bibinfo {author} {\bibfnamefont {T.}~\bibnamefont
  {Becher}}\ and\ \bibinfo {author} {\bibfnamefont {K.}~\bibnamefont
  {Melnikov}},\ }\href {\doibase 10.1088/1126-6708/2007/06/084} {\bibfield
  {journal} {\bibinfo  {journal} {JHEP}\ }\textbf {\bibinfo {volume} {06}},\
  \bibinfo {pages} {084} (\bibinfo {year} {2007})},\ \Eprint
  {http://arxiv.org/abs/0704.3582} {arXiv:0704.3582 [hep-ph]}\BibitemShut
  {NoStop}%
\bibitem [{\citenamefont {Engel}\ \emph {et~al.}(2020)\citenamefont {Engel},
  \citenamefont {Signer},\ and\ \citenamefont {Ulrich}}]{Engel:2019nfw}%
  \BibitemOpen
  \bibfield  {author} {\bibinfo {author} {\bibfnamefont {T.}~\bibnamefont
  {Engel}}, \bibinfo {author} {\bibfnamefont {A.}~\bibnamefont {Signer}}, \
  and\ \bibinfo {author} {\bibfnamefont {Y.}~\bibnamefont {Ulrich}},\ }\href
  {\doibase 10.1007/JHEP01(2020)085} {\bibfield  {journal} {\bibinfo  {journal}
  {JHEP}\ }\textbf {\bibinfo {volume} {01}},\ \bibinfo {pages} {085} (\bibinfo
  {year} {2020})},\ \Eprint {http://arxiv.org/abs/1909.10244} {arXiv:1909.10244
  [hep-ph]}\BibitemShut {NoStop}%
\bibitem [{\citenamefont {Frixione}\ \emph {et~al.}(1996)\citenamefont
  {Frixione}, \citenamefont {Kunszt},\ and\ \citenamefont
  {Signer}}]{Frixione:1995ms}%
  \BibitemOpen
  \bibfield  {author} {\bibinfo {author} {\bibfnamefont {S.}~\bibnamefont
  {Frixione}}, \bibinfo {author} {\bibfnamefont {Z.}~\bibnamefont {Kunszt}}, \
  and\ \bibinfo {author} {\bibfnamefont {A.}~\bibnamefont {Signer}},\ }\href
  {\doibase 10.1016/0550-3213(96)00110-1} {\bibfield  {journal} {\bibinfo
  {journal} {Nucl. Phys. B}\ }\textbf {\bibinfo {volume} {467}},\ \bibinfo
  {pages} {399} (\bibinfo {year} {1996})},\ \Eprint
  {http://arxiv.org/abs/hep-ph/9512328} {arXiv:hep-ph/9512328}\BibitemShut
  {NoStop}%
\bibitem [{\citenamefont {Frederix}\ \emph {et~al.}(2009)\citenamefont
  {Frederix}, \citenamefont {Frixione}, \citenamefont {Maltoni},\ and\
  \citenamefont {Stelzer}}]{Frederix:2009yq}%
  \BibitemOpen
  \bibfield  {author} {\bibinfo {author} {\bibfnamefont {R.}~\bibnamefont
  {Frederix}}, \bibinfo {author} {\bibfnamefont {S.}~\bibnamefont {Frixione}},
  \bibinfo {author} {\bibfnamefont {F.}~\bibnamefont {Maltoni}}, \ and\
  \bibinfo {author} {\bibfnamefont {T.}~\bibnamefont {Stelzer}},\ }\href
  {\doibase 10.1088/1126-6708/2009/10/003} {\bibfield  {journal} {\bibinfo
  {journal} {JHEP}\ }\textbf {\bibinfo {volume} {10}},\ \bibinfo {pages} {003}
  (\bibinfo {year} {2009})},\ \Eprint {http://arxiv.org/abs/0908.4272}
  {arXiv:0908.4272 [hep-ph]}\BibitemShut {NoStop}%
\bibitem [{\citenamefont {Fael}\ and\ \citenamefont
  {Passera}(2019)}]{Fael:2019nsf}%
  \BibitemOpen
  \bibfield  {author} {\bibinfo {author} {\bibfnamefont {M.}~\bibnamefont
  {Fael}}\ and\ \bibinfo {author} {\bibfnamefont {M.}~\bibnamefont {Passera}},\
  }\href {\doibase 10.1103/PhysRevLett.122.192001} {\bibfield  {journal}
  {\bibinfo  {journal} {Phys. Rev. Lett.}\ }\textbf {\bibinfo {volume} {122}},\
  \bibinfo {pages} {192001} (\bibinfo {year} {2019})},\ \Eprint
  {http://arxiv.org/abs/1901.03106} {arXiv:1901.03106 [hep-ph]}\BibitemShut
  {NoStop}%
\bibitem [{\citenamefont {Cabibbo}\ and\ \citenamefont
  {Gatto}(1961)}]{Cabibbo:1961sz}%
  \BibitemOpen
  \bibfield  {author} {\bibinfo {author} {\bibfnamefont {N.}~\bibnamefont
  {Cabibbo}}\ and\ \bibinfo {author} {\bibfnamefont {R.}~\bibnamefont
  {Gatto}},\ }\href {\doibase 10.1103/PhysRev.124.1577} {\bibfield  {journal}
  {\bibinfo  {journal} {Phys. Rev.}\ }\textbf {\bibinfo {volume} {124}},\
  \bibinfo {pages} {1577} (\bibinfo {year} {1961})}\BibitemShut {NoStop}%
\bibitem [{\citenamefont {van Ritbergen}\ and\ \citenamefont
  {Stuart}(1998)}]{vanRitbergen:1998hn}%
  \BibitemOpen
  \bibfield  {author} {\bibinfo {author} {\bibfnamefont {T.}~\bibnamefont {van
  Ritbergen}}\ and\ \bibinfo {author} {\bibfnamefont {R.~G.}\ \bibnamefont
  {Stuart}},\ }\href {\doibase 10.1016/S0370-2693(98)00895-8} {\bibfield
  {journal} {\bibinfo  {journal} {Phys. Lett.}\ }\textbf {\bibinfo {volume}
  {B437}},\ \bibinfo {pages} {201} (\bibinfo {year} {1998})},\ \Eprint
  {http://arxiv.org/abs/hep-ph/9802341} {arXiv:hep-ph/9802341
  [hep-ph]}\BibitemShut {NoStop}%
\bibitem [{\citenamefont {Davydychev}\ \emph {et~al.}(2001)\citenamefont
  {Davydychev}, \citenamefont {Schilcher},\ and\ \citenamefont
  {Spiesberger}}]{Davydychev:2000ee}%
  \BibitemOpen
  \bibfield  {author} {\bibinfo {author} {\bibfnamefont {A.~I.}\ \bibnamefont
  {Davydychev}}, \bibinfo {author} {\bibfnamefont {K.}~\bibnamefont
  {Schilcher}}, \ and\ \bibinfo {author} {\bibfnamefont {H.}~\bibnamefont
  {Spiesberger}},\ }\href {\doibase 10.1007/s100520100577} {\bibfield
  {journal} {\bibinfo  {journal} {Eur. Phys. J.}\ }\textbf {\bibinfo {volume}
  {C19}},\ \bibinfo {pages} {99} (\bibinfo {year} {2001})},\ \Eprint
  {http://arxiv.org/abs/hep-ph/0011221} {arXiv:hep-ph/0011221
  [hep-ph]}\BibitemShut {NoStop}%
\bibitem [{\citenamefont {Actis}\ \emph {et~al.}(2008)\citenamefont {Actis},
  \citenamefont {Czakon}, \citenamefont {Gluza},\ and\ \citenamefont
  {Riemann}}]{Actis:2007fs}%
  \BibitemOpen
  \bibfield  {author} {\bibinfo {author} {\bibfnamefont {S.}~\bibnamefont
  {Actis}}, \bibinfo {author} {\bibfnamefont {M.}~\bibnamefont {Czakon}},
  \bibinfo {author} {\bibfnamefont {J.}~\bibnamefont {Gluza}}, \ and\ \bibinfo
  {author} {\bibfnamefont {T.}~\bibnamefont {Riemann}},\ }\href {\doibase
  10.1103/PhysRevLett.100.131602} {\bibfield  {journal} {\bibinfo  {journal}
  {Phys. Rev. Lett.}\ }\textbf {\bibinfo {volume} {100}},\ \bibinfo {pages}
  {131602} (\bibinfo {year} {2008})},\ \Eprint {http://arxiv.org/abs/0711.3847}
  {arXiv:0711.3847 [hep-ph]}\BibitemShut {NoStop}%
\bibitem [{\citenamefont {K{\"u}hn}\ and\ \citenamefont
  {Uccirati}(2009)}]{Kuhn:2008zs}%
  \BibitemOpen
  \bibfield  {author} {\bibinfo {author} {\bibfnamefont {J.~H.}\ \bibnamefont
  {K{\"u}hn}}\ and\ \bibinfo {author} {\bibfnamefont {S.}~\bibnamefont
  {Uccirati}},\ }\href {\doibase 10.1016/j.nuclphysb.2008.08.002} {\bibfield
  {journal} {\bibinfo  {journal} {Nucl. Phys.}\ }\textbf {\bibinfo {volume}
  {B806}},\ \bibinfo {pages} {300} (\bibinfo {year} {2009})},\ \Eprint
  {http://arxiv.org/abs/0807.1284} {arXiv:0807.1284 [hep-ph]}\BibitemShut
  {NoStop}%
\bibitem [{\citenamefont {Carloni~Calame}\ \emph {et~al.}(2011)\citenamefont
  {Carloni~Calame}, \citenamefont {Czy\.z}, \citenamefont {Gluza},
  \citenamefont {Gunia}, \citenamefont {Montagna}, \citenamefont {Nicrosini},
  \citenamefont {Piccinini}, \citenamefont {Riemann},\ and\ \citenamefont
  {Worek}}]{CarloniCalame:2011zq}%
  \BibitemOpen
  \bibfield  {author} {\bibinfo {author} {\bibfnamefont {C.}~\bibnamefont
  {Carloni~Calame}}, \bibinfo {author} {\bibfnamefont {H.}~\bibnamefont
  {Czy\.z}}, \bibinfo {author} {\bibfnamefont {J.}~\bibnamefont {Gluza}},
  \bibinfo {author} {\bibfnamefont {M.}~\bibnamefont {Gunia}}, \bibinfo
  {author} {\bibfnamefont {G.}~\bibnamefont {Montagna}}, \bibinfo {author}
  {\bibfnamefont {O.}~\bibnamefont {Nicrosini}}, \bibinfo {author}
  {\bibfnamefont {F.}~\bibnamefont {Piccinini}}, \bibinfo {author}
  {\bibfnamefont {T.}~\bibnamefont {Riemann}}, \ and\ \bibinfo {author}
  {\bibfnamefont {M.}~\bibnamefont {Worek}},\ }\href {\doibase
  10.1007/JHEP07(2011)126} {\bibfield  {journal} {\bibinfo  {journal} {JHEP}\
  }\textbf {\bibinfo {volume} {07}},\ \bibinfo {pages} {126} (\bibinfo {year}
  {2011})},\ \Eprint {http://arxiv.org/abs/1106.3178} {arXiv:1106.3178
  [hep-ph]}\BibitemShut {NoStop}%
\bibitem [{\citenamefont {Fael}(2019)}]{Fael:2018dmz}%
  \BibitemOpen
  \bibfield  {author} {\bibinfo {author} {\bibfnamefont {M.}~\bibnamefont
  {Fael}},\ }\href {\doibase 10.1007/JHEP02(2019)027} {\bibfield  {journal}
  {\bibinfo  {journal} {JHEP}\ }\textbf {\bibinfo {volume} {02}},\ \bibinfo
  {pages} {027} (\bibinfo {year} {2019})},\ \Eprint
  {http://arxiv.org/abs/1808.08233} {arXiv:1808.08233 [hep-ph]}\BibitemShut
  {NoStop}%
\bibitem [{\citenamefont {Arbuzov}\ \emph {et~al.}(1995)\citenamefont
  {Arbuzov}, \citenamefont {Kuraev}, \citenamefont {Merenkov},\ and\
  \citenamefont {Trentadue}}]{Arbuzov:1995cn}%
  \BibitemOpen
  \bibfield  {author} {\bibinfo {author} {\bibfnamefont {A.~B.}\ \bibnamefont
  {Arbuzov}}, \bibinfo {author} {\bibfnamefont {E.~A.}\ \bibnamefont {Kuraev}},
  \bibinfo {author} {\bibfnamefont {N.~P.}\ \bibnamefont {Merenkov}}, \ and\
  \bibinfo {author} {\bibfnamefont {L.}~\bibnamefont {Trentadue}},\ }\href
  {http://www.jetp.ac.ru/cgi-bin/dn/e_081_04_0638.pdf} {\bibfield  {journal}
  {\bibinfo  {journal} {J. Exp. Theor. Phys.}\ }\textbf {\bibinfo {volume}
  {81}},\ \bibinfo {pages} {638} (\bibinfo {year} {1995})},\ \bibinfo {note}
  {[Zh. Eksp. Teor. Fiz. {\bf 108}, 1164 (1995)]},\ \Eprint
  {http://arxiv.org/abs/hep-ph/9509405} {arXiv:hep-ph/9509405
  [hep-ph]}\BibitemShut {NoStop}%
\bibitem [{\citenamefont {Arbuzov}\ \emph {et~al.}(1996)\citenamefont
  {Arbuzov}, \citenamefont {Kuraev}, \citenamefont {Merenkov},\ and\
  \citenamefont {Trentadue}}]{Arbuzov:1995vi}%
  \BibitemOpen
  \bibfield  {author} {\bibinfo {author} {\bibfnamefont {A.~B.}\ \bibnamefont
  {Arbuzov}}, \bibinfo {author} {\bibfnamefont {E.~A.}\ \bibnamefont {Kuraev}},
  \bibinfo {author} {\bibfnamefont {N.~P.}\ \bibnamefont {Merenkov}}, \ and\
  \bibinfo {author} {\bibfnamefont {L.}~\bibnamefont {Trentadue}},\ }\href
  {\doibase 10.1016/0550-3213(96)00287-8} {\bibfield  {journal} {\bibinfo
  {journal} {Nucl. Phys.}\ }\textbf {\bibinfo {volume} {B474}},\ \bibinfo
  {pages} {271} (\bibinfo {year} {1996})}\BibitemShut {NoStop}%
\bibitem [{\citenamefont {Catani}\ and\ \citenamefont
  {Trentadue}(1990)}]{Catani:1989et}%
  \BibitemOpen
  \bibfield  {author} {\bibinfo {author} {\bibfnamefont {S.}~\bibnamefont
  {Catani}}\ and\ \bibinfo {author} {\bibfnamefont {L.}~\bibnamefont
  {Trentadue}},\ }\href
  {http://www.jetpletters.ac.ru/ps/1136/article_17189.pdf} {\bibfield
  {journal} {\bibinfo  {journal} {JETP Lett.}\ }\textbf {\bibinfo {volume}
  {51}},\ \bibinfo {pages} {83} (\bibinfo {year} {1990})}\BibitemShut {NoStop}%
\bibitem [{\citenamefont {Skrzypek}(1992)}]{Skrzypek:1992vk}%
  \BibitemOpen
  \bibfield  {author} {\bibinfo {author} {\bibfnamefont {M.}~\bibnamefont
  {Skrzypek}},\ }\href
  {https://s3.cern.ch/inspire-prod-files-9/968b58083e1be6206fb750368b2463b6}
  {\bibfield  {journal} {\bibinfo  {journal} {{Acta Phys. Polon. B}}\ }\textbf
  {\bibinfo {volume} {\textbf{23}}},\ \bibinfo {pages} {135} (\bibinfo {year}
  {1992})}\BibitemShut {NoStop}%
\bibitem [{\citenamefont {Arbuzov}\ \emph {et~al.}(2010)\citenamefont
  {Arbuzov}, \citenamefont {Bytev}, \citenamefont {Kuraev}, \citenamefont
  {Tomasi-Gustafsson},\ and\ \citenamefont {Bystritskiy}}]{Arbuzov:2010zzb}%
  \BibitemOpen
  \bibfield  {author} {\bibinfo {author} {\bibfnamefont {A.}~\bibnamefont
  {Arbuzov}}, \bibinfo {author} {\bibfnamefont {V.}~\bibnamefont {Bytev}},
  \bibinfo {author} {\bibfnamefont {E.}~\bibnamefont {Kuraev}}, \bibinfo
  {author} {\bibfnamefont {E.}~\bibnamefont {Tomasi-Gustafsson}}, \ and\
  \bibinfo {author} {\bibfnamefont {Y.}~\bibnamefont {Bystritskiy}},\ }\href
  {\doibase 10.1134/S1063779610030020} {\bibfield  {journal} {\bibinfo
  {journal} {Phys. Part. Nucl.}\ }\textbf {\bibinfo {volume} {41}},\ \bibinfo
  {pages} {394} (\bibinfo {year} {2010})}\BibitemShut {NoStop}%
\bibitem [{\citenamefont {Montagna}\ \emph {et~al.}(1996)\citenamefont
  {Montagna}, \citenamefont {Nicrosini},\ and\ \citenamefont
  {Piccinini}}]{Montagna:1996gw}%
  \BibitemOpen
  \bibfield  {author} {\bibinfo {author} {\bibfnamefont {G.}~\bibnamefont
  {Montagna}}, \bibinfo {author} {\bibfnamefont {O.}~\bibnamefont {Nicrosini}},
  \ and\ \bibinfo {author} {\bibfnamefont {F.}~\bibnamefont {Piccinini}},\
  }\href {\doibase 10.1016/0370-2693(96)00834-9} {\bibfield  {journal}
  {\bibinfo  {journal} {Phys. Lett.}\ }\textbf {\bibinfo {volume} {B385}},\
  \bibinfo {pages} {348} (\bibinfo {year} {1996})},\ \Eprint
  {http://arxiv.org/abs/hep-ph/9605252} {arXiv:hep-ph/9605252
  [hep-ph]}\BibitemShut {NoStop}%
\bibitem [{\citenamefont {Masiero}\ \emph {et~al.}(2020)\citenamefont
  {Masiero}, \citenamefont {Paradisi},\ and\ \citenamefont
  {Passera}}]{Masiero:2020vxk}%
  \BibitemOpen
  \bibfield  {author} {\bibinfo {author} {\bibfnamefont {A.}~\bibnamefont
  {Masiero}}, \bibinfo {author} {\bibfnamefont {P.}~\bibnamefont {Paradisi}}, \
  and\ \bibinfo {author} {\bibfnamefont {M.}~\bibnamefont {Passera}},\ }\href
  {\doibase 10.1103/PhysRevD.102.075013} {\bibfield  {journal} {\bibinfo
  {journal} {Phys. Rev. D}\ }\textbf {\bibinfo {volume} {102}},\ \bibinfo
  {pages} {075013} (\bibinfo {year} {2020})},\ \Eprint
  {http://arxiv.org/abs/2002.05418} {arXiv:2002.05418 [hep-ph]}\BibitemShut
  {NoStop}%
\bibitem [{\citenamefont {Dev}\ \emph {et~al.}(2020)\citenamefont {Dev},
  \citenamefont {Rodejohann}, \citenamefont {Xu},\ and\ \citenamefont
  {Zhang}}]{Dev:2020drf}%
  \BibitemOpen
  \bibfield  {author} {\bibinfo {author} {\bibfnamefont {P.~B.}\ \bibnamefont
  {Dev}}, \bibinfo {author} {\bibfnamefont {W.}~\bibnamefont {Rodejohann}},
  \bibinfo {author} {\bibfnamefont {X.-J.}\ \bibnamefont {Xu}}, \ and\ \bibinfo
  {author} {\bibfnamefont {Y.}~\bibnamefont {Zhang}},\ }\href {\doibase
  10.1007/JHEP05(2020)053} {\bibfield  {journal} {\bibinfo  {journal} {JHEP}\
  }\textbf {\bibinfo {volume} {05}},\ \bibinfo {pages} {053} (\bibinfo {year}
  {2020})},\ \Eprint {http://arxiv.org/abs/2002.04822} {arXiv:2002.04822
  [hep-ph]}\BibitemShut {NoStop}%
\bibitem [{\citenamefont {Schubert}\ and\ \citenamefont
  {Williams}(2019)}]{Schubert:2019nwm}%
  \BibitemOpen
  \bibfield  {author} {\bibinfo {author} {\bibfnamefont {U.}~\bibnamefont
  {Schubert}}\ and\ \bibinfo {author} {\bibfnamefont {C.}~\bibnamefont
  {Williams}},\ }\href {\doibase 10.1103/PhysRevD.100.035030} {\bibfield
  {journal} {\bibinfo  {journal} {Phys. Rev. D}\ }\textbf {\bibinfo {volume}
  {100}},\ \bibinfo {pages} {035030} (\bibinfo {year} {2019})},\ \Eprint
  {http://arxiv.org/abs/1907.01574} {arXiv:1907.01574 [hep-ph]}\BibitemShut
  {NoStop}%
\bibitem [{\citenamefont {Davier}(2017)}]{michel17}%
  \BibitemOpen
  \bibfield  {author} {\bibinfo {author} {\bibfnamefont {M.}~\bibnamefont
  {Davier}},\ }\href@noop {} {\enquote {\bibinfo {title} {{$e^+e^-$ results
  from BABAR and implications for the muon $g-2$}},}\ }\bibinfo {howpublished}
  {\url{https://indico.fnal.gov/event/13795/session/10/contribution/47/material/slides/0.pdf}}
  (\bibinfo {year} {2017}),\ \bibinfo {note} {{Muon $g-2$ Theory Initiative
  workshop Fermilab}}\BibitemShut {NoStop}%
\bibitem [{\citenamefont {Ignatov}\ \emph {et~al.}(2019)\citenamefont {Ignatov}
  \emph {et~al.}}]{Ignatov:2019omb}%
  \BibitemOpen
  \bibfield  {author} {\bibinfo {author} {\bibfnamefont {F.~V.}\ \bibnamefont
  {Ignatov}} \emph {et~al.} (\bibinfo {collaboration} {CMD-3}),\ }\href
  {\doibase 10.1051/epjconf/201921204001} {\bibfield  {journal} {\bibinfo
  {journal} {EPJ Web Conf.}\ }\textbf {\bibinfo {volume} {212}},\ \bibinfo
  {pages} {04001} (\bibinfo {year} {2019})}\BibitemShut {NoStop}%
\bibitem [{\citenamefont {Logachev}(2020)}]{kedr2020}%
  \BibitemOpen
  \bibfield  {author} {\bibinfo {author} {\bibfnamefont {P.}~\bibnamefont
  {Logachev}},\ }\href@noop {} {\enquote {\bibinfo {title} {{Collider
  experiments at BINP}},}\ }\bibinfo {howpublished}
  {\url{https://indico.inp.nsk.su/event/20/session/0/contribution/209/material/slides/0.pdf}}
  (\bibinfo {year} {2020}),\ \bibinfo {note} {{Conference on Instrumentation
  for Colliding Beam Physics Novosibirsk}}\BibitemShut {NoStop}%
\bibitem [{\citenamefont {Ablikim}\ \emph {et~al.}(2013)\citenamefont {Ablikim}
  \emph {et~al.}}]{Ablikim:2014gna}%
  \BibitemOpen
  \bibfield  {author} {\bibinfo {author} {\bibfnamefont {M.}~\bibnamefont
  {Ablikim}} \emph {et~al.} (\bibinfo {collaboration} {BESIII}),\ }\href
  {\doibase 10.1088/1674-1137/37/12/123001} {\bibfield  {journal} {\bibinfo
  {journal} {Chin. Phys.}\ }\textbf {\bibinfo {volume} {C37}},\ \bibinfo
  {pages} {123001} (\bibinfo {year} {2013})},\ \Eprint
  {http://arxiv.org/abs/1307.2022} {arXiv:1307.2022 [hep-ex]}\BibitemShut
  {NoStop}%
\bibitem [{\citenamefont {Ablikim}\ \emph {et~al.}(2020)\citenamefont {Ablikim}
  \emph {et~al.}}]{Ablikim:2019hff}%
  \BibitemOpen
  \bibfield  {author} {\bibinfo {author} {\bibfnamefont {M.}~\bibnamefont
  {Ablikim}} \emph {et~al.} (\bibinfo {collaboration} {BESIII}),\ }\href
  {\doibase 10.1088/1674-1137/44/4/040001} {\bibfield  {journal} {\bibinfo
  {journal} {Chin. Phys.}\ }\textbf {\bibinfo {volume} {C44}},\ \bibinfo
  {pages} {040001} (\bibinfo {year} {2020})},\ \Eprint
  {http://arxiv.org/abs/1912.05983} {arXiv:1912.05983 [hep-ex]}\BibitemShut
  {NoStop}%
\bibitem [{\citenamefont {Ablikim}\ \emph {et~al.}(2019)\citenamefont {Ablikim}
  \emph {et~al.}}]{Ablikim:2019sjw}%
  \BibitemOpen
  \bibfield  {author} {\bibinfo {author} {\bibfnamefont {M.}~\bibnamefont
  {Ablikim}} \emph {et~al.} (\bibinfo {collaboration} {BESIII}),\ }\href@noop
  {} {\  (\bibinfo {year} {2019})},\ \Eprint {http://arxiv.org/abs/1912.11208}
  {arXiv:1912.11208 [hep-ex]}\BibitemShut {NoStop}%
\bibitem [{\citenamefont {Redmer}(2018{\natexlab{a}})}]{Redmer:2018nxj}%
  \BibitemOpen
  \bibfield  {author} {\bibinfo {author} {\bibfnamefont {C.~F.}\ \bibnamefont
  {Redmer}} (\bibinfo {collaboration} {BESIII}),\ }\href@noop {} {\  (\bibinfo
  {year} {2018}{\natexlab{a}})},\ \Eprint {http://arxiv.org/abs/1810.00643}
  {arXiv:1810.00643 [hep-ex]}\BibitemShut {NoStop}%
\bibitem [{\citenamefont {Akhmetshin}\ \emph {et~al.}(1998)\citenamefont
  {Akhmetshin} \emph {et~al.}}]{Akhmetshin:1998se}%
  \BibitemOpen
  \bibfield  {author} {\bibinfo {author} {\bibfnamefont {R.~R.}\ \bibnamefont
  {Akhmetshin}} \emph {et~al.} (\bibinfo {collaboration} {CMD-2}),\ }\href
  {\doibase 10.1016/S0370-2693(98)00826-0} {\bibfield  {journal} {\bibinfo
  {journal} {Phys. Lett.}\ }\textbf {\bibinfo {volume} {B434}},\ \bibinfo
  {pages} {426} (\bibinfo {year} {1998})}\BibitemShut {NoStop}%
\bibitem [{\citenamefont {Cosme}\ \emph {et~al.}(1976)\citenamefont {Cosme}
  \emph {et~al.}}]{Cosme:1976tf}%
  \BibitemOpen
  \bibfield  {author} {\bibinfo {author} {\bibfnamefont {G.}~\bibnamefont
  {Cosme}} \emph {et~al.},\ }\href {\doibase 10.1016/0370-2693(76)90280-X}
  {\bibfield  {journal} {\bibinfo  {journal} {Phys. Lett.}\ }\textbf {\bibinfo
  {volume} {63B}},\ \bibinfo {pages} {349} (\bibinfo {year}
  {1976})}\BibitemShut {NoStop}%
\bibitem [{\citenamefont {Cosme}\ \emph {et~al.}(1979)\citenamefont {Cosme}
  \emph {et~al.}}]{Cosme:1978qe}%
  \BibitemOpen
  \bibfield  {author} {\bibinfo {author} {\bibfnamefont {G.}~\bibnamefont
  {Cosme}} \emph {et~al.},\ }\href {\doibase 10.1016/0550-3213(79)90100-7}
  {\bibfield  {journal} {\bibinfo  {journal} {Nucl. Phys.}\ }\textbf {\bibinfo
  {volume} {B152}},\ \bibinfo {pages} {215} (\bibinfo {year}
  {1979})}\BibitemShut {NoStop}%
\bibitem [{\citenamefont {Esposito}\ \emph {et~al.}(1981)\citenamefont
  {Esposito} \emph {et~al.}}]{Esposito:1981dv}%
  \BibitemOpen
  \bibfield  {author} {\bibinfo {author} {\bibfnamefont {B.}~\bibnamefont
  {Esposito}} \emph {et~al.} (\bibinfo {collaboration} {MEA}),\ }\href
  {\doibase 10.1007/BF02776174} {\bibfield  {journal} {\bibinfo  {journal}
  {Lett. Nuovo Cim.}\ }\textbf {\bibinfo {volume} {31}},\ \bibinfo {pages}
  {445} (\bibinfo {year} {1981})}\BibitemShut {NoStop}%
\bibitem [{\citenamefont {Bacci}\ \emph {et~al.}(1981)\citenamefont {Bacci}
  \emph {et~al.}}]{Bacci:1980zs}%
  \BibitemOpen
  \bibfield  {author} {\bibinfo {author} {\bibfnamefont {C.}~\bibnamefont
  {Bacci}} \emph {et~al.} (\bibinfo {collaboration} {$\gamma\gamma 2$}),\
  }\href {\doibase 10.1016/0550-3213(81)90208-X} {\bibfield  {journal}
  {\bibinfo  {journal} {Nucl. Phys.}\ }\textbf {\bibinfo {volume} {B184}},\
  \bibinfo {pages} {31} (\bibinfo {year} {1981})}\BibitemShut {NoStop}%
\bibitem [{\citenamefont {Kurdadze}\ \emph {et~al.}(1986)\citenamefont
  {Kurdadze}, \citenamefont {Lelchuk}, \citenamefont {Pakhtusova},
  \citenamefont {Sidorov}, \citenamefont {Skrinsky}, \citenamefont
  {Chilingarov}, \citenamefont {Shatunov}, \citenamefont {Shvarts},\ and\
  \citenamefont {Eidelman}}]{Kurdadze:1986tc}%
  \BibitemOpen
  \bibfield  {author} {\bibinfo {author} {\bibfnamefont {L.~M.}\ \bibnamefont
  {Kurdadze}}, \bibinfo {author} {\bibfnamefont {M.~{\relax Yu}.}\ \bibnamefont
  {Lelchuk}}, \bibinfo {author} {\bibfnamefont {E.~V.}\ \bibnamefont
  {Pakhtusova}}, \bibinfo {author} {\bibfnamefont {V.~A.}\ \bibnamefont
  {Sidorov}}, \bibinfo {author} {\bibfnamefont {A.~N.}\ \bibnamefont
  {Skrinsky}}, \bibinfo {author} {\bibfnamefont {A.~G.}\ \bibnamefont
  {Chilingarov}}, \bibinfo {author} {\bibfnamefont {{\relax Yu}.~M.}\
  \bibnamefont {Shatunov}}, \bibinfo {author} {\bibfnamefont {B.~A.}\
  \bibnamefont {Shvarts}}, \ and\ \bibinfo {author} {\bibfnamefont {S.~I.}\
  \bibnamefont {Eidelman}} (\bibinfo {collaboration} {OLYA}),\ }\href
  {http://www.jetpletters.ac.ru/ps/1409/article_21399.pdf} {\bibfield
  {journal} {\bibinfo  {journal} {JETP Lett.}\ }\textbf {\bibinfo {volume}
  {43}},\ \bibinfo {pages} {643} (\bibinfo {year} {1986})},\ \bibinfo {note}
  {[Pisma Zh. Eksp. Teor. Fiz. {\bf 43}, 497 (1986)]}\BibitemShut {NoStop}%
\bibitem [{\citenamefont {Dolinsky}\ \emph {et~al.}(1991)\citenamefont
  {Dolinsky} \emph {et~al.}}]{Dolinsky:1991vq}%
  \BibitemOpen
  \bibfield  {author} {\bibinfo {author} {\bibfnamefont {S.~I.}\ \bibnamefont
  {Dolinsky}} \emph {et~al.} (\bibinfo {collaboration} {ND}),\ }\href {\doibase
  10.1016/0370-1573(91)90127-8} {\bibfield  {journal} {\bibinfo  {journal}
  {Phys. Rept.}\ }\textbf {\bibinfo {volume} {202}},\ \bibinfo {pages} {99}
  (\bibinfo {year} {1991})}\BibitemShut {NoStop}%
\bibitem [{\citenamefont {Achasov}\ \emph
  {et~al.}(2001{\natexlab{b}})\citenamefont {Achasov} \emph
  {et~al.}}]{Achasov:2001gp}%
  \BibitemOpen
  \bibfield  {author} {\bibinfo {author} {\bibfnamefont {M.~N.}\ \bibnamefont
  {Achasov}} \emph {et~al.} (\bibinfo {collaboration} {SND}),\ }\href@noop {}
  {\enquote {\bibinfo {title} {{$e^+ e^- \to 4\pi$ processes investigation in
  the energy range 0.98\,GeV to 1.38\,GeV with SND detector}},}\ }\bibinfo
  {howpublished}
  {\url{https://s3.cern.ch/inspire-prod-files-e/e0c4030fac25f77794d80ef00073c488}}
  (\bibinfo {year} {2001}{\natexlab{b}})\BibitemShut {NoStop}%
\bibitem [{\citenamefont {Redmer}(2019)}]{Redmer:2019zzr}%
  \BibitemOpen
  \bibfield  {author} {\bibinfo {author} {\bibfnamefont {C.~F.}\ \bibnamefont
  {Redmer}} (\bibinfo {collaboration} {BESIII}),\ }\href {\doibase
  10.1051/epjconf/201921204004} {\bibfield  {journal} {\bibinfo  {journal} {EPJ
  Web Conf.}\ }\textbf {\bibinfo {volume} {212}},\ \bibinfo {pages} {04004}
  (\bibinfo {year} {2019})}\BibitemShut {NoStop}%
\bibitem [{\citenamefont {Maeda}(2018)}]{Maeda:2018}%
  \BibitemOpen
  \bibfield  {author} {\bibinfo {author} {\bibfnamefont {Y.}~\bibnamefont
  {Maeda}},\ }\href@noop {} {\enquote {\bibinfo {title} {{Measurement of cross
  section of light hadron production in $e^+e^-$ collisions in the Belle II
  experiment}},}\ }\bibinfo {howpublished}
  {\url{https://agenda.hepl.phys.nagoya-u.ac.jp/indico/getFile.py/access?contribId=16&sessionId=2&resId=0&materialId=slides&confId=1037}}
  (\bibinfo {year} {2018}),\ \bibinfo {note} {{International workshop Hints for
  New Physics in Heavy Flavor Physics Nagoya}}\BibitemShut {NoStop}%
\bibitem [{\citenamefont {Bernecker}\ and\ \citenamefont
  {Meyer}(2011)}]{Bernecker:2011gh}%
  \BibitemOpen
  \bibfield  {author} {\bibinfo {author} {\bibfnamefont {D.}~\bibnamefont
  {Bernecker}}\ and\ \bibinfo {author} {\bibfnamefont {H.~B.}\ \bibnamefont
  {Meyer}},\ }\href {\doibase 10.1140/epja/i2011-11148-6} {\bibfield  {journal}
  {\bibinfo  {journal} {Eur. Phys. J.}\ }\textbf {\bibinfo {volume} {A47}},\
  \bibinfo {pages} {148} (\bibinfo {year} {2011})},\ \Eprint
  {http://arxiv.org/abs/1107.4388} {arXiv:1107.4388 [hep-lat]}\BibitemShut
  {NoStop}%
\bibitem [{\citenamefont {Aubin}\ \emph {et~al.}(2016)\citenamefont {Aubin},
  \citenamefont {Blum}, \citenamefont {Chau}, \citenamefont {Golterman},
  \citenamefont {Peris},\ and\ \citenamefont {Tu}}]{Aubin:2015rzx}%
  \BibitemOpen
  \bibfield  {author} {\bibinfo {author} {\bibfnamefont {C.}~\bibnamefont
  {Aubin}}, \bibinfo {author} {\bibfnamefont {T.}~\bibnamefont {Blum}},
  \bibinfo {author} {\bibfnamefont {P.}~\bibnamefont {Chau}}, \bibinfo {author}
  {\bibfnamefont {M.}~\bibnamefont {Golterman}}, \bibinfo {author}
  {\bibfnamefont {S.}~\bibnamefont {Peris}}, \ and\ \bibinfo {author}
  {\bibfnamefont {C.}~\bibnamefont {Tu}},\ }\href {\doibase
  10.1103/PhysRevD.93.054508} {\bibfield  {journal} {\bibinfo  {journal} {Phys.
  Rev.}\ }\textbf {\bibinfo {volume} {D93}},\ \bibinfo {pages} {054508}
  (\bibinfo {year} {2016})},\ \Eprint {http://arxiv.org/abs/1512.07555}
  {arXiv:1512.07555 [hep-lat]}\BibitemShut {NoStop}%
\bibitem [{\citenamefont {Lautrup}\ \emph {et~al.}(1972)\citenamefont
  {Lautrup}, \citenamefont {Peterman},\ and\ \citenamefont
  {de~Rafael}}]{Lautrup:1971jf}%
  \BibitemOpen
  \bibfield  {author} {\bibinfo {author} {\bibfnamefont {B.~E.}\ \bibnamefont
  {Lautrup}}, \bibinfo {author} {\bibfnamefont {A.}~\bibnamefont {Peterman}}, \
  and\ \bibinfo {author} {\bibfnamefont {E.}~\bibnamefont {de~Rafael}},\ }\href
  {\doibase 10.1016/0370-1573(72)90011-7} {\bibfield  {journal} {\bibinfo
  {journal} {Phys. Rept.}\ }\textbf {\bibinfo {volume} {3}},\ \bibinfo {pages}
  {193} (\bibinfo {year} {1972})}\BibitemShut {NoStop}%
\bibitem [{\citenamefont {de~Rafael}(1994)}]{deRafael:1993za}%
  \BibitemOpen
  \bibfield  {author} {\bibinfo {author} {\bibfnamefont {E.}~\bibnamefont
  {de~Rafael}},\ }\href {\doibase 10.1016/0370-2693(94)91114-2} {\bibfield
  {journal} {\bibinfo  {journal} {Phys. Lett.}\ }\textbf {\bibinfo {volume}
  {B322}},\ \bibinfo {pages} {239} (\bibinfo {year} {1994})},\ \Eprint
  {http://arxiv.org/abs/hep-ph/9311316} {arXiv:hep-ph/9311316
  [hep-ph]}\BibitemShut {NoStop}%
\bibitem [{\citenamefont {Blum}(2003)}]{Blum:2002ii}%
  \BibitemOpen
  \bibfield  {author} {\bibinfo {author} {\bibfnamefont {T.}~\bibnamefont
  {Blum}},\ }\href {\doibase 10.1103/PhysRevLett.91.052001} {\bibfield
  {journal} {\bibinfo  {journal} {Phys. Rev. Lett.}\ }\textbf {\bibinfo
  {volume} {91}},\ \bibinfo {pages} {052001} (\bibinfo {year} {2003})},\
  \Eprint {http://arxiv.org/abs/hep-lat/0212018} {arXiv:hep-lat/0212018
  [hep-lat]}\BibitemShut {NoStop}%
\bibitem [{\citenamefont {G{\"o}ckeler}\ \emph {et~al.}(2004)\citenamefont
  {G{\"o}ckeler}, \citenamefont {Horsley}, \citenamefont {Kurzinger},
  \citenamefont {Pleiter}, \citenamefont {Rakow},\ and\ \citenamefont
  {Schierholz}}]{Gockeler:2003cw}%
  \BibitemOpen
  \bibfield  {author} {\bibinfo {author} {\bibfnamefont {M.}~\bibnamefont
  {G{\"o}ckeler}}, \bibinfo {author} {\bibfnamefont {R.}~\bibnamefont
  {Horsley}}, \bibinfo {author} {\bibfnamefont {W.}~\bibnamefont {Kurzinger}},
  \bibinfo {author} {\bibfnamefont {D.}~\bibnamefont {Pleiter}}, \bibinfo
  {author} {\bibfnamefont {P.~E.~L.}\ \bibnamefont {Rakow}}, \ and\ \bibinfo
  {author} {\bibfnamefont {G.}~\bibnamefont {Schierholz}} (\bibinfo
  {collaboration} {QCDSF}),\ }\href {\doibase 10.1016/j.nuclphysb.2004.03.026}
  {\bibfield  {journal} {\bibinfo  {journal} {Nucl. Phys.}\ }\textbf {\bibinfo
  {volume} {B688}},\ \bibinfo {pages} {135} (\bibinfo {year} {2004})},\ \Eprint
  {http://arxiv.org/abs/hep-lat/0312032} {arXiv:hep-lat/0312032
  [hep-lat]}\BibitemShut {NoStop}%
\bibitem [{\citenamefont {Della~Morte}\ \emph {et~al.}(2012)\citenamefont
  {Della~Morte}, \citenamefont {J{\"a}ger}, \citenamefont {J{\"u}ttner},\ and\
  \citenamefont {Wittig}}]{DellaMorte:2011aa}%
  \BibitemOpen
  \bibfield  {author} {\bibinfo {author} {\bibfnamefont {M.}~\bibnamefont
  {Della~Morte}}, \bibinfo {author} {\bibfnamefont {B.}~\bibnamefont
  {J{\"a}ger}}, \bibinfo {author} {\bibfnamefont {A.}~\bibnamefont
  {J{\"u}ttner}}, \ and\ \bibinfo {author} {\bibfnamefont {H.}~\bibnamefont
  {Wittig}},\ }\href {\doibase 10.1007/JHEP03(2012)055} {\bibfield  {journal}
  {\bibinfo  {journal} {JHEP}\ }\textbf {\bibinfo {volume} {03}},\ \bibinfo
  {pages} {055} (\bibinfo {year} {2012})},\ \Eprint
  {http://arxiv.org/abs/1112.2894} {arXiv:1112.2894 [hep-lat]}\BibitemShut
  {NoStop}%
\bibitem [{\citenamefont {Aubin}\ \emph {et~al.}(2013)\citenamefont {Aubin},
  \citenamefont {Blum}, \citenamefont {Golterman},\ and\ \citenamefont
  {Peris}}]{Aubin:2013daa}%
  \BibitemOpen
  \bibfield  {author} {\bibinfo {author} {\bibfnamefont {C.}~\bibnamefont
  {Aubin}}, \bibinfo {author} {\bibfnamefont {T.}~\bibnamefont {Blum}},
  \bibinfo {author} {\bibfnamefont {M.}~\bibnamefont {Golterman}}, \ and\
  \bibinfo {author} {\bibfnamefont {S.}~\bibnamefont {Peris}},\ }\href
  {\doibase 10.1103/PhysRevD.88.074505} {\bibfield  {journal} {\bibinfo
  {journal} {Phys. Rev.}\ }\textbf {\bibinfo {volume} {D88}},\ \bibinfo {pages}
  {074505} (\bibinfo {year} {2013})},\ \Eprint {http://arxiv.org/abs/1307.4701}
  {arXiv:1307.4701 [hep-lat]}\BibitemShut {NoStop}%
\bibitem [{\citenamefont {Malak}\ \emph {et~al.}(2015)\citenamefont {Malak},
  \citenamefont {Fodor}, \citenamefont {Hoelbling}, \citenamefont {Lellouch},
  \citenamefont {Sastre},\ and\ \citenamefont {Szabo}}]{Malak:2015sla}%
  \BibitemOpen
  \bibfield  {author} {\bibinfo {author} {\bibfnamefont {R.}~\bibnamefont
  {Malak}}, \bibinfo {author} {\bibfnamefont {Z.}~\bibnamefont {Fodor}},
  \bibinfo {author} {\bibfnamefont {C.}~\bibnamefont {Hoelbling}}, \bibinfo
  {author} {\bibfnamefont {L.}~\bibnamefont {Lellouch}}, \bibinfo {author}
  {\bibfnamefont {A.}~\bibnamefont {Sastre}}, \ and\ \bibinfo {author}
  {\bibfnamefont {K.}~\bibnamefont {Szabo}} (\bibinfo {collaboration}
  {Budapest-Marseille-Wuppertal}),\ }\href {\doibase 10.22323/1.214.0161}
  {\bibfield  {journal} {\bibinfo  {journal} {PoS}\ }\textbf {\bibinfo {volume}
  {LATTICE2014}},\ \bibinfo {pages} {161} (\bibinfo {year} {2015})},\ \Eprint
  {http://arxiv.org/abs/1502.02172} {arXiv:1502.02172 [hep-lat]}\BibitemShut
  {NoStop}%
\bibitem [{\citenamefont {de~Divitiis}\ \emph
  {et~al.}(2012{\natexlab{a}})\citenamefont {de~Divitiis}, \citenamefont
  {Petronzio},\ and\ \citenamefont {Tantalo}}]{deDivitiis:2012vs}%
  \BibitemOpen
  \bibfield  {author} {\bibinfo {author} {\bibfnamefont {G.~M.}\ \bibnamefont
  {de~Divitiis}}, \bibinfo {author} {\bibfnamefont {R.}~\bibnamefont
  {Petronzio}}, \ and\ \bibinfo {author} {\bibfnamefont {N.}~\bibnamefont
  {Tantalo}},\ }\href {\doibase 10.1016/j.physletb.2012.10.035} {\bibfield
  {journal} {\bibinfo  {journal} {Phys. Lett.}\ }\textbf {\bibinfo {volume}
  {B718}},\ \bibinfo {pages} {589} (\bibinfo {year} {2012}{\natexlab{a}})},\
  \Eprint {http://arxiv.org/abs/1208.5914} {arXiv:1208.5914
  [hep-lat]}\BibitemShut {NoStop}%
\bibitem [{\citenamefont {Chakraborty}\ \emph {et~al.}(2014)\citenamefont
  {Chakraborty}, \citenamefont {Davies}, \citenamefont {Donald}, \citenamefont
  {Dowdall}, \citenamefont {Koponen}, \citenamefont {Lepage},\ and\
  \citenamefont {Teubner}}]{Chakraborty:2014mwa}%
  \BibitemOpen
  \bibfield  {author} {\bibinfo {author} {\bibfnamefont {B.}~\bibnamefont
  {Chakraborty}}, \bibinfo {author} {\bibfnamefont {C.~T.~H.}\ \bibnamefont
  {Davies}}, \bibinfo {author} {\bibfnamefont {G.~C.}\ \bibnamefont {Donald}},
  \bibinfo {author} {\bibfnamefont {R.~J.}\ \bibnamefont {Dowdall}}, \bibinfo
  {author} {\bibfnamefont {J.}~\bibnamefont {Koponen}}, \bibinfo {author}
  {\bibfnamefont {G.~P.}\ \bibnamefont {Lepage}}, \ and\ \bibinfo {author}
  {\bibfnamefont {T.}~\bibnamefont {Teubner}} (\bibinfo {collaboration}
  {HPQCD}),\ }\href {\doibase 10.1103/PhysRevD.89.114501} {\bibfield  {journal}
  {\bibinfo  {journal} {Phys. Rev.}\ }\textbf {\bibinfo {volume} {D89}},\
  \bibinfo {pages} {114501} (\bibinfo {year} {2014})},\ \Eprint
  {http://arxiv.org/abs/1403.1778} {arXiv:1403.1778 [hep-lat]}\BibitemShut
  {NoStop}%
\bibitem [{\citenamefont {Bali}\ and\ \citenamefont
  {Endrődi}(2015)}]{Bali:2015msa}%
  \BibitemOpen
  \bibfield  {author} {\bibinfo {author} {\bibfnamefont {G.}~\bibnamefont
  {Bali}}\ and\ \bibinfo {author} {\bibfnamefont {G.}~\bibnamefont
  {Endrődi}},\ }\href {\doibase 10.1103/PhysRevD.92.054506} {\bibfield
  {journal} {\bibinfo  {journal} {Phys. Rev.}\ }\textbf {\bibinfo {volume}
  {D92}},\ \bibinfo {pages} {054506} (\bibinfo {year} {2015})},\ \Eprint
  {http://arxiv.org/abs/1506.08638} {arXiv:1506.08638 [hep-lat]}\BibitemShut
  {NoStop}%
\bibitem [{\citenamefont {Feng}\ \emph {et~al.}(2011)\citenamefont {Feng},
  \citenamefont {Jansen}, \citenamefont {Petschlies},\ and\ \citenamefont
  {Renner}}]{Feng:2011zk}%
  \BibitemOpen
  \bibfield  {author} {\bibinfo {author} {\bibfnamefont {X.}~\bibnamefont
  {Feng}}, \bibinfo {author} {\bibfnamefont {K.}~\bibnamefont {Jansen}},
  \bibinfo {author} {\bibfnamefont {M.}~\bibnamefont {Petschlies}}, \ and\
  \bibinfo {author} {\bibfnamefont {D.~B.}\ \bibnamefont {Renner}} (\bibinfo
  {collaboration} {{ETM}}),\ }\href {\doibase 10.1103/PhysRevLett.107.081802}
  {\bibfield  {journal} {\bibinfo  {journal} {Phys. Rev. Lett.}\ }\textbf
  {\bibinfo {volume} {107}},\ \bibinfo {pages} {081802} (\bibinfo {year}
  {2011})},\ \Eprint {http://arxiv.org/abs/1103.4818} {arXiv:1103.4818
  [hep-lat]}\BibitemShut {NoStop}%
\bibitem [{\citenamefont {Aubin}\ and\ \citenamefont
  {Blum}(2007)}]{Aubin:2006xv}%
  \BibitemOpen
  \bibfield  {author} {\bibinfo {author} {\bibfnamefont {C.}~\bibnamefont
  {Aubin}}\ and\ \bibinfo {author} {\bibfnamefont {T.}~\bibnamefont {Blum}},\
  }\href {\doibase 10.1103/PhysRevD.75.114502} {\bibfield  {journal} {\bibinfo
  {journal} {Phys. Rev.}\ }\textbf {\bibinfo {volume} {D75}},\ \bibinfo {pages}
  {114502} (\bibinfo {year} {2007})},\ \Eprint
  {http://arxiv.org/abs/hep-lat/0608011} {arXiv:hep-lat/0608011
  [hep-lat]}\BibitemShut {NoStop}%
\bibitem [{\citenamefont {Boyle}\ \emph {et~al.}(2012)\citenamefont {Boyle},
  \citenamefont {Del~Debbio}, \citenamefont {Kerrane},\ and\ \citenamefont
  {Zanotti}}]{Boyle:2011hu}%
  \BibitemOpen
  \bibfield  {author} {\bibinfo {author} {\bibfnamefont {P.}~\bibnamefont
  {Boyle}}, \bibinfo {author} {\bibfnamefont {L.}~\bibnamefont {Del~Debbio}},
  \bibinfo {author} {\bibfnamefont {E.}~\bibnamefont {Kerrane}}, \ and\
  \bibinfo {author} {\bibfnamefont {J.}~\bibnamefont {Zanotti}},\ }\href
  {\doibase 10.1103/PhysRevD.85.074504} {\bibfield  {journal} {\bibinfo
  {journal} {Phys. Rev.}\ }\textbf {\bibinfo {volume} {D85}},\ \bibinfo {pages}
  {074504} (\bibinfo {year} {2012})},\ \Eprint {http://arxiv.org/abs/1107.1497}
  {arXiv:1107.1497 [hep-lat]}\BibitemShut {NoStop}%
\bibitem [{\citenamefont {Golterman}\ \emph {et~al.}(2014)\citenamefont
  {Golterman}, \citenamefont {Maltman},\ and\ \citenamefont
  {Peris}}]{Golterman:2014ksa}%
  \BibitemOpen
  \bibfield  {author} {\bibinfo {author} {\bibfnamefont {M.}~\bibnamefont
  {Golterman}}, \bibinfo {author} {\bibfnamefont {K.}~\bibnamefont {Maltman}},
  \ and\ \bibinfo {author} {\bibfnamefont {S.}~\bibnamefont {Peris}},\ }\href
  {\doibase 10.1103/PhysRevD.90.074508} {\bibfield  {journal} {\bibinfo
  {journal} {Phys. Rev.}\ }\textbf {\bibinfo {volume} {D90}},\ \bibinfo {pages}
  {074508} (\bibinfo {year} {2014})},\ \Eprint {http://arxiv.org/abs/1405.2389}
  {arXiv:1405.2389 [hep-lat]}\BibitemShut {NoStop}%
\bibitem [{\citenamefont {Aubin}\ \emph {et~al.}(2012)\citenamefont {Aubin},
  \citenamefont {Blum}, \citenamefont {Golterman},\ and\ \citenamefont
  {Peris}}]{Aubin:2012me}%
  \BibitemOpen
  \bibfield  {author} {\bibinfo {author} {\bibfnamefont {C.}~\bibnamefont
  {Aubin}}, \bibinfo {author} {\bibfnamefont {T.}~\bibnamefont {Blum}},
  \bibinfo {author} {\bibfnamefont {M.}~\bibnamefont {Golterman}}, \ and\
  \bibinfo {author} {\bibfnamefont {S.}~\bibnamefont {Peris}},\ }\href
  {\doibase 10.1103/PhysRevD.86.054509} {\bibfield  {journal} {\bibinfo
  {journal} {Phys. Rev.}\ }\textbf {\bibinfo {volume} {D86}},\ \bibinfo {pages}
  {054509} (\bibinfo {year} {2012})},\ \Eprint {http://arxiv.org/abs/1205.3695}
  {arXiv:1205.3695 [hep-lat]}\BibitemShut {NoStop}%
\bibitem [{\citenamefont {Baker}(1969)}]{Baker:1969sw}%
  \BibitemOpen
  \bibfield  {author} {\bibinfo {author} {\bibfnamefont {G.~A.}\ \bibnamefont
  {Baker}},\ }\href {\doibase 10.1063/1.1664911} {\bibfield  {journal}
  {\bibinfo  {journal} {J. Math. Phys.}\ }\textbf {\bibinfo {volume} {10}},\
  \bibinfo {pages} {814} (\bibinfo {year} {1969})}\BibitemShut {NoStop}%
\bibitem [{\citenamefont {Chakraborty}\ \emph {et~al.}(2016)\citenamefont
  {Chakraborty}, \citenamefont {Davies}, \citenamefont {Koponen}, \citenamefont
  {Lepage}, \citenamefont {Peardon},\ and\ \citenamefont
  {Ryan}}]{Chakraborty:2015ugp}%
  \BibitemOpen
  \bibfield  {author} {\bibinfo {author} {\bibfnamefont {B.}~\bibnamefont
  {Chakraborty}}, \bibinfo {author} {\bibfnamefont {C.~T.~H.}\ \bibnamefont
  {Davies}}, \bibinfo {author} {\bibfnamefont {J.}~\bibnamefont {Koponen}},
  \bibinfo {author} {\bibfnamefont {G.~P.}\ \bibnamefont {Lepage}}, \bibinfo
  {author} {\bibfnamefont {M.~J.}\ \bibnamefont {Peardon}}, \ and\ \bibinfo
  {author} {\bibfnamefont {S.~M.}\ \bibnamefont {Ryan}} (\bibinfo
  {collaboration} {HadSpec and HPQCD}),\ }\href {\doibase
  10.1103/PhysRevD.93.074509} {\bibfield  {journal} {\bibinfo  {journal} {Phys.
  Rev.}\ }\textbf {\bibinfo {volume} {D93}},\ \bibinfo {pages} {074509}
  (\bibinfo {year} {2016})},\ \Eprint {http://arxiv.org/abs/1512.03270}
  {arXiv:1512.03270 [hep-lat]}\BibitemShut {NoStop}%
\bibitem [{\citenamefont {de~Rafael}(2014)}]{deRafael:2014gxa}%
  \BibitemOpen
  \bibfield  {author} {\bibinfo {author} {\bibfnamefont {E.}~\bibnamefont
  {de~Rafael}},\ }\href {\doibase 10.1016/j.physletb.2014.08.003} {\bibfield
  {journal} {\bibinfo  {journal} {Phys.Lett.}\ }\textbf {\bibinfo {volume}
  {B736}},\ \bibinfo {pages} {522} (\bibinfo {year} {2014})},\ \Eprint
  {http://arxiv.org/abs/1406.4671} {arXiv:1406.4671 [hep-lat]}\BibitemShut
  {NoStop}%
\bibitem [{\citenamefont {Charles}\ \emph {et~al.}(2018)\citenamefont
  {Charles}, \citenamefont {de~Rafael},\ and\ \citenamefont
  {Greynat}}]{Charles:2017snx}%
  \BibitemOpen
  \bibfield  {author} {\bibinfo {author} {\bibfnamefont {J.}~\bibnamefont
  {Charles}}, \bibinfo {author} {\bibfnamefont {E.}~\bibnamefont {de~Rafael}},
  \ and\ \bibinfo {author} {\bibfnamefont {D.}~\bibnamefont {Greynat}},\ }\href
  {\doibase 10.1103/PhysRevD.97.076014} {\bibfield  {journal} {\bibinfo
  {journal} {Phys. Rev.}\ }\textbf {\bibinfo {volume} {D97}},\ \bibinfo {pages}
  {076014} (\bibinfo {year} {2018})},\ \Eprint
  {http://arxiv.org/abs/1712.02202} {arXiv:1712.02202 [hep-ph]}\BibitemShut
  {NoStop}%
\bibitem [{\citenamefont {Della~Morte}\ \emph {et~al.}(2017)\citenamefont
  {Della~Morte}, \citenamefont {Francis}, \citenamefont {G{\"u}lpers},
  \citenamefont {Herdo{\'\i}za}, \citenamefont {von Hippel}, \citenamefont
  {Horch}, \citenamefont {J{\"a}ger}, \citenamefont {Meyer}, \citenamefont
  {Nyffeler},\ and\ \citenamefont {Wittig}}]{DellaMorte:2017dyu}%
  \BibitemOpen
  \bibfield  {author} {\bibinfo {author} {\bibfnamefont {M.}~\bibnamefont
  {Della~Morte}}, \bibinfo {author} {\bibfnamefont {A.}~\bibnamefont
  {Francis}}, \bibinfo {author} {\bibfnamefont {V.}~\bibnamefont
  {G{\"u}lpers}}, \bibinfo {author} {\bibfnamefont {G.}~\bibnamefont
  {Herdo{\'\i}za}}, \bibinfo {author} {\bibfnamefont {G.}~\bibnamefont {von
  Hippel}}, \bibinfo {author} {\bibfnamefont {H.}~\bibnamefont {Horch}},
  \bibinfo {author} {\bibfnamefont {B.}~\bibnamefont {J{\"a}ger}}, \bibinfo
  {author} {\bibfnamefont {H.~B.}\ \bibnamefont {Meyer}}, \bibinfo {author}
  {\bibfnamefont {A.}~\bibnamefont {Nyffeler}}, \ and\ \bibinfo {author}
  {\bibfnamefont {H.}~\bibnamefont {Wittig}},\ }\href {\doibase
  10.1007/JHEP10(2017)020} {\bibfield  {journal} {\bibinfo  {journal} {JHEP}\
  }\textbf {\bibinfo {volume} {10}},\ \bibinfo {pages} {020} (\bibinfo {year}
  {2017})},\ \Eprint {http://arxiv.org/abs/1705.01775} {arXiv:1705.01775
  [hep-lat]}\BibitemShut {NoStop}%
\bibitem [{\citenamefont {Parisi}(1984)}]{Parisi:1983ae}%
  \BibitemOpen
  \bibfield  {author} {\bibinfo {author} {\bibfnamefont {G.}~\bibnamefont
  {Parisi}},\ }\href {\doibase 10.1016/0370-1573(84)90081-4} {\bibfield
  {journal} {\bibinfo  {journal} {Phys. Rept.}\ }\textbf {\bibinfo {volume}
  {103}},\ \bibinfo {pages} {203} (\bibinfo {year} {1984})}\BibitemShut
  {NoStop}%
\bibitem [{\citenamefont {Lepage}(1989)}]{Lepage:1989hd}%
  \BibitemOpen
  \bibfield  {author} {\bibinfo {author} {\bibfnamefont {G.~P.}\ \bibnamefont
  {Lepage}},\ }in\ \href
  {https://lib-extopc.kek.jp/preprints/PDF/1990/9003/9003479.pdf} {\emph
  {\bibinfo {booktitle} {{Boulder ASI 1989:97--120}}}}\ (\bibinfo {year}
  {1989})\BibitemShut {NoStop}%
\bibitem [{\citenamefont {C\`e}\ \emph {et~al.}(2016)\citenamefont {C\`e},
  \citenamefont {Giusti},\ and\ \citenamefont {Schaefer}}]{Ce:2016idq}%
  \BibitemOpen
  \bibfield  {author} {\bibinfo {author} {\bibfnamefont {M.}~\bibnamefont
  {C\`e}}, \bibinfo {author} {\bibfnamefont {L.}~\bibnamefont {Giusti}}, \ and\
  \bibinfo {author} {\bibfnamefont {S.}~\bibnamefont {Schaefer}},\ }\href
  {\doibase 10.1103/PhysRevD.93.094507} {\bibfield  {journal} {\bibinfo
  {journal} {Phys. Rev.}\ }\textbf {\bibinfo {volume} {D93}},\ \bibinfo {pages}
  {094507} (\bibinfo {year} {2016})},\ \Eprint
  {http://arxiv.org/abs/1601.04587} {arXiv:1601.04587 [hep-lat]}\BibitemShut
  {NoStop}%
\bibitem [{\citenamefont {C\`e}\ \emph {et~al.}(2017)\citenamefont {C\`e},
  \citenamefont {Giusti},\ and\ \citenamefont {Schaefer}}]{Ce:2016ajy}%
  \BibitemOpen
  \bibfield  {author} {\bibinfo {author} {\bibfnamefont {M.}~\bibnamefont
  {C\`e}}, \bibinfo {author} {\bibfnamefont {L.}~\bibnamefont {Giusti}}, \ and\
  \bibinfo {author} {\bibfnamefont {S.}~\bibnamefont {Schaefer}},\ }\href
  {\doibase 10.1103/PhysRevD.95.034503} {\bibfield  {journal} {\bibinfo
  {journal} {Phys. Rev.}\ }\textbf {\bibinfo {volume} {D95}},\ \bibinfo {pages}
  {034503} (\bibinfo {year} {2017})},\ \Eprint
  {http://arxiv.org/abs/1609.02419} {arXiv:1609.02419 [hep-lat]}\BibitemShut
  {NoStop}%
\bibitem [{\citenamefont {Giusti}\ \emph
  {et~al.}(2018{\natexlab{a}})\citenamefont {Giusti}, \citenamefont {C\`e},\
  and\ \citenamefont {Schaefer}}]{Giusti:2017ksp}%
  \BibitemOpen
  \bibfield  {author} {\bibinfo {author} {\bibfnamefont {L.}~\bibnamefont
  {Giusti}}, \bibinfo {author} {\bibfnamefont {M.}~\bibnamefont {C\`e}}, \ and\
  \bibinfo {author} {\bibfnamefont {S.}~\bibnamefont {Schaefer}},\ }\href
  {\doibase 10.1051/epjconf/201817501003} {\bibfield  {journal} {\bibinfo
  {journal} {EPJ Web Conf.}\ }\textbf {\bibinfo {volume} {175}},\ \bibinfo
  {pages} {01003} (\bibinfo {year} {2018}{\natexlab{a}})},\ \Eprint
  {http://arxiv.org/abs/1710.09212} {arXiv:1710.09212 [hep-lat]}\BibitemShut
  {NoStop}%
\bibitem [{\citenamefont {Giusti}\ \emph
  {et~al.}(2017{\natexlab{a}})\citenamefont {Giusti}, \citenamefont {Lubicz},
  \citenamefont {Martinelli}, \citenamefont {Sanfilippo},\ and\ \citenamefont
  {Simula}}]{Giusti:2017jof}%
  \BibitemOpen
  \bibfield  {author} {\bibinfo {author} {\bibfnamefont {D.}~\bibnamefont
  {Giusti}}, \bibinfo {author} {\bibfnamefont {V.}~\bibnamefont {Lubicz}},
  \bibinfo {author} {\bibfnamefont {G.}~\bibnamefont {Martinelli}}, \bibinfo
  {author} {\bibfnamefont {F.}~\bibnamefont {Sanfilippo}}, \ and\ \bibinfo
  {author} {\bibfnamefont {S.}~\bibnamefont {Simula}},\ }\href {\doibase
  10.1007/JHEP10(2017)157} {\bibfield  {journal} {\bibinfo  {journal} {JHEP}\
  }\textbf {\bibinfo {volume} {10}},\ \bibinfo {pages} {157} (\bibinfo {year}
  {2017}{\natexlab{a}})},\ \Eprint {http://arxiv.org/abs/1707.03019}
  {arXiv:1707.03019 [hep-lat]}\BibitemShut {NoStop}%
\bibitem [{\citenamefont {Chakraborty}\ \emph {et~al.}(2017)\citenamefont
  {Chakraborty}, \citenamefont {Davies}, \citenamefont {de~Oliviera},
  \citenamefont {Koponen}, \citenamefont {Lepage},\ and\ \citenamefont {Van~de
  Water}}]{Chakraborty:2016mwy}%
  \BibitemOpen
  \bibfield  {author} {\bibinfo {author} {\bibfnamefont {B.}~\bibnamefont
  {Chakraborty}}, \bibinfo {author} {\bibfnamefont {C.~T.~H.}\ \bibnamefont
  {Davies}}, \bibinfo {author} {\bibfnamefont {P.~G.}\ \bibnamefont
  {de~Oliviera}}, \bibinfo {author} {\bibfnamefont {J.}~\bibnamefont
  {Koponen}}, \bibinfo {author} {\bibfnamefont {G.~P.}\ \bibnamefont {Lepage}},
  \ and\ \bibinfo {author} {\bibfnamefont {R.~S.}\ \bibnamefont {Van~de Water}}
  (\bibinfo {collaboration} {HPQCD}),\ }\href {\doibase
  10.1103/PhysRevD.96.034516} {\bibfield  {journal} {\bibinfo  {journal} {Phys.
  Rev.}\ }\textbf {\bibinfo {volume} {D96}},\ \bibinfo {pages} {034516}
  (\bibinfo {year} {2017})},\ \Eprint {http://arxiv.org/abs/1601.03071}
  {arXiv:1601.03071 [hep-lat]}\BibitemShut {NoStop}%
\bibitem [{\citenamefont {Giusti}\ \emph
  {et~al.}(2018{\natexlab{b}})\citenamefont {Giusti}, \citenamefont
  {Sanfilippo},\ and\ \citenamefont {Simula}}]{Giusti:2018mdh}%
  \BibitemOpen
  \bibfield  {author} {\bibinfo {author} {\bibfnamefont {D.}~\bibnamefont
  {Giusti}}, \bibinfo {author} {\bibfnamefont {F.}~\bibnamefont {Sanfilippo}},
  \ and\ \bibinfo {author} {\bibfnamefont {S.}~\bibnamefont {Simula}},\ }\href
  {\doibase 10.1103/PhysRevD.98.114504} {\bibfield  {journal} {\bibinfo
  {journal} {Phys. Rev.}\ }\textbf {\bibinfo {volume} {D98}},\ \bibinfo {pages}
  {114504} (\bibinfo {year} {2018}{\natexlab{b}})},\ \Eprint
  {http://arxiv.org/abs/1808.00887} {arXiv:1808.00887 [hep-lat]}\BibitemShut
  {NoStop}%
\bibitem [{\citenamefont {Della~Morte}\ \emph {et~al.}(2018)\citenamefont
  {Della~Morte} \emph {et~al.}}]{DellaMorte:2017khn}%
  \BibitemOpen
  \bibfield  {author} {\bibinfo {author} {\bibfnamefont {M.}~\bibnamefont
  {Della~Morte}} \emph {et~al.},\ }\href {\doibase
  10.1051/epjconf/201817506031} {\bibfield  {journal} {\bibinfo  {journal} {EPJ
  Web Conf.}\ }\textbf {\bibinfo {volume} {175}},\ \bibinfo {pages} {06031}
  (\bibinfo {year} {2018})},\ \Eprint {http://arxiv.org/abs/1710.10072}
  {arXiv:1710.10072 [hep-lat]}\BibitemShut {NoStop}%
\bibitem [{\citenamefont {L{\"u}scher}(1991)}]{Luscher:1991cf}%
  \BibitemOpen
  \bibfield  {author} {\bibinfo {author} {\bibfnamefont {M.}~\bibnamefont
  {L{\"u}scher}},\ }\href {\doibase 10.1016/0550-3213(91)90584-K} {\bibfield
  {journal} {\bibinfo  {journal} {Nucl. Phys.}\ }\textbf {\bibinfo {volume}
  {B364}},\ \bibinfo {pages} {237} (\bibinfo {year} {1991})}\BibitemShut
  {NoStop}%
\bibitem [{\citenamefont {Meyer}(2011)}]{Meyer:2011um}%
  \BibitemOpen
  \bibfield  {author} {\bibinfo {author} {\bibfnamefont {H.~B.}\ \bibnamefont
  {Meyer}},\ }\href {\doibase 10.1103/PhysRevLett.107.072002} {\bibfield
  {journal} {\bibinfo  {journal} {Phys. Rev. Lett.}\ }\textbf {\bibinfo
  {volume} {107}},\ \bibinfo {pages} {072002} (\bibinfo {year} {2011})},\
  \Eprint {http://arxiv.org/abs/1105.1892} {arXiv:1105.1892
  [hep-lat]}\BibitemShut {NoStop}%
\bibitem [{\citenamefont {Francis}\ \emph {et~al.}(2013)\citenamefont
  {Francis}, \citenamefont {J{\"a}ger}, \citenamefont {Meyer},\ and\
  \citenamefont {Wittig}}]{Francis:2013qna}%
  \BibitemOpen
  \bibfield  {author} {\bibinfo {author} {\bibfnamefont {A.}~\bibnamefont
  {Francis}}, \bibinfo {author} {\bibfnamefont {B.}~\bibnamefont {J{\"a}ger}},
  \bibinfo {author} {\bibfnamefont {H.~B.}\ \bibnamefont {Meyer}}, \ and\
  \bibinfo {author} {\bibfnamefont {H.}~\bibnamefont {Wittig}},\ }\href
  {\doibase 10.1103/PhysRevD.88.054502} {\bibfield  {journal} {\bibinfo
  {journal} {Phys. Rev.}\ }\textbf {\bibinfo {volume} {D88}},\ \bibinfo {pages}
  {054502} (\bibinfo {year} {2013})},\ \Eprint {http://arxiv.org/abs/1306.2532}
  {arXiv:1306.2532 [hep-lat]}\BibitemShut {NoStop}%
\bibitem [{\citenamefont {von Hippel}\ \emph {et~al.}(2018)\citenamefont {von
  Hippel} \emph {et~al.}}]{vonHippel:2018talk}%
  \BibitemOpen
  \bibfield  {author} {\bibinfo {author} {\bibfnamefont {G.}~\bibnamefont {von
  Hippel}} \emph {et~al.},\ }\href@noop {} {\enquote {\bibinfo {title}
  {{Lattice results for the hadronic vacuum polarization from Mainz (CLS)}},}\
  }\bibinfo {howpublished}
  {\url{https://kds.kek.jp/indico/event/26780/contributions/88936/attachments/71131/84309/talk.pdf}}
  (\bibinfo {year} {2018}),\ \bibinfo {note} {{Muon $g-2$ Theory Initiative
  workshop KEK}}\BibitemShut {NoStop}%
\bibitem [{\citenamefont {G{\'e}rardin}\ \emph
  {et~al.}(2018{\natexlab{a}})\citenamefont {G{\'e}rardin}, \citenamefont
  {Harris}, \citenamefont {von Hippel}, \citenamefont {H{\"o}rz}, \citenamefont
  {Meyer}, \citenamefont {Mohler}, \citenamefont {Ottnad},\ and\ \citenamefont
  {Wittig}}]{Gerardin:2018sin}%
  \BibitemOpen
  \bibfield  {author} {\bibinfo {author} {\bibfnamefont {A.}~\bibnamefont
  {G{\'e}rardin}}, \bibinfo {author} {\bibfnamefont {T.}~\bibnamefont
  {Harris}}, \bibinfo {author} {\bibfnamefont {G.}~\bibnamefont {von Hippel}},
  \bibinfo {author} {\bibfnamefont {B.}~\bibnamefont {H{\"o}rz}}, \bibinfo
  {author} {\bibfnamefont {H.}~\bibnamefont {Meyer}}, \bibinfo {author}
  {\bibfnamefont {D.}~\bibnamefont {Mohler}}, \bibinfo {author} {\bibfnamefont
  {K.}~\bibnamefont {Ottnad}}, \ and\ \bibinfo {author} {\bibfnamefont
  {H.}~\bibnamefont {Wittig}},\ }\href {\doibase 10.22323/1.334.0139}
  {\bibfield  {journal} {\bibinfo  {journal} {PoS}\ }\textbf {\bibinfo {volume}
  {LATTICE2018}},\ \bibinfo {pages} {139} (\bibinfo {year}
  {2018}{\natexlab{a}})},\ \Eprint {http://arxiv.org/abs/1812.03553}
  {arXiv:1812.03553 [hep-lat]}\BibitemShut {NoStop}%
\bibitem [{\citenamefont {Meyer}\ and\ \citenamefont
  {Wittig}(2019)}]{Meyer:2018til}%
  \BibitemOpen
  \bibfield  {author} {\bibinfo {author} {\bibfnamefont {H.~B.}\ \bibnamefont
  {Meyer}}\ and\ \bibinfo {author} {\bibfnamefont {H.}~\bibnamefont {Wittig}},\
  }\href {\doibase 10.1016/j.ppnp.2018.09.001} {\bibfield  {journal} {\bibinfo
  {journal} {Prog. Part. Nucl. Phys.}\ }\textbf {\bibinfo {volume} {104}},\
  \bibinfo {pages} {46} (\bibinfo {year} {2019})},\ \Eprint
  {http://arxiv.org/abs/1807.09370} {arXiv:1807.09370 [hep-lat]}\BibitemShut
  {NoStop}%
\bibitem [{\citenamefont {Feng}\ \emph {et~al.}(2015)\citenamefont {Feng},
  \citenamefont {Aoki}, \citenamefont {Hashimoto},\ and\ \citenamefont
  {Kaneko}}]{Feng:2014gba}%
  \BibitemOpen
  \bibfield  {author} {\bibinfo {author} {\bibfnamefont {X.}~\bibnamefont
  {Feng}}, \bibinfo {author} {\bibfnamefont {S.}~\bibnamefont {Aoki}}, \bibinfo
  {author} {\bibfnamefont {S.}~\bibnamefont {Hashimoto}}, \ and\ \bibinfo
  {author} {\bibfnamefont {T.}~\bibnamefont {Kaneko}},\ }\href {\doibase
  10.1103/PhysRevD.91.054504} {\bibfield  {journal} {\bibinfo  {journal} {Phys.
  Rev. D}\ }\textbf {\bibinfo {volume} {91}},\ \bibinfo {pages} {054504}
  (\bibinfo {year} {2015})},\ \Eprint {http://arxiv.org/abs/1412.6319}
  {arXiv:1412.6319 [hep-lat]}\BibitemShut {NoStop}%
\bibitem [{\citenamefont {Andersen}\ \emph {et~al.}(2019)\citenamefont
  {Andersen}, \citenamefont {Bulava}, \citenamefont {Hörz},\ and\
  \citenamefont {Morningstar}}]{Andersen:2018mau}%
  \BibitemOpen
  \bibfield  {author} {\bibinfo {author} {\bibfnamefont {C.}~\bibnamefont
  {Andersen}}, \bibinfo {author} {\bibfnamefont {J.}~\bibnamefont {Bulava}},
  \bibinfo {author} {\bibfnamefont {B.}~\bibnamefont {Hörz}}, \ and\ \bibinfo
  {author} {\bibfnamefont {C.}~\bibnamefont {Morningstar}},\ }\href {\doibase
  10.1016/j.nuclphysb.2018.12.018} {\bibfield  {journal} {\bibinfo  {journal}
  {Nucl. Phys. B}\ }\textbf {\bibinfo {volume} {939}},\ \bibinfo {pages} {145}
  (\bibinfo {year} {2019})},\ \Eprint {http://arxiv.org/abs/1808.05007}
  {arXiv:1808.05007 [hep-lat]}\BibitemShut {NoStop}%
\bibitem [{\citenamefont {Erben}\ \emph {et~al.}(2020)\citenamefont {Erben},
  \citenamefont {Green}, \citenamefont {Mohler},\ and\ \citenamefont
  {Wittig}}]{Erben:2019nmx}%
  \BibitemOpen
  \bibfield  {author} {\bibinfo {author} {\bibfnamefont {F.}~\bibnamefont
  {Erben}}, \bibinfo {author} {\bibfnamefont {J.~R.}\ \bibnamefont {Green}},
  \bibinfo {author} {\bibfnamefont {D.}~\bibnamefont {Mohler}}, \ and\ \bibinfo
  {author} {\bibfnamefont {H.}~\bibnamefont {Wittig}},\ }\href {\doibase
  10.1103/PhysRevD.101.054504} {\bibfield  {journal} {\bibinfo  {journal}
  {Phys.\ Rev.\ D}\ }\textbf {\bibinfo {volume} {101}},\ \bibinfo {pages}
  {054504} (\bibinfo {year} {2020})},\ \Eprint
  {http://arxiv.org/abs/1910.01083} {arXiv:1910.01083 [hep-lat]}\BibitemShut
  {NoStop}%
\bibitem [{\citenamefont {Aoki}\ \emph {et~al.}(2017)\citenamefont {Aoki} \emph
  {et~al.}}]{Aoki:2016frl}%
  \BibitemOpen
  \bibfield  {author} {\bibinfo {author} {\bibfnamefont {S.}~\bibnamefont
  {Aoki}} \emph {et~al.},\ }\href {\doibase 10.1140/epjc/s10052-016-4509-7}
  {\bibfield  {journal} {\bibinfo  {journal} {Eur. Phys. J.}\ }\textbf
  {\bibinfo {volume} {C77}},\ \bibinfo {pages} {112} (\bibinfo {year}
  {2017})},\ \Eprint {http://arxiv.org/abs/1607.00299} {arXiv:1607.00299
  [hep-lat]}\BibitemShut {NoStop}%
\bibitem [{\citenamefont {de~Divitiis}\ \emph {et~al.}(2013)\citenamefont
  {de~Divitiis}, \citenamefont {Frezzotti}, \citenamefont {Lubicz},
  \citenamefont {Martinelli}, \citenamefont {Petronzio}, \citenamefont {Rossi},
  \citenamefont {Sanfilippo}, \citenamefont {Simula},\ and\ \citenamefont
  {Tantalo}}]{deDivitiis:2013xla}%
  \BibitemOpen
  \bibfield  {author} {\bibinfo {author} {\bibfnamefont {G.~M.}\ \bibnamefont
  {de~Divitiis}}, \bibinfo {author} {\bibfnamefont {R.}~\bibnamefont
  {Frezzotti}}, \bibinfo {author} {\bibfnamefont {V.}~\bibnamefont {Lubicz}},
  \bibinfo {author} {\bibfnamefont {G.}~\bibnamefont {Martinelli}}, \bibinfo
  {author} {\bibfnamefont {R.}~\bibnamefont {Petronzio}}, \bibinfo {author}
  {\bibfnamefont {G.~C.}\ \bibnamefont {Rossi}}, \bibinfo {author}
  {\bibfnamefont {F.}~\bibnamefont {Sanfilippo}}, \bibinfo {author}
  {\bibfnamefont {S.}~\bibnamefont {Simula}}, \ and\ \bibinfo {author}
  {\bibfnamefont {N.}~\bibnamefont {Tantalo}} (\bibinfo {collaboration}
  {RM123}),\ }\href {\doibase 10.1103/PhysRevD.87.114505} {\bibfield  {journal}
  {\bibinfo  {journal} {Phys. Rev.}\ }\textbf {\bibinfo {volume} {D87}},\
  \bibinfo {pages} {114505} (\bibinfo {year} {2013})},\ \Eprint
  {http://arxiv.org/abs/1303.4896} {arXiv:1303.4896 [hep-lat]}\BibitemShut
  {NoStop}%
\bibitem [{\citenamefont {Borsanyi}\ \emph {et~al.}(2013)\citenamefont
  {Borsanyi} \emph {et~al.}}]{Borsanyi:2013lga}%
  \BibitemOpen
  \bibfield  {author} {\bibinfo {author} {\bibfnamefont {S.}~\bibnamefont
  {Borsanyi}} \emph {et~al.} (\bibinfo {collaboration}
  {Budapest-Marseille-Wuppertal}),\ }\href {\doibase
  10.1103/PhysRevLett.111.252001} {\bibfield  {journal} {\bibinfo  {journal}
  {Phys. Rev. Lett.}\ }\textbf {\bibinfo {volume} {111}},\ \bibinfo {pages}
  {252001} (\bibinfo {year} {2013})},\ \Eprint {http://arxiv.org/abs/1306.2287}
  {arXiv:1306.2287 [hep-lat]}\BibitemShut {NoStop}%
\bibitem [{\citenamefont {Borsanyi}\ \emph {et~al.}(2015)\citenamefont
  {Borsanyi} \emph {et~al.}}]{Borsanyi:2014jba}%
  \BibitemOpen
  \bibfield  {author} {\bibinfo {author} {\bibfnamefont {S.}~\bibnamefont
  {Borsanyi}} \emph {et~al.},\ }\href {\doibase 10.1126/science.1257050}
  {\bibfield  {journal} {\bibinfo  {journal} {Science}\ }\textbf {\bibinfo
  {volume} {347}},\ \bibinfo {pages} {1452} (\bibinfo {year} {2015})},\ \Eprint
  {http://arxiv.org/abs/1406.4088} {arXiv:1406.4088 [hep-lat]}\BibitemShut
  {NoStop}%
\bibitem [{\citenamefont {Fodor}\ \emph {et~al.}(2016)\citenamefont {Fodor},
  \citenamefont {Hoelbling}, \citenamefont {Krieg}, \citenamefont {Lellouch},
  \citenamefont {Lippert}, \citenamefont {Portelli}, \citenamefont {Sastre},
  \citenamefont {Szabo},\ and\ \citenamefont {Varnhorst}}]{Fodor:2016bgu}%
  \BibitemOpen
  \bibfield  {author} {\bibinfo {author} {\bibfnamefont {Z.}~\bibnamefont
  {Fodor}}, \bibinfo {author} {\bibfnamefont {C.}~\bibnamefont {Hoelbling}},
  \bibinfo {author} {\bibfnamefont {S.}~\bibnamefont {Krieg}}, \bibinfo
  {author} {\bibfnamefont {L.}~\bibnamefont {Lellouch}}, \bibinfo {author}
  {\bibfnamefont {T.}~\bibnamefont {Lippert}}, \bibinfo {author} {\bibfnamefont
  {A.}~\bibnamefont {Portelli}}, \bibinfo {author} {\bibfnamefont
  {A.}~\bibnamefont {Sastre}}, \bibinfo {author} {\bibfnamefont
  {K.}~\bibnamefont {Szabo}}, \ and\ \bibinfo {author} {\bibfnamefont
  {L.}~\bibnamefont {Varnhorst}},\ }\href {\doibase
  10.1103/PhysRevLett.117.082001} {\bibfield  {journal} {\bibinfo  {journal}
  {Phys. Rev. Lett.}\ }\textbf {\bibinfo {volume} {117}},\ \bibinfo {pages}
  {082001} (\bibinfo {year} {2016})},\ \Eprint
  {http://arxiv.org/abs/1604.07112} {arXiv:1604.07112 [hep-lat]}\BibitemShut
  {NoStop}%
\bibitem [{\citenamefont {Giusti}\ \emph
  {et~al.}(2017{\natexlab{b}})\citenamefont {Giusti}, \citenamefont {Lubicz},
  \citenamefont {Tarantino}, \citenamefont {Martinelli}, \citenamefont
  {Sanfilippo}, \citenamefont {Simula},\ and\ \citenamefont
  {Tantalo}}]{Giusti:2017dmp}%
  \BibitemOpen
  \bibfield  {author} {\bibinfo {author} {\bibfnamefont {D.}~\bibnamefont
  {Giusti}}, \bibinfo {author} {\bibfnamefont {V.}~\bibnamefont {Lubicz}},
  \bibinfo {author} {\bibfnamefont {C.}~\bibnamefont {Tarantino}}, \bibinfo
  {author} {\bibfnamefont {G.}~\bibnamefont {Martinelli}}, \bibinfo {author}
  {\bibfnamefont {S.}~\bibnamefont {Sanfilippo}}, \bibinfo {author}
  {\bibfnamefont {S.}~\bibnamefont {Simula}}, \ and\ \bibinfo {author}
  {\bibfnamefont {N.}~\bibnamefont {Tantalo}} (\bibinfo {collaboration}
  {RM123}),\ }\href {\doibase 10.1103/PhysRevD.95.114504} {\bibfield  {journal}
  {\bibinfo  {journal} {Phys. Rev.}\ }\textbf {\bibinfo {volume} {D95}},\
  \bibinfo {pages} {114504} (\bibinfo {year} {2017}{\natexlab{b}})},\ \Eprint
  {http://arxiv.org/abs/1704.06561} {arXiv:1704.06561 [hep-lat]}\BibitemShut
  {NoStop}%
\bibitem [{\citenamefont {Basak}\ \emph {et~al.}(2019)\citenamefont {Basak}
  \emph {et~al.}}]{Basak:2018yzz}%
  \BibitemOpen
  \bibfield  {author} {\bibinfo {author} {\bibfnamefont {S.}~\bibnamefont
  {Basak}} \emph {et~al.} (\bibinfo {collaboration} {MILC}),\ }\href {\doibase
  10.1103/PhysRevD.99.034503} {\bibfield  {journal} {\bibinfo  {journal} {Phys.
  Rev.}\ }\textbf {\bibinfo {volume} {D99}},\ \bibinfo {pages} {034503}
  (\bibinfo {year} {2019})},\ \Eprint {http://arxiv.org/abs/1807.05556}
  {arXiv:1807.05556 [hep-lat]}\BibitemShut {NoStop}%
\bibitem [{\citenamefont {Di~Carlo}\ \emph
  {et~al.}(2019{\natexlab{a}})\citenamefont {Di~Carlo}, \citenamefont {Giusti},
  \citenamefont {Lubicz}, \citenamefont {Martinelli}, \citenamefont
  {Sachrajda}, \citenamefont {Sanfilippo}, \citenamefont {Simula},\ and\
  \citenamefont {Tantalo}}]{DiCarlo:2019thl}%
  \BibitemOpen
  \bibfield  {author} {\bibinfo {author} {\bibfnamefont {M.}~\bibnamefont
  {Di~Carlo}}, \bibinfo {author} {\bibfnamefont {D.}~\bibnamefont {Giusti}},
  \bibinfo {author} {\bibfnamefont {V.}~\bibnamefont {Lubicz}}, \bibinfo
  {author} {\bibfnamefont {G.}~\bibnamefont {Martinelli}}, \bibinfo {author}
  {\bibfnamefont {C.~T.}\ \bibnamefont {Sachrajda}}, \bibinfo {author}
  {\bibfnamefont {F.}~\bibnamefont {Sanfilippo}}, \bibinfo {author}
  {\bibfnamefont {S.}~\bibnamefont {Simula}}, \ and\ \bibinfo {author}
  {\bibfnamefont {N.}~\bibnamefont {Tantalo}},\ }\href {\doibase
  10.1103/PhysRevD.100.034514} {\bibfield  {journal} {\bibinfo  {journal}
  {Phys. Rev.}\ }\textbf {\bibinfo {volume} {D100}},\ \bibinfo {pages} {034514}
  (\bibinfo {year} {2019}{\natexlab{a}})},\ \Eprint
  {http://arxiv.org/abs/1904.08731} {arXiv:1904.08731 [hep-lat]}\BibitemShut
  {NoStop}%
\bibitem [{\citenamefont {Borsanyi}\ \emph {et~al.}(2020)\citenamefont
  {Borsanyi} \emph {et~al.}}]{Borsanyi:2020mff}%
  \BibitemOpen
  \bibfield  {author} {\bibinfo {author} {\bibfnamefont {S.}~\bibnamefont
  {Borsanyi}} \emph {et~al.},\ }\href@noop {} {\  (\bibinfo {year} {2020})},\
  \Eprint {http://arxiv.org/abs/2002.12347} {arXiv:2002.12347
  [hep-lat]}\BibitemShut {NoStop}%
\bibitem [{\citenamefont {Aoki}\ \emph {et~al.}(2020)\citenamefont {Aoki} \emph
  {et~al.}}]{Aoki:2019cca}%
  \BibitemOpen
  \bibfield  {author} {\bibinfo {author} {\bibfnamefont {S.}~\bibnamefont
  {Aoki}} \emph {et~al.} (\bibinfo {collaboration} {Flavour Lattice Averaging
  Group}),\ }\href {\doibase 10.1140/epjc/s10052-019-7354-7} {\bibfield
  {journal} {\bibinfo  {journal} {Eur. Phys. J.}\ }\textbf {\bibinfo {volume}
  {C80}},\ \bibinfo {pages} {113} (\bibinfo {year} {2020})},\ \Eprint
  {http://arxiv.org/abs/1902.08191} {arXiv:1902.08191 [hep-lat]}\BibitemShut
  {NoStop}%
\bibitem [{\citenamefont {de~Divitiis}\ \emph
  {et~al.}(2012{\natexlab{b}})\citenamefont {de~Divitiis} \emph
  {et~al.}}]{deDivitiis:2011eh}%
  \BibitemOpen
  \bibfield  {author} {\bibinfo {author} {\bibfnamefont {G.~M.}\ \bibnamefont
  {de~Divitiis}} \emph {et~al.} (\bibinfo {collaboration} {RM123}),\ }\href
  {\doibase 10.1007/JHEP04(2012)124} {\bibfield  {journal} {\bibinfo  {journal}
  {JHEP}\ }\textbf {\bibinfo {volume} {04}},\ \bibinfo {pages} {124} (\bibinfo
  {year} {2012}{\natexlab{b}})},\ \Eprint {http://arxiv.org/abs/1110.6294}
  {arXiv:1110.6294 [hep-lat]}\BibitemShut {NoStop}%
\bibitem [{\citenamefont {Gasser}\ \emph {et~al.}(2003)\citenamefont {Gasser},
  \citenamefont {Rusetsky},\ and\ \citenamefont {Scimemi}}]{Gasser:2003hk}%
  \BibitemOpen
  \bibfield  {author} {\bibinfo {author} {\bibfnamefont {J.}~\bibnamefont
  {Gasser}}, \bibinfo {author} {\bibfnamefont {A.}~\bibnamefont {Rusetsky}}, \
  and\ \bibinfo {author} {\bibfnamefont {I.}~\bibnamefont {Scimemi}},\ }\href
  {\doibase 10.1140/epjc/s2003-01383-1} {\bibfield  {journal} {\bibinfo
  {journal} {Eur. Phys. J.}\ }\textbf {\bibinfo {volume} {C32}},\ \bibinfo
  {pages} {97} (\bibinfo {year} {2003})},\ \Eprint
  {http://arxiv.org/abs/hep-ph/0305260} {arXiv:hep-ph/0305260
  [hep-ph]}\BibitemShut {NoStop}%
\bibitem [{\citenamefont {Cirigliano}\ and\ \citenamefont
  {Neufeld}(2011)}]{Cirigliano:2011tm}%
  \BibitemOpen
  \bibfield  {author} {\bibinfo {author} {\bibfnamefont {V.}~\bibnamefont
  {Cirigliano}}\ and\ \bibinfo {author} {\bibfnamefont {H.}~\bibnamefont
  {Neufeld}},\ }\href {\doibase 10.1016/j.physletb.2011.04.038} {\bibfield
  {journal} {\bibinfo  {journal} {Phys. Lett.}\ }\textbf {\bibinfo {volume}
  {B700}},\ \bibinfo {pages} {7} (\bibinfo {year} {2011})},\ \Eprint
  {http://arxiv.org/abs/1102.0563} {arXiv:1102.0563 [hep-ph]}\BibitemShut
  {NoStop}%
\bibitem [{\citenamefont {Giusti}\ \emph
  {et~al.}(2018{\natexlab{c}})\citenamefont {Giusti}, \citenamefont {Lubicz},
  \citenamefont {Martinelli}, \citenamefont {Sachrajda}, \citenamefont
  {Sanfilippo}, \citenamefont {Simula}, \citenamefont {Tantalo},\ and\
  \citenamefont {Tarantino}}]{Giusti:2017dwk}%
  \BibitemOpen
  \bibfield  {author} {\bibinfo {author} {\bibfnamefont {D.}~\bibnamefont
  {Giusti}}, \bibinfo {author} {\bibfnamefont {V.}~\bibnamefont {Lubicz}},
  \bibinfo {author} {\bibfnamefont {G.}~\bibnamefont {Martinelli}}, \bibinfo
  {author} {\bibfnamefont {C.~T.}\ \bibnamefont {Sachrajda}}, \bibinfo {author}
  {\bibfnamefont {F.}~\bibnamefont {Sanfilippo}}, \bibinfo {author}
  {\bibfnamefont {S.}~\bibnamefont {Simula}}, \bibinfo {author} {\bibfnamefont
  {N.}~\bibnamefont {Tantalo}}, \ and\ \bibinfo {author} {\bibfnamefont
  {C.}~\bibnamefont {Tarantino}},\ }\href {\doibase
  10.1103/PhysRevLett.120.072001} {\bibfield  {journal} {\bibinfo  {journal}
  {Phys. Rev. Lett.}\ }\textbf {\bibinfo {volume} {120}},\ \bibinfo {pages}
  {072001} (\bibinfo {year} {2018}{\natexlab{c}})},\ \Eprint
  {http://arxiv.org/abs/1711.06537} {arXiv:1711.06537 [hep-lat]}\BibitemShut
  {NoStop}%
\bibitem [{\citenamefont {Basak}\ \emph {et~al.}(2016)\citenamefont {Basak}
  \emph {et~al.}}]{Basak:2016jnn}%
  \BibitemOpen
  \bibfield  {author} {\bibinfo {author} {\bibfnamefont {S.}~\bibnamefont
  {Basak}} \emph {et~al.} (\bibinfo {collaboration} {MILC}),\ }\href {\doibase
  10.22323/1.251.0259} {\bibfield  {journal} {\bibinfo  {journal} {PoS}\
  }\textbf {\bibinfo {volume} {LATTICE2015}},\ \bibinfo {pages} {259} (\bibinfo
  {year} {2016})},\ \Eprint {http://arxiv.org/abs/1606.01228} {arXiv:1606.01228
  [hep-lat]}\BibitemShut {NoStop}%
\bibitem [{\citenamefont {G{\"u}lpers}\ \emph {et~al.}(2018)\citenamefont
  {G{\"u}lpers}, \citenamefont {J{\"u}ttner}, \citenamefont {Lehner},\ and\
  \citenamefont {Portelli}}]{Gulpers:2018mim}%
  \BibitemOpen
  \bibfield  {author} {\bibinfo {author} {\bibfnamefont {V.}~\bibnamefont
  {G{\"u}lpers}}, \bibinfo {author} {\bibfnamefont {A.}~\bibnamefont
  {J{\"u}ttner}}, \bibinfo {author} {\bibfnamefont {C.}~\bibnamefont {Lehner}},
  \ and\ \bibinfo {author} {\bibfnamefont {A.}~\bibnamefont {Portelli}},\
  }\href {\doibase 10.22323/1.334.0134} {\bibfield  {journal} {\bibinfo
  {journal} {PoS}\ }\textbf {\bibinfo {volume} {LATTICE2018}},\ \bibinfo
  {pages} {134} (\bibinfo {year} {2018})},\ \Eprint
  {http://arxiv.org/abs/1812.09562} {arXiv:1812.09562 [hep-lat]}\BibitemShut
  {NoStop}%
\bibitem [{\citenamefont {Blum}\ \emph
  {et~al.}(2016{\natexlab{b}})\citenamefont {Blum} \emph
  {et~al.}}]{Blum:2014tka}%
  \BibitemOpen
  \bibfield  {author} {\bibinfo {author} {\bibfnamefont {T.}~\bibnamefont
  {Blum}} \emph {et~al.} (\bibinfo {collaboration} {RBC, UKQCD}),\ }\href
  {\doibase 10.1103/PhysRevD.93.074505} {\bibfield  {journal} {\bibinfo
  {journal} {Phys. Rev.}\ }\textbf {\bibinfo {volume} {D93}},\ \bibinfo {pages}
  {074505} (\bibinfo {year} {2016}{\natexlab{b}})},\ \Eprint
  {http://arxiv.org/abs/1411.7017} {arXiv:1411.7017 [hep-lat]}\BibitemShut
  {NoStop}%
\bibitem [{\citenamefont {Lehner}(2018)}]{Lehner:2018LAT}%
  \BibitemOpen
  \bibfield  {author} {\bibinfo {author} {\bibfnamefont {C.}~\bibnamefont
  {Lehner}},\ }\href@noop {} {\enquote {\bibinfo {title} {{Status of HVP
  calculation by RBC/UKQCD}},}\ }\bibinfo {howpublished}
  {\url{https://indico.fnal.gov/event/15949/session/8/contribution/29/material/slides/0.pdf}}
  (\bibinfo {year} {2018}),\ \bibinfo {note} {{36th International Symposium on
  Lattice Field Theory (Lattice 2018) East Lansing}}\BibitemShut {NoStop}%
\bibitem [{\citenamefont {Bruno}\ \emph {et~al.}(2019)\citenamefont {Bruno},
  \citenamefont {Izubuchi}, \citenamefont {Lehner},\ and\ \citenamefont
  {Meyer}}]{Bruno:2019nzm}%
  \BibitemOpen
  \bibfield  {author} {\bibinfo {author} {\bibfnamefont {M.}~\bibnamefont
  {Bruno}}, \bibinfo {author} {\bibfnamefont {T.}~\bibnamefont {Izubuchi}},
  \bibinfo {author} {\bibfnamefont {C.}~\bibnamefont {Lehner}}, \ and\ \bibinfo
  {author} {\bibfnamefont {A.~S.}\ \bibnamefont {Meyer}},\ }\href {\doibase
  10.22323/1.363.0239} {\bibfield  {journal} {\bibinfo  {journal} {PoS}\
  }\textbf {\bibinfo {volume} {LATTICE2019}},\ \bibinfo {pages} {239} (\bibinfo
  {year} {2019})},\ \Eprint {http://arxiv.org/abs/1910.11745} {arXiv:1910.11745
  [hep-lat]}\BibitemShut {NoStop}%
\bibitem [{\citenamefont {Bijnens}\ and\ \citenamefont
  {Relefors}(2017)}]{Bijnens:2017esv}%
  \BibitemOpen
  \bibfield  {author} {\bibinfo {author} {\bibfnamefont {J.}~\bibnamefont
  {Bijnens}}\ and\ \bibinfo {author} {\bibfnamefont {J.}~\bibnamefont
  {Relefors}},\ }\href {\doibase 10.1007/JHEP12(2017)114} {\bibfield  {journal}
  {\bibinfo  {journal} {JHEP}\ }\textbf {\bibinfo {volume} {12}},\ \bibinfo
  {pages} {114} (\bibinfo {year} {2017})},\ \Eprint
  {http://arxiv.org/abs/1710.04479} {arXiv:1710.04479 [hep-lat]}\BibitemShut
  {NoStop}%
\bibitem [{\citenamefont {Giusti}\ \emph
  {et~al.}(2019{\natexlab{b}})\citenamefont {Giusti}, \citenamefont {Lubicz},
  \citenamefont {Martinelli}, \citenamefont {Sanfilippo},\ and\ \citenamefont
  {Simula}}]{Giusti:2019dmu}%
  \BibitemOpen
  \bibfield  {author} {\bibinfo {author} {\bibfnamefont {D.}~\bibnamefont
  {Giusti}}, \bibinfo {author} {\bibfnamefont {V.}~\bibnamefont {Lubicz}},
  \bibinfo {author} {\bibfnamefont {G.}~\bibnamefont {Martinelli}}, \bibinfo
  {author} {\bibfnamefont {F.}~\bibnamefont {Sanfilippo}}, \ and\ \bibinfo
  {author} {\bibfnamefont {S.}~\bibnamefont {Simula}},\ }\href {\doibase
  10.22323/1.317.0063} {\bibfield  {journal} {\bibinfo  {journal} {PoS}\
  }\textbf {\bibinfo {volume} {CD2018}},\ \bibinfo {pages} {063} (\bibinfo
  {year} {2019}{\natexlab{b}})},\ \Eprint {http://arxiv.org/abs/1909.01962}
  {arXiv:1909.01962 [hep-lat]}\BibitemShut {NoStop}%
\bibitem [{\citenamefont {Izubuchi}\ \emph {et~al.}(2018)\citenamefont
  {Izubuchi}, \citenamefont {Kuramashi}, \citenamefont {Lehner},\ and\
  \citenamefont {Shintani}}]{Izubuchi:2018tdd}%
  \BibitemOpen
  \bibfield  {author} {\bibinfo {author} {\bibfnamefont {T.}~\bibnamefont
  {Izubuchi}}, \bibinfo {author} {\bibfnamefont {Y.}~\bibnamefont {Kuramashi}},
  \bibinfo {author} {\bibfnamefont {C.}~\bibnamefont {Lehner}}, \ and\ \bibinfo
  {author} {\bibfnamefont {E.}~\bibnamefont {Shintani}} (\bibinfo
  {collaboration} {PACS}),\ }\href {\doibase 10.1103/PhysRevD.98.054505}
  {\bibfield  {journal} {\bibinfo  {journal} {Phys. Rev.}\ }\textbf {\bibinfo
  {volume} {D98}},\ \bibinfo {pages} {054505} (\bibinfo {year} {2018})},\
  \Eprint {http://arxiv.org/abs/1805.04250} {arXiv:1805.04250
  [hep-lat]}\BibitemShut {NoStop}%
\bibitem [{\citenamefont {Hansen}\ and\ \citenamefont
  {Patella}(2019)}]{Hansen:2019rbh}%
  \BibitemOpen
  \bibfield  {author} {\bibinfo {author} {\bibfnamefont {M.~T.}\ \bibnamefont
  {Hansen}}\ and\ \bibinfo {author} {\bibfnamefont {A.}~\bibnamefont
  {Patella}},\ }\href {\doibase 10.1103/PhysRevLett.123.172001} {\bibfield
  {journal} {\bibinfo  {journal} {Phys. Rev. Lett.}\ }\textbf {\bibinfo
  {volume} {123}},\ \bibinfo {pages} {172001} (\bibinfo {year} {2019})},\
  \Eprint {http://arxiv.org/abs/1904.10010} {arXiv:1904.10010
  [hep-lat]}\BibitemShut {NoStop}%
\bibitem [{\citenamefont {Burger}\ \emph {et~al.}(2014)\citenamefont {Burger},
  \citenamefont {Feng}, \citenamefont {Hotzel}, \citenamefont {Jansen},
  \citenamefont {Petschlies},\ and\ \citenamefont {Renner}}]{Burger:2013jya}%
  \BibitemOpen
  \bibfield  {author} {\bibinfo {author} {\bibfnamefont {F.}~\bibnamefont
  {Burger}}, \bibinfo {author} {\bibfnamefont {X.}~\bibnamefont {Feng}},
  \bibinfo {author} {\bibfnamefont {G.}~\bibnamefont {Hotzel}}, \bibinfo
  {author} {\bibfnamefont {K.}~\bibnamefont {Jansen}}, \bibinfo {author}
  {\bibfnamefont {M.}~\bibnamefont {Petschlies}}, \ and\ \bibinfo {author}
  {\bibfnamefont {D.~B.}\ \bibnamefont {Renner}} (\bibinfo {collaboration}
  {ETM}),\ }\href {\doibase 10.1007/JHEP02(2014)099} {\bibfield  {journal}
  {\bibinfo  {journal} {JHEP}\ }\textbf {\bibinfo {volume} {02}},\ \bibinfo
  {pages} {099} (\bibinfo {year} {2014})},\ \Eprint
  {http://arxiv.org/abs/1308.4327} {arXiv:1308.4327 [hep-lat]}\BibitemShut
  {NoStop}%
\bibitem [{\citenamefont {Golowich}\ and\ \citenamefont
  {Kambor}(1995)}]{Golowich:1995kd}%
  \BibitemOpen
  \bibfield  {author} {\bibinfo {author} {\bibfnamefont {E.}~\bibnamefont
  {Golowich}}\ and\ \bibinfo {author} {\bibfnamefont {J.}~\bibnamefont
  {Kambor}},\ }\href {\doibase 10.1016/0550-3213(95)00234-J} {\bibfield
  {journal} {\bibinfo  {journal} {Nucl. Phys.}\ }\textbf {\bibinfo {volume}
  {B447}},\ \bibinfo {pages} {373} (\bibinfo {year} {1995})},\ \Eprint
  {http://arxiv.org/abs/hep-ph/9501318} {arXiv:hep-ph/9501318
  [hep-ph]}\BibitemShut {NoStop}%
\bibitem [{\citenamefont {Amoros}\ \emph {et~al.}(2000)\citenamefont {Amoros},
  \citenamefont {Bijnens},\ and\ \citenamefont {Talavera}}]{Amoros:1999dp}%
  \BibitemOpen
  \bibfield  {author} {\bibinfo {author} {\bibfnamefont {G.}~\bibnamefont
  {Amoros}}, \bibinfo {author} {\bibfnamefont {J.}~\bibnamefont {Bijnens}}, \
  and\ \bibinfo {author} {\bibfnamefont {P.}~\bibnamefont {Talavera}},\ }\href
  {\doibase 10.1016/S0550-3213(99)00674-4} {\bibfield  {journal} {\bibinfo
  {journal} {Nucl. Phys.}\ }\textbf {\bibinfo {volume} {B568}},\ \bibinfo
  {pages} {319} (\bibinfo {year} {2000})},\ \Eprint
  {http://arxiv.org/abs/hep-ph/9907264} {arXiv:hep-ph/9907264
  [hep-ph]}\BibitemShut {NoStop}%
\bibitem [{\citenamefont {Bijnens}\ and\ \citenamefont
  {Relefors}(2016{\natexlab{a}})}]{Bijnens:2016ndo}%
  \BibitemOpen
  \bibfield  {author} {\bibinfo {author} {\bibfnamefont {J.}~\bibnamefont
  {Bijnens}}\ and\ \bibinfo {author} {\bibfnamefont {J.}~\bibnamefont
  {Relefors}},\ }\href {\doibase 10.1007/JHEP11(2016)086} {\bibfield  {journal}
  {\bibinfo  {journal} {JHEP}\ }\textbf {\bibinfo {volume} {11}},\ \bibinfo
  {pages} {086} (\bibinfo {year} {2016}{\natexlab{a}})},\ \Eprint
  {http://arxiv.org/abs/1609.01573} {arXiv:1609.01573 [hep-lat]}\BibitemShut
  {NoStop}%
\bibitem [{\citenamefont {Golterman}\ \emph {et~al.}(2017)\citenamefont
  {Golterman}, \citenamefont {Maltman},\ and\ \citenamefont
  {Peris}}]{Golterman:2017njs}%
  \BibitemOpen
  \bibfield  {author} {\bibinfo {author} {\bibfnamefont {M.}~\bibnamefont
  {Golterman}}, \bibinfo {author} {\bibfnamefont {K.}~\bibnamefont {Maltman}},
  \ and\ \bibinfo {author} {\bibfnamefont {S.}~\bibnamefont {Peris}},\ }\href
  {\doibase 10.1103/PhysRevD.95.074509} {\bibfield  {journal} {\bibinfo
  {journal} {Phys. Rev.}\ }\textbf {\bibinfo {volume} {D95}},\ \bibinfo {pages}
  {074509} (\bibinfo {year} {2017})},\ \Eprint
  {http://arxiv.org/abs/1701.08685} {arXiv:1701.08685 [hep-lat]}\BibitemShut
  {NoStop}%
\bibitem [{\citenamefont {Colquhoun}\ \emph {et~al.}(2015)\citenamefont
  {Colquhoun}, \citenamefont {Dowdall}, \citenamefont {Davies}, \citenamefont
  {Hornbostel},\ and\ \citenamefont {Lepage}}]{Colquhoun:2014ica}%
  \BibitemOpen
  \bibfield  {author} {\bibinfo {author} {\bibfnamefont {B.}~\bibnamefont
  {Colquhoun}}, \bibinfo {author} {\bibfnamefont {R.~J.}\ \bibnamefont
  {Dowdall}}, \bibinfo {author} {\bibfnamefont {C.~T.~H.}\ \bibnamefont
  {Davies}}, \bibinfo {author} {\bibfnamefont {K.}~\bibnamefont {Hornbostel}},
  \ and\ \bibinfo {author} {\bibfnamefont {G.~P.}\ \bibnamefont {Lepage}},\
  }\href {\doibase 10.1103/PhysRevD.91.074514} {\bibfield  {journal} {\bibinfo
  {journal} {Phys. Rev.}\ }\textbf {\bibinfo {volume} {D91}},\ \bibinfo {pages}
  {074514} (\bibinfo {year} {2015})},\ \Eprint {http://arxiv.org/abs/1408.5768}
  {arXiv:1408.5768 [hep-lat]}\BibitemShut {NoStop}%
\bibitem [{\citenamefont {Donald}\ \emph {et~al.}(2012)\citenamefont {Donald},
  \citenamefont {Davies}, \citenamefont {Dowdall}, \citenamefont {Follana},
  \citenamefont {Hornbostel}, \citenamefont {Koponen}, \citenamefont {Lepage},\
  and\ \citenamefont {McNeile}}]{Donald:2012ga}%
  \BibitemOpen
  \bibfield  {author} {\bibinfo {author} {\bibfnamefont {G.~C.}\ \bibnamefont
  {Donald}}, \bibinfo {author} {\bibfnamefont {C.~T.~H.}\ \bibnamefont
  {Davies}}, \bibinfo {author} {\bibfnamefont {R.~J.}\ \bibnamefont {Dowdall}},
  \bibinfo {author} {\bibfnamefont {E.}~\bibnamefont {Follana}}, \bibinfo
  {author} {\bibfnamefont {K.}~\bibnamefont {Hornbostel}}, \bibinfo {author}
  {\bibfnamefont {J.}~\bibnamefont {Koponen}}, \bibinfo {author} {\bibfnamefont
  {G.~P.}\ \bibnamefont {Lepage}}, \ and\ \bibinfo {author} {\bibfnamefont
  {C.}~\bibnamefont {McNeile}},\ }\href {\doibase 10.1103/PhysRevD.86.094501}
  {\bibfield  {journal} {\bibinfo  {journal} {Phys. Rev.}\ }\textbf {\bibinfo
  {volume} {D86}},\ \bibinfo {pages} {094501} (\bibinfo {year} {2012})},\
  \Eprint {http://arxiv.org/abs/1208.2855} {arXiv:1208.2855
  [hep-lat]}\BibitemShut {NoStop}%
\bibitem [{\citenamefont {Nakayama}\ \emph {et~al.}(2016)\citenamefont
  {Nakayama}, \citenamefont {Fahy},\ and\ \citenamefont
  {Hashimoto}}]{Nakayama:2016atf}%
  \BibitemOpen
  \bibfield  {author} {\bibinfo {author} {\bibfnamefont {K.}~\bibnamefont
  {Nakayama}}, \bibinfo {author} {\bibfnamefont {B.}~\bibnamefont {Fahy}}, \
  and\ \bibinfo {author} {\bibfnamefont {S.}~\bibnamefont {Hashimoto}},\ }\href
  {\doibase 10.1103/PhysRevD.94.054507} {\bibfield  {journal} {\bibinfo
  {journal} {Phys. Rev.}\ }\textbf {\bibinfo {volume} {D94}},\ \bibinfo {pages}
  {054507} (\bibinfo {year} {2016})},\ \Eprint
  {http://arxiv.org/abs/1606.01002} {arXiv:1606.01002 [hep-lat]}\BibitemShut
  {NoStop}%
\bibitem [{\citenamefont {Blum}\ \emph
  {et~al.}(2016{\natexlab{c}})\citenamefont {Blum} \emph
  {et~al.}}]{Blum:2016xpd}%
  \BibitemOpen
  \bibfield  {author} {\bibinfo {author} {\bibfnamefont {T.}~\bibnamefont
  {Blum}} \emph {et~al.} (\bibinfo {collaboration} {RBC/UKQCD}),\ }\href
  {\doibase 10.1007/JHEP05(2017)034, 10.1007/JHEP04(2016)063} {\bibfield
  {journal} {\bibinfo  {journal} {JHEP}\ }\textbf {\bibinfo {volume} {04}},\
  \bibinfo {pages} {063} (\bibinfo {year} {2016}{\natexlab{c}})},\ \bibinfo
  {note} {[Erratum: JHEP {\bf 05}, 034 (2017)]},\ \Eprint
  {http://arxiv.org/abs/1602.01767} {arXiv:1602.01767 [hep-lat]}\BibitemShut
  {NoStop}%
\bibitem [{\citenamefont {Blum}\ \emph
  {et~al.}(2016{\natexlab{d}})\citenamefont {Blum}, \citenamefont {Boyle},
  \citenamefont {Izubuchi}, \citenamefont {Jin}, \citenamefont {J{\"u}ttner},
  \citenamefont {Lehner}, \citenamefont {Maltman}, \citenamefont
  {Marinkovi{\'c}}, \citenamefont {Portelli},\ and\ \citenamefont
  {Spraggs}}]{Blum:2015you}%
  \BibitemOpen
  \bibfield  {author} {\bibinfo {author} {\bibfnamefont {T.}~\bibnamefont
  {Blum}}, \bibinfo {author} {\bibfnamefont {P.~A.}\ \bibnamefont {Boyle}},
  \bibinfo {author} {\bibfnamefont {T.}~\bibnamefont {Izubuchi}}, \bibinfo
  {author} {\bibfnamefont {L.}~\bibnamefont {Jin}}, \bibinfo {author}
  {\bibfnamefont {A.}~\bibnamefont {J{\"u}ttner}}, \bibinfo {author}
  {\bibfnamefont {C.}~\bibnamefont {Lehner}}, \bibinfo {author} {\bibfnamefont
  {K.}~\bibnamefont {Maltman}}, \bibinfo {author} {\bibfnamefont
  {M.}~\bibnamefont {Marinkovi{\'c}}}, \bibinfo {author} {\bibfnamefont
  {A.}~\bibnamefont {Portelli}}, \ and\ \bibinfo {author} {\bibfnamefont
  {M.}~\bibnamefont {Spraggs}},\ }\href {\doibase
  10.1103/PhysRevLett.116.232002} {\bibfield  {journal} {\bibinfo  {journal}
  {Phys. Rev. Lett.}\ }\textbf {\bibinfo {volume} {116}},\ \bibinfo {pages}
  {232002} (\bibinfo {year} {2016}{\natexlab{d}})},\ \Eprint
  {http://arxiv.org/abs/1512.09054} {arXiv:1512.09054 [hep-lat]}\BibitemShut
  {NoStop}%
\bibitem [{\citenamefont {Borsanyi}\ \emph {et~al.}(2017)\citenamefont
  {Borsanyi}, \citenamefont {Fodor}, \citenamefont {Kawanai}, \citenamefont
  {Krieg}, \citenamefont {Lellouch}, \citenamefont {Malak}, \citenamefont
  {Miura}, \citenamefont {Szabo}, \citenamefont {Torrero},\ and\ \citenamefont
  {Toth}}]{Borsanyi:2016lpl}%
  \BibitemOpen
  \bibfield  {author} {\bibinfo {author} {\bibfnamefont {S.}~\bibnamefont
  {Borsanyi}}, \bibinfo {author} {\bibfnamefont {Z.}~\bibnamefont {Fodor}},
  \bibinfo {author} {\bibfnamefont {T.}~\bibnamefont {Kawanai}}, \bibinfo
  {author} {\bibfnamefont {S.}~\bibnamefont {Krieg}}, \bibinfo {author}
  {\bibfnamefont {L.}~\bibnamefont {Lellouch}}, \bibinfo {author}
  {\bibfnamefont {R.}~\bibnamefont {Malak}}, \bibinfo {author} {\bibfnamefont
  {K.}~\bibnamefont {Miura}}, \bibinfo {author} {\bibfnamefont {K.~K.}\
  \bibnamefont {Szabo}}, \bibinfo {author} {\bibfnamefont {C.}~\bibnamefont
  {Torrero}}, \ and\ \bibinfo {author} {\bibfnamefont {B.}~\bibnamefont
  {Toth}},\ }\href {\doibase 10.1103/PhysRevD.96.074507} {\bibfield  {journal}
  {\bibinfo  {journal} {Phys. Rev.}\ }\textbf {\bibinfo {volume} {D96}},\
  \bibinfo {pages} {074507} (\bibinfo {year} {2017})},\ \Eprint
  {http://arxiv.org/abs/1612.02364} {arXiv:1612.02364 [hep-lat]}\BibitemShut
  {NoStop}%
\bibitem [{\citenamefont {Della~Morte}\ and\ \citenamefont
  {J{\"u}ttner}(2010)}]{DellaMorte:2010aq}%
  \BibitemOpen
  \bibfield  {author} {\bibinfo {author} {\bibfnamefont {M.}~\bibnamefont
  {Della~Morte}}\ and\ \bibinfo {author} {\bibfnamefont {A.}~\bibnamefont
  {J{\"u}ttner}},\ }\href {\doibase 10.1007/JHEP11(2010)154} {\bibfield
  {journal} {\bibinfo  {journal} {JHEP}\ }\textbf {\bibinfo {volume} {11}},\
  \bibinfo {pages} {154} (\bibinfo {year} {2010})},\ \Eprint
  {http://arxiv.org/abs/1009.3783} {arXiv:1009.3783 [hep-lat]}\BibitemShut
  {NoStop}%
\bibitem [{\citenamefont {G{\"u}lpers}\ \emph {et~al.}(2014)\citenamefont
  {G{\"u}lpers}, \citenamefont {Francis}, \citenamefont {J{\"a}ger},
  \citenamefont {Meyer}, \citenamefont {von Hippel},\ and\ \citenamefont
  {Wittig}}]{Francis:2014hoa}%
  \BibitemOpen
  \bibfield  {author} {\bibinfo {author} {\bibfnamefont {V.}~\bibnamefont
  {G{\"u}lpers}}, \bibinfo {author} {\bibfnamefont {A.}~\bibnamefont
  {Francis}}, \bibinfo {author} {\bibfnamefont {B.}~\bibnamefont {J{\"a}ger}},
  \bibinfo {author} {\bibfnamefont {H.}~\bibnamefont {Meyer}}, \bibinfo
  {author} {\bibfnamefont {G.}~\bibnamefont {von Hippel}}, \ and\ \bibinfo
  {author} {\bibfnamefont {H.}~\bibnamefont {Wittig}},\ }\href {\doibase
  10.22323/1.214.0128} {\bibfield  {journal} {\bibinfo  {journal} {PoS}\
  }\textbf {\bibinfo {volume} {LATTICE2014}},\ \bibinfo {pages} {128} (\bibinfo
  {year} {2014})},\ \Eprint {http://arxiv.org/abs/1411.7592} {arXiv:1411.7592
  [hep-lat]}\BibitemShut {NoStop}%
\bibitem [{\citenamefont {Foley}\ \emph {et~al.}(2005)\citenamefont {Foley},
  \citenamefont {Jimmy~Juge}, \citenamefont {O'Cais}, \citenamefont {Peardon},
  \citenamefont {Ryan},\ and\ \citenamefont {Skullerud}}]{Foley:2005ac}%
  \BibitemOpen
  \bibfield  {author} {\bibinfo {author} {\bibfnamefont {J.}~\bibnamefont
  {Foley}}, \bibinfo {author} {\bibfnamefont {K.}~\bibnamefont {Jimmy~Juge}},
  \bibinfo {author} {\bibfnamefont {A.}~\bibnamefont {O'Cais}}, \bibinfo
  {author} {\bibfnamefont {M.}~\bibnamefont {Peardon}}, \bibinfo {author}
  {\bibfnamefont {S.~M.}\ \bibnamefont {Ryan}}, \ and\ \bibinfo {author}
  {\bibfnamefont {J.-I.}\ \bibnamefont {Skullerud}},\ }\href {\doibase
  10.1016/j.cpc.2005.06.008} {\bibfield  {journal} {\bibinfo  {journal}
  {Comput. Phys. Commun.}\ }\textbf {\bibinfo {volume} {172}},\ \bibinfo
  {pages} {145} (\bibinfo {year} {2005})},\ \Eprint
  {http://arxiv.org/abs/hep-lat/0505023} {arXiv:hep-lat/0505023
  [hep-lat]}\BibitemShut {NoStop}%
\bibitem [{\citenamefont {Stathopoulos}\ \emph {et~al.}(2013)\citenamefont
  {Stathopoulos}, \citenamefont {Laeuchli},\ and\ \citenamefont
  {Orginos}}]{Stathopoulos:2013aci}%
  \BibitemOpen
  \bibfield  {author} {\bibinfo {author} {\bibfnamefont {A.}~\bibnamefont
  {Stathopoulos}}, \bibinfo {author} {\bibfnamefont {J.}~\bibnamefont
  {Laeuchli}}, \ and\ \bibinfo {author} {\bibfnamefont {K.}~\bibnamefont
  {Orginos}},\ }\href@noop {} {\  (\bibinfo {year} {2013})},\ \Eprint
  {http://arxiv.org/abs/1302.4018} {arXiv:1302.4018 [hep-lat]}\BibitemShut
  {NoStop}%
\bibitem [{\citenamefont {Bali}\ \emph {et~al.}(2010)\citenamefont {Bali},
  \citenamefont {Collins},\ and\ \citenamefont {Schäfer}}]{Bali:2009hu}%
  \BibitemOpen
  \bibfield  {author} {\bibinfo {author} {\bibfnamefont {G.~S.}\ \bibnamefont
  {Bali}}, \bibinfo {author} {\bibfnamefont {S.}~\bibnamefont {Collins}}, \
  and\ \bibinfo {author} {\bibfnamefont {A.}~\bibnamefont {Schäfer}},\ }\href
  {\doibase 10.1016/j.cpc.2010.05.008} {\bibfield  {journal} {\bibinfo
  {journal} {Comput. Phys. Commun.}\ }\textbf {\bibinfo {volume} {181}},\
  \bibinfo {pages} {1570} (\bibinfo {year} {2010})},\ \Eprint
  {http://arxiv.org/abs/0910.3970} {arXiv:0910.3970 [hep-lat]}\BibitemShut
  {NoStop}%
\bibitem [{\citenamefont {Gülpers}\ \emph {et~al.}(2014)\citenamefont
  {Gülpers}, \citenamefont {von Hippel},\ and\ \citenamefont
  {Wittig}}]{Gulpers:2013uca}%
  \BibitemOpen
  \bibfield  {author} {\bibinfo {author} {\bibfnamefont {V.}~\bibnamefont
  {Gülpers}}, \bibinfo {author} {\bibfnamefont {G.}~\bibnamefont {von
  Hippel}}, \ and\ \bibinfo {author} {\bibfnamefont {H.}~\bibnamefont
  {Wittig}},\ }\href {\doibase 10.1103/PhysRevD.89.094503} {\bibfield
  {journal} {\bibinfo  {journal} {Phys. Rev.}\ }\textbf {\bibinfo {volume}
  {D89}},\ \bibinfo {pages} {094503} (\bibinfo {year} {2014})},\ \Eprint
  {http://arxiv.org/abs/1309.2104} {arXiv:1309.2104 [hep-lat]}\BibitemShut
  {NoStop}%
\bibitem [{\citenamefont {Blum}\ \emph
  {et~al.}(2013{\natexlab{b}})\citenamefont {Blum}, \citenamefont {Izubuchi},\
  and\ \citenamefont {Shintani}}]{Blum:2012uh}%
  \BibitemOpen
  \bibfield  {author} {\bibinfo {author} {\bibfnamefont {T.}~\bibnamefont
  {Blum}}, \bibinfo {author} {\bibfnamefont {T.}~\bibnamefont {Izubuchi}}, \
  and\ \bibinfo {author} {\bibfnamefont {E.}~\bibnamefont {Shintani}},\ }\href
  {\doibase 10.1103/PhysRevD.88.094503} {\bibfield  {journal} {\bibinfo
  {journal} {Phys. Rev.}\ }\textbf {\bibinfo {volume} {D88}},\ \bibinfo {pages}
  {094503} (\bibinfo {year} {2013}{\natexlab{b}})},\ \Eprint
  {http://arxiv.org/abs/1208.4349} {arXiv:1208.4349 [hep-lat]}\BibitemShut
  {NoStop}%
\bibitem [{\citenamefont {Yamamoto}\ \emph {et~al.}(2019)\citenamefont
  {Yamamoto}, \citenamefont {DeTar}, \citenamefont {El-Khadra}, \citenamefont
  {McNeile}, \citenamefont {Van~de Water},\ and\ \citenamefont
  {Vaquero}}]{Yamamoto:2018cqm}%
  \BibitemOpen
  \bibfield  {author} {\bibinfo {author} {\bibfnamefont {S.}~\bibnamefont
  {Yamamoto}}, \bibinfo {author} {\bibfnamefont {C.}~\bibnamefont {DeTar}},
  \bibinfo {author} {\bibfnamefont {A.~X.}\ \bibnamefont {El-Khadra}}, \bibinfo
  {author} {\bibfnamefont {C.}~\bibnamefont {McNeile}}, \bibinfo {author}
  {\bibfnamefont {R.~S.}\ \bibnamefont {Van~de Water}}, \ and\ \bibinfo
  {author} {\bibfnamefont {A.}~\bibnamefont {Vaquero}} (\bibinfo
  {collaboration} {Fermilab Lattice, HPQCD, MILC}),\ }\href {\doibase
  10.22323/1.334.0322} {\bibfield  {journal} {\bibinfo  {journal} {PoS}\
  }\textbf {\bibinfo {volume} {LATTICE2018}},\ \bibinfo {pages} {322} (\bibinfo
  {year} {2019})},\ \Eprint {http://arxiv.org/abs/1811.06058} {arXiv:1811.06058
  [hep-lat]}\BibitemShut {NoStop}%
\bibitem [{\citenamefont {DeTar}\ \emph {et~al.}(2019)\citenamefont {DeTar}
  \emph {et~al.}}]{Davies:2019acq}%
  \BibitemOpen
  \bibfield  {author} {\bibinfo {author} {\bibfnamefont {C.}~\bibnamefont
  {DeTar}} \emph {et~al.} (\bibinfo {collaboration} {Fermilab Lattice, HPQCD,
  MILC}),\ }\href {\doibase 10.22323/1.363.0070} {\bibfield  {journal}
  {\bibinfo  {journal} {PoS}\ }\textbf {\bibinfo {volume} {LATTICE2019}},\
  \bibinfo {pages} {070} (\bibinfo {year} {2019})},\ \Eprint
  {http://arxiv.org/abs/1912.04382} {arXiv:1912.04382 [hep-lat]}\BibitemShut
  {NoStop}%
\bibitem [{\citenamefont {Duncan}\ \emph {et~al.}(1996)\citenamefont {Duncan},
  \citenamefont {Eichten},\ and\ \citenamefont {Thacker}}]{Duncan:1996xy}%
  \BibitemOpen
  \bibfield  {author} {\bibinfo {author} {\bibfnamefont {A.}~\bibnamefont
  {Duncan}}, \bibinfo {author} {\bibfnamefont {E.}~\bibnamefont {Eichten}}, \
  and\ \bibinfo {author} {\bibfnamefont {H.}~\bibnamefont {Thacker}},\ }\href
  {\doibase 10.1103/PhysRevLett.76.3894} {\bibfield  {journal} {\bibinfo
  {journal} {Phys. Rev. Lett.}\ }\textbf {\bibinfo {volume} {76}},\ \bibinfo
  {pages} {3894} (\bibinfo {year} {1996})},\ \Eprint
  {http://arxiv.org/abs/hep-lat/9602005} {arXiv:hep-lat/9602005
  [hep-lat]}\BibitemShut {NoStop}%
\bibitem [{\citenamefont {Boyle}\ \emph {et~al.}(2017)\citenamefont {Boyle},
  \citenamefont {G{\"u}lpers}, \citenamefont {Harrison}, \citenamefont
  {J{\"u}ttner}, \citenamefont {Lehner}, \citenamefont {Portelli},\ and\
  \citenamefont {Sachrajda}}]{Boyle:2017gzv}%
  \BibitemOpen
  \bibfield  {author} {\bibinfo {author} {\bibfnamefont {P.}~\bibnamefont
  {Boyle}}, \bibinfo {author} {\bibfnamefont {V.}~\bibnamefont {G{\"u}lpers}},
  \bibinfo {author} {\bibfnamefont {J.}~\bibnamefont {Harrison}}, \bibinfo
  {author} {\bibfnamefont {A.}~\bibnamefont {J{\"u}ttner}}, \bibinfo {author}
  {\bibfnamefont {C.}~\bibnamefont {Lehner}}, \bibinfo {author} {\bibfnamefont
  {A.}~\bibnamefont {Portelli}}, \ and\ \bibinfo {author} {\bibfnamefont
  {C.~T.}\ \bibnamefont {Sachrajda}},\ }\href {\doibase
  10.1007/JHEP09(2017)153} {\bibfield  {journal} {\bibinfo  {journal} {JHEP}\
  }\textbf {\bibinfo {volume} {09}},\ \bibinfo {pages} {153} (\bibinfo {year}
  {2017})},\ \Eprint {http://arxiv.org/abs/1706.05293} {arXiv:1706.05293
  [hep-lat]}\BibitemShut {NoStop}%
\bibitem [{\citenamefont {Giusti}\ \emph
  {et~al.}(2018{\natexlab{d}})\citenamefont {Giusti}, \citenamefont {Lubicz},
  \citenamefont {Martinelli}, \citenamefont {Sanfilippo},\ and\ \citenamefont
  {Simula}}]{Giusti:2017ier}%
  \BibitemOpen
  \bibfield  {author} {\bibinfo {author} {\bibfnamefont {D.}~\bibnamefont
  {Giusti}}, \bibinfo {author} {\bibfnamefont {V.}~\bibnamefont {Lubicz}},
  \bibinfo {author} {\bibfnamefont {G.}~\bibnamefont {Martinelli}}, \bibinfo
  {author} {\bibfnamefont {F.}~\bibnamefont {Sanfilippo}}, \ and\ \bibinfo
  {author} {\bibfnamefont {S.}~\bibnamefont {Simula}},\ }\href {\doibase
  10.1051/epjconf/201817506006} {\bibfield  {journal} {\bibinfo  {journal} {EPJ
  Web Conf.}\ }\textbf {\bibinfo {volume} {175}},\ \bibinfo {pages} {06006}
  (\bibinfo {year} {2018}{\natexlab{d}})},\ \Eprint
  {http://arxiv.org/abs/1710.06240} {arXiv:1710.06240 [hep-lat]}\BibitemShut
  {NoStop}%
\bibitem [{\citenamefont {Giusti}\ \emph
  {et~al.}(2018{\natexlab{e}})\citenamefont {Giusti}, \citenamefont {Lubicz},
  \citenamefont {Martinelli}, \citenamefont {Sanfilippo}, \citenamefont
  {Simula},\ and\ \citenamefont {Tarantino}}]{Giusti:2018vrc}%
  \BibitemOpen
  \bibfield  {author} {\bibinfo {author} {\bibfnamefont {D.}~\bibnamefont
  {Giusti}}, \bibinfo {author} {\bibfnamefont {V.}~\bibnamefont {Lubicz}},
  \bibinfo {author} {\bibfnamefont {G.}~\bibnamefont {Martinelli}}, \bibinfo
  {author} {\bibfnamefont {F.}~\bibnamefont {Sanfilippo}}, \bibinfo {author}
  {\bibfnamefont {S.}~\bibnamefont {Simula}}, \ and\ \bibinfo {author}
  {\bibfnamefont {C.}~\bibnamefont {Tarantino}},\ }\href {\doibase
  10.22323/1.334.0140} {\bibfield  {journal} {\bibinfo  {journal} {PoS}\
  }\textbf {\bibinfo {volume} {LATTICE2018}},\ \bibinfo {pages} {140} (\bibinfo
  {year} {2018}{\natexlab{e}})},\ \Eprint {http://arxiv.org/abs/1810.05880}
  {arXiv:1810.05880 [hep-lat]}\BibitemShut {NoStop}%
\bibitem [{\citenamefont {Patella}(2017)}]{Patella:2017fgk}%
  \BibitemOpen
  \bibfield  {author} {\bibinfo {author} {\bibfnamefont {A.}~\bibnamefont
  {Patella}},\ }\href {\doibase 10.22323/1.256.0020} {\bibfield  {journal}
  {\bibinfo  {journal} {PoS}\ }\textbf {\bibinfo {volume} {LATTICE2016}},\
  \bibinfo {pages} {020} (\bibinfo {year} {2017})},\ \Eprint
  {http://arxiv.org/abs/1702.03857} {arXiv:1702.03857 [hep-lat]}\BibitemShut
  {NoStop}%
\bibitem [{\citenamefont {Hayakawa}\ and\ \citenamefont
  {Uno}(2008)}]{Hayakawa:2008an}%
  \BibitemOpen
  \bibfield  {author} {\bibinfo {author} {\bibfnamefont {M.}~\bibnamefont
  {Hayakawa}}\ and\ \bibinfo {author} {\bibfnamefont {S.}~\bibnamefont {Uno}},\
  }\href {\doibase 10.1143/PTP.120.413} {\bibfield  {journal} {\bibinfo
  {journal} {Prog. Theor. Phys.}\ }\textbf {\bibinfo {volume} {120}},\ \bibinfo
  {pages} {413} (\bibinfo {year} {2008})},\ \Eprint
  {http://arxiv.org/abs/0804.2044} {arXiv:0804.2044 [hep-ph]}\BibitemShut
  {NoStop}%
\bibitem [{\citenamefont {Westin}\ \emph {et~al.}(2019)\citenamefont {Westin}
  \emph {et~al.}}]{Westin:2019tgc}%
  \BibitemOpen
  \bibfield  {author} {\bibinfo {author} {\bibfnamefont {A.}~\bibnamefont
  {Westin}} \emph {et~al.} (\bibinfo {collaboration} {CSSM/QCDSF/UKQCD}),\
  }\href {\doibase 10.22323/1.334.0136} {\bibfield  {journal} {\bibinfo
  {journal} {PoS}\ }\textbf {\bibinfo {volume} {LATTICE2018}},\ \bibinfo
  {pages} {136} (\bibinfo {year} {2019})},\ \Eprint
  {http://arxiv.org/abs/1902.01518} {arXiv:1902.01518 [hep-lat]}\BibitemShut
  {NoStop}%
\bibitem [{\citenamefont {Di~Carlo}\ \emph
  {et~al.}(2019{\natexlab{b}})\citenamefont {Di~Carlo}, \citenamefont
  {Martinelli}, \citenamefont {Giusti}, \citenamefont {Lubicz}, \citenamefont
  {Sachrajda}, \citenamefont {Sanfilippo}, \citenamefont {Simula},\ and\
  \citenamefont {Tantalo}}]{DiCarlo:2019knp}%
  \BibitemOpen
  \bibfield  {author} {\bibinfo {author} {\bibfnamefont {M.}~\bibnamefont
  {Di~Carlo}}, \bibinfo {author} {\bibfnamefont {G.}~\bibnamefont
  {Martinelli}}, \bibinfo {author} {\bibfnamefont {D.}~\bibnamefont {Giusti}},
  \bibinfo {author} {\bibfnamefont {V.}~\bibnamefont {Lubicz}}, \bibinfo
  {author} {\bibfnamefont {C.~T.}\ \bibnamefont {Sachrajda}}, \bibinfo {author}
  {\bibfnamefont {F.}~\bibnamefont {Sanfilippo}}, \bibinfo {author}
  {\bibfnamefont {S.}~\bibnamefont {Simula}}, \ and\ \bibinfo {author}
  {\bibfnamefont {N.}~\bibnamefont {Tantalo}},\ }\href {\doibase
  10.22323/1.363.0196} {\bibfield  {journal} {\bibinfo  {journal} {PoS}\
  }\textbf {\bibinfo {volume} {LATTICE2019}},\ \bibinfo {pages} {196} (\bibinfo
  {year} {2019}{\natexlab{b}})},\ \Eprint {http://arxiv.org/abs/1911.00938}
  {arXiv:1911.00938 [hep-lat]}\BibitemShut {NoStop}%
\bibitem [{\citenamefont {Martinelli}\ \emph {et~al.}(1995)\citenamefont
  {Martinelli}, \citenamefont {Pittori}, \citenamefont {Sachrajda},
  \citenamefont {Testa},\ and\ \citenamefont {Vladikas}}]{Martinelli:1994ty}%
  \BibitemOpen
  \bibfield  {author} {\bibinfo {author} {\bibfnamefont {G.}~\bibnamefont
  {Martinelli}}, \bibinfo {author} {\bibfnamefont {C.}~\bibnamefont {Pittori}},
  \bibinfo {author} {\bibfnamefont {C.~T.}\ \bibnamefont {Sachrajda}}, \bibinfo
  {author} {\bibfnamefont {M.}~\bibnamefont {Testa}}, \ and\ \bibinfo {author}
  {\bibfnamefont {A.}~\bibnamefont {Vladikas}},\ }\href {\doibase
  10.1016/0550-3213(95)00126-D} {\bibfield  {journal} {\bibinfo  {journal}
  {Nucl. Phys.}\ }\textbf {\bibinfo {volume} {B445}},\ \bibinfo {pages} {81}
  (\bibinfo {year} {1995})},\ \Eprint {http://arxiv.org/abs/hep-lat/9411010}
  {arXiv:hep-lat/9411010 [hep-lat]}\BibitemShut {NoStop}%
\bibitem [{\citenamefont {Bijnens}\ \emph
  {et~al.}(2019{\natexlab{b}})\citenamefont {Bijnens}, \citenamefont
  {Harrison}, \citenamefont {Hermansson-Truedsson}, \citenamefont {Janowski},
  \citenamefont {J{\"u}ttner},\ and\ \citenamefont
  {Portelli}}]{Bijnens:2019ejw}%
  \BibitemOpen
  \bibfield  {author} {\bibinfo {author} {\bibfnamefont {J.}~\bibnamefont
  {Bijnens}}, \bibinfo {author} {\bibfnamefont {J.}~\bibnamefont {Harrison}},
  \bibinfo {author} {\bibfnamefont {N.}~\bibnamefont {Hermansson-Truedsson}},
  \bibinfo {author} {\bibfnamefont {T.}~\bibnamefont {Janowski}}, \bibinfo
  {author} {\bibfnamefont {A.}~\bibnamefont {J{\"u}ttner}}, \ and\ \bibinfo
  {author} {\bibfnamefont {A.}~\bibnamefont {Portelli}},\ }\href {\doibase
  10.1103/PhysRevD.100.014508} {\bibfield  {journal} {\bibinfo  {journal}
  {Phys. Rev.}\ }\textbf {\bibinfo {volume} {D100}},\ \bibinfo {pages} {014508}
  (\bibinfo {year} {2019}{\natexlab{b}})},\ \Eprint
  {http://arxiv.org/abs/1903.10591} {arXiv:1903.10591 [hep-lat]}\BibitemShut
  {NoStop}%
\bibitem [{\citenamefont {Colangelo}\ \emph {et~al.}(2018)\citenamefont
  {Colangelo}, \citenamefont {Hoferichter}, \citenamefont {Procura},\ and\
  \citenamefont {Stoffer}}]{Colangelo:2017urn}%
  \BibitemOpen
  \bibfield  {author} {\bibinfo {author} {\bibfnamefont {G.}~\bibnamefont
  {Colangelo}}, \bibinfo {author} {\bibfnamefont {M.}~\bibnamefont
  {Hoferichter}}, \bibinfo {author} {\bibfnamefont {M.}~\bibnamefont
  {Procura}}, \ and\ \bibinfo {author} {\bibfnamefont {P.}~\bibnamefont
  {Stoffer}},\ }\href {\doibase 10.1051/epjconf/201817501025} {\bibfield
  {journal} {\bibinfo  {journal} {EPJ Web Conf.}\ }\textbf {\bibinfo {volume}
  {175}},\ \bibinfo {pages} {01025} (\bibinfo {year} {2018})},\ \Eprint
  {http://arxiv.org/abs/1711.00281} {arXiv:1711.00281 [hep-ph]}\BibitemShut
  {NoStop}%
\bibitem [{\citenamefont {Giusti}\ \emph
  {et~al.}(2019{\natexlab{c}})\citenamefont {Giusti}, \citenamefont {Lubicz},
  \citenamefont {Martinelli}, \citenamefont {Sachrajda}, \citenamefont
  {Sanfilippo}, \citenamefont {Simula},\ and\ \citenamefont
  {Tantalo}}]{Giusti:2018guw}%
  \BibitemOpen
  \bibfield  {author} {\bibinfo {author} {\bibfnamefont {D.}~\bibnamefont
  {Giusti}}, \bibinfo {author} {\bibfnamefont {V.}~\bibnamefont {Lubicz}},
  \bibinfo {author} {\bibfnamefont {G.}~\bibnamefont {Martinelli}}, \bibinfo
  {author} {\bibfnamefont {C.}~\bibnamefont {Sachrajda}}, \bibinfo {author}
  {\bibfnamefont {F.}~\bibnamefont {Sanfilippo}}, \bibinfo {author}
  {\bibfnamefont {S.}~\bibnamefont {Simula}}, \ and\ \bibinfo {author}
  {\bibfnamefont {N.}~\bibnamefont {Tantalo}},\ }\href {\doibase
  10.22323/1.334.0266} {\bibfield  {journal} {\bibinfo  {journal} {PoS}\
  }\textbf {\bibinfo {volume} {LATTICE2018}},\ \bibinfo {pages} {266} (\bibinfo
  {year} {2019}{\natexlab{c}})},\ \Eprint {http://arxiv.org/abs/1811.06364}
  {arXiv:1811.06364 [hep-lat]}\BibitemShut {NoStop}%
\bibitem [{\citenamefont {Miura}(2019)}]{Miura:2019xtd}%
  \BibitemOpen
  \bibfield  {author} {\bibinfo {author} {\bibfnamefont {K.}~\bibnamefont
  {Miura}},\ }\href {\doibase 10.22323/1.334.0010} {\bibfield  {journal}
  {\bibinfo  {journal} {PoS}\ }\textbf {\bibinfo {volume} {LATTICE2018}},\
  \bibinfo {pages} {010} (\bibinfo {year} {2019})},\ \Eprint
  {http://arxiv.org/abs/1901.09052} {arXiv:1901.09052 [hep-lat]}\BibitemShut
  {NoStop}%
\bibitem [{\citenamefont {Benayoun}\ \emph {et~al.}(2016)\citenamefont
  {Benayoun}, \citenamefont {David}, \citenamefont {DelBuono},\ and\
  \citenamefont {Jegerlehner}}]{Benayoun:2016krn}%
  \BibitemOpen
  \bibfield  {author} {\bibinfo {author} {\bibfnamefont {M.}~\bibnamefont
  {Benayoun}}, \bibinfo {author} {\bibfnamefont {P.}~\bibnamefont {David}},
  \bibinfo {author} {\bibfnamefont {L.}~\bibnamefont {DelBuono}}, \ and\
  \bibinfo {author} {\bibfnamefont {F.}~\bibnamefont {Jegerlehner}},\
  }\href@noop {} {\  (\bibinfo {year} {2016})},\ \Eprint
  {http://arxiv.org/abs/1605.04474} {arXiv:1605.04474 [hep-ph]}\BibitemShut
  {NoStop}%
\bibitem [{\citenamefont {Lehner}()}]{CL:private}%
  \BibitemOpen
  \bibfield  {author} {\bibinfo {author} {\bibfnamefont {C.}~\bibnamefont
  {Lehner}},\ }\href@noop {} {}\bibinfo {howpublished} {{private
  communication}}\BibitemShut {NoStop}%
\bibitem [{\citenamefont {Lehner}\ and\ \citenamefont
  {Meyer}(2020)}]{Lehner:2020crt}%
  \BibitemOpen
  \bibfield  {author} {\bibinfo {author} {\bibfnamefont {C.}~\bibnamefont
  {Lehner}}\ and\ \bibinfo {author} {\bibfnamefont {A.~S.}\ \bibnamefont
  {Meyer}},\ }\href {\doibase 10.1103/PhysRevD.101.074515} {\bibfield
  {journal} {\bibinfo  {journal} {Phys. Rev. D}\ }\textbf {\bibinfo {volume}
  {101}},\ \bibinfo {pages} {074515} (\bibinfo {year} {2020})},\ \Eprint
  {http://arxiv.org/abs/2003.04177} {arXiv:2003.04177 [hep-lat]}\BibitemShut
  {NoStop}%
\bibitem [{\citenamefont {Giusti}\ and\ \citenamefont
  {Simula}(2019{\natexlab{b}})}]{Giusti:2019hoy}%
  \BibitemOpen
  \bibfield  {author} {\bibinfo {author} {\bibfnamefont {D.}~\bibnamefont
  {Giusti}}\ and\ \bibinfo {author} {\bibfnamefont {S.}~\bibnamefont
  {Simula}},\ }\href@noop {} {\  (\bibinfo {year} {2019}{\natexlab{b}})},\
  \Eprint {http://arxiv.org/abs/1910.00611} {arXiv:1910.00611
  [hep-lat]}\BibitemShut {NoStop}%
\bibitem [{\citenamefont {Bruno}\ \emph {et~al.}(2018)\citenamefont {Bruno},
  \citenamefont {Izubuchi}, \citenamefont {Lehner},\ and\ \citenamefont
  {Meyer}}]{Bruno:2018ono}%
  \BibitemOpen
  \bibfield  {author} {\bibinfo {author} {\bibfnamefont {M.}~\bibnamefont
  {Bruno}}, \bibinfo {author} {\bibfnamefont {T.}~\bibnamefont {Izubuchi}},
  \bibinfo {author} {\bibfnamefont {C.}~\bibnamefont {Lehner}}, \ and\ \bibinfo
  {author} {\bibfnamefont {A.}~\bibnamefont {Meyer}},\ }\href {\doibase
  10.22323/1.334.0135} {\bibfield  {journal} {\bibinfo  {journal} {PoS}\
  }\textbf {\bibinfo {volume} {LATTICE2018}},\ \bibinfo {pages} {135} (\bibinfo
  {year} {2018})},\ \Eprint {http://arxiv.org/abs/1811.00508} {arXiv:1811.00508
  [hep-lat]}\BibitemShut {NoStop}%
\bibitem [{\citenamefont {Jegerlehner}(2011)}]{Jegerlehner:2011mw}%
  \BibitemOpen
  \bibfield  {author} {\bibinfo {author} {\bibfnamefont {F.}~\bibnamefont
  {Jegerlehner}},\ }\href {\doibase 10.1393/ncc/i2011-11011-0} {\bibfield
  {journal} {\bibinfo  {journal} {Nuovo Cim.}\ }\textbf {\bibinfo {volume}
  {C034S1}},\ \bibinfo {pages} {31} (\bibinfo {year} {2011})},\ \Eprint
  {http://arxiv.org/abs/1107.4683} {arXiv:1107.4683 [hep-ph]}\BibitemShut
  {NoStop}%
\bibitem [{\citenamefont {Baak}\ \emph {et~al.}(2014)\citenamefont {Baak},
  \citenamefont {C{\'u}th}, \citenamefont {Haller}, \citenamefont {Hoecker},
  \citenamefont {Kogler}, \citenamefont {M{\"o}nig}, \citenamefont {Schott},\
  and\ \citenamefont {Stelzer}}]{Baak:2014ora}%
  \BibitemOpen
  \bibfield  {author} {\bibinfo {author} {\bibfnamefont {M.}~\bibnamefont
  {Baak}}, \bibinfo {author} {\bibfnamefont {J.}~\bibnamefont {C{\'u}th}},
  \bibinfo {author} {\bibfnamefont {J.}~\bibnamefont {Haller}}, \bibinfo
  {author} {\bibfnamefont {A.}~\bibnamefont {Hoecker}}, \bibinfo {author}
  {\bibfnamefont {R.}~\bibnamefont {Kogler}}, \bibinfo {author} {\bibfnamefont
  {K.}~\bibnamefont {M{\"o}nig}}, \bibinfo {author} {\bibfnamefont
  {M.}~\bibnamefont {Schott}}, \ and\ \bibinfo {author} {\bibfnamefont
  {J.}~\bibnamefont {Stelzer}} (\bibinfo {collaboration} {Gfitter Group}),\
  }\href {\doibase 10.1140/epjc/s10052-014-3046-5} {\bibfield  {journal}
  {\bibinfo  {journal} {Eur. Phys. J.}\ }\textbf {\bibinfo {volume} {C74}},\
  \bibinfo {pages} {3046} (\bibinfo {year} {2014})},\ \Eprint
  {http://arxiv.org/abs/1407.3792} {arXiv:1407.3792 [hep-ph]}\BibitemShut
  {NoStop}%
\bibitem [{\citenamefont {Eidelman}\ \emph {et~al.}(1999)\citenamefont
  {Eidelman}, \citenamefont {Jegerlehner}, \citenamefont {Kataev},\ and\
  \citenamefont {Veretin}}]{Eidelman:1998vc}%
  \BibitemOpen
  \bibfield  {author} {\bibinfo {author} {\bibfnamefont {S.}~\bibnamefont
  {Eidelman}}, \bibinfo {author} {\bibfnamefont {F.}~\bibnamefont
  {Jegerlehner}}, \bibinfo {author} {\bibfnamefont {A.~L.}\ \bibnamefont
  {Kataev}}, \ and\ \bibinfo {author} {\bibfnamefont {O.}~\bibnamefont
  {Veretin}},\ }\href {\doibase 10.1016/S0370-2693(99)00389-5} {\bibfield
  {journal} {\bibinfo  {journal} {Phys. Lett.}\ }\textbf {\bibinfo {volume}
  {B454}},\ \bibinfo {pages} {369} (\bibinfo {year} {1999})},\ \Eprint
  {http://arxiv.org/abs/hep-ph/9812521} {arXiv:hep-ph/9812521
  [hep-ph]}\BibitemShut {NoStop}%
\bibitem [{\citenamefont
  {Jegerlehner}(2008{\natexlab{a}})}]{Jegerlehner:2008rs}%
  \BibitemOpen
  \bibfield  {author} {\bibinfo {author} {\bibfnamefont {F.}~\bibnamefont
  {Jegerlehner}},\ }\href {\doibase 10.1016/j.nuclphysbps.2008.09.010}
  {\bibfield  {journal} {\bibinfo  {journal} {Nucl. Phys. Proc. Suppl.}\
  }\textbf {\bibinfo {volume} {181-182}},\ \bibinfo {pages} {135} (\bibinfo
  {year} {2008}{\natexlab{a}})},\ \Eprint {http://arxiv.org/abs/0807.4206}
  {arXiv:0807.4206 [hep-ph]}\BibitemShut {NoStop}%
\bibitem [{\citenamefont {Burger}\ \emph {et~al.}(2015)\citenamefont {Burger},
  \citenamefont {Jansen}, \citenamefont {Petschlies},\ and\ \citenamefont
  {Pientka}}]{Burger:2015lqa}%
  \BibitemOpen
  \bibfield  {author} {\bibinfo {author} {\bibfnamefont {F.}~\bibnamefont
  {Burger}}, \bibinfo {author} {\bibfnamefont {K.}~\bibnamefont {Jansen}},
  \bibinfo {author} {\bibfnamefont {M.}~\bibnamefont {Petschlies}}, \ and\
  \bibinfo {author} {\bibfnamefont {G.}~\bibnamefont {Pientka}},\ }\href
  {\doibase 10.1007/JHEP11(2015)215} {\bibfield  {journal} {\bibinfo  {journal}
  {JHEP}\ }\textbf {\bibinfo {volume} {11}},\ \bibinfo {pages} {215} (\bibinfo
  {year} {2015})},\ \Eprint {http://arxiv.org/abs/1505.03283} {arXiv:1505.03283
  [hep-lat]}\BibitemShut {NoStop}%
\bibitem [{\citenamefont {Francis}\ \emph {et~al.}(2014)\citenamefont
  {Francis}, \citenamefont {Herdo{\'\i}za}, \citenamefont {Horch},
  \citenamefont {J{\"a}ger}, \citenamefont {Meyer},\ and\ \citenamefont
  {Wittig}}]{Francis:2014yga}%
  \BibitemOpen
  \bibfield  {author} {\bibinfo {author} {\bibfnamefont {A.}~\bibnamefont
  {Francis}}, \bibinfo {author} {\bibfnamefont {G.}~\bibnamefont
  {Herdo{\'\i}za}}, \bibinfo {author} {\bibfnamefont {H.}~\bibnamefont
  {Horch}}, \bibinfo {author} {\bibfnamefont {B.}~\bibnamefont {J{\"a}ger}},
  \bibinfo {author} {\bibfnamefont {H.~B.}\ \bibnamefont {Meyer}}, \ and\
  \bibinfo {author} {\bibfnamefont {H.}~\bibnamefont {Wittig}},\ }\href
  {\doibase 10.22323/1.214.0163} {\bibfield  {journal} {\bibinfo  {journal}
  {PoS}\ }\textbf {\bibinfo {volume} {LATTICE2014}},\ \bibinfo {pages} {163}
  (\bibinfo {year} {2014})},\ \Eprint {http://arxiv.org/abs/1412.6934}
  {arXiv:1412.6934 [hep-lat]}\BibitemShut {NoStop}%
\bibitem [{\citenamefont {Francis}\ \emph {et~al.}(2015)\citenamefont
  {Francis}, \citenamefont {G{\"u}lpers}, \citenamefont {Herdo{\'\i}za},
  \citenamefont {Horch}, \citenamefont {J{\"a}ger}, \citenamefont {Meyer},\
  and\ \citenamefont {Wittig}}]{Francis:2015grz}%
  \BibitemOpen
  \bibfield  {author} {\bibinfo {author} {\bibfnamefont {A.}~\bibnamefont
  {Francis}}, \bibinfo {author} {\bibfnamefont {V.}~\bibnamefont
  {G{\"u}lpers}}, \bibinfo {author} {\bibfnamefont {G.}~\bibnamefont
  {Herdo{\'\i}za}}, \bibinfo {author} {\bibfnamefont {H.}~\bibnamefont
  {Horch}}, \bibinfo {author} {\bibfnamefont {B.}~\bibnamefont {J{\"a}ger}},
  \bibinfo {author} {\bibfnamefont {H.~B.}\ \bibnamefont {Meyer}}, \ and\
  \bibinfo {author} {\bibfnamefont {H.}~\bibnamefont {Wittig}},\ }\href
  {\doibase 10.22323/1.251.0110} {\bibfield  {journal} {\bibinfo  {journal}
  {PoS}\ }\textbf {\bibinfo {volume} {LATTICE2015}},\ \bibinfo {pages} {110}
  (\bibinfo {year} {2015})},\ \Eprint {http://arxiv.org/abs/1511.04751}
  {arXiv:1511.04751 [hep-lat]}\BibitemShut {NoStop}%
\bibitem [{\citenamefont {C{\`e}}\ \emph {et~al.}(2019)\citenamefont {C{\`e}},
  \citenamefont {San~Jos{\'e}}, \citenamefont {G{\'e}rardin}, \citenamefont
  {Meyer}, \citenamefont {Miura}, \citenamefont {Ottnad}, \citenamefont
  {Risch}, \citenamefont {Wilhelm},\ and\ \citenamefont {Wittig}}]{Ce:2019imp}%
  \BibitemOpen
  \bibfield  {author} {\bibinfo {author} {\bibfnamefont {M.}~\bibnamefont
  {C{\`e}}}, \bibinfo {author} {\bibfnamefont {T.}~\bibnamefont
  {San~Jos{\'e}}}, \bibinfo {author} {\bibfnamefont {A.}~\bibnamefont
  {G{\'e}rardin}}, \bibinfo {author} {\bibfnamefont {H.~B.}\ \bibnamefont
  {Meyer}}, \bibinfo {author} {\bibfnamefont {K.}~\bibnamefont {Miura}},
  \bibinfo {author} {\bibfnamefont {K.}~\bibnamefont {Ottnad}}, \bibinfo
  {author} {\bibfnamefont {A.}~\bibnamefont {Risch}}, \bibinfo {author}
  {\bibfnamefont {J.}~\bibnamefont {Wilhelm}}, \ and\ \bibinfo {author}
  {\bibfnamefont {H.}~\bibnamefont {Wittig}},\ }\href {\doibase
  10.22323/1.363.0010} {\bibfield  {journal} {\bibinfo  {journal} {PoS}\
  }\textbf {\bibinfo {volume} {LATTICE2019}},\ \bibinfo {pages} {010} (\bibinfo
  {year} {2019})},\ \Eprint {http://arxiv.org/abs/1910.09525} {arXiv:1910.09525
  [hep-lat]}\BibitemShut {NoStop}%
\bibitem [{\citenamefont {Passera}\ \emph {et~al.}(2008)\citenamefont
  {Passera}, \citenamefont {Marciano},\ and\ \citenamefont
  {Sirlin}}]{Passera:2008jk}%
  \BibitemOpen
  \bibfield  {author} {\bibinfo {author} {\bibfnamefont {M.}~\bibnamefont
  {Passera}}, \bibinfo {author} {\bibfnamefont {W.~J.}\ \bibnamefont
  {Marciano}}, \ and\ \bibinfo {author} {\bibfnamefont {A.}~\bibnamefont
  {Sirlin}},\ }\href {\doibase 10.1103/PhysRevD.78.013009} {\bibfield
  {journal} {\bibinfo  {journal} {Phys. Rev.}\ }\textbf {\bibinfo {volume}
  {D78}},\ \bibinfo {pages} {013009} (\bibinfo {year} {2008})},\ \Eprint
  {http://arxiv.org/abs/0804.1142} {arXiv:0804.1142 [hep-ph]}\BibitemShut
  {NoStop}%
\bibitem [{\citenamefont {Crivellin}\ \emph {et~al.}(2020)\citenamefont
  {Crivellin}, \citenamefont {Hoferichter}, \citenamefont {Manzari},\ and\
  \citenamefont {Montull}}]{Crivellin:2020zul}%
  \BibitemOpen
  \bibfield  {author} {\bibinfo {author} {\bibfnamefont {A.}~\bibnamefont
  {Crivellin}}, \bibinfo {author} {\bibfnamefont {M.}~\bibnamefont
  {Hoferichter}}, \bibinfo {author} {\bibfnamefont {C.~A.}\ \bibnamefont
  {Manzari}}, \ and\ \bibinfo {author} {\bibfnamefont {M.}~\bibnamefont
  {Montull}},\ }\href {\doibase 10.1103/PhysRevLett.125.091801} {\bibfield
  {journal} {\bibinfo  {journal} {Phys. Rev. Lett.}\ }\textbf {\bibinfo
  {volume} {125}},\ \bibinfo {pages} {091801} (\bibinfo {year} {2020})},\
  \Eprint {http://arxiv.org/abs/2003.04886} {arXiv:2003.04886
  [hep-ph]}\BibitemShut {NoStop}%
\bibitem [{\citenamefont {Haller}\ \emph {et~al.}(2018)\citenamefont {Haller},
  \citenamefont {Hoecker}, \citenamefont {Kogler}, \citenamefont {M{\"o}nig},
  \citenamefont {Peiffer},\ and\ \citenamefont {Stelzer}}]{Haller:2018nnx}%
  \BibitemOpen
  \bibfield  {author} {\bibinfo {author} {\bibfnamefont {J.}~\bibnamefont
  {Haller}}, \bibinfo {author} {\bibfnamefont {A.}~\bibnamefont {Hoecker}},
  \bibinfo {author} {\bibfnamefont {R.}~\bibnamefont {Kogler}}, \bibinfo
  {author} {\bibfnamefont {K.}~\bibnamefont {M{\"o}nig}}, \bibinfo {author}
  {\bibfnamefont {T.}~\bibnamefont {Peiffer}}, \ and\ \bibinfo {author}
  {\bibfnamefont {J.}~\bibnamefont {Stelzer}},\ }\href {\doibase
  10.1140/epjc/s10052-018-6131-3} {\bibfield  {journal} {\bibinfo  {journal}
  {Eur. Phys. J.}\ }\textbf {\bibinfo {volume} {C78}},\ \bibinfo {pages} {675}
  (\bibinfo {year} {2018})},\ \Eprint {http://arxiv.org/abs/1803.01853}
  {arXiv:1803.01853 [hep-ph]}\BibitemShut {NoStop}%
\bibitem [{\citenamefont {De~Blas}\ \emph {et~al.}(2020)\citenamefont {De~Blas}
  \emph {et~al.}}]{deBlas:2019okz}%
  \BibitemOpen
  \bibfield  {author} {\bibinfo {author} {\bibfnamefont {J.}~\bibnamefont
  {De~Blas}} \emph {et~al.},\ }\href {\doibase 10.1140/epjc/s10052-020-7904-z}
  {\bibfield  {journal} {\bibinfo  {journal} {Eur. Phys. J. C}\ }\textbf
  {\bibinfo {volume} {80}},\ \bibinfo {pages} {456} (\bibinfo {year} {2020})},\
  \Eprint {http://arxiv.org/abs/1910.14012} {arXiv:1910.14012
  [hep-ph]}\BibitemShut {NoStop}%
\bibitem [{\citenamefont {Erler}\ and\ \citenamefont
  {Schott}(2019)}]{Erler:2019hds}%
  \BibitemOpen
  \bibfield  {author} {\bibinfo {author} {\bibfnamefont {J.}~\bibnamefont
  {Erler}}\ and\ \bibinfo {author} {\bibfnamefont {M.}~\bibnamefont {Schott}},\
  }\href {\doibase 10.1016/j.ppnp.2019.02.007} {\bibfield  {journal} {\bibinfo
  {journal} {Prog. Part. Nucl. Phys.}\ }\textbf {\bibinfo {volume} {106}},\
  \bibinfo {pages} {68} (\bibinfo {year} {2019})},\ \Eprint
  {http://arxiv.org/abs/1902.05142} {arXiv:1902.05142 [hep-ph]}\BibitemShut
  {NoStop}%
\bibitem [{\citenamefont {Erler}\ and\ \citenamefont
  {Ferro-Hern{\'a}ndez}(2018)}]{Erler:2017knj}%
  \BibitemOpen
  \bibfield  {author} {\bibinfo {author} {\bibfnamefont {J.}~\bibnamefont
  {Erler}}\ and\ \bibinfo {author} {\bibfnamefont {R.}~\bibnamefont
  {Ferro-Hern{\'a}ndez}},\ }\href {\doibase 10.1007/JHEP03(2018)196} {\bibfield
   {journal} {\bibinfo  {journal} {JHEP}\ }\textbf {\bibinfo {volume} {03}},\
  \bibinfo {pages} {196} (\bibinfo {year} {2018})},\ \Eprint
  {http://arxiv.org/abs/1712.09146} {arXiv:1712.09146 [hep-ph]}\BibitemShut
  {NoStop}%
\bibitem [{\citenamefont {G{\"u}lpers}\ \emph {et~al.}(2016)\citenamefont
  {G{\"u}lpers}, \citenamefont {Meyer}, \citenamefont {von Hippel},\ and\
  \citenamefont {Wittig}}]{Guelpers:2015nfb}%
  \BibitemOpen
  \bibfield  {author} {\bibinfo {author} {\bibfnamefont {V.}~\bibnamefont
  {G{\"u}lpers}}, \bibinfo {author} {\bibfnamefont {H.}~\bibnamefont {Meyer}},
  \bibinfo {author} {\bibfnamefont {G.}~\bibnamefont {von Hippel}}, \ and\
  \bibinfo {author} {\bibfnamefont {H.}~\bibnamefont {Wittig}},\ }\href
  {\doibase 10.22323/1.251.0263} {\bibfield  {journal} {\bibinfo  {journal}
  {PoS}\ }\textbf {\bibinfo {volume} {LATTICE2015}},\ \bibinfo {pages} {263}
  (\bibinfo {year} {2016})}\BibitemShut {NoStop}%
\bibitem [{\citenamefont {C{\`e}}\ \emph {et~al.}(2018)\citenamefont {C{\`e}},
  \citenamefont {G{\'e}rardin}, \citenamefont {Ottnad},\ and\ \citenamefont
  {Meyer}}]{Ce:2018ziv}%
  \BibitemOpen
  \bibfield  {author} {\bibinfo {author} {\bibfnamefont {M.}~\bibnamefont
  {C{\`e}}}, \bibinfo {author} {\bibfnamefont {A.}~\bibnamefont
  {G{\'e}rardin}}, \bibinfo {author} {\bibfnamefont {K.}~\bibnamefont
  {Ottnad}}, \ and\ \bibinfo {author} {\bibfnamefont {H.~B.}\ \bibnamefont
  {Meyer}},\ }\href {\doibase 10.22323/1.334.0137} {\bibfield  {journal}
  {\bibinfo  {journal} {PoS}\ }\textbf {\bibinfo {volume} {LATTICE2018}},\
  \bibinfo {pages} {137} (\bibinfo {year} {2018})},\ \Eprint
  {http://arxiv.org/abs/1811.08669} {arXiv:1811.08669 [hep-lat]}\BibitemShut
  {NoStop}%
\bibitem [{\citenamefont {Giusti}\ and\ \citenamefont
  {Simula}(2020)}]{Giusti:2020efo}%
  \BibitemOpen
  \bibfield  {author} {\bibinfo {author} {\bibfnamefont {D.}~\bibnamefont
  {Giusti}}\ and\ \bibinfo {author} {\bibfnamefont {S.}~\bibnamefont
  {Simula}},\ }\href {\doibase 10.1103/PhysRevD.102.054503} {\bibfield
  {journal} {\bibinfo  {journal} {Phys. Rev. D}\ }\textbf {\bibinfo {volume}
  {102}},\ \bibinfo {pages} {054503} (\bibinfo {year} {2020})},\ \Eprint
  {http://arxiv.org/abs/2003.12086} {arXiv:2003.12086 [hep-lat]}\BibitemShut
  {NoStop}%
\bibitem [{\citenamefont {Lehner}\ \emph {et~al.}(2019)\citenamefont {Lehner}
  \emph {et~al.}}]{Lehner:2019wvv}%
  \BibitemOpen
  \bibfield  {author} {\bibinfo {author} {\bibfnamefont {C.}~\bibnamefont
  {Lehner}} \emph {et~al.} (\bibinfo {collaboration} {USQCD}),\ }\href
  {\doibase 10.1140/epja/i2019-12891-2} {\bibfield  {journal} {\bibinfo
  {journal} {Eur. Phys. J. A}\ }\textbf {\bibinfo {volume} {55}},\ \bibinfo
  {pages} {195} (\bibinfo {year} {2019})},\ \Eprint
  {http://arxiv.org/abs/1904.09479} {arXiv:1904.09479 [hep-lat]}\BibitemShut
  {NoStop}%
\bibitem [{\citenamefont {Giusti}\ \emph {et~al.}(2004)\citenamefont {Giusti},
  \citenamefont {Hernandez}, \citenamefont {Laine}, \citenamefont {Weisz},\
  and\ \citenamefont {Wittig}}]{Giusti:2004yp}%
  \BibitemOpen
  \bibfield  {author} {\bibinfo {author} {\bibfnamefont {L.}~\bibnamefont
  {Giusti}}, \bibinfo {author} {\bibfnamefont {P.}~\bibnamefont {Hernandez}},
  \bibinfo {author} {\bibfnamefont {M.}~\bibnamefont {Laine}}, \bibinfo
  {author} {\bibfnamefont {P.}~\bibnamefont {Weisz}}, \ and\ \bibinfo {author}
  {\bibfnamefont {H.}~\bibnamefont {Wittig}},\ }\href {\doibase
  10.1088/1126-6708/2004/04/013} {\bibfield  {journal} {\bibinfo  {journal}
  {JHEP}\ }\textbf {\bibinfo {volume} {04}},\ \bibinfo {pages} {013} (\bibinfo
  {year} {2004})},\ \Eprint {http://arxiv.org/abs/hep-lat/0402002}
  {arXiv:hep-lat/0402002}\BibitemShut {NoStop}%
\bibitem [{\citenamefont {DeGrand}\ and\ \citenamefont
  {Schaefer}(2004)}]{DeGrand:2004qw}%
  \BibitemOpen
  \bibfield  {author} {\bibinfo {author} {\bibfnamefont {T.~A.}\ \bibnamefont
  {DeGrand}}\ and\ \bibinfo {author} {\bibfnamefont {S.}~\bibnamefont
  {Schaefer}},\ }\href {\doibase 10.1016/j.cpc.2004.02.006} {\bibfield
  {journal} {\bibinfo  {journal} {Comput. Phys. Commun.}\ }\textbf {\bibinfo
  {volume} {159}},\ \bibinfo {pages} {185} (\bibinfo {year} {2004})},\ \Eprint
  {http://arxiv.org/abs/hep-lat/0401011} {arXiv:hep-lat/0401011}\BibitemShut
  {NoStop}%
\bibitem [{\citenamefont {Neff}\ \emph {et~al.}(2001)\citenamefont {Neff},
  \citenamefont {Eicker}, \citenamefont {Lippert}, \citenamefont {Negele},\
  and\ \citenamefont {Schilling}}]{Neff:2001zr}%
  \BibitemOpen
  \bibfield  {author} {\bibinfo {author} {\bibfnamefont {H.}~\bibnamefont
  {Neff}}, \bibinfo {author} {\bibfnamefont {N.}~\bibnamefont {Eicker}},
  \bibinfo {author} {\bibfnamefont {T.}~\bibnamefont {Lippert}}, \bibinfo
  {author} {\bibfnamefont {J.~W.}\ \bibnamefont {Negele}}, \ and\ \bibinfo
  {author} {\bibfnamefont {K.}~\bibnamefont {Schilling}},\ }\href {\doibase
  10.1103/PhysRevD.64.114509} {\bibfield  {journal} {\bibinfo  {journal} {Phys.
  Rev. D}\ }\textbf {\bibinfo {volume} {64}},\ \bibinfo {pages} {114509}
  (\bibinfo {year} {2001})},\ \Eprint {http://arxiv.org/abs/hep-lat/0106016}
  {arXiv:hep-lat/0106016}\BibitemShut {NoStop}%
\bibitem [{\citenamefont {Erben}\ \emph {et~al.}(2018)\citenamefont {Erben},
  \citenamefont {Green}, \citenamefont {Mohler},\ and\ \citenamefont
  {Wittig}}]{Erben:2017hvr}%
  \BibitemOpen
  \bibfield  {author} {\bibinfo {author} {\bibfnamefont {F.}~\bibnamefont
  {Erben}}, \bibinfo {author} {\bibfnamefont {J.}~\bibnamefont {Green}},
  \bibinfo {author} {\bibfnamefont {D.}~\bibnamefont {Mohler}}, \ and\ \bibinfo
  {author} {\bibfnamefont {H.}~\bibnamefont {Wittig}},\ }\href {\doibase
  10.1051/epjconf/201817505027} {\bibfield  {journal} {\bibinfo  {journal} {EPJ
  Web Conf.}\ }\textbf {\bibinfo {volume} {175}},\ \bibinfo {pages} {05027}
  (\bibinfo {year} {2018})},\ \Eprint {http://arxiv.org/abs/1710.03529}
  {arXiv:1710.03529 [hep-lat]}\BibitemShut {NoStop}%
\bibitem [{\citenamefont {Nyffeler}(2019)}]{Nyffeler:2017ohp}%
  \BibitemOpen
  \bibfield  {author} {\bibinfo {author} {\bibfnamefont {A.}~\bibnamefont
  {Nyffeler}},\ }\href {\doibase 10.1051/epjconf/201921801001} {\bibfield
  {journal} {\bibinfo  {journal} {EPJ Web Conf.}\ }\textbf {\bibinfo {volume}
  {218}},\ \bibinfo {pages} {01001} (\bibinfo {year} {2019})},\ \Eprint
  {http://arxiv.org/abs/1710.09742} {arXiv:1710.09742 [hep-ph]}\BibitemShut
  {NoStop}%
\bibitem [{\citenamefont {Melnikov}(2001)}]{Melnikov:2001uw}%
  \BibitemOpen
  \bibfield  {author} {\bibinfo {author} {\bibfnamefont {K.}~\bibnamefont
  {Melnikov}},\ }\href {\doibase 10.1142/S0217751X01005602} {\bibfield
  {journal} {\bibinfo  {journal} {Int. J. Mod. Phys.}\ }\textbf {\bibinfo
  {volume} {A16}},\ \bibinfo {pages} {4591} (\bibinfo {year} {2001})},\ \Eprint
  {http://arxiv.org/abs/hep-ph/0105267} {arXiv:hep-ph/0105267
  [hep-ph]}\BibitemShut {NoStop}%
\bibitem [{\citenamefont {Knecht}\ and\ \citenamefont
  {Nyffeler}(2002)}]{Knecht:2001qf}%
  \BibitemOpen
  \bibfield  {author} {\bibinfo {author} {\bibfnamefont {M.}~\bibnamefont
  {Knecht}}\ and\ \bibinfo {author} {\bibfnamefont {A.}~\bibnamefont
  {Nyffeler}},\ }\href {\doibase 10.1103/PhysRevD.65.073034} {\bibfield
  {journal} {\bibinfo  {journal} {Phys. Rev.}\ }\textbf {\bibinfo {volume}
  {D65}},\ \bibinfo {pages} {073034} (\bibinfo {year} {2002})},\ \Eprint
  {http://arxiv.org/abs/hep-ph/0111058} {arXiv:hep-ph/0111058
  [hep-ph]}\BibitemShut {NoStop}%
\bibitem [{\citenamefont {Knecht}\ \emph
  {et~al.}(2002{\natexlab{a}})\citenamefont {Knecht}, \citenamefont {Nyffeler},
  \citenamefont {Perrottet},\ and\ \citenamefont {de~Rafael}}]{Knecht:2001qg}%
  \BibitemOpen
  \bibfield  {author} {\bibinfo {author} {\bibfnamefont {M.}~\bibnamefont
  {Knecht}}, \bibinfo {author} {\bibfnamefont {A.}~\bibnamefont {Nyffeler}},
  \bibinfo {author} {\bibfnamefont {M.}~\bibnamefont {Perrottet}}, \ and\
  \bibinfo {author} {\bibfnamefont {E.}~\bibnamefont {de~Rafael}},\ }\href
  {\doibase 10.1103/PhysRevLett.88.071802} {\bibfield  {journal} {\bibinfo
  {journal} {Phys. Rev. Lett.}\ }\textbf {\bibinfo {volume} {88}},\ \bibinfo
  {pages} {071802} (\bibinfo {year} {2002}{\natexlab{a}})},\ \Eprint
  {http://arxiv.org/abs/hep-ph/0111059} {arXiv:hep-ph/0111059
  [hep-ph]}\BibitemShut {NoStop}%
\bibitem [{\citenamefont {Prades}\ \emph {et~al.}(2009)\citenamefont {Prades},
  \citenamefont {de~Rafael},\ and\ \citenamefont {Vainshtein}}]{Prades:2009tw}%
  \BibitemOpen
  \bibfield  {author} {\bibinfo {author} {\bibfnamefont {J.}~\bibnamefont
  {Prades}}, \bibinfo {author} {\bibfnamefont {E.}~\bibnamefont {de~Rafael}}, \
  and\ \bibinfo {author} {\bibfnamefont {A.}~\bibnamefont {Vainshtein}},\
  }\href {\doibase 10.1142/9789814271844_0009} {\bibfield  {journal} {\bibinfo
  {journal} {Adv. Ser. Direct. High Energy Phys.}\ }\textbf {\bibinfo {volume}
  {20}},\ \bibinfo {pages} {303} (\bibinfo {year} {2009})},\ \Eprint
  {http://arxiv.org/abs/0901.0306} {arXiv:0901.0306 [hep-ph]}\BibitemShut
  {NoStop}%
\bibitem [{\citenamefont {Jegerlehner}\ and\ \citenamefont
  {Nyffeler}(2009)}]{Jegerlehner:2009ry}%
  \BibitemOpen
  \bibfield  {author} {\bibinfo {author} {\bibfnamefont {F.}~\bibnamefont
  {Jegerlehner}}\ and\ \bibinfo {author} {\bibfnamefont {A.}~\bibnamefont
  {Nyffeler}},\ }\href {\doibase 10.1016/j.physrep.2009.04.003} {\bibfield
  {journal} {\bibinfo  {journal} {Phys. Rept.}\ }\textbf {\bibinfo {volume}
  {477}},\ \bibinfo {pages} {1} (\bibinfo {year} {2009})},\ \Eprint
  {http://arxiv.org/abs/0902.3360} {arXiv:0902.3360 [hep-ph]}\BibitemShut
  {NoStop}%
\bibitem [{\citenamefont {Colangelo}\ \emph {et~al.}(2015)\citenamefont
  {Colangelo}, \citenamefont {Hoferichter}, \citenamefont {Procura},\ and\
  \citenamefont {Stoffer}}]{Colangelo:2015ama}%
  \BibitemOpen
  \bibfield  {author} {\bibinfo {author} {\bibfnamefont {G.}~\bibnamefont
  {Colangelo}}, \bibinfo {author} {\bibfnamefont {M.}~\bibnamefont
  {Hoferichter}}, \bibinfo {author} {\bibfnamefont {M.}~\bibnamefont
  {Procura}}, \ and\ \bibinfo {author} {\bibfnamefont {P.}~\bibnamefont
  {Stoffer}},\ }\href {\doibase 10.1007/JHEP09(2015)074} {\bibfield  {journal}
  {\bibinfo  {journal} {JHEP}\ }\textbf {\bibinfo {volume} {09}},\ \bibinfo
  {pages} {074} (\bibinfo {year} {2015})},\ \Eprint
  {http://arxiv.org/abs/1506.01386} {arXiv:1506.01386 [hep-ph]}\BibitemShut
  {NoStop}%
\bibitem [{\citenamefont {Colangelo}\ \emph
  {et~al.}(2014{\natexlab{b}})\citenamefont {Colangelo}, \citenamefont
  {Hoferichter}, \citenamefont {Procura},\ and\ \citenamefont
  {Stoffer}}]{Colangelo:2014dfa}%
  \BibitemOpen
  \bibfield  {author} {\bibinfo {author} {\bibfnamefont {G.}~\bibnamefont
  {Colangelo}}, \bibinfo {author} {\bibfnamefont {M.}~\bibnamefont
  {Hoferichter}}, \bibinfo {author} {\bibfnamefont {M.}~\bibnamefont
  {Procura}}, \ and\ \bibinfo {author} {\bibfnamefont {P.}~\bibnamefont
  {Stoffer}},\ }\href {\doibase 10.1007/JHEP09(2014)091} {\bibfield  {journal}
  {\bibinfo  {journal} {JHEP}\ }\textbf {\bibinfo {volume} {09}},\ \bibinfo
  {pages} {091} (\bibinfo {year} {2014}{\natexlab{b}})},\ \Eprint
  {http://arxiv.org/abs/1402.7081} {arXiv:1402.7081 [hep-ph]}\BibitemShut
  {NoStop}%
\bibitem [{\citenamefont {Pauk}\ and\ \citenamefont
  {Vanderhaeghen}(2014{\natexlab{b}})}]{Pauk:2014rfa}%
  \BibitemOpen
  \bibfield  {author} {\bibinfo {author} {\bibfnamefont {V.}~\bibnamefont
  {Pauk}}\ and\ \bibinfo {author} {\bibfnamefont {M.}~\bibnamefont
  {Vanderhaeghen}},\ }\href {\doibase 10.1103/PhysRevD.90.113012} {\bibfield
  {journal} {\bibinfo  {journal} {Phys. Rev.}\ }\textbf {\bibinfo {volume}
  {D90}},\ \bibinfo {pages} {113012} (\bibinfo {year} {2014}{\natexlab{b}})},\
  \Eprint {http://arxiv.org/abs/1409.0819} {arXiv:1409.0819
  [hep-ph]}\BibitemShut {NoStop}%
\bibitem [{\citenamefont {Aldins}\ \emph {et~al.}(1970)\citenamefont {Aldins},
  \citenamefont {Kinoshita}, \citenamefont {Brodsky},\ and\ \citenamefont
  {Dufner}}]{Aldins:1970id}%
  \BibitemOpen
  \bibfield  {author} {\bibinfo {author} {\bibfnamefont {J.}~\bibnamefont
  {Aldins}}, \bibinfo {author} {\bibfnamefont {T.}~\bibnamefont {Kinoshita}},
  \bibinfo {author} {\bibfnamefont {S.~J.}\ \bibnamefont {Brodsky}}, \ and\
  \bibinfo {author} {\bibfnamefont {A.~J.}\ \bibnamefont {Dufner}},\ }\href
  {\doibase 10.1103/PhysRevD.1.2378} {\bibfield  {journal} {\bibinfo  {journal}
  {Phys. Rev.}\ }\textbf {\bibinfo {volume} {D1}},\ \bibinfo {pages} {2378}
  (\bibinfo {year} {1970})}\BibitemShut {NoStop}%
\bibitem [{\citenamefont
  {Jegerlehner}(2008{\natexlab{b}})}]{Jegerlehner:2008zza}%
  \BibitemOpen
  \bibfield  {author} {\bibinfo {author} {\bibfnamefont {F.}~\bibnamefont
  {Jegerlehner}},\ }\href {\doibase 10.1007/978-3-540-72634-0} {\bibfield
  {journal} {\bibinfo  {journal} {Springer Tracts Mod. Phys.}\ }\textbf
  {\bibinfo {volume} {226}},\ \bibinfo {pages} {1} (\bibinfo {year}
  {2008}{\natexlab{b}})}\BibitemShut {NoStop}%
\bibitem [{\citenamefont {Pauk}\ and\ \citenamefont
  {Vanderhaeghen}(2014{\natexlab{c}})}]{Pauk:2014jza}%
  \BibitemOpen
  \bibfield  {author} {\bibinfo {author} {\bibfnamefont {V.}~\bibnamefont
  {Pauk}}\ and\ \bibinfo {author} {\bibfnamefont {M.}~\bibnamefont
  {Vanderhaeghen}},\ }\href@noop {} {\  (\bibinfo {year}
  {2014}{\natexlab{c}})},\ \Eprint {http://arxiv.org/abs/1403.7503}
  {arXiv:1403.7503 [hep-ph]}\BibitemShut {NoStop}%
\bibitem [{\citenamefont {Schwinger}(1975{\natexlab{a}})}]{Schwinger:1975ti}%
  \BibitemOpen
  \bibfield  {author} {\bibinfo {author} {\bibfnamefont {J.~S.}\ \bibnamefont
  {Schwinger}},\ }\href {\doibase 10.1007/978-3-7091-8424-0_9,
  10.1073/pnas.72.1.1} {\bibfield  {journal} {\bibinfo  {journal} {Proc. Nat.
  Acad. Sci.}\ }\textbf {\bibinfo {volume} {72}},\ \bibinfo {pages} {1}
  (\bibinfo {year} {1975}{\natexlab{a}})},\ \bibinfo {note} {[Acta Phys.
  Austriaca Suppl. {\bf 14}, 471 (1975)]}\BibitemShut {NoStop}%
\bibitem [{\citenamefont {Schwinger}(1975{\natexlab{b}})}]{Schwinger:1975uq}%
  \BibitemOpen
  \bibfield  {author} {\bibinfo {author} {\bibfnamefont {J.~S.}\ \bibnamefont
  {Schwinger}},\ }\href {\doibase 10.1073/pnas.72.11.4216} {\bibfield
  {journal} {\bibinfo  {journal} {Proc. Nat. Acad. Sci.}\ }\textbf {\bibinfo
  {volume} {72}},\ \bibinfo {pages} {1559} (\bibinfo {year}
  {1975}{\natexlab{b}})}\BibitemShut {NoStop}%
\bibitem [{\citenamefont {Hagelstein}\ and\ \citenamefont
  {Pascalutsa}(2018)}]{Hagelstein:2017obr}%
  \BibitemOpen
  \bibfield  {author} {\bibinfo {author} {\bibfnamefont {F.}~\bibnamefont
  {Hagelstein}}\ and\ \bibinfo {author} {\bibfnamefont {V.}~\bibnamefont
  {Pascalutsa}},\ }\href {\doibase 10.1103/PhysRevLett.120.072002} {\bibfield
  {journal} {\bibinfo  {journal} {Phys. Rev. Lett.}\ }\textbf {\bibinfo
  {volume} {120}},\ \bibinfo {pages} {072002} (\bibinfo {year} {2018})},\
  \Eprint {http://arxiv.org/abs/1710.04571} {arXiv:1710.04571
  [hep-ph]}\BibitemShut {NoStop}%
\bibitem [{\citenamefont {Karplus}\ and\ \citenamefont
  {Neuman}(1950)}]{Karplus:1950zza}%
  \BibitemOpen
  \bibfield  {author} {\bibinfo {author} {\bibfnamefont {R.}~\bibnamefont
  {Karplus}}\ and\ \bibinfo {author} {\bibfnamefont {M.}~\bibnamefont
  {Neuman}},\ }\href {\doibase 10.1103/PhysRev.80.380} {\bibfield  {journal}
  {\bibinfo  {journal} {Phys. Rev.}\ }\textbf {\bibinfo {volume} {80}},\
  \bibinfo {pages} {380} (\bibinfo {year} {1950})}\BibitemShut {NoStop}%
\bibitem [{\citenamefont {Leo}\ \emph {et~al.}(1975)\citenamefont {Leo},
  \citenamefont {Minguzzi},\ and\ \citenamefont {Soliani}}]{Leo:1975fb}%
  \BibitemOpen
  \bibfield  {author} {\bibinfo {author} {\bibfnamefont {R.~A.}\ \bibnamefont
  {Leo}}, \bibinfo {author} {\bibfnamefont {A.}~\bibnamefont {Minguzzi}}, \
  and\ \bibinfo {author} {\bibfnamefont {G.}~\bibnamefont {Soliani}},\ }\href
  {\doibase 10.1007/BF02730173} {\bibfield  {journal} {\bibinfo  {journal}
  {Nuovo Cim.}\ }\textbf {\bibinfo {volume} {A30}},\ \bibinfo {pages} {270}
  (\bibinfo {year} {1975})}\BibitemShut {NoStop}%
\bibitem [{\citenamefont {Bijnens}\ \emph {et~al.}(1996)\citenamefont
  {Bijnens}, \citenamefont {Pallante},\ and\ \citenamefont
  {Prades}}]{Bijnens:1995xf}%
  \BibitemOpen
  \bibfield  {author} {\bibinfo {author} {\bibfnamefont {J.}~\bibnamefont
  {Bijnens}}, \bibinfo {author} {\bibfnamefont {E.}~\bibnamefont {Pallante}}, \
  and\ \bibinfo {author} {\bibfnamefont {J.}~\bibnamefont {Prades}},\ }\href
  {\doibase 10.1016/0550-3213(96)00288-X} {\bibfield  {journal} {\bibinfo
  {journal} {Nucl. Phys.}\ }\textbf {\bibinfo {volume} {B474}},\ \bibinfo
  {pages} {379} (\bibinfo {year} {1996})}\BibitemShut {NoStop}%
\bibitem [{\citenamefont {Eichmann}\ \emph {et~al.}(2016)\citenamefont
  {Eichmann}, \citenamefont {Fischer}, \citenamefont {Heupel},\ and\
  \citenamefont {Williams}}]{Eichmann:2014ooa}%
  \BibitemOpen
  \bibfield  {author} {\bibinfo {author} {\bibfnamefont {G.}~\bibnamefont
  {Eichmann}}, \bibinfo {author} {\bibfnamefont {C.~S.}\ \bibnamefont
  {Fischer}}, \bibinfo {author} {\bibfnamefont {W.}~\bibnamefont {Heupel}}, \
  and\ \bibinfo {author} {\bibfnamefont {R.}~\bibnamefont {Williams}},\ }\href
  {\doibase 10.1063/1.4938621} {\bibfield  {journal} {\bibinfo  {journal} {AIP
  Conf. Proc.}\ }\textbf {\bibinfo {volume} {1701}},\ \bibinfo {pages} {040004}
  (\bibinfo {year} {2016})},\ \Eprint {http://arxiv.org/abs/1411.7876}
  {arXiv:1411.7876 [hep-ph]}\BibitemShut {NoStop}%
\bibitem [{\citenamefont {Bardeen}\ and\ \citenamefont
  {Tung}(1968)}]{Bardeen:1969aw}%
  \BibitemOpen
  \bibfield  {author} {\bibinfo {author} {\bibfnamefont {W.~A.}\ \bibnamefont
  {Bardeen}}\ and\ \bibinfo {author} {\bibfnamefont {W.~K.}\ \bibnamefont
  {Tung}},\ }\href {\doibase 10.1103/PhysRev.173.1423} {\bibfield  {journal}
  {\bibinfo  {journal} {Phys. Rev.}\ }\textbf {\bibinfo {volume} {173}},\
  \bibinfo {pages} {1423} (\bibinfo {year} {1968})},\ \bibinfo {note}
  {[Erratum: Phys. Rev. {\bf D4}, 3229 (1971)]}\BibitemShut {NoStop}%
\bibitem [{\citenamefont {Tarrach}(1975)}]{Tarrach:1975tu}%
  \BibitemOpen
  \bibfield  {author} {\bibinfo {author} {\bibfnamefont {R.}~\bibnamefont
  {Tarrach}},\ }\href {\doibase 10.1007/BF02894857} {\bibfield  {journal}
  {\bibinfo  {journal} {Nuovo Cim.}\ }\textbf {\bibinfo {volume} {A28}},\
  \bibinfo {pages} {409} (\bibinfo {year} {1975})}\BibitemShut {NoStop}%
\bibitem [{\citenamefont {Rosner}(1967)}]{Rosner:1967zz}%
  \BibitemOpen
  \bibfield  {author} {\bibinfo {author} {\bibfnamefont {J.~L.}\ \bibnamefont
  {Rosner}},\ }\href {\doibase 10.1016/0003-4916(67)90262-X} {\bibfield
  {journal} {\bibinfo  {journal} {Annals Phys.}\ }\textbf {\bibinfo {volume}
  {44}},\ \bibinfo {pages} {11} (\bibinfo {year} {1967})}\BibitemShut {NoStop}%
\bibitem [{\citenamefont {Levine}\ and\ \citenamefont
  {Roskies}(1974)}]{Levine:1974xh}%
  \BibitemOpen
  \bibfield  {author} {\bibinfo {author} {\bibfnamefont {M.~J.}\ \bibnamefont
  {Levine}}\ and\ \bibinfo {author} {\bibfnamefont {R.}~\bibnamefont
  {Roskies}},\ }\href {\doibase 10.1103/PhysRevD.9.421} {\bibfield  {journal}
  {\bibinfo  {journal} {Phys. Rev.}\ }\textbf {\bibinfo {volume} {D9}},\
  \bibinfo {pages} {421} (\bibinfo {year} {1974})}\BibitemShut {NoStop}%
\bibitem [{\citenamefont {Hoferichter}\ \emph
  {et~al.}(2014{\natexlab{a}})\citenamefont {Hoferichter}, \citenamefont
  {Colangelo}, \citenamefont {Procura},\ and\ \citenamefont
  {Stoffer}}]{Hoferichter:2013ama}%
  \BibitemOpen
  \bibfield  {author} {\bibinfo {author} {\bibfnamefont {M.}~\bibnamefont
  {Hoferichter}}, \bibinfo {author} {\bibfnamefont {G.}~\bibnamefont
  {Colangelo}}, \bibinfo {author} {\bibfnamefont {M.}~\bibnamefont {Procura}},
  \ and\ \bibinfo {author} {\bibfnamefont {P.}~\bibnamefont {Stoffer}},\ }\href
  {\doibase 10.1142/S2010194514604001} {\bibfield  {journal} {\bibinfo
  {journal} {Int. J. Mod. Phys. Conf. Ser.}\ }\textbf {\bibinfo {volume}
  {35}},\ \bibinfo {pages} {1460400} (\bibinfo {year} {2014}{\natexlab{a}})},\
  \Eprint {http://arxiv.org/abs/1309.6877} {arXiv:1309.6877
  [hep-ph]}\BibitemShut {NoStop}%
\bibitem [{\citenamefont {Colangelo}\ \emph
  {et~al.}(2014{\natexlab{c}})\citenamefont {Colangelo}, \citenamefont
  {Hoferichter}, \citenamefont {Kubis}, \citenamefont {Procura},\ and\
  \citenamefont {Stoffer}}]{Colangelo:2014pva}%
  \BibitemOpen
  \bibfield  {author} {\bibinfo {author} {\bibfnamefont {G.}~\bibnamefont
  {Colangelo}}, \bibinfo {author} {\bibfnamefont {M.}~\bibnamefont
  {Hoferichter}}, \bibinfo {author} {\bibfnamefont {B.}~\bibnamefont {Kubis}},
  \bibinfo {author} {\bibfnamefont {M.}~\bibnamefont {Procura}}, \ and\
  \bibinfo {author} {\bibfnamefont {P.}~\bibnamefont {Stoffer}},\ }\href
  {\doibase 10.1016/j.physletb.2014.09.021} {\bibfield  {journal} {\bibinfo
  {journal} {Phys. Lett.}\ }\textbf {\bibinfo {volume} {B738}},\ \bibinfo
  {pages} {6} (\bibinfo {year} {2014}{\natexlab{c}})},\ \Eprint
  {http://arxiv.org/abs/1408.2517} {arXiv:1408.2517 [hep-ph]}\BibitemShut
  {NoStop}%
\bibitem [{\citenamefont {Colangelo}\ \emph
  {et~al.}(2017{\natexlab{b}})\citenamefont {Colangelo}, \citenamefont
  {Hoferichter}, \citenamefont {Procura},\ and\ \citenamefont
  {Stoffer}}]{Colangelo:2017qdm}%
  \BibitemOpen
  \bibfield  {author} {\bibinfo {author} {\bibfnamefont {G.}~\bibnamefont
  {Colangelo}}, \bibinfo {author} {\bibfnamefont {M.}~\bibnamefont
  {Hoferichter}}, \bibinfo {author} {\bibfnamefont {M.}~\bibnamefont
  {Procura}}, \ and\ \bibinfo {author} {\bibfnamefont {P.}~\bibnamefont
  {Stoffer}},\ }\href {\doibase 10.1103/PhysRevLett.118.232001} {\bibfield
  {journal} {\bibinfo  {journal} {Phys. Rev. Lett.}\ }\textbf {\bibinfo
  {volume} {118}},\ \bibinfo {pages} {232001} (\bibinfo {year}
  {2017}{\natexlab{b}})},\ \Eprint {http://arxiv.org/abs/1701.06554}
  {arXiv:1701.06554 [hep-ph]}\BibitemShut {NoStop}%
\bibitem [{\citenamefont {Hoferichter}\ \emph
  {et~al.}(2018{\natexlab{b}})\citenamefont {Hoferichter}, \citenamefont
  {Hoid}, \citenamefont {Kubis}, \citenamefont {Leupold},\ and\ \citenamefont
  {Schneider}}]{Hoferichter:2018dmo}%
  \BibitemOpen
  \bibfield  {author} {\bibinfo {author} {\bibfnamefont {M.}~\bibnamefont
  {Hoferichter}}, \bibinfo {author} {\bibfnamefont {B.-L.}\ \bibnamefont
  {Hoid}}, \bibinfo {author} {\bibfnamefont {B.}~\bibnamefont {Kubis}},
  \bibinfo {author} {\bibfnamefont {S.}~\bibnamefont {Leupold}}, \ and\
  \bibinfo {author} {\bibfnamefont {S.~P.}\ \bibnamefont {Schneider}},\ }\href
  {\doibase 10.1103/PhysRevLett.121.112002} {\bibfield  {journal} {\bibinfo
  {journal} {Phys. Rev. Lett.}\ }\textbf {\bibinfo {volume} {121}},\ \bibinfo
  {pages} {112002} (\bibinfo {year} {2018}{\natexlab{b}})},\ \Eprint
  {http://arxiv.org/abs/1805.01471} {arXiv:1805.01471 [hep-ph]}\BibitemShut
  {NoStop}%
\bibitem [{\citenamefont {Kinoshita}\ \emph {et~al.}(1985)\citenamefont
  {Kinoshita}, \citenamefont {Nizic},\ and\ \citenamefont
  {Okamoto}}]{Kinoshita:1984it}%
  \BibitemOpen
  \bibfield  {author} {\bibinfo {author} {\bibfnamefont {T.}~\bibnamefont
  {Kinoshita}}, \bibinfo {author} {\bibfnamefont {B.}~\bibnamefont {Nizic}}, \
  and\ \bibinfo {author} {\bibfnamefont {Y.}~\bibnamefont {Okamoto}},\ }\href
  {\doibase 10.1103/PhysRevD.31.2108} {\bibfield  {journal} {\bibinfo
  {journal} {Phys. Rev.}\ }\textbf {\bibinfo {volume} {D31}},\ \bibinfo {pages}
  {2108} (\bibinfo {year} {1985})}\BibitemShut {NoStop}%
\bibitem [{\citenamefont {Hayakawa}\ \emph {et~al.}(1995)\citenamefont
  {Hayakawa}, \citenamefont {Kinoshita},\ and\ \citenamefont
  {Sanda}}]{Hayakawa:1995ps}%
  \BibitemOpen
  \bibfield  {author} {\bibinfo {author} {\bibfnamefont {M.}~\bibnamefont
  {Hayakawa}}, \bibinfo {author} {\bibfnamefont {T.}~\bibnamefont {Kinoshita}},
  \ and\ \bibinfo {author} {\bibfnamefont {A.~I.}\ \bibnamefont {Sanda}},\
  }\href {\doibase 10.1103/PhysRevLett.75.790} {\bibfield  {journal} {\bibinfo
  {journal} {Phys. Rev. Lett.}\ }\textbf {\bibinfo {volume} {75}},\ \bibinfo
  {pages} {790} (\bibinfo {year} {1995})},\ \Eprint
  {http://arxiv.org/abs/hep-ph/9503463} {arXiv:hep-ph/9503463
  [hep-ph]}\BibitemShut {NoStop}%
\bibitem [{\citenamefont {Hayakawa}\ \emph {et~al.}(1996)\citenamefont
  {Hayakawa}, \citenamefont {Kinoshita},\ and\ \citenamefont
  {Sanda}}]{Hayakawa:1996ki}%
  \BibitemOpen
  \bibfield  {author} {\bibinfo {author} {\bibfnamefont {M.}~\bibnamefont
  {Hayakawa}}, \bibinfo {author} {\bibfnamefont {T.}~\bibnamefont {Kinoshita}},
  \ and\ \bibinfo {author} {\bibfnamefont {A.~I.}\ \bibnamefont {Sanda}},\
  }\href {\doibase 10.1103/PhysRevD.54.3137} {\bibfield  {journal} {\bibinfo
  {journal} {Phys. Rev.}\ }\textbf {\bibinfo {volume} {D54}},\ \bibinfo {pages}
  {3137} (\bibinfo {year} {1996})},\ \Eprint
  {http://arxiv.org/abs/hep-ph/9601310} {arXiv:hep-ph/9601310
  [hep-ph]}\BibitemShut {NoStop}%
\bibitem [{\citenamefont {Hayakawa}\ and\ \citenamefont
  {Kinoshita}(1998)}]{Hayakawa:1997rq}%
  \BibitemOpen
  \bibfield  {author} {\bibinfo {author} {\bibfnamefont {M.}~\bibnamefont
  {Hayakawa}}\ and\ \bibinfo {author} {\bibfnamefont {T.}~\bibnamefont
  {Kinoshita}},\ }\href {\doibase 10.1103/PhysRevD.57.465} {\bibfield
  {journal} {\bibinfo  {journal} {Phys. Rev.}\ }\textbf {\bibinfo {volume}
  {D57}},\ \bibinfo {pages} {465} (\bibinfo {year} {1998})},\ \bibinfo {note}
  {[Erratum: Phys. Rev. {\bf D66}, 019902 (2002)]},\ \Eprint
  {http://arxiv.org/abs/hep-ph/9708227} {arXiv:hep-ph/9708227
  [hep-ph]}\BibitemShut {NoStop}%
\bibitem [{\citenamefont {Bijnens}\ \emph {et~al.}(1995)\citenamefont
  {Bijnens}, \citenamefont {Pallante},\ and\ \citenamefont
  {Prades}}]{Bijnens:1995cc}%
  \BibitemOpen
  \bibfield  {author} {\bibinfo {author} {\bibfnamefont {J.}~\bibnamefont
  {Bijnens}}, \bibinfo {author} {\bibfnamefont {E.}~\bibnamefont {Pallante}}, \
  and\ \bibinfo {author} {\bibfnamefont {J.}~\bibnamefont {Prades}},\ }\href
  {\doibase 10.1103/PhysRevLett.75.1447} {\bibfield  {journal} {\bibinfo
  {journal} {Phys. Rev. Lett.}\ }\textbf {\bibinfo {volume} {75}},\ \bibinfo
  {pages} {1447} (\bibinfo {year} {1995})},\ \bibinfo {note} {[Erratum: Phys.
  Rev. Lett. {\bf 75}, 3781 (1995)]},\ \Eprint
  {http://arxiv.org/abs/hep-ph/9505251} {arXiv:hep-ph/9505251
  [hep-ph]}\BibitemShut {NoStop}%
\bibitem [{\citenamefont {Bijnens}\ \emph {et~al.}(2002)\citenamefont
  {Bijnens}, \citenamefont {Pallante},\ and\ \citenamefont
  {Prades}}]{Bijnens:2001cq}%
  \BibitemOpen
  \bibfield  {author} {\bibinfo {author} {\bibfnamefont {J.}~\bibnamefont
  {Bijnens}}, \bibinfo {author} {\bibfnamefont {E.}~\bibnamefont {Pallante}}, \
  and\ \bibinfo {author} {\bibfnamefont {J.}~\bibnamefont {Prades}},\ }\href
  {\doibase 10.1016/S0550-3213(02)00074-3} {\bibfield  {journal} {\bibinfo
  {journal} {Nucl. Phys.}\ }\textbf {\bibinfo {volume} {B626}},\ \bibinfo
  {pages} {410} (\bibinfo {year} {2002})},\ \Eprint
  {http://arxiv.org/abs/hep-ph/0112255} {arXiv:hep-ph/0112255
  [hep-ph]}\BibitemShut {NoStop}%
\bibitem [{\citenamefont {Ramsey-Musolf}\ and\ \citenamefont
  {Wise}(2002)}]{RamseyMusolf:2002cy}%
  \BibitemOpen
  \bibfield  {author} {\bibinfo {author} {\bibfnamefont {M.~J.}\ \bibnamefont
  {Ramsey-Musolf}}\ and\ \bibinfo {author} {\bibfnamefont {M.~B.}\ \bibnamefont
  {Wise}},\ }\href {\doibase 10.1103/PhysRevLett.89.041601} {\bibfield
  {journal} {\bibinfo  {journal} {Phys. Rev. Lett.}\ }\textbf {\bibinfo
  {volume} {89}},\ \bibinfo {pages} {041601} (\bibinfo {year} {2002})},\
  \Eprint {http://arxiv.org/abs/hep-ph/0201297} {arXiv:hep-ph/0201297
  [hep-ph]}\BibitemShut {NoStop}%
\bibitem [{\citenamefont {Bijnens}\ and\ \citenamefont
  {Prades}(2007)}]{Bijnens:2007pz}%
  \BibitemOpen
  \bibfield  {author} {\bibinfo {author} {\bibfnamefont {J.}~\bibnamefont
  {Bijnens}}\ and\ \bibinfo {author} {\bibfnamefont {J.}~\bibnamefont
  {Prades}},\ }\href {\doibase 10.1142/S0217732307022992} {\bibfield  {journal}
  {\bibinfo  {journal} {Mod. Phys. Lett.}\ }\textbf {\bibinfo {volume} {A22}},\
  \bibinfo {pages} {767} (\bibinfo {year} {2007})},\ \Eprint
  {http://arxiv.org/abs/hep-ph/0702170} {arXiv:hep-ph/0702170
  [HEP-PH]}\BibitemShut {NoStop}%
\bibitem [{\citenamefont {Bijnens}\ and\ \citenamefont
  {Relefors}(2016{\natexlab{b}})}]{Bijnens:2016hgx}%
  \BibitemOpen
  \bibfield  {author} {\bibinfo {author} {\bibfnamefont {J.}~\bibnamefont
  {Bijnens}}\ and\ \bibinfo {author} {\bibfnamefont {J.}~\bibnamefont
  {Relefors}},\ }\href {\doibase 10.1007/JHEP09(2016)113} {\bibfield  {journal}
  {\bibinfo  {journal} {JHEP}\ }\textbf {\bibinfo {volume} {09}},\ \bibinfo
  {pages} {113} (\bibinfo {year} {2016}{\natexlab{b}})},\ \Eprint
  {http://arxiv.org/abs/1608.01454} {arXiv:1608.01454 [hep-ph]}\BibitemShut
  {NoStop}%
\bibitem [{\citenamefont {Engel}\ \emph {et~al.}(2012)\citenamefont {Engel},
  \citenamefont {Patel},\ and\ \citenamefont {Ramsey-Musolf}}]{Engel:2012xb}%
  \BibitemOpen
  \bibfield  {author} {\bibinfo {author} {\bibfnamefont {K.~T.}\ \bibnamefont
  {Engel}}, \bibinfo {author} {\bibfnamefont {H.~H.}\ \bibnamefont {Patel}}, \
  and\ \bibinfo {author} {\bibfnamefont {M.~J.}\ \bibnamefont
  {Ramsey-Musolf}},\ }\href {\doibase 10.1103/PhysRevD.86.037502} {\bibfield
  {journal} {\bibinfo  {journal} {Phys. Rev. D}\ }\textbf {\bibinfo {volume}
  {86}},\ \bibinfo {pages} {037502} (\bibinfo {year} {2012})},\ \Eprint
  {http://arxiv.org/abs/1201.0809} {arXiv:1201.0809 [hep-ph]}\BibitemShut
  {NoStop}%
\bibitem [{\citenamefont {Engel}\ and\ \citenamefont
  {Ramsey-Musolf}(2014)}]{Engel:2013kda}%
  \BibitemOpen
  \bibfield  {author} {\bibinfo {author} {\bibfnamefont {K.~T.}\ \bibnamefont
  {Engel}}\ and\ \bibinfo {author} {\bibfnamefont {M.~J.}\ \bibnamefont
  {Ramsey-Musolf}},\ }\href {\doibase 10.1016/j.physletb.2014.09.006}
  {\bibfield  {journal} {\bibinfo  {journal} {Phys. Lett.}\ }\textbf {\bibinfo
  {volume} {B738}},\ \bibinfo {pages} {123} (\bibinfo {year} {2014})},\ \Eprint
  {http://arxiv.org/abs/1309.2225} {arXiv:1309.2225 [hep-ph]}\BibitemShut
  {NoStop}%
\bibitem [{\citenamefont {Greynat}\ and\ \citenamefont
  {de~Rafael}(2012)}]{Greynat:2012ww}%
  \BibitemOpen
  \bibfield  {author} {\bibinfo {author} {\bibfnamefont {D.}~\bibnamefont
  {Greynat}}\ and\ \bibinfo {author} {\bibfnamefont {E.}~\bibnamefont
  {de~Rafael}},\ }\href {\doibase 10.1007/JHEP07(2012)020} {\bibfield
  {journal} {\bibinfo  {journal} {JHEP}\ }\textbf {\bibinfo {volume} {07}},\
  \bibinfo {pages} {020} (\bibinfo {year} {2012})},\ \Eprint
  {http://arxiv.org/abs/1204.3029} {arXiv:1204.3029 [hep-ph]}\BibitemShut
  {NoStop}%
\bibitem [{\citenamefont {Dorokhov}\ \emph {et~al.}(2012)\citenamefont
  {Dorokhov}, \citenamefont {Radzhabov},\ and\ \citenamefont
  {Zhevlakov}}]{Dorokhov:2012qa}%
  \BibitemOpen
  \bibfield  {author} {\bibinfo {author} {\bibfnamefont {A.~E.}\ \bibnamefont
  {Dorokhov}}, \bibinfo {author} {\bibfnamefont {A.~E.}\ \bibnamefont
  {Radzhabov}}, \ and\ \bibinfo {author} {\bibfnamefont {A.~S.}\ \bibnamefont
  {Zhevlakov}},\ }\href {\doibase 10.1140/epjc/s10052-012-2227-3} {\bibfield
  {journal} {\bibinfo  {journal} {Eur. Phys. J.}\ }\textbf {\bibinfo {volume}
  {C72}},\ \bibinfo {pages} {2227} (\bibinfo {year} {2012})},\ \Eprint
  {http://arxiv.org/abs/1204.3729} {arXiv:1204.3729 [hep-ph]}\BibitemShut
  {NoStop}%
\bibitem [{\citenamefont {Dorokhov}\ \emph {et~al.}(2015)\citenamefont
  {Dorokhov}, \citenamefont {Radzhabov},\ and\ \citenamefont
  {Zhevlakov}}]{Dorokhov:2015psa}%
  \BibitemOpen
  \bibfield  {author} {\bibinfo {author} {\bibfnamefont {A.~E.}\ \bibnamefont
  {Dorokhov}}, \bibinfo {author} {\bibfnamefont {A.~E.}\ \bibnamefont
  {Radzhabov}}, \ and\ \bibinfo {author} {\bibfnamefont {A.~S.}\ \bibnamefont
  {Zhevlakov}},\ }\href {\doibase 10.1140/epjc/s10052-015-3577-4} {\bibfield
  {journal} {\bibinfo  {journal} {Eur. Phys. J.}\ }\textbf {\bibinfo {volume}
  {C75}},\ \bibinfo {pages} {417} (\bibinfo {year} {2015})},\ \Eprint
  {http://arxiv.org/abs/1502.04487} {arXiv:1502.04487 [hep-ph]}\BibitemShut
  {NoStop}%
\bibitem [{\citenamefont {Boughezal}\ and\ \citenamefont
  {Melnikov}(2011)}]{Boughezal:2011vw}%
  \BibitemOpen
  \bibfield  {author} {\bibinfo {author} {\bibfnamefont {R.}~\bibnamefont
  {Boughezal}}\ and\ \bibinfo {author} {\bibfnamefont {K.}~\bibnamefont
  {Melnikov}},\ }\href {\doibase 10.1016/j.physletb.2011.09.001} {\bibfield
  {journal} {\bibinfo  {journal} {Phys. Lett.}\ }\textbf {\bibinfo {volume}
  {B704}},\ \bibinfo {pages} {193} (\bibinfo {year} {2011})},\ \Eprint
  {http://arxiv.org/abs/1104.4510} {arXiv:1104.4510 [hep-ph]}\BibitemShut
  {NoStop}%
\bibitem [{\citenamefont {Masjuan}\ and\ \citenamefont
  {Vanderhaeghen}(2015)}]{Masjuan:2012qn}%
  \BibitemOpen
  \bibfield  {author} {\bibinfo {author} {\bibfnamefont {P.}~\bibnamefont
  {Masjuan}}\ and\ \bibinfo {author} {\bibfnamefont {M.}~\bibnamefont
  {Vanderhaeghen}},\ }\href {\doibase 10.1088/0954-3899/42/12/125004}
  {\bibfield  {journal} {\bibinfo  {journal} {J. Phys.}\ }\textbf {\bibinfo
  {volume} {G42}},\ \bibinfo {pages} {125004} (\bibinfo {year} {2015})},\
  \Eprint {http://arxiv.org/abs/1212.0357} {arXiv:1212.0357
  [hep-ph]}\BibitemShut {NoStop}%
\bibitem [{\citenamefont {Goecke}\ \emph {et~al.}(2011)\citenamefont {Goecke},
  \citenamefont {Fischer},\ and\ \citenamefont {Williams}}]{Goecke:2010if}%
  \BibitemOpen
  \bibfield  {author} {\bibinfo {author} {\bibfnamefont {T.}~\bibnamefont
  {Goecke}}, \bibinfo {author} {\bibfnamefont {C.~S.}\ \bibnamefont {Fischer}},
  \ and\ \bibinfo {author} {\bibfnamefont {R.}~\bibnamefont {Williams}},\
  }\href {\doibase 10.1103/PhysRevD.83.094006} {\bibfield  {journal} {\bibinfo
  {journal} {Phys. Rev.}\ }\textbf {\bibinfo {volume} {D83}},\ \bibinfo {pages}
  {094006} (\bibinfo {year} {2011})},\ \bibinfo {note} {[Erratum: Phys. Rev.
  {\bf D86}, 099901 (2012)]},\ \Eprint {http://arxiv.org/abs/1012.3886}
  {arXiv:1012.3886 [hep-ph]}\BibitemShut {NoStop}%
\bibitem [{\citenamefont {Abegg}\ \emph {et~al.}(1994)\citenamefont {Abegg}
  \emph {et~al.}}]{Abegg:1994wx}%
  \BibitemOpen
  \bibfield  {author} {\bibinfo {author} {\bibfnamefont {R.}~\bibnamefont
  {Abegg}} \emph {et~al.},\ }\href {\doibase 10.1103/PhysRevD.50.92} {\bibfield
   {journal} {\bibinfo  {journal} {Phys. Rev.}\ }\textbf {\bibinfo {volume}
  {D50}},\ \bibinfo {pages} {92} (\bibinfo {year} {1994})}\BibitemShut
  {NoStop}%
\bibitem [{\citenamefont {Abouzaid}\ \emph {et~al.}(2007)\citenamefont
  {Abouzaid} \emph {et~al.}}]{Abouzaid:2006kk}%
  \BibitemOpen
  \bibfield  {author} {\bibinfo {author} {\bibfnamefont {E.}~\bibnamefont
  {Abouzaid}} \emph {et~al.} (\bibinfo {collaboration} {KTeV}),\ }\href
  {\doibase 10.1103/PhysRevD.75.012004} {\bibfield  {journal} {\bibinfo
  {journal} {Phys. Rev.}\ }\textbf {\bibinfo {volume} {D75}},\ \bibinfo {pages}
  {012004} (\bibinfo {year} {2007})},\ \Eprint
  {http://arxiv.org/abs/hep-ex/0610072} {arXiv:hep-ex/0610072
  [hep-ex]}\BibitemShut {NoStop}%
\bibitem [{\citenamefont {Akhmetshin}\ \emph {et~al.}(2015)\citenamefont
  {Akhmetshin} \emph {et~al.}}]{Akhmetshin:2014hxv}%
  \BibitemOpen
  \bibfield  {author} {\bibinfo {author} {\bibfnamefont {R.~R.}\ \bibnamefont
  {Akhmetshin}} \emph {et~al.} (\bibinfo {collaboration} {CMD-3}),\ }\href
  {\doibase 10.1016/j.physletb.2014.11.056} {\bibfield  {journal} {\bibinfo
  {journal} {Phys. Lett.}\ }\textbf {\bibinfo {volume} {B740}},\ \bibinfo
  {pages} {273} (\bibinfo {year} {2015})},\ \Eprint
  {http://arxiv.org/abs/1409.1664} {arXiv:1409.1664 [hep-ex]}\BibitemShut
  {NoStop}%
\bibitem [{\citenamefont {Achasov}\ \emph {et~al.}(2015)\citenamefont {Achasov}
  \emph {et~al.}}]{Achasov:2015mek}%
  \BibitemOpen
  \bibfield  {author} {\bibinfo {author} {\bibfnamefont {M.~N.}\ \bibnamefont
  {Achasov}} \emph {et~al.} (\bibinfo {collaboration} {SND}),\ }\href {\doibase
  10.1103/PhysRevD.91.092010} {\bibfield  {journal} {\bibinfo  {journal} {Phys.
  Rev.}\ }\textbf {\bibinfo {volume} {D91}},\ \bibinfo {pages} {092010}
  (\bibinfo {year} {2015})},\ \Eprint {http://arxiv.org/abs/1504.01245}
  {arXiv:1504.01245 [hep-ex]}\BibitemShut {NoStop}%
\bibitem [{\citenamefont {Achasov}\ \emph
  {et~al.}(2018{\natexlab{d}})\citenamefont {Achasov} \emph
  {et~al.}}]{Achasov:2018idb}%
  \BibitemOpen
  \bibfield  {author} {\bibinfo {author} {\bibfnamefont {M.~N.}\ \bibnamefont
  {Achasov}} \emph {et~al.} (\bibinfo {collaboration} {SND}),\ }\href {\doibase
  10.1103/PhysRevD.98.052007} {\bibfield  {journal} {\bibinfo  {journal} {Phys.
  Rev.}\ }\textbf {\bibinfo {volume} {D98}},\ \bibinfo {pages} {052007}
  (\bibinfo {year} {2018}{\natexlab{d}})},\ \Eprint
  {http://arxiv.org/abs/1806.07609} {arXiv:1806.07609 [hep-ex]}\BibitemShut
  {NoStop}%
\bibitem [{\citenamefont {Achasov}\ \emph
  {et~al.}(2020{\natexlab{b}})\citenamefont {Achasov} \emph
  {et~al.}}]{Achasov:2019wtd}%
  \BibitemOpen
  \bibfield  {author} {\bibinfo {author} {\bibfnamefont {M.~N.}\ \bibnamefont
  {Achasov}} \emph {et~al.} (\bibinfo {collaboration} {SND}),\ }\href {\doibase
  10.1016/j.physletb.2019.135074} {\bibfield  {journal} {\bibinfo  {journal}
  {Phys. Lett.}\ }\textbf {\bibinfo {volume} {B800}},\ \bibinfo {pages}
  {135074} (\bibinfo {year} {2020}{\natexlab{b}})},\ \Eprint
  {http://arxiv.org/abs/1906.03838} {arXiv:1906.03838 [hep-ex]}\BibitemShut
  {NoStop}%
\bibitem [{\citenamefont {Adler}(1969)}]{Adler:1969gk}%
  \BibitemOpen
  \bibfield  {author} {\bibinfo {author} {\bibfnamefont {S.~L.}\ \bibnamefont
  {Adler}},\ }\href {\doibase 10.1103/PhysRev.177.2426} {\bibfield  {journal}
  {\bibinfo  {journal} {Phys. Rev.}\ }\textbf {\bibinfo {volume} {177}},\
  \bibinfo {pages} {2426} (\bibinfo {year} {1969})}\BibitemShut {NoStop}%
\bibitem [{\citenamefont {Bell}\ and\ \citenamefont
  {Jackiw}(1969)}]{Bell:1969ts}%
  \BibitemOpen
  \bibfield  {author} {\bibinfo {author} {\bibfnamefont {J.~S.}\ \bibnamefont
  {Bell}}\ and\ \bibinfo {author} {\bibfnamefont {R.}~\bibnamefont {Jackiw}},\
  }\href {\doibase 10.1007/BF02823296} {\bibfield  {journal} {\bibinfo
  {journal} {Nuovo Cim.}\ }\textbf {\bibinfo {volume} {A60}},\ \bibinfo {pages}
  {47} (\bibinfo {year} {1969})}\BibitemShut {NoStop}%
\bibitem [{\citenamefont {Moussallam}(1995)}]{Moussallam:1994xp}%
  \BibitemOpen
  \bibfield  {author} {\bibinfo {author} {\bibfnamefont {B.}~\bibnamefont
  {Moussallam}},\ }\href {\doibase 10.1103/PhysRevD.51.4939} {\bibfield
  {journal} {\bibinfo  {journal} {Phys. Rev.}\ }\textbf {\bibinfo {volume}
  {D51}},\ \bibinfo {pages} {4939} (\bibinfo {year} {1995})},\ \Eprint
  {http://arxiv.org/abs/hep-ph/9407402} {arXiv:hep-ph/9407402
  [hep-ph]}\BibitemShut {NoStop}%
\bibitem [{\citenamefont {Goity}\ \emph {et~al.}(2002)\citenamefont {Goity},
  \citenamefont {Bernstein},\ and\ \citenamefont {Holstein}}]{Goity:2002nn}%
  \BibitemOpen
  \bibfield  {author} {\bibinfo {author} {\bibfnamefont {J.~L.}\ \bibnamefont
  {Goity}}, \bibinfo {author} {\bibfnamefont {A.~M.}\ \bibnamefont
  {Bernstein}}, \ and\ \bibinfo {author} {\bibfnamefont {B.~R.}\ \bibnamefont
  {Holstein}},\ }\href {\doibase 10.1103/PhysRevD.66.076014} {\bibfield
  {journal} {\bibinfo  {journal} {Phys. Rev.}\ }\textbf {\bibinfo {volume}
  {D66}},\ \bibinfo {pages} {076014} (\bibinfo {year} {2002})},\ \Eprint
  {http://arxiv.org/abs/hep-ph/0206007} {arXiv:hep-ph/0206007
  [hep-ph]}\BibitemShut {NoStop}%
\bibitem [{\citenamefont {Ananthanarayan}\ and\ \citenamefont
  {Moussallam}(2002)}]{Ananthanarayan:2002kj}%
  \BibitemOpen
  \bibfield  {author} {\bibinfo {author} {\bibfnamefont {B.}~\bibnamefont
  {Ananthanarayan}}\ and\ \bibinfo {author} {\bibfnamefont {B.}~\bibnamefont
  {Moussallam}},\ }\href {\doibase 10.1088/1126-6708/2002/05/052} {\bibfield
  {journal} {\bibinfo  {journal} {JHEP}\ }\textbf {\bibinfo {volume} {05}},\
  \bibinfo {pages} {052} (\bibinfo {year} {2002})},\ \Eprint
  {http://arxiv.org/abs/hep-ph/0205232} {arXiv:hep-ph/0205232
  [hep-ph]}\BibitemShut {NoStop}%
\bibitem [{\citenamefont {Kampf}\ and\ \citenamefont
  {Moussallam}(2009)}]{Kampf:2009tk}%
  \BibitemOpen
  \bibfield  {author} {\bibinfo {author} {\bibfnamefont {K.}~\bibnamefont
  {Kampf}}\ and\ \bibinfo {author} {\bibfnamefont {B.}~\bibnamefont
  {Moussallam}},\ }\href {\doibase 10.1103/PhysRevD.79.076005} {\bibfield
  {journal} {\bibinfo  {journal} {Phys. Rev.}\ }\textbf {\bibinfo {volume}
  {D79}},\ \bibinfo {pages} {076005} (\bibinfo {year} {2009})},\ \Eprint
  {http://arxiv.org/abs/0901.4688} {arXiv:0901.4688 [hep-ph]}\BibitemShut
  {NoStop}%
\bibitem [{\citenamefont {Ioffe}\ and\ \citenamefont
  {Oganesian}(2007)}]{Ioffe:2007pl}%
  \BibitemOpen
  \bibfield  {author} {\bibinfo {author} {\bibfnamefont {B.~L.}\ \bibnamefont
  {Ioffe}}\ and\ \bibinfo {author} {\bibfnamefont {A.~G.}\ \bibnamefont
  {Oganesian}},\ }\href {\doibase 10.1016/j.physletb.2007.02.021} {\bibfield
  {journal} {\bibinfo  {journal} {Phys. Lett.}\ }\textbf {\bibinfo {volume}
  {B647}},\ \bibinfo {pages} {389} (\bibinfo {year} {2007})},\ \Eprint
  {http://arxiv.org/abs/hep-ph/0701077} {arXiv:hep-ph/0701077
  [hep-ph]}\BibitemShut {NoStop}%
\bibitem [{\citenamefont {Larin}\ \emph {et~al.}(2011)\citenamefont {Larin}
  \emph {et~al.}}]{Larin:2010kq}%
  \BibitemOpen
  \bibfield  {author} {\bibinfo {author} {\bibfnamefont {I.}~\bibnamefont
  {Larin}} \emph {et~al.} (\bibinfo {collaboration} {PrimEx}),\ }\href
  {\doibase 10.1103/PhysRevLett.106.162303} {\bibfield  {journal} {\bibinfo
  {journal} {Phys. Rev. Lett.}\ }\textbf {\bibinfo {volume} {106}},\ \bibinfo
  {pages} {162303} (\bibinfo {year} {2011})},\ \Eprint
  {http://arxiv.org/abs/1009.1681} {arXiv:1009.1681 [nucl-ex]}\BibitemShut
  {NoStop}%
\bibitem [{\citenamefont {Redmer}(2018{\natexlab{b}})}]{Redmer:2018uew}%
  \BibitemOpen
  \bibfield  {author} {\bibinfo {author} {\bibfnamefont {C.~F.}\ \bibnamefont
  {Redmer}} (\bibinfo {collaboration} {BESIII}),\ }\href@noop {} {\  (\bibinfo
  {year} {2018}{\natexlab{b}})},\ \Eprint {http://arxiv.org/abs/1810.00654}
  {arXiv:1810.00654 [hep-ex]}\BibitemShut {NoStop}%
\bibitem [{\citenamefont {Farzanpay}\ \emph {et~al.}(1992)\citenamefont
  {Farzanpay} \emph {et~al.}}]{Farzanpay:1992pz}%
  \BibitemOpen
  \bibfield  {author} {\bibinfo {author} {\bibfnamefont {F.}~\bibnamefont
  {Farzanpay}} \emph {et~al.},\ }\href {\doibase 10.1016/0370-2693(92)90577-Q}
  {\bibfield  {journal} {\bibinfo  {journal} {Phys. Lett.}\ }\textbf {\bibinfo
  {volume} {B278}},\ \bibinfo {pages} {413} (\bibinfo {year}
  {1992})}\BibitemShut {NoStop}%
\bibitem [{\citenamefont {Meijer~Drees}\ \emph {et~al.}(1992)\citenamefont
  {Meijer~Drees} \emph {et~al.}}]{MeijerDrees:1992qb}%
  \BibitemOpen
  \bibfield  {author} {\bibinfo {author} {\bibfnamefont {R.}~\bibnamefont
  {Meijer~Drees}} \emph {et~al.} (\bibinfo {collaboration} {SINDRUM-I}),\
  }\href {\doibase 10.1103/PhysRevD.45.1439} {\bibfield  {journal} {\bibinfo
  {journal} {Phys. Rev.}\ }\textbf {\bibinfo {volume} {D45}},\ \bibinfo {pages}
  {1439} (\bibinfo {year} {1992})}\BibitemShut {NoStop}%
\bibitem [{\citenamefont {Husek}\ \emph {et~al.}(2015)\citenamefont {Husek},
  \citenamefont {Kampf},\ and\ \citenamefont {Novotn\'y}}]{Husek:2015sma}%
  \BibitemOpen
  \bibfield  {author} {\bibinfo {author} {\bibfnamefont {T.}~\bibnamefont
  {Husek}}, \bibinfo {author} {\bibfnamefont {K.}~\bibnamefont {Kampf}}, \ and\
  \bibinfo {author} {\bibfnamefont {J.}~\bibnamefont {Novotn\'y}},\ }\href
  {\doibase 10.1103/PhysRevD.92.054027} {\bibfield  {journal} {\bibinfo
  {journal} {Phys. Rev.}\ }\textbf {\bibinfo {volume} {D92}},\ \bibinfo {pages}
  {054027} (\bibinfo {year} {2015})},\ \Eprint
  {http://arxiv.org/abs/1504.06178} {arXiv:1504.06178 [hep-ph]}\BibitemShut
  {NoStop}%
\bibitem [{\citenamefont {Husek}\ \emph {et~al.}(2018)\citenamefont {Husek},
  \citenamefont {Kampf}, \citenamefont {Leupold},\ and\ \citenamefont
  {Novotn\'y}}]{Husek:2017vmo}%
  \BibitemOpen
  \bibfield  {author} {\bibinfo {author} {\bibfnamefont {T.}~\bibnamefont
  {Husek}}, \bibinfo {author} {\bibfnamefont {K.}~\bibnamefont {Kampf}},
  \bibinfo {author} {\bibfnamefont {S.}~\bibnamefont {Leupold}}, \ and\
  \bibinfo {author} {\bibfnamefont {J.}~\bibnamefont {Novotn\'y}},\ }\href
  {\doibase 10.1103/PhysRevD.97.096013} {\bibfield  {journal} {\bibinfo
  {journal} {Phys. Rev.}\ }\textbf {\bibinfo {volume} {D97}},\ \bibinfo {pages}
  {096013} (\bibinfo {year} {2018})},\ \Eprint
  {http://arxiv.org/abs/1711.11001} {arXiv:1711.11001 [hep-ph]}\BibitemShut
  {NoStop}%
\bibitem [{\citenamefont {Landsberg}(1985)}]{Landsberg:1986fd}%
  \BibitemOpen
  \bibfield  {author} {\bibinfo {author} {\bibfnamefont {L.~G.}\ \bibnamefont
  {Landsberg}},\ }\href {\doibase 10.1016/0370-1573(85)90129-2} {\bibfield
  {journal} {\bibinfo  {journal} {Phys. Rept.}\ }\textbf {\bibinfo {volume}
  {128}},\ \bibinfo {pages} {301} (\bibinfo {year} {1985})}\BibitemShut
  {NoStop}%
\bibitem [{\citenamefont {Abouzaid}\ \emph {et~al.}(2008)\citenamefont
  {Abouzaid} \emph {et~al.}}]{Abouzaid:2008cd}%
  \BibitemOpen
  \bibfield  {author} {\bibinfo {author} {\bibfnamefont {E.}~\bibnamefont
  {Abouzaid}} \emph {et~al.} (\bibinfo {collaboration} {KTeV}),\ }\href
  {\doibase 10.1103/PhysRevLett.100.182001} {\bibfield  {journal} {\bibinfo
  {journal} {Phys. Rev. Lett.}\ }\textbf {\bibinfo {volume} {100}},\ \bibinfo
  {pages} {182001} (\bibinfo {year} {2008})},\ \Eprint
  {http://arxiv.org/abs/0802.2064} {arXiv:0802.2064 [hep-ex]}\BibitemShut
  {NoStop}%
\bibitem [{\citenamefont {Ambrosino}\ \emph
  {et~al.}(2011{\natexlab{b}})\citenamefont {Ambrosino} \emph
  {et~al.}}]{KLOE2:2011aa}%
  \BibitemOpen
  \bibfield  {author} {\bibinfo {author} {\bibfnamefont {F.}~\bibnamefont
  {Ambrosino}} \emph {et~al.} (\bibinfo {collaboration} {KLOE, KLOE-2}),\
  }\href {\doibase 10.1016/j.physletb.2011.07.033} {\bibfield  {journal}
  {\bibinfo  {journal} {Phys. Lett.}\ }\textbf {\bibinfo {volume} {B702}},\
  \bibinfo {pages} {324} (\bibinfo {year} {2011}{\natexlab{b}})},\ \Eprint
  {http://arxiv.org/abs/1105.6067} {arXiv:1105.6067 [hep-ex]}\BibitemShut
  {NoStop}%
\bibitem [{\citenamefont {Akhmetshin}\ \emph {et~al.}(2005)\citenamefont
  {Akhmetshin} \emph {et~al.}}]{Akhmetshin:2004gw}%
  \BibitemOpen
  \bibfield  {author} {\bibinfo {author} {\bibfnamefont {R.~R.}\ \bibnamefont
  {Akhmetshin}} \emph {et~al.} (\bibinfo {collaboration} {CMD-2}),\ }\href
  {\doibase 10.1016/j.physletb.2004.11.020} {\bibfield  {journal} {\bibinfo
  {journal} {Phys. Lett.}\ }\textbf {\bibinfo {volume} {B605}},\ \bibinfo
  {pages} {26} (\bibinfo {year} {2005})},\ \Eprint
  {http://arxiv.org/abs/hep-ex/0409030} {arXiv:hep-ex/0409030
  [hep-ex]}\BibitemShut {NoStop}%
\bibitem [{\citenamefont {Achasov}\ \emph
  {et~al.}(2006{\natexlab{b}})\citenamefont {Achasov} \emph
  {et~al.}}]{Achasov:2006dv}%
  \BibitemOpen
  \bibfield  {author} {\bibinfo {author} {\bibfnamefont {M.~N.}\ \bibnamefont
  {Achasov}} \emph {et~al.} (\bibinfo {collaboration} {SND}),\ }\href {\doibase
  10.1103/PhysRevD.74.014016} {\bibfield  {journal} {\bibinfo  {journal} {Phys.
  Rev.}\ }\textbf {\bibinfo {volume} {D74}},\ \bibinfo {pages} {014016}
  (\bibinfo {year} {2006}{\natexlab{b}})},\ \Eprint
  {http://arxiv.org/abs/hep-ex/0605109} {arXiv:hep-ex/0605109
  [hep-ex]}\BibitemShut {NoStop}%
\bibitem [{\citenamefont {Achasov}\ \emph
  {et~al.}(2014{\natexlab{c}})\citenamefont {Achasov} \emph
  {et~al.}}]{Achasov:2013eli}%
  \BibitemOpen
  \bibfield  {author} {\bibinfo {author} {\bibfnamefont {M.~N.}\ \bibnamefont
  {Achasov}} \emph {et~al.} (\bibinfo {collaboration} {SND}),\ }\href {\doibase
  10.1103/PhysRevD.90.032002} {\bibfield  {journal} {\bibinfo  {journal} {Phys.
  Rev.}\ }\textbf {\bibinfo {volume} {D90}},\ \bibinfo {pages} {032002}
  (\bibinfo {year} {2014}{\natexlab{c}})},\ \Eprint
  {http://arxiv.org/abs/1312.7078} {arXiv:1312.7078 [hep-ex]}\BibitemShut
  {NoStop}%
\bibitem [{\citenamefont {Pedlar}\ \emph {et~al.}(2009)\citenamefont {Pedlar}
  \emph {et~al.}}]{Pedlar:2009aa}%
  \BibitemOpen
  \bibfield  {author} {\bibinfo {author} {\bibfnamefont {T.~K.}\ \bibnamefont
  {Pedlar}} \emph {et~al.} (\bibinfo {collaboration} {CLEO}),\ }\href {\doibase
  10.1103/PhysRevD.79.111101} {\bibfield  {journal} {\bibinfo  {journal} {Phys.
  Rev.}\ }\textbf {\bibinfo {volume} {D79}},\ \bibinfo {pages} {111101}
  (\bibinfo {year} {2009})},\ \Eprint {http://arxiv.org/abs/0904.1394}
  {arXiv:0904.1394 [hep-ex]}\BibitemShut {NoStop}%
\bibitem [{\citenamefont {Aubert}\ \emph
  {et~al.}(2006{\natexlab{c}})\citenamefont {Aubert} \emph
  {et~al.}}]{Aubert:2006cy}%
  \BibitemOpen
  \bibfield  {author} {\bibinfo {author} {\bibfnamefont {B.}~\bibnamefont
  {Aubert}} \emph {et~al.} (\bibinfo {collaboration} {BABAR}),\ }\href
  {\doibase 10.1103/PhysRevD.74.012002} {\bibfield  {journal} {\bibinfo
  {journal} {Phys. Rev.}\ }\textbf {\bibinfo {volume} {D74}},\ \bibinfo {pages}
  {012002} (\bibinfo {year} {2006}{\natexlab{c}})},\ \Eprint
  {http://arxiv.org/abs/hep-ex/0605018} {arXiv:hep-ex/0605018
  [hep-ex]}\BibitemShut {NoStop}%
\bibitem [{\citenamefont {Akhmetshin}\ \emph {et~al.}(2001)\citenamefont
  {Akhmetshin} \emph {et~al.}}]{Akhmetshin:2001hm}%
  \BibitemOpen
  \bibfield  {author} {\bibinfo {author} {\bibfnamefont {R.~R.}\ \bibnamefont
  {Akhmetshin}} \emph {et~al.} (\bibinfo {collaboration} {CMD-2}),\ }\href
  {\doibase 10.1016/S0370-2693(01)00567-6} {\bibfield  {journal} {\bibinfo
  {journal} {Phys. Lett.}\ }\textbf {\bibinfo {volume} {B509}},\ \bibinfo
  {pages} {217} (\bibinfo {year} {2001})},\ \Eprint
  {http://arxiv.org/abs/hep-ex/0103043} {arXiv:hep-ex/0103043
  [hep-ex]}\BibitemShut {NoStop}%
\bibitem [{\citenamefont {Ambrosino}\ \emph
  {et~al.}(2007{\natexlab{a}})\citenamefont {Ambrosino} \emph
  {et~al.}}]{Ambrosino:2006gk}%
  \BibitemOpen
  \bibfield  {author} {\bibinfo {author} {\bibfnamefont {F.}~\bibnamefont
  {Ambrosino}} \emph {et~al.} (\bibinfo {collaboration} {KLOE}),\ }\href
  {\doibase 10.1016/j.physletb.2007.03.032} {\bibfield  {journal} {\bibinfo
  {journal} {Phys. Lett. B}\ }\textbf {\bibinfo {volume} {648}},\ \bibinfo
  {pages} {267} (\bibinfo {year} {2007}{\natexlab{a}})},\ \Eprint
  {http://arxiv.org/abs/hep-ex/0612029} {arXiv:hep-ex/0612029}\BibitemShut
  {NoStop}%
\bibitem [{\citenamefont {Aulchenko}\ \emph {et~al.}(2003)\citenamefont
  {Aulchenko} \emph {et~al.}}]{Aulchenko:2003wv}%
  \BibitemOpen
  \bibfield  {author} {\bibinfo {author} {\bibfnamefont {V.~M.}\ \bibnamefont
  {Aulchenko}} \emph {et~al.} (\bibinfo {collaboration} {SND}),\ }\href
  {\doibase 10.1134/1.1600793} {\bibfield  {journal} {\bibinfo  {journal} {J.
  Exp. Theor. Phys.}\ }\textbf {\bibinfo {volume} {97}},\ \bibinfo {pages} {24}
  (\bibinfo {year} {2003})},\ \bibinfo {note} {[Zh. Eksp. Teor. Fiz. {\bf 124},
  28 (2003)]}\BibitemShut {NoStop}%
\bibitem [{\citenamefont {Akhmetshin}\ \emph
  {et~al.}(2000{\natexlab{c}})\citenamefont {Akhmetshin} \emph
  {et~al.}}]{Akhmetshin:2000hp}%
  \BibitemOpen
  \bibfield  {author} {\bibinfo {author} {\bibfnamefont {R.}~\bibnamefont
  {Akhmetshin}} \emph {et~al.} (\bibinfo {collaboration} {CMD-2}),\ }\href
  {\doibase 10.1016/S0370-2693(00)01188-6} {\bibfield  {journal} {\bibinfo
  {journal} {Phys. Lett. B}\ }\textbf {\bibinfo {volume} {494}},\ \bibinfo
  {pages} {26} (\bibinfo {year} {2000}{\natexlab{c}})},\ \Eprint
  {http://arxiv.org/abs/hep-ex/0010018} {arXiv:hep-ex/0010018}\BibitemShut
  {NoStop}%
\bibitem [{\citenamefont {Anastasi}\ \emph {et~al.}(2016)\citenamefont
  {Anastasi} \emph {et~al.}}]{Anastasi:2016qga}%
  \BibitemOpen
  \bibfield  {author} {\bibinfo {author} {\bibfnamefont {A.}~\bibnamefont
  {Anastasi}} \emph {et~al.} (\bibinfo {collaboration} {KLOE-2}),\ }\href
  {\doibase 10.1016/j.physletb.2016.04.015} {\bibfield  {journal} {\bibinfo
  {journal} {Phys. Lett.}\ }\textbf {\bibinfo {volume} {B757}},\ \bibinfo
  {pages} {362} (\bibinfo {year} {2016})},\ \Eprint
  {http://arxiv.org/abs/1601.06565} {arXiv:1601.06565 [hep-ex]}\BibitemShut
  {NoStop}%
\bibitem [{\citenamefont {Babusci}\ \emph {et~al.}(2015)\citenamefont {Babusci}
  \emph {et~al.}}]{Babusci:2014ldz}%
  \BibitemOpen
  \bibfield  {author} {\bibinfo {author} {\bibfnamefont {D.}~\bibnamefont
  {Babusci}} \emph {et~al.} (\bibinfo {collaboration} {KLOE-2}),\ }\href
  {\doibase 10.1016/j.physletb.2015.01.011} {\bibfield  {journal} {\bibinfo
  {journal} {Phys. Lett.}\ }\textbf {\bibinfo {volume} {B742}},\ \bibinfo
  {pages} {1} (\bibinfo {year} {2015})},\ \Eprint
  {http://arxiv.org/abs/1409.4582} {arXiv:1409.4582 [hep-ex]}\BibitemShut
  {NoStop}%
\bibitem [{\citenamefont {Babusci}\ \emph {et~al.}(2012)\citenamefont
  {Babusci}, \citenamefont {Czy\.z}, \citenamefont {Gonnella}, \citenamefont
  {Ivashyn}, \citenamefont {Mascolo}, \citenamefont {Messi}, \citenamefont
  {Moricciani}, \citenamefont {Nyffeler},\ and\ \citenamefont
  {Venanzoni}}]{Babusci:2011bg}%
  \BibitemOpen
  \bibfield  {author} {\bibinfo {author} {\bibfnamefont {D.}~\bibnamefont
  {Babusci}}, \bibinfo {author} {\bibfnamefont {H.}~\bibnamefont {Czy\.z}},
  \bibinfo {author} {\bibfnamefont {F.}~\bibnamefont {Gonnella}}, \bibinfo
  {author} {\bibfnamefont {S.}~\bibnamefont {Ivashyn}}, \bibinfo {author}
  {\bibfnamefont {M.}~\bibnamefont {Mascolo}}, \bibinfo {author} {\bibfnamefont
  {R.}~\bibnamefont {Messi}}, \bibinfo {author} {\bibfnamefont
  {D.}~\bibnamefont {Moricciani}}, \bibinfo {author} {\bibfnamefont
  {A.}~\bibnamefont {Nyffeler}}, \ and\ \bibinfo {author} {\bibfnamefont
  {G.}~\bibnamefont {Venanzoni}},\ }\href {\doibase
  10.1140/epjc/s10052-012-1917-1} {\bibfield  {journal} {\bibinfo  {journal}
  {Eur. Phys. J.}\ }\textbf {\bibinfo {volume} {C72}},\ \bibinfo {pages} {1917}
  (\bibinfo {year} {2012})},\ \Eprint {http://arxiv.org/abs/1109.2461}
  {arXiv:1109.2461 [hep-ph]}\BibitemShut {NoStop}%
\bibitem [{\citenamefont {Knecht}\ and\ \citenamefont
  {Nyffeler}(2001)}]{Knecht:2001xc}%
  \BibitemOpen
  \bibfield  {author} {\bibinfo {author} {\bibfnamefont {M.}~\bibnamefont
  {Knecht}}\ and\ \bibinfo {author} {\bibfnamefont {A.}~\bibnamefont
  {Nyffeler}},\ }\href {\doibase 10.1007/s100520100755} {\bibfield  {journal}
  {\bibinfo  {journal} {Eur. Phys. J.}\ }\textbf {\bibinfo {volume} {C21}},\
  \bibinfo {pages} {659} (\bibinfo {year} {2001})},\ \Eprint
  {http://arxiv.org/abs/hep-ph/0106034} {arXiv:hep-ph/0106034
  [hep-ph]}\BibitemShut {NoStop}%
\bibitem [{\citenamefont {Curciarello}(2019)}]{Curciarello:2019}%
  \BibitemOpen
  \bibfield  {author} {\bibinfo {author} {\bibfnamefont {F.}~\bibnamefont
  {Curciarello}} (\bibinfo {collaboration} {KLOE-2}),\ }\href@noop {} {\enquote
  {\bibinfo {title} {{Recent results on hadron physics at KLOE-2}},}\ }\bibinfo
  {howpublished}
  {\url{https://indico.inp.nsk.su/event/15/session/3/contribution/48/material/slides/0.pdf}}
  (\bibinfo {year} {2019}),\ \bibinfo {note} {talk at the International
  Workshop on $e^+e^-$ collisions from $\phi$ to $\psi$ (PhiPsi19)
  Novosibirsk}\BibitemShut {NoStop}%
\bibitem [{\citenamefont {Archilli}\ \emph {et~al.}(2010)\citenamefont
  {Archilli}, \citenamefont {Babusci}, \citenamefont {Badoni}, \citenamefont
  {Beretta}, \citenamefont {Gonnella}, \citenamefont {Iafolla}, \citenamefont
  {Messi}, \citenamefont {Moricciani},\ and\ \citenamefont
  {Quintieri}}]{Archilli:2010zza}%
  \BibitemOpen
  \bibfield  {author} {\bibinfo {author} {\bibfnamefont {F.}~\bibnamefont
  {Archilli}}, \bibinfo {author} {\bibfnamefont {D.}~\bibnamefont {Babusci}},
  \bibinfo {author} {\bibfnamefont {D.}~\bibnamefont {Badoni}}, \bibinfo
  {author} {\bibfnamefont {M.}~\bibnamefont {Beretta}}, \bibinfo {author}
  {\bibfnamefont {F.}~\bibnamefont {Gonnella}}, \bibinfo {author}
  {\bibfnamefont {L.}~\bibnamefont {Iafolla}}, \bibinfo {author} {\bibfnamefont
  {R.}~\bibnamefont {Messi}}, \bibinfo {author} {\bibfnamefont
  {D.}~\bibnamefont {Moricciani}}, \ and\ \bibinfo {author} {\bibfnamefont
  {L.}~\bibnamefont {Quintieri}},\ }\href {\doibase 10.1016/j.nima.2009.06.082}
  {\bibfield  {journal} {\bibinfo  {journal} {Nucl. Instrum. Meth.}\ }\textbf
  {\bibinfo {volume} {A617}},\ \bibinfo {pages} {266} (\bibinfo {year}
  {2010})}\BibitemShut {NoStop}%
\bibitem [{JLa(2004)}]{JLab12pCDR:2014}%
  \BibitemOpen
  \href@noop {} {\enquote {\bibinfo {title} {{Pre-Conceptual Design Report
  (pCDR) for the Science and Experimental Equipment for the 12 GeV Upgrade of
  CEBAF}},}\ }\bibinfo {howpublished}
  {\url{https://www.jlab.org/div_dept/physics_division/pCDR_public/pCDR_final/pCDR_final.pdf}}
  (\bibinfo {year} {2004})\BibitemShut {NoStop}%
\bibitem [{\citenamefont {Gan}(2015)}]{Gan:2015nyc}%
  \BibitemOpen
  \bibfield  {author} {\bibinfo {author} {\bibfnamefont {L.}~\bibnamefont
  {Gan}},\ }\href {\doibase 10.22323/1.253.0017} {\bibfield  {journal}
  {\bibinfo  {journal} {PoS}\ }\textbf {\bibinfo {volume} {CD15}},\ \bibinfo
  {pages} {017} (\bibinfo {year} {2015})}\BibitemShut {NoStop}%
\bibitem [{\citenamefont {Redmer}(2018{\natexlab{c}})}]{Redmer:2018gah}%
  \BibitemOpen
  \bibfield  {author} {\bibinfo {author} {\bibfnamefont {C.~F.}\ \bibnamefont
  {Redmer}} (\bibinfo {collaboration} {BESIII}),\ }\href {\doibase
  10.1051/epjconf/201816600017} {\bibfield  {journal} {\bibinfo  {journal} {EPJ
  Web Conf.}\ }\textbf {\bibinfo {volume} {166}},\ \bibinfo {pages} {00017}
  (\bibinfo {year} {2018}{\natexlab{c}})}\BibitemShut {NoStop}%
\bibitem [{\citenamefont {Nyffeler}(2016)}]{Nyffeler:2016gnb}%
  \BibitemOpen
  \bibfield  {author} {\bibinfo {author} {\bibfnamefont {A.}~\bibnamefont
  {Nyffeler}},\ }\href {\doibase 10.1103/PhysRevD.94.053006} {\bibfield
  {journal} {\bibinfo  {journal} {Phys. Rev.}\ }\textbf {\bibinfo {volume}
  {D94}},\ \bibinfo {pages} {053006} (\bibinfo {year} {2016})},\ \Eprint
  {http://arxiv.org/abs/1602.03398} {arXiv:1602.03398 [hep-ph]}\BibitemShut
  {NoStop}%
\bibitem [{\citenamefont {Morgan}\ \emph {et~al.}(1994)\citenamefont {Morgan},
  \citenamefont {Pennington},\ and\ \citenamefont {Whalley}}]{Morgan:1994ip}%
  \BibitemOpen
  \bibfield  {author} {\bibinfo {author} {\bibfnamefont {D.}~\bibnamefont
  {Morgan}}, \bibinfo {author} {\bibfnamefont {M.~R.}\ \bibnamefont
  {Pennington}}, \ and\ \bibinfo {author} {\bibfnamefont {M.~R.}\ \bibnamefont
  {Whalley}},\ }\href {\doibase 10.1088/0954-3899/20/8a/001} {\bibfield
  {journal} {\bibinfo  {journal} {J. Phys.}\ }\textbf {\bibinfo {volume}
  {G20}},\ \bibinfo {pages} {A1} (\bibinfo {year} {1994})}\BibitemShut
  {NoStop}%
\bibitem [{\citenamefont {Whalley}(2001)}]{Whalley:2001mk}%
  \BibitemOpen
  \bibfield  {author} {\bibinfo {author} {\bibfnamefont {M.~R.}\ \bibnamefont
  {Whalley}},\ }\href {\doibase 10.1088/0954-3899/27/12A/301} {\bibfield
  {journal} {\bibinfo  {journal} {J. Phys.}\ }\textbf {\bibinfo {volume}
  {G27}},\ \bibinfo {pages} {A1} (\bibinfo {year} {2001})}\BibitemShut
  {NoStop}%
\bibitem [{\citenamefont {Uehara}\ \emph {et~al.}(2008)\citenamefont {Uehara}
  \emph {et~al.}}]{Uehara:2008ep}%
  \BibitemOpen
  \bibfield  {author} {\bibinfo {author} {\bibfnamefont {S.}~\bibnamefont
  {Uehara}} \emph {et~al.} (\bibinfo {collaboration} {Belle}),\ }\href
  {\doibase 10.1103/PhysRevD.78.052004} {\bibfield  {journal} {\bibinfo
  {journal} {Phys. Rev.}\ }\textbf {\bibinfo {volume} {D78}},\ \bibinfo {pages}
  {052004} (\bibinfo {year} {2008})},\ \Eprint {http://arxiv.org/abs/0805.3387}
  {arXiv:0805.3387 [hep-ex]}\BibitemShut {NoStop}%
\bibitem [{\citenamefont {Uehara}\ \emph
  {et~al.}(2009{\natexlab{a}})\citenamefont {Uehara} \emph
  {et~al.}}]{Uehara:2009cka}%
  \BibitemOpen
  \bibfield  {author} {\bibinfo {author} {\bibfnamefont {S.}~\bibnamefont
  {Uehara}} \emph {et~al.} (\bibinfo {collaboration} {Belle}),\ }\href
  {\doibase 10.1103/PhysRevD.79.052009} {\bibfield  {journal} {\bibinfo
  {journal} {Phys. Rev.}\ }\textbf {\bibinfo {volume} {D79}},\ \bibinfo {pages}
  {052009} (\bibinfo {year} {2009}{\natexlab{a}})},\ \Eprint
  {http://arxiv.org/abs/0903.3697} {arXiv:0903.3697 [hep-ex]}\BibitemShut
  {NoStop}%
\bibitem [{\citenamefont {Nakazawa}\ \emph {et~al.}(2005)\citenamefont
  {Nakazawa} \emph {et~al.}}]{Nakazawa:2004gu}%
  \BibitemOpen
  \bibfield  {author} {\bibinfo {author} {\bibfnamefont {H.}~\bibnamefont
  {Nakazawa}} \emph {et~al.} (\bibinfo {collaboration} {Belle}),\ }\href
  {\doibase 10.1016/j.physletb.2005.03.067} {\bibfield  {journal} {\bibinfo
  {journal} {Phys. Lett.}\ }\textbf {\bibinfo {volume} {B615}},\ \bibinfo
  {pages} {39} (\bibinfo {year} {2005})},\ \Eprint
  {http://arxiv.org/abs/hep-ex/0412058} {arXiv:hep-ex/0412058
  [hep-ex]}\BibitemShut {NoStop}%
\bibitem [{\citenamefont {Mori}\ \emph {et~al.}(2007)\citenamefont {Mori} \emph
  {et~al.}}]{Mori:2007bu}%
  \BibitemOpen
  \bibfield  {author} {\bibinfo {author} {\bibfnamefont {T.}~\bibnamefont
  {Mori}} \emph {et~al.} (\bibinfo {collaboration} {Belle}),\ }\href {\doibase
  10.1143/JPSJ.76.074102} {\bibfield  {journal} {\bibinfo  {journal} {J. Phys.
  Soc. Jap.}\ }\textbf {\bibinfo {volume} {76}},\ \bibinfo {pages} {074102}
  (\bibinfo {year} {2007})},\ \Eprint {http://arxiv.org/abs/0704.3538}
  {arXiv:0704.3538 [hep-ex]}\BibitemShut {NoStop}%
\bibitem [{\citenamefont {Uehara}\ \emph
  {et~al.}(2009{\natexlab{b}})\citenamefont {Uehara} \emph
  {et~al.}}]{Uehara:2009cf}%
  \BibitemOpen
  \bibfield  {author} {\bibinfo {author} {\bibfnamefont {S.}~\bibnamefont
  {Uehara}} \emph {et~al.} (\bibinfo {collaboration} {Belle}),\ }\href
  {\doibase 10.1103/PhysRevD.80.032001} {\bibfield  {journal} {\bibinfo
  {journal} {Phys. Rev.}\ }\textbf {\bibinfo {volume} {D80}},\ \bibinfo {pages}
  {032001} (\bibinfo {year} {2009}{\natexlab{b}})},\ \Eprint
  {http://arxiv.org/abs/0906.1464} {arXiv:0906.1464 [hep-ex]}\BibitemShut
  {NoStop}%
\bibitem [{\citenamefont {Uehara}\ \emph {et~al.}(2010)\citenamefont {Uehara}
  \emph {et~al.}}]{Uehara:2010mq}%
  \BibitemOpen
  \bibfield  {author} {\bibinfo {author} {\bibfnamefont {S.}~\bibnamefont
  {Uehara}} \emph {et~al.} (\bibinfo {collaboration} {Belle}),\ }\href
  {\doibase 10.1103/PhysRevD.82.114031} {\bibfield  {journal} {\bibinfo
  {journal} {Phys. Rev.}\ }\textbf {\bibinfo {volume} {D82}},\ \bibinfo {pages}
  {114031} (\bibinfo {year} {2010})},\ \Eprint {http://arxiv.org/abs/1007.3779}
  {arXiv:1007.3779 [hep-ex]}\BibitemShut {NoStop}%
\bibitem [{\citenamefont {Abe}\ \emph {et~al.}(2003)\citenamefont {Abe} \emph
  {et~al.}}]{Abe:2003vn}%
  \BibitemOpen
  \bibfield  {author} {\bibinfo {author} {\bibfnamefont {K.}~\bibnamefont
  {Abe}} \emph {et~al.} (\bibinfo {collaboration} {Belle}),\ }\href {\doibase
  10.1140/epjc/s2003-01468-9} {\bibfield  {journal} {\bibinfo  {journal} {Eur.
  Phys. J.}\ }\textbf {\bibinfo {volume} {C32}},\ \bibinfo {pages} {323}
  (\bibinfo {year} {2003})},\ \Eprint {http://arxiv.org/abs/hep-ex/0309077}
  {arXiv:hep-ex/0309077 [hep-ex]}\BibitemShut {NoStop}%
\bibitem [{\citenamefont {Uehara}\ \emph {et~al.}(2013)\citenamefont {Uehara}
  \emph {et~al.}}]{Uehara:2013mbo}%
  \BibitemOpen
  \bibfield  {author} {\bibinfo {author} {\bibfnamefont {S.}~\bibnamefont
  {Uehara}} \emph {et~al.} (\bibinfo {collaboration} {Belle}),\ }\href
  {\doibase 10.1093/ptep/ptt097} {\bibfield  {journal} {\bibinfo  {journal}
  {PTEP}\ }\textbf {\bibinfo {volume} {2013}},\ \bibinfo {pages} {123C01}
  (\bibinfo {year} {2013})},\ \Eprint {http://arxiv.org/abs/1307.7457}
  {arXiv:1307.7457 [hep-ex]}\BibitemShut {NoStop}%
\bibitem [{\citenamefont {Masuda}\ \emph {et~al.}(2018)\citenamefont {Masuda}
  \emph {et~al.}}]{Masuda:2017rhm}%
  \BibitemOpen
  \bibfield  {author} {\bibinfo {author} {\bibfnamefont {M.}~\bibnamefont
  {Masuda}} \emph {et~al.} (\bibinfo {collaboration} {Belle}),\ }\href
  {\doibase 10.1103/PhysRevD.97.052003} {\bibfield  {journal} {\bibinfo
  {journal} {Phys. Rev.}\ }\textbf {\bibinfo {volume} {D97}},\ \bibinfo {pages}
  {052003} (\bibinfo {year} {2018})},\ \Eprint
  {http://arxiv.org/abs/1712.02145} {arXiv:1712.02145 [hep-ex]}\BibitemShut
  {NoStop}%
\bibitem [{\citenamefont {Achasov}\ \emph
  {et~al.}(2000{\natexlab{a}})\citenamefont {Achasov} \emph
  {et~al.}}]{Achasov:2000zr}%
  \BibitemOpen
  \bibfield  {author} {\bibinfo {author} {\bibfnamefont {M.~N.}\ \bibnamefont
  {Achasov}} \emph {et~al.} (\bibinfo {collaboration} {SND}),\ }\href {\doibase
  10.1134/1.568352} {\bibfield  {journal} {\bibinfo  {journal} {JETP Lett.}\
  }\textbf {\bibinfo {volume} {71}},\ \bibinfo {pages} {355} (\bibinfo {year}
  {2000}{\natexlab{a}})},\ \bibinfo {note} {[Pisma Zh. Eksp. Teor. Fiz. {\bf
  71}, 519 (2000)]}\BibitemShut {NoStop}%
\bibitem [{\citenamefont {Achasov}\ \emph
  {et~al.}(2002{\natexlab{b}})\citenamefont {Achasov} \emph
  {et~al.}}]{Achasov:2002jv}%
  \BibitemOpen
  \bibfield  {author} {\bibinfo {author} {\bibfnamefont {M.~N.}\ \bibnamefont
  {Achasov}} \emph {et~al.} (\bibinfo {collaboration} {SND}),\ }\href {\doibase
  10.1016/S0370-2693(02)01825-7} {\bibfield  {journal} {\bibinfo  {journal}
  {Phys. Lett.}\ }\textbf {\bibinfo {volume} {B537}},\ \bibinfo {pages} {201}
  (\bibinfo {year} {2002}{\natexlab{b}})},\ \Eprint
  {http://arxiv.org/abs/hep-ex/0205068} {arXiv:hep-ex/0205068
  [hep-ex]}\BibitemShut {NoStop}%
\bibitem [{\citenamefont {Akhmetshin}\ \emph {et~al.}(2003)\citenamefont
  {Akhmetshin} \emph {et~al.}}]{Akhmetshin:2003ag}%
  \BibitemOpen
  \bibfield  {author} {\bibinfo {author} {\bibfnamefont {R.~R.}\ \bibnamefont
  {Akhmetshin}} \emph {et~al.} (\bibinfo {collaboration} {CMD-2}),\ }\href
  {\doibase 10.1016/S0370-2693(03)00595-1} {\bibfield  {journal} {\bibinfo
  {journal} {Phys. Lett.}\ }\textbf {\bibinfo {volume} {B562}},\ \bibinfo
  {pages} {173} (\bibinfo {year} {2003})},\ \Eprint
  {http://arxiv.org/abs/hep-ex/0304009} {arXiv:hep-ex/0304009
  [hep-ex]}\BibitemShut {NoStop}%
\bibitem [{\citenamefont {Akhmetshin}\ \emph
  {et~al.}(2004{\natexlab{c}})\citenamefont {Akhmetshin} \emph
  {et~al.}}]{Akhmetshin:2003rg}%
  \BibitemOpen
  \bibfield  {author} {\bibinfo {author} {\bibfnamefont {R.~R.}\ \bibnamefont
  {Akhmetshin}} \emph {et~al.} (\bibinfo {collaboration} {CMD-2}),\ }\href
  {\doibase 10.1016/j.physletb.2003.11.046} {\bibfield  {journal} {\bibinfo
  {journal} {Phys. Lett.}\ }\textbf {\bibinfo {volume} {B580}},\ \bibinfo
  {pages} {119} (\bibinfo {year} {2004}{\natexlab{c}})},\ \Eprint
  {http://arxiv.org/abs/hep-ex/0310012} {arXiv:hep-ex/0310012
  [hep-ex]}\BibitemShut {NoStop}%
\bibitem [{\citenamefont {Ambrosino}\ \emph
  {et~al.}(2007{\natexlab{b}})\citenamefont {Ambrosino} \emph
  {et~al.}}]{Ambrosino:2006hb}%
  \BibitemOpen
  \bibfield  {author} {\bibinfo {author} {\bibfnamefont {F.}~\bibnamefont
  {Ambrosino}} \emph {et~al.} (\bibinfo {collaboration} {KLOE}),\ }\href
  {\doibase 10.1140/epjc/s10052-006-0157-7} {\bibfield  {journal} {\bibinfo
  {journal} {Eur. Phys. J.}\ }\textbf {\bibinfo {volume} {C49}},\ \bibinfo
  {pages} {473} (\bibinfo {year} {2007}{\natexlab{b}})},\ \Eprint
  {http://arxiv.org/abs/hep-ex/0609009} {arXiv:hep-ex/0609009
  [hep-ex]}\BibitemShut {NoStop}%
\bibitem [{\citenamefont {Colangelo}\ \emph
  {et~al.}(2020{\natexlab{b}})\citenamefont {Colangelo}, \citenamefont
  {Hagelstein}, \citenamefont {Hoferichter}, \citenamefont {Laub},\ and\
  \citenamefont {Stoffer}}]{Colangelo:2019lpu}%
  \BibitemOpen
  \bibfield  {author} {\bibinfo {author} {\bibfnamefont {G.}~\bibnamefont
  {Colangelo}}, \bibinfo {author} {\bibfnamefont {F.}~\bibnamefont
  {Hagelstein}}, \bibinfo {author} {\bibfnamefont {M.}~\bibnamefont
  {Hoferichter}}, \bibinfo {author} {\bibfnamefont {L.}~\bibnamefont {Laub}}, \
  and\ \bibinfo {author} {\bibfnamefont {P.}~\bibnamefont {Stoffer}},\ }\href
  {\doibase 10.1103/PhysRevD.101.051501} {\bibfield  {journal} {\bibinfo
  {journal} {Phys. Rev.}\ }\textbf {\bibinfo {volume} {D101}},\ \bibinfo
  {pages} {051501} (\bibinfo {year} {2020}{\natexlab{b}})},\ \Eprint
  {http://arxiv.org/abs/1910.11881} {arXiv:1910.11881 [hep-ph]}\BibitemShut
  {NoStop}%
\bibitem [{\citenamefont {Landau}(1948)}]{Landau:1948kw}%
  \BibitemOpen
  \bibfield  {author} {\bibinfo {author} {\bibfnamefont {L.~D.}\ \bibnamefont
  {Landau}},\ }\href {\doibase 10.1016/B978-0-08-010586-4.50070-5} {\bibfield
  {journal} {\bibinfo  {journal} {Dokl. Akad. Nauk Ser. Fiz.}\ }\textbf
  {\bibinfo {volume} {60}},\ \bibinfo {pages} {207} (\bibinfo {year}
  {1948})}\BibitemShut {NoStop}%
\bibitem [{\citenamefont {Yang}(1950)}]{Yang:1950rg}%
  \BibitemOpen
  \bibfield  {author} {\bibinfo {author} {\bibfnamefont {C.-N.}\ \bibnamefont
  {Yang}},\ }\href {\doibase 10.1103/PhysRev.77.242} {\bibfield  {journal}
  {\bibinfo  {journal} {Phys. Rev.}\ }\textbf {\bibinfo {volume} {77}},\
  \bibinfo {pages} {242} (\bibinfo {year} {1950})}\BibitemShut {NoStop}%
\bibitem [{\citenamefont {Achard}\ \emph {et~al.}(2003)\citenamefont {Achard}
  \emph {et~al.}}]{Achard:2003qa}%
  \BibitemOpen
  \bibfield  {author} {\bibinfo {author} {\bibfnamefont {P.}~\bibnamefont
  {Achard}} \emph {et~al.} (\bibinfo {collaboration} {L3}),\ }\href {\doibase
  10.1016/j.physletb.2003.05.003} {\bibfield  {journal} {\bibinfo  {journal}
  {Phys. Lett.}\ }\textbf {\bibinfo {volume} {B568}},\ \bibinfo {pages} {11}
  (\bibinfo {year} {2003})},\ \Eprint {http://arxiv.org/abs/hep-ex/0305082}
  {arXiv:hep-ex/0305082 [hep-ex]}\BibitemShut {NoStop}%
\bibitem [{\citenamefont {Achard}\ \emph
  {et~al.}(2004{\natexlab{a}})\citenamefont {Achard} \emph
  {et~al.}}]{Achard:2004ux}%
  \BibitemOpen
  \bibfield  {author} {\bibinfo {author} {\bibfnamefont {P.}~\bibnamefont
  {Achard}} \emph {et~al.} (\bibinfo {collaboration} {L3}),\ }\href {\doibase
  10.1016/j.physletb.2004.07.013} {\bibfield  {journal} {\bibinfo  {journal}
  {Phys. Lett.}\ }\textbf {\bibinfo {volume} {B597}},\ \bibinfo {pages} {26}
  (\bibinfo {year} {2004}{\natexlab{a}})},\ \Eprint
  {http://arxiv.org/abs/hep-ex/0407020} {arXiv:hep-ex/0407020
  [hep-ex]}\BibitemShut {NoStop}%
\bibitem [{\citenamefont {Achard}\ \emph
  {et~al.}(2004{\natexlab{b}})\citenamefont {Achard} \emph
  {et~al.}}]{Achard:2004us}%
  \BibitemOpen
  \bibfield  {author} {\bibinfo {author} {\bibfnamefont {P.}~\bibnamefont
  {Achard}} \emph {et~al.} (\bibinfo {collaboration} {L3}),\ }\href {\doibase
  10.1016/j.physletb.2004.10.049} {\bibfield  {journal} {\bibinfo  {journal}
  {Phys. Lett.}\ }\textbf {\bibinfo {volume} {B604}},\ \bibinfo {pages} {48}
  (\bibinfo {year} {2004}{\natexlab{b}})},\ \Eprint
  {http://arxiv.org/abs/hep-ex/0410073} {arXiv:hep-ex/0410073
  [hep-ex]}\BibitemShut {NoStop}%
\bibitem [{\citenamefont {Achard}\ \emph {et~al.}(2005)\citenamefont {Achard}
  \emph {et~al.}}]{Achard:2005pb}%
  \BibitemOpen
  \bibfield  {author} {\bibinfo {author} {\bibfnamefont {P.}~\bibnamefont
  {Achard}} \emph {et~al.} (\bibinfo {collaboration} {L3}),\ }\href {\doibase
  10.1016/j.physletb.2005.04.011} {\bibfield  {journal} {\bibinfo  {journal}
  {Phys. Lett.}\ }\textbf {\bibinfo {volume} {B615}},\ \bibinfo {pages} {19}
  (\bibinfo {year} {2005})},\ \Eprint {http://arxiv.org/abs/hep-ex/0504016}
  {arXiv:hep-ex/0504016 [hep-ex]}\BibitemShut {NoStop}%
\bibitem [{\citenamefont {Berger}\ \emph {et~al.}(1984)\citenamefont {Berger}
  \emph {et~al.}}]{Berger:1984uk}%
  \BibitemOpen
  \bibfield  {author} {\bibinfo {author} {\bibfnamefont {C.}~\bibnamefont
  {Berger}} \emph {et~al.} (\bibinfo {collaboration} {PLUTO}),\ }\href
  {\doibase 10.1016/0370-2693(84)90437-4} {\bibfield  {journal} {\bibinfo
  {journal} {Phys. Lett.}\ }\textbf {\bibinfo {volume} {149B}},\ \bibinfo
  {pages} {421} (\bibinfo {year} {1984})}\BibitemShut {NoStop}%
\bibitem [{\citenamefont {Bintinger}\ \emph {et~al.}(1985)\citenamefont
  {Bintinger} \emph {et~al.}}]{Bintinger:1984as}%
  \BibitemOpen
  \bibfield  {author} {\bibinfo {author} {\bibfnamefont {D.}~\bibnamefont
  {Bintinger}} \emph {et~al.} (\bibinfo {collaboration} {TPC/$2\gamma$}),\
  }\href {\doibase 10.1103/PhysRevLett.54.763} {\bibfield  {journal} {\bibinfo
  {journal} {Phys. Rev. Lett.}\ }\textbf {\bibinfo {volume} {54}},\ \bibinfo
  {pages} {763} (\bibinfo {year} {1985})}\BibitemShut {NoStop}%
\bibitem [{\citenamefont {Aihara}\ \emph {et~al.}(1990)\citenamefont {Aihara}
  \emph {et~al.}}]{Aihara:1989kr}%
  \BibitemOpen
  \bibfield  {author} {\bibinfo {author} {\bibfnamefont {H.}~\bibnamefont
  {Aihara}} \emph {et~al.} (\bibinfo {collaboration} {TPC/$2\gamma$}),\ }\href
  {\doibase 10.1103/PhysRevD.41.2667} {\bibfield  {journal} {\bibinfo
  {journal} {Phys. Rev.}\ }\textbf {\bibinfo {volume} {D41}},\ \bibinfo {pages}
  {2667} (\bibinfo {year} {1990})}\BibitemShut {NoStop}%
\bibitem [{\citenamefont {Baru}\ \emph {et~al.}(1992)\citenamefont {Baru} \emph
  {et~al.}}]{Baru:1991bt}%
  \BibitemOpen
  \bibfield  {author} {\bibinfo {author} {\bibfnamefont {S.~E.}\ \bibnamefont
  {Baru}} \emph {et~al.} (\bibinfo {collaboration} {MD-1}),\ }\href {\doibase
  10.1007/BF01597557} {\bibfield  {journal} {\bibinfo  {journal} {Z. Phys.}\
  }\textbf {\bibinfo {volume} {C53}},\ \bibinfo {pages} {219} (\bibinfo {year}
  {1992})}\BibitemShut {NoStop}%
\bibitem [{\citenamefont {Kubis}(2018)}]{Kubis:2018bej}%
  \BibitemOpen
  \bibfield  {author} {\bibinfo {author} {\bibfnamefont {B.}~\bibnamefont
  {Kubis}},\ }\href {\doibase 10.1051/epjconf/201816600012} {\bibfield
  {journal} {\bibinfo  {journal} {EPJ Web Conf.}\ }\textbf {\bibinfo {volume}
  {166}},\ \bibinfo {pages} {00012} (\bibinfo {year} {2018})}\BibitemShut
  {NoStop}%
\bibitem [{\citenamefont {Leupold}\ \emph {et~al.}(2018)\citenamefont
  {Leupold}, \citenamefont {Hoferichter}, \citenamefont {Kubis}, \citenamefont
  {Niecknig},\ and\ \citenamefont {Schneider}}]{Leupold:2018mgr}%
  \BibitemOpen
  \bibfield  {author} {\bibinfo {author} {\bibfnamefont {S.}~\bibnamefont
  {Leupold}}, \bibinfo {author} {\bibfnamefont {M.}~\bibnamefont
  {Hoferichter}}, \bibinfo {author} {\bibfnamefont {B.}~\bibnamefont {Kubis}},
  \bibinfo {author} {\bibfnamefont {F.}~\bibnamefont {Niecknig}}, \ and\
  \bibinfo {author} {\bibfnamefont {S.~P.}\ \bibnamefont {Schneider}},\ }\href
  {\doibase 10.1051/epjconf/201816600013} {\bibfield  {journal} {\bibinfo
  {journal} {EPJ Web Conf.}\ }\textbf {\bibinfo {volume} {166}},\ \bibinfo
  {pages} {00013} (\bibinfo {year} {2018})}\BibitemShut {NoStop}%
\bibitem [{\citenamefont {Aloisio}\ \emph {et~al.}(2003)\citenamefont {Aloisio}
  \emph {et~al.}}]{Aloisio:2003ur}%
  \BibitemOpen
  \bibfield  {author} {\bibinfo {author} {\bibfnamefont {A.}~\bibnamefont
  {Aloisio}} \emph {et~al.} (\bibinfo {collaboration} {KLOE}),\ }\href
  {\doibase 10.1016/S0370-2693(03)00402-7} {\bibfield  {journal} {\bibinfo
  {journal} {Phys. Lett.}\ }\textbf {\bibinfo {volume} {B561}},\ \bibinfo
  {pages} {55} (\bibinfo {year} {2003})},\ \bibinfo {note} {[Erratum: Phys.
  Lett. {\bf B609}, 449 (2005)]},\ \Eprint
  {http://arxiv.org/abs/hep-ex/0303016} {arXiv:hep-ex/0303016
  [hep-ex]}\BibitemShut {NoStop}%
\bibitem [{\citenamefont {Adlarson}\ \emph
  {et~al.}(2017{\natexlab{c}})\citenamefont {Adlarson} \emph
  {et~al.}}]{Adlarson:2016wkw}%
  \BibitemOpen
  \bibfield  {author} {\bibinfo {author} {\bibfnamefont {P.}~\bibnamefont
  {Adlarson}} \emph {et~al.} (\bibinfo {collaboration} {WASA-at-COSY}),\ }\href
  {\doibase 10.1016/j.physletb.2017.03.050} {\bibfield  {journal} {\bibinfo
  {journal} {Phys. Lett.}\ }\textbf {\bibinfo {volume} {B770}},\ \bibinfo
  {pages} {418} (\bibinfo {year} {2017}{\natexlab{c}})},\ \Eprint
  {http://arxiv.org/abs/1610.02187} {arXiv:1610.02187 [nucl-ex]}\BibitemShut
  {NoStop}%
\bibitem [{\citenamefont {Ablikim}\ \emph
  {et~al.}(2018{\natexlab{a}})\citenamefont {Ablikim} \emph
  {et~al.}}]{Ablikim:2018yen}%
  \BibitemOpen
  \bibfield  {author} {\bibinfo {author} {\bibfnamefont {M.}~\bibnamefont
  {Ablikim}} \emph {et~al.} (\bibinfo {collaboration} {BESIII}),\ }\href
  {\doibase 10.1103/PhysRevD.98.112007} {\bibfield  {journal} {\bibinfo
  {journal} {Phys. Rev.}\ }\textbf {\bibinfo {volume} {D98}},\ \bibinfo {pages}
  {112007} (\bibinfo {year} {2018}{\natexlab{a}})},\ \Eprint
  {http://arxiv.org/abs/1811.03817} {arXiv:1811.03817 [hep-ex]}\BibitemShut
  {NoStop}%
\bibitem [{\citenamefont {Niecknig}\ \emph {et~al.}(2012)\citenamefont
  {Niecknig}, \citenamefont {Kubis},\ and\ \citenamefont
  {Schneider}}]{Niecknig:2012sj}%
  \BibitemOpen
  \bibfield  {author} {\bibinfo {author} {\bibfnamefont {F.}~\bibnamefont
  {Niecknig}}, \bibinfo {author} {\bibfnamefont {B.}~\bibnamefont {Kubis}}, \
  and\ \bibinfo {author} {\bibfnamefont {S.~P.}\ \bibnamefont {Schneider}},\
  }\href {\doibase 10.1140/epjc/s10052-012-2014-1} {\bibfield  {journal}
  {\bibinfo  {journal} {Eur. Phys. J.}\ }\textbf {\bibinfo {volume} {C72}},\
  \bibinfo {pages} {2014} (\bibinfo {year} {2012})},\ \Eprint
  {http://arxiv.org/abs/1203.2501} {arXiv:1203.2501 [hep-ph]}\BibitemShut
  {NoStop}%
\bibitem [{\citenamefont {Danilkin}\ \emph {et~al.}(2015)\citenamefont
  {Danilkin}, \citenamefont {Fern{\'a}ndez-Ram{\'i}rez}, \citenamefont {Guo},
  \citenamefont {Mathieu}, \citenamefont {Schott}, \citenamefont {Shi},\ and\
  \citenamefont {Szczepaniak}}]{Danilkin:2014cra}%
  \BibitemOpen
  \bibfield  {author} {\bibinfo {author} {\bibfnamefont {I.~V.}\ \bibnamefont
  {Danilkin}}, \bibinfo {author} {\bibfnamefont {C.}~\bibnamefont
  {Fern{\'a}ndez-Ram{\'i}rez}}, \bibinfo {author} {\bibfnamefont
  {P.}~\bibnamefont {Guo}}, \bibinfo {author} {\bibfnamefont {V.}~\bibnamefont
  {Mathieu}}, \bibinfo {author} {\bibfnamefont {D.}~\bibnamefont {Schott}},
  \bibinfo {author} {\bibfnamefont {M.}~\bibnamefont {Shi}}, \ and\ \bibinfo
  {author} {\bibfnamefont {A.~P.}\ \bibnamefont {Szczepaniak}},\ }\href
  {\doibase 10.1103/PhysRevD.91.094029} {\bibfield  {journal} {\bibinfo
  {journal} {Phys. Rev.}\ }\textbf {\bibinfo {volume} {D91}},\ \bibinfo {pages}
  {094029} (\bibinfo {year} {2015})},\ \Eprint {http://arxiv.org/abs/1409.7708}
  {arXiv:1409.7708 [hep-ph]}\BibitemShut {NoStop}%
\bibitem [{\citenamefont {Guo}(2019)}]{Guo:2019gjf}%
  \BibitemOpen
  \bibfield  {author} {\bibinfo {author} {\bibfnamefont {Y.}~\bibnamefont
  {Guo}} (\bibinfo {collaboration} {BESIII}),\ }\href {\doibase
  10.1088/1742-6596/1137/1/012008} {\bibfield  {journal} {\bibinfo  {journal}
  {J. Phys. Conf. Ser.}\ }\textbf {\bibinfo {volume} {1137}},\ \bibinfo {pages}
  {012008} (\bibinfo {year} {2019})}\BibitemShut {NoStop}%
\bibitem [{\citenamefont {Uehara}(1996)}]{Uehara:2013dna}%
  \BibitemOpen
  \bibfield  {author} {\bibinfo {author} {\bibfnamefont {S.}~\bibnamefont
  {Uehara}},\ }\href@noop {} {\  (\bibinfo {year} {1996})},\ \Eprint
  {http://arxiv.org/abs/1310.0157} {arXiv:1310.0157 [hep-ph]}\BibitemShut
  {NoStop}%
\bibitem [{\citenamefont {Druzhinin}\ \emph {et~al.}(2014)\citenamefont
  {Druzhinin}, \citenamefont {Kardapoltsev},\ and\ \citenamefont
  {Tayursky}}]{Druzhinin:2014sba}%
  \BibitemOpen
  \bibfield  {author} {\bibinfo {author} {\bibfnamefont {V.~P.}\ \bibnamefont
  {Druzhinin}}, \bibinfo {author} {\bibfnamefont {L.~V.}\ \bibnamefont
  {Kardapoltsev}}, \ and\ \bibinfo {author} {\bibfnamefont {V.~A.}\
  \bibnamefont {Tayursky}},\ }\href {\doibase 10.1016/j.cpc.2013.07.017}
  {\bibfield  {journal} {\bibinfo  {journal} {Comput. Phys. Commun.}\ }\textbf
  {\bibinfo {volume} {185}},\ \bibinfo {pages} {236} (\bibinfo {year}
  {2014})}\BibitemShut {NoStop}%
\bibitem [{\citenamefont {Czy\.z}\ and\ \citenamefont
  {Nowak-Kubat}(2006)}]{Czyz:2006dm}%
  \BibitemOpen
  \bibfield  {author} {\bibinfo {author} {\bibfnamefont {H.}~\bibnamefont
  {Czy\.z}}\ and\ \bibinfo {author} {\bibfnamefont {E.}~\bibnamefont
  {Nowak-Kubat}},\ }\href {\doibase 10.1016/j.physletb.2006.02.024} {\bibfield
  {journal} {\bibinfo  {journal} {Phys. Lett.}\ }\textbf {\bibinfo {volume}
  {B634}},\ \bibinfo {pages} {493} (\bibinfo {year} {2006})},\ \Eprint
  {http://arxiv.org/abs/hep-ph/0601169} {arXiv:hep-ph/0601169
  [hep-ph]}\BibitemShut {NoStop}%
\bibitem [{\citenamefont {Czy\.z}\ and\ \citenamefont
  {Ivashyn}(2011)}]{Czyz:2010sp}%
  \BibitemOpen
  \bibfield  {author} {\bibinfo {author} {\bibfnamefont {H.}~\bibnamefont
  {Czy\.z}}\ and\ \bibinfo {author} {\bibfnamefont {S.}~\bibnamefont
  {Ivashyn}},\ }\href {\doibase 10.1016/j.cpc.2011.01.029} {\bibfield
  {journal} {\bibinfo  {journal} {Comput. Phys. Commun.}\ }\textbf {\bibinfo
  {volume} {182}},\ \bibinfo {pages} {1338} (\bibinfo {year} {2011})},\ \Eprint
  {http://arxiv.org/abs/1009.1881} {arXiv:1009.1881 [hep-ph]}\BibitemShut
  {NoStop}%
\bibitem [{\citenamefont {Czy\.z}\ and\ \citenamefont
  {Kisza}(2019)}]{Czyz:2018jpp}%
  \BibitemOpen
  \bibfield  {author} {\bibinfo {author} {\bibfnamefont {H.}~\bibnamefont
  {Czy\.z}}\ and\ \bibinfo {author} {\bibfnamefont {P.}~\bibnamefont {Kisza}},\
  }\href {\doibase 10.1016/j.cpc.2018.07.021} {\bibfield  {journal} {\bibinfo
  {journal} {Comput. Phys. Commun.}\ }\textbf {\bibinfo {volume} {234}},\
  \bibinfo {pages} {245} (\bibinfo {year} {2019})},\ \Eprint
  {http://arxiv.org/abs/1805.07756} {arXiv:1805.07756 [hep-ph]}\BibitemShut
  {NoStop}%
\bibitem [{\citenamefont {Nyffeler}(2009)}]{Nyffeler:2009tw}%
  \BibitemOpen
  \bibfield  {author} {\bibinfo {author} {\bibfnamefont {A.}~\bibnamefont
  {Nyffeler}},\ }\href {\doibase 10.1103/PhysRevD.79.073012} {\bibfield
  {journal} {\bibinfo  {journal} {Phys. Rev.}\ }\textbf {\bibinfo {volume}
  {D79}},\ \bibinfo {pages} {073012} (\bibinfo {year} {2009})},\ \Eprint
  {http://arxiv.org/abs/0901.1172} {arXiv:0901.1172 [hep-ph]}\BibitemShut
  {NoStop}%
\bibitem [{\citenamefont {Bardeen}(1969)}]{Bardeen:1969md}%
  \BibitemOpen
  \bibfield  {author} {\bibinfo {author} {\bibfnamefont {W.~A.}\ \bibnamefont
  {Bardeen}},\ }\href {\doibase 10.1103/PhysRev.184.1848} {\bibfield  {journal}
  {\bibinfo  {journal} {Phys. Rev.}\ }\textbf {\bibinfo {volume} {184}},\
  \bibinfo {pages} {1848} (\bibinfo {year} {1969})}\BibitemShut {NoStop}%
\bibitem [{\citenamefont {Feldmann}(2000)}]{Feldmann:1999uf}%
  \BibitemOpen
  \bibfield  {author} {\bibinfo {author} {\bibfnamefont {T.}~\bibnamefont
  {Feldmann}},\ }\href {\doibase 10.1142/S0217751X00000082} {\bibfield
  {journal} {\bibinfo  {journal} {Int. J. Mod. Phys.}\ }\textbf {\bibinfo
  {volume} {A15}},\ \bibinfo {pages} {159} (\bibinfo {year} {2000})},\ \Eprint
  {http://arxiv.org/abs/hep-ph/9907491} {arXiv:hep-ph/9907491
  [hep-ph]}\BibitemShut {NoStop}%
\bibitem [{\citenamefont {Escribano}\ \emph {et~al.}(2016)\citenamefont
  {Escribano}, \citenamefont {Gonz{\`a}lez-Sol{\'i}s}, \citenamefont
  {Masjuan},\ and\ \citenamefont {S{\'a}nchez-Puertas}}]{Escribano:2015yup}%
  \BibitemOpen
  \bibfield  {author} {\bibinfo {author} {\bibfnamefont {R.}~\bibnamefont
  {Escribano}}, \bibinfo {author} {\bibfnamefont {S.}~\bibnamefont
  {Gonz{\`a}lez-Sol{\'i}s}}, \bibinfo {author} {\bibfnamefont {P.}~\bibnamefont
  {Masjuan}}, \ and\ \bibinfo {author} {\bibfnamefont {P.}~\bibnamefont
  {S{\'a}nchez-Puertas}},\ }\href {\doibase 10.1103/PhysRevD.94.054033}
  {\bibfield  {journal} {\bibinfo  {journal} {Phys. Rev.}\ }\textbf {\bibinfo
  {volume} {D94}},\ \bibinfo {pages} {054033} (\bibinfo {year} {2016})},\
  \Eprint {http://arxiv.org/abs/1512.07520} {arXiv:1512.07520
  [hep-ph]}\BibitemShut {NoStop}%
\bibitem [{\citenamefont {Lepage}\ and\ \citenamefont
  {Brodsky}(1979)}]{Lepage:1979zb}%
  \BibitemOpen
  \bibfield  {author} {\bibinfo {author} {\bibfnamefont {G.~P.}\ \bibnamefont
  {Lepage}}\ and\ \bibinfo {author} {\bibfnamefont {S.~J.}\ \bibnamefont
  {Brodsky}},\ }\href {\doibase 10.1016/0370-2693(79)90554-9} {\bibfield
  {journal} {\bibinfo  {journal} {Phys. Lett.}\ }\textbf {\bibinfo {volume}
  {87B}},\ \bibinfo {pages} {359} (\bibinfo {year} {1979})}\BibitemShut
  {NoStop}%
\bibitem [{\citenamefont {Lepage}\ and\ \citenamefont
  {Brodsky}(1980)}]{Lepage:1980fj}%
  \BibitemOpen
  \bibfield  {author} {\bibinfo {author} {\bibfnamefont {G.~P.}\ \bibnamefont
  {Lepage}}\ and\ \bibinfo {author} {\bibfnamefont {S.~J.}\ \bibnamefont
  {Brodsky}},\ }\href {\doibase 10.1103/PhysRevD.22.2157} {\bibfield  {journal}
  {\bibinfo  {journal} {Phys. Rev.}\ }\textbf {\bibinfo {volume} {D22}},\
  \bibinfo {pages} {2157} (\bibinfo {year} {1980})}\BibitemShut {NoStop}%
\bibitem [{\citenamefont {del Aguila}\ and\ \citenamefont
  {Chase}(1981)}]{delAguila:1981nk}%
  \BibitemOpen
  \bibfield  {author} {\bibinfo {author} {\bibfnamefont {F.}~\bibnamefont {del
  Aguila}}\ and\ \bibinfo {author} {\bibfnamefont {M.~K.}\ \bibnamefont
  {Chase}},\ }\href {\doibase 10.1016/0550-3213(81)90344-8} {\bibfield
  {journal} {\bibinfo  {journal} {Nucl. Phys.}\ }\textbf {\bibinfo {volume}
  {B193}},\ \bibinfo {pages} {517} (\bibinfo {year} {1981})}\BibitemShut
  {NoStop}%
\bibitem [{\citenamefont {Braaten}(1983)}]{Braaten:1982yp}%
  \BibitemOpen
  \bibfield  {author} {\bibinfo {author} {\bibfnamefont {E.}~\bibnamefont
  {Braaten}},\ }\href {\doibase 10.1103/PhysRevD.28.524} {\bibfield  {journal}
  {\bibinfo  {journal} {Phys. Rev.}\ }\textbf {\bibinfo {volume} {D28}},\
  \bibinfo {pages} {524} (\bibinfo {year} {1983})}\BibitemShut {NoStop}%
\bibitem [{\citenamefont {Braun}\ \emph {et~al.}(2003)\citenamefont {Braun},
  \citenamefont {Korchemsky},\ and\ \citenamefont {M{\"u}ller}}]{Braun:2003rp}%
  \BibitemOpen
  \bibfield  {author} {\bibinfo {author} {\bibfnamefont {V.}~\bibnamefont
  {Braun}}, \bibinfo {author} {\bibfnamefont {G.}~\bibnamefont {Korchemsky}}, \
  and\ \bibinfo {author} {\bibfnamefont {D.}~\bibnamefont {M{\"u}ller}},\
  }\href {\doibase 10.1016/S0146-6410(03)90004-4} {\bibfield  {journal}
  {\bibinfo  {journal} {Prog. Part. Nucl. Phys.}\ }\textbf {\bibinfo {volume}
  {51}},\ \bibinfo {pages} {311} (\bibinfo {year} {2003})},\ \Eprint
  {http://arxiv.org/abs/hep-ph/0306057} {arXiv:hep-ph/0306057}\BibitemShut
  {NoStop}%
\bibitem [{\citenamefont {Nesterenko}\ and\ \citenamefont
  {Radyushkin}(1983)}]{Nesterenko:1982dn}%
  \BibitemOpen
  \bibfield  {author} {\bibinfo {author} {\bibfnamefont {V.~A.}\ \bibnamefont
  {Nesterenko}}\ and\ \bibinfo {author} {\bibfnamefont {A.~V.}\ \bibnamefont
  {Radyushkin}},\ }\href@noop {} {\bibfield  {journal} {\bibinfo  {journal}
  {Sov. J. Nucl. Phys.}\ }\textbf {\bibinfo {volume} {38}},\ \bibinfo {pages}
  {284} (\bibinfo {year} {1983})},\ \bibinfo {note} {[Yad. Fiz. {\bf 38}, 476
  (1983)]}\BibitemShut {NoStop}%
\bibitem [{\citenamefont {Novikov}\ \emph {et~al.}(1984)\citenamefont
  {Novikov}, \citenamefont {Shifman}, \citenamefont {Vainshtein}, \citenamefont
  {Voloshin},\ and\ \citenamefont {Zakharov}}]{Novikov:1983jt}%
  \BibitemOpen
  \bibfield  {author} {\bibinfo {author} {\bibfnamefont {V.~A.}\ \bibnamefont
  {Novikov}}, \bibinfo {author} {\bibfnamefont {M.~A.}\ \bibnamefont
  {Shifman}}, \bibinfo {author} {\bibfnamefont {A.~I.}\ \bibnamefont
  {Vainshtein}}, \bibinfo {author} {\bibfnamefont {M.~B.}\ \bibnamefont
  {Voloshin}}, \ and\ \bibinfo {author} {\bibfnamefont {V.~I.}\ \bibnamefont
  {Zakharov}},\ }\href {\doibase 10.1016/0550-3213(84)90006-3} {\bibfield
  {journal} {\bibinfo  {journal} {Nucl. Phys.}\ }\textbf {\bibinfo {volume}
  {B237}},\ \bibinfo {pages} {525} (\bibinfo {year} {1984})}\BibitemShut
  {NoStop}%
\bibitem [{\citenamefont {Gorsky}(1987)}]{Gorsky:1987mu}%
  \BibitemOpen
  \bibfield  {author} {\bibinfo {author} {\bibfnamefont {A.~S.}\ \bibnamefont
  {Gorsky}},\ }\href@noop {} {\bibfield  {journal} {\bibinfo  {journal} {Sov.
  J. Nucl. Phys.}\ }\textbf {\bibinfo {volume} {46}},\ \bibinfo {pages} {537}
  (\bibinfo {year} {1987})},\ \bibinfo {note} {[Yad. Fiz. {\bf 46}, 938
  (1987)]}\BibitemShut {NoStop}%
\bibitem [{\citenamefont {Manohar}(1990)}]{Manohar:1990hu}%
  \BibitemOpen
  \bibfield  {author} {\bibinfo {author} {\bibfnamefont {A.~V.}\ \bibnamefont
  {Manohar}},\ }\href {\doibase 10.1016/0370-2693(90)90276-C} {\bibfield
  {journal} {\bibinfo  {journal} {Phys. Lett.}\ }\textbf {\bibinfo {volume}
  {B244}},\ \bibinfo {pages} {101} (\bibinfo {year} {1990})}\BibitemShut
  {NoStop}%
\bibitem [{\citenamefont {Leutwyler}(1998)}]{Leutwyler:1997yr}%
  \BibitemOpen
  \bibfield  {author} {\bibinfo {author} {\bibfnamefont {H.}~\bibnamefont
  {Leutwyler}},\ }\href {\doibase 10.1016/S0920-5632(97)01065-7} {\bibfield
  {journal} {\bibinfo  {journal} {Nucl. Phys. Proc. Suppl.}\ }\textbf {\bibinfo
  {volume} {64}},\ \bibinfo {pages} {223} (\bibinfo {year} {1998})},\ \Eprint
  {http://arxiv.org/abs/hep-ph/9709408} {arXiv:hep-ph/9709408
  [hep-ph]}\BibitemShut {NoStop}%
\bibitem [{\citenamefont {Agaev}\ \emph {et~al.}(2014)\citenamefont {Agaev},
  \citenamefont {Braun}, \citenamefont {Offen}, \citenamefont {Porkert},\ and\
  \citenamefont {Sch{\"a}fer}}]{Agaev:2014wna}%
  \BibitemOpen
  \bibfield  {author} {\bibinfo {author} {\bibfnamefont {S.~S.}\ \bibnamefont
  {Agaev}}, \bibinfo {author} {\bibfnamefont {V.~M.}\ \bibnamefont {Braun}},
  \bibinfo {author} {\bibfnamefont {N.}~\bibnamefont {Offen}}, \bibinfo
  {author} {\bibfnamefont {F.~A.}\ \bibnamefont {Porkert}}, \ and\ \bibinfo
  {author} {\bibfnamefont {A.}~\bibnamefont {Sch{\"a}fer}},\ }\href {\doibase
  10.1103/PhysRevD.90.074019} {\bibfield  {journal} {\bibinfo  {journal} {Phys.
  Rev.}\ }\textbf {\bibinfo {volume} {D90}},\ \bibinfo {pages} {074019}
  (\bibinfo {year} {2014})},\ \Eprint {http://arxiv.org/abs/1409.4311}
  {arXiv:1409.4311 [hep-ph]}\BibitemShut {NoStop}%
\bibitem [{\citenamefont {Alte}\ \emph {et~al.}(2016)\citenamefont {Alte},
  \citenamefont {K{\"o}nig},\ and\ \citenamefont {Neubert}}]{Alte:2015dpo}%
  \BibitemOpen
  \bibfield  {author} {\bibinfo {author} {\bibfnamefont {S.}~\bibnamefont
  {Alte}}, \bibinfo {author} {\bibfnamefont {M.}~\bibnamefont {K{\"o}nig}}, \
  and\ \bibinfo {author} {\bibfnamefont {M.}~\bibnamefont {Neubert}},\ }\href
  {\doibase 10.1007/JHEP02(2016)162} {\bibfield  {journal} {\bibinfo  {journal}
  {JHEP}\ }\textbf {\bibinfo {volume} {02}},\ \bibinfo {pages} {162} (\bibinfo
  {year} {2016})},\ \Eprint {http://arxiv.org/abs/1512.09135} {arXiv:1512.09135
  [hep-ph]}\BibitemShut {NoStop}%
\bibitem [{\citenamefont {Bali}\ \emph {et~al.}(2018)\citenamefont {Bali},
  \citenamefont {Braun}, \citenamefont {Gl{\"a}{\ss}le}, \citenamefont
  {G{\"o}ckeler}, \citenamefont {Gruber}, \citenamefont {Hutzler},
  \citenamefont {Korcyl}, \citenamefont {Sch{\"a}fer}, \citenamefont {Wein},\
  and\ \citenamefont {Zhang}}]{Bali:2018spj}%
  \BibitemOpen
  \bibfield  {author} {\bibinfo {author} {\bibfnamefont {G.~S.}\ \bibnamefont
  {Bali}}, \bibinfo {author} {\bibfnamefont {V.~M.}\ \bibnamefont {Braun}},
  \bibinfo {author} {\bibfnamefont {B.}~\bibnamefont {Gl{\"a}{\ss}le}},
  \bibinfo {author} {\bibfnamefont {M.}~\bibnamefont {G{\"o}ckeler}}, \bibinfo
  {author} {\bibfnamefont {M.}~\bibnamefont {Gruber}}, \bibinfo {author}
  {\bibfnamefont {F.}~\bibnamefont {Hutzler}}, \bibinfo {author} {\bibfnamefont
  {P.}~\bibnamefont {Korcyl}}, \bibinfo {author} {\bibfnamefont
  {A.}~\bibnamefont {Sch{\"a}fer}}, \bibinfo {author} {\bibfnamefont
  {P.}~\bibnamefont {Wein}}, \ and\ \bibinfo {author} {\bibfnamefont {J.-H.}\
  \bibnamefont {Zhang}},\ }\href {\doibase 10.1103/PhysRevD.98.094507}
  {\bibfield  {journal} {\bibinfo  {journal} {Phys. Rev.}\ }\textbf {\bibinfo
  {volume} {D98}},\ \bibinfo {pages} {094507} (\bibinfo {year} {2018})},\
  \Eprint {http://arxiv.org/abs/1807.06671} {arXiv:1807.06671
  [hep-lat]}\BibitemShut {NoStop}%
\bibitem [{\citenamefont {Hoferichter}\ \emph
  {et~al.}(2014{\natexlab{b}})\citenamefont {Hoferichter}, \citenamefont
  {Kubis}, \citenamefont {Leupold}, \citenamefont {Niecknig},\ and\
  \citenamefont {Schneider}}]{Hoferichter:2014vra}%
  \BibitemOpen
  \bibfield  {author} {\bibinfo {author} {\bibfnamefont {M.}~\bibnamefont
  {Hoferichter}}, \bibinfo {author} {\bibfnamefont {B.}~\bibnamefont {Kubis}},
  \bibinfo {author} {\bibfnamefont {S.}~\bibnamefont {Leupold}}, \bibinfo
  {author} {\bibfnamefont {F.}~\bibnamefont {Niecknig}}, \ and\ \bibinfo
  {author} {\bibfnamefont {S.~P.}\ \bibnamefont {Schneider}},\ }\href {\doibase
  10.1140/epjc/s10052-014-3180-0} {\bibfield  {journal} {\bibinfo  {journal}
  {Eur. Phys. J.}\ }\textbf {\bibinfo {volume} {C74}},\ \bibinfo {pages} {3180}
  (\bibinfo {year} {2014}{\natexlab{b}})},\ \Eprint
  {http://arxiv.org/abs/1410.4691} {arXiv:1410.4691 [hep-ph]}\BibitemShut
  {NoStop}%
\bibitem [{\citenamefont {Hoferichter}\ \emph {et~al.}(2012)\citenamefont
  {Hoferichter}, \citenamefont {Kubis},\ and\ \citenamefont
  {Sakkas}}]{Hoferichter:2012pm}%
  \BibitemOpen
  \bibfield  {author} {\bibinfo {author} {\bibfnamefont {M.}~\bibnamefont
  {Hoferichter}}, \bibinfo {author} {\bibfnamefont {B.}~\bibnamefont {Kubis}},
  \ and\ \bibinfo {author} {\bibfnamefont {D.}~\bibnamefont {Sakkas}},\ }\href
  {\doibase 10.1103/PhysRevD.86.116009} {\bibfield  {journal} {\bibinfo
  {journal} {Phys. Rev.}\ }\textbf {\bibinfo {volume} {D86}},\ \bibinfo {pages}
  {116009} (\bibinfo {year} {2012})},\ \Eprint {http://arxiv.org/abs/1210.6793}
  {arXiv:1210.6793 [hep-ph]}\BibitemShut {NoStop}%
\bibitem [{\citenamefont {Hoferichter}\ \emph {et~al.}(2017)\citenamefont
  {Hoferichter}, \citenamefont {Kubis},\ and\ \citenamefont
  {Zanke}}]{Hoferichter:2017ftn}%
  \BibitemOpen
  \bibfield  {author} {\bibinfo {author} {\bibfnamefont {M.}~\bibnamefont
  {Hoferichter}}, \bibinfo {author} {\bibfnamefont {B.}~\bibnamefont {Kubis}},
  \ and\ \bibinfo {author} {\bibfnamefont {M.}~\bibnamefont {Zanke}},\ }\href
  {\doibase 10.1103/PhysRevD.96.114016} {\bibfield  {journal} {\bibinfo
  {journal} {Phys. Rev.}\ }\textbf {\bibinfo {volume} {D96}},\ \bibinfo {pages}
  {114016} (\bibinfo {year} {2017})},\ \Eprint
  {http://arxiv.org/abs/1710.00824} {arXiv:1710.00824 [hep-ph]}\BibitemShut
  {NoStop}%
\bibitem [{\citenamefont {Achasov}\ \emph
  {et~al.}(2003{\natexlab{b}})\citenamefont {Achasov} \emph
  {et~al.}}]{Achasov:2003ir}%
  \BibitemOpen
  \bibfield  {author} {\bibinfo {author} {\bibfnamefont {M.~N.}\ \bibnamefont
  {Achasov}} \emph {et~al.} (\bibinfo {collaboration} {SND}),\ }\href {\doibase
  10.1103/PhysRevD.68.052006} {\bibfield  {journal} {\bibinfo  {journal} {Phys.
  Rev.}\ }\textbf {\bibinfo {volume} {D68}},\ \bibinfo {pages} {052006}
  (\bibinfo {year} {2003}{\natexlab{b}})},\ \Eprint
  {http://arxiv.org/abs/hep-ex/0305049} {arXiv:hep-ex/0305049
  [hep-ex]}\BibitemShut {NoStop}%
\bibitem [{\citenamefont {Schneider}\ \emph {et~al.}(2012)\citenamefont
  {Schneider}, \citenamefont {Kubis},\ and\ \citenamefont
  {Niecknig}}]{Schneider:2012ez}%
  \BibitemOpen
  \bibfield  {author} {\bibinfo {author} {\bibfnamefont {S.~P.}\ \bibnamefont
  {Schneider}}, \bibinfo {author} {\bibfnamefont {B.}~\bibnamefont {Kubis}}, \
  and\ \bibinfo {author} {\bibfnamefont {F.}~\bibnamefont {Niecknig}},\ }\href
  {\doibase 10.1103/PhysRevD.86.054013} {\bibfield  {journal} {\bibinfo
  {journal} {Phys. Rev.}\ }\textbf {\bibinfo {volume} {D86}},\ \bibinfo {pages}
  {054013} (\bibinfo {year} {2012})},\ \Eprint {http://arxiv.org/abs/1206.3098}
  {arXiv:1206.3098 [hep-ph]}\BibitemShut {NoStop}%
\bibitem [{\citenamefont {Achasov}\ \emph
  {et~al.}(2000{\natexlab{b}})\citenamefont {Achasov} \emph
  {et~al.}}]{Achasov:2000zd}%
  \BibitemOpen
  \bibfield  {author} {\bibinfo {author} {\bibfnamefont {M.~N.}\ \bibnamefont
  {Achasov}} \emph {et~al.} (\bibinfo {collaboration} {SND}),\ }\href {\doibase
  10.1007/s100529900222} {\bibfield  {journal} {\bibinfo  {journal} {Eur. Phys.
  J.}\ }\textbf {\bibinfo {volume} {C12}},\ \bibinfo {pages} {25} (\bibinfo
  {year} {2000}{\natexlab{b}})}\BibitemShut {NoStop}%
\bibitem [{\citenamefont {Achasov}\ \emph
  {et~al.}(2003{\natexlab{c}})\citenamefont {Achasov} \emph
  {et~al.}}]{Achasov:2003ed}%
  \BibitemOpen
  \bibfield  {author} {\bibinfo {author} {\bibfnamefont {M.~N.}\ \bibnamefont
  {Achasov}} \emph {et~al.} (\bibinfo {collaboration} {SND}),\ }\href {\doibase
  10.1016/S0370-2693(03)00336-8} {\bibfield  {journal} {\bibinfo  {journal}
  {Phys. Lett.}\ }\textbf {\bibinfo {volume} {B559}},\ \bibinfo {pages} {171}
  (\bibinfo {year} {2003}{\natexlab{c}})},\ \Eprint
  {http://arxiv.org/abs/hep-ex/0302004} {arXiv:hep-ex/0302004
  [hep-ex]}\BibitemShut {NoStop}%
\bibitem [{\citenamefont {Baker}\ and\ \citenamefont
  {Graves-Morris}(1996)}]{BakerMorris}%
  \BibitemOpen
  \bibfield  {author} {\bibinfo {author} {\bibfnamefont {G.~A.}\ \bibnamefont
  {Baker}}\ and\ \bibinfo {author} {\bibfnamefont {P.}~\bibnamefont
  {Graves-Morris}},\ }\href@noop {} {\emph {\bibinfo {title} {Pad{\'e}
  Approximants}}},\ \bibinfo {edition} {2nd}\ ed.,\ \bibinfo {series}
  {Enciclopedia of Mathematics and its Applications}\ No.~\bibinfo {number}
  {59}\ (\bibinfo  {publisher} {Cambridge University Press},\ \bibinfo
  {address} {New York},\ \bibinfo {year} {1996})\BibitemShut {NoStop}%
\bibitem [{\citenamefont {Baker}(1975)}]{Baker}%
  \BibitemOpen
  \bibfield  {author} {\bibinfo {author} {\bibfnamefont {G.~A.}\ \bibnamefont
  {Baker}},\ }\href@noop {} {\emph {\bibinfo {title} {Essentials of Pad{\'e}
  Approximants}}},\ \bibinfo {edition} {1st}\ ed.\ (\bibinfo  {publisher}
  {Academic Press},\ \bibinfo {address} {New York},\ \bibinfo {year}
  {1975})\BibitemShut {NoStop}%
\bibitem [{\citenamefont {Peris}(2006)}]{Peris:2006ds}%
  \BibitemOpen
  \bibfield  {author} {\bibinfo {author} {\bibfnamefont {S.}~\bibnamefont
  {Peris}},\ }\href {\doibase 10.1103/PhysRevD.74.054013} {\bibfield  {journal}
  {\bibinfo  {journal} {Phys. Rev.}\ }\textbf {\bibinfo {volume} {D74}},\
  \bibinfo {pages} {054013} (\bibinfo {year} {2006})},\ \Eprint
  {http://arxiv.org/abs/hep-ph/0603190} {arXiv:hep-ph/0603190
  [hep-ph]}\BibitemShut {NoStop}%
\bibitem [{\citenamefont {Golterman}\ and\ \citenamefont
  {Peris}(2006)}]{Golterman:2006gv}%
  \BibitemOpen
  \bibfield  {author} {\bibinfo {author} {\bibfnamefont {M.}~\bibnamefont
  {Golterman}}\ and\ \bibinfo {author} {\bibfnamefont {S.}~\bibnamefont
  {Peris}},\ }\href {\doibase 10.1103/PhysRevD.74.096002} {\bibfield  {journal}
  {\bibinfo  {journal} {Phys. Rev.}\ }\textbf {\bibinfo {volume} {D74}},\
  \bibinfo {pages} {096002} (\bibinfo {year} {2006})},\ \Eprint
  {http://arxiv.org/abs/hep-ph/0607152} {arXiv:hep-ph/0607152
  [hep-ph]}\BibitemShut {NoStop}%
\bibitem [{\citenamefont {Masjuan}\ and\ \citenamefont
  {Peris}(2007)}]{Masjuan:2007ay}%
  \BibitemOpen
  \bibfield  {author} {\bibinfo {author} {\bibfnamefont {P.}~\bibnamefont
  {Masjuan}}\ and\ \bibinfo {author} {\bibfnamefont {S.}~\bibnamefont
  {Peris}},\ }\href {\doibase 10.1088/1126-6708/2007/05/040} {\bibfield
  {journal} {\bibinfo  {journal} {JHEP}\ }\textbf {\bibinfo {volume} {05}},\
  \bibinfo {pages} {040} (\bibinfo {year} {2007})},\ \Eprint
  {http://arxiv.org/abs/0704.1247} {arXiv:0704.1247 [hep-ph]}\BibitemShut
  {NoStop}%
\bibitem [{\citenamefont {Masjuan}\ and\ \citenamefont
  {Peris}(2008)}]{Masjuan:2008fr}%
  \BibitemOpen
  \bibfield  {author} {\bibinfo {author} {\bibfnamefont {P.}~\bibnamefont
  {Masjuan}}\ and\ \bibinfo {author} {\bibfnamefont {S.}~\bibnamefont
  {Peris}},\ }\href {\doibase 10.1016/j.physletb.2008.03.040} {\bibfield
  {journal} {\bibinfo  {journal} {Phys. Lett.}\ }\textbf {\bibinfo {volume}
  {B663}},\ \bibinfo {pages} {61} (\bibinfo {year} {2008})},\ \Eprint
  {http://arxiv.org/abs/0801.3558} {arXiv:0801.3558 [hep-ph]}\BibitemShut
  {NoStop}%
\bibitem [{\citenamefont {Masjuan}\ and\ \citenamefont
  {Peris}(2010)}]{Masjuan:2009wy}%
  \BibitemOpen
  \bibfield  {author} {\bibinfo {author} {\bibfnamefont {P.}~\bibnamefont
  {Masjuan}}\ and\ \bibinfo {author} {\bibfnamefont {S.}~\bibnamefont
  {Peris}},\ }\href {\doibase 10.1016/j.physletb.2010.02.069} {\bibfield
  {journal} {\bibinfo  {journal} {Phys. Lett.}\ }\textbf {\bibinfo {volume}
  {B686}},\ \bibinfo {pages} {307} (\bibinfo {year} {2010})},\ \Eprint
  {http://arxiv.org/abs/0903.0294} {arXiv:0903.0294 [hep-ph]}\BibitemShut
  {NoStop}%
\bibitem [{\citenamefont {Masjuan}(2012)}]{Masjuan:2012wy}%
  \BibitemOpen
  \bibfield  {author} {\bibinfo {author} {\bibfnamefont {P.}~\bibnamefont
  {Masjuan}},\ }\href {\doibase 10.1103/PhysRevD.86.094021} {\bibfield
  {journal} {\bibinfo  {journal} {Phys. Rev.}\ }\textbf {\bibinfo {volume}
  {D86}},\ \bibinfo {pages} {094021} (\bibinfo {year} {2012})},\ \Eprint
  {http://arxiv.org/abs/1206.2549} {arXiv:1206.2549 [hep-ph]}\BibitemShut
  {NoStop}%
\bibitem [{\citenamefont {Escribano}\ \emph {et~al.}(2014)\citenamefont
  {Escribano}, \citenamefont {Masjuan},\ and\ \citenamefont
  {S{\'a}nchez-Puertas}}]{Escribano:2013kba}%
  \BibitemOpen
  \bibfield  {author} {\bibinfo {author} {\bibfnamefont {R.}~\bibnamefont
  {Escribano}}, \bibinfo {author} {\bibfnamefont {P.}~\bibnamefont {Masjuan}},
  \ and\ \bibinfo {author} {\bibfnamefont {P.}~\bibnamefont
  {S{\'a}nchez-Puertas}},\ }\href {\doibase 10.1103/PhysRevD.89.034014}
  {\bibfield  {journal} {\bibinfo  {journal} {Phys. Rev.}\ }\textbf {\bibinfo
  {volume} {D89}},\ \bibinfo {pages} {034014} (\bibinfo {year} {2014})},\
  \Eprint {http://arxiv.org/abs/1307.2061} {arXiv:1307.2061
  [hep-ph]}\BibitemShut {NoStop}%
\bibitem [{\citenamefont {Escribano}\ \emph {et~al.}(2015)\citenamefont
  {Escribano}, \citenamefont {Masjuan},\ and\ \citenamefont
  {S{\'a}nchez-Puertas}}]{Escribano:2015nra}%
  \BibitemOpen
  \bibfield  {author} {\bibinfo {author} {\bibfnamefont {R.}~\bibnamefont
  {Escribano}}, \bibinfo {author} {\bibfnamefont {P.}~\bibnamefont {Masjuan}},
  \ and\ \bibinfo {author} {\bibfnamefont {P.}~\bibnamefont
  {S{\'a}nchez-Puertas}},\ }\href {\doibase 10.1140/epjc/s10052-015-3642-z}
  {\bibfield  {journal} {\bibinfo  {journal} {Eur. Phys. J.}\ }\textbf
  {\bibinfo {volume} {C75}},\ \bibinfo {pages} {414} (\bibinfo {year}
  {2015})},\ \Eprint {http://arxiv.org/abs/1504.07742} {arXiv:1504.07742
  [hep-ph]}\BibitemShut {NoStop}%
\bibitem [{\citenamefont {Masjuan~Queralt}(2010)}]{Queralt:2010sv}%
  \BibitemOpen
  \bibfield  {author} {\bibinfo {author} {\bibfnamefont {P.}~\bibnamefont
  {Masjuan~Queralt}},\ }\emph {\bibinfo {title} {{Rational Approximations in
  Quantum Chromodynamics}}},\ \href@noop {} {Ph.D. thesis},\ \bibinfo  {school}
  {Barcelona, IFAE} (\bibinfo {year} {2010}),\ \Eprint
  {http://arxiv.org/abs/1005.5683} {arXiv:1005.5683 [hep-ph]}\BibitemShut
  {NoStop}%
\bibitem [{\citenamefont {Kubis}\ and\ \citenamefont
  {Plenter}(2015)}]{Kubis:2015sga}%
  \BibitemOpen
  \bibfield  {author} {\bibinfo {author} {\bibfnamefont {B.}~\bibnamefont
  {Kubis}}\ and\ \bibinfo {author} {\bibfnamefont {J.}~\bibnamefont
  {Plenter}},\ }\href {\doibase 10.1140/epjc/s10052-015-3495-5} {\bibfield
  {journal} {\bibinfo  {journal} {Eur. Phys. J.}\ }\textbf {\bibinfo {volume}
  {C75}},\ \bibinfo {pages} {283} (\bibinfo {year} {2015})},\ \Eprint
  {http://arxiv.org/abs/1504.02588} {arXiv:1504.02588 [hep-ph]}\BibitemShut
  {NoStop}%
\bibitem [{\citenamefont {Masjuan}\ \emph {et~al.}(2008)\citenamefont
  {Masjuan}, \citenamefont {Sanz-Cillero},\ and\ \citenamefont
  {Virto}}]{Masjuan:2008cp}%
  \BibitemOpen
  \bibfield  {author} {\bibinfo {author} {\bibfnamefont {P.}~\bibnamefont
  {Masjuan}}, \bibinfo {author} {\bibfnamefont {J.~J.}\ \bibnamefont
  {Sanz-Cillero}}, \ and\ \bibinfo {author} {\bibfnamefont {J.}~\bibnamefont
  {Virto}},\ }\href {\doibase 10.1016/j.physletb.2008.07.106} {\bibfield
  {journal} {\bibinfo  {journal} {Phys. Lett.}\ }\textbf {\bibinfo {volume}
  {B668}},\ \bibinfo {pages} {14} (\bibinfo {year} {2008})},\ \Eprint
  {http://arxiv.org/abs/0805.3291} {arXiv:0805.3291 [hep-ph]}\BibitemShut
  {NoStop}%
\bibitem [{\citenamefont {Chisholm}(1973)}]{Chisholm:1973}%
  \BibitemOpen
  \bibfield  {author} {\bibinfo {author} {\bibfnamefont {J.~S.~R.}\
  \bibnamefont {Chisholm}},\ }\href {\doibase
  10.1090/S0025-5718-1973-0382928-6} {\bibfield  {journal} {\bibinfo  {journal}
  {Math. Comp.}\ }\textbf {\bibinfo {volume} {27}},\ \bibinfo {pages} {841}
  (\bibinfo {year} {1973})}\BibitemShut {NoStop}%
\bibitem [{\citenamefont {Chisholm}\ and\ \citenamefont
  {McEwan}(1974)}]{Chisholm:1974}%
  \BibitemOpen
  \bibfield  {author} {\bibinfo {author} {\bibfnamefont {J.~S.~R.}\
  \bibnamefont {Chisholm}}\ and\ \bibinfo {author} {\bibfnamefont
  {J.}~\bibnamefont {McEwan}},\ }\href {\doibase 10.1098/rspa.1974.0028}
  {\bibfield  {journal} {\bibinfo  {journal} {Proc. R. Soc. Lond.}\ }\textbf
  {\bibinfo {volume} {A336}},\ \bibinfo {pages} {421} (\bibinfo {year}
  {1974})}\BibitemShut {NoStop}%
\bibitem [{\citenamefont {Hughes~Jones}(1976)}]{Jones:1976}%
  \BibitemOpen
  \bibfield  {author} {\bibinfo {author} {\bibfnamefont {R.}~\bibnamefont
  {Hughes~Jones}},\ }\href {\doibase 10.1016/0021-9045(76)90050-2} {\bibfield
  {journal} {\bibinfo  {journal} {Journal of Approximation Theory}\ }\textbf
  {\bibinfo {volume} {16}},\ \bibinfo {pages} {201} (\bibinfo {year}
  {1976})}\BibitemShut {NoStop}%
\bibitem [{\citenamefont {Masjuan}\ and\ \citenamefont
  {S{\'a}nchez-Puertas}(2015)}]{Masjuan:2015lca}%
  \BibitemOpen
  \bibfield  {author} {\bibinfo {author} {\bibfnamefont {P.}~\bibnamefont
  {Masjuan}}\ and\ \bibinfo {author} {\bibfnamefont {P.}~\bibnamefont
  {S{\'a}nchez-Puertas}},\ }\href@noop {} {\  (\bibinfo {year} {2015})},\
  \Eprint {http://arxiv.org/abs/1504.07001} {arXiv:1504.07001
  [hep-ph]}\BibitemShut {NoStop}%
\bibitem [{\citenamefont
  {S{\'a}nchez-Puertas}(2016)}]{Sanchez-Puertas:2017sih}%
  \BibitemOpen
  \bibfield  {author} {\bibinfo {author} {\bibfnamefont {P.}~\bibnamefont
  {S{\'a}nchez-Puertas}},\ }\emph {\bibinfo {title} {{A theoretical study of
  meson transition form factors}}},\ \href@noop {} {Ph.D. thesis},\ \bibinfo
  {school} {Mainz U., Inst. Phys.} (\bibinfo {year} {2016}),\ \Eprint
  {http://arxiv.org/abs/1709.04792} {arXiv:1709.04792 [hep-ph]}\BibitemShut
  {NoStop}%
\bibitem [{\citenamefont {Alabiso}\ and\ \citenamefont
  {Butera}(1975)}]{Alabiso:1974vk}%
  \BibitemOpen
  \bibfield  {author} {\bibinfo {author} {\bibfnamefont {C.}~\bibnamefont
  {Alabiso}}\ and\ \bibinfo {author} {\bibfnamefont {P.}~\bibnamefont
  {Butera}},\ }\href {\doibase 10.1063/1.522617} {\bibfield  {journal}
  {\bibinfo  {journal} {J. Math. Phys.}\ }\textbf {\bibinfo {volume} {16}},\
  \bibinfo {pages} {840} (\bibinfo {year} {1975})}\BibitemShut {NoStop}%
\bibitem [{\citenamefont {Raya}\ \emph {et~al.}(2016)\citenamefont {Raya},
  \citenamefont {Chang}, \citenamefont {Bashir}, \citenamefont
  {Cobos-Martinez}, \citenamefont {Guti{\'e}rrez-Guerrero}, \citenamefont
  {Roberts},\ and\ \citenamefont {Tandy}}]{Raya:2015gva}%
  \BibitemOpen
  \bibfield  {author} {\bibinfo {author} {\bibfnamefont {K.}~\bibnamefont
  {Raya}}, \bibinfo {author} {\bibfnamefont {L.}~\bibnamefont {Chang}},
  \bibinfo {author} {\bibfnamefont {A.}~\bibnamefont {Bashir}}, \bibinfo
  {author} {\bibfnamefont {J.~J.}\ \bibnamefont {Cobos-Martinez}}, \bibinfo
  {author} {\bibfnamefont {L.~X.}\ \bibnamefont {Guti{\'e}rrez-Guerrero}},
  \bibinfo {author} {\bibfnamefont {C.~D.}\ \bibnamefont {Roberts}}, \ and\
  \bibinfo {author} {\bibfnamefont {P.~C.}\ \bibnamefont {Tandy}},\ }\href
  {\doibase 10.1103/PhysRevD.93.074017} {\bibfield  {journal} {\bibinfo
  {journal} {Phys. Rev.}\ }\textbf {\bibinfo {volume} {D93}},\ \bibinfo {pages}
  {074017} (\bibinfo {year} {2016})},\ \Eprint
  {http://arxiv.org/abs/1510.02799} {arXiv:1510.02799 [nucl-th]}\BibitemShut
  {NoStop}%
\bibitem [{\citenamefont {Eichmann}\ \emph {et~al.}(2017)\citenamefont
  {Eichmann}, \citenamefont {Fischer}, \citenamefont {Weil},\ and\
  \citenamefont {Williams}}]{Eichmann:2017wil}%
  \BibitemOpen
  \bibfield  {author} {\bibinfo {author} {\bibfnamefont {G.}~\bibnamefont
  {Eichmann}}, \bibinfo {author} {\bibfnamefont {C.~S.}\ \bibnamefont
  {Fischer}}, \bibinfo {author} {\bibfnamefont {E.}~\bibnamefont {Weil}}, \
  and\ \bibinfo {author} {\bibfnamefont {R.}~\bibnamefont {Williams}},\ }\href
  {\doibase 10.1016/j.physletb.2017.10.002} {\bibfield  {journal} {\bibinfo
  {journal} {Phys. Lett.}\ }\textbf {\bibinfo {volume} {B774}},\ \bibinfo
  {pages} {425} (\bibinfo {year} {2017})},\ \Eprint
  {http://arxiv.org/abs/1704.05774} {arXiv:1704.05774 [hep-ph]}\BibitemShut
  {NoStop}%
\bibitem [{\citenamefont {Weil}\ \emph {et~al.}(2017)\citenamefont {Weil},
  \citenamefont {Eichmann}, \citenamefont {Fischer},\ and\ \citenamefont
  {Williams}}]{Weil:2017knt}%
  \BibitemOpen
  \bibfield  {author} {\bibinfo {author} {\bibfnamefont {E.}~\bibnamefont
  {Weil}}, \bibinfo {author} {\bibfnamefont {G.}~\bibnamefont {Eichmann}},
  \bibinfo {author} {\bibfnamefont {C.~S.}\ \bibnamefont {Fischer}}, \ and\
  \bibinfo {author} {\bibfnamefont {R.}~\bibnamefont {Williams}},\ }\href
  {\doibase 10.1103/PhysRevD.96.014021} {\bibfield  {journal} {\bibinfo
  {journal} {Phys. Rev.}\ }\textbf {\bibinfo {volume} {D96}},\ \bibinfo {pages}
  {014021} (\bibinfo {year} {2017})},\ \Eprint
  {http://arxiv.org/abs/1704.06046} {arXiv:1704.06046 [hep-ph]}\BibitemShut
  {NoStop}%
\bibitem [{\citenamefont {Eichmann}\ \emph {et~al.}(2019)\citenamefont
  {Eichmann}, \citenamefont {Fischer}, \citenamefont {Weil},\ and\
  \citenamefont {Williams}}]{Eichmann:2019tjk}%
  \BibitemOpen
  \bibfield  {author} {\bibinfo {author} {\bibfnamefont {G.}~\bibnamefont
  {Eichmann}}, \bibinfo {author} {\bibfnamefont {C.~S.}\ \bibnamefont
  {Fischer}}, \bibinfo {author} {\bibfnamefont {E.}~\bibnamefont {Weil}}, \
  and\ \bibinfo {author} {\bibfnamefont {R.}~\bibnamefont {Williams}},\ }\href
  {\doibase 10.1016/j.physletb.2019.134855} {\bibfield  {journal} {\bibinfo
  {journal} {Phys. Lett.}\ }\textbf {\bibinfo {volume} {B797}},\ \bibinfo
  {pages} {134855} (\bibinfo {year} {2019})},\ \Eprint
  {http://arxiv.org/abs/1903.10844} {arXiv:1903.10844 [hep-ph]}\BibitemShut
  {NoStop}%
\bibitem [{\citenamefont {Raya}\ \emph {et~al.}(2020)\citenamefont {Raya},
  \citenamefont {Bashir},\ and\ \citenamefont {Roig}}]{Raya:2019dnh}%
  \BibitemOpen
  \bibfield  {author} {\bibinfo {author} {\bibfnamefont {K.}~\bibnamefont
  {Raya}}, \bibinfo {author} {\bibfnamefont {A.}~\bibnamefont {Bashir}}, \ and\
  \bibinfo {author} {\bibfnamefont {P.}~\bibnamefont {Roig}},\ }\href {\doibase
  10.1103/PhysRevD.101.074021} {\bibfield  {journal} {\bibinfo  {journal}
  {Phys. Rev.}\ }\textbf {\bibinfo {volume} {D101}},\ \bibinfo {pages} {074021}
  (\bibinfo {year} {2020})},\ \Eprint {http://arxiv.org/abs/1910.05960}
  {arXiv:1910.05960 [hep-ph]}\BibitemShut {NoStop}%
\bibitem [{\citenamefont {Chang}\ \emph {et~al.}(2013)\citenamefont {Chang},
  \citenamefont {Clo{\"e}t}, \citenamefont {Cobos-Martinez}, \citenamefont
  {Roberts}, \citenamefont {Schmidt},\ and\ \citenamefont
  {Tandy}}]{Chang:2013pq}%
  \BibitemOpen
  \bibfield  {author} {\bibinfo {author} {\bibfnamefont {L.}~\bibnamefont
  {Chang}}, \bibinfo {author} {\bibfnamefont {I.~C.}\ \bibnamefont
  {Clo{\"e}t}}, \bibinfo {author} {\bibfnamefont {J.~J.}\ \bibnamefont
  {Cobos-Martinez}}, \bibinfo {author} {\bibfnamefont {C.~D.}\ \bibnamefont
  {Roberts}}, \bibinfo {author} {\bibfnamefont {S.~M.}\ \bibnamefont
  {Schmidt}}, \ and\ \bibinfo {author} {\bibfnamefont {P.~C.}\ \bibnamefont
  {Tandy}},\ }\href {\doibase 10.1103/PhysRevLett.110.132001} {\bibfield
  {journal} {\bibinfo  {journal} {Phys. Rev. Lett.}\ }\textbf {\bibinfo
  {volume} {110}},\ \bibinfo {pages} {132001} (\bibinfo {year} {2013})},\
  \Eprint {http://arxiv.org/abs/1301.0324} {arXiv:1301.0324
  [nucl-th]}\BibitemShut {NoStop}%
\bibitem [{\citenamefont {Bali}\ \emph {et~al.}(2019)\citenamefont {Bali},
  \citenamefont {Braun}, \citenamefont {B{\"u}rger}, \citenamefont
  {G{\"o}ckeler}, \citenamefont {Gruber}, \citenamefont {Hutzler},
  \citenamefont {Korcyl}, \citenamefont {Sch{\"a}fer}, \citenamefont
  {Sternbeck},\ and\ \citenamefont {Wein}}]{Bali:2019dqc}%
  \BibitemOpen
  \bibfield  {author} {\bibinfo {author} {\bibfnamefont {G.~S.}\ \bibnamefont
  {Bali}}, \bibinfo {author} {\bibfnamefont {V.~M.}\ \bibnamefont {Braun}},
  \bibinfo {author} {\bibfnamefont {S.}~\bibnamefont {B{\"u}rger}}, \bibinfo
  {author} {\bibfnamefont {M.}~\bibnamefont {G{\"o}ckeler}}, \bibinfo {author}
  {\bibfnamefont {M.}~\bibnamefont {Gruber}}, \bibinfo {author} {\bibfnamefont
  {F.}~\bibnamefont {Hutzler}}, \bibinfo {author} {\bibfnamefont
  {P.}~\bibnamefont {Korcyl}}, \bibinfo {author} {\bibfnamefont
  {A.}~\bibnamefont {Sch{\"a}fer}}, \bibinfo {author} {\bibfnamefont
  {A.}~\bibnamefont {Sternbeck}}, \ and\ \bibinfo {author} {\bibfnamefont
  {P.}~\bibnamefont {Wein}} (\bibinfo {collaboration} {RQCD}),\ }\href
  {\doibase 10.1007/JHEP08(2019)065} {\bibfield  {journal} {\bibinfo  {journal}
  {JHEP}\ }\textbf {\bibinfo {volume} {08}},\ \bibinfo {pages} {065} (\bibinfo
  {year} {2019})},\ \Eprint {http://arxiv.org/abs/1903.08038} {arXiv:1903.08038
  [hep-lat]}\BibitemShut {NoStop}%
\bibitem [{\citenamefont {Hong}\ and\ \citenamefont {Kim}(2009)}]{Hong:2009zw}%
  \BibitemOpen
  \bibfield  {author} {\bibinfo {author} {\bibfnamefont {D.~K.}\ \bibnamefont
  {Hong}}\ and\ \bibinfo {author} {\bibfnamefont {D.}~\bibnamefont {Kim}},\
  }\href {\doibase 10.1016/j.physletb.2009.09.026} {\bibfield  {journal}
  {\bibinfo  {journal} {Phys. Lett.}\ }\textbf {\bibinfo {volume} {B680}},\
  \bibinfo {pages} {480} (\bibinfo {year} {2009})},\ \Eprint
  {http://arxiv.org/abs/0904.4042} {arXiv:0904.4042 [hep-ph]}\BibitemShut
  {NoStop}%
\bibitem [{\citenamefont {Cappiello}\ \emph {et~al.}(2011)\citenamefont
  {Cappiello}, \citenamefont {Cat\`a},\ and\ \citenamefont
  {D'Ambrosio}}]{Cappiello:2010uy}%
  \BibitemOpen
  \bibfield  {author} {\bibinfo {author} {\bibfnamefont {L.}~\bibnamefont
  {Cappiello}}, \bibinfo {author} {\bibfnamefont {O.}~\bibnamefont {Cat\`a}}, \
  and\ \bibinfo {author} {\bibfnamefont {G.}~\bibnamefont {D'Ambrosio}},\
  }\href {\doibase 10.1103/PhysRevD.83.093006} {\bibfield  {journal} {\bibinfo
  {journal} {Phys. Rev.}\ }\textbf {\bibinfo {volume} {D83}},\ \bibinfo {pages}
  {093006} (\bibinfo {year} {2011})},\ \Eprint {http://arxiv.org/abs/1009.1161}
  {arXiv:1009.1161 [hep-ph]}\BibitemShut {NoStop}%
\bibitem [{\citenamefont {Leutgeb}\ \emph {et~al.}(2019)\citenamefont
  {Leutgeb}, \citenamefont {Mager},\ and\ \citenamefont
  {Rebhan}}]{Leutgeb:2019zpq}%
  \BibitemOpen
  \bibfield  {author} {\bibinfo {author} {\bibfnamefont {J.}~\bibnamefont
  {Leutgeb}}, \bibinfo {author} {\bibfnamefont {J.}~\bibnamefont {Mager}}, \
  and\ \bibinfo {author} {\bibfnamefont {A.}~\bibnamefont {Rebhan}},\ }\href
  {\doibase 10.1103/PhysRevD.100.094038} {\bibfield  {journal} {\bibinfo
  {journal} {Phys. Rev. D}\ }\textbf {\bibinfo {volume} {100}},\ \bibinfo
  {pages} {094038} (\bibinfo {year} {2019})},\ \Eprint
  {http://arxiv.org/abs/1906.11795} {arXiv:1906.11795 [hep-ph]}\BibitemShut
  {NoStop}%
\bibitem [{\citenamefont {Czy{\.z}}\ \emph {et~al.}(2018)\citenamefont
  {Czy{\.z}}, \citenamefont {Kisza},\ and\ \citenamefont
  {Tracz}}]{Czyz:2017veo}%
  \BibitemOpen
  \bibfield  {author} {\bibinfo {author} {\bibfnamefont {H.}~\bibnamefont
  {Czy{\.z}}}, \bibinfo {author} {\bibfnamefont {P.}~\bibnamefont {Kisza}}, \
  and\ \bibinfo {author} {\bibfnamefont {S.}~\bibnamefont {Tracz}},\ }\href
  {\doibase 10.1103/PhysRevD.97.016006} {\bibfield  {journal} {\bibinfo
  {journal} {Phys. Rev.}\ }\textbf {\bibinfo {volume} {D97}},\ \bibinfo {pages}
  {016006} (\bibinfo {year} {2018})},\ \Eprint
  {http://arxiv.org/abs/1711.00820} {arXiv:1711.00820 [hep-ph]}\BibitemShut
  {NoStop}%
\bibitem [{\citenamefont {Guevara}\ \emph {et~al.}(2018)\citenamefont
  {Guevara}, \citenamefont {Roig},\ and\ \citenamefont
  {Sanz-Cillero}}]{Guevara:2018rhj}%
  \BibitemOpen
  \bibfield  {author} {\bibinfo {author} {\bibfnamefont {A.}~\bibnamefont
  {Guevara}}, \bibinfo {author} {\bibfnamefont {P.}~\bibnamefont {Roig}}, \
  and\ \bibinfo {author} {\bibfnamefont {J.~J.}\ \bibnamefont {Sanz-Cillero}},\
  }\href {\doibase 10.1007/JHEP06(2018)160} {\bibfield  {journal} {\bibinfo
  {journal} {JHEP}\ }\textbf {\bibinfo {volume} {06}},\ \bibinfo {pages} {160}
  (\bibinfo {year} {2018})},\ \Eprint {http://arxiv.org/abs/1803.08099}
  {arXiv:1803.08099 [hep-ph]}\BibitemShut {NoStop}%
\bibitem [{\citenamefont {Brodsky}\ and\ \citenamefont
  {Lepage}(1981)}]{Brodsky:1981rp}%
  \BibitemOpen
  \bibfield  {author} {\bibinfo {author} {\bibfnamefont {S.~J.}\ \bibnamefont
  {Brodsky}}\ and\ \bibinfo {author} {\bibfnamefont {G.~P.}\ \bibnamefont
  {Lepage}},\ }\href {\doibase 10.1103/PhysRevD.24.1808} {\bibfield  {journal}
  {\bibinfo  {journal} {Phys. Rev.}\ }\textbf {\bibinfo {volume} {D24}},\
  \bibinfo {pages} {1808} (\bibinfo {year} {1981})}\BibitemShut {NoStop}%
\bibitem [{\citenamefont {Danilkin}\ \emph {et~al.}(2019)\citenamefont
  {Danilkin}, \citenamefont {Redmer},\ and\ \citenamefont
  {Vanderhaeghen}}]{Danilkin:2019mhd}%
  \BibitemOpen
  \bibfield  {author} {\bibinfo {author} {\bibfnamefont {I.}~\bibnamefont
  {Danilkin}}, \bibinfo {author} {\bibfnamefont {C.~F.}\ \bibnamefont
  {Redmer}}, \ and\ \bibinfo {author} {\bibfnamefont {M.}~\bibnamefont
  {Vanderhaeghen}},\ }\href {\doibase 10.1016/j.ppnp.2019.05.002} {\bibfield
  {journal} {\bibinfo  {journal} {Prog. Part. Nucl. Phys.}\ }\textbf {\bibinfo
  {volume} {107}},\ \bibinfo {pages} {20} (\bibinfo {year} {2019})},\ \Eprint
  {http://arxiv.org/abs/1901.10346} {arXiv:1901.10346 [hep-ph]}\BibitemShut
  {NoStop}%
\bibitem [{\citenamefont {Bartos}\ \emph {et~al.}(2002)\citenamefont {Bartos},
  \citenamefont {Dubnickova}, \citenamefont {Dubnicka}, \citenamefont
  {Kuraev},\ and\ \citenamefont {Zemlyanaya}}]{Bartos:2001pg}%
  \BibitemOpen
  \bibfield  {author} {\bibinfo {author} {\bibfnamefont {E.}~\bibnamefont
  {Bartos}}, \bibinfo {author} {\bibfnamefont {A.~Z.}\ \bibnamefont
  {Dubnickova}}, \bibinfo {author} {\bibfnamefont {S.}~\bibnamefont
  {Dubnicka}}, \bibinfo {author} {\bibfnamefont {E.~A.}\ \bibnamefont
  {Kuraev}}, \ and\ \bibinfo {author} {\bibfnamefont {E.}~\bibnamefont
  {Zemlyanaya}},\ }\href {\doibase 10.1016/S0550-3213(02)00242-0} {\bibfield
  {journal} {\bibinfo  {journal} {Nucl. Phys.}\ }\textbf {\bibinfo {volume}
  {B632}},\ \bibinfo {pages} {330} (\bibinfo {year} {2002})},\ \Eprint
  {http://arxiv.org/abs/hep-ph/0106084} {arXiv:hep-ph/0106084
  [hep-ph]}\BibitemShut {NoStop}%
\bibitem [{\citenamefont {Dorokhov}\ and\ \citenamefont
  {Broniowski}(2008)}]{Dorokhov:2008pw}%
  \BibitemOpen
  \bibfield  {author} {\bibinfo {author} {\bibfnamefont {A.~E.}\ \bibnamefont
  {Dorokhov}}\ and\ \bibinfo {author} {\bibfnamefont {W.}~\bibnamefont
  {Broniowski}},\ }\href {\doibase 10.1103/PhysRevD.78.073011} {\bibfield
  {journal} {\bibinfo  {journal} {Phys. Rev.}\ }\textbf {\bibinfo {volume}
  {D78}},\ \bibinfo {pages} {073011} (\bibinfo {year} {2008})},\ \Eprint
  {http://arxiv.org/abs/0805.0760} {arXiv:0805.0760 [hep-ph]}\BibitemShut
  {NoStop}%
\bibitem [{\citenamefont {Dorokhov}\ \emph {et~al.}(2011)\citenamefont
  {Dorokhov}, \citenamefont {Radzhabov},\ and\ \citenamefont
  {Zhevlakov}}]{Dorokhov:2011zf}%
  \BibitemOpen
  \bibfield  {author} {\bibinfo {author} {\bibfnamefont {A.~E.}\ \bibnamefont
  {Dorokhov}}, \bibinfo {author} {\bibfnamefont {A.~E.}\ \bibnamefont
  {Radzhabov}}, \ and\ \bibinfo {author} {\bibfnamefont {A.~S.}\ \bibnamefont
  {Zhevlakov}},\ }\href {\doibase 10.1140/epjc/s10052-011-1702-6} {\bibfield
  {journal} {\bibinfo  {journal} {Eur. Phys. J.}\ }\textbf {\bibinfo {volume}
  {C71}},\ \bibinfo {pages} {1702} (\bibinfo {year} {2011})},\ \Eprint
  {http://arxiv.org/abs/1103.2042} {arXiv:1103.2042 [hep-ph]}\BibitemShut
  {NoStop}%
\bibitem [{\citenamefont {Kampf}\ and\ \citenamefont
  {Novotn\'y}(2011)}]{Kampf:2011ty}%
  \BibitemOpen
  \bibfield  {author} {\bibinfo {author} {\bibfnamefont {K.}~\bibnamefont
  {Kampf}}\ and\ \bibinfo {author} {\bibfnamefont {J.}~\bibnamefont
  {Novotn\'y}},\ }\href {\doibase 10.1103/PhysRevD.84.014036} {\bibfield
  {journal} {\bibinfo  {journal} {Phys. Rev.}\ }\textbf {\bibinfo {volume}
  {D84}},\ \bibinfo {pages} {014036} (\bibinfo {year} {2011})},\ \Eprint
  {http://arxiv.org/abs/1104.3137} {arXiv:1104.3137 [hep-ph]}\BibitemShut
  {NoStop}%
\bibitem [{\citenamefont {Roig}\ \emph {et~al.}(2014)\citenamefont {Roig},
  \citenamefont {Guevara},\ and\ \citenamefont
  {L{\'o}pez~Castro}}]{Roig:2014uja}%
  \BibitemOpen
  \bibfield  {author} {\bibinfo {author} {\bibfnamefont {P.}~\bibnamefont
  {Roig}}, \bibinfo {author} {\bibfnamefont {A.}~\bibnamefont {Guevara}}, \
  and\ \bibinfo {author} {\bibfnamefont {G.}~\bibnamefont {L{\'o}pez~Castro}},\
  }\href {\doibase 10.1103/PhysRevD.89.073016} {\bibfield  {journal} {\bibinfo
  {journal} {Phys. Rev.}\ }\textbf {\bibinfo {volume} {D89}},\ \bibinfo {pages}
  {073016} (\bibinfo {year} {2014})},\ \Eprint {http://arxiv.org/abs/1401.4099}
  {arXiv:1401.4099 [hep-ph]}\BibitemShut {NoStop}%
\bibitem [{\citenamefont {Hanhart}\ \emph {et~al.}(2013)\citenamefont
  {Hanhart}, \citenamefont {Kup{\'s}{\'c}}, \citenamefont {Mei{\ss}ner},
  \citenamefont {Stollenwerk},\ and\ \citenamefont {Wirzba}}]{Hanhart:2013vba}%
  \BibitemOpen
  \bibfield  {author} {\bibinfo {author} {\bibfnamefont {C.}~\bibnamefont
  {Hanhart}}, \bibinfo {author} {\bibfnamefont {A.}~\bibnamefont
  {Kup{\'s}{\'c}}}, \bibinfo {author} {\bibfnamefont {U.-G.}\ \bibnamefont
  {Mei{\ss}ner}}, \bibinfo {author} {\bibfnamefont {F.}~\bibnamefont
  {Stollenwerk}}, \ and\ \bibinfo {author} {\bibfnamefont {A.}~\bibnamefont
  {Wirzba}},\ }\href {\doibase 10.1140/epjc/s10052-013-2668-3} {\bibfield
  {journal} {\bibinfo  {journal} {Eur. Phys. J.}\ }\textbf {\bibinfo {volume}
  {C73}},\ \bibinfo {pages} {2668} (\bibinfo {year} {2013})},\ \bibinfo {note}
  {[Erratum: Eur.\ Phys.\ J.\ {\bf C75}, 242 (2015)]},\ \Eprint
  {http://arxiv.org/abs/1307.5654} {arXiv:1307.5654 [hep-ph]}\BibitemShut
  {NoStop}%
\bibitem [{\citenamefont {Adlarson}\ \emph {et~al.}(2012)\citenamefont
  {Adlarson} \emph {et~al.}}]{Adlarson:2011xb}%
  \BibitemOpen
  \bibfield  {author} {\bibinfo {author} {\bibfnamefont {P.}~\bibnamefont
  {Adlarson}} \emph {et~al.} (\bibinfo {collaboration} {WASA-at-COSY}),\ }\href
  {\doibase 10.1016/j.physletb.2011.12.027} {\bibfield  {journal} {\bibinfo
  {journal} {Phys. Lett.}\ }\textbf {\bibinfo {volume} {B707}},\ \bibinfo
  {pages} {243} (\bibinfo {year} {2012})},\ \Eprint
  {http://arxiv.org/abs/1107.5277} {arXiv:1107.5277 [nucl-ex]}\BibitemShut
  {NoStop}%
\bibitem [{\citenamefont {Babusci}\ \emph
  {et~al.}(2013{\natexlab{c}})\citenamefont {Babusci} \emph
  {et~al.}}]{Babusci:2012ft}%
  \BibitemOpen
  \bibfield  {author} {\bibinfo {author} {\bibfnamefont {D.}~\bibnamefont
  {Babusci}} \emph {et~al.} (\bibinfo {collaboration} {KLOE}),\ }\href
  {\doibase 10.1016/j.physletb.2012.11.032} {\bibfield  {journal} {\bibinfo
  {journal} {Phys. Lett.}\ }\textbf {\bibinfo {volume} {B718}},\ \bibinfo
  {pages} {910} (\bibinfo {year} {2013}{\natexlab{c}})},\ \Eprint
  {http://arxiv.org/abs/1209.4611} {arXiv:1209.4611 [hep-ex]}\BibitemShut
  {NoStop}%
\bibitem [{\citenamefont {Ablikim}\ \emph
  {et~al.}(2018{\natexlab{b}})\citenamefont {Ablikim} \emph
  {et~al.}}]{Ablikim:2017fll}%
  \BibitemOpen
  \bibfield  {author} {\bibinfo {author} {\bibfnamefont {M.}~\bibnamefont
  {Ablikim}} \emph {et~al.} (\bibinfo {collaboration} {BESIII}),\ }\href
  {\doibase 10.1103/PhysRevLett.120.242003} {\bibfield  {journal} {\bibinfo
  {journal} {Phys. Rev. Lett.}\ }\textbf {\bibinfo {volume} {120}},\ \bibinfo
  {pages} {242003} (\bibinfo {year} {2018}{\natexlab{b}})},\ \Eprint
  {http://arxiv.org/abs/1712.01525} {arXiv:1712.01525 [hep-ex]}\BibitemShut
  {NoStop}%
\bibitem [{\citenamefont {Stollenwerk}\ \emph {et~al.}(2012)\citenamefont
  {Stollenwerk}, \citenamefont {Hanhart}, \citenamefont {Kup{\'s}{\'c}},
  \citenamefont {Mei{\ss}ner},\ and\ \citenamefont
  {Wirzba}}]{Stollenwerk:2011zz}%
  \BibitemOpen
  \bibfield  {author} {\bibinfo {author} {\bibfnamefont {F.}~\bibnamefont
  {Stollenwerk}}, \bibinfo {author} {\bibfnamefont {C.}~\bibnamefont
  {Hanhart}}, \bibinfo {author} {\bibfnamefont {A.}~\bibnamefont
  {Kup{\'s}{\'c}}}, \bibinfo {author} {\bibfnamefont {U.-G.}\ \bibnamefont
  {Mei{\ss}ner}}, \ and\ \bibinfo {author} {\bibfnamefont {A.}~\bibnamefont
  {Wirzba}},\ }\href {\doibase 10.1016/j.physletb.2011.12.008} {\bibfield
  {journal} {\bibinfo  {journal} {Phys. Lett.}\ }\textbf {\bibinfo {volume}
  {B707}},\ \bibinfo {pages} {184} (\bibinfo {year} {2012})},\ \Eprint
  {http://arxiv.org/abs/1108.2419} {arXiv:1108.2419 [nucl-th]}\BibitemShut
  {NoStop}%
\bibitem [{\citenamefont {Xiao}\ \emph {et~al.}(2015)\citenamefont {Xiao},
  \citenamefont {Dato}, \citenamefont {Hanhart}, \citenamefont {Kubis},
  \citenamefont {Mei{\ss}ner},\ and\ \citenamefont {Wirzba}}]{Xiao:2015uva}%
  \BibitemOpen
  \bibfield  {author} {\bibinfo {author} {\bibfnamefont {C.-W.}\ \bibnamefont
  {Xiao}}, \bibinfo {author} {\bibfnamefont {T.}~\bibnamefont {Dato}}, \bibinfo
  {author} {\bibfnamefont {C.}~\bibnamefont {Hanhart}}, \bibinfo {author}
  {\bibfnamefont {B.}~\bibnamefont {Kubis}}, \bibinfo {author} {\bibfnamefont
  {U.-G.}\ \bibnamefont {Mei{\ss}ner}}, \ and\ \bibinfo {author} {\bibfnamefont
  {A.}~\bibnamefont {Wirzba}},\ }\href@noop {} {\  (\bibinfo {year} {2015})},\
  \Eprint {http://arxiv.org/abs/1509.02194} {arXiv:1509.02194
  [hep-ph]}\BibitemShut {NoStop}%
\bibitem [{\citenamefont {Guo}\ \emph {et~al.}(2012)\citenamefont {Guo},
  \citenamefont {Kubis},\ and\ \citenamefont {Wirzba}}]{Guo:2011ir}%
  \BibitemOpen
  \bibfield  {author} {\bibinfo {author} {\bibfnamefont {F.-K.}\ \bibnamefont
  {Guo}}, \bibinfo {author} {\bibfnamefont {B.}~\bibnamefont {Kubis}}, \ and\
  \bibinfo {author} {\bibfnamefont {A.}~\bibnamefont {Wirzba}},\ }\href
  {\doibase 10.1103/PhysRevD.85.014014} {\bibfield  {journal} {\bibinfo
  {journal} {Phys. Rev.}\ }\textbf {\bibinfo {volume} {D85}},\ \bibinfo {pages}
  {014014} (\bibinfo {year} {2012})},\ \Eprint {http://arxiv.org/abs/1111.5949}
  {arXiv:1111.5949 [hep-ph]}\BibitemShut {NoStop}%
\bibitem [{\citenamefont {Ablikim}\ \emph {et~al.}(2014)\citenamefont {Ablikim}
  \emph {et~al.}}]{Ablikim:2014eoc}%
  \BibitemOpen
  \bibfield  {author} {\bibinfo {author} {\bibfnamefont {M.}~\bibnamefont
  {Ablikim}} \emph {et~al.} (\bibinfo {collaboration} {BESIII}),\ }\href
  {\doibase 10.1103/PhysRevLett.112.251801} {\bibfield  {journal} {\bibinfo
  {journal} {Phys. Rev. Lett.}\ }\textbf {\bibinfo {volume} {112}},\ \bibinfo
  {pages} {251801} (\bibinfo {year} {2014})},\ \bibinfo {note} {[Addendum:
  Phys.\ Rev.\ Lett.\ {\bf 113}, 039903 (2014)]},\ \Eprint
  {http://arxiv.org/abs/1404.0096} {arXiv:1404.0096 [hep-ex]}\BibitemShut
  {NoStop}%
\bibitem [{\citenamefont {Ding}\ \emph {et~al.}(2019)\citenamefont {Ding},
  \citenamefont {Raya}, \citenamefont {Bashir}, \citenamefont {Binosi},
  \citenamefont {Chang}, \citenamefont {Chen},\ and\ \citenamefont
  {Roberts}}]{Ding:2018xwy}%
  \BibitemOpen
  \bibfield  {author} {\bibinfo {author} {\bibfnamefont {M.}~\bibnamefont
  {Ding}}, \bibinfo {author} {\bibfnamefont {K.}~\bibnamefont {Raya}}, \bibinfo
  {author} {\bibfnamefont {A.}~\bibnamefont {Bashir}}, \bibinfo {author}
  {\bibfnamefont {D.}~\bibnamefont {Binosi}}, \bibinfo {author} {\bibfnamefont
  {L.}~\bibnamefont {Chang}}, \bibinfo {author} {\bibfnamefont
  {M.}~\bibnamefont {Chen}}, \ and\ \bibinfo {author} {\bibfnamefont {C.~D.}\
  \bibnamefont {Roberts}},\ }\href {\doibase 10.1103/PhysRevD.99.014014}
  {\bibfield  {journal} {\bibinfo  {journal} {Phys. Rev.}\ }\textbf {\bibinfo
  {volume} {D99}},\ \bibinfo {pages} {014014} (\bibinfo {year} {2019})},\
  \Eprint {http://arxiv.org/abs/1810.12313} {arXiv:1810.12313
  [nucl-th]}\BibitemShut {NoStop}%
\bibitem [{\citenamefont {Mandelstam}(1958)}]{Mandelstam:1958xc}%
  \BibitemOpen
  \bibfield  {author} {\bibinfo {author} {\bibfnamefont {S.}~\bibnamefont
  {Mandelstam}},\ }\href {\doibase 10.1103/PhysRev.112.1344} {\bibfield
  {journal} {\bibinfo  {journal} {Phys. Rev.}\ }\textbf {\bibinfo {volume}
  {112}},\ \bibinfo {pages} {1344} (\bibinfo {year} {1958})}\BibitemShut
  {NoStop}%
\bibitem [{\citenamefont {Pascalutsa}\ \emph {et~al.}(2012)\citenamefont
  {Pascalutsa}, \citenamefont {Pauk},\ and\ \citenamefont
  {Vanderhaeghen}}]{Pascalutsa:2012pr}%
  \BibitemOpen
  \bibfield  {author} {\bibinfo {author} {\bibfnamefont {V.}~\bibnamefont
  {Pascalutsa}}, \bibinfo {author} {\bibfnamefont {V.}~\bibnamefont {Pauk}}, \
  and\ \bibinfo {author} {\bibfnamefont {M.}~\bibnamefont {Vanderhaeghen}},\
  }\href {\doibase 10.1103/PhysRevD.85.116001} {\bibfield  {journal} {\bibinfo
  {journal} {Phys. Rev.}\ }\textbf {\bibinfo {volume} {D85}},\ \bibinfo {pages}
  {116001} (\bibinfo {year} {2012})},\ \Eprint {http://arxiv.org/abs/1204.0740}
  {arXiv:1204.0740 [hep-ph]}\BibitemShut {NoStop}%
\bibitem [{\citenamefont {Danilkin}\ and\ \citenamefont
  {Vanderhaeghen}(2019)}]{Danilkin:2018qfn}%
  \BibitemOpen
  \bibfield  {author} {\bibinfo {author} {\bibfnamefont {I.}~\bibnamefont
  {Danilkin}}\ and\ \bibinfo {author} {\bibfnamefont {M.}~\bibnamefont
  {Vanderhaeghen}},\ }\href {\doibase 10.1016/j.physletb.2018.12.047}
  {\bibfield  {journal} {\bibinfo  {journal} {Phys. Lett.}\ }\textbf {\bibinfo
  {volume} {B789}},\ \bibinfo {pages} {366} (\bibinfo {year} {2019})},\ \Eprint
  {http://arxiv.org/abs/1810.03669} {arXiv:1810.03669 [hep-ph]}\BibitemShut
  {NoStop}%
\bibitem [{\citenamefont {Hoferichter}\ and\ \citenamefont
  {Stoffer}(2019)}]{Hoferichter:2019nlq}%
  \BibitemOpen
  \bibfield  {author} {\bibinfo {author} {\bibfnamefont {M.}~\bibnamefont
  {Hoferichter}}\ and\ \bibinfo {author} {\bibfnamefont {P.}~\bibnamefont
  {Stoffer}},\ }\href {\doibase 10.1007/JHEP07(2019)073} {\bibfield  {journal}
  {\bibinfo  {journal} {JHEP}\ }\textbf {\bibinfo {volume} {07}},\ \bibinfo
  {pages} {073} (\bibinfo {year} {2019})},\ \Eprint
  {http://arxiv.org/abs/1905.13198} {arXiv:1905.13198 [hep-ph]}\BibitemShut
  {NoStop}%
\bibitem [{\citenamefont {Boyer}\ \emph {et~al.}(1990)\citenamefont {Boyer}
  \emph {et~al.}}]{Boyer:1990vu}%
  \BibitemOpen
  \bibfield  {author} {\bibinfo {author} {\bibfnamefont {J.}~\bibnamefont
  {Boyer}} \emph {et~al.} (\bibinfo {collaboration} {Mark II}),\ }\href
  {\doibase 10.1103/PhysRevD.42.1350} {\bibfield  {journal} {\bibinfo
  {journal} {Phys. Rev.}\ }\textbf {\bibinfo {volume} {D42}},\ \bibinfo {pages}
  {1350} (\bibinfo {year} {1990})}\BibitemShut {NoStop}%
\bibitem [{\citenamefont {Behrend}\ \emph {et~al.}(1992)\citenamefont {Behrend}
  \emph {et~al.}}]{Behrend:1992hy}%
  \BibitemOpen
  \bibfield  {author} {\bibinfo {author} {\bibfnamefont {H.~J.}\ \bibnamefont
  {Behrend}} \emph {et~al.} (\bibinfo {collaboration} {CELLO}),\ }\href
  {\doibase 10.1007/BF01565945} {\bibfield  {journal} {\bibinfo  {journal} {Z.
  Phys.}\ }\textbf {\bibinfo {volume} {C56}},\ \bibinfo {pages} {381} (\bibinfo
  {year} {1992})}\BibitemShut {NoStop}%
\bibitem [{\citenamefont {Marsiske}\ \emph {et~al.}(1990)\citenamefont
  {Marsiske} \emph {et~al.}}]{Marsiske:1990hx}%
  \BibitemOpen
  \bibfield  {author} {\bibinfo {author} {\bibfnamefont {H.}~\bibnamefont
  {Marsiske}} \emph {et~al.} (\bibinfo {collaboration} {Crystal Ball}),\ }\href
  {\doibase 10.1103/PhysRevD.41.3324} {\bibfield  {journal} {\bibinfo
  {journal} {Phys. Rev.}\ }\textbf {\bibinfo {volume} {D41}},\ \bibinfo {pages}
  {3324} (\bibinfo {year} {1990})}\BibitemShut {NoStop}%
\bibitem [{\citenamefont {Drechsel}\ \emph {et~al.}(1998)\citenamefont
  {Drechsel}, \citenamefont {Kn{\"o}chlein}, \citenamefont {Korchin},
  \citenamefont {Metz},\ and\ \citenamefont {Scherer}}]{Drechsel:1997xv}%
  \BibitemOpen
  \bibfield  {author} {\bibinfo {author} {\bibfnamefont {D.}~\bibnamefont
  {Drechsel}}, \bibinfo {author} {\bibfnamefont {G.}~\bibnamefont
  {Kn{\"o}chlein}}, \bibinfo {author} {\bibfnamefont {A.~{\relax Yu}.}\
  \bibnamefont {Korchin}}, \bibinfo {author} {\bibfnamefont {A.}~\bibnamefont
  {Metz}}, \ and\ \bibinfo {author} {\bibfnamefont {S.}~\bibnamefont
  {Scherer}},\ }\href {\doibase 10.1103/PhysRevC.57.941} {\bibfield  {journal}
  {\bibinfo  {journal} {Phys. Rev.}\ }\textbf {\bibinfo {volume} {C57}},\
  \bibinfo {pages} {941} (\bibinfo {year} {1998})},\ \Eprint
  {http://arxiv.org/abs/nucl-th/9704064} {arXiv:nucl-th/9704064
  [nucl-th]}\BibitemShut {NoStop}%
\bibitem [{\citenamefont {G\'omez~Nicola}\ \emph {et~al.}(2008)\citenamefont
  {G\'omez~Nicola}, \citenamefont {Pel\'aez},\ and\ \citenamefont
  {R\'ios}}]{GomezNicola:2007qj}%
  \BibitemOpen
  \bibfield  {author} {\bibinfo {author} {\bibfnamefont {A.}~\bibnamefont
  {G\'omez~Nicola}}, \bibinfo {author} {\bibfnamefont {J.~R.}\ \bibnamefont
  {Pel\'aez}}, \ and\ \bibinfo {author} {\bibfnamefont {G.}~\bibnamefont
  {R\'ios}},\ }\href {\doibase 10.1103/PhysRevD.77.056006} {\bibfield
  {journal} {\bibinfo  {journal} {Phys. Rev.}\ }\textbf {\bibinfo {volume}
  {D77}},\ \bibinfo {pages} {056006} (\bibinfo {year} {2008})},\ \Eprint
  {http://arxiv.org/abs/0712.2763} {arXiv:0712.2763 [hep-ph]}\BibitemShut
  {NoStop}%
\bibitem [{\citenamefont {Colangelo}\ \emph {et~al.}(2001)\citenamefont
  {Colangelo}, \citenamefont {Gasser},\ and\ \citenamefont
  {Leutwyler}}]{Colangelo:2001df}%
  \BibitemOpen
  \bibfield  {author} {\bibinfo {author} {\bibfnamefont {G.}~\bibnamefont
  {Colangelo}}, \bibinfo {author} {\bibfnamefont {J.}~\bibnamefont {Gasser}}, \
  and\ \bibinfo {author} {\bibfnamefont {H.}~\bibnamefont {Leutwyler}},\ }\href
  {\doibase 10.1016/S0550-3213(01)00147-X} {\bibfield  {journal} {\bibinfo
  {journal} {Nucl. Phys.}\ }\textbf {\bibinfo {volume} {B603}},\ \bibinfo
  {pages} {125} (\bibinfo {year} {2001})},\ \Eprint
  {http://arxiv.org/abs/hep-ph/0103088} {arXiv:hep-ph/0103088
  [hep-ph]}\BibitemShut {NoStop}%
\bibitem [{\citenamefont {Danilkin}\ \emph {et~al.}(2020)\citenamefont
  {Danilkin}, \citenamefont {Deineka},\ and\ \citenamefont
  {Vanderhaeghen}}]{Danilkin:2019opj}%
  \BibitemOpen
  \bibfield  {author} {\bibinfo {author} {\bibfnamefont {I.}~\bibnamefont
  {Danilkin}}, \bibinfo {author} {\bibfnamefont {O.}~\bibnamefont {Deineka}}, \
  and\ \bibinfo {author} {\bibfnamefont {M.}~\bibnamefont {Vanderhaeghen}},\
  }\href {\doibase 10.1103/PhysRevD.101.054008} {\bibfield  {journal} {\bibinfo
   {journal} {Phys. Rev.}\ }\textbf {\bibinfo {volume} {D101}},\ \bibinfo
  {pages} {054008} (\bibinfo {year} {2020})},\ \Eprint
  {http://arxiv.org/abs/1909.04158} {arXiv:1909.04158 [hep-ph]}\BibitemShut
  {NoStop}%
\bibitem [{\citenamefont {Dai}\ and\ \citenamefont
  {Pennington}(2014)}]{Dai:2014zta}%
  \BibitemOpen
  \bibfield  {author} {\bibinfo {author} {\bibfnamefont {L.-Y.}\ \bibnamefont
  {Dai}}\ and\ \bibinfo {author} {\bibfnamefont {M.~R.}\ \bibnamefont
  {Pennington}},\ }\href {\doibase 10.1103/PhysRevD.90.036004} {\bibfield
  {journal} {\bibinfo  {journal} {Phys. Rev.}\ }\textbf {\bibinfo {volume}
  {D90}},\ \bibinfo {pages} {036004} (\bibinfo {year} {2014})},\ \Eprint
  {http://arxiv.org/abs/1404.7524} {arXiv:1404.7524 [hep-ph]}\BibitemShut
  {NoStop}%
\bibitem [{\citenamefont {Garc\'ia-Mart\'in}\ and\ \citenamefont
  {Moussallam}(2010)}]{GarciaMartin:2010cw}%
  \BibitemOpen
  \bibfield  {author} {\bibinfo {author} {\bibfnamefont {R.}~\bibnamefont
  {Garc\'ia-Mart\'in}}\ and\ \bibinfo {author} {\bibfnamefont {B.}~\bibnamefont
  {Moussallam}},\ }\href {\doibase 10.1140/epjc/s10052-010-1471-7} {\bibfield
  {journal} {\bibinfo  {journal} {Eur. Phys. J.}\ }\textbf {\bibinfo {volume}
  {C70}},\ \bibinfo {pages} {155} (\bibinfo {year} {2010})},\ \Eprint
  {http://arxiv.org/abs/1006.5373} {arXiv:1006.5373 [hep-ph]}\BibitemShut
  {NoStop}%
\bibitem [{\citenamefont {Hoferichter}\ \emph {et~al.}(2011)\citenamefont
  {Hoferichter}, \citenamefont {Phillips},\ and\ \citenamefont
  {Schat}}]{Hoferichter:2011wk}%
  \BibitemOpen
  \bibfield  {author} {\bibinfo {author} {\bibfnamefont {M.}~\bibnamefont
  {Hoferichter}}, \bibinfo {author} {\bibfnamefont {D.~R.}\ \bibnamefont
  {Phillips}}, \ and\ \bibinfo {author} {\bibfnamefont {C.}~\bibnamefont
  {Schat}},\ }\href {\doibase 10.1140/epjc/s10052-011-1743-x} {\bibfield
  {journal} {\bibinfo  {journal} {Eur. Phys. J.}\ }\textbf {\bibinfo {volume}
  {C71}},\ \bibinfo {pages} {1743} (\bibinfo {year} {2011})},\ \Eprint
  {http://arxiv.org/abs/1106.4147} {arXiv:1106.4147 [hep-ph]}\BibitemShut
  {NoStop}%
\bibitem [{\citenamefont {Moussallam}(2013)}]{Moussallam:2013una}%
  \BibitemOpen
  \bibfield  {author} {\bibinfo {author} {\bibfnamefont {B.}~\bibnamefont
  {Moussallam}},\ }\href {\doibase 10.1140/epjc/s10052-013-2539-y} {\bibfield
  {journal} {\bibinfo  {journal} {Eur. Phys. J.}\ }\textbf {\bibinfo {volume}
  {C73}},\ \bibinfo {pages} {2539} (\bibinfo {year} {2013})},\ \Eprint
  {http://arxiv.org/abs/1305.3143} {arXiv:1305.3143 [hep-ph]}\BibitemShut
  {NoStop}%
\bibitem [{\citenamefont {Gasparyan}\ and\ \citenamefont
  {Lutz}(2010)}]{Gasparyan:2010xz}%
  \BibitemOpen
  \bibfield  {author} {\bibinfo {author} {\bibfnamefont {A.}~\bibnamefont
  {Gasparyan}}\ and\ \bibinfo {author} {\bibfnamefont {M.~F.~M.}\ \bibnamefont
  {Lutz}},\ }\href {\doibase 10.1016/j.nuclphysa.2010.08.006} {\bibfield
  {journal} {\bibinfo  {journal} {Nucl. Phys.}\ }\textbf {\bibinfo {volume}
  {A848}},\ \bibinfo {pages} {126} (\bibinfo {year} {2010})},\ \Eprint
  {http://arxiv.org/abs/1003.3426} {arXiv:1003.3426 [hep-ph]}\BibitemShut
  {NoStop}%
\bibitem [{\citenamefont {Danilkin}\ \emph
  {et~al.}(2011{\natexlab{a}})\citenamefont {Danilkin}, \citenamefont
  {Gasparyan},\ and\ \citenamefont {Lutz}}]{Danilkin:2010xd}%
  \BibitemOpen
  \bibfield  {author} {\bibinfo {author} {\bibfnamefont {I.~V.}\ \bibnamefont
  {Danilkin}}, \bibinfo {author} {\bibfnamefont {A.~M.}\ \bibnamefont
  {Gasparyan}}, \ and\ \bibinfo {author} {\bibfnamefont {M.~F.~M.}\
  \bibnamefont {Lutz}},\ }\href {\doibase 10.1016/j.physletb.2011.01.036}
  {\bibfield  {journal} {\bibinfo  {journal} {Phys. Lett.}\ }\textbf {\bibinfo
  {volume} {B697}},\ \bibinfo {pages} {147} (\bibinfo {year}
  {2011}{\natexlab{a}})},\ \Eprint {http://arxiv.org/abs/1009.5928}
  {arXiv:1009.5928 [hep-ph]}\BibitemShut {NoStop}%
\bibitem [{\citenamefont {Danilkin}\ \emph
  {et~al.}(2011{\natexlab{b}})\citenamefont {Danilkin}, \citenamefont {Gil},\
  and\ \citenamefont {Lutz}}]{Danilkin:2011fz}%
  \BibitemOpen
  \bibfield  {author} {\bibinfo {author} {\bibfnamefont {I.~V.}\ \bibnamefont
  {Danilkin}}, \bibinfo {author} {\bibfnamefont {L.~I.~R.}\ \bibnamefont
  {Gil}}, \ and\ \bibinfo {author} {\bibfnamefont {M.~F.~M.}\ \bibnamefont
  {Lutz}},\ }\href {\doibase 10.1016/j.physletb.2011.08.001} {\bibfield
  {journal} {\bibinfo  {journal} {Phys. Lett.}\ }\textbf {\bibinfo {volume}
  {B703}},\ \bibinfo {pages} {504} (\bibinfo {year} {2011}{\natexlab{b}})},\
  \Eprint {http://arxiv.org/abs/1106.2230} {arXiv:1106.2230
  [hep-ph]}\BibitemShut {NoStop}%
\bibitem [{\citenamefont {Chew}\ and\ \citenamefont
  {Mandelstam}(1960)}]{Chew:1960iv}%
  \BibitemOpen
  \bibfield  {author} {\bibinfo {author} {\bibfnamefont {G.~F.}\ \bibnamefont
  {Chew}}\ and\ \bibinfo {author} {\bibfnamefont {S.}~\bibnamefont
  {Mandelstam}},\ }\href {\doibase 10.1103/PhysRev.119.467} {\bibfield
  {journal} {\bibinfo  {journal} {Phys. Rev.}\ }\textbf {\bibinfo {volume}
  {119}},\ \bibinfo {pages} {467} (\bibinfo {year} {1960})}\BibitemShut
  {NoStop}%
\bibitem [{\citenamefont {B\"uttiker}\ \emph {et~al.}(2004)\citenamefont
  {B\"uttiker}, \citenamefont {Descotes-Genon},\ and\ \citenamefont
  {Moussallam}}]{Buettiker:2003pp}%
  \BibitemOpen
  \bibfield  {author} {\bibinfo {author} {\bibfnamefont {P.}~\bibnamefont
  {B\"uttiker}}, \bibinfo {author} {\bibfnamefont {S.}~\bibnamefont
  {Descotes-Genon}}, \ and\ \bibinfo {author} {\bibfnamefont {B.}~\bibnamefont
  {Moussallam}},\ }\href {\doibase 10.1140/epjc/s2004-01591-1} {\bibfield
  {journal} {\bibinfo  {journal} {Eur. Phys. J.}\ }\textbf {\bibinfo {volume}
  {C33}},\ \bibinfo {pages} {409} (\bibinfo {year} {2004})},\ \Eprint
  {http://arxiv.org/abs/hep-ph/0310283} {arXiv:hep-ph/0310283
  [hep-ph]}\BibitemShut {NoStop}%
\bibitem [{\citenamefont {Pel\'aez}\ and\ \citenamefont
  {Rodas}(2018)}]{Pelaez:2018qny}%
  \BibitemOpen
  \bibfield  {author} {\bibinfo {author} {\bibfnamefont {J.~R.}\ \bibnamefont
  {Pel\'aez}}\ and\ \bibinfo {author} {\bibfnamefont {A.}~\bibnamefont
  {Rodas}},\ }\href {\doibase 10.1140/epjc/s10052-018-6296-9} {\bibfield
  {journal} {\bibinfo  {journal} {Eur. Phys. J.}\ }\textbf {\bibinfo {volume}
  {C78}},\ \bibinfo {pages} {897} (\bibinfo {year} {2018})},\ \Eprint
  {http://arxiv.org/abs/1807.04543} {arXiv:1807.04543 [hep-ph]}\BibitemShut
  {NoStop}%
\bibitem [{\citenamefont {Budnev}\ \emph {et~al.}(1975)\citenamefont {Budnev},
  \citenamefont {Ginzburg}, \citenamefont {Meledin},\ and\ \citenamefont
  {Serbo}}]{Budnev:1974de}%
  \BibitemOpen
  \bibfield  {author} {\bibinfo {author} {\bibfnamefont {V.~M.}\ \bibnamefont
  {Budnev}}, \bibinfo {author} {\bibfnamefont {I.~F.}\ \bibnamefont
  {Ginzburg}}, \bibinfo {author} {\bibfnamefont {G.~V.}\ \bibnamefont
  {Meledin}}, \ and\ \bibinfo {author} {\bibfnamefont {V.~G.}\ \bibnamefont
  {Serbo}},\ }\href {\doibase 10.1016/0370-1573(75)90009-5} {\bibfield
  {journal} {\bibinfo  {journal} {Phys. Rept.}\ }\textbf {\bibinfo {volume}
  {15}},\ \bibinfo {pages} {181} (\bibinfo {year} {1975})}\BibitemShut
  {NoStop}%
\bibitem [{\citenamefont {Redmer}(2017)}]{Redmer:2017fhg}%
  \BibitemOpen
  \bibfield  {author} {\bibinfo {author} {\bibfnamefont {C.~F.}\ \bibnamefont
  {Redmer}} (\bibinfo {collaboration} {BESIII}),\ }\href {\doibase
  10.1016/j.nuclphysbps.2017.03.053} {\bibfield  {journal} {\bibinfo  {journal}
  {Nucl. Part. Phys. Proc.}\ }\textbf {\bibinfo {volume} {287-288}},\ \bibinfo
  {pages} {99} (\bibinfo {year} {2017})}\BibitemShut {NoStop}%
\bibitem [{\citenamefont {Bijnens}()}]{Bijnens-private}%
  \BibitemOpen
  \bibfield  {author} {\bibinfo {author} {\bibfnamefont {J.}~\bibnamefont
  {Bijnens}},\ }\href@noop {} {}\bibinfo {howpublished} {private
  communication}\BibitemShut {NoStop}%
\bibitem [{\citenamefont {Blatnik}\ \emph {et~al.}(1979)\citenamefont
  {Blatnik}, \citenamefont {Stahov},\ and\ \citenamefont
  {Lang}}]{Blatnik:1978wj}%
  \BibitemOpen
  \bibfield  {author} {\bibinfo {author} {\bibfnamefont {S.}~\bibnamefont
  {Blatnik}}, \bibinfo {author} {\bibfnamefont {J.}~\bibnamefont {Stahov}}, \
  and\ \bibinfo {author} {\bibfnamefont {C.~B.}\ \bibnamefont {Lang}},\ }\href
  {\doibase 10.1007/BF02725742} {\bibfield  {journal} {\bibinfo  {journal}
  {Lett. Nuovo Cim.}\ }\textbf {\bibinfo {volume} {24}},\ \bibinfo {pages} {39}
  (\bibinfo {year} {1979})}\BibitemShut {NoStop}%
\bibitem [{\citenamefont {Ecker}\ \emph
  {et~al.}(1989{\natexlab{b}})\citenamefont {Ecker}, \citenamefont {Gasser},
  \citenamefont {Pich},\ and\ \citenamefont {de~Rafael}}]{Ecker:1988te}%
  \BibitemOpen
  \bibfield  {author} {\bibinfo {author} {\bibfnamefont {G.}~\bibnamefont
  {Ecker}}, \bibinfo {author} {\bibfnamefont {J.}~\bibnamefont {Gasser}},
  \bibinfo {author} {\bibfnamefont {A.}~\bibnamefont {Pich}}, \ and\ \bibinfo
  {author} {\bibfnamefont {E.}~\bibnamefont {de~Rafael}},\ }\href {\doibase
  10.1016/0550-3213(89)90346-5} {\bibfield  {journal} {\bibinfo  {journal}
  {Nucl. Phys.}\ }\textbf {\bibinfo {volume} {B321}},\ \bibinfo {pages} {311}
  (\bibinfo {year} {1989}{\natexlab{b}})}\BibitemShut {NoStop}%
\bibitem [{\citenamefont {Oller}\ and\ \citenamefont
  {Oset}(1998)}]{Oller:1997yg}%
  \BibitemOpen
  \bibfield  {author} {\bibinfo {author} {\bibfnamefont {J.~A.}\ \bibnamefont
  {Oller}}\ and\ \bibinfo {author} {\bibfnamefont {E.}~\bibnamefont {Oset}},\
  }\href {\doibase 10.1016/S0375-9474(97)00649-0} {\bibfield  {journal}
  {\bibinfo  {journal} {Nucl. Phys.}\ }\textbf {\bibinfo {volume} {A629}},\
  \bibinfo {pages} {739} (\bibinfo {year} {1998})},\ \Eprint
  {http://arxiv.org/abs/hep-ph/9706487} {arXiv:hep-ph/9706487
  [hep-ph]}\BibitemShut {NoStop}%
\bibitem [{\citenamefont {Danilkin}\ \emph {et~al.}(2013)\citenamefont
  {Danilkin}, \citenamefont {Lutz}, \citenamefont {Leupold},\ and\
  \citenamefont {Terschl{\"u}sen}}]{Danilkin:2012ua}%
  \BibitemOpen
  \bibfield  {author} {\bibinfo {author} {\bibfnamefont {I.~V.}\ \bibnamefont
  {Danilkin}}, \bibinfo {author} {\bibfnamefont {M.~F.~M.}\ \bibnamefont
  {Lutz}}, \bibinfo {author} {\bibfnamefont {S.}~\bibnamefont {Leupold}}, \
  and\ \bibinfo {author} {\bibfnamefont {C.}~\bibnamefont {Terschl{\"u}sen}},\
  }\href {\doibase 10.1140/epjc/s10052-013-2358-1} {\bibfield  {journal}
  {\bibinfo  {journal} {Eur. Phys. J.}\ }\textbf {\bibinfo {volume} {C73}},\
  \bibinfo {pages} {2358} (\bibinfo {year} {2013})},\ \Eprint
  {http://arxiv.org/abs/1211.1503} {arXiv:1211.1503 [hep-ph]}\BibitemShut
  {NoStop}%
\bibitem [{\citenamefont {Danilkin}\ \emph {et~al.}(2017)\citenamefont
  {Danilkin}, \citenamefont {Deineka},\ and\ \citenamefont
  {Vanderhaeghen}}]{Danilkin:2017lyn}%
  \BibitemOpen
  \bibfield  {author} {\bibinfo {author} {\bibfnamefont {I.}~\bibnamefont
  {Danilkin}}, \bibinfo {author} {\bibfnamefont {O.}~\bibnamefont {Deineka}}, \
  and\ \bibinfo {author} {\bibfnamefont {M.}~\bibnamefont {Vanderhaeghen}},\
  }\href {\doibase 10.1103/PhysRevD.96.114018} {\bibfield  {journal} {\bibinfo
  {journal} {Phys. Rev.}\ }\textbf {\bibinfo {volume} {D96}},\ \bibinfo {pages}
  {114018} (\bibinfo {year} {2017})},\ \Eprint
  {http://arxiv.org/abs/1709.08595} {arXiv:1709.08595 [hep-ph]}\BibitemShut
  {NoStop}%
\bibitem [{\citenamefont {Deineka}\ \emph {et~al.}(2019)\citenamefont
  {Deineka}, \citenamefont {Danilkin},\ and\ \citenamefont
  {Vanderhaeghen}}]{Deineka:2018nuh}%
  \BibitemOpen
  \bibfield  {author} {\bibinfo {author} {\bibfnamefont {O.}~\bibnamefont
  {Deineka}}, \bibinfo {author} {\bibfnamefont {I.}~\bibnamefont {Danilkin}}, \
  and\ \bibinfo {author} {\bibfnamefont {M.}~\bibnamefont {Vanderhaeghen}},\
  }\href {\doibase 10.1051/epjconf/201919902005} {\bibfield  {journal}
  {\bibinfo  {journal} {EPJ Web Conf.}\ }\textbf {\bibinfo {volume} {199}},\
  \bibinfo {pages} {02005} (\bibinfo {year} {2019})},\ \Eprint
  {http://arxiv.org/abs/1808.04117} {arXiv:1808.04117 [hep-ph]}\BibitemShut
  {NoStop}%
\bibitem [{\citenamefont {Colangelo}\ \emph {et~al.}()\citenamefont
  {Colangelo}, \citenamefont {Hoferichter}, \citenamefont {Procura},\ and\
  \citenamefont {Stoffer}}]{CHPSinProgress}%
  \BibitemOpen
  \bibfield  {author} {\bibinfo {author} {\bibfnamefont {G.}~\bibnamefont
  {Colangelo}}, \bibinfo {author} {\bibfnamefont {M.}~\bibnamefont
  {Hoferichter}}, \bibinfo {author} {\bibfnamefont {M.}~\bibnamefont
  {Procura}}, \ and\ \bibinfo {author} {\bibfnamefont {P.}~\bibnamefont
  {Stoffer}},\ }\href@noop {} {}\bibinfo {howpublished} {{in
  preparation}}\BibitemShut {NoStop}%
\bibitem [{\citenamefont {Hoferichter}(2019)}]{Hoferichter-Seattle}%
  \BibitemOpen
  \bibfield  {author} {\bibinfo {author} {\bibfnamefont {M.}~\bibnamefont
  {Hoferichter}},\ }\href@noop {} {\enquote {\bibinfo {title} {{Axial vectors
  and transversal short-distance constraints}},}\ }\bibinfo {howpublished}
  {\url{https://indico.fnal.gov/event/21626/session/9/contribution/49/material/slides/0.pdf}}
  (\bibinfo {year} {2019}),\ \bibinfo {note} {{Muon $g-2$ Theory Initiative
  workshop Seattle}}\BibitemShut {NoStop}%
\bibitem [{\citenamefont {Hoferichter}\ and\ \citenamefont
  {Stoffer}(2020)}]{Hoferichter:2020lap}%
  \BibitemOpen
  \bibfield  {author} {\bibinfo {author} {\bibfnamefont {M.}~\bibnamefont
  {Hoferichter}}\ and\ \bibinfo {author} {\bibfnamefont {P.}~\bibnamefont
  {Stoffer}},\ }\href {\doibase 10.1007/JHEP05(2020)159} {\bibfield  {journal}
  {\bibinfo  {journal} {JHEP}\ }\textbf {\bibinfo {volume} {05}},\ \bibinfo
  {pages} {159} (\bibinfo {year} {2020})},\ \Eprint
  {http://arxiv.org/abs/2004.06127} {arXiv:2004.06127 [hep-ph]}\BibitemShut
  {NoStop}%
\bibitem [{\citenamefont {Leutgeb}\ and\ \citenamefont
  {Rebhan}(2020)}]{Leutgeb:2019gbz}%
  \BibitemOpen
  \bibfield  {author} {\bibinfo {author} {\bibfnamefont {J.}~\bibnamefont
  {Leutgeb}}\ and\ \bibinfo {author} {\bibfnamefont {A.}~\bibnamefont
  {Rebhan}},\ }\href {\doibase 10.1103/PhysRevD.101.114015} {\bibfield
  {journal} {\bibinfo  {journal} {Phys. Rev. D}\ }\textbf {\bibinfo {volume}
  {101}},\ \bibinfo {pages} {114015} (\bibinfo {year} {2020})},\ \Eprint
  {http://arxiv.org/abs/1912.01596} {arXiv:1912.01596 [hep-ph]}\BibitemShut
  {NoStop}%
\bibitem [{\citenamefont {Cappiello}\ \emph {et~al.}(2020)\citenamefont
  {Cappiello}, \citenamefont {Cat{\`a}}, \citenamefont {D'Ambrosio},
  \citenamefont {Greynat},\ and\ \citenamefont {Iyer}}]{Cappiello:2019hwh}%
  \BibitemOpen
  \bibfield  {author} {\bibinfo {author} {\bibfnamefont {L.}~\bibnamefont
  {Cappiello}}, \bibinfo {author} {\bibfnamefont {O.}~\bibnamefont {Cat{\`a}}},
  \bibinfo {author} {\bibfnamefont {G.}~\bibnamefont {D'Ambrosio}}, \bibinfo
  {author} {\bibfnamefont {D.}~\bibnamefont {Greynat}}, \ and\ \bibinfo
  {author} {\bibfnamefont {A.}~\bibnamefont {Iyer}},\ }\href {\doibase
  10.1103/PhysRevD.102.016009} {\bibfield  {journal} {\bibinfo  {journal}
  {Phys. Rev. D}\ }\textbf {\bibinfo {volume} {102}},\ \bibinfo {pages}
  {016009} (\bibinfo {year} {2020})},\ \Eprint
  {http://arxiv.org/abs/1912.02779} {arXiv:1912.02779 [hep-ph]}\BibitemShut
  {NoStop}%
\bibitem [{\citenamefont {Bijnens}\ \emph {et~al.}(2003)\citenamefont
  {Bijnens}, \citenamefont {G{\'a}miz}, \citenamefont {Lipartia},\ and\
  \citenamefont {Prades}}]{Bijnens:2003rc}%
  \BibitemOpen
  \bibfield  {author} {\bibinfo {author} {\bibfnamefont {J.}~\bibnamefont
  {Bijnens}}, \bibinfo {author} {\bibfnamefont {E.}~\bibnamefont {G{\'a}miz}},
  \bibinfo {author} {\bibfnamefont {E.}~\bibnamefont {Lipartia}}, \ and\
  \bibinfo {author} {\bibfnamefont {J.}~\bibnamefont {Prades}},\ }\href
  {\doibase 10.1088/1126-6708/2003/04/055} {\bibfield  {journal} {\bibinfo
  {journal} {JHEP}\ }\textbf {\bibinfo {volume} {04}},\ \bibinfo {pages} {055}
  (\bibinfo {year} {2003})},\ \Eprint {http://arxiv.org/abs/hep-ph/0304222}
  {arXiv:hep-ph/0304222 [hep-ph]}\BibitemShut {NoStop}%
\bibitem [{\citenamefont {Bali}\ \emph {et~al.}(2012)\citenamefont {Bali},
  \citenamefont {Bruckmann}, \citenamefont {Constantinou}, \citenamefont
  {Costa}, \citenamefont {Endrődi}, \citenamefont {Katz}, \citenamefont
  {Panagopoulos},\ and\ \citenamefont {Sch{\"a}fer}}]{Bali:2012jv}%
  \BibitemOpen
  \bibfield  {author} {\bibinfo {author} {\bibfnamefont {G.~S.}\ \bibnamefont
  {Bali}}, \bibinfo {author} {\bibfnamefont {F.}~\bibnamefont {Bruckmann}},
  \bibinfo {author} {\bibfnamefont {M.}~\bibnamefont {Constantinou}}, \bibinfo
  {author} {\bibfnamefont {M.}~\bibnamefont {Costa}}, \bibinfo {author}
  {\bibfnamefont {G.}~\bibnamefont {Endrődi}}, \bibinfo {author}
  {\bibfnamefont {S.~D.}\ \bibnamefont {Katz}}, \bibinfo {author}
  {\bibfnamefont {H.}~\bibnamefont {Panagopoulos}}, \ and\ \bibinfo {author}
  {\bibfnamefont {A.}~\bibnamefont {Sch{\"a}fer}},\ }\href {\doibase
  10.1103/PhysRevD.86.094512} {\bibfield  {journal} {\bibinfo  {journal} {Phys.
  Rev.}\ }\textbf {\bibinfo {volume} {D86}},\ \bibinfo {pages} {094512}
  (\bibinfo {year} {2012})},\ \Eprint {http://arxiv.org/abs/1209.6015}
  {arXiv:1209.6015 [hep-lat]}\BibitemShut {NoStop}%
\bibitem [{\citenamefont {Knecht}\ \emph
  {et~al.}(2002{\natexlab{b}})\citenamefont {Knecht}, \citenamefont {Peris},
  \citenamefont {Perrottet},\ and\ \citenamefont {de~Rafael}}]{Knecht:2002hr}%
  \BibitemOpen
  \bibfield  {author} {\bibinfo {author} {\bibfnamefont {M.}~\bibnamefont
  {Knecht}}, \bibinfo {author} {\bibfnamefont {S.}~\bibnamefont {Peris}},
  \bibinfo {author} {\bibfnamefont {M.}~\bibnamefont {Perrottet}}, \ and\
  \bibinfo {author} {\bibfnamefont {E.}~\bibnamefont {de~Rafael}},\ }\href
  {\doibase 10.1088/1126-6708/2002/11/003} {\bibfield  {journal} {\bibinfo
  {journal} {JHEP}\ }\textbf {\bibinfo {volume} {11}},\ \bibinfo {pages} {003}
  (\bibinfo {year} {2002}{\natexlab{b}})},\ \Eprint
  {http://arxiv.org/abs/hep-ph/0205102} {arXiv:hep-ph/0205102
  [hep-ph]}\BibitemShut {NoStop}%
\bibitem [{\citenamefont {Vainshtein}(2003)}]{Vainshtein:2002nv}%
  \BibitemOpen
  \bibfield  {author} {\bibinfo {author} {\bibfnamefont {A.}~\bibnamefont
  {Vainshtein}},\ }\href {\doibase 10.1016/j.physletb.2003.07.038} {\bibfield
  {journal} {\bibinfo  {journal} {Phys. Lett.}\ }\textbf {\bibinfo {volume}
  {B569}},\ \bibinfo {pages} {187} (\bibinfo {year} {2003})},\ \Eprint
  {http://arxiv.org/abs/hep-ph/0212231} {arXiv:hep-ph/0212231
  [hep-ph]}\BibitemShut {NoStop}%
\bibitem [{\citenamefont {Knecht}\ \emph {et~al.}(2004)\citenamefont {Knecht},
  \citenamefont {Peris}, \citenamefont {Perrottet},\ and\ \citenamefont
  {de~Rafael}}]{Knecht:2003xy}%
  \BibitemOpen
  \bibfield  {author} {\bibinfo {author} {\bibfnamefont {M.}~\bibnamefont
  {Knecht}}, \bibinfo {author} {\bibfnamefont {S.}~\bibnamefont {Peris}},
  \bibinfo {author} {\bibfnamefont {M.}~\bibnamefont {Perrottet}}, \ and\
  \bibinfo {author} {\bibfnamefont {E.}~\bibnamefont {de~Rafael}},\ }\href
  {\doibase 10.1088/1126-6708/2004/03/035} {\bibfield  {journal} {\bibinfo
  {journal} {JHEP}\ }\textbf {\bibinfo {volume} {03}},\ \bibinfo {pages} {035}
  (\bibinfo {year} {2004})},\ \Eprint {http://arxiv.org/abs/hep-ph/0311100}
  {arXiv:hep-ph/0311100 [hep-ph]}\BibitemShut {NoStop}%
\bibitem [{\citenamefont {Melnikov}\ and\ \citenamefont
  {Vainshtein}(2019)}]{Melnikov:2019xkq}%
  \BibitemOpen
  \bibfield  {author} {\bibinfo {author} {\bibfnamefont {K.}~\bibnamefont
  {Melnikov}}\ and\ \bibinfo {author} {\bibfnamefont {A.}~\bibnamefont
  {Vainshtein}},\ }\href@noop {} {\  (\bibinfo {year} {2019})},\ \Eprint
  {http://arxiv.org/abs/1911.05874} {arXiv:1911.05874 [hep-ph]}\BibitemShut
  {NoStop}%
\bibitem [{\citenamefont {Ruiz~Arriola}\ and\ \citenamefont
  {Broniowski}(2006)}]{RuizArriola:2006jge}%
  \BibitemOpen
  \bibfield  {author} {\bibinfo {author} {\bibfnamefont {E.}~\bibnamefont
  {Ruiz~Arriola}}\ and\ \bibinfo {author} {\bibfnamefont {W.}~\bibnamefont
  {Broniowski}},\ }\href {\doibase 10.1103/PhysRevD.74.034008} {\bibfield
  {journal} {\bibinfo  {journal} {Phys. Rev.}\ }\textbf {\bibinfo {volume}
  {D74}},\ \bibinfo {pages} {034008} (\bibinfo {year} {2006})},\ \Eprint
  {http://arxiv.org/abs/hep-ph/0605318} {arXiv:hep-ph/0605318
  [hep-ph]}\BibitemShut {NoStop}%
\bibitem [{\citenamefont {Ruiz~Arriola}\ and\ \citenamefont
  {Broniowski}(2010)}]{Arriola:2010aq}%
  \BibitemOpen
  \bibfield  {author} {\bibinfo {author} {\bibfnamefont {E.}~\bibnamefont
  {Ruiz~Arriola}}\ and\ \bibinfo {author} {\bibfnamefont {W.}~\bibnamefont
  {Broniowski}},\ }\href {\doibase 10.1103/PhysRevD.81.094021} {\bibfield
  {journal} {\bibinfo  {journal} {Phys. Rev.}\ }\textbf {\bibinfo {volume}
  {D81}},\ \bibinfo {pages} {094021} (\bibinfo {year} {2010})},\ \Eprint
  {http://arxiv.org/abs/1004.0837} {arXiv:1004.0837 [hep-ph]}\BibitemShut
  {NoStop}%
\bibitem [{\citenamefont {Acciarri}\ \emph {et~al.}(1997)\citenamefont
  {Acciarri} \emph {et~al.}}]{Acciarri:1997rb}%
  \BibitemOpen
  \bibfield  {author} {\bibinfo {author} {\bibfnamefont {M.}~\bibnamefont
  {Acciarri}} \emph {et~al.} (\bibinfo {collaboration} {L3}),\ }\href {\doibase
  10.1016/S0370-2693(97)01091-5} {\bibfield  {journal} {\bibinfo  {journal}
  {Phys. Lett.}\ }\textbf {\bibinfo {volume} {B413}},\ \bibinfo {pages} {147}
  (\bibinfo {year} {1997})}\BibitemShut {NoStop}%
\bibitem [{\citenamefont {Acciarri}\ \emph {et~al.}(2001)\citenamefont
  {Acciarri} \emph {et~al.}}]{Acciarri:2000ev}%
  \BibitemOpen
  \bibfield  {author} {\bibinfo {author} {\bibfnamefont {M.}~\bibnamefont
  {Acciarri}} \emph {et~al.} (\bibinfo {collaboration} {L3}),\ }\href {\doibase
  10.1016/S0370-2693(01)00102-2} {\bibfield  {journal} {\bibinfo  {journal}
  {Phys. Lett.}\ }\textbf {\bibinfo {volume} {B501}},\ \bibinfo {pages} {1}
  (\bibinfo {year} {2001})},\ \Eprint {http://arxiv.org/abs/hep-ex/0011035}
  {arXiv:hep-ex/0011035 [hep-ex]}\BibitemShut {NoStop}%
\bibitem [{\citenamefont {Ahohe}\ \emph {et~al.}(2005)\citenamefont {Ahohe}
  \emph {et~al.}}]{Ahohe:2005ug}%
  \BibitemOpen
  \bibfield  {author} {\bibinfo {author} {\bibfnamefont {R.}~\bibnamefont
  {Ahohe}} \emph {et~al.} (\bibinfo {collaboration} {CLEO}),\ }\href {\doibase
  10.1103/PhysRevD.71.072001} {\bibfield  {journal} {\bibinfo  {journal} {Phys.
  Rev.}\ }\textbf {\bibinfo {volume} {D71}},\ \bibinfo {pages} {072001}
  (\bibinfo {year} {2005})},\ \Eprint {http://arxiv.org/abs/hep-ex/0501026}
  {arXiv:hep-ex/0501026 [hep-ex]}\BibitemShut {NoStop}%
\bibitem [{\citenamefont {Zhang}\ \emph {et~al.}(2012)\citenamefont {Zhang}
  \emph {et~al.}}]{Zhang:2012tj}%
  \BibitemOpen
  \bibfield  {author} {\bibinfo {author} {\bibfnamefont {C.~C.}\ \bibnamefont
  {Zhang}} \emph {et~al.} (\bibinfo {collaboration} {Belle}),\ }\href {\doibase
  10.1103/PhysRevD.86.052002} {\bibfield  {journal} {\bibinfo  {journal} {Phys.
  Rev.}\ }\textbf {\bibinfo {volume} {D86}},\ \bibinfo {pages} {052002}
  (\bibinfo {year} {2012})},\ \Eprint {http://arxiv.org/abs/1206.5087}
  {arXiv:1206.5087 [hep-ex]}\BibitemShut {NoStop}%
\bibitem [{\citenamefont {Ablikim}\ \emph
  {et~al.}(2018{\natexlab{c}})\citenamefont {Ablikim} \emph
  {et~al.}}]{Ablikim:2018ajr}%
  \BibitemOpen
  \bibfield  {author} {\bibinfo {author} {\bibfnamefont {M.}~\bibnamefont
  {Ablikim}} \emph {et~al.} (\bibinfo {collaboration} {BESIII}),\ }\href
  {\doibase 10.1103/PhysRevD.97.072014} {\bibfield  {journal} {\bibinfo
  {journal} {Phys. Rev.}\ }\textbf {\bibinfo {volume} {D97}},\ \bibinfo {pages}
  {072014} (\bibinfo {year} {2018}{\natexlab{c}})},\ \Eprint
  {http://arxiv.org/abs/1802.09854} {arXiv:1802.09854 [hep-ex]}\BibitemShut
  {NoStop}%
\bibitem [{\citenamefont {Raya}\ \emph {et~al.}(2017)\citenamefont {Raya},
  \citenamefont {Ding}, \citenamefont {Bashir}, \citenamefont {Chang},\ and\
  \citenamefont {Roberts}}]{Raya:2016yuj}%
  \BibitemOpen
  \bibfield  {author} {\bibinfo {author} {\bibfnamefont {K.}~\bibnamefont
  {Raya}}, \bibinfo {author} {\bibfnamefont {M.}~\bibnamefont {Ding}}, \bibinfo
  {author} {\bibfnamefont {A.}~\bibnamefont {Bashir}}, \bibinfo {author}
  {\bibfnamefont {L.}~\bibnamefont {Chang}}, \ and\ \bibinfo {author}
  {\bibfnamefont {C.~D.}\ \bibnamefont {Roberts}},\ }\href {\doibase
  10.1103/PhysRevD.95.074014} {\bibfield  {journal} {\bibinfo  {journal} {Phys.
  Rev.}\ }\textbf {\bibinfo {volume} {D95}},\ \bibinfo {pages} {074014}
  (\bibinfo {year} {2017})},\ \Eprint {http://arxiv.org/abs/1610.06575}
  {arXiv:1610.06575 [nucl-th]}\BibitemShut {NoStop}%
\bibitem [{\citenamefont {Lautrup}\ and\ \citenamefont
  {de~Rafael}(1974)}]{Lautrup:1974ic}%
  \BibitemOpen
  \bibfield  {author} {\bibinfo {author} {\bibfnamefont {B.~E.}\ \bibnamefont
  {Lautrup}}\ and\ \bibinfo {author} {\bibfnamefont {E.}~\bibnamefont
  {de~Rafael}},\ }\href {\doibase 10.1016/0550-3213(74)90481-7} {\bibfield
  {journal} {\bibinfo  {journal} {Nucl. Phys.}\ }\textbf {\bibinfo {volume}
  {B70}},\ \bibinfo {pages} {317} (\bibinfo {year} {1974})},\ \bibinfo {note}
  {[Erratum: Nucl.\ Phys.\ {\bf B78}, 576 (1974)]}\BibitemShut {NoStop}%
\bibitem [{Gla(2007)}]{Glasgow2007}%
  \BibitemOpen
  \href@noop {} {\enquote {\bibinfo {title} {{Topical Workshop on The Muon
  Magnetic Dipole Moment $(g-2)_\mu$}},}\ }\bibinfo {howpublished}
  {\url{http://www.maths.liv.ac.uk/TheorPhys/people/staff/teubner/GlasgowGm2_2007/}}
  (\bibinfo {year} {2007}),\ \bibinfo {note} {{held at the University of
  Glasgow, Glasgow, United Kingdom, October 25--26.}}\BibitemShut {Stop}%
\bibitem [{\citenamefont {Green}\ \emph {et~al.}(2015)\citenamefont {Green},
  \citenamefont {Gryniuk}, \citenamefont {von Hippel}, \citenamefont {Meyer},\
  and\ \citenamefont {Pascalutsa}}]{Green:2015sra}%
  \BibitemOpen
  \bibfield  {author} {\bibinfo {author} {\bibfnamefont {J.}~\bibnamefont
  {Green}}, \bibinfo {author} {\bibfnamefont {O.}~\bibnamefont {Gryniuk}},
  \bibinfo {author} {\bibfnamefont {G.}~\bibnamefont {von Hippel}}, \bibinfo
  {author} {\bibfnamefont {H.~B.}\ \bibnamefont {Meyer}}, \ and\ \bibinfo
  {author} {\bibfnamefont {V.}~\bibnamefont {Pascalutsa}},\ }\href {\doibase
  10.1103/PhysRevLett.115.222003} {\bibfield  {journal} {\bibinfo  {journal}
  {Phys. Rev. Lett.}\ }\textbf {\bibinfo {volume} {115}},\ \bibinfo {pages}
  {222003} (\bibinfo {year} {2015})},\ \Eprint
  {http://arxiv.org/abs/1507.01577} {arXiv:1507.01577 [hep-lat]}\BibitemShut
  {NoStop}%
\bibitem [{\citenamefont {Asmussen}\ \emph
  {et~al.}(2019{\natexlab{b}})\citenamefont {Asmussen}, \citenamefont
  {Gérardin}, \citenamefont {Nyffeler},\ and\ \citenamefont
  {Meyer}}]{Asmussen:2018oip}%
  \BibitemOpen
  \bibfield  {author} {\bibinfo {author} {\bibfnamefont {N.}~\bibnamefont
  {Asmussen}}, \bibinfo {author} {\bibfnamefont {A.}~\bibnamefont {Gérardin}},
  \bibinfo {author} {\bibfnamefont {A.}~\bibnamefont {Nyffeler}}, \ and\
  \bibinfo {author} {\bibfnamefont {H.~B.}\ \bibnamefont {Meyer}},\ }\href
  {\doibase 10.21468/SciPostPhysProc.1.031} {\bibfield  {journal} {\bibinfo
  {journal} {SciPost Phys. Proc.}\ }\textbf {\bibinfo {volume} {1}},\ \bibinfo
  {pages} {031} (\bibinfo {year} {2019}{\natexlab{b}})},\ \Eprint
  {http://arxiv.org/abs/1811.08320} {arXiv:1811.08320 [hep-lat]}\BibitemShut
  {NoStop}%
\bibitem [{\citenamefont {Asmussen}\ \emph
  {et~al.}(2018{\natexlab{a}})\citenamefont {Asmussen}, \citenamefont
  {Gérardin}, \citenamefont {Meyer},\ and\ \citenamefont
  {Nyffeler}}]{Asmussen:2017bup}%
  \BibitemOpen
  \bibfield  {author} {\bibinfo {author} {\bibfnamefont {N.}~\bibnamefont
  {Asmussen}}, \bibinfo {author} {\bibfnamefont {A.}~\bibnamefont {Gérardin}},
  \bibinfo {author} {\bibfnamefont {H.~B.}\ \bibnamefont {Meyer}}, \ and\
  \bibinfo {author} {\bibfnamefont {A.}~\bibnamefont {Nyffeler}},\ }\href
  {\doibase 10.1051/epjconf/201817506023} {\bibfield  {journal} {\bibinfo
  {journal} {EPJ Web Conf.}\ }\textbf {\bibinfo {volume} {175}},\ \bibinfo
  {pages} {06023} (\bibinfo {year} {2018}{\natexlab{a}})},\ \Eprint
  {http://arxiv.org/abs/1711.02466} {arXiv:1711.02466 [hep-lat]}\BibitemShut
  {NoStop}%
\bibitem [{\citenamefont {Laporta}\ and\ \citenamefont
  {Remiddi}(1991)}]{Laporta:1991zw}%
  \BibitemOpen
  \bibfield  {author} {\bibinfo {author} {\bibfnamefont {S.}~\bibnamefont
  {Laporta}}\ and\ \bibinfo {author} {\bibfnamefont {E.}~\bibnamefont
  {Remiddi}},\ }\href {\doibase 10.1016/0370-2693(91)90036-P} {\bibfield
  {journal} {\bibinfo  {journal} {Phys. Lett.}\ }\textbf {\bibinfo {volume}
  {B265}},\ \bibinfo {pages} {182} (\bibinfo {year} {1991})}\BibitemShut
  {NoStop}%
\bibitem [{\citenamefont {Brodsky}\ and\ \citenamefont
  {Sullivan}(1967)}]{Brodsky:1966mv}%
  \BibitemOpen
  \bibfield  {author} {\bibinfo {author} {\bibfnamefont {S.~J.}\ \bibnamefont
  {Brodsky}}\ and\ \bibinfo {author} {\bibfnamefont {J.~D.}\ \bibnamefont
  {Sullivan}},\ }\href {\doibase 10.1103/PhysRev.156.1644} {\bibfield
  {journal} {\bibinfo  {journal} {Phys. Rev.}\ }\textbf {\bibinfo {volume}
  {156}},\ \bibinfo {pages} {1644} (\bibinfo {year} {1967})}\BibitemShut
  {NoStop}%
\bibitem [{\citenamefont {Asmussen}\ \emph {et~al.}()\citenamefont {Asmussen},
  \citenamefont {G\'erardin}, \citenamefont {Green}, \citenamefont {Meyer},\
  and\ \citenamefont {Nyffeler}}]{MainzPrep}%
  \BibitemOpen
  \bibfield  {author} {\bibinfo {author} {\bibfnamefont {N.}~\bibnamefont
  {Asmussen}}, \bibinfo {author} {\bibfnamefont {A.}~\bibnamefont
  {G\'erardin}}, \bibinfo {author} {\bibfnamefont {J.}~\bibnamefont {Green}},
  \bibinfo {author} {\bibfnamefont {H.~B.}\ \bibnamefont {Meyer}}, \ and\
  \bibinfo {author} {\bibfnamefont {A.}~\bibnamefont {Nyffeler}},\ }\href@noop
  {} {}\bibinfo {howpublished} {{in preparation}}\BibitemShut {NoStop}%
\bibitem [{\citenamefont {Asmussen}\ \emph
  {et~al.}(2018{\natexlab{b}})\citenamefont {Asmussen}, \citenamefont
  {G{\'e}rardin}, \citenamefont {Green}, \citenamefont {Gryniuk}, \citenamefont
  {von Hippel}, \citenamefont {Meyer}, \citenamefont {Nyffeler}, \citenamefont
  {Pascalutsa},\ and\ \citenamefont {Wittig}}]{Asmussen:2018lcw}%
  \BibitemOpen
  \bibfield  {author} {\bibinfo {author} {\bibfnamefont {N.}~\bibnamefont
  {Asmussen}}, \bibinfo {author} {\bibfnamefont {A.}~\bibnamefont
  {G{\'e}rardin}}, \bibinfo {author} {\bibfnamefont {J.}~\bibnamefont {Green}},
  \bibinfo {author} {\bibfnamefont {O.}~\bibnamefont {Gryniuk}}, \bibinfo
  {author} {\bibfnamefont {G.}~\bibnamefont {von Hippel}}, \bibinfo {author}
  {\bibfnamefont {H.~B.}\ \bibnamefont {Meyer}}, \bibinfo {author}
  {\bibfnamefont {A.}~\bibnamefont {Nyffeler}}, \bibinfo {author}
  {\bibfnamefont {V.}~\bibnamefont {Pascalutsa}}, \ and\ \bibinfo {author}
  {\bibfnamefont {H.}~\bibnamefont {Wittig}},\ }\href {\doibase
  10.1051/epjconf/201817901017} {\bibfield  {journal} {\bibinfo  {journal} {EPJ
  Web Conf.}\ }\textbf {\bibinfo {volume} {179}},\ \bibinfo {pages} {01017}
  (\bibinfo {year} {2018}{\natexlab{b}})},\ \Eprint
  {http://arxiv.org/abs/1801.04238} {arXiv:1801.04238 [hep-lat]}\BibitemShut
  {NoStop}%
\bibitem [{\citenamefont {Laporta}\ and\ \citenamefont
  {Remiddi}(1993)}]{Laporta:1992pa}%
  \BibitemOpen
  \bibfield  {author} {\bibinfo {author} {\bibfnamefont {S.}~\bibnamefont
  {Laporta}}\ and\ \bibinfo {author} {\bibfnamefont {E.}~\bibnamefont
  {Remiddi}},\ }\href {\doibase 10.1016/0370-2693(93)91176-N} {\bibfield
  {journal} {\bibinfo  {journal} {Phys. Lett.}\ }\textbf {\bibinfo {volume}
  {B301}},\ \bibinfo {pages} {440} (\bibinfo {year} {1993})}\BibitemShut
  {NoStop}%
\bibitem [{\citenamefont {Passera}()}]{Passera:PC}%
  \BibitemOpen
  \bibfield  {author} {\bibinfo {author} {\bibfnamefont {M.}~\bibnamefont
  {Passera}},\ }\href@noop {} {}\bibinfo {howpublished} {private
  communication}\BibitemShut {NoStop}%
\bibitem [{\citenamefont {Jin}\ \emph {et~al.}(2016)\citenamefont {Jin},
  \citenamefont {Blum}, \citenamefont {Christ}, \citenamefont {Hayakawa},
  \citenamefont {Izubuchi}, \citenamefont {Jung},\ and\ \citenamefont
  {Lehner}}]{Jin:2016rmu}%
  \BibitemOpen
  \bibfield  {author} {\bibinfo {author} {\bibfnamefont {L.}~\bibnamefont
  {Jin}}, \bibinfo {author} {\bibfnamefont {T.}~\bibnamefont {Blum}}, \bibinfo
  {author} {\bibfnamefont {N.}~\bibnamefont {Christ}}, \bibinfo {author}
  {\bibfnamefont {M.}~\bibnamefont {Hayakawa}}, \bibinfo {author}
  {\bibfnamefont {T.}~\bibnamefont {Izubuchi}}, \bibinfo {author}
  {\bibfnamefont {C.}~\bibnamefont {Jung}}, \ and\ \bibinfo {author}
  {\bibfnamefont {C.}~\bibnamefont {Lehner}},\ }\href {\doibase
  10.22323/1.256.0181} {\bibfield  {journal} {\bibinfo  {journal} {PoS}\
  }\textbf {\bibinfo {volume} {LATTICE2016}},\ \bibinfo {pages} {181} (\bibinfo
  {year} {2016})},\ \Eprint {http://arxiv.org/abs/1611.08685} {arXiv:1611.08685
  [hep-lat]}\BibitemShut {NoStop}%
\bibitem [{\citenamefont {G{\'e}rardin}\ \emph
  {et~al.}(2018{\natexlab{b}})\citenamefont {G{\'e}rardin}, \citenamefont
  {Green}, \citenamefont {Gryniuk}, \citenamefont {von Hippel}, \citenamefont
  {Meyer}, \citenamefont {Pascalutsa},\ and\ \citenamefont
  {Wittig}}]{Gerardin:2017ryf}%
  \BibitemOpen
  \bibfield  {author} {\bibinfo {author} {\bibfnamefont {A.}~\bibnamefont
  {G{\'e}rardin}}, \bibinfo {author} {\bibfnamefont {J.}~\bibnamefont {Green}},
  \bibinfo {author} {\bibfnamefont {O.}~\bibnamefont {Gryniuk}}, \bibinfo
  {author} {\bibfnamefont {G.}~\bibnamefont {von Hippel}}, \bibinfo {author}
  {\bibfnamefont {H.~B.}\ \bibnamefont {Meyer}}, \bibinfo {author}
  {\bibfnamefont {V.}~\bibnamefont {Pascalutsa}}, \ and\ \bibinfo {author}
  {\bibfnamefont {H.}~\bibnamefont {Wittig}},\ }\href {\doibase
  10.1103/PhysRevD.98.074501} {\bibfield  {journal} {\bibinfo  {journal} {Phys.
  Rev.}\ }\textbf {\bibinfo {volume} {D98}},\ \bibinfo {pages} {074501}
  (\bibinfo {year} {2018}{\natexlab{b}})},\ \Eprint
  {http://arxiv.org/abs/1712.00421} {arXiv:1712.00421 [hep-lat]}\BibitemShut
  {NoStop}%
\bibitem [{\citenamefont {Ji}\ and\ \citenamefont {Jung}(2001)}]{Ji:2001wha}%
  \BibitemOpen
  \bibfield  {author} {\bibinfo {author} {\bibfnamefont {X.-d.}\ \bibnamefont
  {Ji}}\ and\ \bibinfo {author} {\bibfnamefont {C.-w.}\ \bibnamefont {Jung}},\
  }\href {\doibase 10.1103/PhysRevLett.86.208} {\bibfield  {journal} {\bibinfo
  {journal} {Phys. Rev. Lett.}\ }\textbf {\bibinfo {volume} {86}},\ \bibinfo
  {pages} {208} (\bibinfo {year} {2001})},\ \Eprint
  {http://arxiv.org/abs/hep-lat/0101014} {arXiv:hep-lat/0101014
  [hep-lat]}\BibitemShut {NoStop}%
\bibitem [{\citenamefont {G{\'e}rardin}\ \emph {et~al.}(2016)\citenamefont
  {G{\'e}rardin}, \citenamefont {Meyer},\ and\ \citenamefont
  {Nyffeler}}]{Gerardin:2016cqj}%
  \BibitemOpen
  \bibfield  {author} {\bibinfo {author} {\bibfnamefont {A.}~\bibnamefont
  {G{\'e}rardin}}, \bibinfo {author} {\bibfnamefont {H.~B.}\ \bibnamefont
  {Meyer}}, \ and\ \bibinfo {author} {\bibfnamefont {A.}~\bibnamefont
  {Nyffeler}},\ }\href {\doibase 10.1103/PhysRevD.94.074507} {\bibfield
  {journal} {\bibinfo  {journal} {Phys. Rev.}\ }\textbf {\bibinfo {volume}
  {D94}},\ \bibinfo {pages} {074507} (\bibinfo {year} {2016})},\ \Eprint
  {http://arxiv.org/abs/1607.08174} {arXiv:1607.08174 [hep-lat]}\BibitemShut
  {NoStop}%
\bibitem [{\citenamefont {G{\'e}rardin}\ \emph
  {et~al.}(2019{\natexlab{b}})\citenamefont {G{\'e}rardin}, \citenamefont
  {Harris},\ and\ \citenamefont {Meyer}}]{Gerardin:2018kpy}%
  \BibitemOpen
  \bibfield  {author} {\bibinfo {author} {\bibfnamefont {A.}~\bibnamefont
  {G{\'e}rardin}}, \bibinfo {author} {\bibfnamefont {T.}~\bibnamefont
  {Harris}}, \ and\ \bibinfo {author} {\bibfnamefont {H.~B.}\ \bibnamefont
  {Meyer}},\ }\href {\doibase 10.1103/PhysRevD.99.014519} {\bibfield  {journal}
  {\bibinfo  {journal} {Phys. Rev.}\ }\textbf {\bibinfo {volume} {D99}},\
  \bibinfo {pages} {014519} (\bibinfo {year} {2019}{\natexlab{b}})},\ \Eprint
  {http://arxiv.org/abs/1811.08209} {arXiv:1811.08209 [hep-lat]}\BibitemShut
  {NoStop}%
\bibitem [{\citenamefont {Donoghue}\ \emph {et~al.}(1985)\citenamefont
  {Donoghue}, \citenamefont {Holstein},\ and\ \citenamefont
  {Lin}}]{Donoghue:1986wv}%
  \BibitemOpen
  \bibfield  {author} {\bibinfo {author} {\bibfnamefont {J.~F.}\ \bibnamefont
  {Donoghue}}, \bibinfo {author} {\bibfnamefont {B.~R.}\ \bibnamefont
  {Holstein}}, \ and\ \bibinfo {author} {\bibfnamefont {Y.~C.~R.}\ \bibnamefont
  {Lin}},\ }\href {\doibase 10.1103/PhysRevLett.55.2766} {\bibfield  {journal}
  {\bibinfo  {journal} {Phys. Rev. Lett.}\ }\textbf {\bibinfo {volume} {55}},\
  \bibinfo {pages} {2766} (\bibinfo {year} {1985})},\ \bibinfo {note}
  {[Erratum: Phys. Rev. Lett. {\bf 61}, 1527 (1988)]}\BibitemShut {NoStop}%
\bibitem [{\citenamefont {Bijnens}\ \emph {et~al.}(1988)\citenamefont
  {Bijnens}, \citenamefont {Bramon},\ and\ \citenamefont
  {Cornet}}]{Bijnens:1988kx}%
  \BibitemOpen
  \bibfield  {author} {\bibinfo {author} {\bibfnamefont {J.}~\bibnamefont
  {Bijnens}}, \bibinfo {author} {\bibfnamefont {A.}~\bibnamefont {Bramon}}, \
  and\ \bibinfo {author} {\bibfnamefont {F.}~\bibnamefont {Cornet}},\ }\href
  {\doibase 10.1103/PhysRevLett.61.1453} {\bibfield  {journal} {\bibinfo
  {journal} {Phys. Rev. Lett.}\ }\textbf {\bibinfo {volume} {61}},\ \bibinfo
  {pages} {1453} (\bibinfo {year} {1988})}\BibitemShut {NoStop}%
\bibitem [{\citenamefont {Pascalutsa}\ and\ \citenamefont
  {Vanderhaeghen}(2010)}]{Pascalutsa:2010sj}%
  \BibitemOpen
  \bibfield  {author} {\bibinfo {author} {\bibfnamefont {V.}~\bibnamefont
  {Pascalutsa}}\ and\ \bibinfo {author} {\bibfnamefont {M.}~\bibnamefont
  {Vanderhaeghen}},\ }\href {\doibase 10.1103/PhysRevLett.105.201603}
  {\bibfield  {journal} {\bibinfo  {journal} {Phys. Rev. Lett.}\ }\textbf
  {\bibinfo {volume} {105}},\ \bibinfo {pages} {201603} (\bibinfo {year}
  {2010})},\ \Eprint {http://arxiv.org/abs/1008.1088} {arXiv:1008.1088
  [hep-ph]}\BibitemShut {NoStop}%
\bibitem [{\citenamefont {Aoyama}\ \emph
  {et~al.}(2012{\natexlab{b}})\citenamefont {Aoyama}, \citenamefont {Hayakawa},
  \citenamefont {Kinoshita},\ and\ \citenamefont {Nio}}]{Aoyama:2012qma}%
  \BibitemOpen
  \bibfield  {author} {\bibinfo {author} {\bibfnamefont {T.}~\bibnamefont
  {Aoyama}}, \bibinfo {author} {\bibfnamefont {M.}~\bibnamefont {Hayakawa}},
  \bibinfo {author} {\bibfnamefont {T.}~\bibnamefont {Kinoshita}}, \ and\
  \bibinfo {author} {\bibfnamefont {M.}~\bibnamefont {Nio}},\ }\href {\doibase
  10.1093/ptep/pts030} {\bibfield  {journal} {\bibinfo  {journal} {PTEP}\
  }\textbf {\bibinfo {volume} {2012}},\ \bibinfo {pages} {01A107} (\bibinfo
  {year} {2012}{\natexlab{b}})}\BibitemShut {NoStop}%
\bibitem [{\citenamefont {Schwinger}(1948)}]{Schwinger:1948iu}%
  \BibitemOpen
  \bibfield  {author} {\bibinfo {author} {\bibfnamefont {J.~S.}\ \bibnamefont
  {Schwinger}},\ }\href {\doibase 10.1103/PhysRev.73.416} {\bibfield  {journal}
  {\bibinfo  {journal} {Phys. Rev.}\ }\textbf {\bibinfo {volume} {73}},\
  \bibinfo {pages} {416} (\bibinfo {year} {1948})}\BibitemShut {NoStop}%
\bibitem [{\citenamefont {Petermann}(1957{\natexlab{a}})}]{Petermann:1957hs}%
  \BibitemOpen
  \bibfield  {author} {\bibinfo {author} {\bibfnamefont {A.}~\bibnamefont
  {Petermann}},\ }\href {\doibase 10.5169/seals-112823} {\bibfield  {journal}
  {\bibinfo  {journal} {Helv. Phys. Acta}\ }\textbf {\bibinfo {volume} {30}},\
  \bibinfo {pages} {407} (\bibinfo {year} {1957}{\natexlab{a}})}\BibitemShut
  {NoStop}%
\bibitem [{\citenamefont {Sommerfield}(1958)}]{Sommerfield:1958}%
  \BibitemOpen
  \bibfield  {author} {\bibinfo {author} {\bibfnamefont {C.~M.}\ \bibnamefont
  {Sommerfield}},\ }\href {\doibase 10.1016/0003-4916(58)90003-4} {\bibfield
  {journal} {\bibinfo  {journal} {Ann. Phys. (N.Y.)}\ }\textbf {\bibinfo
  {volume} {5}},\ \bibinfo {pages} {26} (\bibinfo {year} {1958})}\BibitemShut
  {NoStop}%
\bibitem [{\citenamefont {Laporta}\ and\ \citenamefont
  {Remiddi}(1996)}]{Laporta:1996mq}%
  \BibitemOpen
  \bibfield  {author} {\bibinfo {author} {\bibfnamefont {S.}~\bibnamefont
  {Laporta}}\ and\ \bibinfo {author} {\bibfnamefont {E.}~\bibnamefont
  {Remiddi}},\ }\href {\doibase 10.1016/0370-2693(96)00439-X} {\bibfield
  {journal} {\bibinfo  {journal} {Phys. Lett.}\ }\textbf {\bibinfo {volume}
  {B379}},\ \bibinfo {pages} {283} (\bibinfo {year} {1996})},\ \Eprint
  {http://arxiv.org/abs/hep-ph/9602417} {arXiv:hep-ph/9602417
  [hep-ph]}\BibitemShut {NoStop}%
\bibitem [{\citenamefont {Karplus}\ and\ \citenamefont
  {Kroll}(1950)}]{Karplus:1950zzb}%
  \BibitemOpen
  \bibfield  {author} {\bibinfo {author} {\bibfnamefont {R.}~\bibnamefont
  {Karplus}}\ and\ \bibinfo {author} {\bibfnamefont {N.~M.}\ \bibnamefont
  {Kroll}},\ }\href {\doibase 10.1103/PhysRev.77.536} {\bibfield  {journal}
  {\bibinfo  {journal} {Phys. Rev.}\ }\textbf {\bibinfo {volume} {77}},\
  \bibinfo {pages} {536} (\bibinfo {year} {1950})}\BibitemShut {NoStop}%
\bibitem [{\citenamefont {Kinoshita}\ and\ \citenamefont
  {Cvitanovic}(1972)}]{Kinoshita:1973ph}%
  \BibitemOpen
  \bibfield  {author} {\bibinfo {author} {\bibfnamefont {T.}~\bibnamefont
  {Kinoshita}}\ and\ \bibinfo {author} {\bibfnamefont {P.}~\bibnamefont
  {Cvitanovic}},\ }\href {\doibase 10.1103/PhysRevLett.29.1534} {\bibfield
  {journal} {\bibinfo  {journal} {Phys. Rev. Lett.}\ }\textbf {\bibinfo
  {volume} {29}},\ \bibinfo {pages} {1534} (\bibinfo {year}
  {1972})}\BibitemShut {NoStop}%
\bibitem [{\citenamefont {Cvitanovi{\'c}}\ and\ \citenamefont
  {Kinoshita}(1974)}]{Cvitanovic:1974um}%
  \BibitemOpen
  \bibfield  {author} {\bibinfo {author} {\bibfnamefont {P.}~\bibnamefont
  {Cvitanovi{\'c}}}\ and\ \bibinfo {author} {\bibfnamefont {T.}~\bibnamefont
  {Kinoshita}},\ }\href {\doibase 10.1103/PhysRevD.10.4007} {\bibfield
  {journal} {\bibinfo  {journal} {Phys. Rev.}\ }\textbf {\bibinfo {volume}
  {D10}},\ \bibinfo {pages} {4007} (\bibinfo {year} {1974})}\BibitemShut
  {NoStop}%
\bibitem [{\citenamefont {Levine}\ and\ \citenamefont
  {Wright}(1973)}]{Levine:1974cb}%
  \BibitemOpen
  \bibfield  {author} {\bibinfo {author} {\bibfnamefont {M.~J.}\ \bibnamefont
  {Levine}}\ and\ \bibinfo {author} {\bibfnamefont {J.}~\bibnamefont
  {Wright}},\ }\href {\doibase 10.1103/PhysRevD.8.3171} {\bibfield  {journal}
  {\bibinfo  {journal} {Phys. Rev.}\ }\textbf {\bibinfo {volume} {D8}},\
  \bibinfo {pages} {3171} (\bibinfo {year} {1973})}\BibitemShut {NoStop}%
\bibitem [{\citenamefont {Carroll}\ and\ \citenamefont
  {Yao}(1974)}]{Carroll:1974bu}%
  \BibitemOpen
  \bibfield  {author} {\bibinfo {author} {\bibfnamefont {R.}~\bibnamefont
  {Carroll}}\ and\ \bibinfo {author} {\bibfnamefont {Y.~P.}\ \bibnamefont
  {Yao}},\ }\href {\doibase 10.1016/0370-2693(74)90659-5} {\bibfield  {journal}
  {\bibinfo  {journal} {Phys. Lett.}\ }\textbf {\bibinfo {volume} {48B}},\
  \bibinfo {pages} {125} (\bibinfo {year} {1974})}\BibitemShut {NoStop}%
\bibitem [{\citenamefont {Carroll}(1975)}]{Carroll:1975jf}%
  \BibitemOpen
  \bibfield  {author} {\bibinfo {author} {\bibfnamefont {R.}~\bibnamefont
  {Carroll}},\ }\href {\doibase 10.1103/PhysRevD.12.2344} {\bibfield  {journal}
  {\bibinfo  {journal} {Phys. Rev.}\ }\textbf {\bibinfo {volume} {D12}},\
  \bibinfo {pages} {2344} (\bibinfo {year} {1975})}\BibitemShut {NoStop}%
\bibitem [{\citenamefont {Kinoshita}(1995)}]{Kinoshita:1995ym}%
  \BibitemOpen
  \bibfield  {author} {\bibinfo {author} {\bibfnamefont {T.}~\bibnamefont
  {Kinoshita}},\ }\href {\doibase 10.1103/PhysRevLett.75.4728} {\bibfield
  {journal} {\bibinfo  {journal} {Phys. Rev. Lett.}\ }\textbf {\bibinfo
  {volume} {75}},\ \bibinfo {pages} {4728} (\bibinfo {year}
  {1995})}\BibitemShut {NoStop}%
\bibitem [{\citenamefont {Laporta}(2017)}]{Laporta:2017okg}%
  \BibitemOpen
  \bibfield  {author} {\bibinfo {author} {\bibfnamefont {S.}~\bibnamefont
  {Laporta}},\ }\href {\doibase 10.1016/j.physletb.2017.06.056} {\bibfield
  {journal} {\bibinfo  {journal} {Phys. Lett.}\ }\textbf {\bibinfo {volume}
  {B772}},\ \bibinfo {pages} {232} (\bibinfo {year} {2017})},\ \Eprint
  {http://arxiv.org/abs/1704.06996} {arXiv:1704.06996 [hep-ph]}\BibitemShut
  {NoStop}%
\bibitem [{\citenamefont {Aoyama}\ \emph {et~al.}(2015)\citenamefont {Aoyama},
  \citenamefont {Hayakawa}, \citenamefont {Kinoshita},\ and\ \citenamefont
  {Nio}}]{Aoyama:2014sxa}%
  \BibitemOpen
  \bibfield  {author} {\bibinfo {author} {\bibfnamefont {T.}~\bibnamefont
  {Aoyama}}, \bibinfo {author} {\bibfnamefont {M.}~\bibnamefont {Hayakawa}},
  \bibinfo {author} {\bibfnamefont {T.}~\bibnamefont {Kinoshita}}, \ and\
  \bibinfo {author} {\bibfnamefont {M.}~\bibnamefont {Nio}},\ }\href {\doibase
  10.1103/PhysRevD.91.033006} {\bibfield  {journal} {\bibinfo  {journal} {Phys.
  Rev.}\ }\textbf {\bibinfo {volume} {D91}},\ \bibinfo {pages} {033006}
  (\bibinfo {year} {2015})},\ \bibinfo {note} {[Erratum: Phys. Rev. {\bf D96},
  019901 (2017)]},\ \Eprint {http://arxiv.org/abs/1412.8284} {arXiv:1412.8284
  [hep-ph]}\BibitemShut {NoStop}%
\bibitem [{\citenamefont {Kinoshita}\ and\ \citenamefont
  {Lindquist}(1981)}]{Kinoshita:1981vs}%
  \BibitemOpen
  \bibfield  {author} {\bibinfo {author} {\bibfnamefont {T.}~\bibnamefont
  {Kinoshita}}\ and\ \bibinfo {author} {\bibfnamefont {W.~B.}\ \bibnamefont
  {Lindquist}},\ }\href {\doibase 10.1103/PhysRevLett.47.1573} {\bibfield
  {journal} {\bibinfo  {journal} {Phys. Rev. Lett.}\ }\textbf {\bibinfo
  {volume} {47}},\ \bibinfo {pages} {1573} (\bibinfo {year}
  {1981})}\BibitemShut {NoStop}%
\bibitem [{\citenamefont {Kinoshita}\ and\ \citenamefont
  {Lindquist}(1983{\natexlab{a}})}]{Kinoshita:1979dy}%
  \BibitemOpen
  \bibfield  {author} {\bibinfo {author} {\bibfnamefont {T.}~\bibnamefont
  {Kinoshita}}\ and\ \bibinfo {author} {\bibfnamefont {W.~B.}\ \bibnamefont
  {Lindquist}},\ }\href {\doibase 10.1103/PhysRevD.27.853} {\bibfield
  {journal} {\bibinfo  {journal} {Phys. Rev. D}\ }\textbf {\bibinfo {volume}
  {27}},\ \bibinfo {pages} {853} (\bibinfo {year}
  {1983}{\natexlab{a}})}\BibitemShut {NoStop}%
\bibitem [{\citenamefont {Kinoshita}\ and\ \citenamefont
  {Lindquist}(1983{\natexlab{b}})}]{Kinoshita:1979ej}%
  \BibitemOpen
  \bibfield  {author} {\bibinfo {author} {\bibfnamefont {T.}~\bibnamefont
  {Kinoshita}}\ and\ \bibinfo {author} {\bibfnamefont {W.~B.}\ \bibnamefont
  {Lindquist}},\ }\href {\doibase 10.1103/PhysRevD.27.867} {\bibfield
  {journal} {\bibinfo  {journal} {Phys. Rev. D}\ }\textbf {\bibinfo {volume}
  {27}},\ \bibinfo {pages} {867} (\bibinfo {year}
  {1983}{\natexlab{b}})}\BibitemShut {NoStop}%
\bibitem [{\citenamefont {Kinoshita}\ and\ \citenamefont
  {Lindquist}(1983{\natexlab{c}})}]{Kinoshita:1979ei}%
  \BibitemOpen
  \bibfield  {author} {\bibinfo {author} {\bibfnamefont {T.}~\bibnamefont
  {Kinoshita}}\ and\ \bibinfo {author} {\bibfnamefont {W.~B.}\ \bibnamefont
  {Lindquist}},\ }\href {\doibase 10.1103/PhysRevD.27.877} {\bibfield
  {journal} {\bibinfo  {journal} {Phys. Rev. D}\ }\textbf {\bibinfo {volume}
  {27}},\ \bibinfo {pages} {877} (\bibinfo {year}
  {1983}{\natexlab{c}})}\BibitemShut {NoStop}%
\bibitem [{\citenamefont {Kinoshita}\ and\ \citenamefont
  {Lindquist}(1983{\natexlab{d}})}]{Kinoshita:1981wx}%
  \BibitemOpen
  \bibfield  {author} {\bibinfo {author} {\bibfnamefont {T.}~\bibnamefont
  {Kinoshita}}\ and\ \bibinfo {author} {\bibfnamefont {W.~B.}\ \bibnamefont
  {Lindquist}},\ }\href {\doibase 10.1103/PhysRevD.27.886} {\bibfield
  {journal} {\bibinfo  {journal} {Phys. Rev. D}\ }\textbf {\bibinfo {volume}
  {27}},\ \bibinfo {pages} {886} (\bibinfo {year}
  {1983}{\natexlab{d}})}\BibitemShut {NoStop}%
\bibitem [{\citenamefont {Kinoshita}\ and\ \citenamefont
  {Lindquist}(1989)}]{Kinoshita:1981ww}%
  \BibitemOpen
  \bibfield  {author} {\bibinfo {author} {\bibfnamefont {T.}~\bibnamefont
  {Kinoshita}}\ and\ \bibinfo {author} {\bibfnamefont {W.~B.}\ \bibnamefont
  {Lindquist}},\ }\href {\doibase 10.1103/PhysRevD.39.2407} {\bibfield
  {journal} {\bibinfo  {journal} {Phys. Rev. D}\ }\textbf {\bibinfo {volume}
  {39}},\ \bibinfo {pages} {2407} (\bibinfo {year} {1989})}\BibitemShut
  {NoStop}%
\bibitem [{\citenamefont {Kinoshita}\ and\ \citenamefont
  {Lindquist}(1990)}]{Kinoshita:1981wm}%
  \BibitemOpen
  \bibfield  {author} {\bibinfo {author} {\bibfnamefont {T.}~\bibnamefont
  {Kinoshita}}\ and\ \bibinfo {author} {\bibfnamefont {W.~B.}\ \bibnamefont
  {Lindquist}},\ }\href {\doibase 10.1103/PhysRevD.42.636} {\bibfield
  {journal} {\bibinfo  {journal} {Phys. Rev. D}\ }\textbf {\bibinfo {volume}
  {42}},\ \bibinfo {pages} {636} (\bibinfo {year} {1990})}\BibitemShut
  {NoStop}%
\bibitem [{\citenamefont {Kinoshita}\ and\ \citenamefont
  {Nio}(2003)}]{Kinoshita:2002ns}%
  \BibitemOpen
  \bibfield  {author} {\bibinfo {author} {\bibfnamefont {T.}~\bibnamefont
  {Kinoshita}}\ and\ \bibinfo {author} {\bibfnamefont {M.}~\bibnamefont
  {Nio}},\ }\href {\doibase 10.1103/PhysRevLett.90.021803} {\bibfield
  {journal} {\bibinfo  {journal} {Phys. Rev. Lett.}\ }\textbf {\bibinfo
  {volume} {90}},\ \bibinfo {pages} {021803} (\bibinfo {year} {2003})},\
  \Eprint {http://arxiv.org/abs/hep-ph/0210322} {arXiv:hep-ph/0210322
  [hep-ph]}\BibitemShut {NoStop}%
\bibitem [{\citenamefont {Kinoshita}\ and\ \citenamefont
  {Nio}(2006{\natexlab{a}})}]{Kinoshita:2005zr}%
  \BibitemOpen
  \bibfield  {author} {\bibinfo {author} {\bibfnamefont {T.}~\bibnamefont
  {Kinoshita}}\ and\ \bibinfo {author} {\bibfnamefont {M.}~\bibnamefont
  {Nio}},\ }\href {\doibase 10.1103/PhysRevD.73.013003} {\bibfield  {journal}
  {\bibinfo  {journal} {Phys. Rev.}\ }\textbf {\bibinfo {volume} {D73}},\
  \bibinfo {pages} {013003} (\bibinfo {year} {2006}{\natexlab{a}})},\ \Eprint
  {http://arxiv.org/abs/hep-ph/0507249} {arXiv:hep-ph/0507249
  [hep-ph]}\BibitemShut {NoStop}%
\bibitem [{\citenamefont {Aoyama}\ \emph {et~al.}(2007)\citenamefont {Aoyama},
  \citenamefont {Hayakawa}, \citenamefont {Kinoshita},\ and\ \citenamefont
  {Nio}}]{Aoyama:2007dv}%
  \BibitemOpen
  \bibfield  {author} {\bibinfo {author} {\bibfnamefont {T.}~\bibnamefont
  {Aoyama}}, \bibinfo {author} {\bibfnamefont {M.}~\bibnamefont {Hayakawa}},
  \bibinfo {author} {\bibfnamefont {T.}~\bibnamefont {Kinoshita}}, \ and\
  \bibinfo {author} {\bibfnamefont {M.}~\bibnamefont {Nio}},\ }\href {\doibase
  10.1103/PhysRevLett.99.110406} {\bibfield  {journal} {\bibinfo  {journal}
  {Phys. Rev. Lett.}\ }\textbf {\bibinfo {volume} {99}},\ \bibinfo {pages}
  {110406} (\bibinfo {year} {2007})},\ \Eprint {http://arxiv.org/abs/0706.3496}
  {arXiv:0706.3496 [hep-ph]}\BibitemShut {NoStop}%
\bibitem [{\citenamefont {Aoyama}\ \emph
  {et~al.}(2008{\natexlab{a}})\citenamefont {Aoyama}, \citenamefont {Hayakawa},
  \citenamefont {Kinoshita},\ and\ \citenamefont {Nio}}]{Aoyama:2007mn}%
  \BibitemOpen
  \bibfield  {author} {\bibinfo {author} {\bibfnamefont {T.}~\bibnamefont
  {Aoyama}}, \bibinfo {author} {\bibfnamefont {M.}~\bibnamefont {Hayakawa}},
  \bibinfo {author} {\bibfnamefont {T.}~\bibnamefont {Kinoshita}}, \ and\
  \bibinfo {author} {\bibfnamefont {M.}~\bibnamefont {Nio}},\ }\href {\doibase
  10.1103/PhysRevD.77.053012} {\bibfield  {journal} {\bibinfo  {journal} {Phys.
  Rev.}\ }\textbf {\bibinfo {volume} {D77}},\ \bibinfo {pages} {053012}
  (\bibinfo {year} {2008}{\natexlab{a}})},\ \Eprint
  {http://arxiv.org/abs/0712.2607} {arXiv:0712.2607 [hep-ph]}\BibitemShut
  {NoStop}%
\bibitem [{\citenamefont {Marquard}\ \emph {et~al.}(2019)\citenamefont
  {Marquard}, \citenamefont {Smirnov}, \citenamefont {Smirnov}, \citenamefont
  {Steinhauser},\ and\ \citenamefont {Wellmann}}]{Marquard:2017iib}%
  \BibitemOpen
  \bibfield  {author} {\bibinfo {author} {\bibfnamefont {P.}~\bibnamefont
  {Marquard}}, \bibinfo {author} {\bibfnamefont {A.~V.}\ \bibnamefont
  {Smirnov}}, \bibinfo {author} {\bibfnamefont {V.~A.}\ \bibnamefont
  {Smirnov}}, \bibinfo {author} {\bibfnamefont {M.}~\bibnamefont
  {Steinhauser}}, \ and\ \bibinfo {author} {\bibfnamefont {D.}~\bibnamefont
  {Wellmann}},\ }\href {\doibase 10.1051/epjconf/201921801004} {\bibfield
  {journal} {\bibinfo  {journal} {EPJ Web Conf.}\ }\textbf {\bibinfo {volume}
  {218}},\ \bibinfo {pages} {01004} (\bibinfo {year} {2019})},\ \Eprint
  {http://arxiv.org/abs/1708.07138} {arXiv:1708.07138 [hep-ph]}\BibitemShut
  {NoStop}%
\bibitem [{\citenamefont {Volkov}(2017)}]{Volkov:2017xaq}%
  \BibitemOpen
  \bibfield  {author} {\bibinfo {author} {\bibfnamefont {S.}~\bibnamefont
  {Volkov}},\ }\href {\doibase 10.1103/PhysRevD.96.096018} {\bibfield
  {journal} {\bibinfo  {journal} {Phys. Rev.}\ }\textbf {\bibinfo {volume}
  {D96}},\ \bibinfo {pages} {096018} (\bibinfo {year} {2017})},\ \Eprint
  {http://arxiv.org/abs/1705.05800} {arXiv:1705.05800 [hep-ph]}\BibitemShut
  {NoStop}%
\bibitem [{\citenamefont {Volkov}(2018)}]{Volkov:2018jhy}%
  \BibitemOpen
  \bibfield  {author} {\bibinfo {author} {\bibfnamefont {S.}~\bibnamefont
  {Volkov}},\ }\href {\doibase 10.1103/PhysRevD.98.076018} {\bibfield
  {journal} {\bibinfo  {journal} {Phys. Rev.}\ }\textbf {\bibinfo {volume}
  {D98}},\ \bibinfo {pages} {076018} (\bibinfo {year} {2018})},\ \Eprint
  {http://arxiv.org/abs/1807.05281} {arXiv:1807.05281 [hep-ph]}\BibitemShut
  {NoStop}%
\bibitem [{\citenamefont {Aoyama}\ \emph {et~al.}(2018)\citenamefont {Aoyama},
  \citenamefont {Kinoshita},\ and\ \citenamefont {Nio}}]{Aoyama:2017uqe}%
  \BibitemOpen
  \bibfield  {author} {\bibinfo {author} {\bibfnamefont {T.}~\bibnamefont
  {Aoyama}}, \bibinfo {author} {\bibfnamefont {T.}~\bibnamefont {Kinoshita}}, \
  and\ \bibinfo {author} {\bibfnamefont {M.}~\bibnamefont {Nio}},\ }\href
  {\doibase 10.1103/PhysRevD.97.036001} {\bibfield  {journal} {\bibinfo
  {journal} {Phys. Rev.}\ }\textbf {\bibinfo {volume} {D97}},\ \bibinfo {pages}
  {036001} (\bibinfo {year} {2018})},\ \Eprint
  {http://arxiv.org/abs/1712.06060} {arXiv:1712.06060 [hep-ph]}\BibitemShut
  {NoStop}%
\bibitem [{\citenamefont {Aoyama}\ \emph
  {et~al.}(2008{\natexlab{b}})\citenamefont {Aoyama}, \citenamefont {Hayakawa},
  \citenamefont {Kinoshita}, \citenamefont {Nio},\ and\ \citenamefont
  {Watanabe}}]{Aoyama:2008gy}%
  \BibitemOpen
  \bibfield  {author} {\bibinfo {author} {\bibfnamefont {T.}~\bibnamefont
  {Aoyama}}, \bibinfo {author} {\bibfnamefont {M.}~\bibnamefont {Hayakawa}},
  \bibinfo {author} {\bibfnamefont {T.}~\bibnamefont {Kinoshita}}, \bibinfo
  {author} {\bibfnamefont {M.}~\bibnamefont {Nio}}, \ and\ \bibinfo {author}
  {\bibfnamefont {N.}~\bibnamefont {Watanabe}},\ }\href {\doibase
  10.1103/PhysRevD.78.053005} {\bibfield  {journal} {\bibinfo  {journal} {Phys.
  Rev.}\ }\textbf {\bibinfo {volume} {D78}},\ \bibinfo {pages} {053005}
  (\bibinfo {year} {2008}{\natexlab{b}})},\ \Eprint
  {http://arxiv.org/abs/0806.3390} {arXiv:0806.3390 [hep-ph]}\BibitemShut
  {NoStop}%
\bibitem [{\citenamefont {Aoyama}\ \emph
  {et~al.}(2008{\natexlab{c}})\citenamefont {Aoyama}, \citenamefont {Hayakawa},
  \citenamefont {Kinoshita},\ and\ \citenamefont {Nio}}]{Aoyama:2008hz}%
  \BibitemOpen
  \bibfield  {author} {\bibinfo {author} {\bibfnamefont {T.}~\bibnamefont
  {Aoyama}}, \bibinfo {author} {\bibfnamefont {M.}~\bibnamefont {Hayakawa}},
  \bibinfo {author} {\bibfnamefont {T.}~\bibnamefont {Kinoshita}}, \ and\
  \bibinfo {author} {\bibfnamefont {M.}~\bibnamefont {Nio}},\ }\href {\doibase
  10.1103/PhysRevD.78.113006} {\bibfield  {journal} {\bibinfo  {journal} {Phys.
  Rev.}\ }\textbf {\bibinfo {volume} {D78}},\ \bibinfo {pages} {113006}
  (\bibinfo {year} {2008}{\natexlab{c}})},\ \Eprint
  {http://arxiv.org/abs/0810.5208} {arXiv:0810.5208 [hep-ph]}\BibitemShut
  {NoStop}%
\bibitem [{\citenamefont {Aoyama}\ \emph
  {et~al.}(2010{\natexlab{a}})\citenamefont {Aoyama}, \citenamefont {Asano},
  \citenamefont {Hayakawa}, \citenamefont {Kinoshita}, \citenamefont {Nio},\
  and\ \citenamefont {Watanabe}}]{Aoyama:2010yt}%
  \BibitemOpen
  \bibfield  {author} {\bibinfo {author} {\bibfnamefont {T.}~\bibnamefont
  {Aoyama}}, \bibinfo {author} {\bibfnamefont {K.}~\bibnamefont {Asano}},
  \bibinfo {author} {\bibfnamefont {M.}~\bibnamefont {Hayakawa}}, \bibinfo
  {author} {\bibfnamefont {T.}~\bibnamefont {Kinoshita}}, \bibinfo {author}
  {\bibfnamefont {M.}~\bibnamefont {Nio}}, \ and\ \bibinfo {author}
  {\bibfnamefont {N.}~\bibnamefont {Watanabe}},\ }\href {\doibase
  10.1103/PhysRevD.81.053009} {\bibfield  {journal} {\bibinfo  {journal} {Phys.
  Rev.}\ }\textbf {\bibinfo {volume} {D81}},\ \bibinfo {pages} {053009}
  (\bibinfo {year} {2010}{\natexlab{a}})},\ \Eprint
  {http://arxiv.org/abs/1001.3704} {arXiv:1001.3704 [hep-ph]}\BibitemShut
  {NoStop}%
\bibitem [{\citenamefont {Aoyama}\ \emph
  {et~al.}(2010{\natexlab{b}})\citenamefont {Aoyama}, \citenamefont {Hayakawa},
  \citenamefont {Kinoshita},\ and\ \citenamefont {Nio}}]{Aoyama:2010pk}%
  \BibitemOpen
  \bibfield  {author} {\bibinfo {author} {\bibfnamefont {T.}~\bibnamefont
  {Aoyama}}, \bibinfo {author} {\bibfnamefont {M.}~\bibnamefont {Hayakawa}},
  \bibinfo {author} {\bibfnamefont {T.}~\bibnamefont {Kinoshita}}, \ and\
  \bibinfo {author} {\bibfnamefont {M.}~\bibnamefont {Nio}},\ }\href {\doibase
  10.1103/PhysRevD.82.113004} {\bibfield  {journal} {\bibinfo  {journal} {Phys.
  Rev.}\ }\textbf {\bibinfo {volume} {D82}},\ \bibinfo {pages} {113004}
  (\bibinfo {year} {2010}{\natexlab{b}})},\ \Eprint
  {http://arxiv.org/abs/1009.3077} {arXiv:1009.3077 [hep-ph]}\BibitemShut
  {NoStop}%
\bibitem [{\citenamefont {Aoyama}\ \emph
  {et~al.}(2011{\natexlab{a}})\citenamefont {Aoyama}, \citenamefont {Hayakawa},
  \citenamefont {Kinoshita},\ and\ \citenamefont {Nio}}]{Aoyama:2010zp}%
  \BibitemOpen
  \bibfield  {author} {\bibinfo {author} {\bibfnamefont {T.}~\bibnamefont
  {Aoyama}}, \bibinfo {author} {\bibfnamefont {M.}~\bibnamefont {Hayakawa}},
  \bibinfo {author} {\bibfnamefont {T.}~\bibnamefont {Kinoshita}}, \ and\
  \bibinfo {author} {\bibfnamefont {M.}~\bibnamefont {Nio}},\ }\href {\doibase
  10.1103/PhysRevD.83.053003} {\bibfield  {journal} {\bibinfo  {journal} {Phys.
  Rev.}\ }\textbf {\bibinfo {volume} {D83}},\ \bibinfo {pages} {053003}
  (\bibinfo {year} {2011}{\natexlab{a}})},\ \Eprint
  {http://arxiv.org/abs/1012.5569} {arXiv:1012.5569 [hep-ph]}\BibitemShut
  {NoStop}%
\bibitem [{\citenamefont {Aoyama}\ \emph
  {et~al.}(2011{\natexlab{b}})\citenamefont {Aoyama}, \citenamefont {Hayakawa},
  \citenamefont {Kinoshita},\ and\ \citenamefont {Nio}}]{Aoyama:2011rm}%
  \BibitemOpen
  \bibfield  {author} {\bibinfo {author} {\bibfnamefont {T.}~\bibnamefont
  {Aoyama}}, \bibinfo {author} {\bibfnamefont {M.}~\bibnamefont {Hayakawa}},
  \bibinfo {author} {\bibfnamefont {T.}~\bibnamefont {Kinoshita}}, \ and\
  \bibinfo {author} {\bibfnamefont {M.}~\bibnamefont {Nio}},\ }\href {\doibase
  10.1103/PhysRevD.83.053002} {\bibfield  {journal} {\bibinfo  {journal} {Phys.
  Rev.}\ }\textbf {\bibinfo {volume} {D83}},\ \bibinfo {pages} {053002}
  (\bibinfo {year} {2011}{\natexlab{b}})},\ \Eprint
  {http://arxiv.org/abs/1101.0459} {arXiv:1101.0459 [hep-ph]}\BibitemShut
  {NoStop}%
\bibitem [{\citenamefont {Aoyama}\ \emph
  {et~al.}(2011{\natexlab{c}})\citenamefont {Aoyama}, \citenamefont {Hayakawa},
  \citenamefont {Kinoshita},\ and\ \citenamefont {Nio}}]{Aoyama:2011zy}%
  \BibitemOpen
  \bibfield  {author} {\bibinfo {author} {\bibfnamefont {T.}~\bibnamefont
  {Aoyama}}, \bibinfo {author} {\bibfnamefont {M.}~\bibnamefont {Hayakawa}},
  \bibinfo {author} {\bibfnamefont {T.}~\bibnamefont {Kinoshita}}, \ and\
  \bibinfo {author} {\bibfnamefont {M.}~\bibnamefont {Nio}},\ }\href {\doibase
  10.1103/PhysRevD.84.053003} {\bibfield  {journal} {\bibinfo  {journal} {Phys.
  Rev.}\ }\textbf {\bibinfo {volume} {D84}},\ \bibinfo {pages} {053003}
  (\bibinfo {year} {2011}{\natexlab{c}})},\ \Eprint
  {http://arxiv.org/abs/1105.5200} {arXiv:1105.5200 [hep-ph]}\BibitemShut
  {NoStop}%
\bibitem [{\citenamefont {Aoyama}\ \emph
  {et~al.}(2012{\natexlab{c}})\citenamefont {Aoyama}, \citenamefont {Hayakawa},
  \citenamefont {Kinoshita},\ and\ \citenamefont {Nio}}]{Aoyama:2011dy}%
  \BibitemOpen
  \bibfield  {author} {\bibinfo {author} {\bibfnamefont {T.}~\bibnamefont
  {Aoyama}}, \bibinfo {author} {\bibfnamefont {M.}~\bibnamefont {Hayakawa}},
  \bibinfo {author} {\bibfnamefont {T.}~\bibnamefont {Kinoshita}}, \ and\
  \bibinfo {author} {\bibfnamefont {M.}~\bibnamefont {Nio}},\ }\href {\doibase
  10.1103/PhysRevD.85.033007} {\bibfield  {journal} {\bibinfo  {journal} {Phys.
  Rev.}\ }\textbf {\bibinfo {volume} {D85}},\ \bibinfo {pages} {033007}
  (\bibinfo {year} {2012}{\natexlab{c}})},\ \Eprint
  {http://arxiv.org/abs/1110.2826} {arXiv:1110.2826 [hep-ph]}\BibitemShut
  {NoStop}%
\bibitem [{\citenamefont {Aoyama}\ \emph
  {et~al.}(2012{\natexlab{d}})\citenamefont {Aoyama}, \citenamefont {Hayakawa},
  \citenamefont {Kinoshita},\ and\ \citenamefont {Nio}}]{Aoyama:2012fc}%
  \BibitemOpen
  \bibfield  {author} {\bibinfo {author} {\bibfnamefont {T.}~\bibnamefont
  {Aoyama}}, \bibinfo {author} {\bibfnamefont {M.}~\bibnamefont {Hayakawa}},
  \bibinfo {author} {\bibfnamefont {T.}~\bibnamefont {Kinoshita}}, \ and\
  \bibinfo {author} {\bibfnamefont {M.}~\bibnamefont {Nio}},\ }\href {\doibase
  10.1103/PhysRevD.85.093013} {\bibfield  {journal} {\bibinfo  {journal} {Phys.
  Rev.}\ }\textbf {\bibinfo {volume} {D85}},\ \bibinfo {pages} {093013}
  (\bibinfo {year} {2012}{\natexlab{d}})},\ \Eprint
  {http://arxiv.org/abs/1201.2461} {arXiv:1201.2461 [hep-ph]}\BibitemShut
  {NoStop}%
\bibitem [{\citenamefont {Aoyama}\ \emph
  {et~al.}(2012{\natexlab{e}})\citenamefont {Aoyama}, \citenamefont {Hayakawa},
  \citenamefont {Kinoshita},\ and\ \citenamefont {Nio}}]{Aoyama:2012wj}%
  \BibitemOpen
  \bibfield  {author} {\bibinfo {author} {\bibfnamefont {T.}~\bibnamefont
  {Aoyama}}, \bibinfo {author} {\bibfnamefont {M.}~\bibnamefont {Hayakawa}},
  \bibinfo {author} {\bibfnamefont {T.}~\bibnamefont {Kinoshita}}, \ and\
  \bibinfo {author} {\bibfnamefont {M.}~\bibnamefont {Nio}},\ }\href {\doibase
  10.1103/PhysRevLett.109.111807} {\bibfield  {journal} {\bibinfo  {journal}
  {Phys. Rev. Lett.}\ }\textbf {\bibinfo {volume} {109}},\ \bibinfo {pages}
  {111807} (\bibinfo {year} {2012}{\natexlab{e}})},\ \Eprint
  {http://arxiv.org/abs/1205.5368} {arXiv:1205.5368 [hep-ph]}\BibitemShut
  {NoStop}%
\bibitem [{\citenamefont {Volkov}(2019)}]{Volkov:2019phy}%
  \BibitemOpen
  \bibfield  {author} {\bibinfo {author} {\bibfnamefont {S.}~\bibnamefont
  {Volkov}},\ }\href {\doibase 10.1103/PhysRevD.100.096004} {\bibfield
  {journal} {\bibinfo  {journal} {Phys. Rev.}\ }\textbf {\bibinfo {volume}
  {D100}},\ \bibinfo {pages} {096004} (\bibinfo {year} {2019})},\ \Eprint
  {http://arxiv.org/abs/1909.08015} {arXiv:1909.08015 [hep-ph]}\BibitemShut
  {NoStop}%
\bibitem [{\citenamefont {Mohr}\ \emph {et~al.}(2016)\citenamefont {Mohr},
  \citenamefont {Newell},\ and\ \citenamefont {Taylor}}]{Mohr:2015ccw}%
  \BibitemOpen
  \bibfield  {author} {\bibinfo {author} {\bibfnamefont {P.~J.}\ \bibnamefont
  {Mohr}}, \bibinfo {author} {\bibfnamefont {D.~B.}\ \bibnamefont {Newell}}, \
  and\ \bibinfo {author} {\bibfnamefont {B.~N.}\ \bibnamefont {Taylor}},\
  }\href {\doibase 10.1103/RevModPhys.88.035009} {\bibfield  {journal}
  {\bibinfo  {journal} {Rev. Mod. Phys.}\ }\textbf {\bibinfo {volume} {88}},\
  \bibinfo {pages} {035009} (\bibinfo {year} {2016})},\ \Eprint
  {http://arxiv.org/abs/1507.07956} {arXiv:1507.07956
  [physics.atom-ph]}\BibitemShut {NoStop}%
\bibitem [{\citenamefont {Suura}\ and\ \citenamefont
  {Wichmann}(1957)}]{Suura:1957zz}%
  \BibitemOpen
  \bibfield  {author} {\bibinfo {author} {\bibfnamefont {H.}~\bibnamefont
  {Suura}}\ and\ \bibinfo {author} {\bibfnamefont {E.~H.}\ \bibnamefont
  {Wichmann}},\ }\href {\doibase 10.1103/PhysRev.105.1930} {\bibfield
  {journal} {\bibinfo  {journal} {Phys. Rev.}\ }\textbf {\bibinfo {volume}
  {105}},\ \bibinfo {pages} {1930} (\bibinfo {year} {1957})}\BibitemShut
  {NoStop}%
\bibitem [{\citenamefont {Petermann}(1957{\natexlab{b}})}]{Petermann:1957ir}%
  \BibitemOpen
  \bibfield  {author} {\bibinfo {author} {\bibfnamefont {A.}~\bibnamefont
  {Petermann}},\ }\href {\doibase 10.1103/PhysRev.105.1931} {\bibfield
  {journal} {\bibinfo  {journal} {Phys. Rev.}\ }\textbf {\bibinfo {volume}
  {105}},\ \bibinfo {pages} {1931} (\bibinfo {year}
  {1957}{\natexlab{b}})}\BibitemShut {NoStop}%
\bibitem [{\citenamefont {Elend}(1966)}]{Elend:1966a}%
  \BibitemOpen
  \bibfield  {author} {\bibinfo {author} {\bibfnamefont {H.~H.}\ \bibnamefont
  {Elend}},\ }\href {\doibase 10.1016/0031-9163(66)91171-1} {\bibfield
  {journal} {\bibinfo  {journal} {Phys. Lett.}\ }\textbf {\bibinfo {volume}
  {20}},\ \bibinfo {pages} {682} (\bibinfo {year} {1966})},\ \bibinfo {note}
  {[Erratum: Phys. Lett. {\bf 21}, 720 (1966)]}\BibitemShut {NoStop}%
\bibitem [{\citenamefont {Li}\ \emph {et~al.}(1993)\citenamefont {Li},
  \citenamefont {Mendel},\ and\ \citenamefont {Samuel}}]{Li:1992xf}%
  \BibitemOpen
  \bibfield  {author} {\bibinfo {author} {\bibfnamefont {G.}~\bibnamefont
  {Li}}, \bibinfo {author} {\bibfnamefont {R.}~\bibnamefont {Mendel}}, \ and\
  \bibinfo {author} {\bibfnamefont {M.~A.}\ \bibnamefont {Samuel}},\ }\href
  {\doibase 10.1103/PhysRevD.47.1723} {\bibfield  {journal} {\bibinfo
  {journal} {Phys. Rev.}\ }\textbf {\bibinfo {volume} {D47}},\ \bibinfo {pages}
  {1723} (\bibinfo {year} {1993})}\BibitemShut {NoStop}%
\bibitem [{\citenamefont {Passera}(2007)}]{Passera:2006gc}%
  \BibitemOpen
  \bibfield  {author} {\bibinfo {author} {\bibfnamefont {M.}~\bibnamefont
  {Passera}},\ }\href {\doibase 10.1103/PhysRevD.75.013002} {\bibfield
  {journal} {\bibinfo  {journal} {Phys. Rev.}\ }\textbf {\bibinfo {volume}
  {D75}},\ \bibinfo {pages} {013002} (\bibinfo {year} {2007})},\ \Eprint
  {http://arxiv.org/abs/hep-ph/0606174} {arXiv:hep-ph/0606174
  [hep-ph]}\BibitemShut {NoStop}%
\bibitem [{\citenamefont {Laporta}(1993{\natexlab{a}})}]{Laporta:1993ju}%
  \BibitemOpen
  \bibfield  {author} {\bibinfo {author} {\bibfnamefont {S.}~\bibnamefont
  {Laporta}},\ }\href {\doibase 10.1007/BF02787236} {\bibfield  {journal}
  {\bibinfo  {journal} {Nuovo Cim.}\ }\textbf {\bibinfo {volume} {A106}},\
  \bibinfo {pages} {675} (\bibinfo {year} {1993}{\natexlab{a}})}\BibitemShut
  {NoStop}%
\bibitem [{\citenamefont {Kinoshita}(1967)}]{Kinoshita:1967}%
  \BibitemOpen
  \bibfield  {author} {\bibinfo {author} {\bibfnamefont {T.}~\bibnamefont
  {Kinoshita}},\ }\href {\doibase 10.1007/BF02712327} {\bibfield  {journal}
  {\bibinfo  {journal} {Nuovo Cimento}\ }\textbf {\bibinfo {volume} {B51}},\
  \bibinfo {pages} {140} (\bibinfo {year} {1967})}\BibitemShut {NoStop}%
\bibitem [{\citenamefont {Lautrup}(1969)}]{Lautrup:1969uk}%
  \BibitemOpen
  \bibfield  {author} {\bibinfo {author} {\bibfnamefont {B.~E.}\ \bibnamefont
  {Lautrup}},\ }\href {\doibase 10.1007/BF02754894} {\bibfield  {journal}
  {\bibinfo  {journal} {Nuovo Cim.}\ }\textbf {\bibinfo {volume} {A64}},\
  \bibinfo {pages} {322} (\bibinfo {year} {1969})}\BibitemShut {NoStop}%
\bibitem [{\citenamefont {Lautrup}\ \emph {et~al.}(1971)\citenamefont
  {Lautrup}, \citenamefont {Peterman},\ and\ \citenamefont
  {de~Rafael}}]{Lautrup:1971yp}%
  \BibitemOpen
  \bibfield  {author} {\bibinfo {author} {\bibfnamefont {B.~E.}\ \bibnamefont
  {Lautrup}}, \bibinfo {author} {\bibfnamefont {A.}~\bibnamefont {Peterman}}, \
  and\ \bibinfo {author} {\bibfnamefont {E.}~\bibnamefont {de~Rafael}},\ }\href
  {\doibase 10.1007/BF02722666} {\bibfield  {journal} {\bibinfo  {journal}
  {Nuovo Cim.}\ }\textbf {\bibinfo {volume} {A1}},\ \bibinfo {pages} {238}
  (\bibinfo {year} {1971})}\BibitemShut {NoStop}%
\bibitem [{\citenamefont {Lautrup}\ and\ \citenamefont
  {Samuel}(1977)}]{Lautrup:1977tc}%
  \BibitemOpen
  \bibfield  {author} {\bibinfo {author} {\bibfnamefont {B.~E.}\ \bibnamefont
  {Lautrup}}\ and\ \bibinfo {author} {\bibfnamefont {M.~A.}\ \bibnamefont
  {Samuel}},\ }\href {\doibase 10.1016/0370-2693(77)90075-2} {\bibfield
  {journal} {\bibinfo  {journal} {Phys. Lett.}\ }\textbf {\bibinfo {volume}
  {72B}},\ \bibinfo {pages} {114} (\bibinfo {year} {1977})}\BibitemShut
  {NoStop}%
\bibitem [{\citenamefont {Samuel}\ and\ \citenamefont
  {Chlouber}(1976)}]{Samuel:1976yt}%
  \BibitemOpen
  \bibfield  {author} {\bibinfo {author} {\bibfnamefont {M.~A.}\ \bibnamefont
  {Samuel}}\ and\ \bibinfo {author} {\bibfnamefont {C.}~\bibnamefont
  {Chlouber}},\ }\href {\doibase 10.1103/PhysRevLett.36.442} {\bibfield
  {journal} {\bibinfo  {journal} {Phys. Rev. Lett.}\ }\textbf {\bibinfo
  {volume} {36}},\ \bibinfo {pages} {442} (\bibinfo {year} {1976})}\BibitemShut
  {NoStop}%
\bibitem [{\citenamefont {Kinoshita}(1989)}]{Kinoshita:1989mf}%
  \BibitemOpen
  \bibfield  {author} {\bibinfo {author} {\bibfnamefont {T.}~\bibnamefont
  {Kinoshita}},\ }\href {\doibase 10.1103/PhysRevD.40.1323} {\bibfield
  {journal} {\bibinfo  {journal} {Phys. Rev.}\ }\textbf {\bibinfo {volume}
  {D40}},\ \bibinfo {pages} {1323} (\bibinfo {year} {1989})}\BibitemShut
  {NoStop}%
\bibitem [{\citenamefont {Kinoshita}\ \emph {et~al.}(1990)\citenamefont
  {Kinoshita}, \citenamefont {Nizic},\ and\ \citenamefont
  {Okamoto}}]{Kinoshita:1990wp}%
  \BibitemOpen
  \bibfield  {author} {\bibinfo {author} {\bibfnamefont {T.}~\bibnamefont
  {Kinoshita}}, \bibinfo {author} {\bibfnamefont {B.}~\bibnamefont {Nizic}}, \
  and\ \bibinfo {author} {\bibfnamefont {Y.}~\bibnamefont {Okamoto}},\ }\href
  {\doibase 10.1103/PhysRevD.41.593} {\bibfield  {journal} {\bibinfo  {journal}
  {Phys. Rev. D}\ }\textbf {\bibinfo {volume} {41}},\ \bibinfo {pages} {593}
  (\bibinfo {year} {1990})}\BibitemShut {NoStop}%
\bibitem [{\citenamefont {Samuel}\ and\ \citenamefont
  {Li}(1991)}]{Samuel:1990qf}%
  \BibitemOpen
  \bibfield  {author} {\bibinfo {author} {\bibfnamefont {M.~A.}\ \bibnamefont
  {Samuel}}\ and\ \bibinfo {author} {\bibfnamefont {G.-w.}\ \bibnamefont
  {Li}},\ }\href {\doibase 10.1103/PhysRevD.44.3935} {\bibfield  {journal}
  {\bibinfo  {journal} {Phys. Rev.}\ }\textbf {\bibinfo {volume} {D44}},\
  \bibinfo {pages} {3935} (\bibinfo {year} {1991})},\ \bibinfo {note} {[Errata:
  Phys. Rev. {\bf D46}, 4782 (1992); Phys. Rev. {\bf D48}, 1879
  (1993)]}\BibitemShut {NoStop}%
\bibitem [{\citenamefont {Czarnecki}\ and\ \citenamefont
  {Skrzypek}(1999)}]{Czarnecki:1998rc}%
  \BibitemOpen
  \bibfield  {author} {\bibinfo {author} {\bibfnamefont {A.}~\bibnamefont
  {Czarnecki}}\ and\ \bibinfo {author} {\bibfnamefont {M.}~\bibnamefont
  {Skrzypek}},\ }\href {\doibase 10.1016/S0370-2693(99)00076-3} {\bibfield
  {journal} {\bibinfo  {journal} {Phys. Lett.}\ }\textbf {\bibinfo {volume}
  {B449}},\ \bibinfo {pages} {354} (\bibinfo {year} {1999})},\ \Eprint
  {http://arxiv.org/abs/hep-ph/9812394} {arXiv:hep-ph/9812394
  [hep-ph]}\BibitemShut {NoStop}%
\bibitem [{\citenamefont {Friot}\ \emph {et~al.}(2005)\citenamefont {Friot},
  \citenamefont {Greynat},\ and\ \citenamefont {de~Rafael}}]{Friot:2005cu}%
  \BibitemOpen
  \bibfield  {author} {\bibinfo {author} {\bibfnamefont {S.}~\bibnamefont
  {Friot}}, \bibinfo {author} {\bibfnamefont {D.}~\bibnamefont {Greynat}}, \
  and\ \bibinfo {author} {\bibfnamefont {E.}~\bibnamefont {de~Rafael}},\ }\href
  {\doibase 10.1016/j.physletb.2005.08.126} {\bibfield  {journal} {\bibinfo
  {journal} {Phys. Lett.}\ }\textbf {\bibinfo {volume} {B628}},\ \bibinfo
  {pages} {73} (\bibinfo {year} {2005})},\ \Eprint
  {http://arxiv.org/abs/hep-ph/0505038} {arXiv:hep-ph/0505038
  [hep-ph]}\BibitemShut {NoStop}%
\bibitem [{\citenamefont {Ananthanarayan}\ \emph {et~al.}(2020)\citenamefont
  {Ananthanarayan}, \citenamefont {Friot},\ and\ \citenamefont
  {Ghosh}}]{Ananthanarayan:2020acj}%
  \BibitemOpen
  \bibfield  {author} {\bibinfo {author} {\bibfnamefont {B.}~\bibnamefont
  {Ananthanarayan}}, \bibinfo {author} {\bibfnamefont {S.}~\bibnamefont
  {Friot}}, \ and\ \bibinfo {author} {\bibfnamefont {S.}~\bibnamefont
  {Ghosh}},\ }\href {\doibase 10.1103/PhysRevD.101.116008} {\bibfield
  {journal} {\bibinfo  {journal} {Phys. Rev.}\ }\textbf {\bibinfo {volume}
  {D101}},\ \bibinfo {pages} {116008} (\bibinfo {year} {2020})},\ \Eprint
  {http://arxiv.org/abs/2003.12030} {arXiv:2003.12030 [hep-ph]}\BibitemShut
  {NoStop}%
\bibitem [{\citenamefont {Lautrup}(1972)}]{Lautrup:1972iw}%
  \BibitemOpen
  \bibfield  {author} {\bibinfo {author} {\bibfnamefont {B.~E.}\ \bibnamefont
  {Lautrup}},\ }\href {\doibase 10.1016/0370-2693(72)90168-2} {\bibfield
  {journal} {\bibinfo  {journal} {Phys. Lett.}\ }\textbf {\bibinfo {volume}
  {38B}},\ \bibinfo {pages} {408} (\bibinfo {year} {1972})}\BibitemShut
  {NoStop}%
\bibitem [{\citenamefont {Laporta}(1993{\natexlab{b}})}]{Laporta:1993ds}%
  \BibitemOpen
  \bibfield  {author} {\bibinfo {author} {\bibfnamefont {S.}~\bibnamefont
  {Laporta}},\ }\href {\doibase 10.1016/0370-2693(93)90988-T} {\bibfield
  {journal} {\bibinfo  {journal} {Phys. Lett.}\ }\textbf {\bibinfo {volume}
  {B312}},\ \bibinfo {pages} {495} (\bibinfo {year} {1993}{\natexlab{b}})},\
  \Eprint {http://arxiv.org/abs/hep-ph/9306324} {arXiv:hep-ph/9306324
  [hep-ph]}\BibitemShut {NoStop}%
\bibitem [{\citenamefont {Lepage}(1978)}]{Lepage:1977sw}%
  \BibitemOpen
  \bibfield  {author} {\bibinfo {author} {\bibfnamefont {G.~P.}\ \bibnamefont
  {Lepage}},\ }\href {\doibase 10.1016/0021-9991(78)90004-9} {\bibfield
  {journal} {\bibinfo  {journal} {J. Comput. Phys.}\ }\textbf {\bibinfo
  {volume} {27}},\ \bibinfo {pages} {192} (\bibinfo {year} {1978})}\BibitemShut
  {NoStop}%
\bibitem [{\citenamefont {Kurz}\ \emph {et~al.}(2016)\citenamefont {Kurz},
  \citenamefont {Liu}, \citenamefont {Marquard}, \citenamefont {Smirnov},
  \citenamefont {Smirnov},\ and\ \citenamefont {Steinhauser}}]{Kurz:2016bau}%
  \BibitemOpen
  \bibfield  {author} {\bibinfo {author} {\bibfnamefont {A.}~\bibnamefont
  {Kurz}}, \bibinfo {author} {\bibfnamefont {T.}~\bibnamefont {Liu}}, \bibinfo
  {author} {\bibfnamefont {P.}~\bibnamefont {Marquard}}, \bibinfo {author}
  {\bibfnamefont {A.}~\bibnamefont {Smirnov}}, \bibinfo {author} {\bibfnamefont
  {V.}~\bibnamefont {Smirnov}}, \ and\ \bibinfo {author} {\bibfnamefont
  {M.}~\bibnamefont {Steinhauser}},\ }\href {\doibase
  10.1103/PhysRevD.93.053017} {\bibfield  {journal} {\bibinfo  {journal} {Phys.
  Rev.}\ }\textbf {\bibinfo {volume} {D93}},\ \bibinfo {pages} {053017}
  (\bibinfo {year} {2016})},\ \Eprint {http://arxiv.org/abs/1602.02785}
  {arXiv:1602.02785 [hep-ph]}\BibitemShut {NoStop}%
\bibitem [{\citenamefont {Kurz}\ \emph
  {et~al.}(2014{\natexlab{b}})\citenamefont {Kurz}, \citenamefont {Liu},
  \citenamefont {Marquard},\ and\ \citenamefont {Steinhauser}}]{Kurz:2013exa}%
  \BibitemOpen
  \bibfield  {author} {\bibinfo {author} {\bibfnamefont {A.}~\bibnamefont
  {Kurz}}, \bibinfo {author} {\bibfnamefont {T.}~\bibnamefont {Liu}}, \bibinfo
  {author} {\bibfnamefont {P.}~\bibnamefont {Marquard}}, \ and\ \bibinfo
  {author} {\bibfnamefont {M.}~\bibnamefont {Steinhauser}},\ }\href {\doibase
  10.1016/j.nuclphysb.2013.11.018} {\bibfield  {journal} {\bibinfo  {journal}
  {Nucl. Phys.}\ }\textbf {\bibinfo {volume} {B879}},\ \bibinfo {pages} {1}
  (\bibinfo {year} {2014}{\natexlab{b}})},\ \Eprint
  {http://arxiv.org/abs/1311.2471} {arXiv:1311.2471 [hep-ph]}\BibitemShut
  {NoStop}%
\bibitem [{\citenamefont {Kataev}(2012)}]{Kataev:2012kn}%
  \BibitemOpen
  \bibfield  {author} {\bibinfo {author} {\bibfnamefont {A.~L.}\ \bibnamefont
  {Kataev}},\ }\href {\doibase 10.1103/PhysRevD.86.013010} {\bibfield
  {journal} {\bibinfo  {journal} {Phys. Rev.}\ }\textbf {\bibinfo {volume}
  {D86}},\ \bibinfo {pages} {013010} (\bibinfo {year} {2012})},\ \Eprint
  {http://arxiv.org/abs/1205.6191} {arXiv:1205.6191 [hep-ph]}\BibitemShut
  {NoStop}%
\bibitem [{\citenamefont {Laporta}(1994)}]{Laporta:1994md}%
  \BibitemOpen
  \bibfield  {author} {\bibinfo {author} {\bibfnamefont {S.}~\bibnamefont
  {Laporta}},\ }\href {\doibase 10.1016/0370-2693(94)91513-X} {\bibfield
  {journal} {\bibinfo  {journal} {Phys. Lett.}\ }\textbf {\bibinfo {volume}
  {B328}},\ \bibinfo {pages} {522} (\bibinfo {year} {1994})},\ \Eprint
  {http://arxiv.org/abs/hep-ph/9404204} {arXiv:hep-ph/9404204
  [hep-ph]}\BibitemShut {NoStop}%
\bibitem [{\citenamefont {Kinoshita}\ and\ \citenamefont
  {Nio}(2006{\natexlab{b}})}]{Kinoshita:2005sm}%
  \BibitemOpen
  \bibfield  {author} {\bibinfo {author} {\bibfnamefont {T.}~\bibnamefont
  {Kinoshita}}\ and\ \bibinfo {author} {\bibfnamefont {M.}~\bibnamefont
  {Nio}},\ }\href {\doibase 10.1103/PhysRevD.73.053007} {\bibfield  {journal}
  {\bibinfo  {journal} {Phys. Rev.}\ }\textbf {\bibinfo {volume} {D73}},\
  \bibinfo {pages} {053007} (\bibinfo {year} {2006}{\natexlab{b}})},\ \Eprint
  {http://arxiv.org/abs/hep-ph/0512330} {arXiv:hep-ph/0512330
  [hep-ph]}\BibitemShut {NoStop}%
\bibitem [{\citenamefont {Kataev}(1992)}]{Kataev:1991cp}%
  \BibitemOpen
  \bibfield  {author} {\bibinfo {author} {\bibfnamefont {A.~L.}\ \bibnamefont
  {Kataev}},\ }\href {\doibase 10.1016/0370-2693(92)90452-A} {\bibfield
  {journal} {\bibinfo  {journal} {Phys. Lett.}\ }\textbf {\bibinfo {volume}
  {B284}},\ \bibinfo {pages} {401} (\bibinfo {year} {1992})},\ \bibinfo {note}
  {[Erratum: Phys. Lett. {\bf B710}, 710 (2012)]}\BibitemShut {NoStop}%
\bibitem [{\citenamefont {Kataev}\ and\ \citenamefont
  {Starshenko}(1995)}]{Kataev:1994rw}%
  \BibitemOpen
  \bibfield  {author} {\bibinfo {author} {\bibfnamefont {A.~L.}\ \bibnamefont
  {Kataev}}\ and\ \bibinfo {author} {\bibfnamefont {V.~V.}\ \bibnamefont
  {Starshenko}},\ }\href {\doibase 10.1103/PhysRevD.52.402} {\bibfield
  {journal} {\bibinfo  {journal} {Phys. Rev.}\ }\textbf {\bibinfo {volume}
  {D52}},\ \bibinfo {pages} {402} (\bibinfo {year} {1995})},\ \Eprint
  {http://arxiv.org/abs/hep-ph/9412305} {arXiv:hep-ph/9412305
  [hep-ph]}\BibitemShut {NoStop}%
\bibitem [{\citenamefont {Kataev}(2006)}]{Kataev:2006yh}%
  \BibitemOpen
  \bibfield  {author} {\bibinfo {author} {\bibfnamefont {A.~L.}\ \bibnamefont
  {Kataev}},\ }\href {\doibase 10.1103/PhysRevD.74.073011} {\bibfield
  {journal} {\bibinfo  {journal} {Phys. Rev.}\ }\textbf {\bibinfo {volume}
  {D74}},\ \bibinfo {pages} {073011} (\bibinfo {year} {2006})},\ \Eprint
  {http://arxiv.org/abs/hep-ph/0608120} {arXiv:hep-ph/0608120
  [hep-ph]}\BibitemShut {NoStop}%
\bibitem [{\citenamefont {Karshenboim}(1993)}]{Karshenboim:1993rt}%
  \BibitemOpen
  \bibfield  {author} {\bibinfo {author} {\bibfnamefont {S.~G.}\ \bibnamefont
  {Karshenboim}},\ }\href@noop {} {\bibfield  {journal} {\bibinfo  {journal}
  {Phys. Atom. Nucl.}\ }\textbf {\bibinfo {volume} {56}},\ \bibinfo {pages}
  {857} (\bibinfo {year} {1993})},\ \bibinfo {note} {[Yad. Fiz. {\bf 56N6}, 252
  (1993)]}\BibitemShut {NoStop}%
\bibitem [{\citenamefont {Baikov}\ \emph {et~al.}(2013)\citenamefont {Baikov},
  \citenamefont {Maier},\ and\ \citenamefont {Marquard}}]{Baikov:2013ula}%
  \BibitemOpen
  \bibfield  {author} {\bibinfo {author} {\bibfnamefont {P.~A.}\ \bibnamefont
  {Baikov}}, \bibinfo {author} {\bibfnamefont {A.}~\bibnamefont {Maier}}, \
  and\ \bibinfo {author} {\bibfnamefont {P.}~\bibnamefont {Marquard}},\ }\href
  {\doibase 10.1016/j.nuclphysb.2013.10.020} {\bibfield  {journal} {\bibinfo
  {journal} {Nucl. Phys.}\ }\textbf {\bibinfo {volume} {B877}},\ \bibinfo
  {pages} {647} (\bibinfo {year} {2013})},\ \Eprint
  {http://arxiv.org/abs/1307.6105} {arXiv:1307.6105 [hep-ph]}\BibitemShut
  {NoStop}%
\bibitem [{\citenamefont {Baikov}\ and\ \citenamefont
  {Broadhurst}(1995)}]{Baikov:1995ui}%
  \BibitemOpen
  \bibfield  {author} {\bibinfo {author} {\bibfnamefont {P.~A.}\ \bibnamefont
  {Baikov}}\ and\ \bibinfo {author} {\bibfnamefont {D.~J.}\ \bibnamefont
  {Broadhurst}},\ }\href@noop {} {\  (\bibinfo {year} {1995})},\ \Eprint
  {http://arxiv.org/abs/hep-ph/9504398} {arXiv:hep-ph/9504398
  [hep-ph]}\BibitemShut {NoStop}%
\bibitem [{\citenamefont {Bouchendira}\ \emph {et~al.}(2011)\citenamefont
  {Bouchendira}, \citenamefont {Clade}, \citenamefont {Guellati-Khelifa},
  \citenamefont {Nez},\ and\ \citenamefont {Biraben}}]{Bouchendira:2010es}%
  \BibitemOpen
  \bibfield  {author} {\bibinfo {author} {\bibfnamefont {R.}~\bibnamefont
  {Bouchendira}}, \bibinfo {author} {\bibfnamefont {P.}~\bibnamefont {Clade}},
  \bibinfo {author} {\bibfnamefont {S.}~\bibnamefont {Guellati-Khelifa}},
  \bibinfo {author} {\bibfnamefont {F.}~\bibnamefont {Nez}}, \ and\ \bibinfo
  {author} {\bibfnamefont {F.}~\bibnamefont {Biraben}},\ }\href {\doibase
  10.1103/PhysRevLett.106.080801} {\bibfield  {journal} {\bibinfo  {journal}
  {Phys. Rev. Lett.}\ }\textbf {\bibinfo {volume} {106}},\ \bibinfo {pages}
  {080801} (\bibinfo {year} {2011})},\ \Eprint {http://arxiv.org/abs/1012.3627}
  {arXiv:1012.3627 [physics.atom-ph]}\BibitemShut {NoStop}%
\bibitem [{\citenamefont {Beyer}\ \emph {et~al.}(2017)\citenamefont {Beyer},
  \citenamefont {Maisenbacher}, \citenamefont {Matveev}, \citenamefont {Pohl},
  \citenamefont {Khabarova}, \citenamefont {Grinin}, \citenamefont {Lamour},
  \citenamefont {Yost}, \citenamefont {H{\"a}nsch}, \citenamefont
  {Kolachevsky},\ and\ \citenamefont {Udem}}]{Beyer:2017}%
  \BibitemOpen
  \bibfield  {author} {\bibinfo {author} {\bibfnamefont {A.}~\bibnamefont
  {Beyer}}, \bibinfo {author} {\bibfnamefont {L.}~\bibnamefont {Maisenbacher}},
  \bibinfo {author} {\bibfnamefont {A.}~\bibnamefont {Matveev}}, \bibinfo
  {author} {\bibfnamefont {R.}~\bibnamefont {Pohl}}, \bibinfo {author}
  {\bibfnamefont {K.}~\bibnamefont {Khabarova}}, \bibinfo {author}
  {\bibfnamefont {A.}~\bibnamefont {Grinin}}, \bibinfo {author} {\bibfnamefont
  {T.}~\bibnamefont {Lamour}}, \bibinfo {author} {\bibfnamefont {D.~C.}\
  \bibnamefont {Yost}}, \bibinfo {author} {\bibfnamefont {T.~W.}\ \bibnamefont
  {H{\"a}nsch}}, \bibinfo {author} {\bibfnamefont {N.}~\bibnamefont
  {Kolachevsky}}, \ and\ \bibinfo {author} {\bibfnamefont {T.}~\bibnamefont
  {Udem}},\ }\href {\doibase 10.1126/science.aah6677} {\bibfield  {journal}
  {\bibinfo  {journal} {Science}\ }\textbf {\bibinfo {volume} {358}},\ \bibinfo
  {pages} {79} (\bibinfo {year} {2017})}\BibitemShut {NoStop}%
\bibitem [{\citenamefont {Fleurbaey}\ \emph {et~al.}(2018)\citenamefont
  {Fleurbaey}, \citenamefont {Galtier}, \citenamefont {Thomas}, \citenamefont
  {Bonnaud}, \citenamefont {Julien}, \citenamefont {Biraben}, \citenamefont
  {Nez}, \citenamefont {Abgrall},\ and\ \citenamefont
  {Gu{\'e}na}}]{Fleurbaey:2018fih}%
  \BibitemOpen
  \bibfield  {author} {\bibinfo {author} {\bibfnamefont {H.}~\bibnamefont
  {Fleurbaey}}, \bibinfo {author} {\bibfnamefont {S.}~\bibnamefont {Galtier}},
  \bibinfo {author} {\bibfnamefont {S.}~\bibnamefont {Thomas}}, \bibinfo
  {author} {\bibfnamefont {M.}~\bibnamefont {Bonnaud}}, \bibinfo {author}
  {\bibfnamefont {L.}~\bibnamefont {Julien}}, \bibinfo {author} {\bibfnamefont
  {F.}~\bibnamefont {Biraben}}, \bibinfo {author} {\bibfnamefont
  {F.}~\bibnamefont {Nez}}, \bibinfo {author} {\bibfnamefont {M.}~\bibnamefont
  {Abgrall}}, \ and\ \bibinfo {author} {\bibfnamefont {J.}~\bibnamefont
  {Gu{\'e}na}},\ }\href {\doibase 10.1103/PhysRevLett.120.183001} {\bibfield
  {journal} {\bibinfo  {journal} {Phys. Rev. Lett.}\ }\textbf {\bibinfo
  {volume} {120}},\ \bibinfo {pages} {183001} (\bibinfo {year} {2018})},\
  \Eprint {http://arxiv.org/abs/1801.08816} {arXiv:1801.08816
  [physics.atom-ph]}\BibitemShut {NoStop}%
\bibitem [{\citenamefont {Hanneke}\ \emph {et~al.}(2008)\citenamefont
  {Hanneke}, \citenamefont {Fogwell},\ and\ \citenamefont
  {Gabrielse}}]{Hanneke:2008tm}%
  \BibitemOpen
  \bibfield  {author} {\bibinfo {author} {\bibfnamefont {D.}~\bibnamefont
  {Hanneke}}, \bibinfo {author} {\bibfnamefont {S.}~\bibnamefont {Fogwell}}, \
  and\ \bibinfo {author} {\bibfnamefont {G.}~\bibnamefont {Gabrielse}},\ }\href
  {\doibase 10.1103/PhysRevLett.100.120801} {\bibfield  {journal} {\bibinfo
  {journal} {Phys. Rev. Lett.}\ }\textbf {\bibinfo {volume} {100}},\ \bibinfo
  {pages} {120801} (\bibinfo {year} {2008})},\ \Eprint
  {http://arxiv.org/abs/0801.1134} {arXiv:0801.1134
  [physics.atom-ph]}\BibitemShut {NoStop}%
\bibitem [{\citenamefont {Czarnecki}\ and\ \citenamefont
  {Marciano}(2001)}]{Czarnecki:2001pv}%
  \BibitemOpen
  \bibfield  {author} {\bibinfo {author} {\bibfnamefont {A.}~\bibnamefont
  {Czarnecki}}\ and\ \bibinfo {author} {\bibfnamefont {W.~J.}\ \bibnamefont
  {Marciano}},\ }\href {\doibase 10.1103/PhysRevD.64.013014} {\bibfield
  {journal} {\bibinfo  {journal} {Phys. Rev.}\ }\textbf {\bibinfo {volume}
  {D64}},\ \bibinfo {pages} {013014} (\bibinfo {year} {2001})},\ \Eprint
  {http://arxiv.org/abs/hep-ph/0102122} {arXiv:hep-ph/0102122
  [hep-ph]}\BibitemShut {NoStop}%
\bibitem [{\citenamefont {St{\"o}ckinger}(2009)}]{Stockinger:1900zz}%
  \BibitemOpen
  \bibfield  {author} {\bibinfo {author} {\bibfnamefont {D.}~\bibnamefont
  {St{\"o}ckinger}},\ }\href {\doibase 10.1142/9789814271844\_0012} {\bibfield
  {journal} {\bibinfo  {journal} {Adv. Ser. Direct. High Energy Phys.}\
  }\textbf {\bibinfo {volume} {20}},\ \bibinfo {pages} {393} (\bibinfo {year}
  {2009})}\BibitemShut {NoStop}%
\bibitem [{\citenamefont {Giudice}\ \emph {et~al.}(2012)\citenamefont
  {Giudice}, \citenamefont {Paradisi},\ and\ \citenamefont
  {Passera}}]{Giudice:2012ms}%
  \BibitemOpen
  \bibfield  {author} {\bibinfo {author} {\bibfnamefont {G.}~\bibnamefont
  {Giudice}}, \bibinfo {author} {\bibfnamefont {P.}~\bibnamefont {Paradisi}}, \
  and\ \bibinfo {author} {\bibfnamefont {M.}~\bibnamefont {Passera}},\ }\href
  {\doibase 10.1007/JHEP11(2012)113} {\bibfield  {journal} {\bibinfo  {journal}
  {JHEP}\ }\textbf {\bibinfo {volume} {11}},\ \bibinfo {pages} {113} (\bibinfo
  {year} {2012})},\ \Eprint {http://arxiv.org/abs/1208.6583} {arXiv:1208.6583
  [hep-ph]}\BibitemShut {NoStop}%
\bibitem [{\citenamefont {Crivellin}\ \emph {et~al.}(2018)\citenamefont
  {Crivellin}, \citenamefont {Hoferichter},\ and\ \citenamefont
  {Schmidt-Wellenburg}}]{Crivellin:2018qmi}%
  \BibitemOpen
  \bibfield  {author} {\bibinfo {author} {\bibfnamefont {A.}~\bibnamefont
  {Crivellin}}, \bibinfo {author} {\bibfnamefont {M.}~\bibnamefont
  {Hoferichter}}, \ and\ \bibinfo {author} {\bibfnamefont {P.}~\bibnamefont
  {Schmidt-Wellenburg}},\ }\href {\doibase 10.1103/PhysRevD.98.113002}
  {\bibfield  {journal} {\bibinfo  {journal} {Phys. Rev. D}\ }\textbf {\bibinfo
  {volume} {98}},\ \bibinfo {pages} {113002} (\bibinfo {year} {2018})},\
  \Eprint {http://arxiv.org/abs/1807.11484} {arXiv:1807.11484
  [hep-ph]}\BibitemShut {NoStop}%
\bibitem [{\citenamefont {Kukhto}\ \emph {et~al.}(1992)\citenamefont {Kukhto},
  \citenamefont {Kuraev}, \citenamefont {Silagadze},\ and\ \citenamefont
  {Schiller}}]{Kukhto:1992qv}%
  \BibitemOpen
  \bibfield  {author} {\bibinfo {author} {\bibfnamefont {T.}~\bibnamefont
  {Kukhto}}, \bibinfo {author} {\bibfnamefont {E.}~\bibnamefont {Kuraev}},
  \bibinfo {author} {\bibfnamefont {Z.}~\bibnamefont {Silagadze}}, \ and\
  \bibinfo {author} {\bibfnamefont {A.}~\bibnamefont {Schiller}},\ }\href
  {\doibase 10.1016/0550-3213(92)90687-7} {\bibfield  {journal} {\bibinfo
  {journal} {Nucl. Phys. B}\ }\textbf {\bibinfo {volume} {371}},\ \bibinfo
  {pages} {567} (\bibinfo {year} {1992})}\BibitemShut {NoStop}%
\bibitem [{\citenamefont {Czarnecki}\ \emph {et~al.}(1995)\citenamefont
  {Czarnecki}, \citenamefont {Krause},\ and\ \citenamefont
  {Marciano}}]{Czarnecki:1995wq}%
  \BibitemOpen
  \bibfield  {author} {\bibinfo {author} {\bibfnamefont {A.}~\bibnamefont
  {Czarnecki}}, \bibinfo {author} {\bibfnamefont {B.}~\bibnamefont {Krause}}, \
  and\ \bibinfo {author} {\bibfnamefont {W.~J.}\ \bibnamefont {Marciano}},\
  }\href {\doibase 10.1103/PhysRevD.52.R2619} {\bibfield  {journal} {\bibinfo
  {journal} {Phys. Rev. D}\ }\textbf {\bibinfo {volume} {52}},\ \bibinfo
  {pages} {2619} (\bibinfo {year} {1995})},\ \Eprint
  {http://arxiv.org/abs/hep-ph/9506256} {arXiv:hep-ph/9506256}\BibitemShut
  {NoStop}%
\bibitem [{\citenamefont {Czarnecki}\ \emph {et~al.}(1996)\citenamefont
  {Czarnecki}, \citenamefont {Krause},\ and\ \citenamefont
  {Marciano}}]{Czarnecki:1995sz}%
  \BibitemOpen
  \bibfield  {author} {\bibinfo {author} {\bibfnamefont {A.}~\bibnamefont
  {Czarnecki}}, \bibinfo {author} {\bibfnamefont {B.}~\bibnamefont {Krause}}, \
  and\ \bibinfo {author} {\bibfnamefont {W.~J.}\ \bibnamefont {Marciano}},\
  }\href {\doibase 10.1103/PhysRevLett.76.3267} {\bibfield  {journal} {\bibinfo
   {journal} {Phys. Rev. Lett.}\ }\textbf {\bibinfo {volume} {76}},\ \bibinfo
  {pages} {3267} (\bibinfo {year} {1996})},\ \Eprint
  {http://arxiv.org/abs/hep-ph/9512369} {arXiv:hep-ph/9512369
  [hep-ph]}\BibitemShut {NoStop}%
\bibitem [{\citenamefont {Degrassi}\ and\ \citenamefont
  {Giudice}(1998)}]{Degrassi:1998es}%
  \BibitemOpen
  \bibfield  {author} {\bibinfo {author} {\bibfnamefont {G.}~\bibnamefont
  {Degrassi}}\ and\ \bibinfo {author} {\bibfnamefont {G.}~\bibnamefont
  {Giudice}},\ }\href {\doibase 10.1103/PhysRevD.58.053007} {\bibfield
  {journal} {\bibinfo  {journal} {Phys. Rev. D}\ }\textbf {\bibinfo {volume}
  {58}},\ \bibinfo {pages} {053007} (\bibinfo {year} {1998})},\ \Eprint
  {http://arxiv.org/abs/hep-ph/9803384} {arXiv:hep-ph/9803384}\BibitemShut
  {NoStop}%
\bibitem [{\citenamefont {von Weitershausen}\ \emph {et~al.}(2010)\citenamefont
  {von Weitershausen}, \citenamefont {Sch{\"a}fer}, \citenamefont
  {St{\"o}ckinger-Kim},\ and\ \citenamefont
  {St{\"o}ckinger}}]{vonWeitershausen:2010zr}%
  \BibitemOpen
  \bibfield  {author} {\bibinfo {author} {\bibfnamefont {P.}~\bibnamefont {von
  Weitershausen}}, \bibinfo {author} {\bibfnamefont {M.}~\bibnamefont
  {Sch{\"a}fer}}, \bibinfo {author} {\bibfnamefont {H.}~\bibnamefont
  {St{\"o}ckinger-Kim}}, \ and\ \bibinfo {author} {\bibfnamefont
  {D.}~\bibnamefont {St{\"o}ckinger}},\ }\href {\doibase
  10.1103/PhysRevD.81.093004} {\bibfield  {journal} {\bibinfo  {journal} {Phys.
  Rev. D}\ }\textbf {\bibinfo {volume} {81}},\ \bibinfo {pages} {093004}
  (\bibinfo {year} {2010})},\ \Eprint {http://arxiv.org/abs/1003.5820}
  {arXiv:1003.5820 [hep-ph]}\BibitemShut {NoStop}%
\bibitem [{\citenamefont {Peris}\ \emph {et~al.}(1995)\citenamefont {Peris},
  \citenamefont {Perrottet},\ and\ \citenamefont {de~Rafael}}]{Peris:1995bb}%
  \BibitemOpen
  \bibfield  {author} {\bibinfo {author} {\bibfnamefont {S.}~\bibnamefont
  {Peris}}, \bibinfo {author} {\bibfnamefont {M.}~\bibnamefont {Perrottet}}, \
  and\ \bibinfo {author} {\bibfnamefont {E.}~\bibnamefont {de~Rafael}},\ }\href
  {\doibase 10.1016/0370-2693(95)00768-G} {\bibfield  {journal} {\bibinfo
  {journal} {Phys. Lett. B}\ }\textbf {\bibinfo {volume} {355}},\ \bibinfo
  {pages} {523} (\bibinfo {year} {1995})},\ \Eprint
  {http://arxiv.org/abs/hep-ph/9505405} {arXiv:hep-ph/9505405}\BibitemShut
  {NoStop}%
\bibitem [{\citenamefont {Heinemeyer}\ \emph {et~al.}(2004)\citenamefont
  {Heinemeyer}, \citenamefont {St{\"o}ckinger},\ and\ \citenamefont
  {Weiglein}}]{Heinemeyer:2004yq}%
  \BibitemOpen
  \bibfield  {author} {\bibinfo {author} {\bibfnamefont {S.}~\bibnamefont
  {Heinemeyer}}, \bibinfo {author} {\bibfnamefont {D.}~\bibnamefont
  {St{\"o}ckinger}}, \ and\ \bibinfo {author} {\bibfnamefont {G.}~\bibnamefont
  {Weiglein}},\ }\href {\doibase 10.1016/j.nuclphysb.2004.08.014} {\bibfield
  {journal} {\bibinfo  {journal} {Nucl. Phys. B}\ }\textbf {\bibinfo {volume}
  {699}},\ \bibinfo {pages} {103} (\bibinfo {year} {2004})},\ \Eprint
  {http://arxiv.org/abs/hep-ph/0405255} {arXiv:hep-ph/0405255}\BibitemShut
  {NoStop}%
\bibitem [{\citenamefont {Gribouk}\ and\ \citenamefont
  {Czarnecki}(2005)}]{Gribouk:2005ee}%
  \BibitemOpen
  \bibfield  {author} {\bibinfo {author} {\bibfnamefont {T.}~\bibnamefont
  {Gribouk}}\ and\ \bibinfo {author} {\bibfnamefont {A.}~\bibnamefont
  {Czarnecki}},\ }\href {\doibase 10.1103/PhysRevD.72.053016} {\bibfield
  {journal} {\bibinfo  {journal} {Phys. Rev. D}\ }\textbf {\bibinfo {volume}
  {72}},\ \bibinfo {pages} {053016} (\bibinfo {year} {2005})},\ \Eprint
  {http://arxiv.org/abs/hep-ph/0509205} {arXiv:hep-ph/0509205}\BibitemShut
  {NoStop}%
\bibitem [{\citenamefont {Ishikawa}\ \emph {et~al.}(2019)\citenamefont
  {Ishikawa}, \citenamefont {Nakazawa},\ and\ \citenamefont
  {Yasui}}]{Ishikawa:2018rlv}%
  \BibitemOpen
  \bibfield  {author} {\bibinfo {author} {\bibfnamefont {T.}~\bibnamefont
  {Ishikawa}}, \bibinfo {author} {\bibfnamefont {N.}~\bibnamefont {Nakazawa}},
  \ and\ \bibinfo {author} {\bibfnamefont {Y.}~\bibnamefont {Yasui}},\ }\href
  {\doibase 10.1103/PhysRevD.99.073004} {\bibfield  {journal} {\bibinfo
  {journal} {Phys. Rev. D}\ }\textbf {\bibinfo {volume} {99}},\ \bibinfo
  {pages} {073004} (\bibinfo {year} {2019})},\ \Eprint
  {http://arxiv.org/abs/1810.13445} {arXiv:1810.13445 [hep-ph]}\BibitemShut
  {NoStop}%
\bibitem [{\citenamefont {Czarnecki}\ and\ \citenamefont
  {Marciano}(2017)}]{Czarnecki:2017rlm}%
  \BibitemOpen
  \bibfield  {author} {\bibinfo {author} {\bibfnamefont {A.}~\bibnamefont
  {Czarnecki}}\ and\ \bibinfo {author} {\bibfnamefont {W.~J.}\ \bibnamefont
  {Marciano}},\ }\href {\doibase 10.1103/PhysRevD.96.113001} {\bibfield
  {journal} {\bibinfo  {journal} {Phys. Rev. D}\ }\textbf {\bibinfo {volume}
  {96}},\ \bibinfo {pages} {113001} (\bibinfo {year} {2017})},\ \bibinfo {note}
  {[Erratum: Phys. Rev. {\bf D97}, 019901 (2018)]},\ \Eprint
  {http://arxiv.org/abs/1711.00550} {arXiv:1711.00550 [hep-ph]}\BibitemShut
  {NoStop}%
\end{thebibliography}%

\end{document}